\documentclass[structabstract]{aa}  
\usepackage[round]{natbib}
\usepackage{graphicx}
\usepackage{txfonts}
\usepackage{aalongtable}

\begin{document}
\title{Star forming galaxies in the AKARI Deep Field South: identifications and SEDs}
 
 \titlerunning{Star forming galaxies in the ADF-S}
 \authorrunning{Ma{\l}ek, Pollo, Takeuchi et al.}


\author{K.\ Ma{\l}ek
	\inst{1}, A.\ Pollo \inst{2,3}, T.\ T.\ Takeuchi \inst{4}, P.\ Bienias \inst{5}, 
	M.\ Shirahata \inst{6}, S.\ Matsuura \inst{6} \and M.\ Kawada \inst{7}
 }

\institute{
	Center for Theoretical Physics of the Polish Academy of Sciences, Al. Lotnik\'ow 32/46, 02-668 Warsaw,  Poland \\ 
  \email{
		malek@cft.edu.pl}
\and
	  The Andrzej So{\l}tan Institute for Nuclear Studies, 69, ul.\ Ho\.{z}a, 00-681 Warsaw, Poland  
\and
 The Astronomical Observatory of the Jagiellonian University, ul.\ Orla 171, 30-244 Krak\'{o}w, Poland 
	\and
	Institute for Advanced Research, Nagoya University, Furo-cho, Chikusa-ku, Nagoya 464-8601, Japan 
	\and
College of Inter-Faculty Individual Studies in Mathematics and Natural
Sciences, University of Warsaw, ul.\ \.{Z}wirki i Wigury 93, 02-089 Warsaw, Poland
	\and
	Institute of Space and Astronautical Science, JAXA, 3-1-1 Yoshinodai, Sagamihara, Kanagawa 229-8510, Japan
	\and
	Division of Particle and Astrophysical Science, Nagoya University, Furo-cho, Chikusa-ku, Nagoya 464-8602, Japan
 }

\date{Received ; accepted }

 
  \abstract
	{}
{We investigate the nature and properties of far-infrared (FIR) 
sources from the AKARI Deep Field South (ADF-S).}
{We performed an extensive search for the counterparts of $1000$  
ADF-S objects brighter than 0.0301 Jy in the {{\it WIDE-S}} (90 $\mu$m) 
AKARI band in the public databases (NED and SIMBAD). 
We analyzed 
the properties of the resulting sample: statistic of the identified objects, 
quality of position determination of the ADF-S sources, 
their number counts, redshift distribution and comparison of morphological 
types, when the corresponding information was available. 
We also made a crude analysis of the clustering properties 
of the ADF-S sources and we constructed spectral energy distributions 
(SEDs) of selected objects with the best photometry, using three 
different models. }
{Among 1000 investigated ADF-S sources, 
545 were identified at other wavelengths using the public databases.
From them, 518 are known galaxies, and 343 of these galaxies were not known previously as infra-red sources.
Among the remaining 
sources there are two quasars and infrared and radio sources of unknown origin. 
Among six stellar identifications, at least five is probably an effect 
of contamination. 
We found redshifts of 48 extragalactic objects and morphological types of 
77 galaxies. 
We present models of SEDs of 47 sources with sufficiently good photometric data.  }
{We conclude that the bright FIR point sources observed in the ADF-S are mostly nearby galaxies. 
Their properties are very 
similar to properties of the local population of optically bright galaxies, except an unusually high ratio of peculiar or interacting 
objects and a smaller percentage of elliptical galaxies.  
The percentage of lenticular galaxies is the same as in the optically bright population which 
suggests that galaxies of this type may frequently contain a significant amount of cool dust.
It is possible that source confusion plays a significant role in more than 34~\% of measurements. 
The SEDs display a variety of galaxy types, from very
actively star forming to very quiescent. 
Thanks to the AKARI long wavelength bands it was revealed for the first time that these galaxies form a population of objects with very cool dust.}


\maketitle
%

\section{Introduction}\label{sec:introduction}

Active star formation (SF) is followed by heavy element production from the
birth and death of stars. 
Several of the heavy elements produced by stars end up substantially depleted into dust grains. These dust grains in galaxies tend to absorb ultraviolet (UV) light, emitted by young stars, and re-emit it in the far infrared (FIR).
Indeed, there is an extreme category of galaxies which have large
amount of dust and are extremely luminous in the FIR and submillimeter
(submm) wavelengths. 
Heavily hidden SF is suggested in these galaxies. 

By examining the luminosity functions (LFs) at UV and FIR from GALEX and IRAS/Spitzer, 
\citet{takeuchi05a} proved that the FIR LF shows much stronger evolution than that 
of UV, though both evolve very strongly. 
This indicates that the fraction of hidden SF rapidly increases toward higher redshifts 
up to $z < 1$. 
There is another important observable closely related to the dust emission from 
galaxies: the cosmic IR background (CIB). 
Recently, \citet{takeuchi06} constructed the IR spectral energy distribution (SED) 
of the Local Universe. 
The energy emitted in the IR is 25--30~\% of the total energy budget. 
In contrast, the IR (from near/mid-IR to millimeter) contribution is roughly 
(or even more than) a half in the CIB spectrum \citep[e.g.,][]{dole06}. 
This also suggests a strong evolution of the IR contribution to the cosmic SED 
in the Universe. 

Thus, understanding the radiative physics of dust is a fundamental task to have 
an unbiased view of the cosmic SF history.
Especially, exploring the evolution of SEDs of galaxies at each epoch is a key to have 
a unified view of the SF history.
First step to this is to know the properties of Local galaxies.
Vast amount of new knowledge about the IR universe was provided by IRAS
\citep[e.g.,][]{soifer87}, followed by MSX \citep[e.g.,][]{egan03}, ISO \citep[e.g.,][]{genzel00,verma05} and 
Spitzer \citep[e.g.,][]{soifer08}. 

Recently, after IRAS, a Japanese IR satellite AKARI has performed an all-sky survey
\citep{murakami} and various smaller but deeper surveys at different IR wavelengths. 
In particular, 
with the aid of the Far-Infrared Surveyor \cite[FIS:][]{kawada}, observations in four FIR bands were possible.
Among 
the observed fields, the lowest Galactic cirrus emission density region near the South Ecliptic Pole 
was selected 
for observations since it can provide the best FIR extragalactic image of the Universe.
This field is referred to as the AKARI Deep Field South (ADF-S). 
This survey is unique in having continuous wavelength coverage with four photometric bands 
(65, 90, 140 and 160 $\mu$m) mapped over a
 wide area ($\sim$12 square degrees). 
2268 infrared sources were detected, down to $\sim 20$~mJy at 90~$\mu$m, and 
infrared colors for about 400 of these were measured.

By the advent of AKARI surveys, a new generation of large database of the Local Universe
will be available.
In this paper, we present first result of our cross identification (cross-ID) of the
sources in the ADF-S.
After showing the process of identification of the ADF-S objects, we discuss 
their statistical properties.
Further, we show the UV-optical-FIR SEDs of selected nearby star-forming 
galaxies from our sample with the best photometric data. 
We have made a analysis of these SEDs by fitting a few simple model of dust emission from
galaxies.

The paper is organized as follows: in Section~\ref{sec:data}, we present the data.
Section~\ref{sec:basic} is devoted to the basic analysis of the properties of 
identified sources, their distribution on the sky, and the quality of 
the catalog and possible biases, e.g. related to the source confusion.
Then we discuss the statistical properties of the obtained sample, e.g., the number counts, 
redshift distribution, galaxy morphologies, and other properties. 
In Section~\ref{sec:sed} we present SEDs of the identified galaxies 
and we attempt to model the dust emission from them. 
We present our conclusions in Section~\ref{sec:conclusion}.

\section{The data}\label{sec:data}

We cross-identified the ADF-S point source catalog (based on $90\;\mu$m) with
publicly available databases, mainly the SIMBAD\footnote{URL: {\tt http://simbad.u-strasbg.fr/simbad/.}} and 
NED\footnote{URL: {\tt http://nedwww.ipac.caltech.edu/}.}. 
We performed this search in two stages: first for 500 ADF-S sources brighter than 0.0482 $\mu$Jy 
in the {\it WIDE-S} AKARI band (90 $\mu$m), which corresponds roughly to $\sim 10\sigma$ detection, 
and then for the next 500 ADF-S sources, brighter than 0.0301 $\mu$Jy in the {\it WIDE-S} band, 
which corresponds approximately to $\sim 6 \sigma$ detection. 
In the following sections, we will refer to these two data sets as $10 \sigma$ and $6 \sigma$ catalogs. 
We will present the properties of sources in these catalogs. 

The search for counterparts was performed in the radius of $40''$ around each source.  The ADF-S images were obtained by the slow-scan mode of FIS.
The synthesized point spread functions (PSFs) of the slow-scanned image were presented 
and examined extensively in \citet[][]{shirahata09}.
They showed that the PSF at each band is well represented by a ``double-Gaussian profile'',
i.e., a superposition of two 2-dimensional Gaussian profiles with different standard deviations.
The standard deviations of the narrower component are $32''\pm 1''$ (for the 65 $\mu$m band), $30''\pm 1''$ (the 90 $\mu$m band), $41'' \pm 1''$ (the 140 $\mu$m band), and $38''\pm 1''$ (the 160 $\mu$m band). 
They have also shown that approximately 80~\% of the flux power is included in this component.
Hence, practically this size can be regarded as a reasonable counterpart search radius.
In the case of ADF-S, nearby galaxies have more extended profile than a simple point source, 
and the reduction method and scan speed are slightly different from the early PSF analysis.
In addition, we have used all the four FIS bands.
Hence, we have chosen the largest size $40''$ as the tolerance
radius of the counterpart search.

For sources from the $10 \sigma$ catalog in the search radii we found in total $500$ counterparts, 
corresponding to 330 ADF-S sources. 
Among them there are two cases of two ADF-S sources corresponding to the same counterpart 
(one star and one galaxy). 
For 170 sources (34~\%) no counterparts were found. 208 sources (42~\%) have one possible 
counterpart, 114 sources (23~\%) have two or three counterparts and 
for 8 sources (1.6~\%) more than 3 possible counterparts were found.

Extending the identification process to the $6 \sigma$ level, we found 284 more counterparts, 
corresponding to 215 sources fainter than 0.0482~$\mu$Jy and brighter then 0.0329~$\mu$Jy 
in the {\it WIDE-S} band. 
Among them were 49 cases (10~\% of this fainter part of the sample) of a double and 10 cases 
(2~\%) of a triple counterpart. 
Two sources correspond to the same counterpart (Seyfert-1 galaxy  located at at $z \sim 0.24$). 
One of the sources corresponds to the extragalactic X-ray source which is related to a starburst 
galaxy NGC~1705, identified as the fourth brightest source in the $10 \sigma$ sample. 
For 156 sources (31~\%), we found just one counterpart. 
No counterparts were found for 285 sources (57~\%).
Thus, the probability of finding a counterpart decreases with the 90~$\mu$m brightness of 
sources by a factor of 1.5. 
However, in the same time the probability of a source having multiple counterparts decreases 
much faster, by a factor of 2. 
There are two possible reasons for this behavior: 1) in ADF-S data, a source confusion 
would give a more significant contribution to the flux of the brighter sources.
This is usually not the case \citep[e.g.,][]{takeuchi04}, but because of the special way of 
image construction in the ADF-S data, the confusion effect might play some role,
2) interaction between physically close galaxies  (and, consequently, on the sky) may increase their 
intrinsic IR luminosity significantly.
Actual situation would be a mixture of these two effects.

For the purpose of this work we assume, unless specified otherwise, that the most nearby 
counterpart of an ADF-S source are real ones. 
However, as stated above, we are aware that the source confusion plays an important role 
at least in a part of measurements of the multiply identified sources. 
We will try to address this issue in Section~\ref{subsec:CF}.

\section{Basic Analysis}\label{sec:basic}

\subsection{Accuracy of the position determination of the ADF-S sources}\label{subsec:position_accuracy}

\begin{figure}[t]
\centering
\includegraphics[width=8.5cm]{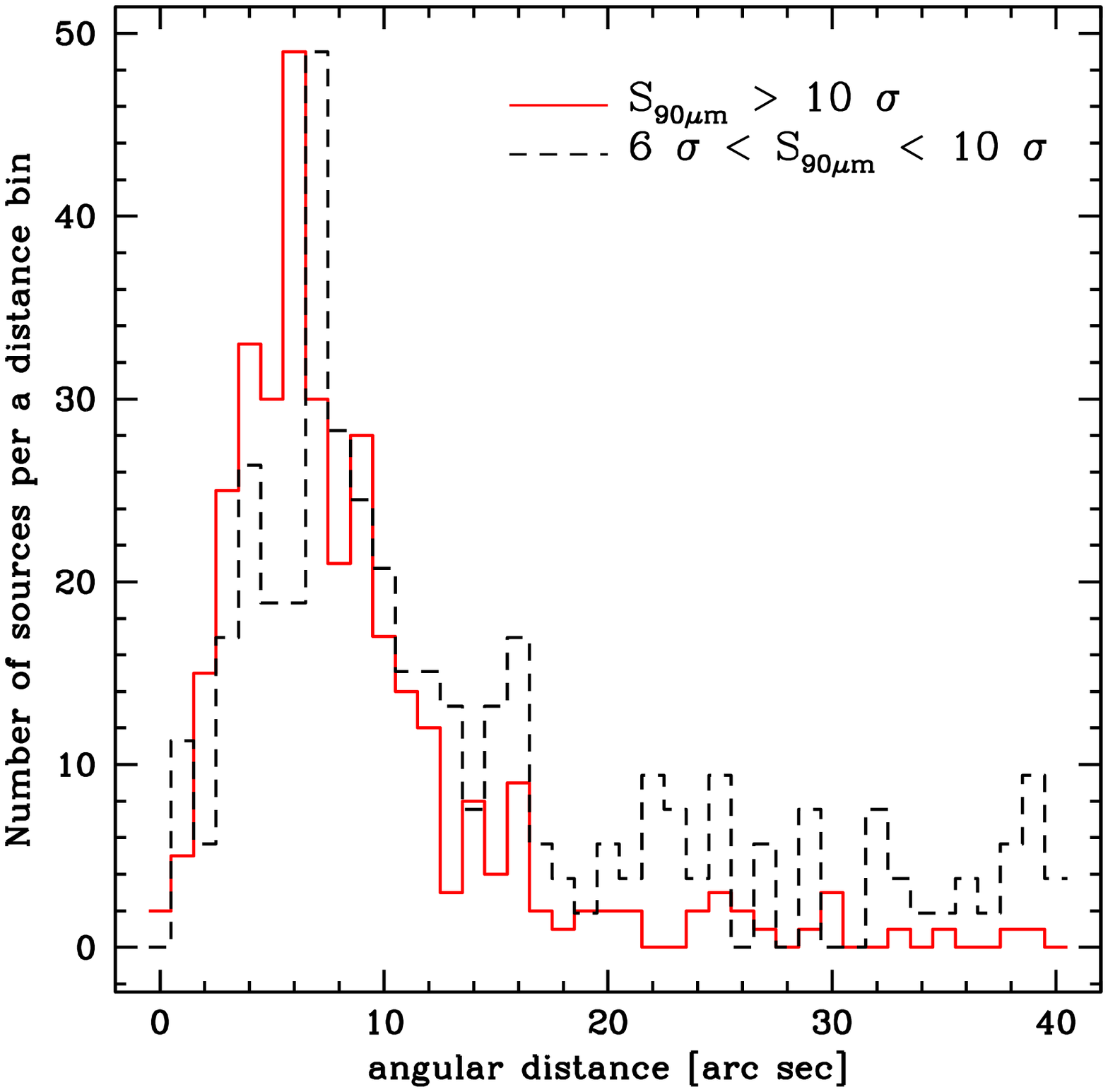}
\caption{The distribution of the angular deviations of the nearest counterparts from the ADF-S sources. 
A solid line corresponds to the $10 \sigma$ catalog, while a dashed line corresponds to
the faint part of the $6 \sigma$ sample. 
In both cases most of the identified sources have counterparts closer than 
$20 ''$. 
The precision of the position determination is decreasing with the source 
flux density - for the $10 \sigma$ sample the median value of angular distance of the 
source from a counterpart is 6.9 arc sec (6.45 below $20 ''$), while for the faint part 
of the $6 \sigma$ sample the median value of angular distance of the source from 
a counterpart is 9.7 arc sec (8.2 below $20 ''$). 
The identifications at angular distance $>20''$ might be an effect of contamination.  
}\label{angdist}
\end{figure}

\begin{figure}[tb]
\centering
\includegraphics[width=8.5cm, height=8.5cm]{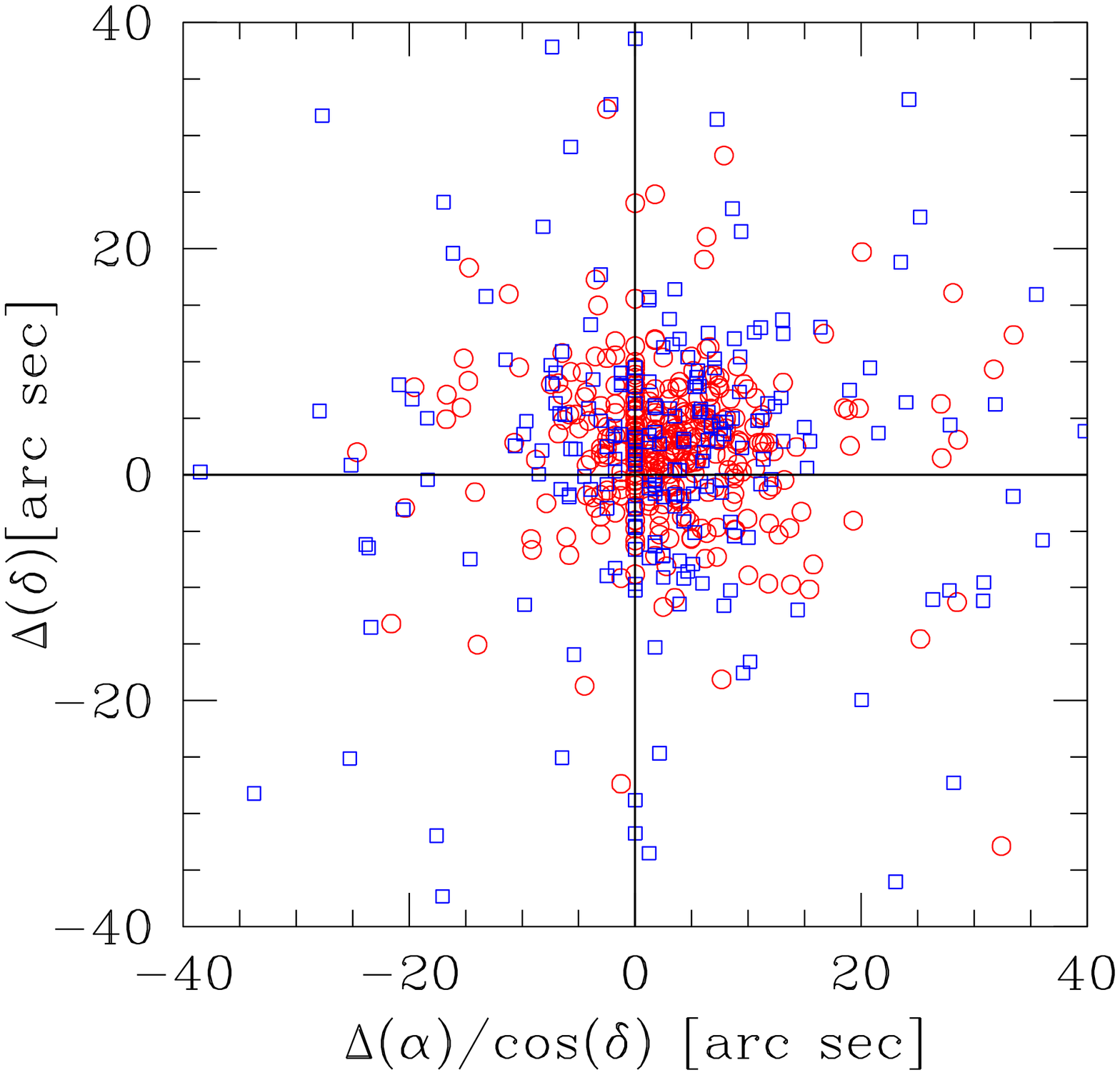}
\caption{ 
Scatter plot of the deviation of counterparts from the ADF-S sources in right ascension and declination. 
Open circles denote sources from the $10 \sigma$ catalog, open squares - remaining 
objects from the faint part of the $6 \sigma$ catalog. 
We can observe a small but systematic bias, similar for brighter and fainter part of the sample, 
even if the scatter for the fainter objects is larger.
}\label{scatter}
\end{figure}

As shown in Figure~\ref{angdist}, the angular separation between the ADF-S source 
and its counterpart is smaller than 20$''$ in most cases. 
It suggests that the actual resolution of the ADF-S map roughly corresponds to 
the pixel size of the FIS detector. 
It is plausible that the more distant identifications are caused by the contamination. 
However, since they are very few, their presence should not affect the quality of 
our sample and we decided to keep our original criterion.

\begin{figure*}[t]
\centering
\includegraphics[width=8.5cm]{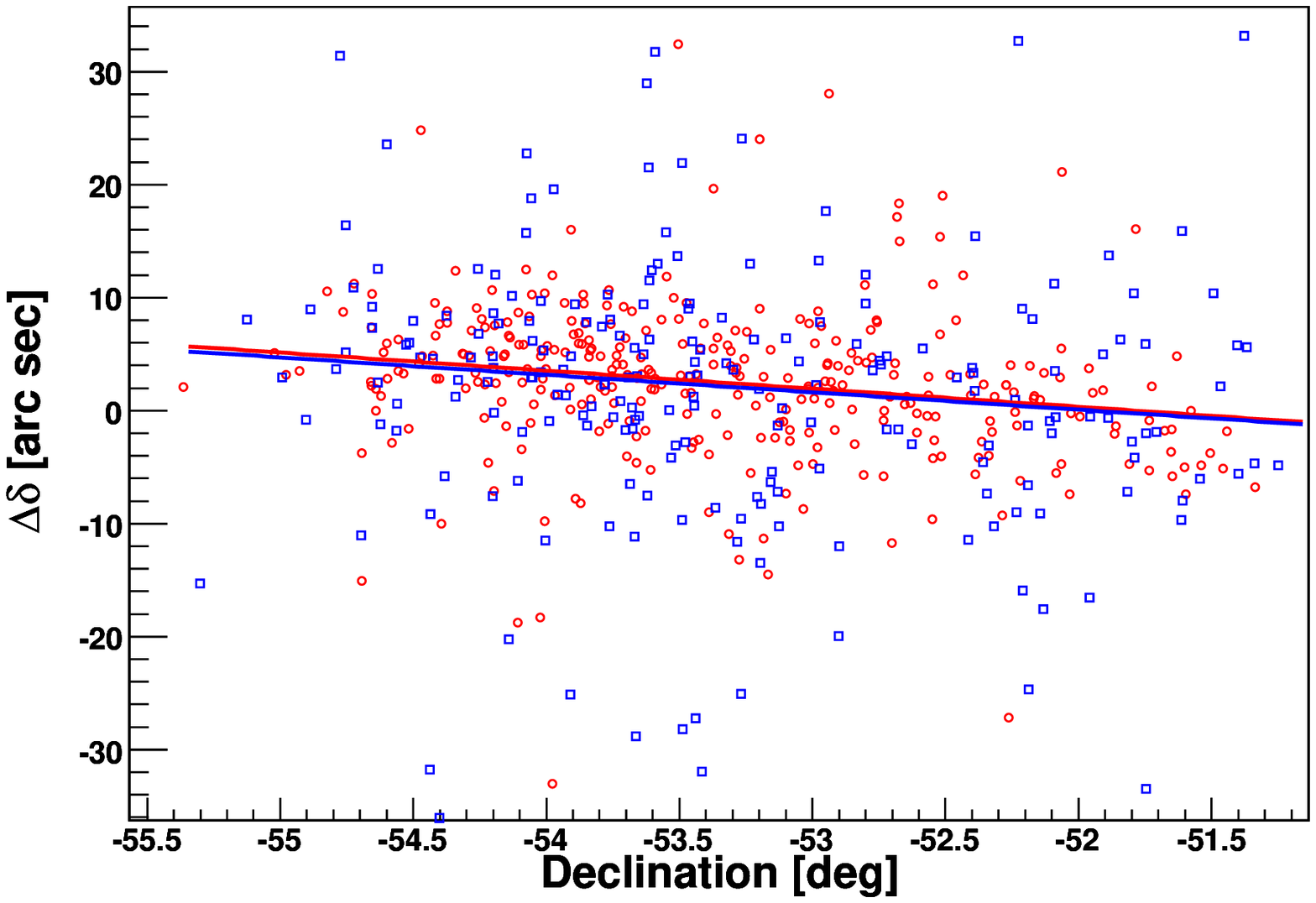}
\includegraphics[width=8.5cm]{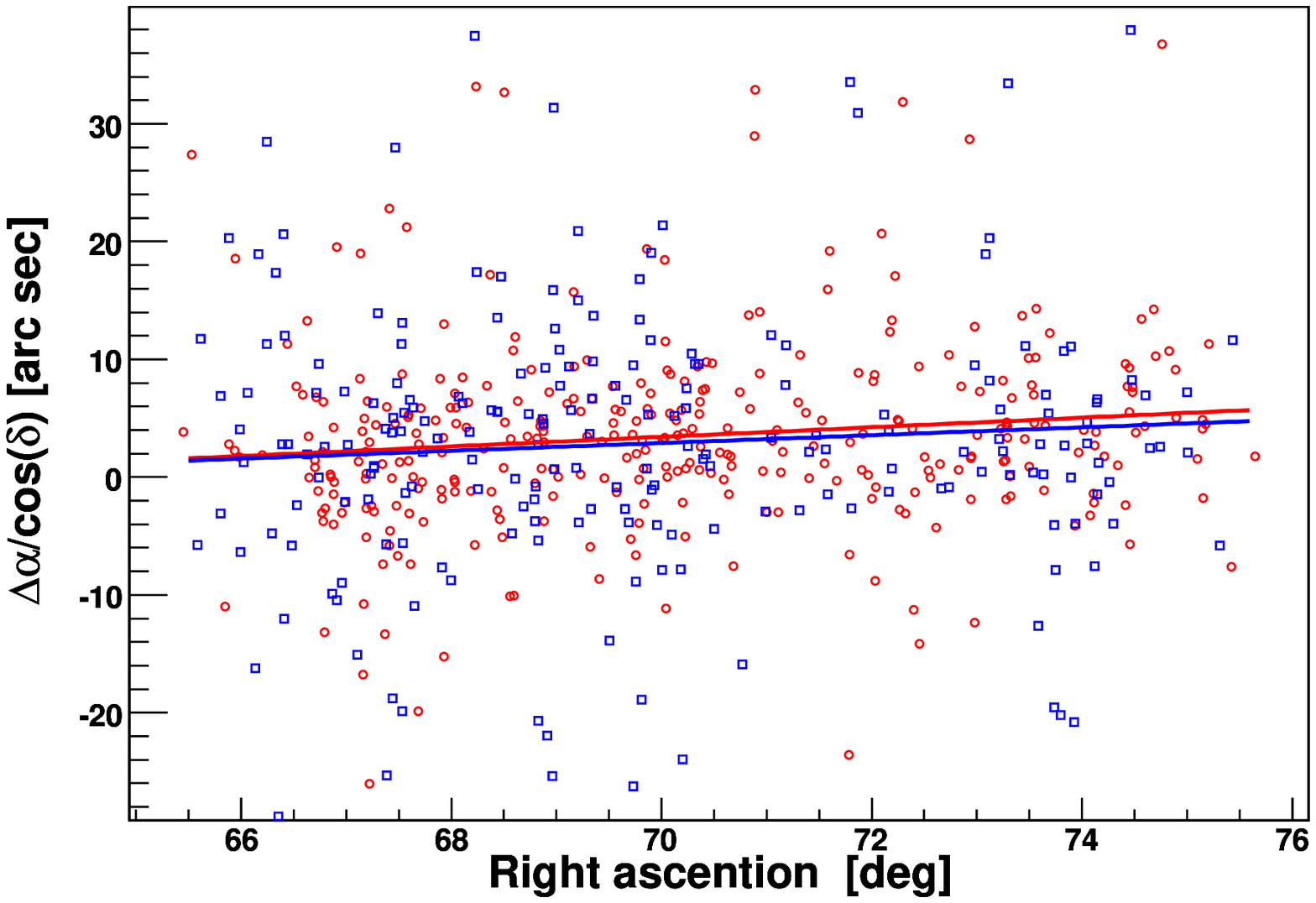}
\includegraphics[width=8.5cm]{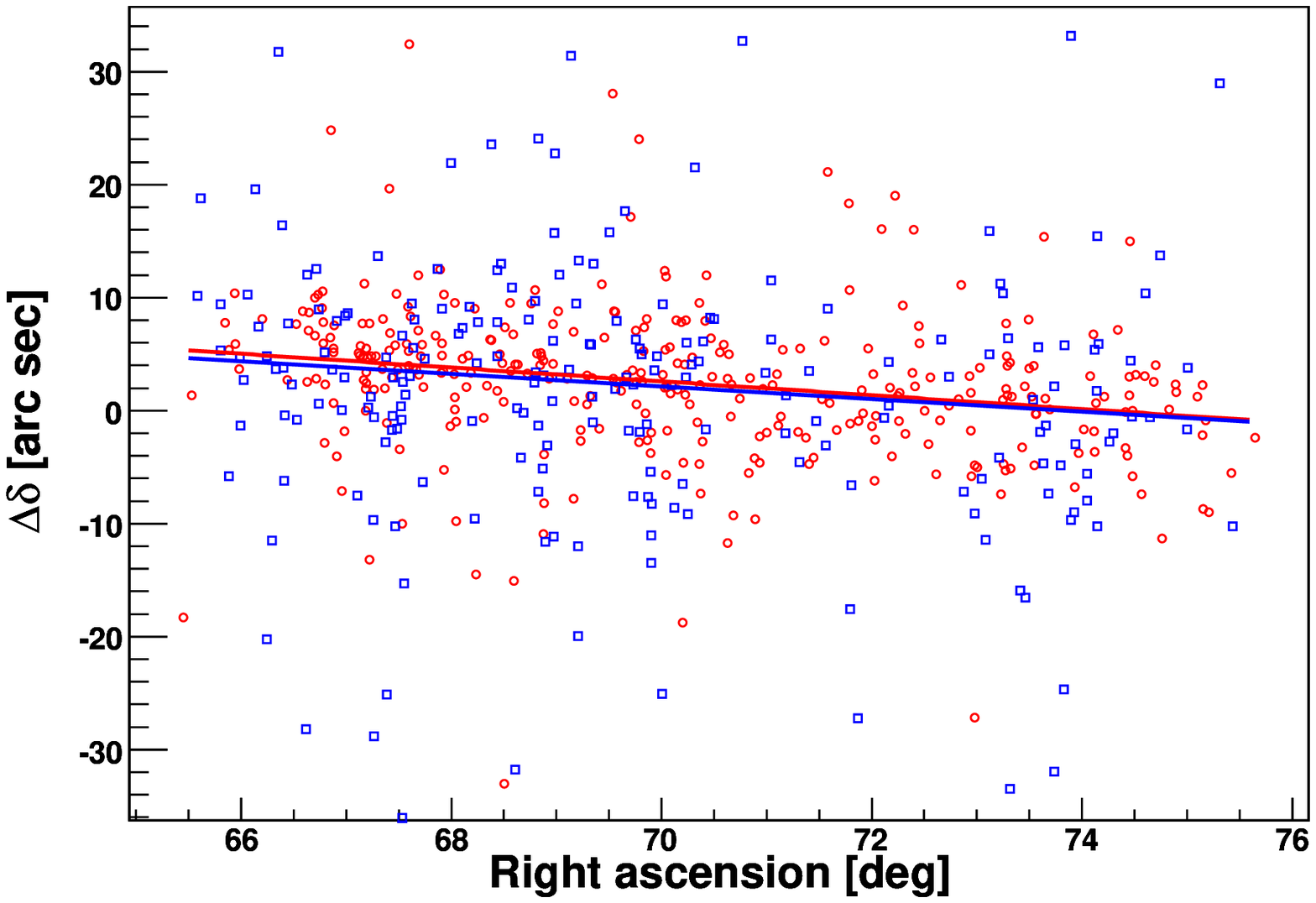}
\includegraphics[width=8.5cm]{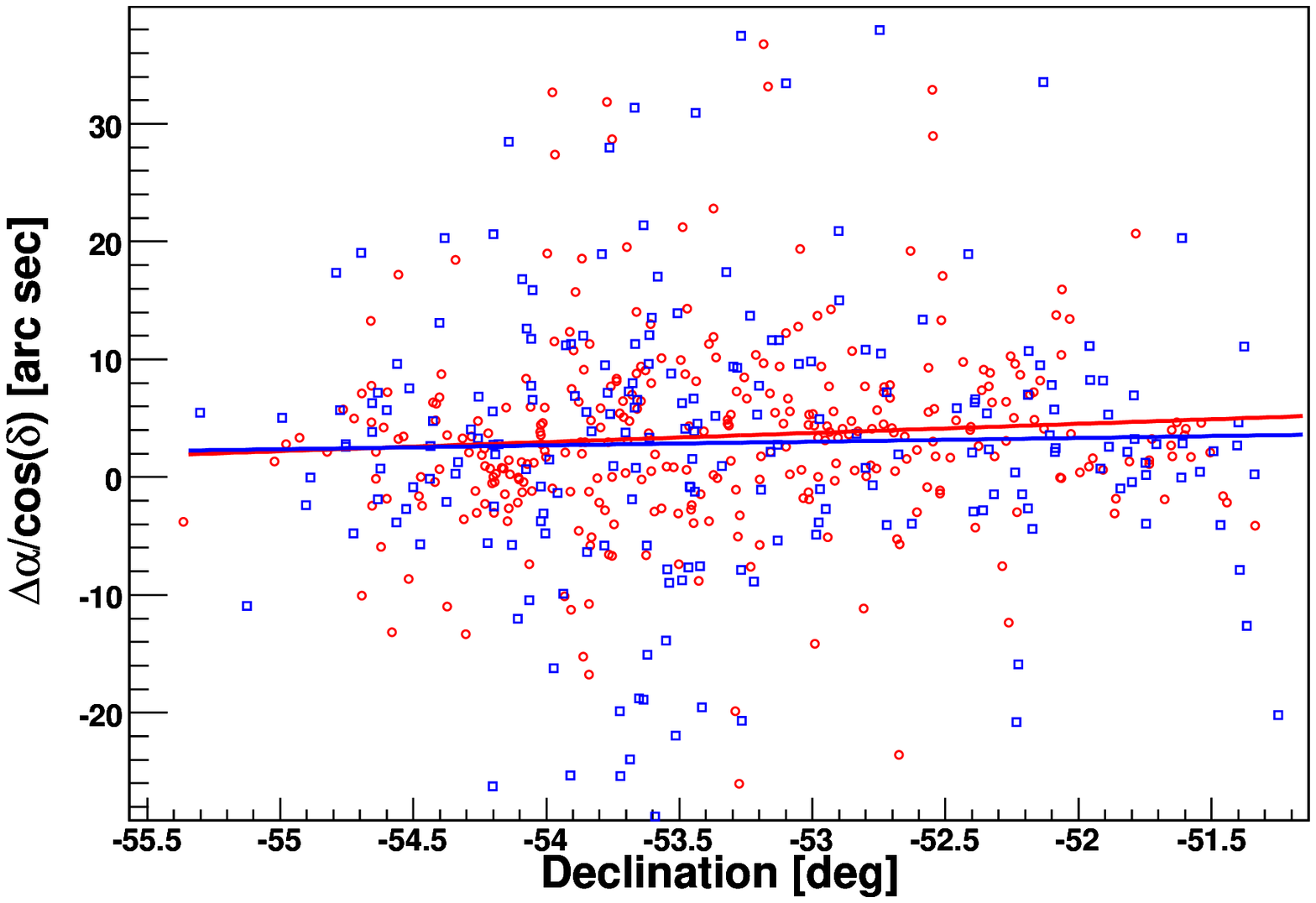}
\caption{ 
Dependence of the deviation of the position of counterparts of the ADF-S sources in right ascension 
and declination, as a function of right ascension and declination. 
As in Figure~\ref{scatter}, open circles correspond to sources from the $10 \sigma$ catalog and open squares - to 
sources from the faint part of the $6 \sigma$ catalog. 
The solid lines correspond to the best $rms$ fits for the bright 
and faint samples. 
}\label{radec}
\end{figure*}

\begin{table}[ht]
\caption{Best $rms$ fit of the positional deviation}
\label{tabrms}
\centering 
\begin{tabular} {lcc}
\hline
\hline
 Deviation & $10 \sigma$ catalog & $6 \sigma$ catalog  \\
\hline
\hline
 & slope& slope  \\
\hline
$\Delta\alpha$/cos($\delta$) vs $\alpha$ & +0.4$\pm$0.2 & +0.3$\pm$0.2\\
$\Delta\delta$ vs $\delta$ & -1.6$\pm$0.5 & -1.6$\pm$0.5 \\
$\Delta\delta$ vs $\alpha$ & -0.6$\pm$0.2 & -0.6$\pm$0.2 \\
$\Delta\alpha$/cos($\delta$) vs $\delta$ & +0.8$\pm$0.5& +0.3$\pm$0.5\\ 
\hline
  &  amplitude &  amplitude \\
\hline
$\Delta\alpha$/cos($\delta$) vs $\alpha$ & -25$\pm$12&-20$\pm$12\\
$\Delta\delta$ vs $\delta$ &-83$\pm$24&-80$\pm$24\\
$\Delta\delta$ vs $\alpha$&+45$\pm$10&+41$\pm$11\\
$\Delta\alpha$/cos($\delta$) vs $\delta$ &+45$\pm$28&+20$\pm$26\\ 
\hline
\hline
\end{tabular}
\end{table}

Positional scatter map, shown in Figure~\ref{scatter}, displays a small but systematic 
bias of $\sim 4''$ in declination of the ADF-S positions with respect to counterparts.
Dependence of a positional deviation in the right ascension $\alpha$ and declination 
$\delta$ as a function of $\alpha$ and $\delta$ is presented in the four panels of Figure~\ref{radec}. 
We can observe that the bias of the position determination depends on the source position, 
being the strongest in the south-west part of the field. 
However, it should be noted that in all cases the bias is much smaller than the uncertainty of 
the determined position resulting from the AKARI FIS detector resolution. 

The parameters of the the best $rms$ fits to the data in Figure~\ref{radec} are listed in 
Table~\ref{tabrms}. 
The results are presented both for the 330 identified sources from the $10 \sigma$ catalog 
and for all 545 identified sources from the full $6 \sigma$ sample. 
We can observe that including the fainter sources into the sample increases the observed scatter, 
however, at the same time it reduces the bias in the determination of $\alpha$ and its 
dependence on the source position in the ADF-S. 
Thus, we should be aware that the positions of brightest ADF-S sources are the most biased, 
even if the bias itself remains small. 
This bias may result from the complex method of source extraction from the slow-scan 
data in the ADF-S \citep[see][for details]{shirahata}.

\subsection{Classification of objects identified in the ADF-S}\label{subsec:classification}

In the $10 \sigma$ catalog, among 330 identified objects, 314 are known galaxies (one appearing twice). 
173 (55~\%) of these galaxies were previously observed in the infrared (either by IRAS or 2MASS) 
but 141 (45~\%) of them were not known previously as infrared sources. 
Remaining 16 objects include radio and infrared sources of unknown nature, 
five stars and one quasar at $z = 1.053$. 
However, the stars in this sample belong to a sparse group of objects with the distance of 
a counterpart from the ADF-S object close to $40''$. 
A careful examination has revealed that they are most probably falsely identified 
because of the contamination (M.\ Fukagawa, private communication). 

In the fainter half of the $6 \sigma$ catalog, i.e. fainter than $10 \sigma$, among 215 identified sources, 205 are known galaxies 
(one appearing twice). 
However, only 32 (16~\%) of them were previously known as infrared sources.
The remaining 173 galaxies are observed in infrared for the first time. 
Among the remaining objects there are 7 radio sources of an unknown nature, 
one quasar at z $\sim 1.23$, one X-ray source related to the starburst galaxy 
NGC~1705, which is the fourth brightest source in our $10 \sigma$ data set, and 
one double stellar system of $B \sim 6$. 
Summary of the properties of identified sources is given in Table \ref{tabgal}. 

\begin{table}[ht]
\caption{Classification of identified ADF-S sources}
\label{tabgal}
\centering 
\begin{tabular} {llcc}
\hline
\hline
Catalogs: & & $10 \sigma$ & $6 \sigma$ \\
\hline
\hline
Galaxies & & 314 & 518 \\
\hline
& infrared galaxies & 144  & 176 \\
& radio-loud galaxies & 12  & 13\\
\hline
& galaxies in a cluster of galaxies & 33 & 40 \\
& galaxies in a pair of galaxies & 2  & 2\\
& interacting galaxies & 2  & 2\\
\hline
& low surface brightness galaxies & 1 & 1\\
& Seyfert-1 galaxies & 1  & 2\\
& starburst galaxies & 1  & 1\\
\hline
\hline
Stars & & 5  & 6\\
IR sources & & 3 & 3 \\
Radio sources & & 6 & 13 \\
Quasars & & 1  & 2\\
X-ray sources & & - & 1 \\
\hline
\hline
\end{tabular}
\end{table}

The data related to the nearest counterparts of the ADF-S sources are summarized in the tables which are available online. 
The names of counterparts (in case of different naming conventions, we use the primary name given by the NED, when available; otherwise the primary name given by the SIMBAD), the positions of the ADF-S sources and corresponding nearest counterparts, the angular distances of counterparts from the ADF-S sources, as well as their redshifts, 
when available, are given in the online Tables~\ref{ADFScount} and~\ref{ADFScount2}. 
The exemplary part of this table, containing data for the 10 brightest 
ADF-S sources, is shown in the text as Table~\ref{exADFScount}.
Flux densities of ADF-S sources with the identified counterparts, in four 
FIR bands, are given in online Tables~\ref{ADFSmeasurements} and~\ref{ADFSmeasurements2}. 
Again, an exemplary part containing data for the 10 brightest ADF-S sources, is shown in the text as Table~\ref{exADFSmeasurements}.
The flux densities of counterparts in all the other wavebands, always in the units of Janskys, are given in online Tables~\ref{measurements1},~\ref{measurements2},~\ref{measurements3},~\ref{measurements4},~\ref{measurements1a}, and~\ref{measurements2a}. 
An example of such a table, showing a part of the data for the 10 brightest ADF-S sources, is shown in the text as Table~\ref{exmeasurements}.

\begin{figure*}[t]
\centering
\includegraphics[width=17.0cm]{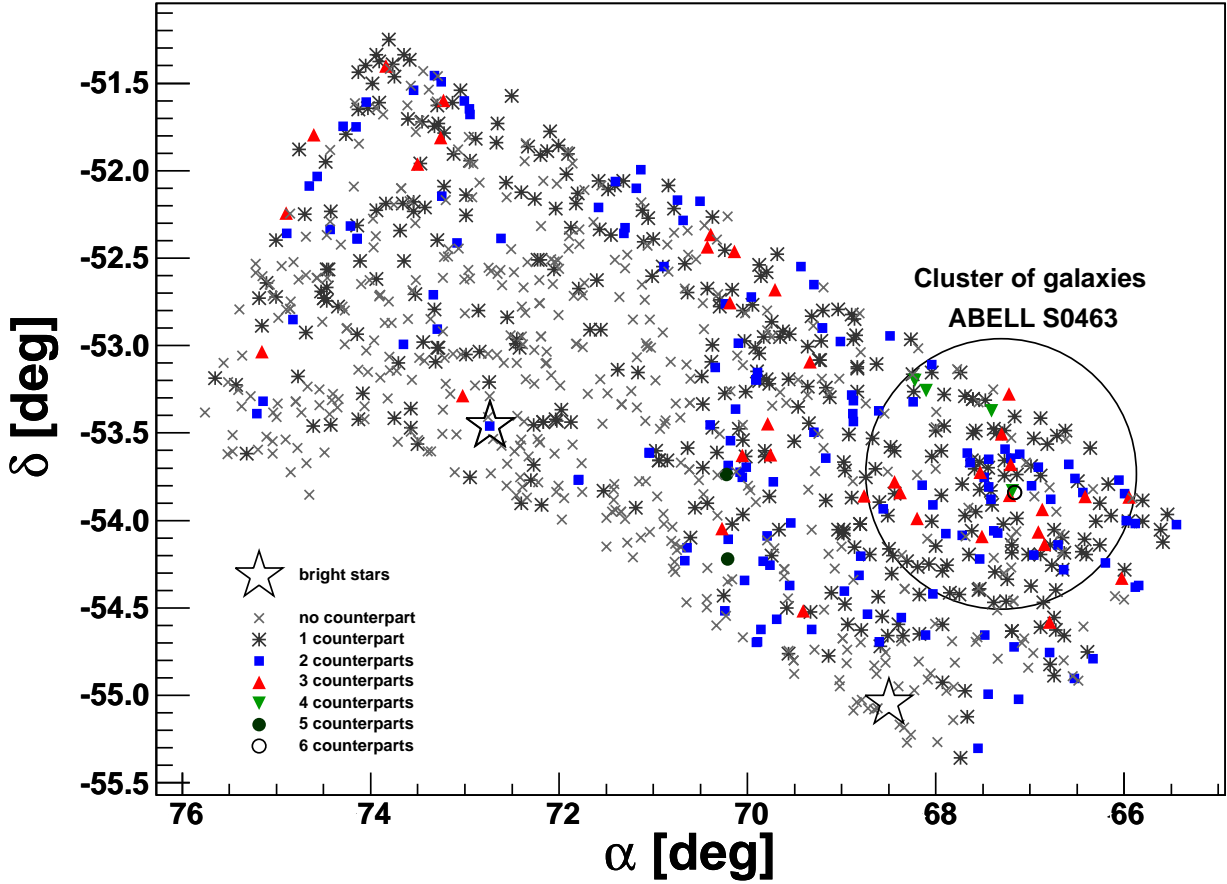}
  \caption{
The map of the ADF-S with the positions of the identified and unidentified 
objects marked. 
Sources for which no identification was found (455 sources) are marked as x-s. 
Sources with one counterpart (362) as shown as small stars, sources with two 
counterparts (139) are marked as full squares, those with three counterparts (36) triangles, 
with four counterparts (4) -- as upside-down triangles, with five counterparts (2) as full 
circles and with six counterparts (2) as open circles. 
The positions of two optically bright stars (around 3 and 6~mag) present in the ADF-S 
are indicated by big empty stars. 
One of these stars is the double system and an X-ray source; we have also identified it as 
one of the ADF-S sources. 
The position of the cluster of galaxies Abell S0463 at $z \sim 0.04$ is indicated by a big empty circle. 
}\label{counterparts}
\end{figure*}

As shown in Figure~\ref{counterparts}, the sources with multiple counterparts appear 
mainly in a particular part of the ADF-S, where the nearby cluster of galaxies 
Abell~S0463 is located. 
We can also observe that in the region corresponding to this cluster there are very few
unidentified sources. 
This facts can be partially related to the higher local density of galaxies 
in this part of the field, which increases the risk of the chance coincidence of the angular 
position of the ADF-S source with one of the cluster galaxies. 
On the other hand, however, this cluster has been intensely observed in the past, 
and the optically bright galaxies in this area are sampled much better and deeper 
than in the rest of the ADF-S. 
Then, if we assume that most of the identifications in the cluster area correct, 
we can deduce that the unidentified objects in the other part of the field are 
similar nearby galaxies, not observed before because of their relatively low surface brightness.

\subsection{Number counts}\label{subsec:nc}

\begin{figure}[t]
\centering
\includegraphics[width=8.5cm]{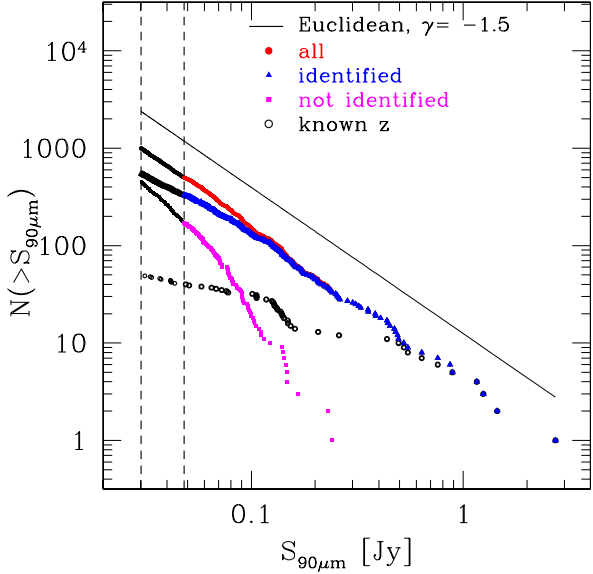}
\caption{Integral number counts of objects in the the $10 \sigma$ 
and $6 \sigma$ ADF-S samples in 90 $\mu$m. 
Filled circles correspond to all the 6 and $10\sigma$ datasets, triangles - 
to the identified objects, squares - to not identified objects and open circles - to objects with known $z$. 
The dashed vertical lines mark the limiting luminosities of $10 \sigma$ and $6 \sigma$ catalogs. 
For comparison, the ideal Euclidean case, with an arbitrary amplitude, has been shown as a solid line.}
\label{ncounts}
\end{figure}

A similar conclusion that the bright ADF-S sources are mostly nearby galaxies can 
be deduced by the analysis of number counts of our sample, which is presented in 
Figure~\ref{ncounts} \citep[for a more detailed analysis of the number counts of 
the ADF-S galaxies, see][]{shirahata}. 
In a homogeneous Euclidean Universe it is expected that the cumulative number counts 
of sources follow a power law $N(>S) \propto S^{-3/2}$ 
\citep[for cosmological discussions, see e.g.,][]{peebles93}. 
This relation does not depend on the luminosity function of the sources. 
Indeed, the number counts of nearby galaxies usually can be well fitted by the power 
law with a slope $-1.5$. 
This relation deviates from the Euclidean slope when the cosmological and evolutionary 
effects become important \citep[e.g.,][]{yoshii88, metcalfe96}. 
Thus, the deviation of the measured number counts of the sample from the 
Euclidean slope at the bright flux densities may be a simple test of the completeness 
of the sample at a first approximation.

\begin{table}[ht]
\caption{Slope of number counts of ADF-S sources}
\label{tabnc}
\centering
\begin{tabular} {lcc}
\hline
\hline
Objects & $10\sigma$ sample & $6\sigma$ sample \\
	& $S_{90\mu{\rm m}}>0.0482$ Jy & $S_{90\mu{\rm m}}>0.0301$ Jy \\
\hline
All & -1.63$\pm$0.03 & -1.58$\pm$0.02 \\
Identified & -1.45$\pm$0.03 & -1.34$\pm$0.02 \\
Not identified & -3.15$\pm$0.09 & -2.55$\pm$0.04 \\
Known $z$ & -0.84$\pm$0.05 & -0.77$\pm$0.04 \\
\hline
\hline
\end{tabular}
\end{table}

As shown in Figure~\ref{ncounts}, the number counts for all the FIR-bright 
ADF-S sample, from the 90 $\mu$m {\it WIDE-S} measurement, 
are quite well fitted to the Euclidean number counts, being only slightly steeper. 
It implies that these objects are mainly nearby galaxies. 
The number counts of identified sources are less steep, while the number counts of 
unidentified sources are steeper than Euclidean.
This, together with the above discussion, implies that the unidentified sources are less 
luminous (and, therefore, more difficult to observe also optically) nearby population 
of galaxies. 
As expected, galaxies with known redshifts form the brightest sample with 
the flattest number counts. 

For the twice larger $6 \sigma$ sample the situation changes only slightly. 
The comparison of slopes of the best $rms$ fits to the observed number counts for 
both samples are given in Table~\ref{tabnc}. 

In total, we conclude that the bright part of the FIR-selected sample of celestial 
objects in the ADF-S consists mainly of nearby galaxies. 
The average properties of these galaxies are examined in the following.

\subsection{Redshift distribution}\label{subsec:redshift}

\begin{figure}
\centering
\includegraphics[width=8.5cm]{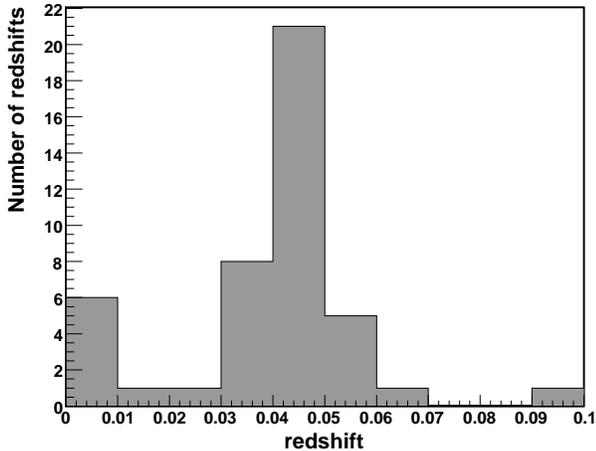}
  \caption{
The redshift histogram of 44 counterparts of the ADF-S objects with known 
redshifts in 0.01 bins. 
Four objects with redshifts higher than 0.2 are not shown here. 
These are: one galaxy at $z=0.2591$, one Seyfert-1 galaxy at $z=0.243$ and 
two quasars: HE $0435-5304$, located at $z=1.232$ and VV2006 J044011.9-524818, 
located at $z=1.053$.
}\label{zhist}
\end{figure}

The redshift information is available for 48 galaxies from the full $6 \sigma$ sample. 
The only two high redshift sources are quasars: VV2006 J044011.9-524818, 
located at $z=1.053$ and HE $0435-5304$, located at $z=1.232$, and
two more galaxies are located at $z \sim 0.25$. 
All the other sources are nearby galaxies at $z < 0.1$. 
The redshift distribution of these galaxies, shown in Figure~\ref{zhist},
demonstrates that a large part of them belongs to a cluster Abell S0463 at z~0.04. 

\subsection{Completeness of identifications}\label{subsec:completeness}

\begin{figure}[t]
\centering
\includegraphics[width=8.5cm]{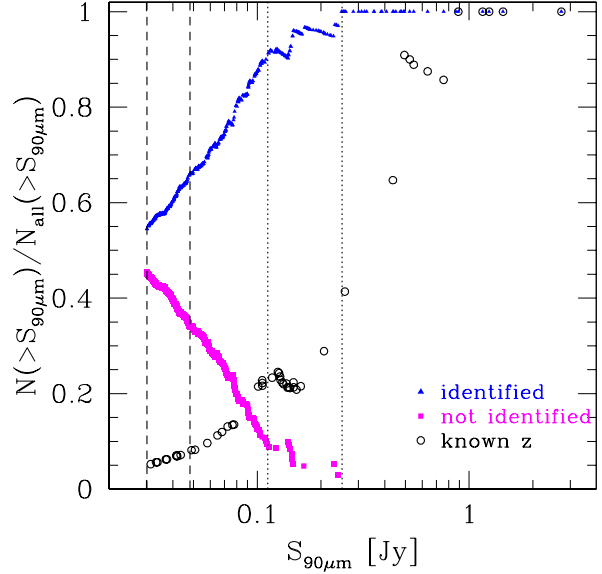}
\caption{
Ratios of identified and unidentified sources, and of sources with a known
redshift, with respect to the total number of sources in the subsamples
of a different 90 $\mu$m limiting luminosity. Similarly to
Figure~\ref{ncounts},
triangles correspond
to the identified objects, squares - to not identified objects and open circles - to objects with known $z$.
The dashed vertical lines mark the limiting luminosities of $10 \sigma$ and $6 \sigma$ catalogs. The dotted vertical lines mark the limits of 100~\% and 90~\% completeness of the catalog of counterparts.
}
\label{perc}
\end{figure}

In Figure~\ref{perc} we present the ratio of
identified and not identified sources, as well as that of objects with
measured redshifts, with respect to the total number of sources in
subsamples with different limiting flux density at 90 $\mu$m.
It shows that the sample of identified sources is 100~\%
complete until $S_{90} = 0.25$ Jy, which corresponds to
first 32 sources and remains more than 90~\% complete until 
$S_{90} = 0.105$ Jy, which corresponds to the first 140 sources. 
For sources fainter than 0.105 Jy, the completeness of the catalog
starts to fall down rapidly to 66~\% for the $10 \sigma$
and 55~\% for the complete $6 \sigma$ sample. 

This incompleteness has to be taken into account if we try 
to apply the conclusions from the analysis of the identified sample
to to whole FIR-bright ADF-S dataset.

\subsection{Galaxy morphologies and environment}\label{subsec:morphology}

Among the identified galaxies, 67 in the $10 \sigma$ sample and 10 in the fainter 
part of the $6 \sigma$ sample have determined morphologies. 
Most of them (but not all) belong to the cluster Abell~S0463 and were identified by 
\cite{dressler80a}. The redshift of the cluster is $z \sim 0.0394$ \citep{abell89}. 
It is a regular (type I-II in the Bautz-Morgan classification), moderately rich (population 84), 
lenticular-rich galaxy cluster and it was used as a typical regular cluster in 
a sample investigated by \cite{dressler80b}. 
We can assume, then, that our morphological sample can be - in a large part - representative 
for the nearby bright galaxy population. However, the presence of a rich cluster may introduce some bias for towards dense environments. 
In Table \ref{morph} we present the statistics of the galaxy types in our sample 
compared to frequencies of different types usually found in an optically bright 
galaxy population in a nearby Universe \citep{deVac}. 
The first, not unexpected but striking observation is a high percentage of peculiar 
galaxies in our sample. 
The high FIR luminosity of dust in these objects reflects on-going star forming 
processes induced in them by interactions with other galaxies. 
We observe slightly more spiral galaxies than expected in an optically-selected 
sample but - given the number statistics - this over-representation of spirals is not significant. 
This is compensated by a significantly (even given a small number statistics) 
lower amount of elliptical galaxies in the sample. 
Moreover, among five found elliptical galaxies found one belongs to a pair 
of interacting galaxies, and one is a Seyfert-1 active galaxy. 
These particularities probably explain the unusual FIR luminosity of these 
two galaxies and make a ratio of seemingly normal elliptical galaxies in our sample even lower. 

The amount of lenticular (S0) galaxies in our sample turned out to be practically identical
 among normal optically bright galaxies. 
The issue of dust in lenticular galaxies has been already discussed in the literature 
for years. 
Lanes of dust and gas were found in many objects of this type \citep[e.g.][]{danks79,silchenko2004}.
The IRAS observations have revealed that $\sim 68$~\% lenticular galaxies, compared to
$\sim 45$~\% ellipticals, contain cool 
dust \citep{knapp89} and remain visible in the infrared. 
Our results are consistent with the conclusion of \citet{robhay94} that, taking the FIR detection 
as a criterion, the most prominent distinction can be made between ellipticals and 
spirals, with lenticulars remaining an intermediate (and slightly elusive) type.  
The Spitzer detailed observations of three lenticular galaxies has revealed that 
even if their bulge-to-disk ratios support their classification as lenticulars, they 
contain warm dust forming a structure similar to spiral arms \citep{pahre}. 
The ratio of lenticular galaxies in the FIR selected sample being the same as 
in the optically selected sample of bright galaxies may be a further evidence 
that that the presence of warm dust in lenticular galaxies may be their common feature.

Although lenticulars were originally introduced as an intermediate transition class between elliptical and spiral galaxies \citep{hubble}, recently there is a mounting evidence that they are probably a much more complex class of objects. 
It has been suggested \citep{vandenbergh} that lenticulars can be divided into two subpopulations, with different formation histories. 
Some faint lenticulars could indeed form via secular formation processes at early epochs or by slow stripping of gas from spirals in the cluster environment \citep{abadi}. 
The luminous ones could be rather an effect of mergers of spiral galaxies \citep{bekki}. 
Most probably also other processes, like gas starvation, or gas ejection by active nuclei, should be taken into account to describe in a satisfactory way the evolution of these class of galaxies \citep{vandenbergh09}.
This more complex scenarios seem to be supported by recent observations of lenticulars e.g. in the near infrared \citep{barway}.
Detection of a significant population of lenticular galaxies in FIR proves that presence of substantial amounts of cold dust is a quite normal feature for this class of galaxies and may be a further evidence for their complex formation process.

%
\begin{table*}[htb]
\begin{minipage}[t]{\columnwidth}
\caption{77 ADF-S galaxies with known morphological types} 
\label{morph}
\centering  
\renewcommand{\footnoterule}{} 
\begin{tabular}{l r r r r c}  
\hline\hline  
 & \multicolumn{2}{c}{$10 \sigma$ catalog} & \multicolumn{2}{c}{$6 \sigma$ catalog} & \\
\hline
 & number & frequency & number & frequency & usual frequency  \\
& & & & & among bright galaxies\footnote{as in \cite{deVac}} \\ 
\hline
spiral & 45 & 67~\% $\pm$ 10~\% & 50 & 65~\% $\pm$ 9~\% &  61~\% \\
lenticular & 15 & 22.5~\% $\pm$ 6~\% & 17 & 22~\% $\pm$ 5~\% & 22~\% \\
elliptical & 4 & 6~\% $\pm$ 3~\% & 5 & 6.5~\% $\pm$ 3~\% & 13~\%\\
irregular & 1 & 1.5~\% $\pm$ 1.5~\% & 3  & 4~\% $\pm$ 2~\% &3.5~\%  \\
compact & 2 & 3~\% $\pm$ 2~\% & 2  & 2.5~\% $\pm$ 2~\% &-- \\
\hline  
peculiar & 8 & 12~\% $\pm$ 4~\%  & 8 & 10~\% $\pm$ 4~\% & 1~\%  \\
\hline 
\hline
\end{tabular}
\end{minipage}
\end{table*}

\begin{figure*}[t]
\centering
\includegraphics[width=14cm,angle=270]{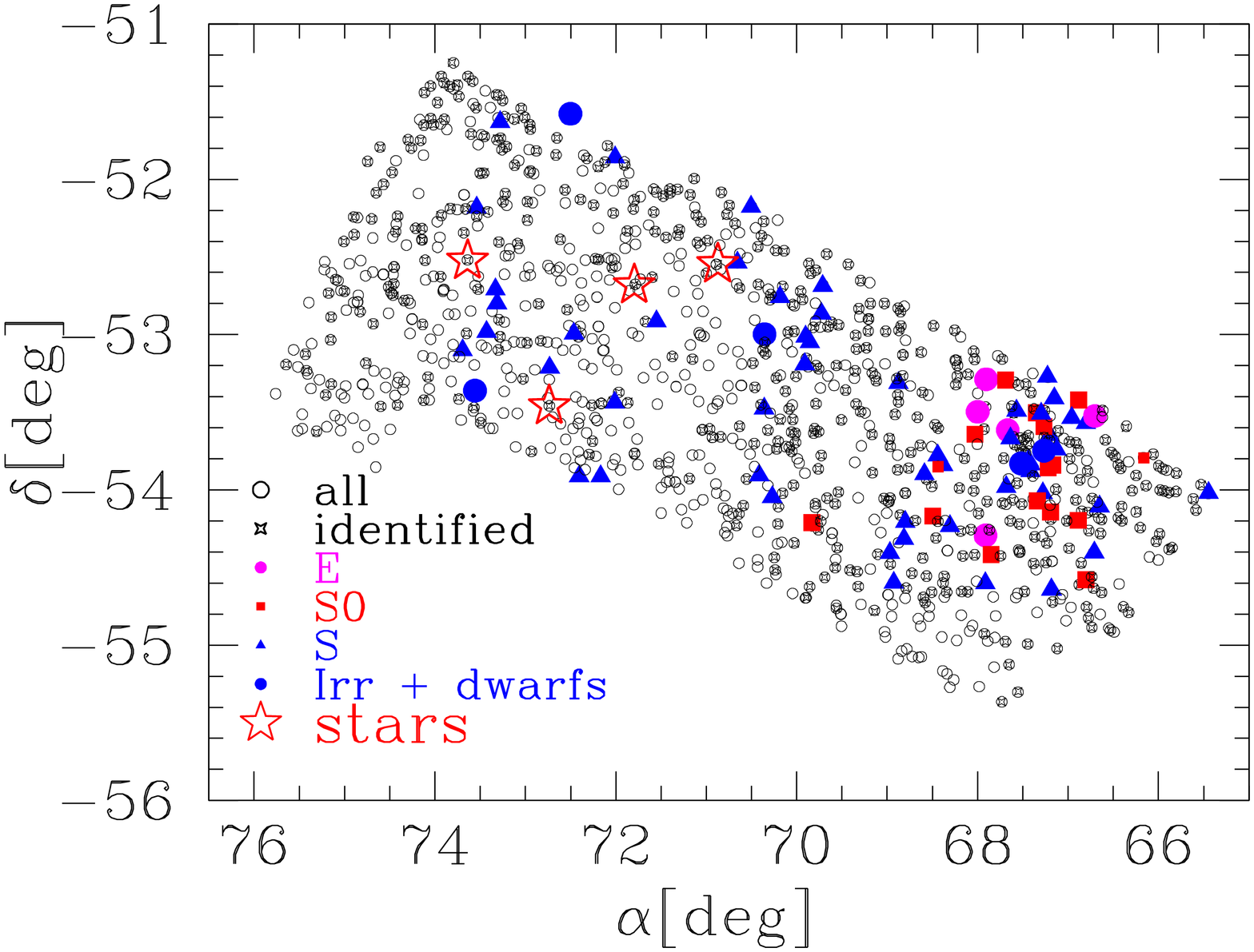}
  \caption{
Positions of galaxies with known morphological types in the ADF-S. 
In this plot, 
positions of all the ADF-S sources are marked by open circles. Identified 
sources are shown as small stars. The elliptical galaxies are shown as full 
circles, lenticular galaxies - as full squares, spiral galaxies - 
as full triangles and 
irregular or dwarf galaxies - as full circles. 
Positions of identified 
stars are shown as big empty stars; however, at least three of these identifications are most probably the effect of contamination.
We can observe that all the elliptical and lenticular galaxies are located in the region of the galaxy 
cluster Abell~S0463 at $z \sim 0.04$. Their morphologies were mostly determined 
by \cite{dressler80a}. The spiral and irregular galaxies, however, are distributed in 
all the field. 
}\label{mapmorph}
\end{figure*}

\subsection{Correlation function}\label{subsec:CF}

A careful analysis of the correlation function and clustering properties of 
galaxies in the ADF-S data will be given in~\cite{kawada09}. 
Here we present a simple analysis of clustering of galaxies from our sample, 
aiming at examining the quality of the data and possible biases.

The correlation function is the simplest statistical measurement of clustering, 
as a function of scale (angular or spatial). 
It corresponds to the second moment of the galaxy distribution. 
{}To compute the angular correlation function $\omega$ of 
the ADF-S galaxies as 
a function of the angular scale $\theta$, we adopted the Landy-Szalay estimator 
\citep{lansal}, that expresses $\omega(\theta)$ as:
\begin{equation}\label{lseq}
\omega(\theta) = \frac{N_R(N_R-1)}{N_G(N_G-1)} \frac{GG(\theta)}{RR(\theta)} 
  - \frac{N_R-1}{N_G} \frac{GR(\theta)}{RR(\theta)} + 1 \;  .
\end{equation}
In this expression $N_G$ and $N_R$ are the mean densities (or, equivalently, 
the total numbers) of objects, respectively, in the galaxy sample and in a catalog 
of random points. 
The random points are distributed within the same survey area and 
with the same 
angular selection biases as galaxies in the ADF-S catalog.  
$GG(\theta)$ is the number of independent galaxy-galaxy pairs with
separation between $\theta$ and $\theta+d\theta$; $RR(\theta)$ is the number 
of independent random-random pairs within the same interval of separations 
and $GR(\theta)$ represents the number of galaxy-random pairs.

In the nearby Universe, the angular correlation function of galaxies usually can be fitted by a power law 
$\omega(\theta) = A_{\omega} \theta^{1-\gamma}$, where an amplitude $A_{\omega}$ 
is a measure of a clustering strength and $\gamma$ informs about its scale dependence. 
In practice, because of the finite size of the survey, the measured $\omega(\theta)$ 
is a biased estimator of the real correlation function and becomes underestimated 
on the large scale. 
The correction factor which needs to be applied is related to as integral constrained 
IC, \citep{pg76},
\begin{equation}
  IC = \frac{1}{\Omega^2} \int \int \omega(\theta) d\Omega_1 d\Omega_2,
\end{equation}
where $\Omega$ is the area of the observed field.
Then, the measured correlation function can be written as:
\begin{equation}
  \omega(\theta) = A_{\omega} (\theta^{1-\gamma} - \mbox{IC}).
\end{equation}

\begin{figure*}[t]
\centering
\includegraphics[width=8.5cm]{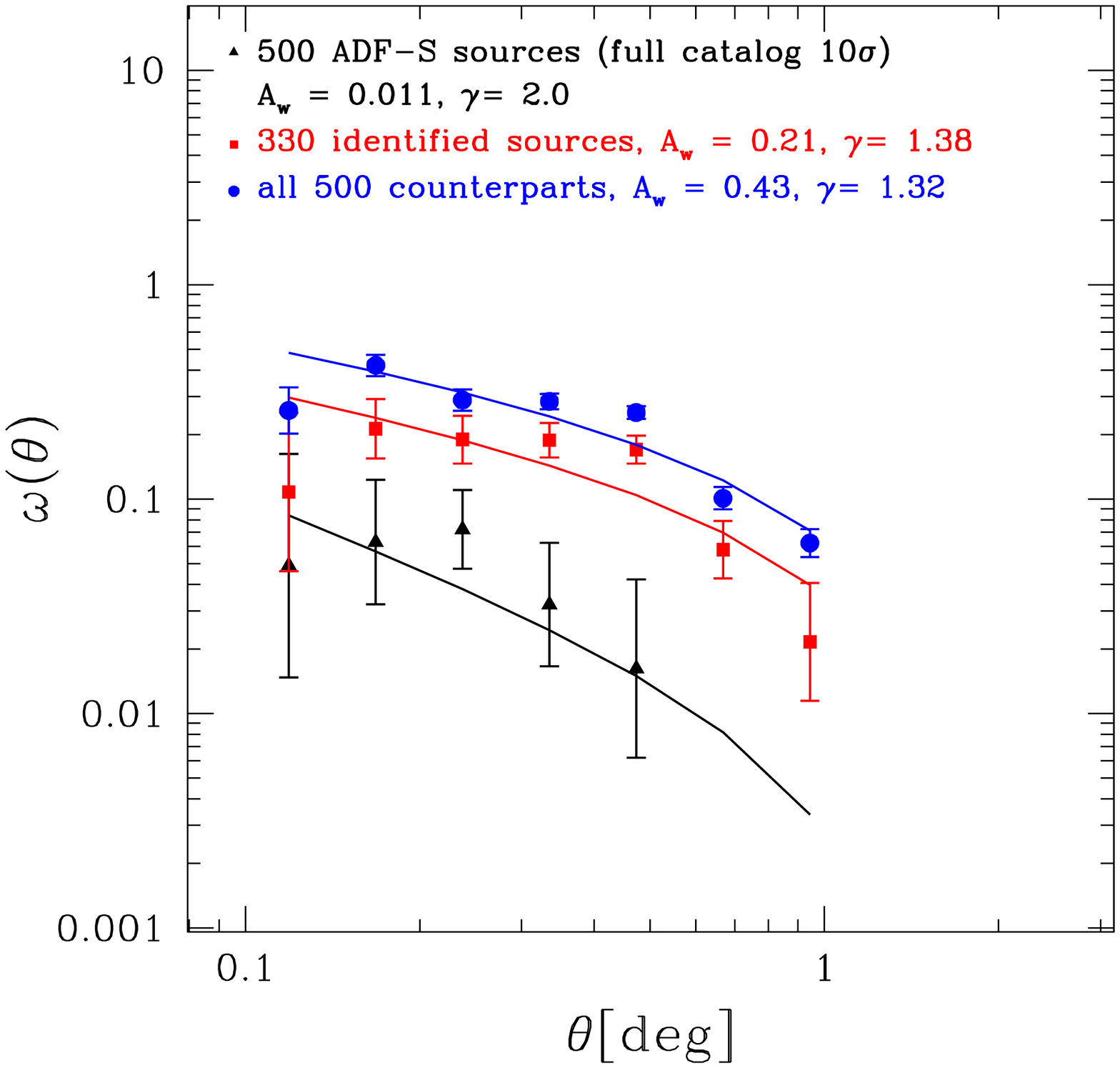}
\includegraphics[width=8.5cm]{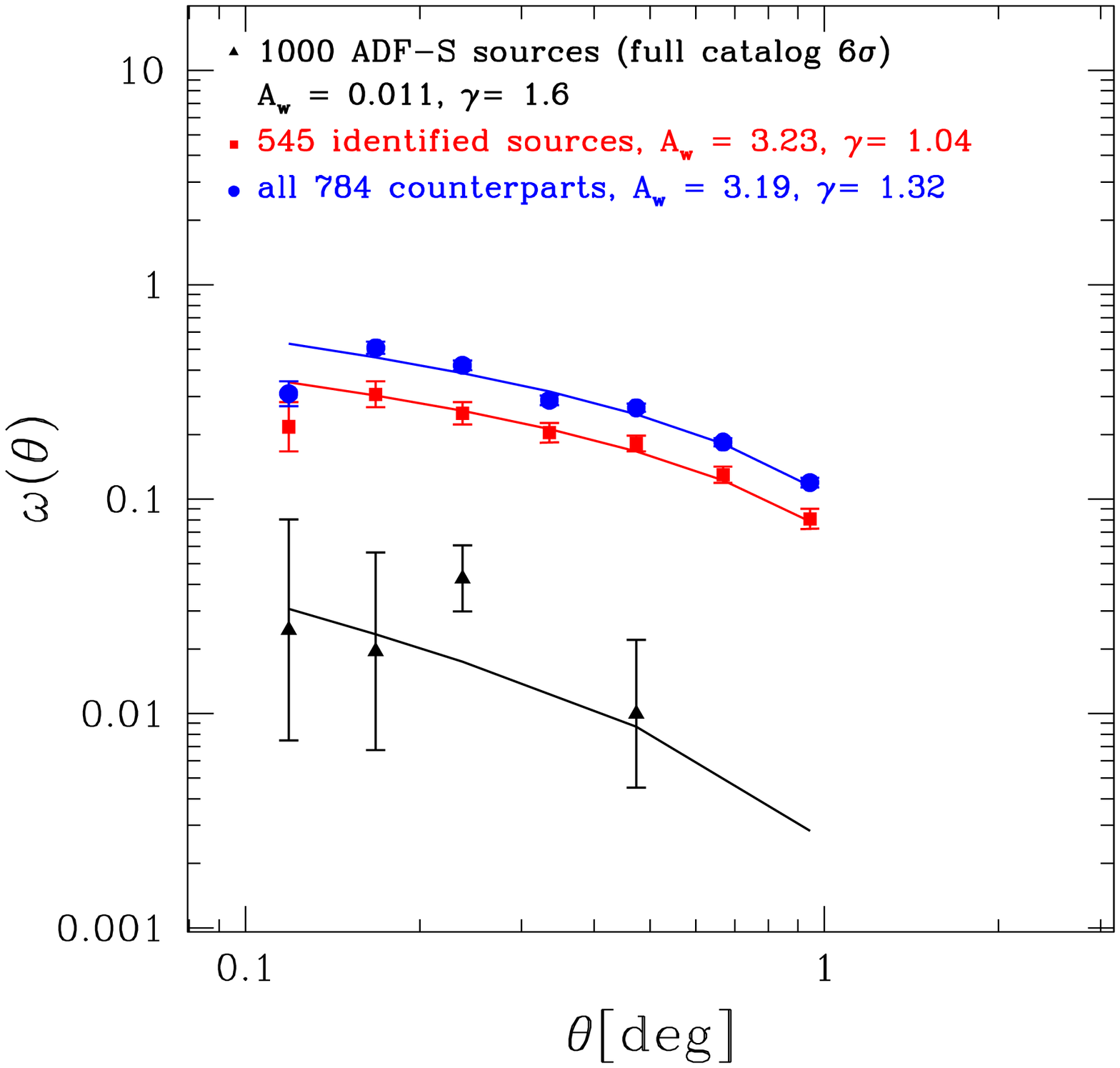}
\caption{
Correlation function of the $10 \sigma$ (left panel) and 
$6 \sigma$ (right panel) ADF-S samples. 
Triangles correspond to the full $10 \sigma$ and $6 \sigma$ 
dataset of ADF-S sources, squares to the samples of the most nearby 
counterparts of identified sources, and circles to samples constructed 
from all the counterparts of the identified ADF-S sources at 
the angular distance smaller than 40'' from the ADF-S source. 
The corresponding best-fitting power-law with an integral 
constraint $\omega(\theta)= A\theta^{-\gamma}-IC$ is shown in each case.
We can observe a lower clustering of the full sample and a higher clustering 
of the identified sample, comparing to the $10 \sigma$ catalog. Correlation 
function of all the counterparts found shows in both cases a deficit of pairs
on small scales which suggests that the level of source confusion 
is higher than the ratio of secondary counterparts, i.e. 34\% of the sample.
}
\label{cf_6s}
\end{figure*}

Correlation functions measured for the $10 \sigma$ and $6 \sigma$ ADF-S samples are shown in 
Figure~\ref{cf_6s}. 
We plot there the best fitted power law with the IC applied. 
In this case we do not use the covariance matrix to correct 
for the correlation between the bins, since the precise measurement of 
the clustering parameters was not our goal. 
Also, for the same reason, the error bars are simple Poissonian errors. 
Thus, both the fits and the errors in our calculation are no more than indicative. 
However, they give a sufficient information to compare the properties 
of different subsets of our data. 
The clustering amplitude for the full $10 \sigma$ ADF-S $90\;\mu$m 
sample is equal to $1.1\times 10^{-3}$ with the slope $\gamma = 2$, which 
is not very 
different from what was found for other infrared galaxy surveys, 
for example SWIRE \citep{oliver,delatorre}. 
In case of the $6 \sigma$ 
catalog we observe a lower clustering amplitude of the whole sample,
which may be attributed to the fact that we are dealing with fainter
 and hence less massive sources.
Computing the correlation function only for
identified sources, in both cases we have a good clustering signal.
The correlation function becomes less steep, and the clustering
amplitude much higher. 
This confirms that the identified sources on the sky
belong to the brighter local population of galaxies which resides in
the denser environments.

In case of the $6 \sigma$ sample, the clustering amplitude of 
identified sources is even higher than for the brighter $10 \sigma$ sample. 
This increase of the clustering amplitude probably reflects the fact 
that the identification process, more in case of the fainter sources 
than in case of brighter ones, was biased because of the presence 
of the cluster of galaxies Abell~S0463 in the ADF-S. 
The objects from the region occupied by this cluster were observed in the past 
more intensely and, as a result, they are much better sampled in the 
public databases. 
Thus, we should be aware that all the conclusions drawn from the catalog of counterparts 
of $6 \sigma$ sources are biased towards the dense environment, comparing
to the $10 \sigma$ catalog. Also, the high clustering amplitude of the
counterparts of the $10 \sigma$ sources should be probably partially 
attributed to this bias, even if the effect is clearly smaller. 

Deriving real-space clustering parameters requires more 
detailed estimation of the redshift distribution of 
our sources, as well as the proper treatment of the possible biases,
and will be a subject of future studies. 

What is visible for $\omega(\theta)$ is a lack of pairs on 
scales smaller than $\sim 0.2$ deg. 
It is a common feature of measurements in IR data, and is caused mainly 
by source confusion. 
The scale where the deficit of pairs occurs in case of 
a sample of nearest counterparts remains, expectantly, the same as 
in case of the full dataset. 
This observation assures us that it is really caused only by a source confusion,
and not by some incompleteness in the point source identification
process, for instance.

{}To examine the possible impact of the source confusion on our data, we made 
an experiment, assuming that all the counterparts found in the $40''$ range, 
not only the nearest ones, can contribute to the FIR flux and, therefore, to the 
clustering signal. As expected, the clustering amplitude of a sample constructed in such a way is even higher. 
The deficit of pairs on the small scales decreases significantly but still remains. 
This result may suggest that the amount of sources missing because of the source confusion in 
our sample is even larger than the amount of the found secondary counterparts, i.e., 34~\% of the sample. 

It requires more careful examination of the data to identify what percentage of 
secondary counterparts actually contributes to the IR flux of ADF-S sources. 
In the same time, as mentioned above, even the assumption 
that all the found secondary counterparts could contribute to the clustering signal does not 
assure a sufficiently large number of close galaxy pairs in our data to obtain a power-law 
shape of the correlation function. 
Then, it is very possible that also some other unresolved (and not  identified) extragalactic IR sources 
contribute to the flux of the identified sources.  
Additionally, since many IR-bright objects found in our sample are peculiar or interacting, we may  
suspect that even in a greater amount of cases we do not know nor see interacting partners 
which actually exist and might contribute to the IR flux. 
The conclusion is that the small-scale environment of the galaxies observed in the infrared 
should be studied in future even more carefully. 

\section{Spectral Energy Distributions}\label{sec:sed}

\begin{figure}[tb]
\includegraphics[width=8.5cm]{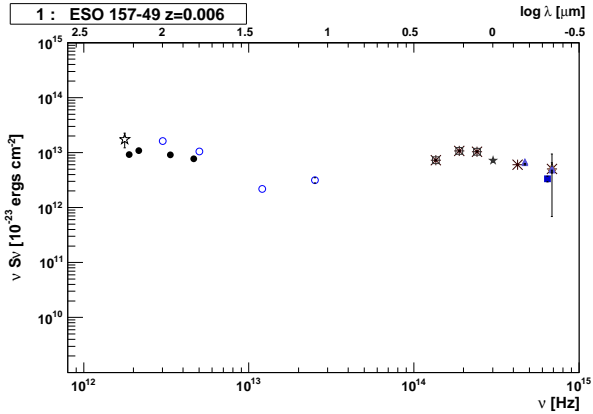}
\caption {An example of an SED: here we show the data points for the 
brightest 
ADF-S
source, plotted using all the information listed in online Tables~\ref{ADFSmeasurements}-\ref{measurements2a}.
The data points from AKARI Deep Field South (full circles), 2MASS (open squares),
SIMBAD database (eight pointed stars), IRAS (open circles), ESO/Uppsala
(full triangles), APM (full squares), RC3 (open triangles), ISOPHOT
(five pointed stars) are shown. SEDs of all the identified ADF-S sources are
included in the online material. Since the redshift of this 
object is known, the SED is presented in the rest frame.
}
\label{expoints}
\end{figure}

As mentioned in Introduction, SEDs give first important clue to the physics
of radiation of the sources.
The deep image at the AKARI filter bands gives us for the first time the opportunity
to analyze SEDs and update the models of their interstellar dust emission.
Online Figures 1 -- 16
present SEDs of 518 identified infrared ADF-S sources. In the text we show
only one example of such SED, for the brightest ADF-S source in 90 $\mu$m, 
in Figure~\ref{expoints}.

\subsection{SED models}\label{subsec:sed_model}

From the $10\sigma$ sample we have selected 
47 galaxies 
with the best available photometry to fit the models of their SEDs. 
The results are presented in Figures~\ref{sed}--\ref{sed6}. 
The main selection criterion were at least three data points in the infrared range of the spectrum, in order to 
be able to model 
dust properties sensibly. 
In the selected sample 
23 galaxies have their morphologies determined. There is
one elliptical galaxy, two compact galaxies (one of them being a starburst blue compact dwarf), 
five lenticular galaxies and 15 spiral
 galaxies.

{}To fit the SEDs of these galaxies we use three methods. 
First we applied a modified blackbody to the dust emission part, and a blackbody to the stellar emission 
part in the galaxy SEDs. 
The results are presented in Section~\ref{subsubsec:modifiedbb}.
Since the galaxies in our sample are often evolved, 
a single blackbody often gives a poor fit to the observed SEDs for the stellar emission part. 
In a future work, we plan to use a more sophisticated stellar population synthesis model with
realistic star formation history to model their stellar component. 

It is widely known that some 
dust components 
in galaxies
cannot establish an equilibrium with  ambient
radiation field.
These components produce strong mid-IR (MIR) emission which extends in a wide range of wavelengths
and cannot be well fitted by the modified blackbody \citep[e.g.,][]{purcell76,draine85,li2001,draine01,takeuchi03,takeuchi05c}.  
In order to deal with this MIR emission, we should use more sophisticated models for the fit.
Then, in addition to the modified blackbody model, we used models of 
\citet{dale2002} and \citet{li2001}. 
These more refined models in most cases succeeded in reproducing the MIR part 
of the dust emission. 
The results are presented Sections~\ref{subsubsec:dh} and \ref{subsubsec:draine}. 

To fit SEDs of galaxies we tried to use all available measurements
from the ADF-S catalog (listed in the online
Tables~\ref{ADFSmeasurements} and~\ref{ADFSmeasurements2}) 
and data from public databases (listed in the online 
Tables~\ref{measurements1},~\ref{measurements2},~\ref{measurements3} and
~\ref{measurements4}), excluding only several most dubious measurements,
as discussed below.

To fit \citet{dale2002} and \citet{li2001} models we used all four bands 
from the ADF-S measurements, when available, i.e. N60 (65 $\mu$m), 
WIDE-S (90 $\mu$m), WIDE-L (140 $\mu$m) and  N160 (160 $\mu$m), 
four IRAS bands (12 $\mu$m, 25 $\mu$m, 60 $\mu$m and 
100 $\mu$m)\footnote{Note that in some cases IRAS provides only the upper 
limits on the IR flux.}, in one case of a galaxy NGC~1705 seven 
Spitzer bands (3.6 $\mu$m, 4.5$\mu$m, 5.8 $\mu$m, 8 $\mu$m, 
24 $\mu$m, 70 $\mu$m and 160 $\mu$m), and in one case of a galaxy 
ESO 157-49 one ISOPHOT band (170 $\mu$m). We included the uncertainties 
of all 
these data points in the fitting process. Data from the 
ADF-S were
not treated in any preferential way.

To find the best fitting models we used a $\chi^2$ test; the details
of this procedure in case of simple blackbody and dust models are 
explained in the corresponding sections below. 

Among considered 47 galaxies, there are 15 objects with known redshifts. 
SEDs of these 15 galaxies are fitted and presented in the rest frame. 
SEDs of the remaining objects are fitted and presented in the observed frame.

The parameters resulting from all the three methods 
are summarized in Table~\ref{MODELtable}. As the complementary information, 
we summarize morphological and environmental properties of 
galaxies used 
for the fitting in Table~\ref{morphological_table}.

\subsubsection{Modified blackbody model}\label{subsubsec:modifiedbb}

As the simplest approach, we modeled the stellar component of galaxies 
$\nu = 5\times 10^{13}$ -- $10^{15}$ Hz, i.e. $\lambda = 0.3$ -- $6$ $\mu$m)

with the blackbody spectrum
\begin{equation}\label{eq:bb}
  B_{\nu}(T) = \frac{2h\nu^{2}}{c^{2}}\frac{1}{e^\frac{h\nu}{kT}-1}
\end{equation}
and their dust emission ($\nu = 10^{12}$ -- $10^{13}$ Hz, 
i.e. $\lambda = 30$ -- $300$ $\mu$m) with a modified blackbody 
\begin{equation}\label{eq:modbb}
  B'_{\nu}(T) = \nu^\gamma B_\nu(T) 
\end{equation}

Since we treat these models as indicative only, we looked for
the best fitting parameters of the blackbody spectra by a $\chi^2$ 
minimization, without taking errors into account. Since the 
blackbody spectra provide rather a poor fit to the data, we found 
that using the error information does not improve the fitting, and 
in some cases even makes the blackbody models more difficult to fit.  

We found the best fitting parameters looking for a minimal
\begin{equation}
\chi^2 = \sum_{i=1}^n \frac{\left( \nu S_{\nu,i}-B'_{\nu}(T)_{i} \right)^2}{B'_{\nu}(T)_{i}},
\end{equation}
where $n$ is a number of data points, $\nu S_{\nu,i}$ is a measured flux and $B'_{\nu}(T)_{i}$  is a corresponding theoretical value given by a modified blackbody model.

If the absolute difference between data and the best fit was larger 
than $2\sigma$, we performed the fitting process again, but without taking 
into account the point for which the relative difference was the highest,
providing that there were still at least three data points left. 
The result of the rejection of the least fitting point in case of the poor 
fit can be seen e.g. in Figure~\ref{sed} in case of a galaxy number 8: the
data point in the B filter (the last point in the optical range) had to 
be removed to provide a reasonable fit. 
Perhaps not accidentally, the data points which are most often rejected
from the fitting procedure are measurements in the optical B filter. 
In fact,
usually they are the measurements of the poorest or most dubious 
quality. For instance, in case of 4 galaxies (sources 7, 8 in Figure~\ref{sed}, source 9 in Figure~\ref{sed2} and source 45 in
Figure~\ref{sed5}),
these measurements, acquired from the SIMBAD database, originally come 
from the Dressler's catalog of galaxies in clusters 
\citep{dressler80b}. In the original paper, they are described as 
the "estimated total apparent visual magnitude" and their accuracy is
about one magnitude. Then, they are obviously not suitable for 
the SED fitting.
Other often rejected points, also in the B band (e.g. galaxy number 6 in
Figure~\ref{sed}), come from the APM survey \citep{maddox}, whose 
photometry is known to suffer a systematic bias 
\citep[e.g.,][]{metcalfe95, bertin97}.

Having in mind all that was said about 
the small reliability of the stellar blackbody spectra, we 
should note that the ``temperature'' of stellar component works only as 
a symbolic index of the hardness of the SEDs, and does not have an absolute sense
of the temperature of their stellar population.
The temperature is used
here only for an internal comparison of the analyzed ADF-S sources.

We did not take the into account dust attenuation in the fitting
in this analysis.
For this kind of analysis, we need to use a more sophisticated
stellar spectral synthesis model to extract the information on the
dust attenuation assuming a certain dust attenuation curve
\citep[e.g.,][]{mathis90,calzetti1994, calzetti2000}.
However, in addition to the adopted rough approximation of stellar spectra
by blackbody, almost all our SEDs lack UV--optical photometry which hampers 
a precise determination of extinction.
If we had a FIR/UV flux ratio, we would be able to estimate a correlation between
FIR/UV flux ratio and luminosity from newly forming stars \citep{buat05,buat07}.
Unfortunately, the UV photometry or distance of a source is only available for a 
small number of sources in our sample. 
Hence, we only discuss a possible effect of dust attenuation qualitatively.

As stated before, dust scatters and absorbs UV--optical light from stars and re-emits
the energy at IR.
Since usual attenuation curves of galaxies have a steep rise toward UV, dust 
attenuation causes reddening of galaxy stellar spectra. 
Then, this causes a significant underestimation of stellar temperature by 
blackbody fitting.

\begin{figure*}[t]
\centering
\includegraphics[width=8.5cm]{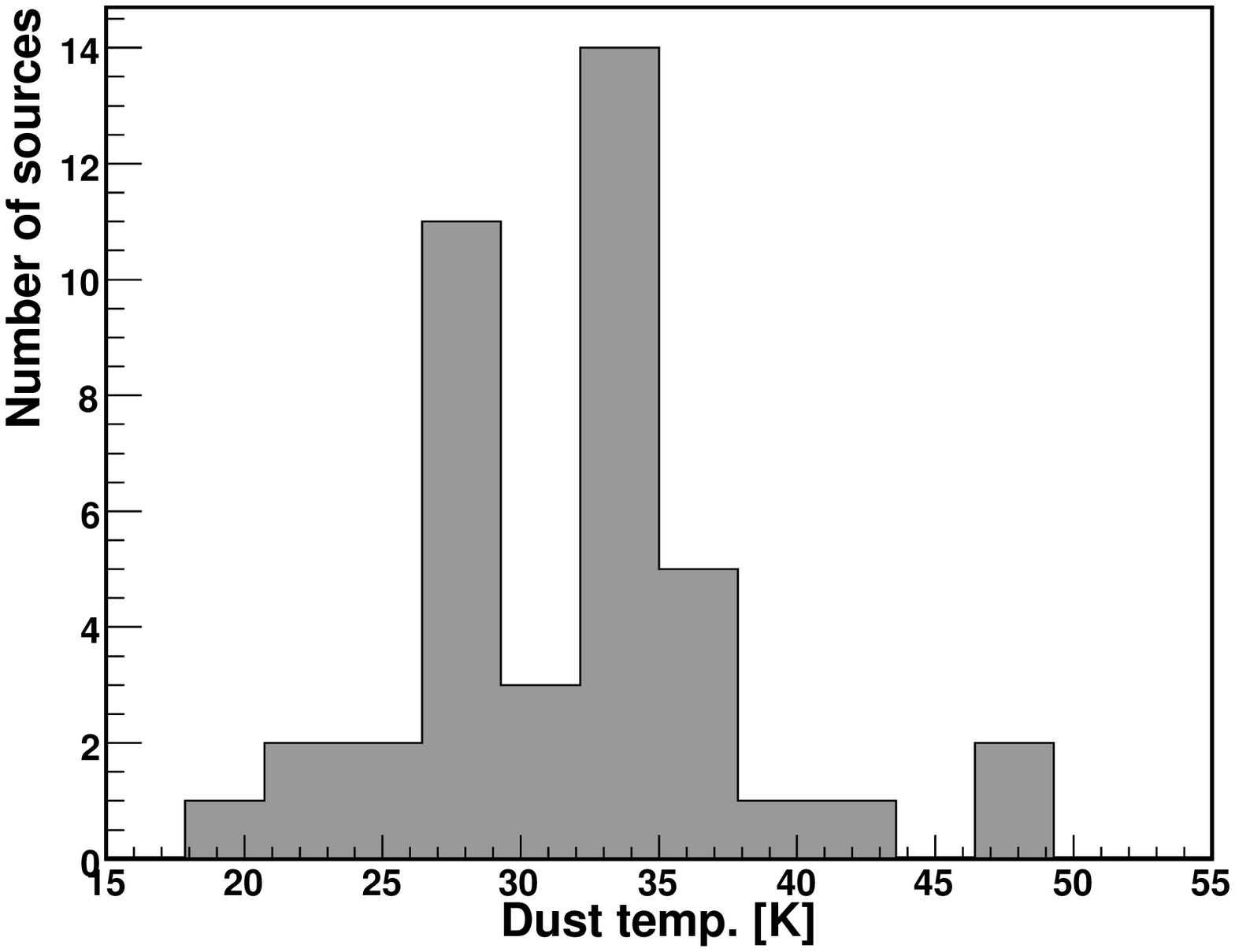}
\includegraphics[width=8.5cm]{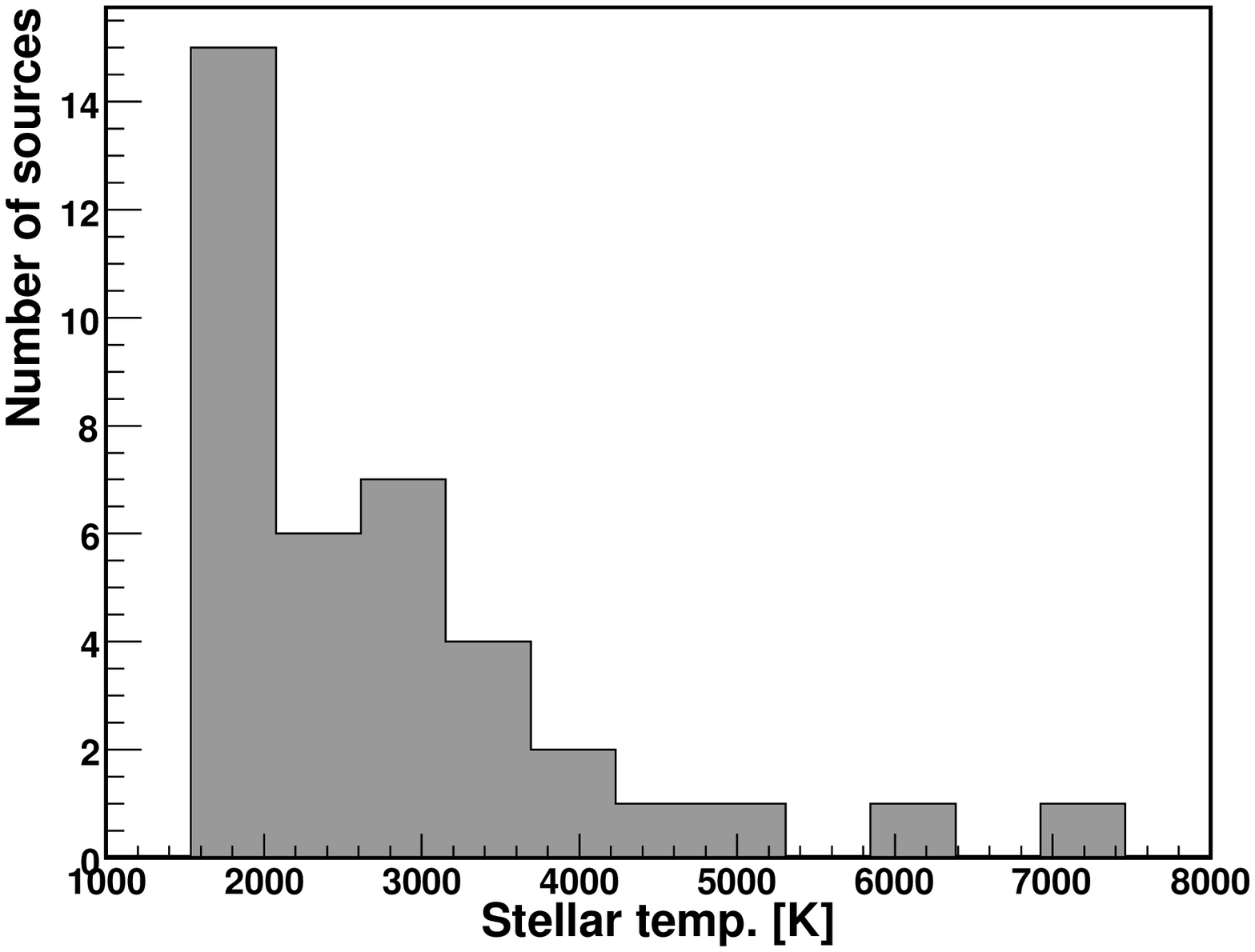}
\caption{
Histograms of temperatures of dust and stellar components of galaxies 
in the best fit modified blackbody models.
}\label{temp}
\end{figure*}

The best fitted values of temperature of stars and dust in the modeled galaxies 
are listed in the second and third column of Table~\ref{MODELtable}. 
The histograms of these two temperatures for all 47 modeled galaxies are presented in Figure~\ref{temp}. 
According to this measurement, the both the median and mean temperature of dust is around $32\pm4$~K. The median temperature of the stellar component is 
$2240\pm360$~K, while its mean is $2830\pm410$~K.
In the latter case, the discrepancy between mean and median values 
is related to the fact that the temperature distribution is far from 
Gaussian.

The obtained values of stellar temperatures are generally significantly too low for stars emitting UV.
As we mentioned above, the estimated low temperature of our sample would be partially 
due to the dust attenuation. Here we use it only to make an internal comparison of our sample.

The highest stellar temperature we have fitted for a starburst 
blue compact dwarf NGC~1705. The dust temperature of this galaxy
is also significantly higher than average. However, as it can be 
seen from the fourth panel in Figure~\ref{sed}, a simple blackbody gives
a very poor fit to its stellar population.

The elliptical galaxy ESO 157-IGA040 is significantly warmer than the 
median stellar temperature, with $T_* = 3060$~K, and its dust 
temperature is also relatively high, at $T_{dust} \approx 38$~K. In its
case, as seen from the second panel of Figure~\ref{sed}, blackbody models
seem to work quite well. A particular activity of this elliptical galaxy
should be probably related to the fact that it is a member of a pair 
of interacting galaxies. 

Even if 15 spiral and five lenticular galaxies make a rather poor statistical sample, we attempted some simple comparison between these two groups.

Rather surprisingly, spiral and lenticular galaxies in our sample
seem to have very similar properties, and spirals typically have even lower
dust and stellar temperatures than lenticulars. The
mean dust temperature of spirals ($\sim 30 \pm 5$~K) is 5~K 
lower than that of 
lenticulars ($\sim 35 \pm 17$~K), 
while the average temperature of a stellar component in spirals 
is $880\pm200$~K lower than in case of lenticulars. The difference in the
dust temperatures is mainly due to the fact that the two outliers with
the highest dust temperatures are lenticulars and a galaxy with the lowest
fitted dust temperature is a spiral. However, the black bodies of these
objects seem to be fitted reasonably well.
Consequently, looking at the median
values, the difference is less pronounced: the median dust temperature is 
$29\pm4$~K for lenticulars and $28\pm 4$~K for spirals. Also the difference 
median between stellar temperatures for spirals and lenticulars is still 
significant ($570\pm130$~K) but smaller.

Of course, given a small number statistics, this difference 
between spiral and lenticular galaxies is not 
 very significant in a statistical sense. 
It might be interpreted as lenticulars detected in FIR being particularly 
active in star formation, at least at the same level as, 
average spirals.
However, it is more plausible that the simplest models are not particularly accurate 
in reproducing the dust and stellar emission in this type of galaxies.

\subsubsection{Dale and Helou model}\label{subsubsec:dh}

As mentioned above, modified blackbody gives a poor fit the MIR part of dust spectrum.
{}To fit the dust emission of our galaxies in a whole range 
of wavelengths 
between 3 and 1100 $\mu$m, we applied a model proposed by \cite{dale2002}. 
It was developed to model a wide range of interstellar 
environments in normal star-forming galaxies for different heating intensity levels. 
In this model the local SED of dust is given by a power-law distribution: 
\begin{equation}
  {\rm d} M_{\rm d}(U) \propto U^{-\alpha_{\rm SED}}\, {\rm d}U,
\end{equation}
where $M_{\rm d}(U)$ is the 
mass of dust heated by a radiation field $U$; $\alpha_{\rm SED}$ is 
a parameter that represents relative contributions of active and quiescent regions from 
different galaxies. 
Values of $\alpha_{\rm SED} \leq 1$ represent actively star-forming
galaxies. 
The lower value of $\alpha_{\rm SED}$, the more actively star forming is a galaxy.
More quiescent cooler galaxies have $1 < \alpha_{\rm SED}< 2.6$. 
The case of $\alpha_{\rm SED}=2.6$ corresponds to the most quiescent dust model. 
Values of $\alpha_{\rm SED}$ between 2.6 and 4 are fitted for galaxies where the FIR
emission peak appears at even longer wavelengths than in case of 
most quiescent well examined galaxies 
\citep{dale2005}. 
This region is still not very well examined, but it appears to correspond to 
the 
coolest and most quiescent galaxies.

If enough data are available, it is possible to calculate the expected value of $\alpha_{\rm SED}$ 
e.g., from $IRAS$ $f_{\nu}(60\mu\mbox{m})$ and $f_{\nu}(100\mu \mbox{m})$ 
measurements 
\citep[\citet{moshir}; this method was presented e.g. in][]{dale2001}. 
However, in our case, a relatively small amount of data and the fact that we are exploring 
a rather poorly known regime of FIR have prompted us to use a minimal $\chi^{2}$-fitting. 
We took into account uncertainties of photometric 
measurements. To compare our data points to the models effectively, 
the modeled spectra have been convolved with the AKARI, IRAS, and Spitzer
photometric bands. Using a grid of models with all the parameters available,
we looked for the best fitting model with a minimal
$\chi^2$ defined as:
\begin{equation}
\chi^2 = \sum_{i=1}^n \frac{\left( (\nu S_{\nu,i}+K)-F_{\nu,i} \right)^2}{\sigma_{\nu,i}^{2}},
\end{equation}
where $n$ is a number of points, $K$ - a normalization constant, used 
as a free parameter,
$\nu S_{\nu,i}$  is a measured flux, $F_{\nu,i}$ is a corresponding theoretical value of flux from the dust model, $\sigma_{\nu,i}$ is a corresponding measurement error.

The resulting $\alpha_{\rm SED}$ for all modeled galaxies are listed in column~4 
of Table~\ref{MODELtable}. 

\begin{figure}[tb]
\centering
\includegraphics[width=8.5cm]{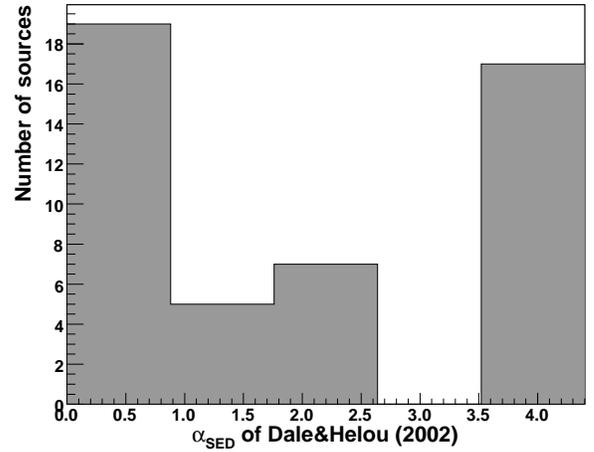}
  \caption{The histogram of the parameter $\alpha$ of the model by \cite{dale2002}. 
}\label{dale_alpha}
\end{figure}

Histogram of $\alpha_{\rm SED}$, presented in Figure~\ref{dale_alpha},  shows that
the distribution of $\alpha_{\rm SED}$ is quite discontinuous. 
Most of
the analyzed sources have extreme values of $\alpha_{\rm SED}$: less than 0.9 (19 sources) or more than 3.5 (17 sources). 
We have found seven galaxies with values of $\alpha_{\rm SED}$ between 
1.7 and 2.1 and five with values $\alpha_{\rm SED}$ between 
0.9 and 1.7. 
No galaxies were fitted by models with $\alpha_{\rm SED}$ between 2.2 and 3.5.

Among galaxies with known morphological types,
80~\% of lenticular galaxies were fitted as the warmest, 
actively star forming galaxies with $\alpha_{\rm SED}<1$ and 
only one of them, 2MASX J$04292360-5330114$, was placed in a 
different regime, as 
the most quiescent galaxy with $\alpha_{\rm SED} = 4$. 
No lenticular galaxy was found to be 
in 
the $\alpha_{\rm SED} = 1$ -- $2.6$ regime. 

The elliptical and 4 of the spiral galaxies were
classified as cool and quiescent. 
One of the compact galaxies, the starburst 
blue compact dwarf NGC~1705, similarly to 60~\% of spiral galaxies, is 
classified as warm and actively star forming. 
ESO 157-51, a second compact galaxy, and the remaining 40~\% of spiral galaxies were placed 
in poorly examined and, in all probability, the most quiescent regime with $\alpha_{\rm SED}>3$. 
This suggests that the peak of their dust emission is located at yet longer wavelengths. 

We have compared our conclusions with the results 
of \cite{dale2005}, 
who constructed SEDs for 71 nearby galaxies from the SINGS sample \citep{kennicut} 
in the range of $\lambda$ from $1\; \mu$m to $850\;\mu$m. 
Among them there are five elliptical galaxies with $\alpha_{\rm SED}$ between 1.32 and 2.14 and 
one with $\alpha_{\rm SED}= 4$. 
This range of $\alpha_{\rm SED}$ indicates that all elliptical galaxies in the SINGS sample are cold 
and quiescent. 
The only elliptical galaxy in our sample seems to belong to the same 
type. This is expected, however, not consistent with the results of the
simple blackbody fitting, which may point out the insufficiency of the latter. 

In the sample used by 
\cite{dale2005} there were also two lenticular galaxies, one of them warm and active in star formation
($\alpha_{\rm SED}=0.6$) and one cold ($\alpha_{\rm SED}=2.67$). 
In this case, because of a poor number statistics, it is difficult to make any comparison, 
but our modeling confirms that lenticulars can be very warm as well as extremely quiescent. 

The only 
discrepancy concerns spiral galaxies. 
In the data used by \cite{dale2005}, 54 spiral 
galaxies were included, and none of them was classified as 
actively star forming:
54~\%  of them were cold 
and quiescent with $\alpha_{\rm SED}$ between 1.68 and 2.6 and the remaining 45~\% were described 
as even more quiescent with $\alpha_{\rm SED} > 2.6$. 
In our case a half of spiral galaxies is very warm and active in star formation, there are no quiescent ones with $\alpha_{\rm SED}$ between 1.68 and 2.6 %
but 
the other half are spiral galaxies with $\alpha_{\rm SED} = 4$, i.e., extremely quiescent.
This finding is consistent with the fact that our sample is based on a genuine FIR ($90\;\mu$m)
selection.
On one hand, this sample is sensitive to pick up dusty IR galaxies, but on the other hand, it can
also select galaxies which are quiescent because of the long wavelength comparing with IRAS-based
studies (usually based on $60\;\mu$m selected).

\subsubsection{Li \& Draine model}\label{subsubsec:draine}

As yet another more refined model of the infrared emission from dust grains heated by 
starlight in galaxies we have chosen the model proposed by \citet{draine07}, which is 
an improved version of \citet{li2001}. 
This model assumes that the dust heated by starlight consists of a mixture of amorphous 
silicate and carbonaceous grains. 
Each molecules have a wide size distribution ranging from molecules containing tens of 
atom to large grains $\ge 1\; \mu$m in diameter \citep{draine07}. 
In this paper authors have calculated the emission spectrum for dust heated by the stellar 
light and parametrized this model using three parameters: the fraction of the total dust mass 
that is contributed by the polycyclic aromatic hydrocarbon (PAH) particles ($q_{\rm PAH}$),
lower ($U_{\rm min}$) and higher ($U_{\rm max}$) cutoff of the starlight intensity distribution and 
the fraction of a dust heated by the starlight ($\gamma$).

As before, 
the modeled spectra have been first convolved with the AKARI, IRAS, and Spitzer
photometric bands.
We have fitted \cite{draine07} using a $\chi^{2}$ test including - 
as in case of the Dale \& Helou model - the information about the 
photometric errors.

\begin{figure}[tb]
\centering \includegraphics[width=8.5cm]{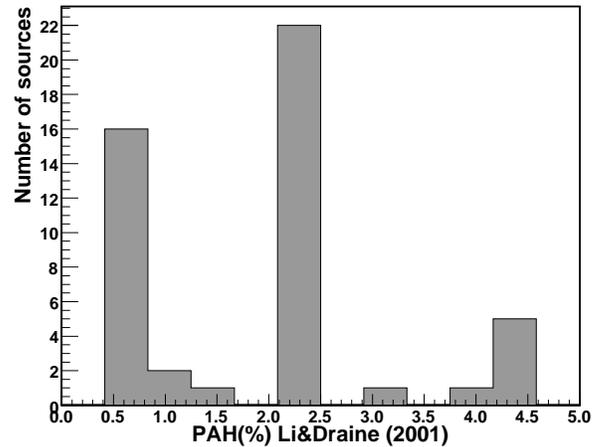}
\caption{The histogram of the amount of polycyclic aromatic hydrocarbons (PAH) 
in the dust of the analyzed galaxies, according to the model by \cite{li2001}.}\label{pah}
\end{figure}

\begin{figure*}[t]
\centering

\includegraphics[width=9cm]{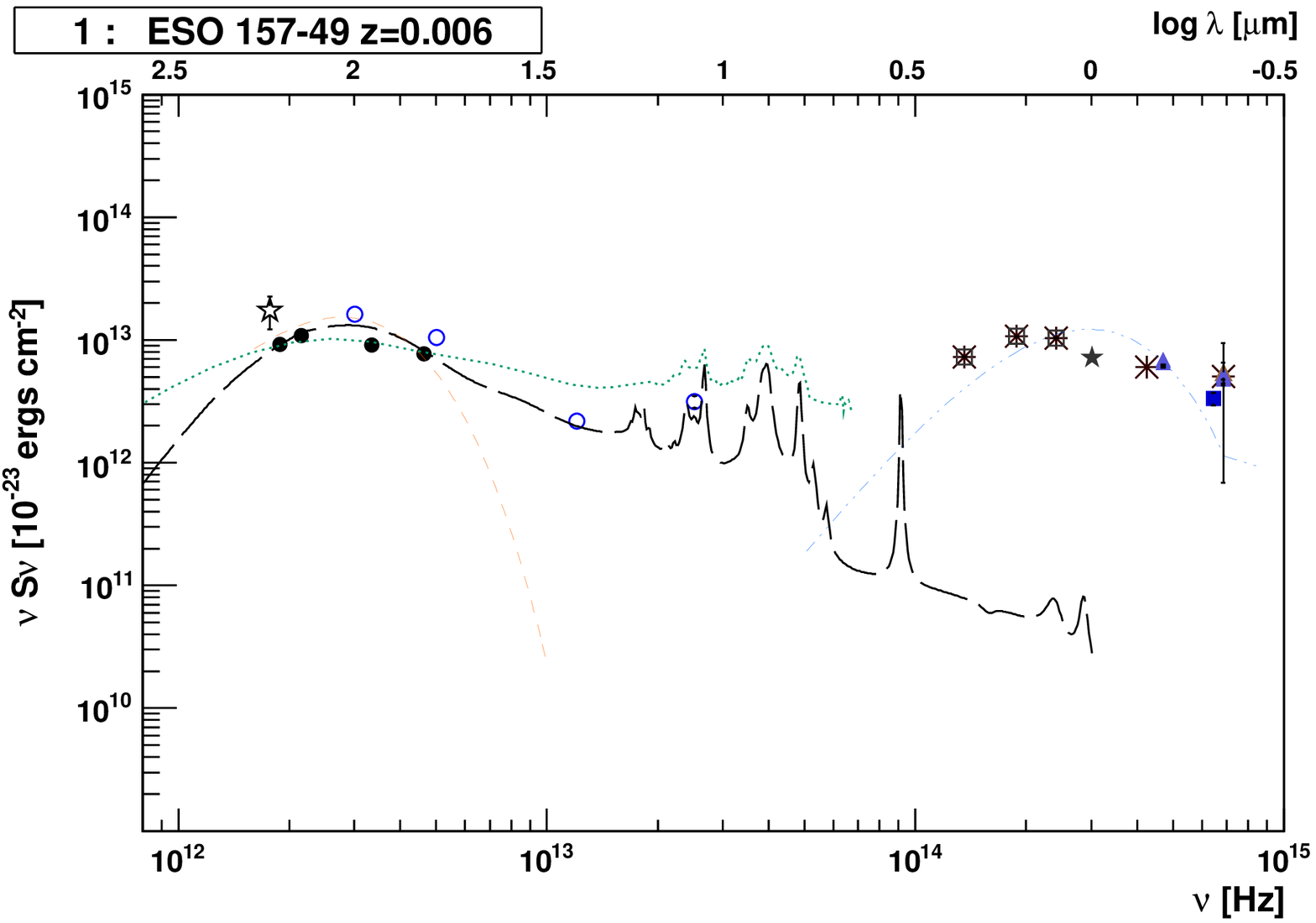}
\includegraphics[width=9cm]{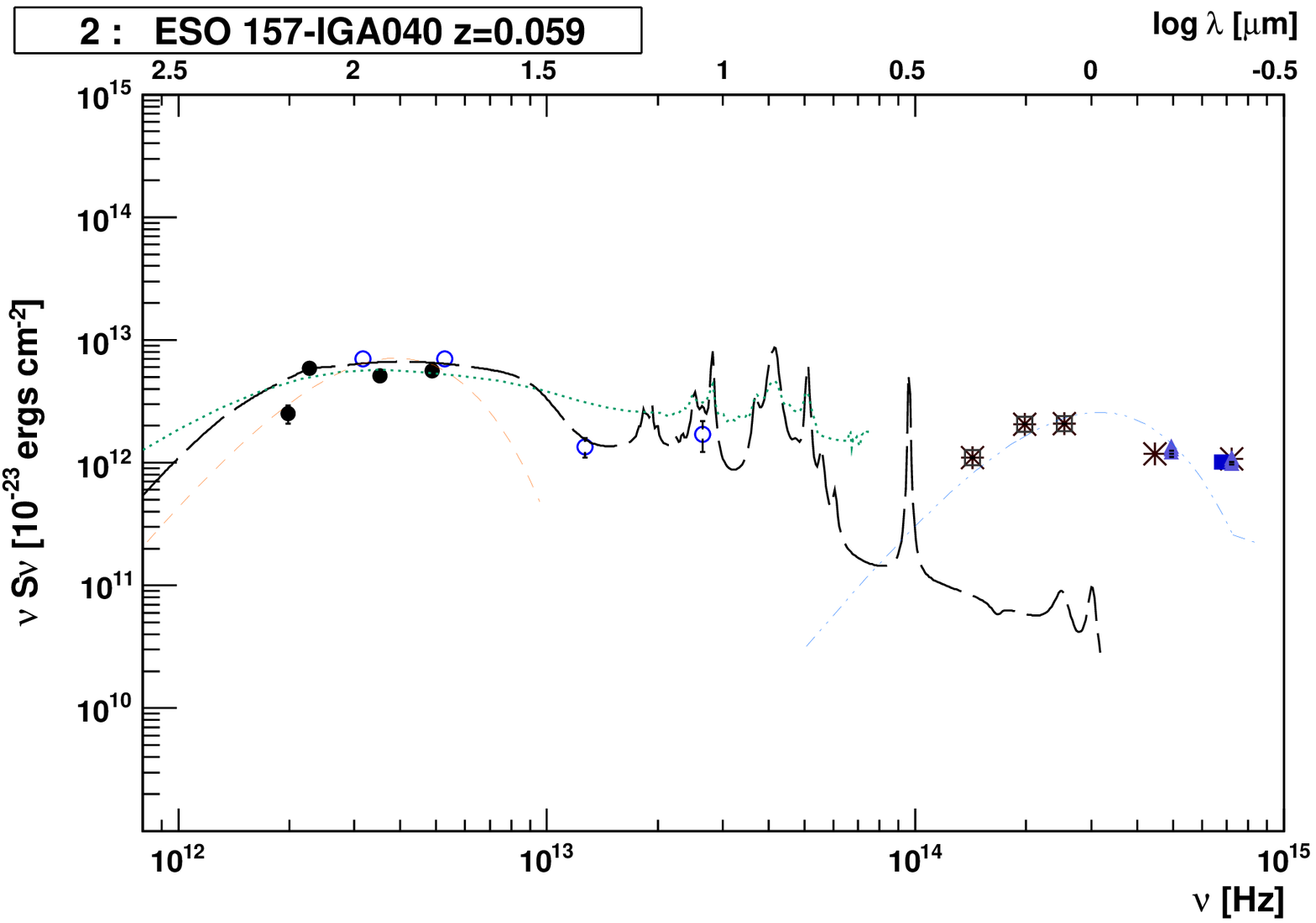}
\includegraphics[width=9cm]{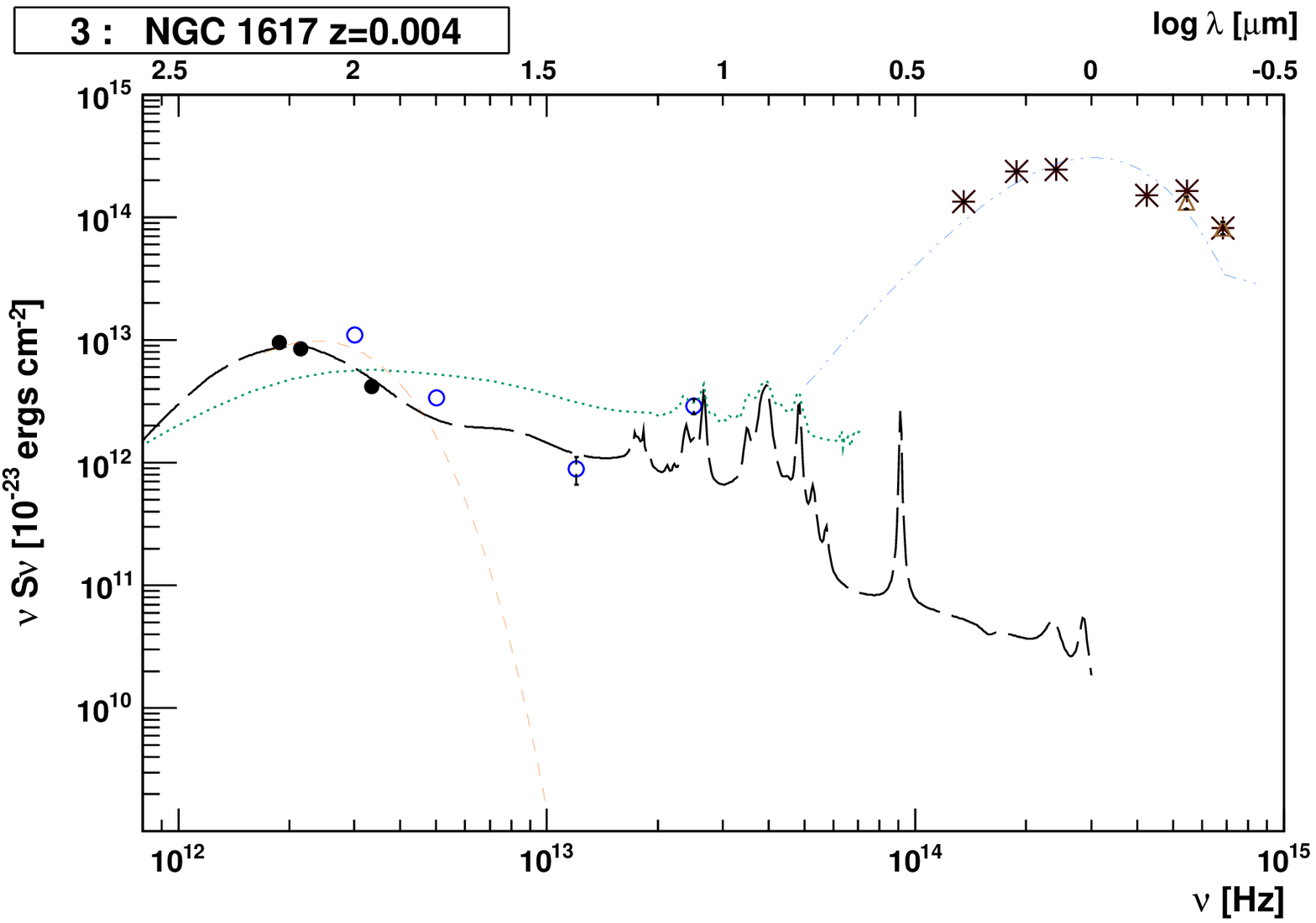}
\includegraphics[width=9cm]{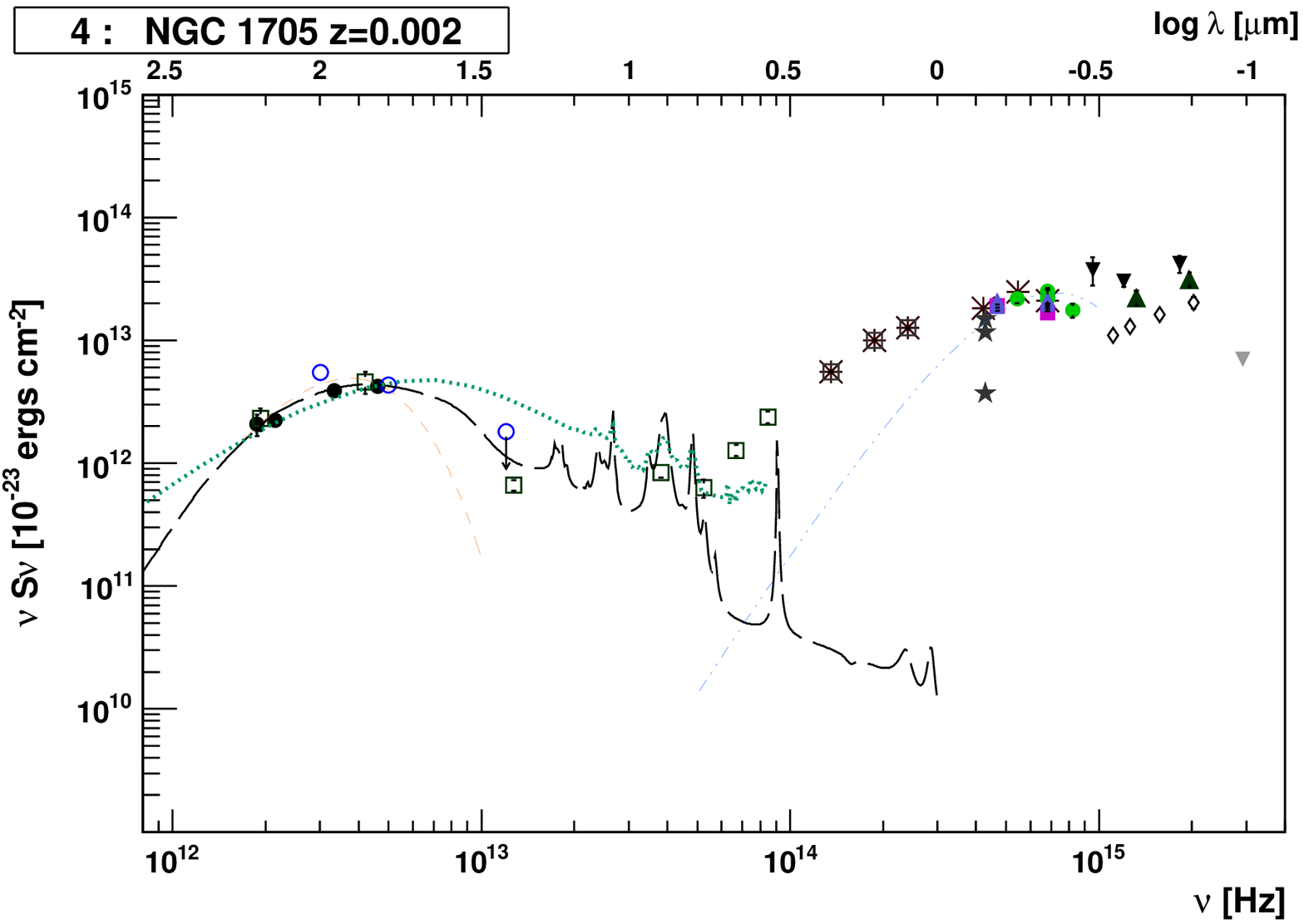}
\includegraphics[width=9cm]{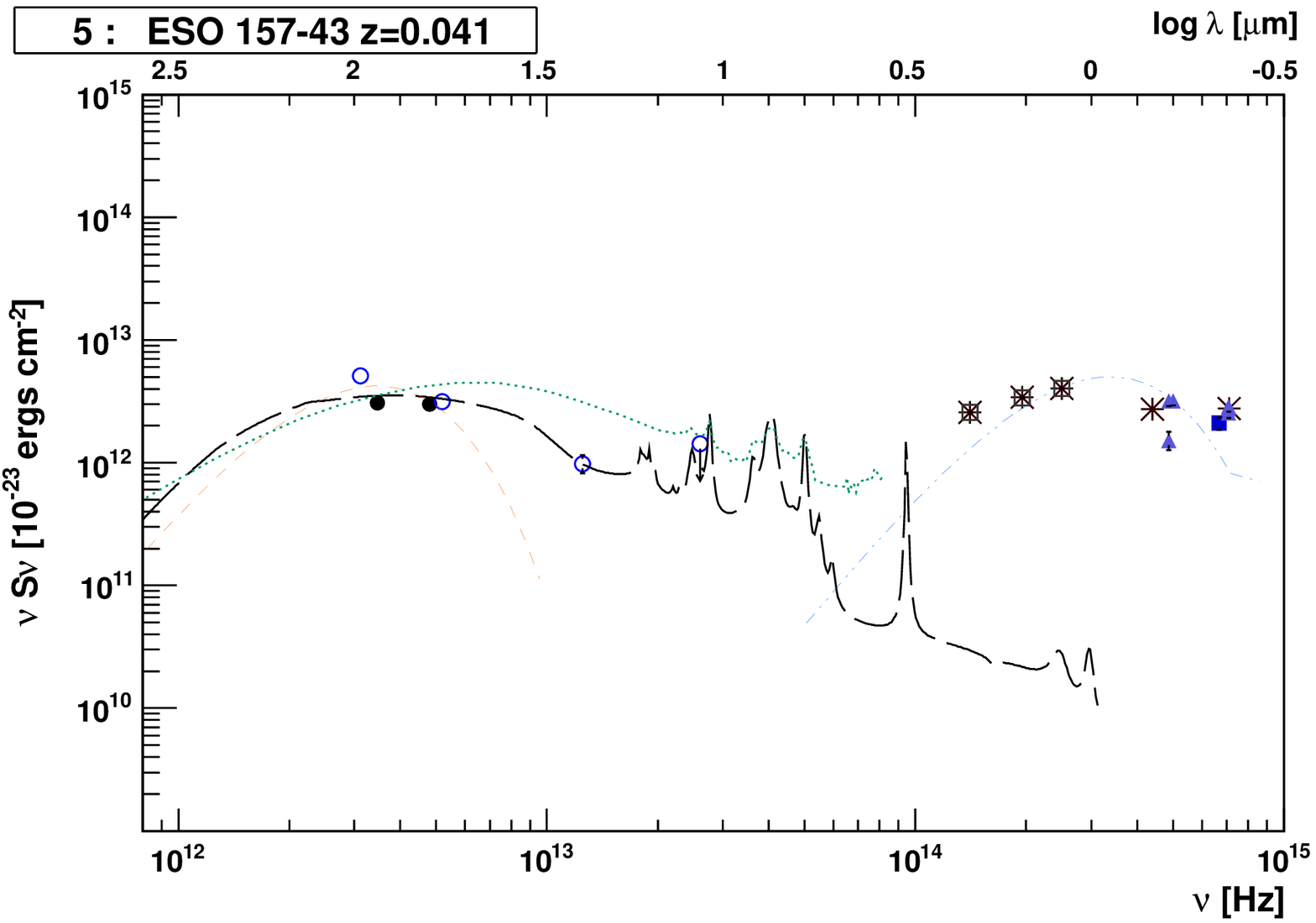}
\includegraphics[width=9cm]{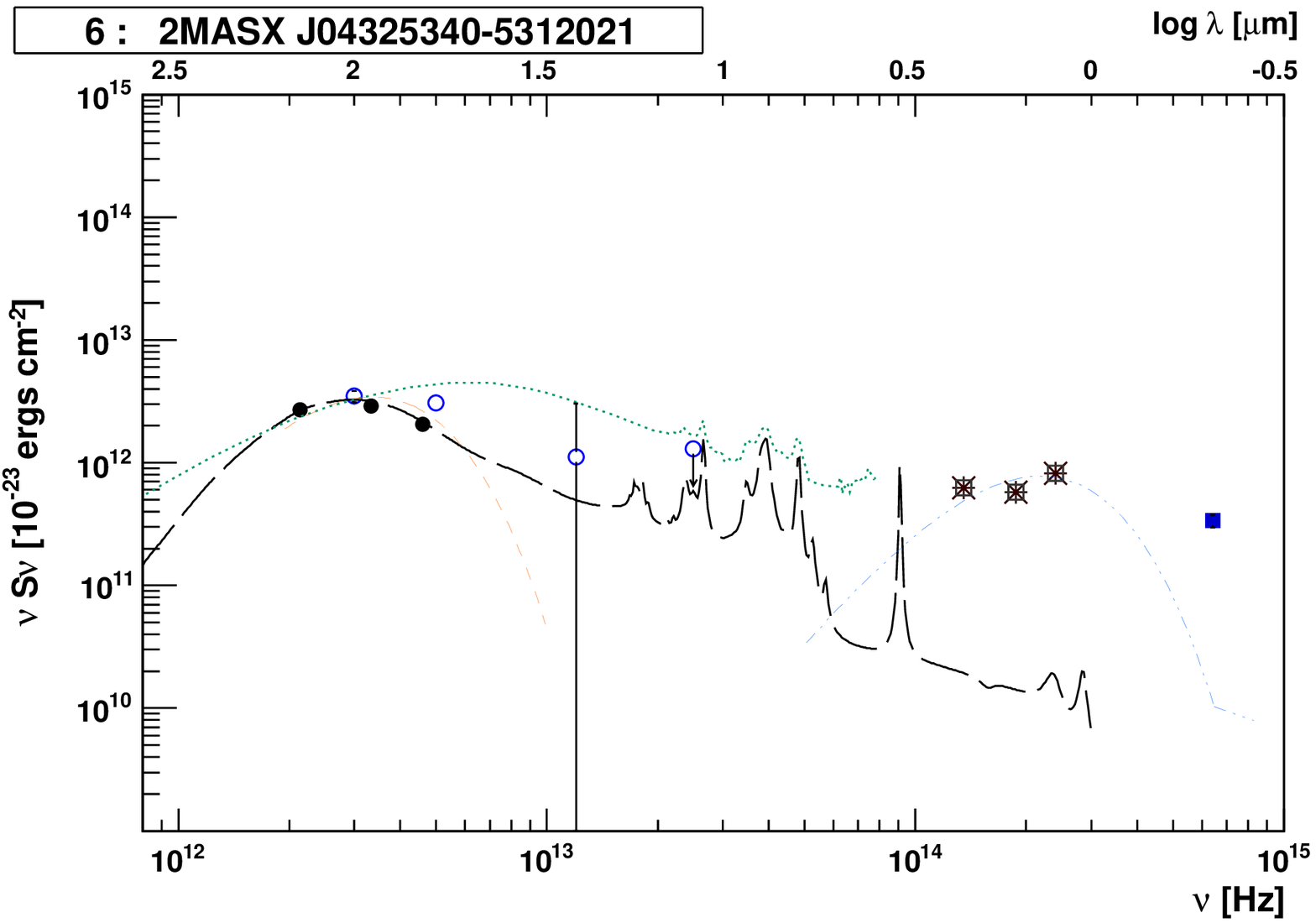}
\includegraphics[width=9cm]{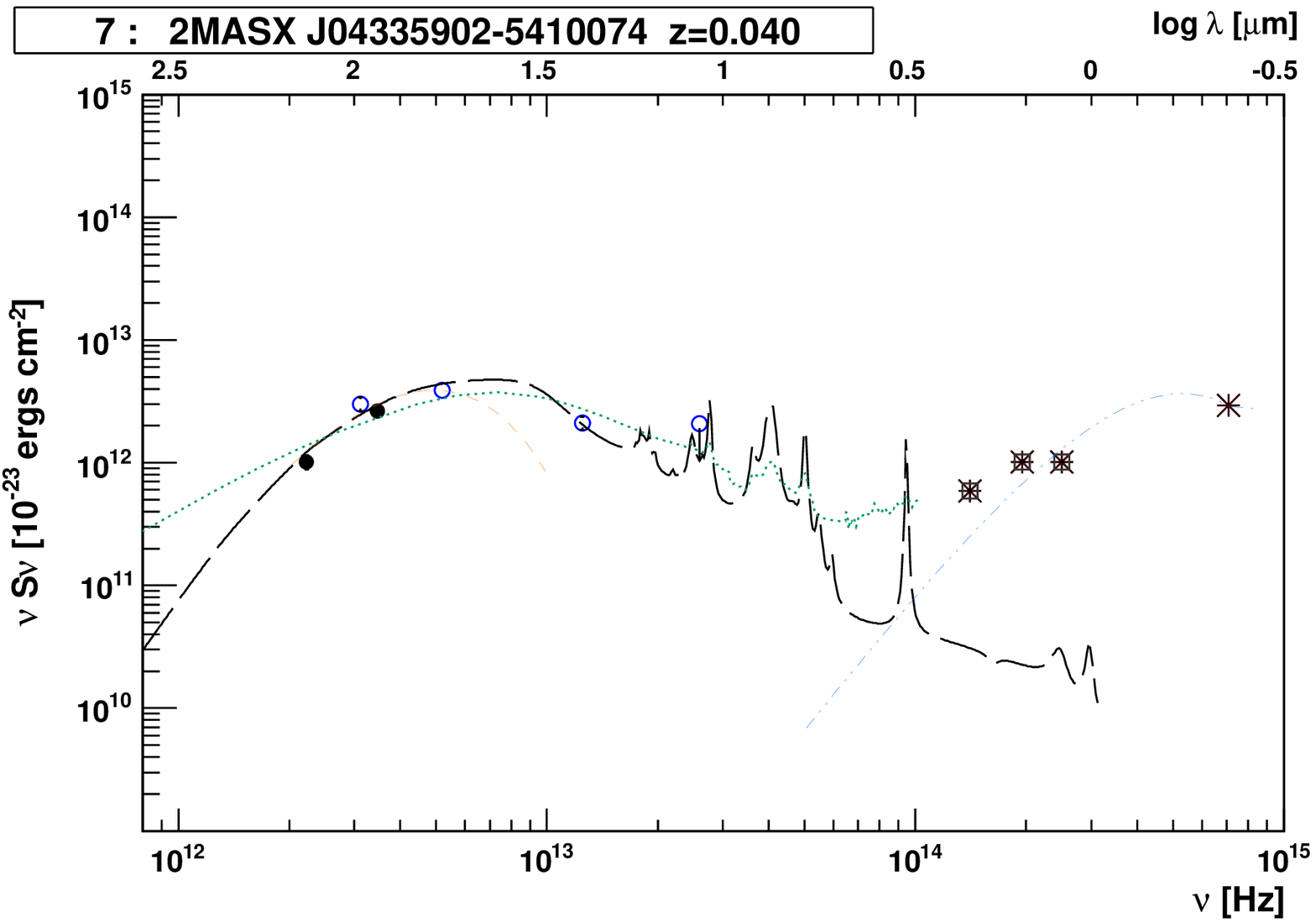}
\includegraphics[width=9cm]{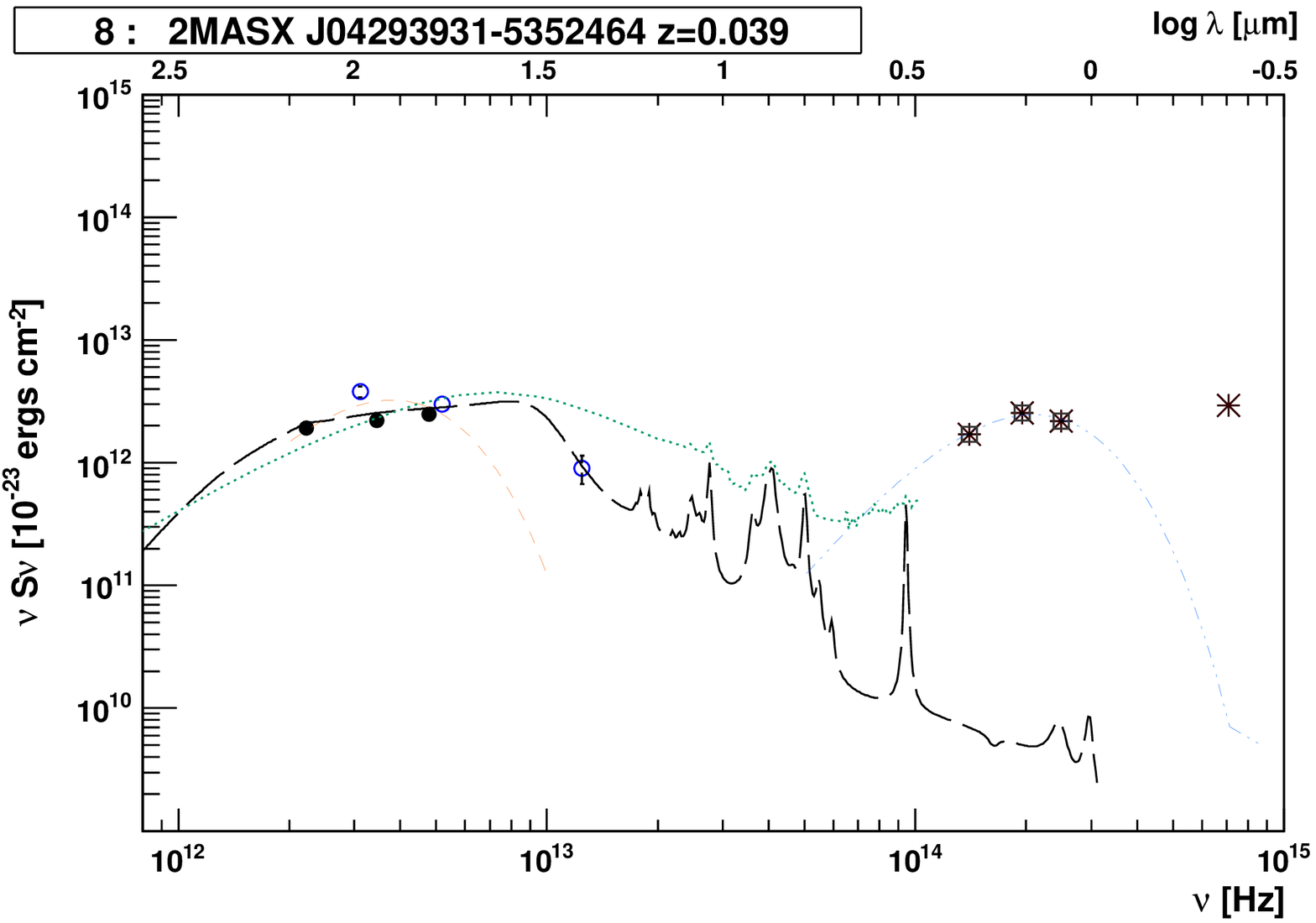}
\caption{The SEDs of ADF-S galaxies with the best photometry and available 
data from other catalogs. 
The data points from AKARI Deep Field South (full circles), 2MASS (open squares),
SIMBAD database (eight pointed stars), IRAS (open circles), ESO/Uppsala
(full triangles), APM (full squares), RC3 (full triangles), ISOPHOT
(five pointed stars), Siding Spring Observatory (five pointed stars), GALEX
(full triangles), HIPASS catalogue (full circles), Palomar/Las Campanas Imaging
Atlas of Blue Compact Dwarf Galaxies
(full squares), IUE (open diamonds), Spitzer 
(open squares), FUSE (upside-down light triangles) and UV: 1650, 2500, 315 (upside-down dark triangles) were fitted by
three different models of dust emission: modified blackbody 
 (short-dashed line) model of \citet{dale2002} 
(dotted line), model of \citet{li2001} (long-dashed line) and stellar 
emission: modified blackbody (dot-dot-dashed line). 
SEDs of galaxies with a given redshift (objects number 1, 2, 3, 
4, 5, 7, 8) are fitted after shifting to the rest frame and presented 
in the rest frame. Galaxy number 6, whose redshift is not known, is 
shown in the observed frame.}
\label{sed}
\end{figure*}

\begin{figure*}[t]
 \centering
\includegraphics[width=9cm]{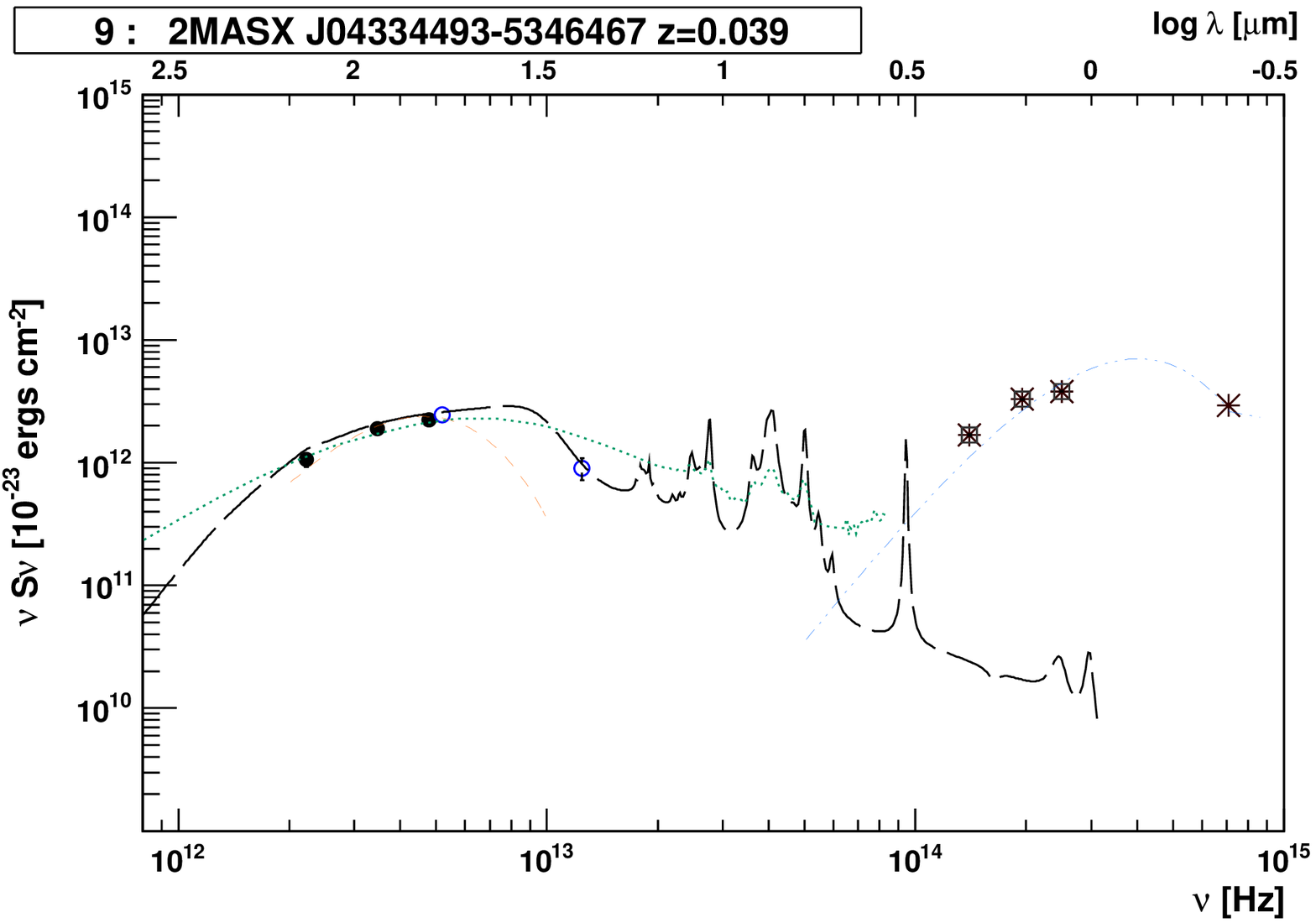}
\includegraphics[width=9cm]{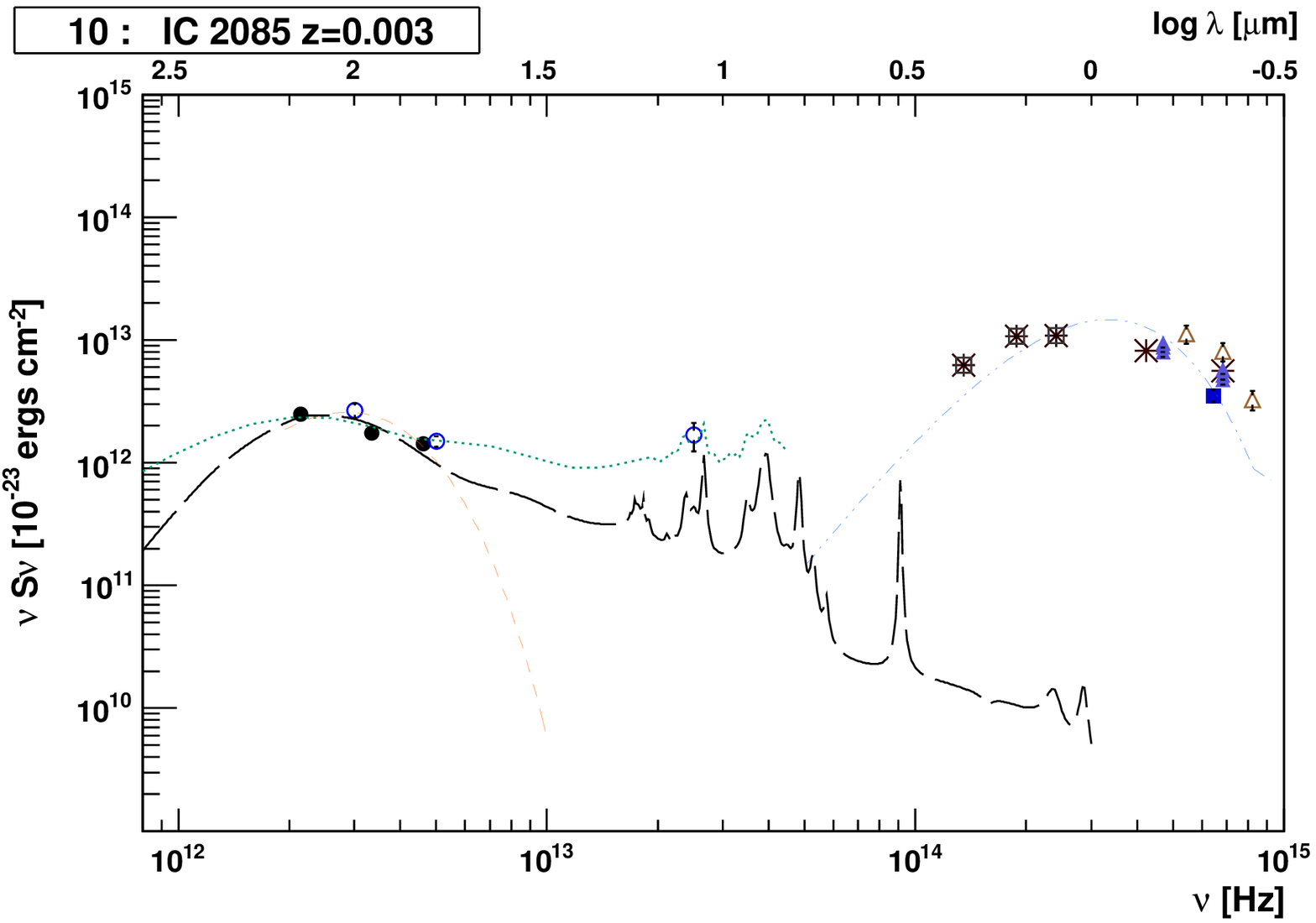}
\includegraphics[width=9cm]{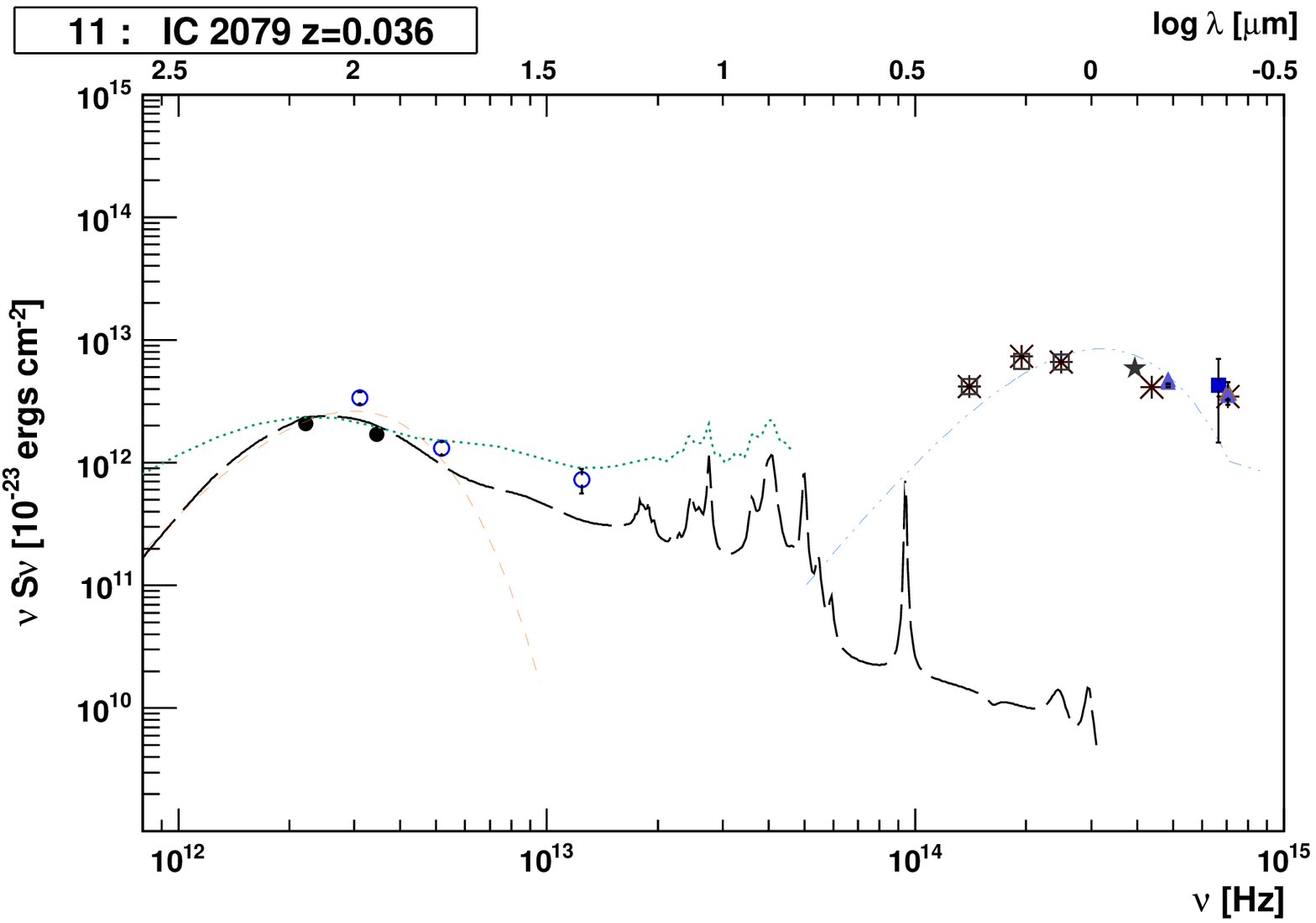}
\includegraphics[width=9cm]{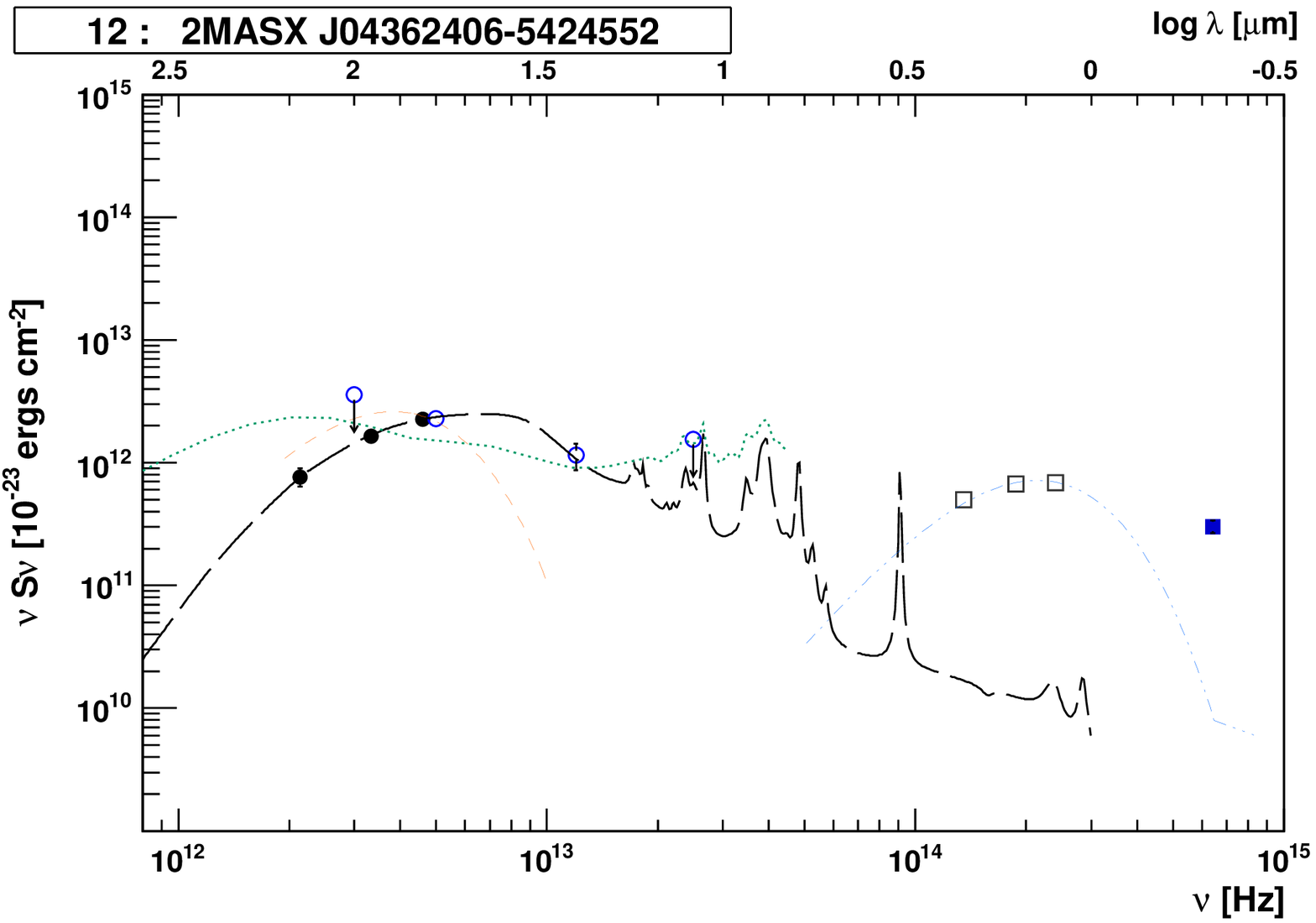}
\includegraphics[width=9cm]{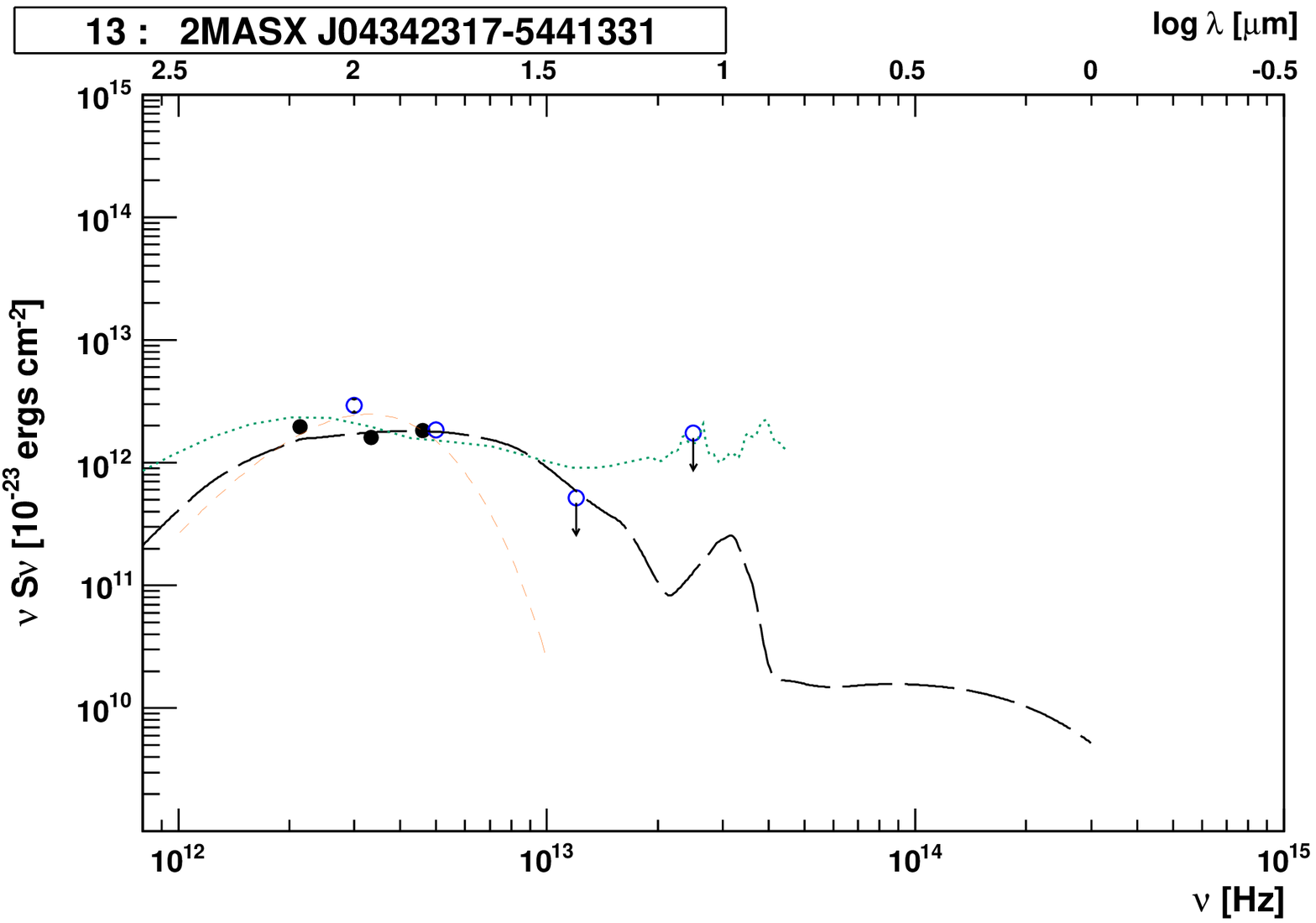}
\includegraphics[width=9cm]{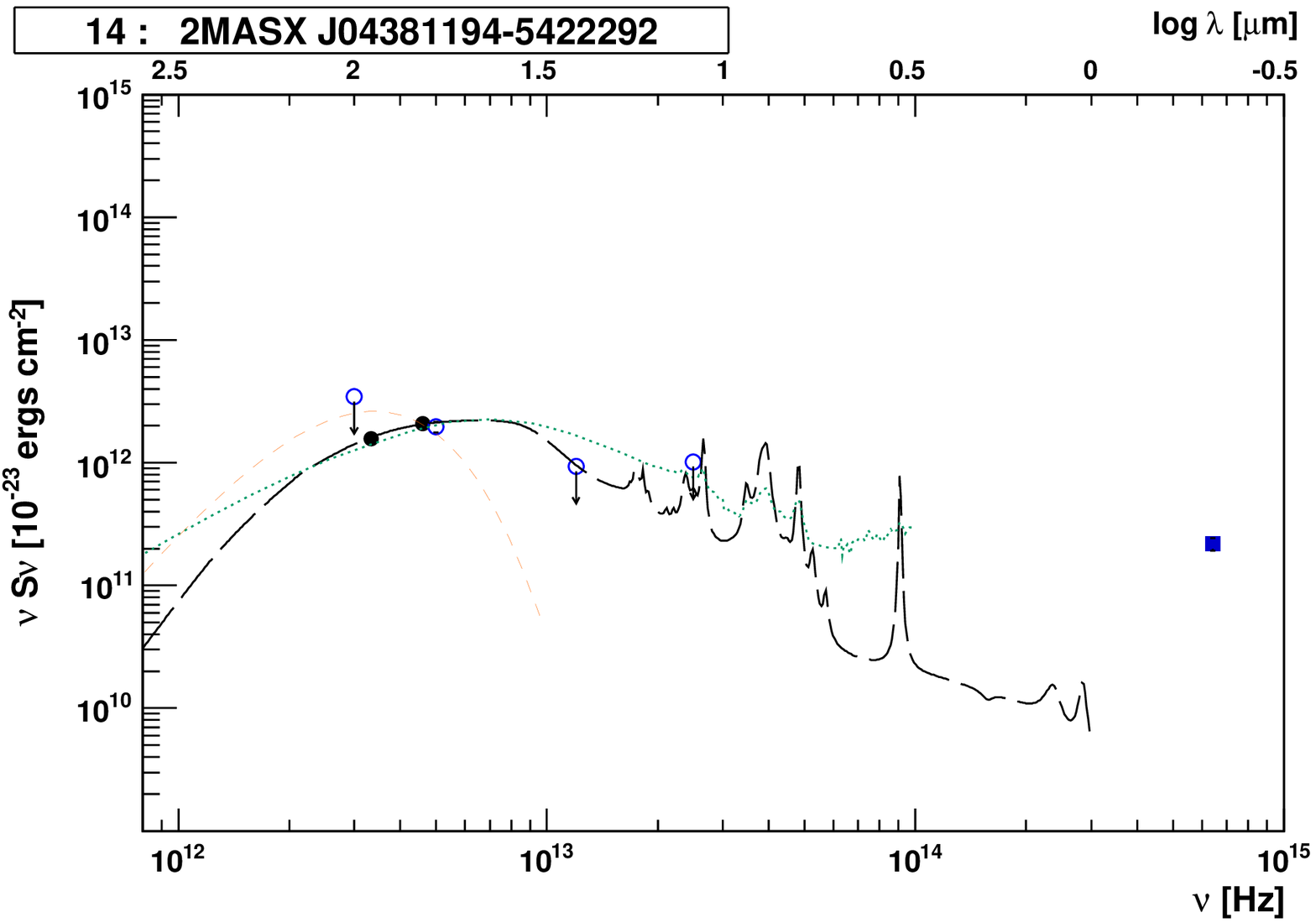}
\includegraphics[width=9cm]{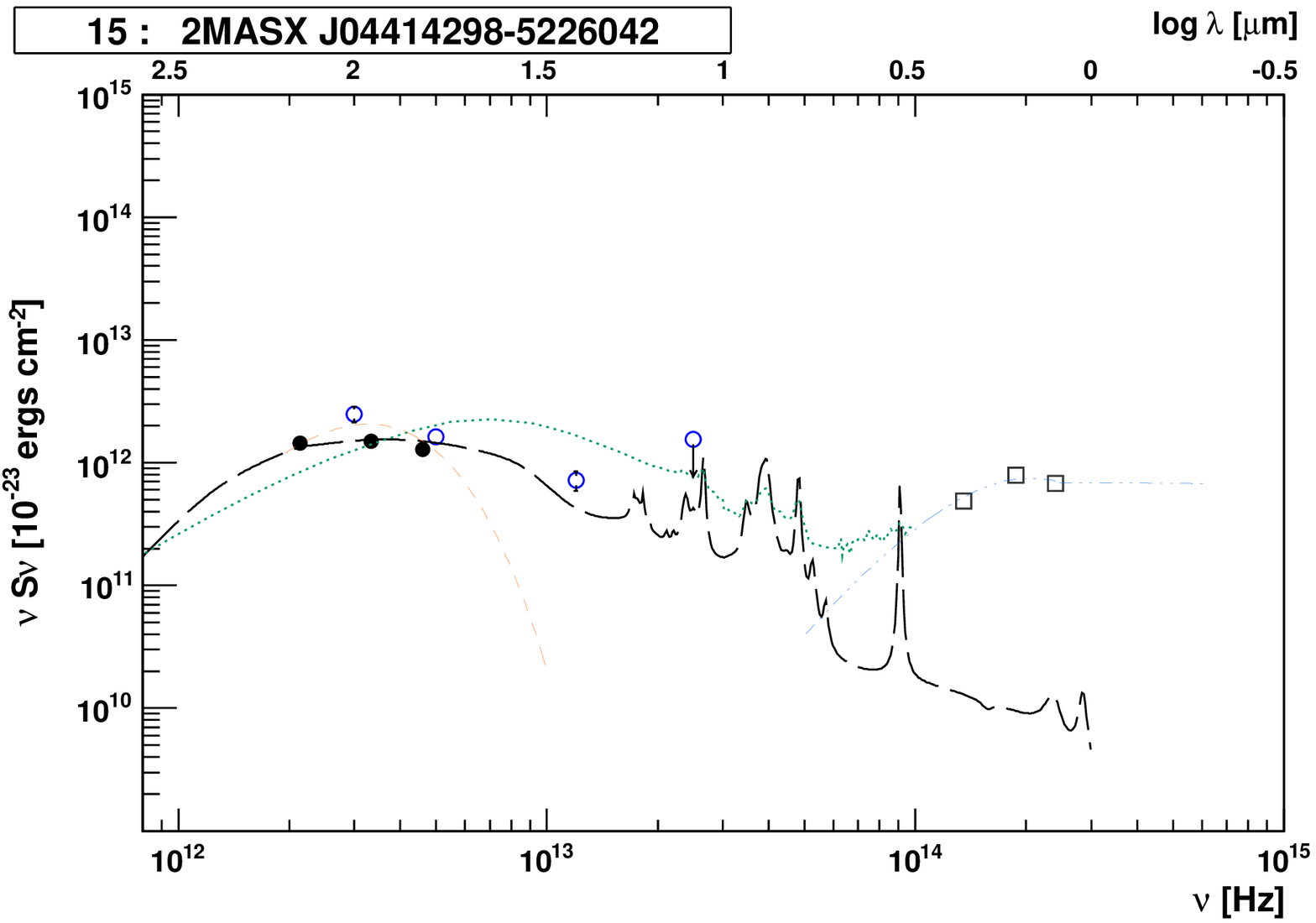}
\includegraphics[width=9cm]{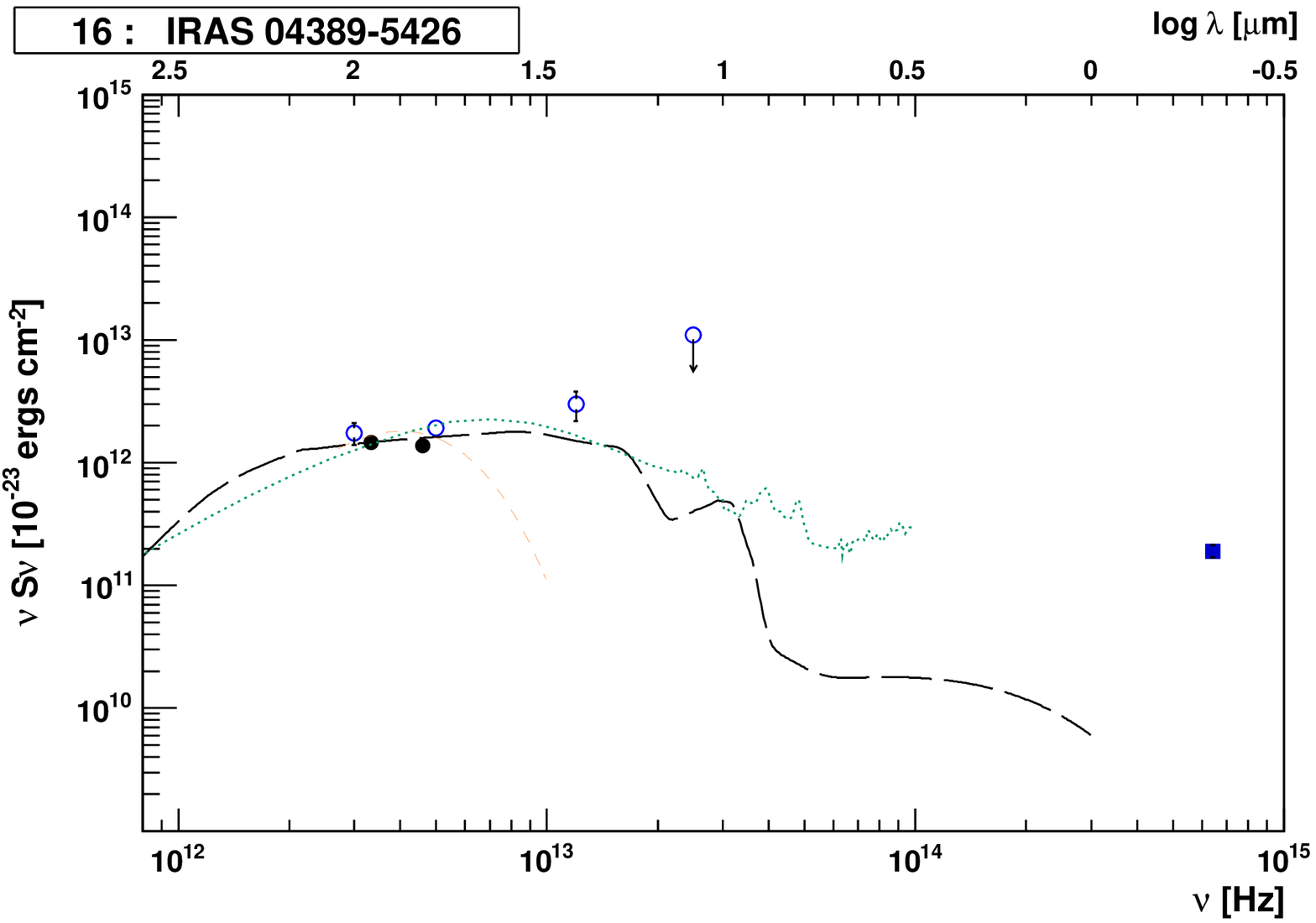}
\caption{Next 8 SEDs of ADF-S galaxies, with symbols as in Figure~\ref{sed}. 
SEDs of galaxies with a given redshift (objects number 9, 10, 11) 
are fitted after shifting to the rest frame and presented 
in the rest frame. The remaining objects are 
shown in the observed frame.}
  \label{sed2}%
 \end{figure*}

\begin{figure*}[t]
 \centering
\includegraphics[width=9cm]{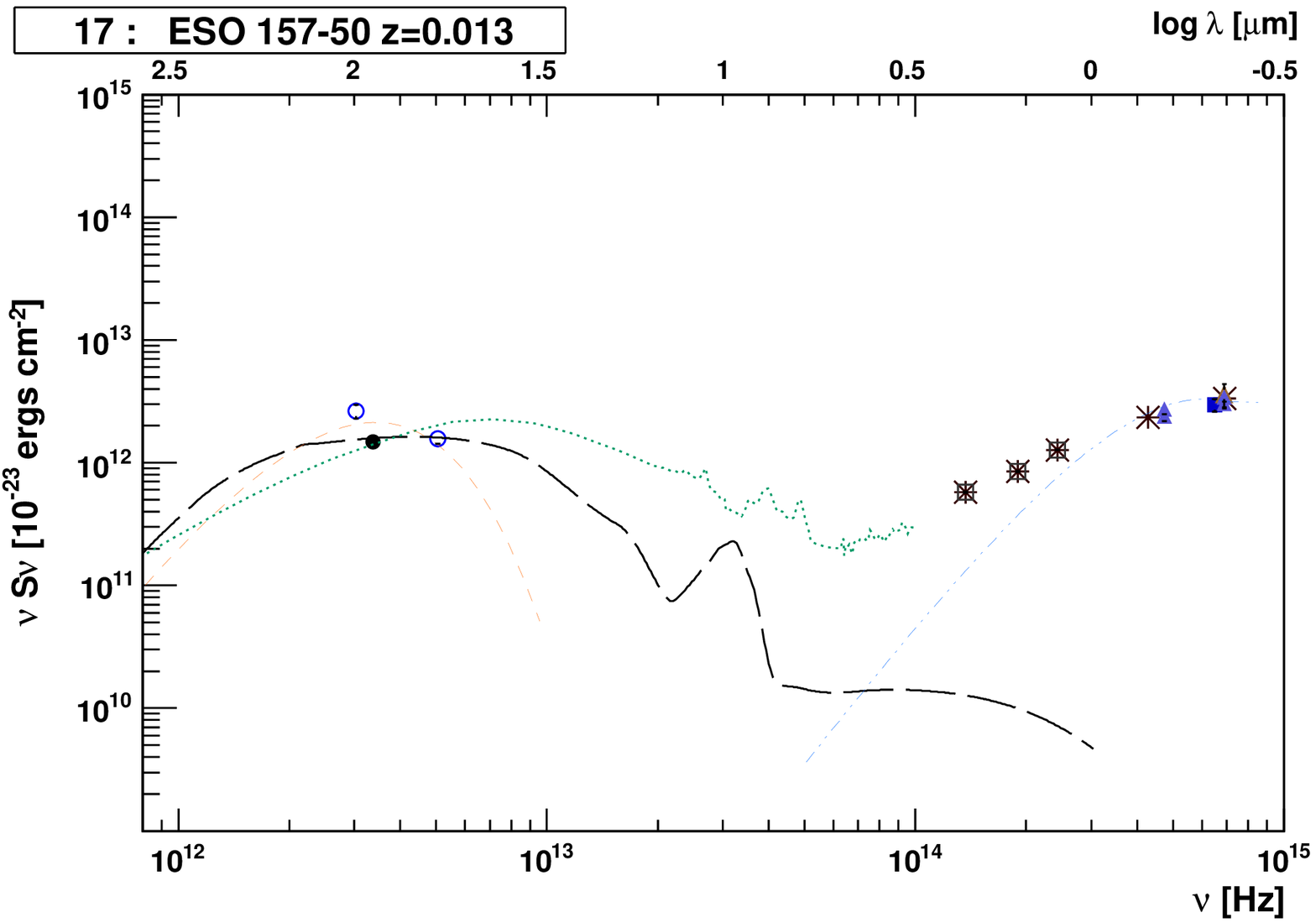}
\includegraphics[width=9cm]{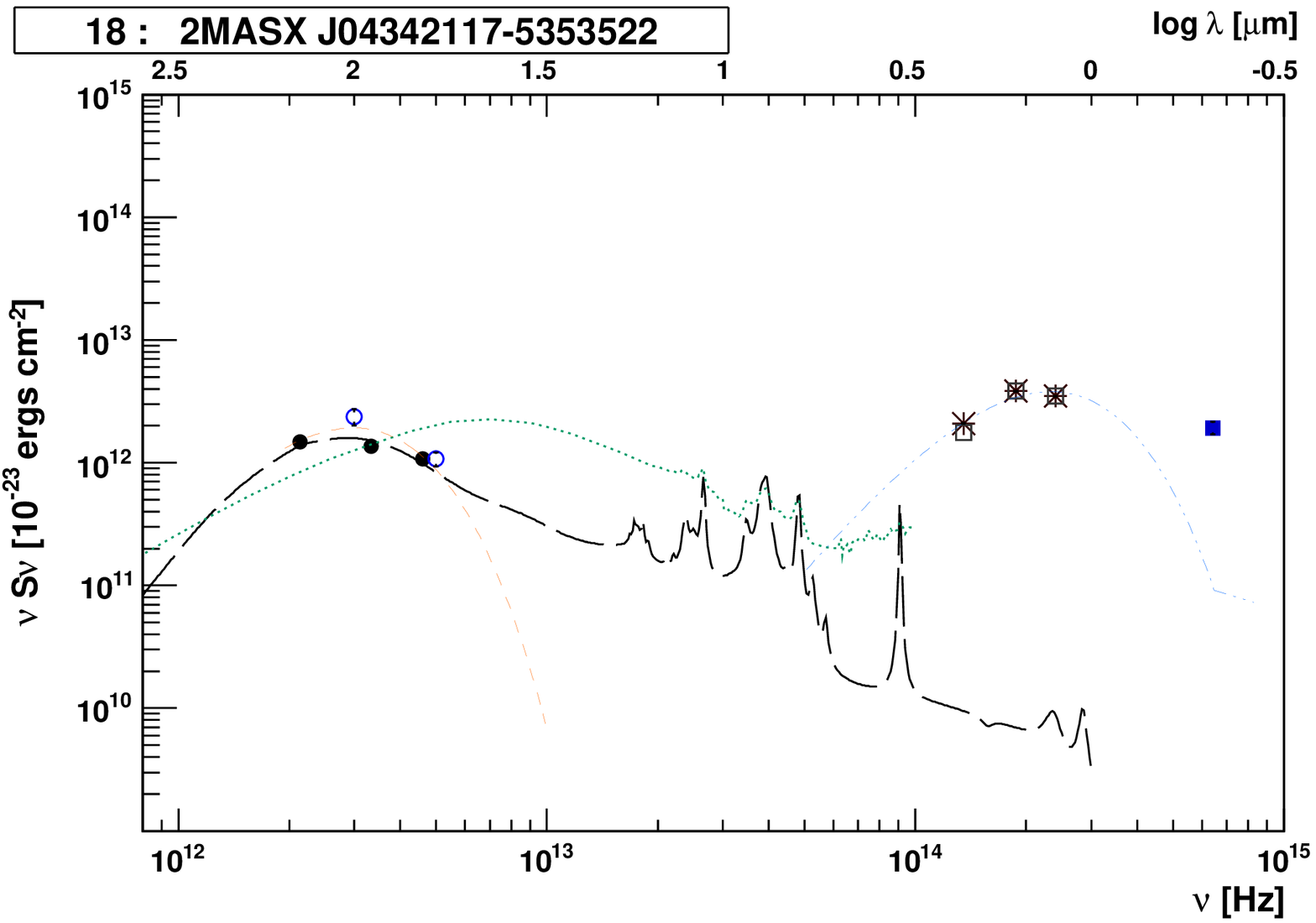}
\includegraphics[width=9cm]{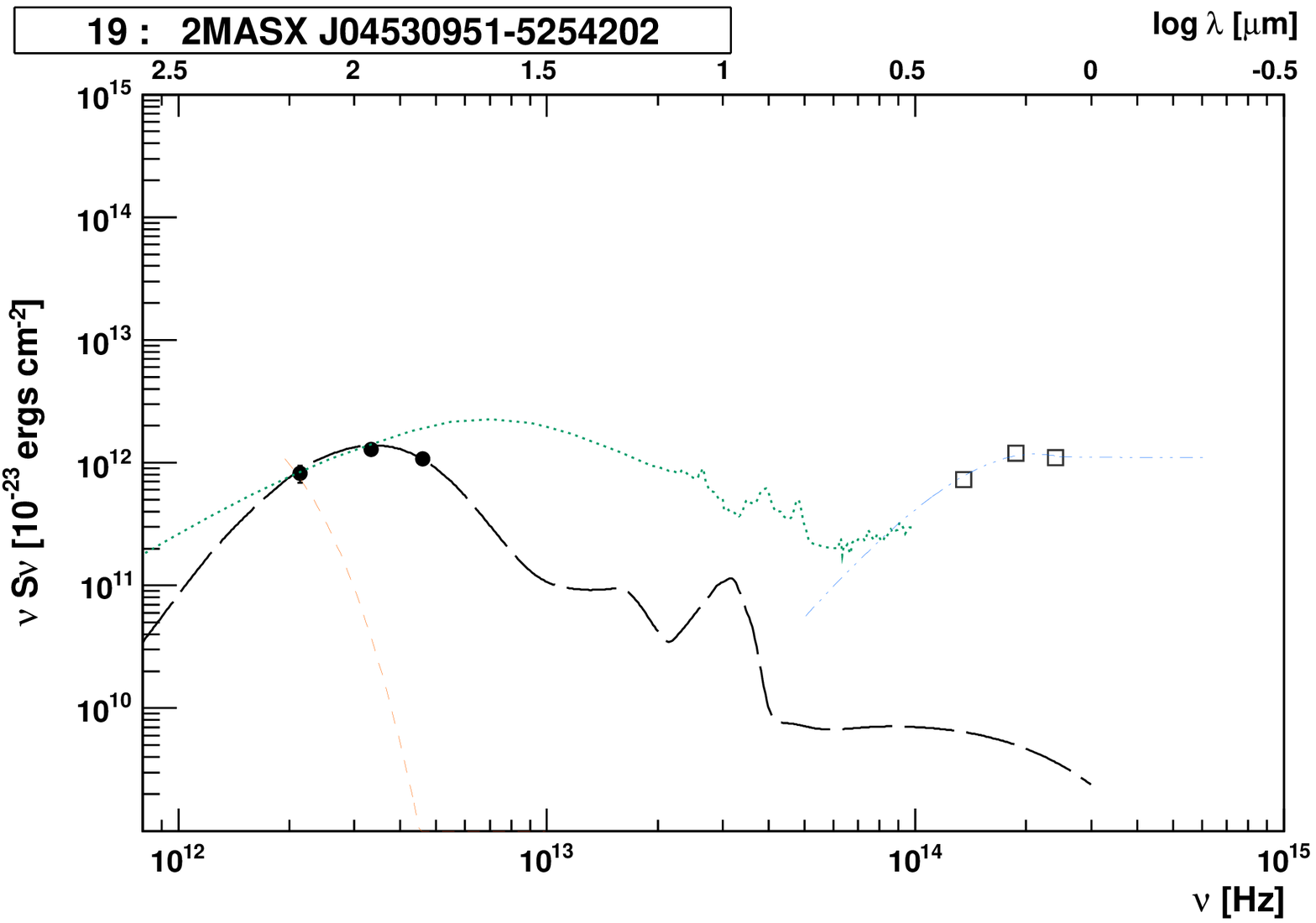}
\includegraphics[width=9cm]{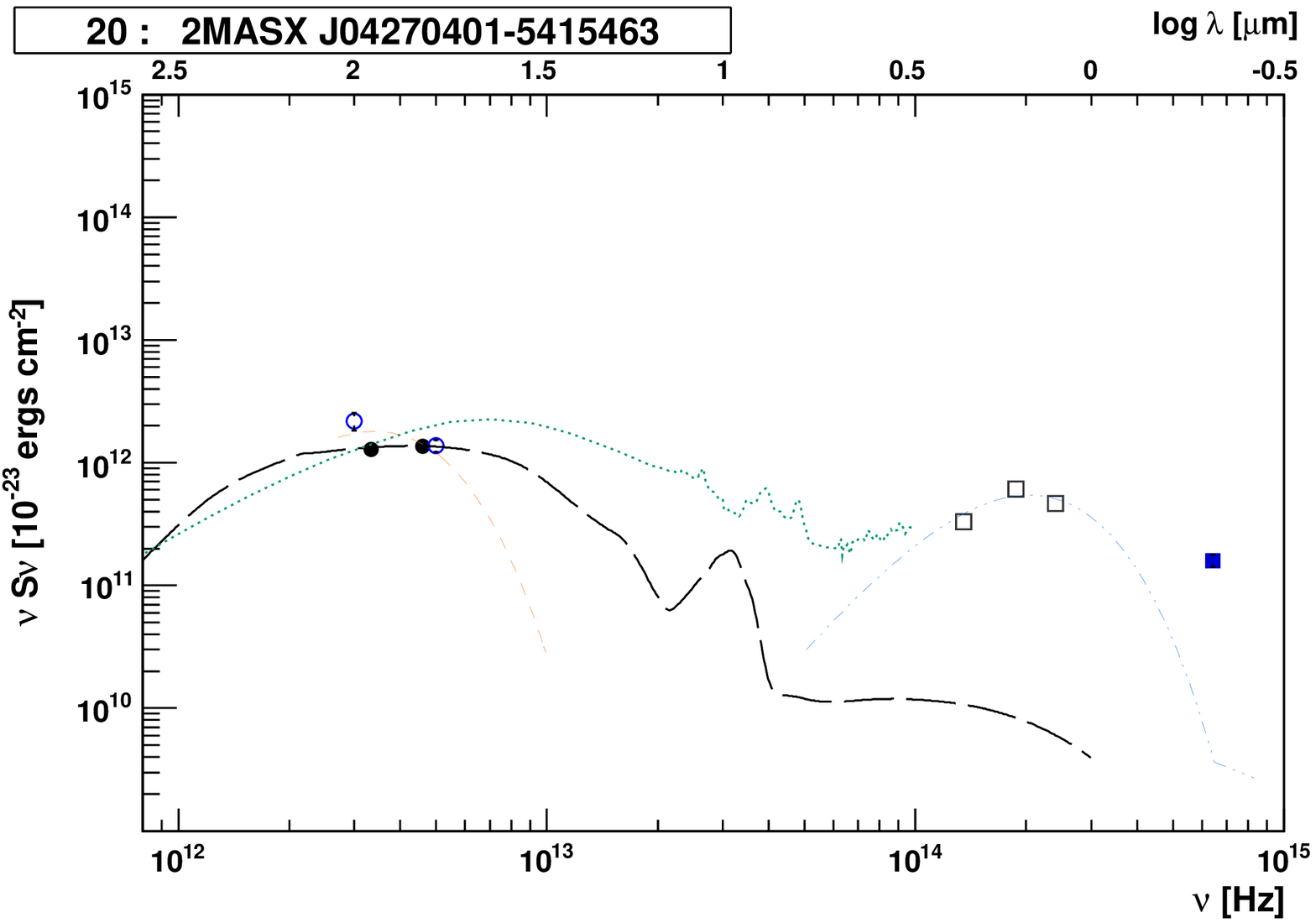}
\includegraphics[width=9cm]{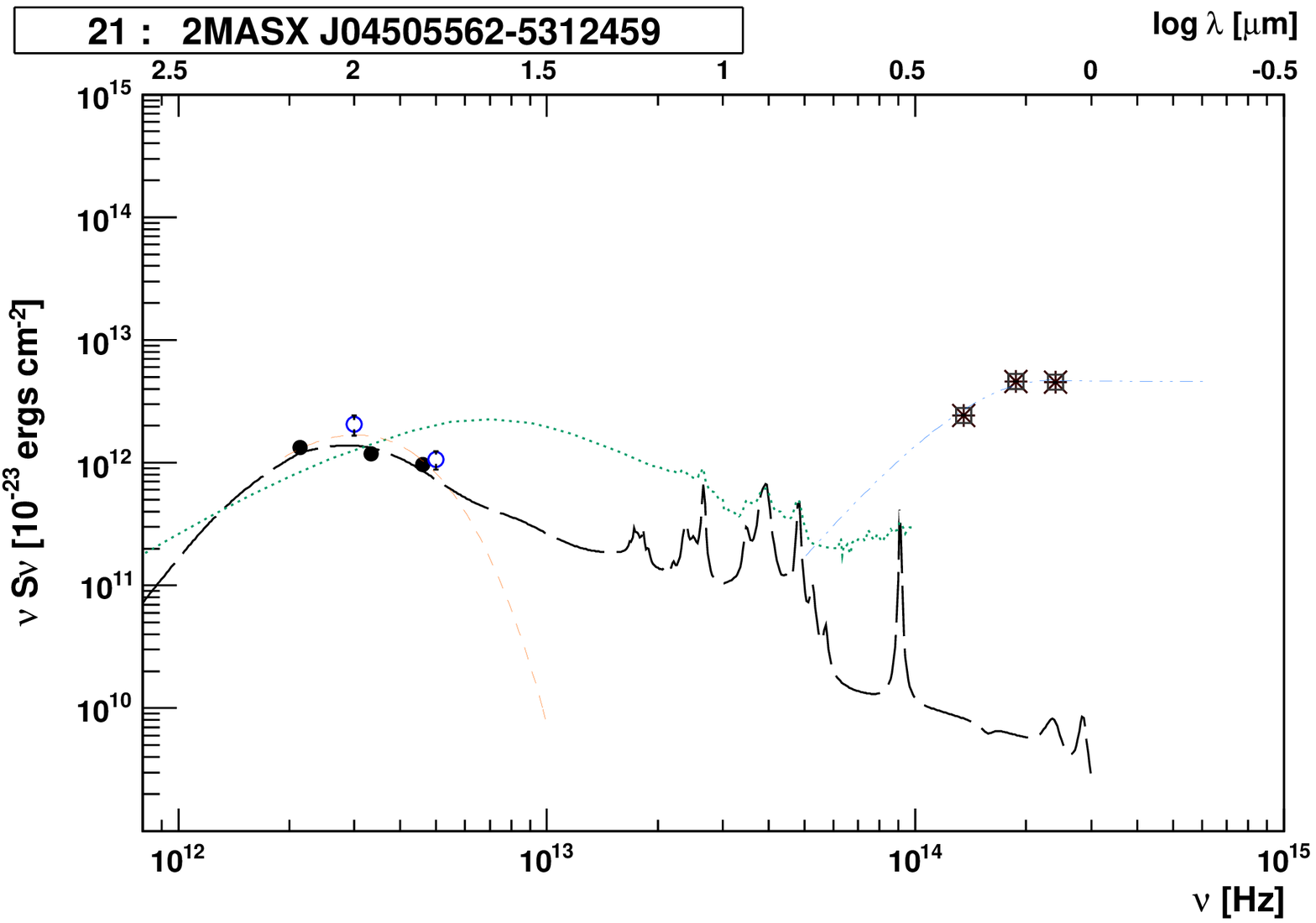}
\includegraphics[width=9cm]{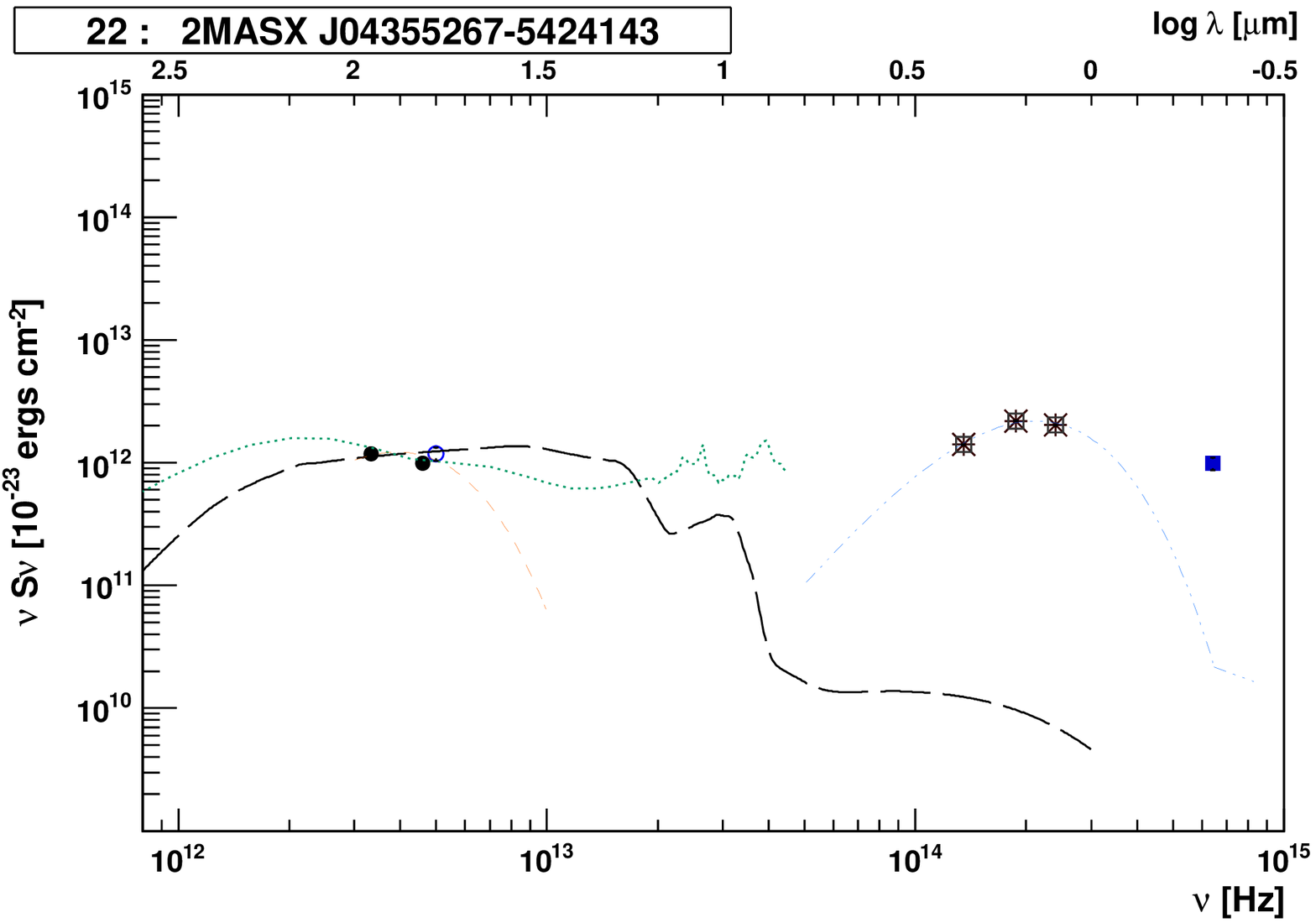}
\includegraphics[width=9cm]{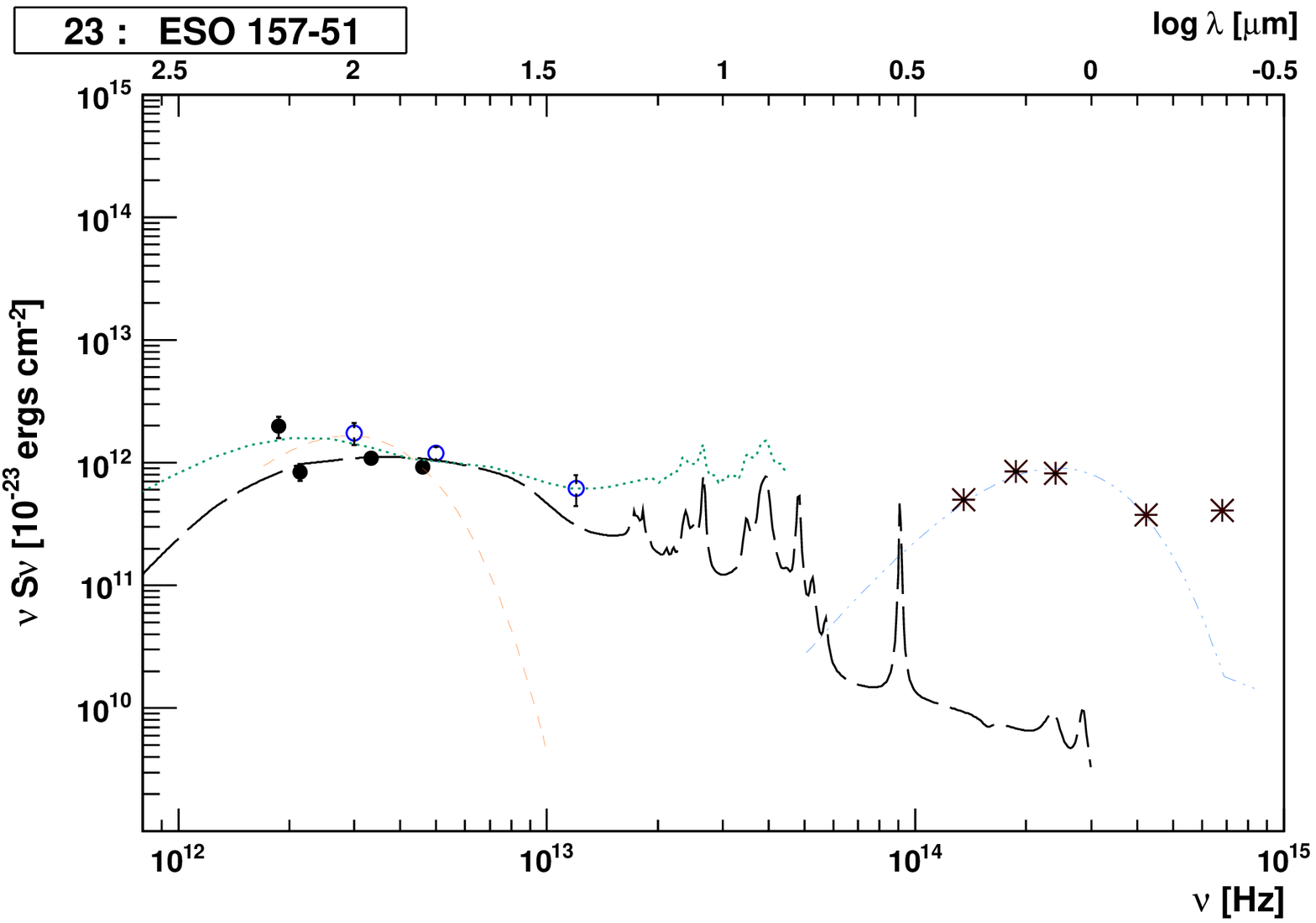}
  \includegraphics[width=9cm]{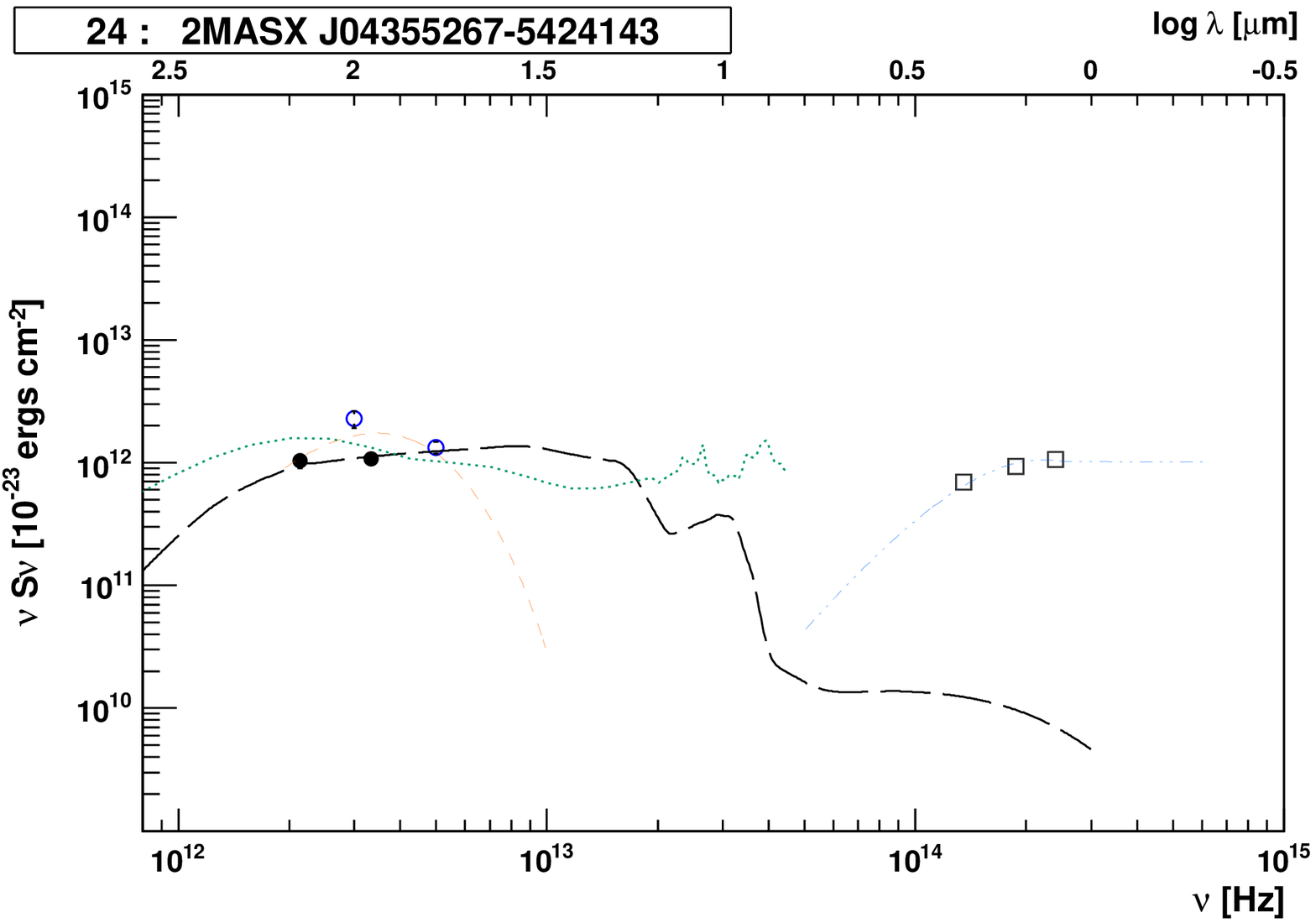}
\caption{Next 8 SEDs of ADF-S galaxies, with symbols as in Figure~\ref{sed}.
SED of a galaxy number 17, for which the redshift is known,
is fitted after shifting to the rest frame and presented 
in the rest frame. The remaining objects are
shown in the observed frame.
}
  \label{sed3}%
 \end{figure*}

\begin{figure*}[t]
 \centering
\includegraphics[width=9cm]{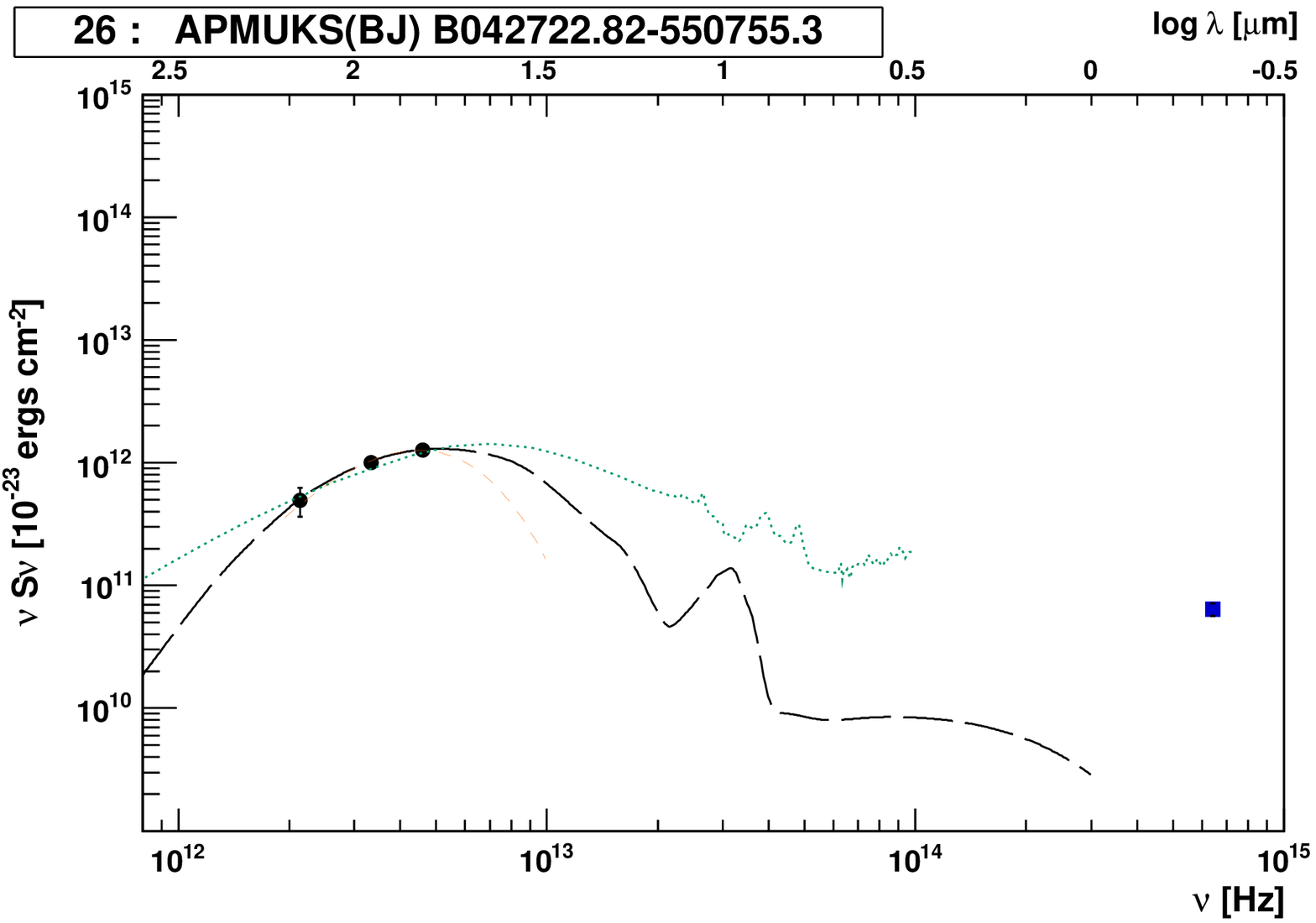}
\includegraphics[width=9cm]{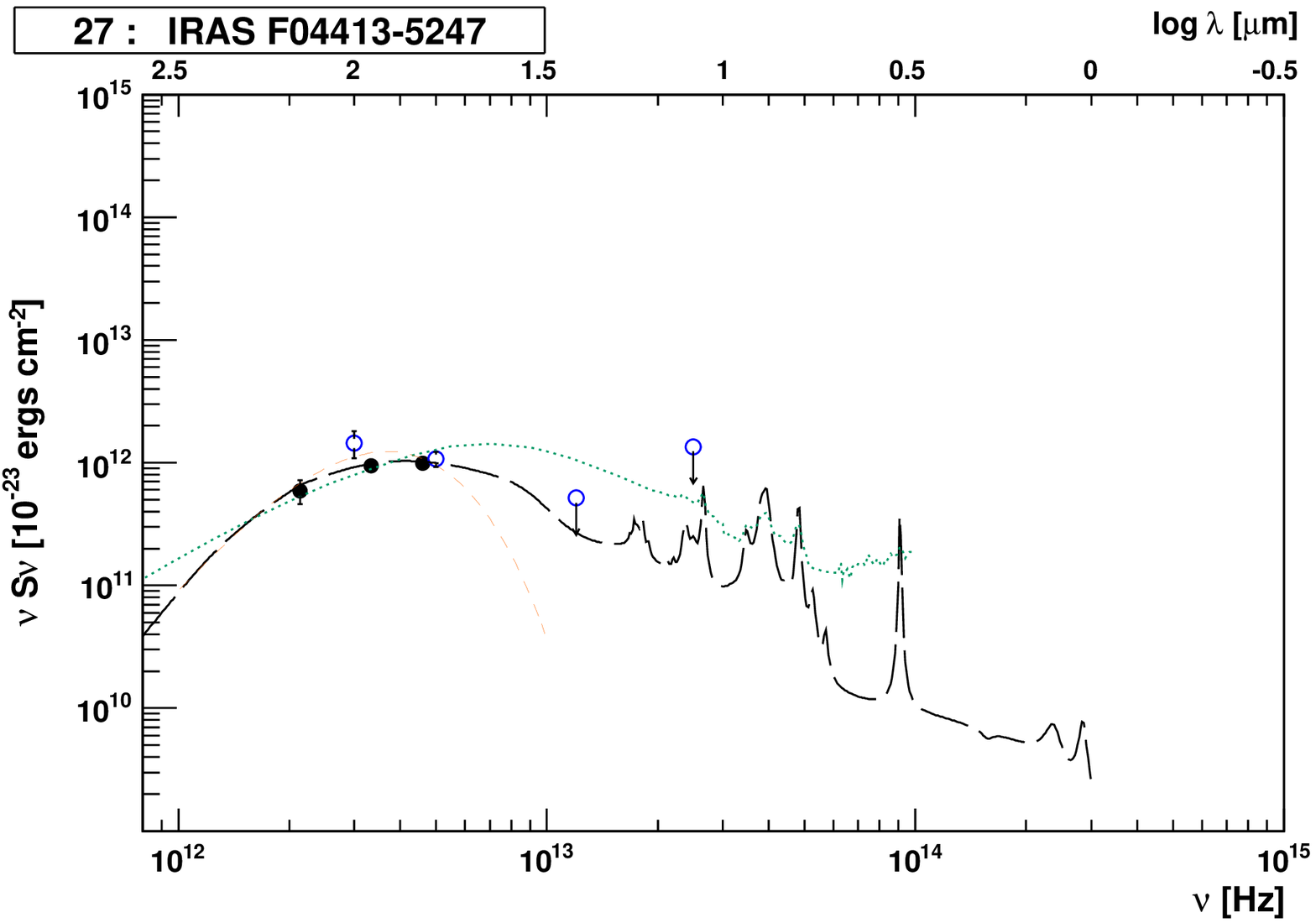}
\includegraphics[width=9cm]{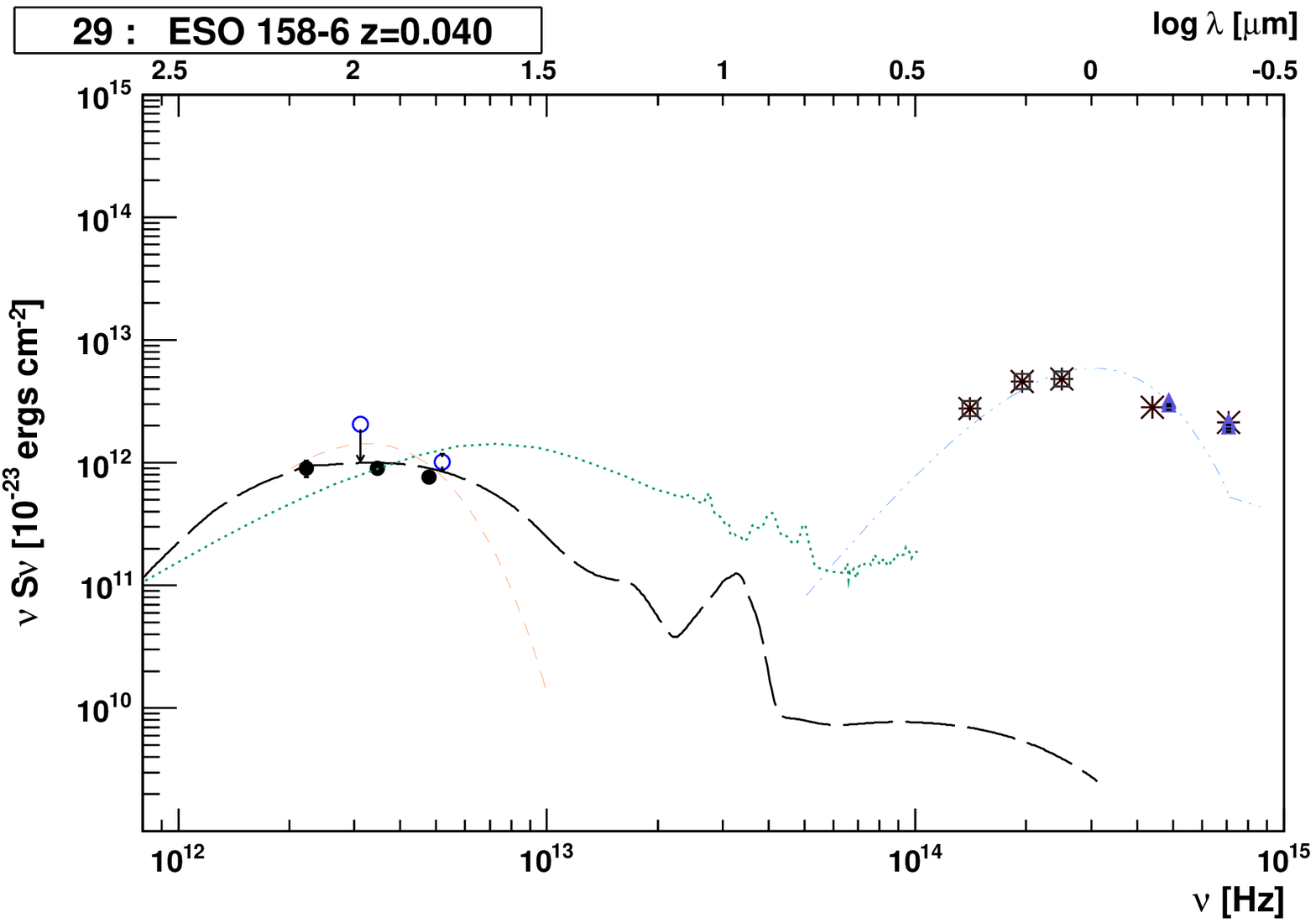}
\includegraphics[width=9cm]{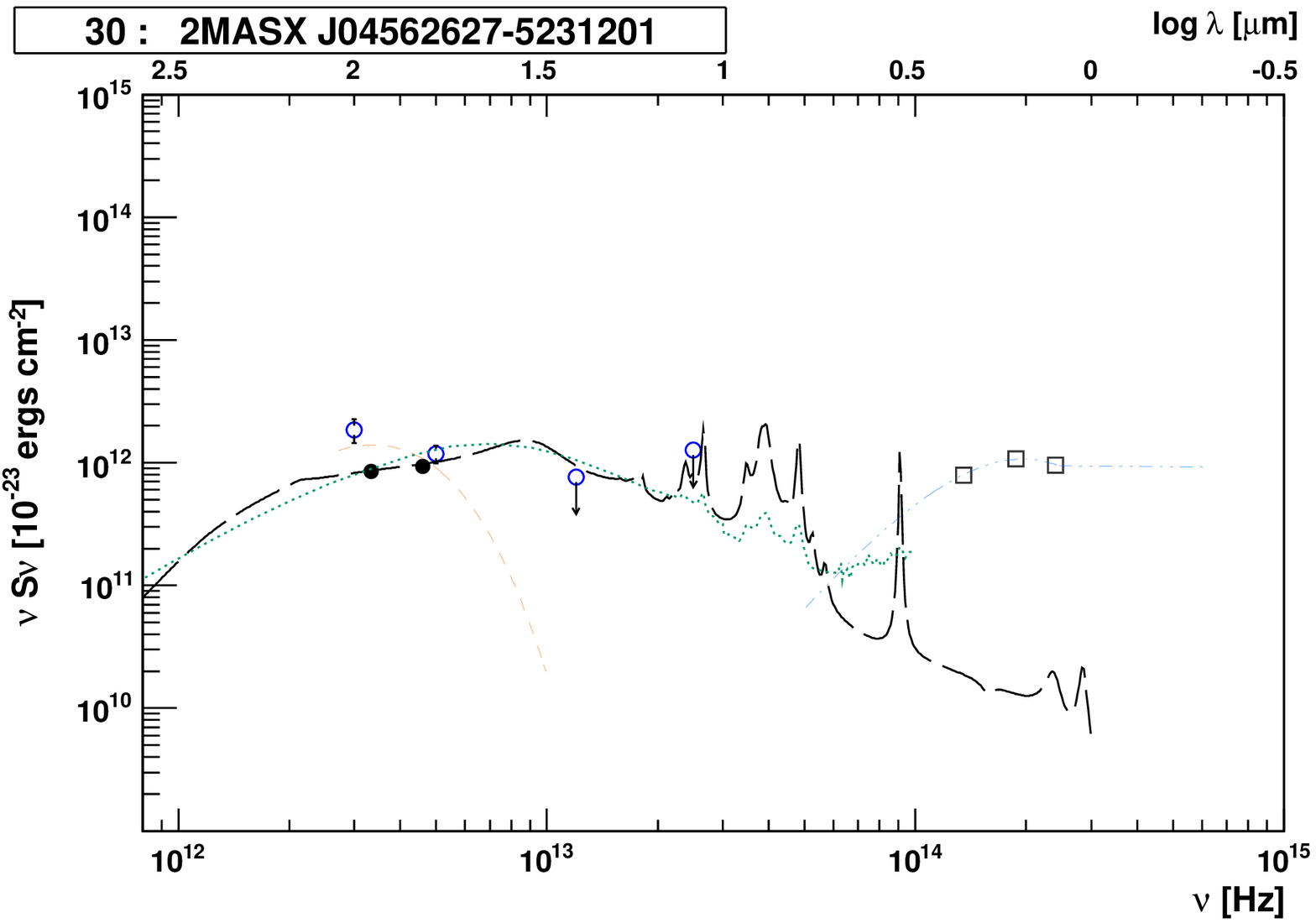}
\includegraphics[width=9cm]{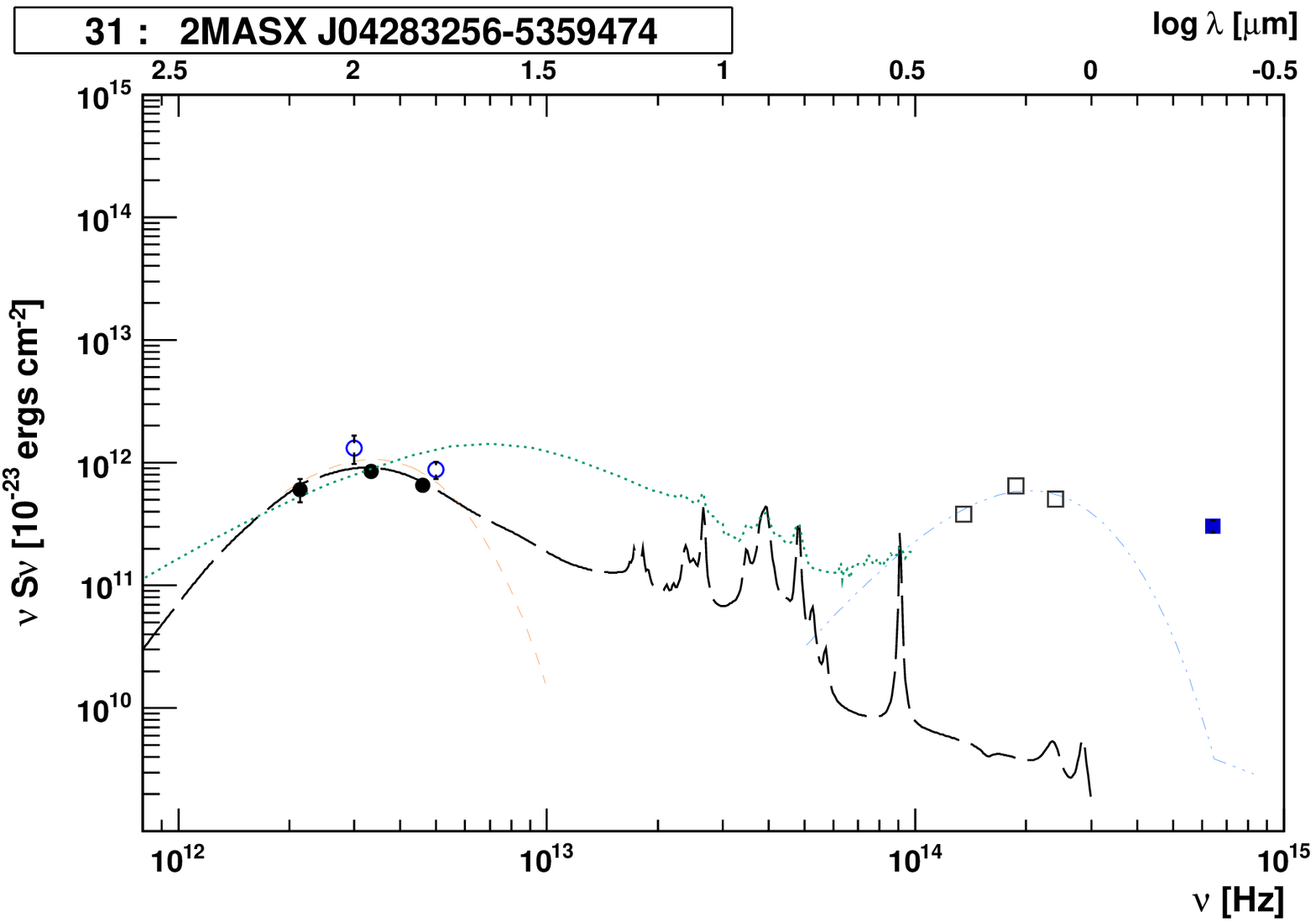}
\includegraphics[width=9cm]{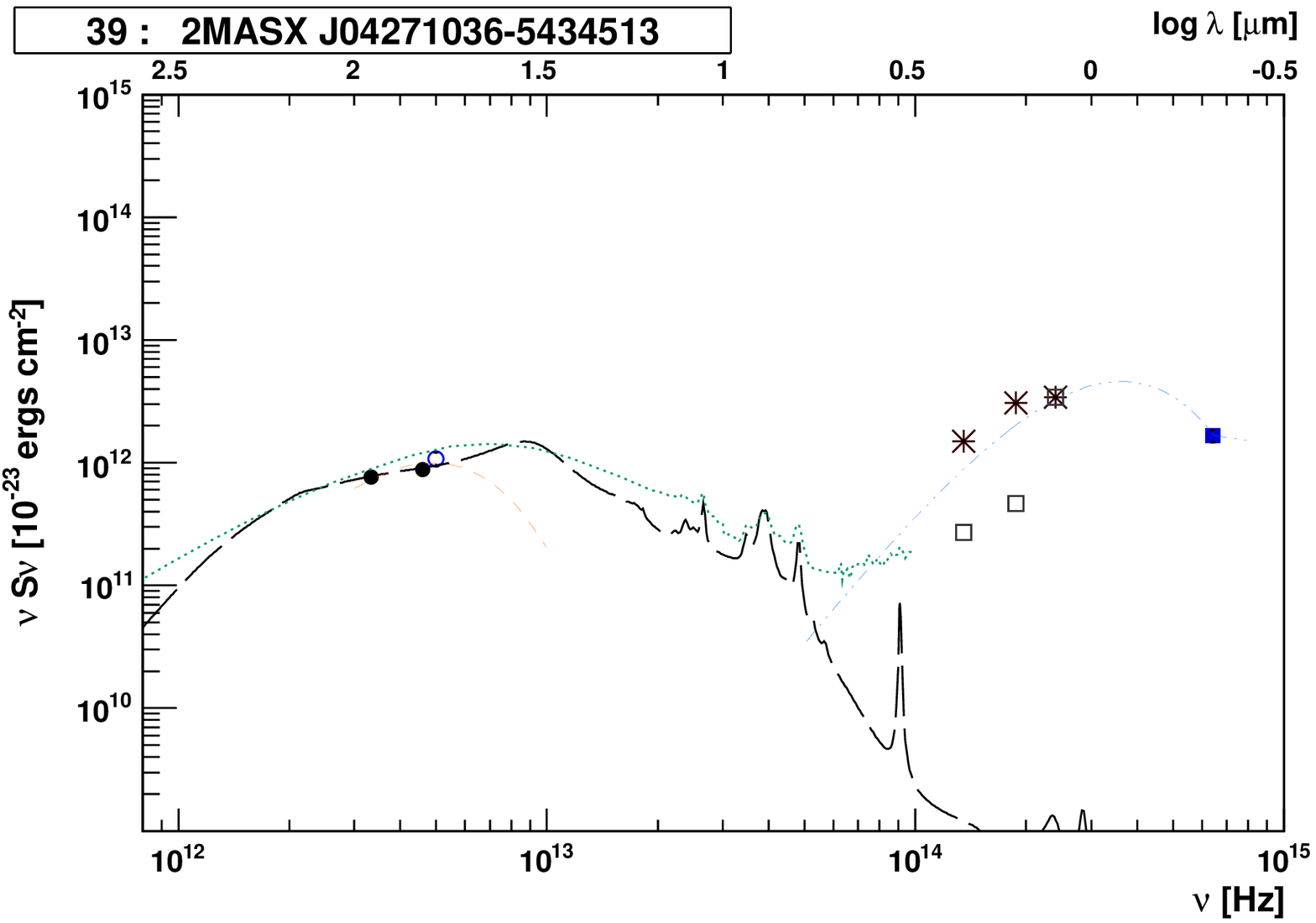}
\includegraphics[width=9cm]{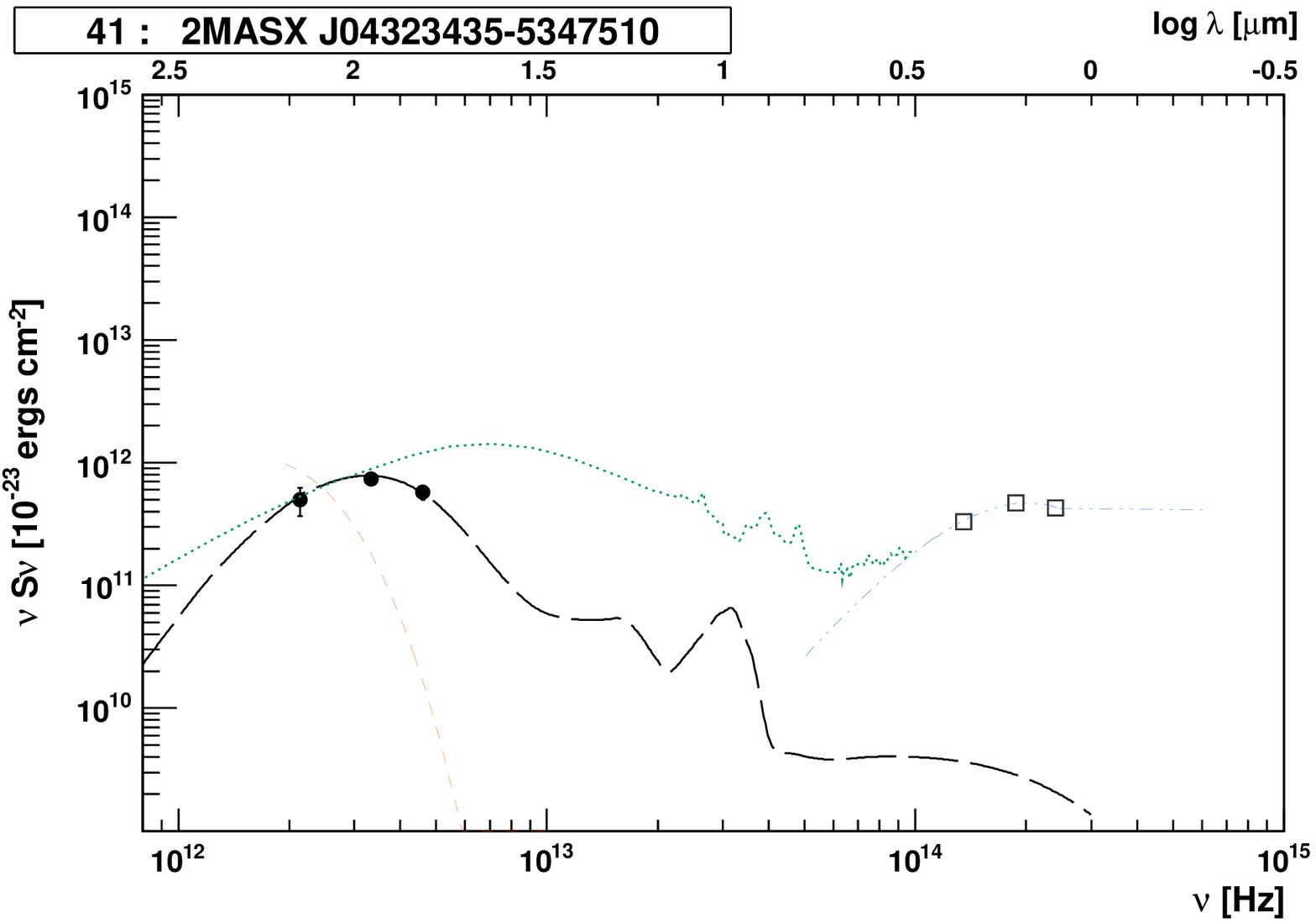}
\includegraphics[width=9cm]{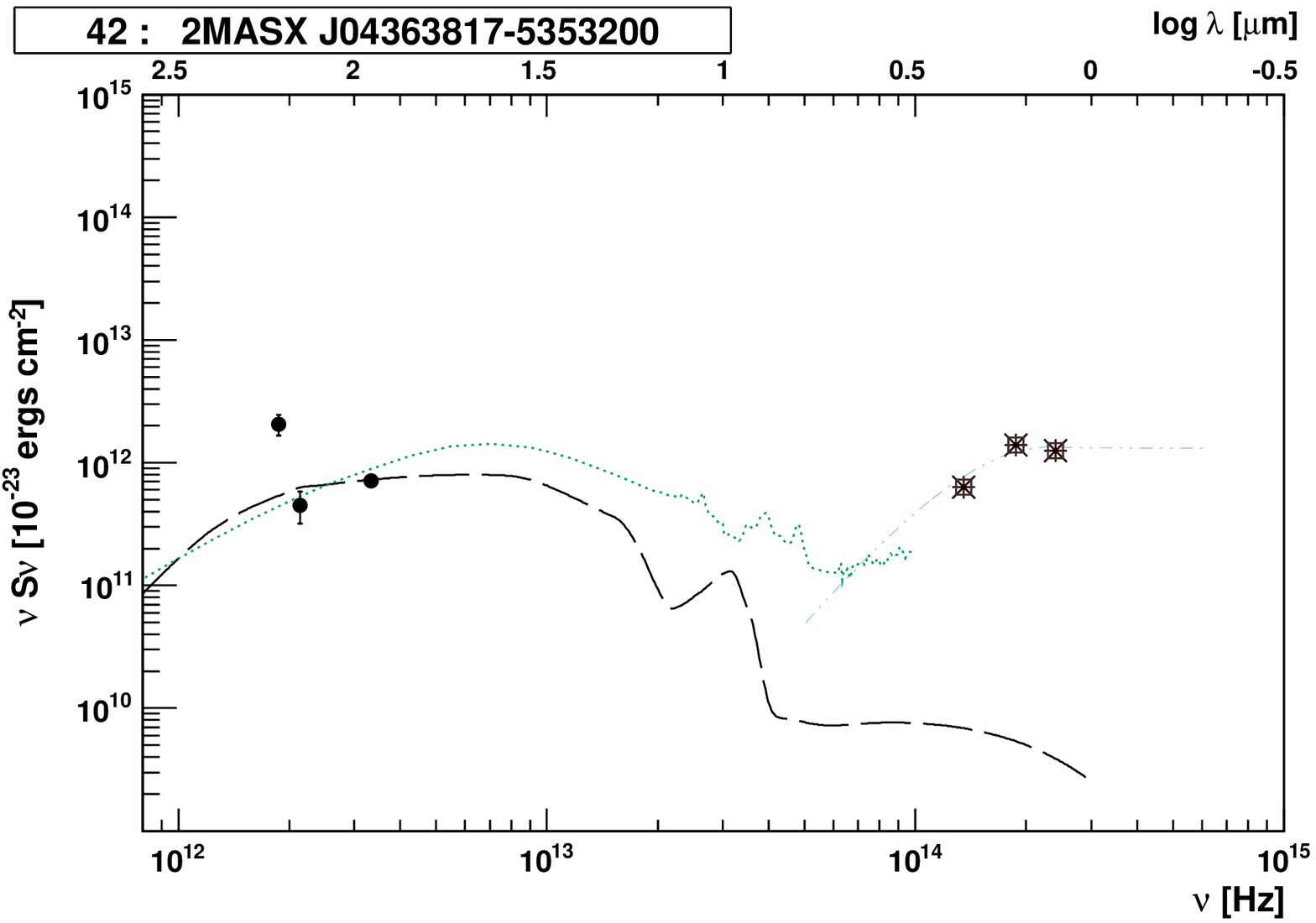}
\caption{Next 8 SEDs of ADF-S galaxies, with symbols as in Figure~\ref{sed}. 
SED of a galaxy number 29, for which the redshift is known, 
is fitted after shifting to the rest frame and presented
in the rest frame. The remaining objects are
shown in the observed frame.
}

  \label{sed4}%
 \end{figure*}

\begin{figure*}[t]
 \centering
\includegraphics[width=9cm]{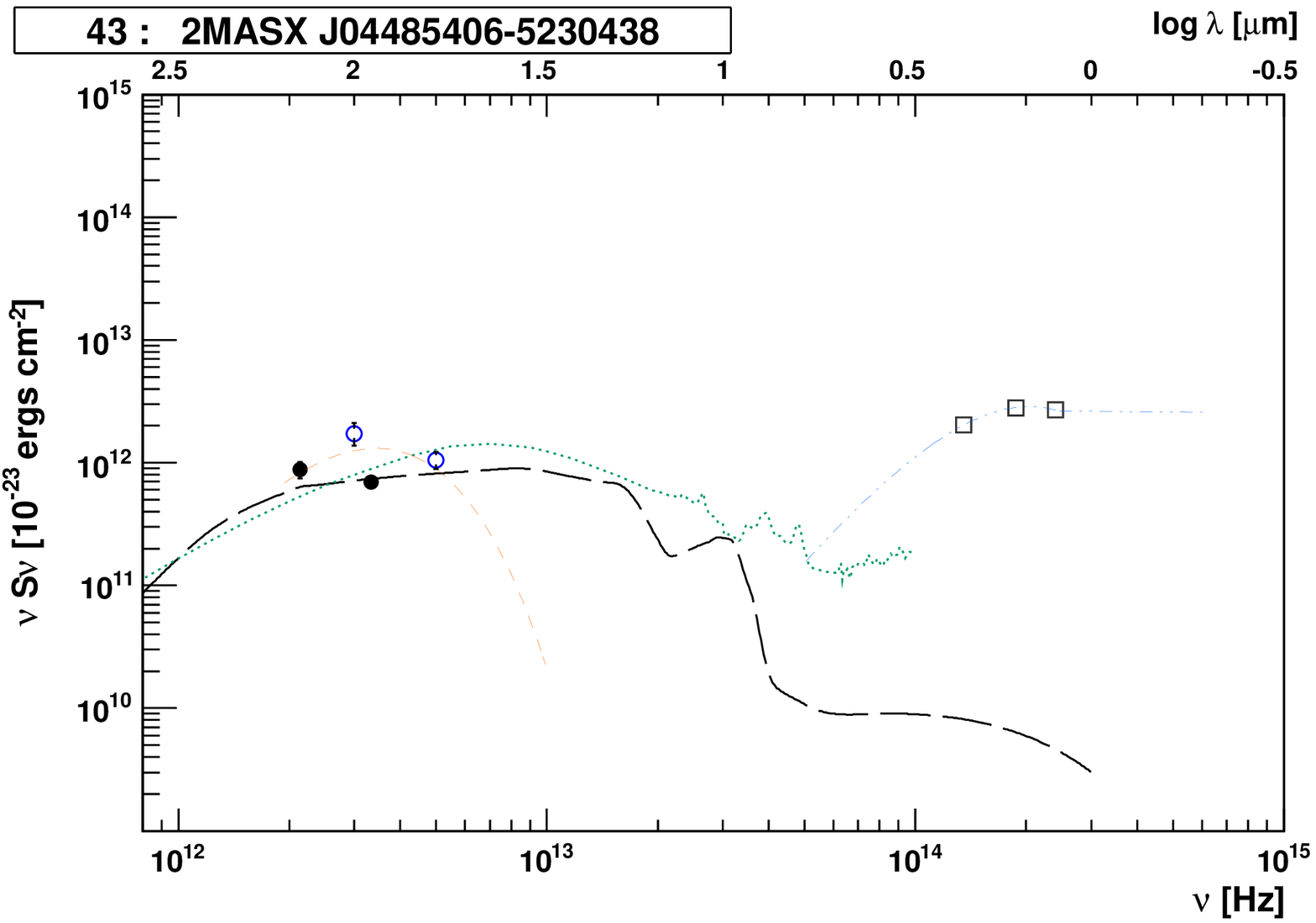}
\includegraphics[width=9cm]{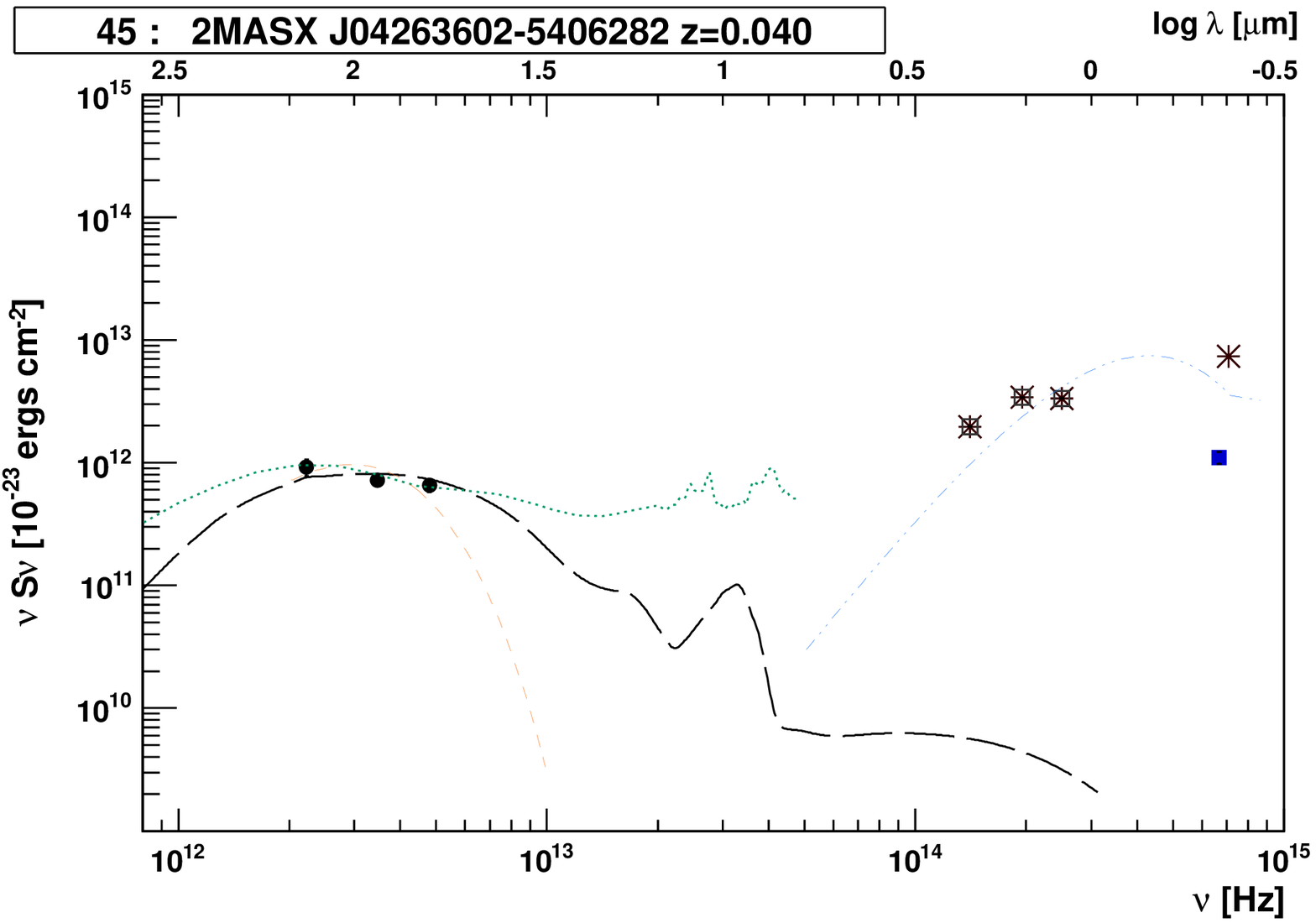}
\includegraphics[width=9cm]{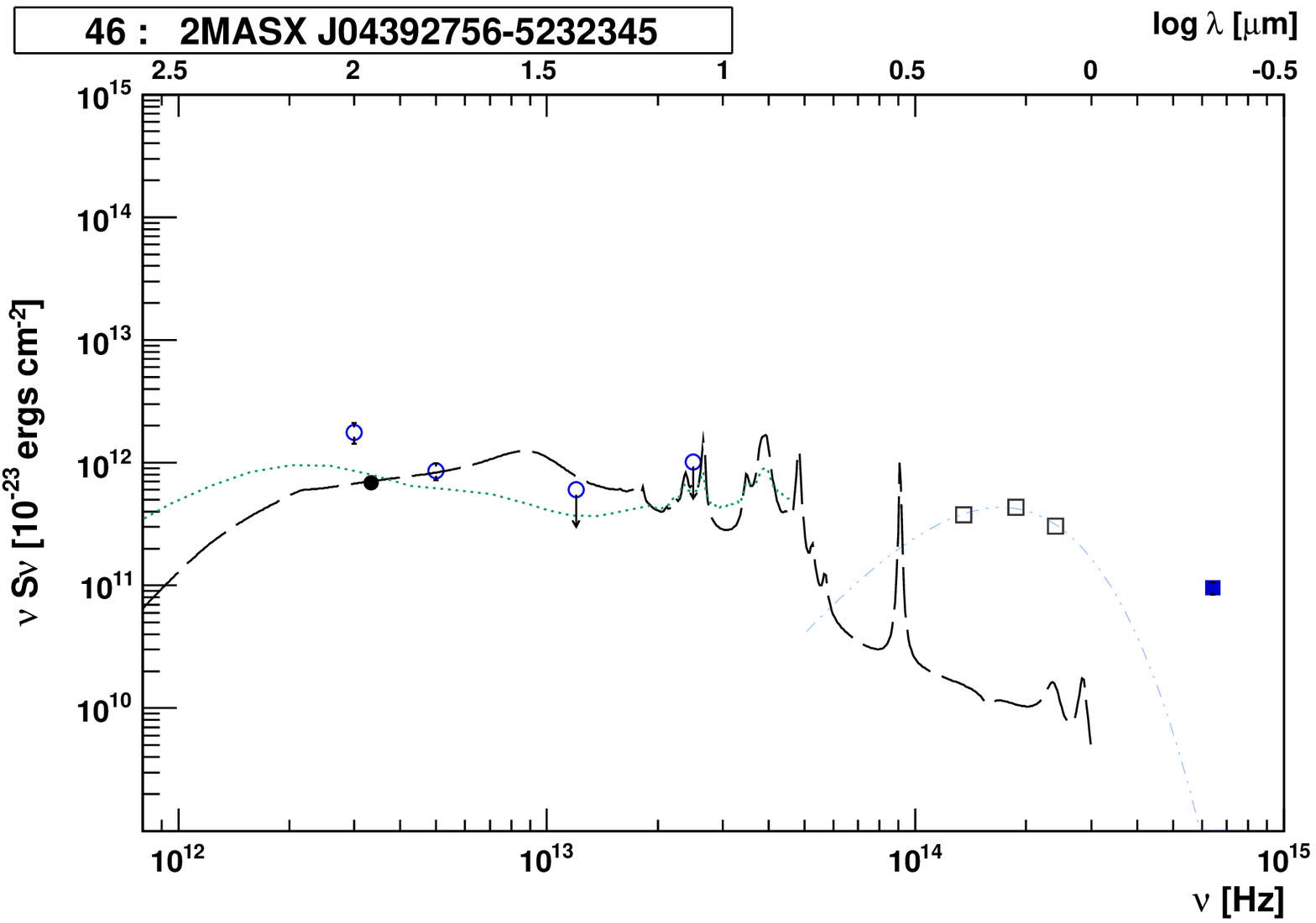}
\includegraphics[width=9cm]{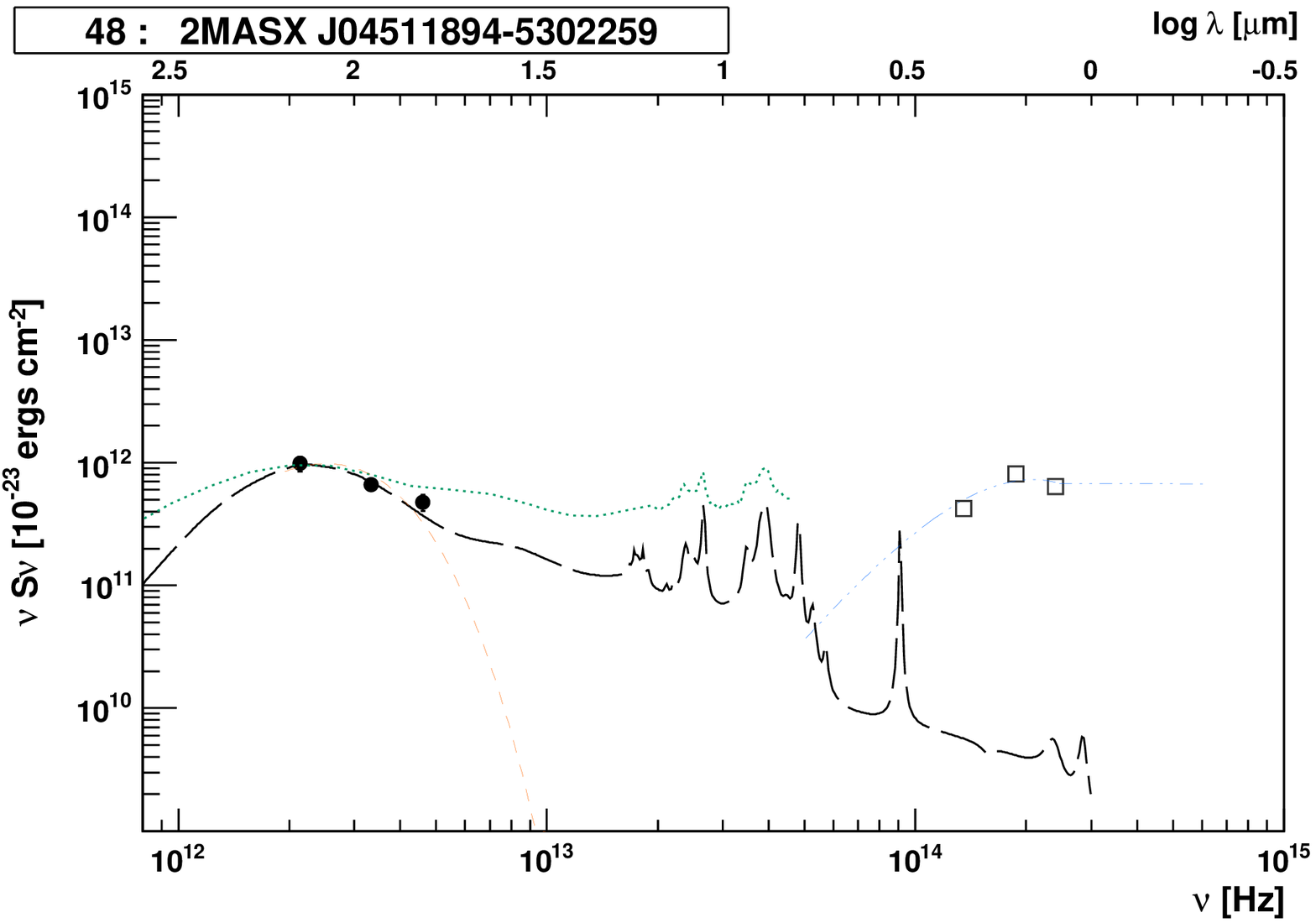}
\includegraphics[width=9cm]{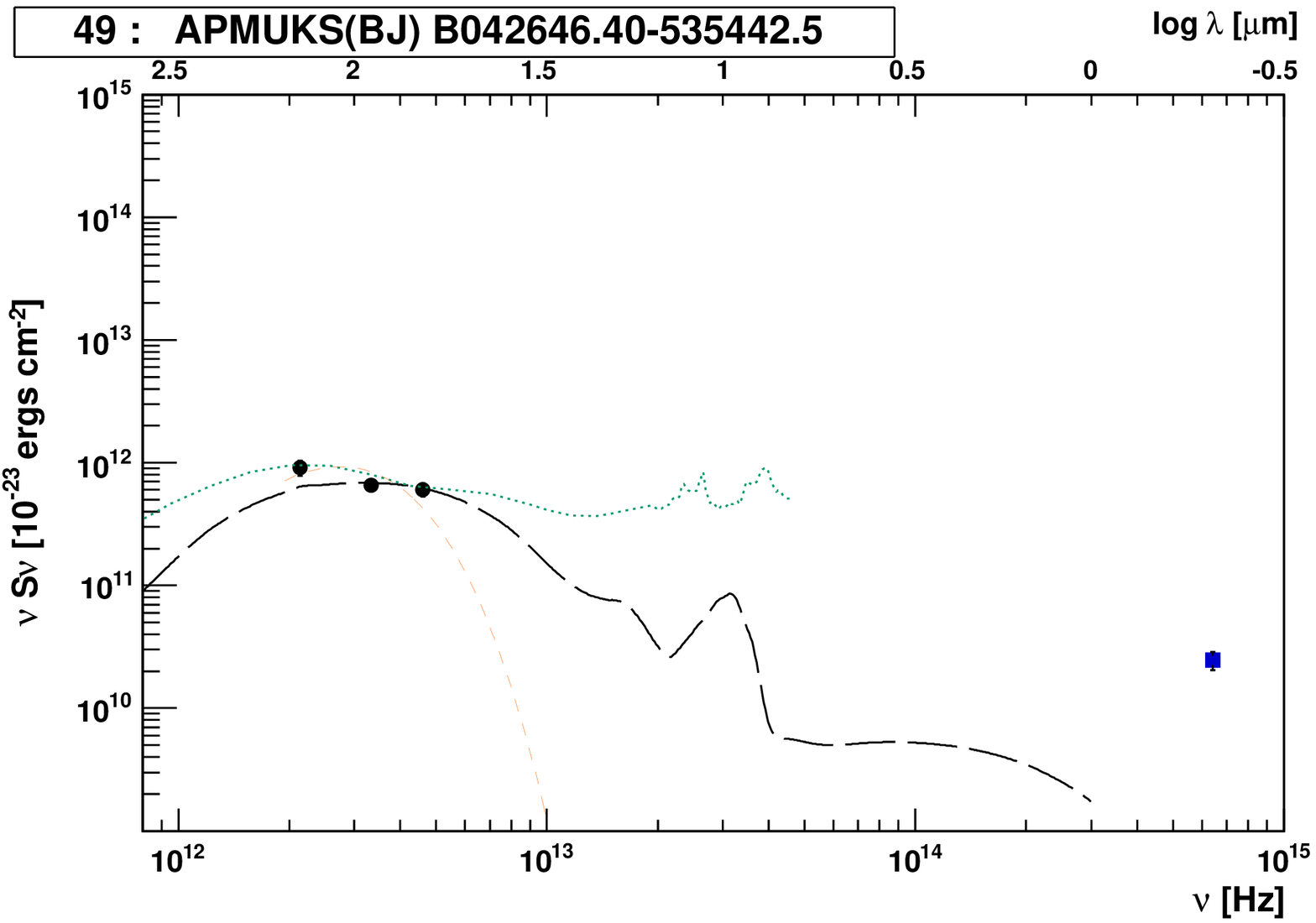}
\includegraphics[width=9cm]{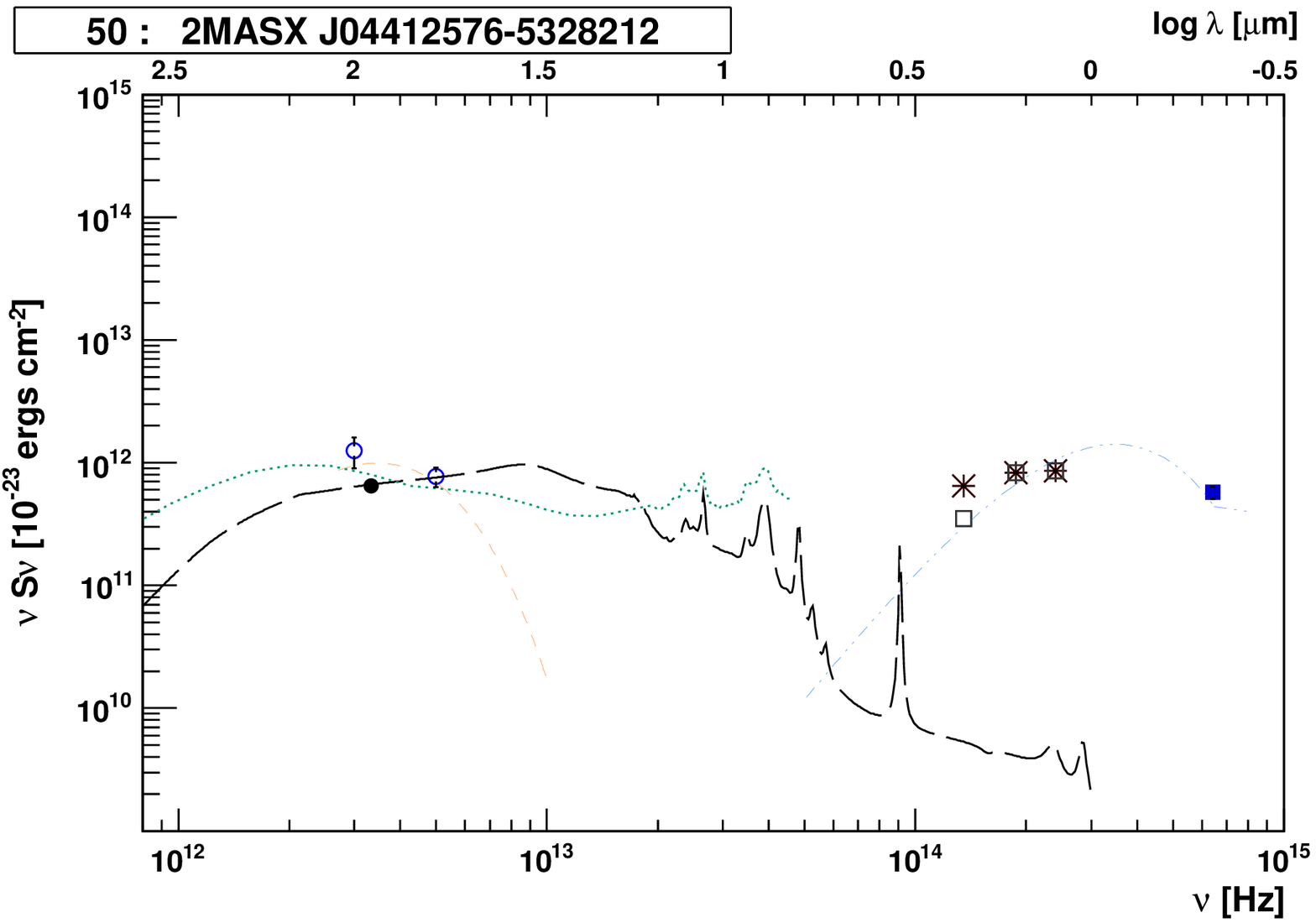}
\includegraphics[width=9cm]{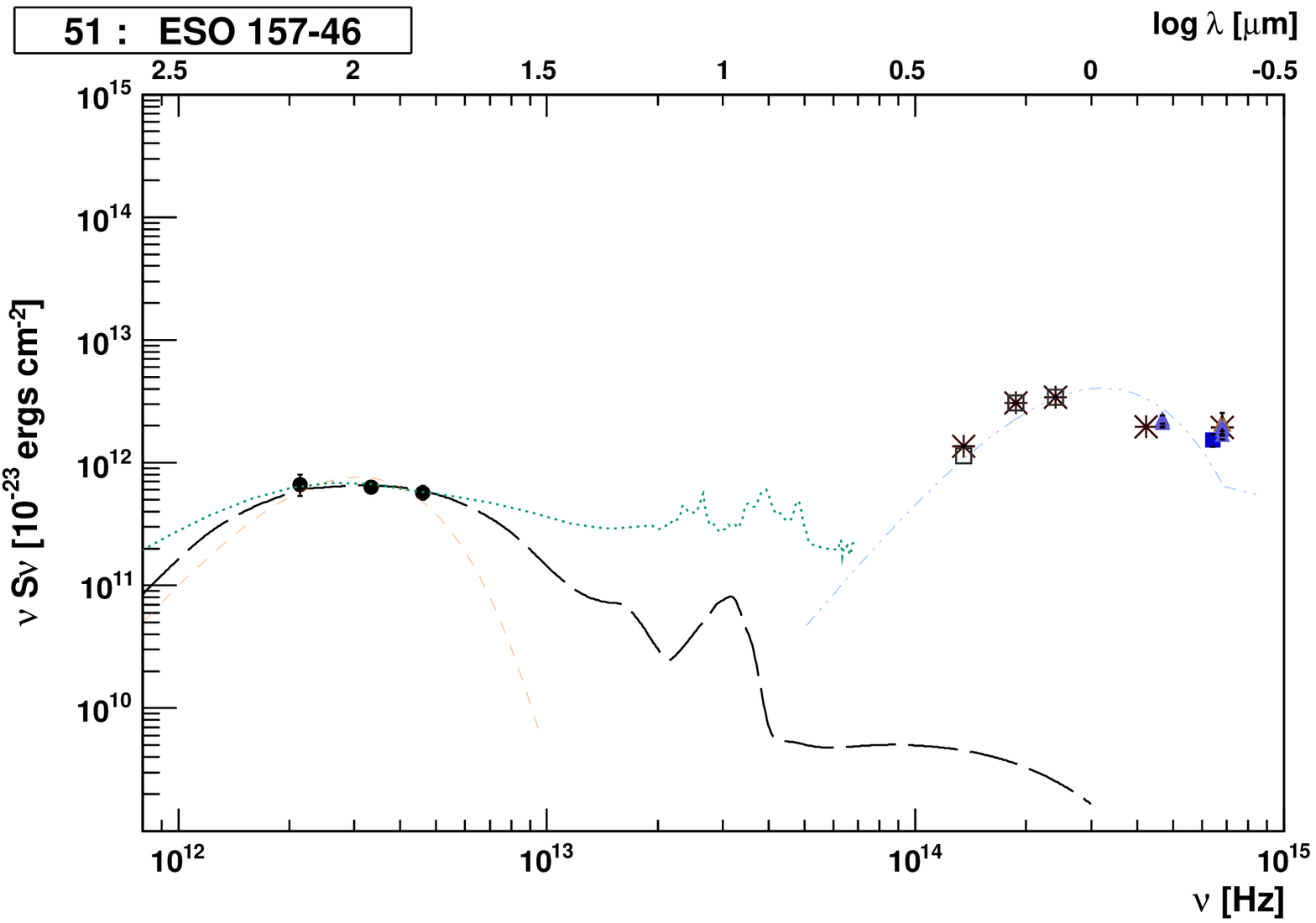}
\includegraphics[width=9cm]{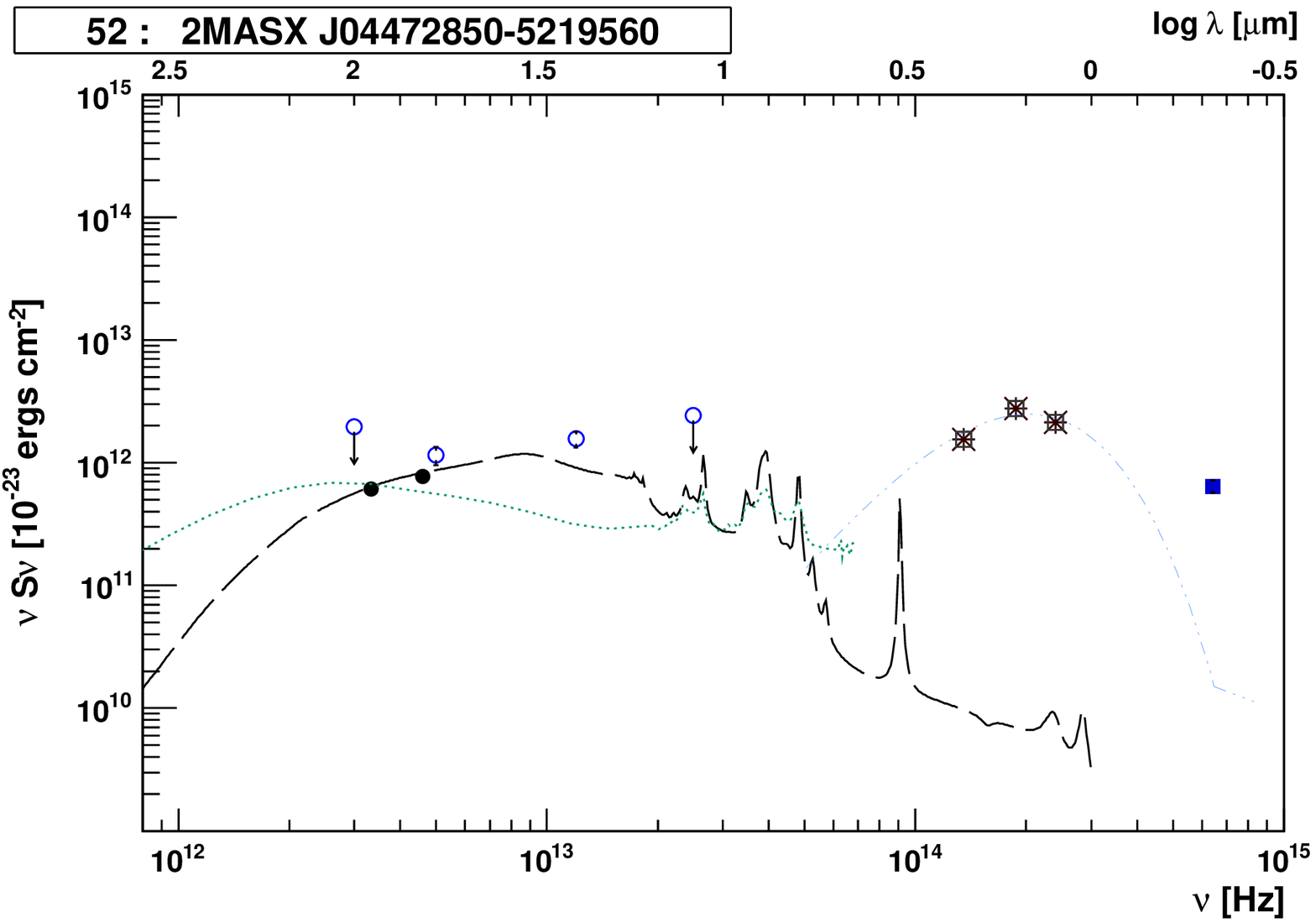}
\caption{Next 8 SEDs of ADF-S galaxies, with symbols as in Figure~\ref{sed}.
SED of a galaxy number 45, for which the redshift is known,
is fitted after shifting to the rest frame and presented
in the rest frame. The remaining objects are
shown in the observed frame.
}

  \label{sed5}%
 \end{figure*}

\begin{figure*}[t]
\centering
\includegraphics[width=9cm]{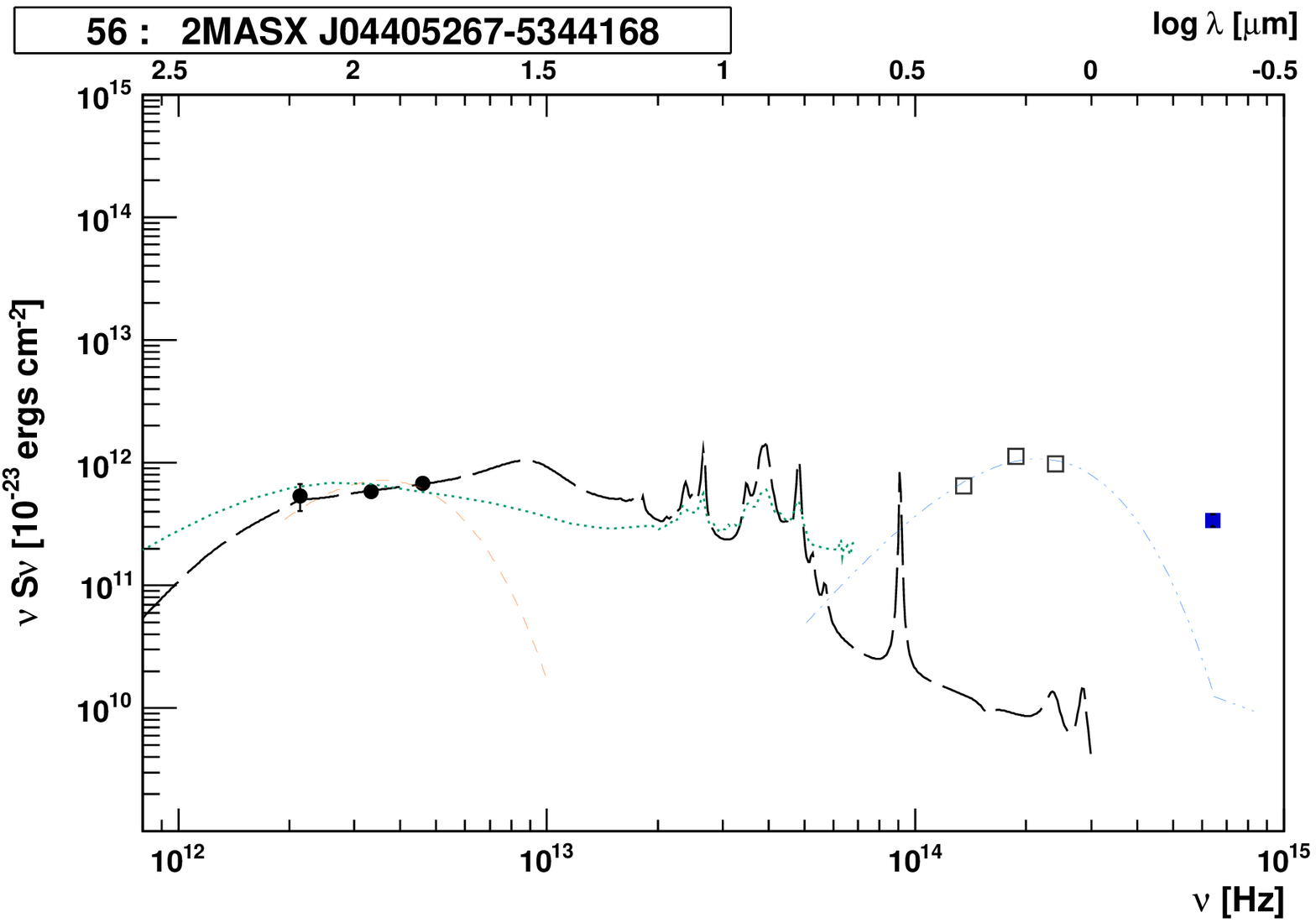}
\includegraphics[width=9cm]{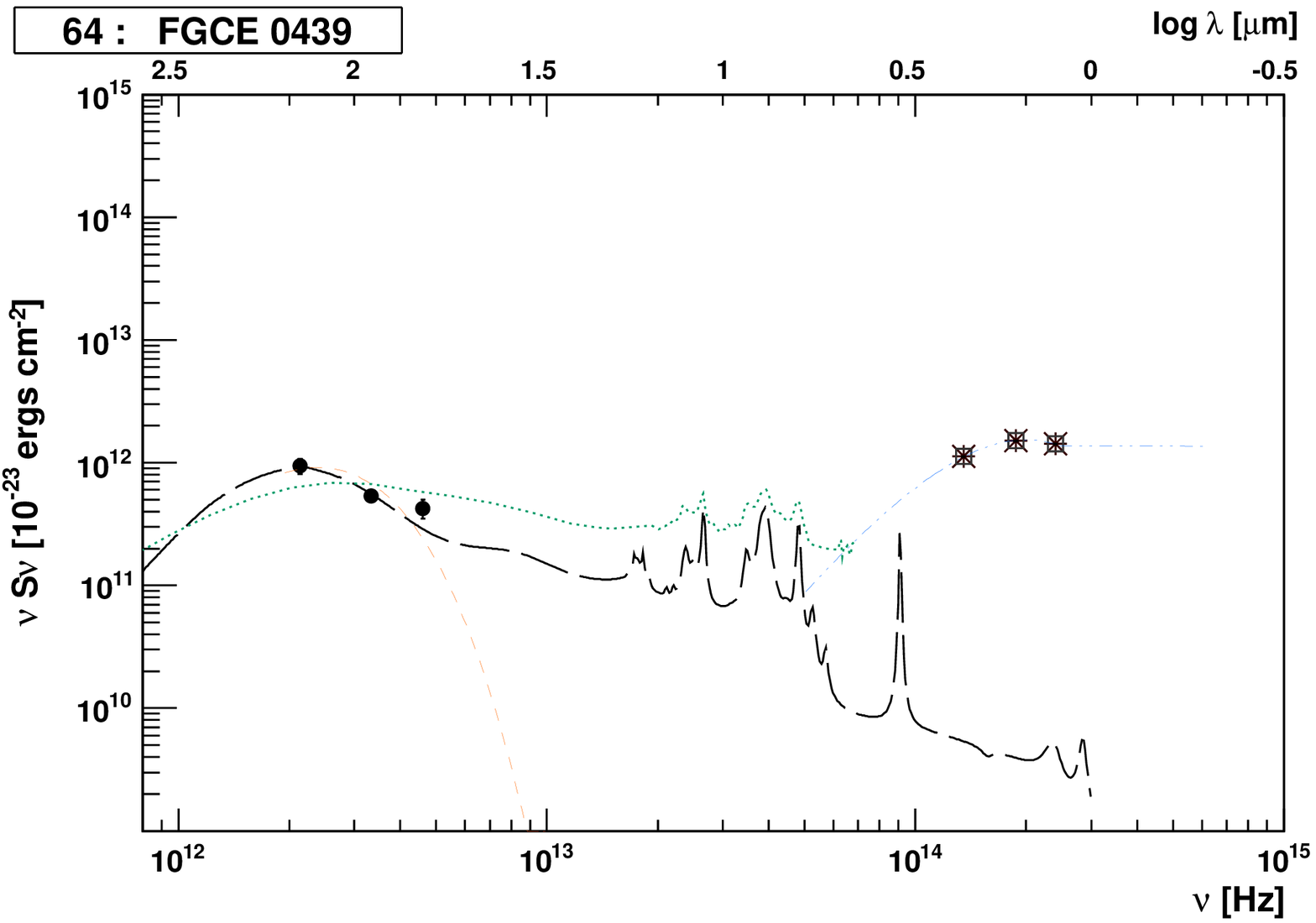}
\includegraphics[width=9cm]{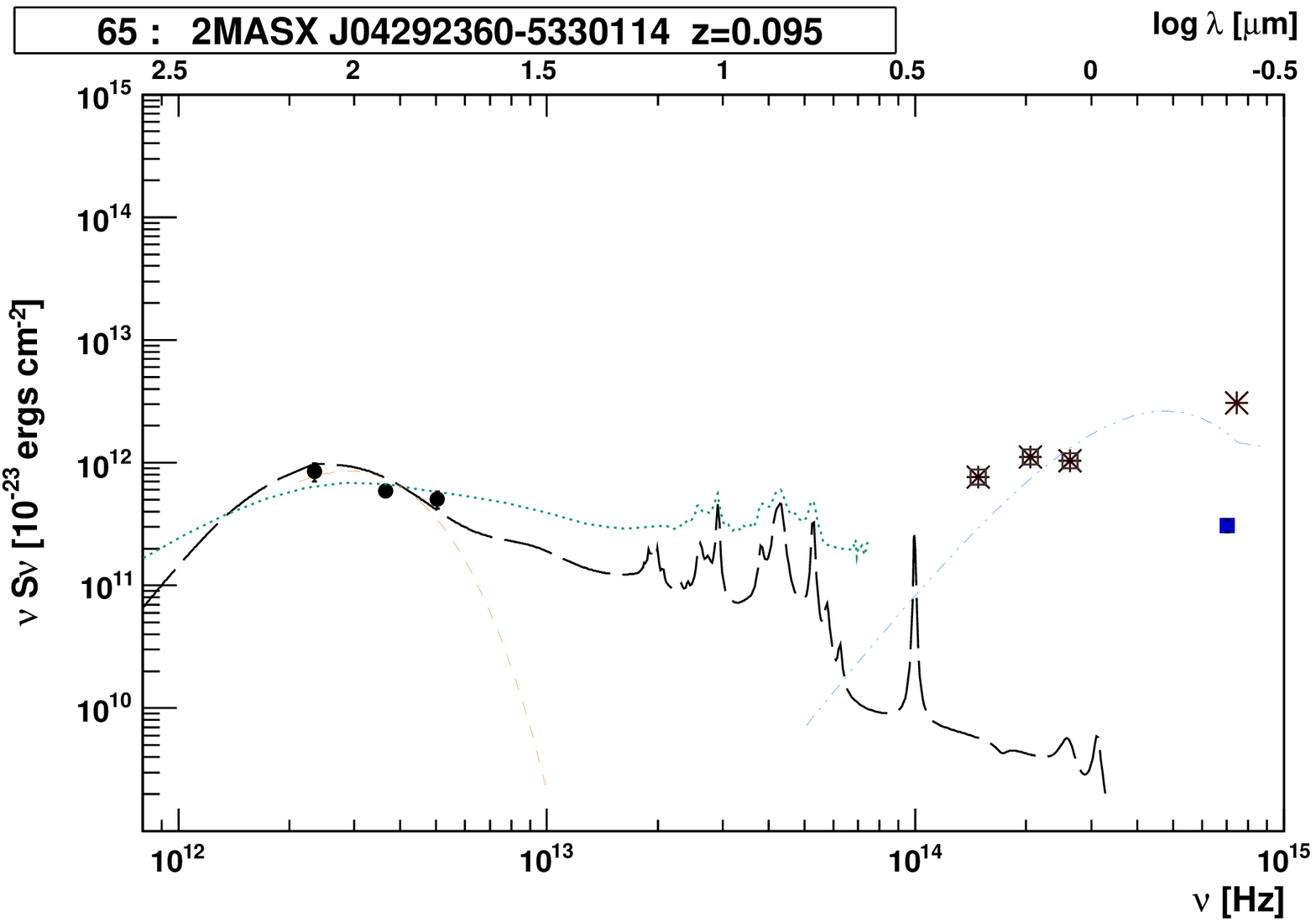}
\includegraphics[width=9cm]{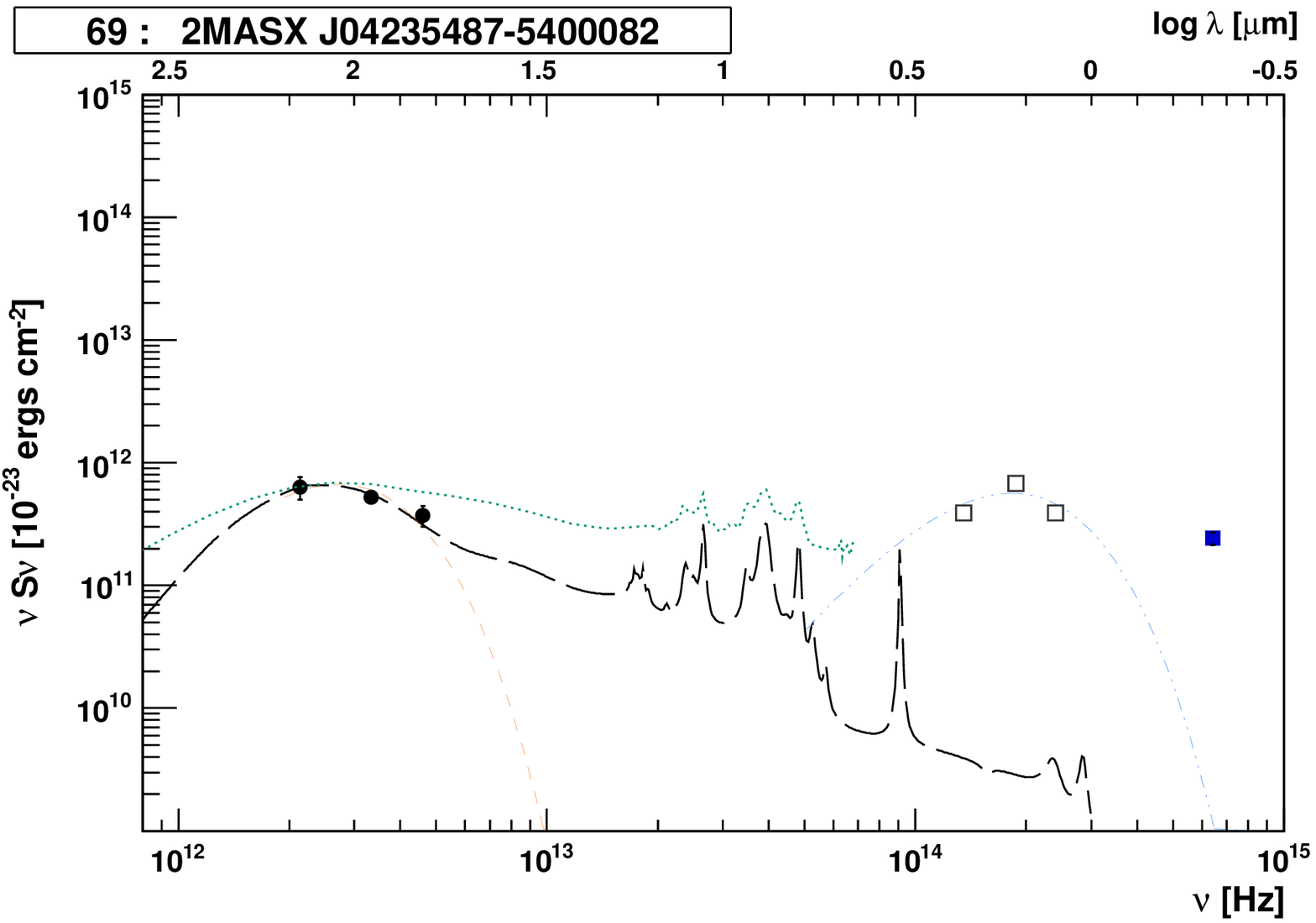}
\includegraphics[width=9cm]{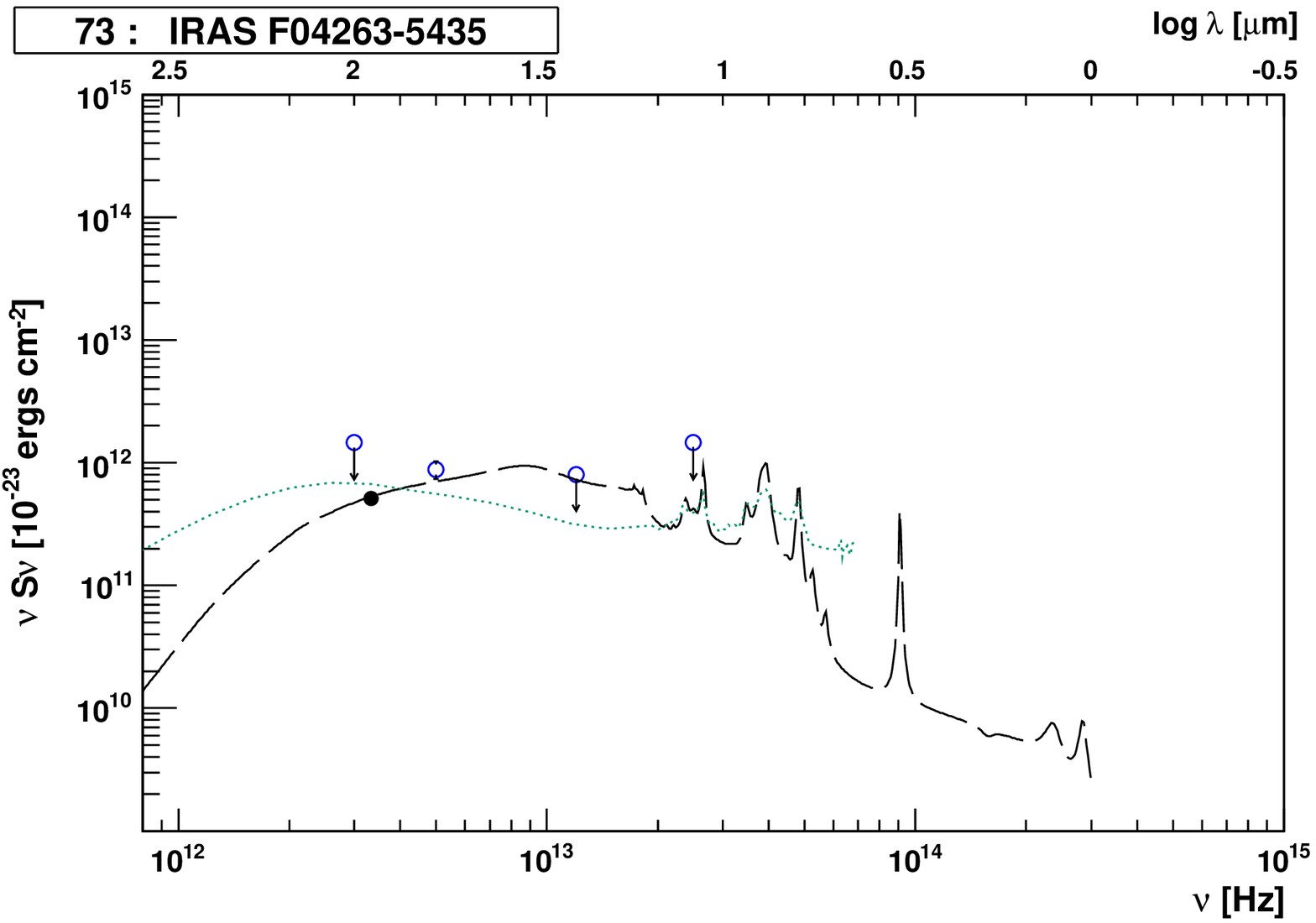}
\includegraphics[width=9cm]{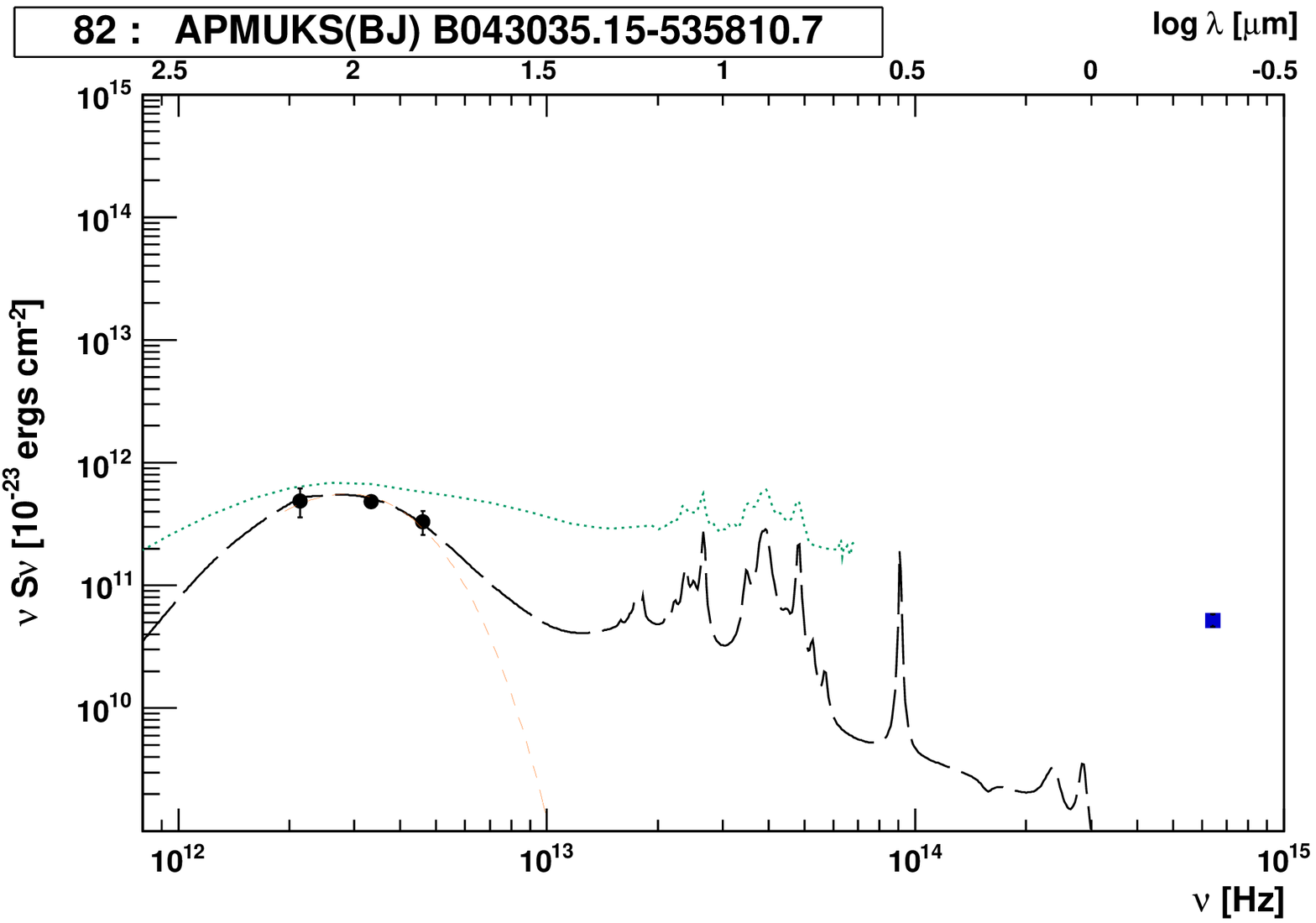}
\includegraphics[width=9cm]{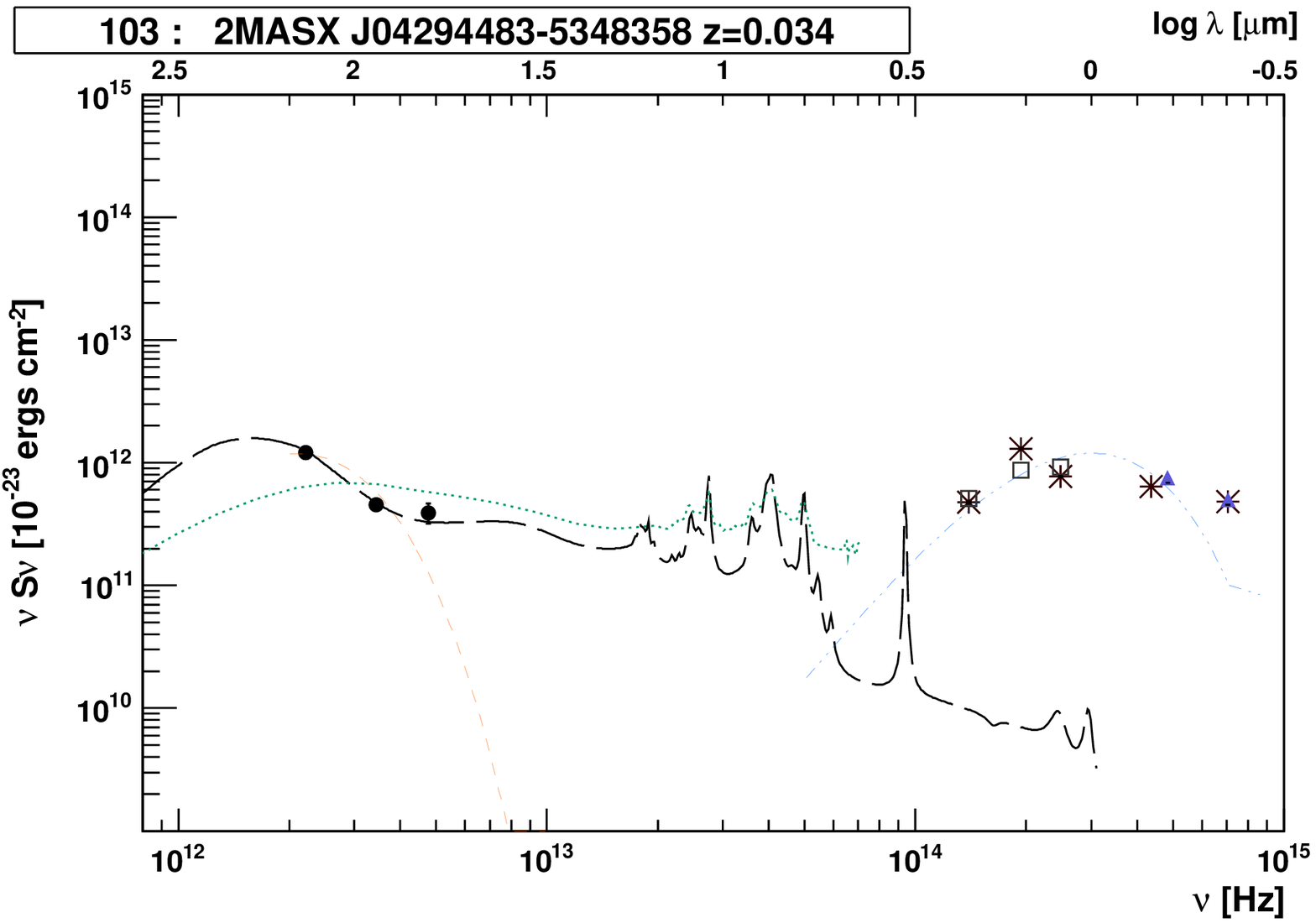}
\caption{Next 7 SEDs of ADF-S galaxies, with symbols as in Figure~\ref{sed}. 
SEDs of galaxies number 65 and 103, with known redshifts,
are fitted after shifting to the rest frame and presented
in the rest frame. The remaining objects are
shown in the observed frame.
}

  \label{sed6}%
 \end{figure*}

\citet{draine07} calculate the parameters 
directly from Spitzer data which was impossible in our case, but 
may be a part of a future project.
The resulting values of $q_{\rm PAH}$ are listed in column~7 of Table~\ref{MODELtable}. 
The histogram of $q_{\rm PAH}$, shown in Figure~\ref{pah}, 
is
very discrete, with peaks around three specific values: $q_{\rm PAH}=0.75, 2.37$ 
and 4.58, which correspond to the 
Large Magellanic Cloud models.  

Spectra presented by \cite{draine07} correspond to 
11 dust models with $q_{\rm PAH}$ ranging from 
$0.1\;\%$ to $4.58\;\%$. 
The lowest value of $q_{\rm PAH}$ corresponds 
to the Small Magellanic Cloud (SMC) model; $q_{\rm PAH} =$ $0.75\;\%$, $1.49\;\%$ and $2.37~\%$ 
correspond to the  
Large Magellanic Cloud (LMC) and the remaining 7 models 
are related to the Milky Way (MW)-like galaxies. 
81~\% from 47 modeled ADF-S sources 
were 
identified as LMC-like galaxies and 9~\% corresponded to
MW-type galaxies. No SMC-like galaxy was found. 
In all the sample,
the median $q_{\rm PAH} = 2.37~\%$.

All five lenticular galaxies, as well as both compact dwarfs
in our sample are best fitted 
by LMC-like models with values of $q_{\rm PAH} \leq 2.37$. 
Also the majority (87.5~\%) of spiral galaxies is best fitted by
LMC-like models, while for the rest MW-like models are applied.
Maximal $q_{\rm PAH}$ for spiral galaxies is 3.9~\%.  
Then, spiral galaxies tend to be bigger and richer in PAH, but
the difference with lenticulars is not statistically significant.
The
elliptical galaxy has the highest $q_{\rm PAH}=4.55$~\% 
among modeled galaxies and is classified as MW-like galaxy. The
discrepant conclusions about this particular galaxy probably reflect
its more complex structure and processes, related to its interactions
with another galaxy.

\section{Conclusions}\label{sec:conclusion}

\begin{enumerate}
  \item In this paper, we presented the catalog of counterparts of the ADF-S 
90~$\mu$m sources, detected at the $>6 \sigma$ level. 
We found counterparts for 545 among 1000 sources from the analyzed catalog. 
We discussed the properties of these sources and tried to derive conclusions 
about the average properties of the sample. 
 \item The point source ADF-S catalog itself appeared to be quite reliable 
in the position determination, with most of the counterparts located 
at the angular distance significantly lower than the nominal resolution 
of the FIS detector. A small number of counterparts detected at higher 
angular distances can be possibly an effect of contamination. In 
the position determination we observe a small (of an order of 4'', but 
systematic bias, which depends on the location in the ADF-S. 
This small systematics should be taken into account in the future works 
on these data.
  \item  We conclude that FIR sources detected in the ADF-S 
are mostly nearby galaxies. This conclusion is supported, first of all, by source 
number counts in the FIR, but also a large number of bright optical 
counterparts and the redshift distribution of counterparts. It should be 
noted, however, that the completeness of the identified sample, close 
to 100\% for sources brighter than 0.1 Jy, falls down rather rapidly 
for fainter sources, to 55~\% for the whole 10$\sigma$ catalog.  
  \item The population of identified galaxies appears surprisingly "normal", 
similar to one expected for local optically bright galaxies. The main differences 
are: \\ {\it a)}
a significantly lower percentage of elliptical galaxies, which can 
be explained by the fact they are less dusty than other galaxies, \\{\it b)} much 
higher percentage of peculiar galaxies, which can be attributed to strong star-forming 
activity  of these objects, related to intergalactic interactions, and hence stronger 
radiation of cold dust in them). \\ c) We detected also a fraction of lenticular galaxies 
(all from the cluster of galaxies Abell~S0463) which is practically the same as expected 
in the optically bright galaxy population. It suggests that these galaxies contain 
a significant amount of cold dust and supports more complex models of their formation 
than a simple secular evolution.  
  \item The estimated source confusion is on the level higher than 34~\% of the number of identified sources. 
It suggests that to remove the effect of the source confusion from the IR flux measurements and for a proper estimation of local environment of the FIR-bright galaxies dedicated 
observations of their closest neighborhood will be required.
  \item The SEDs of the identified sources display a variety of properties. 
In the first approach, the examined galaxies seem to be either extremely quiescent or 
very active in star formation. The lenticular galaxies usually belong to the 
actively star forming group. The analysis suggests that to explain well 
the FIR properties of otherwise "normal" galaxies from our sample new updated 
models should be developed.
\end{enumerate}

\begin{acknowledgements}
 We thank the anonymous referee for her/his very careful reading of the manuscript 
and giving useful comments which significantly improved the clarity of this paper.
This work is based on observations with AKARI, a JAXA project with the participation of ESA. 
This research has made use of the NASA/IPAC Extragalactic Database (NED) which is operated by 
the Jet Propulsion Laboratory, California Institute of Technology, under contract with 
the National Aeronautics and Space Administration, and the SIMBAD database, operated at CDS, Strasbourg, France
We thank Misato Fukagawa for sending the information about Vega-like star candidates.
This work has been supported in part by the Polish Astroparticle Physics Network.
AP was financed by the research grant of the Polish Ministry of Science
PBZ/MNiSW/07/2006/34A.
TTT has been supported by Program for Improvement of Research
Environment for Young Researchers from Special Coordination Funds for
Promoting Science and Technology, and the Grant-in-Aid for the Scientific
Research Fund (20740105) commissioned by the Ministry of Education, Culture,
Sports, Science and Technology (MEXT) of Japan.
TTT has been partially supported from the Grand-in-Aid for the Global
COE Program ``Quest for Fundamental Principles in the Universe: from
Particles to the Solar System and the Cosmos'' from the MEXT.
\end{acknowledgements}

\begin{table*}[ht]
\caption{Parameters of the fitted models of SEDs of 47 ADF-S galaxies 
with the best photometric data. The table contains ADF-S 
identification 
numbers (first column), indicative temperatures of dust and stellar 
component resulting from the best fitted modified blackbody models, 
the parameter $\alpha$ of the \citet{dale2001} model, three 
parameters and the name of the best fitted \citet{li2001} model.}
\label{MODELtable}
\centering
\begin{tabular}{ccccccccccc} 
\hline\hline
ID  &  \multicolumn{2}{c}{Modified blackbody} & \multicolumn{1}{c}{Dale \& Helou} & \multicolumn{4}{c}{Li \& Draine} \\
(ADF-S)  &  Dust temp.\ [K] &  Stellar temp.\ [K]  & $\alpha$ & U${\rm min}$ & U${\rm max}$ & ${\rm PAH}$ (\%) & name\\
 \hline
1 & 26.9 $\pm$ 0.2 & 2800 $\pm$ 40 & 2.4375 & 4.00 & 4.00 & 2.37 & LMC2 \\  
2 & 37.8$\pm$  0.3 & 3060 $\pm$ 30 & 2.1250 & 0.10 & 100 & 4.58 & MW3.1 \\  
3 & 22.9 $\pm$ 0.2 & 2950 $\pm$ 30 & 2.1250 & 0.50 & 0.50 & 2.37 & LMC2 \\  
4 & 36.3 $\pm$ 0.1 & 7000 $\pm$ 120& 1.0000 & 1.20 & 100 & 2.37 & LMC2    \\   
5 & 33.6 $\pm$ 0.3 & 3260 $\pm$ 50 & 1.3125 & 0.10 & 100 & 2.37 & LMC2 \\  
6 & 32.6 $\pm$ 0.3 & 2110 $\pm$ 70 & 1.3125 & 5.00 & 5.00 & 2.37 & LMC2 \\  
7 & 47.8 $\pm$ 0.4 & 5110 $\pm$ 50 & 0.0625 & 8.00 & 1e4 & 2.37 & LMC2 \\  
8 & 36.8 $\pm$ 0.4 & 2020 $\pm$ 70 & 0.0625 & 0.10 & 1e4 & 1.12 & MW3.1 \\  
9 & - & 3890 $\pm$ 50 & 1.1875 & 0.80 & 1e4 & 3.90 & MW3.1 \\  
10 & 27.8 $\pm$ 0.2 & 3230 $\pm$ 30 & 4.0000 & 2.00 & 2.00 & 2.37 & LMC2 \\  
11 & 28.9 $\pm$ 0.3 & 3030 $\pm$ 40 & 4.0000 & 2.00 & 2.00 & 2.37 & LMC2 \\  
12 & 36.0 $\pm$ 0.6 & 2070 $\pm$ 120 & 4.0000 & 5.00 & 1e4 & 2.37 & LMC2 \\  
13 & 33.2 $\pm$ 0.4 & - & 4.0000 & 0.10 & 100 & 0.75 & LMC2 \\  
14 & 32.4 $\pm$ 0.4 & - & 0.0625 & 8.00 & 100 & 2.37 & LMC2 \\  
15 & 32.1 $\pm$ 0.2 & 1980 $\pm$ 10 & 0.0625 & 0.10 & 100 & 0.75 & LMC2 \\  
16 & 28.2 $\pm$ 0.7 & - & 0.0625 & 0.10 & 100 & 4.58 & MW3.1 \\  
17 & 33.2 $\pm$ 0.6 & 5890 $\pm$ 60 & 0.0625 & 0.10 & 1e4 & 0.75 & LMC2 \\  
18 & 28.8 $\pm$ 0.2 & 2260 $\pm$ 70 & 0.0625 & 4.00 & 4.00 & 2.37 & LMC2 \\  
19 & 33.8 $\pm$ 0.1 & 2050 $\pm$ 10 & 0.0625 & 15.0 & 15.0 & 0.75 & LMC2 \\  
20 & 33.0 $\pm$ 0.5 & 1980 $\pm$ 70 & 0.0625 & 0.10 & 1e4 & 0.75 & LMC2 \\  
21 & 29.4 $\pm$ 0.2 & 2210 $\pm$ 10 & 0.0625 & 4.00 & 4.00 & 2.37 & LMC2 \\  
22 & 38.0 $\pm$ 0.1 & 2060 $\pm$ 40 & 4.0000 & 0.10 & 100 & 0.75 & LMC2 \\  
23 & 28.1 $\pm$ 0.3 & 2340 $\pm$ 70 & 4.0000 & 0.10 & 100 & 2.37 & LMC2 \\  
24 & 33.4 $\pm$ 0.6 & 2140 $\pm$ 10 & 4.0000 & 0.10 & 1e6 & 0.75 & LMC2 \\  
26 & 43.5 $\pm$ 0.1 & - & 0.0625 & 4.00 & 1e4 & 0.75 & LMC2 \\  
27 & 35.7 $\pm$ 0.3 & - & 0.0625 & 0.80 & 100 & 2.37 & LMC2 \\  
29 & 27.5 $\pm$ 0.8 & 2950 $\pm$ 20 & 0.0625 & 0.10 & 100 & 0.75 & LMC2 \\  
30 & 32.7 $\pm$ 0.7 & 1900 $\pm$ 10 & 0.0625 & 0.10 & 1e6 & 4.58 & MW3.1 \\  
31 & 32.8 $\pm$ 0.3 & 1970 $\pm$ 120 & 0.0625 & 8.00 & 8.00 & 2.37 & LMC2 \\  
39 & 47.4 $\pm$ 0.4 & 3500 $\pm$ 60 & 0.0625 & 0.20 & 1e6 & 0.47 & MW3.1 \\  
41 & 32.5 $\pm$ 0.1 & 1950 $\pm$ 10 & 0.0625 & 12.0 & 12.0 & 0.75 & LMC2 \\  
42 & - & 2210 $\pm$ 30 & 0.0625 & 0.10 & 1e5 & 0.75 & LMC2 \\  
43 & 33.3 $\pm$ 0.7 & 1970 $\pm$ 10 & 0.0625 & 0.10 & 1e6 & 0.75 & LMC2 \\  
45 & 28.5 $\pm$ 0.1 & 4160 $\pm$ 80 & 3.8750 & 0.10 & 100 & 0.75 & LMC2 \\  
46 & - & 1650 $\pm$ 20 & 3.8750 & 0.10 & 1e6 & 4.58 & MW3.1 \\  
48 & 24.3 $\pm$ 0.1 & 2010 $\pm$ 10 & 3.8750 & 1.20 & 1.20 & 2.37 & LMC2 \\  
49 & 26.6 $\pm$ 0.1 & - & 3.8750 & 0.10 & 100 & 0.75 & LMC2 \\  
50 & 33.6 $\pm$ 0.1 & 3380 $\pm$ 60 & 3.8750 & 0.10 & 1e6 & 1.49 & LMC2 \\  
51 & 29.7 $\pm$ 0.2 & 3120 $\pm$ 30 & 2.2500 & 0.10 & 100 & 0.75 & LMC2 \\  
52 & - & 2000 $\pm$ 30 & 2.2500 & 2.00 & 1e6 & 2.37 & LMC2 \\  
56 & 34.8 $\pm$ 0.1 & - & 2.2500 & 0.10 & 1e6 & 4.58 & MW3.1 \\  
64 & 23.1 $\pm$ 0.1 & 1940 $\pm$ 10 & 2.2500 & 0.70 & 0.70 & 2.37 & LMC2 \\  
65 & 28.0 $\pm$ 0.1 & 4600 $\pm$ 110 & 4.0000 & 1.50 & 1.50 & 2.37 & LMC2 \\  
69 & 26.4 $\pm$ 0.1 & 1770 $\pm$ 80 & 4.0000 & 2.00 & 2.00 & 2.37 & LMC2 \\  

73 & - & - & 4.0000 & 1.50 & 1e6 & 2.37 & LMC2 \\  
82 & 27.8 $\pm$  0.1 & - & 4.0000 & 2.00 & 2.00 & 3.19 & MW3.1 \\  
103 & 20.2 $\pm$ 0.1 & 2910 $\pm$ 50 & 4.0000 & 0.10 & 0.10 & 2.37 & LMC2 \\

\hline
\end{tabular}
\end{table*}

\begin{table*}[ht]
\caption{Morphological and environmental properties of 
47 ADF-S galaxies used for fitting of SED models. 
After the ADF-S identification number, in the second column we give 
the object's classification: "Galaxy" means that it has been identified 
as a galaxy, IRS means that it has been previously identified as the 
infra-red source, RS means that it has been identified as the radio source, 
"other" means that it has been observed in some other (UV or X-ray) 
wavelengths. This column contains also the available information about 
the object's environment - whether it is an object from the cluster 
of galaxies (in this case it always applies to Abell~S0463) or an 
interacting group of galaxies. The third column gives the object's 
indicative morphological type, as provided by the NED. The galaxies
whose morphological types were determined by \citet{dressler80a} are 
indicated. The last column gives the number of counterparts found 
for each object in our 40'' search range, which may be treated 
as a very crude indicator of a local environment of an investigated object. }
\label{morphological_table}
\centering
\begin{tabular}{cllc} 
\hline\hline
ID  &  Type of source & Morphological type & Number of  \\
(ADFS)  &&  & counterparts\\
\hline
\renewcommand{\footnoterule}{}
1 & Galaxy/IRS/RS  & Sb  & 1 \\  
2 & Galaxy in Pair of Interacting Galaxies/IRS/RS  & E  & 1 \\  
3 & Galaxy/IRS/RS/other  & Sa  & 1 \\  
4 & Stardust Galaxy IR emission line up with  & S0, SA0- pec: HII  BCDG  & 1\\  
  & Super Star Cluster/IRS/RS/other  & &\\
5 & Galaxy/IRS/RS  & S  & 2 \\
\hline
\end{tabular}
\begin{list}{}{}
\item[$^{\mathrm{a}}$] \cite{dressler80a}
\end{list}
\end{table*}

\begin{table*}[ht]
\caption{Nearest counterparts of 10 brightest ADF-S sources (full version of the table in the online material). In this table we present: the ADF-S identification number, the name of the nearest identified counterpart, right ascension and declination of both ADF-S source and the counterpart, the angular distance between them and the redshift of the counterpart, when available.}
\label{exADFScount}
\centering
\begin{tabular} {clccccccl}
\hline\hline
{ID}  &  {Name} & {$\alpha_{\rm ADF-S}$} & {$\delta_{\rm ADF-S}$} & {$\alpha_{\rm counterpart}$} & {$\delta_{\rm counterpart}$} & {Dist [arc sec]} &  {redshift}  \\
{(ADF-S)}  &  {} & {(J2000)} & {(J2000)} & {(J2000)} & {(J2000)} & {} & {} \\
\hline
\hline
1 & ESO~157-49 & 69.90448 & -53.01348 & 69.90300 & -53.01294 & 3.74 & 0.00578$^{\mathrm{1}}$  \\
2 & ESO~157-IGA040 & 67.90437 & -53.28733 & 67.90467 & -53.28825 & 3.37 & 0.05881$^{\mathrm{2}}$  \\
3 & NGC~1617 & 67.91312 & -54.60001 & 67.91417 & -54.60167 & 6.35 & 0.00355$^{\mathrm{3}}$\\
4 & NGC~1705 & 73.55990 & -53.36119 & 73.55708 & -53.36111 & 6.06 & 0.00210$^{\mathrm{1}}$ \\
5 & ESO~157-43 & 68.81647 & -54.31440 & 68.81460 & -54.31580 & 6.51 & 0.04104$^{\mathrm{1}}$\\
6 & 2MASX~J04325340-5312021 & 68.22050 & -53.19939 & 68.22210 & -53.20190 & 9.81 & \\
7 & 2MASX~J04335902-5410074  & 68.49623 & -54.16840 & 68.49579 & -54.16861 & 2.52 & 0.04041$^{\mathrm{1}}$ \\
8 & 2MASX~J04293931-5352464 & 67.41413 & -53.87810 & 67.41540 & -53.88000 & 7.36 & 0.03910$^{\mathrm{1}}$ \\
9 & 2MASX~J04334493-5346467 & 68.43882 & -53.77891 & 68.43960 & -53.77940 & 2.52 & 0.03944$^{\mathrm{4}}$\\
10 & IC~2085 & 67.85253 & -54.41524 & 67.84979 & -54.41708 & 8.77 & 0.00328$^{\mathrm{5}}$ \\
\hline \hline
\end{tabular}
\begin{list}{}{}
\item{\textit{1} - \cite{lawrence01}} 
\item{\textit{2} - \cite{karl95}} 
\item{\textit{3} - \cite{deVac91}} 
\item{\textit{4} - \cite{dressler88}} 
\item{\textit{5} - \cite{mathewson92}} 
\end{list}
\end{table*}

\begin{table*}[ht]
\caption{Flux densities of 10 brightest ADF-S objects with identified counterparts, measured by AKARI (the full table is available in the online material).}
\label{exADFSmeasurements}
\centering
\begin{tabular} {cccccc}
\hline\hline
{ID}  &\multicolumn{4}{c}{Flux densities [Jy]} \\
{(ADF-S)} & {N60 (65 $\mu$m)} & {WIDE-S (90 $\mu$m)} &{WIDE-L (140 $\mu$m)} & {N160 (160 $\mu$m)} \\
\hline
1 &  1.65E+00 $\pm$ 5.83E-02 & 2.72E+00 $\pm$ 9.14E-02 & 5.04E+00 $\pm$ 1.53E-01 & 4.87E+00 $\pm$ 3.42E-01    \\  
2 &  1.15E+00 $\pm$ 4.18E-02 & 1.45E+00 $\pm$ 4.79E-02 & 2.59E+00 $\pm$ 9.43E-02 & 1.26E+00 $\pm$ 2.12E-01    \\  
3 &  - & 1.24E+00 $\pm$ 4.12E-02 & 3.93E+00 $\pm$ 1.25E-01 & 5.09E+00 $\pm$ 3.52E-01    \\  
4 &  9.12E-01 $\pm$ 3.44E-02 & 1.16E+00 $\pm$ 3.84E-02 & 1.05E+00 $\pm$ 6.74E-02 & 6.08E-01 $\pm$ 2.00E+00    \\  
5 &  6.22E-01 $\pm$ 2.60E-02 & 8.89E-01 $\pm$ 2.96E-02 & - & -    \\  
6 &  4.46E-01 $\pm$ 2.16E-02 & 8.66E-01 $\pm$ 2.89E-02 & 1.25E+00 $\pm$ 7.01E-02 & -    \\  
7 &  - & 7.58E-01 $\pm$ 2.54E-02 & 4.56E-01 $\pm$ 6.21E-02 & -    \\  
8 &  5.18E-01 $\pm$ 2.33E-02 & 6.37E-01 $\pm$ 2.15E-02 & 8.65E-01 $\pm$ 6.54E-02 & -    \\  
9 &  4.66E-01 $\pm$ 2.20E-02 & 5.48E-01 $\pm$ 1.87E-02 & 4.77E-01 $\pm$ 6.22E-02 & -    \\ 
\hline \hline
\end{tabular}
\end{table*}

\begin{table*}[ht]
\caption{Photometric data for 10 brightest ADF-S sources with identified counterparts, collected from public databases (full version of the corresponding tables are available in the online material).}
\label{exmeasurements}
\centering
\begin{tabular} {cccccc}
\hline\hline
{ID}  &  {0.44 $\mu$m} & {0.47 $\mu$m$^C$} & {0.55 $\mu$m}  & {0.64 $\mu$m}  & {0.71 $\mu$m} \\
{(ADF-S)}  & {} & {} & {} & {} & {} \\
\hline
1&7.05E-03$\pm$6.09E-04$^A$ & 5.16E-03$\pm$6.03E-04& - &1.39E-02$\pm$1.20E-03$^A$ &1.41E-02$^D$ \\
& 7.35E-03$\pm$6.35E-03$^A$ & &  &1.44E-02$\pm$1.25E-03$^A$&  \\
 & 7.59E-03$\pm$6.56E-04$^A$ & & &1.47E-02$\pm$1.27E-03$^D$  & \\
& 7.90E-03$\pm$1.69E-03$^B$& & & & \\
& 7.38E-03$^D$& & & & \\
2&1.38E-03$\pm$1.19E-14$^A$ & 1.50E-03$\pm$1.23E-04& - &2.46E-03$\pm$2.13E-04$^A$ &2.63E-03$^D$ \\
& 1.46E-03$\pm$1.27E-04$^A$ & &  &2.62E-03$\pm$2.26E-04$^A$&  \\
& 1.53E-03$\pm$1.32E-04$^A$ & & &2.76E-03$\pm$2.38E-04$^D$  & \\
& 1.49E-03$^D$  & & & & \\
3& 1.19E-01$^D$  & - &2.43E-01$\pm$2.85E-02$^B$ & - &3.54E-01$^D$ \\
&  1.20E-01$\pm$1.40E-02$^B$& &2.98E-01$^D$  & - &  \\
4&2.89E-02$\pm$2.50E-03$^A$ & 2.47E-02$\pm$2.29E-04&4.03E-02$\pm$3.49E-03$^B$ &4.04E-02$\pm$3.49E-03$^A$ &4.33E-02$^D$ \\
 & 3.03E-02$\pm$2.62E-03$^A$ & &4.55E-02$^D$  &4.31E-02$\pm$3.73E-03$^A$& 3.50E-02$\pm$9.81E-04$^F$  \\
 & 3.16E-02$\pm$2.74E-03$^A$ & & &4.55E-02$\pm$3.93E-03$^D$  & 8.63E-03$\pm$2.42E-04$^F$\\
 & 3.08E-02$^D$  & & & 4.10E-02$\pm$2.33E-03$^E$&2.73E-02$\pm$7.65E-04$^F$ \\
 & 3.20E-02$\pm$6.83E-03$^B$ & & & & \\
 & 3.32E-02$\pm$4.23E-03$^B$ & & & & \\
 & 3.69E-02$^B$ & & & & \\
 & 2.47E-02$\pm$2.29E-04$^E$ & & & & \\
5&3.58E-03$\pm$3.09E-04$^A$ & 3.14E-03$\pm$3.67E-04& - &3.12E-03$\pm$5.29E-04$^A$ &6.20E-03$^D$ \\ & 3.91E-03$\pm$3.09E-04$^A$ & & &6.38E-03$\pm$5.51E-04$^A$&  \\  & 4.04E-03$\pm$3.49E-04$^A$ & & &6.51E-03$\pm$5.63E-04$^D$  & \\ & 3.91E-03$^D$  & & & & \\
6& - & 5.26E-04$\pm$6.14E-05& - & - & - \\
7& 4.13E-03$^D$  & -&- &- & -\\
8& 4.13E-03$^D$  &- &- & -& -\\
9& 4.13E-03$^D$  & -& -& -& -\\
10&6.95E-03$\pm$6.01E-04$^A$ & 5.45E-03$\pm$6.37E-04&2.06E-02$\pm$3.54E-03$^B$ &1.70E-02$\pm$1.47E-03$^A$ &1.93E-02$^D$ \\
\hline \hline
\end{tabular}
\begin{list}{}{}
\item{\textit{A} - \cite{lauberts}}\\
\item{\textit{B} - \cite{deVac91}} \\
\item{\textit{C} - \cite{maddox}}\\
\item{\textit{D} - The SIMBAD astronomical database, \cite{wenger}}\\
\item{\textit{E} - \cite{gildepaz}} \\
\end{list}
\end{table*}
\clearpage

\onltab{1}{
\begin{longtable}{clccccccl} 
\caption{\label{ADFScount} The list of 330 nearest counterparts of the ADF-S sources from the $10 \sigma$ catalog ($S_{90\mu m} >$ 0.0482 Jy). In this table we present: the ADF-S identification number, the name of the nearest identified counterpart, right ascension and declination of both ADF-S source and the identified counterpart, the angular distance between them and the redshift of the counterpart, when available.}\\
\hline\hline
{ID}  &  {Name} & {$\alpha_{\rm ADF-S}$} & {$\delta_{\rm ADF-S}$} & {$\alpha_{\rm counterpart}$} & {$\delta_{\rm counterpart}$} & {Dist [arc sec]} &  {redshift}  \\
{(ADF-S)}  &  {} & {(J2000)} & {(J2000)} & {(J2000)} & {(J2000)} & {} & {} \\ 
\hline
\endfirsthead
\caption{continued.}\\
\hline\hline
{ID}  &  {Name} & {$\alpha_{\rm ADF-S}$} & {$\delta_{\rm ADF-S}$} & {$\alpha_{\rm counterpart}$} & {$\delta_{\rm counterpart}$} & {Dist [arc sec]} &  {redshift}  \\
{(ADF-S)}  &  {} & {(J2000)} & {(J2000)} & {(J2000)} & {(J2000)} & {} & {}\\
\hline
\endhead
\hline
\endfoot
\hline
\renewcommand{\footnoterule}{}
1 & ESO~157-49 & 69.90448 & -53.01348 & 69.90300 & -53.01294 & 3.74 & 0.00578$^{\mathrm{1}}$  \\
2 & ESO~157-IGA040 & 67.90437 & -53.28733 & 67.90467 & -53.28825 & 3.37 & 0.05881$^{\mathrm{2}}$  \\
3 & NGC~1617 & 67.91312 & -54.60001 & 67.91417 & -54.60167 & 6.35 & 0.00355$^{\mathrm{3}}$\\
4 & NGC~1705 & 73.55990 & -53.36119 & 73.55708 & -53.36111 & 6.06 & 0.00210$^{\mathrm{1}}$ \\
5 & ESO~157-43 & 68.81647 & -54.31440 & 68.81460 & -54.31580 & 6.51 & 0.04104$^{\mathrm{1}}$\\
6 & 2MASX~J04325340-5312021 & 68.22050 & -53.19939 & 68.22210 & -53.20190 & 9.81 & \\
7 & 2MASX~J04335902-5410074  & 68.49623 & -54.16840 & 68.49579 & -54.16861 & 2.52 & 0.04041$^{\mathrm{1}}$ \\
8 & 2MASX~J04293931-5352464 & 67.41413 & -53.87810 & 67.41540 & -53.88000 & 7.36 & 0.03910$^{\mathrm{1}}$ \\
\hline
\end{longtable}
\begin{list}{}{}
\item{\textit{1} - \cite{lawrence01}} 
\item{\textit{2} - \cite{karl95}} 
\item{\textit{3} - \cite{deVac91}} 
\item{\textit{4} - \cite{dressler88}} 
\item{\textit{5} - \cite{mathewson92}} 
\item{\textit{6} - \cite{mathewson}}  
\item{\textit{7} - \cite{dinella96}}  
\item{\textit{8} - \cite{desouza97}}  
\item{\textit{9} - \cite{dacosta91}}  
\item{\textit{10} - \cite{paturel02}}  
\item{\textit{11} - \cite{fairall84}}  
\item{\textit{12} - \cite{ellis84}}  
\item{\textit{13} - \cite{loveday96}}  
\item{\textit{14} - \cite{wisotzki00}} 
\end{list}
}

\clearpage

\onltab{2}{
\begin{longtable}{clccccccl} 

\caption{\label{ADFScount2} The list of 218 nearest counterparts of the ADF-S sources from the fainter part of the $6 \sigma$ catalog (0.0.82 Jy $< S_{90 \mu m} <$ 0.0301 Jy). All the columns are ordered as in Table~\ref{ADFScount}.}\\
\hline\hline
{ID}  &  {Name} & {$\alpha_{\rm ADF-S}$} & {$\delta_{\rm ADF-S}$} & {$\alpha_{\rm counterpart}$} & {$\delta_{\rm counterpart}$} & {Dist [arc sec]} &  {redshift}  \\
{(ADF-S)}  &  {} & {(J2000)} & {(J2000)} & {(J2000)} & {(J2000)} & {} & {} \\ 
\hline
\endfirsthead
\caption{continued.}\\
\hline\hline
{ID}  &  {Name} & {$\alpha_{\rm ADF-S}$} & {$\delta_{\rm ADF-S}$} & {$\alpha_{\rm counterpart}$} & {$\delta_{\rm counterpart}$} & {Dist [arc sec]} &  {redshift}  \\
{(ADF-S)}  &  {} & {(J2000)} & {(J2000)} & {(J2000)} & {(J2000)} & {} & {}\\
\hline
\endhead
\hline
\endfoot

\hline
\renewcommand{\footnoterule}{}
503 &APMUKS(BJ) B042654.63-541839.4  & 67.01398 & -54.19913 & 67.01242 & -54.20153 & 9.24 &  \\
504 &APMUKS(BJ) B042901.44-541940.2  & 67.53602 & -54.22007 & 67.53917 & -54.22078 & 7.1 &  \\
507 &APMUKS(BJ) B045318.69-512458.0  & 73.63448 & -51.33819 & 73.63429 & -51.33689 & 4.7 &  \\
513 &2MASX J04483850-5326377& 72.16142 & -53.44373 & 72.16033 & -53.44386 & 2.38 &  \\
514 &APMUKS(BJ) B042858.99-542959.8  & 67.53412 & -54.40284 & 67.52671 & -54.39283 & 39.23 &  \\
519 &2MASX J04554545-5237330& 73.93758 & -52.62664 & 73.93946 & -52.62581 & 5.09 &  \\
523 &APMUKS(BJ) B043801.14-541109.3  & 69.79368 & -54.08956 & 69.78417 & -54.08903 & 20.18 &  \\
524 &APMUKS(BJ) B043420.17-533205.3  & 68.87540 & -53.43288 & 68.87413 & -53.43375 & 4.16 &  \\
\hline
\end{longtable}

\begin{list}{}{}
\item{\textit{1} - \textit{14} As in Table~\ref{ADFScount}} 
\item{\textit{15} - \cite{thomas}}
\end{list}
}

\clearpage

\onltab{3}{
\begin{longtable}{cccccc} 

\caption{\label{ADFSmeasurements} Flux densities of 330 ADF-S sources from the 10$\sigma$ catalog with identified counterparts, measured by AKARI.}\\
\hline\hline
{ID}  &\multicolumn{4}{c}{Flux densities [Jy]} \\
{(ADF-S)} & {N60 (65 $\mu$m)} & {WIDE-S (90 $\mu$m)} &{WIDE-L (140 $\mu$m)} & {N160 (160 $\mu$m)} \\ 
\hline
\endfirsthead
\caption{continued.}\\
\hline\hline
{ID}  &\multicolumn{4}{c}{Flux densities [Jy]} \\
{(ADFS)}& {N60 (65 $\mu$m)} & {WIDE-S (90 $\mu$m)} &{WIDE-L (140 $\mu$m)} & {N160 (160 $\mu$m)} \\ 
\hline
\endhead
\hline
\endfoot

\hline
\renewcommand{\footnoterule}{}
1 &  1.65E+00 $\pm$ 5.83E-02 & 2.72E+00 $\pm$ 9.14E-02 & 5.04E+00 $\pm$ 1.53E-01 & 4.87E+00 $\pm$ 3.42E-01    \\  
2 &  1.15E+00 $\pm$ 4.18E-02 & 1.45E+00 $\pm$ 4.79E-02 & 2.59E+00 $\pm$ 9.43E-02 & 1.26E+00 $\pm$ 2.12E-01    \\  
3 &  - & 1.24E+00 $\pm$ 4.12E-02 & 3.93E+00 $\pm$ 1.25E-01 & 5.09E+00 $\pm$ 3.52E-01    \\  
4 &  9.12E-01 $\pm$ 3.44E-02 & 1.16E+00 $\pm$ 3.84E-02 & 1.05E+00 $\pm$ 6.74E-02 & 6.08E-01 $\pm$ 2.00E+00    \\  
5 &  6.22E-01 $\pm$ 2.60E-02 & 8.89E-01 $\pm$ 2.96E-02 & - & -    \\  
6 &  4.46E-01 $\pm$ 2.16E-02 & 8.66E-01 $\pm$ 2.89E-02 & 1.25E+00 $\pm$ 7.01E-02 & -    \\  
7 &  - & 7.58E-01 $\pm$ 2.54E-02 & 4.56E-01 $\pm$ 6.21E-02 & -    \\  
8 &  5.18E-01 $\pm$ 2.33E-02 & 6.37E-01 $\pm$ 2.15E-02 & 8.65E-01 $\pm$ 6.54E-02 & -    \\
\hline
\end{longtable}
}

\clearpage

\onltab{4}{
\begin{longtable}{cccccc} 

\caption{\label{ADFSmeasurements2} Flux densities of 218 ADF-S sources from the fainter part of the 6$\sigma$ catalog, measured by AKARI.. }\\
\hline\hline
{ID}  &\multicolumn{4}{c}{Flux densities [Jy]} \\
{(ADF-S)} & {N60 (65 $\mu$m)} & {WIDE-S (90 $\mu$m)} &{WIDE-L (140 $\mu$m)} & {N160 (160 $\mu$m)} \\ 
\hline
\endfirsthead
\caption{continued.}\\
\hline\hline
{ID}  &\multicolumn{4}{c}{Flux densities [Jy]} \\
{(ADFS)}& {N60 (65 $\mu$m)} & {WIDE-S (90 $\mu$m)} &{WIDE-L (140 $\mu$m)} & {N160 (160 $\mu$m)} \\ 
\hline
\endhead
\hline
\endfoot

\hline
\renewcommand{\footnoterule}{}

503 & - & 4.79E-02 $\pm$ 5.45E-03 & - & -    \\  
504 & - & 4.78E-02 $\pm$ 5.45E-03 & - & -    \\  
507 & - & 4.75E-02 $\pm$ 5.45E-03 & - & -    \\  
513 & - & 4.70E-02 $\pm$ 5.44E-03 & - & -    \\  
514 & - & 4.69E-02 $\pm$ 5.44E-03 & - & -    \\  
519 & - & 4.68E-02 $\pm$ 5.44E-03 & - & -    \\  
523 & - & 4.65E-02 $\pm$ 5.44E-03 & - & -    \\  
524 & - & 4.65E-02 $\pm$ 5.44E-03 & - & -    \\
\hline
\end{longtable}
}

\clearpage

\onltab{5}{
\begin{longtable}{ccccccc} 

\caption{\label{measurements1} Flux densities in other wavelengths, found for the counterparts of the ADFS sources in the public databases, for the 10$\sigma$ catalog - part 1.}\\ 

\hline\hline
{ID}  &  {0.44 $\mu$m} & {0.47 $\mu$m$^C$} & {0.55 $\mu$m}  & {0.64 $\mu$m}  & {0.71 $\mu$m} \\ 
{(ADF-S)}  & {} & {} & {} & {} & {} \\
\hline
\endfirsthead
\caption{continued.}\\
\hline\hline
{ID}  &  {0.44 $\mu$m} & {0.47 $\mu$m$^C$} & {0.55 $\mu$m}  & {0.64 $\mu$m}  & {0.71 $\mu$m} \\ 
{(ADF-S)} &{} & {} & {} & {} & {}\\
\hline
\endhead
\hline
\endfoot

\hline

1&7.05E-03$\pm$6.09E-04$^A$ & 5.16E-03$\pm$6.03E-04& - &1.39E-02$\pm$1.20E-03$^A$ &1.41E-02$^D$ \\ 
& 7.35E-03$\pm$6.35E-03$^A$ & &  &1.44E-02$\pm$1.25E-03$^A$&  \\ 
 & 7.59E-03$\pm$6.56E-04$^A$ & & &1.47E-02$\pm$1.27E-03$^D$  & \\ 
& 7.90E-03$\pm$1.69E-03$^B$& & & & \\  
& 7.38E-03$^D$& & & & \\

2&1.38E-03$\pm$1.19E-14$^A$ & 1.50E-03$\pm$1.23E-04& - &2.46E-03$\pm$2.13E-04$^A$ &2.63E-03$^D$ \\ 
& 1.46E-03$\pm$1.27E-04$^A$ & &  &2.62E-03$\pm$2.26E-04$^A$&  \\  
& 1.53E-03$\pm$1.32E-04$^A$ & & &2.76E-03$\pm$2.38E-04$^D$  & \\
\hline
\end{longtable}

\begin{list}{}{}
\item{\textit{A} - \cite{lauberts}}\\
\item{\textit{B} - \cite{deVac91}} \\
\item{\textit{C} - \cite{maddox}}\\
\item{\textit{D} - The SIMBAD astronomical database, \cite{wenger}}\\
\item{\textit{E} - \cite{gildepaz}} \\
\item{\textit{F} - \cite{sandage}} \\
\end{list}
}%

\clearpage

\onltab{6}{
\begin{longtable}{ccccccc} 

\caption{\label{measurements2} Flux densities in other wavelengths found for the counterparts of the ADFS sources in the public databases, for the 10$\sigma$ catalog - part 2.}\\ 

\hline\hline
{ID}  &  {0.79 $\mu$m} & {1.25 $\mu$m} & {1.65 $\mu$m}  & {2.22 $\mu$m}  & {12 $\mu$m$^I$} \\ 
{(ADF-S)}  & {} & {} & {} & {} & {} \\
\hline
\endfirsthead
\caption{continued.}\\
\hline\hline
{ID}  &  {0.79 $\mu$m} & {1.25 $\mu$m} & {1.65 $\mu$m}  & {2.22 $\mu$m}  & {12 $\mu$m$^I$} \\ 
{(ADF-S)} &{} & {} & {} & {} & {}\\
\hline
\endhead
\hline
\endfoot

\hline

1&2.38E-02$\pm$1.19E-03$^G$ &4.31E-02$\pm$1.13E-04$^H$ & 5.66E-02$\pm$1.85E-04$^H$ &5.37E-02$\pm$2.31E-04$^H$ &1.25E-01$\pm$1.38E-02  \\ & &4.31E-02$\pm$1.13E-04$^D$ &5.66E-02$\pm$1.85E-04$^D$ &5.37E-02$\pm$2.31E-04$^D$ & & \\

2& - &8.17E-03$\pm$3.21E-05$^H$ & 1.04E-02$\pm$6.71E-05$^H$ &7.71E-03$\pm$6.75E-05$^H$ &6.45E-02$\pm$1.81E-01  \\ & &8.17E-03$\pm$3.89E-05$^D$ &1.04E-02$\pm$6.71E-05$^D$ &7.71E-03$\pm$6.75E-05$^D$ & & \\
3& -&1.02E+00$\pm$2.17E-03$^D$ &1.26E+00$\pm$3.27E-03$^D$ &9.85E-01$\pm$3.06E-03$^D$&- \\

4&2.02E+15$\pm$1.24E-02$^J$  &5.24E-02$\pm$1.54E-04$^H$ & 3.30E-02$\pm$5.34E-02$^H$ &4.50E-02$\pm$4.11E-02$^H$ &2.38E-04$\pm$2.50E-01  \\ & &5.24E-02$\pm$1.54E-02$^D$ &5.34E-02$\pm$2.23E-04$^D$ &4.11E-04$\pm$2.38E-04$^D$ & & \\
5& - &1.62E-02$\pm$6.35E-05$^H$ & 1.73E-02$\pm$1.05E-04$^H$ &1.83E-02$\pm$1.10E-04$^H$ & $<$ 5.48E-02 \\
\hline
\end{longtable}
\begin{list}{}{}
\item{\textit{A} - \textit{F} - as in table~\ref{measurements1}}\\
\item{\textit{G} - \cite{mathewson} }\\
\item{\textit{H} - \cite{2mass} }\\
\item{\textit{I} - \cite{moshir} }\\
\item{\textit{J} - \cite{kinney} }
\end{list}
}

\clearpage

\onltab{7}{  
\begin{longtable}{ccccccc} 

\caption{\label{measurements3} Flux densities in other wavelengths found for the counterparts of the ADFS sources in the public databases, for the 10$\sigma$ catalog - part 3.}\\ 

\hline\hline
{ID}  &  {25 $\mu$m$^I$} & {60 $\mu$m$^I$} & {100 $\mu$m$^I$}  & {35.69 cm $^K$} \\ 
{(ADF-S)}  & {} & {} & {} & {(843 MHz)}  \\
\hline
\endfirsthead
\caption{continued.}\\
\hline\hline
{ID}  &  {25 $\mu$m$^I$} & {60 $\mu$m$^I$} & {100 $\mu$m$^I$}  & {356895.8 $\mu$m$^K$} \\ 
{(ADF-S)} &{} & {} & {} & {} \\
\hline
\endhead
\hline
\endfoot

\hline

2&1.06E-01$\pm$1.90E-02&1.32E+00$\pm$6.60E-02& 2.20E+00$\pm$1.32E-01&1.06E-02$\pm$1.00E-03\\
3&7.36E-02$\pm$1.84E-02&6.70E-01$\pm$3.35E-02& 3.65E+00$\pm$1.83E-01&-\\
4&1.50E-01$\pm$1.50E-01&8.68E-01$\pm$6.07E-02& 1.82E+00$\pm$1.45E-01&-\\
5&7.87E-02$\pm$1.34E-02&6.07E-01$\pm$3.64E-02& 1.64E+00$\pm$1.31E-01&1.06E-02$\pm$1.00E-03\\
6&9.28E-02$\pm$1.60E-01&6.16E-01$\pm$4.31E-02& 1.16E+00$\pm$1.27E-01&-\\
7&1.69E-01$\pm$2.03E-02&7.47E-01$\pm$3.73E-02& 9.65E-01$\pm$1.25E-01&-\\
8&7.25E-02$\pm$1.88E-02&5.81E-01$\pm$3.49E-02& 1.22E+00$\pm$1.22E-01&-\\
9&7.24E-02$\pm$1.45E-02&4.75E-01$\pm$3.32E-02&-&-\\
\hline
\end{longtable}

\begin{list}{}{}
\item{\textit{I} - as in table~\ref{measurements2}}\\
\item{\textit{K} - \cite{mauch} }\\
\end{list}
}

\clearpage

\onltab{8}{
\begin{table*}[ht]
\caption{\label{measurements4}Flux densities in other wavelengths found for the counterparts of the ADFS sources in the public databases, for the 10$\sigma$ catalog - part 4. }
\label{other}
\centering  
\begin{tabular}{l c c c }  
\hline\hline  
{ID}  & $\lambda$ [$\mu$m] & Measurement &{Passband} \\ 
{AKARI}  & & \\ 
\hline\hline

4&0.10&2.41E-03&1030 A $^L$\\
4&0.15&1.60E-02$\pm$2.22E-03&GALEX FUV$^M$\\
4&0.16&2.30E-02$\pm$3.66E-03& UV 1650$^N$\\
4&0.18& 1.68E-02$\pm$2.32E-03&GALEX NUV$^M$\\
4&0.24&1.02E-02$\pm$7.90E-04&2373 A$^J$\\
4&0.25&2.52E-02$\pm$2.42E-03 &UV 2500$^N$\\
4&0.27&9.93E-03$\pm$3.65E-04&2700 A$^J$\\
4&0.31&3.94E-02$\pm$1.02E-02&UV3150$^N$ \\
4&0.36&2.14E-02$\pm$2.73E-03&B\_J$^O$\\
4&1.90&1.03E-02$\pm$6.11E-04&1913 A$^J$\\
4&3.60&2.60E-02$\pm$4.00E-03&3.6 IRAC$^P$\\
4&3.60&2.81E-02$\pm$2.00E-04&3.6 IRAC$^P$\\
4&4.50&1.80E-02$\pm$3.00E-03&4.5 IRAC$^P$\\
4&4.50&1.99E-02$\pm$4.00E-04&4.5 IRAC$^P$\\
4&4.50&1.79E-02$\pm$5.80E-03&4.5 IRAC$^P$\\
4&5.80&1.00E-02$\pm$2.00E-03&5.8 IRAC$^P$\\
4&8.00&1.70E-02$\pm$2.00E-03&8 IRAC$^P$\\
4&8.00&2.44E-02$\pm$1.40E-03&8 IRAC$^P$\\
4&24.0&5.20E-02$\pm$5.00E-03&24 MIPS$^P$\\
4&24.0&5.60E-02$\pm$2.00E-03&24 MIPS$^P$\\
4&24.0&5.27E-02$\pm$2.20E-03&24 MIPS$^P$\\
4&70.0&1.38E+00$\pm$1.00E-01&70 MIPS$^P$\\
4&70.0&1.09E+00$\pm$2.20E-01&24 MIPS$^P$\\
4&160.0&1.20E+00$\pm$2.50E-01&160 MIPS$^P$\\
4&160.0&1.66E+00$\pm$2.10E-01&160 MIPS$^P$\\
\hline
10&0.36&3.96E-03$\pm$7.02E-04&b\_j$^B$\\
\hline
\end{tabular}

\begin{list}{}{}
\item{\textit{B} - as in table~\ref{measurements1}}\\
\item{\textit{J} - as in table~\ref{measurements2}}\\
\item{\textit{L} - \cite{fox} }\\
\item{\textit{M} - \cite{dale2007}}\\
\item{\textit{N} - \cite{rifatto}}\\
\item{\textit{O} - \cite{doyle}}\\
\item{\textit{P} - \cite{dale2005}}
\end{list}
\end{table*}
 }

\clearpage

\onltab{9}{
\begin{longtable}{ccccccc} 

\caption{\label{measurements1a} Flux densities in other wavelengths found for the counterparts of the ADFS sources, in the public databases, for the fainer part of the 6$\sigma$ catalog - part 1.}\\ 

\hline\hline
{ID}& {0.47 $\mu$m$^C$}   & {1.25 $\mu$m} & {1.65 $\mu$m}  & {2.22 $\mu$m}    \\ 
{(ADF-S)}  & {} & {} & {} & {} \\
\hline
\endfirsthead
\caption{continued.}\\
\hline\hline
{ID}& {0.47 $\mu$m$^C$}   & {1.25 $\mu$m} & {1.65 $\mu$m}  & {2.22 $\mu$m}    \\ 
{(ADF-S)}  & {} & {} & {} & {} \\
\hline
\endhead
\hline
\endfoot

\hline
503 &   3.17E-05 $\pm$ 5.37E-06  &  -  &  -  &  -  \\
504 &   1.26E-04 $\pm$ 1.47E-05  &  -  &  -  &  -  \\
507 &   8.48E-05 $\pm$ 9.91E-06  &  -  &  -  &  -  \\
513 &   -  &  2.25E-03 $\pm$ 2.63E-04$^H$  &   3.14E-03 $\pm$ 3.73E-04$^H$  &   2.49E-03 $\pm$ 4.31E-04$^H$  \\
514 &   2.97E-05 $\pm$ 5.03E-06  &  -  &  -  &  -  \\
519 &   -  &  1.65E-03 $\pm$ 3.60E-04$^H$  &   2.68E-03 $\pm$ 5.01E-04$^H$  &   1.94E-03 $\pm$ 6.37E-04$^H$  \\
523 &   2.68E-05 $\pm$ 4.55E-06  &  -  &  -  &  -  \\
524 &   1.86E-04 $\pm$ 2.17E-05  &  -  &  -  &  -  \\
\hline
\end{longtable}
\begin{list}{}{}
\item{\textit{C}, \textit{D} - as in table~\ref{measurements1}}\\
\item{\textit{H} - as in table~\ref{measurements2}}\\
\end{list}

}

\onltab{10}{

\begin{longtable}{ccccccc} 

\caption{\label{measurements2a} Flux densities in other wavelengths found for the counterparts of the ADFS sources in the public databases, for the fainter part of the 6$\sigma$ catalog - part 2.}\\ 

\hline\hline
{ID}   &  {0.44 $\mu$m$^D$}&{0.55 $\mu$m$^D$}&  {35.69 cm$^K$}  \\ 
{(ADF-S)}   & {} & {} & {(843 MHz)} \\
\hline
\endfirsthead
\caption{continued.}\\
\hline\hline
\hline
\endhead
\hline
\endfoot
\hline
541&	1.94E-03&-&-\\
574&-&	 	 	1.25E-03&-\\
585& -&-&5.88E-02 $\pm$ 2.00E-03 \\
604& -&-&6.80E-03 $\pm$ 8.00E-04 \\
610 &-&1.25E-03&-&-\\
621&	4.13E-03&-&-\\
624& 1.64E-03&-&-\\
699& -&-&1.32E-02 $\pm$ 1.20E-03 \\
760& -&-&1.11E-02 $\pm$ 1.10E-03 \\
794&1.64E-03&-&-\\
806&1.64E-03&-&-\\
831& -&-&1.73E-02 $\pm$ 1.60E-03 \\
837&-&1.04E-03&-\\
856	&2.38E-01&3.09E-01&-&-\\
857& -&-&6.31E-02 $\pm$ 2.30E-03 \\
892&1.64E-03&-&-\\
945	&8.63E+000&1.02E+001&-\\
955& -&-&5.04E-02 $\pm$ 1.70E-03 \\
989	&5.44E-01&1.81E+000&-&-\\
\hline
\end{longtable}
\begin{list}{}{}
\item{\textit{D} - as in table~\ref{measurements1}}\\
\item{\textit{K} - as in table~\ref{measurements3}}\\
\end{list}

}

\onlfig{1}{
\begin{figure*}[t]
\centering
\includegraphics[width=4cm]{points/kmalek_1.eps}
\includegraphics[width=4cm]{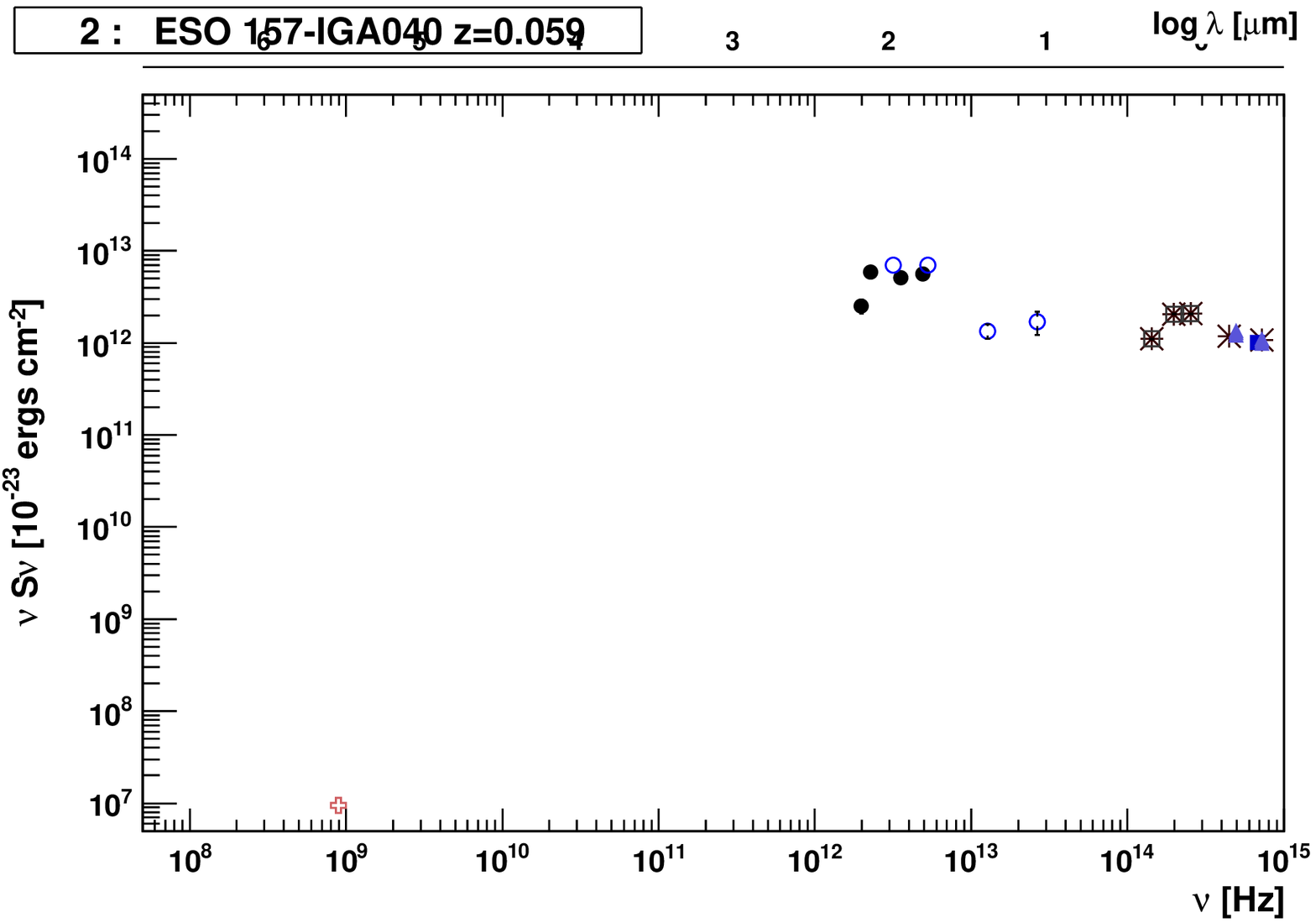}
\includegraphics[width=4cm]{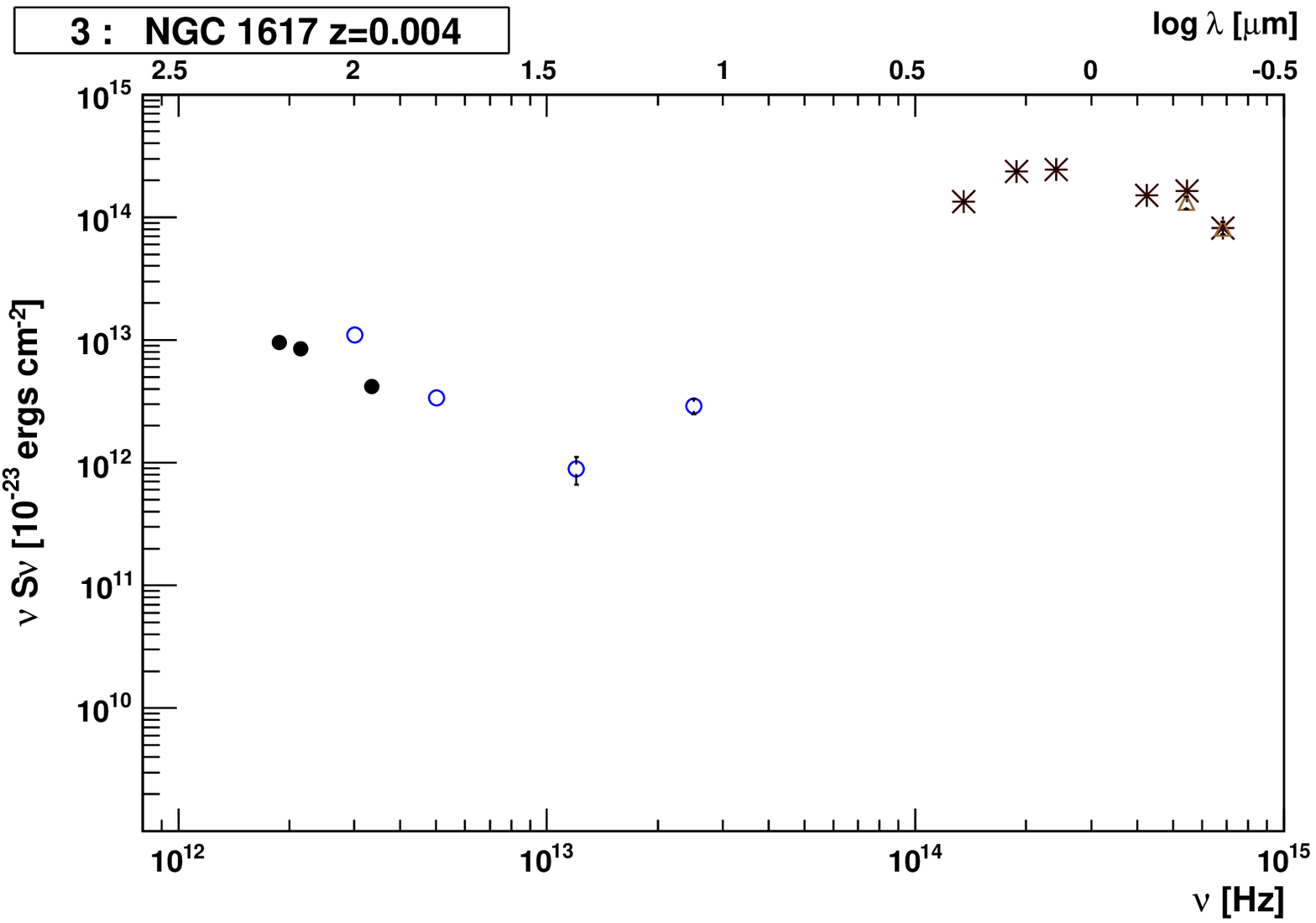}
\includegraphics[width=4cm]{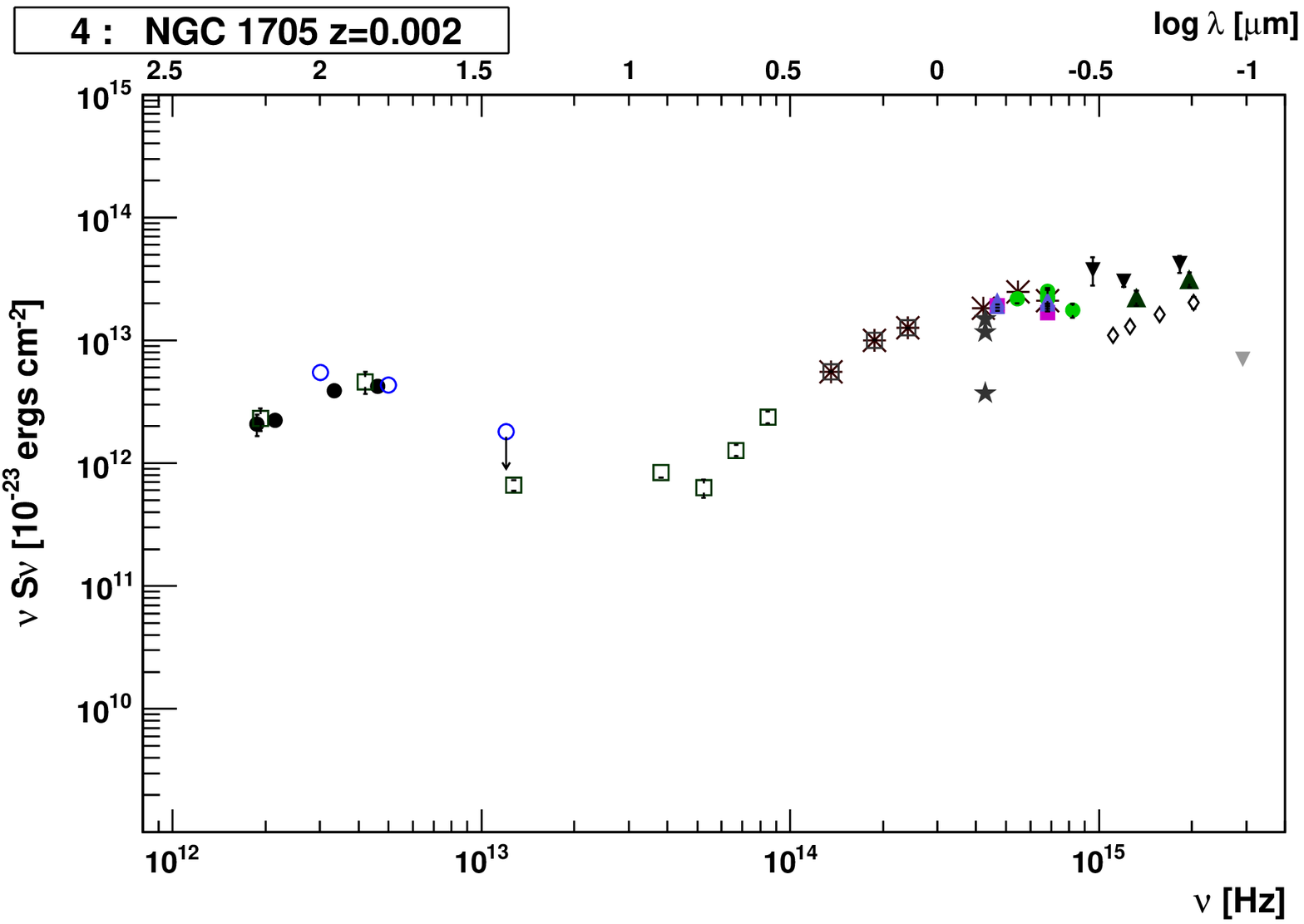}
\includegraphics[width=4cm]{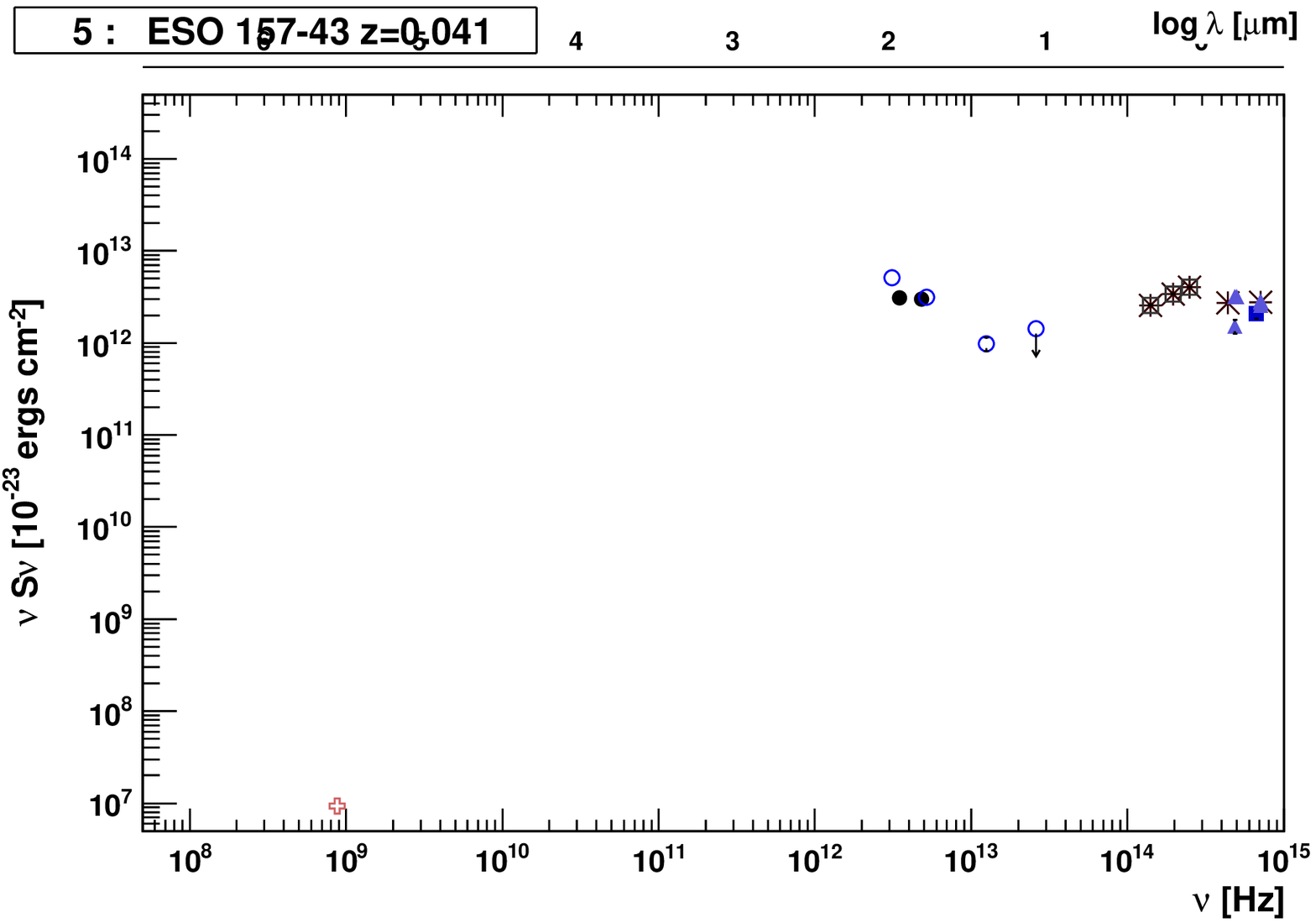}
\includegraphics[width=4cm]{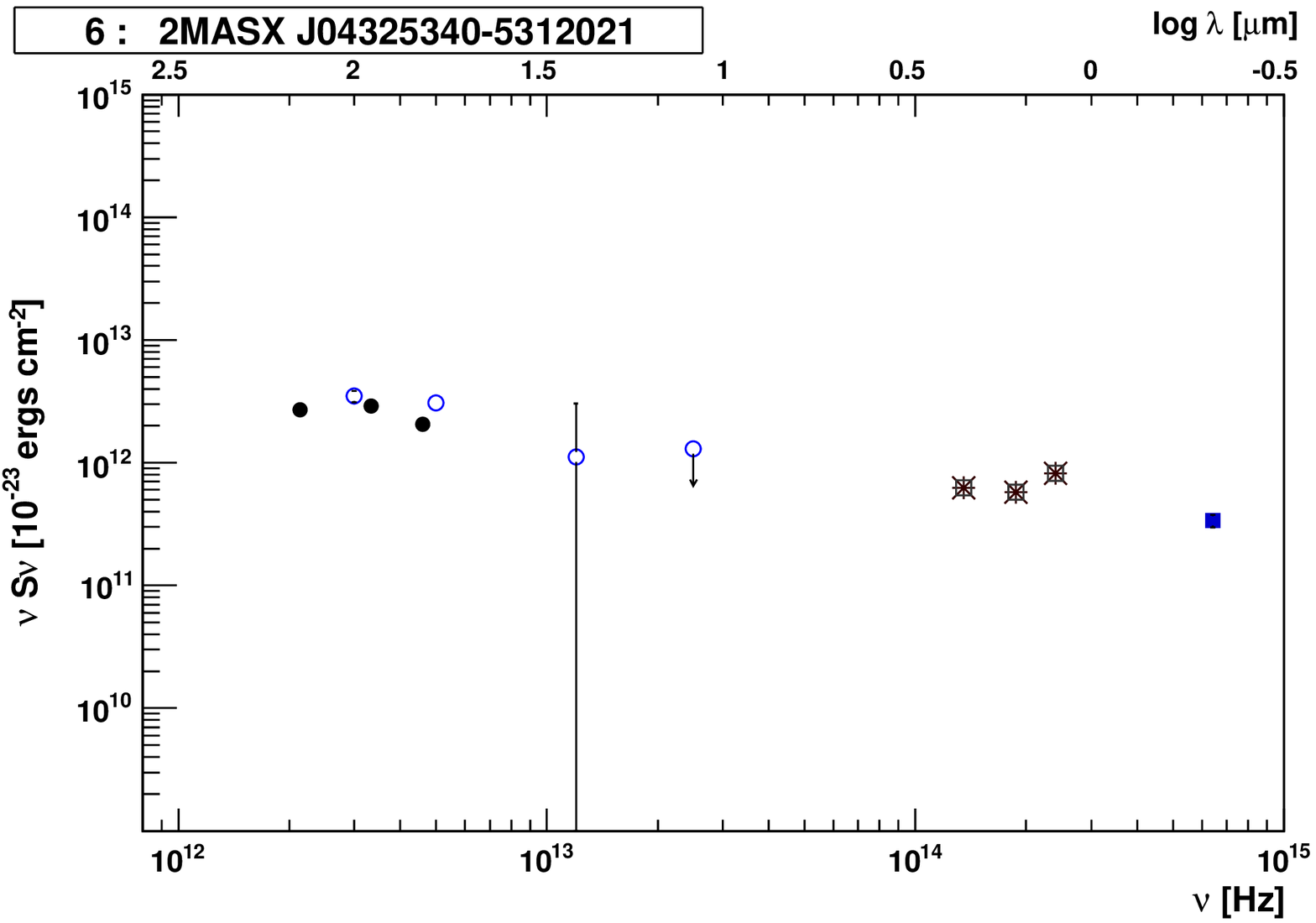}
\includegraphics[width=4cm]{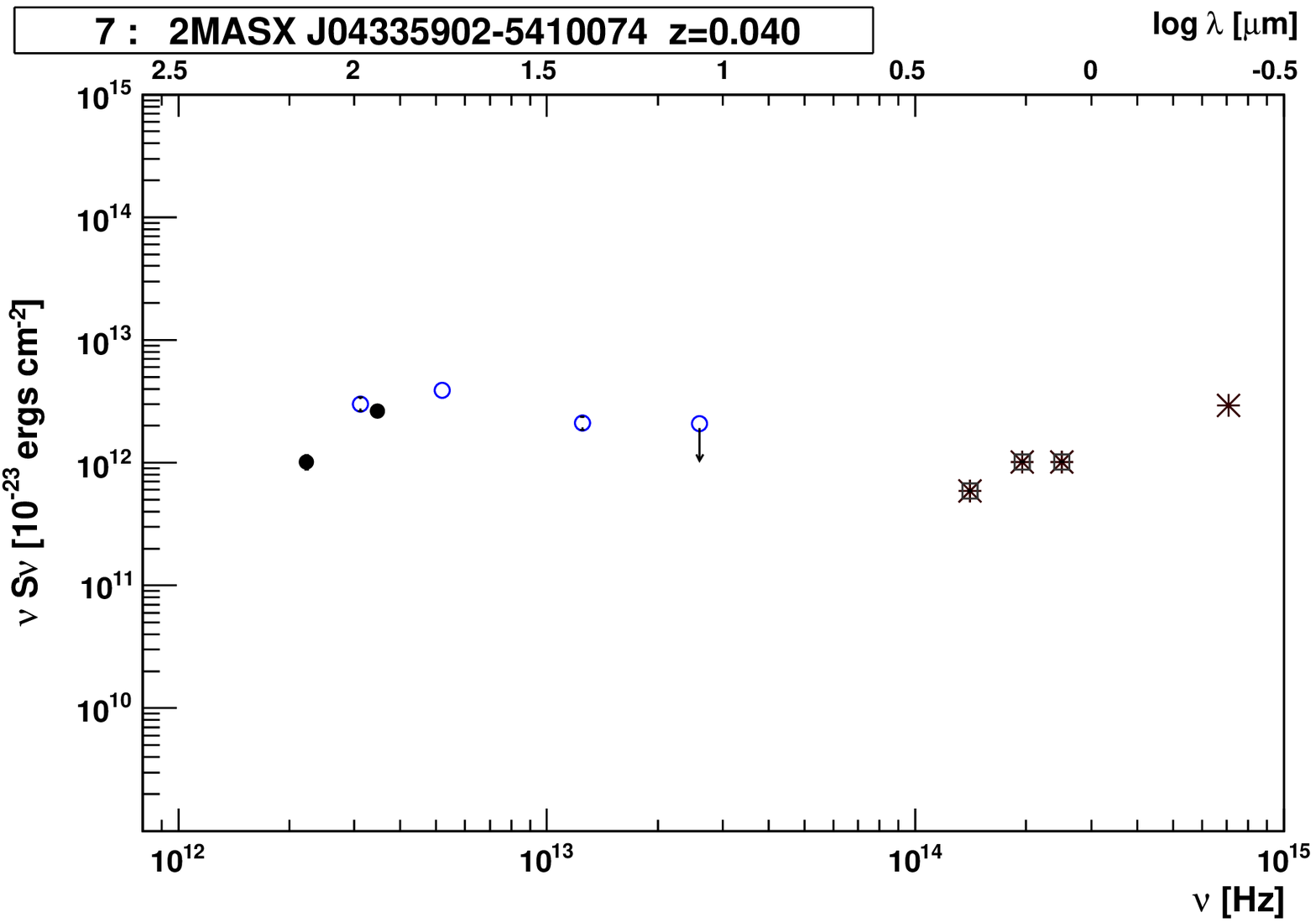}
\includegraphics[width=4cm]{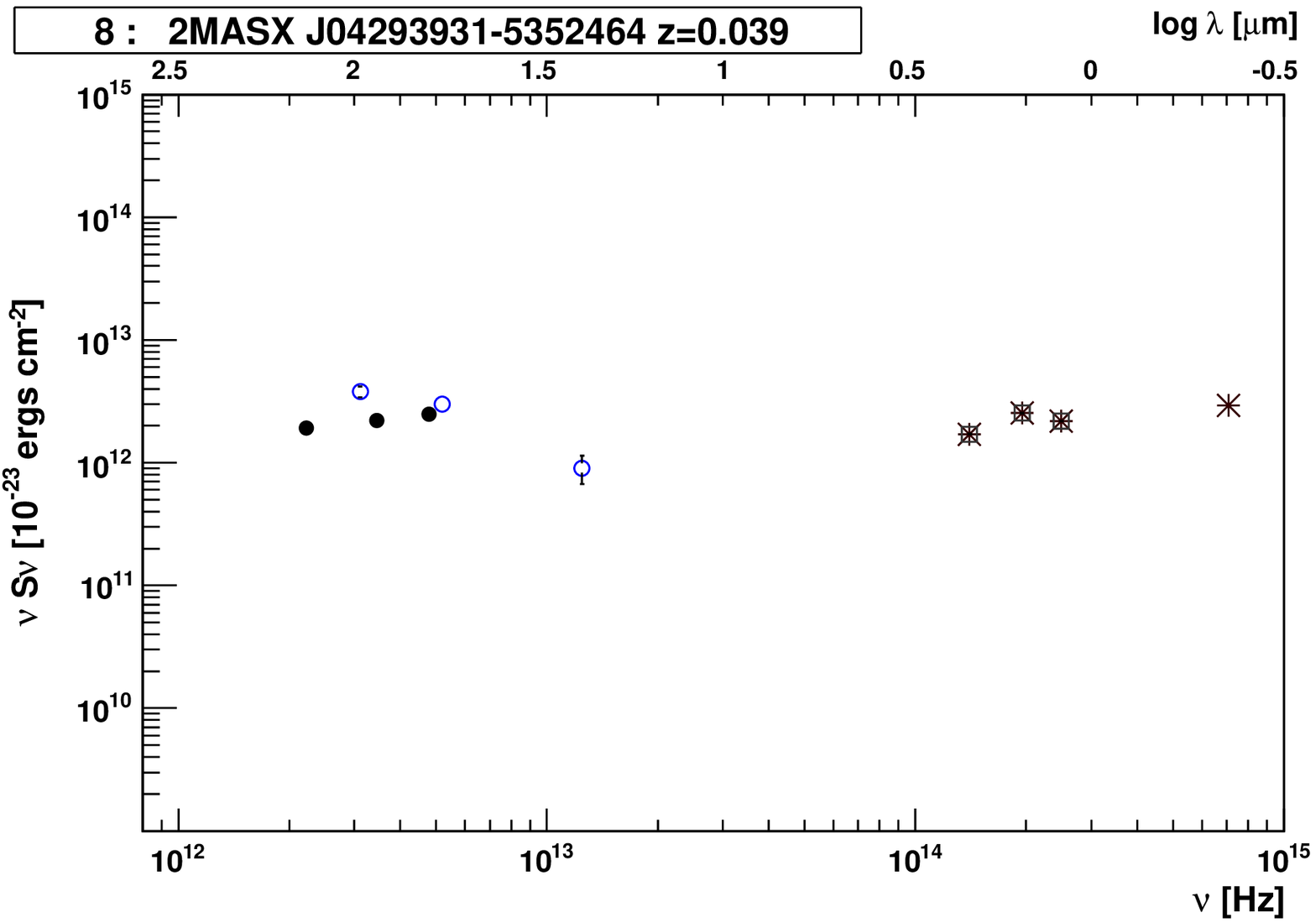}
\includegraphics[width=4cm]{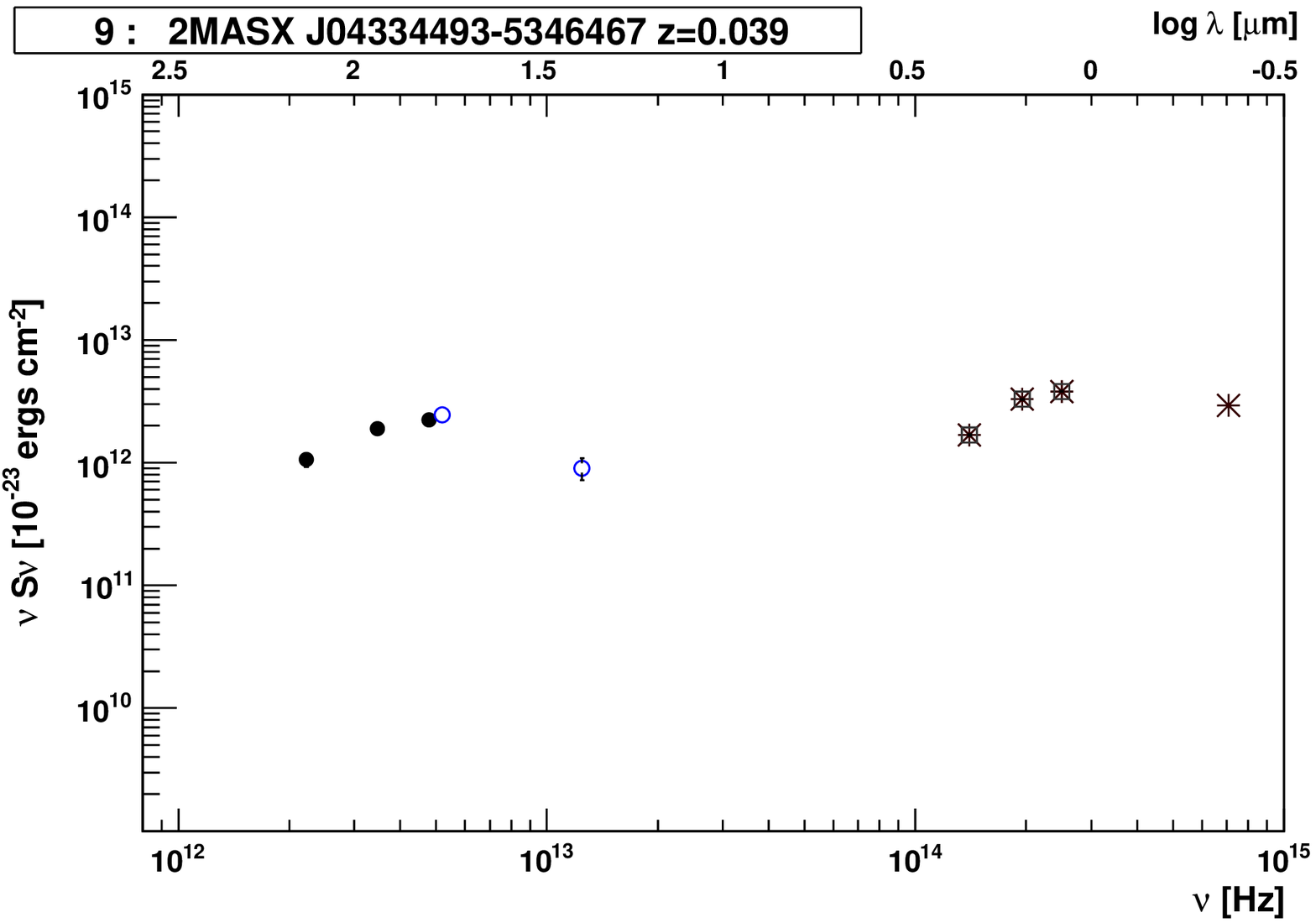}
\includegraphics[width=4cm]{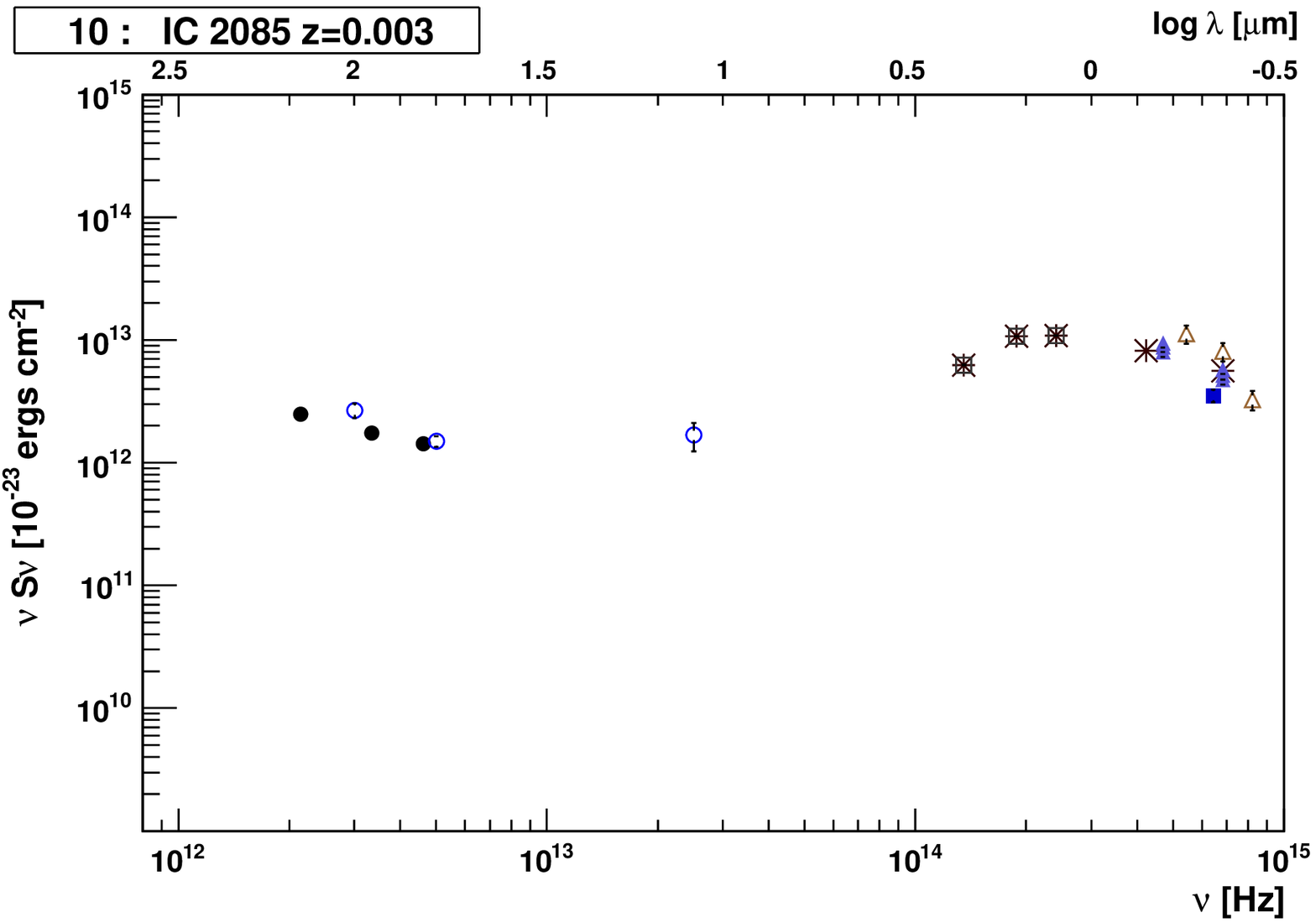}
\includegraphics[width=4cm]{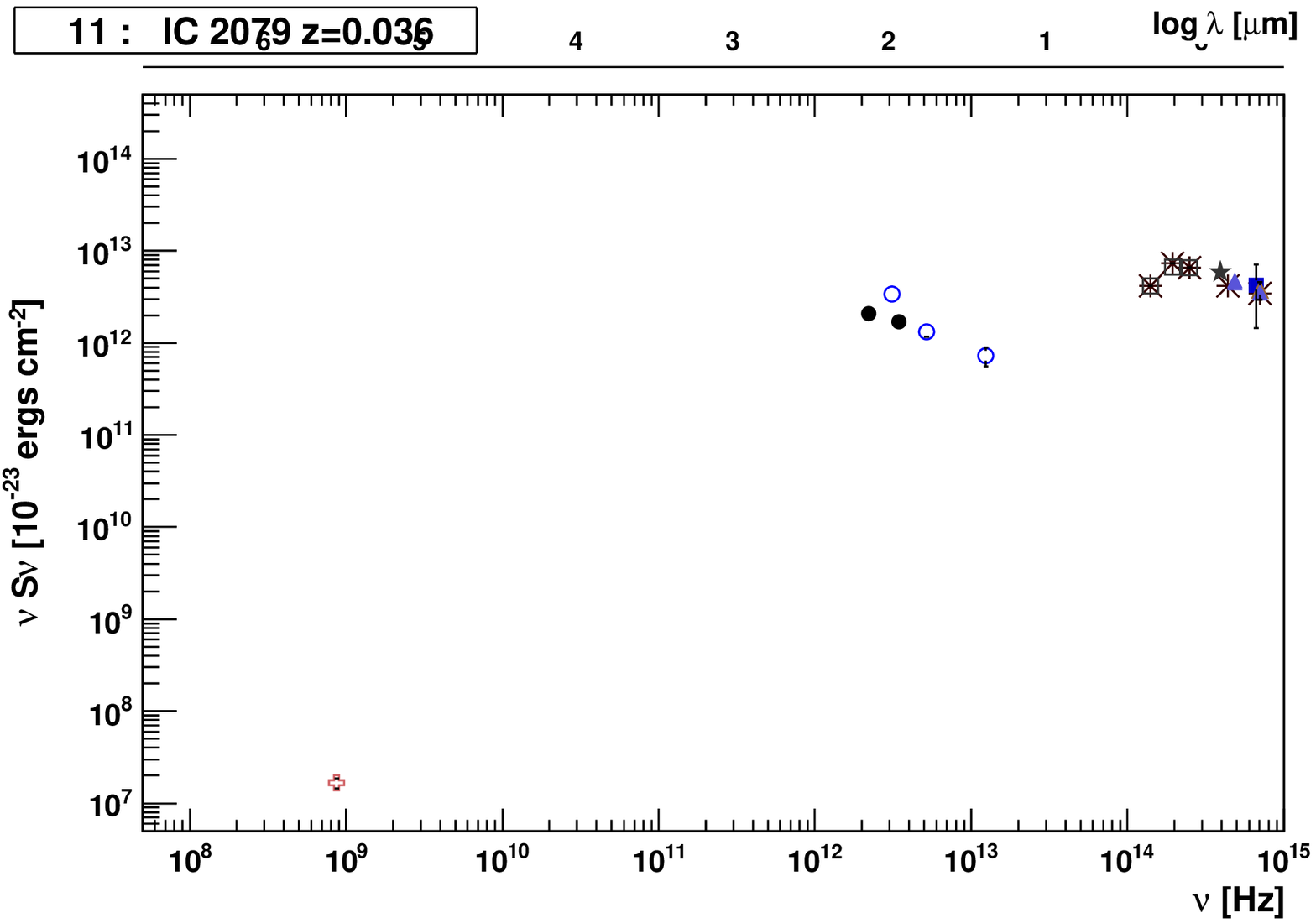}
\includegraphics[width=4cm]{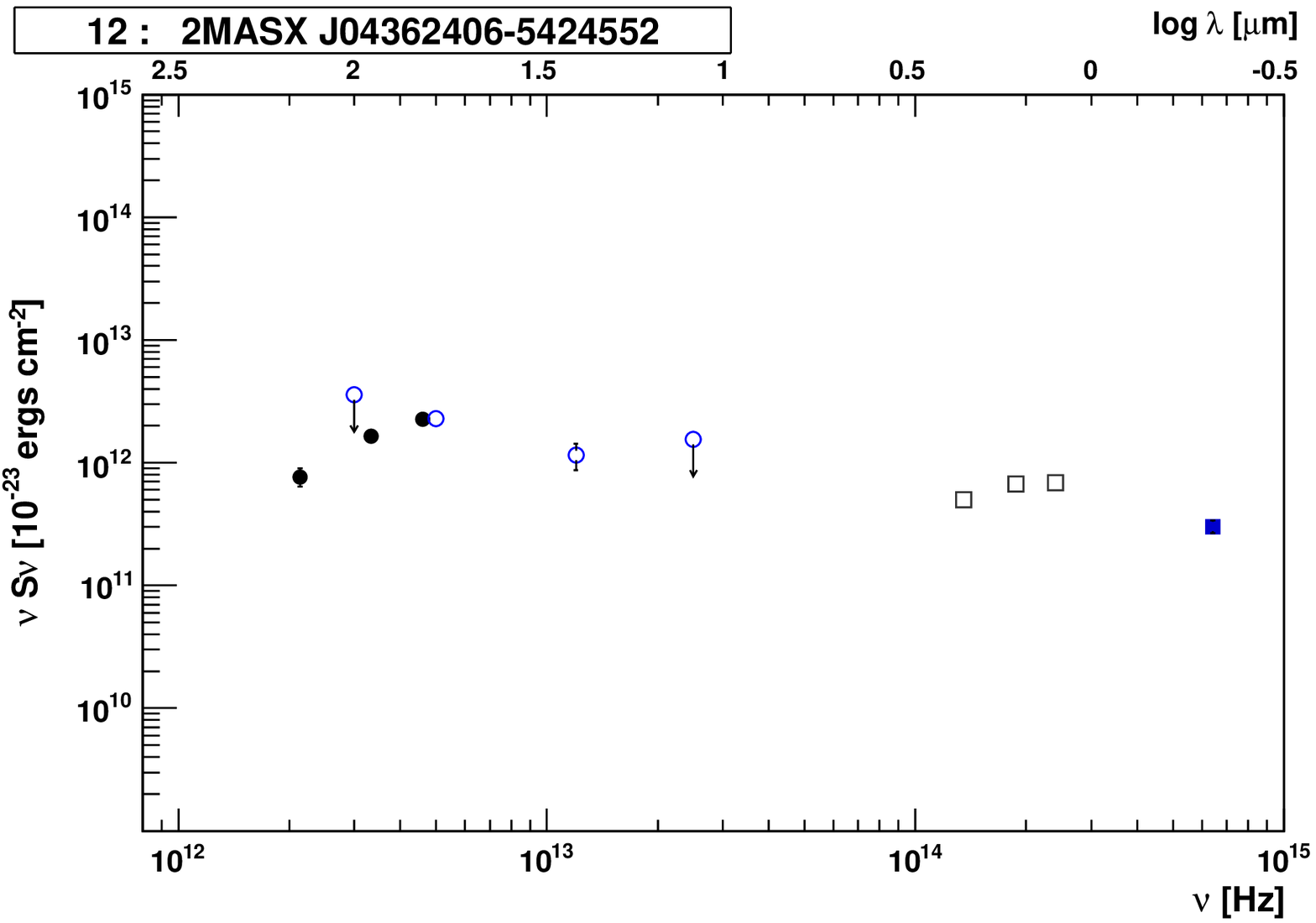}
\includegraphics[width=4cm]{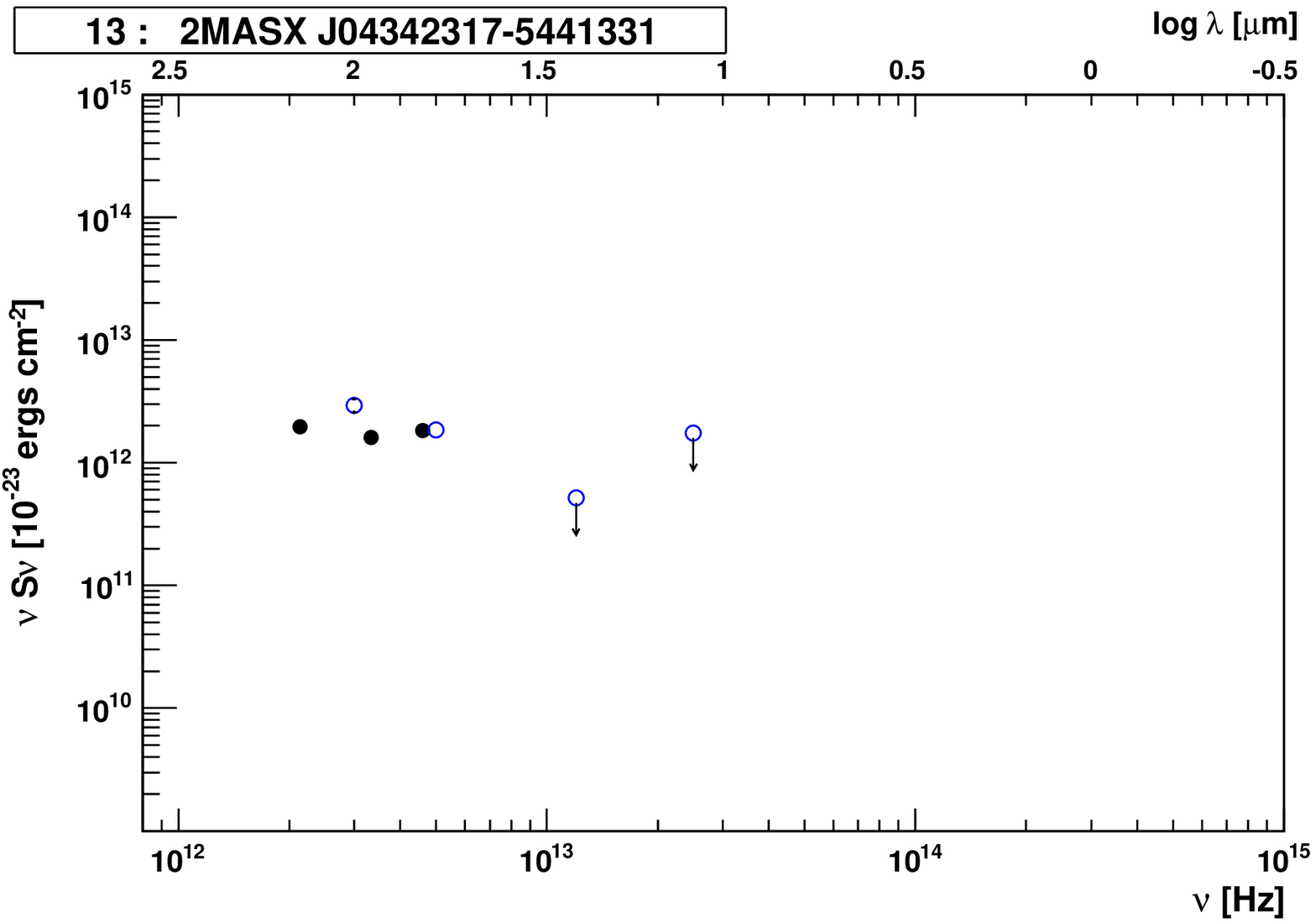}
\includegraphics[width=4cm]{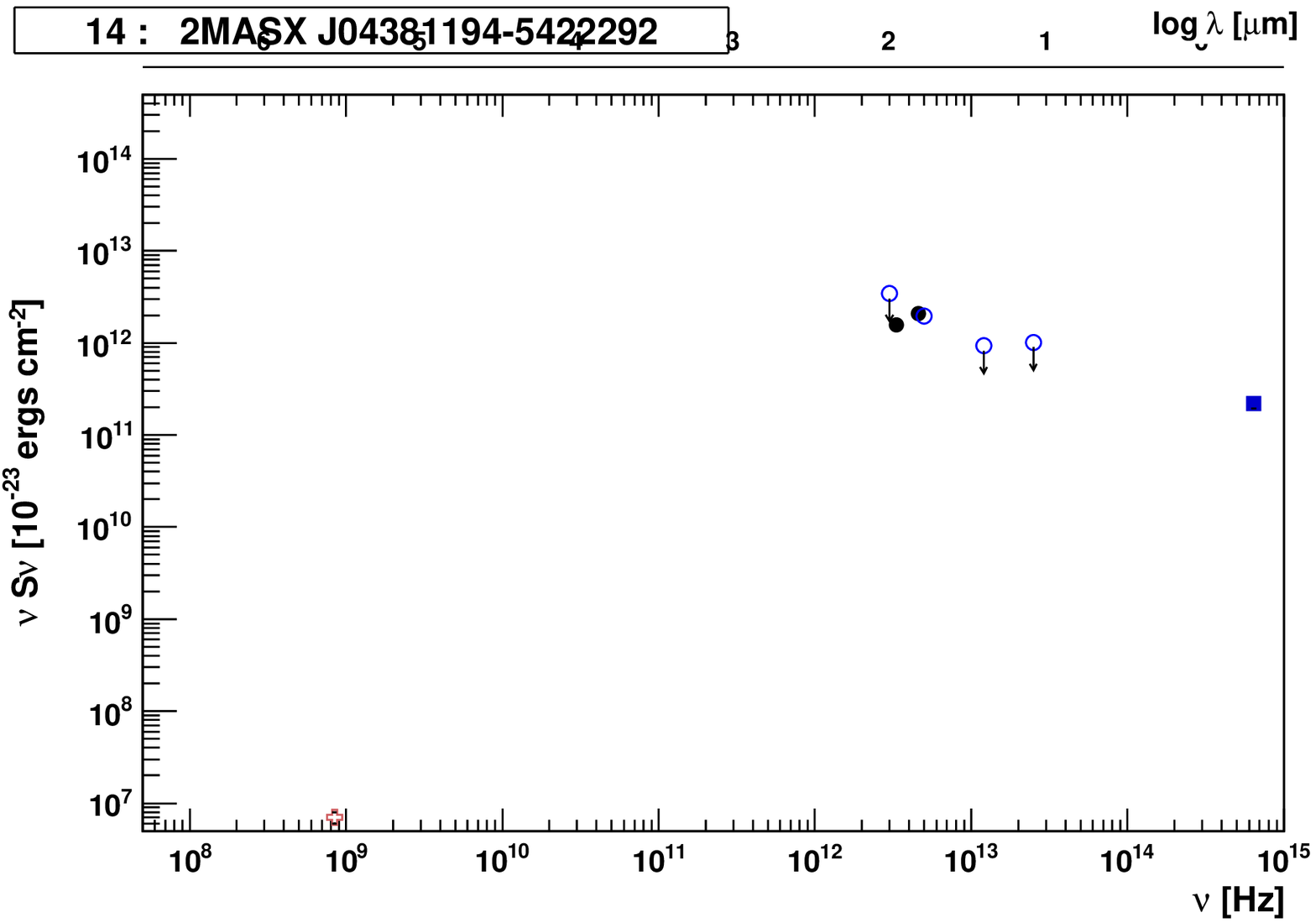}
\includegraphics[width=4cm]{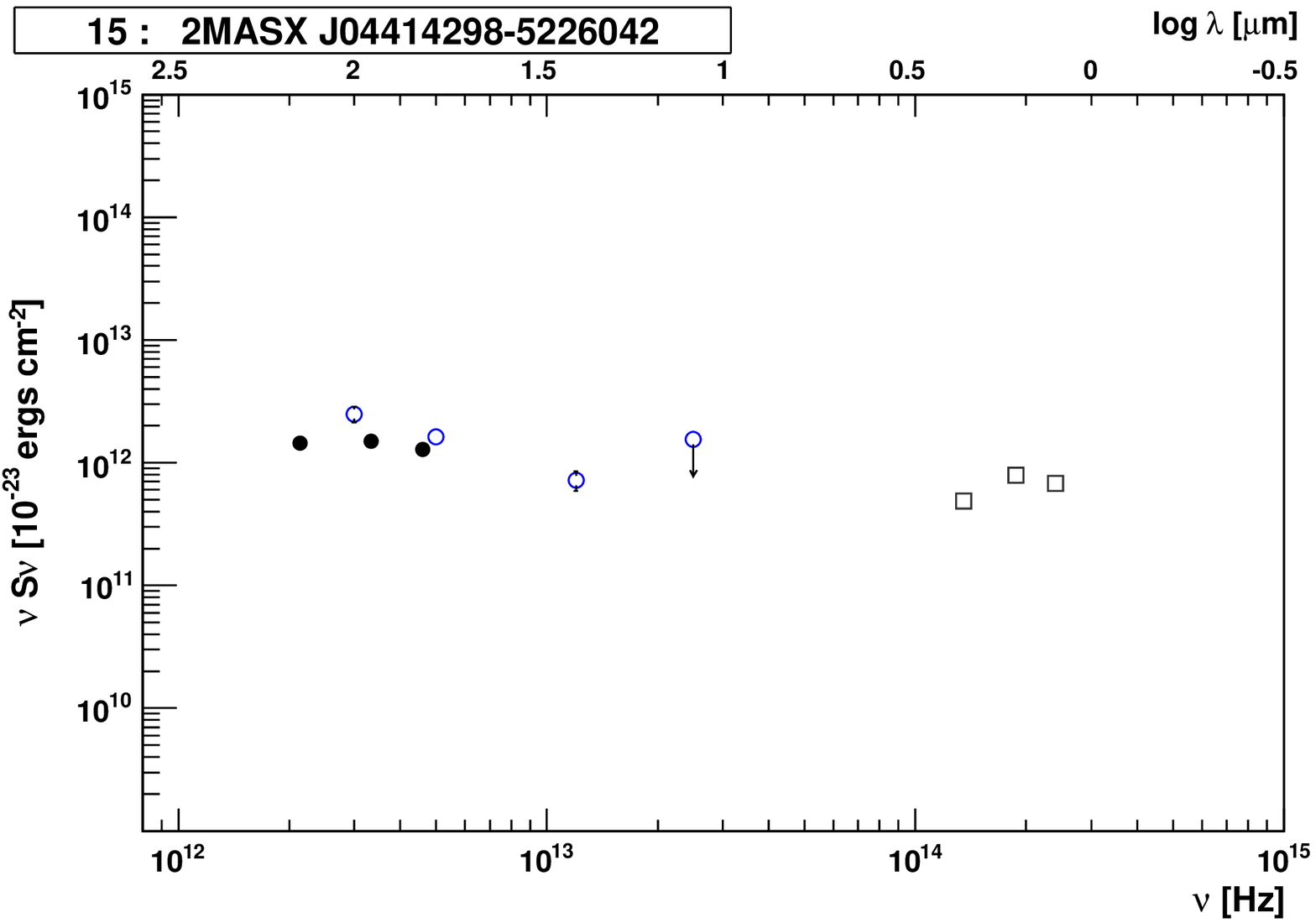}
\includegraphics[width=4cm]{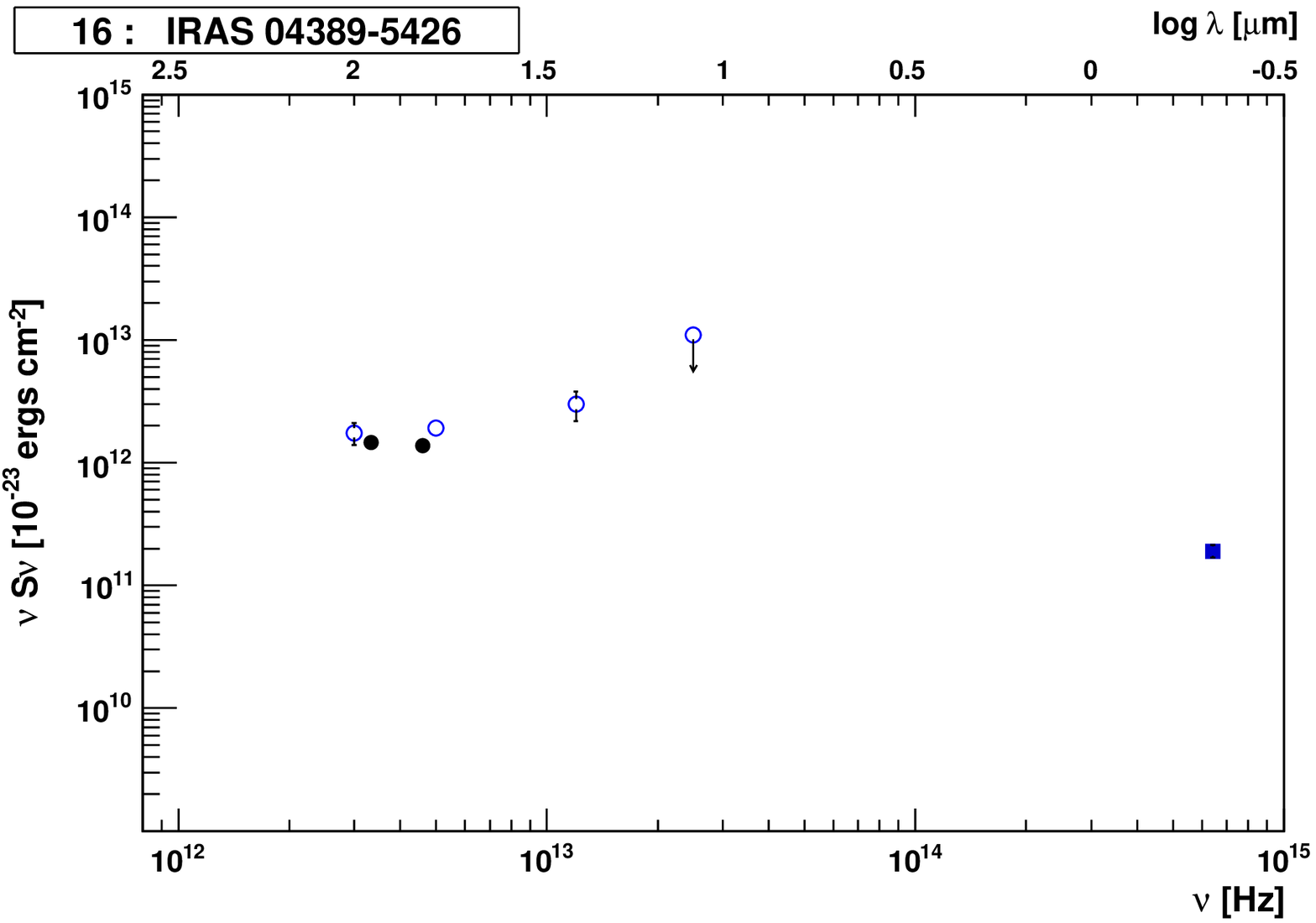}
\includegraphics[width=4cm]{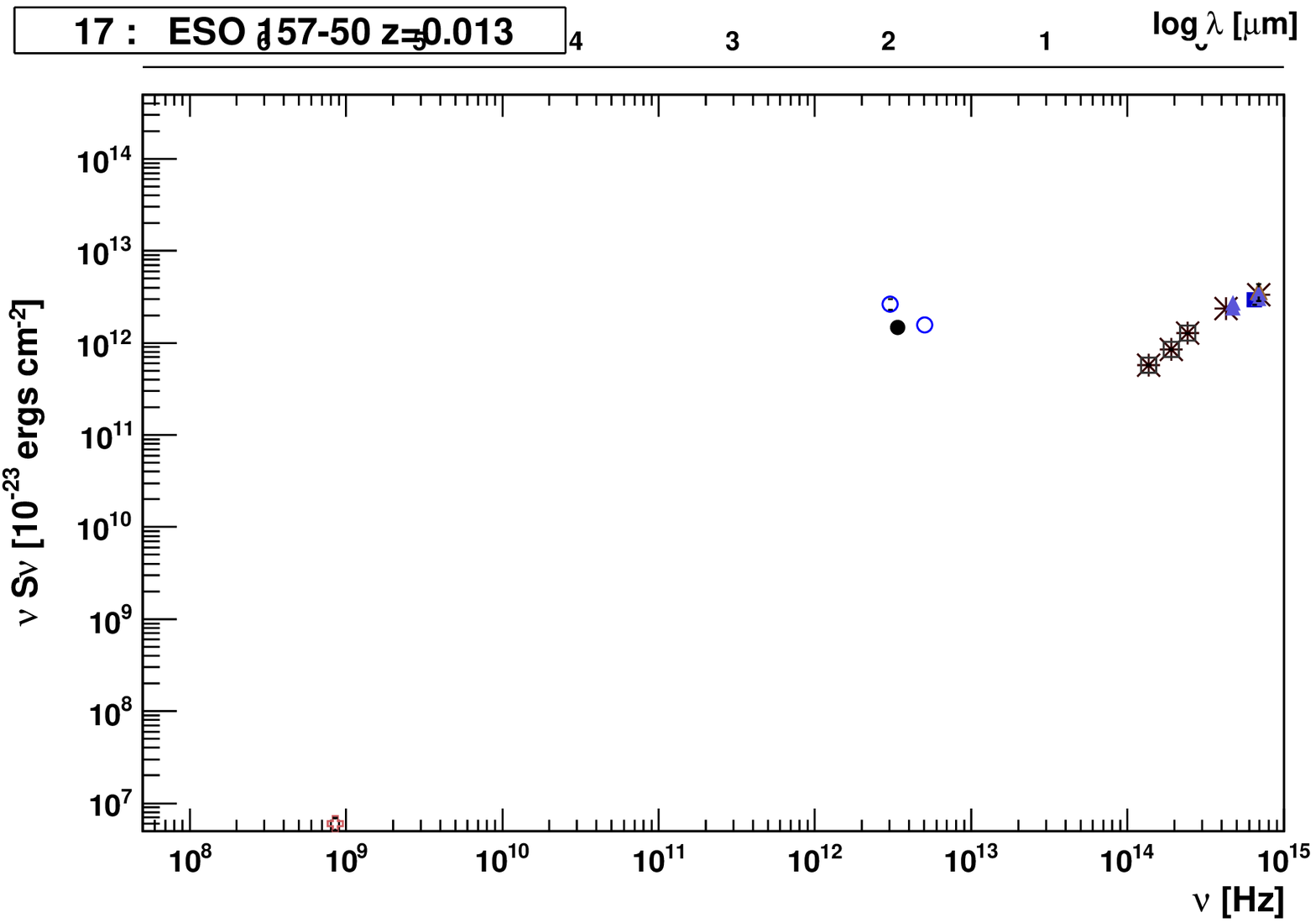}
\includegraphics[width=4cm]{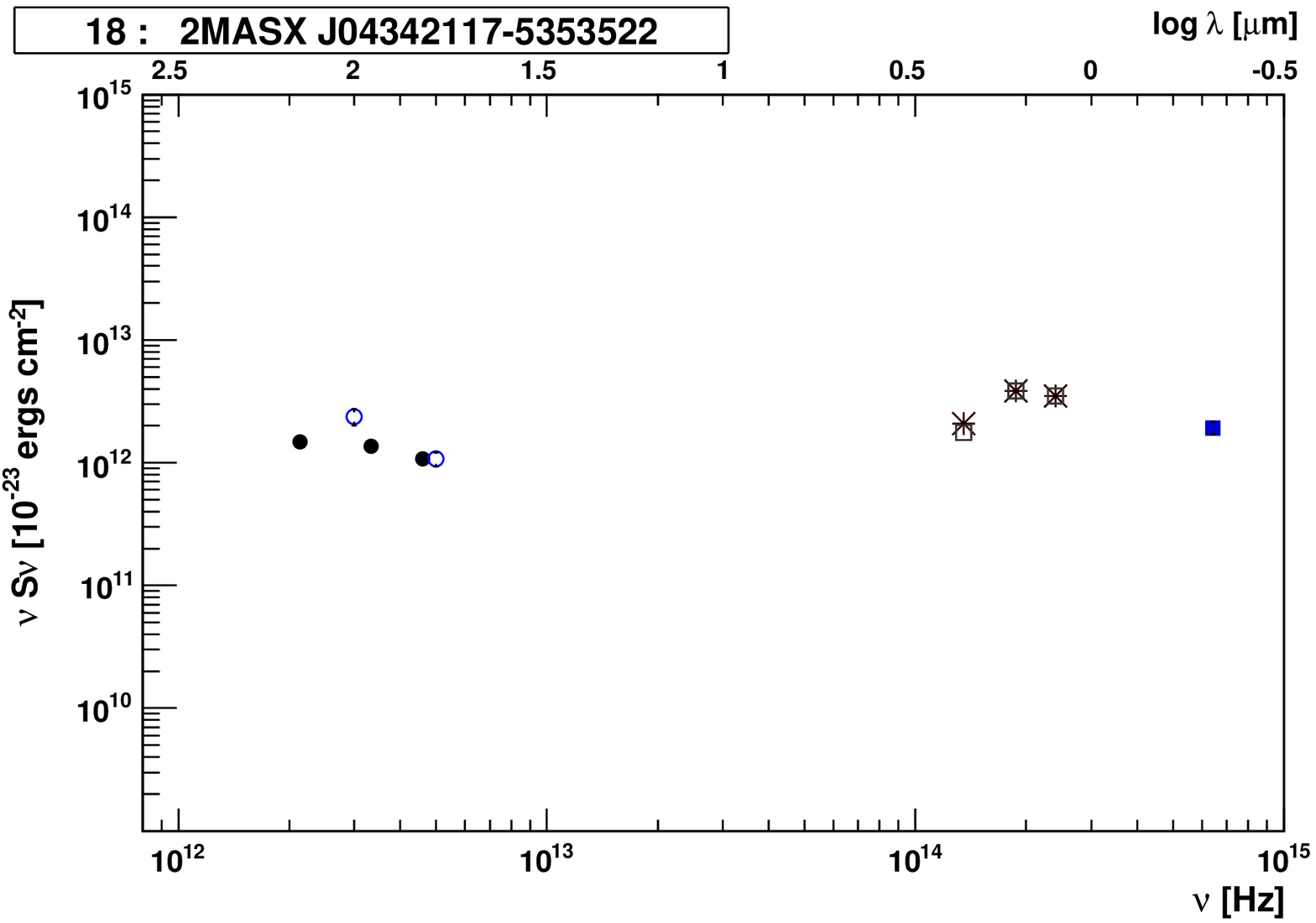}
\includegraphics[width=4cm]{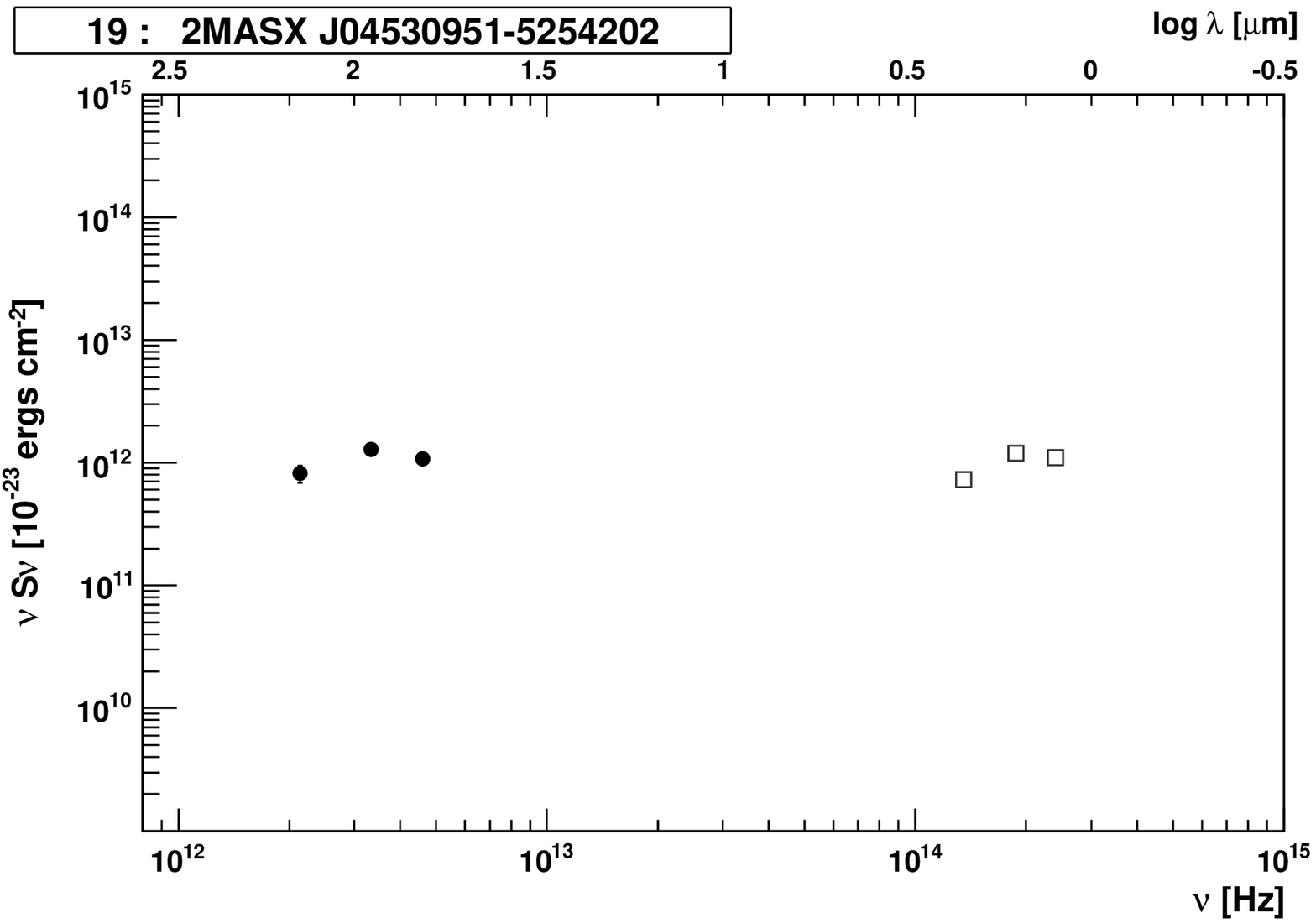}
\includegraphics[width=4cm]{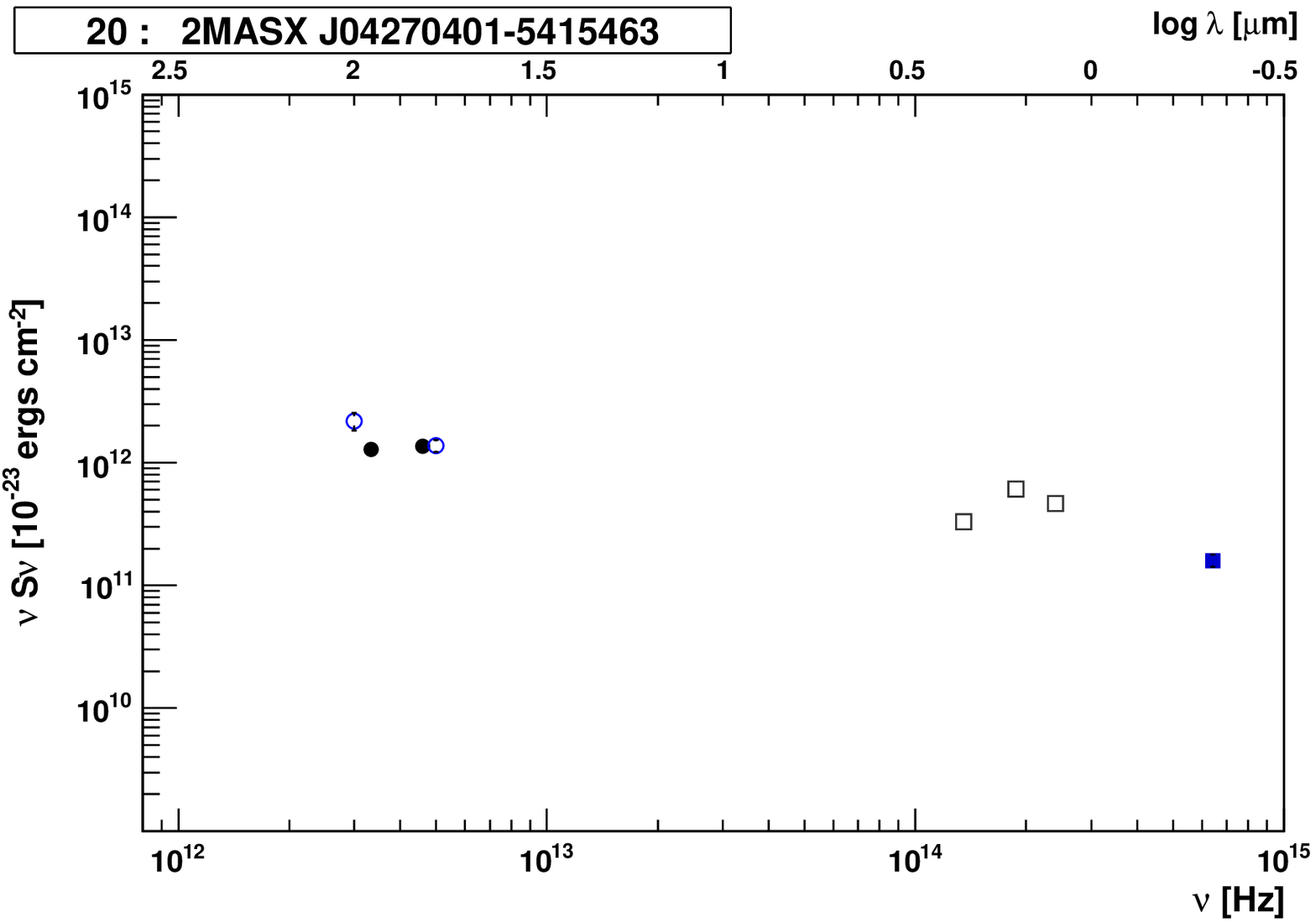}
\includegraphics[width=4cm]{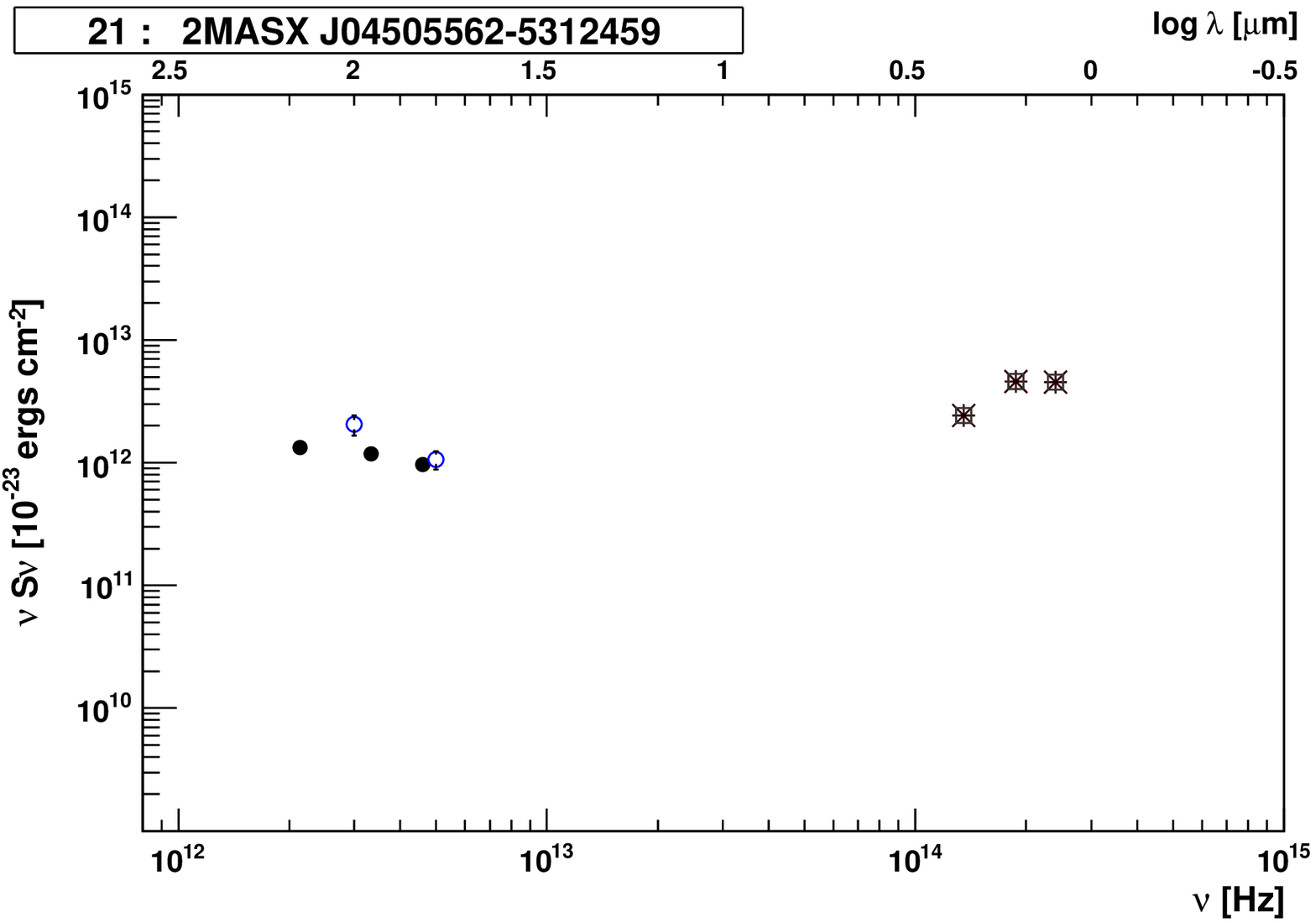}
\includegraphics[width=4cm]{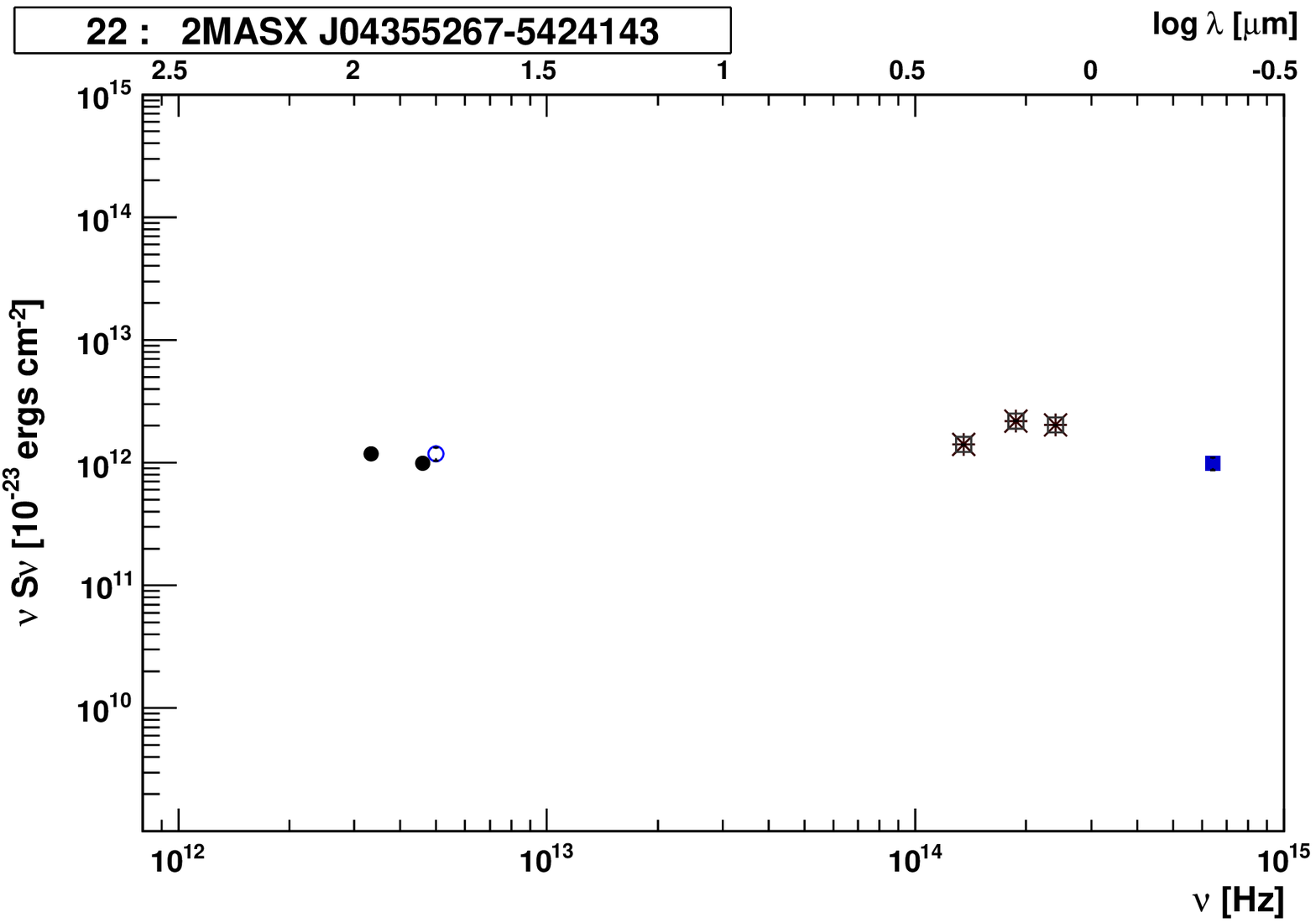}
\includegraphics[width=4cm]{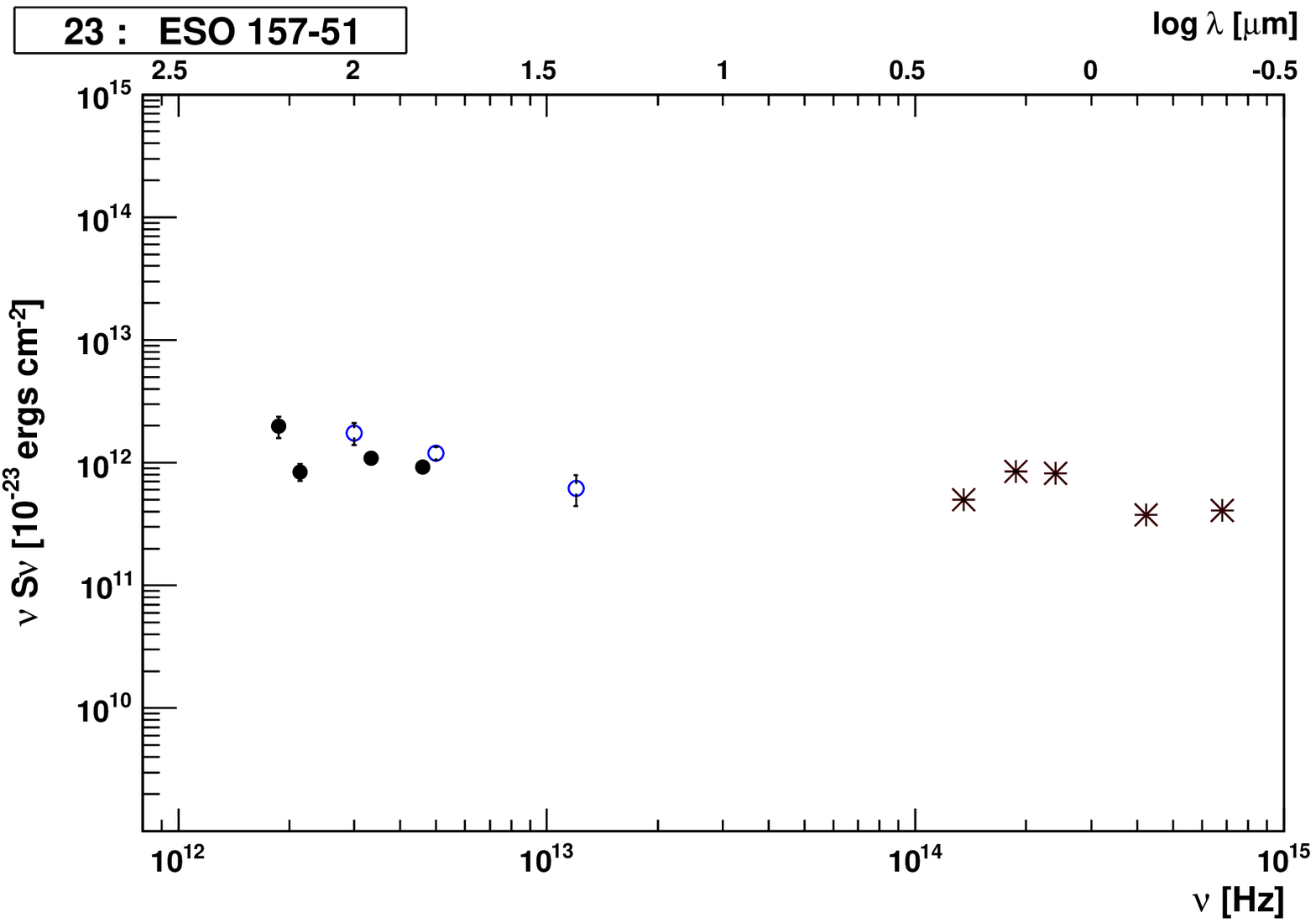}
\includegraphics[width=4cm]{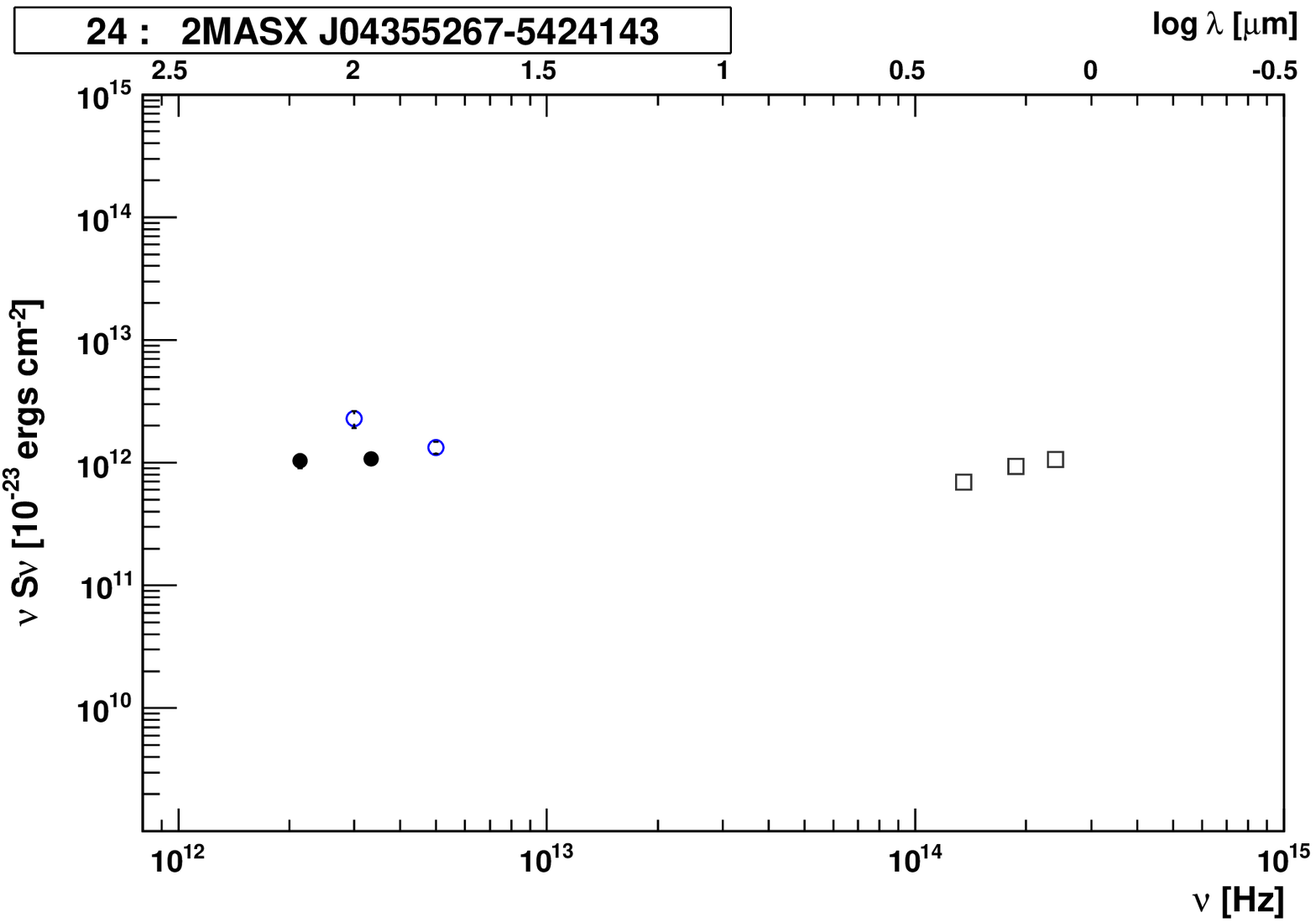}
\includegraphics[width=4cm]{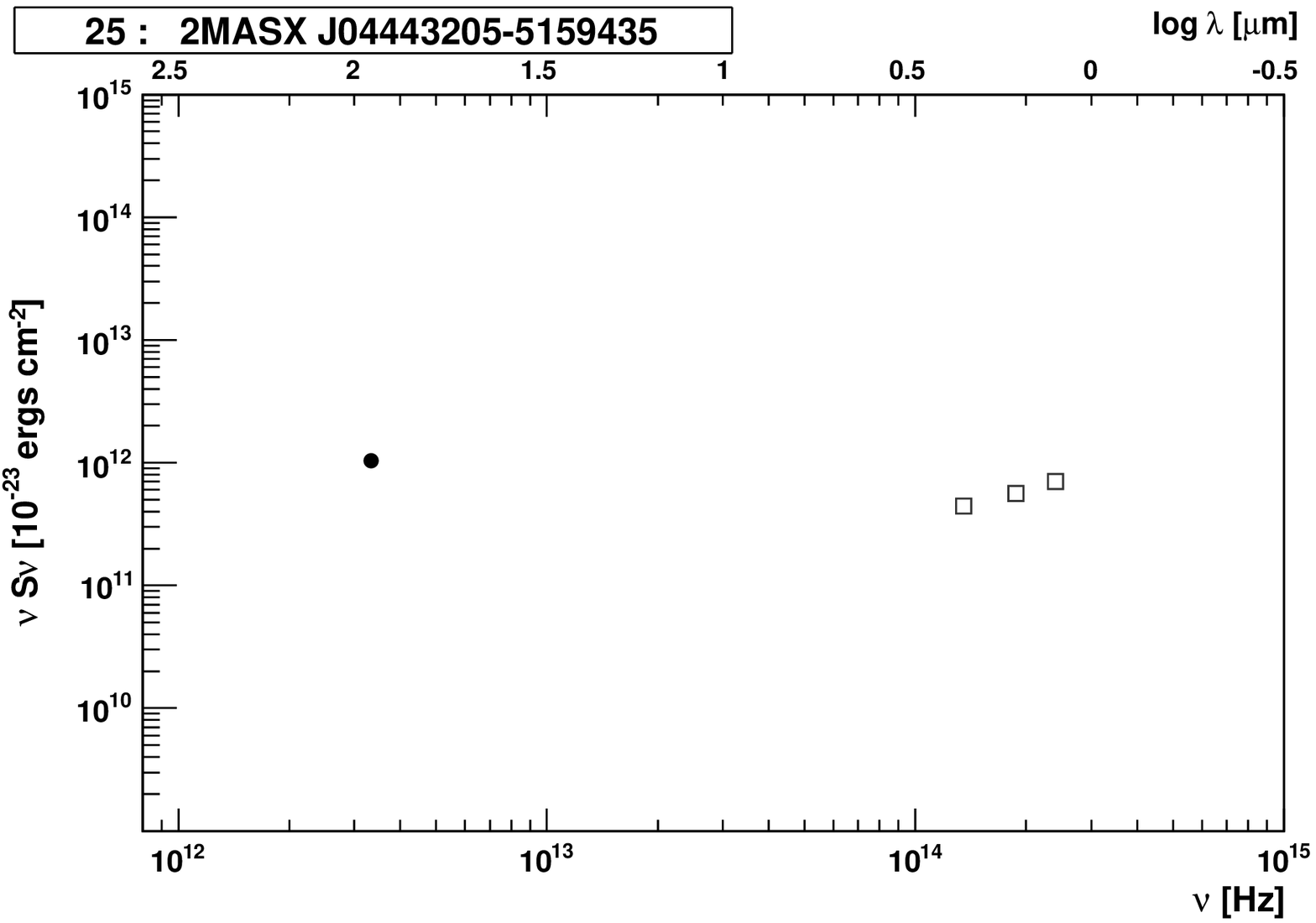}
\includegraphics[width=4cm]{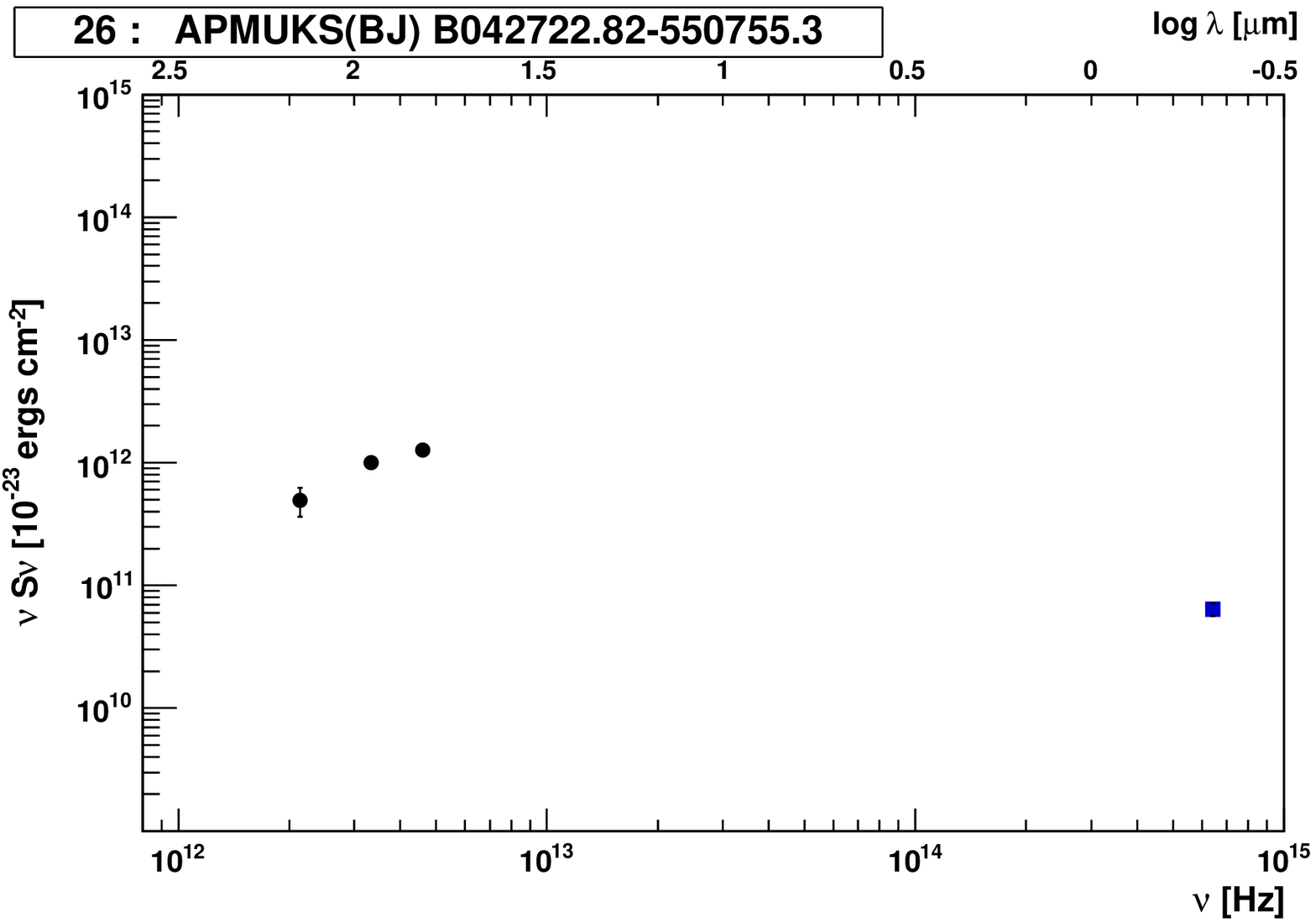}
\includegraphics[width=4cm]{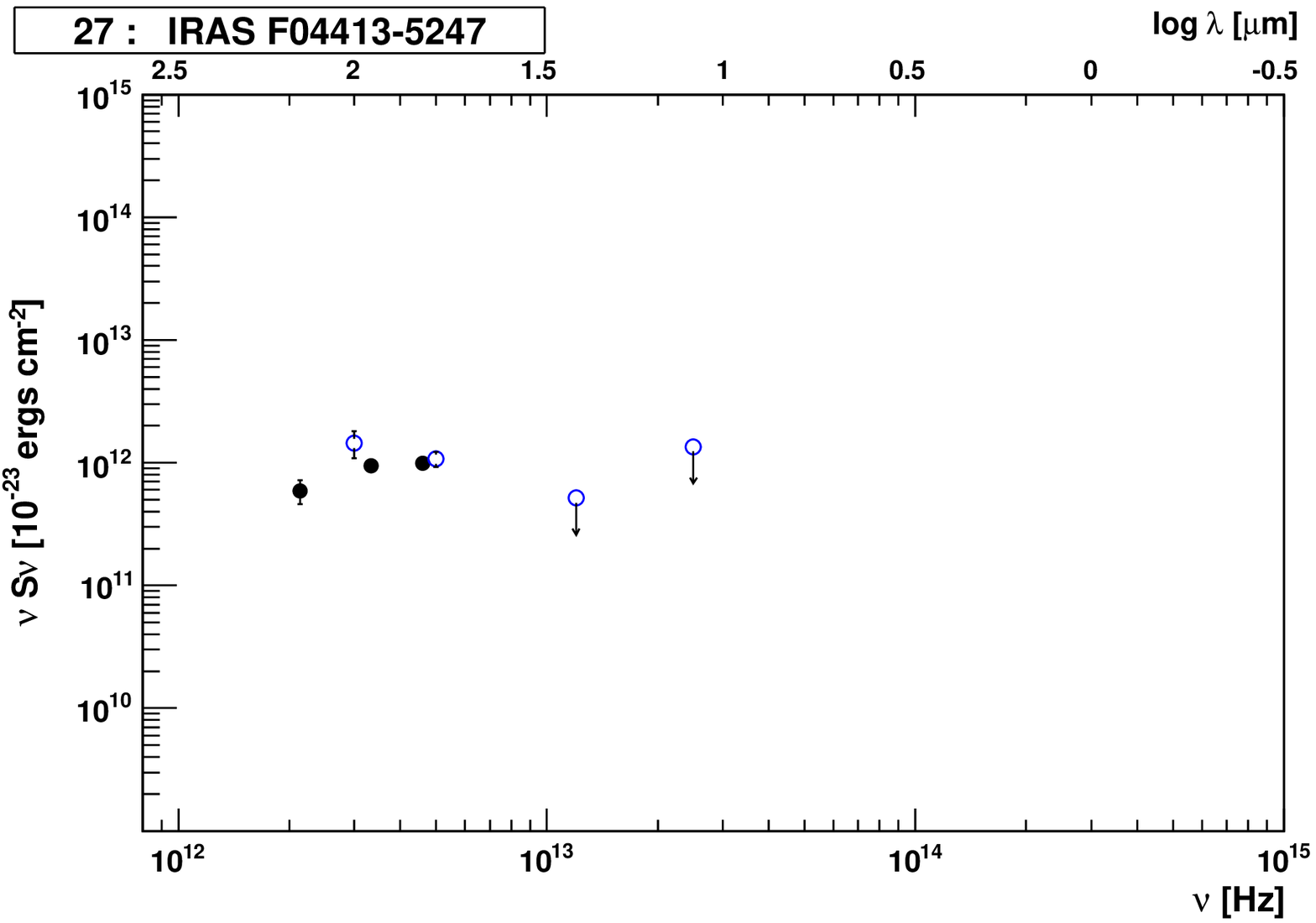}
\includegraphics[width=4cm]{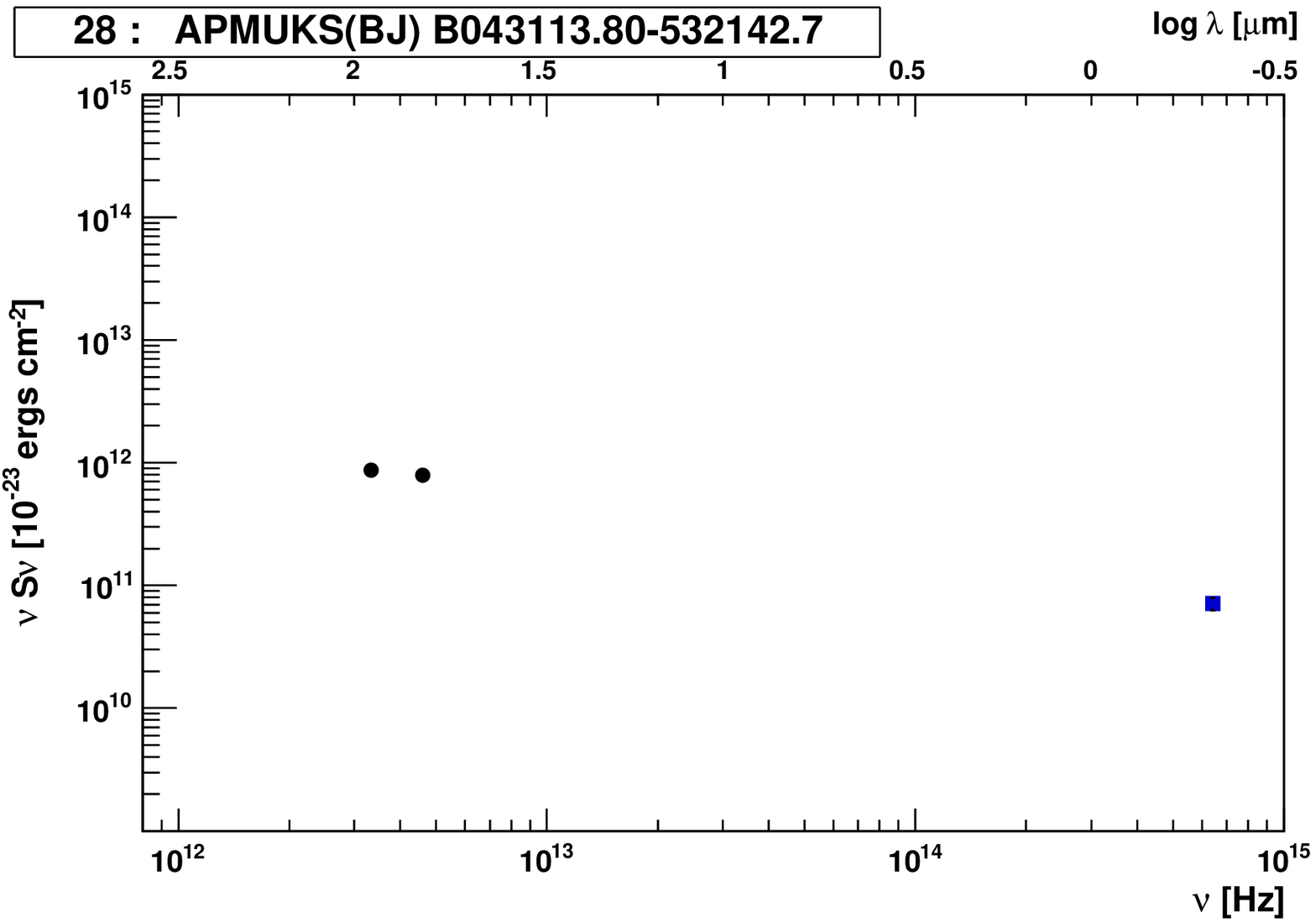}
\includegraphics[width=4cm]{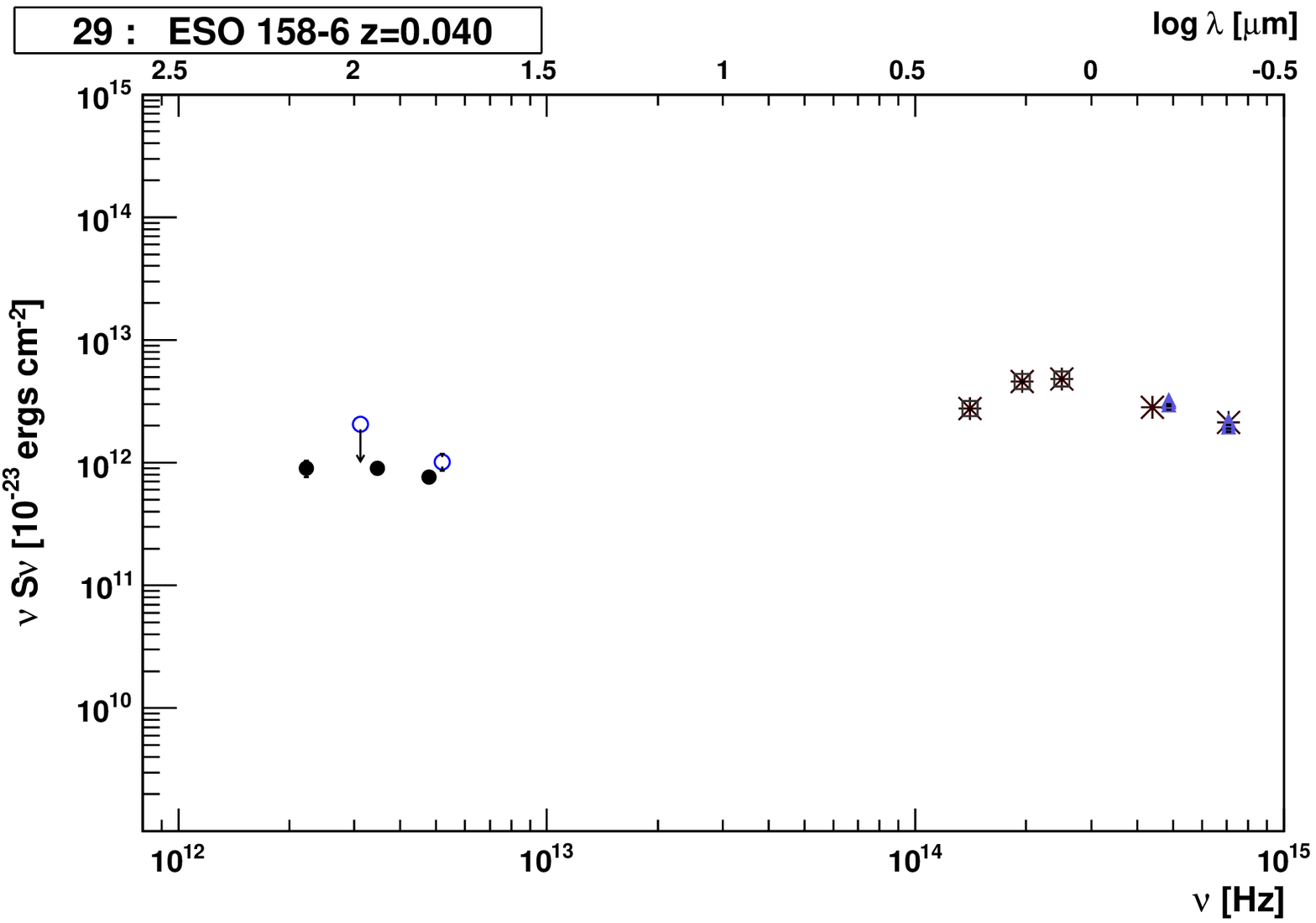}
\includegraphics[width=4cm]{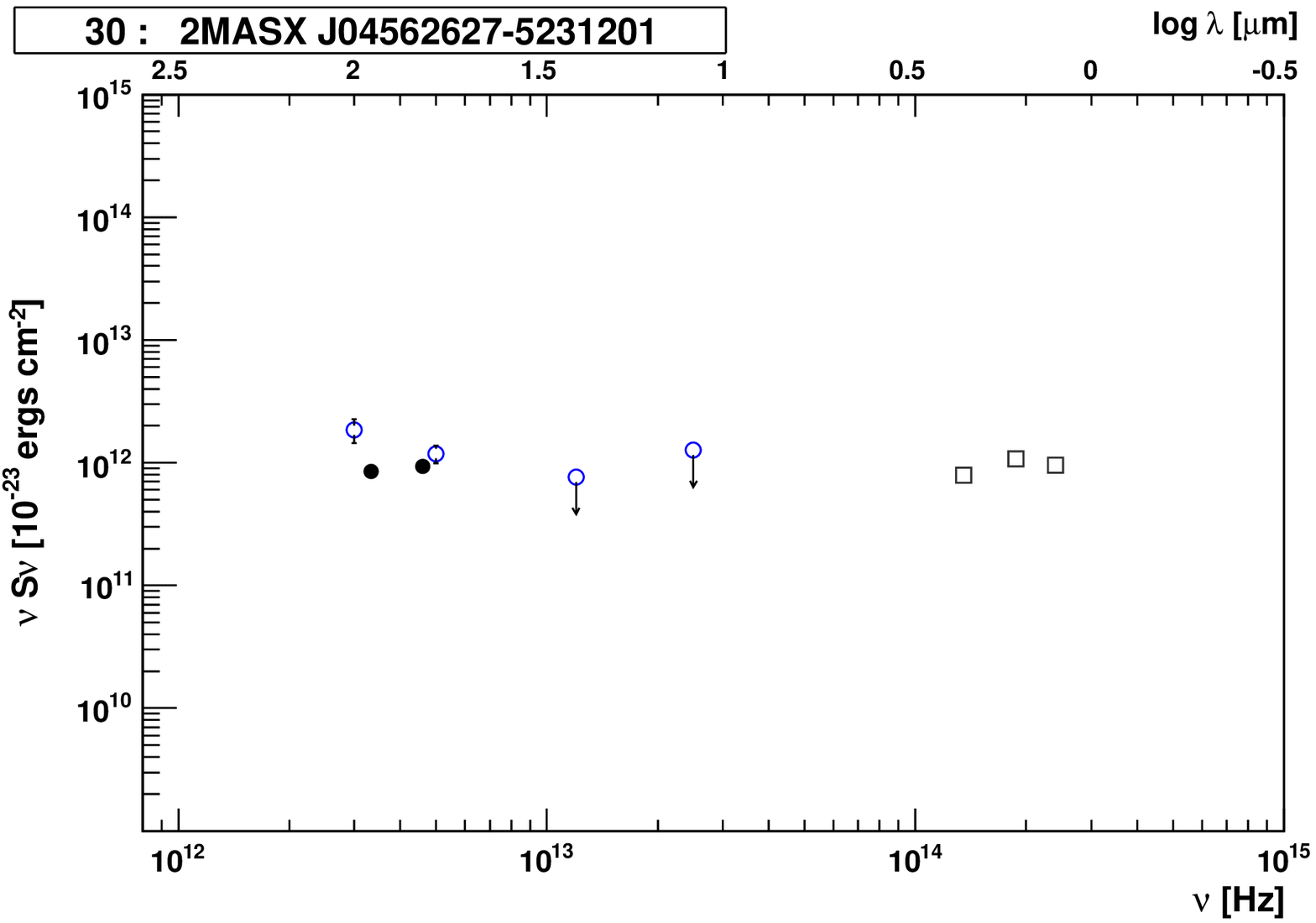}
\includegraphics[width=4cm]{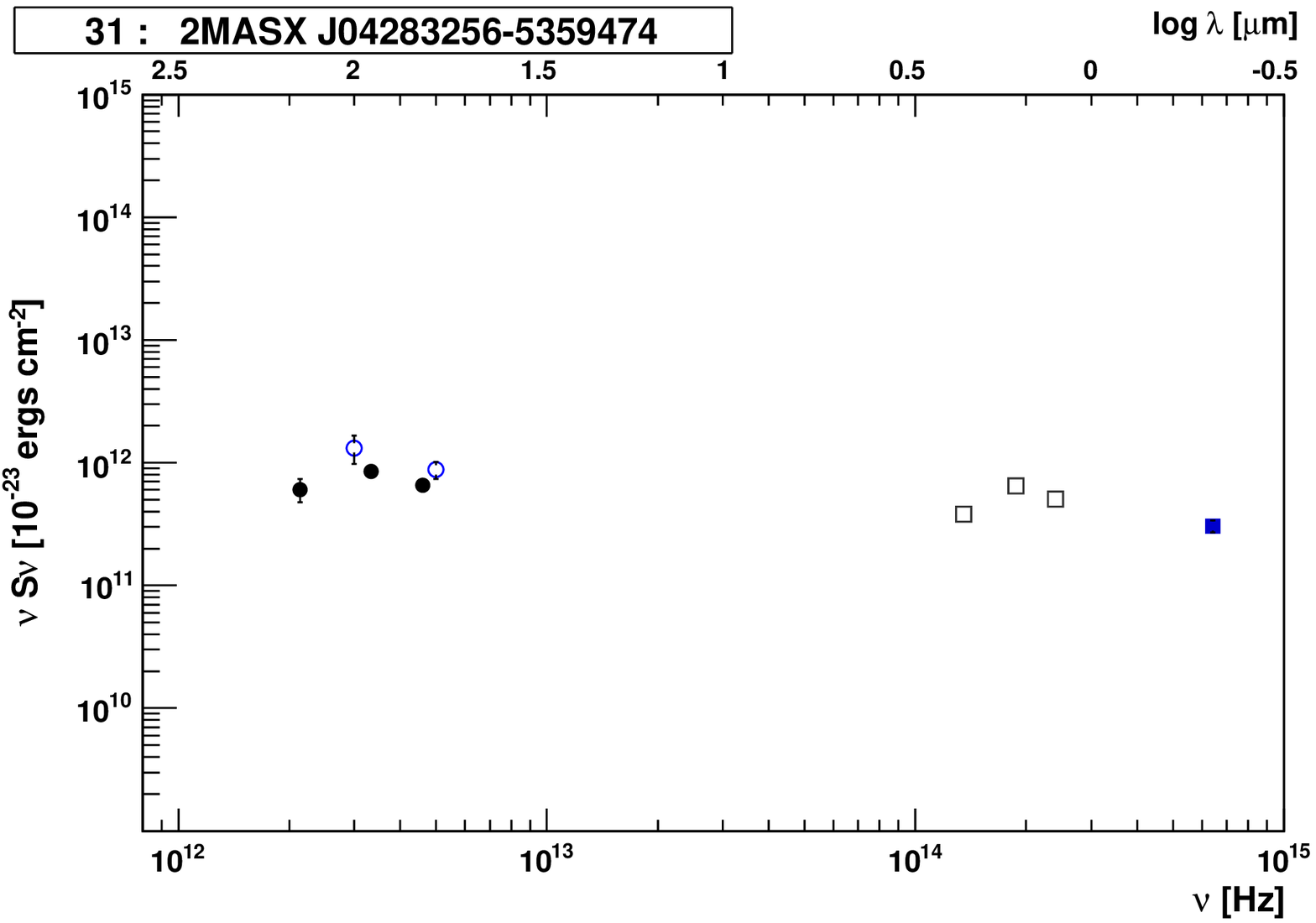}
\includegraphics[width=4cm]{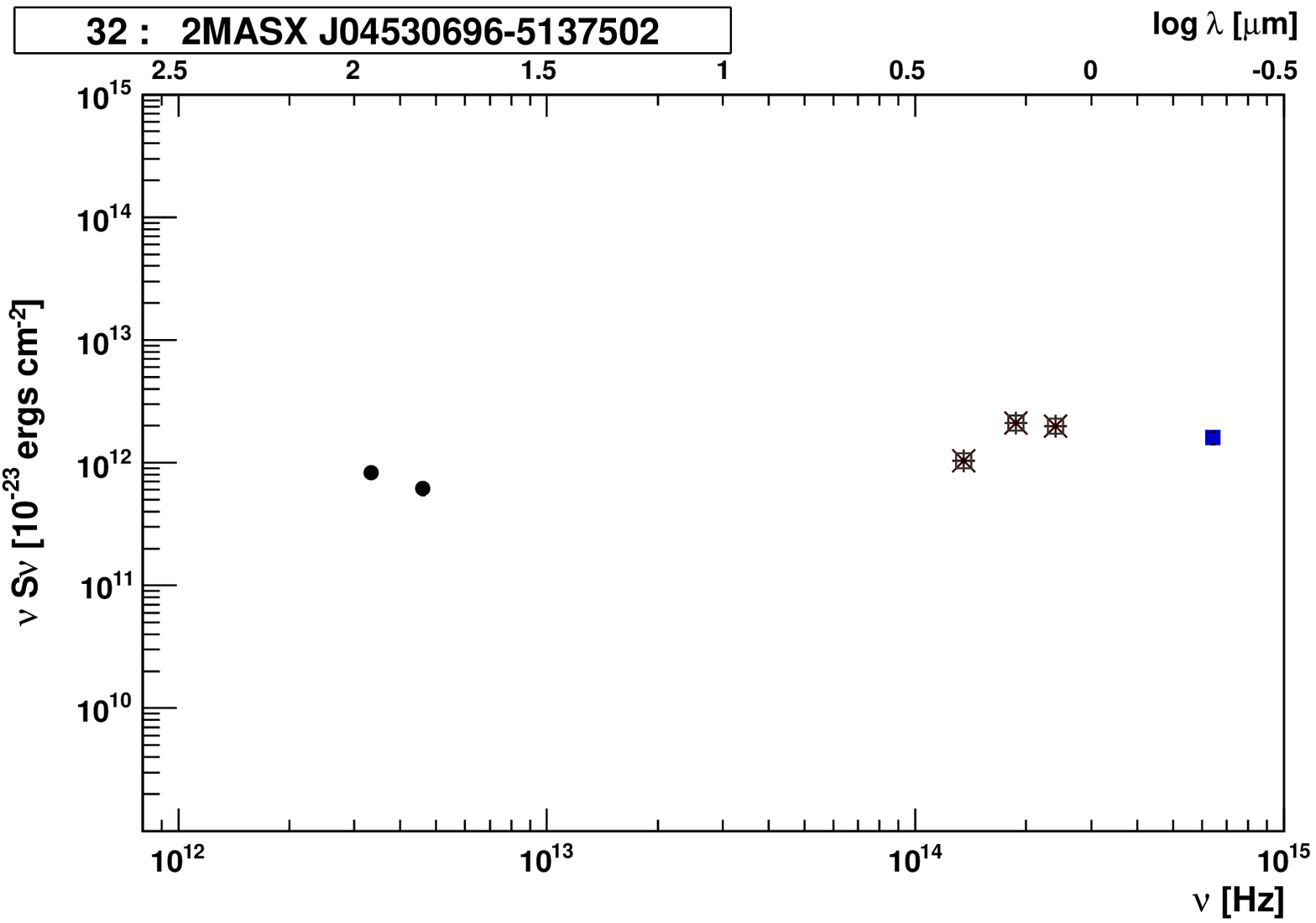}
\includegraphics[width=4cm]{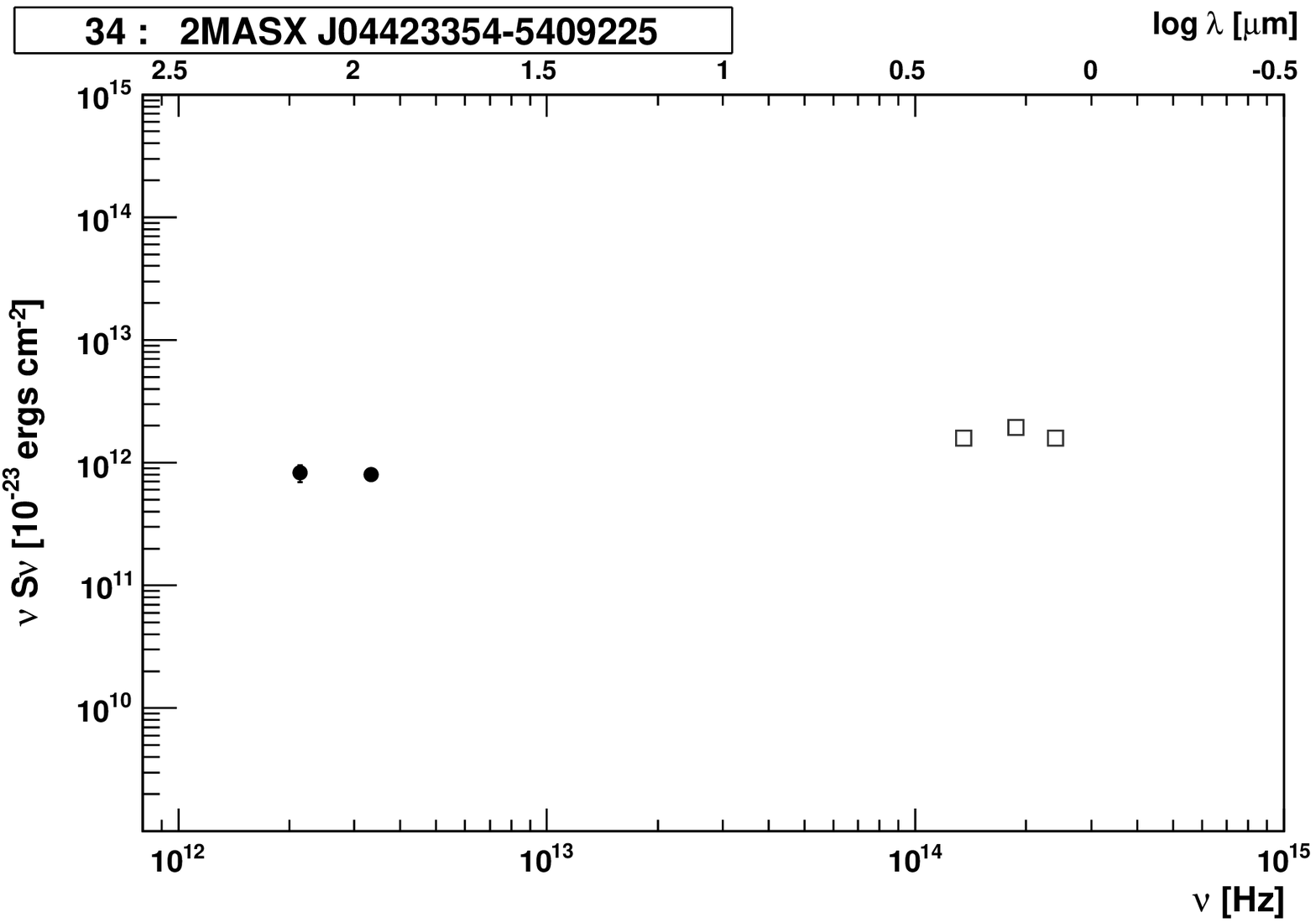}
\includegraphics[width=4cm]{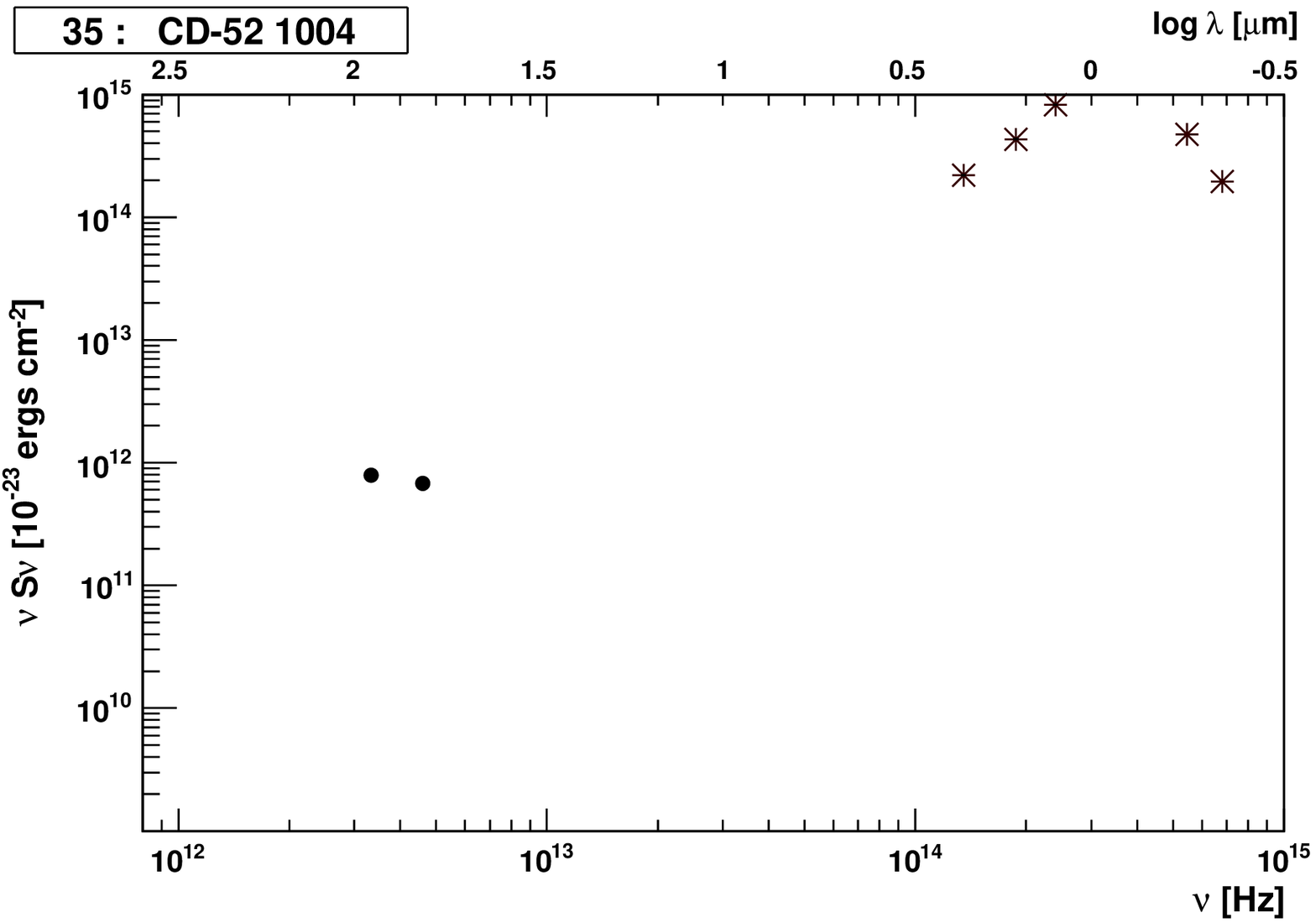}
\includegraphics[width=4cm]{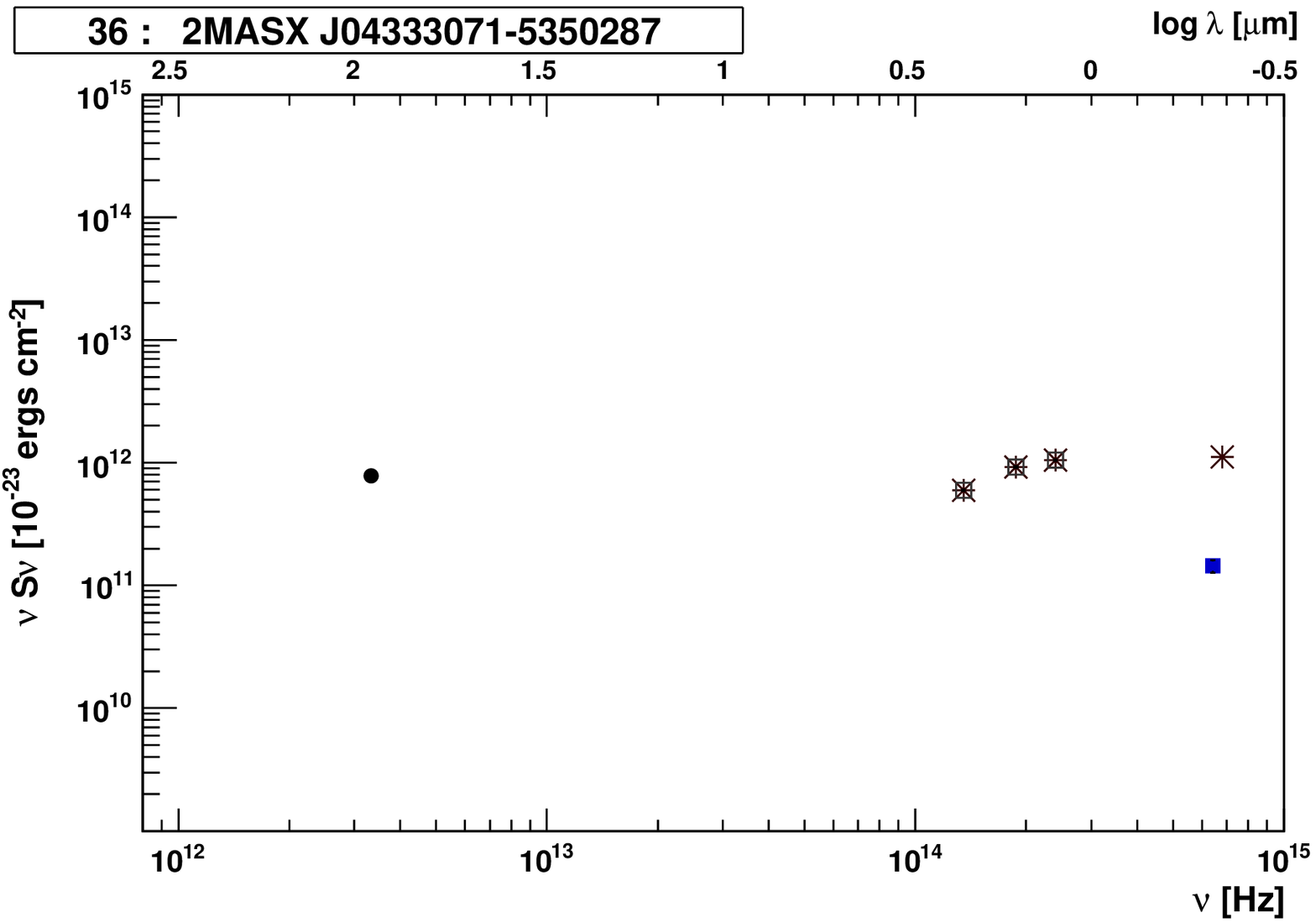}
\includegraphics[width=4cm]{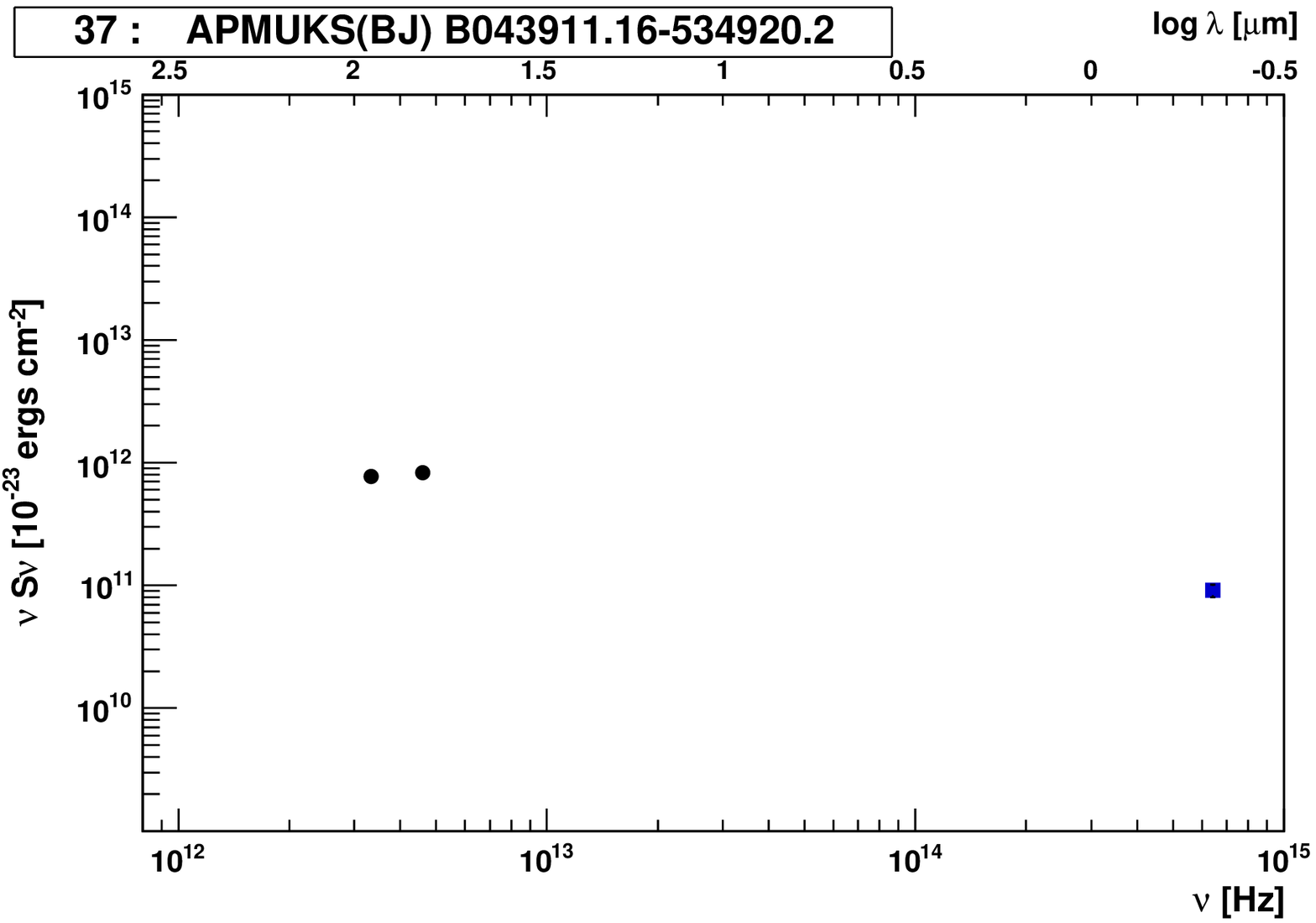}
\caption {SEDs for 
first 36 infrared ADF-S identified sources, using all the information listed in online Tables~\ref{ADFSmeasurements}-\ref{measurements2a}. 
The data points from AKARI Deep Field South (full black circles), 2MASS (open gray squares),
SIMBAD database (black eight pointed stars), IRAS (blue open circles), 
ESO/Uppsala
(full blue triangles), APM (full navy blue squares), RC3 (open brown  triangles), ISOPHOT
(five pointed black stars), 
Siding Spring Observatory (five pointed gray stars), GALEX
(full green triangles), HIPASS catalogue (full green circles), Palomar/Las Campanas Imaging
Atlas of Blue Compact Dwarf Galaxies
(full magenta squares), IUE (open black diamonds), Spitzer (full red squares), 
FUSE
(upside-down red triangles) and UV: 1650, 2500, 3150 nm (full navy blue upside-down triangles) and SUMMS at $843MHz$ (magenta open crosses) are shown. 
Whenever the redshift is available (and shown in the plot), 
the data are presented in the rest frame. In other cases the observed frame is used.} 
\label{points1}
\end{figure*}
}

\clearpage

 \onlfig{2}{
 \begin{figure*}[t]
\centering
\includegraphics[width=4cm]{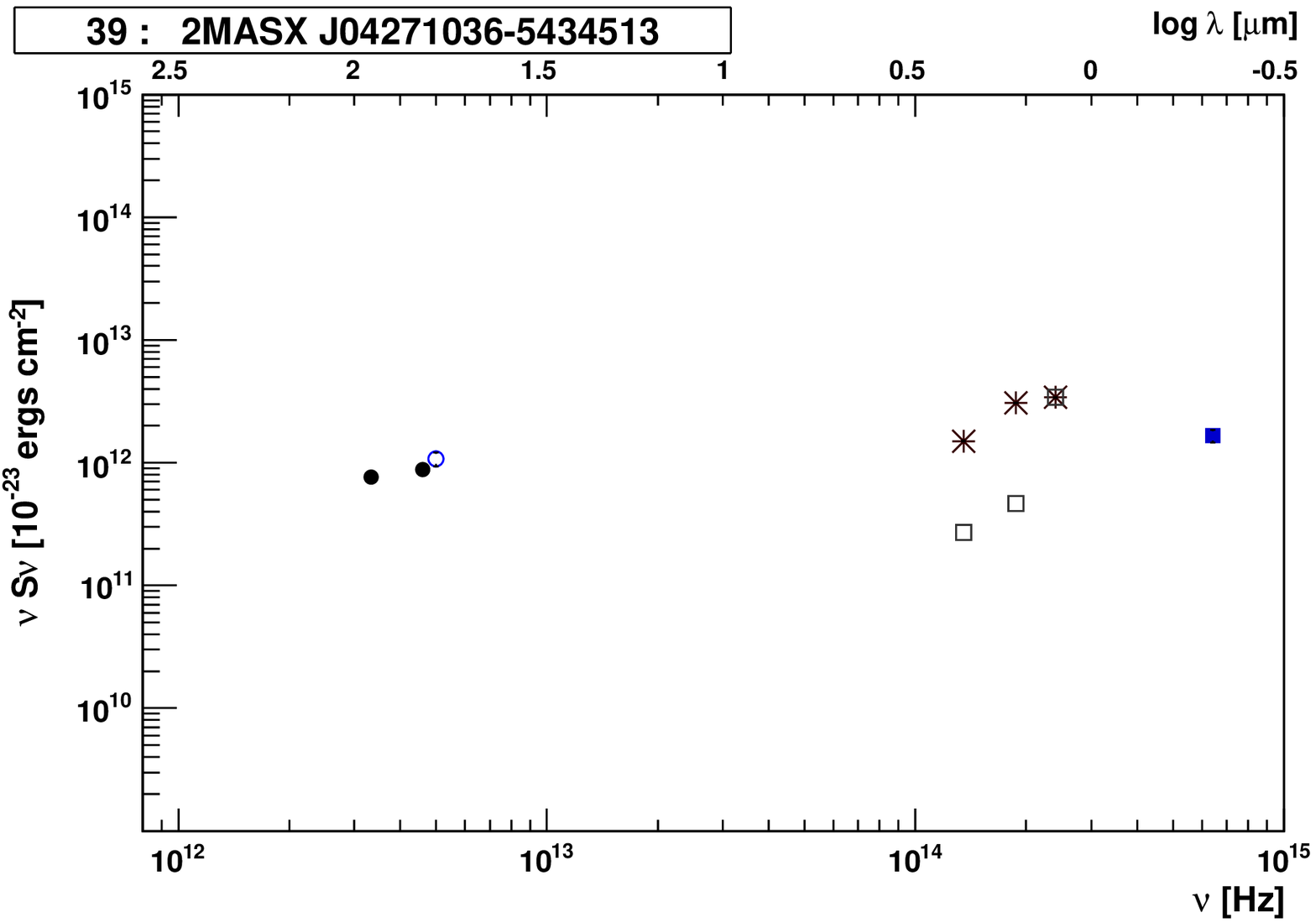}
\includegraphics[width=4cm]{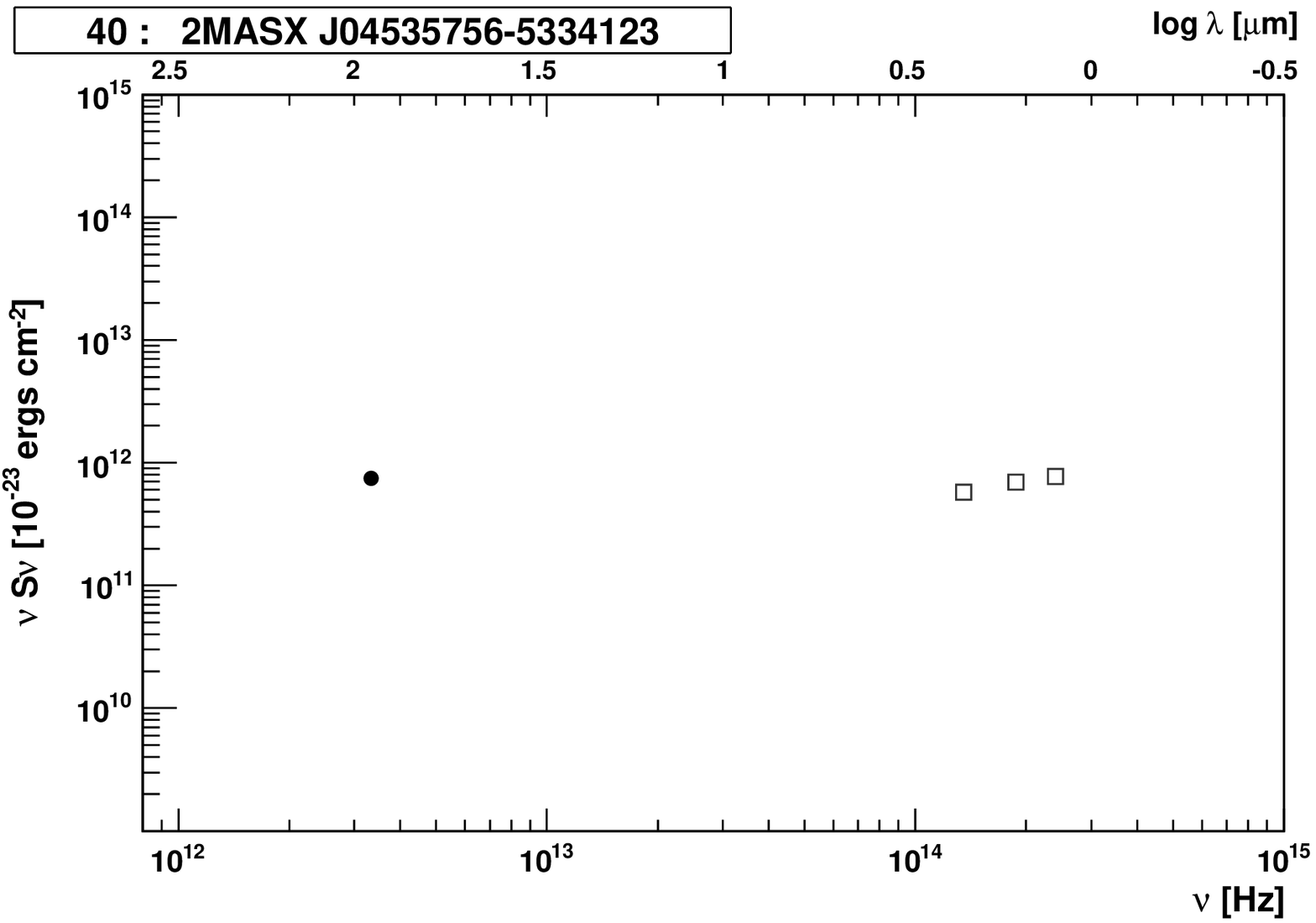}
\includegraphics[width=4cm]{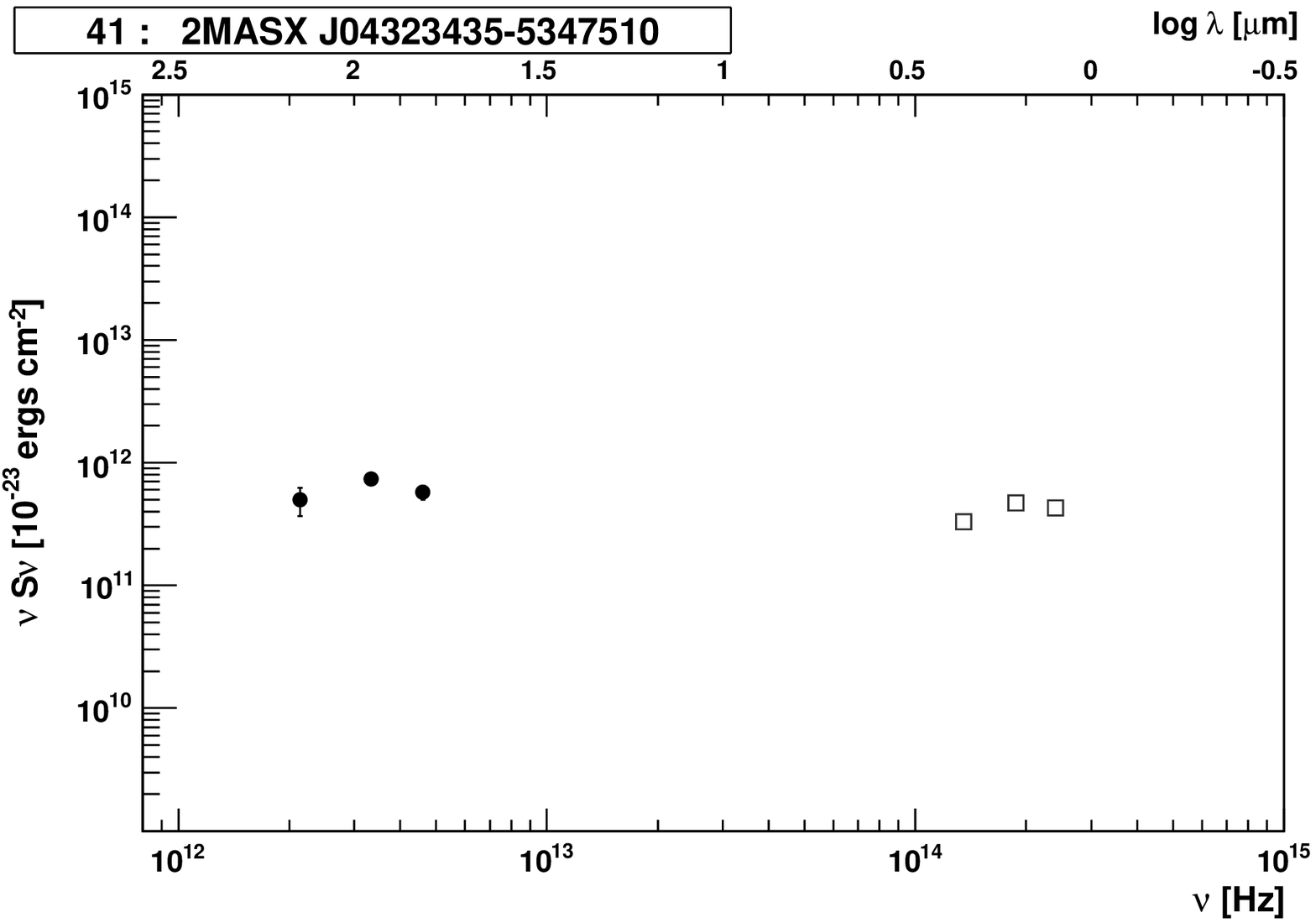}
\includegraphics[width=4cm]{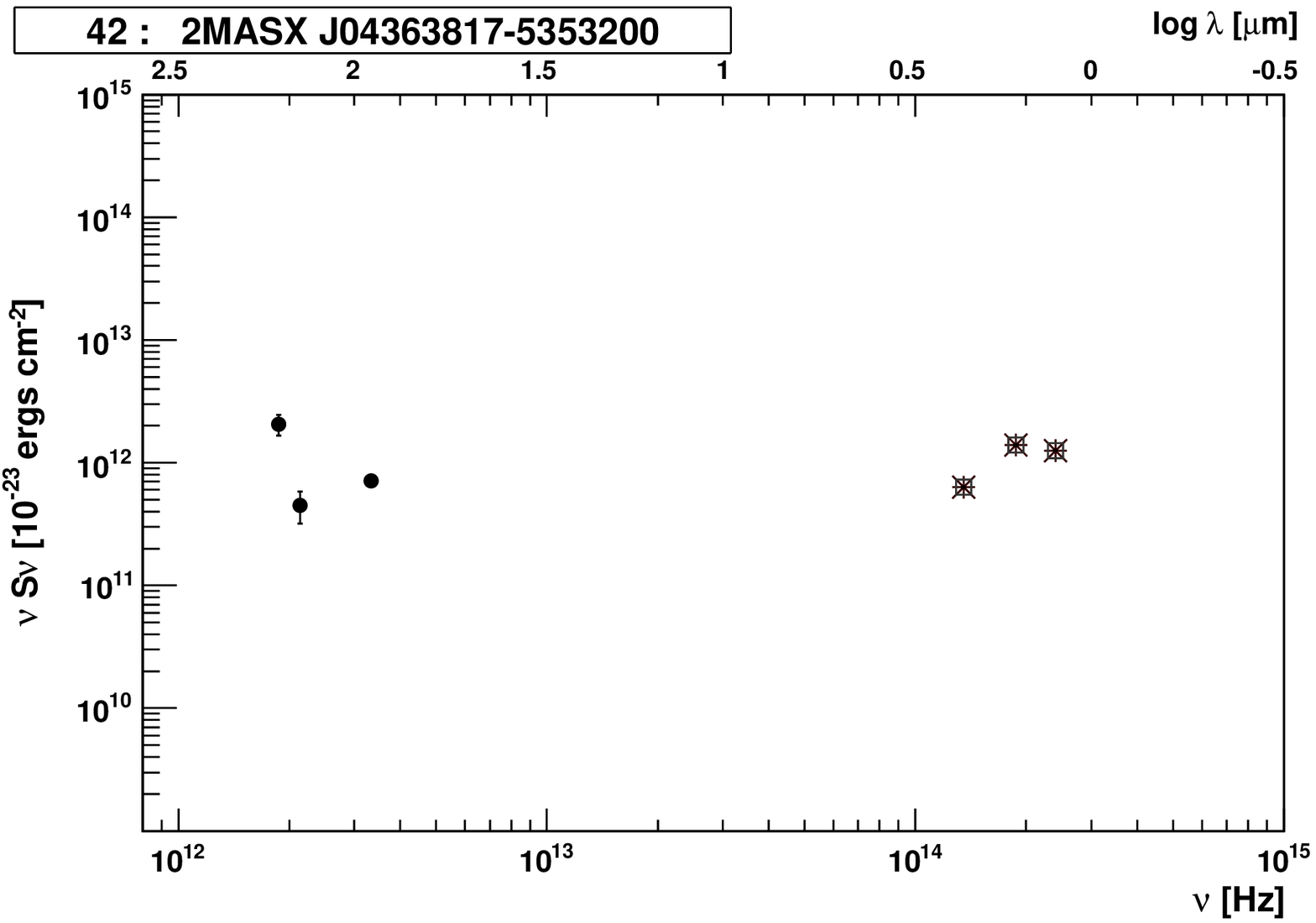}
\includegraphics[width=4cm]{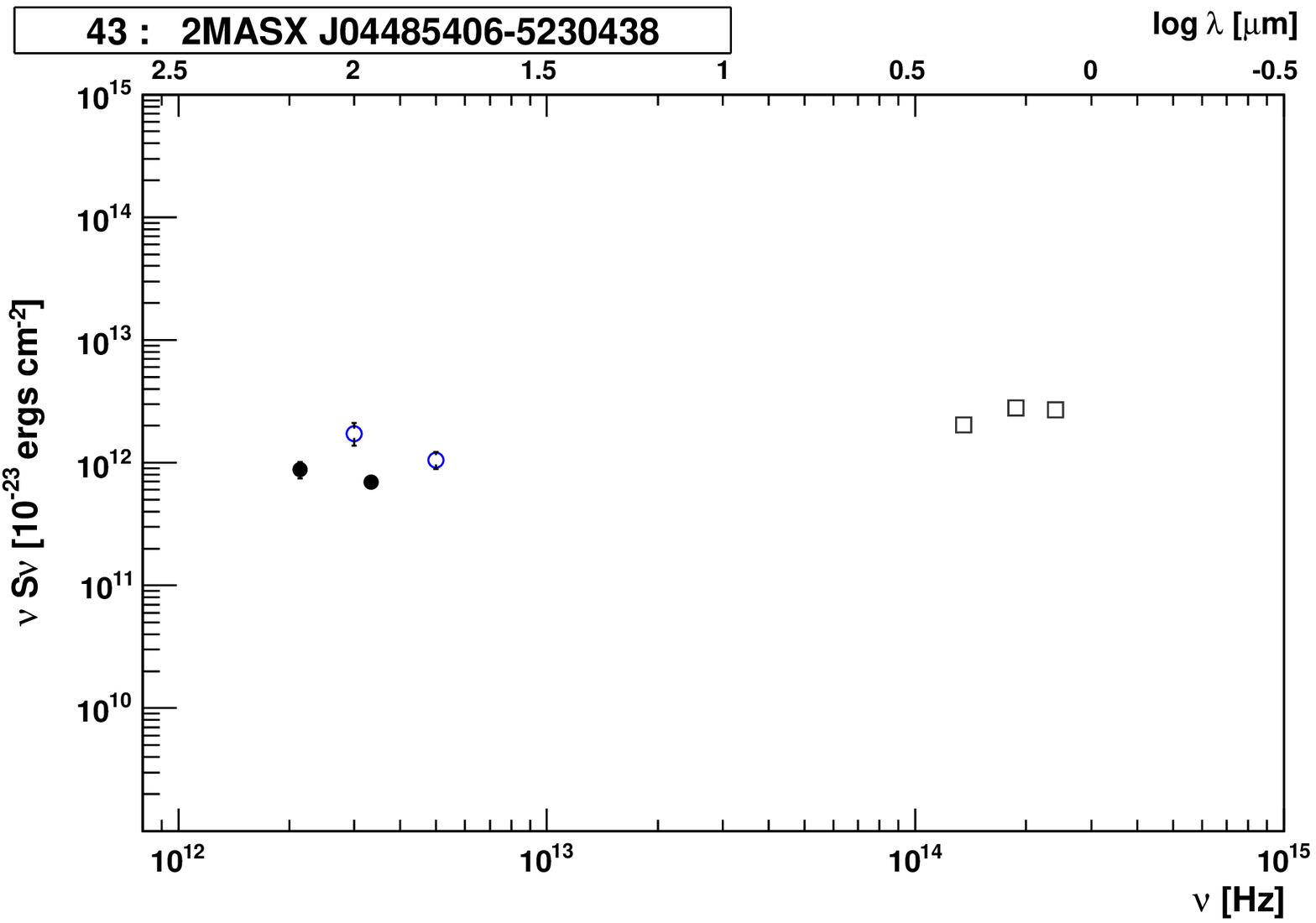}
\includegraphics[width=4cm]{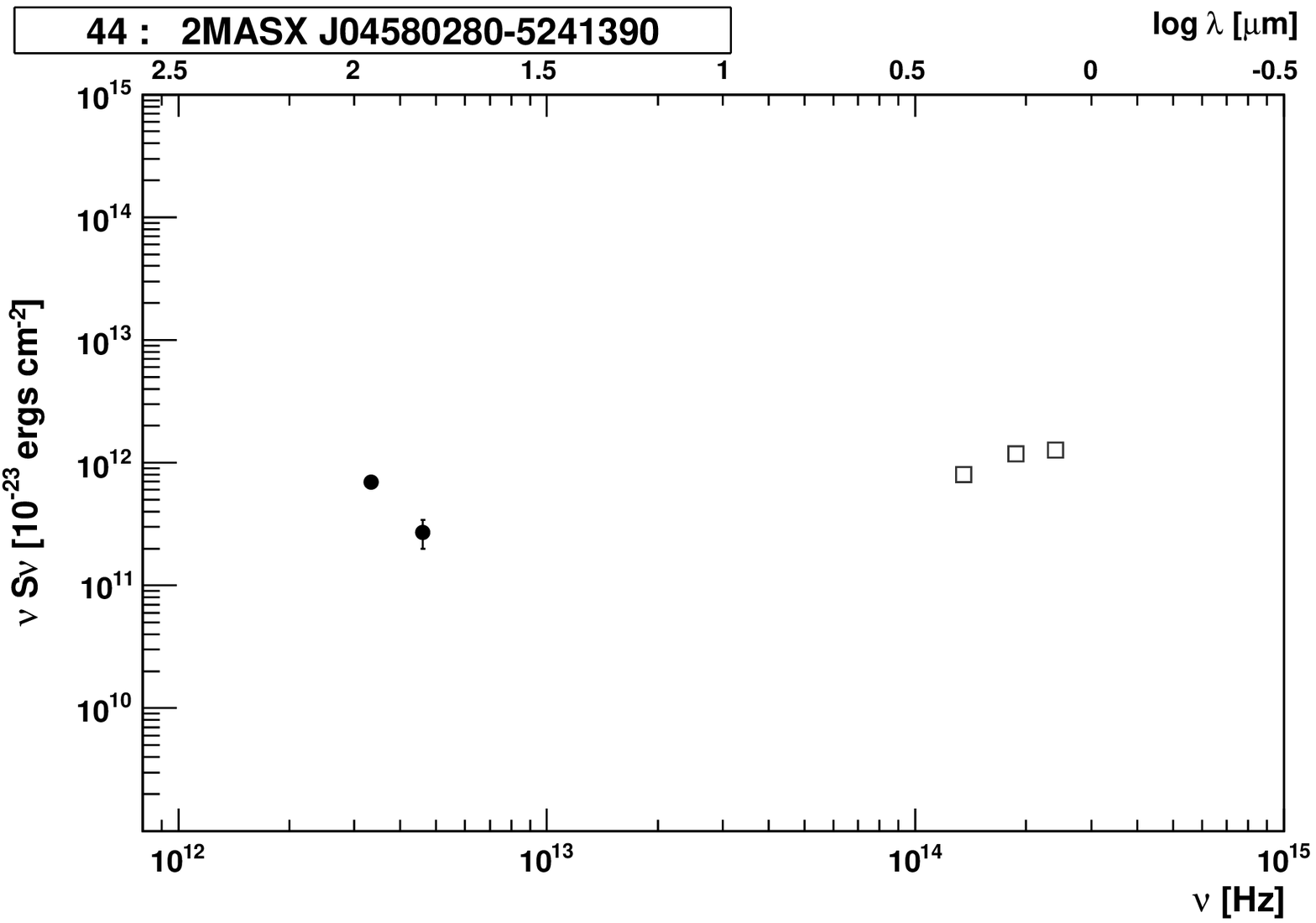}
\includegraphics[width=4cm]{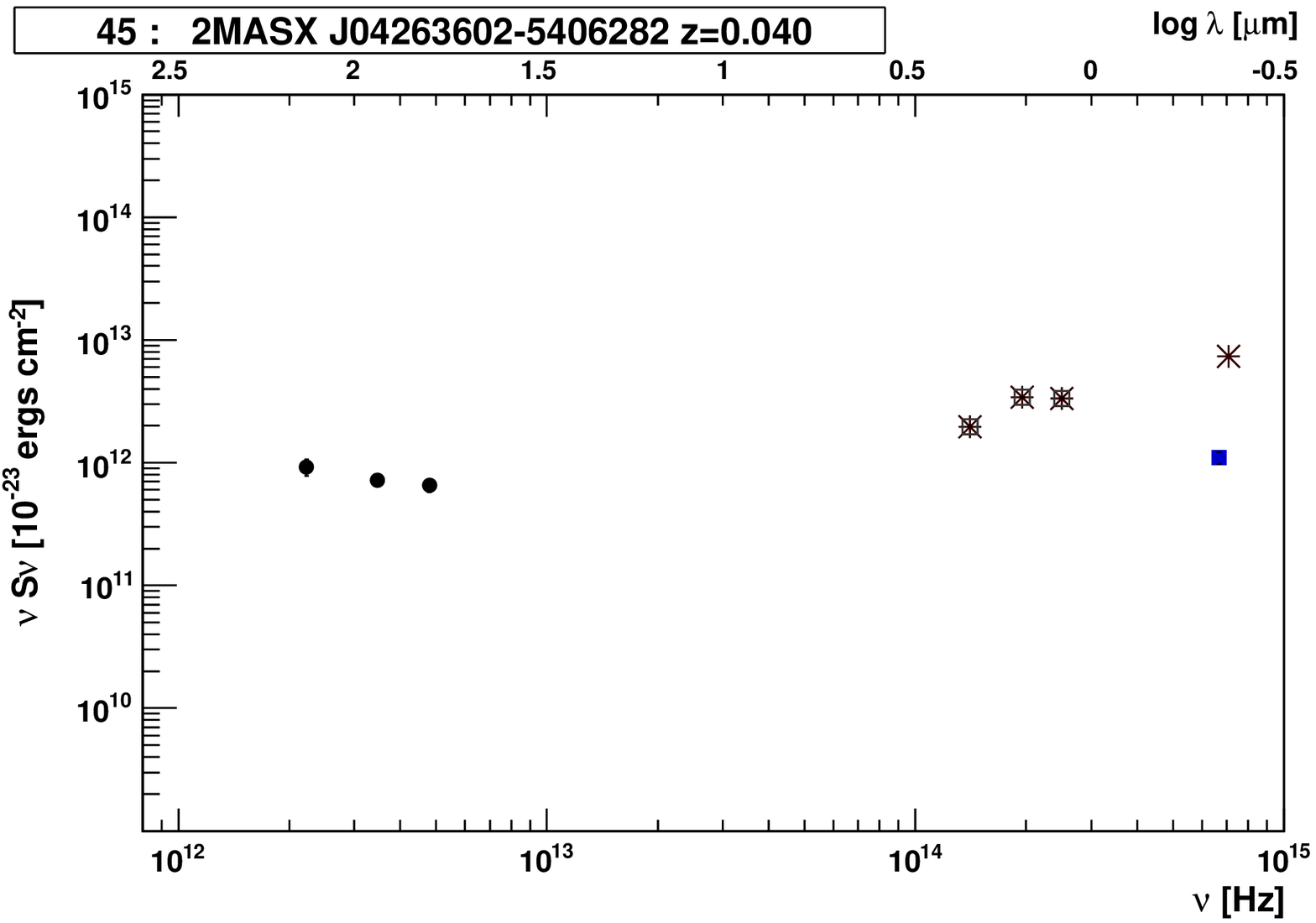}
\includegraphics[width=4cm]{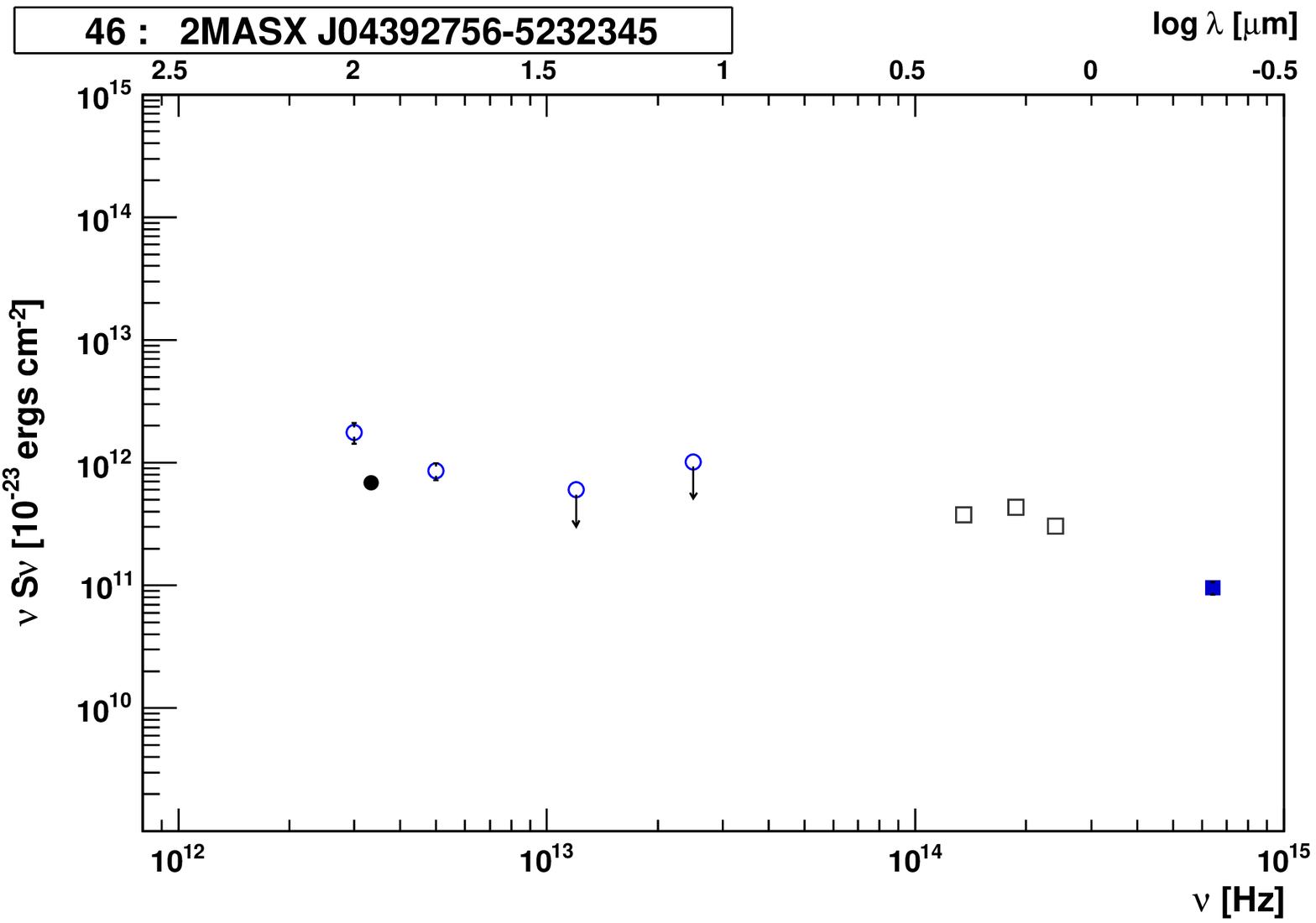}
\includegraphics[width=4cm]{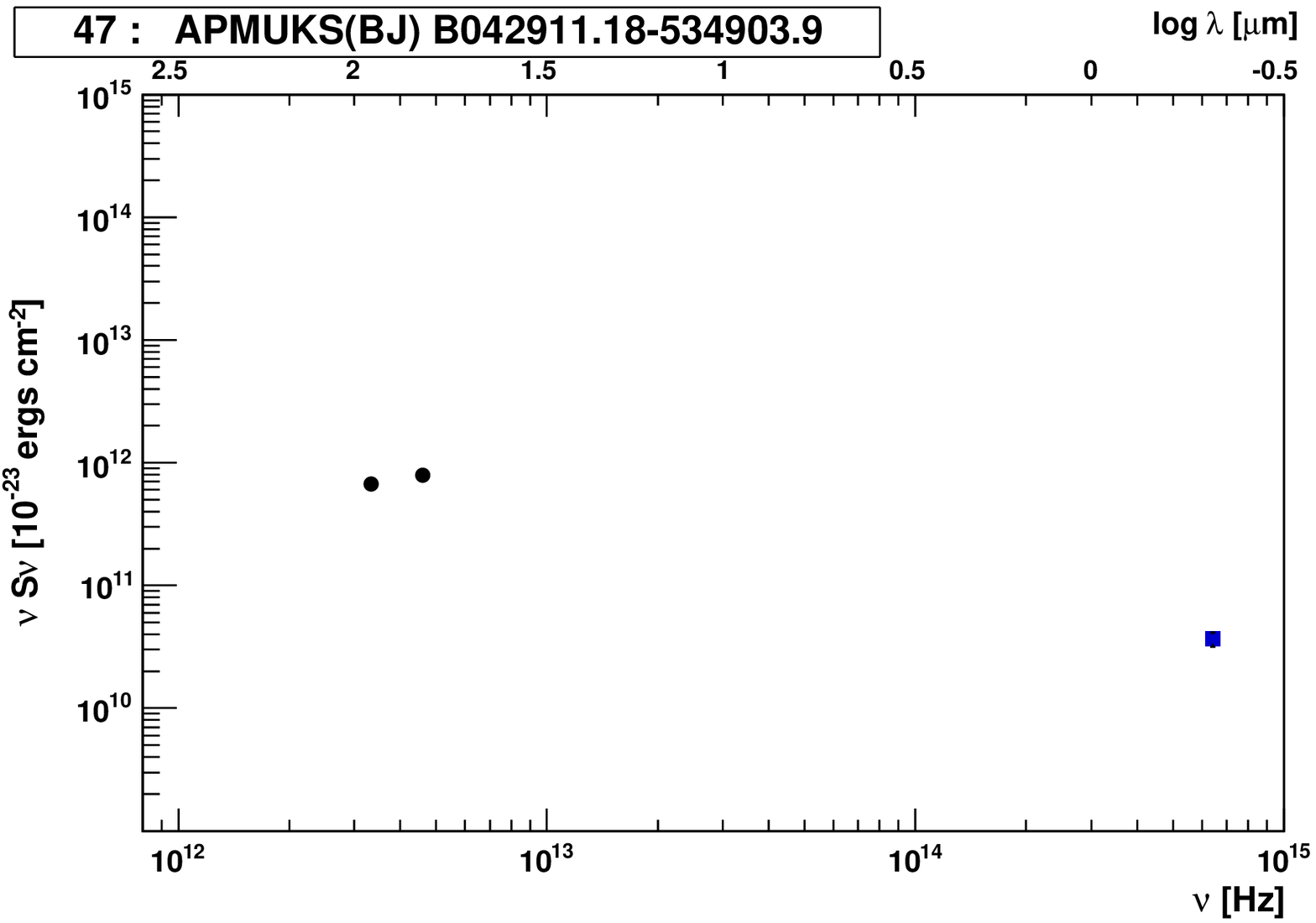}
\includegraphics[width=4cm]{points/kmalek_44.eps}
\includegraphics[width=4cm]{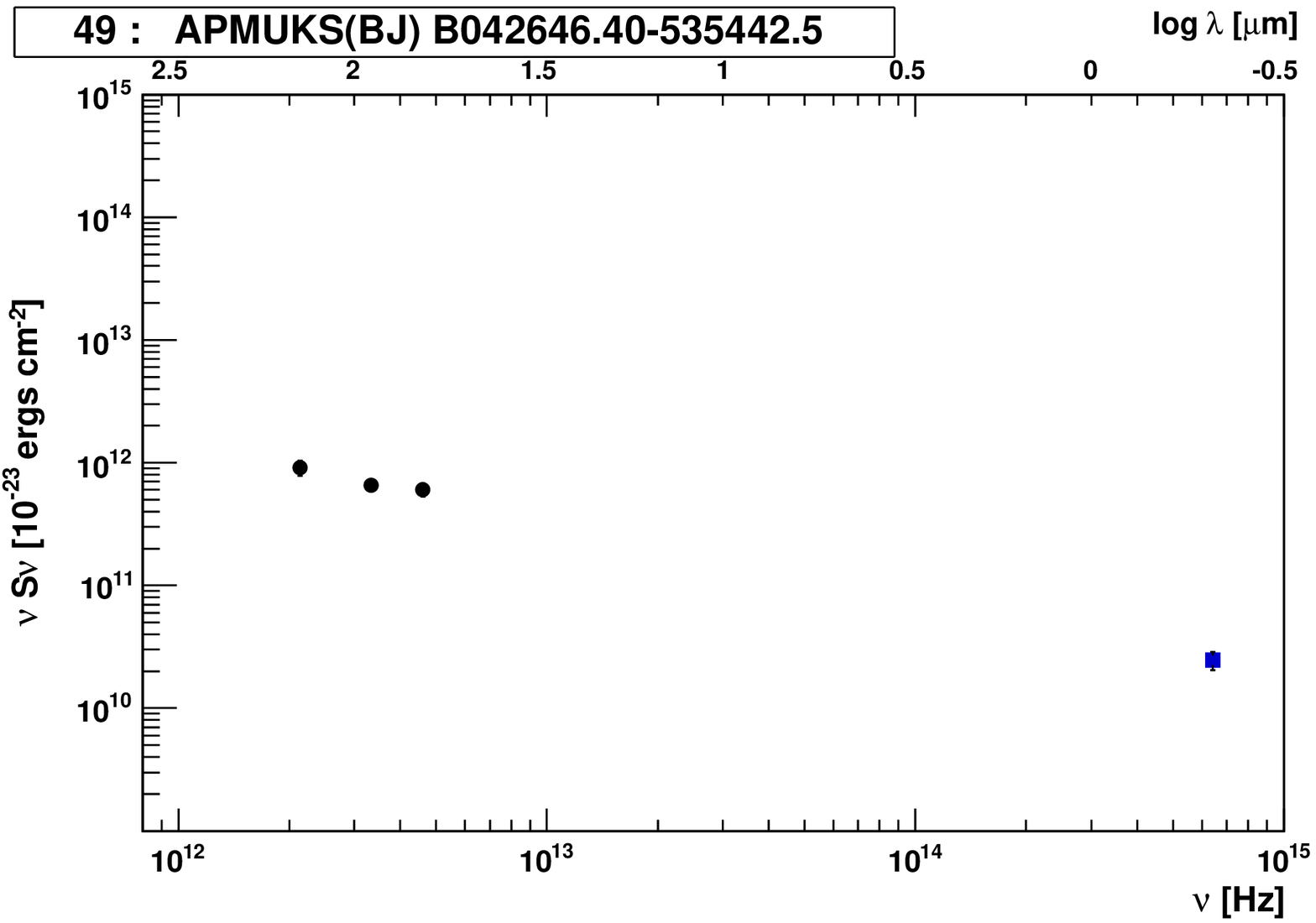}
\includegraphics[width=4cm]{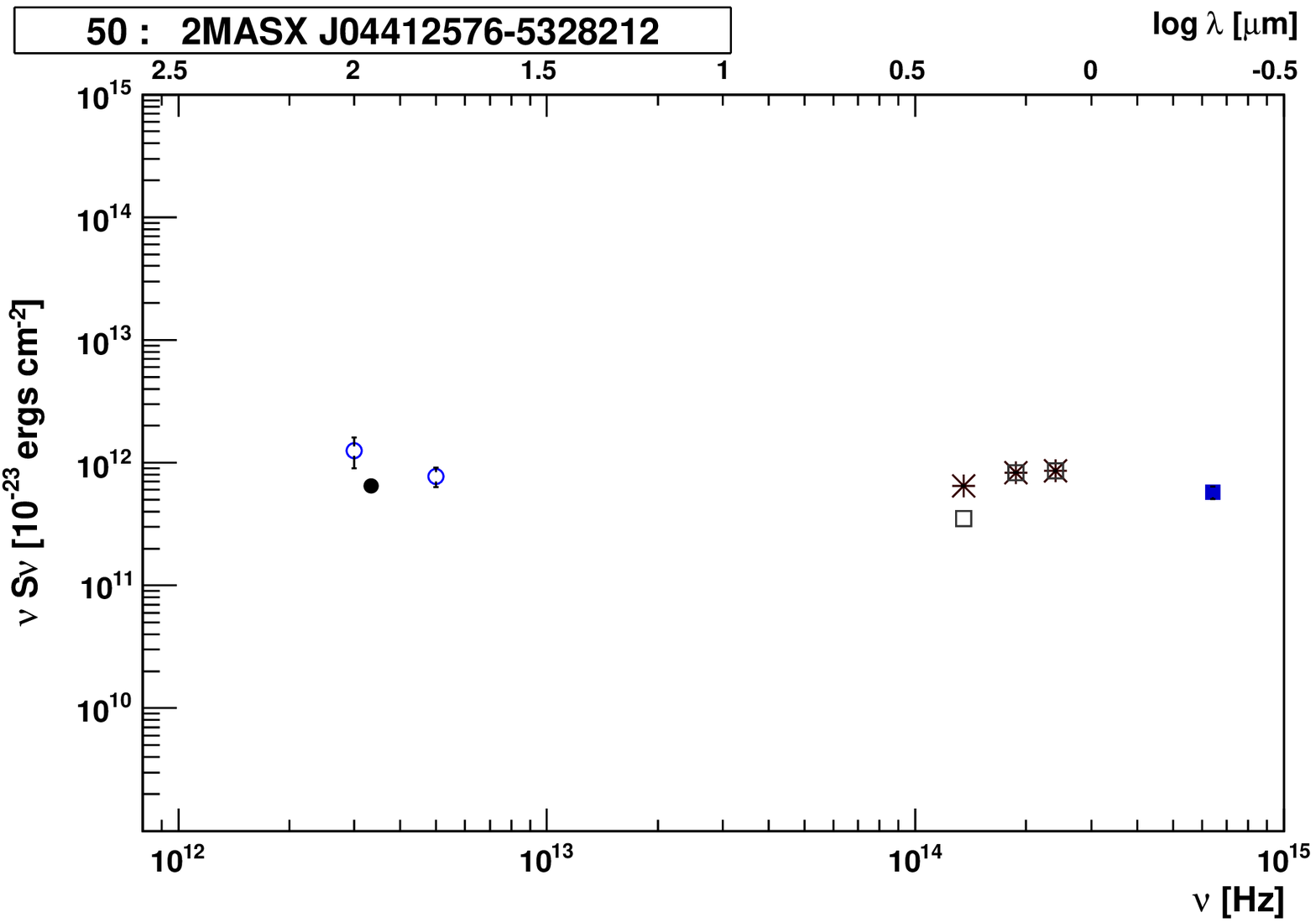}
\includegraphics[width=4cm]{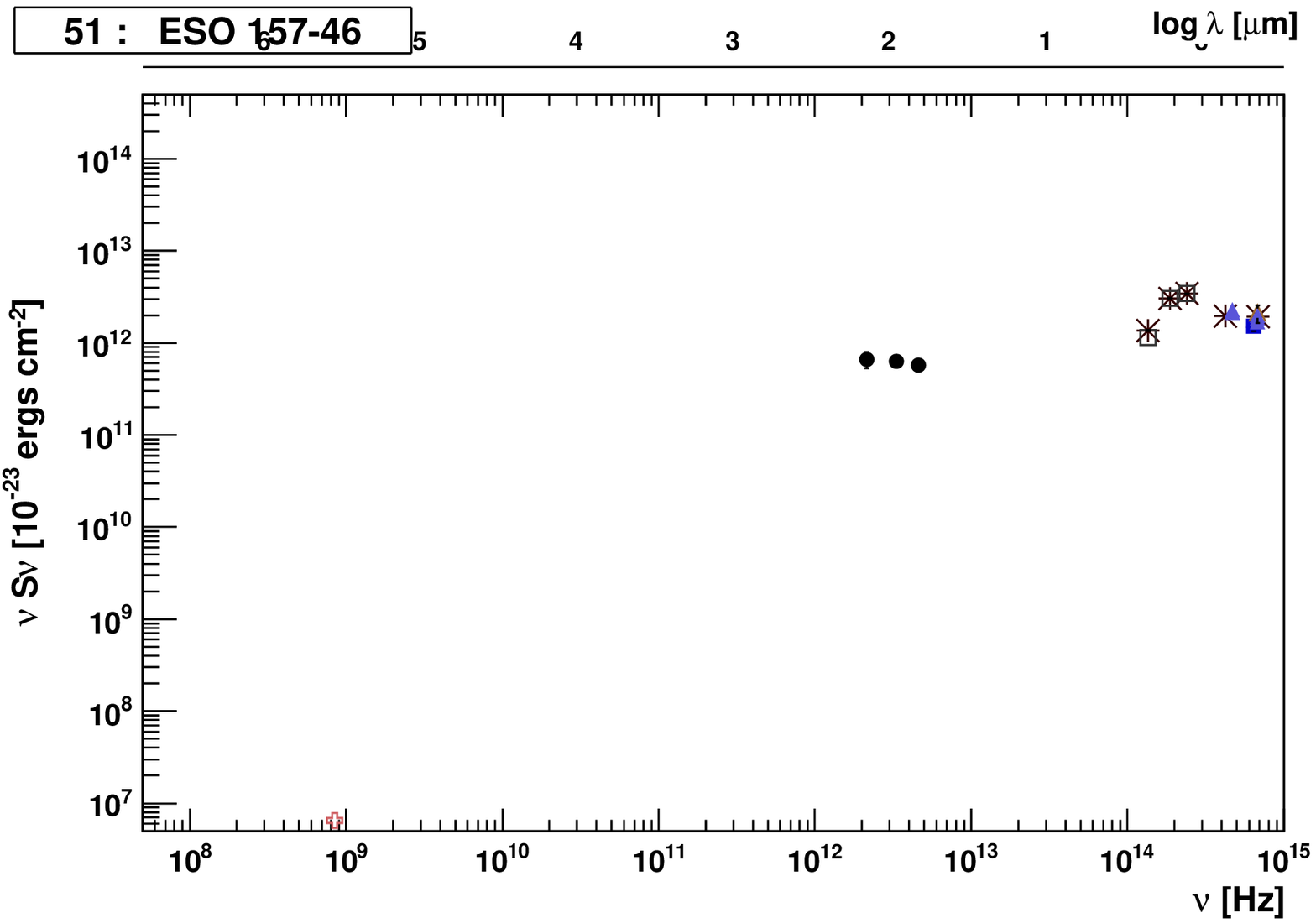}
\includegraphics[width=4cm]{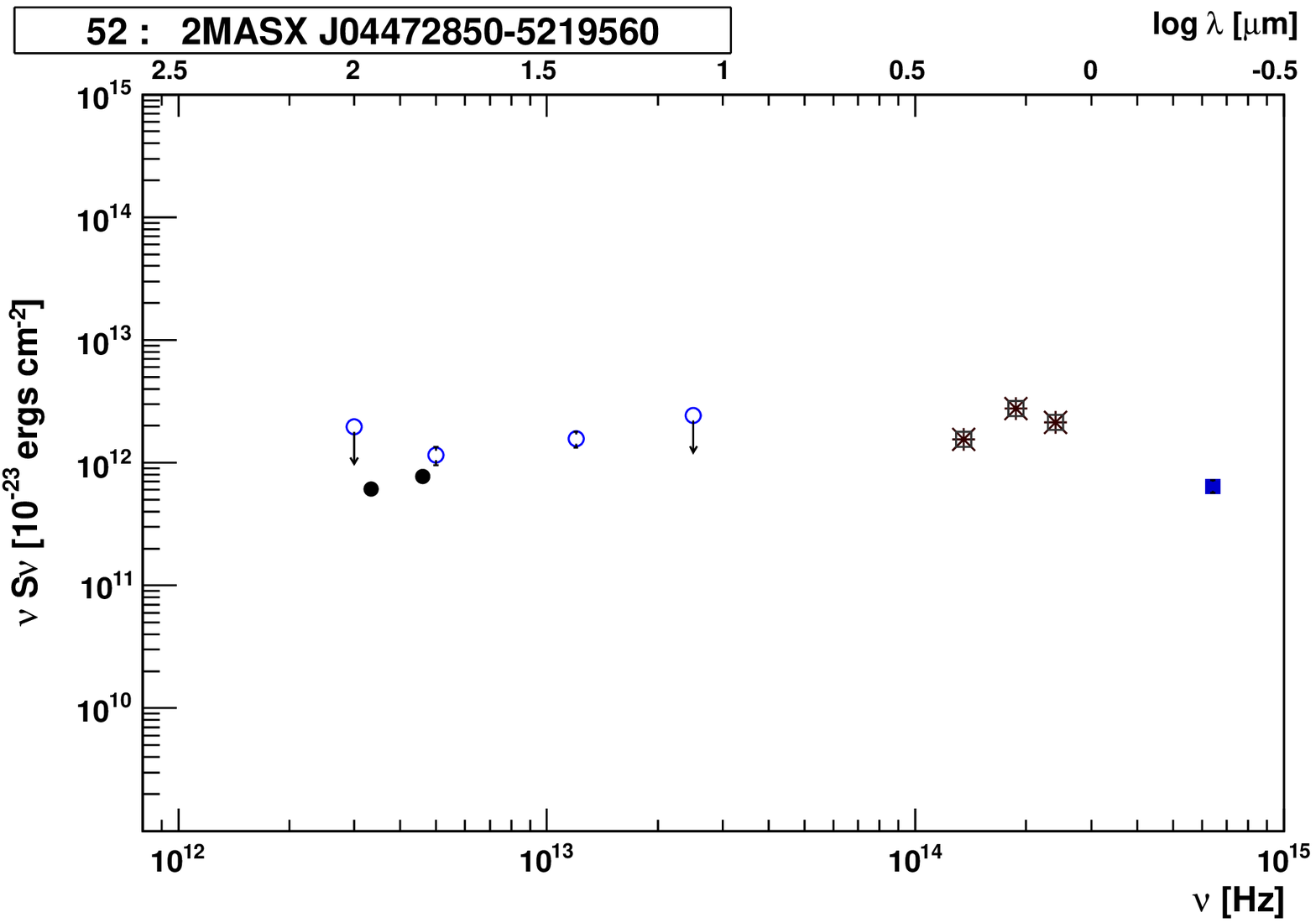}
\includegraphics[width=4cm]{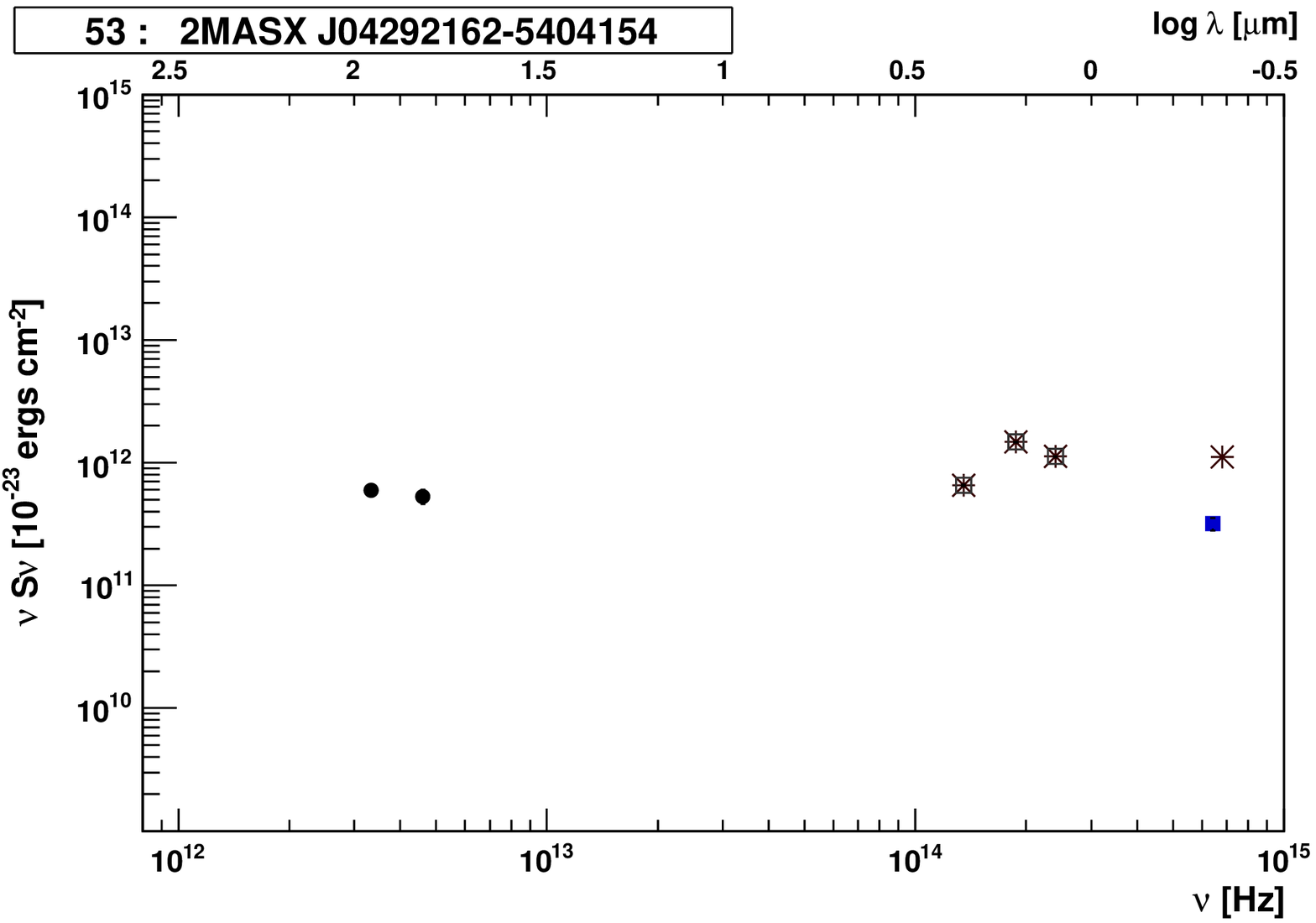}
\includegraphics[width=4cm]{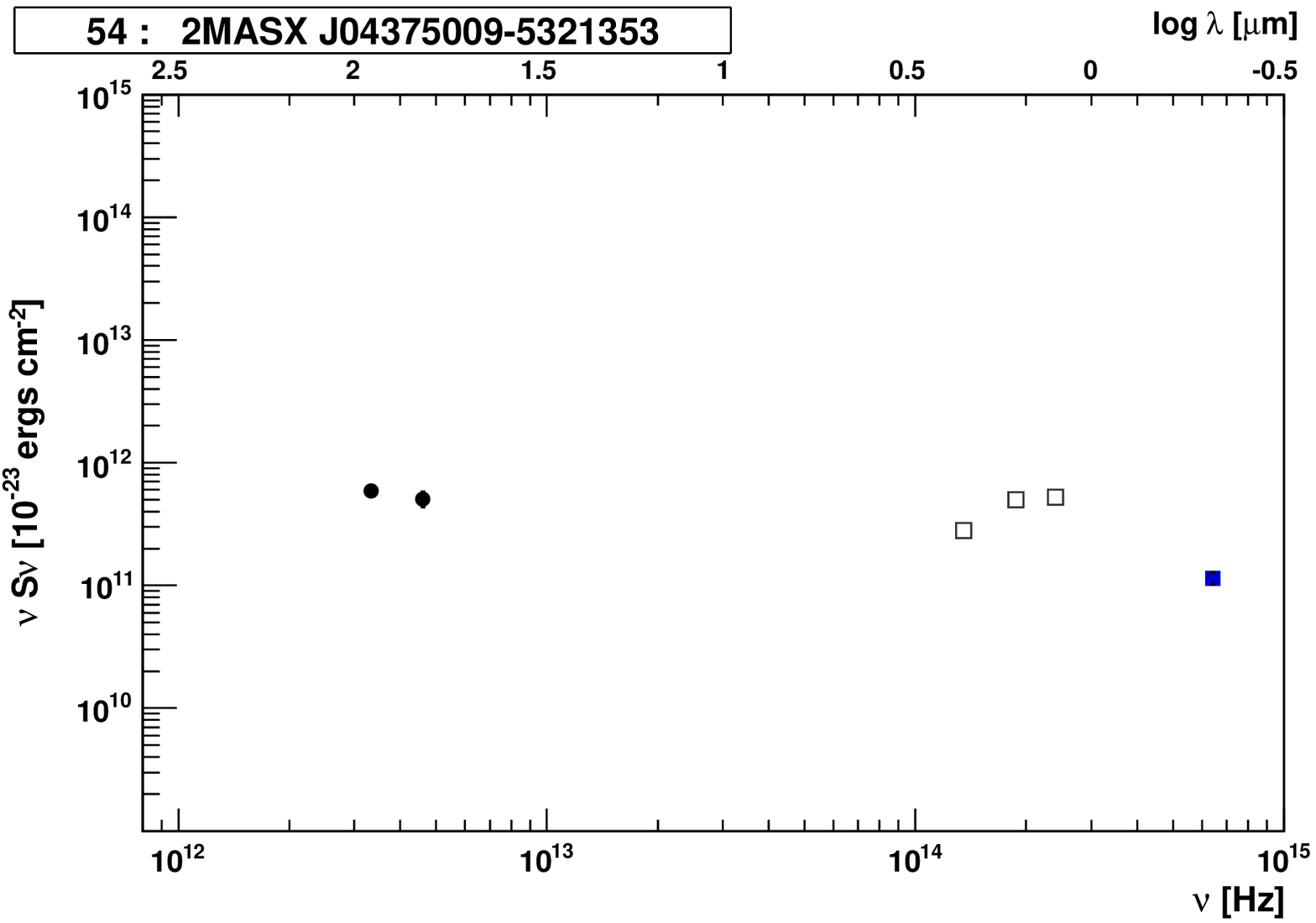}
\includegraphics[width=4cm]{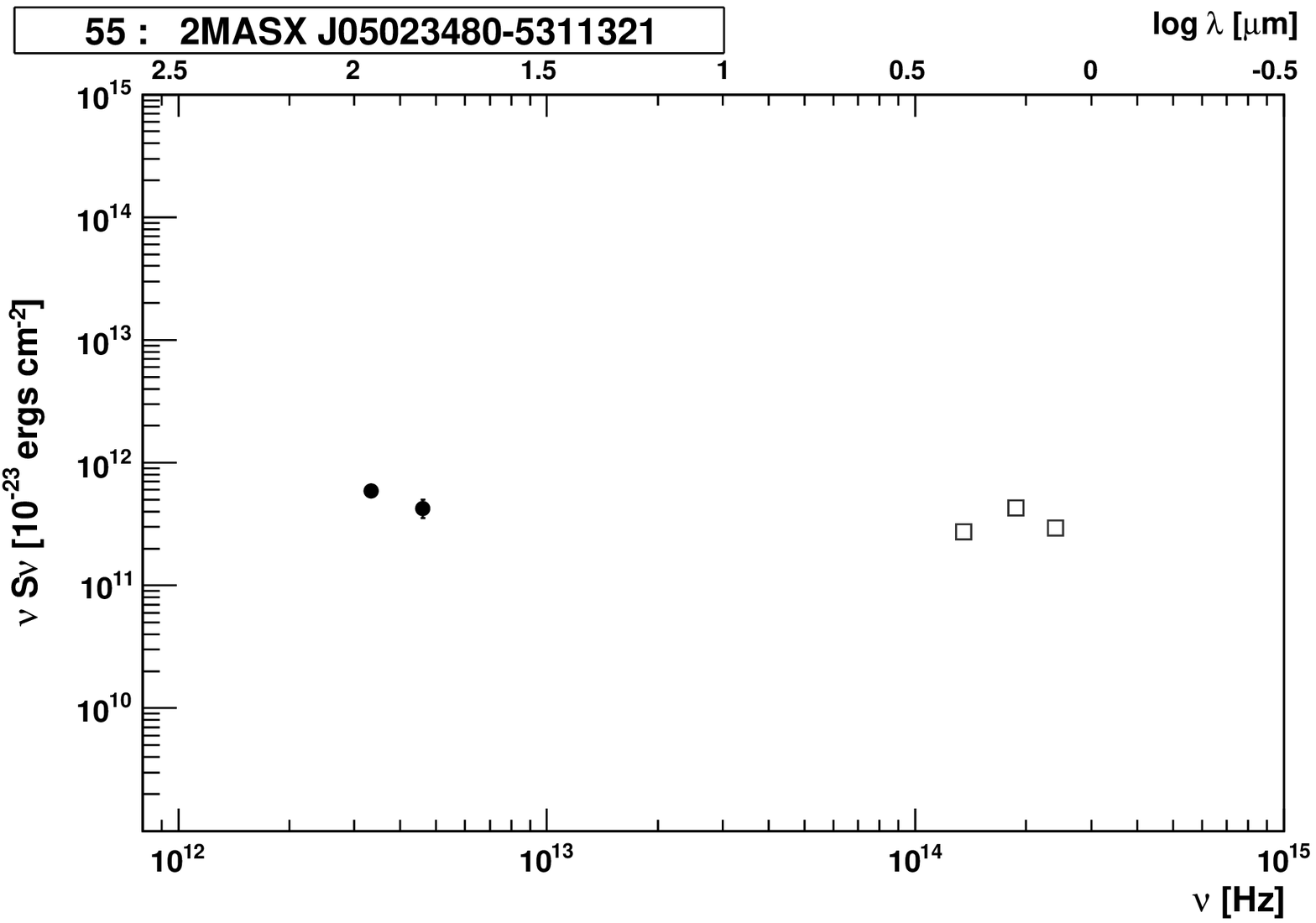}
\includegraphics[width=4cm]{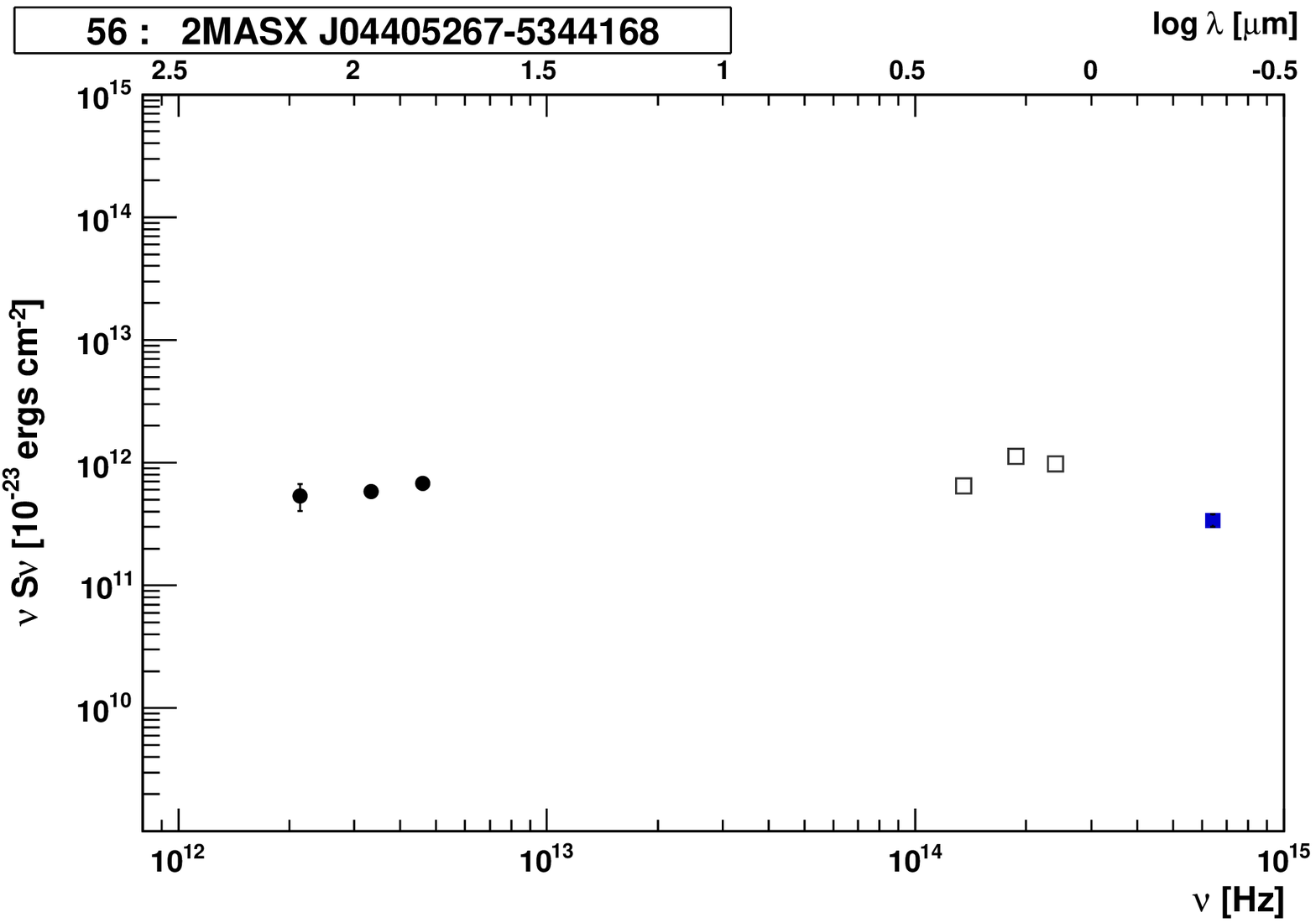}
\includegraphics[width=4cm]{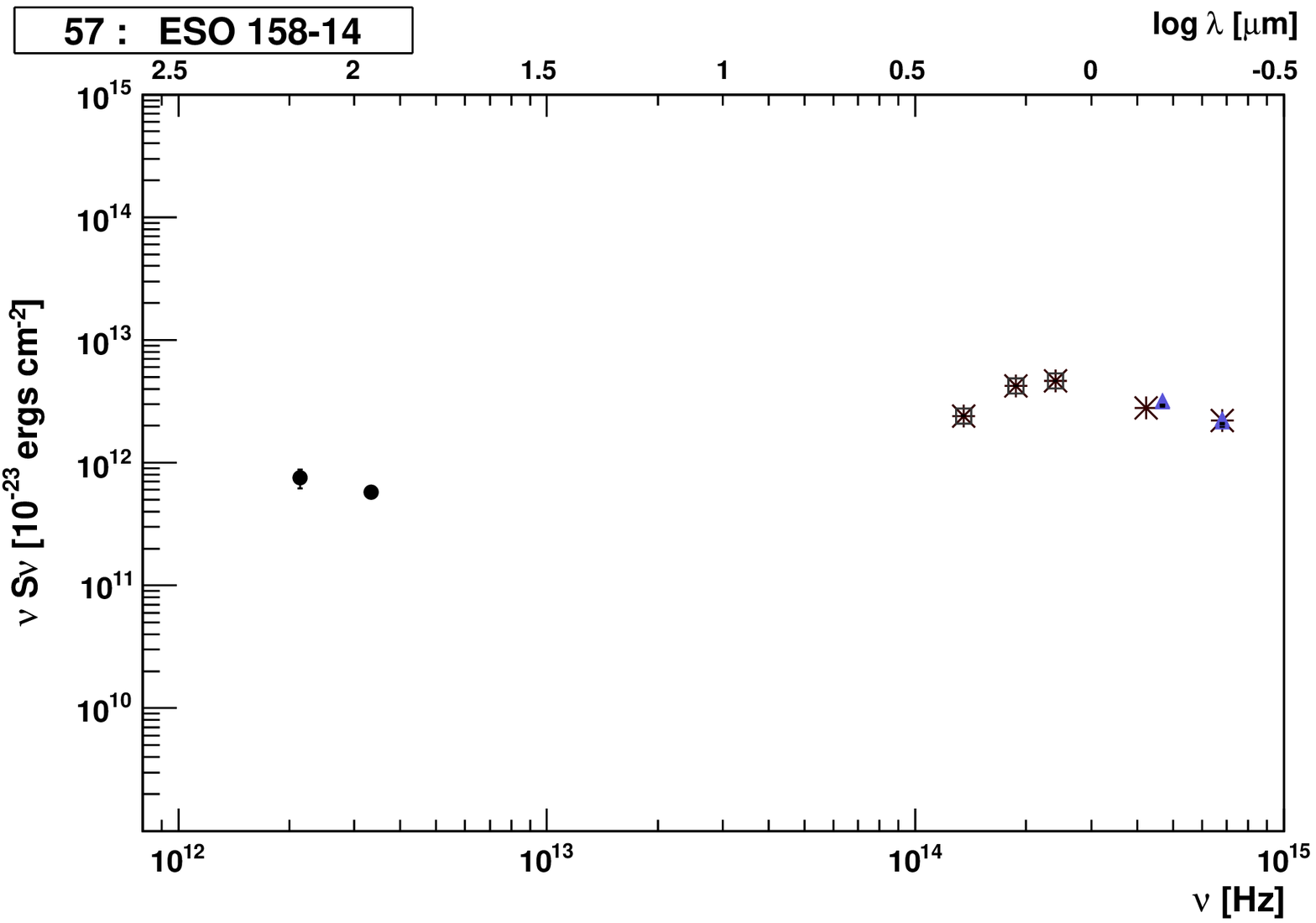}
\includegraphics[width=4cm]{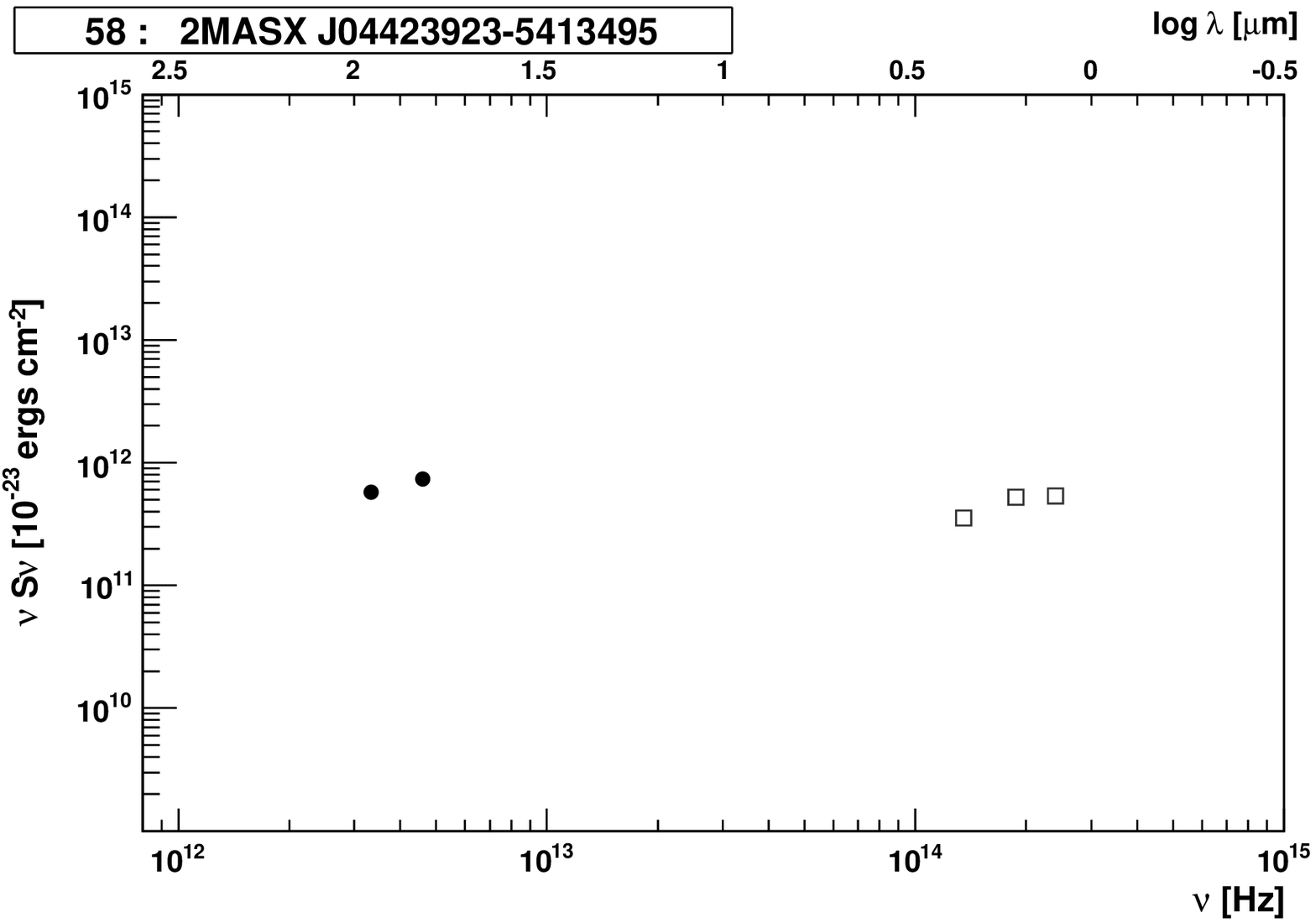}
\includegraphics[width=4cm]{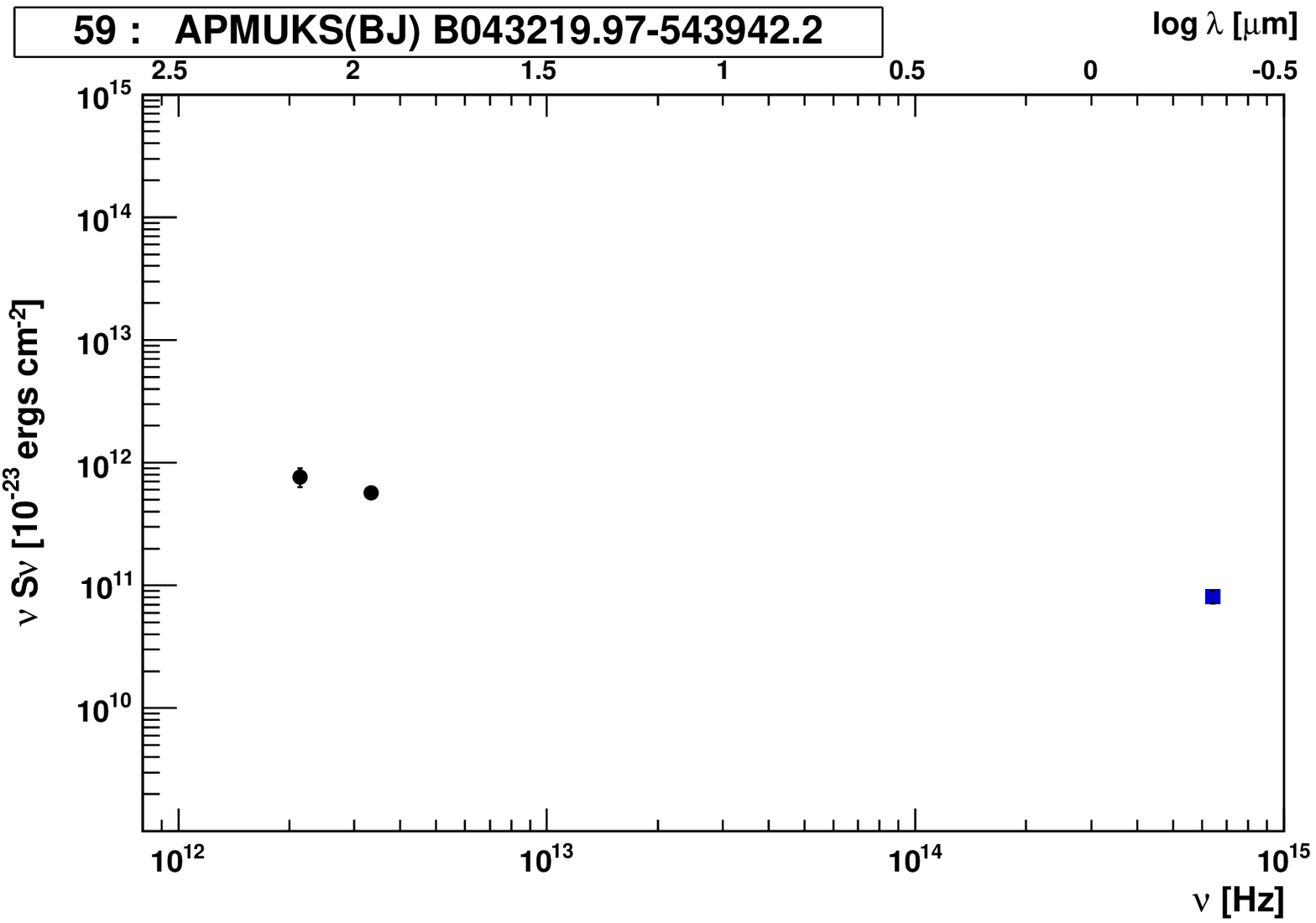}
\includegraphics[width=4cm]{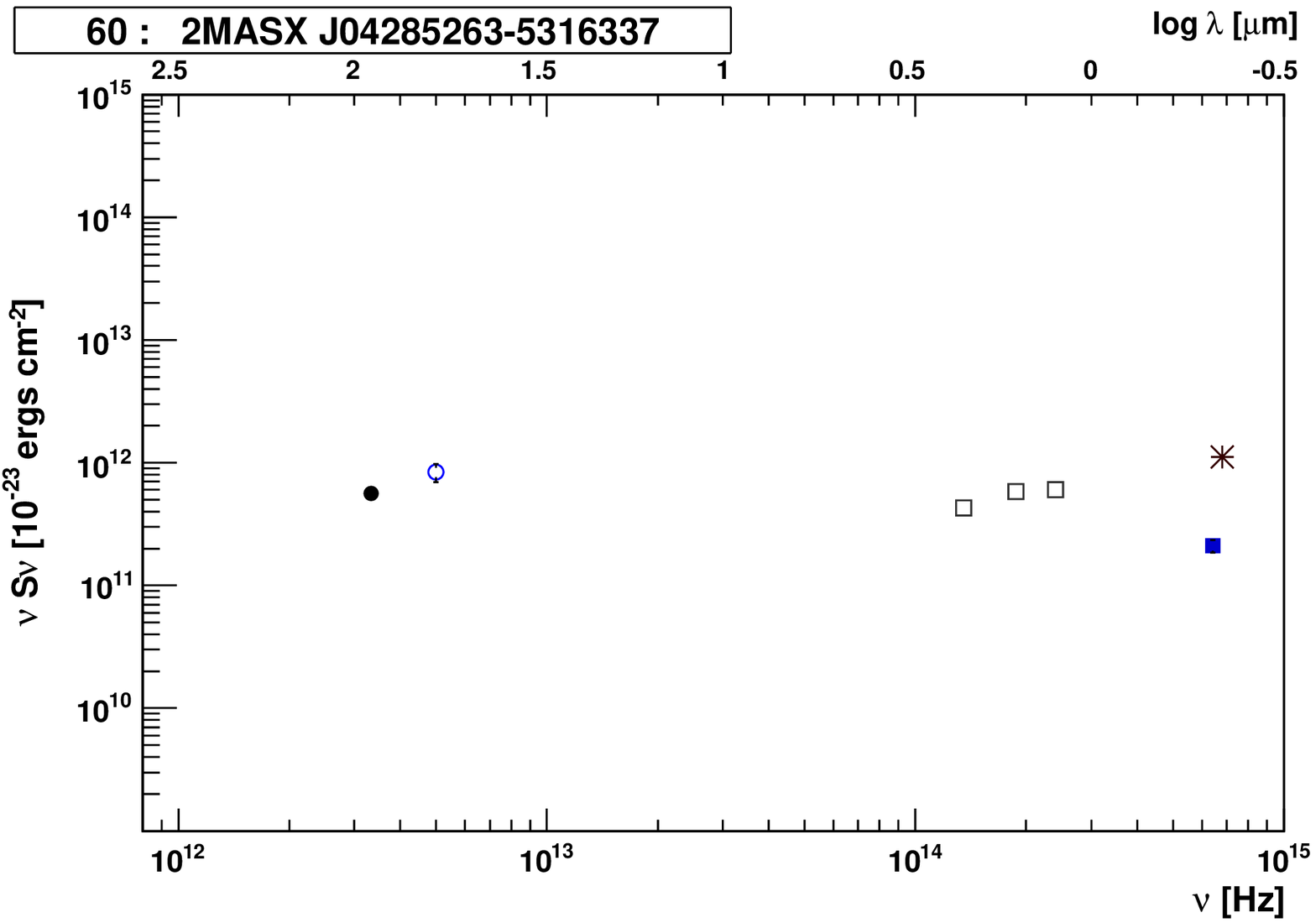}
\includegraphics[width=4cm]{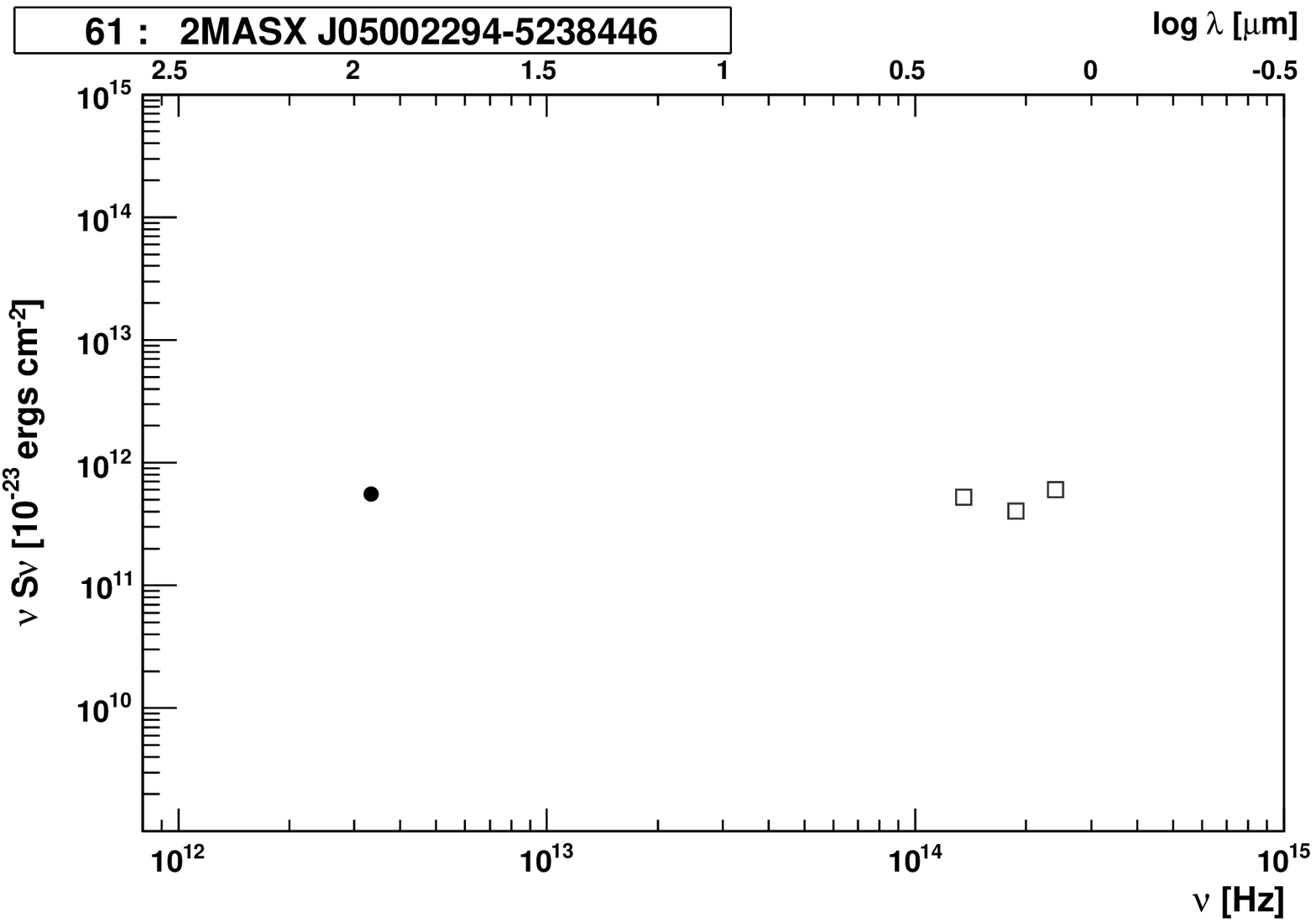}
\includegraphics[width=4cm]{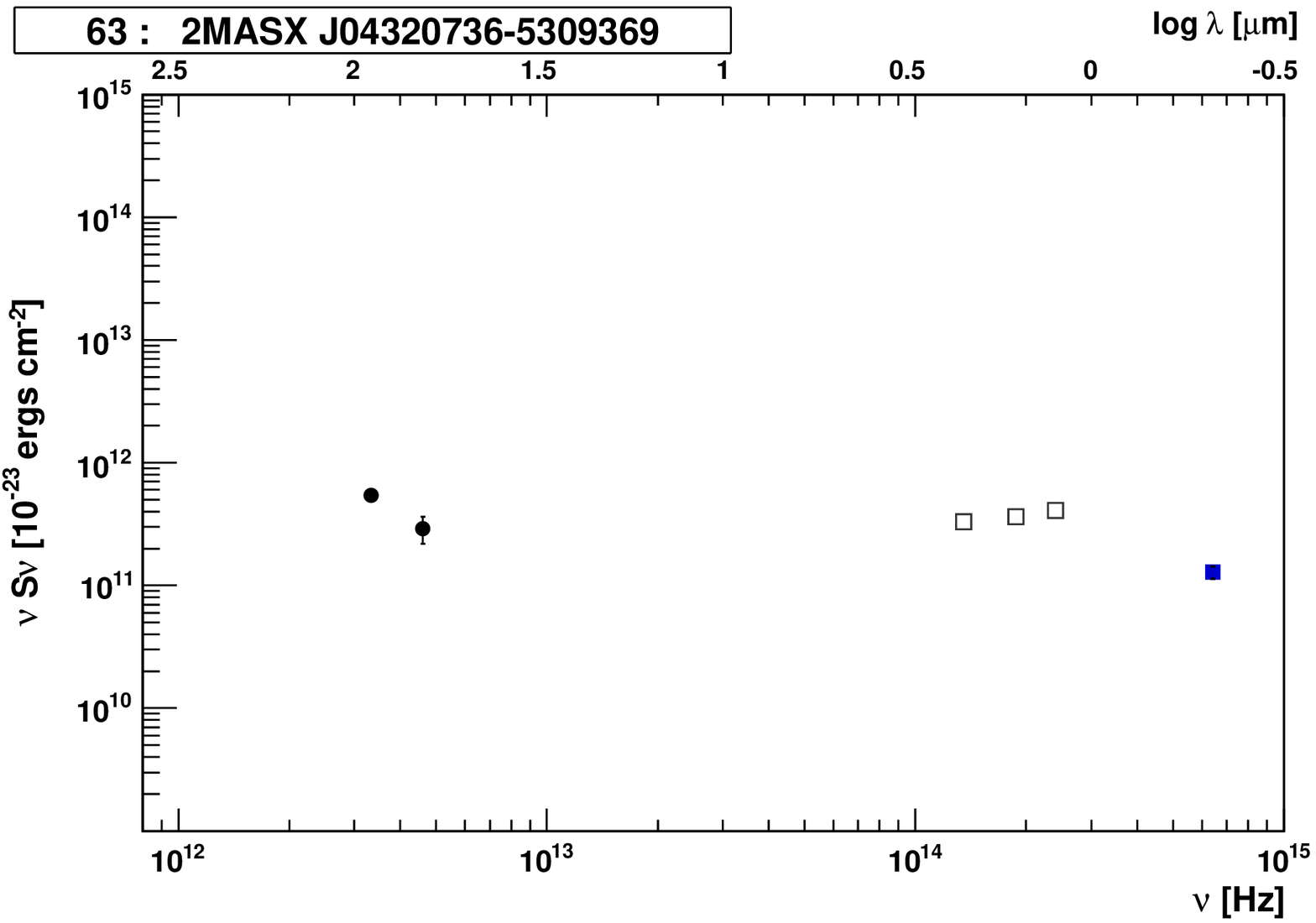}
\includegraphics[width=4cm]{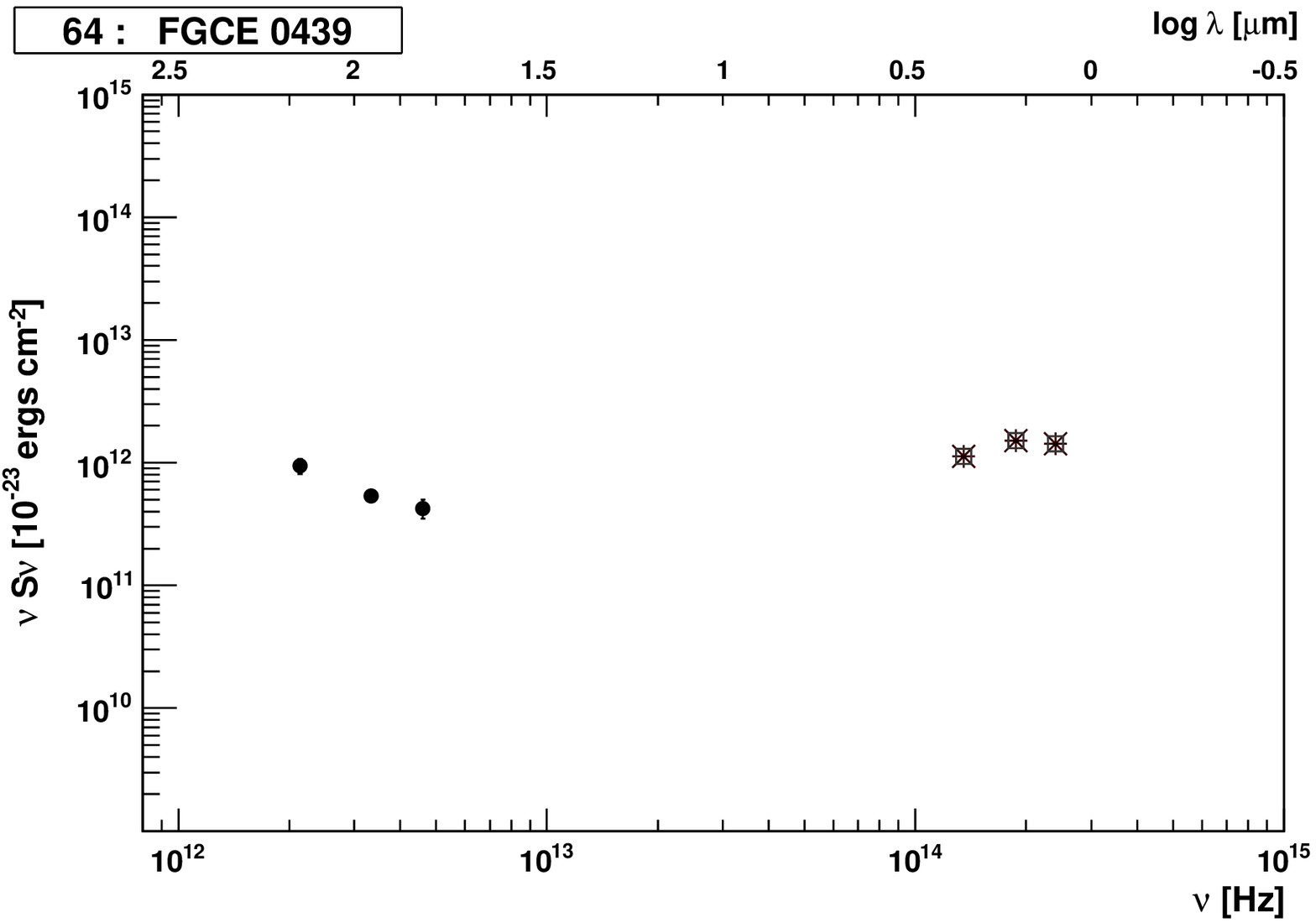}
\includegraphics[width=4cm]{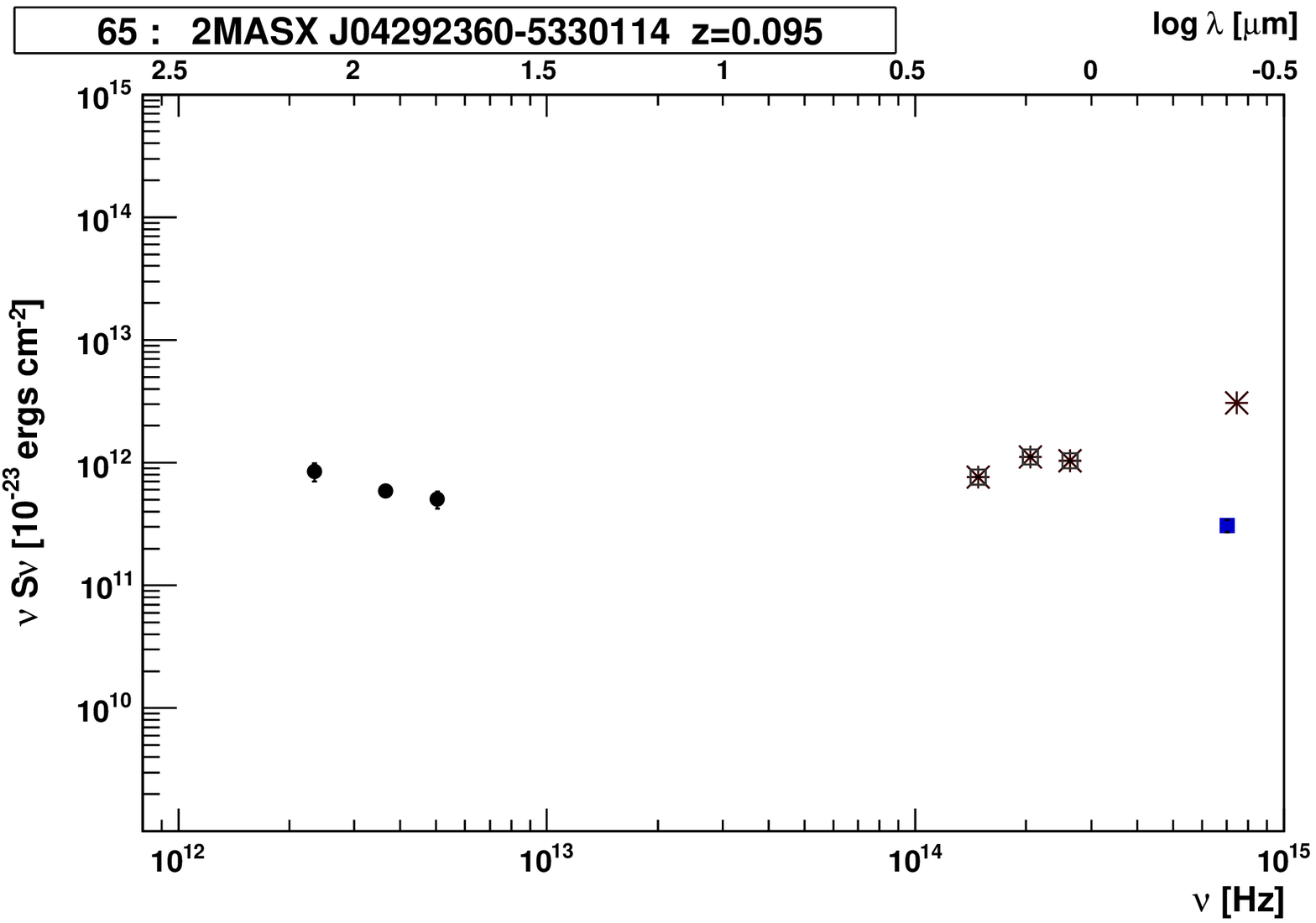}
\includegraphics[width=4cm]{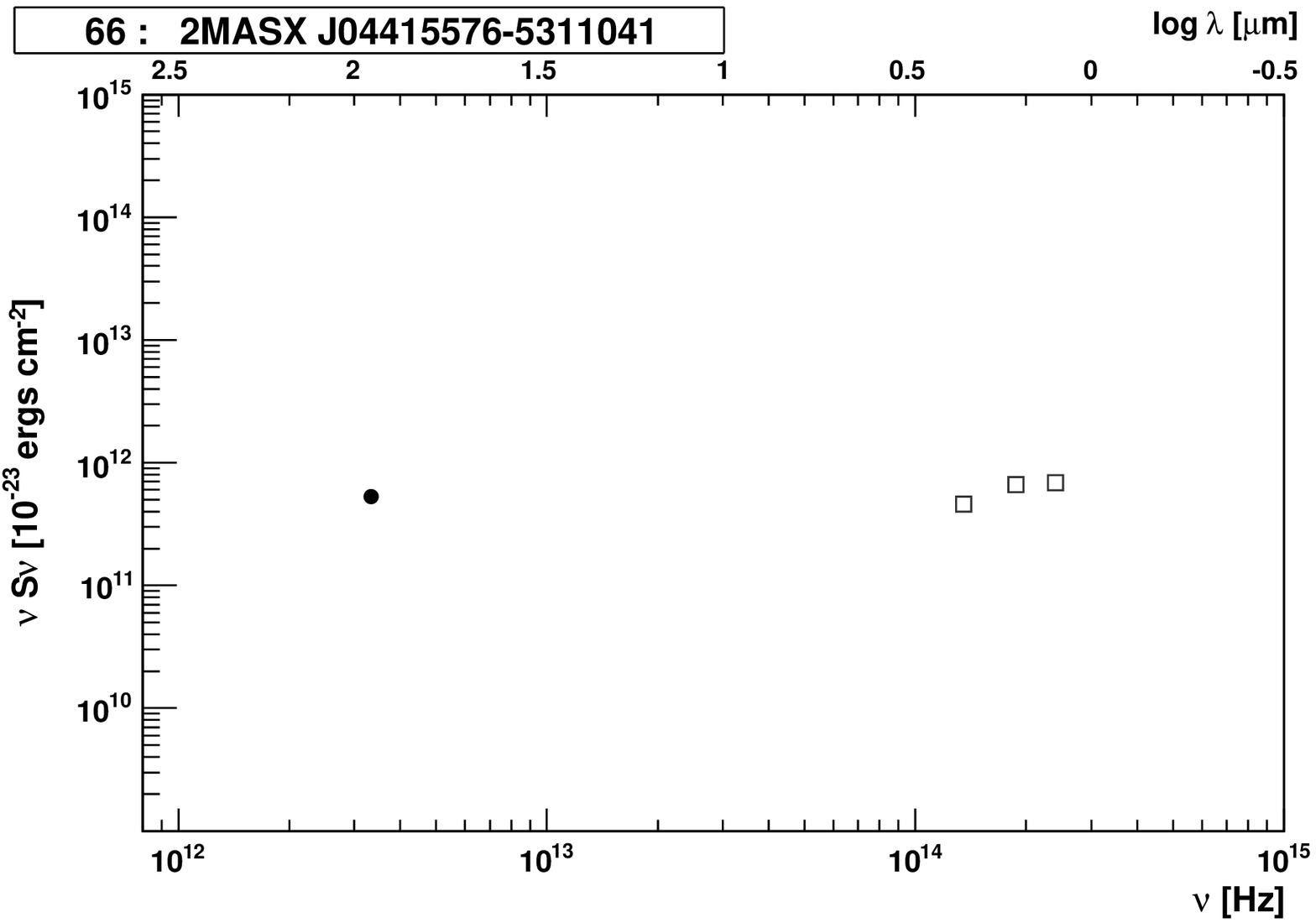}
\includegraphics[width=4cm]{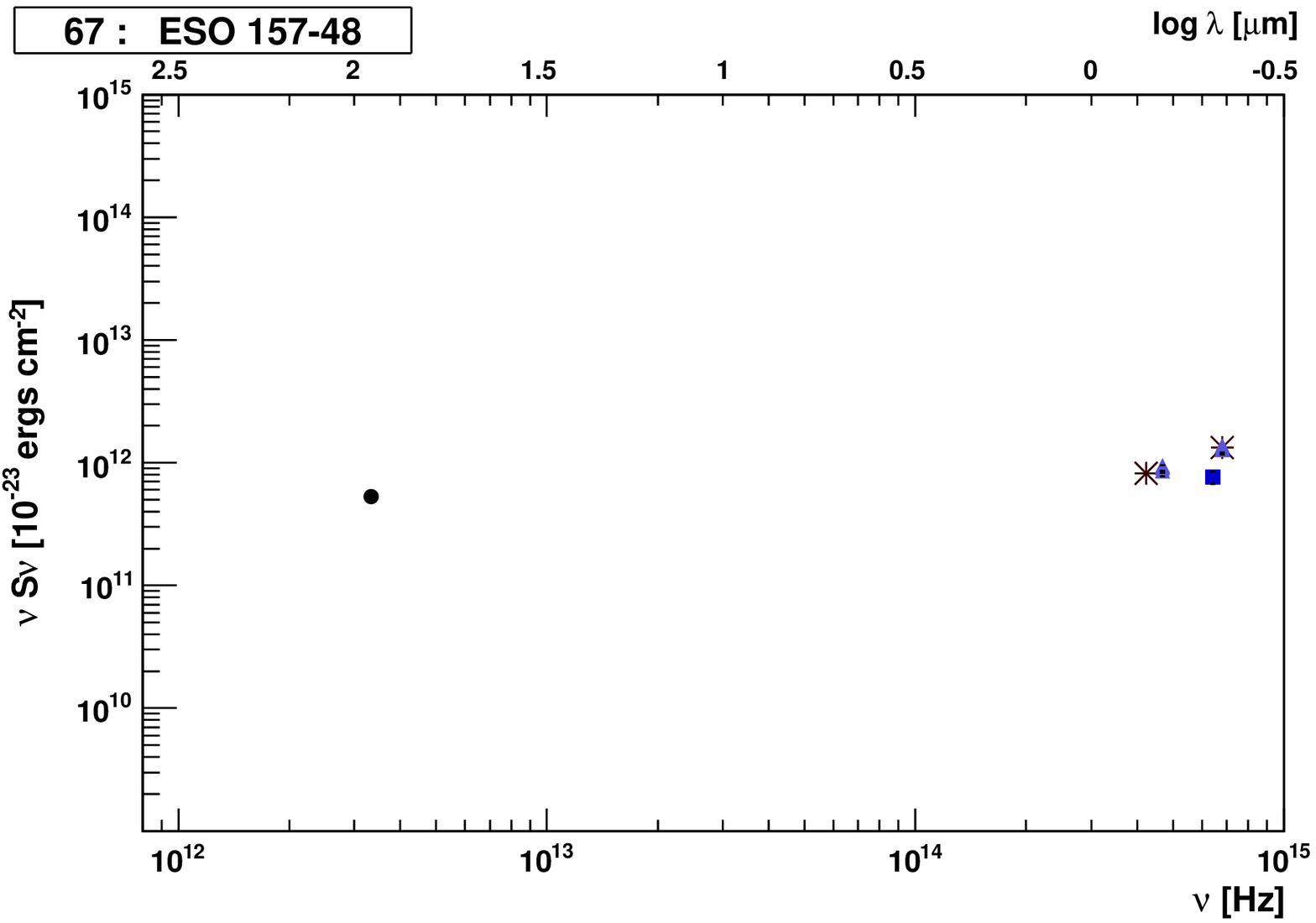}
\includegraphics[width=4cm]{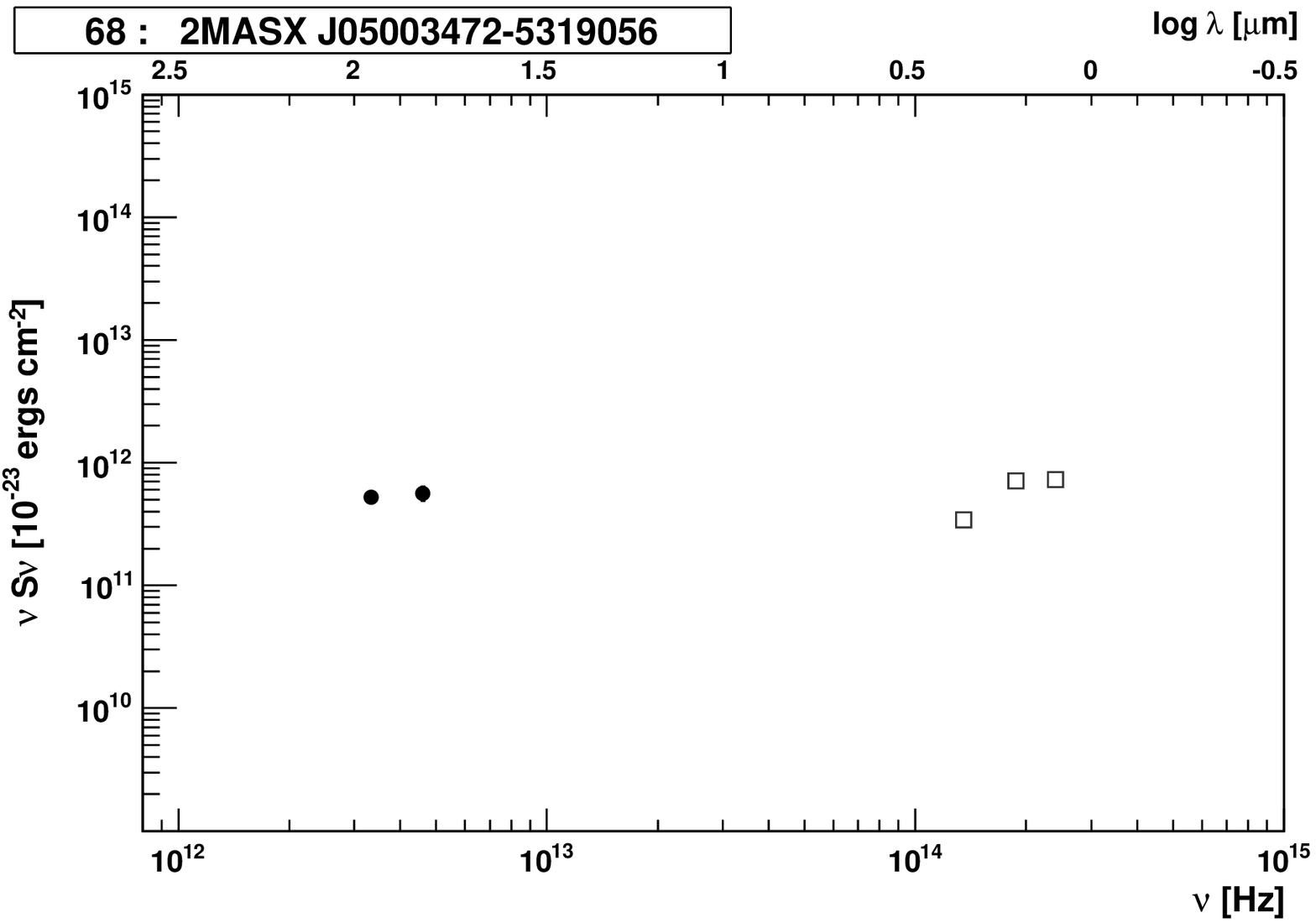}
\includegraphics[width=4cm]{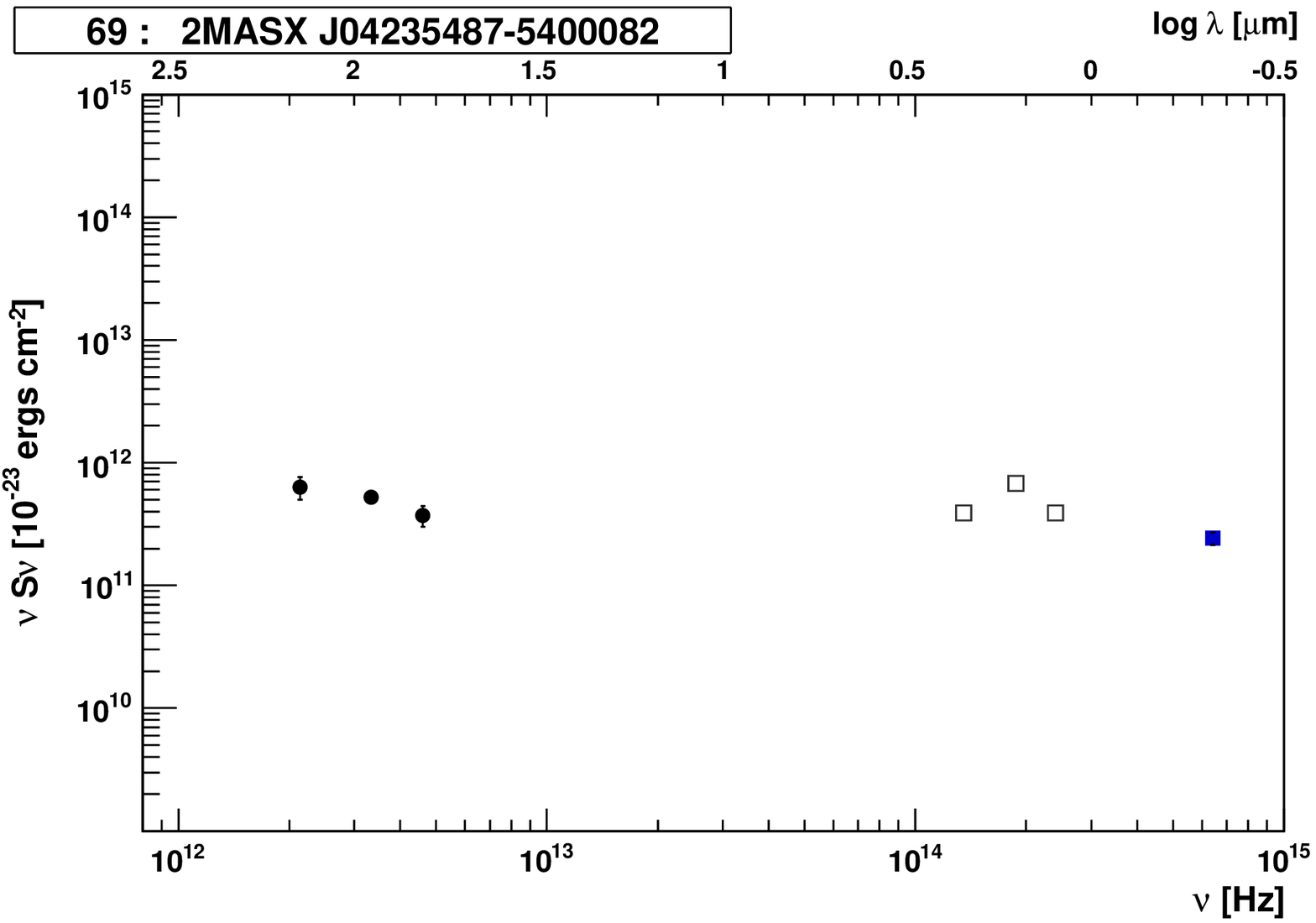}
\includegraphics[width=4cm]{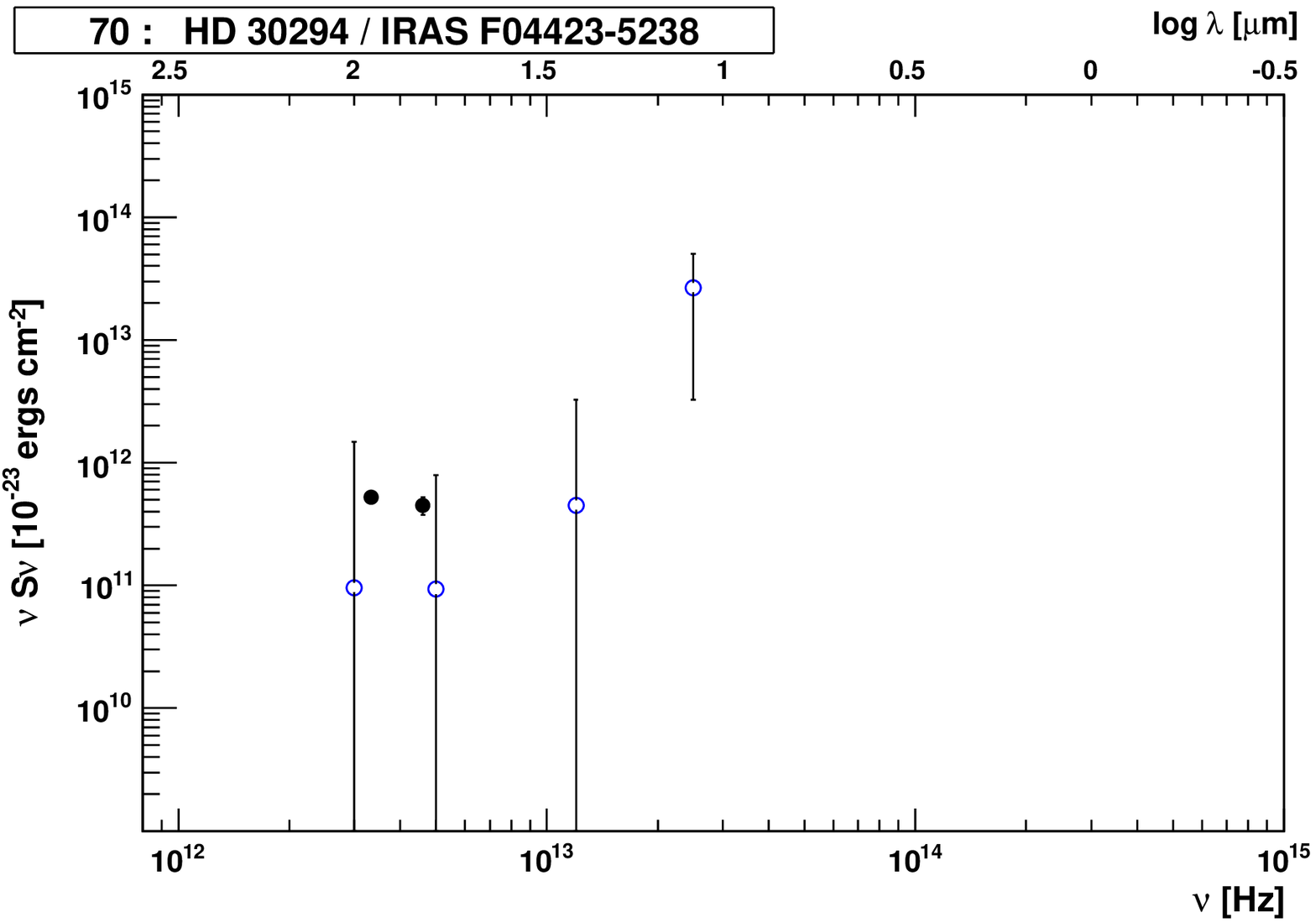}
\includegraphics[width=4cm]{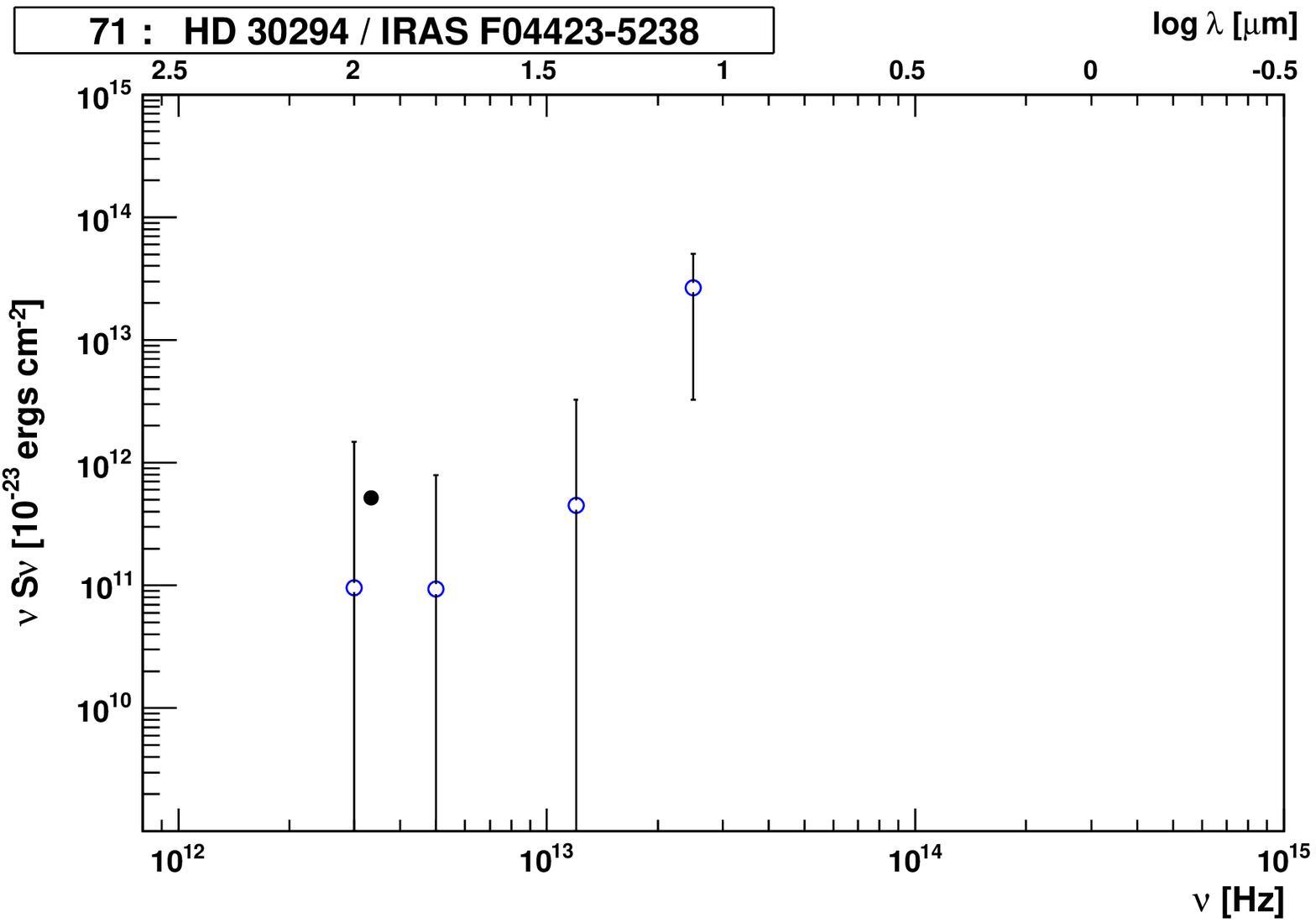}
\includegraphics[width=4cm]{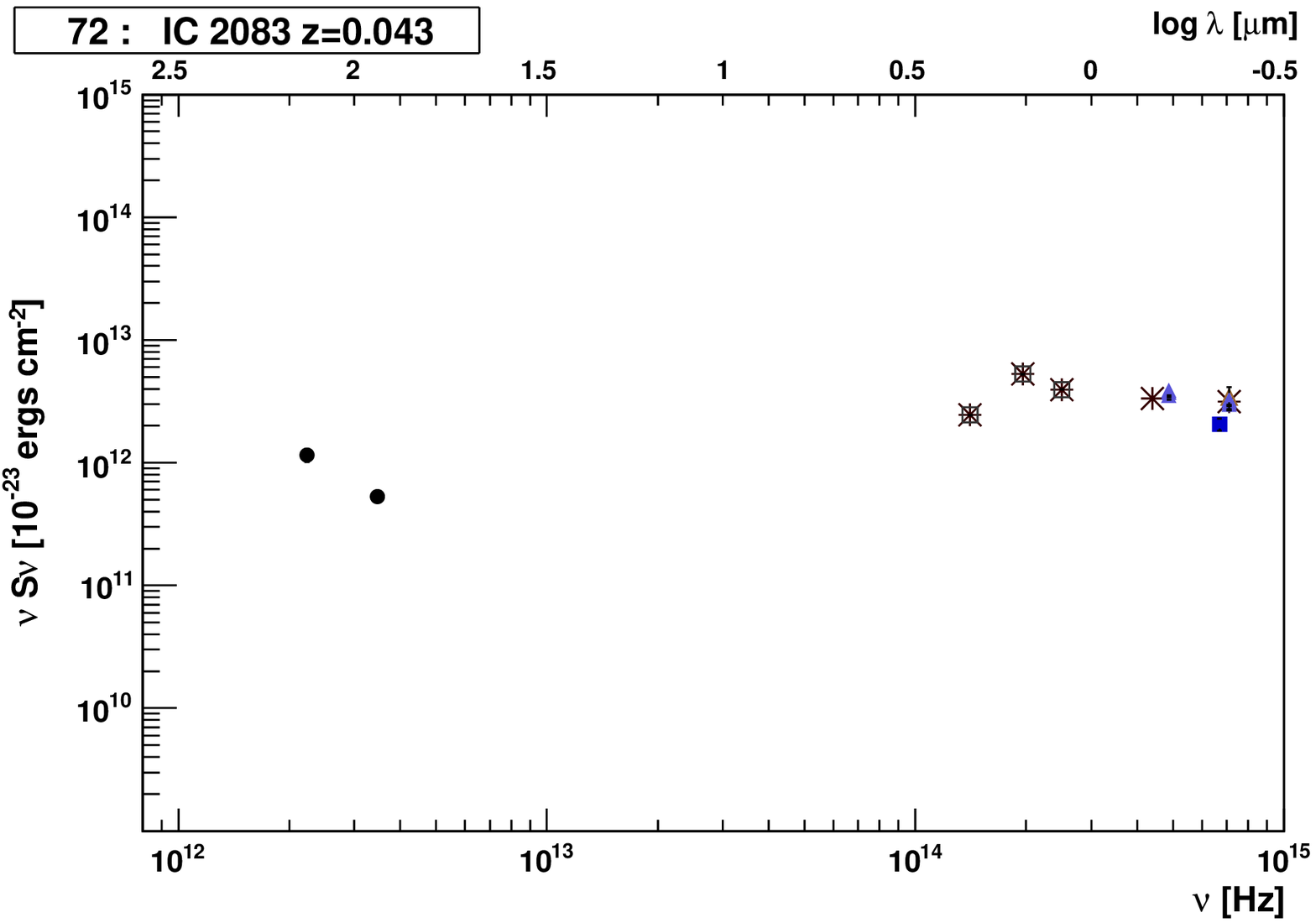}
\includegraphics[width=4cm]{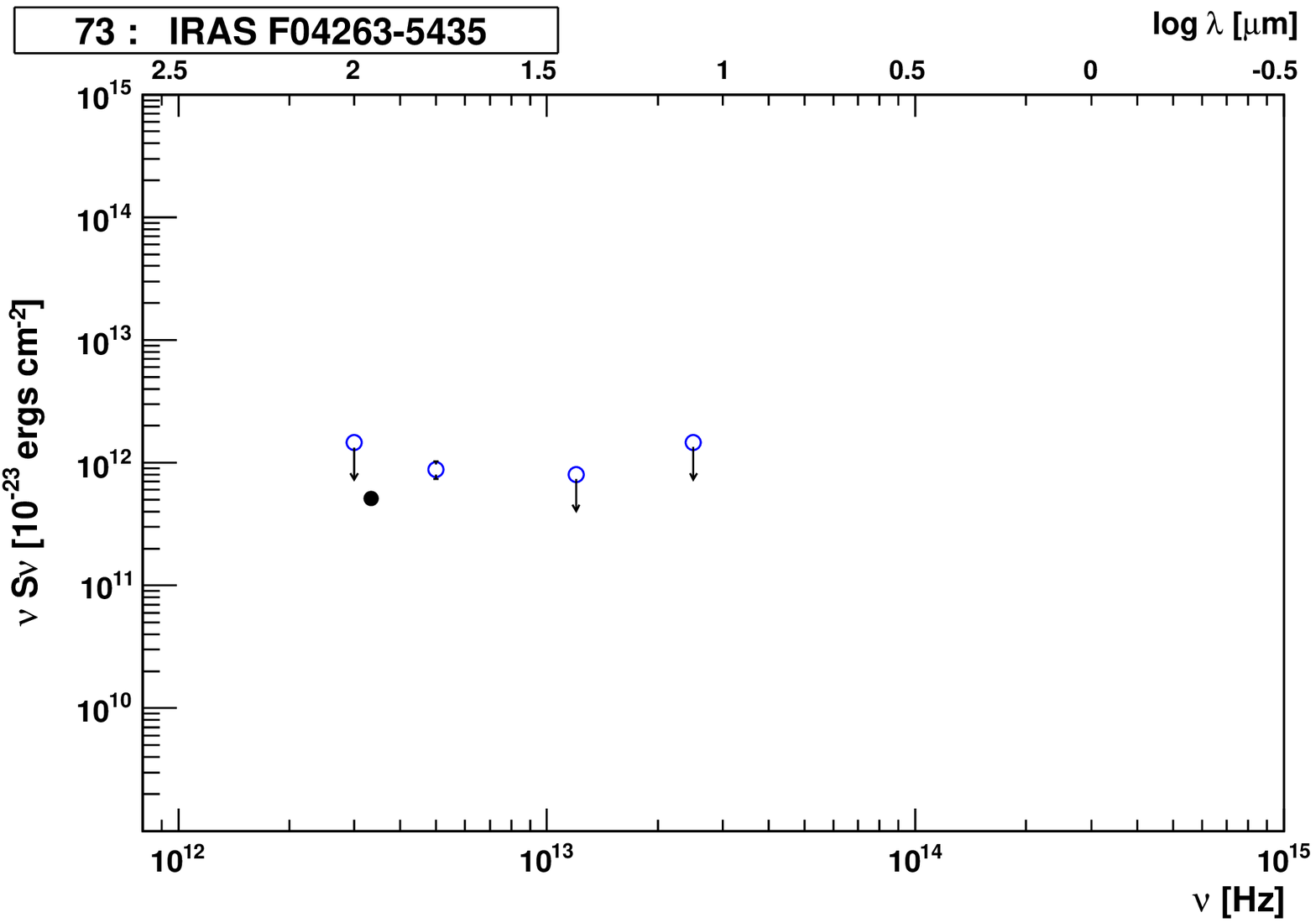}
\includegraphics[width=4cm]{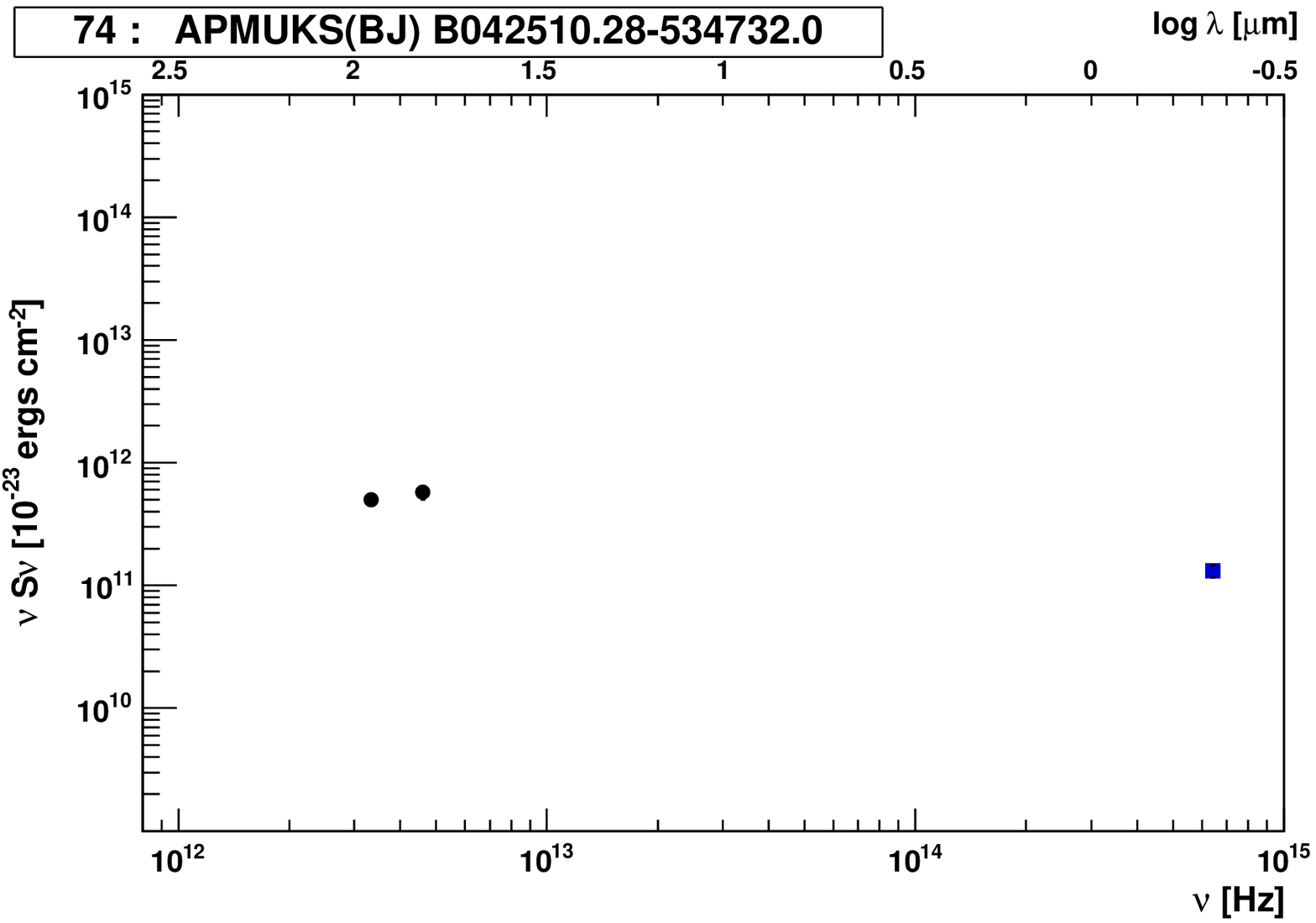}
\includegraphics[width=4cm]{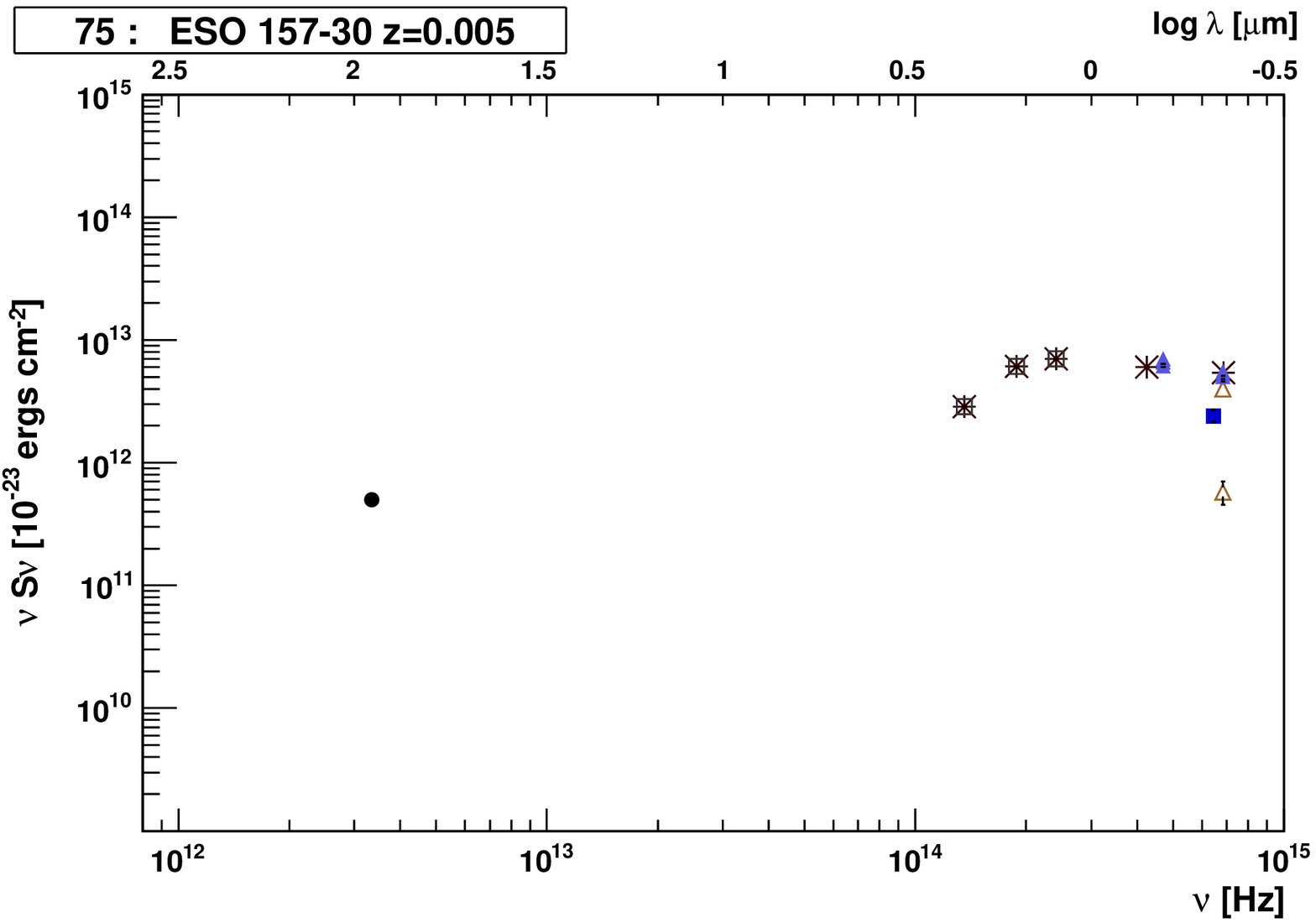}
 \label{points2}
 \caption {SEDs for the next 36 ADF-S identified sources, with symbols as in Figure~\ref{points1}.}
 \end{figure*}
 }

\clearpage

\onlfig{3}{
\begin{figure*}[t]
\centering
\includegraphics[width=4cm]{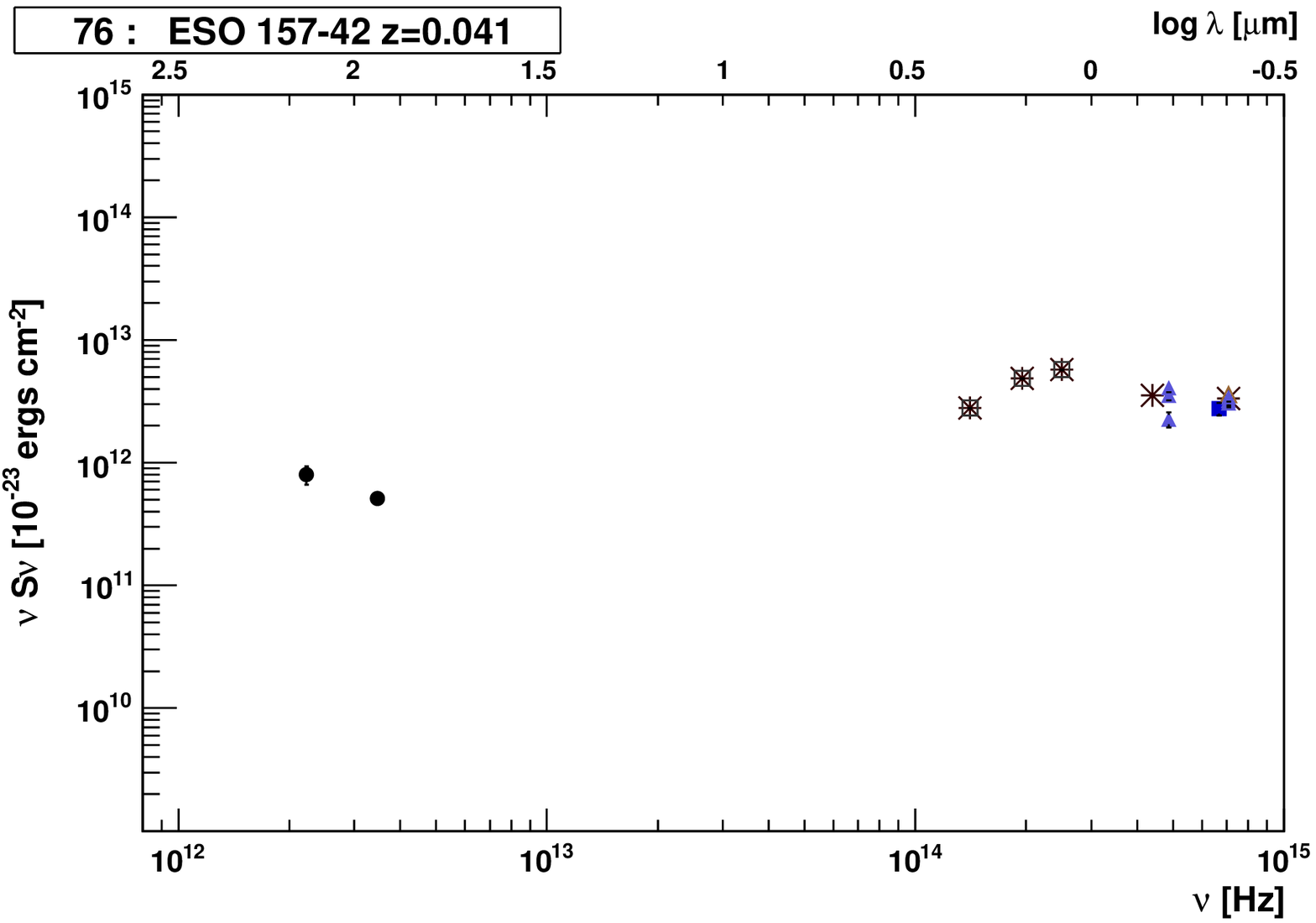}
\includegraphics[width=4cm]{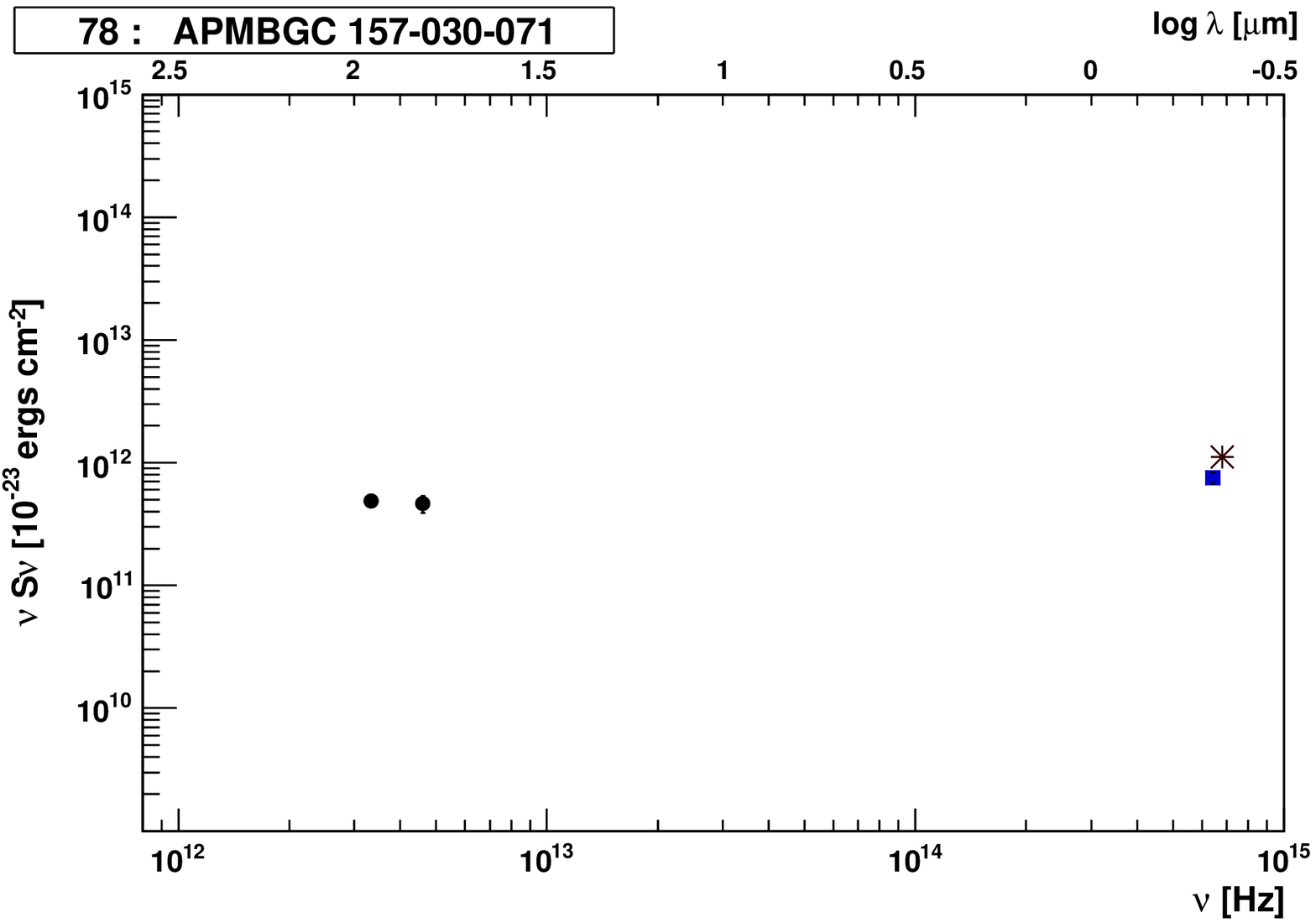}
\includegraphics[width=4cm]{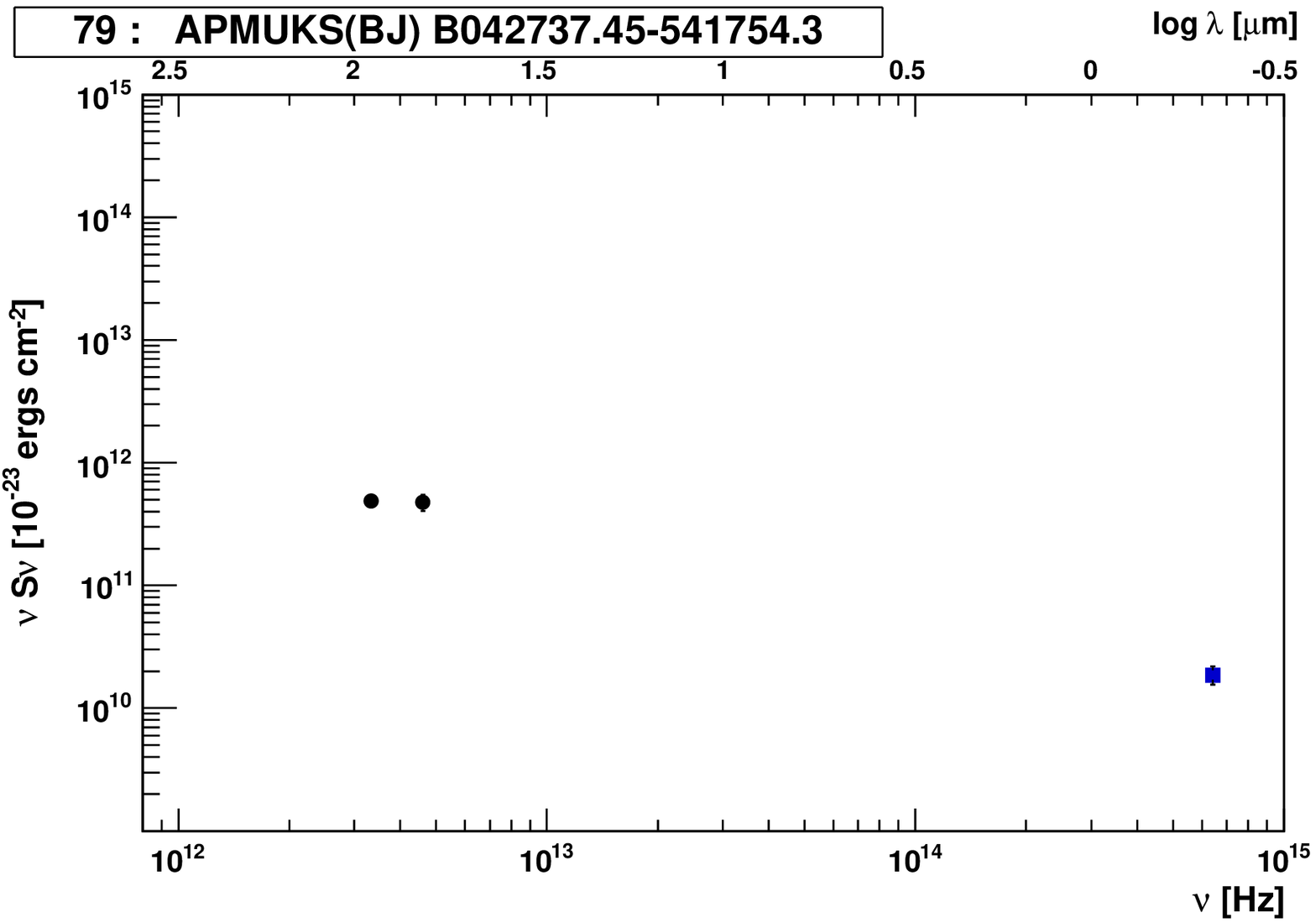}
\includegraphics[width=4cm]{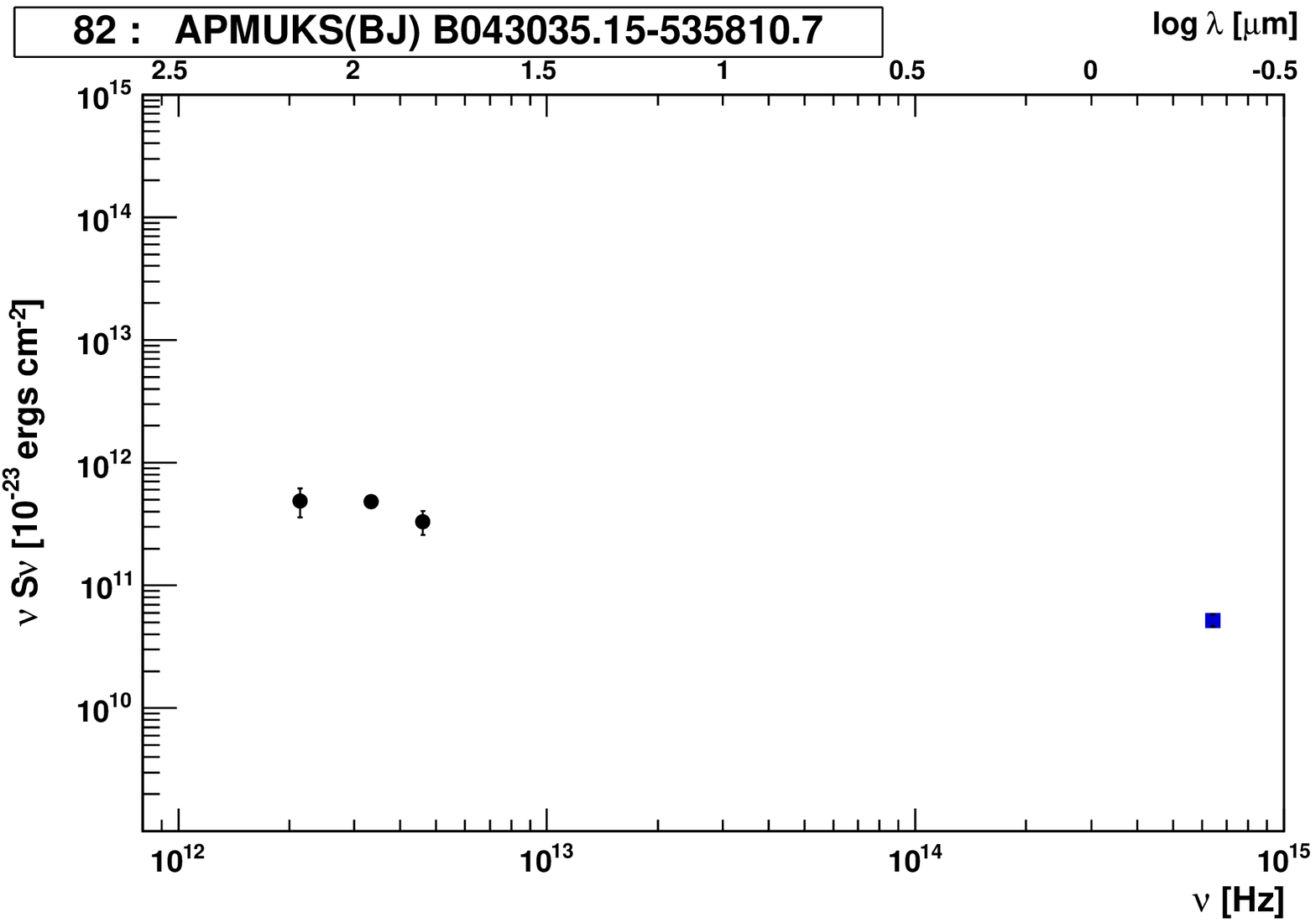}
\includegraphics[width=4cm]{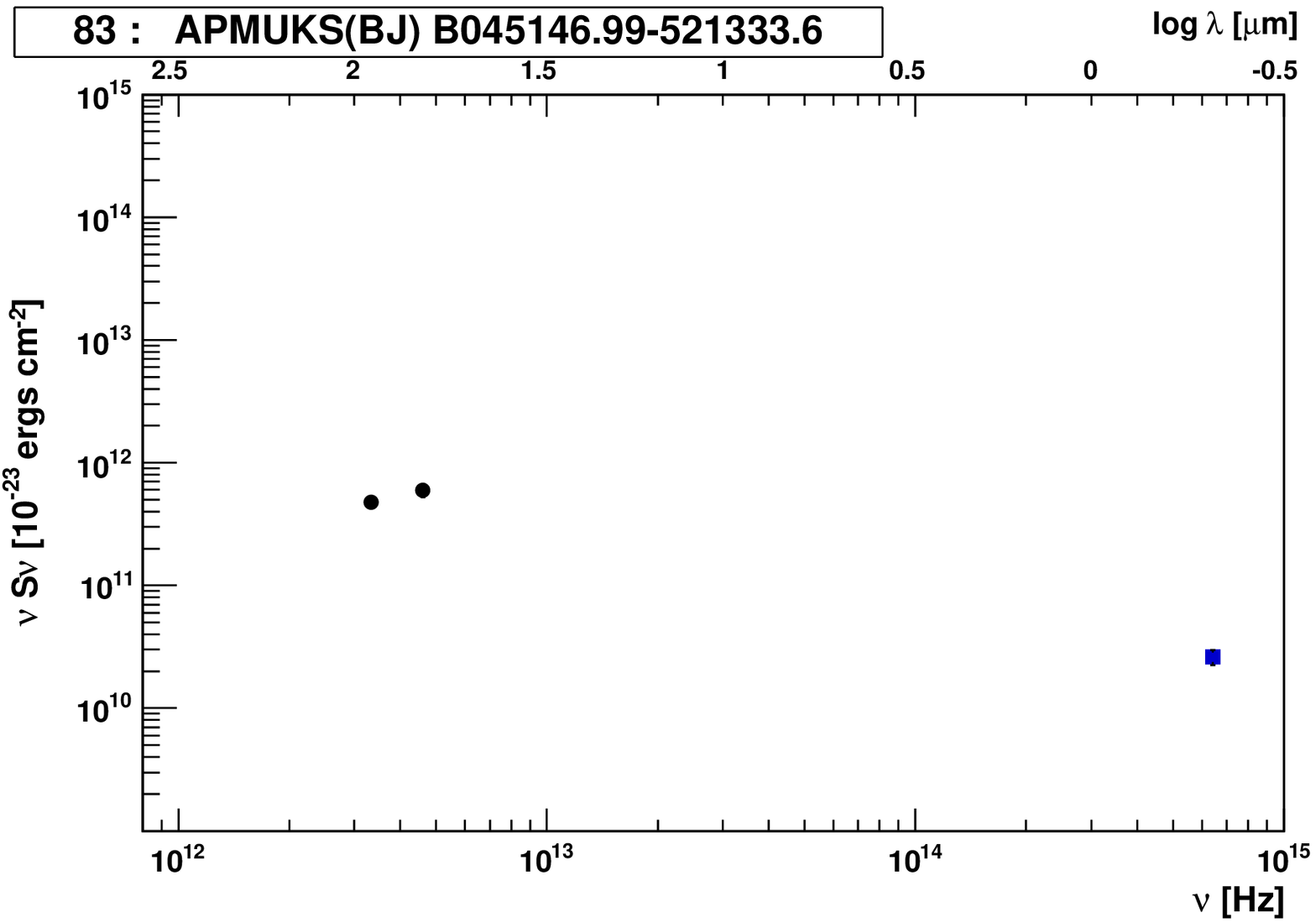}
\includegraphics[width=4cm]{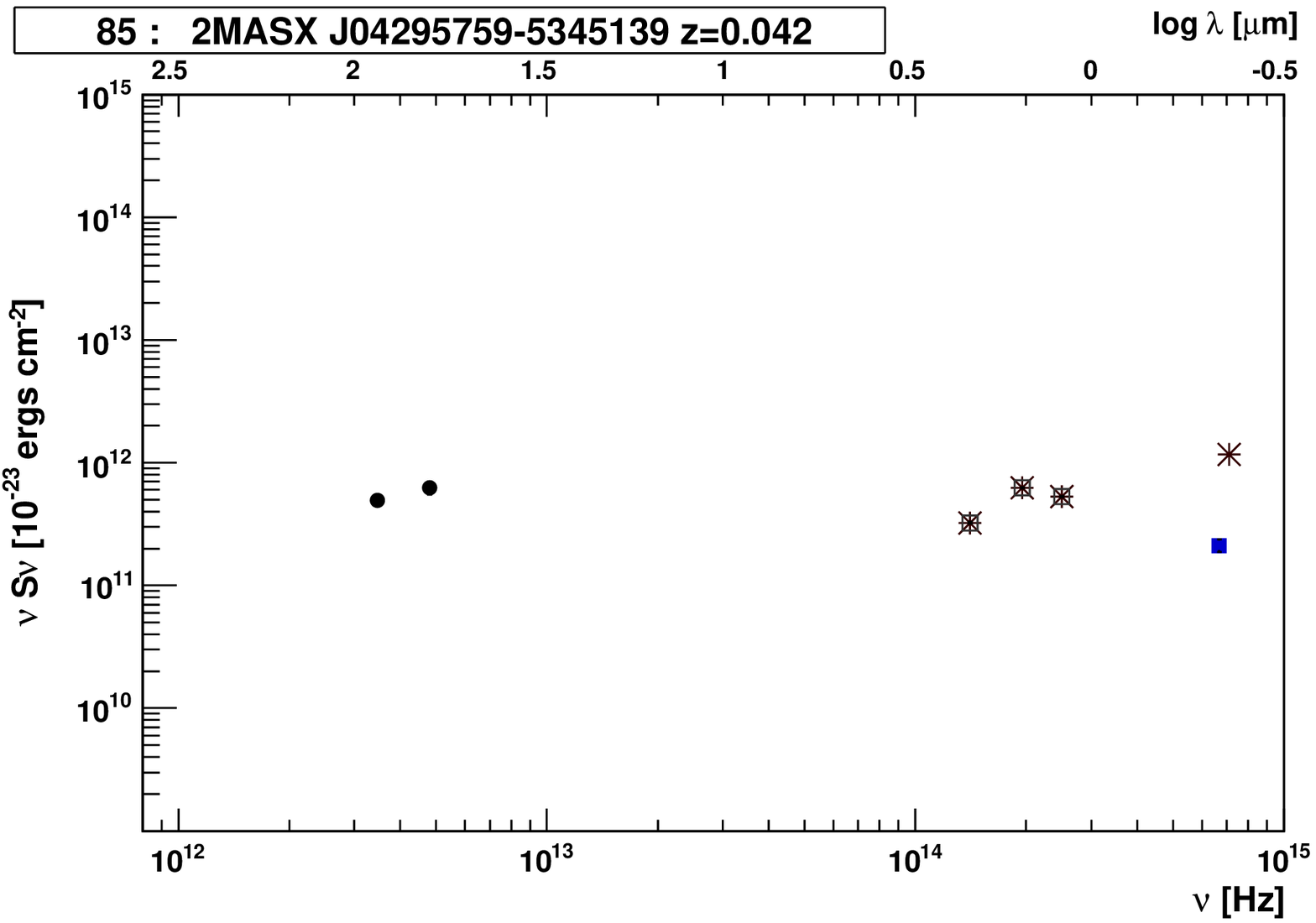}
\includegraphics[width=4cm]{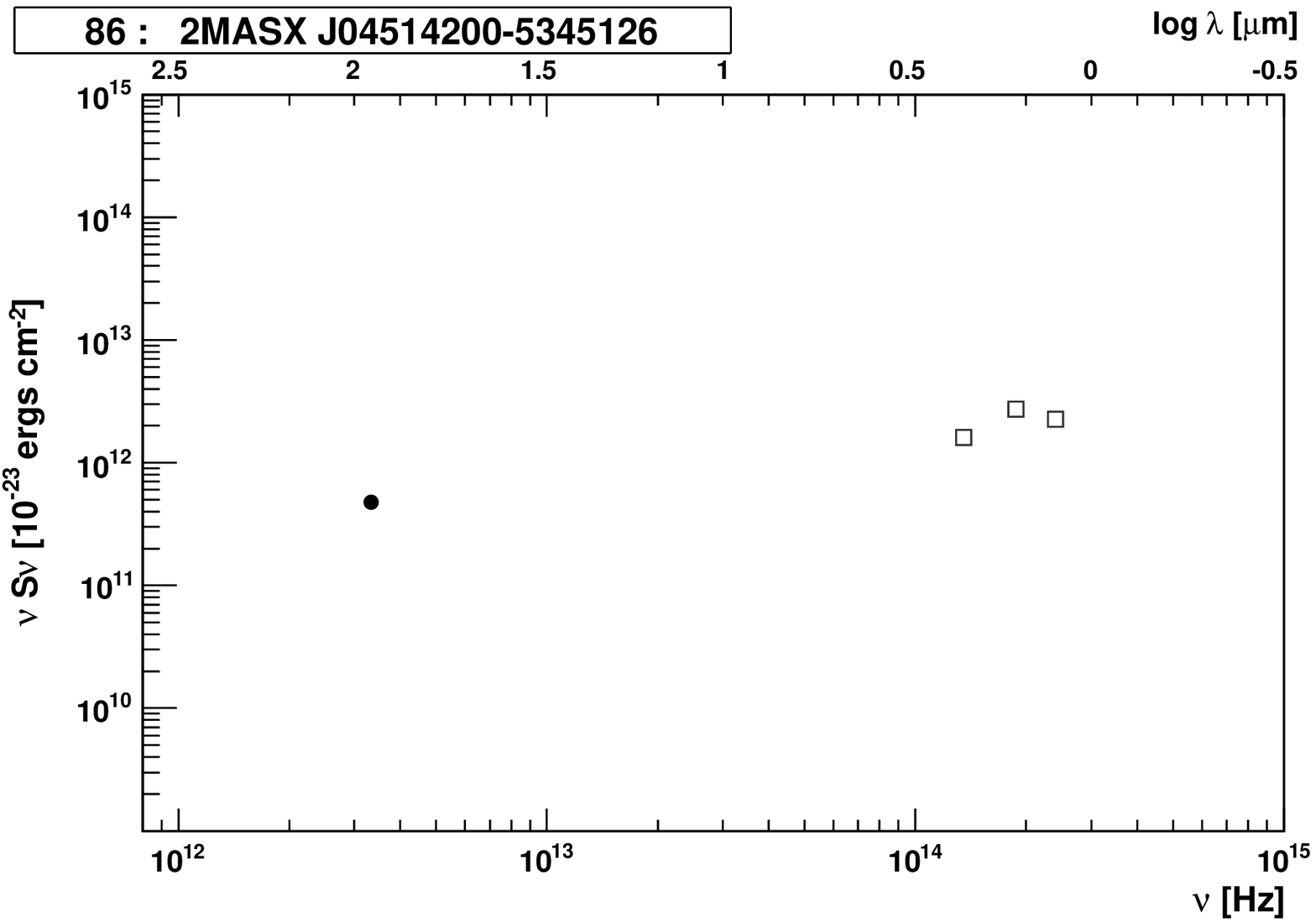}
\includegraphics[width=4cm]{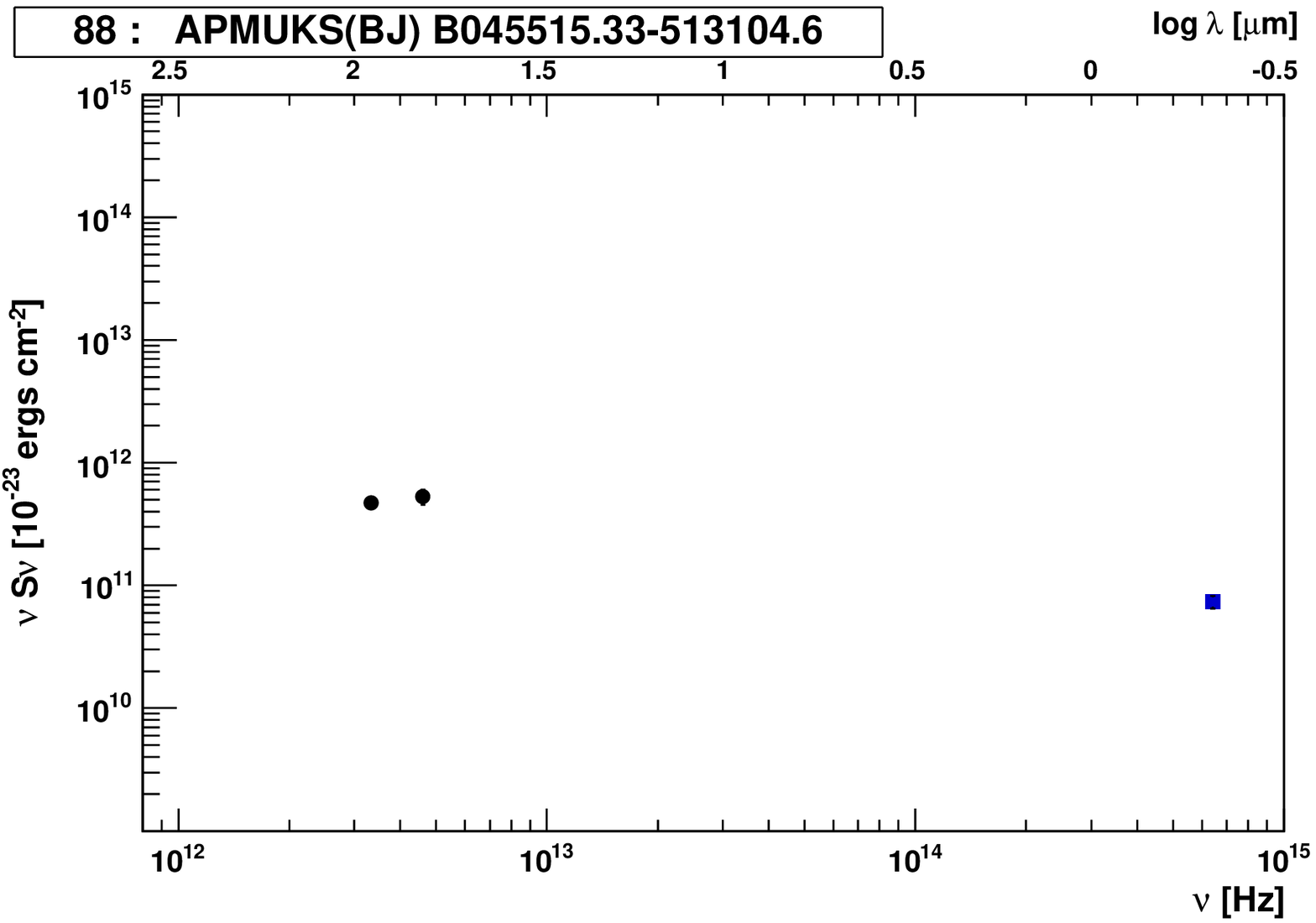}
\includegraphics[width=4cm]{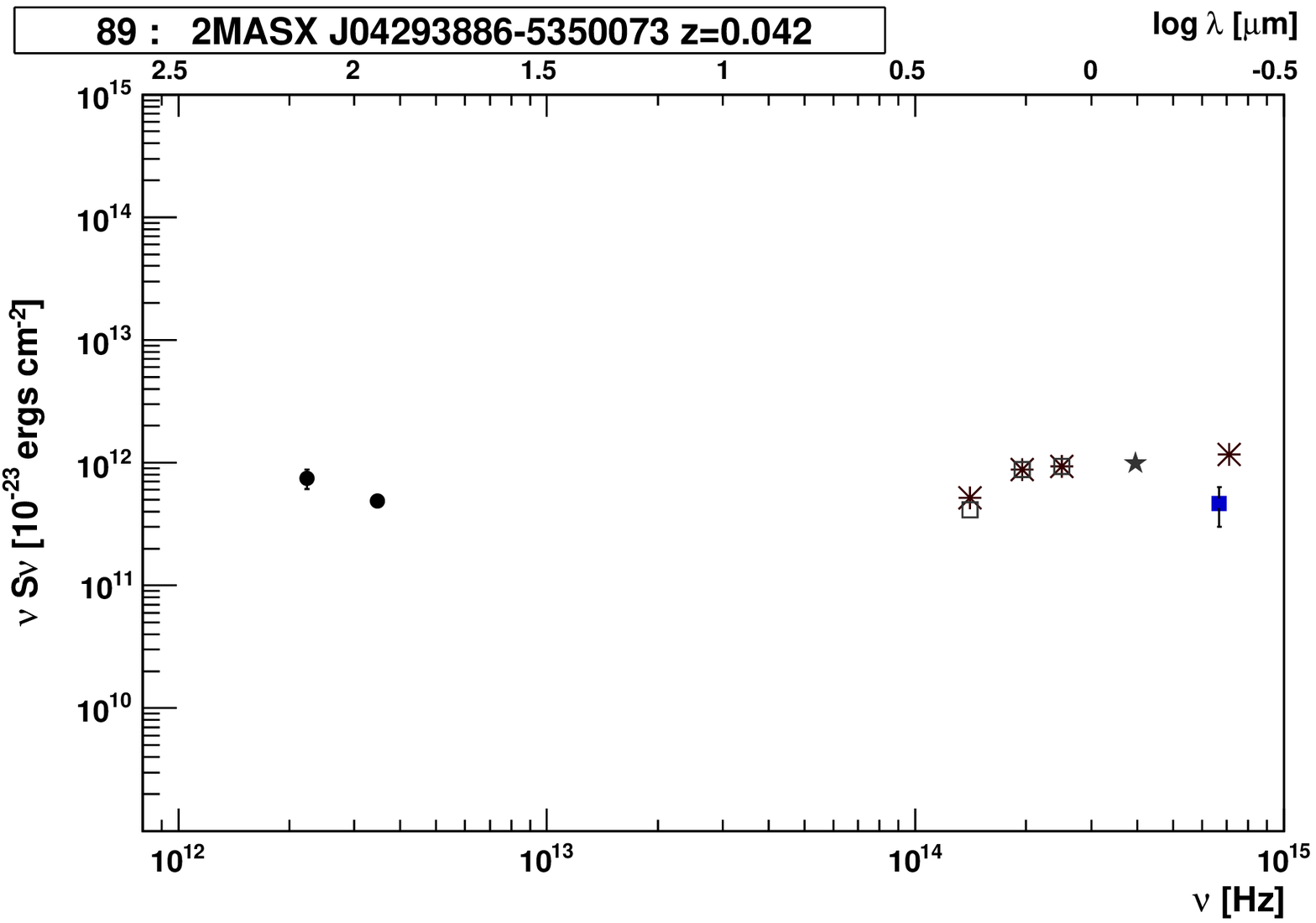}
\includegraphics[width=4cm]{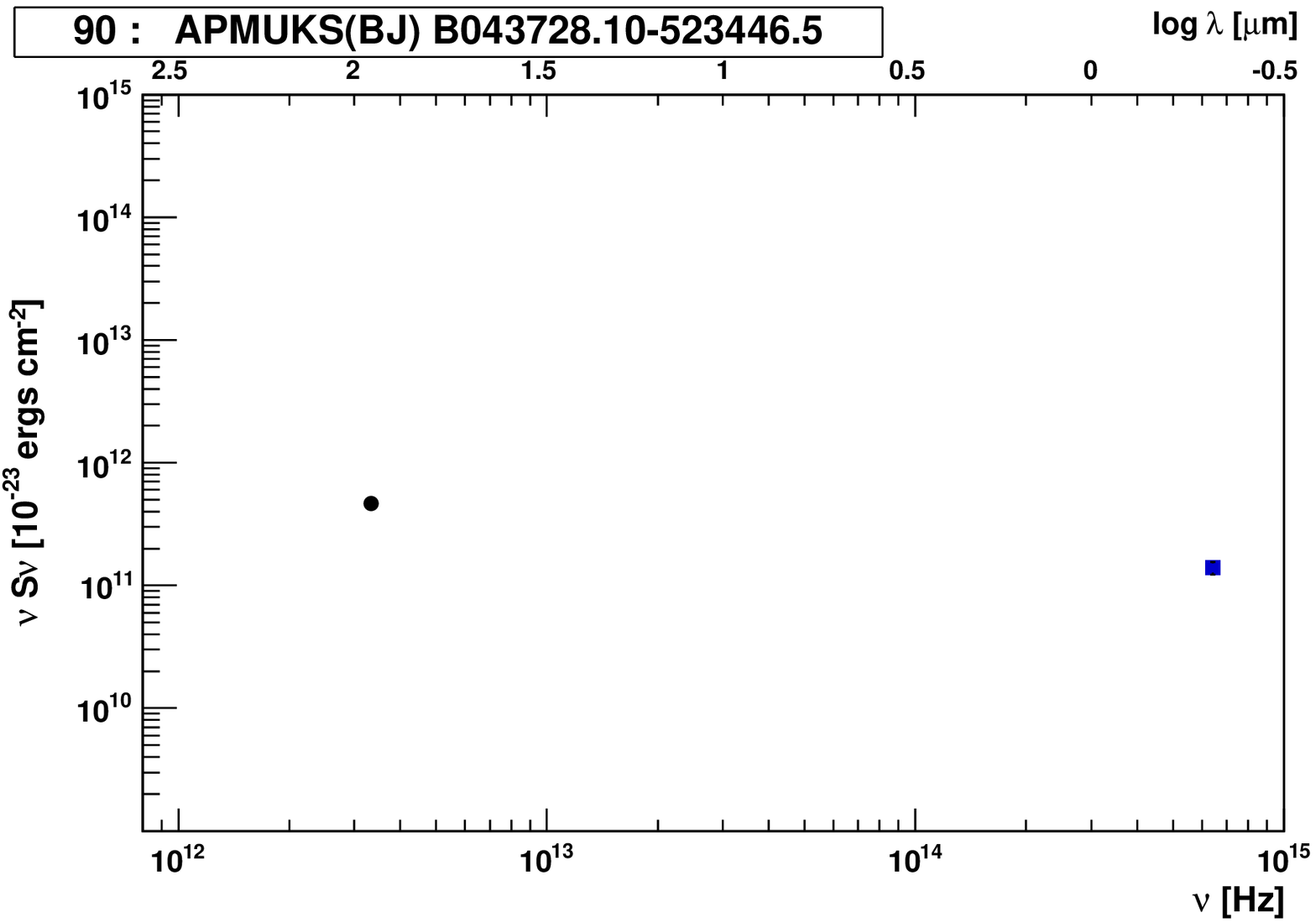}
\includegraphics[width=4cm]{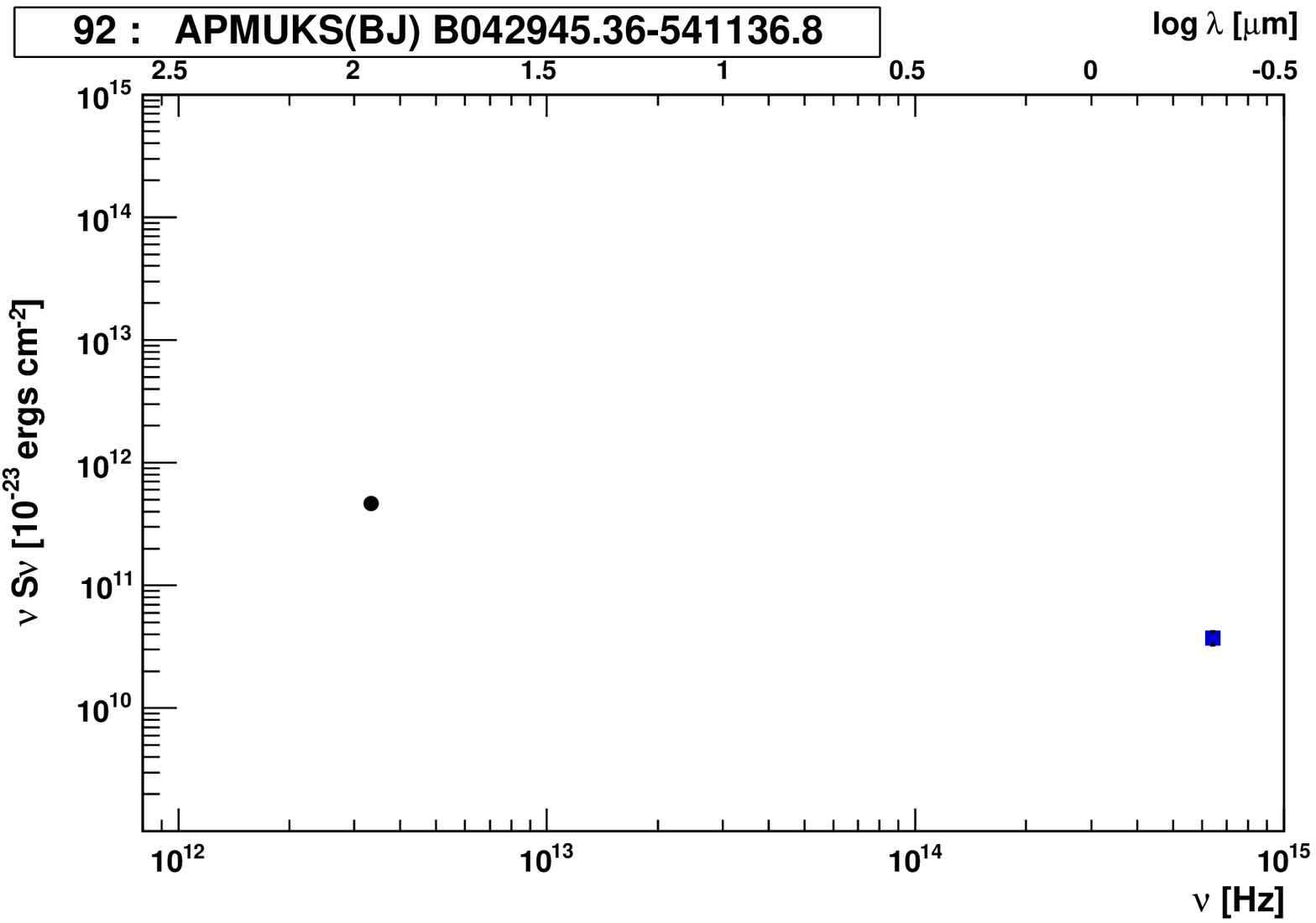}
\includegraphics[width=4cm]{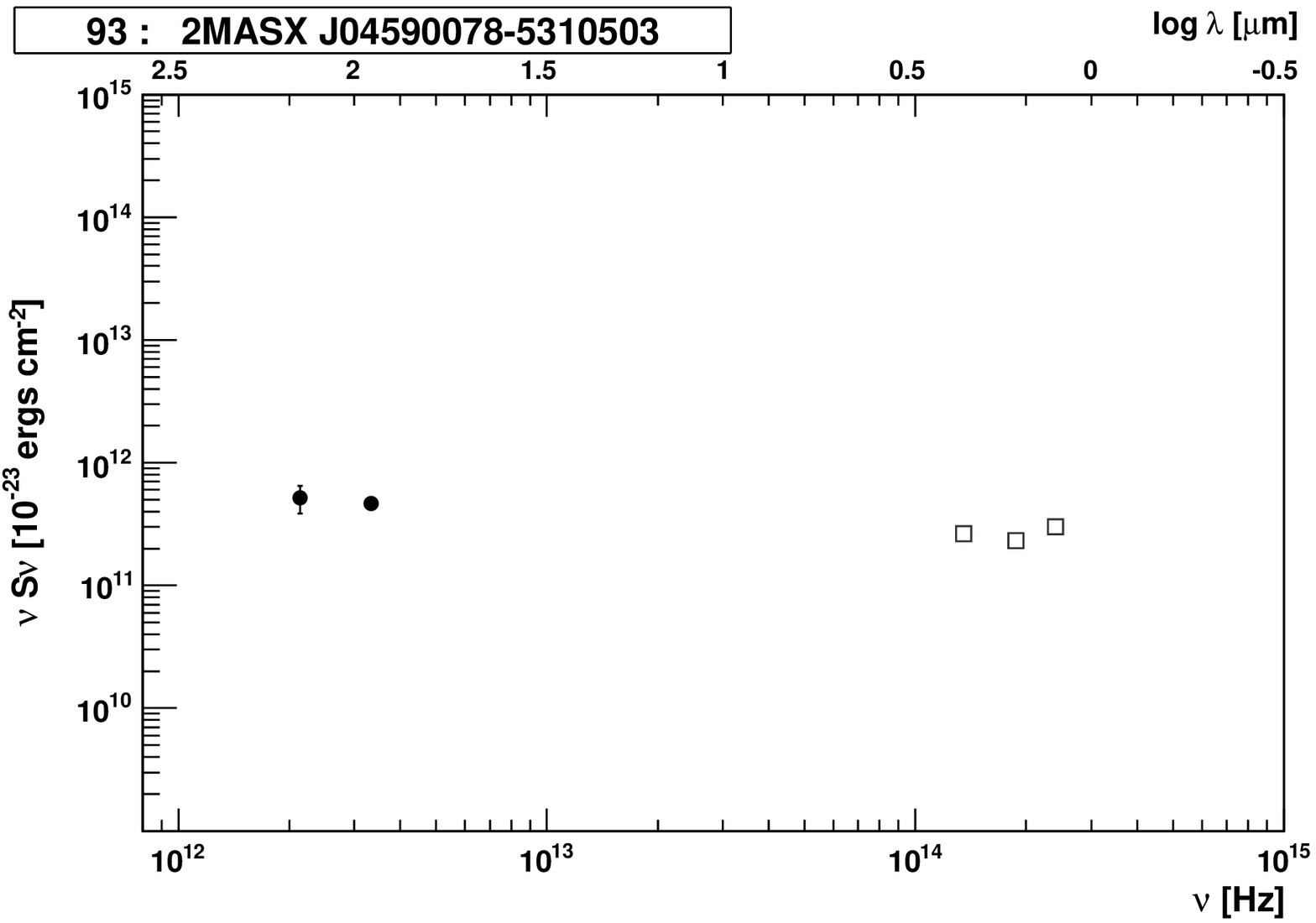}
\includegraphics[width=4cm]{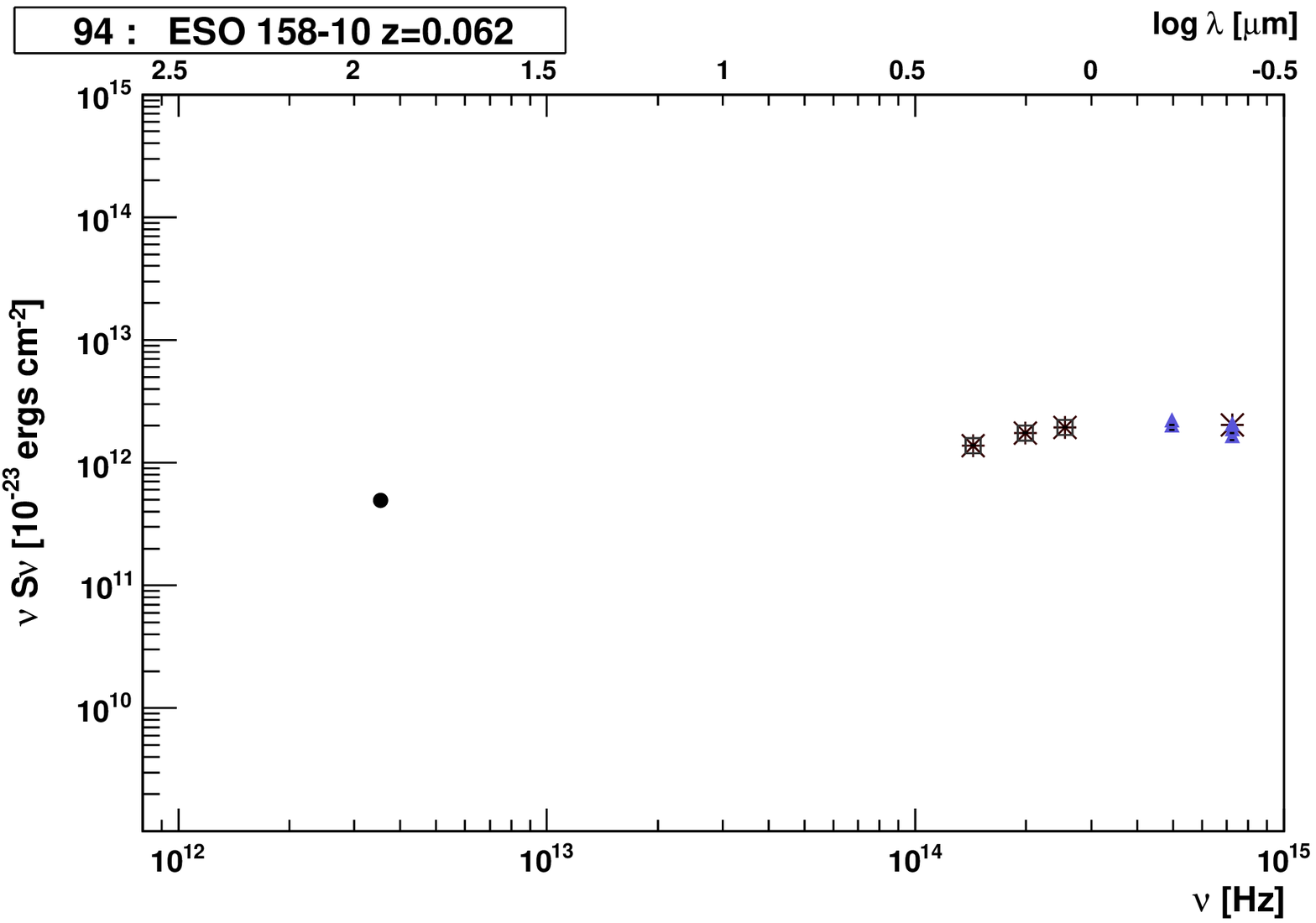}
\includegraphics[width=4cm]{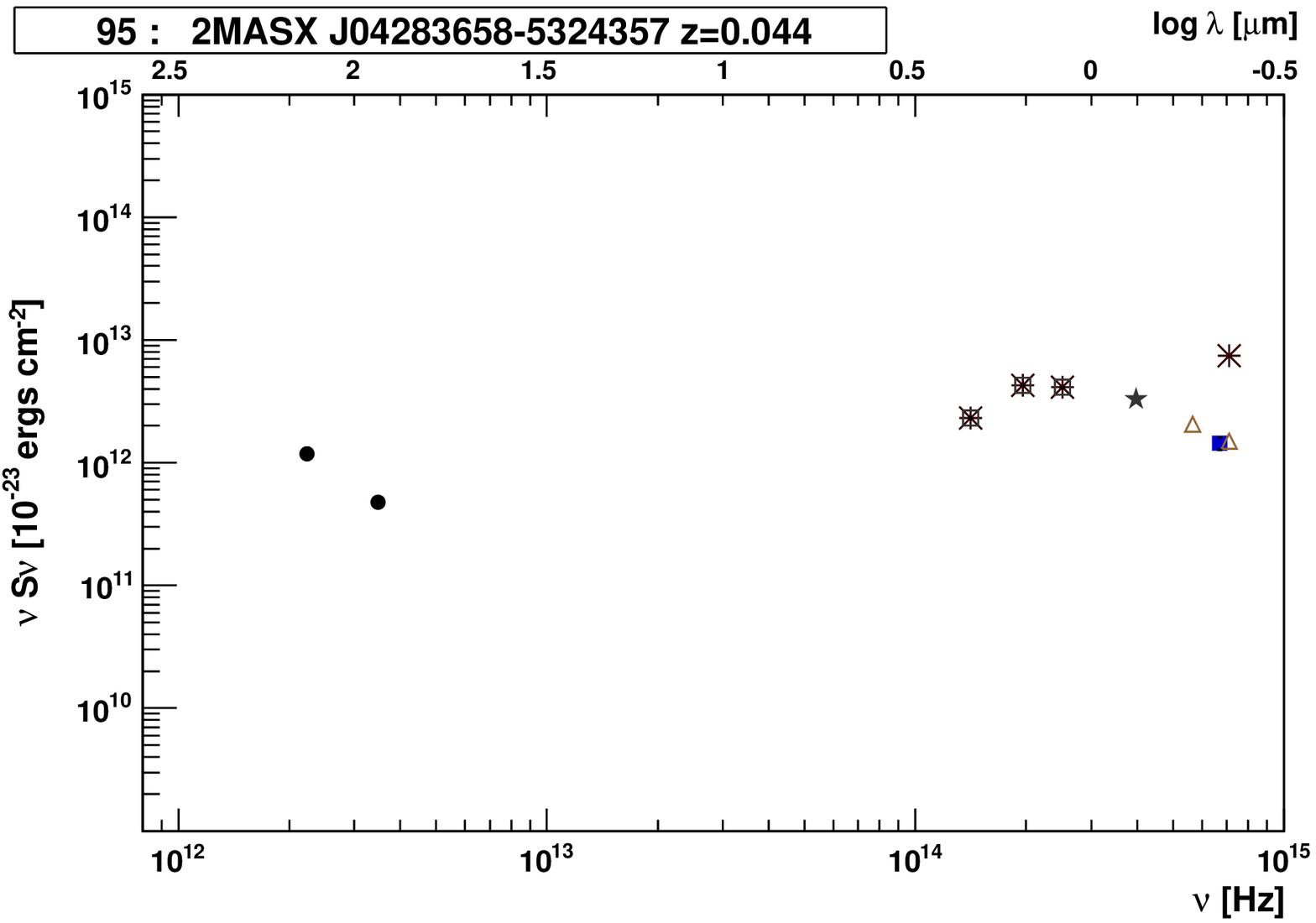}
\includegraphics[width=4cm]{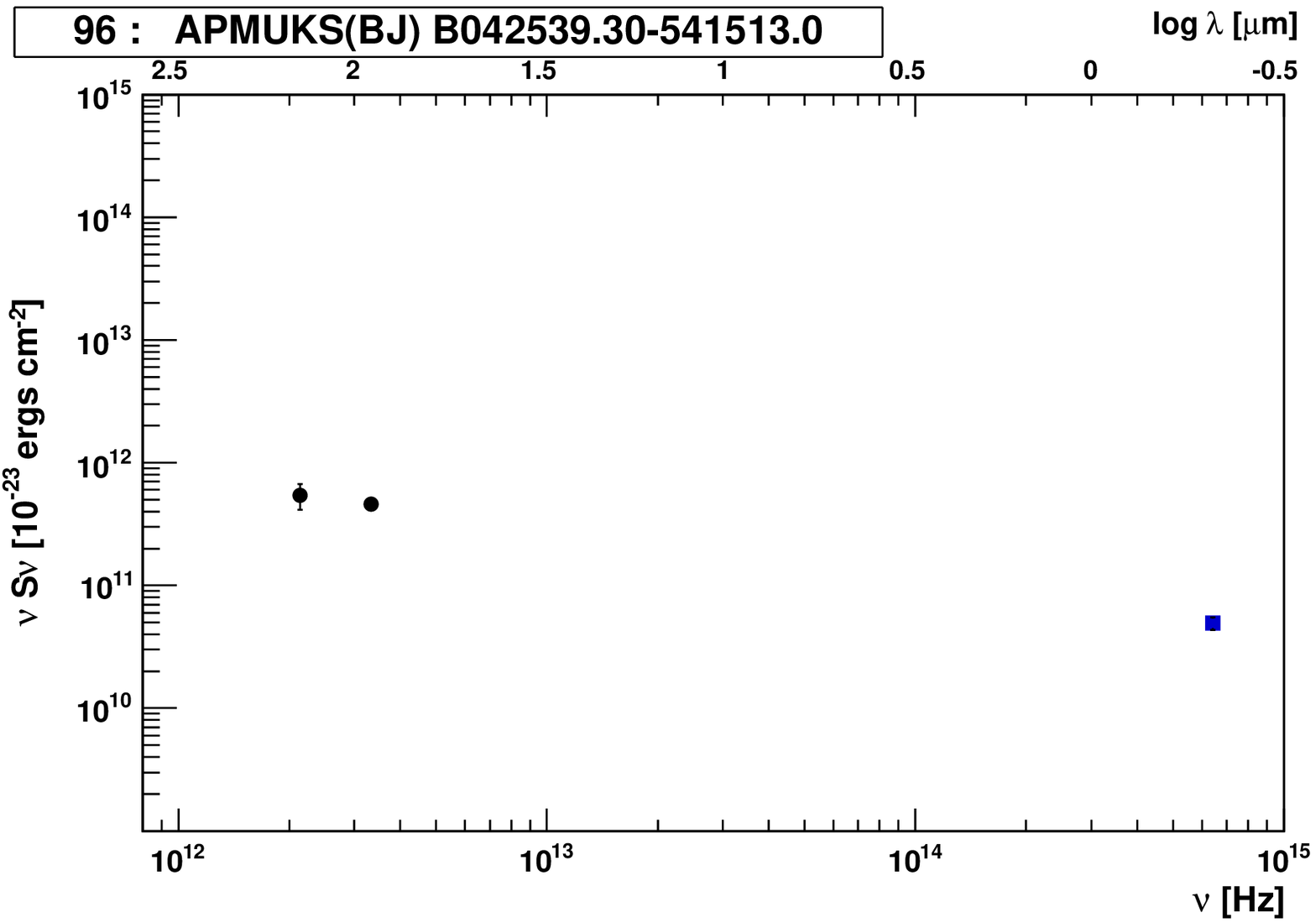}
\includegraphics[width=4cm]{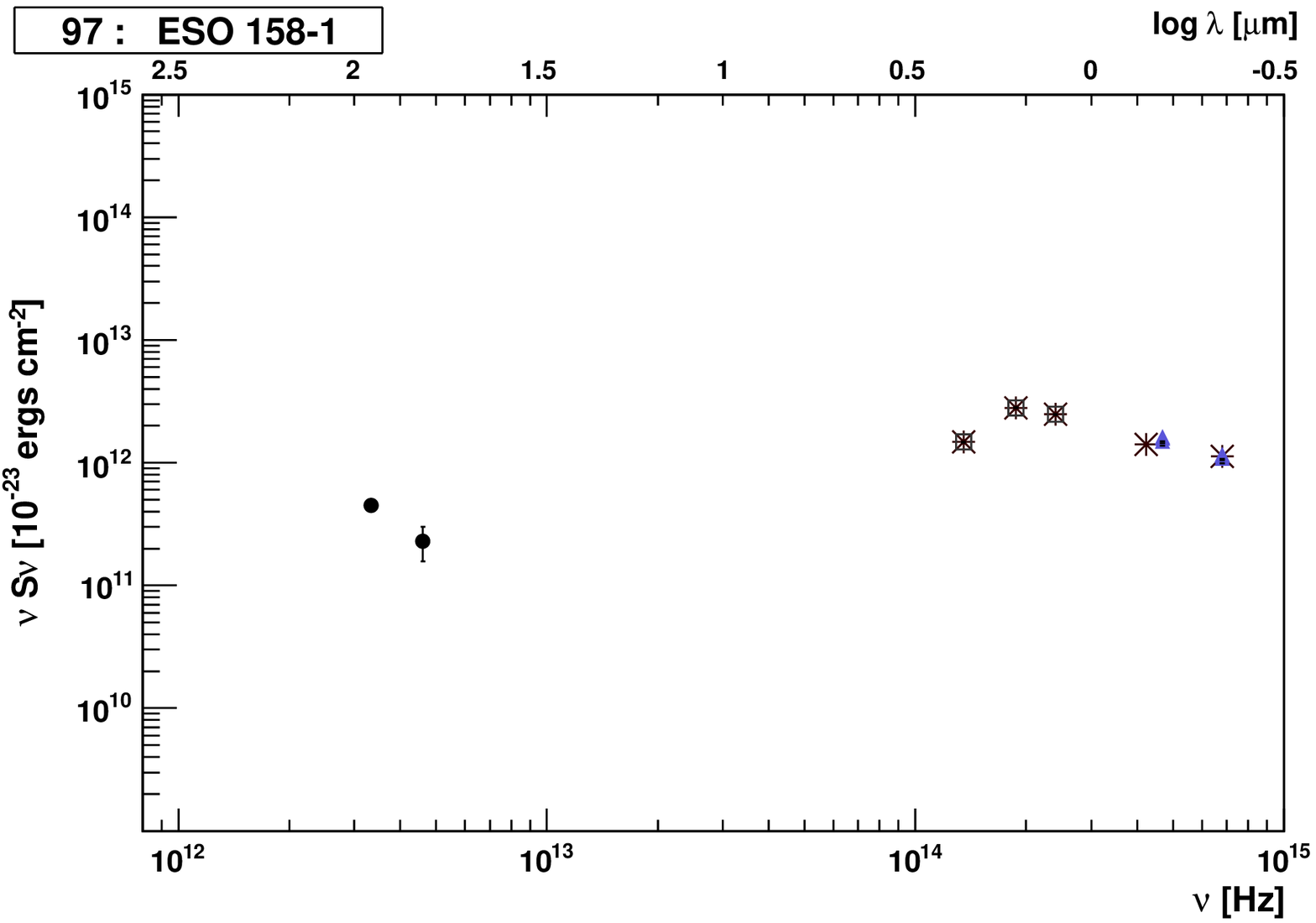}
\includegraphics[width=4cm]{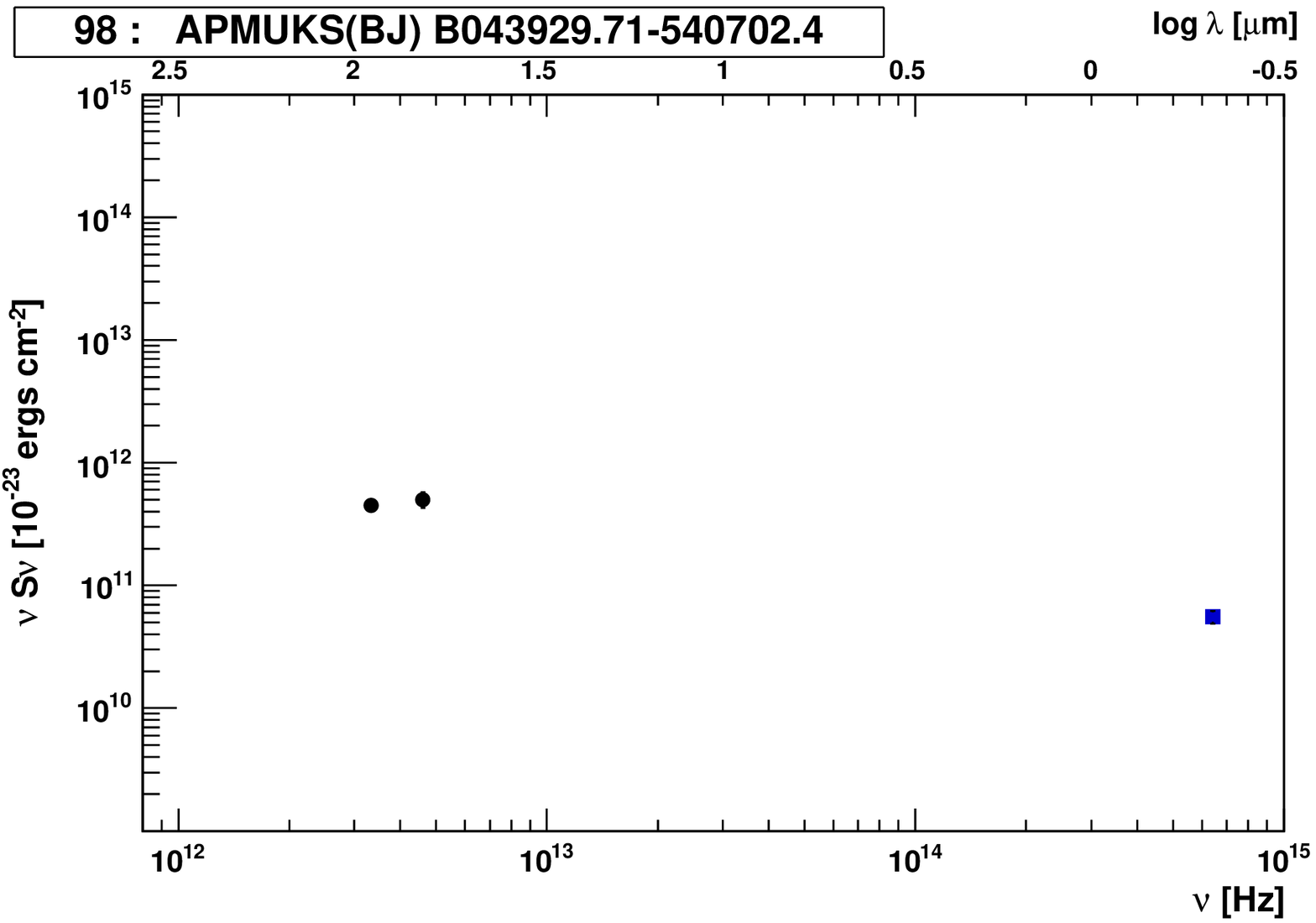}
\includegraphics[width=4cm]{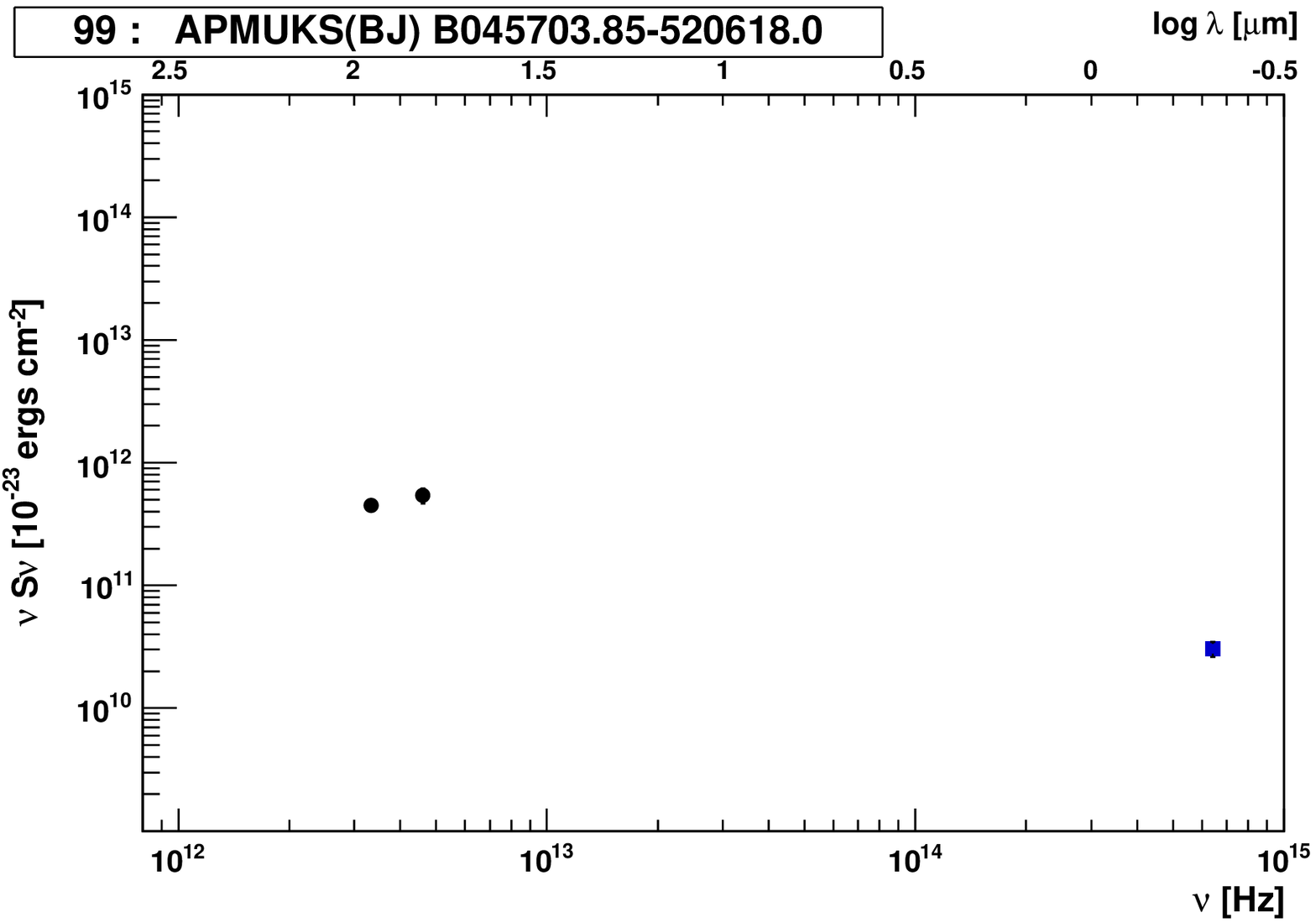}
\includegraphics[width=4cm]{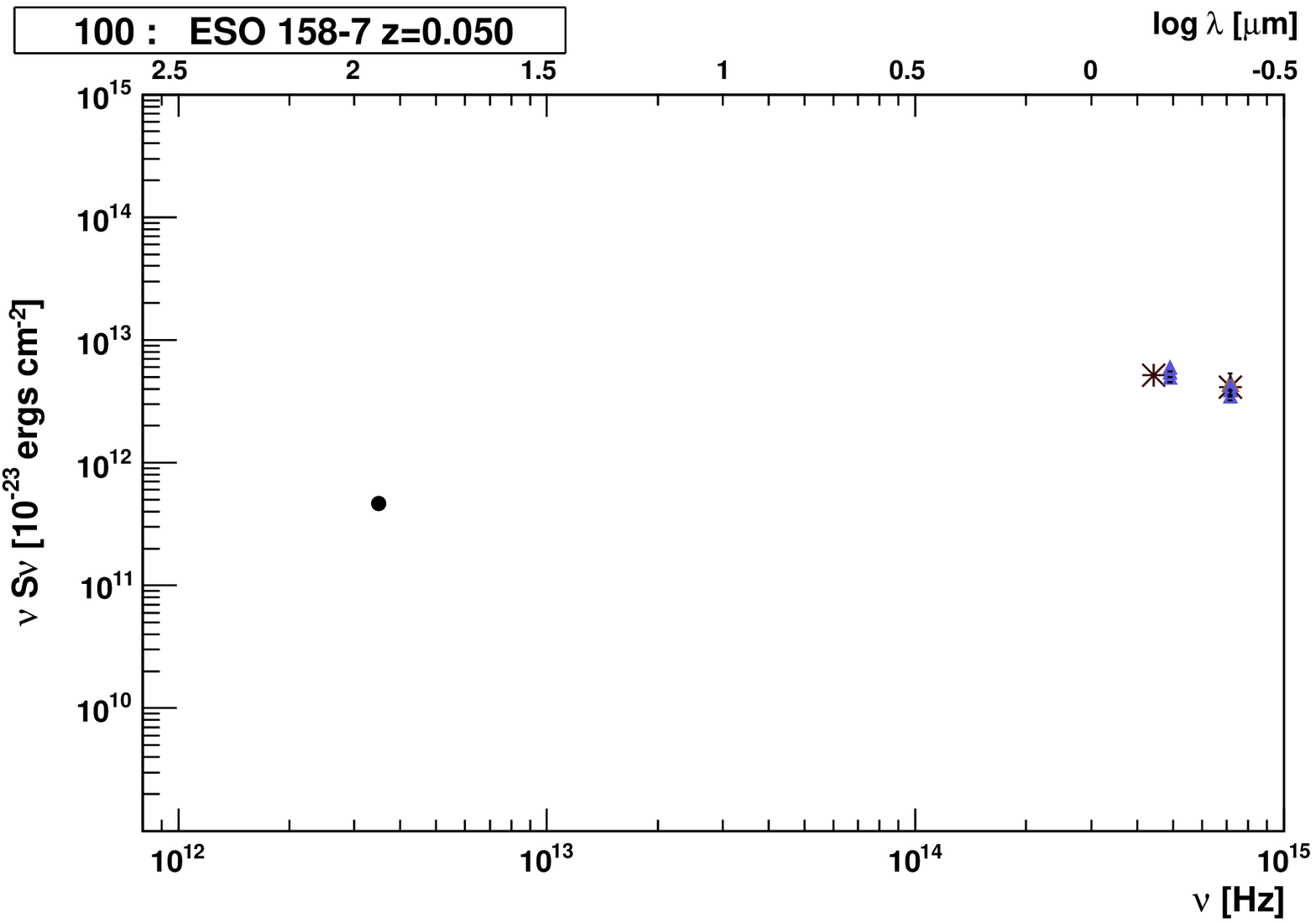}
\includegraphics[width=4cm]{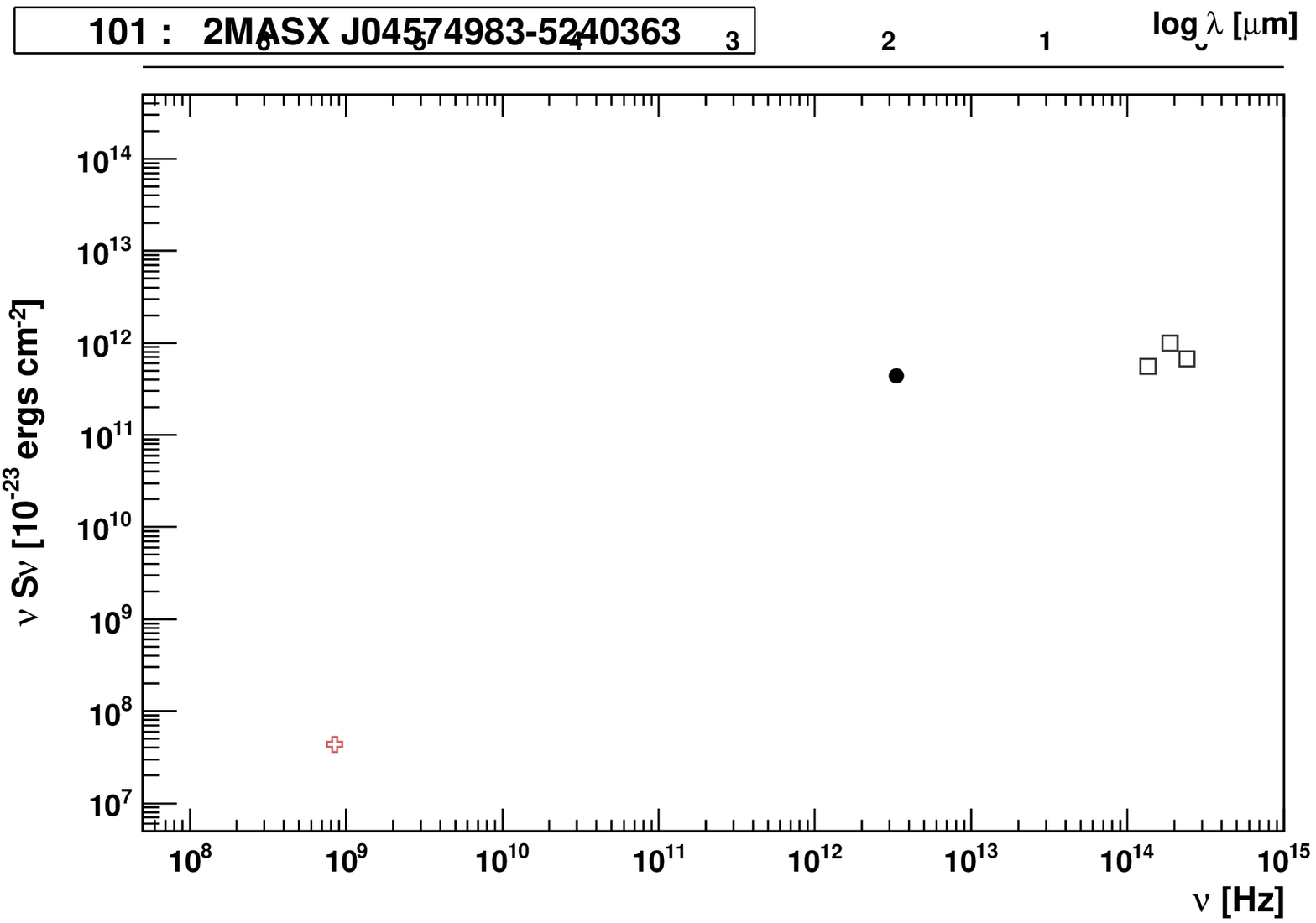}
\includegraphics[width=4cm]{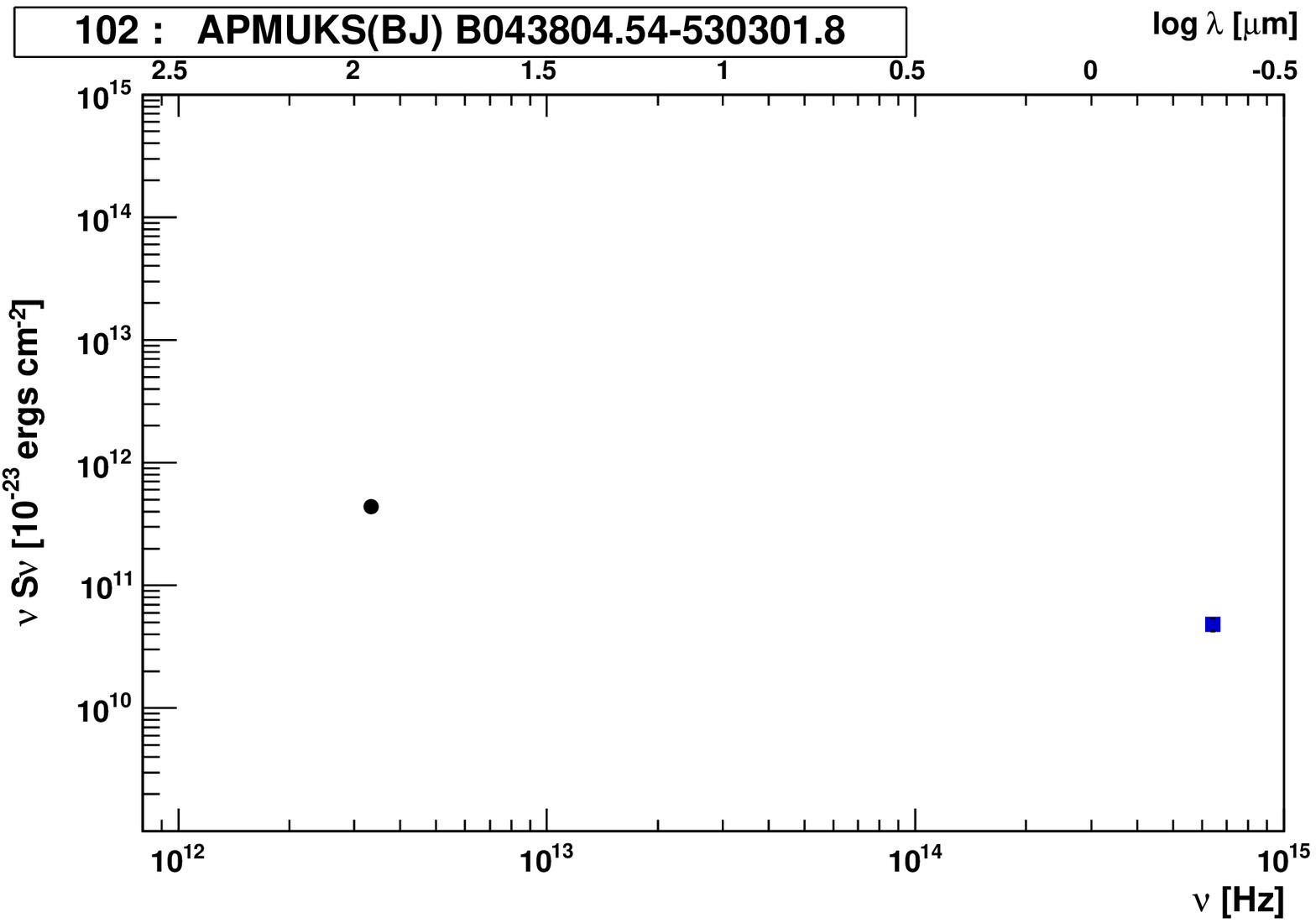}
\includegraphics[width=4cm]{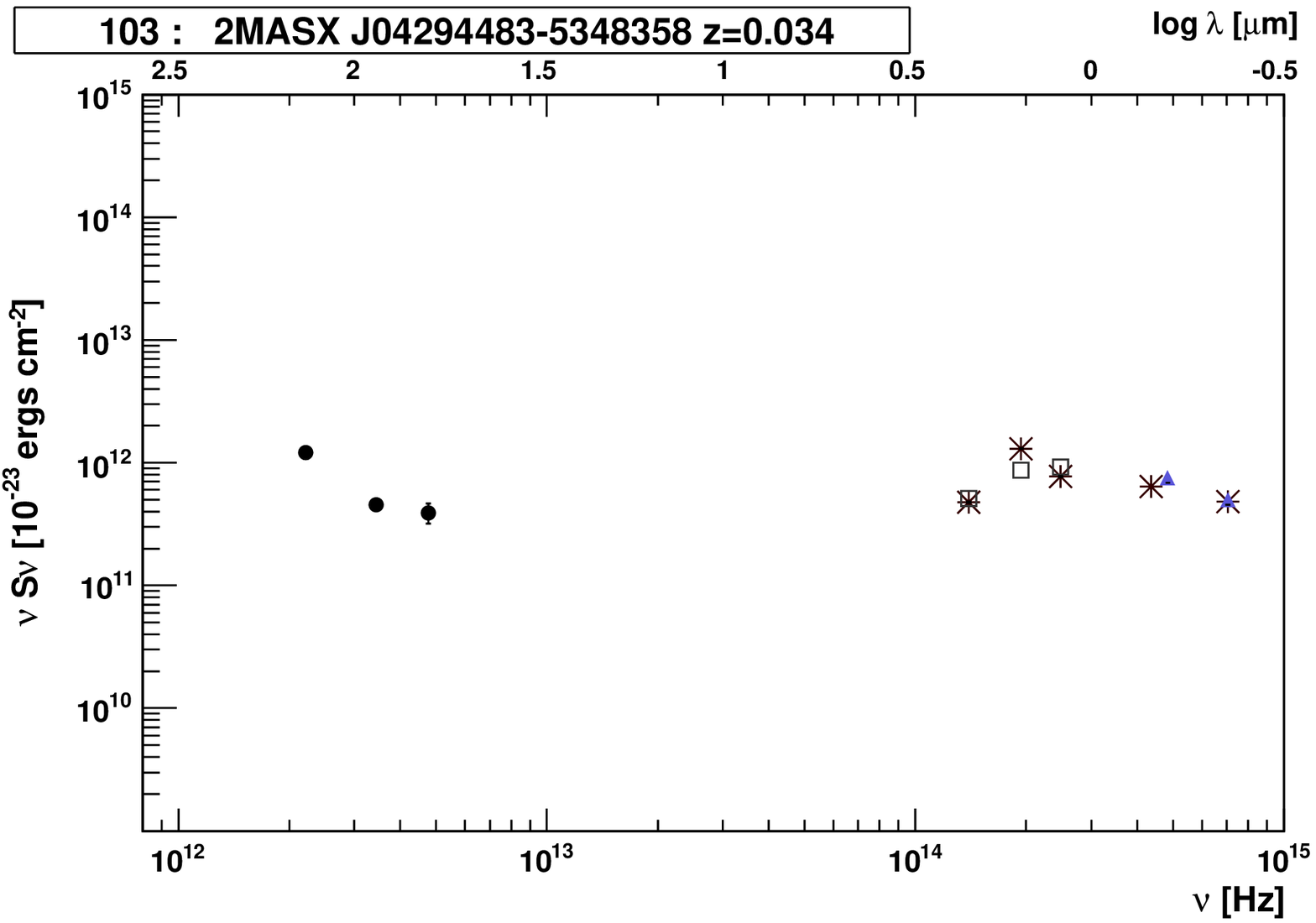}
\includegraphics[width=4cm]{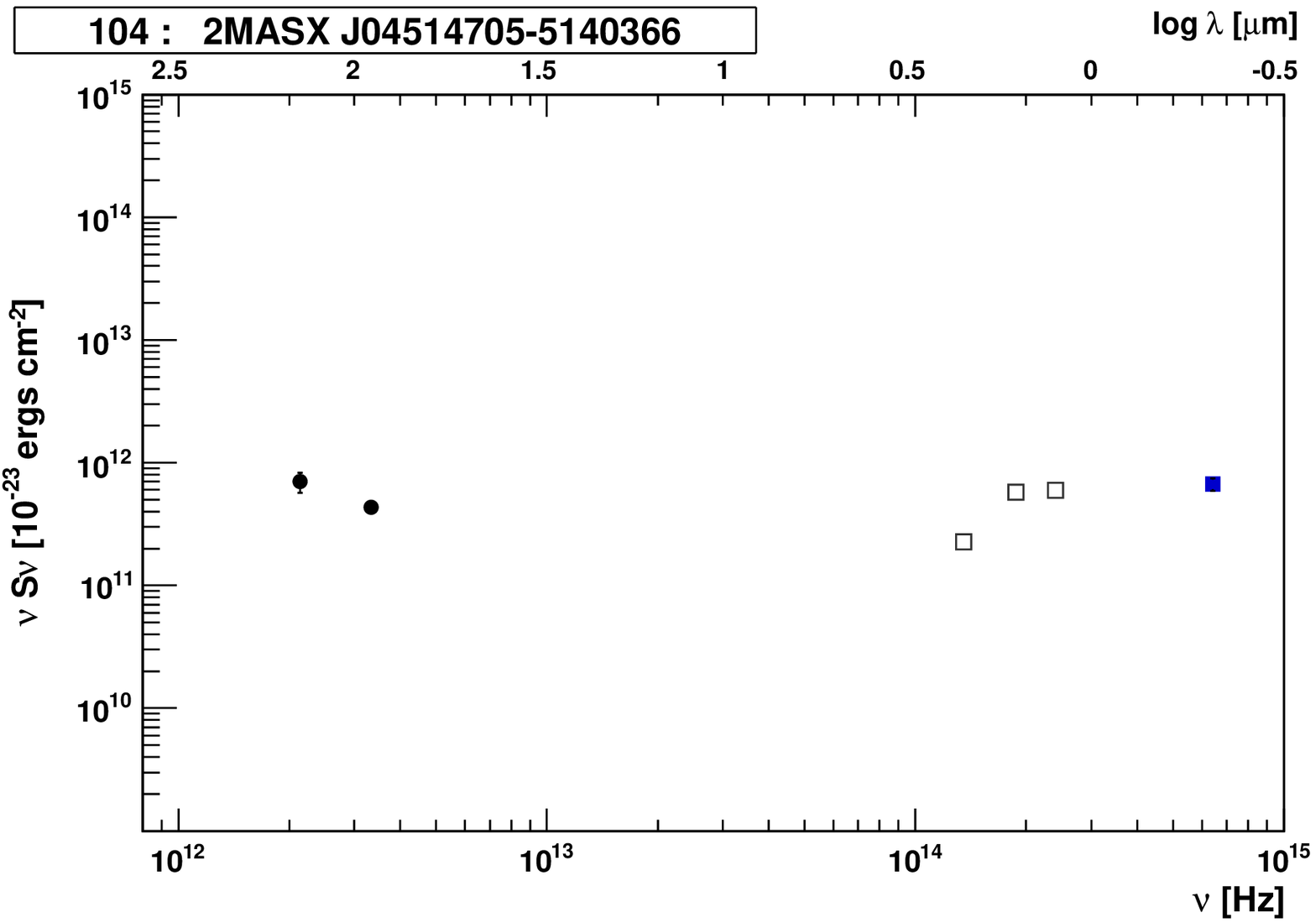}
\includegraphics[width=4cm]{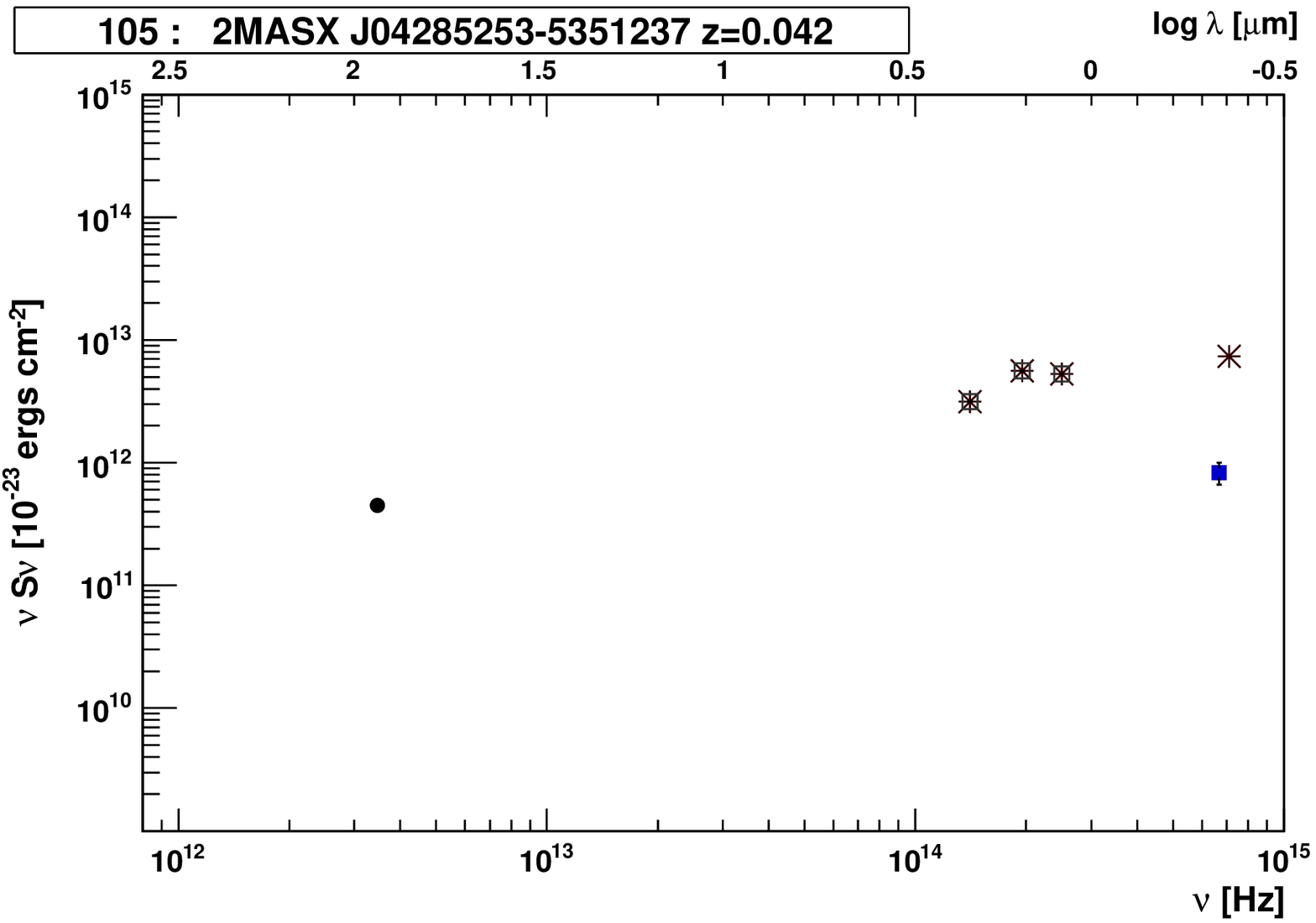}
\includegraphics[width=4cm]{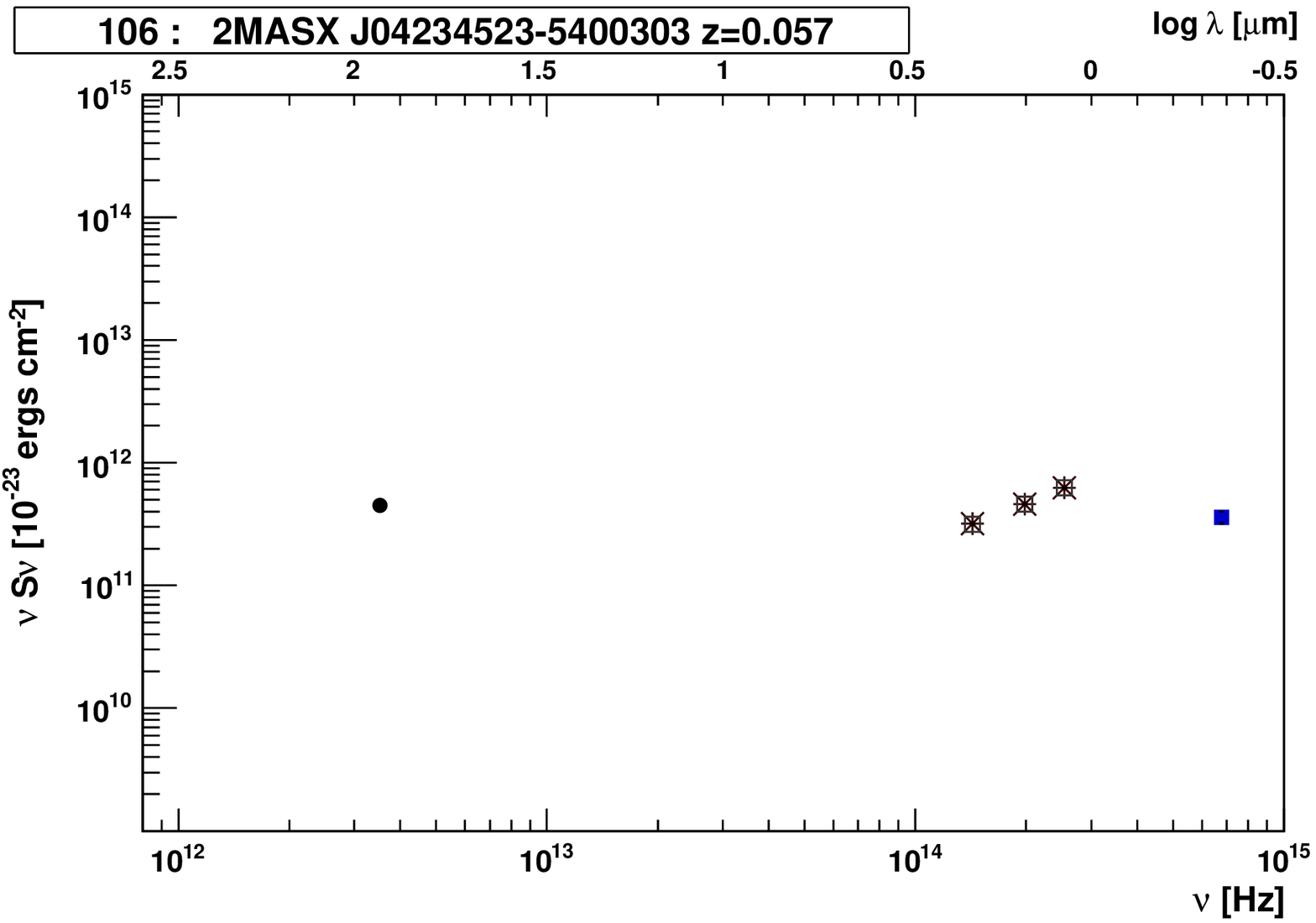}
\includegraphics[width=4cm]{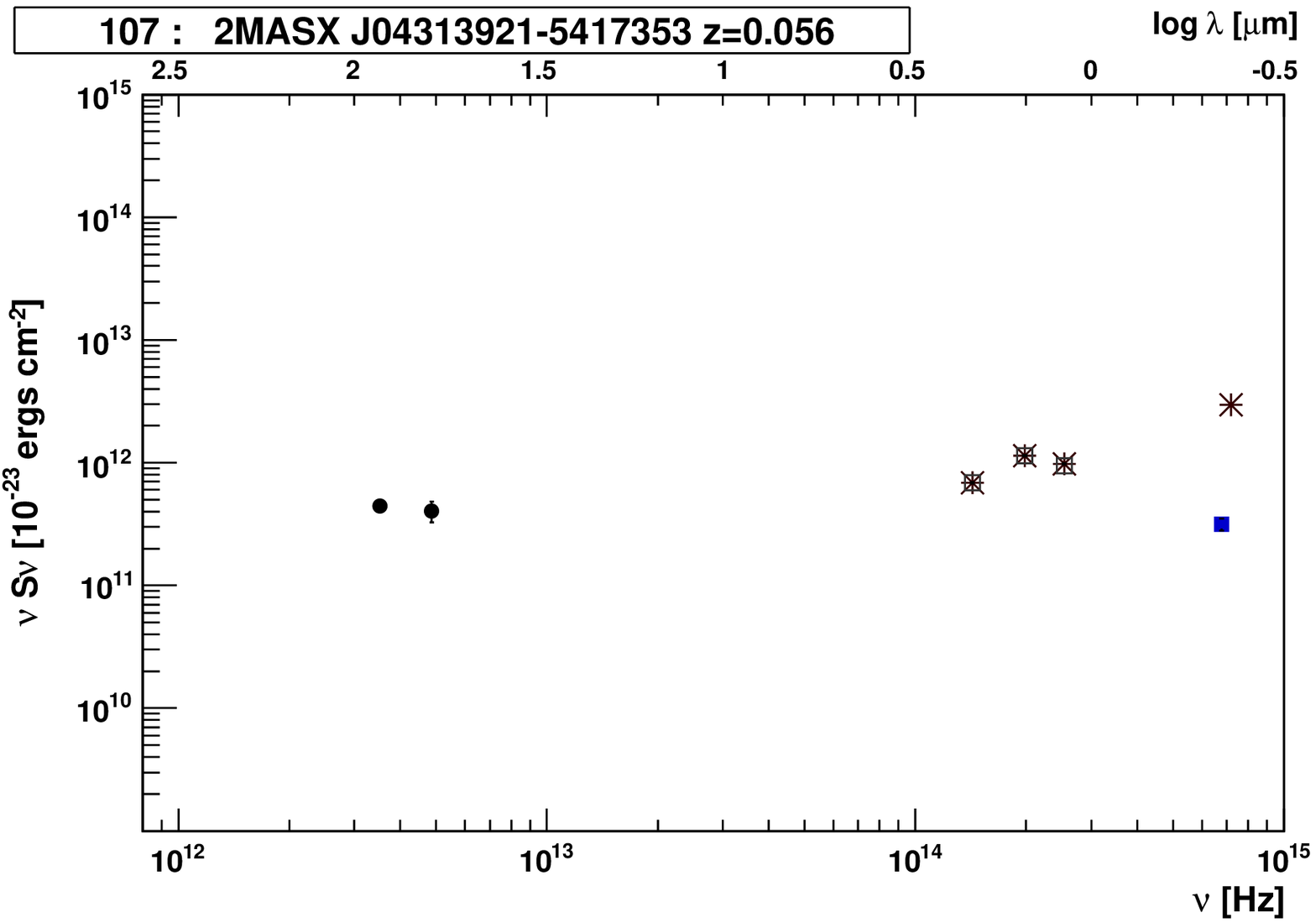}
\includegraphics[width=4cm]{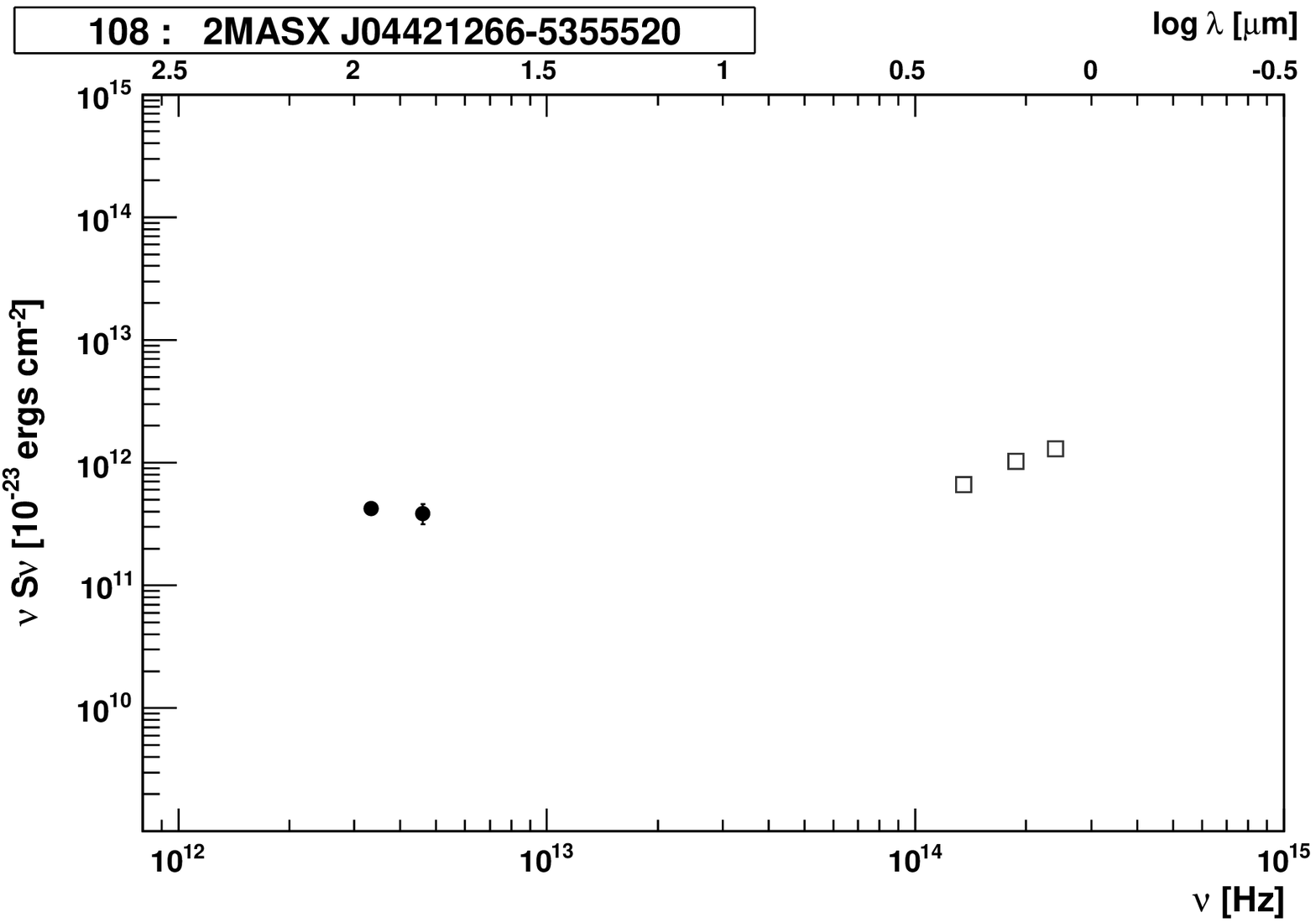}
\includegraphics[width=4cm]{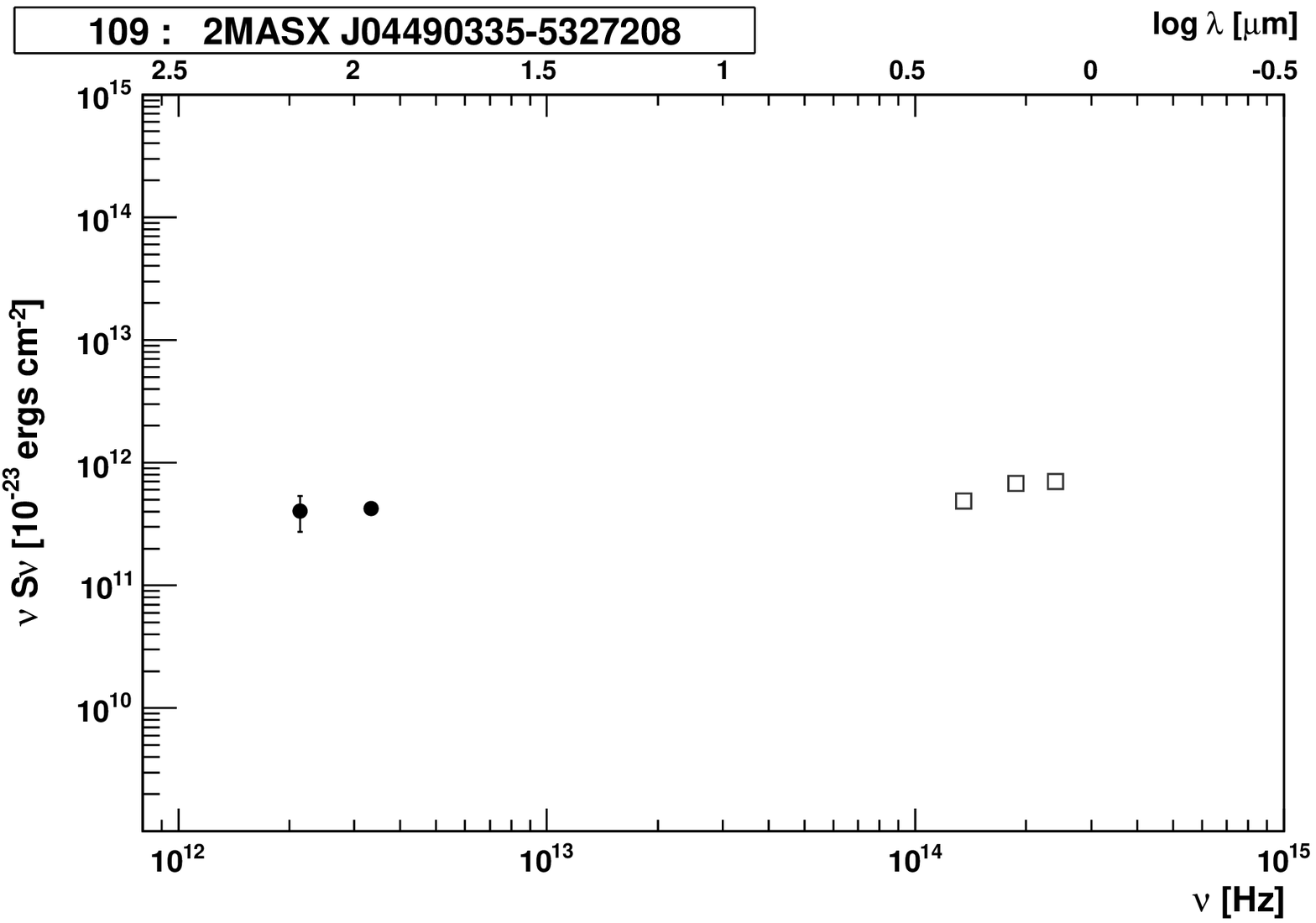}
\includegraphics[width=4cm]{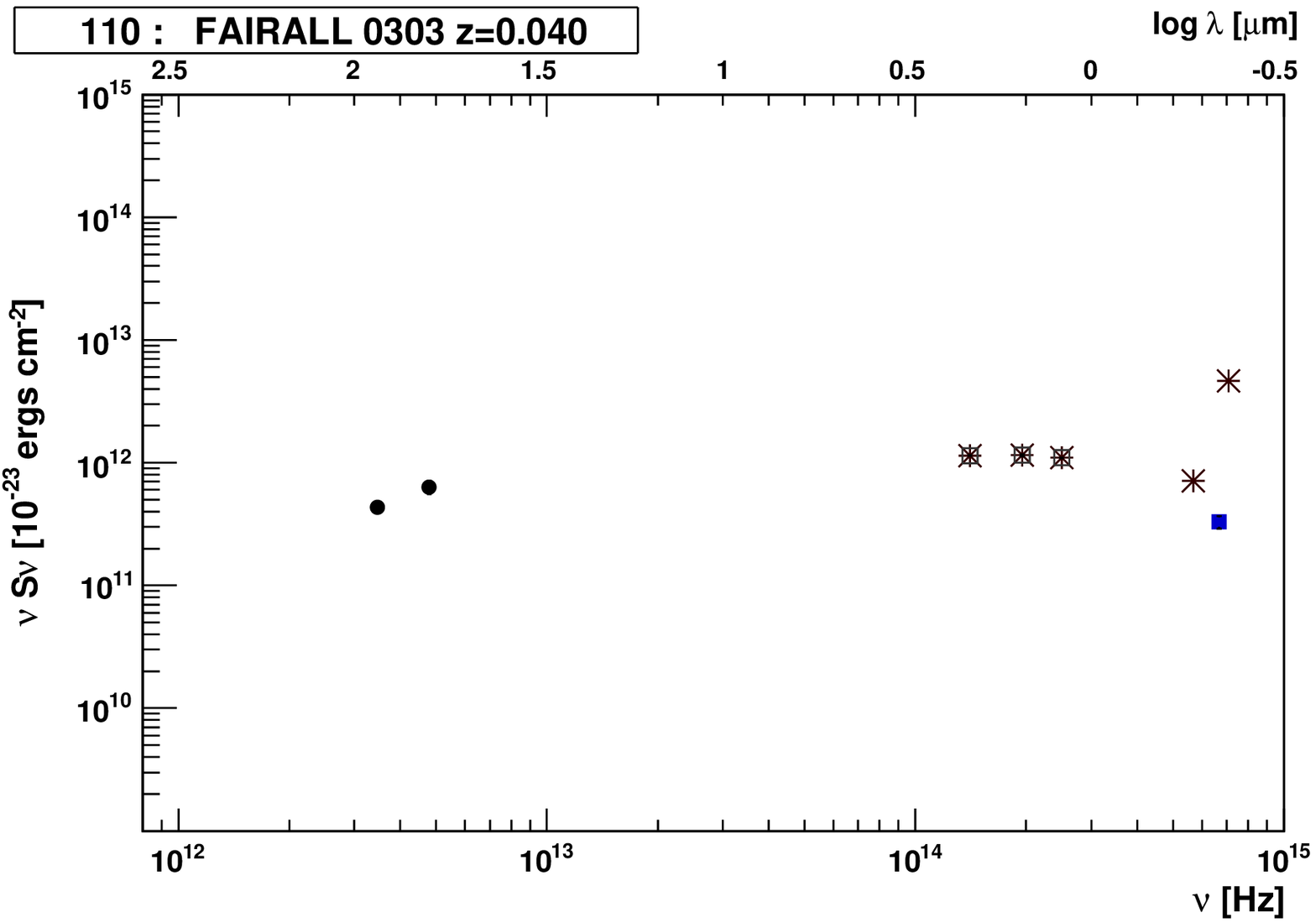}
\includegraphics[width=4cm]{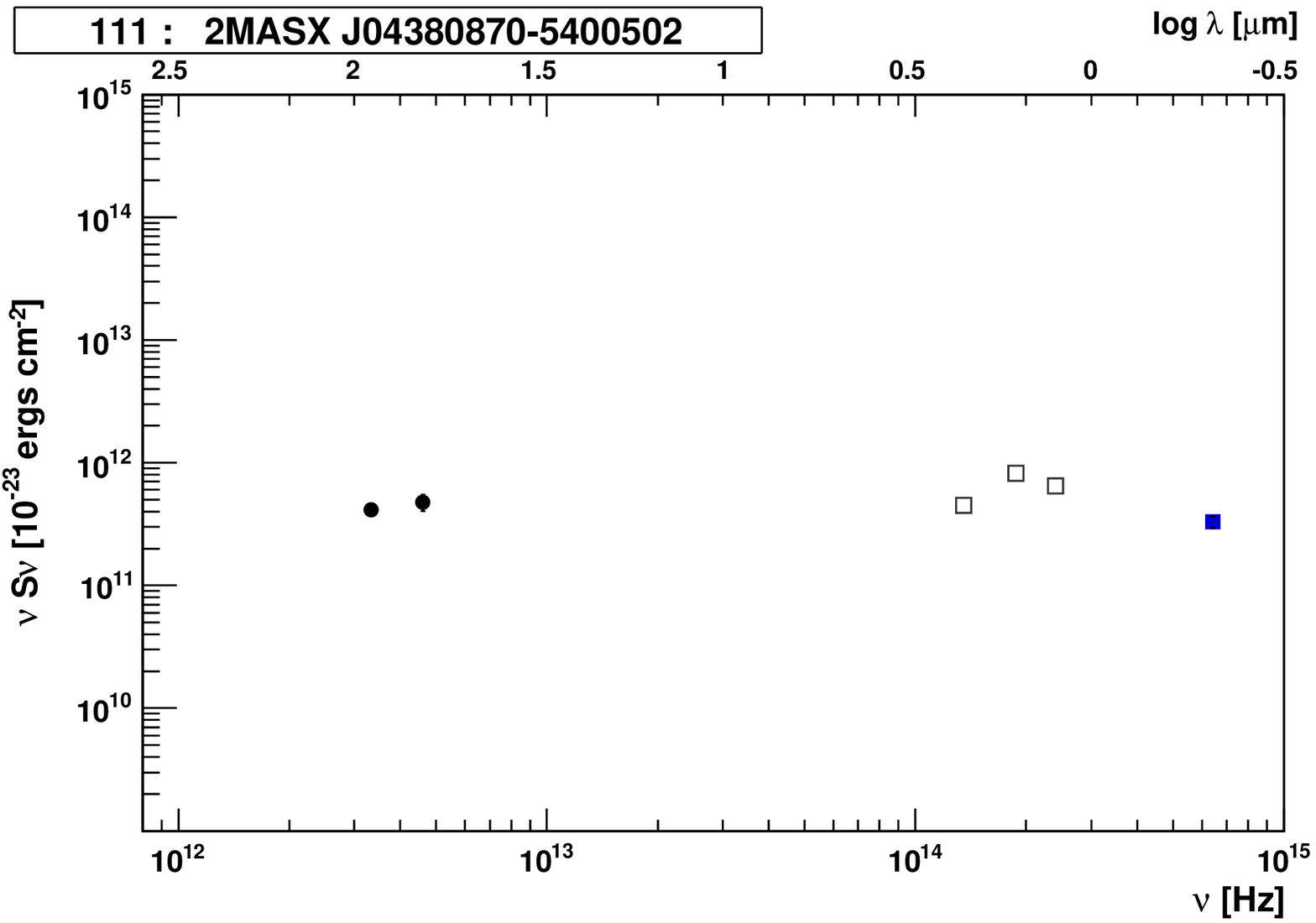}
\includegraphics[width=4cm]{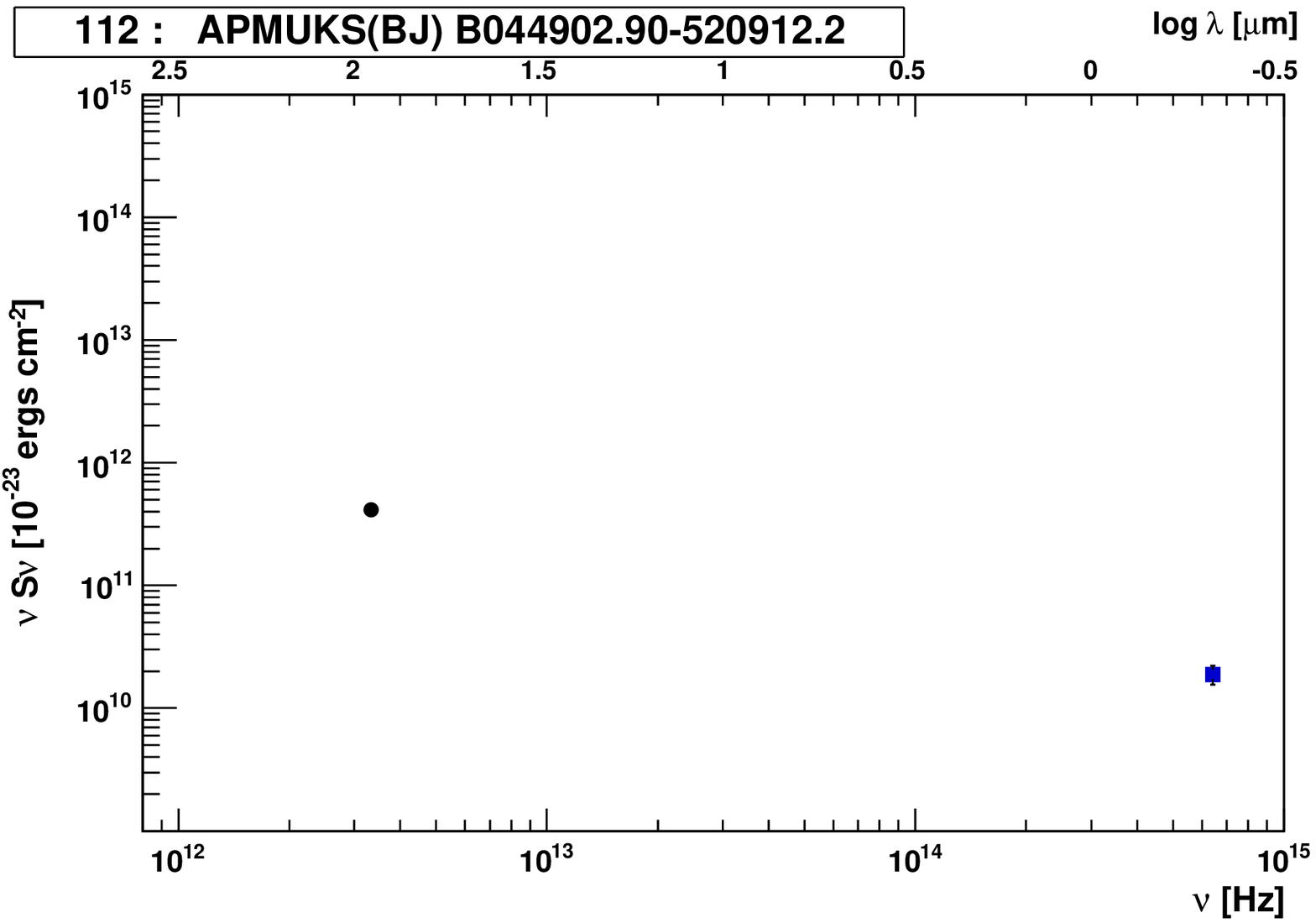}
\includegraphics[width=4cm]{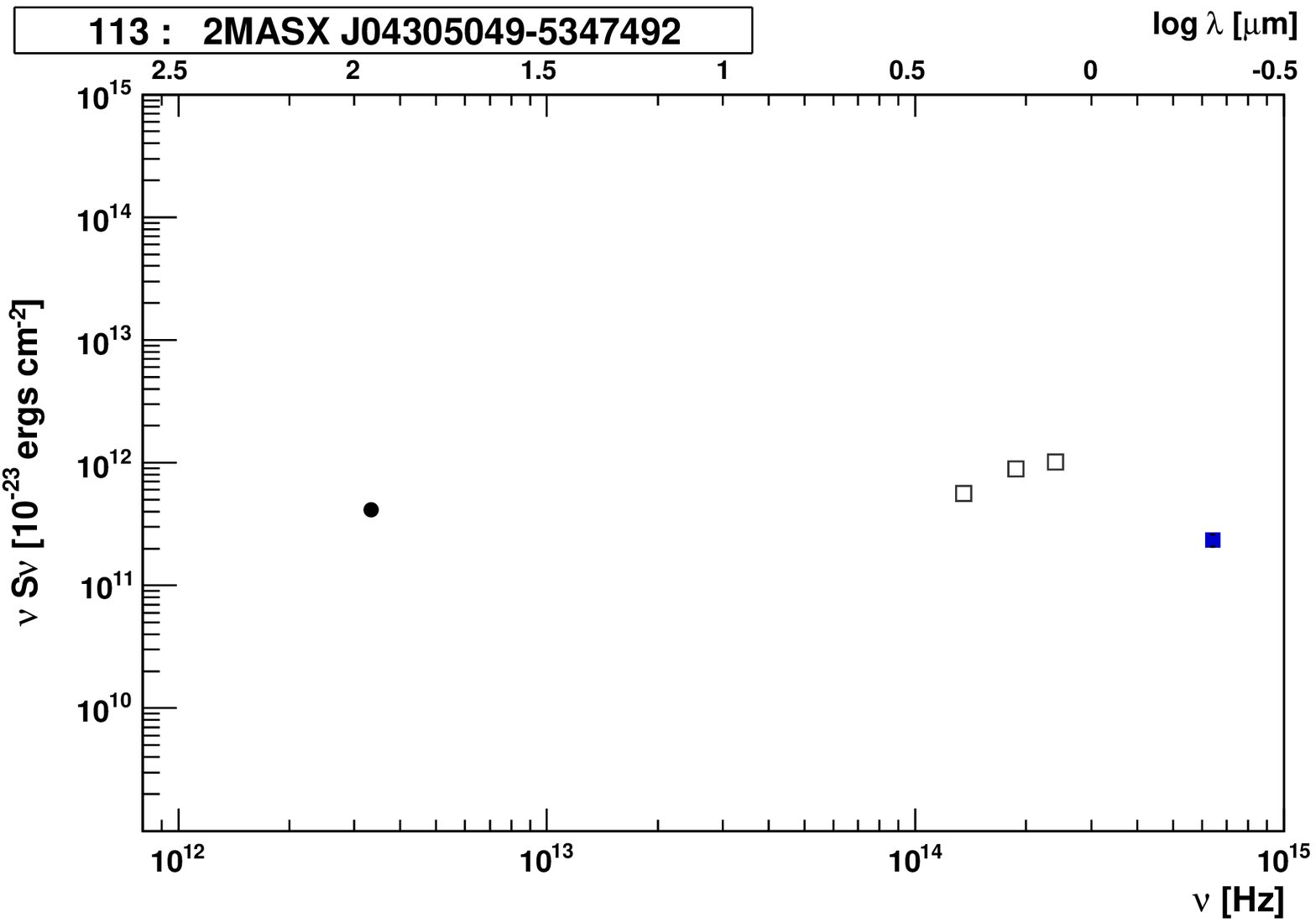}
\includegraphics[width=4cm]{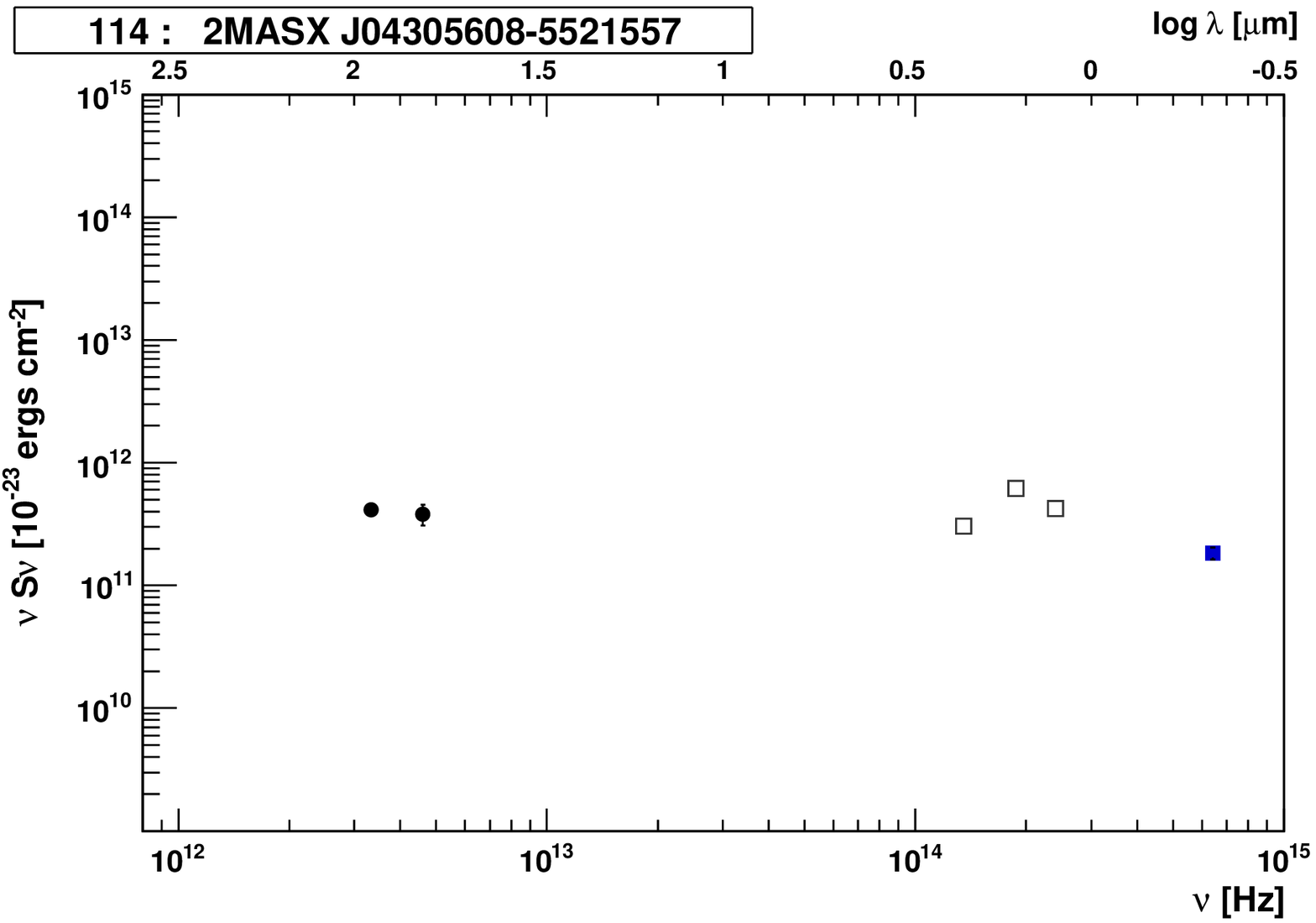}
\includegraphics[width=4cm]{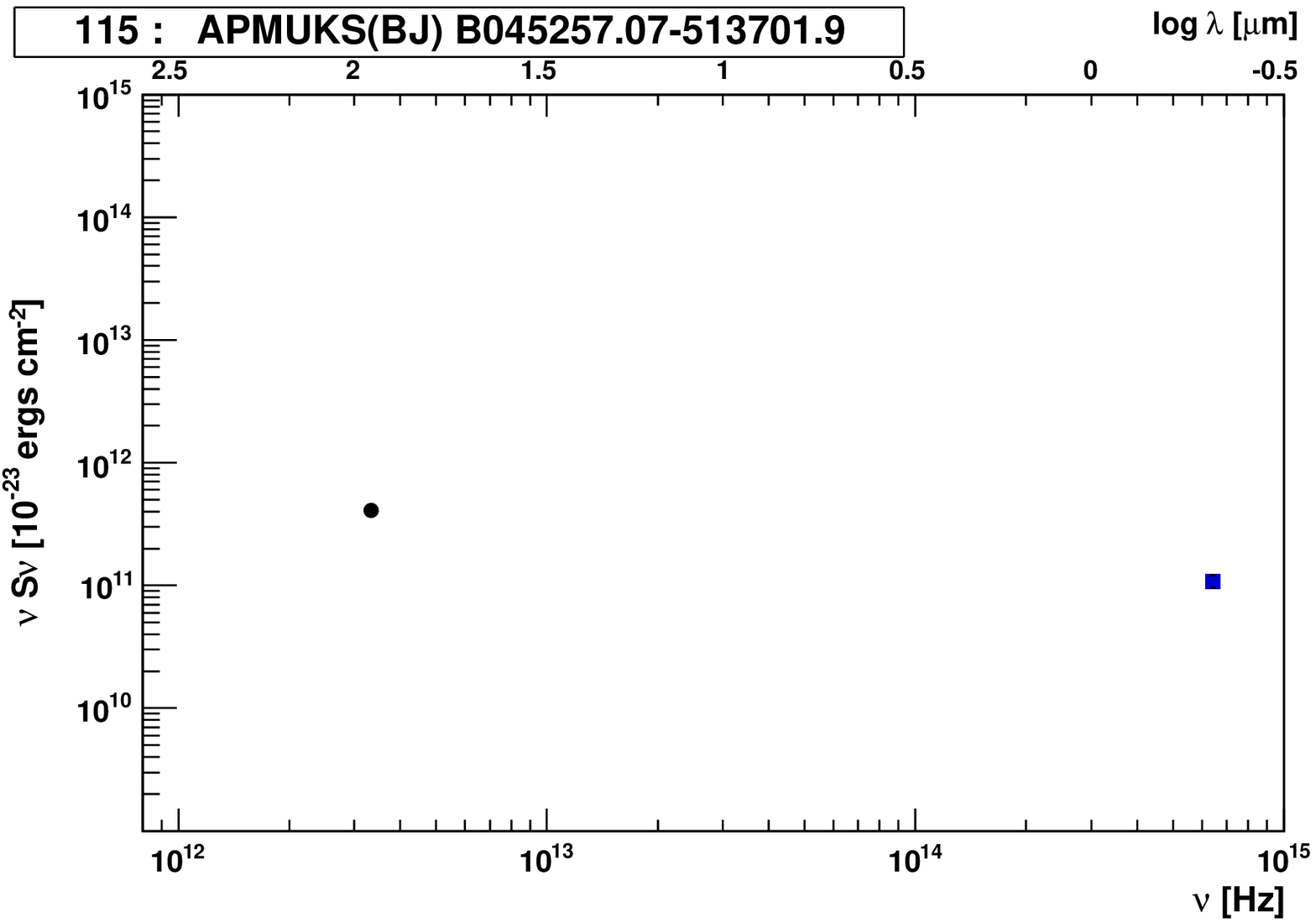}
\includegraphics[width=4cm]{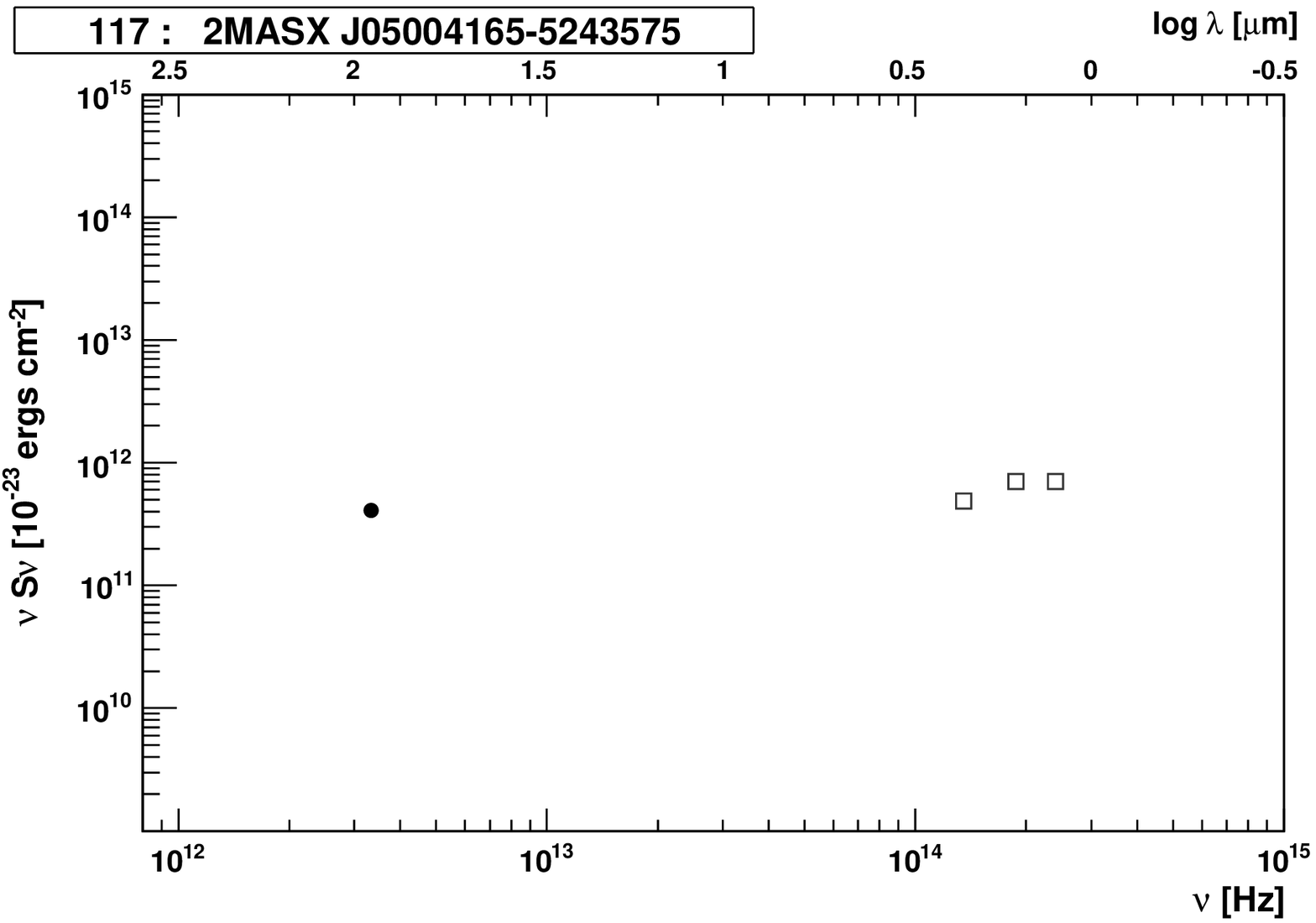}
\includegraphics[width=4cm]{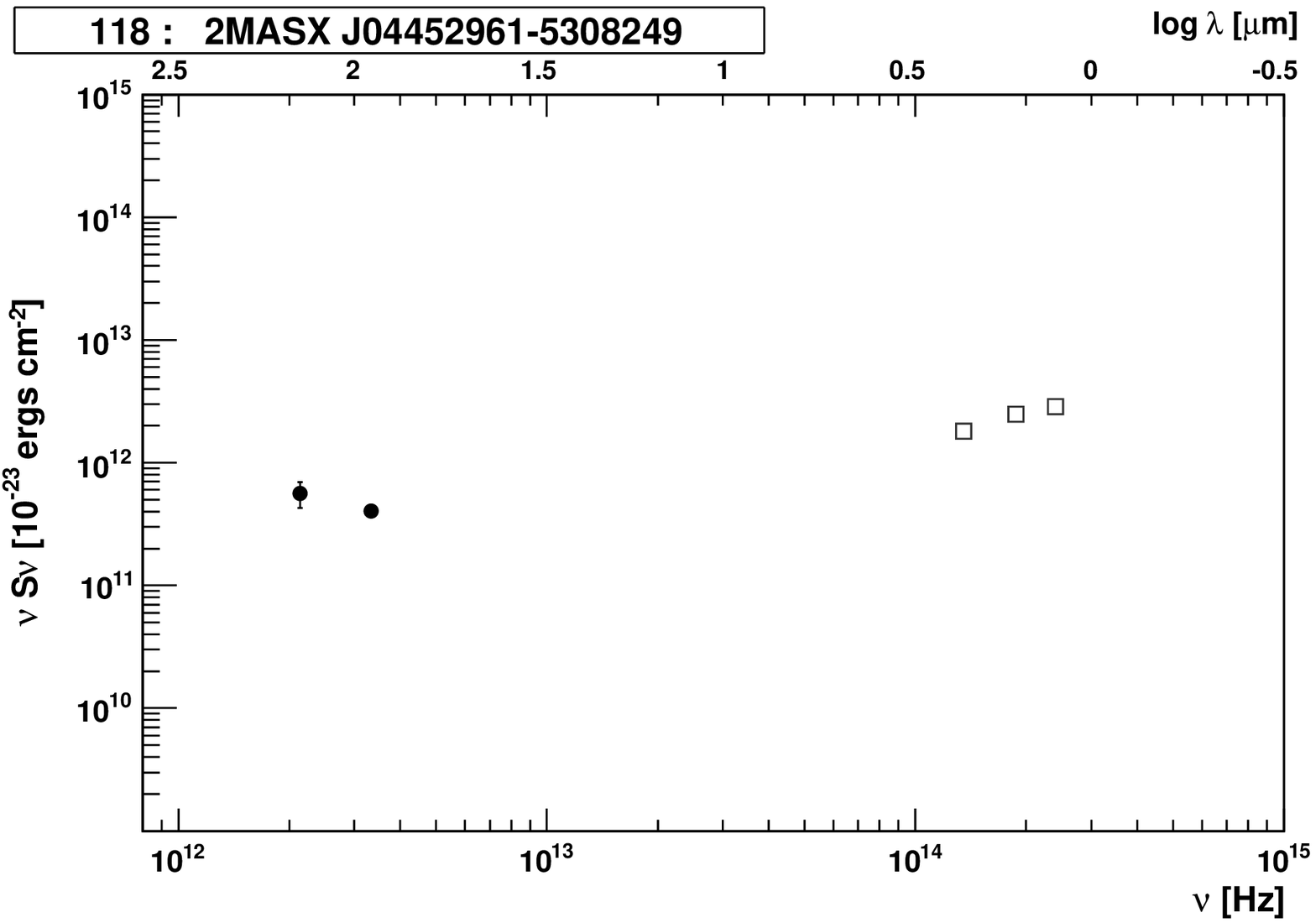}
\label{points3}
\caption {SEDs for the next 36 ADF-S identified sources, with symbols as in Figure~\ref{points1}.}
\end{figure*}
}

\clearpage

\onlfig{4}{
\begin{figure*}[t]
\centering

\includegraphics[width=4cm]{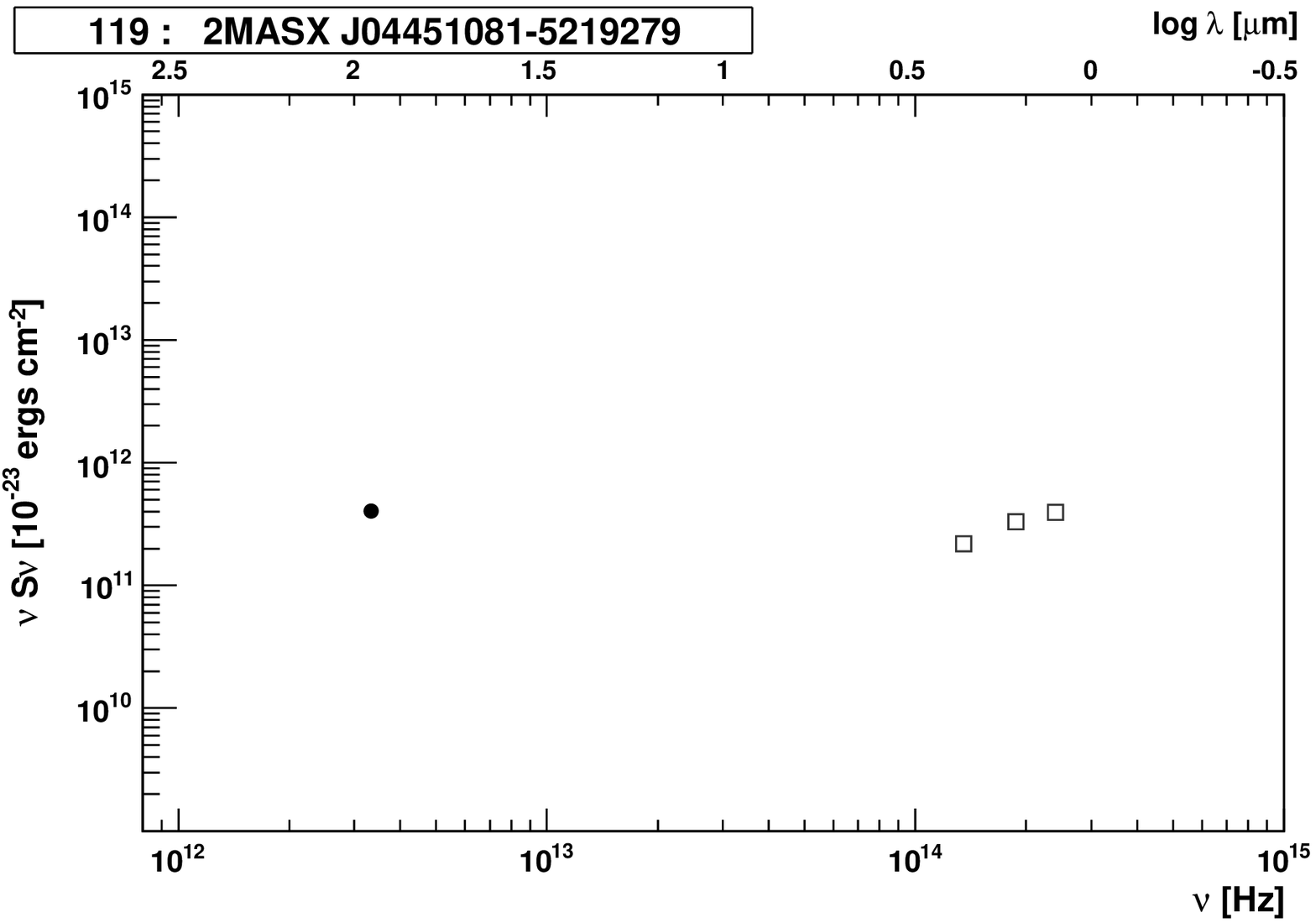}
\includegraphics[width=4cm]{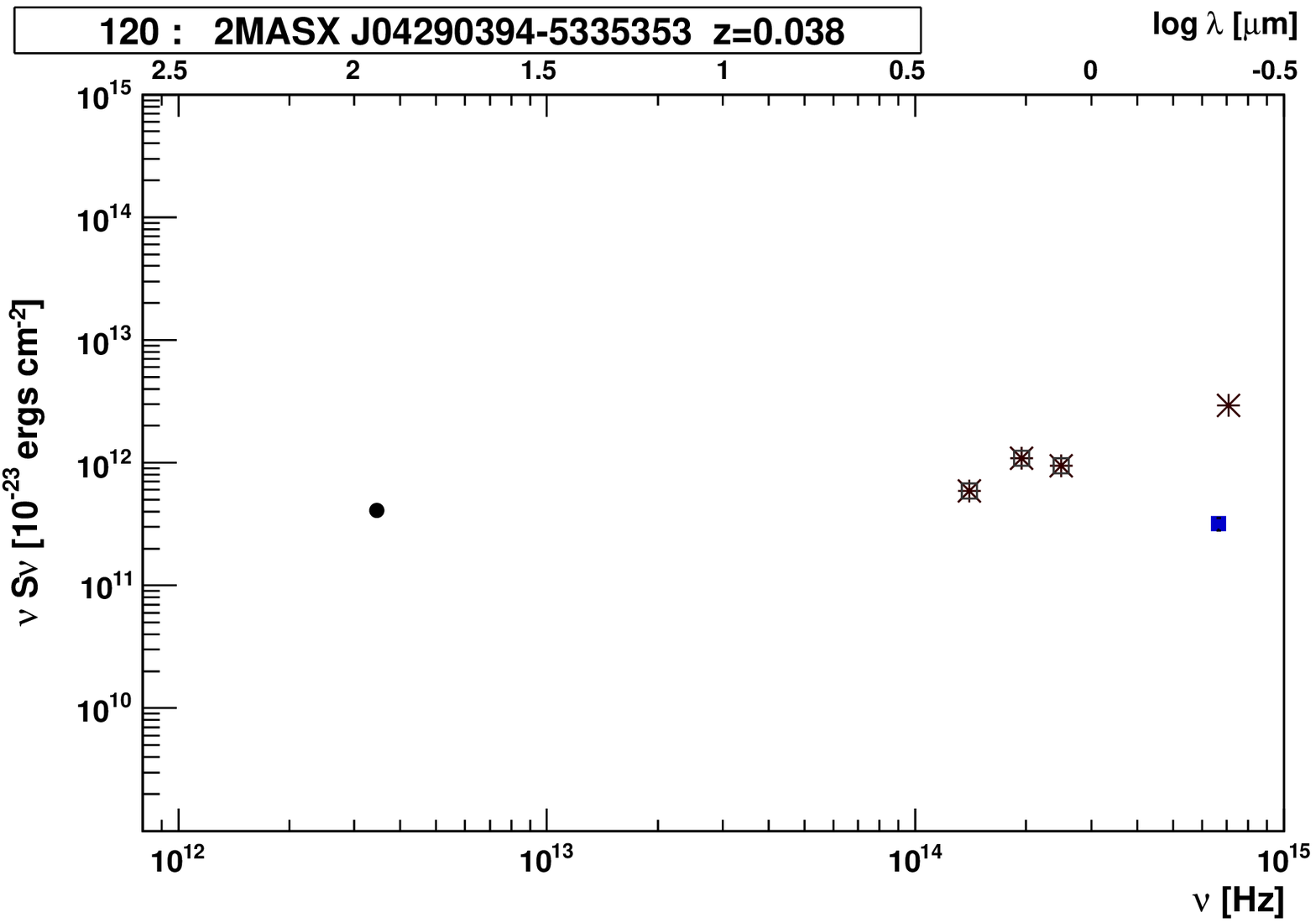}
\includegraphics[width=4cm]{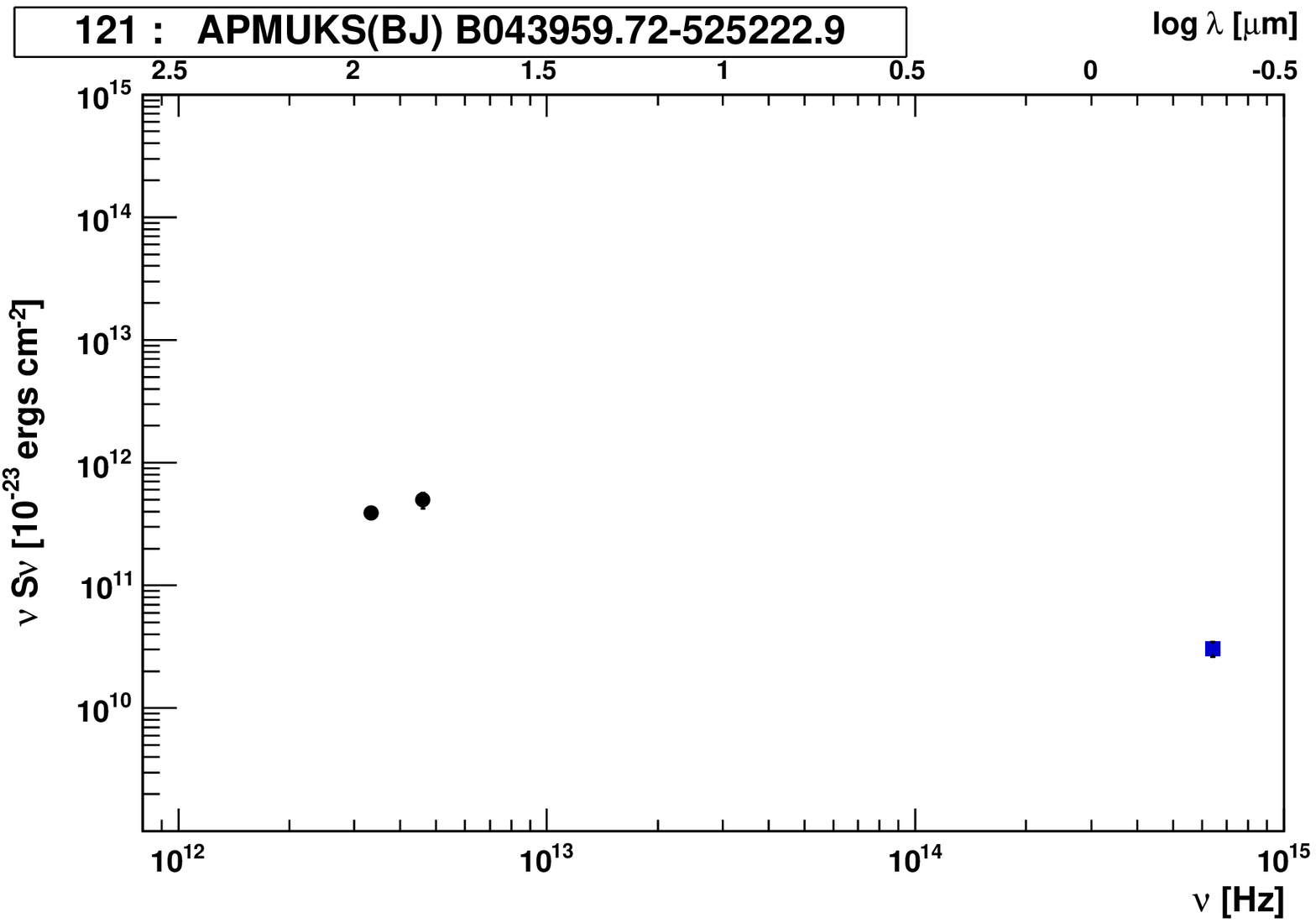}
\includegraphics[width=4cm]{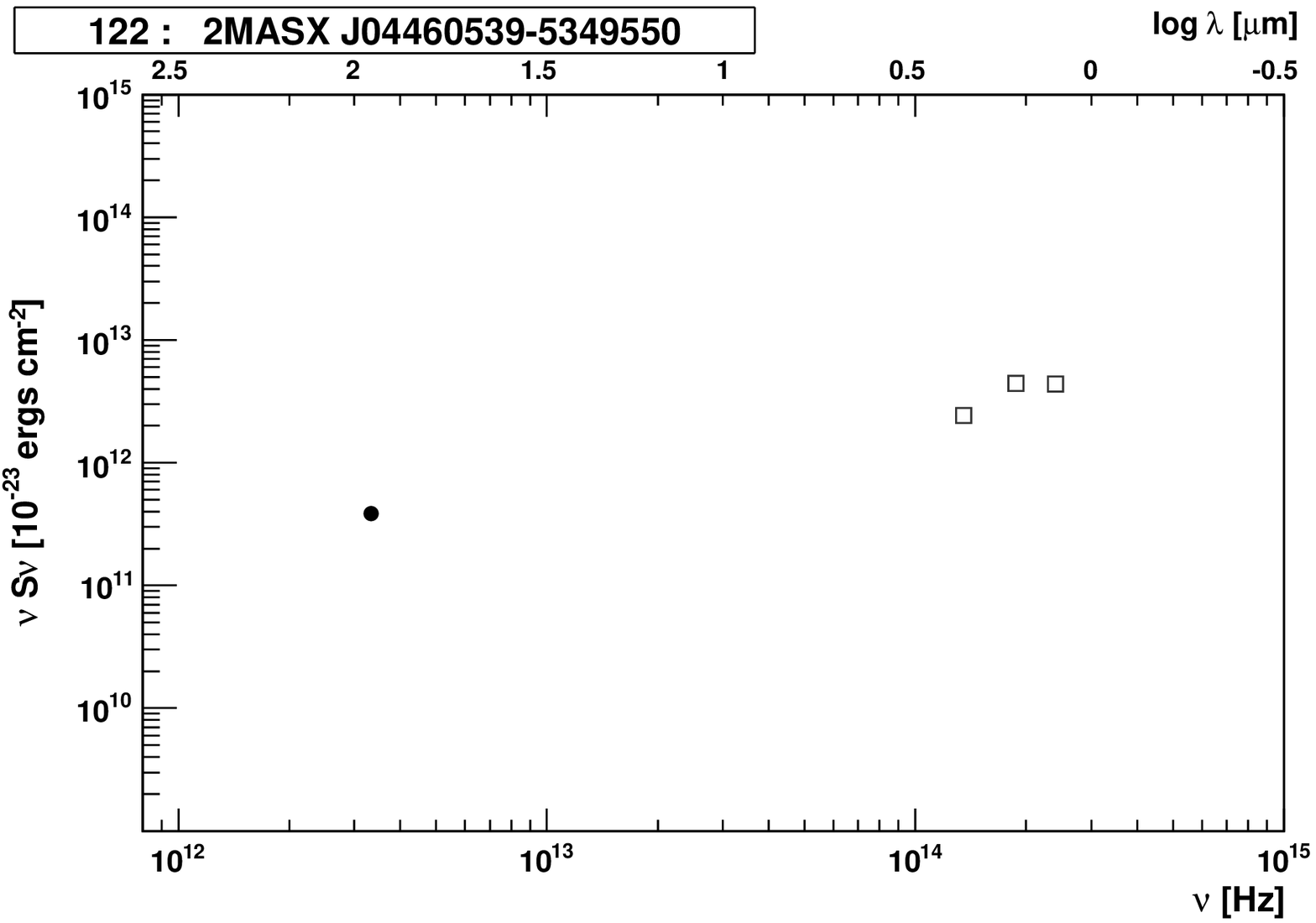}
\includegraphics[width=4cm]{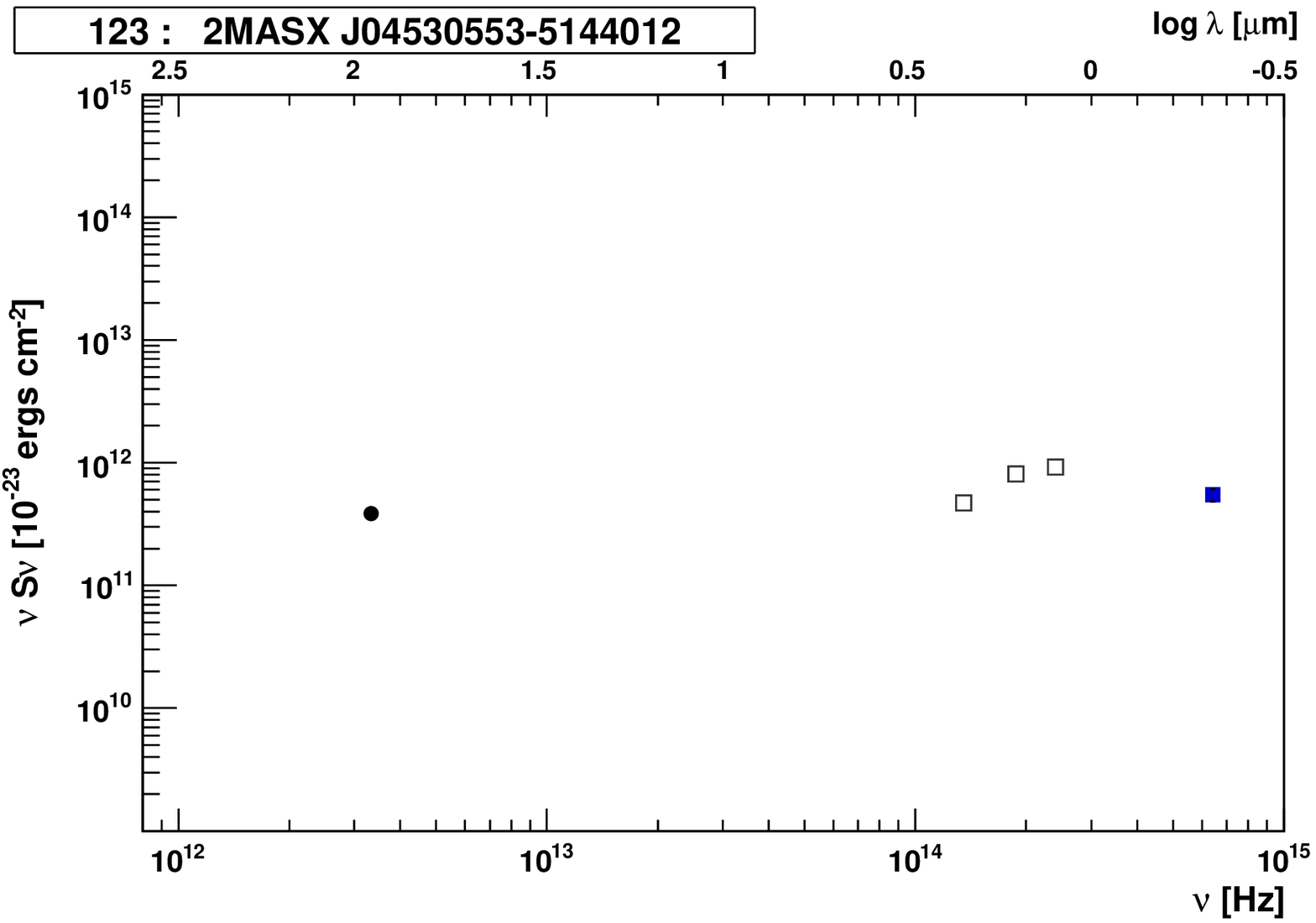}
\includegraphics[width=4cm]{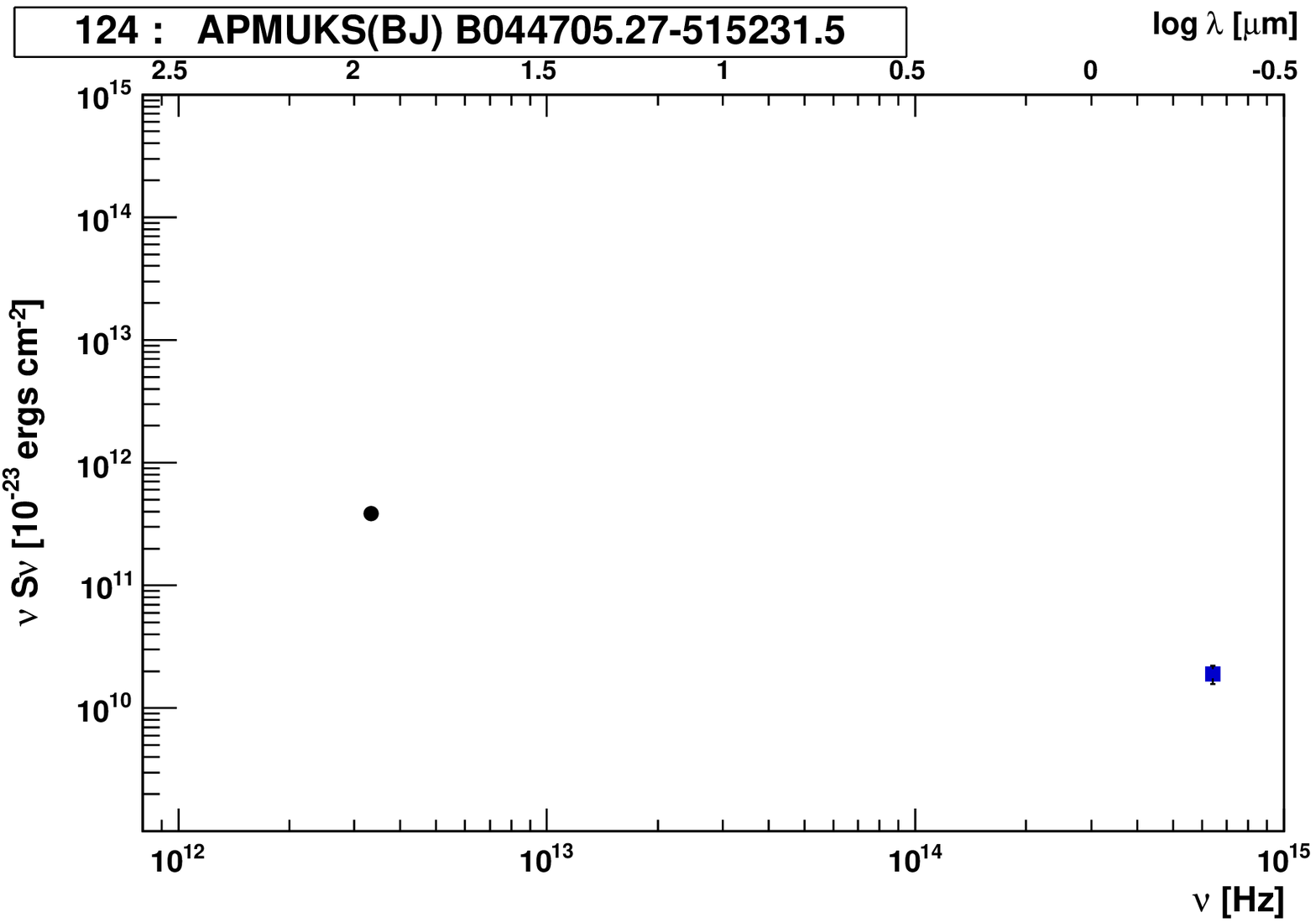}
\includegraphics[width=4cm]{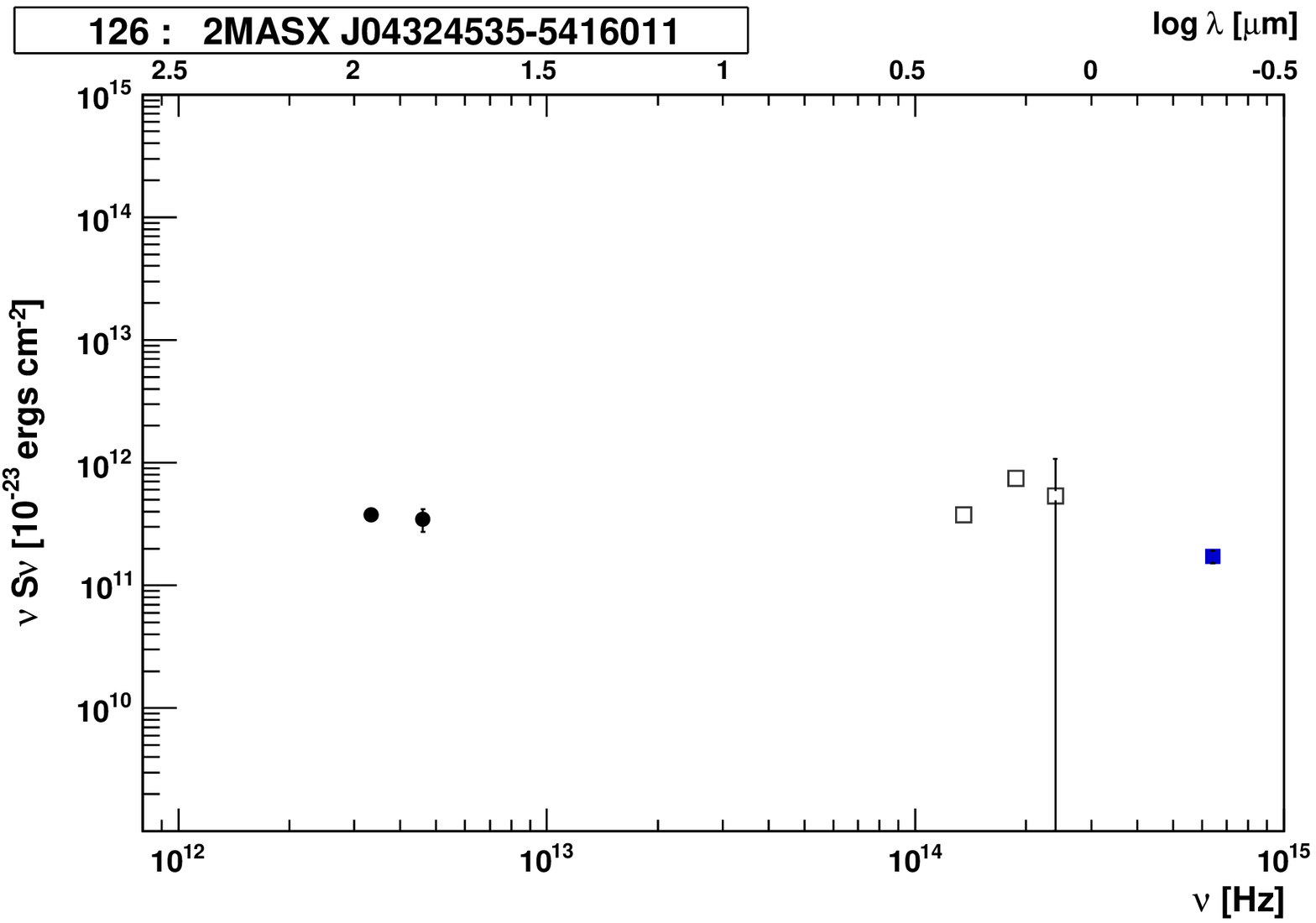}
\includegraphics[width=4cm]{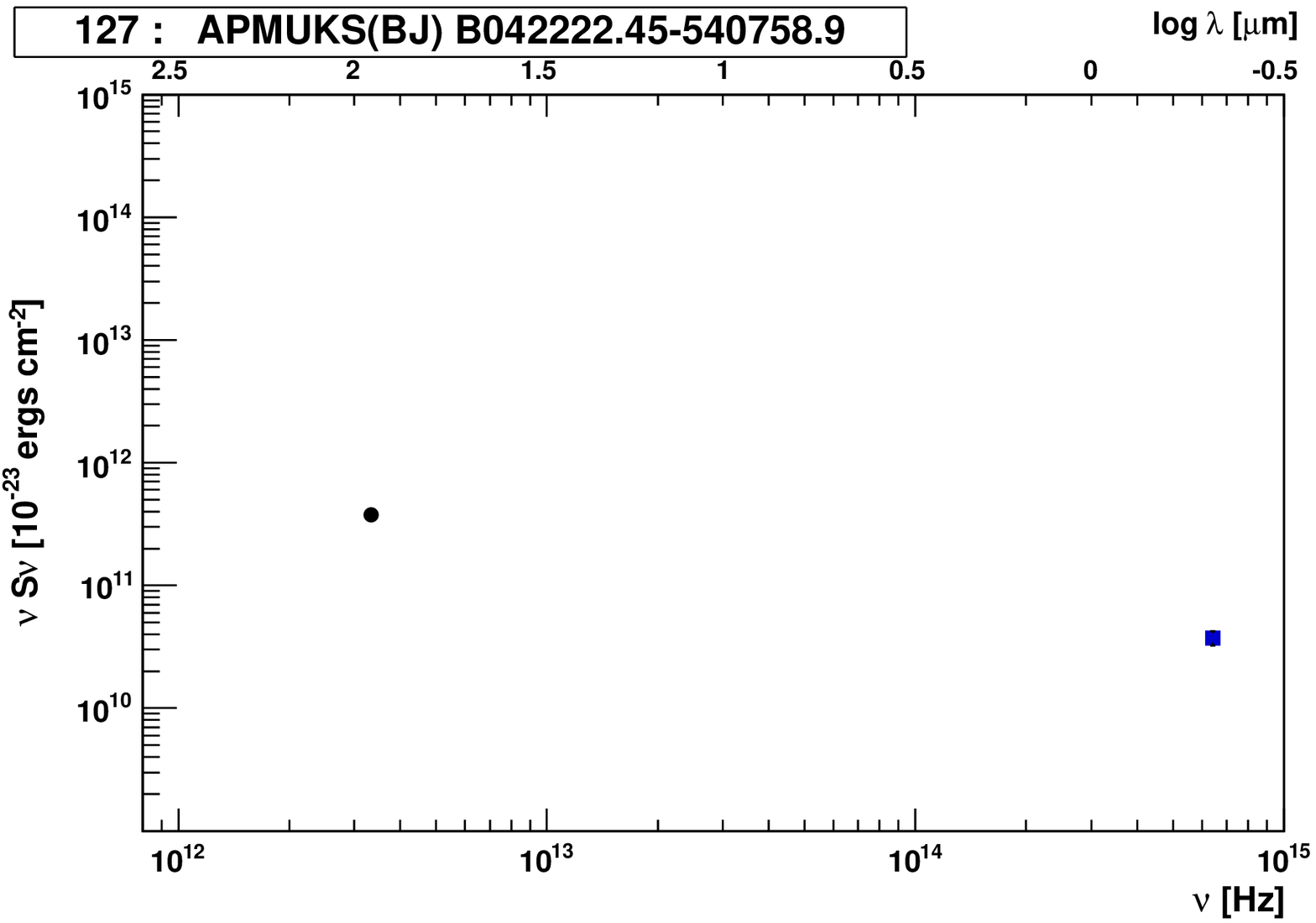}
\includegraphics[width=4cm]{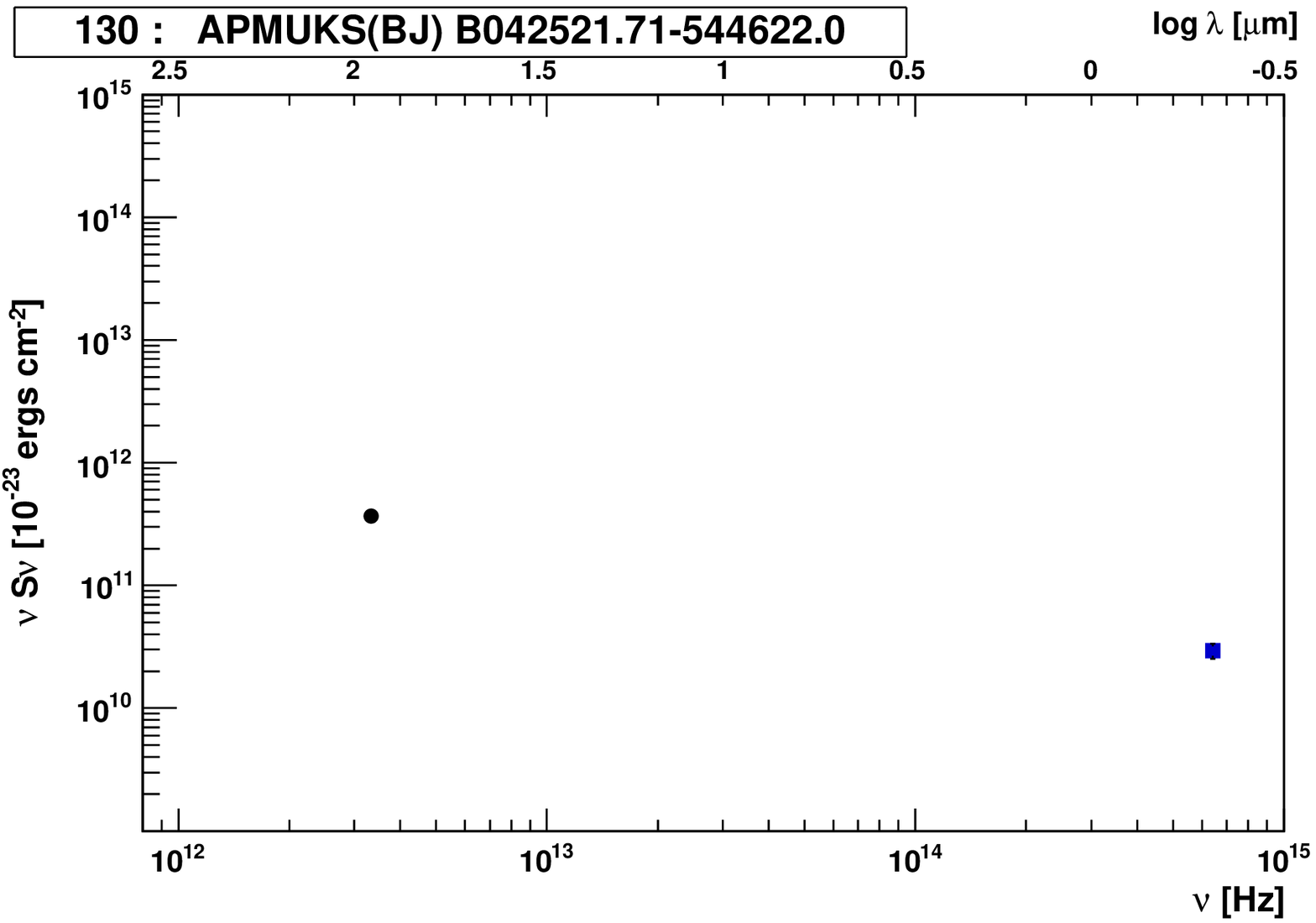}
\includegraphics[width=4cm]{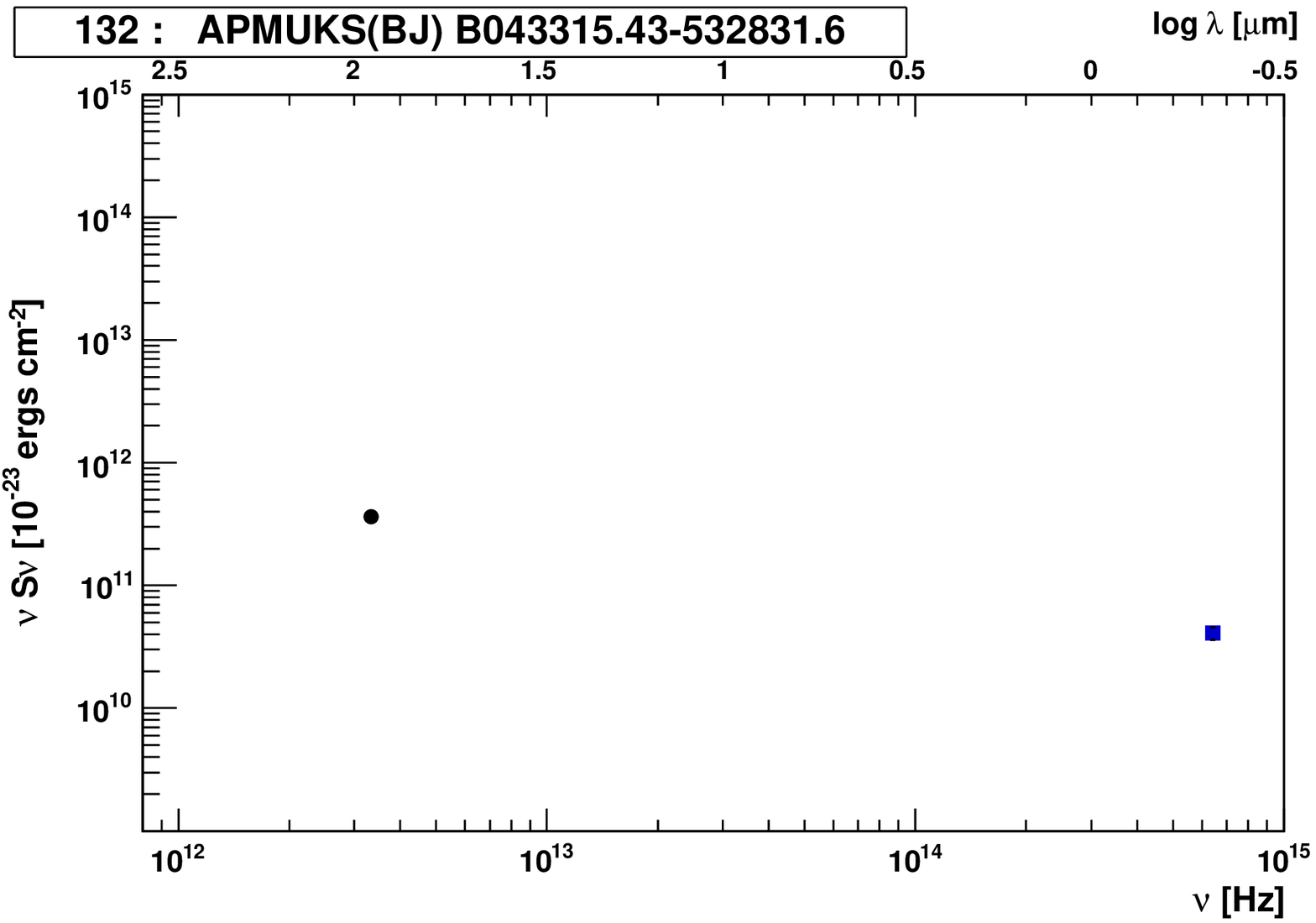}
\includegraphics[width=4cm]{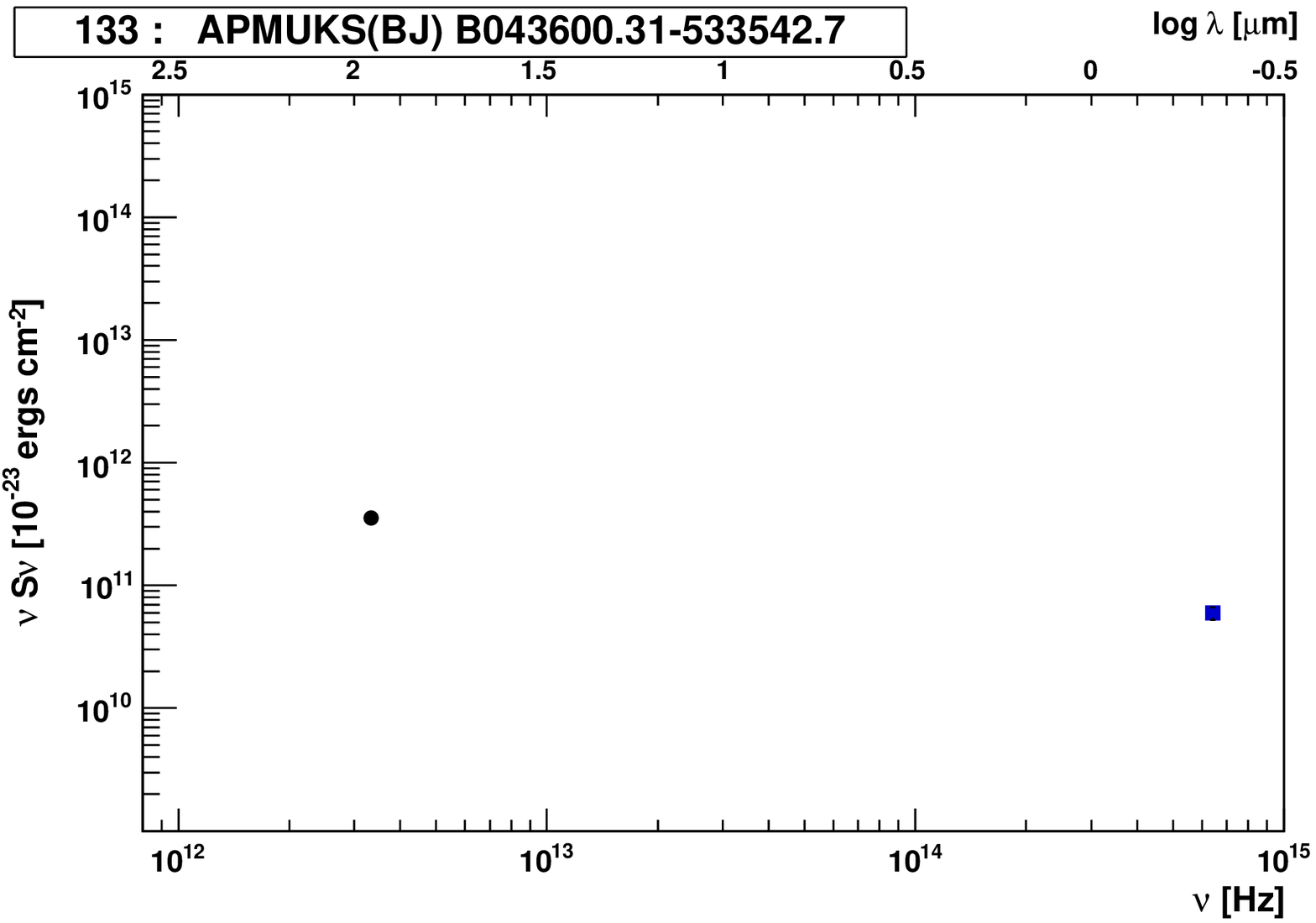}
\includegraphics[width=4cm]{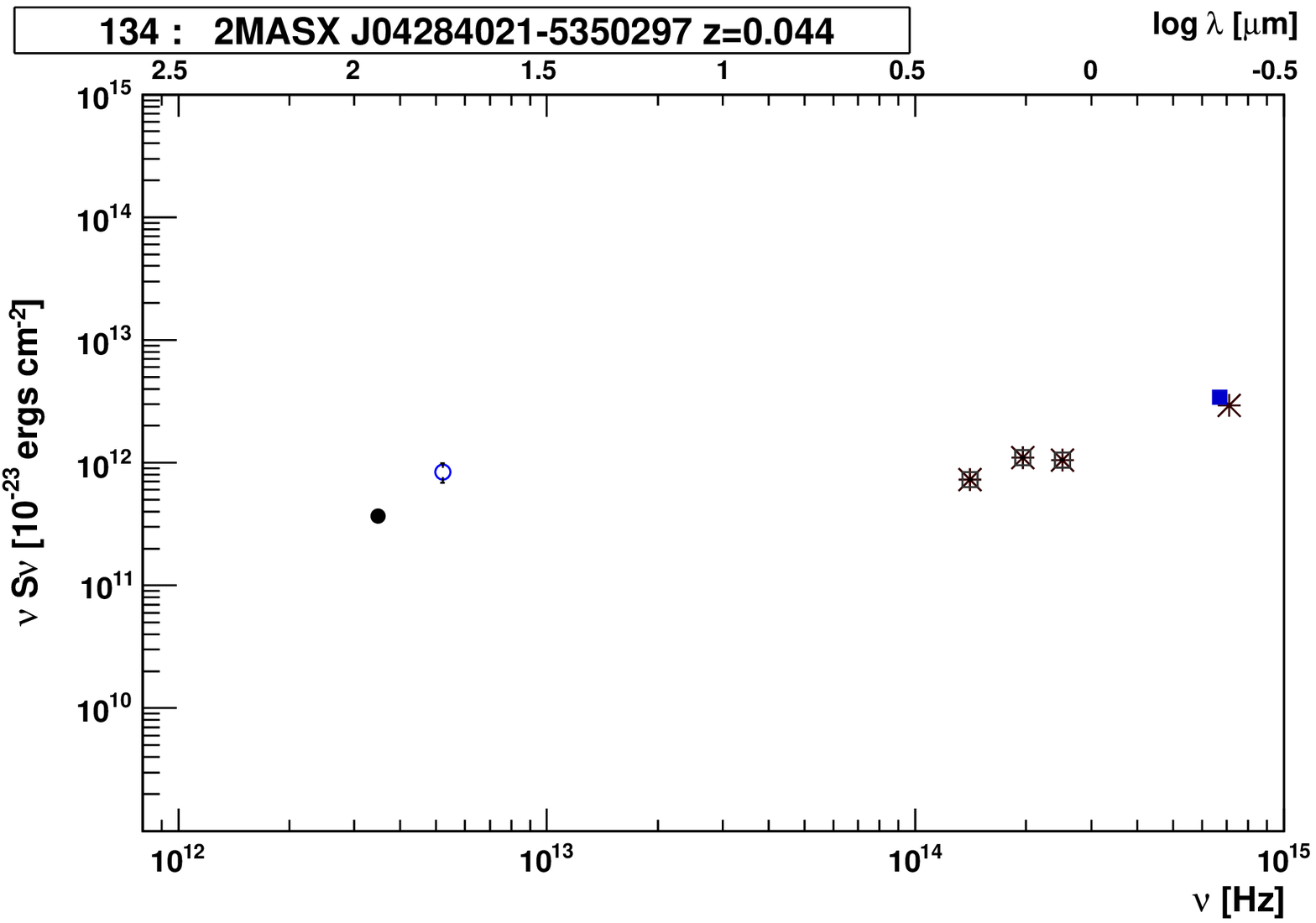}
\includegraphics[width=4cm]{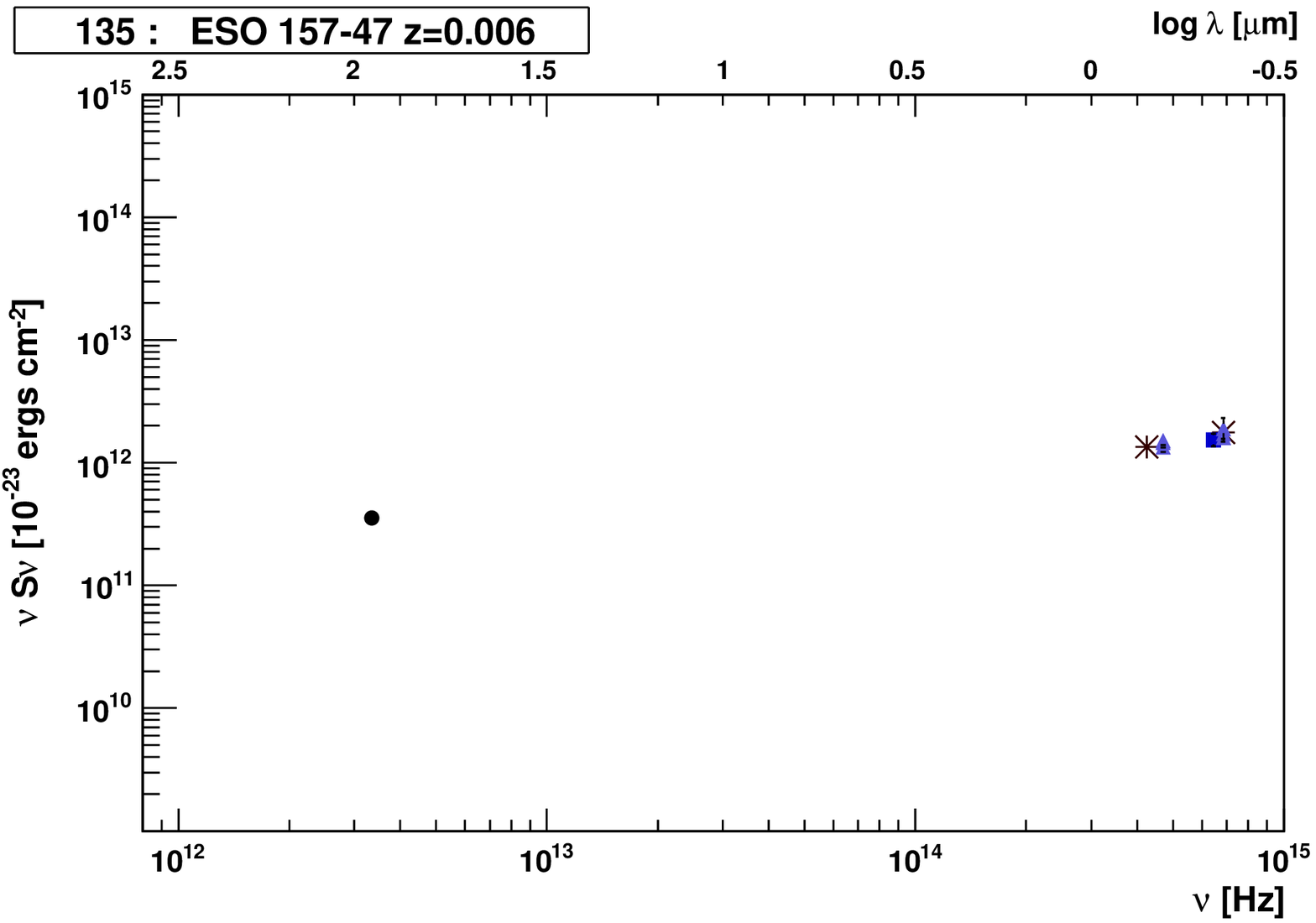}
\includegraphics[width=4cm]{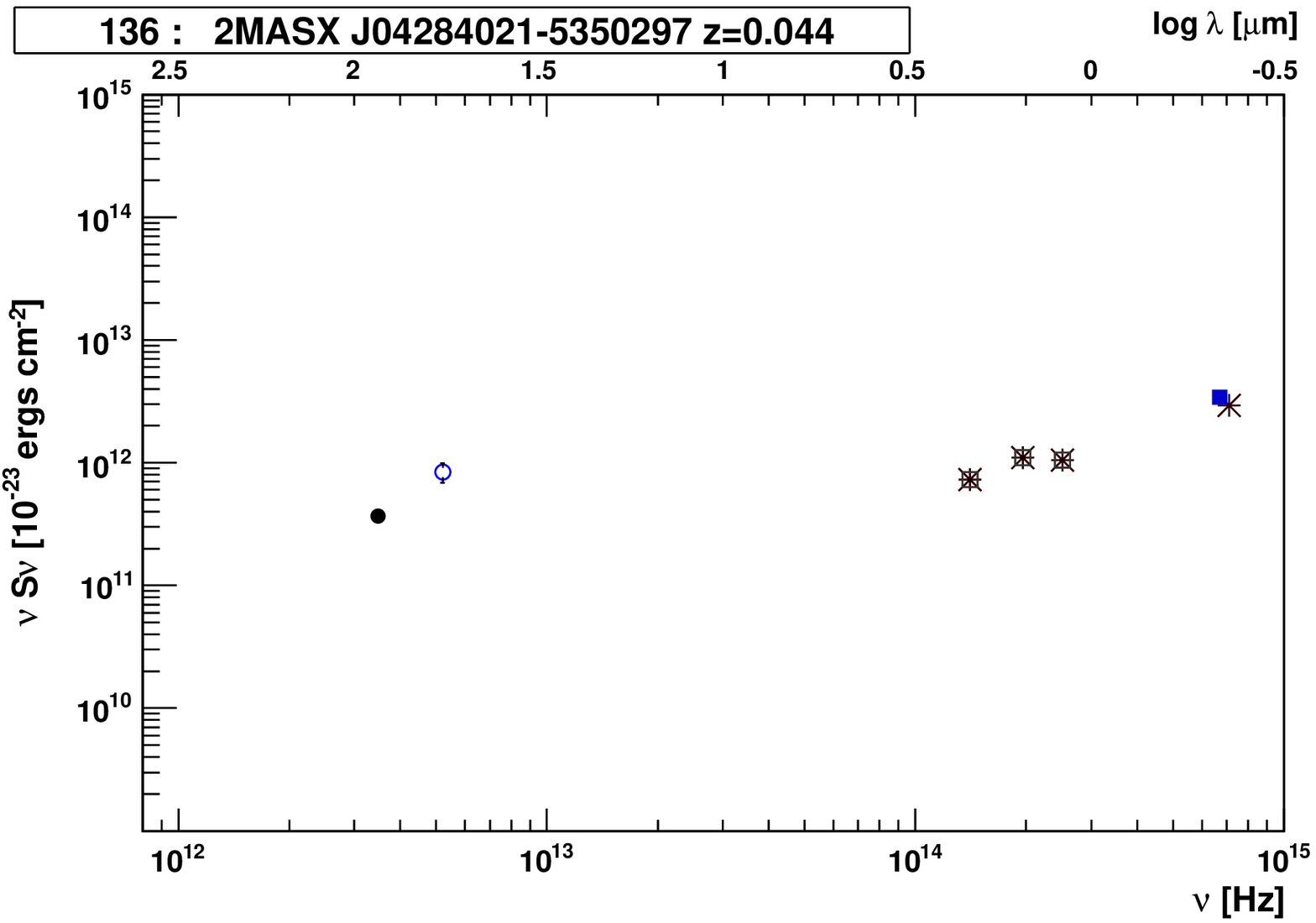}
\includegraphics[width=4cm]{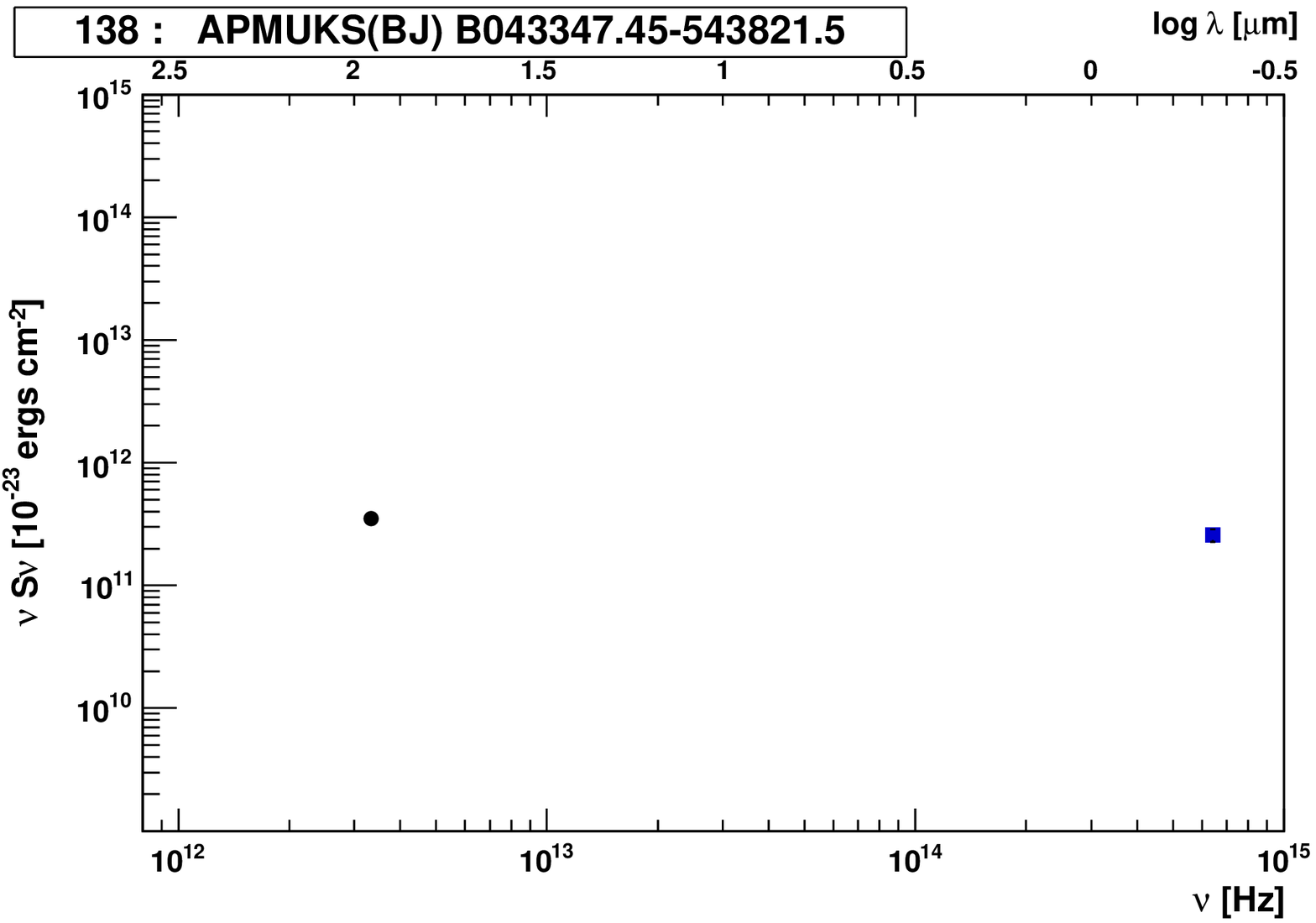}
\includegraphics[width=4cm]{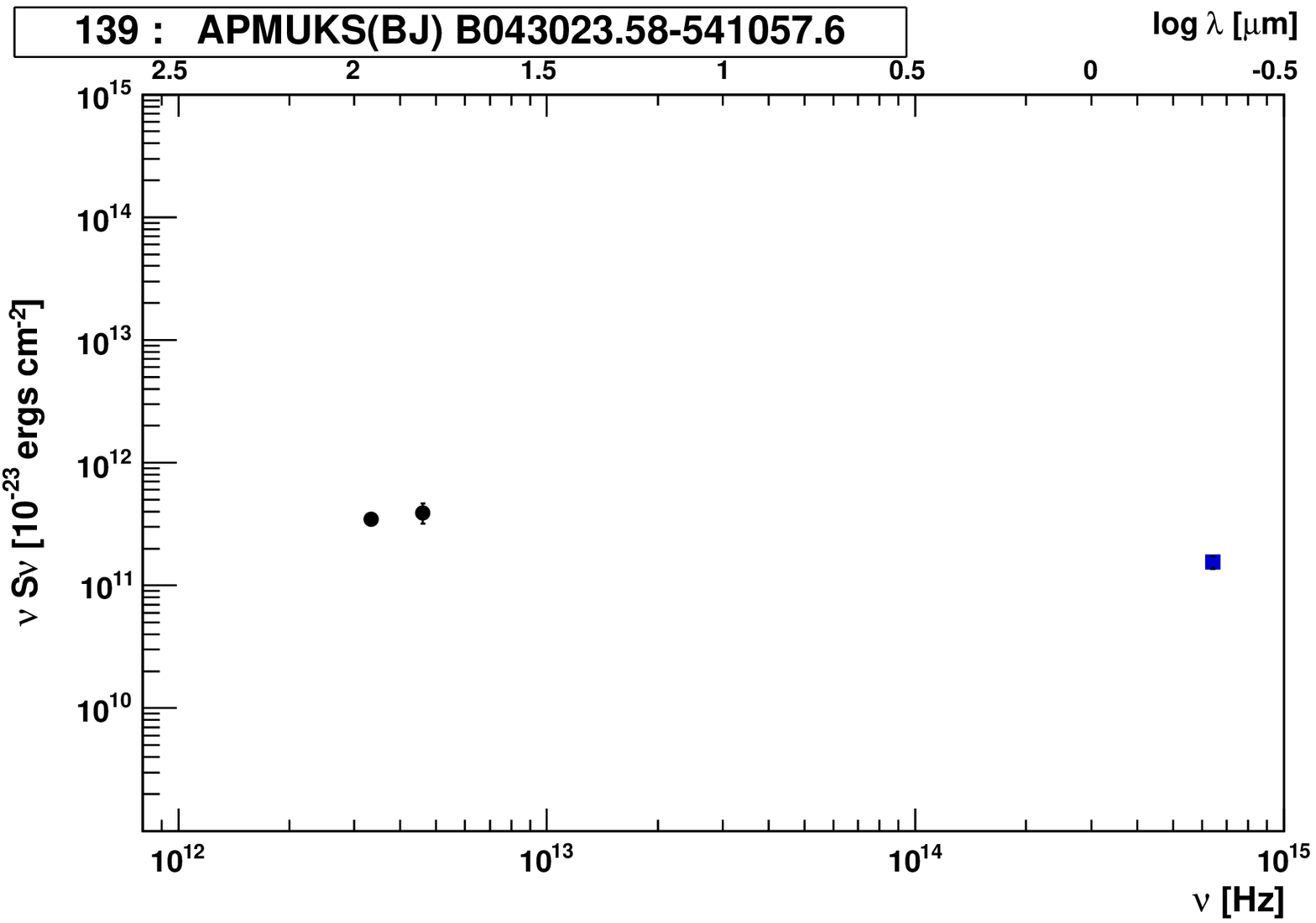}
\includegraphics[width=4cm]{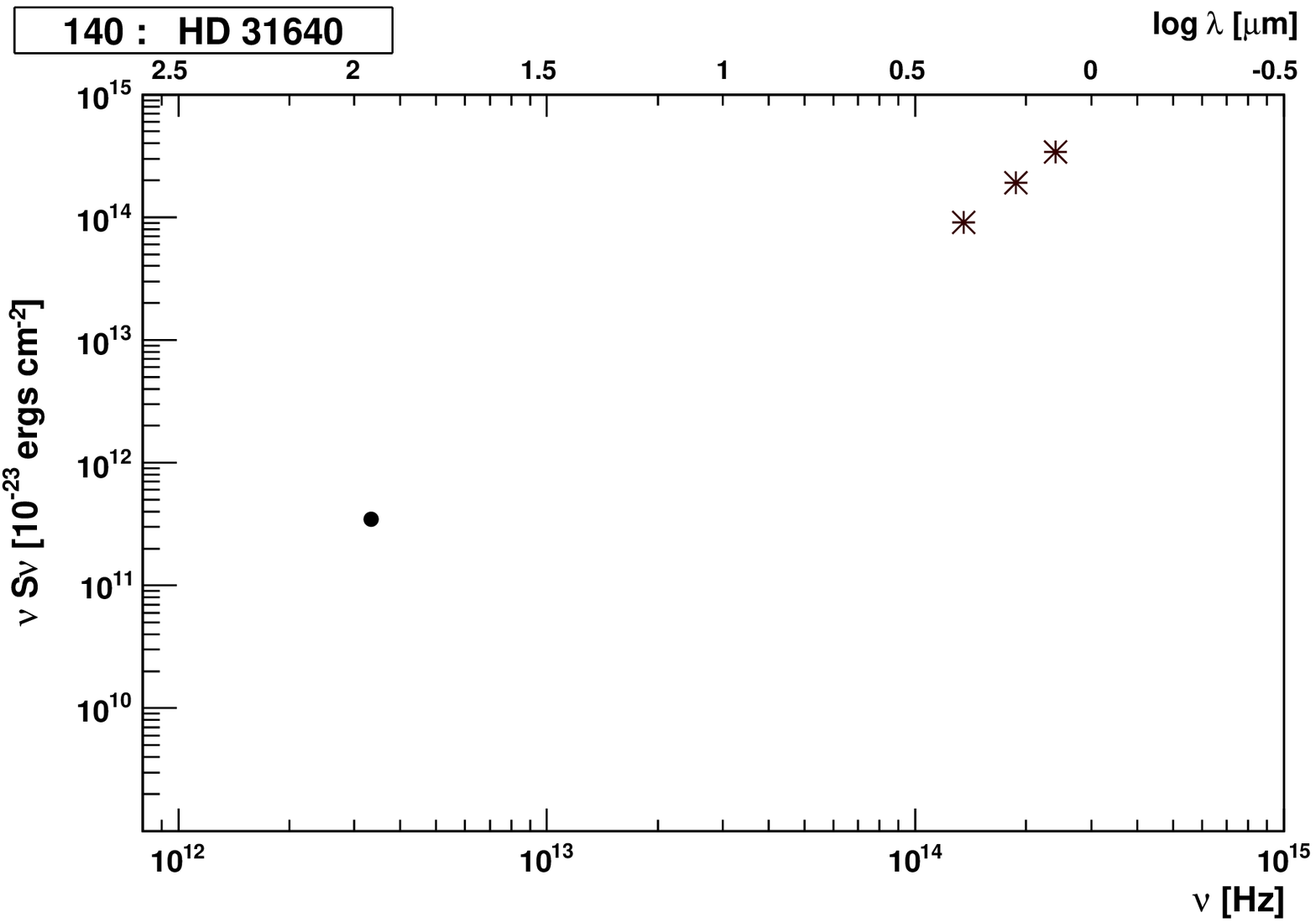}
\includegraphics[width=4cm]{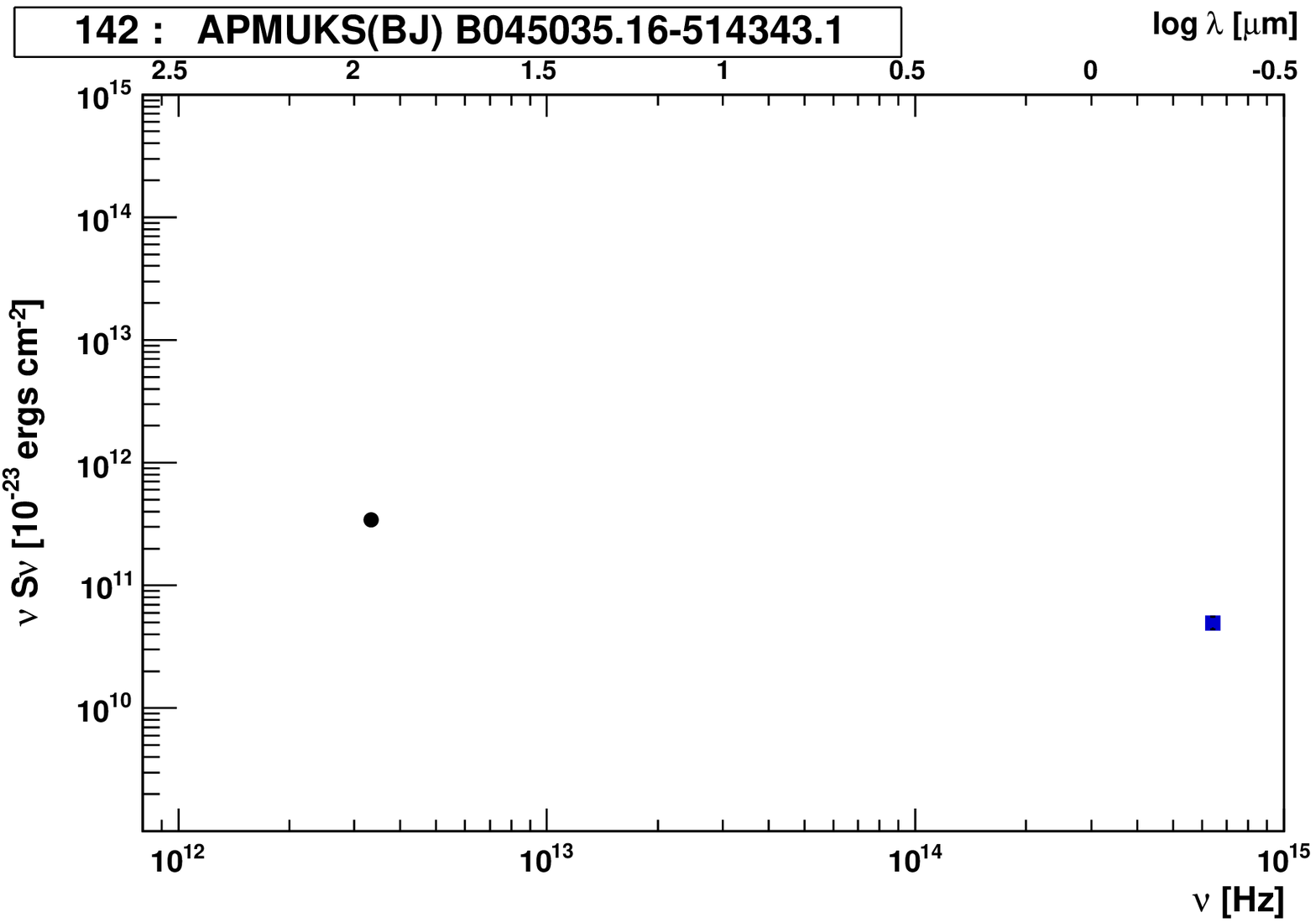}
\includegraphics[width=4cm]{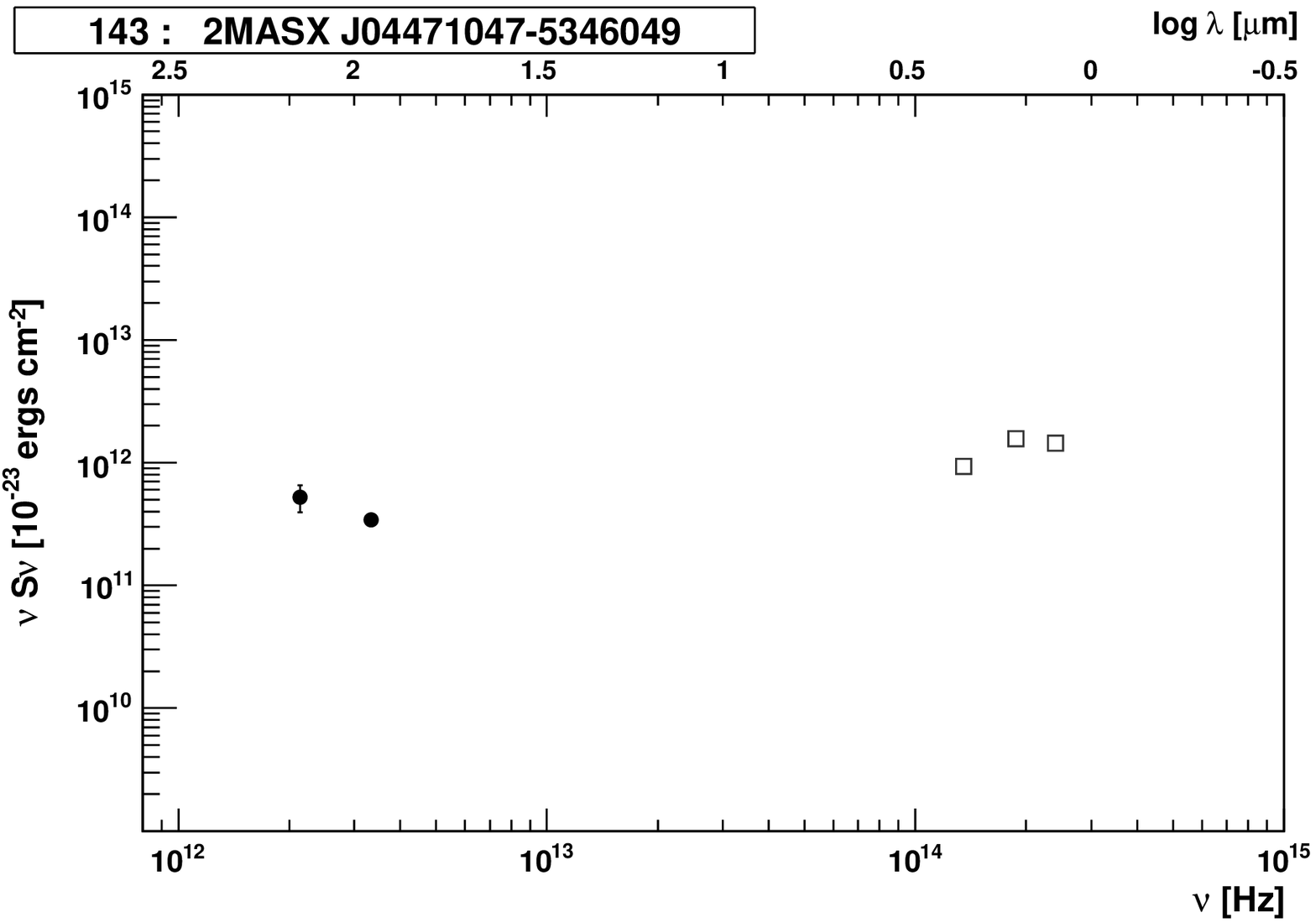}
\includegraphics[width=4cm]{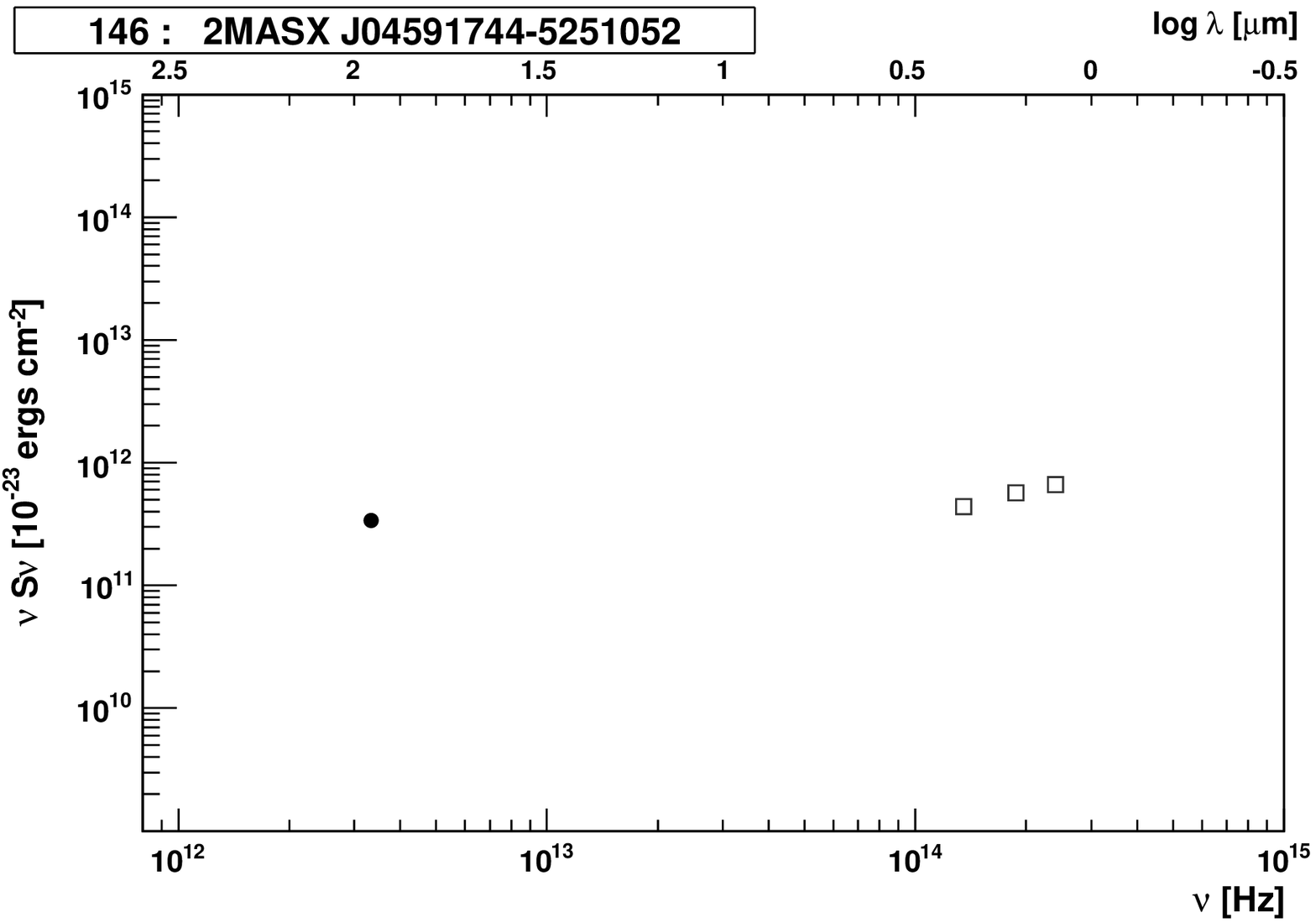}
\includegraphics[width=4cm]{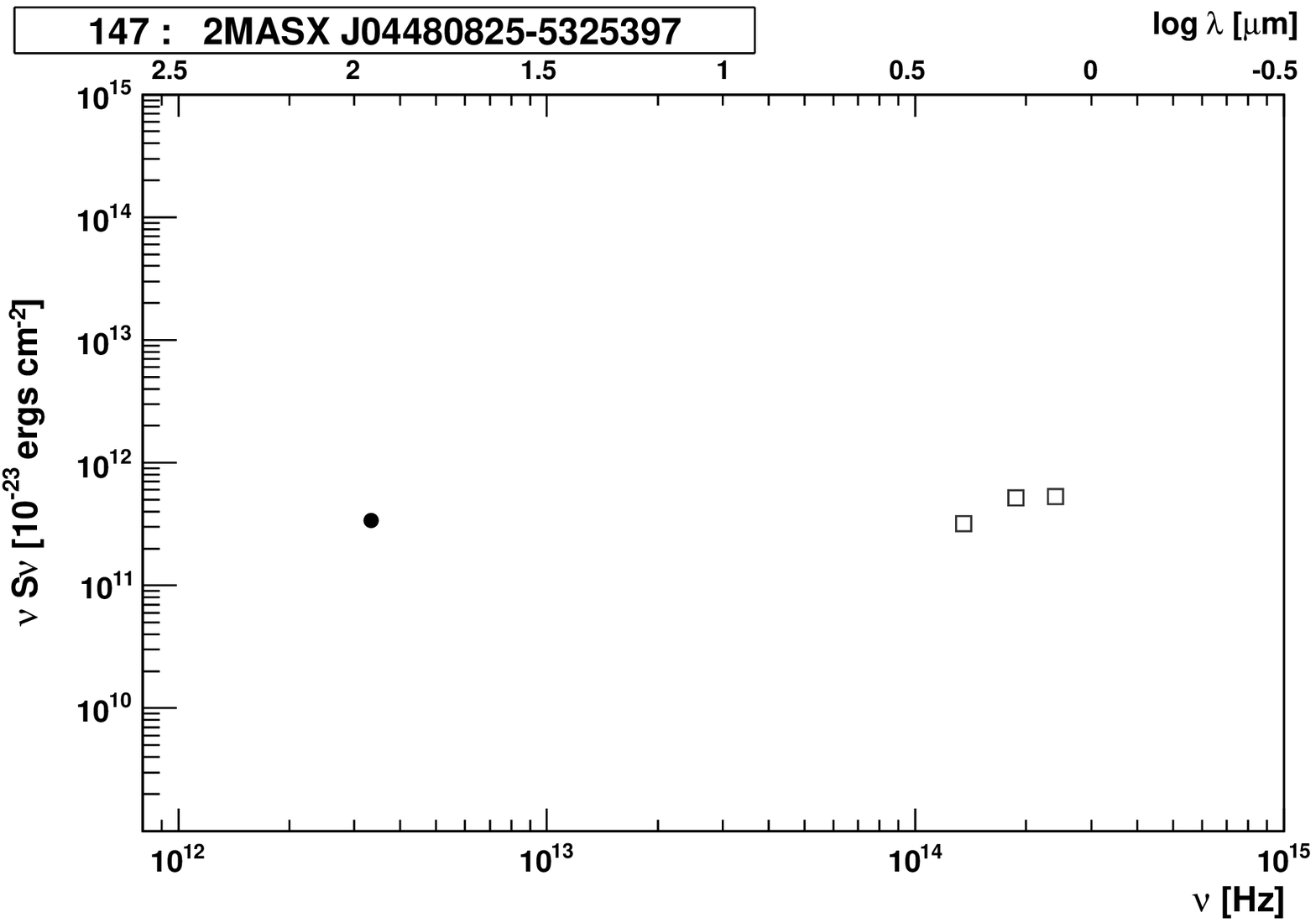}
\includegraphics[width=4cm]{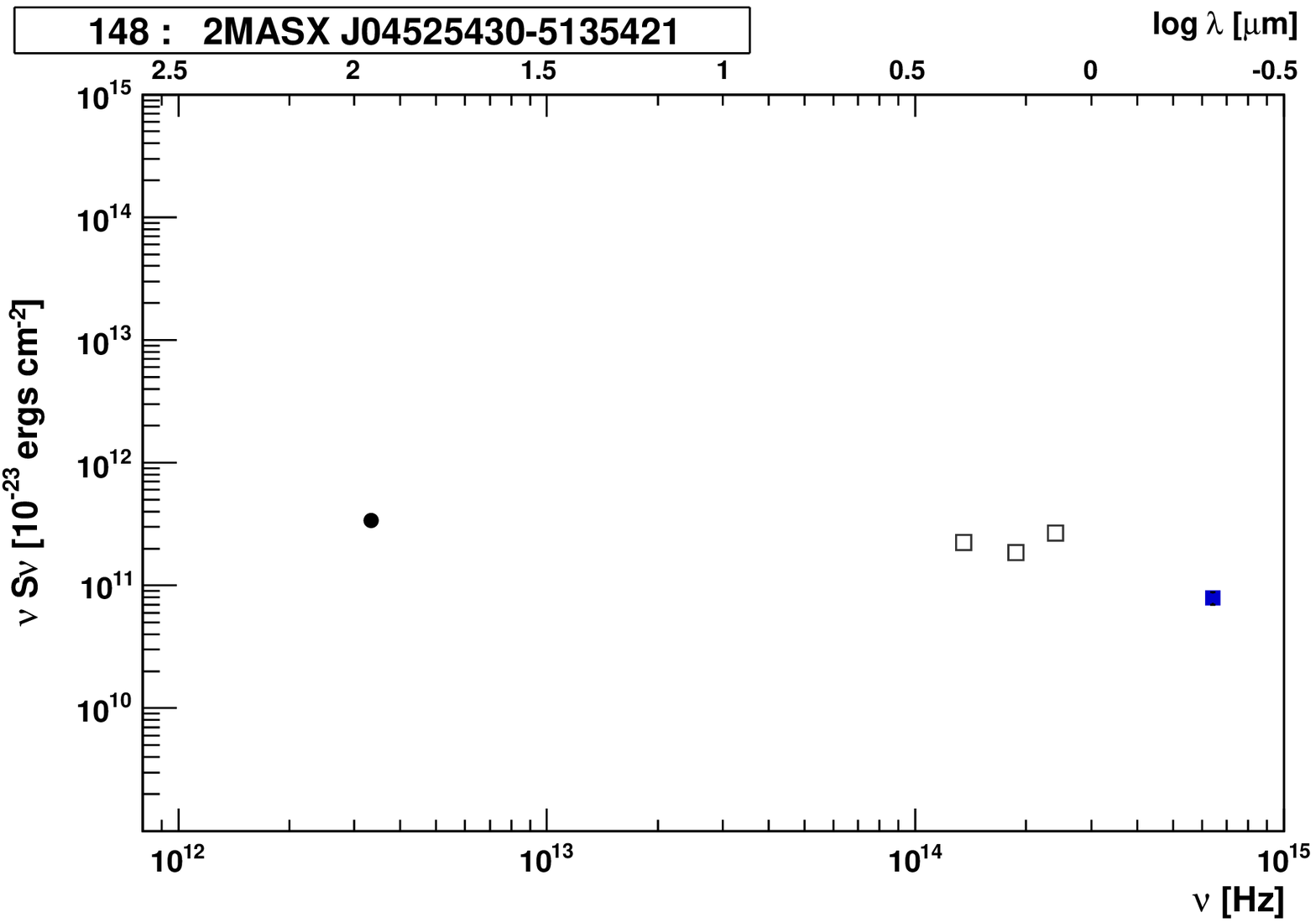}
\includegraphics[width=4cm]{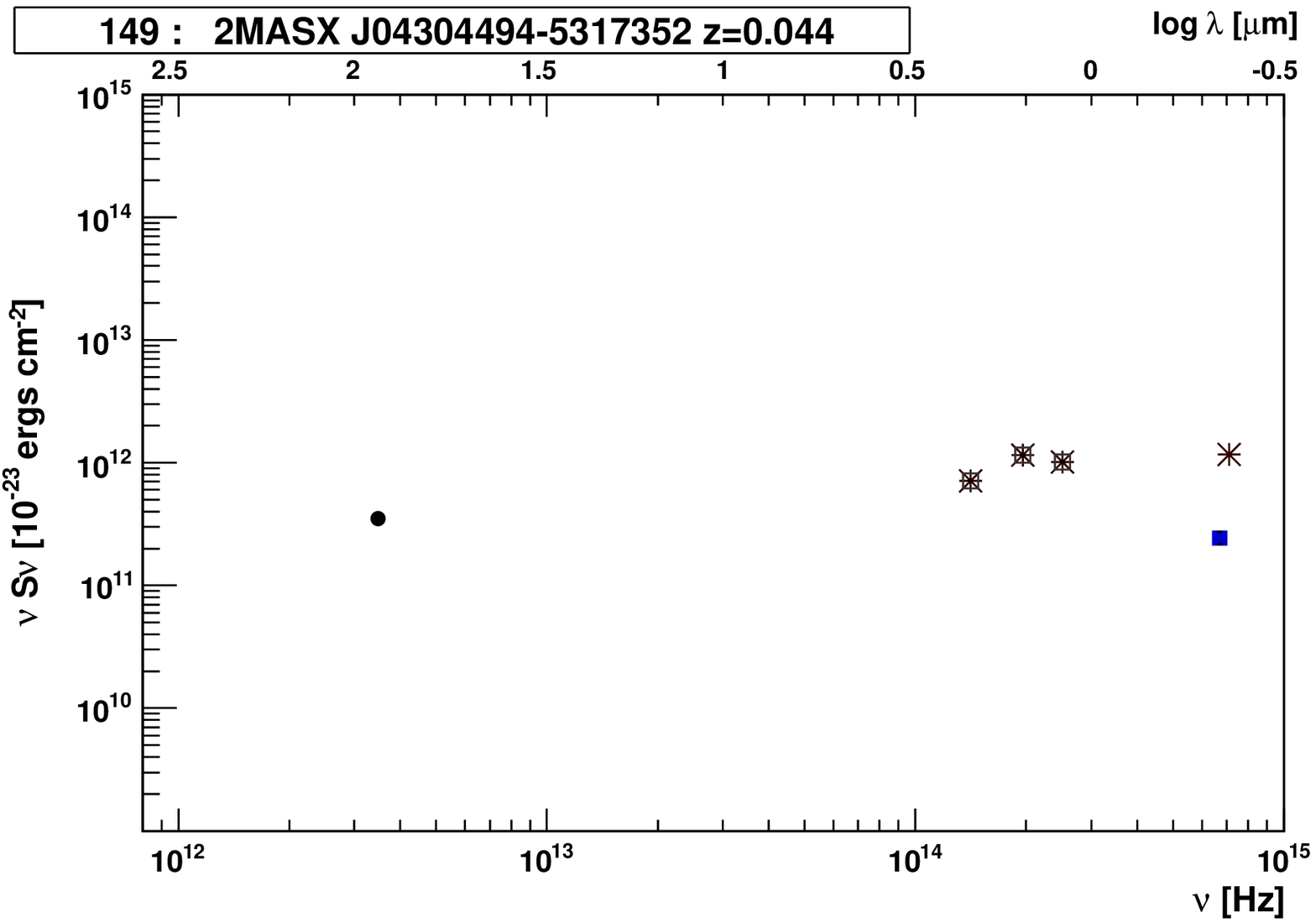}
\includegraphics[width=4cm]{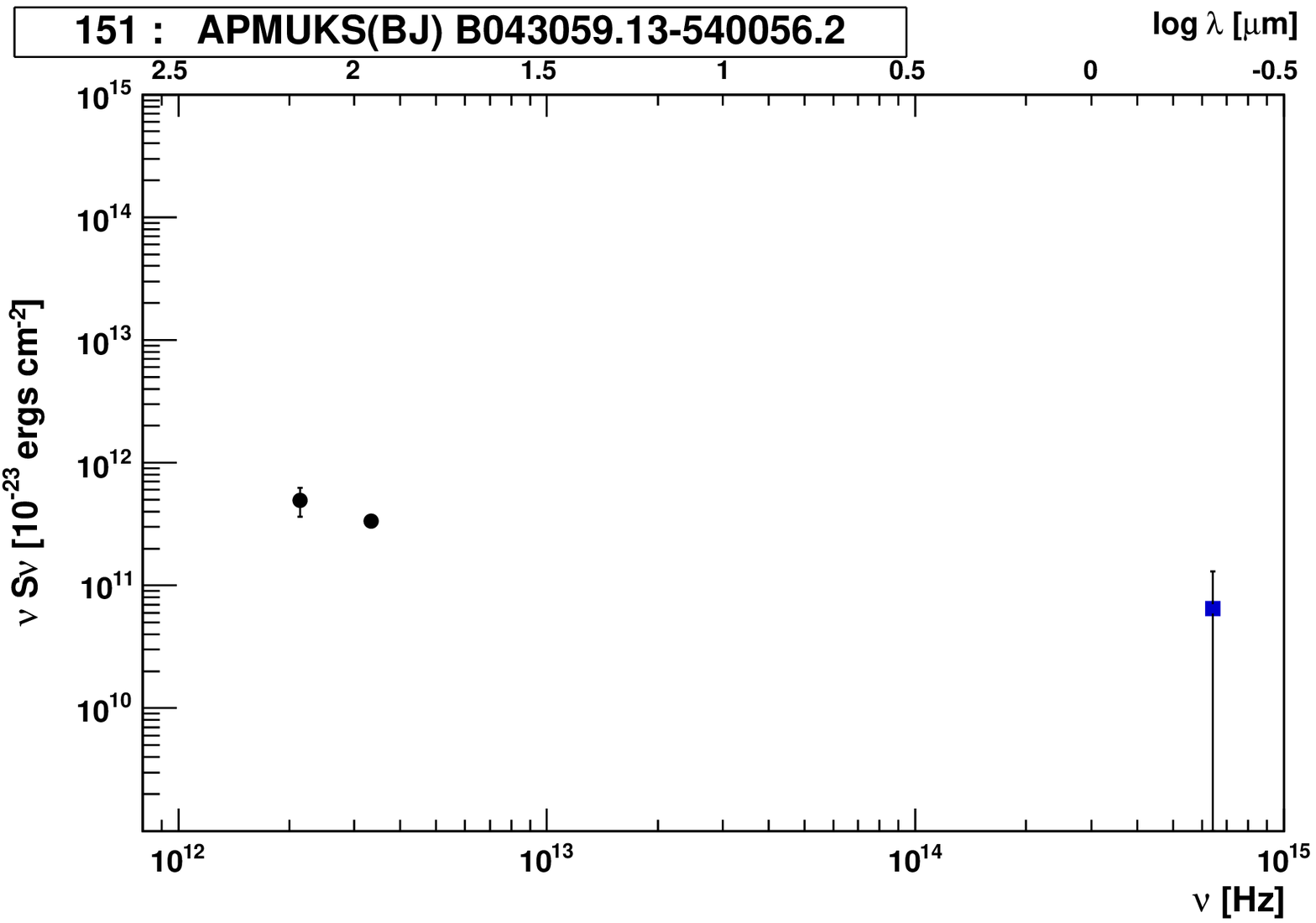}
\includegraphics[width=4cm]{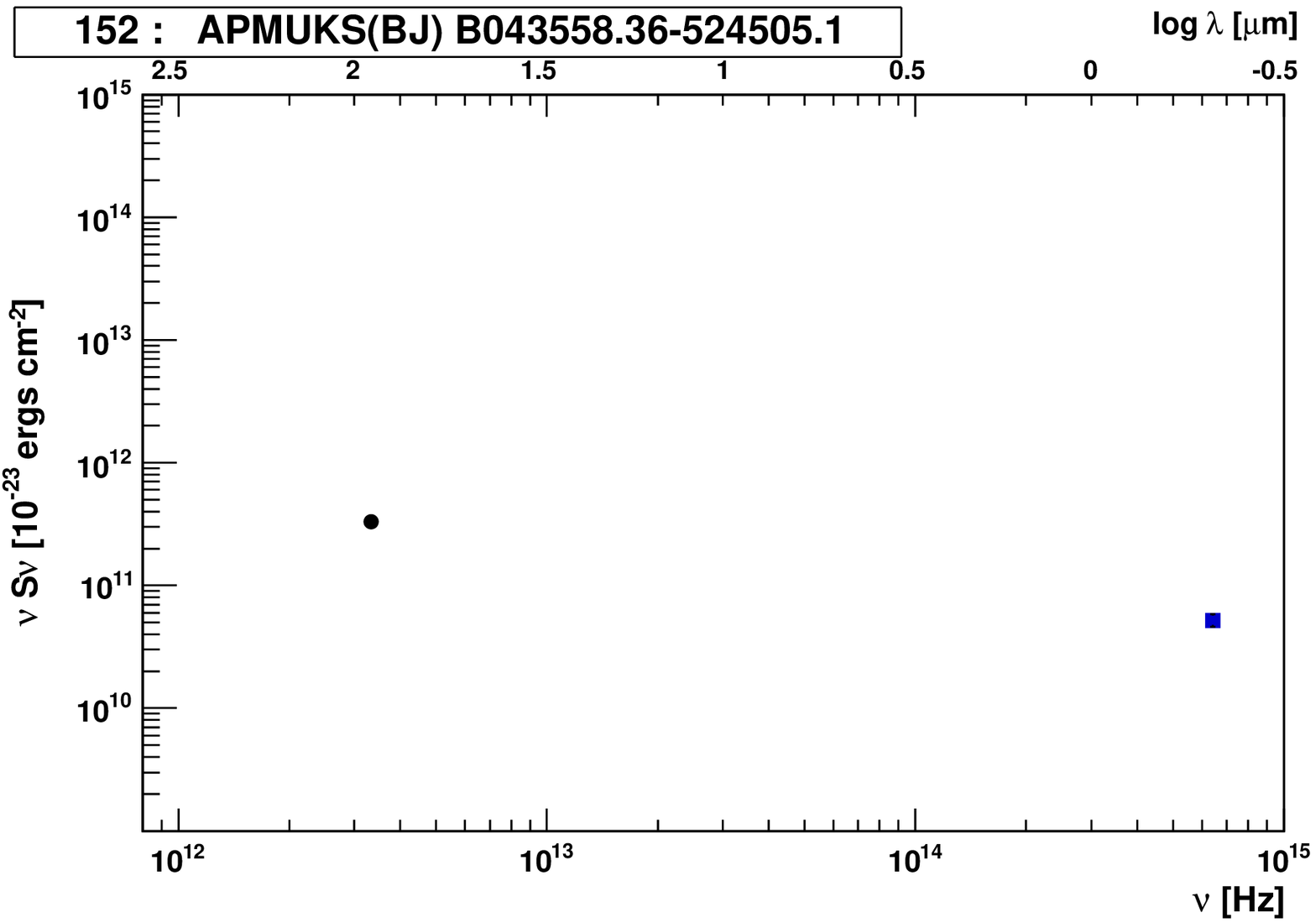}
\includegraphics[width=4cm]{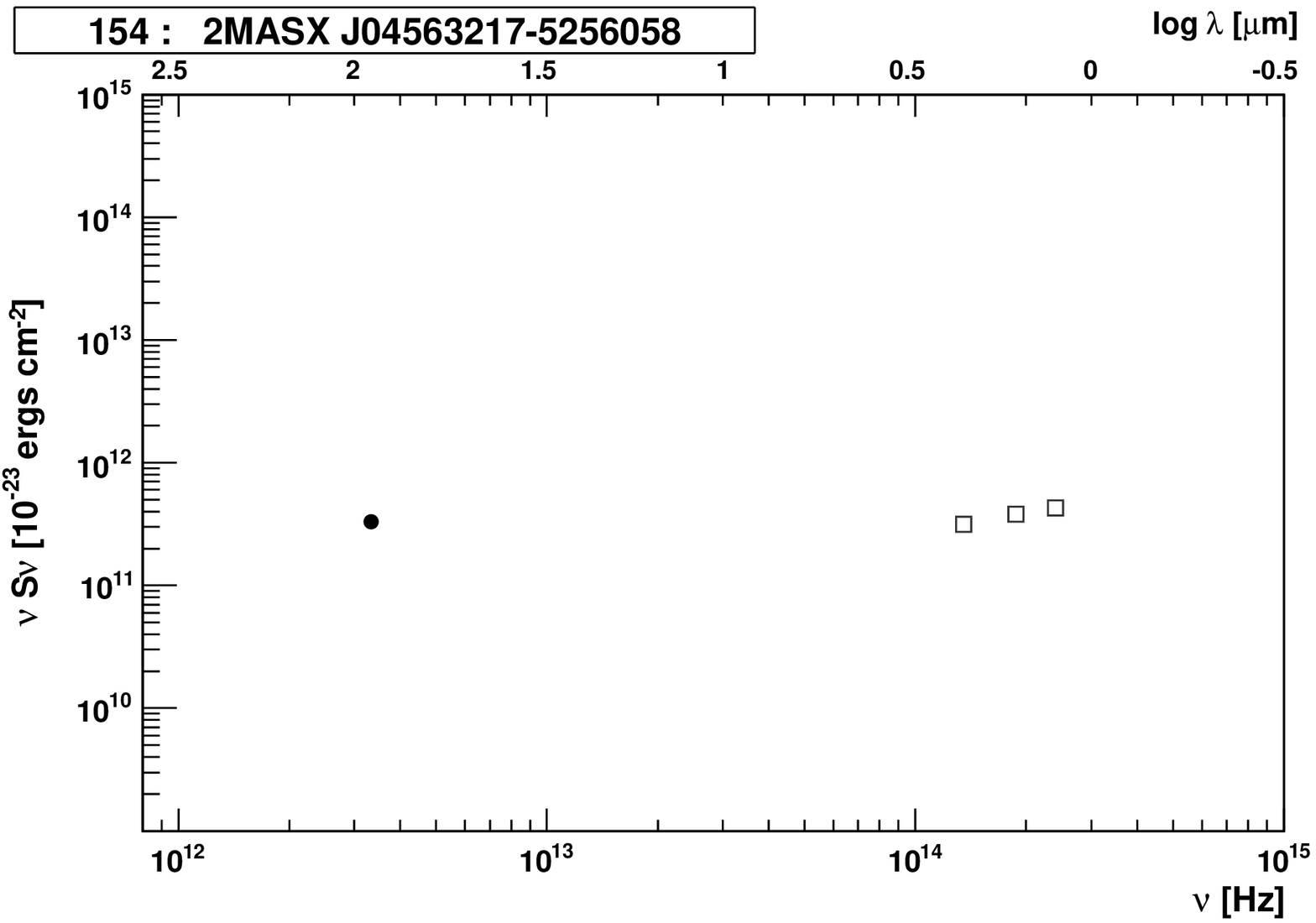}
\includegraphics[width=4cm]{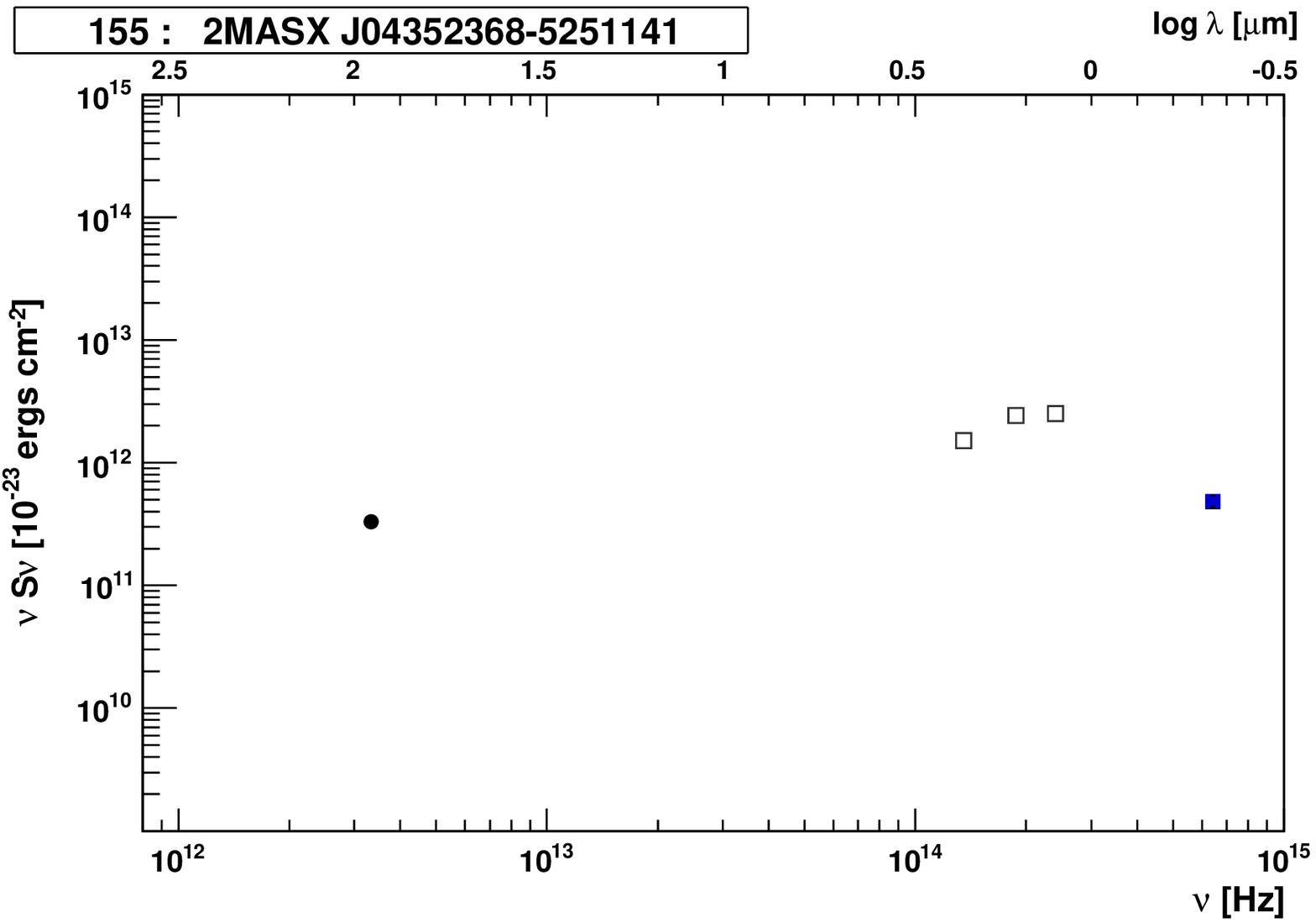}
\includegraphics[width=4cm]{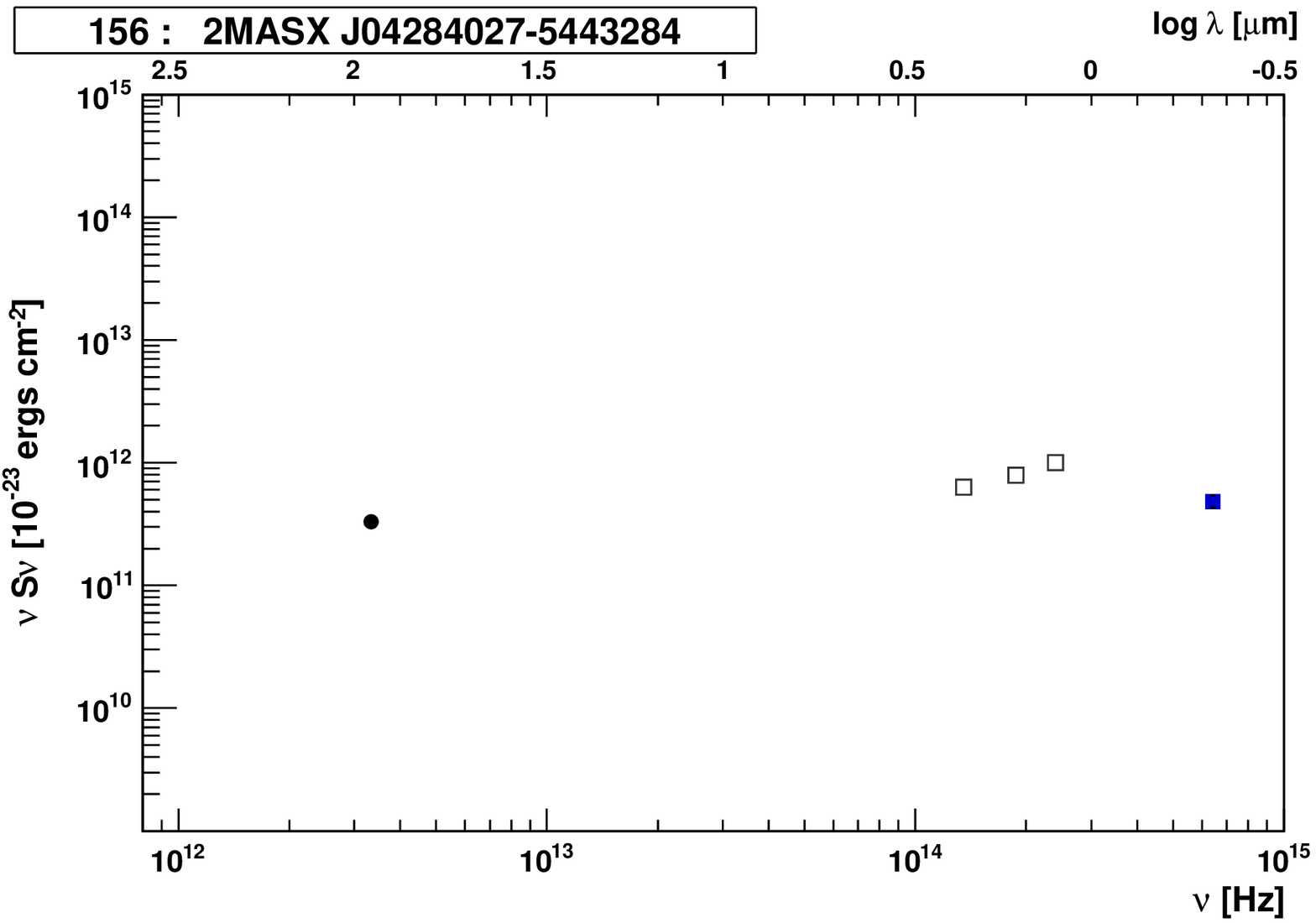}
\includegraphics[width=4cm]{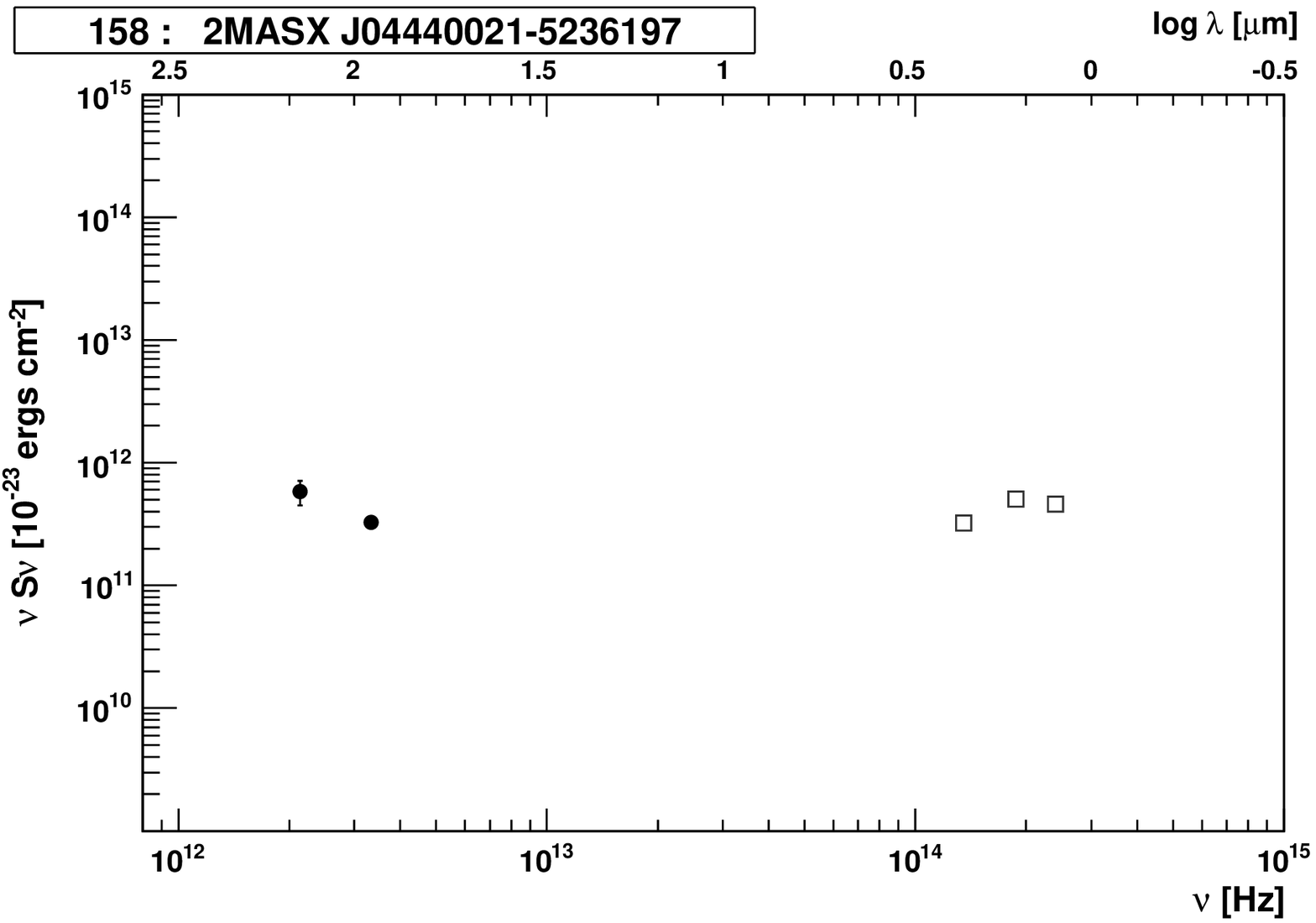}
\includegraphics[width=4cm]{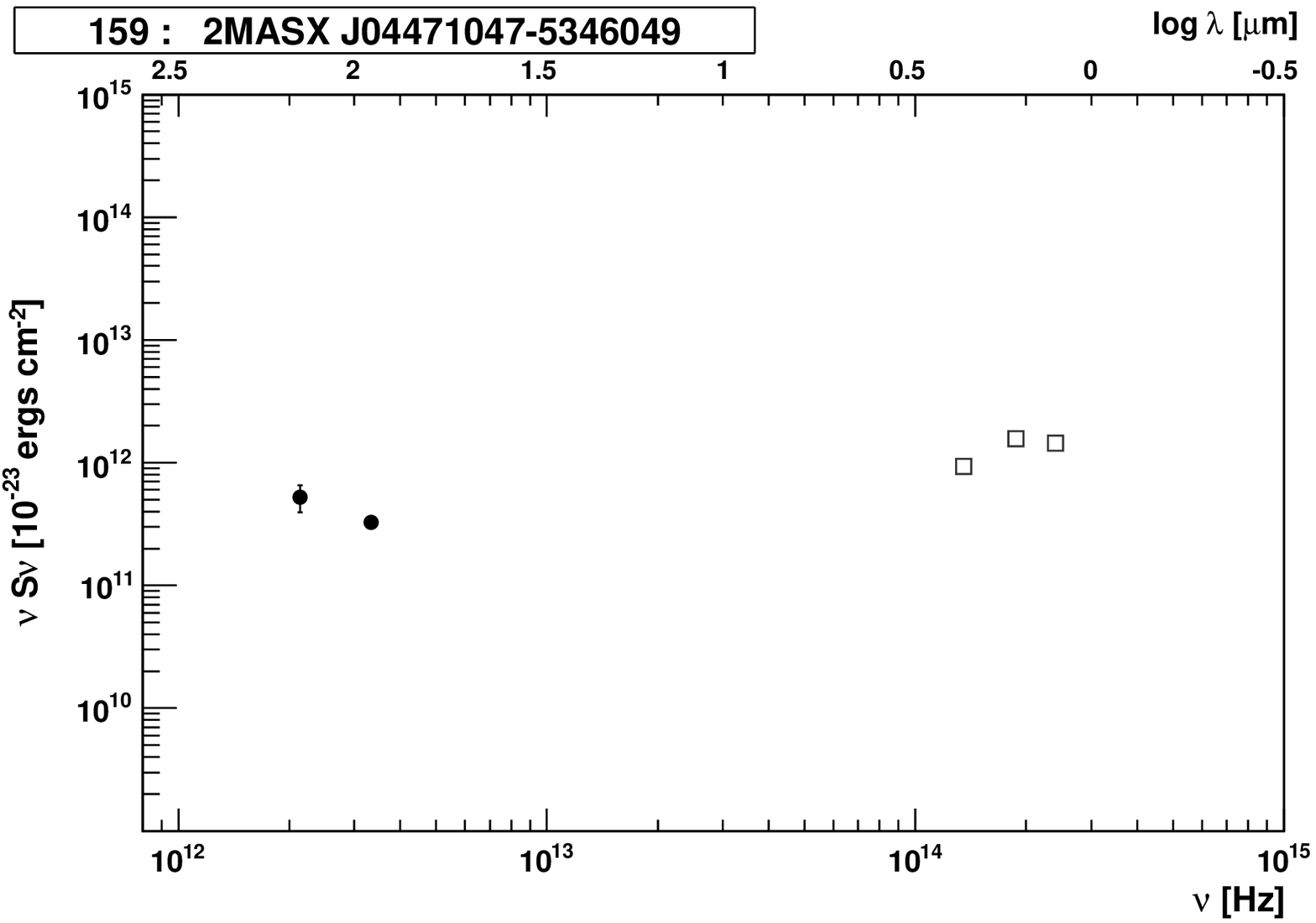}
\includegraphics[width=4cm]{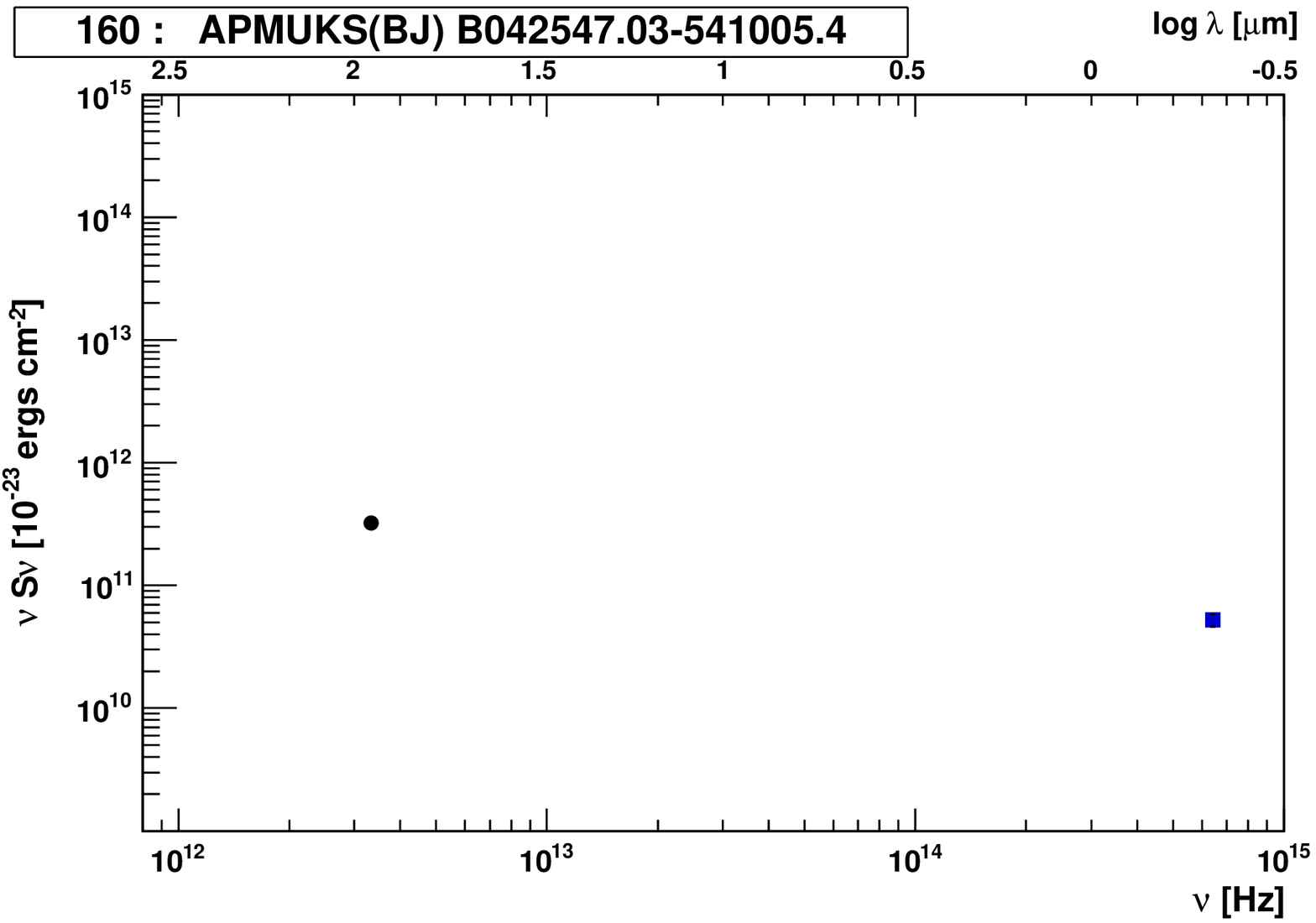}
\includegraphics[width=4cm]{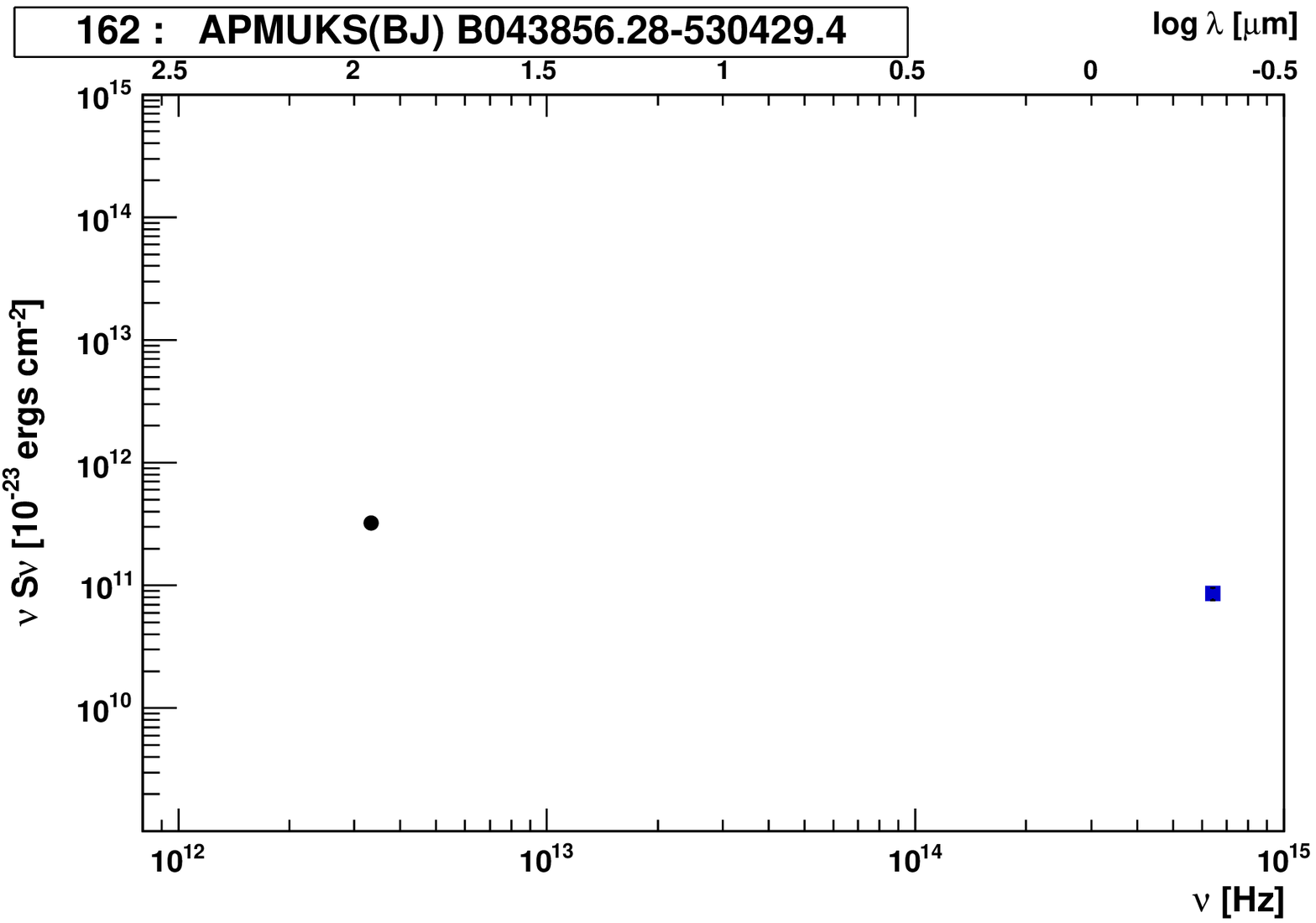}
\includegraphics[width=4cm]{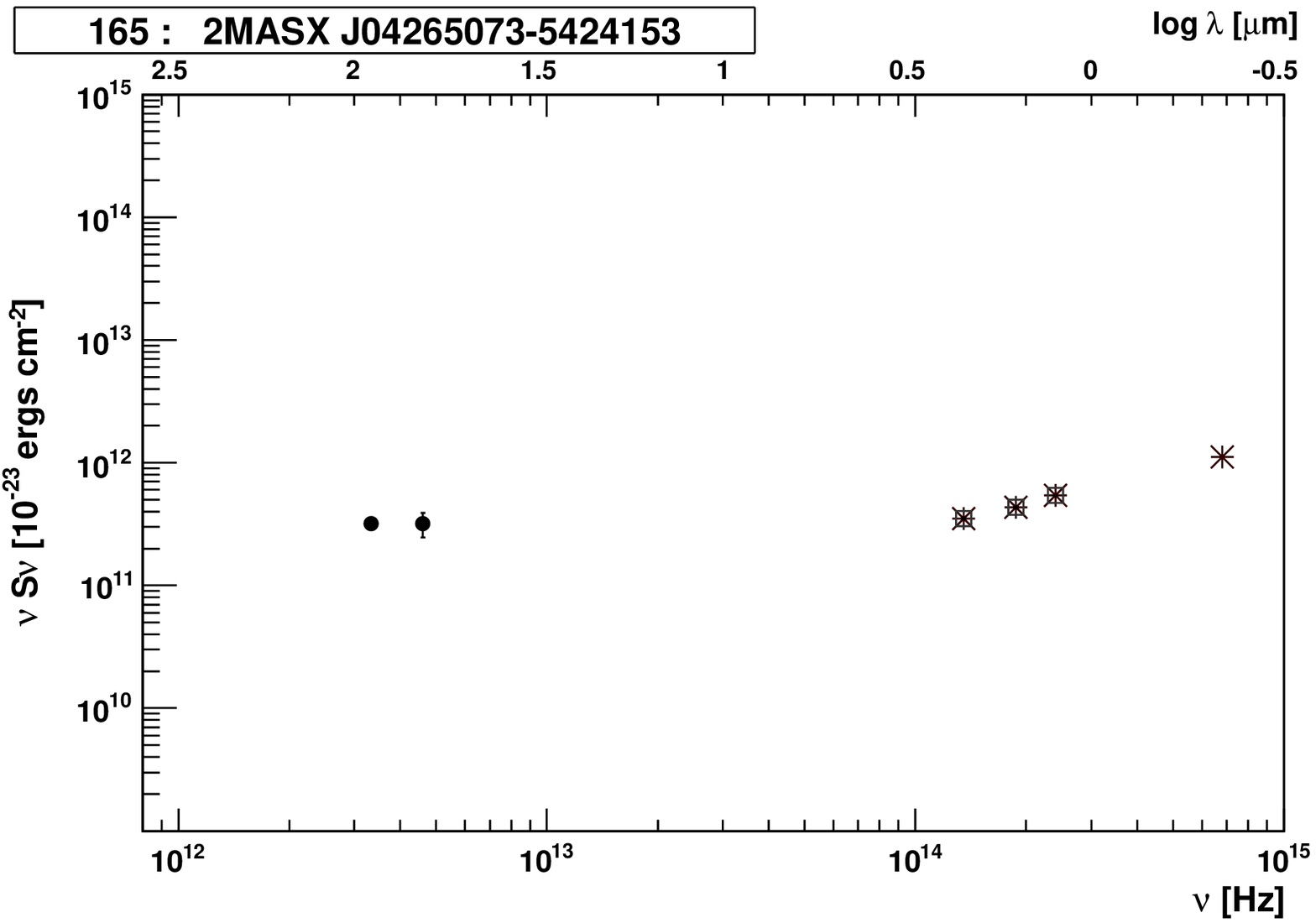}
\includegraphics[width=4cm]{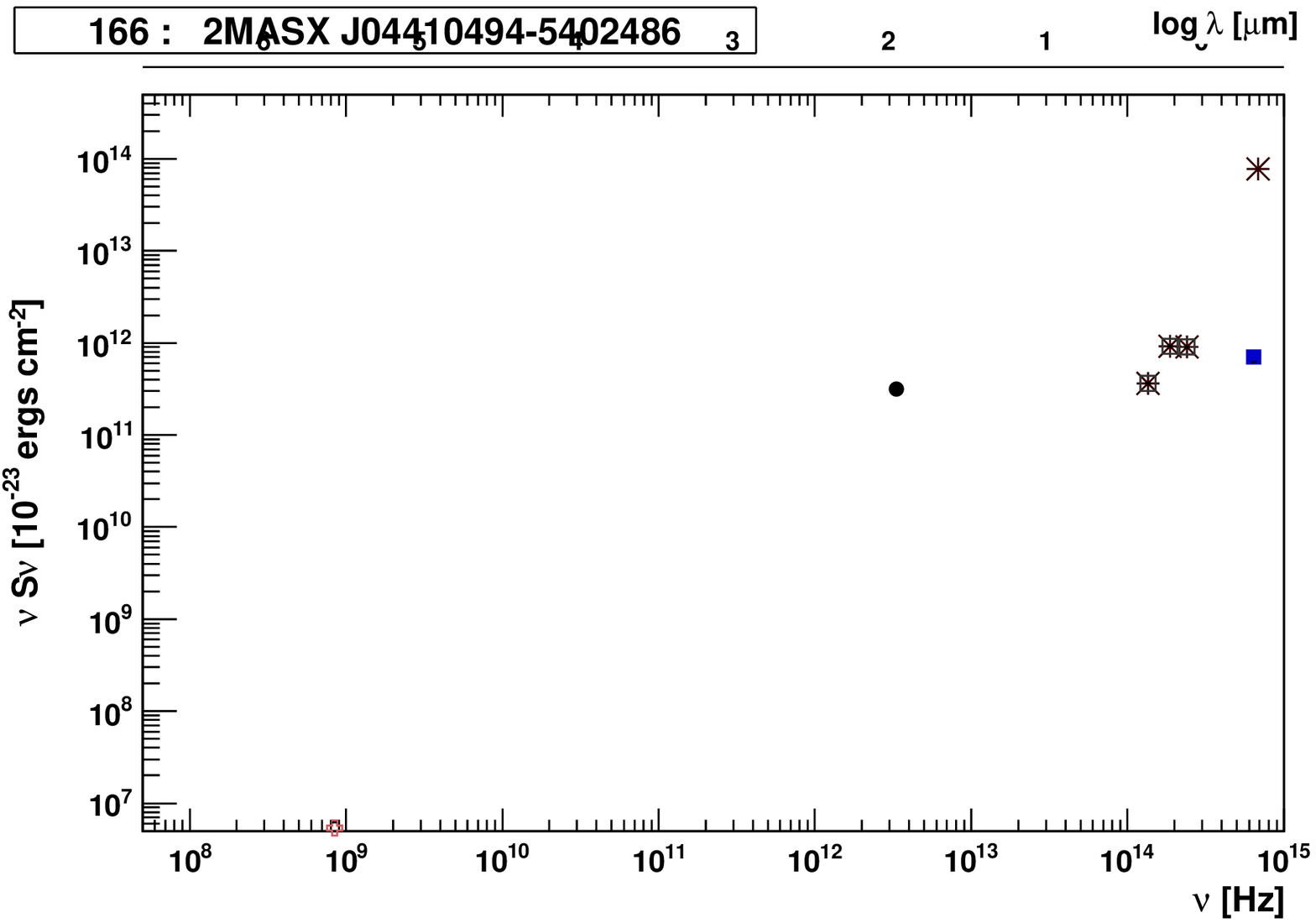}
\includegraphics[width=4cm]{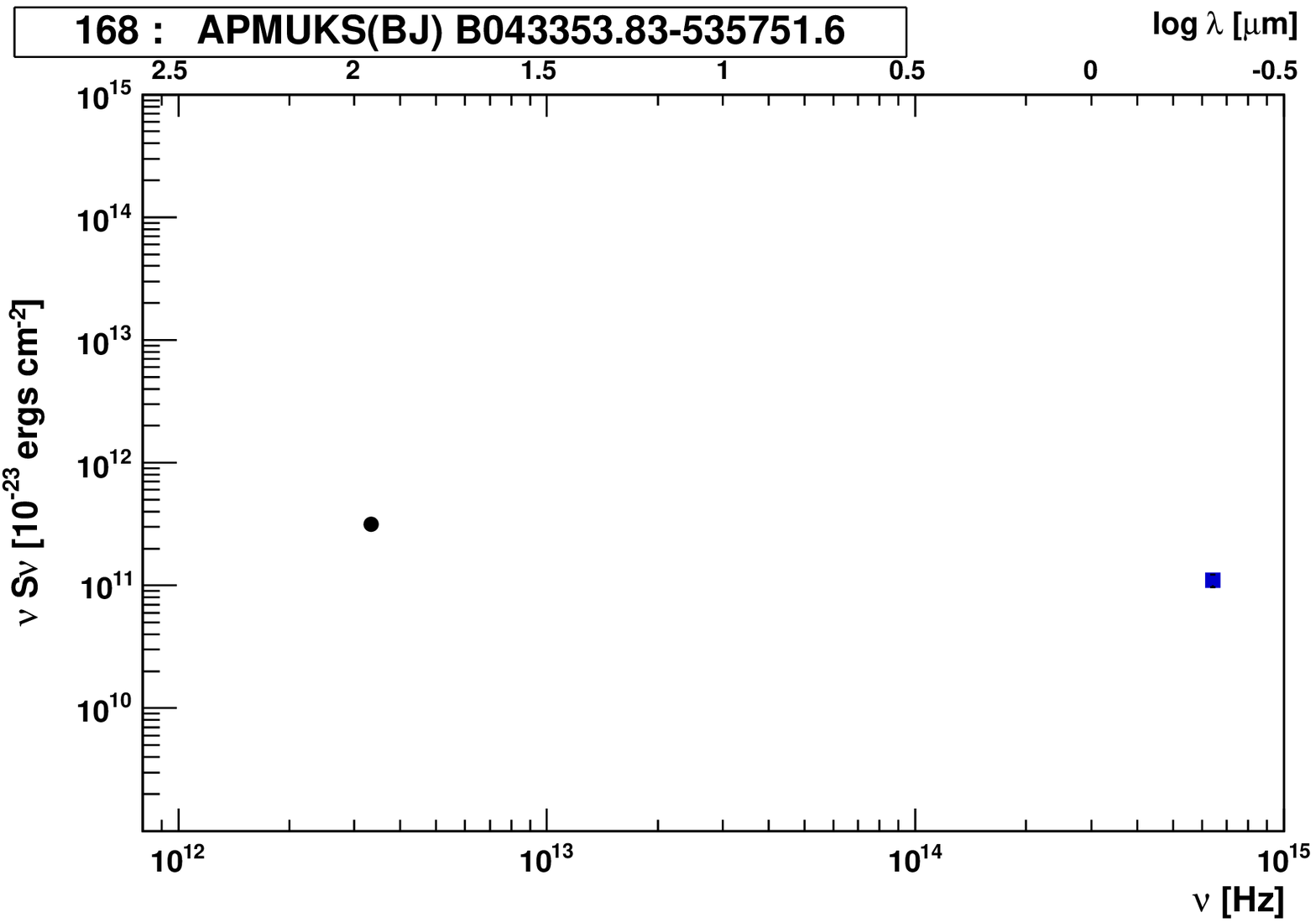}
\includegraphics[width=4cm]{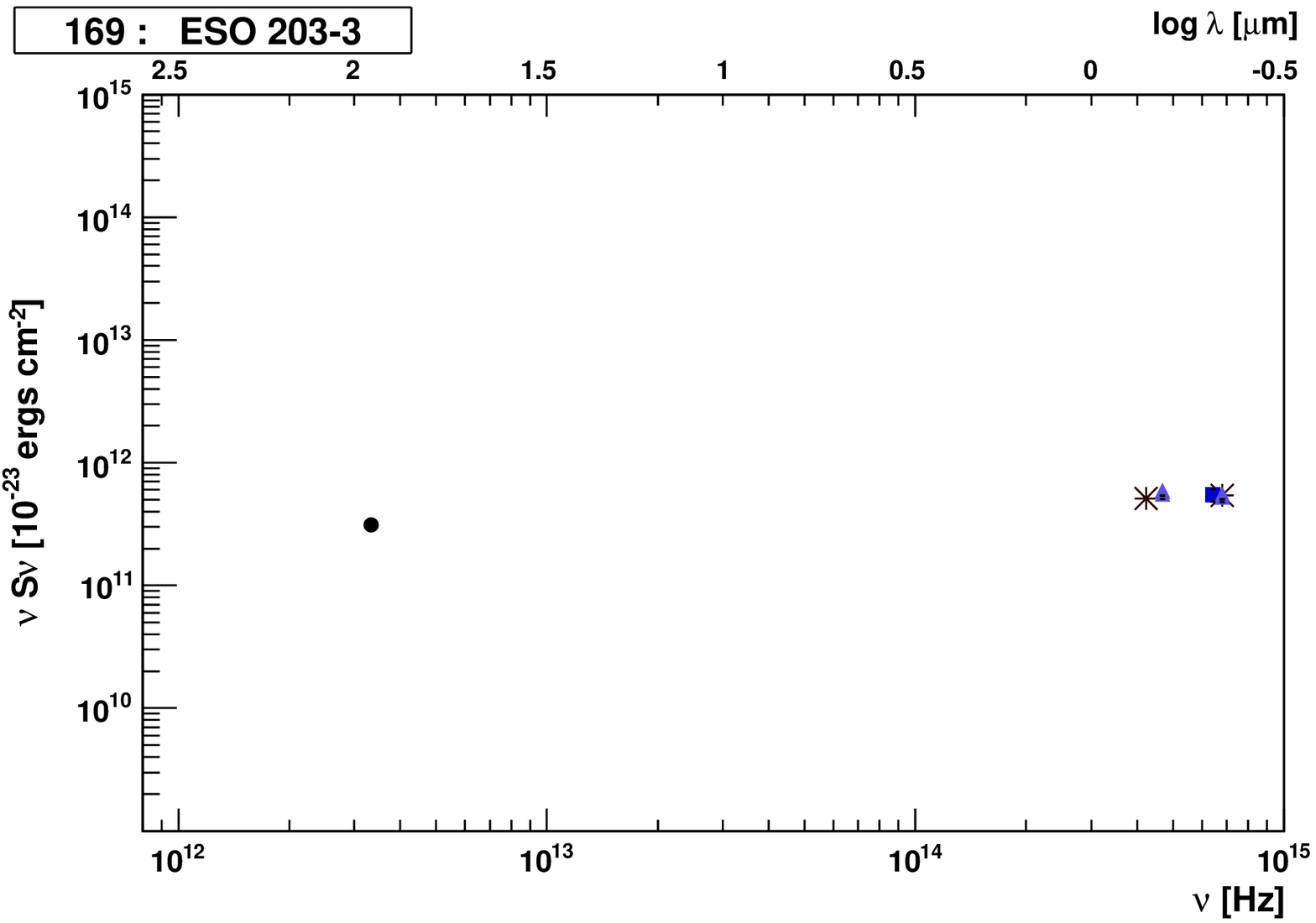}
\label{points4}
\caption {SEDs for the next 36 ADF-S identified sources, with symbols as in Figure~\ref{points1}.}
\end{figure*}
}

\clearpage

\onlfig{5}{
\begin{figure*}[t]
\centering
\includegraphics[width=4cm]{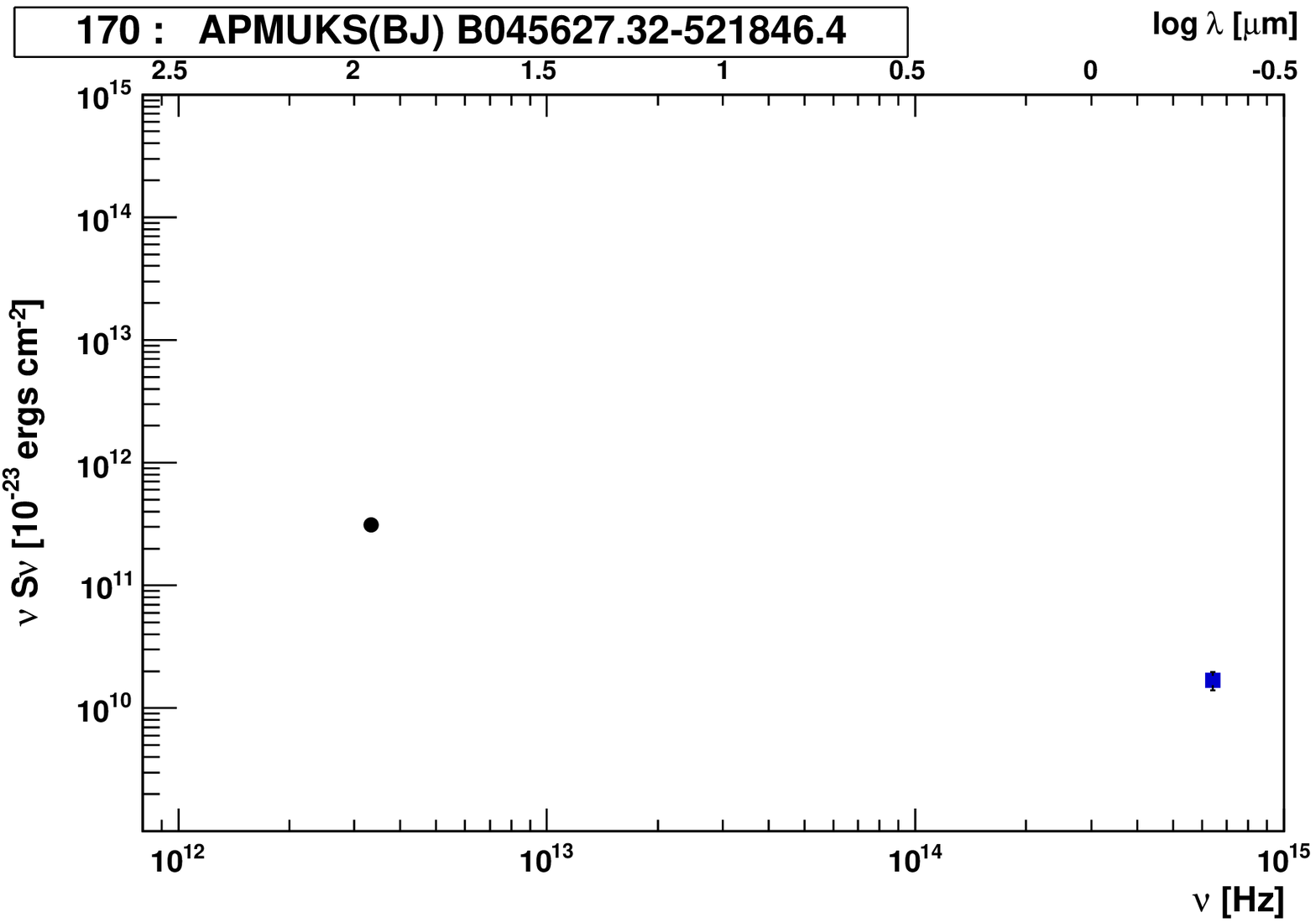}
\includegraphics[width=4cm]{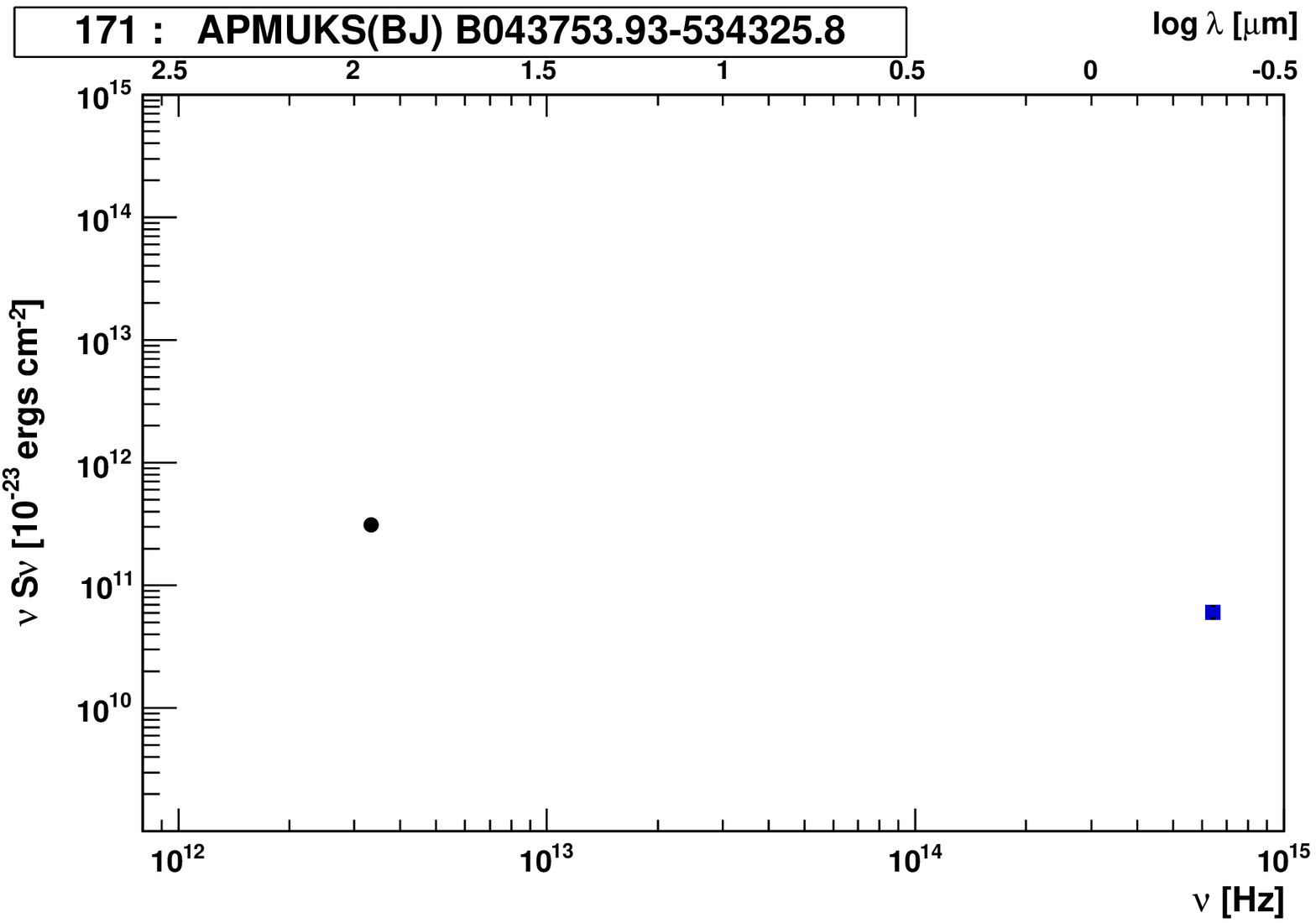}
\includegraphics[width=4cm]{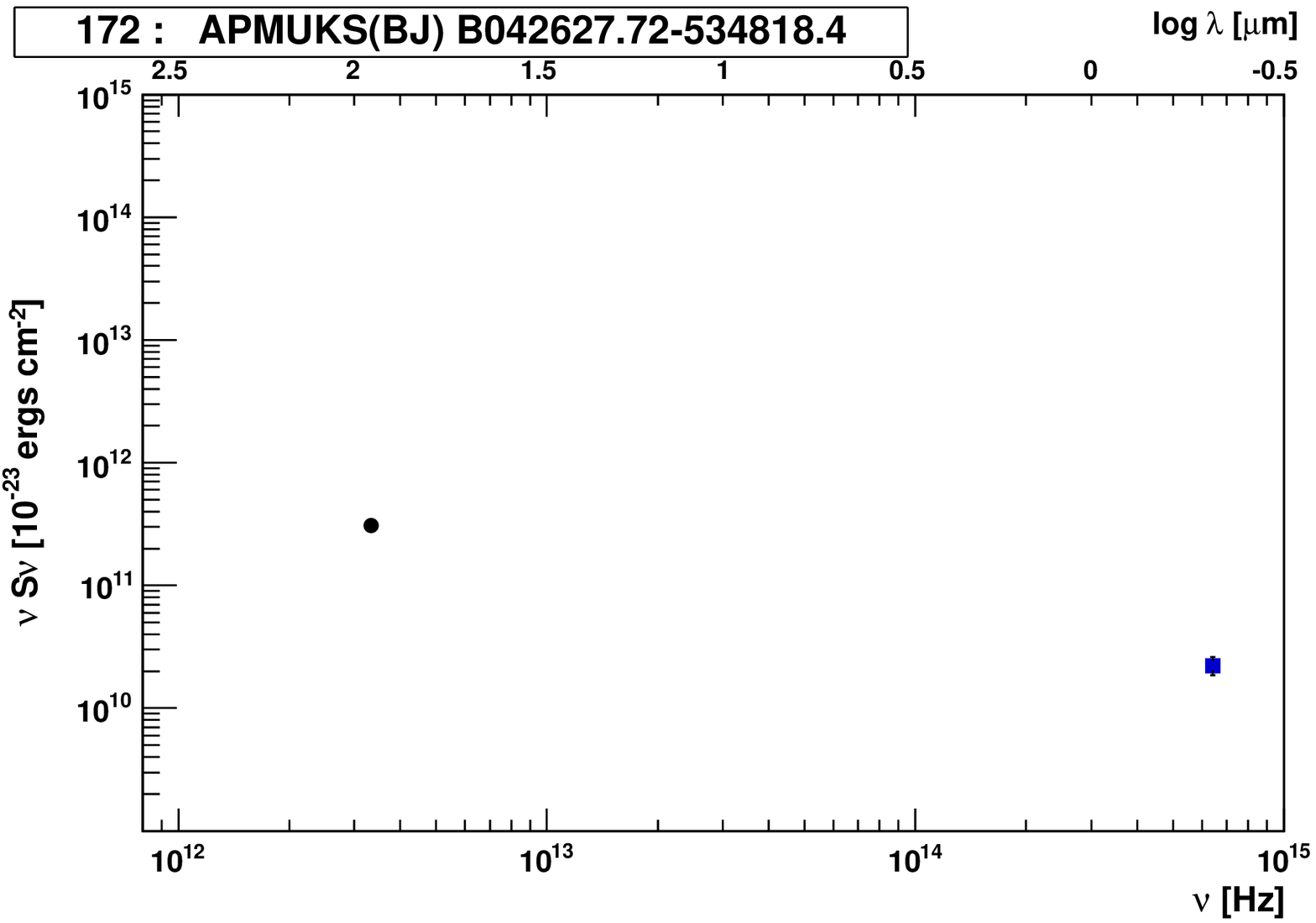}
\includegraphics[width=4cm]{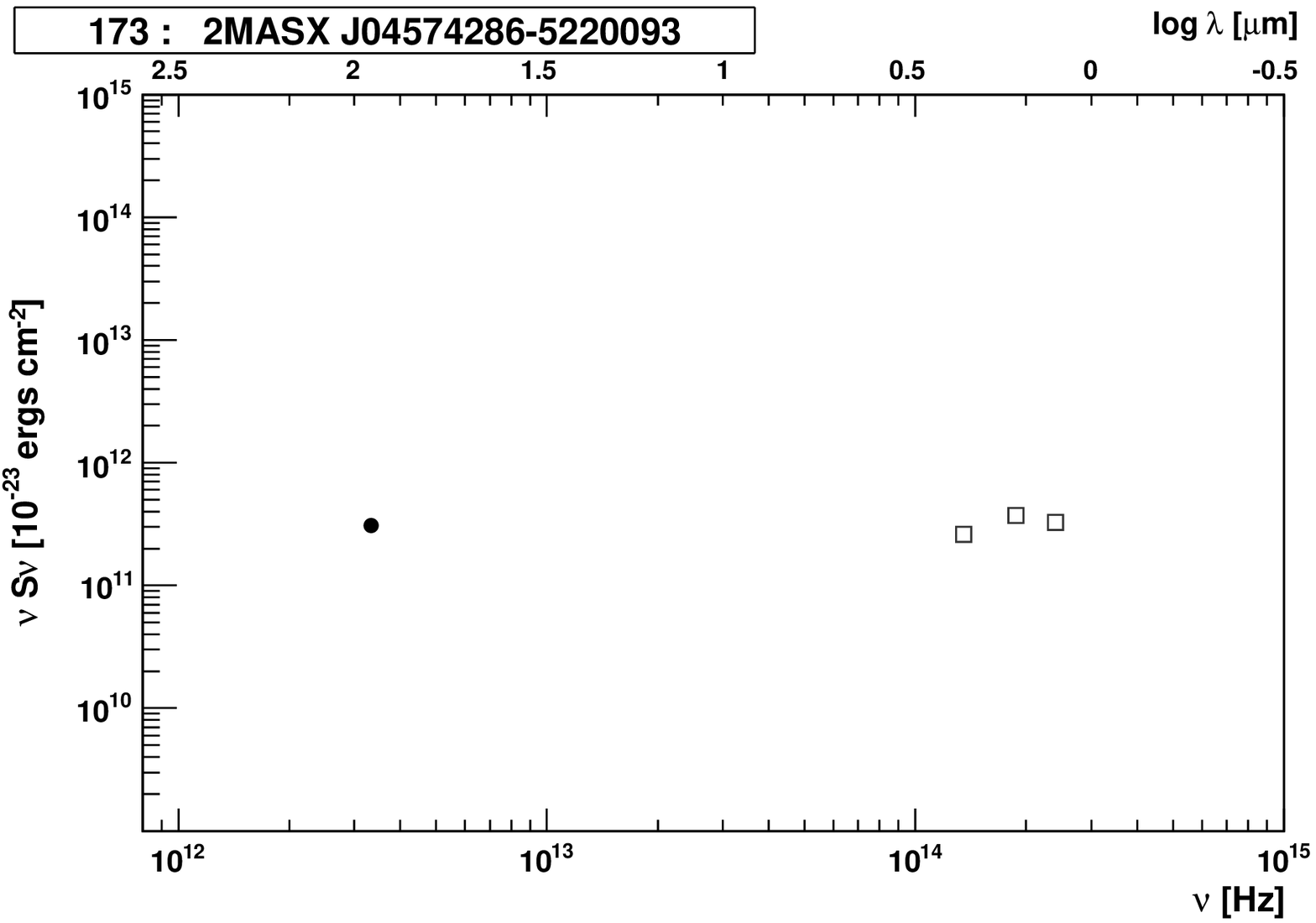}
\includegraphics[width=4cm]{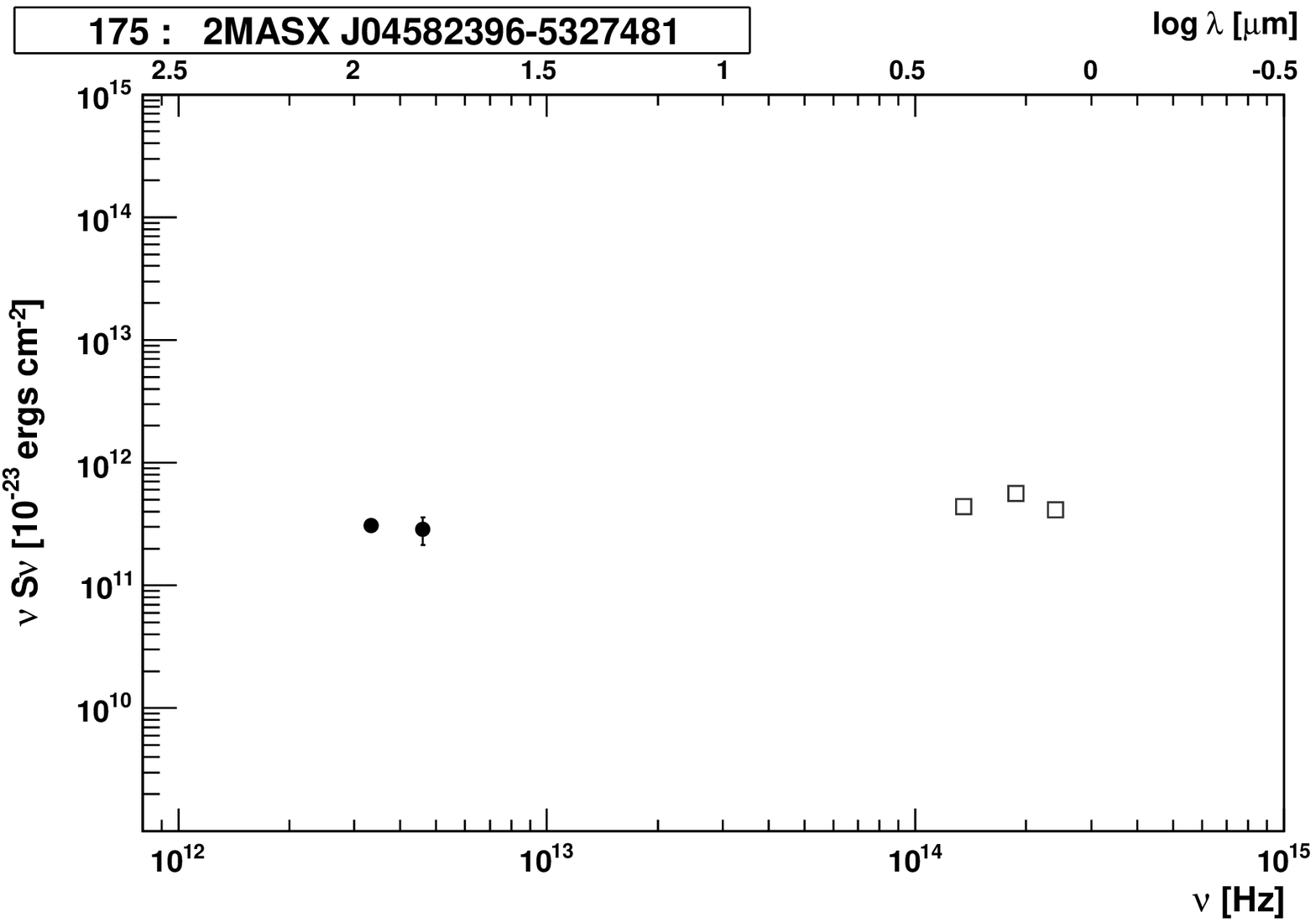}
\includegraphics[width=4cm]{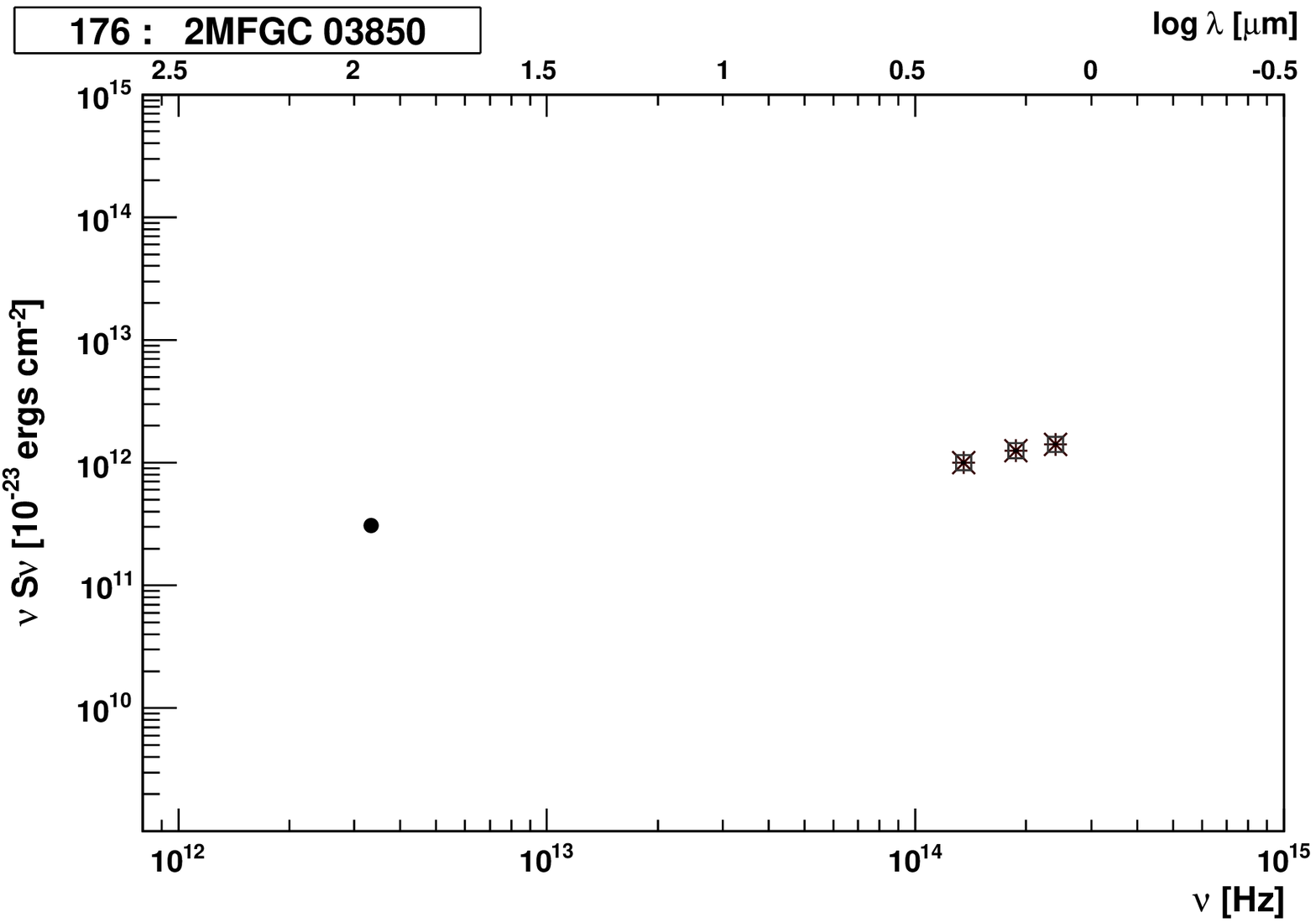}
\includegraphics[width=4cm]{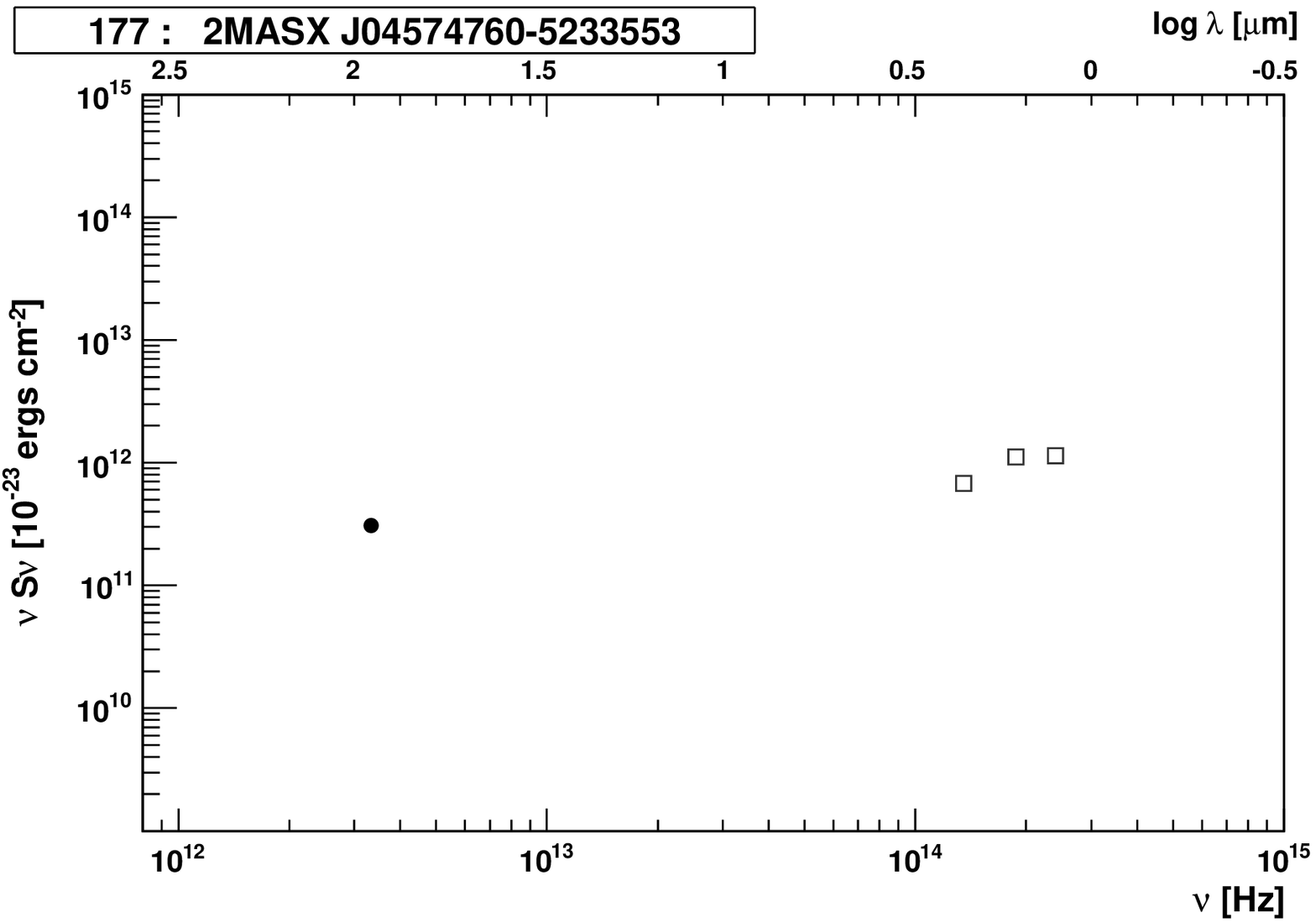}
\includegraphics[width=4cm]{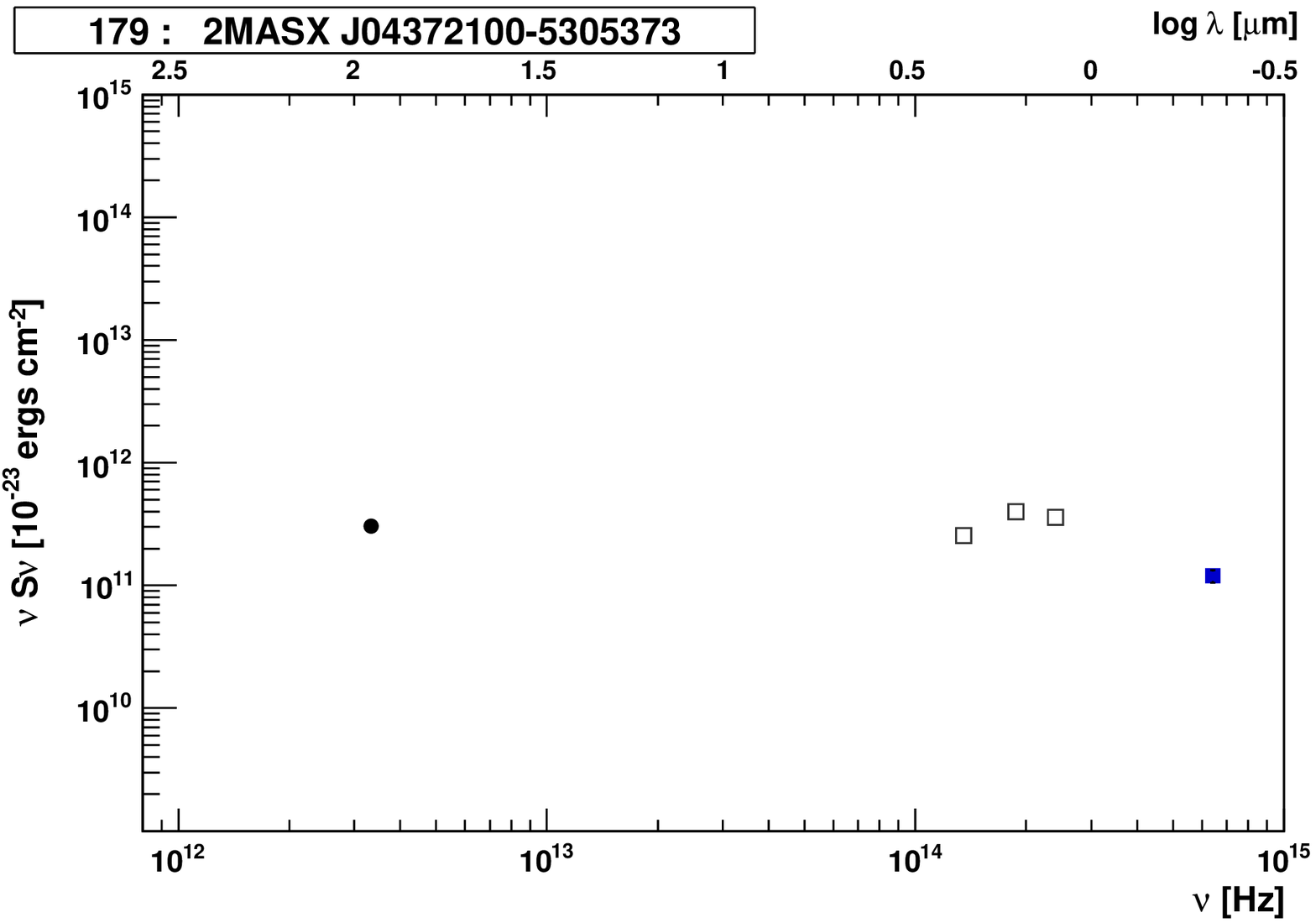}
\includegraphics[width=4cm]{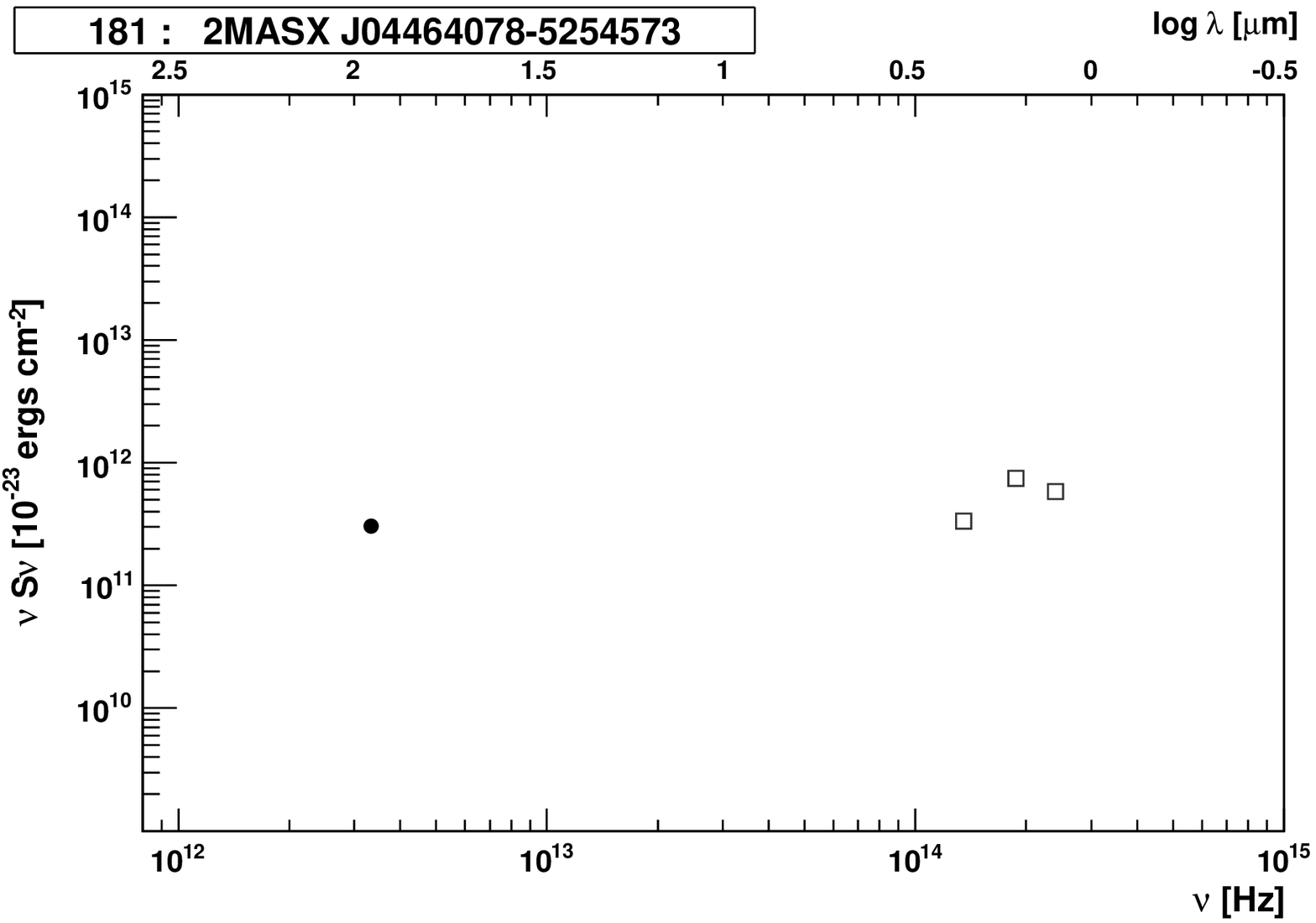}
\includegraphics[width=4cm]{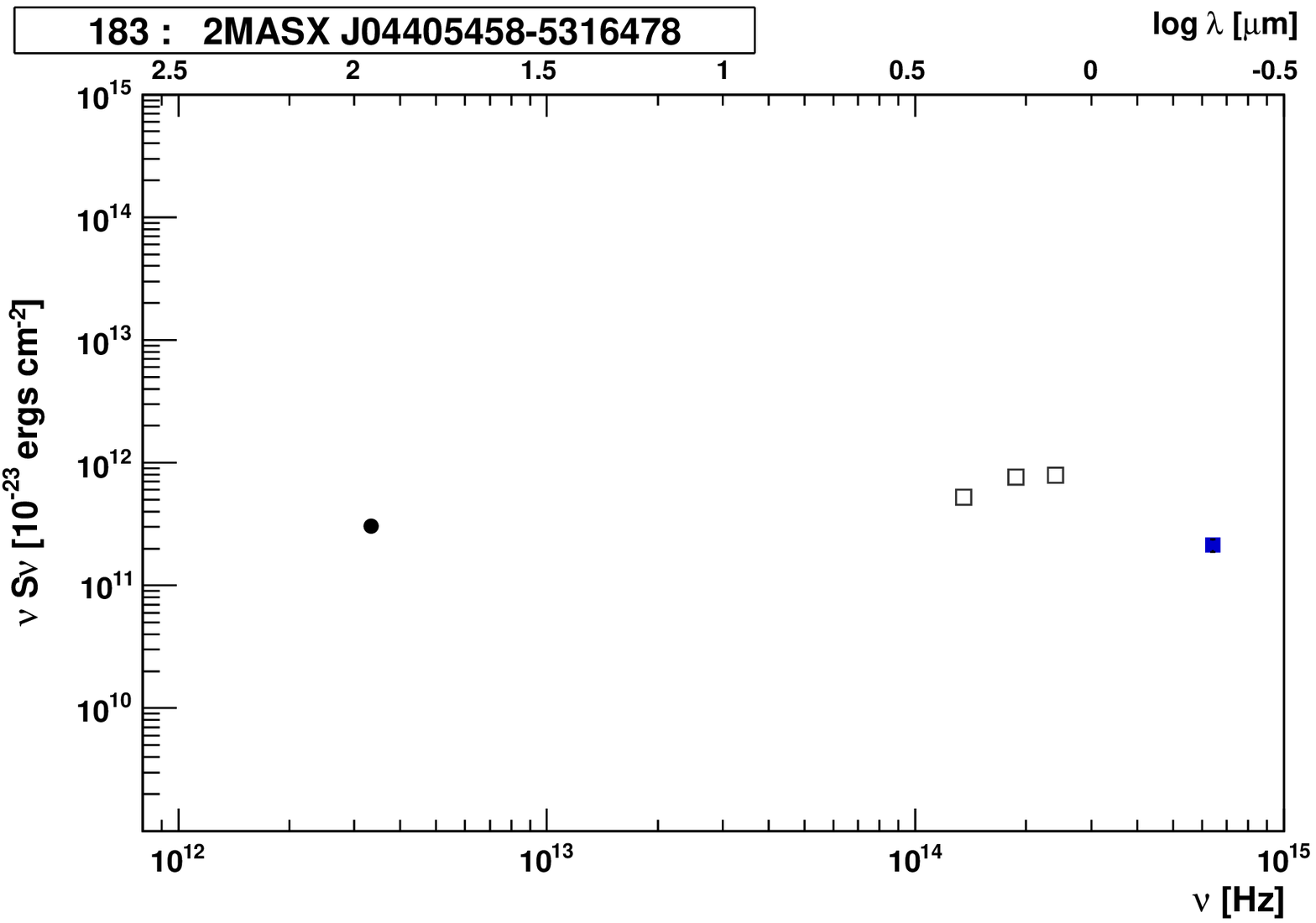}
\includegraphics[width=4cm]{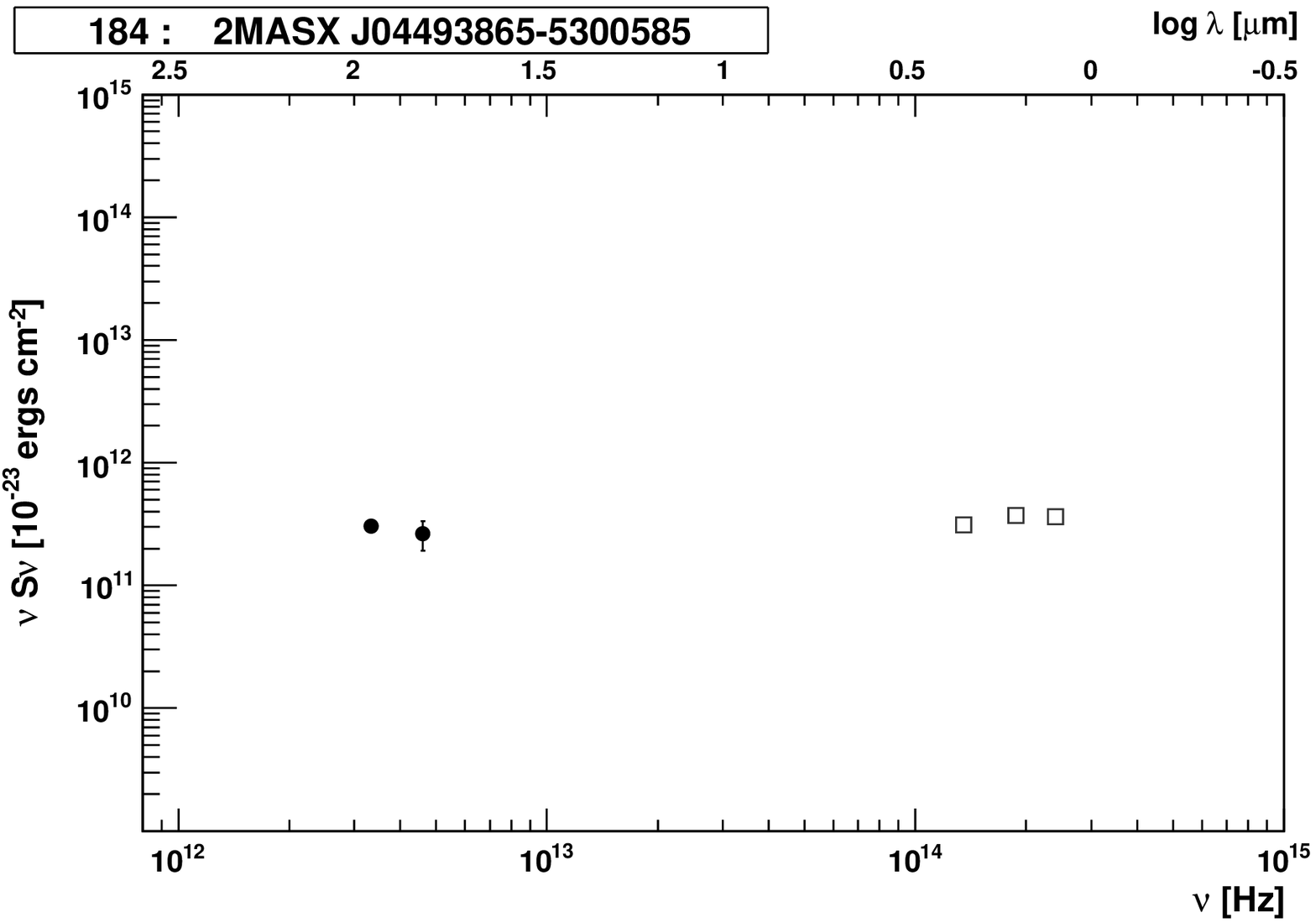}
\includegraphics[width=4cm]{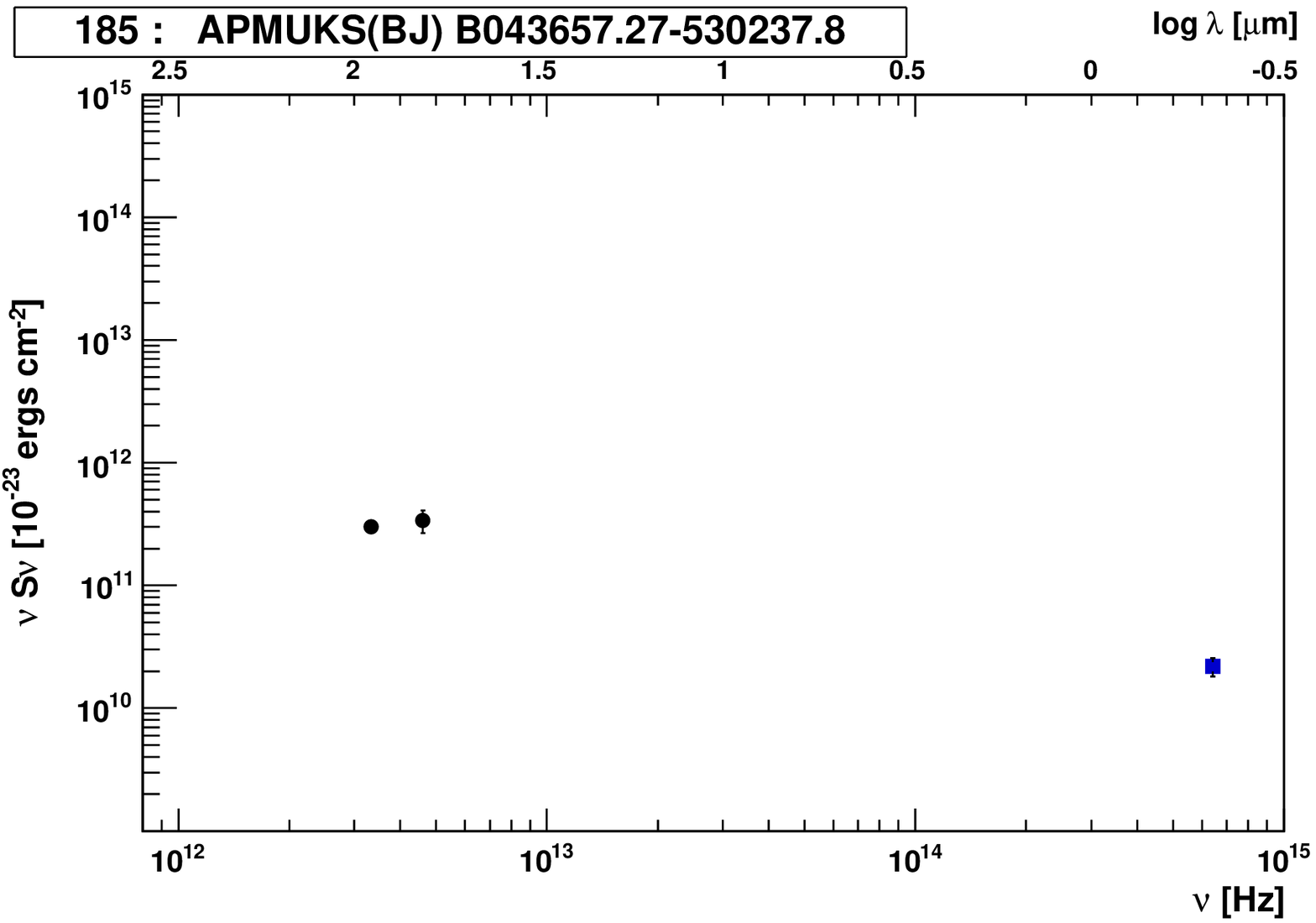}
\includegraphics[width=4cm]{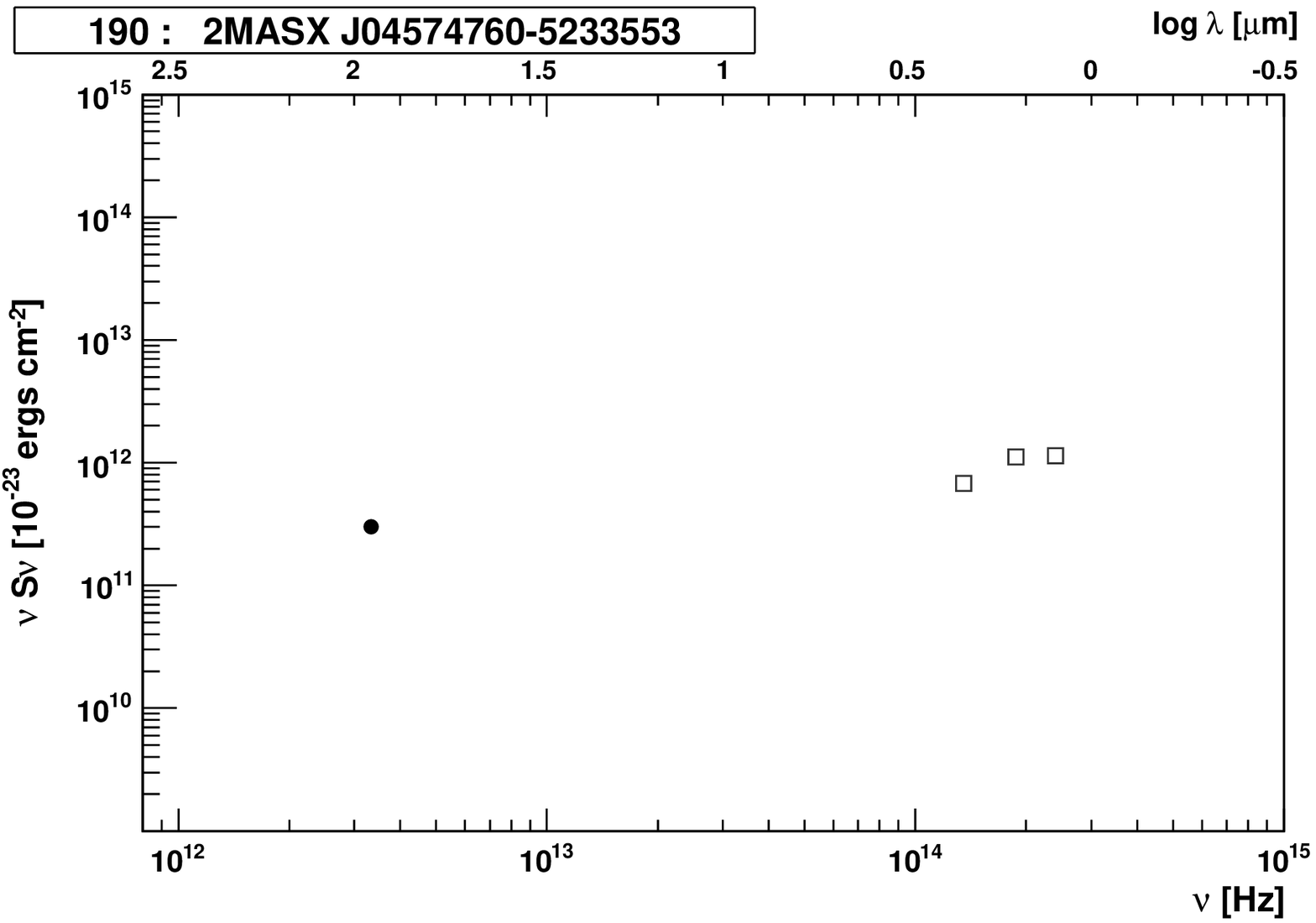}
\includegraphics[width=4cm]{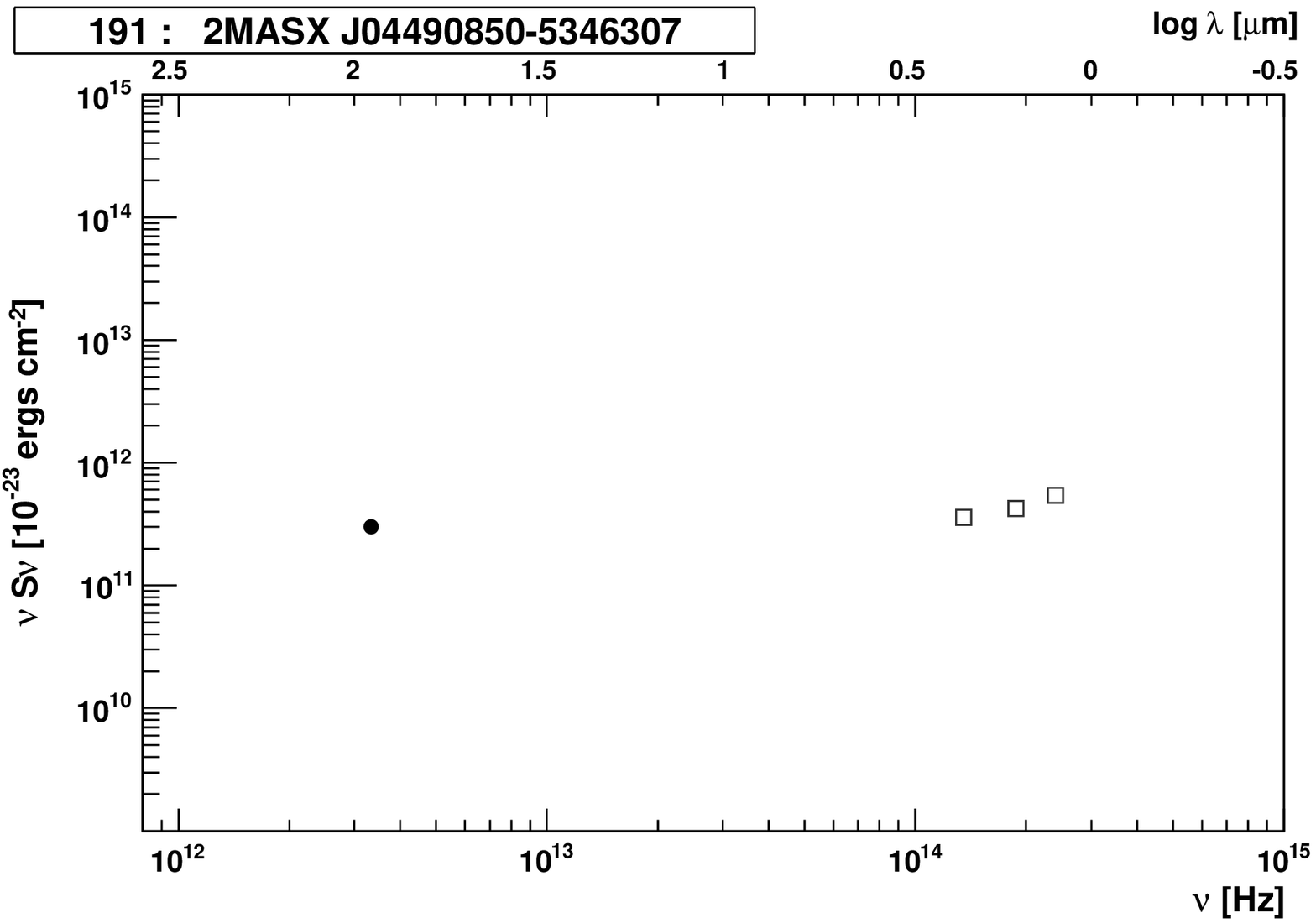}
\includegraphics[width=4cm]{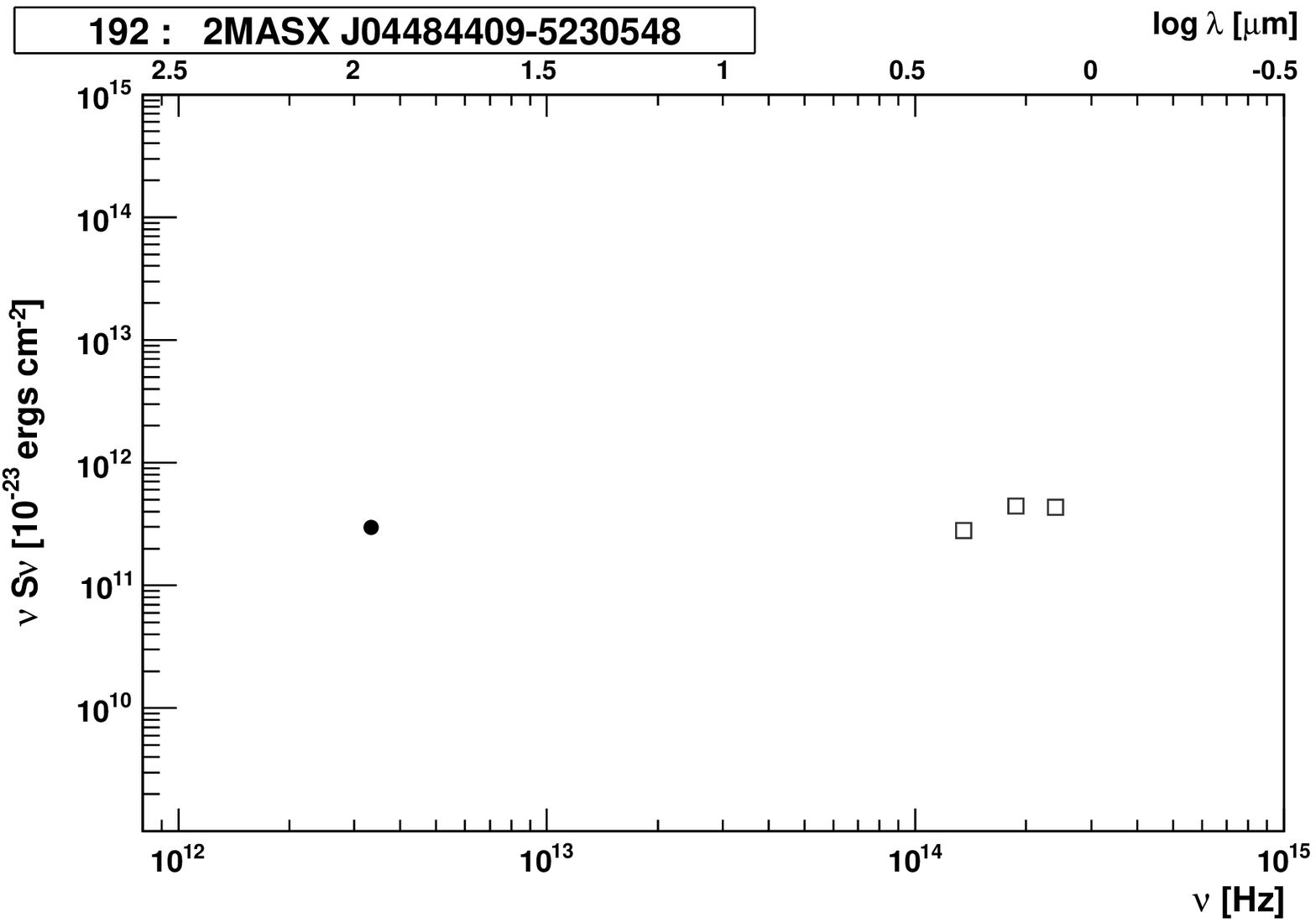}
\includegraphics[width=4cm]{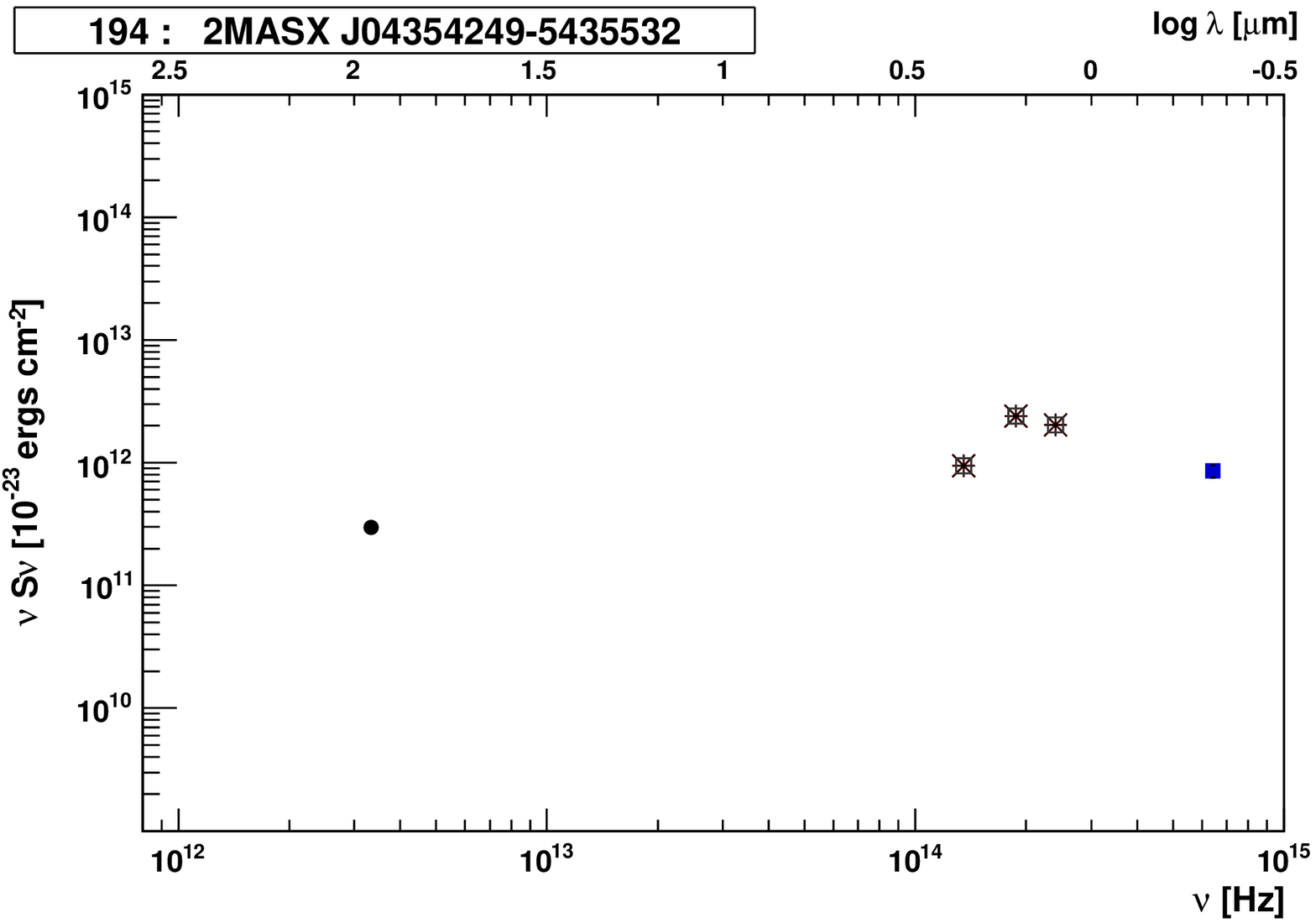}
\includegraphics[width=4cm]{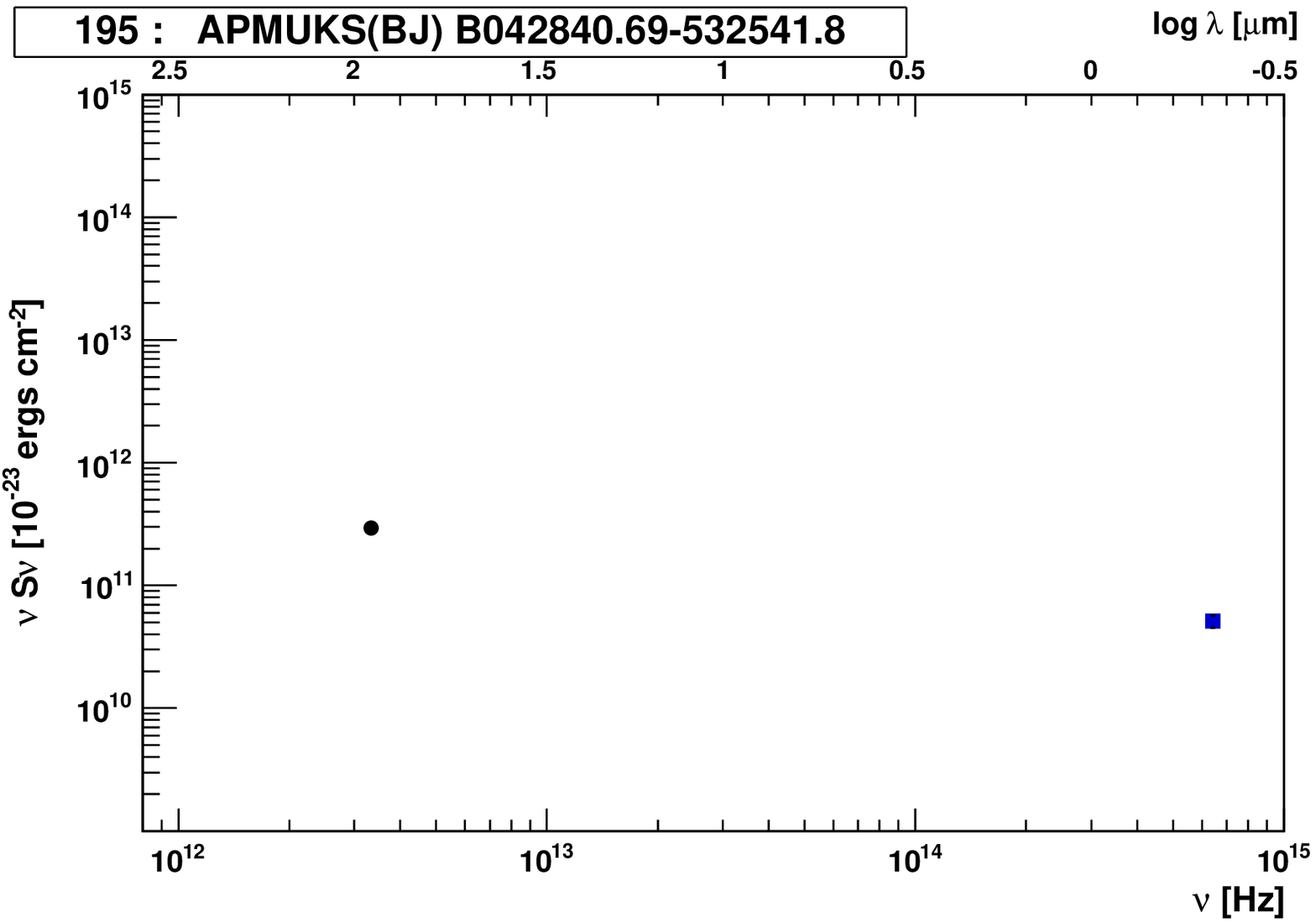}
\includegraphics[width=4cm]{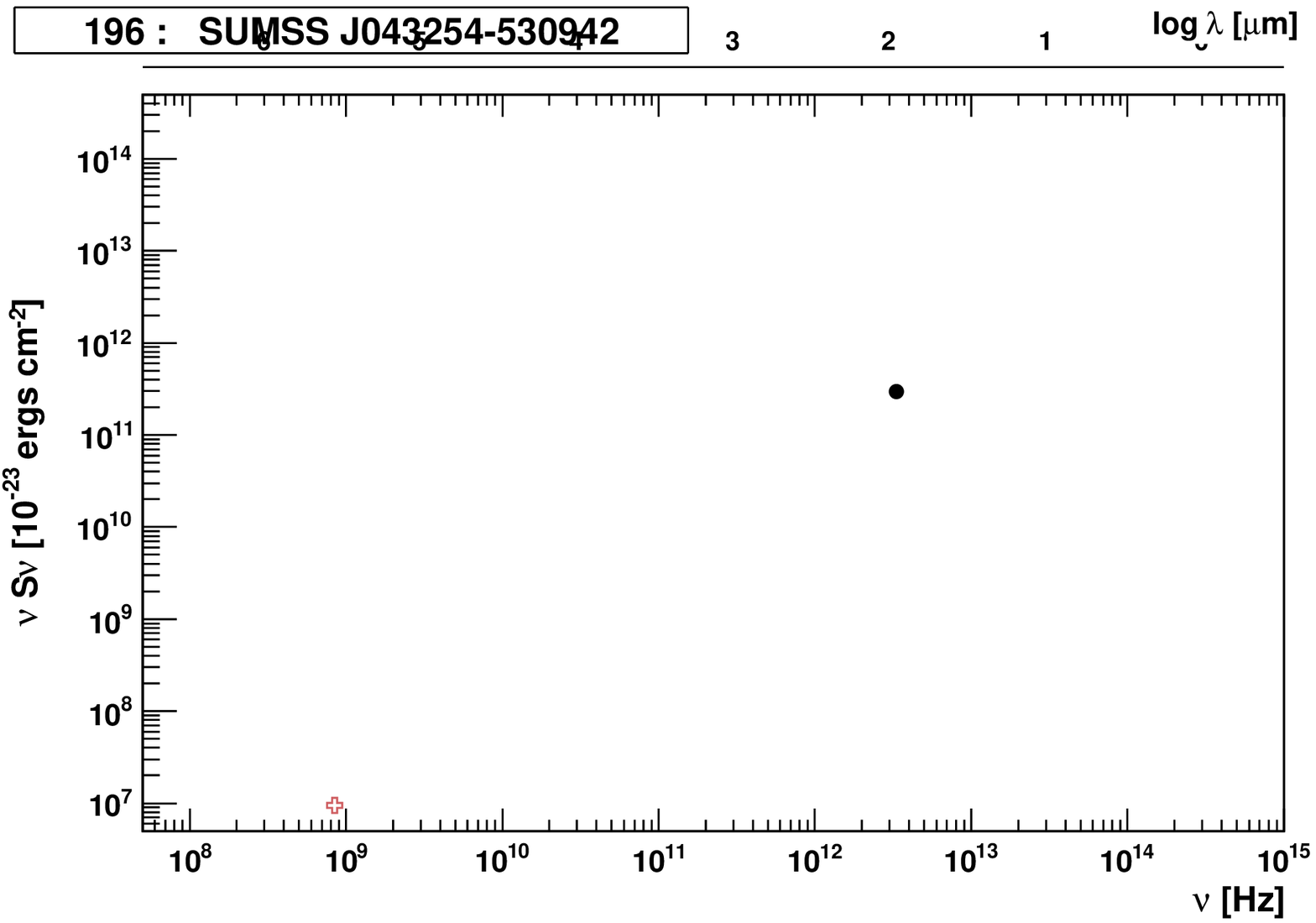}
\includegraphics[width=4cm]{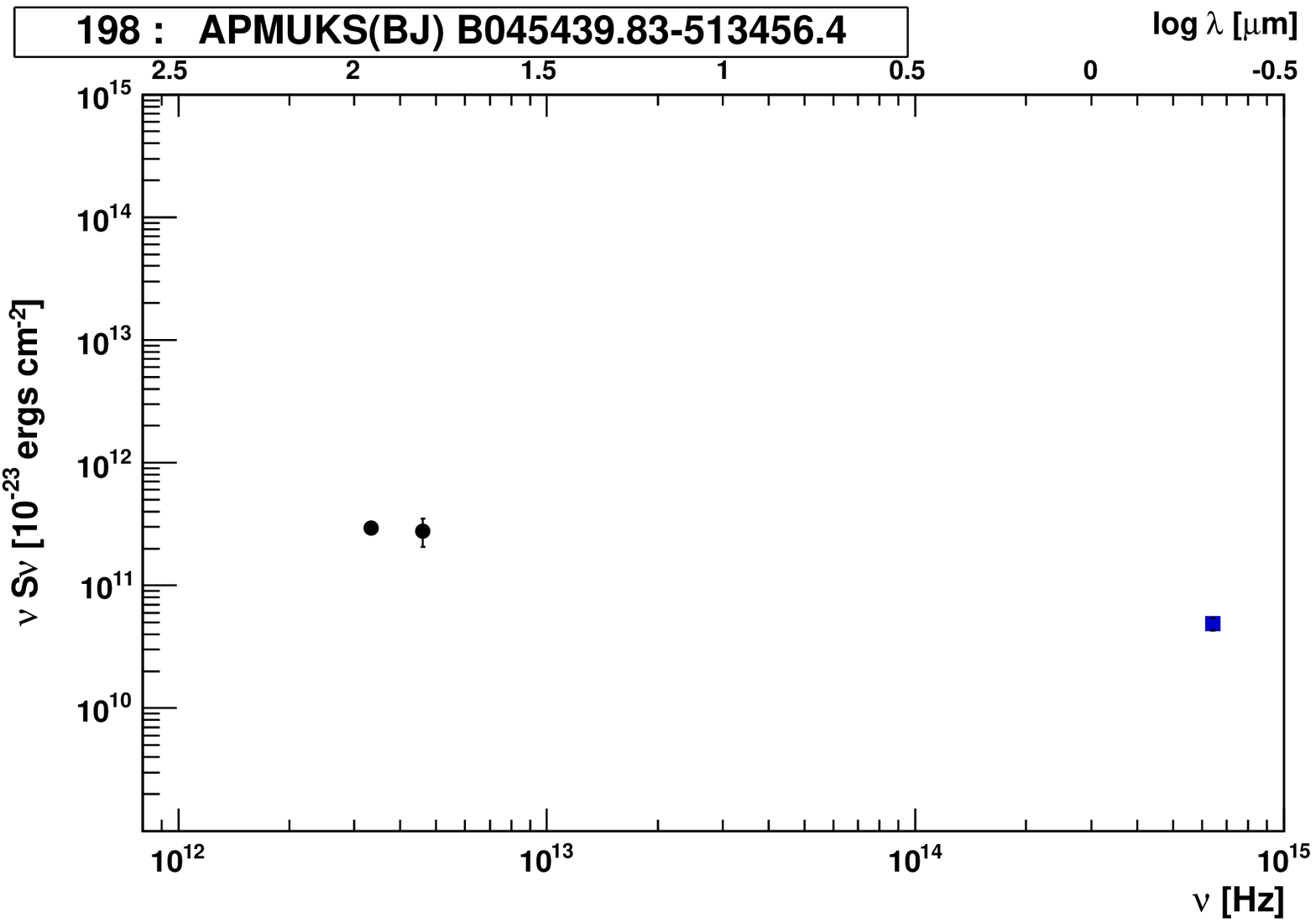}
\includegraphics[width=4cm]{points/kmalek_170.eps}
\includegraphics[width=4cm]{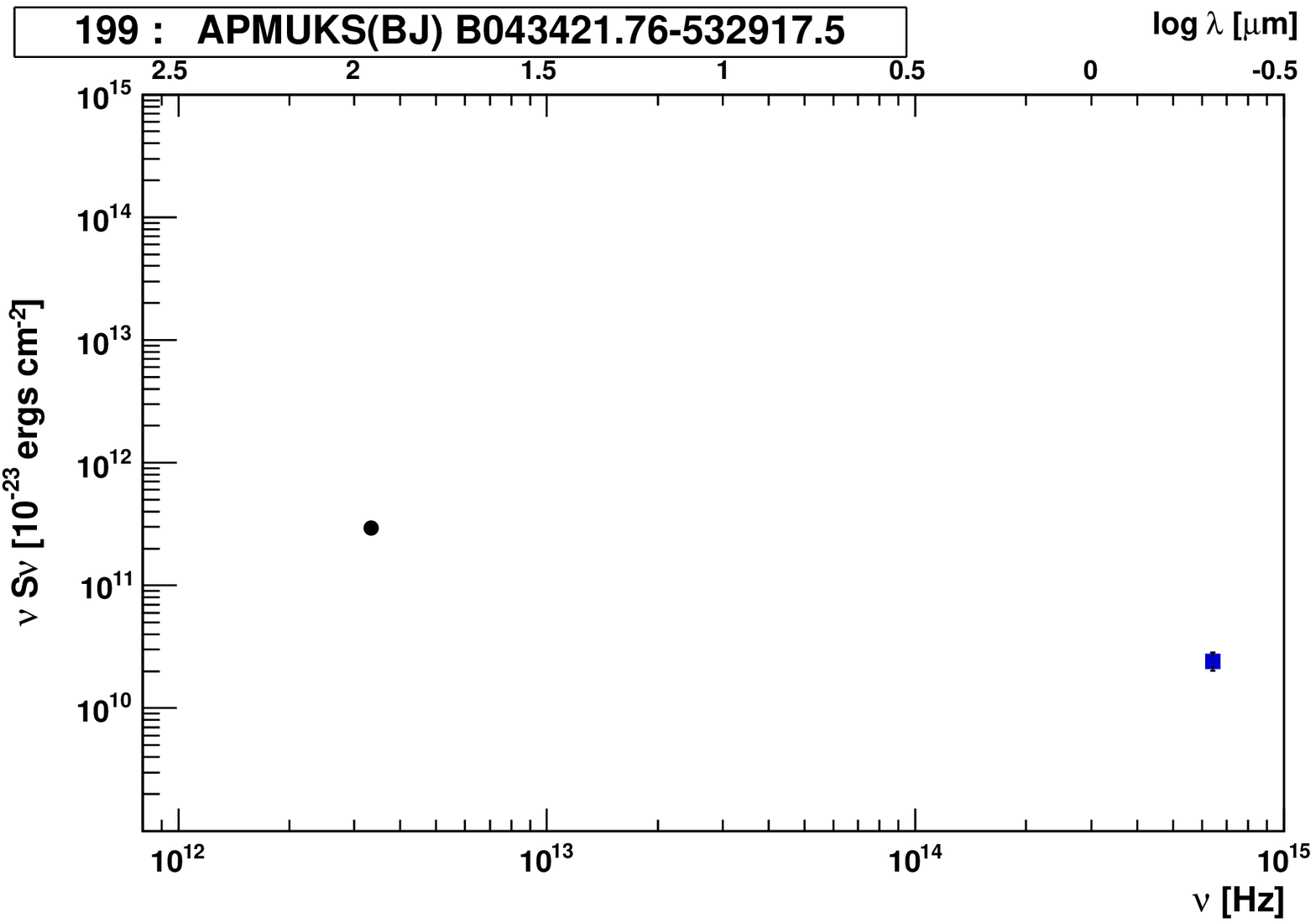}
\includegraphics[width=4cm]{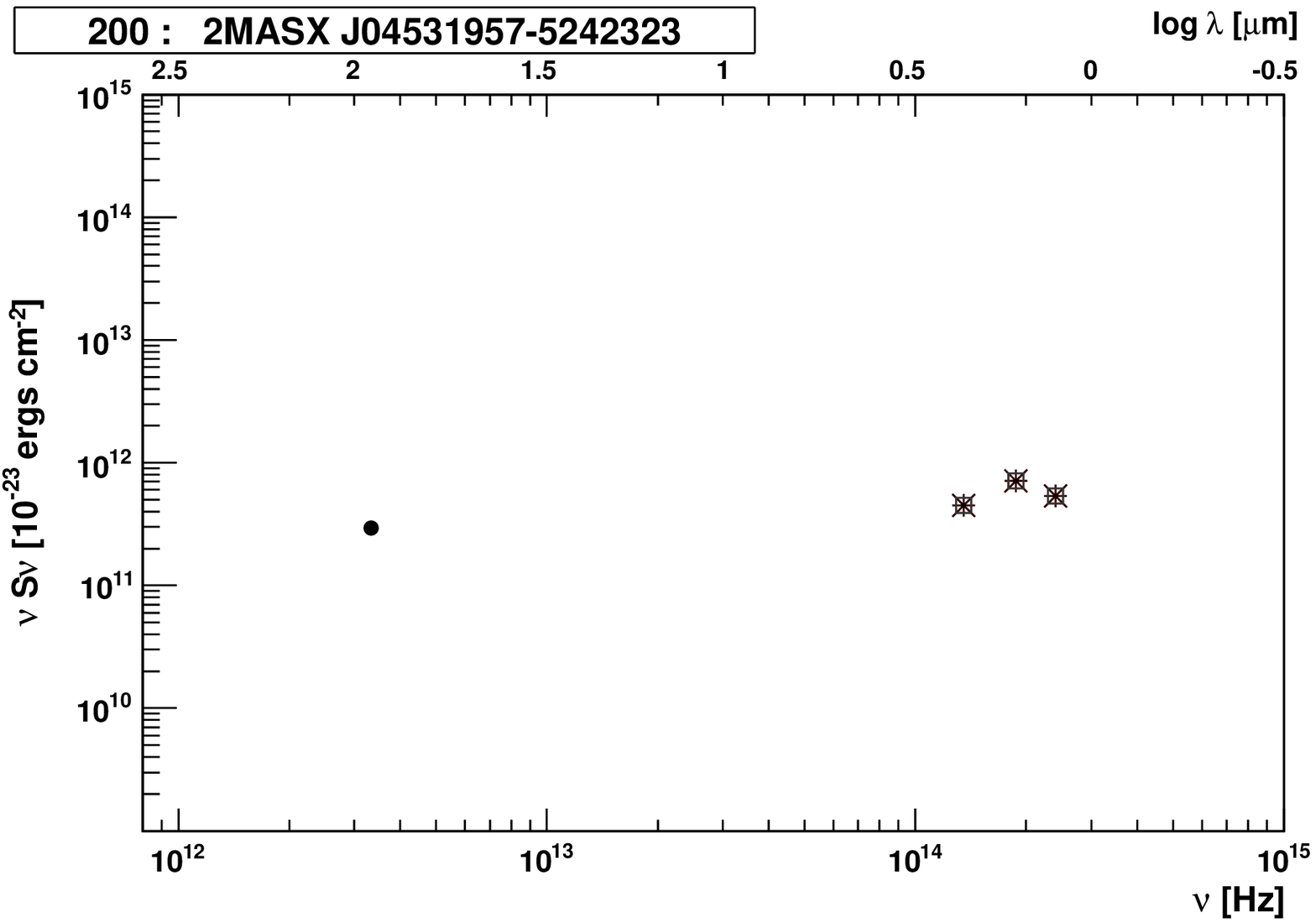}
\includegraphics[width=4cm]{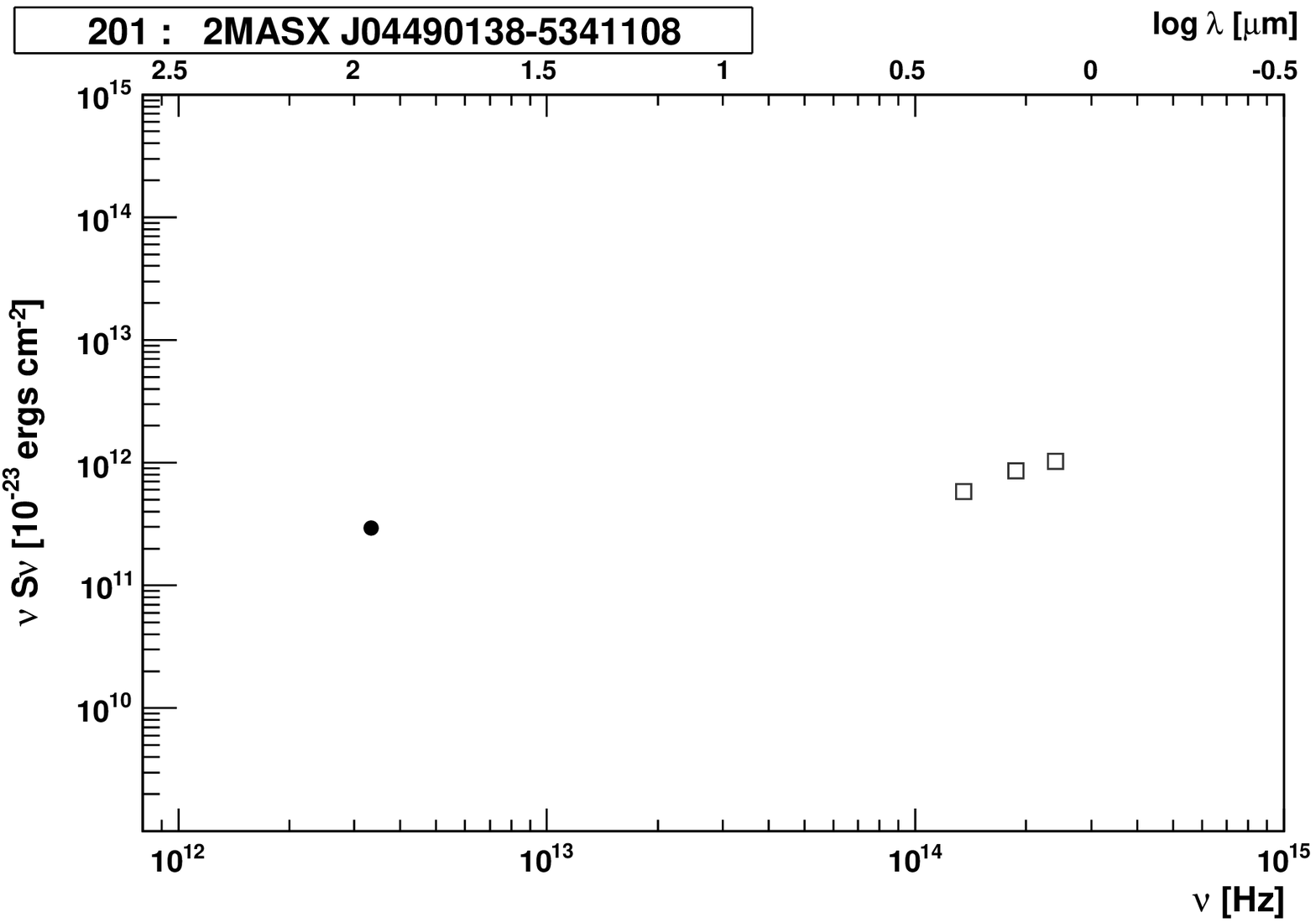}
\includegraphics[width=4cm]{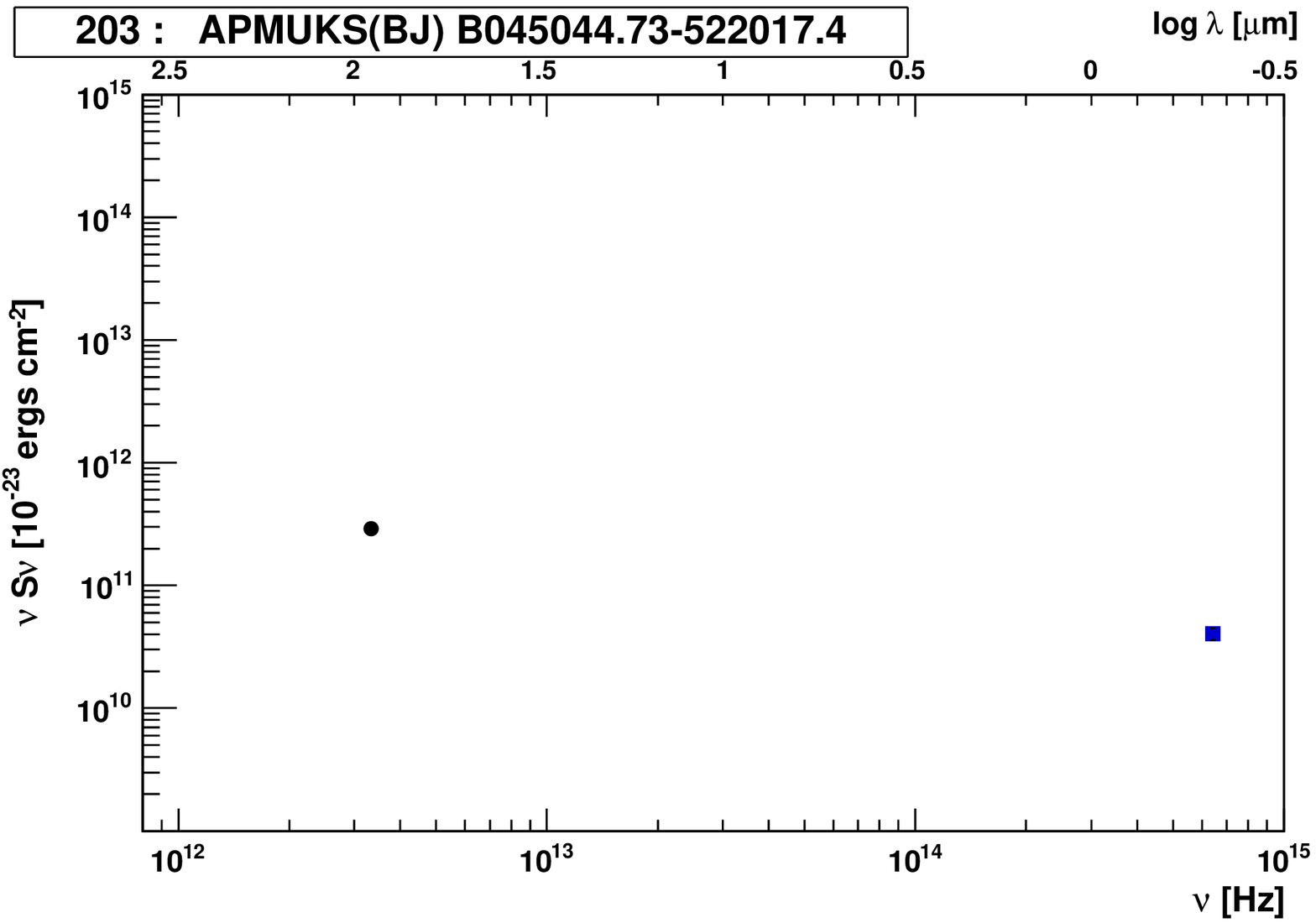}
\includegraphics[width=4cm]{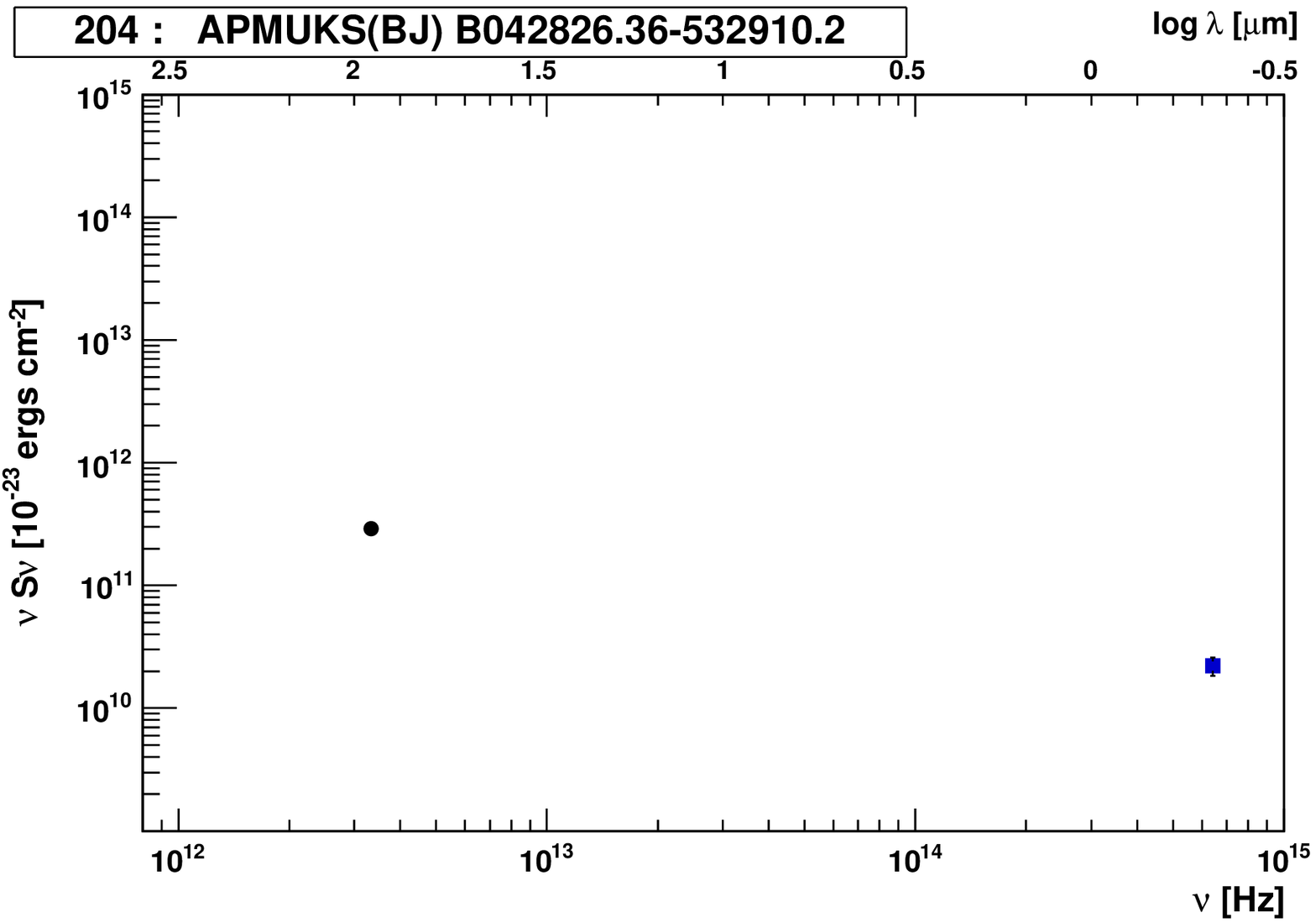}
\includegraphics[width=4cm]{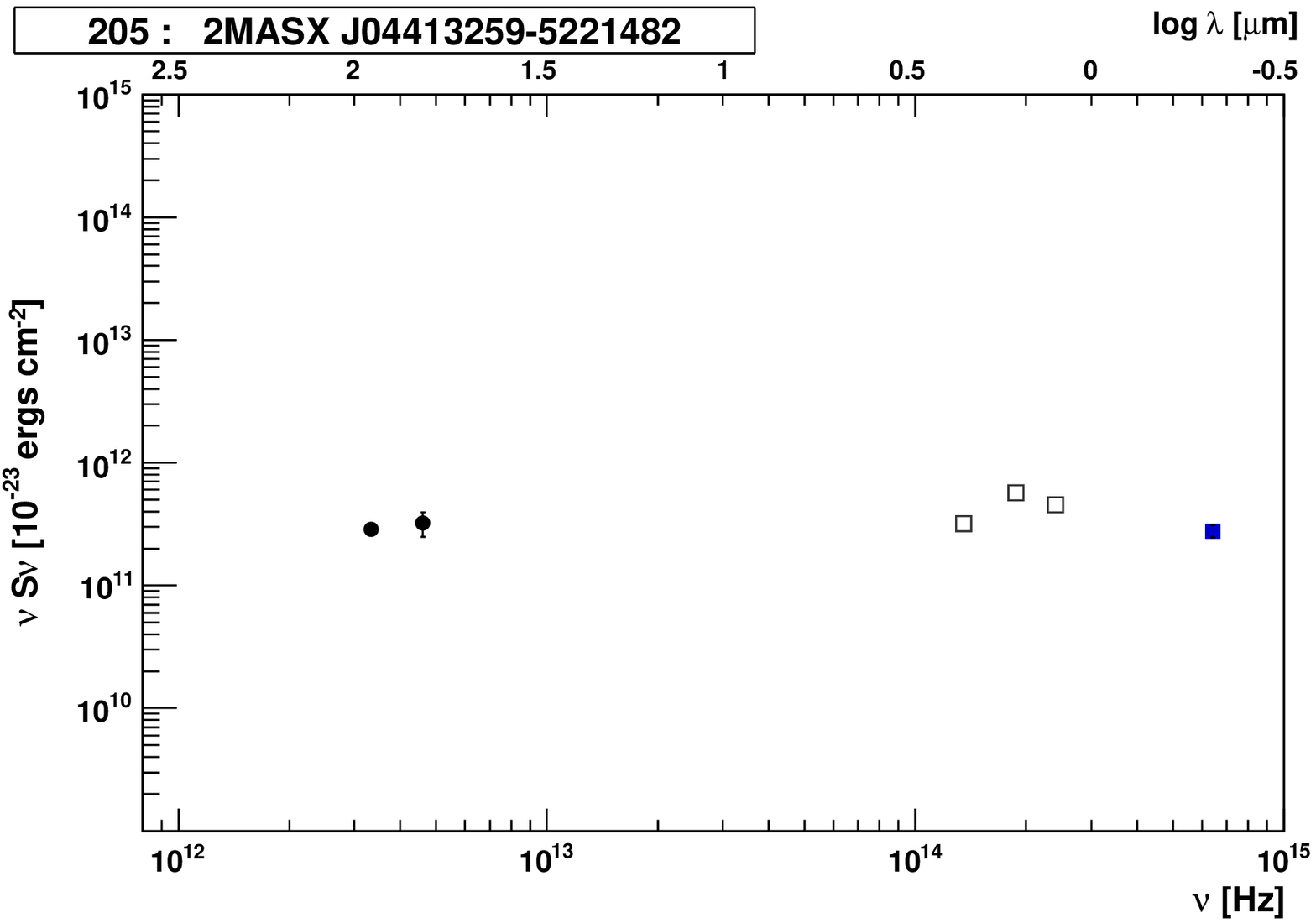}
\includegraphics[width=4cm]{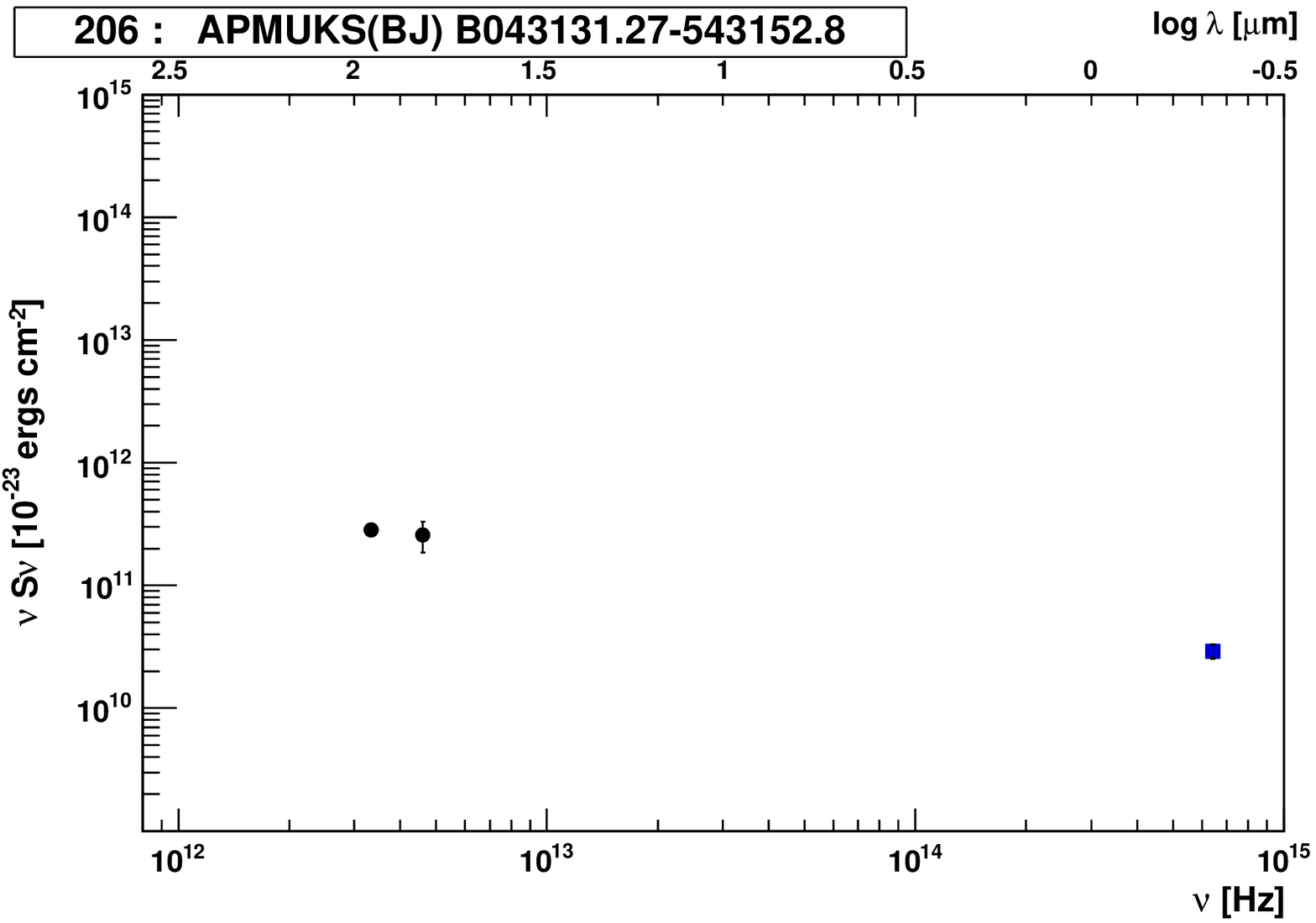}
\includegraphics[width=4cm]{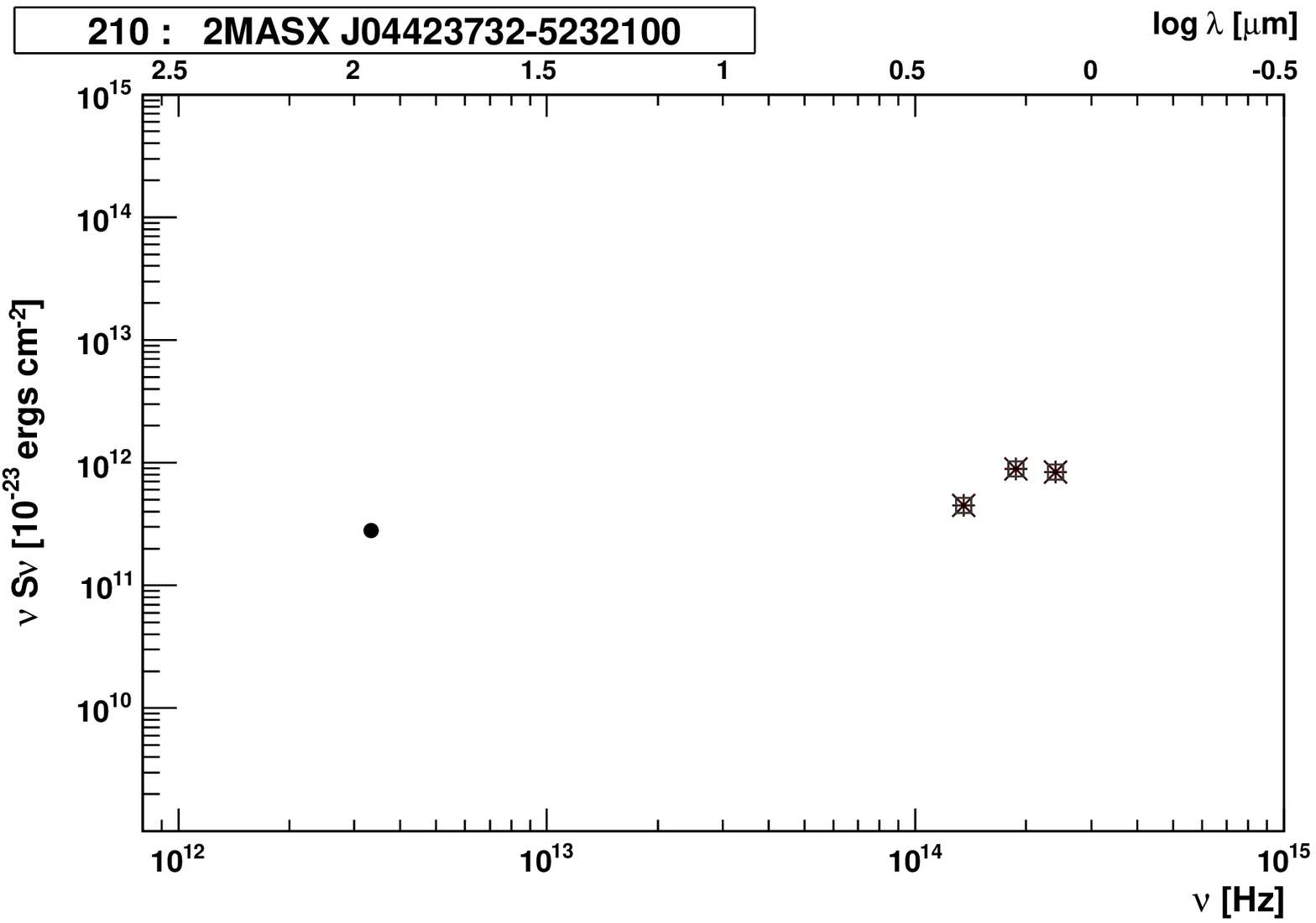}
\includegraphics[width=4cm]{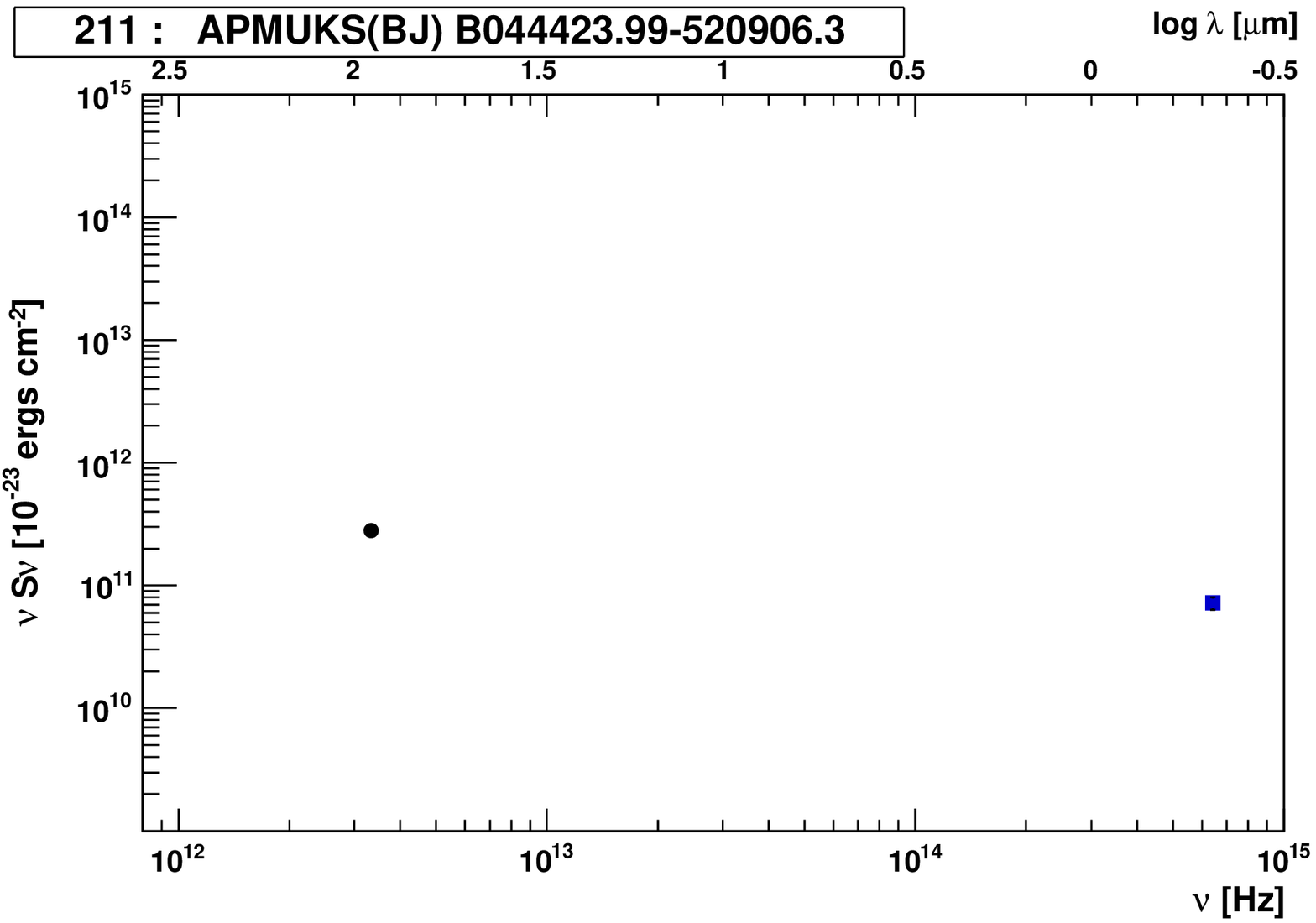}
\includegraphics[width=4cm]{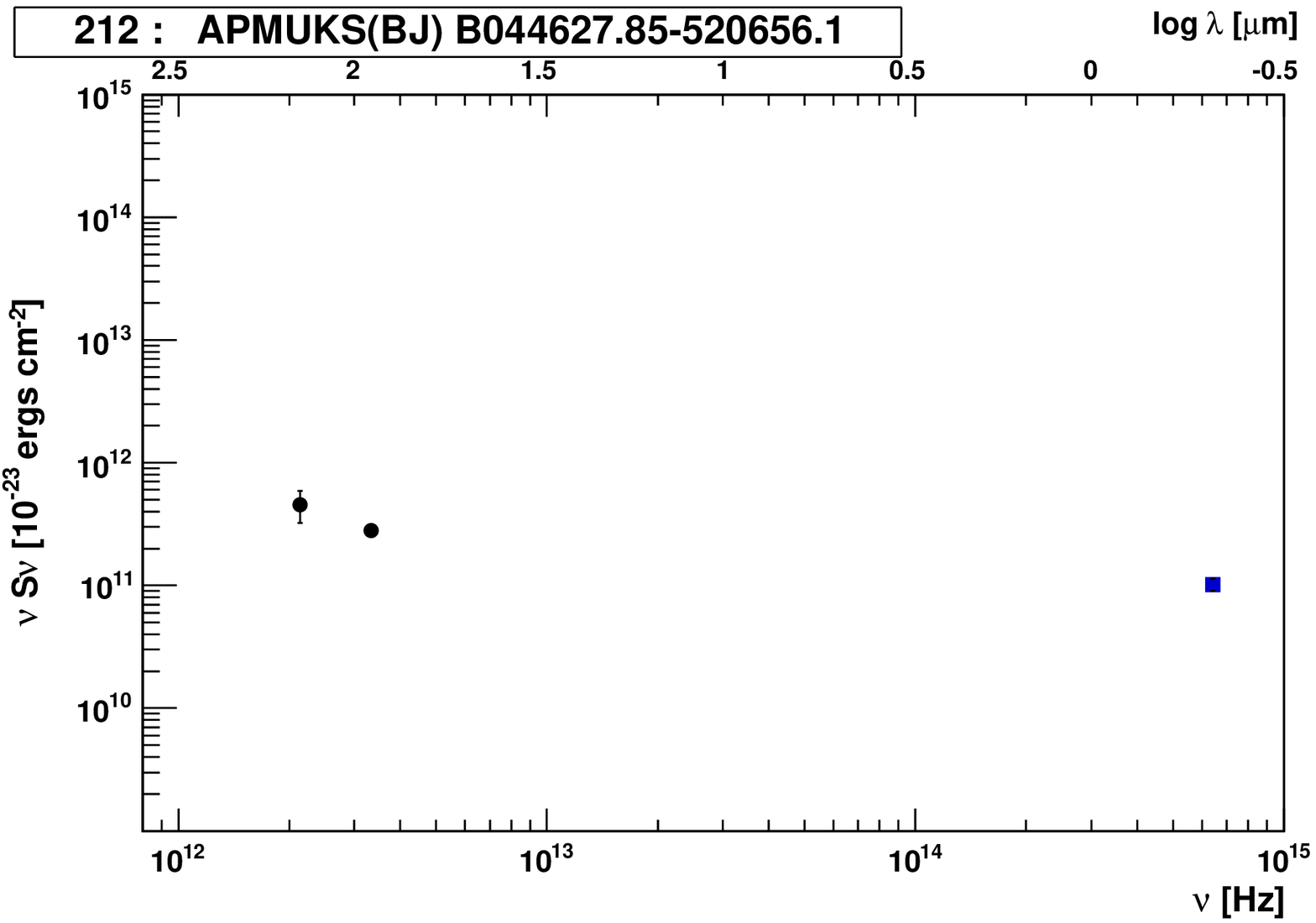}
\includegraphics[width=4cm]{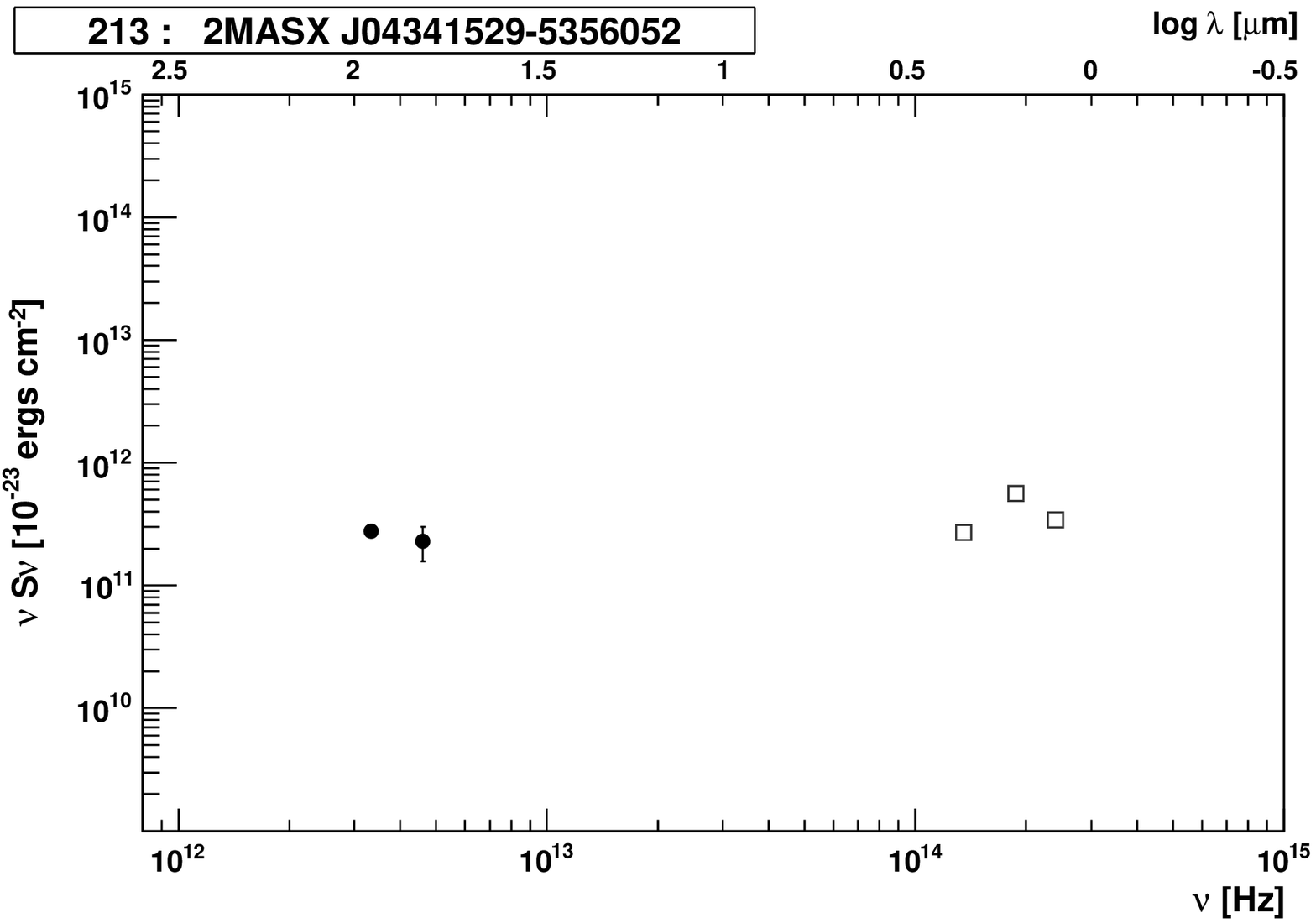}
\includegraphics[width=4cm]{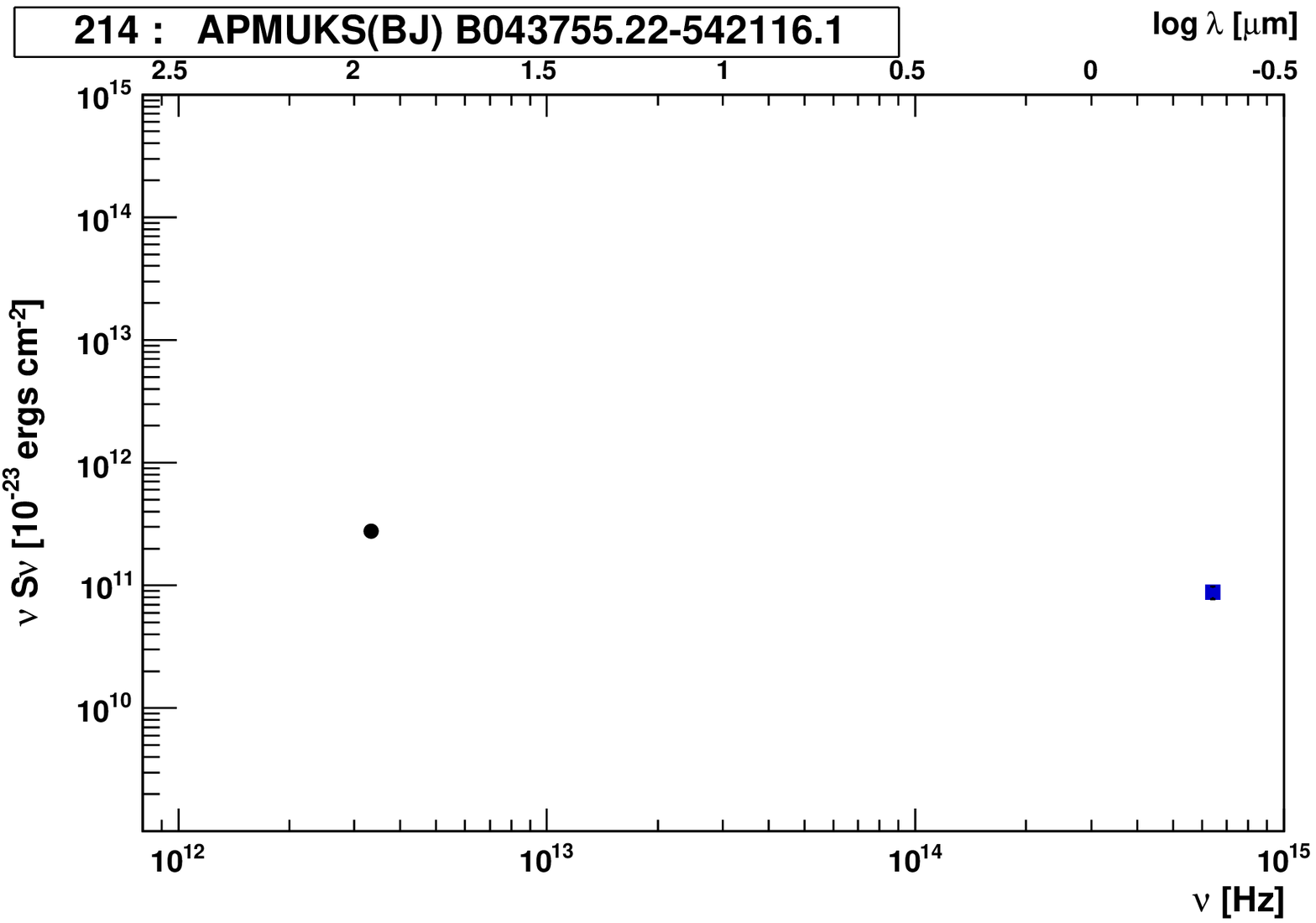}
\includegraphics[width=4cm]{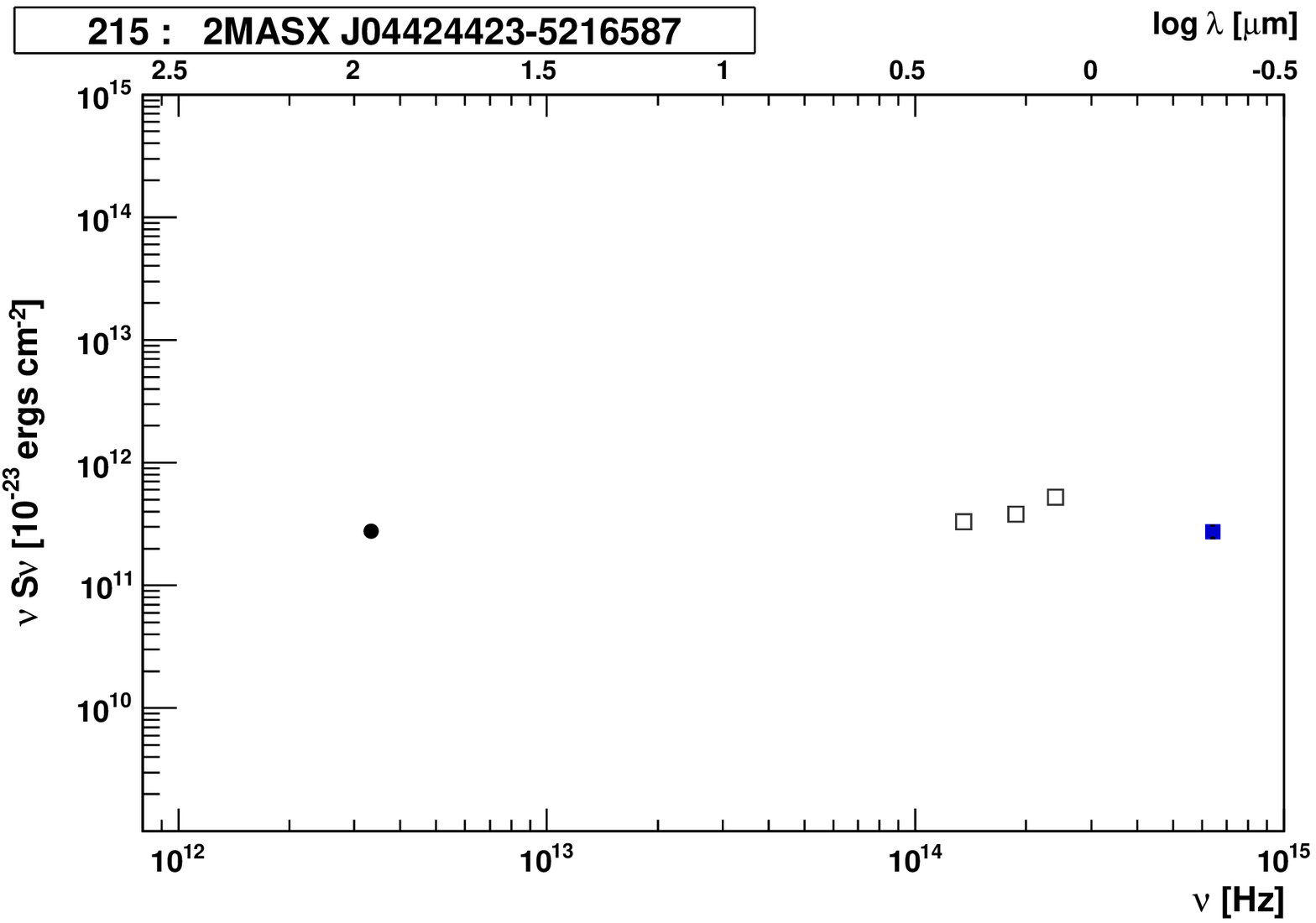}
\includegraphics[width=4cm]{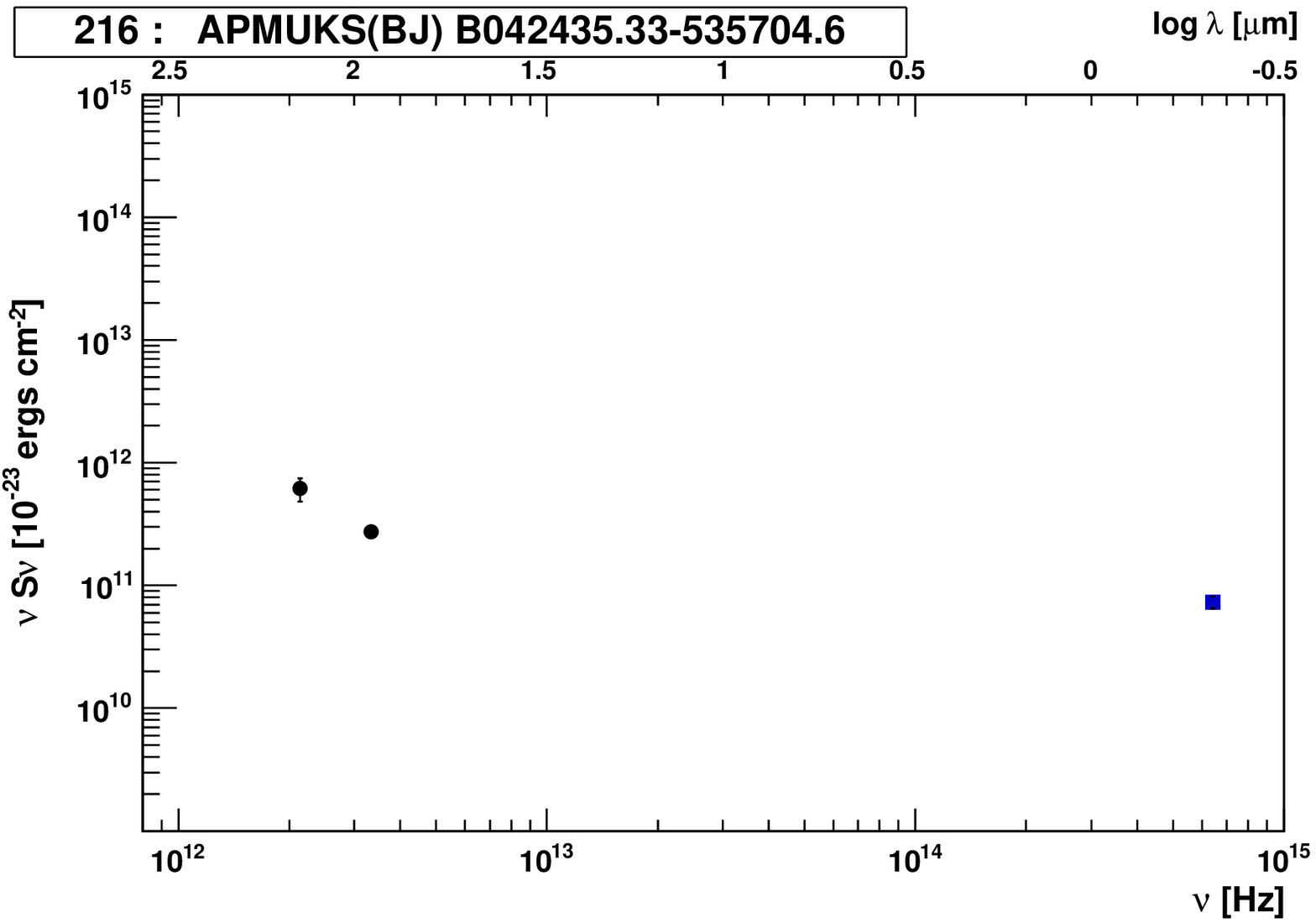}
\includegraphics[width=4cm]{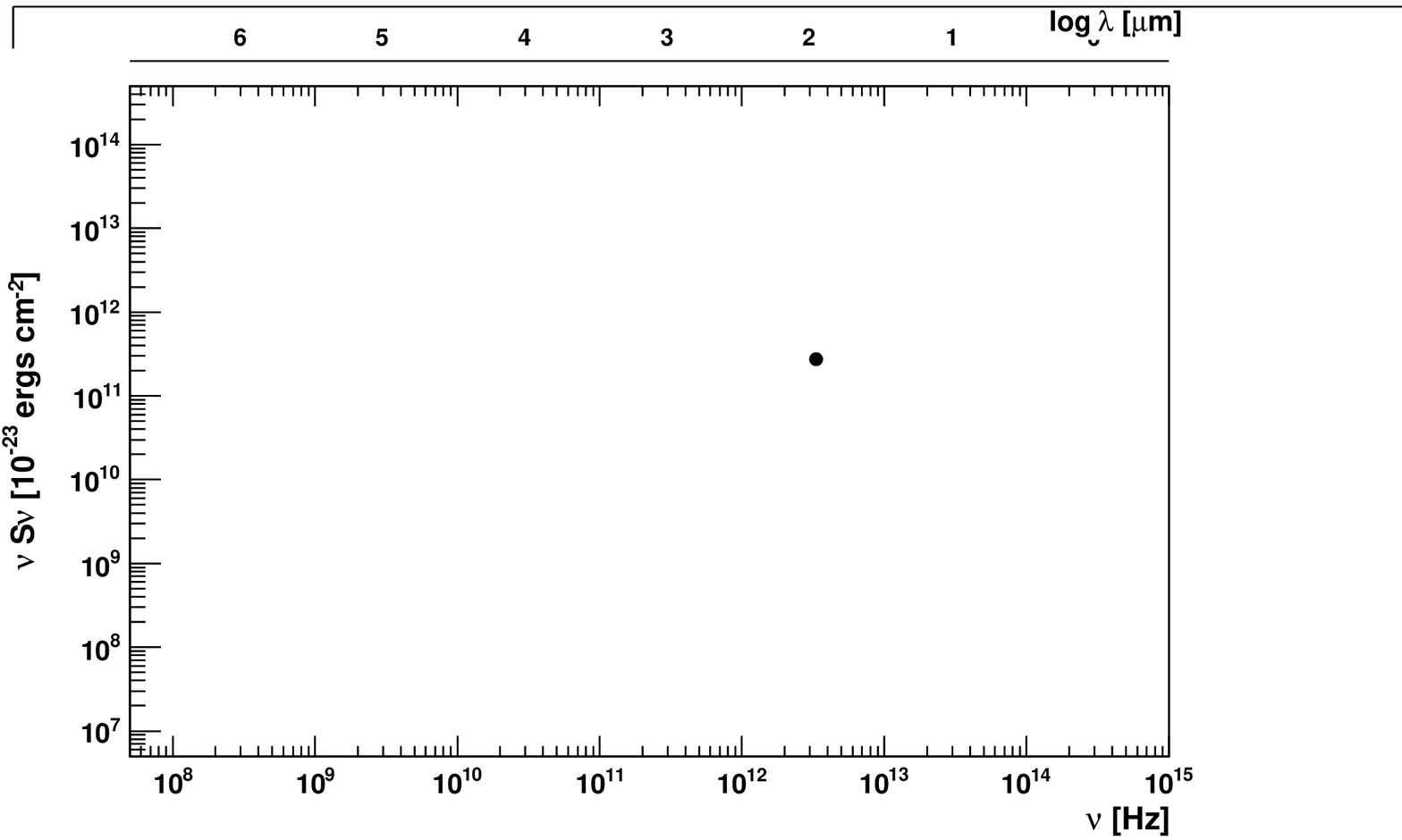}
\includegraphics[width=4cm]{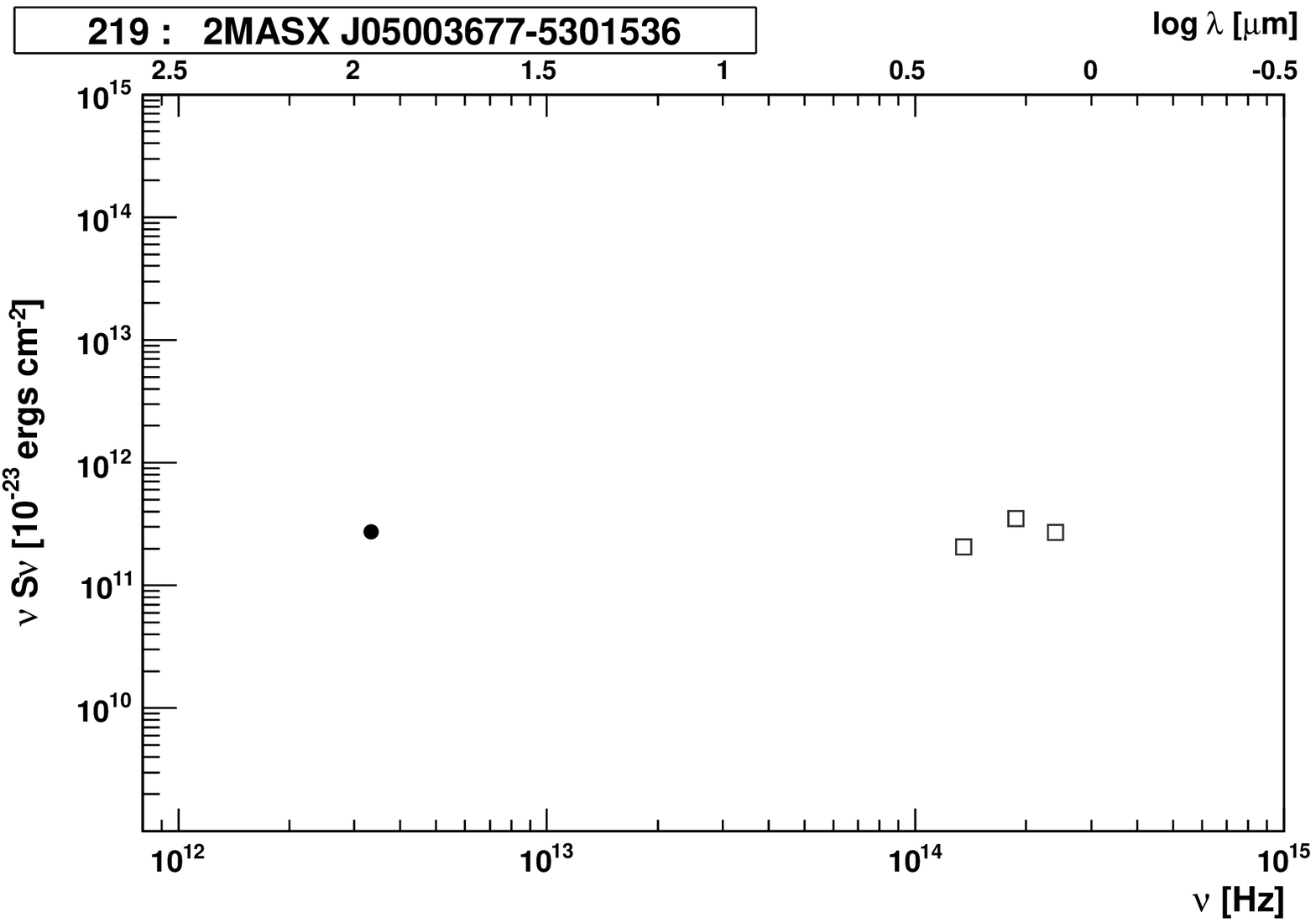}
\label{points5}
\caption {SEDs for the next 36 ADF-S identified sources, with symbols as in Figure~\ref{points1}.}
\end{figure*}
}

\clearpage

\onlfig{6}{
\begin{figure*}[t]
\centering
\includegraphics[width=4cm]{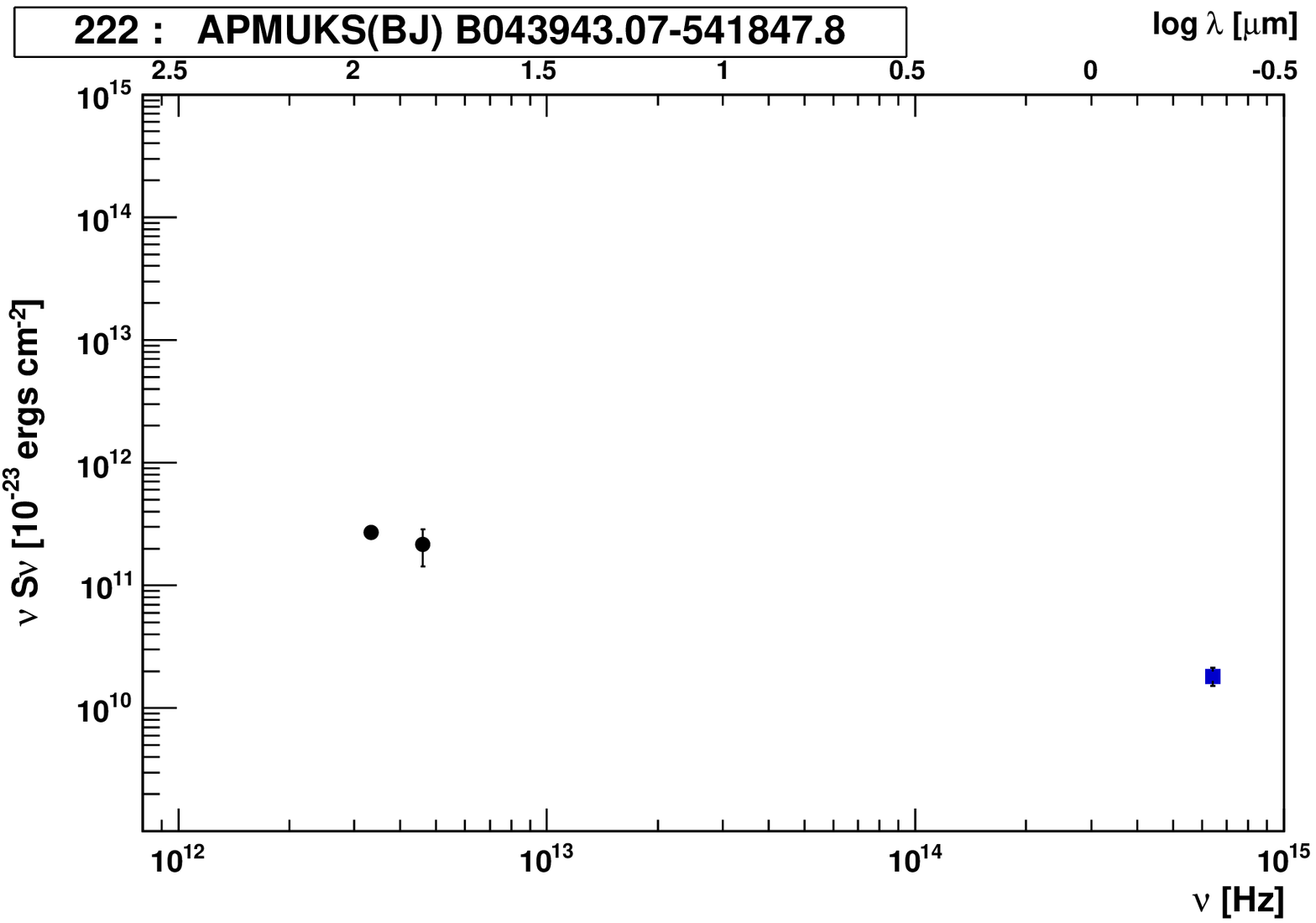}
\includegraphics[width=4cm]{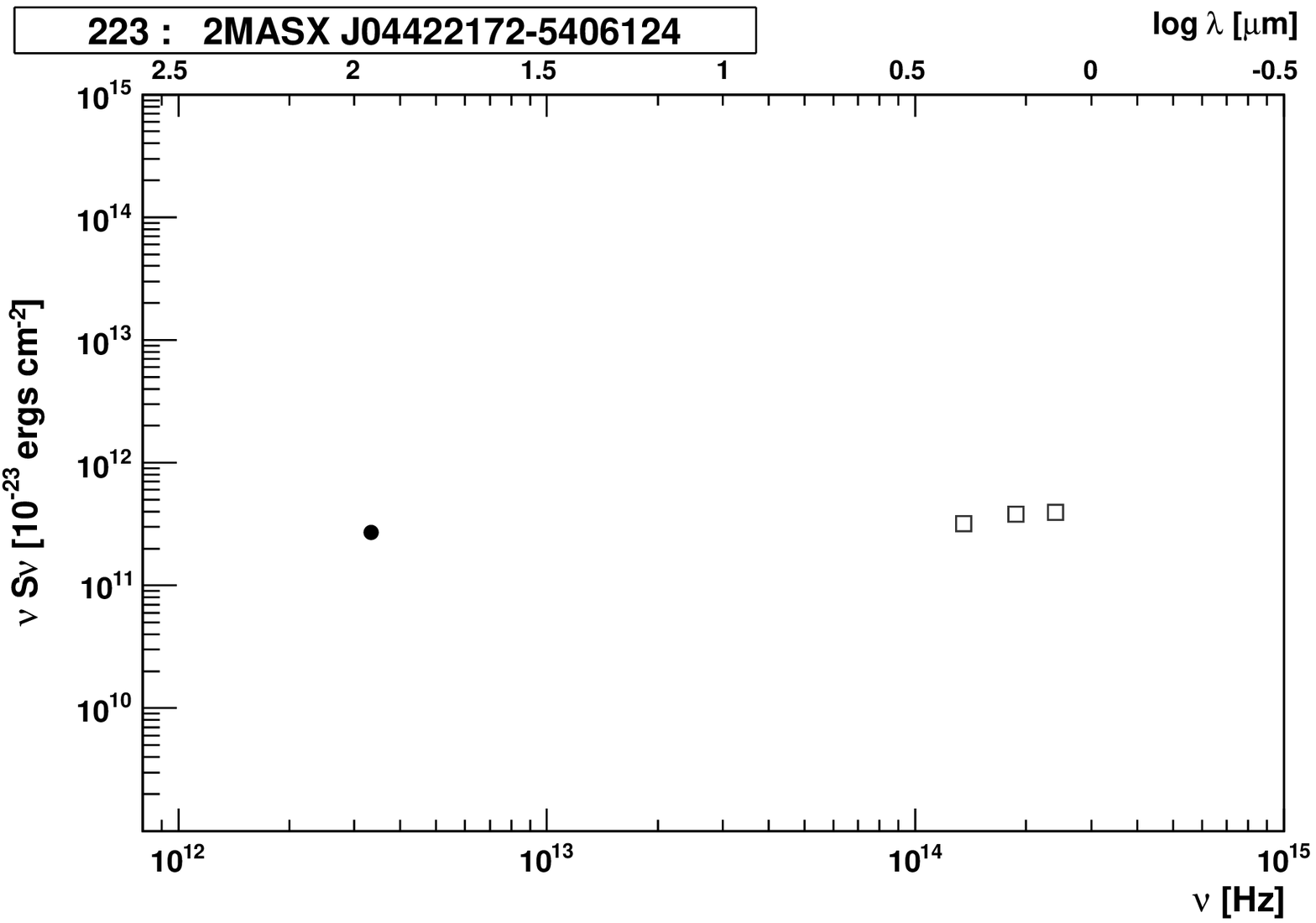}
\includegraphics[width=4cm]{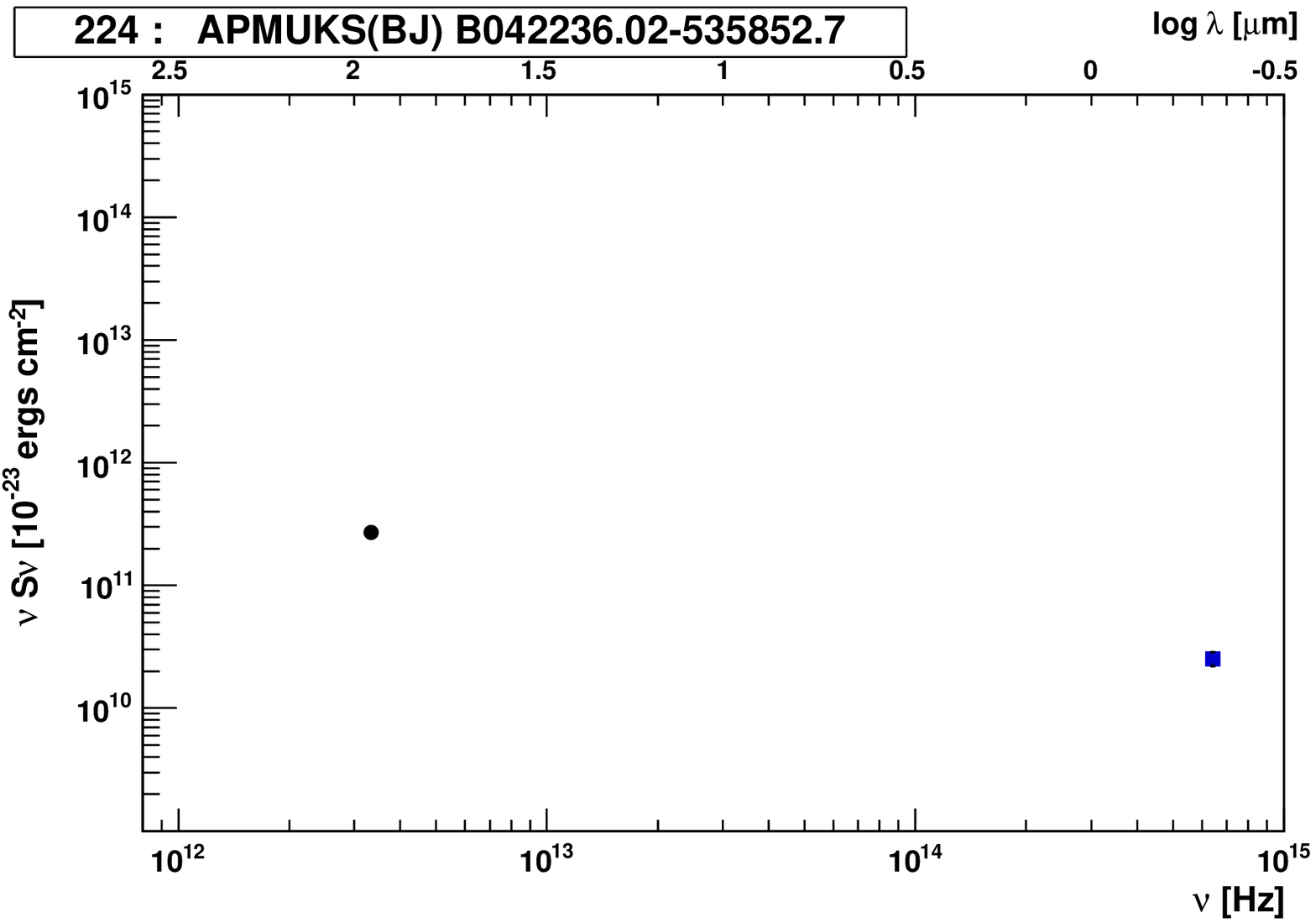}
\includegraphics[width=4cm]{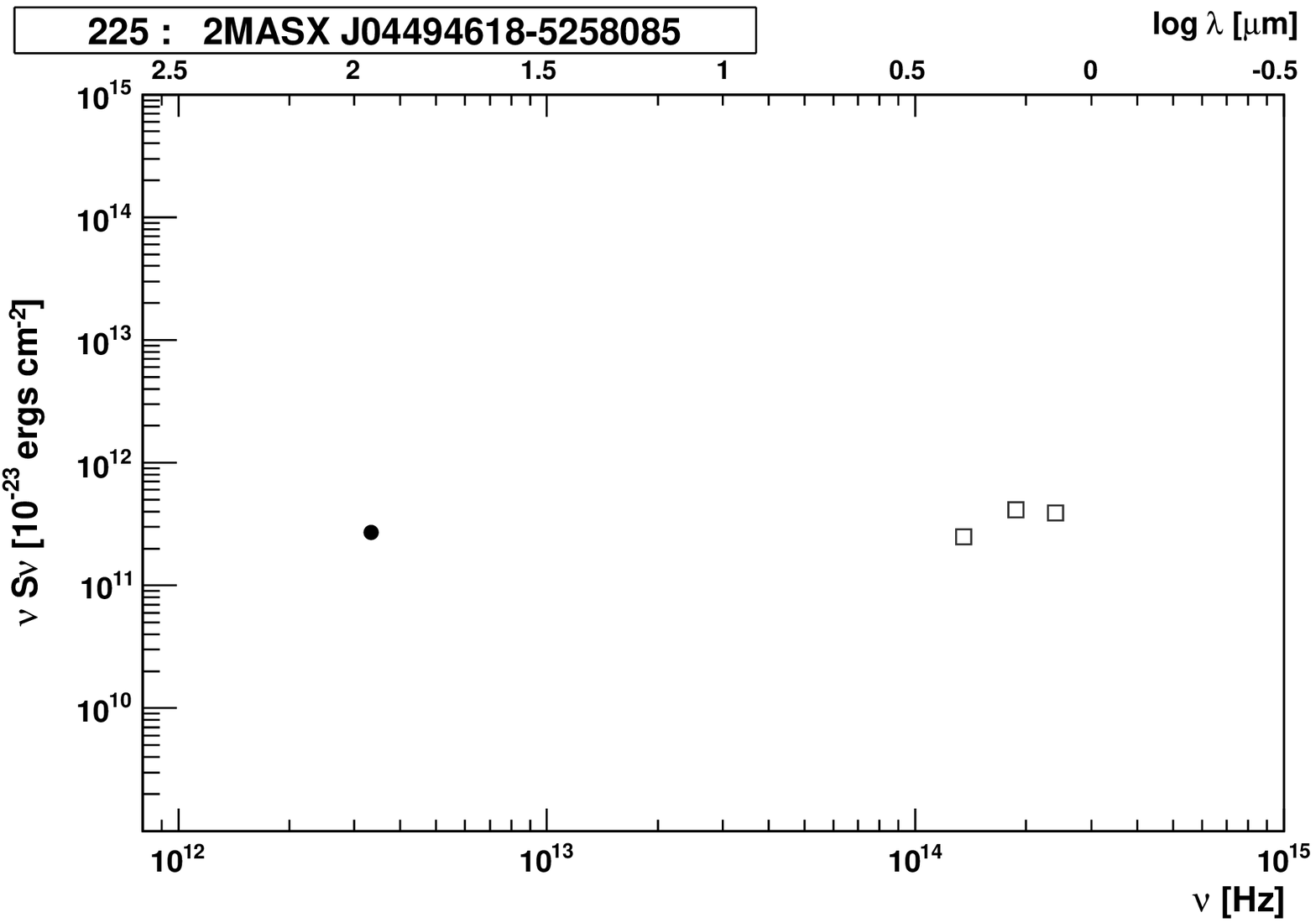}
\includegraphics[width=4cm]{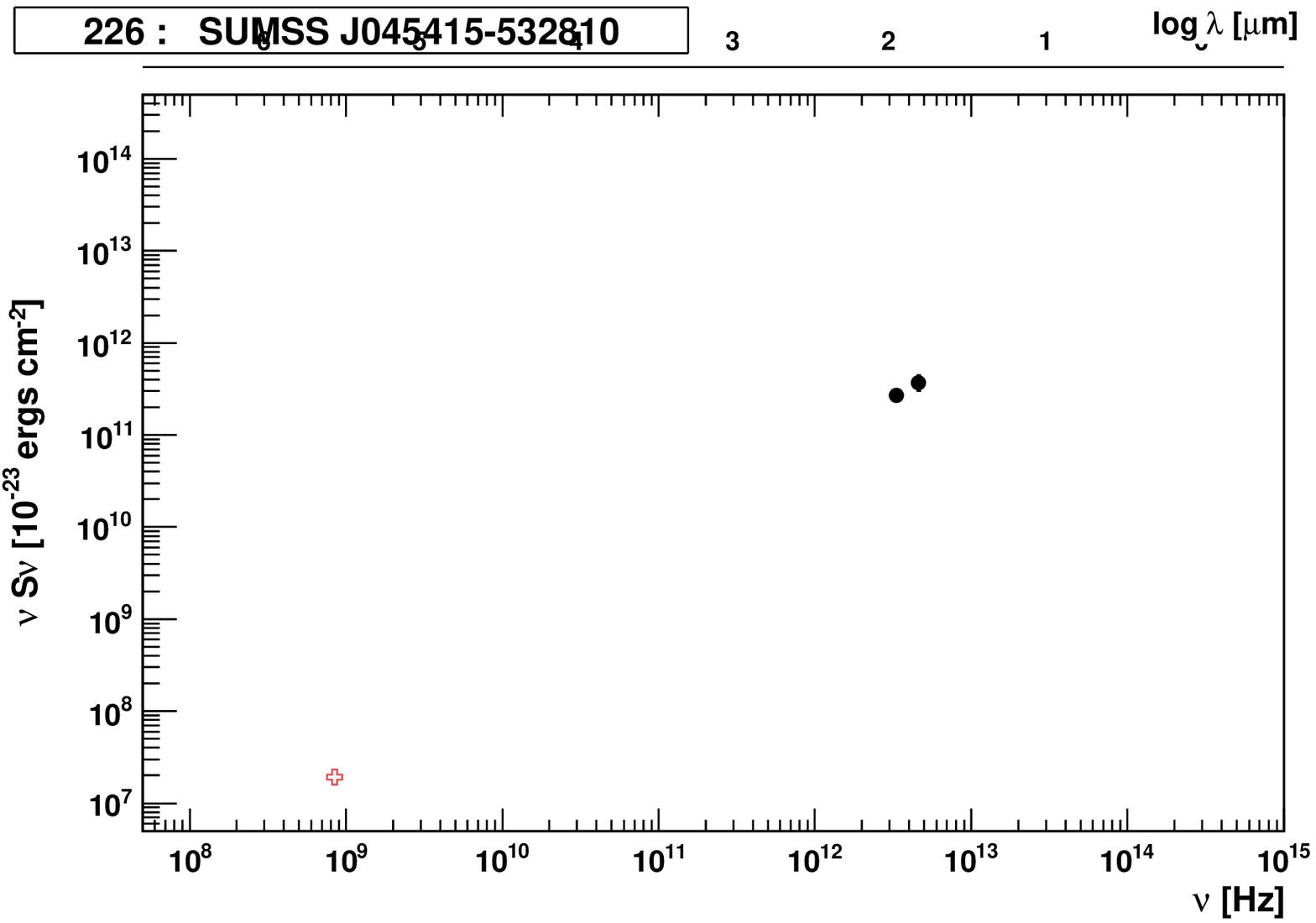}
\includegraphics[width=4cm]{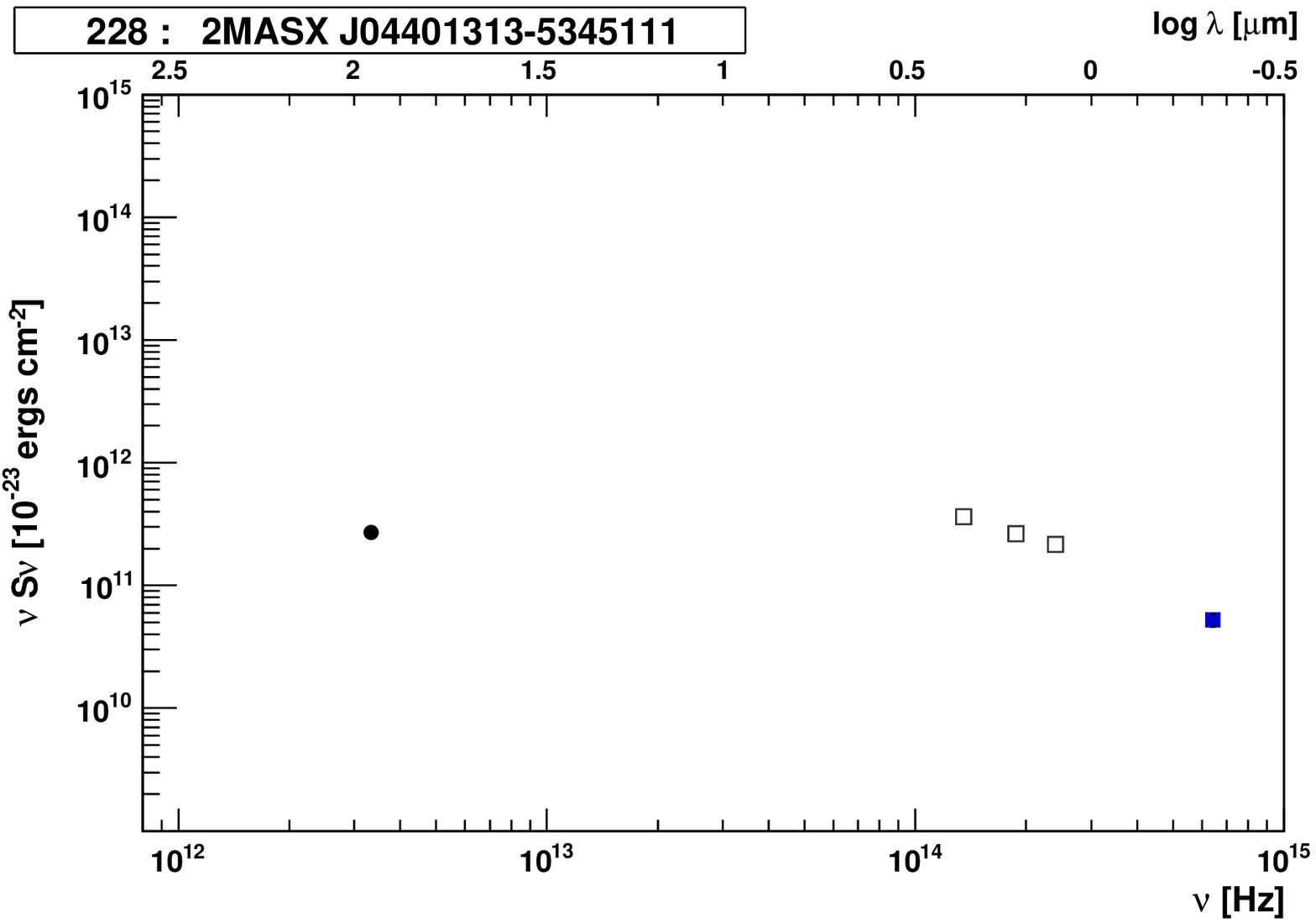}
\includegraphics[width=4cm]{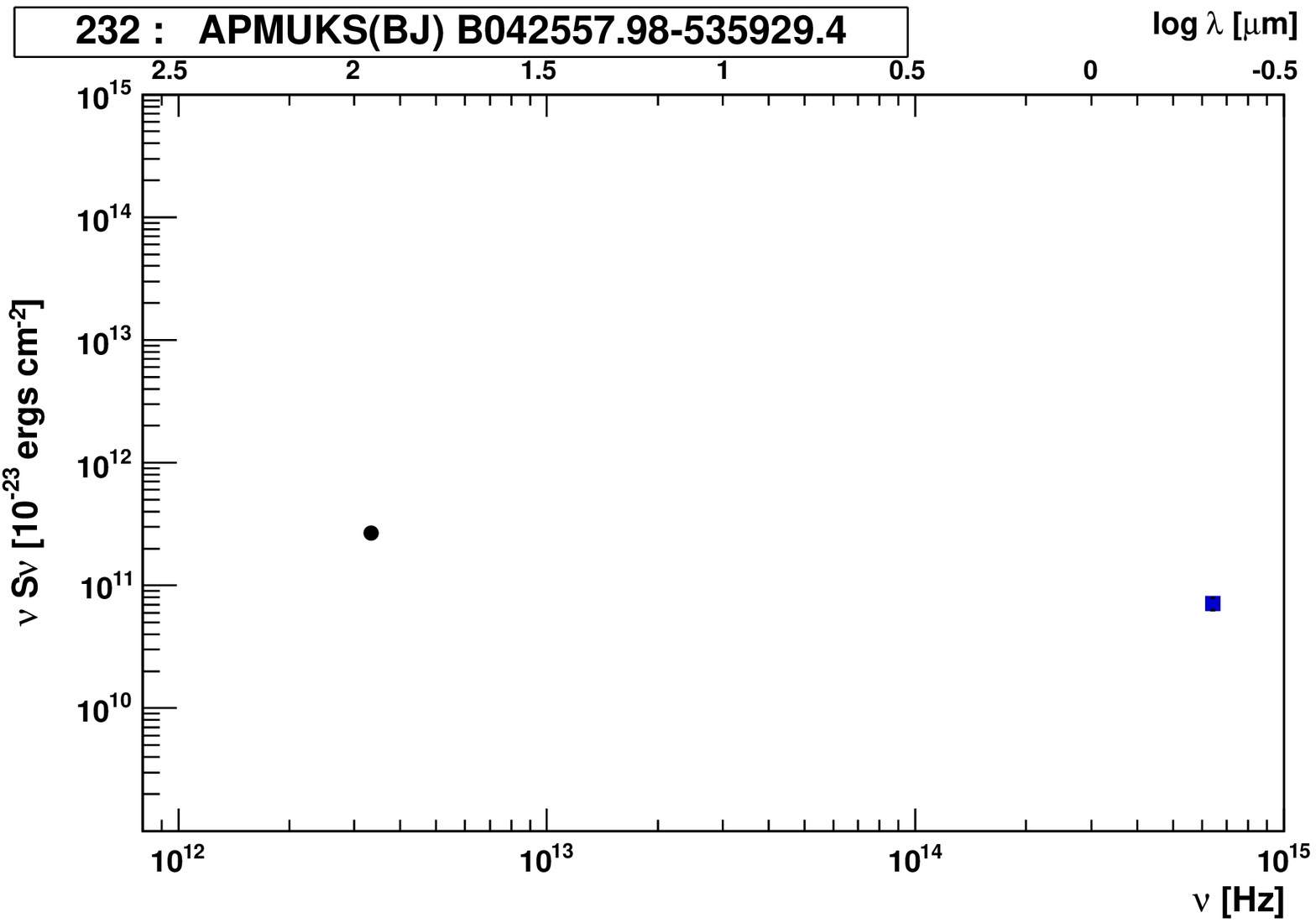}
\includegraphics[width=4cm]{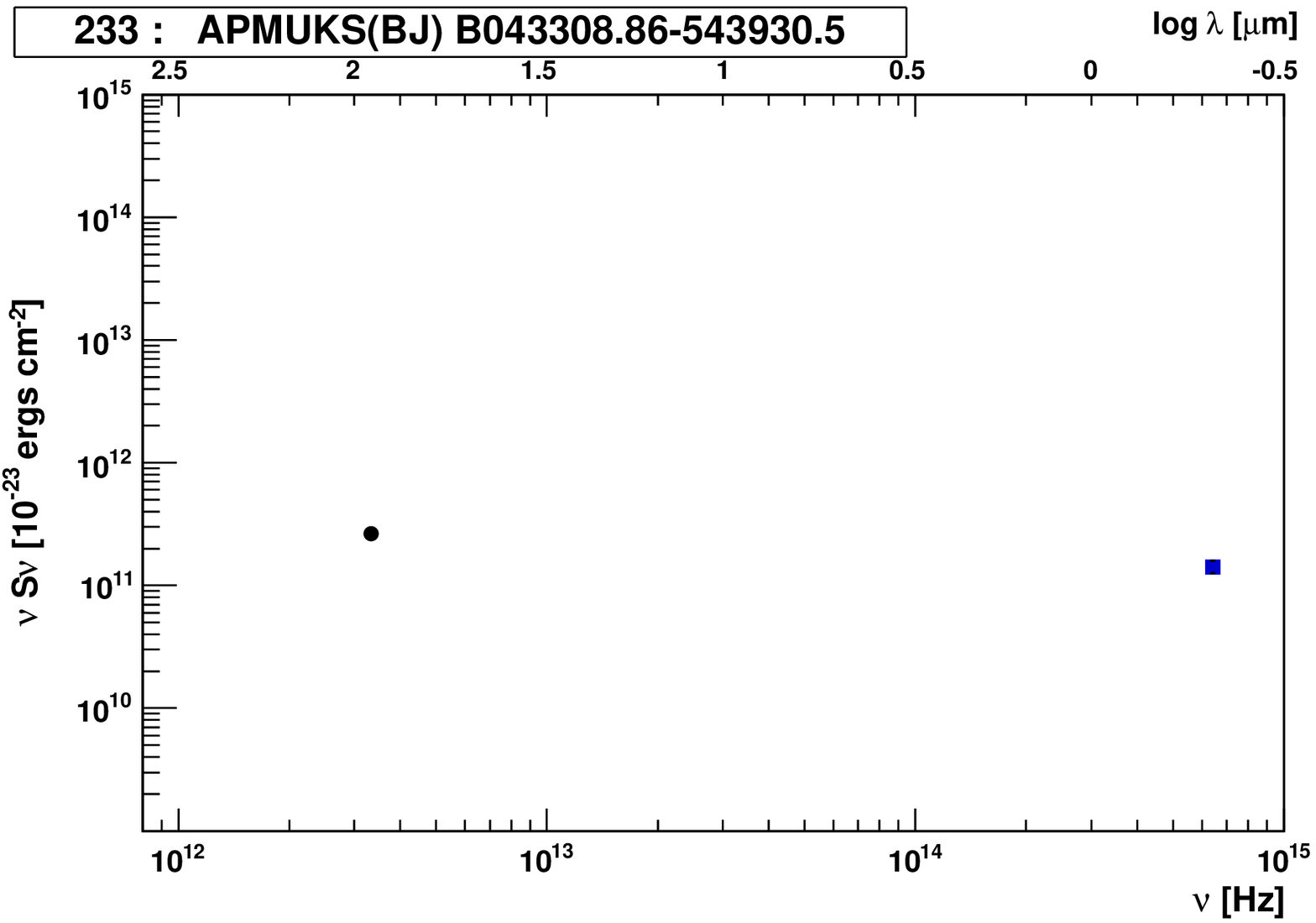}
\includegraphics[width=4cm]{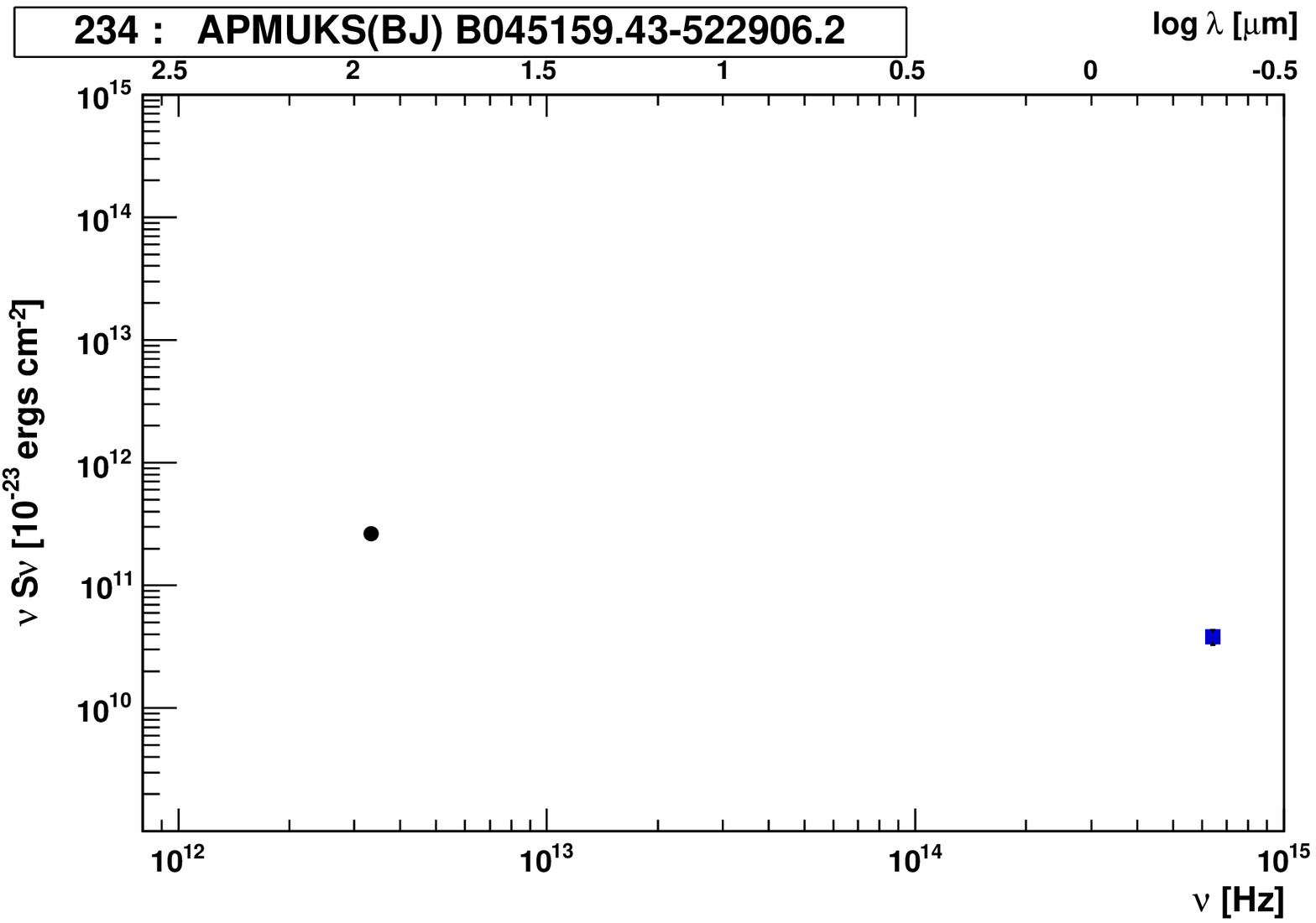}
\includegraphics[width=4cm]{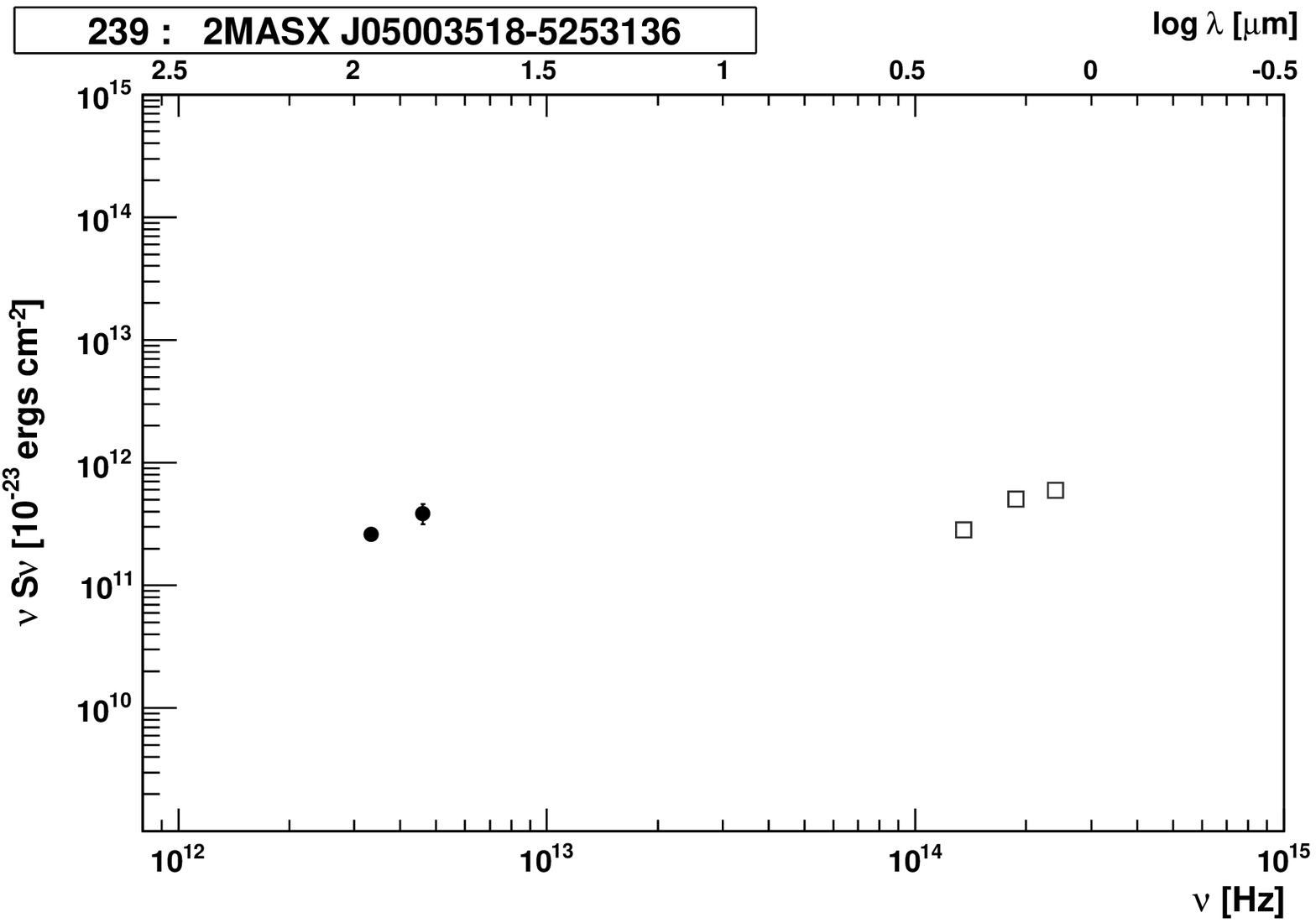}
\includegraphics[width=4cm]{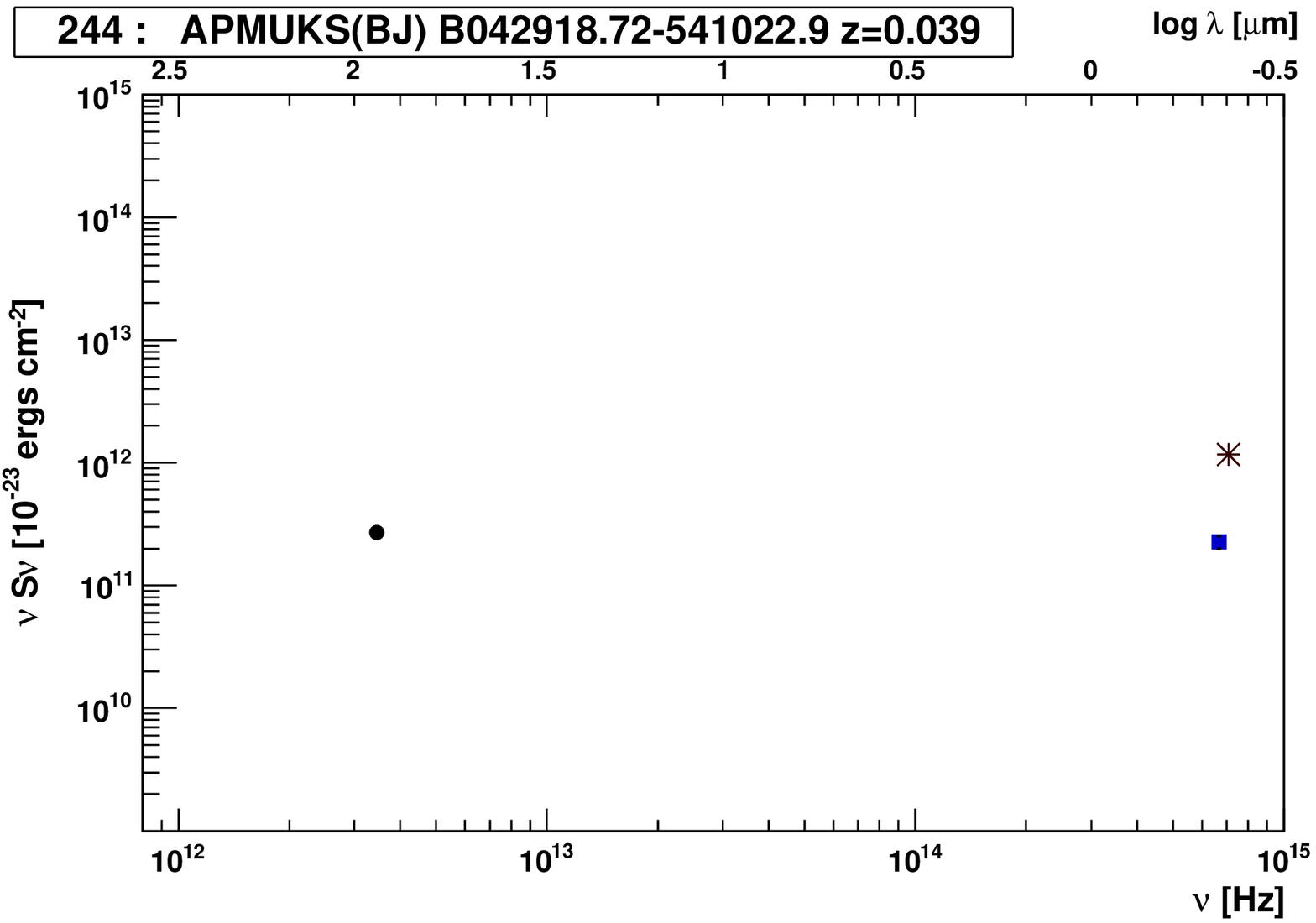}
\includegraphics[width=4cm]{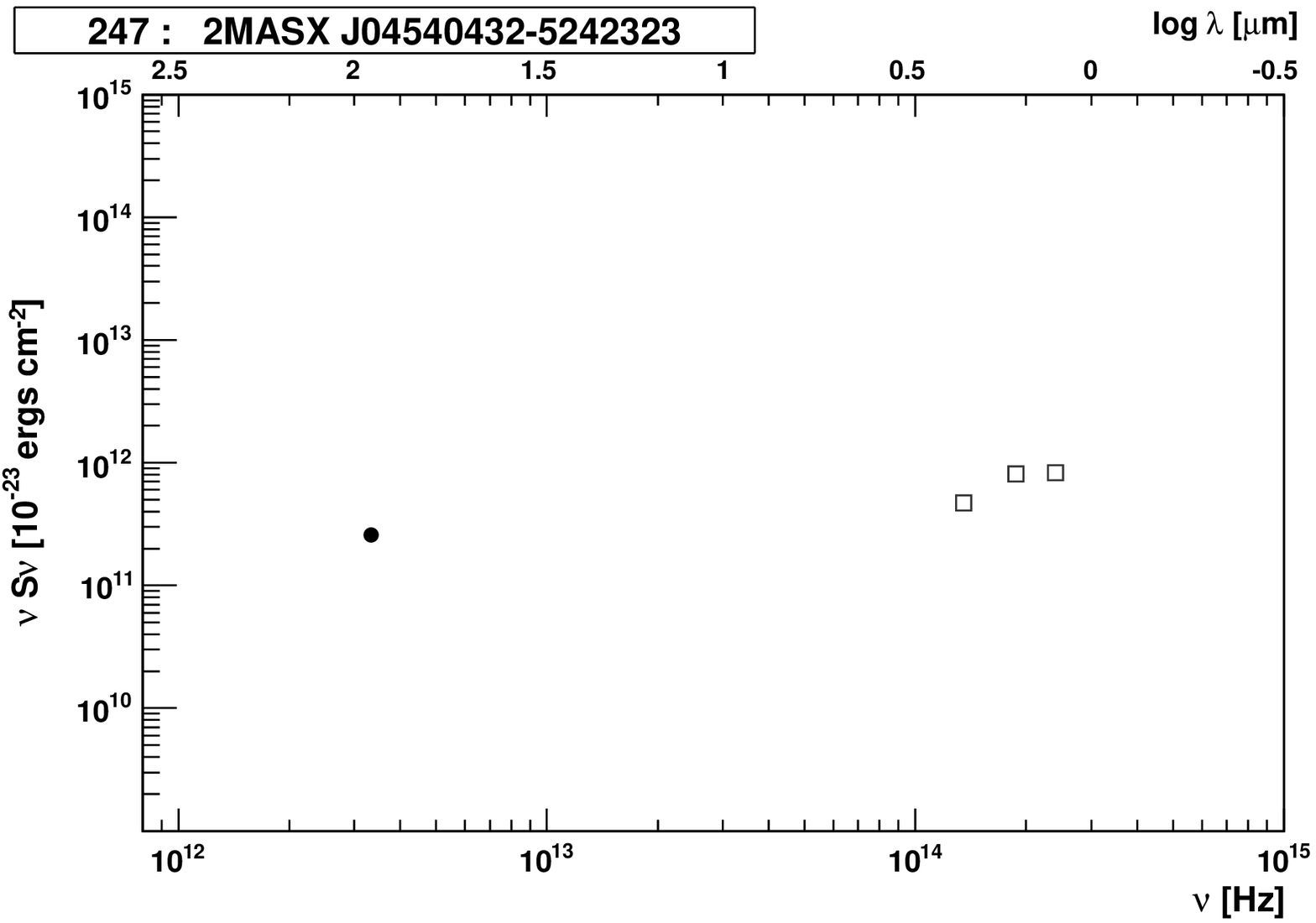}
\includegraphics[width=4cm]{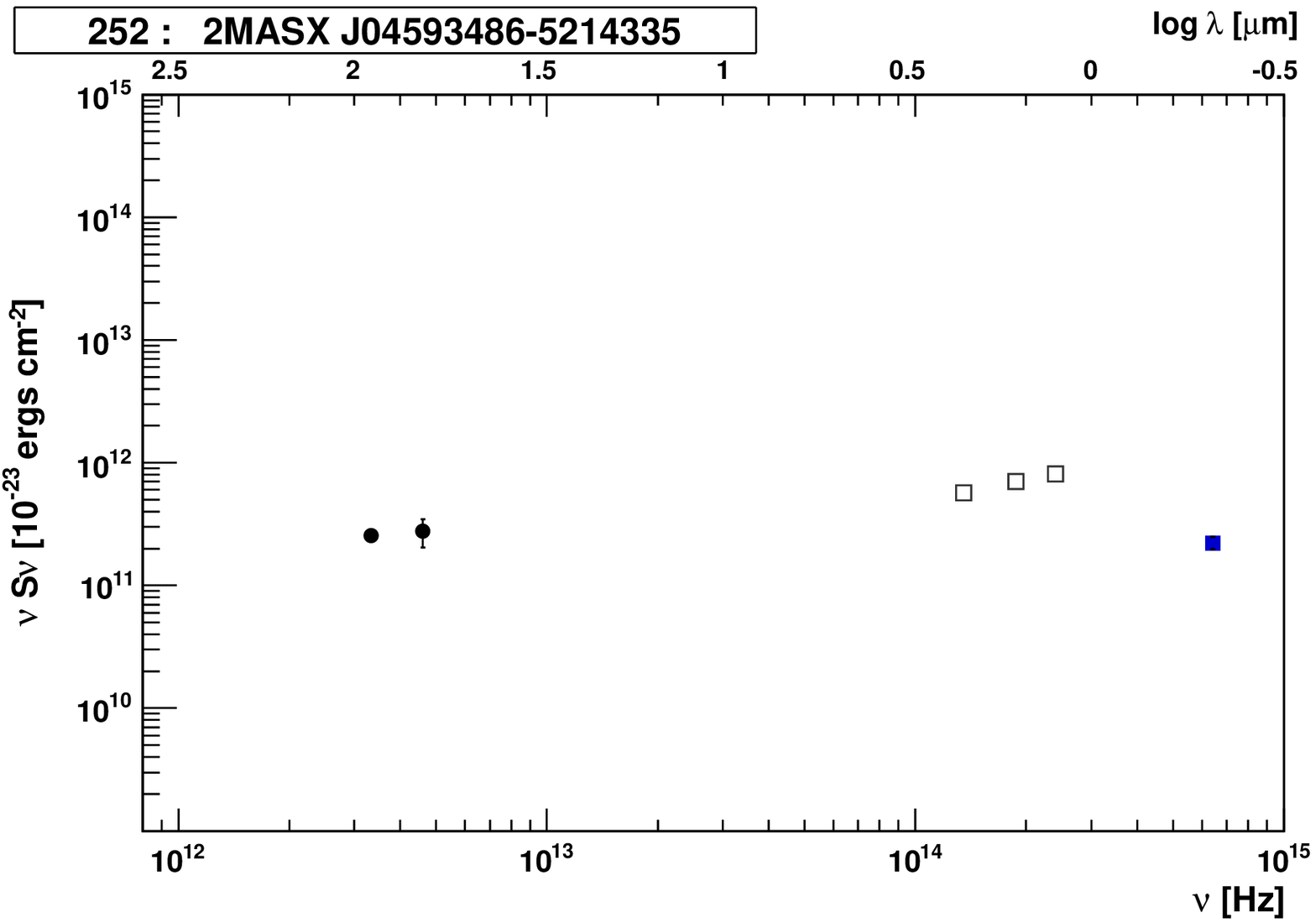}
\includegraphics[width=4cm]{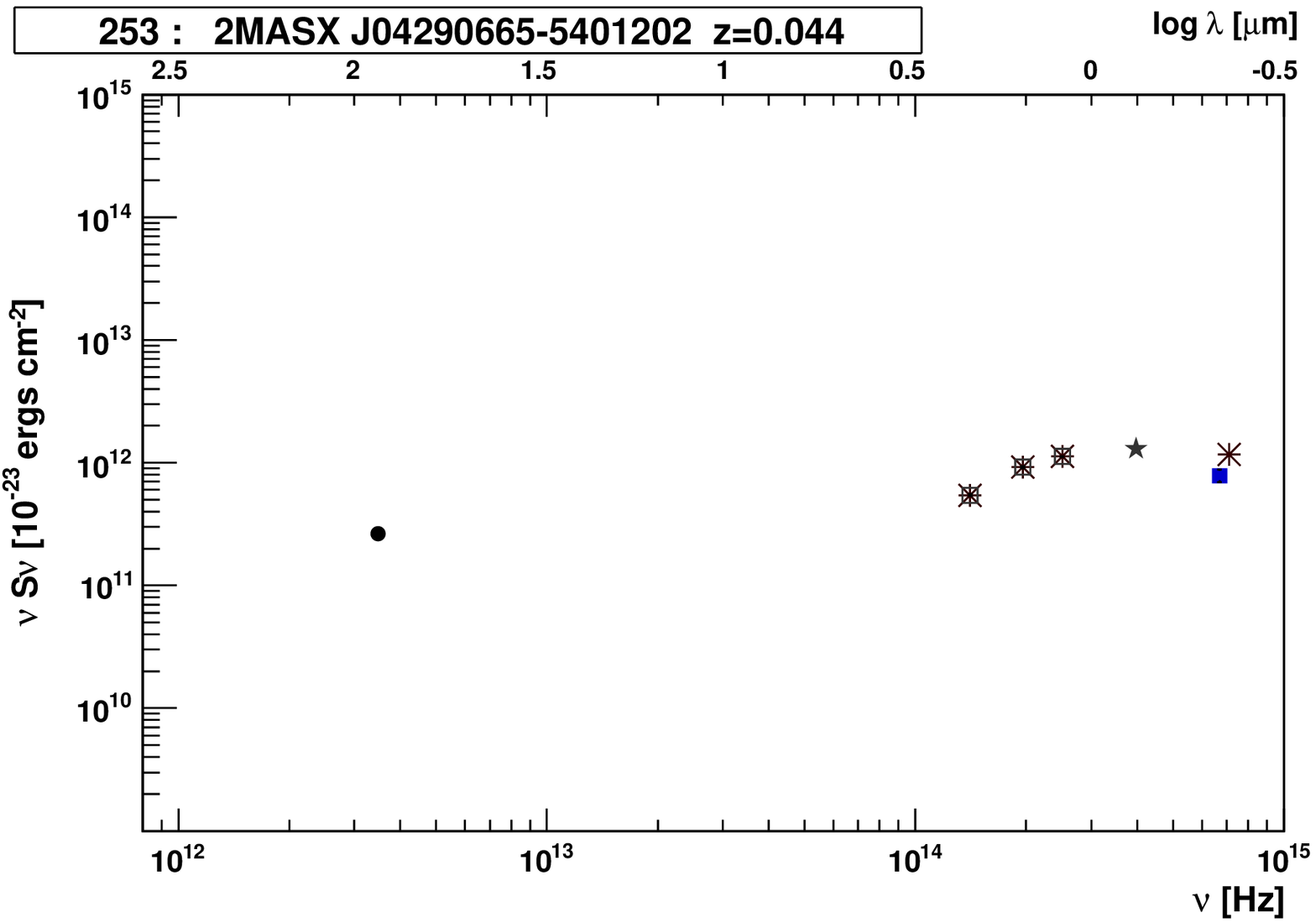}
\includegraphics[width=4cm]{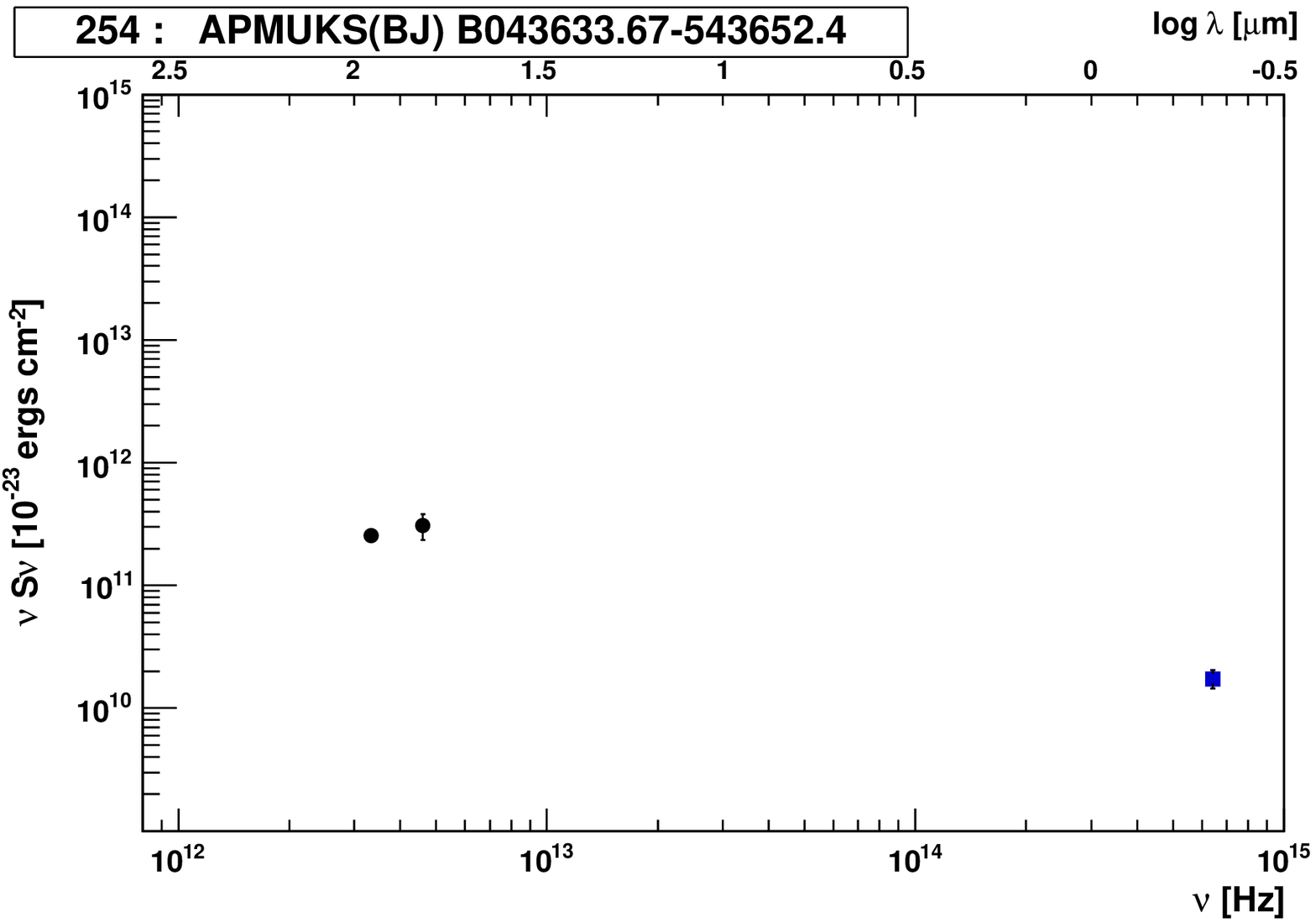}
\includegraphics[width=4cm]{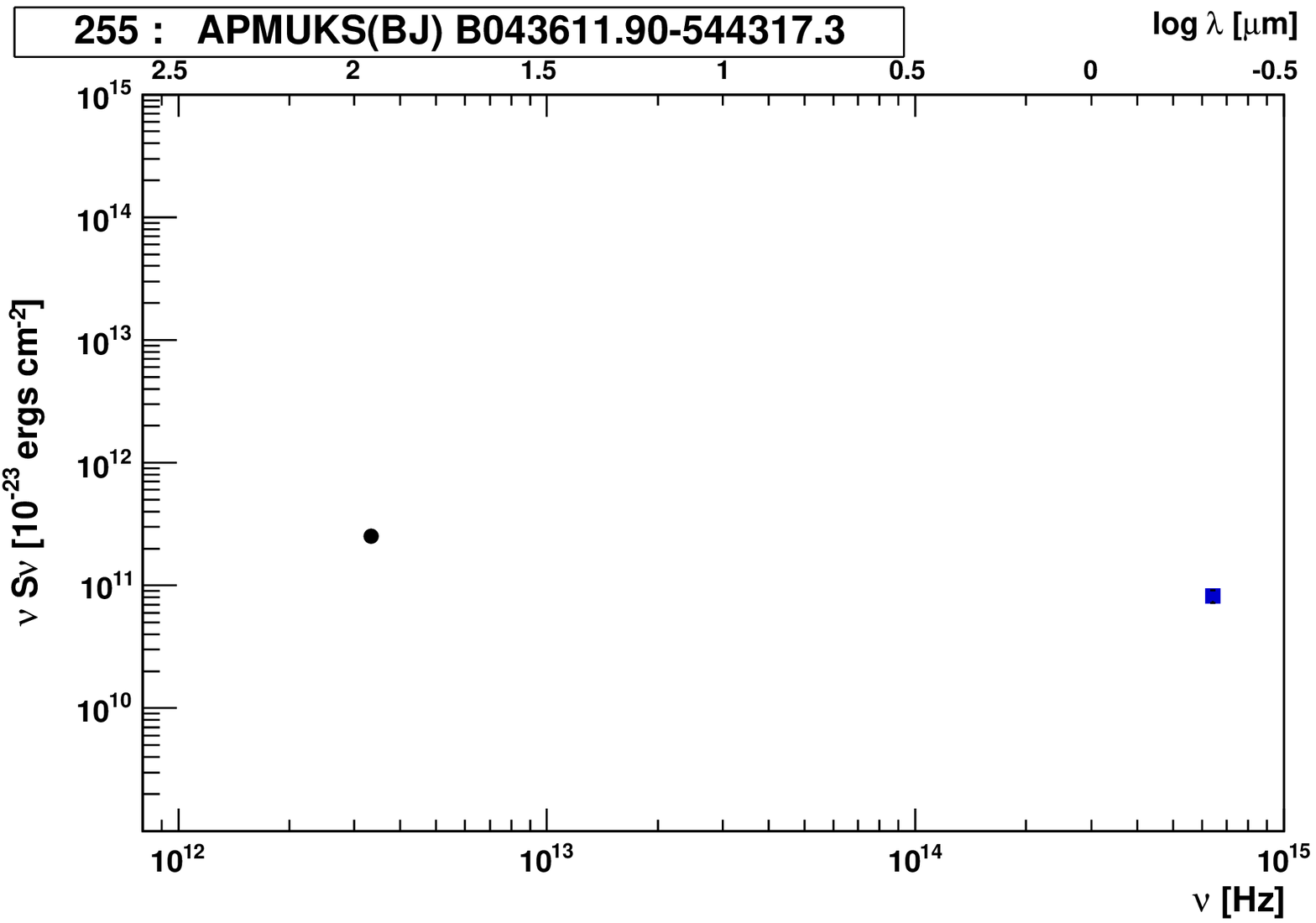}
\includegraphics[width=4cm]{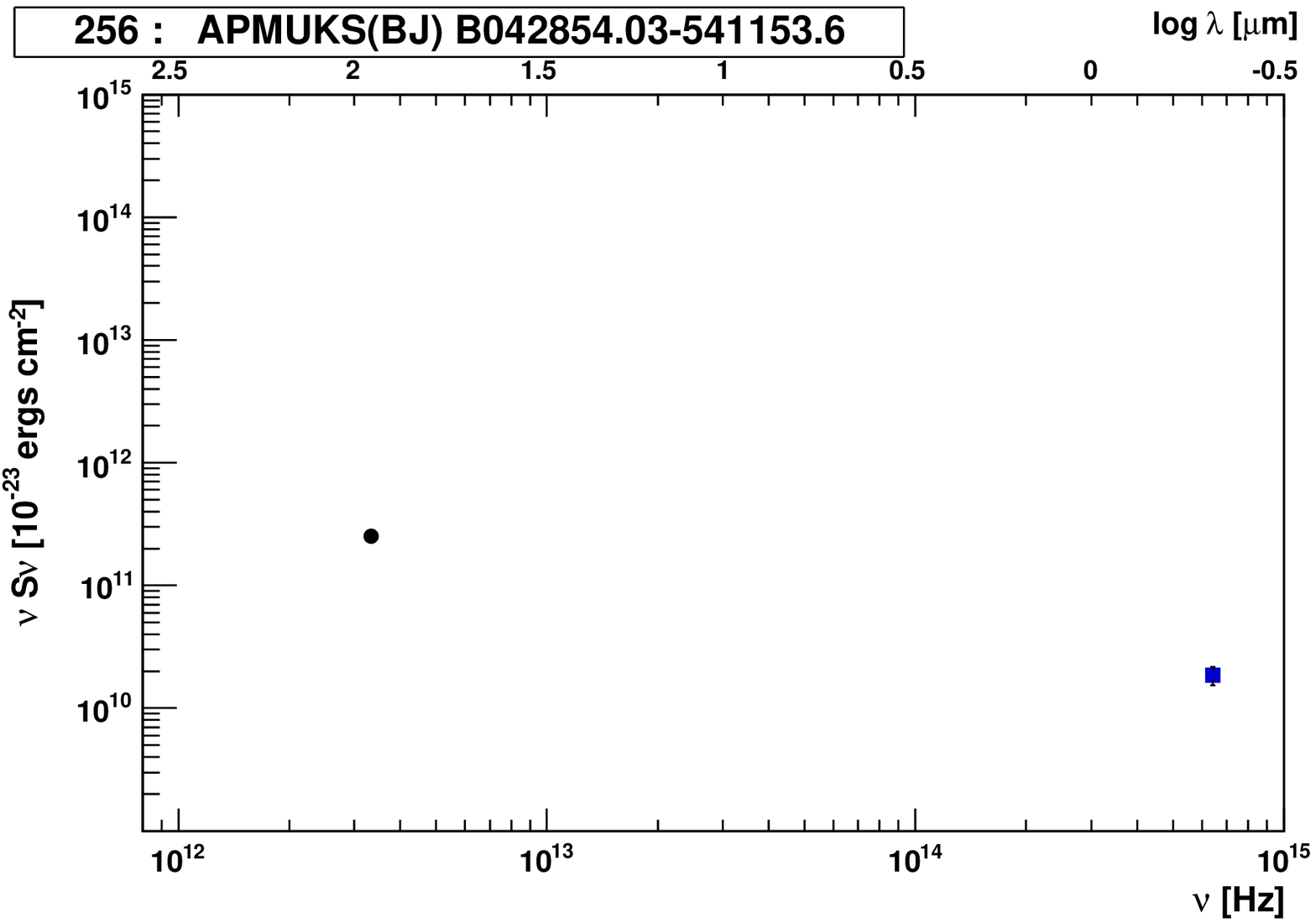}
\includegraphics[width=4cm]{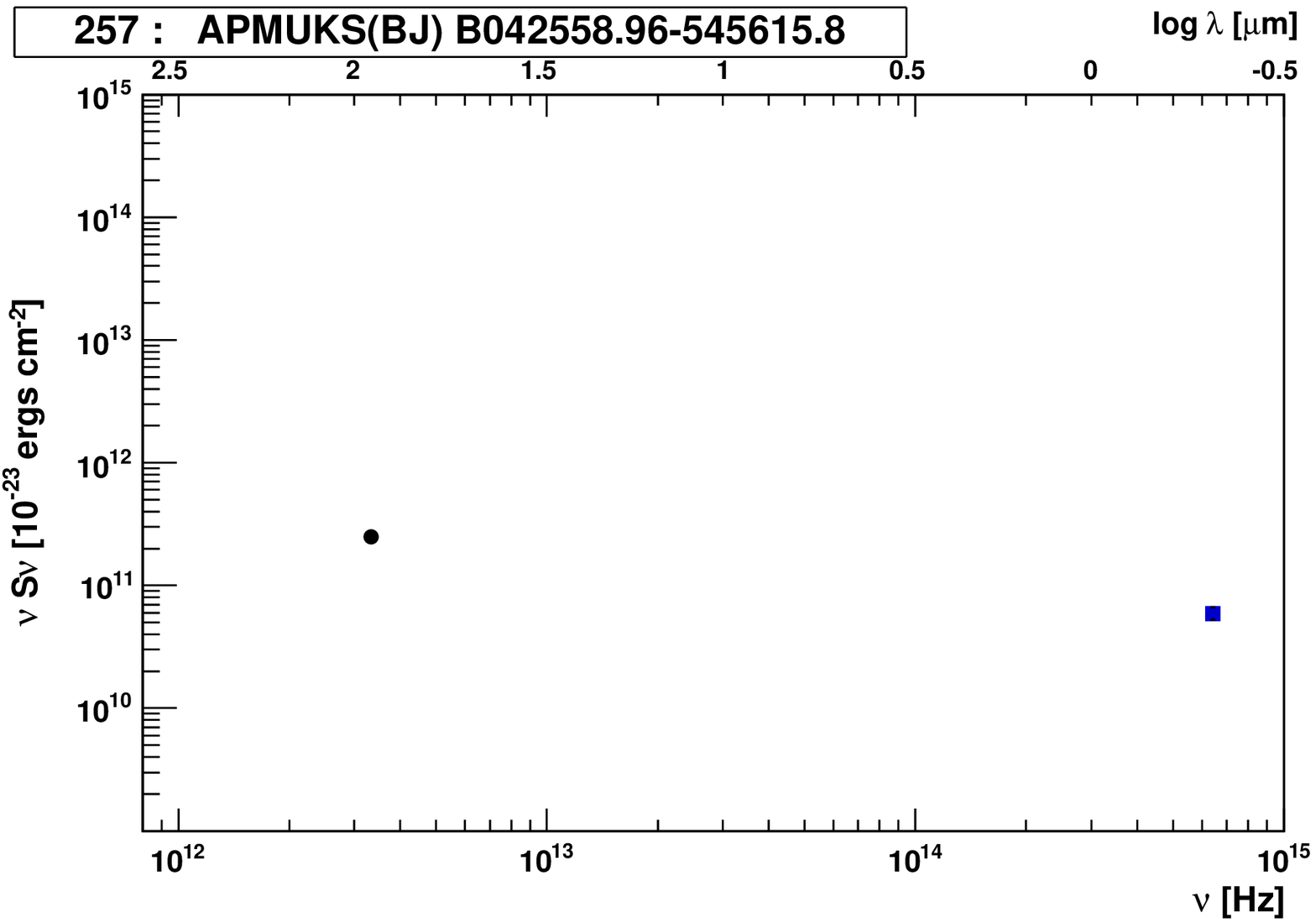}
\includegraphics[width=4cm]{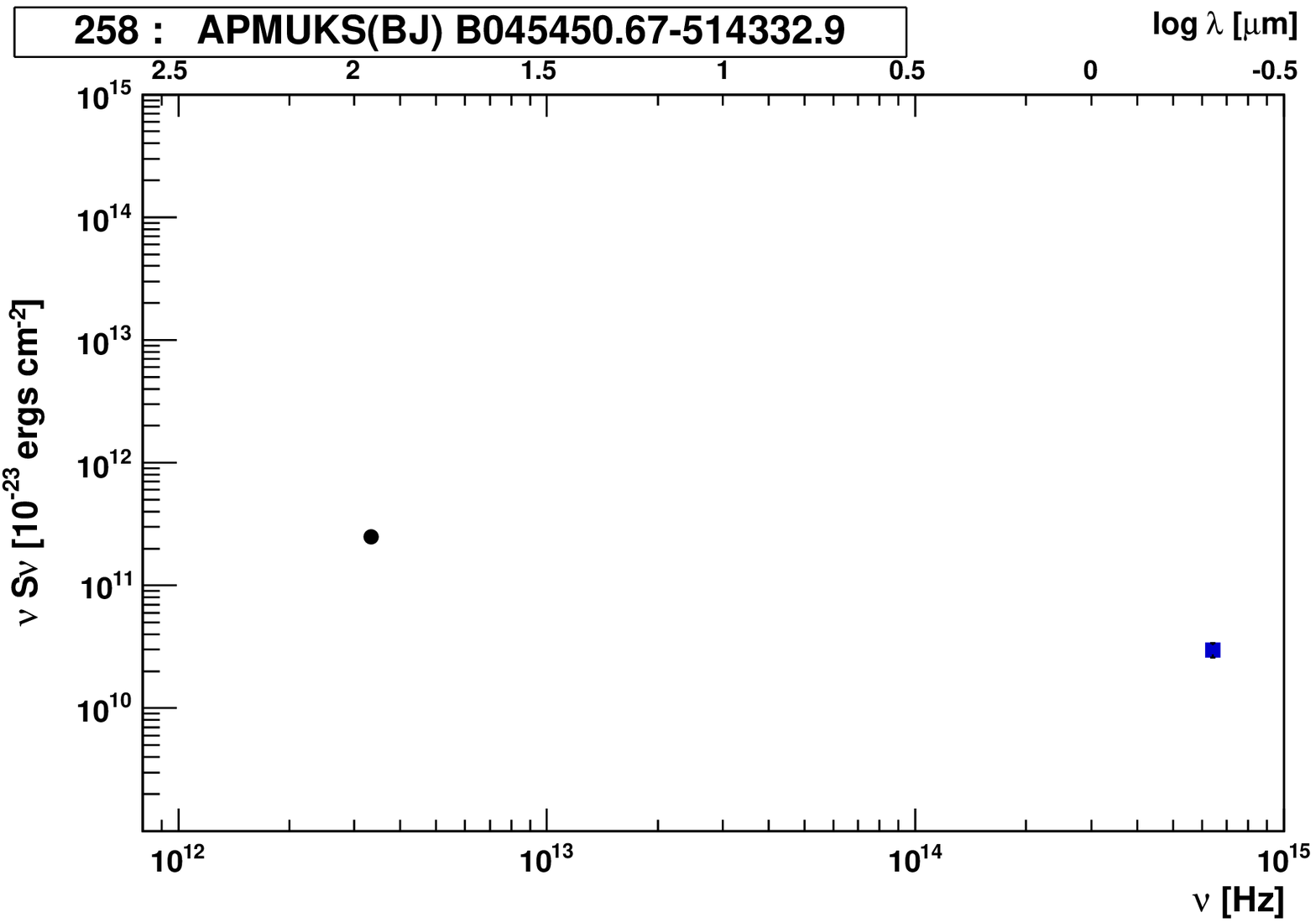}
\includegraphics[width=4cm]{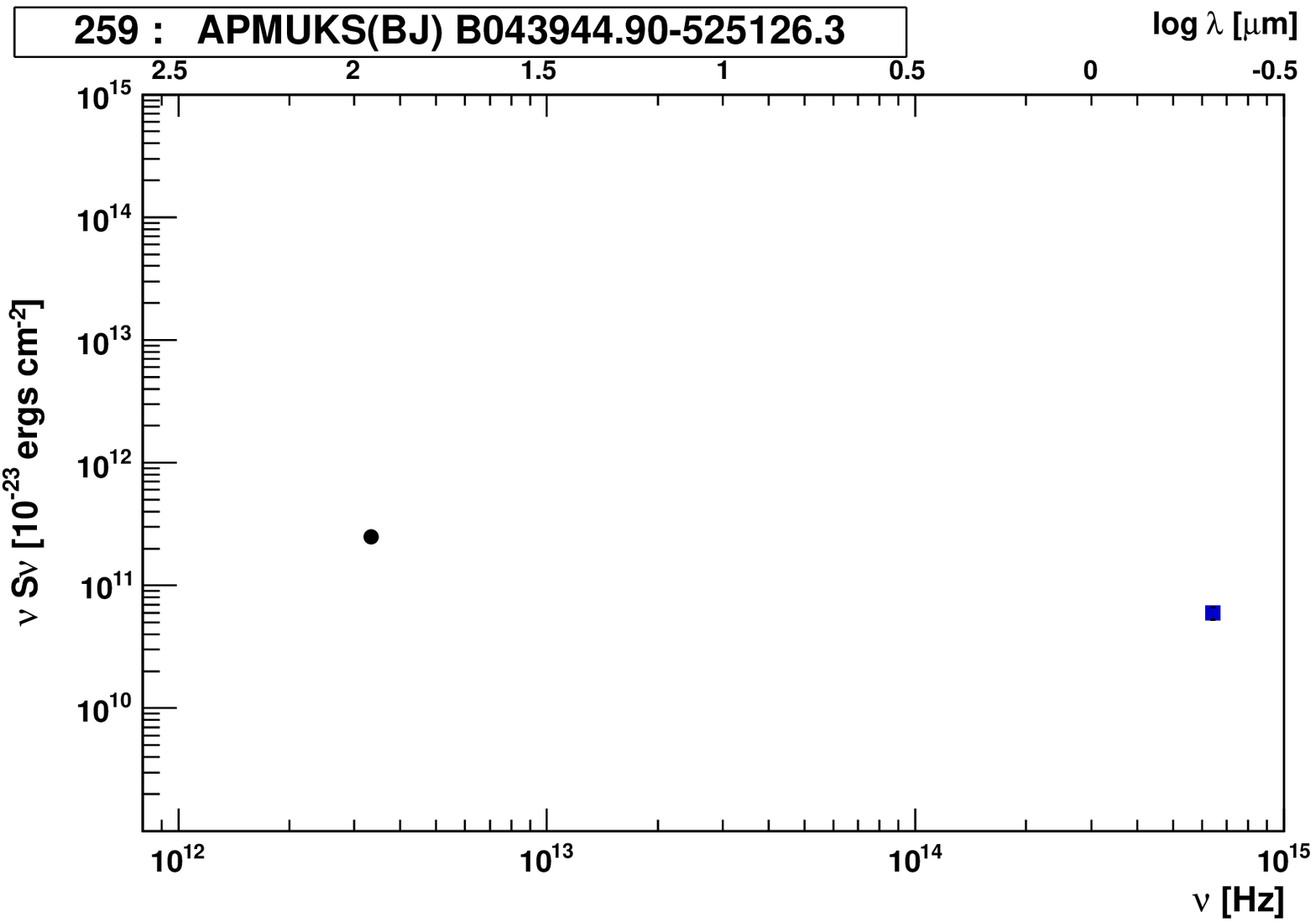}
\includegraphics[width=4cm]{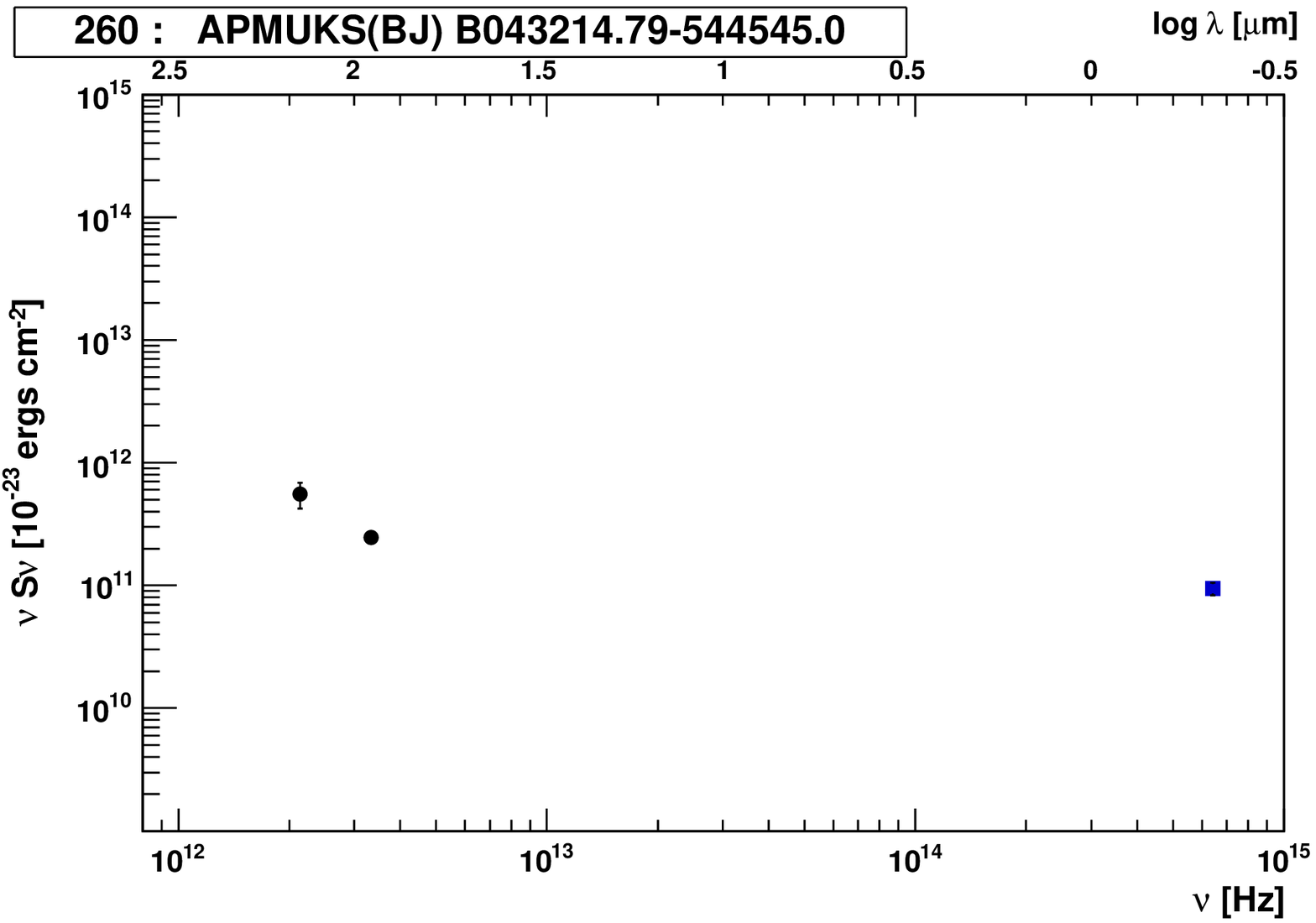}
\includegraphics[width=4cm]{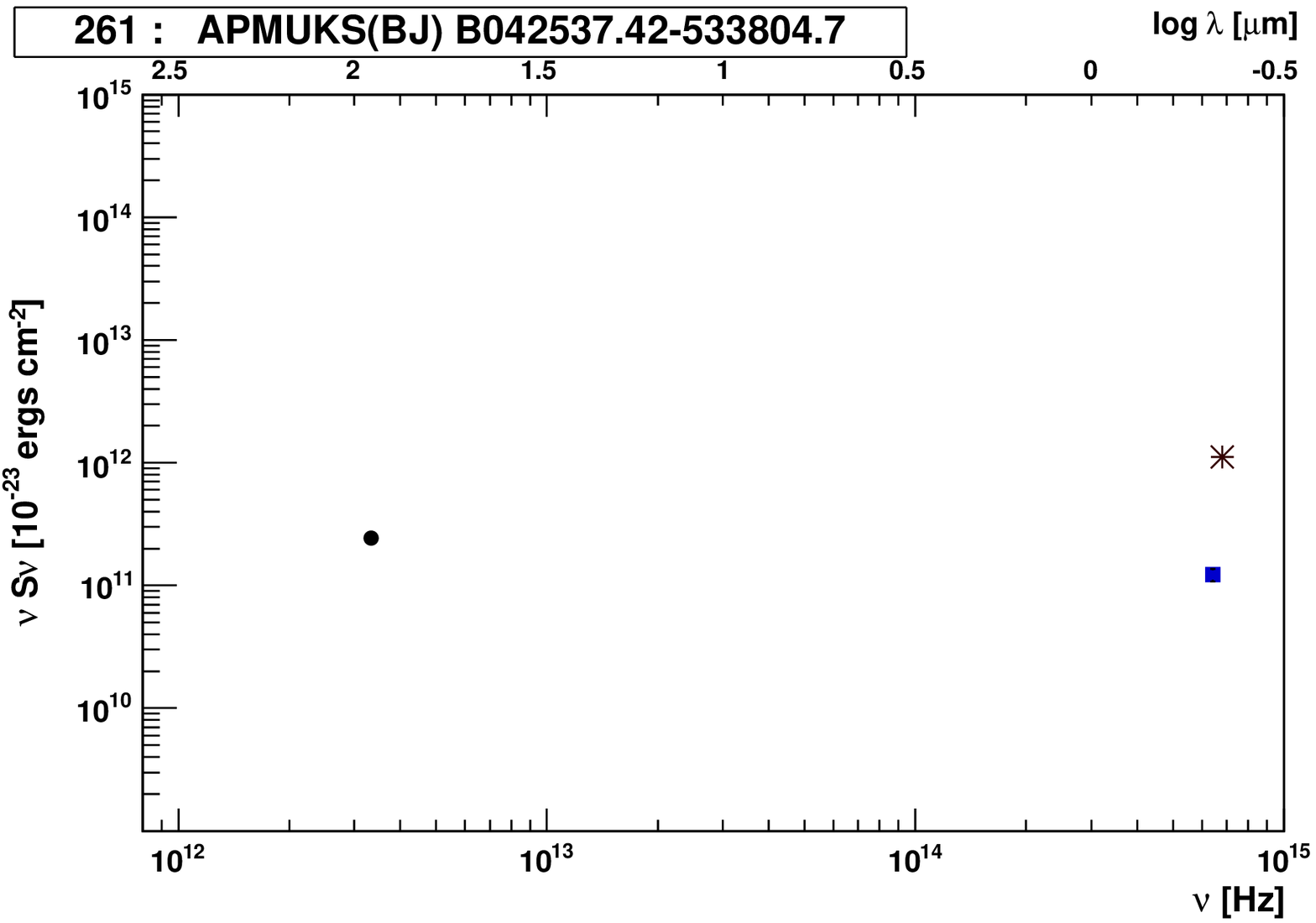}
\includegraphics[width=4cm]{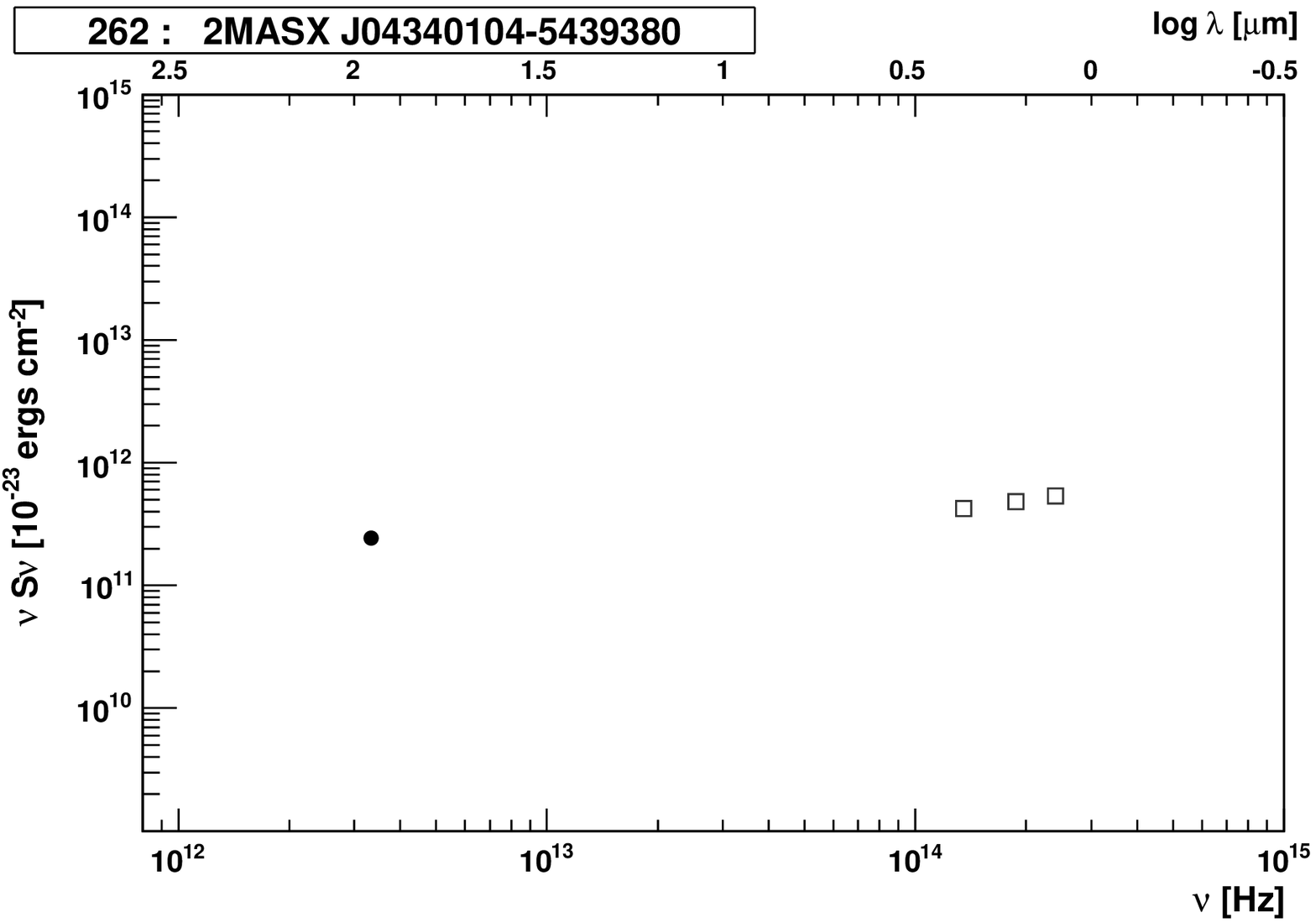}
\includegraphics[width=4cm]{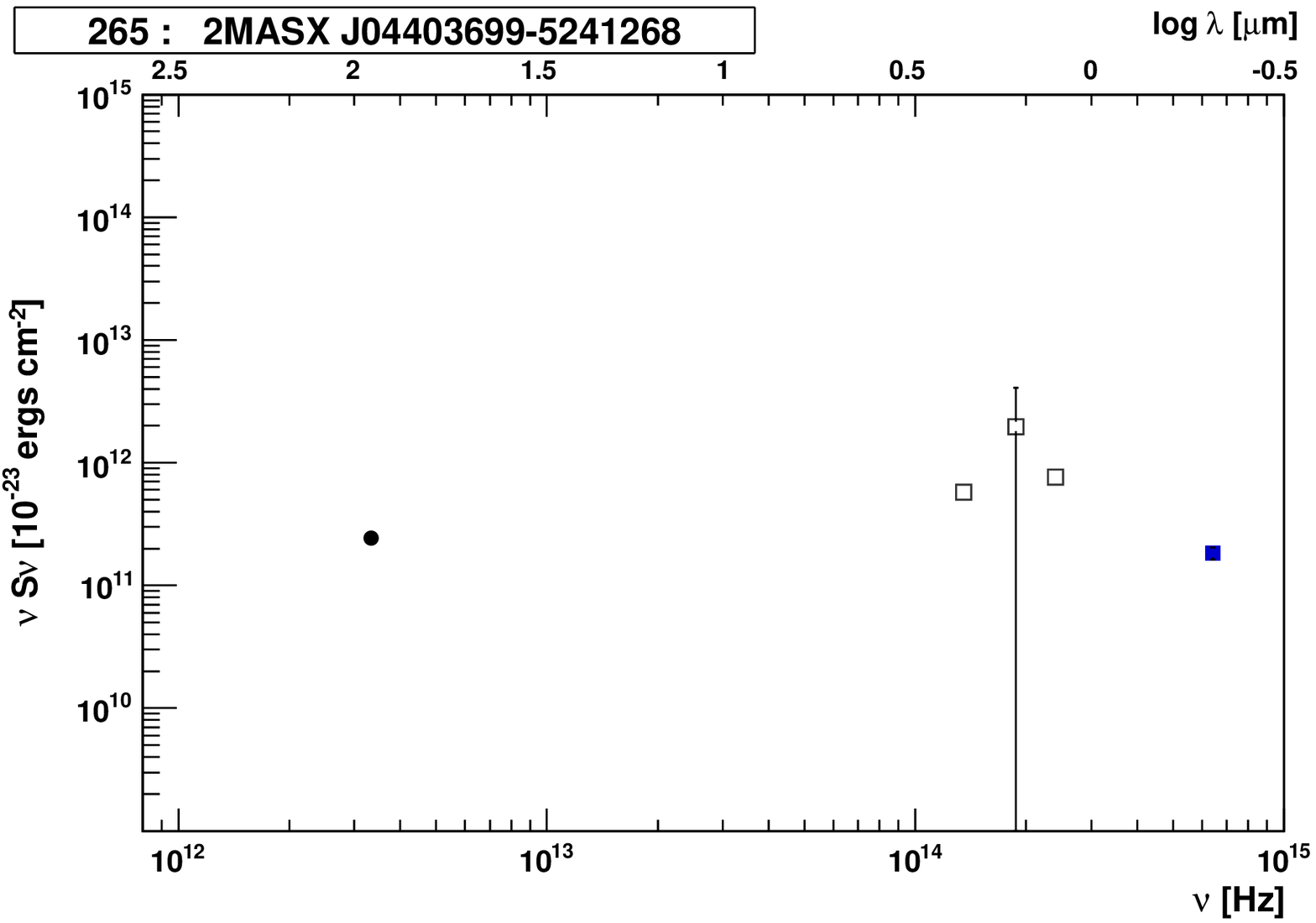}
\includegraphics[width=4cm]{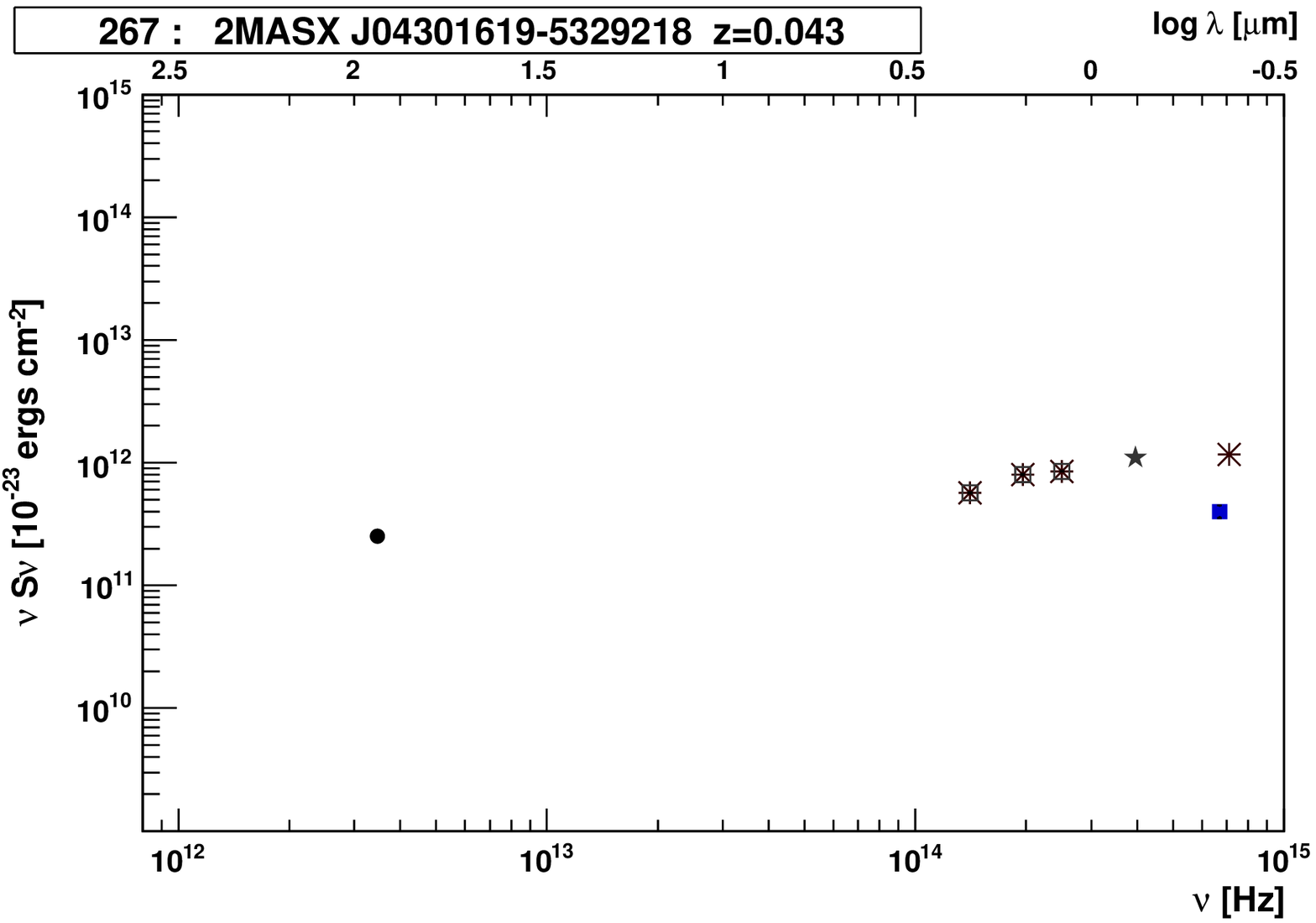}
\includegraphics[width=4cm]{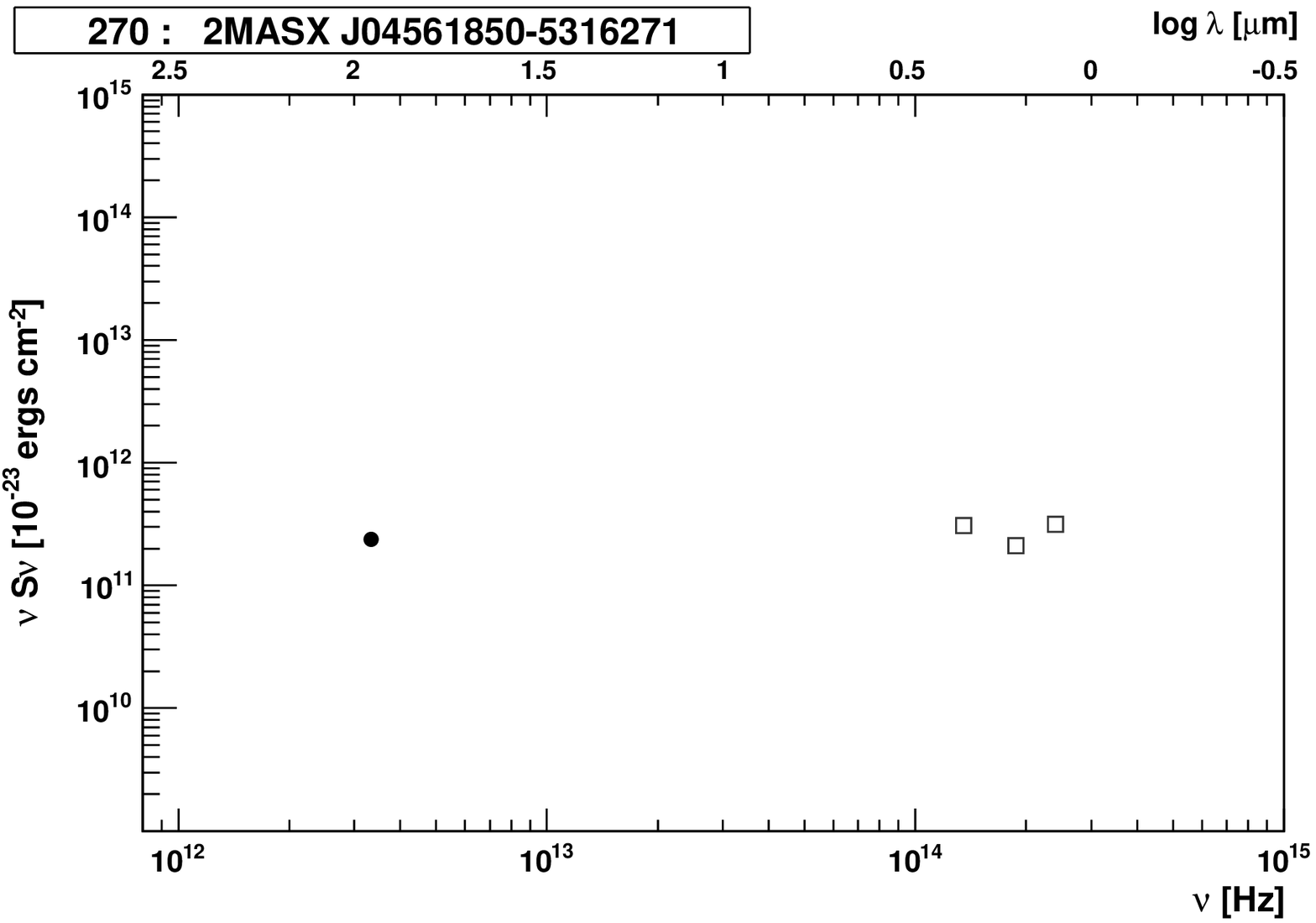}
\includegraphics[width=4cm]{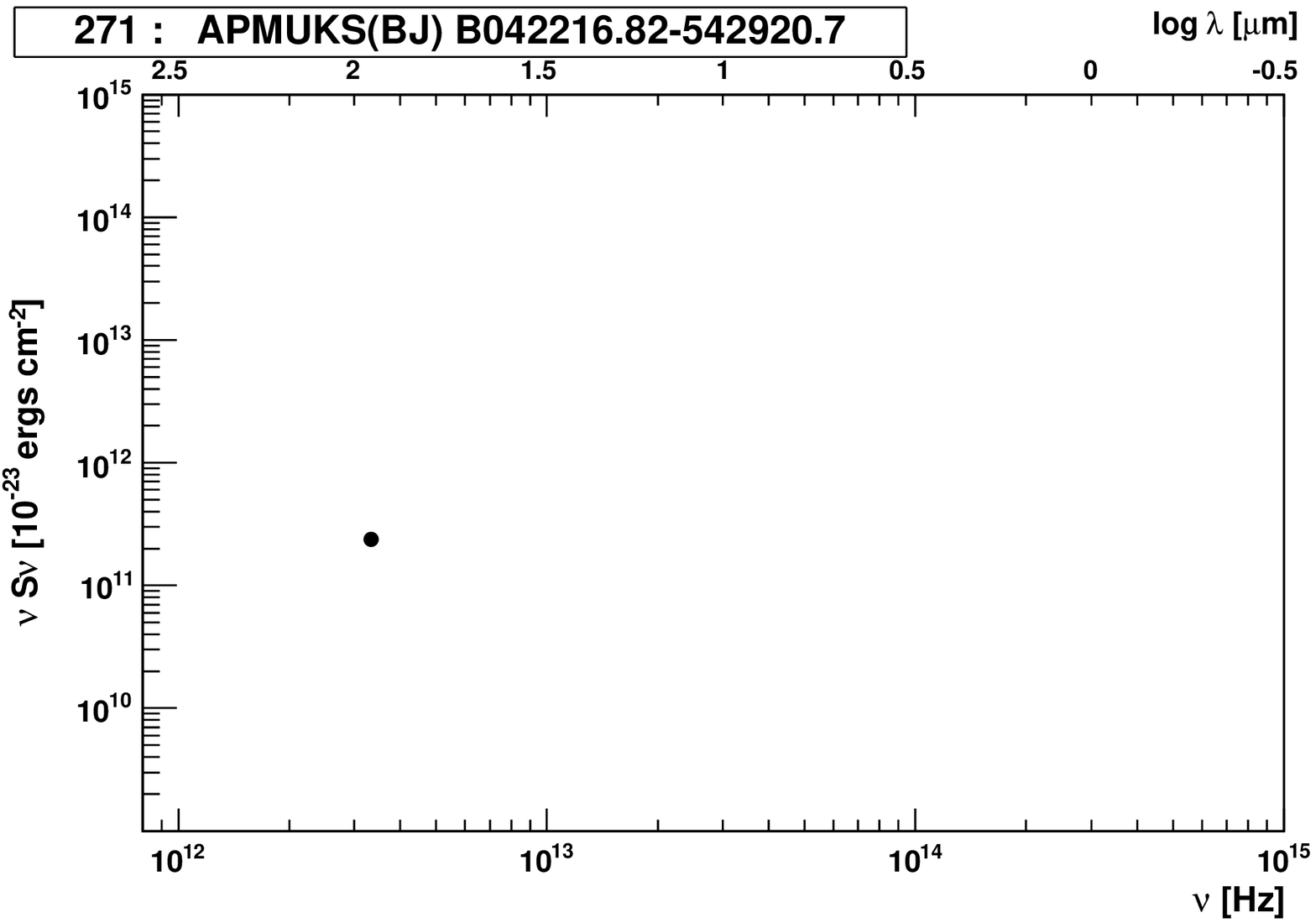}
\includegraphics[width=4cm]{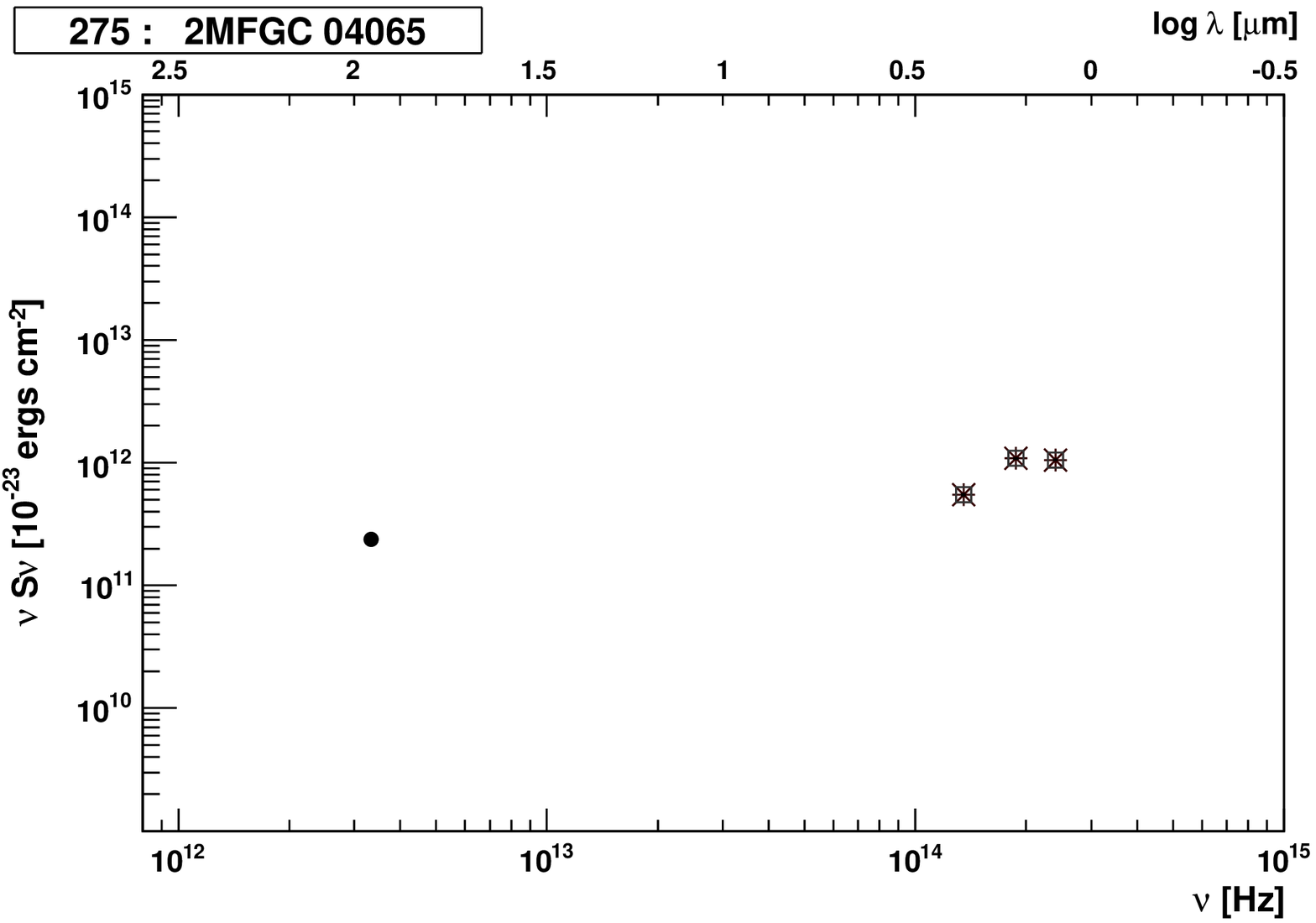}
\includegraphics[width=4cm]{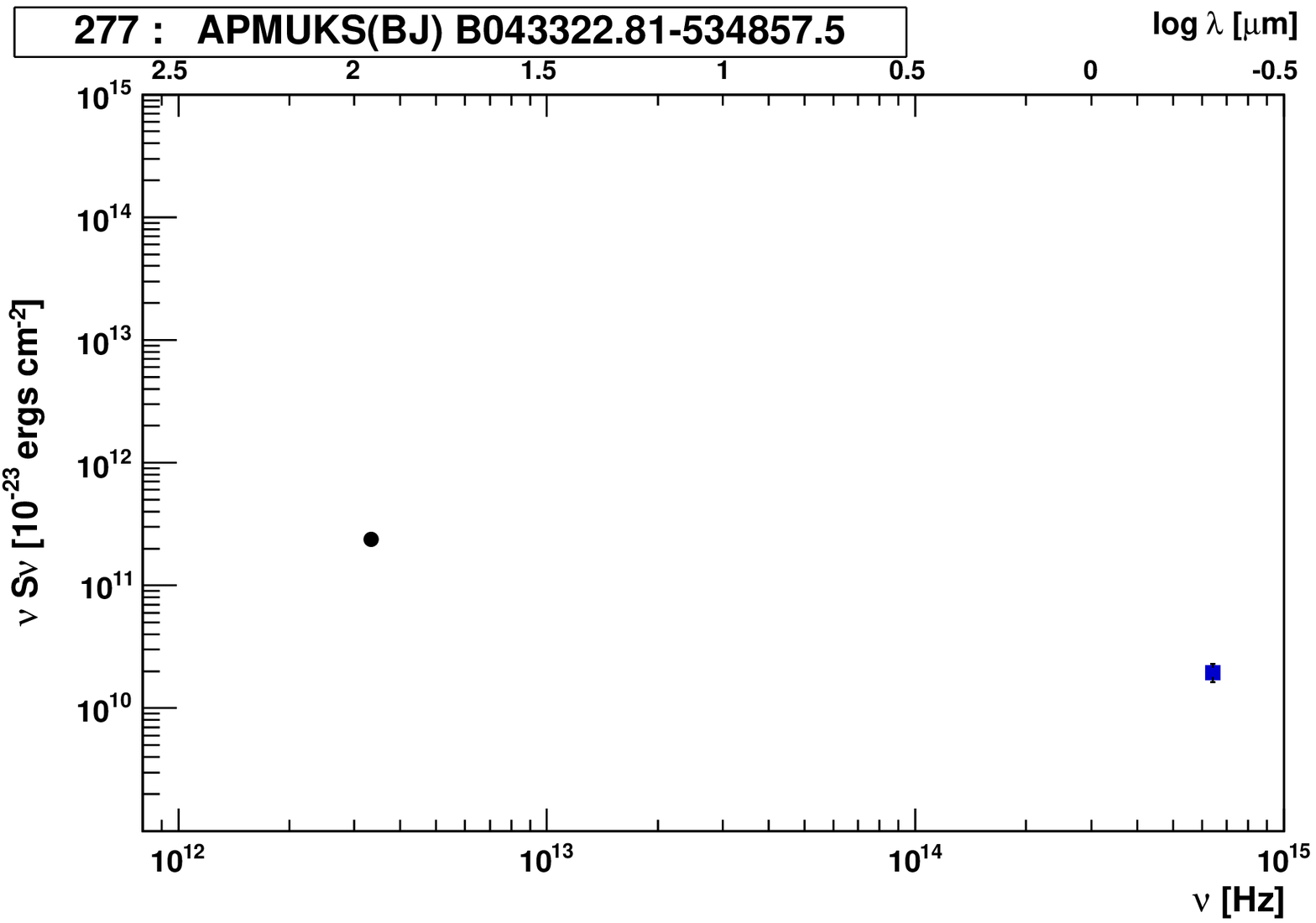}
\includegraphics[width=4cm]{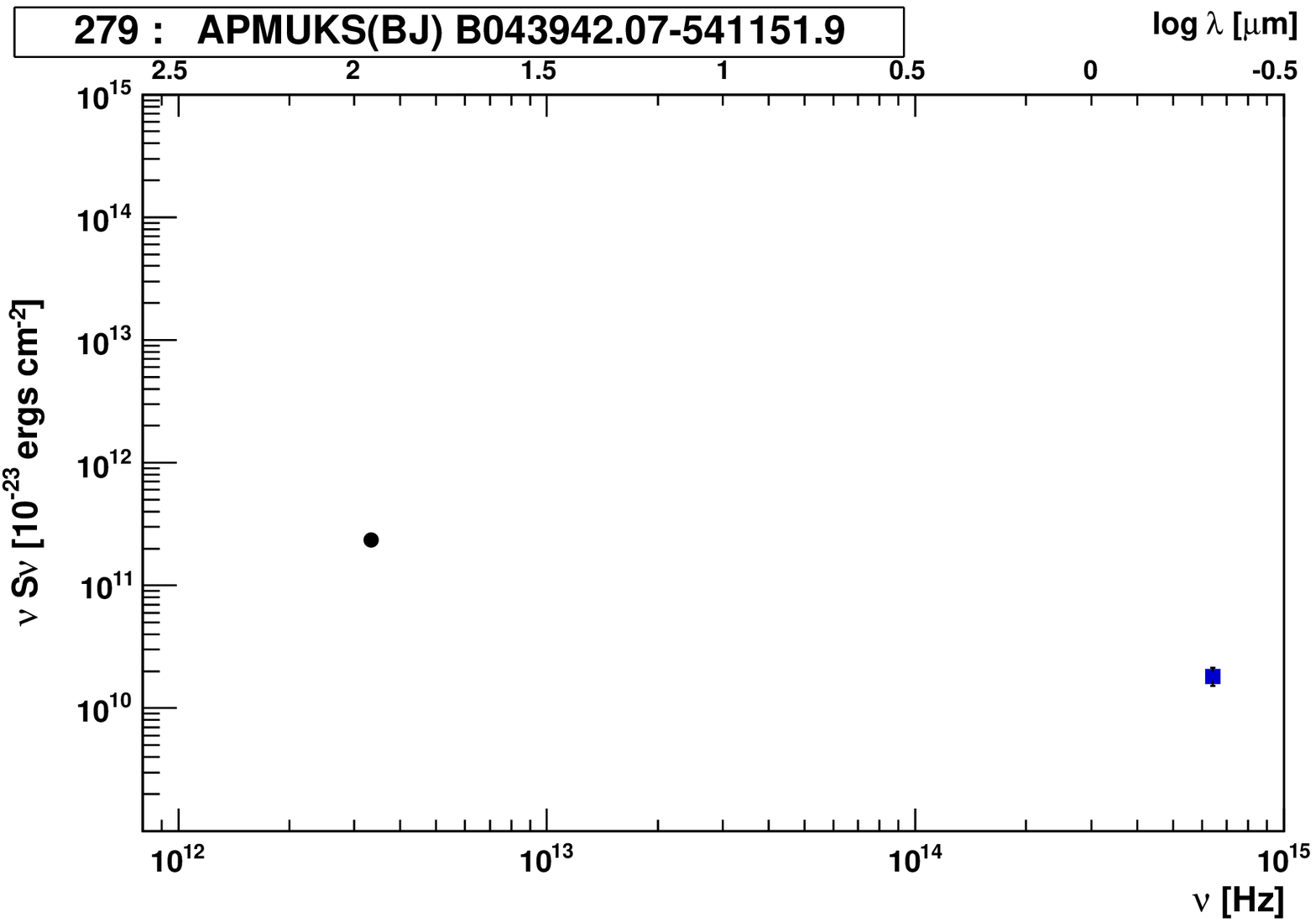}
\includegraphics[width=4cm]{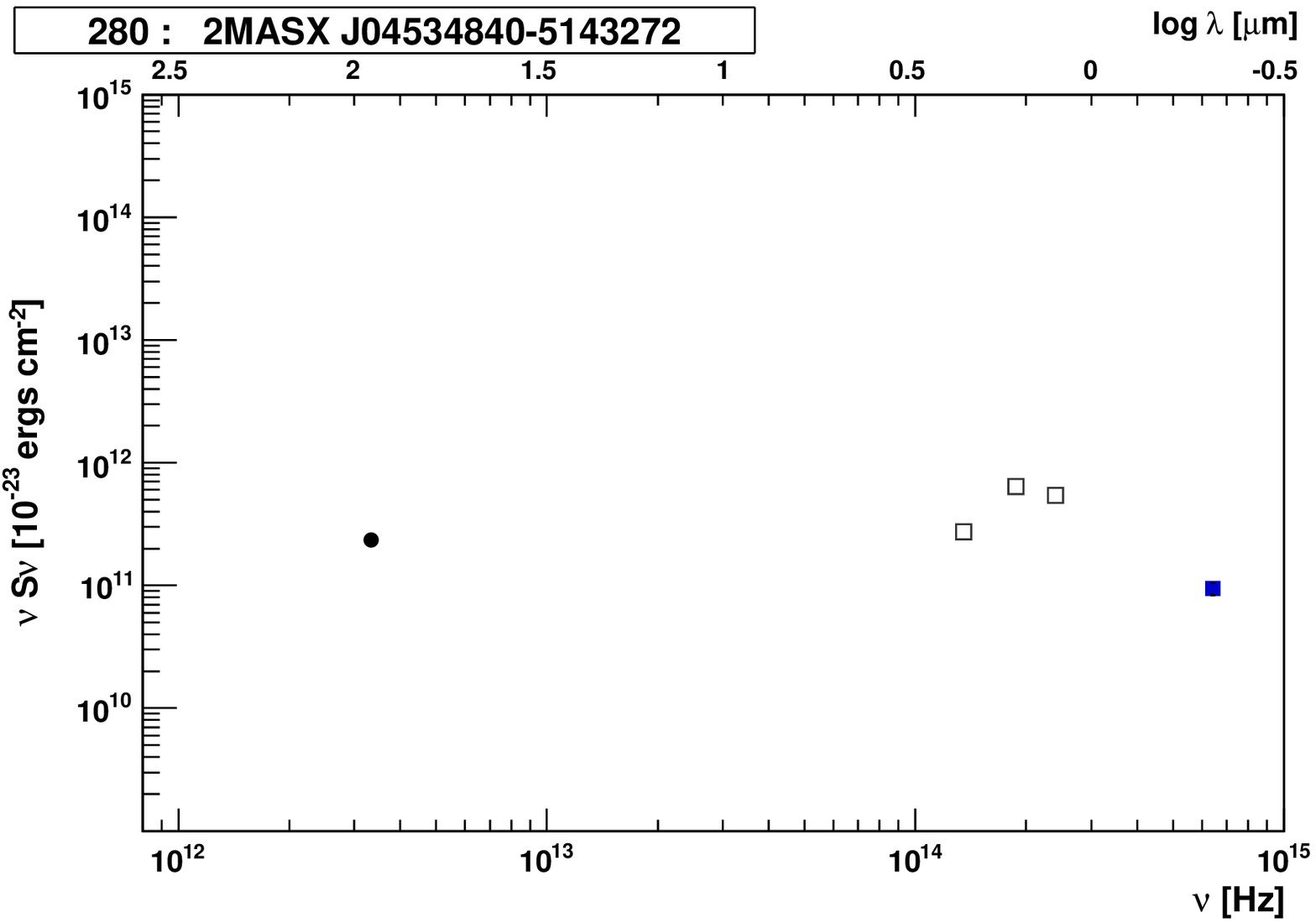}
\includegraphics[width=4cm]{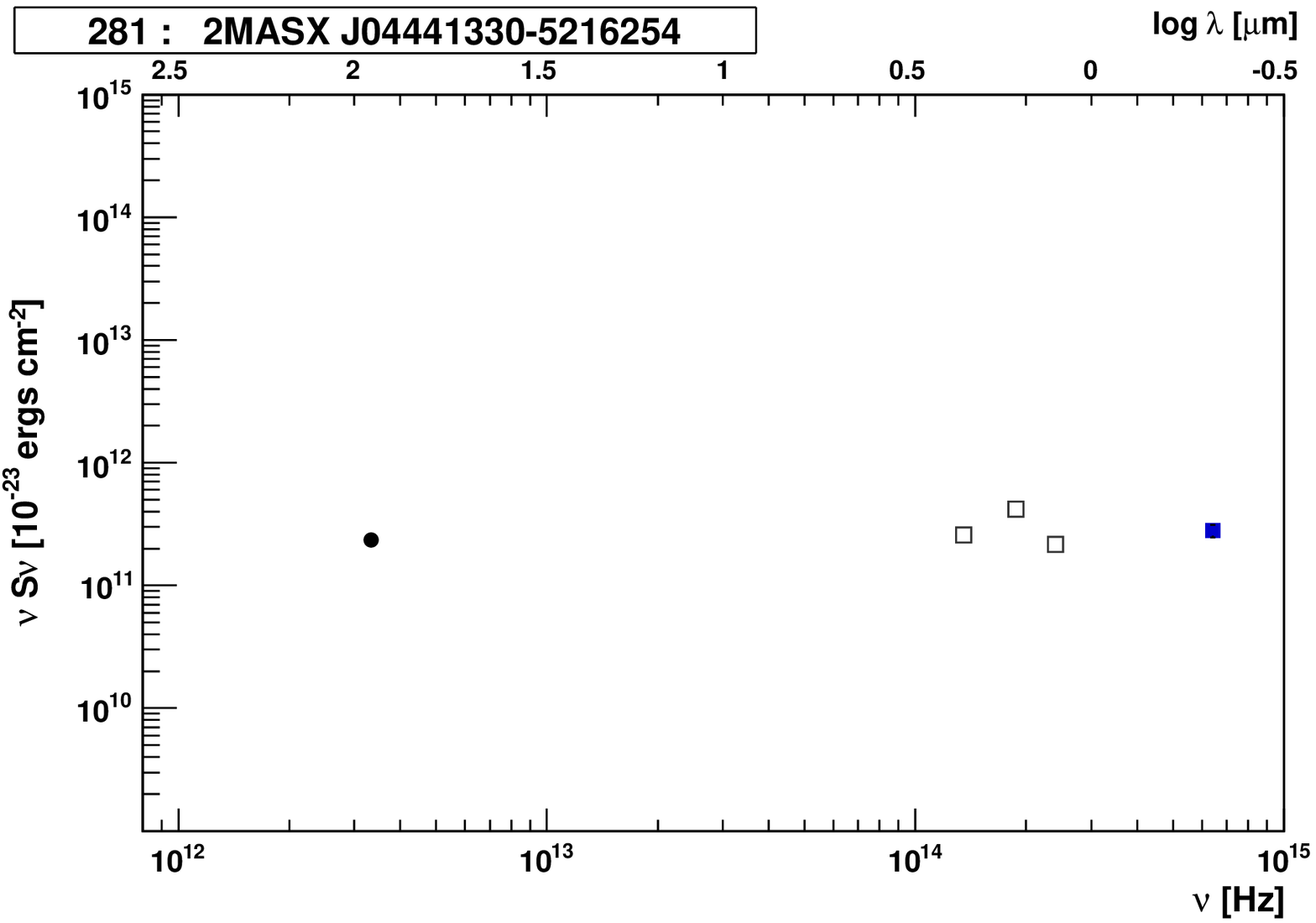}
\includegraphics[width=4cm]{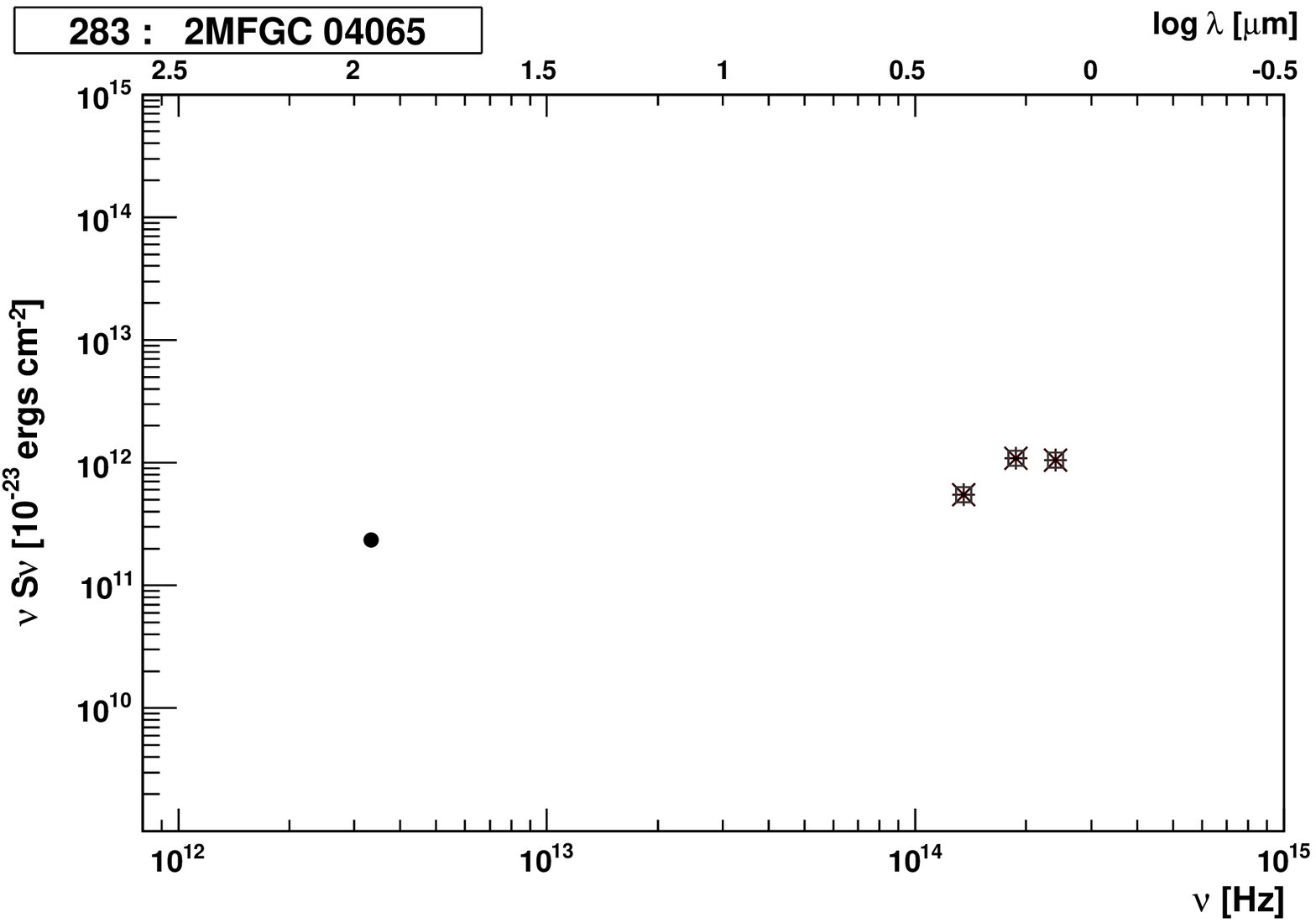}
\includegraphics[width=4cm]{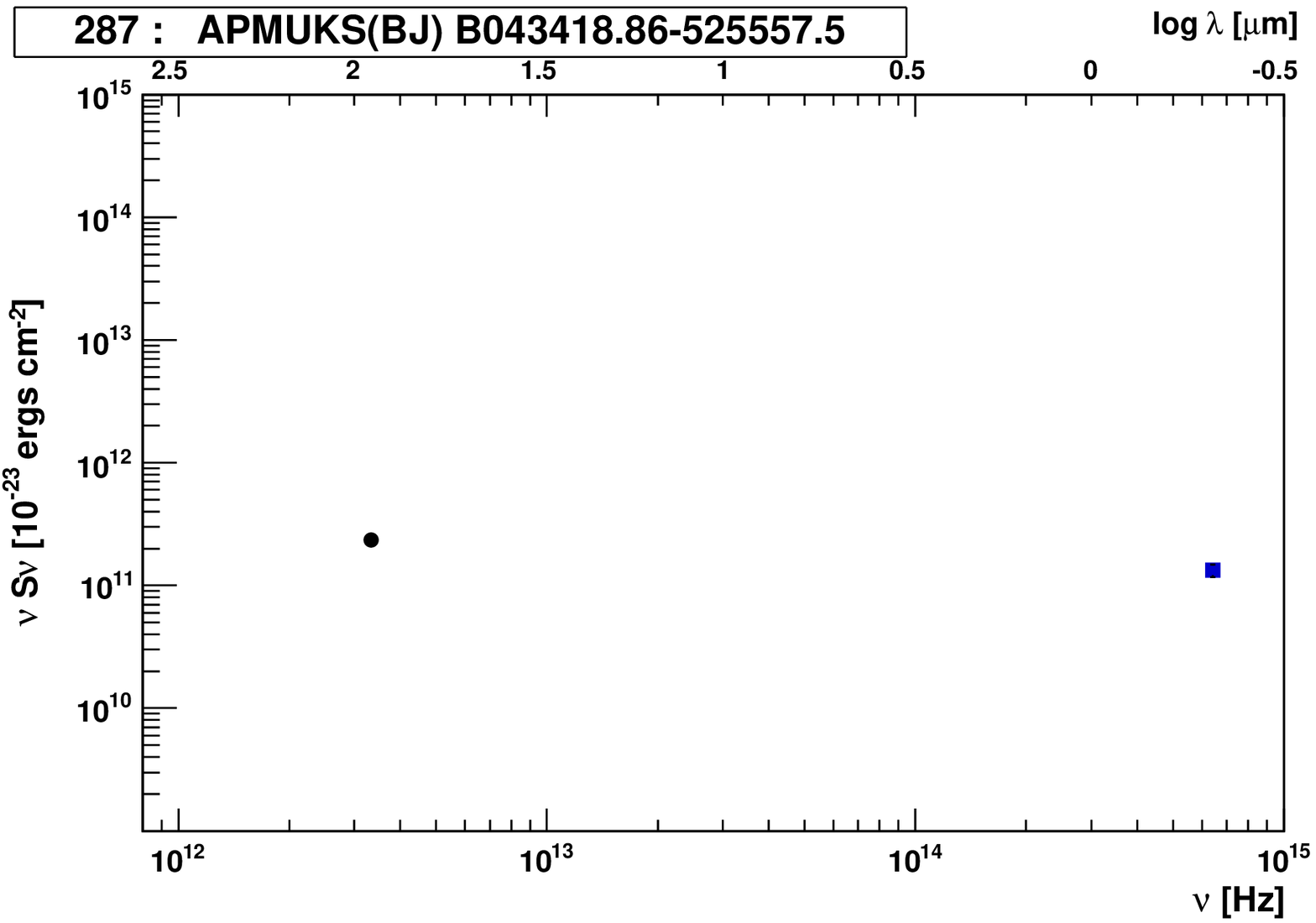}
\includegraphics[width=4cm]{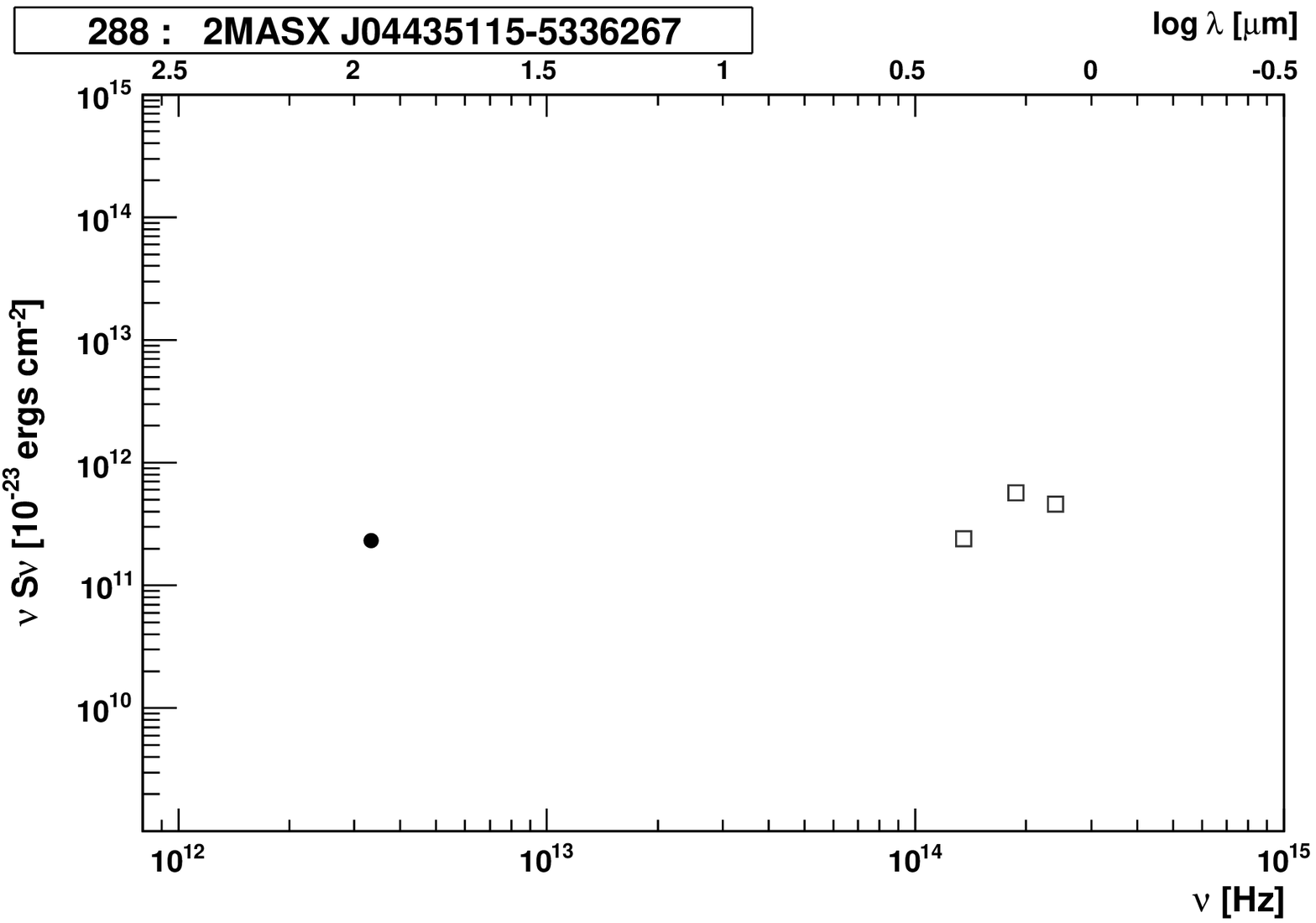}
\includegraphics[width=4cm]{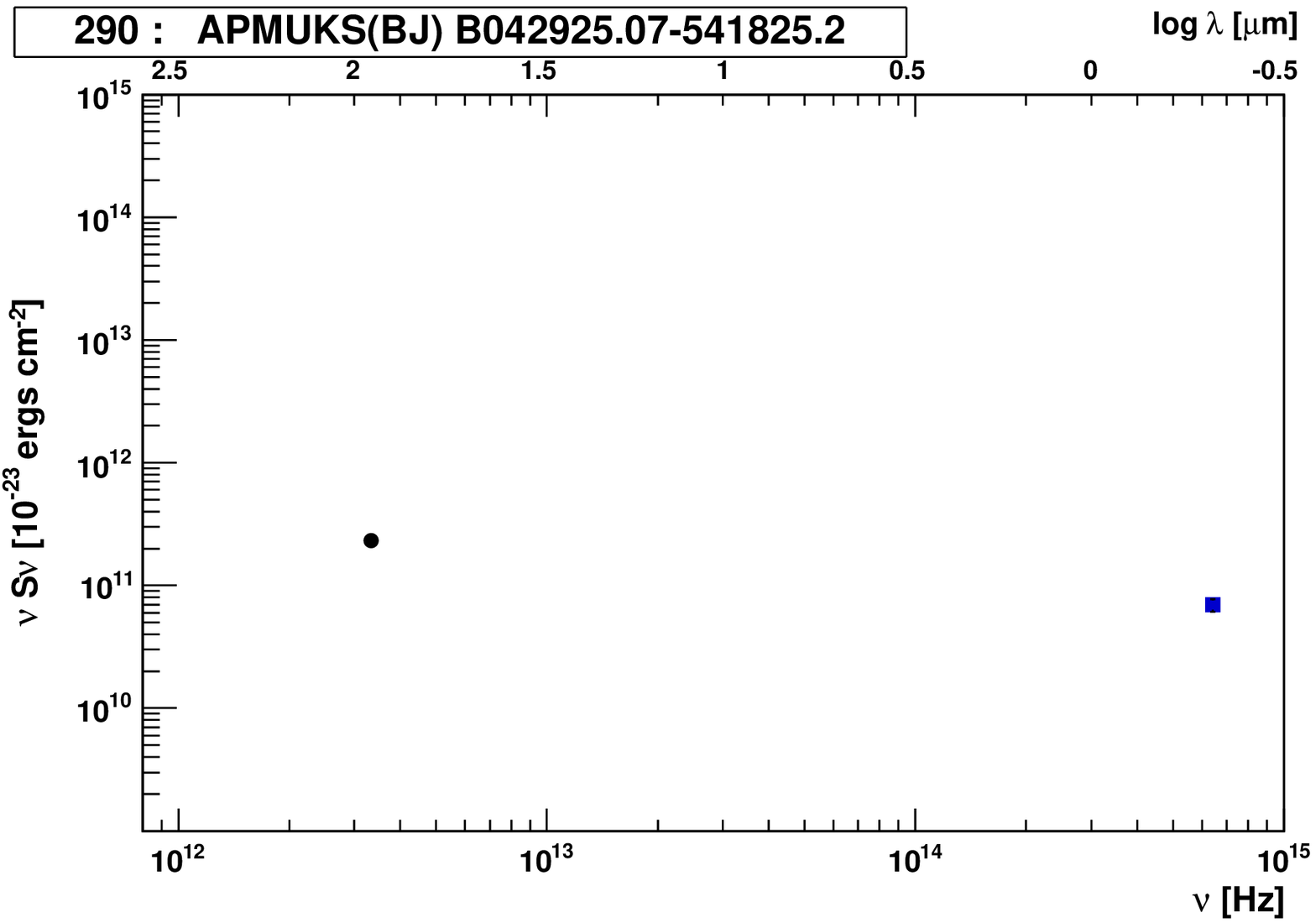}
\label{points6}
\caption {SEDs for the next 36 ADF-S identified sources, with symbols as in Figure~\ref{points1}.}
\end{figure*}
}

\clearpage

\onlfig{7}{
\begin{figure*}[t]
\centering

\includegraphics[width=4cm]{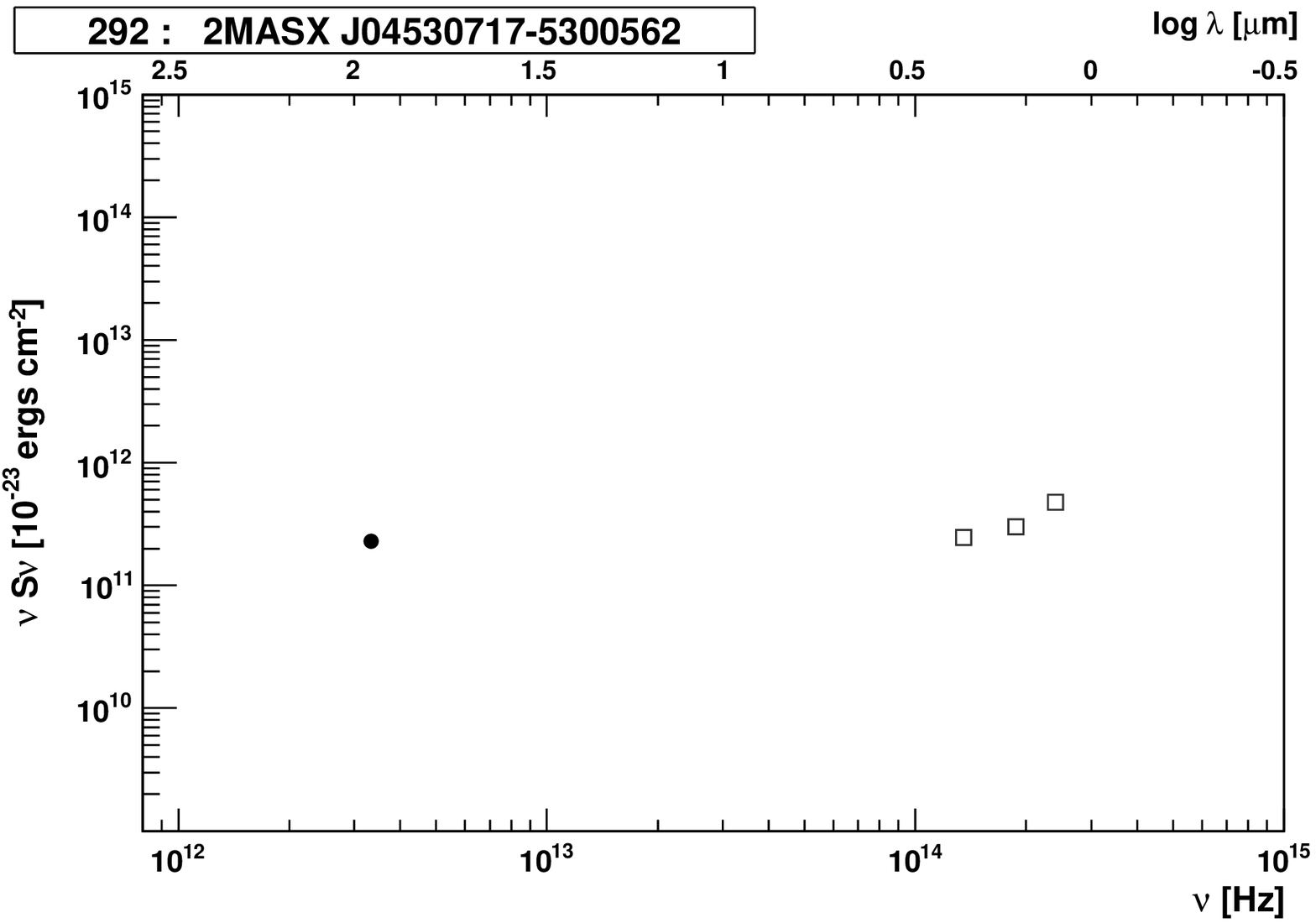}
\includegraphics[width=4cm]{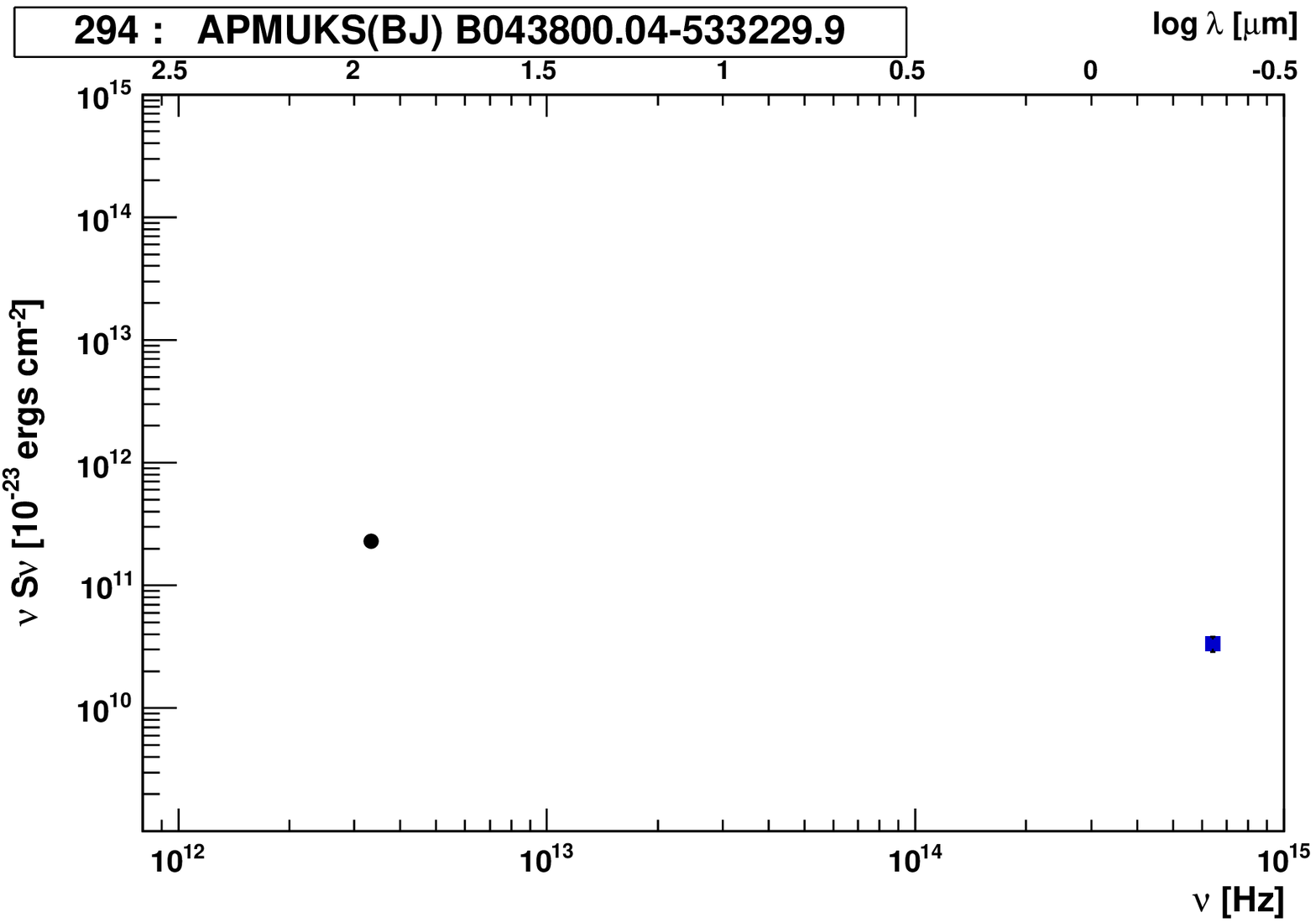}
\includegraphics[width=4cm]{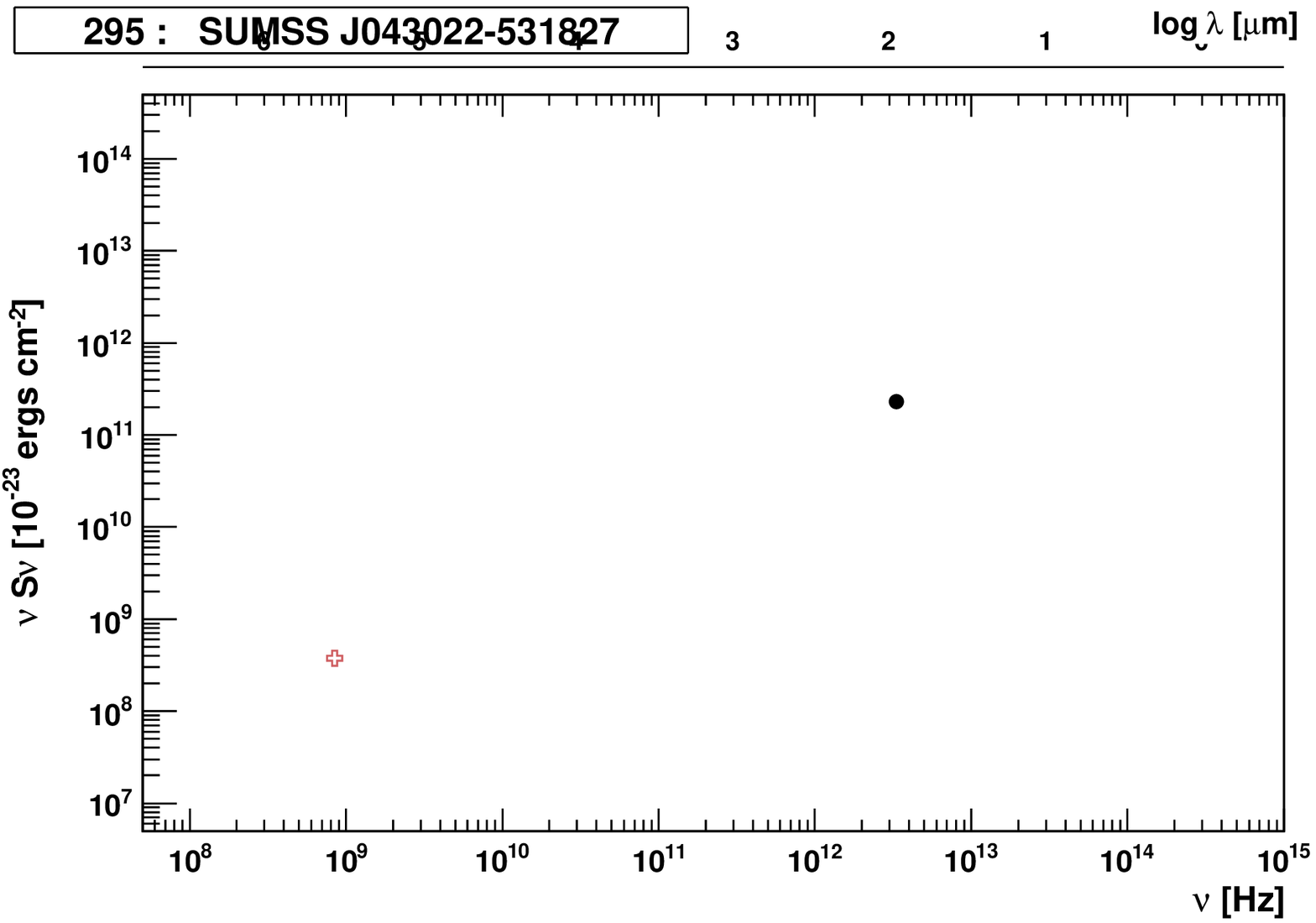}
\includegraphics[width=4cm]{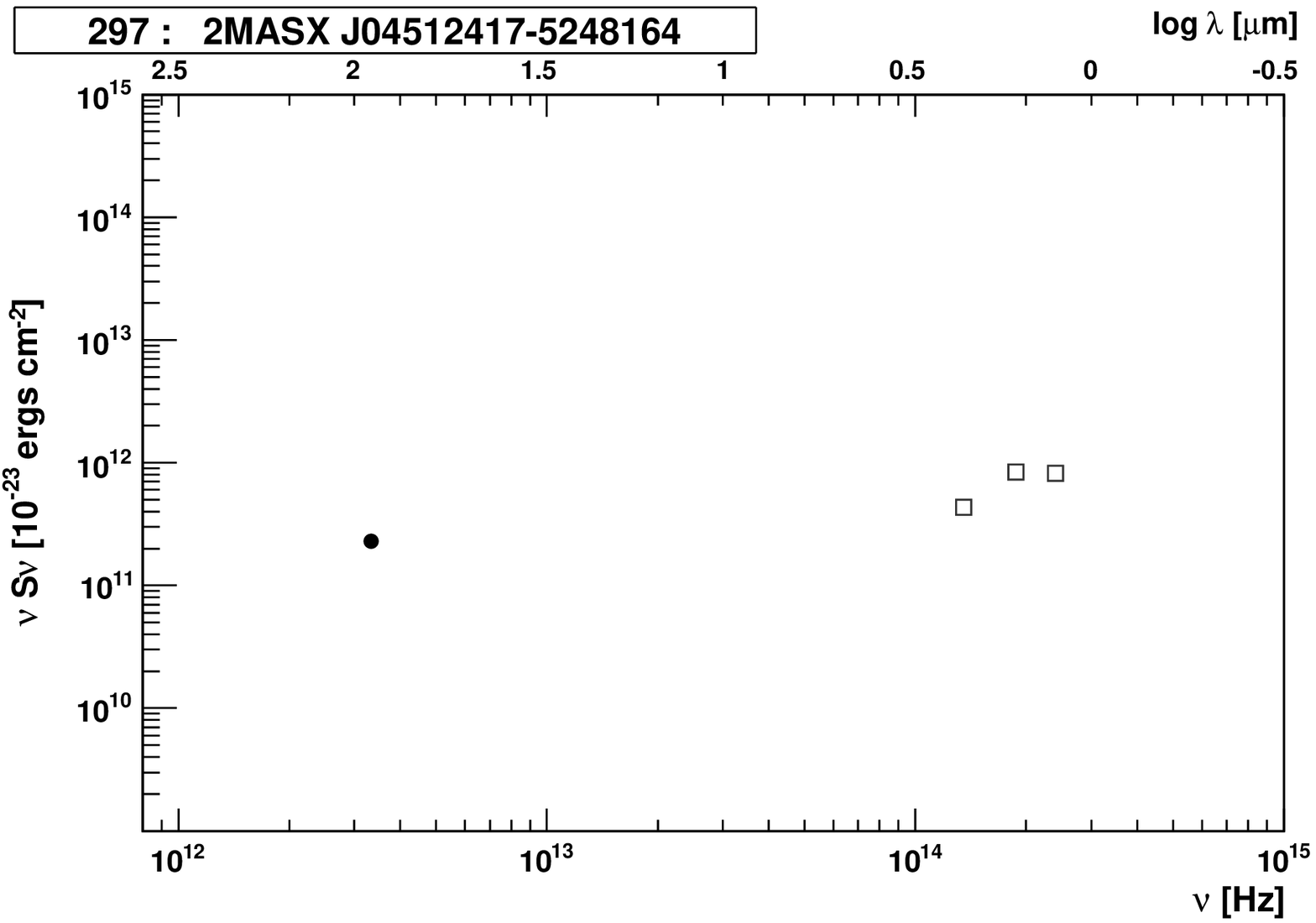}
\includegraphics[width=4cm]{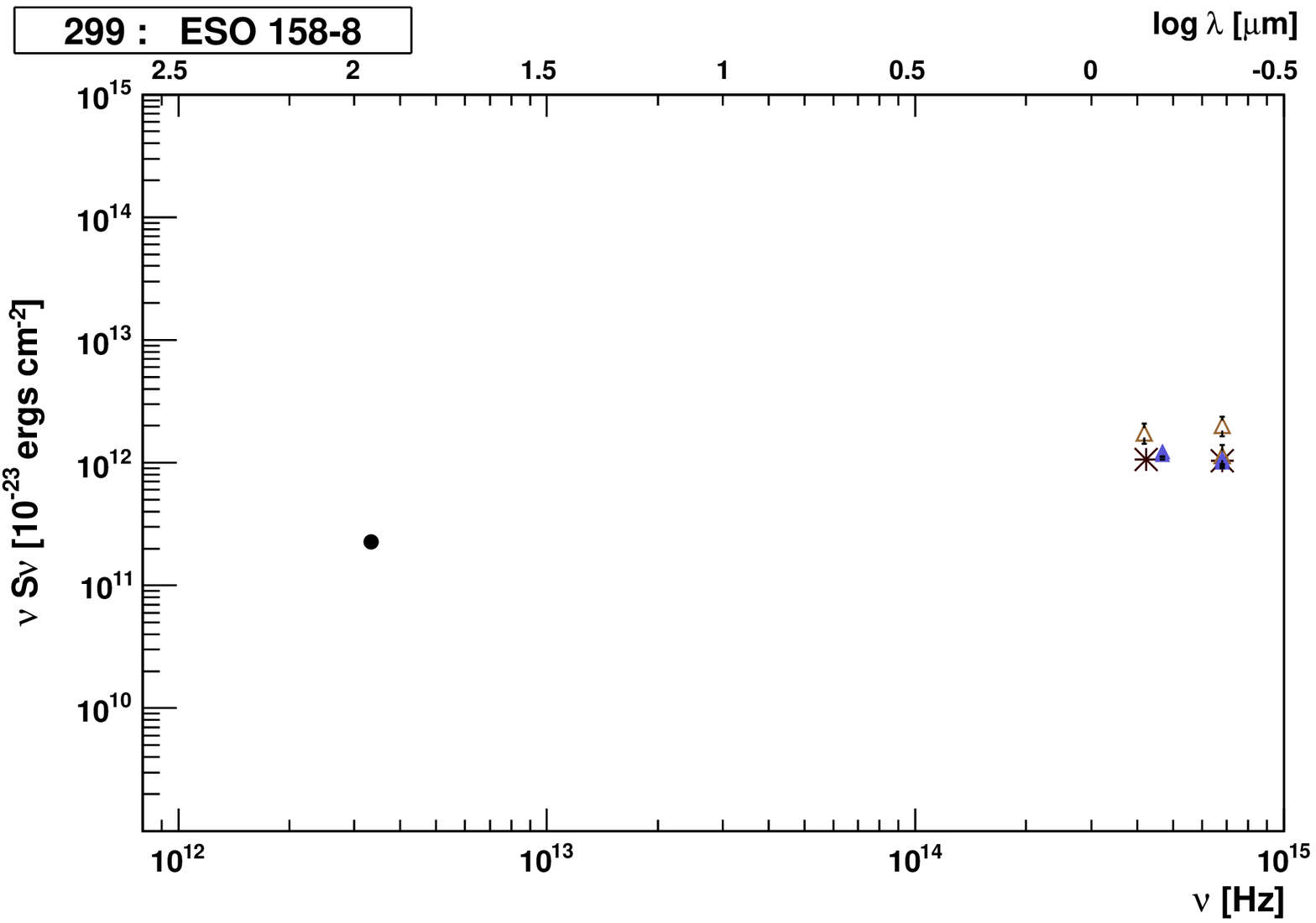}
\includegraphics[width=4cm]{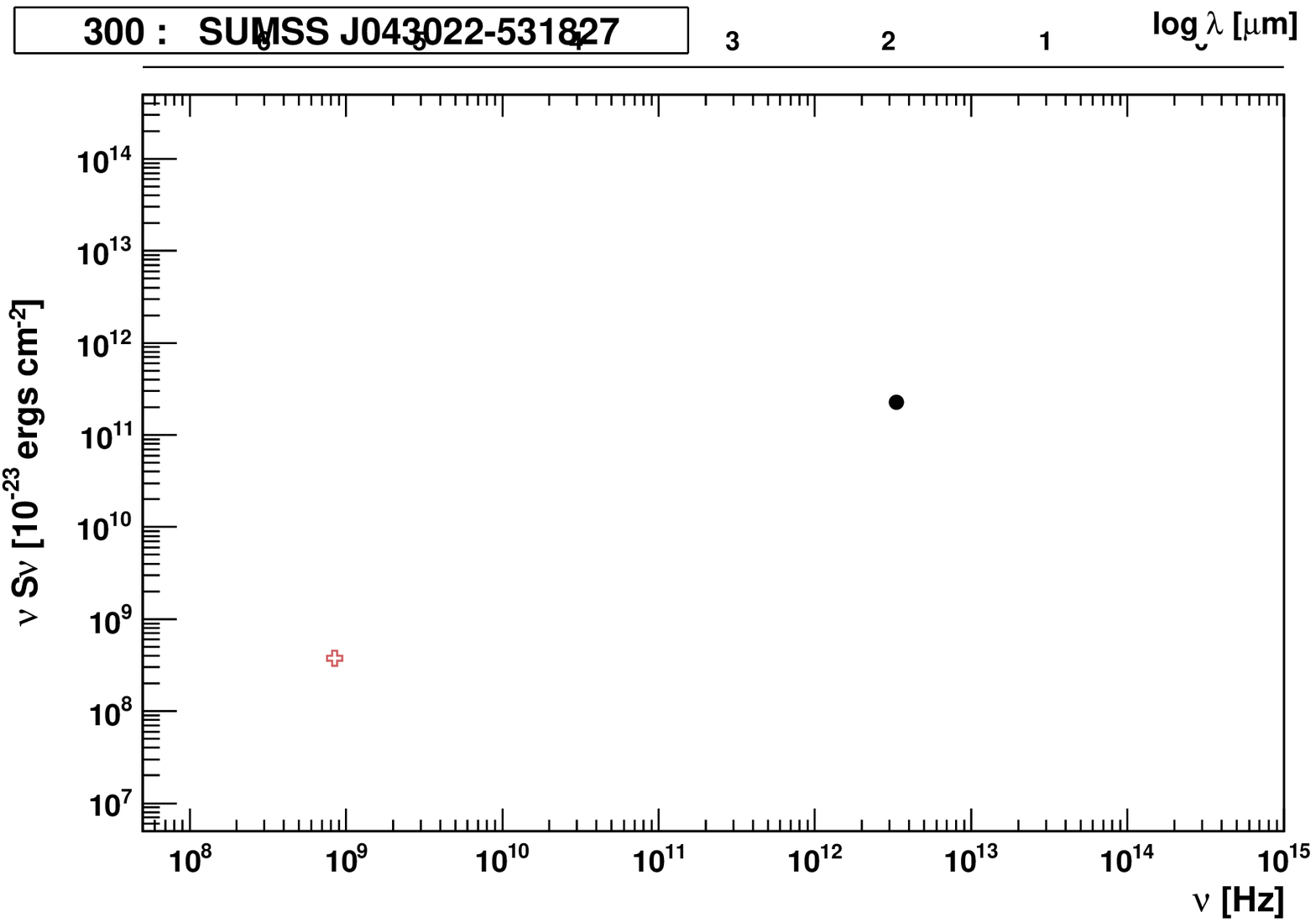}
\includegraphics[width=4cm]{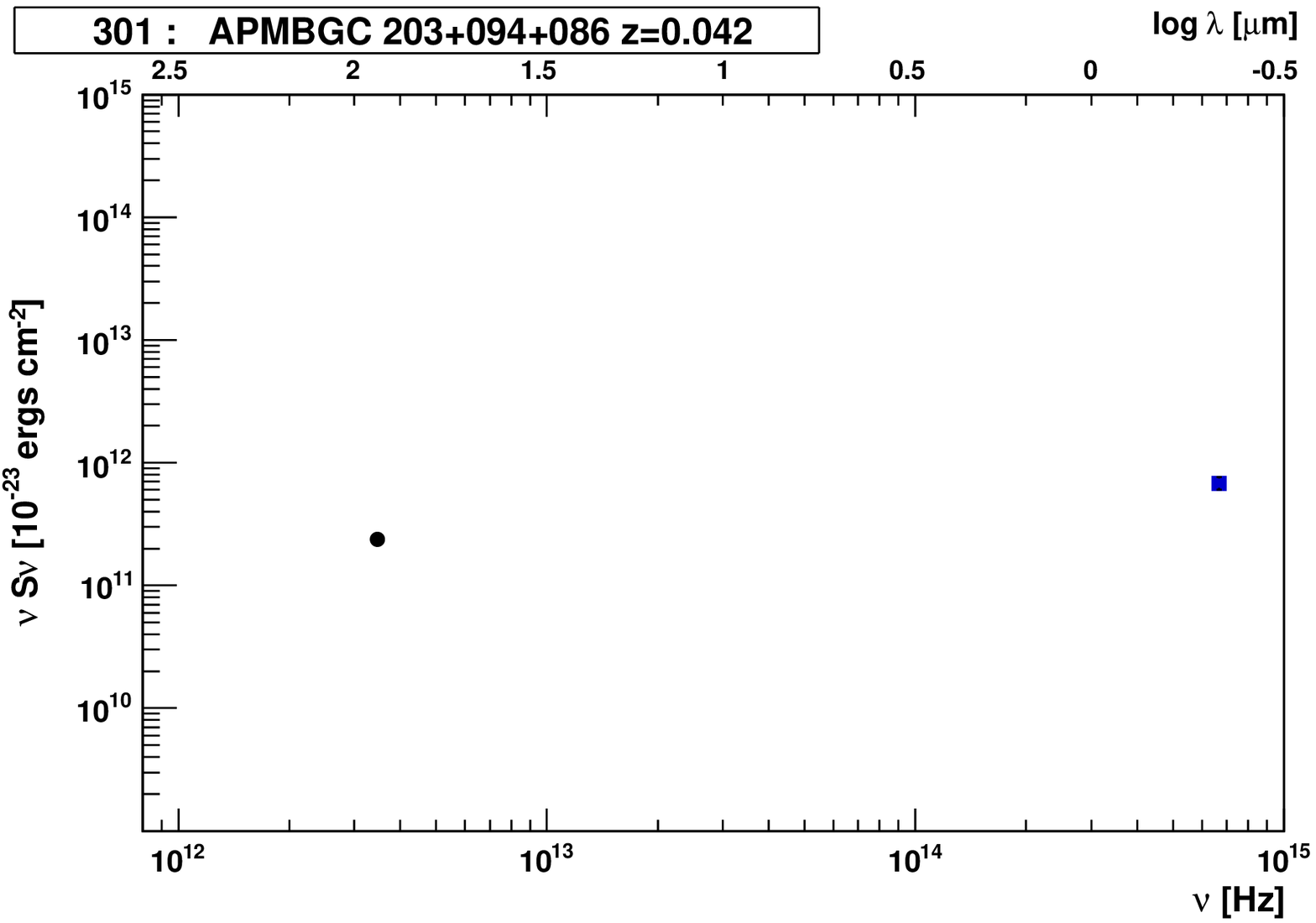}
\includegraphics[width=4cm]{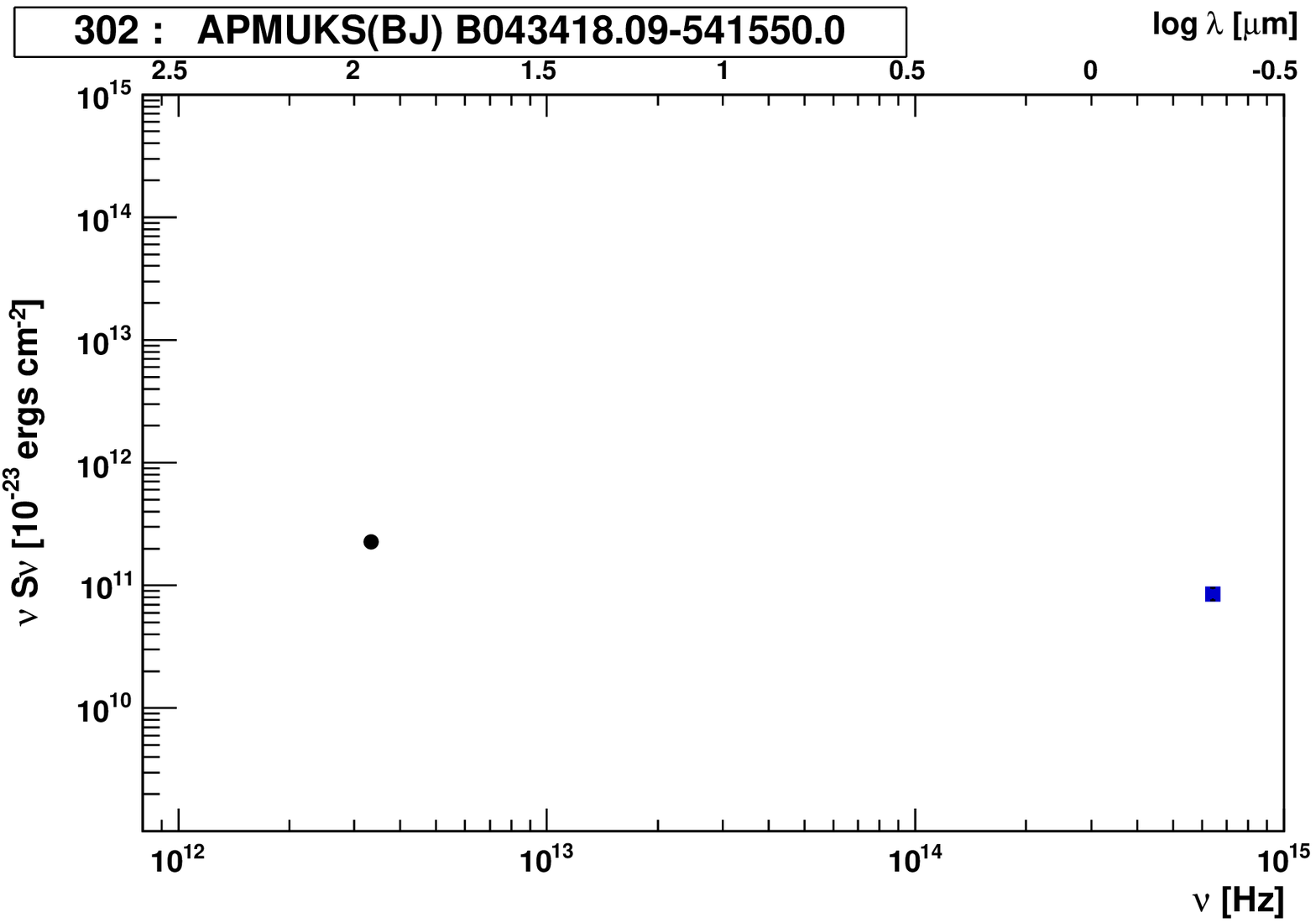}
\includegraphics[width=4cm]{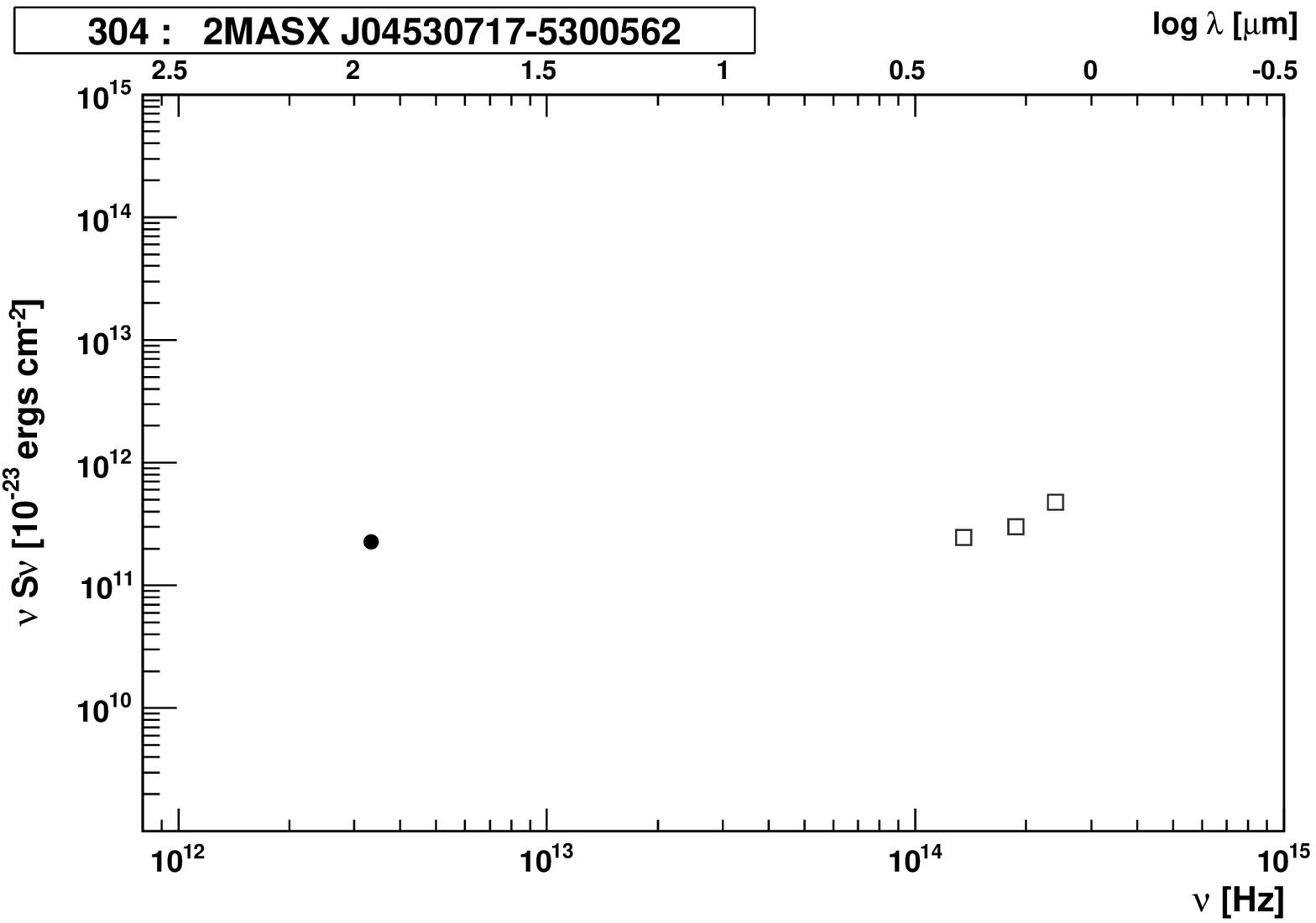}
\includegraphics[width=4cm]{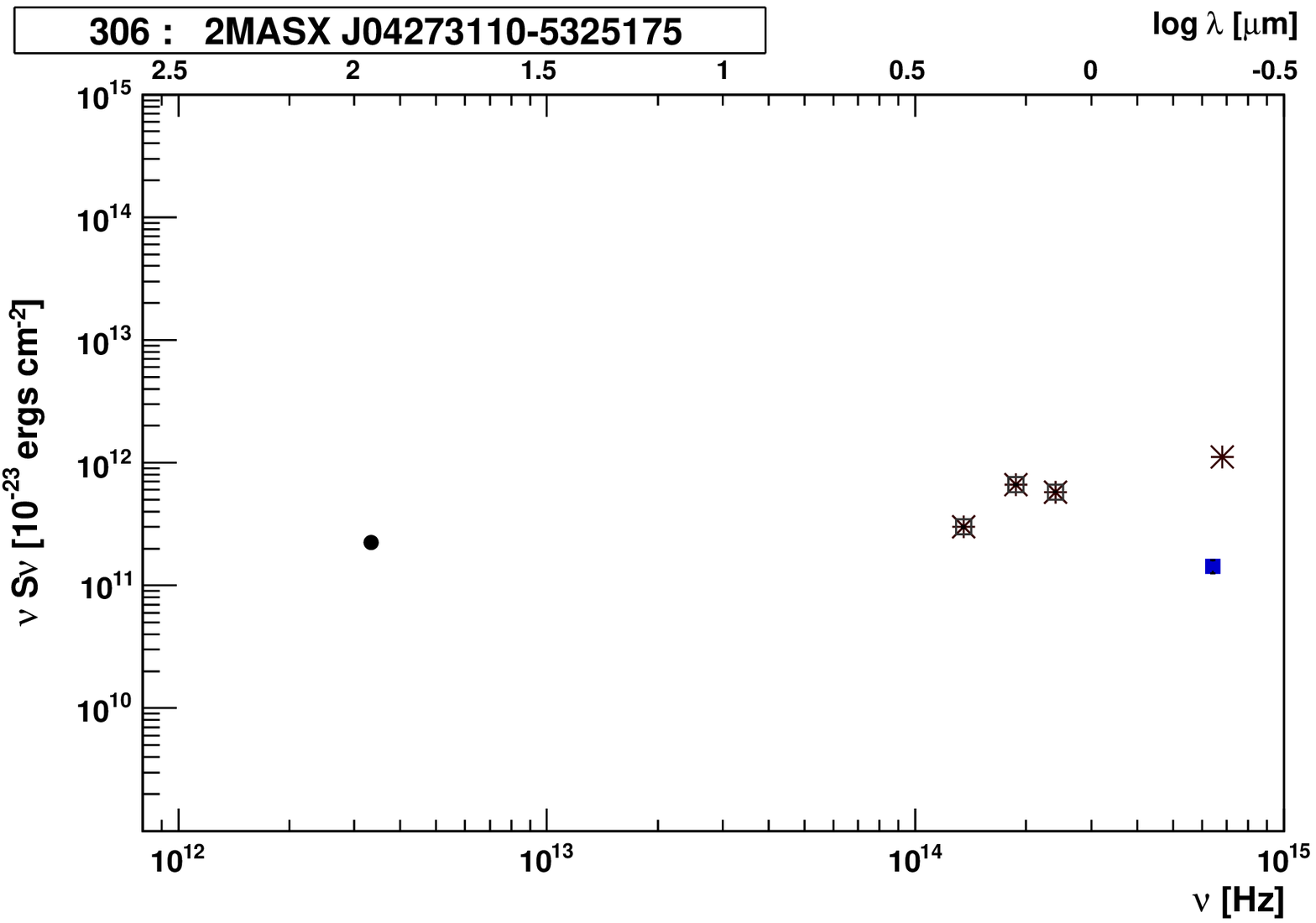}
\includegraphics[width=4cm]{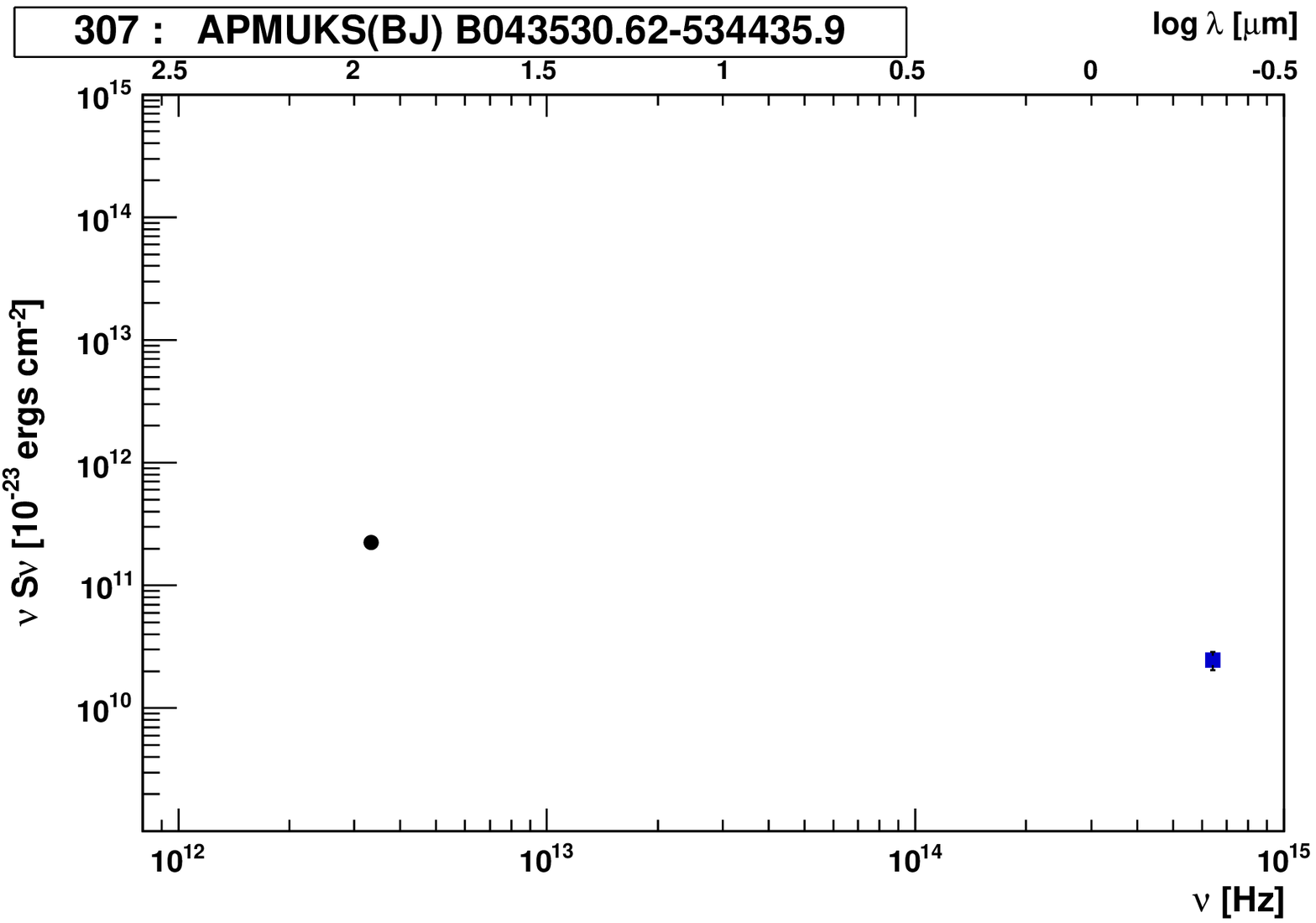}
\includegraphics[width=4cm]{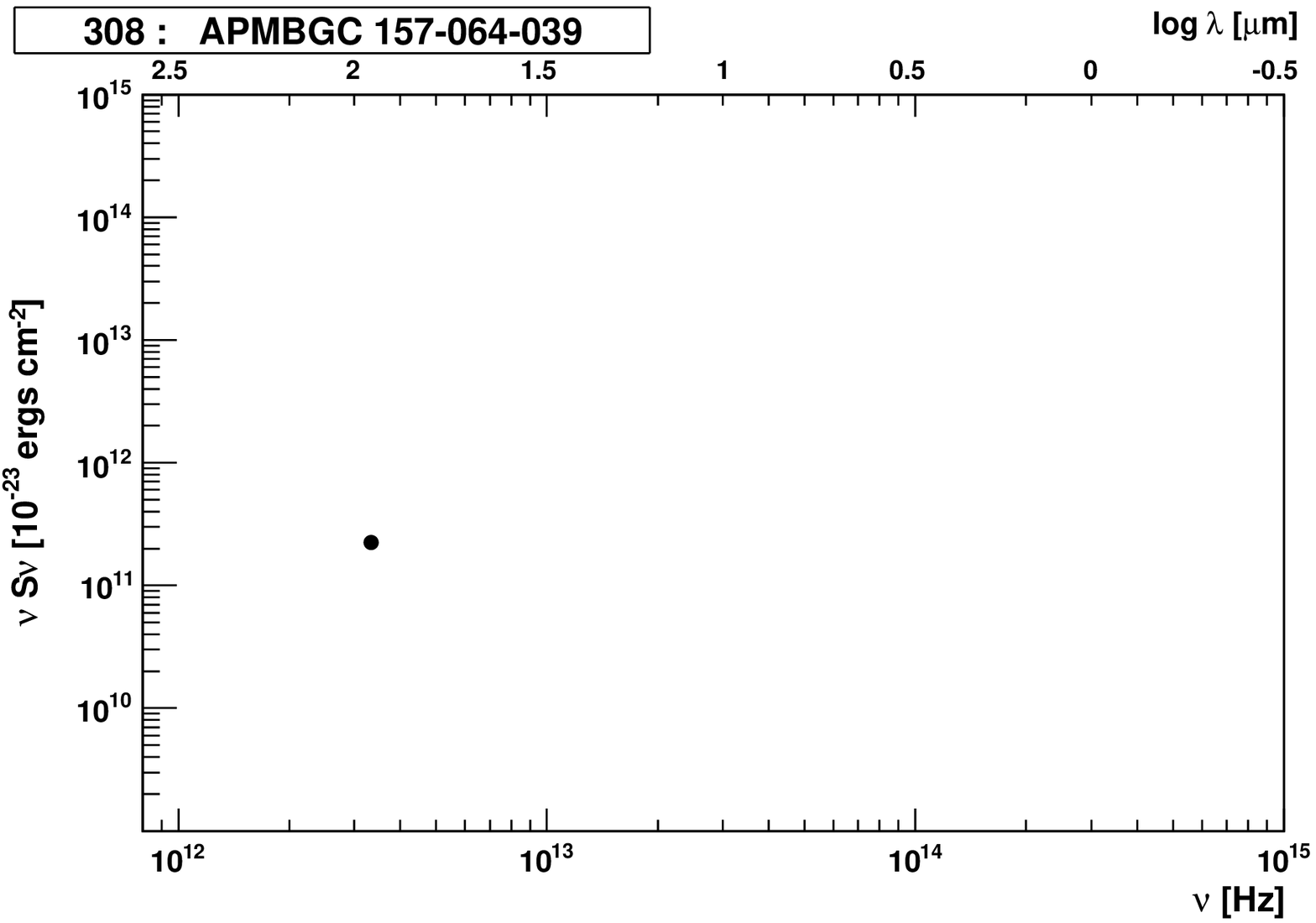}
\includegraphics[width=4cm]{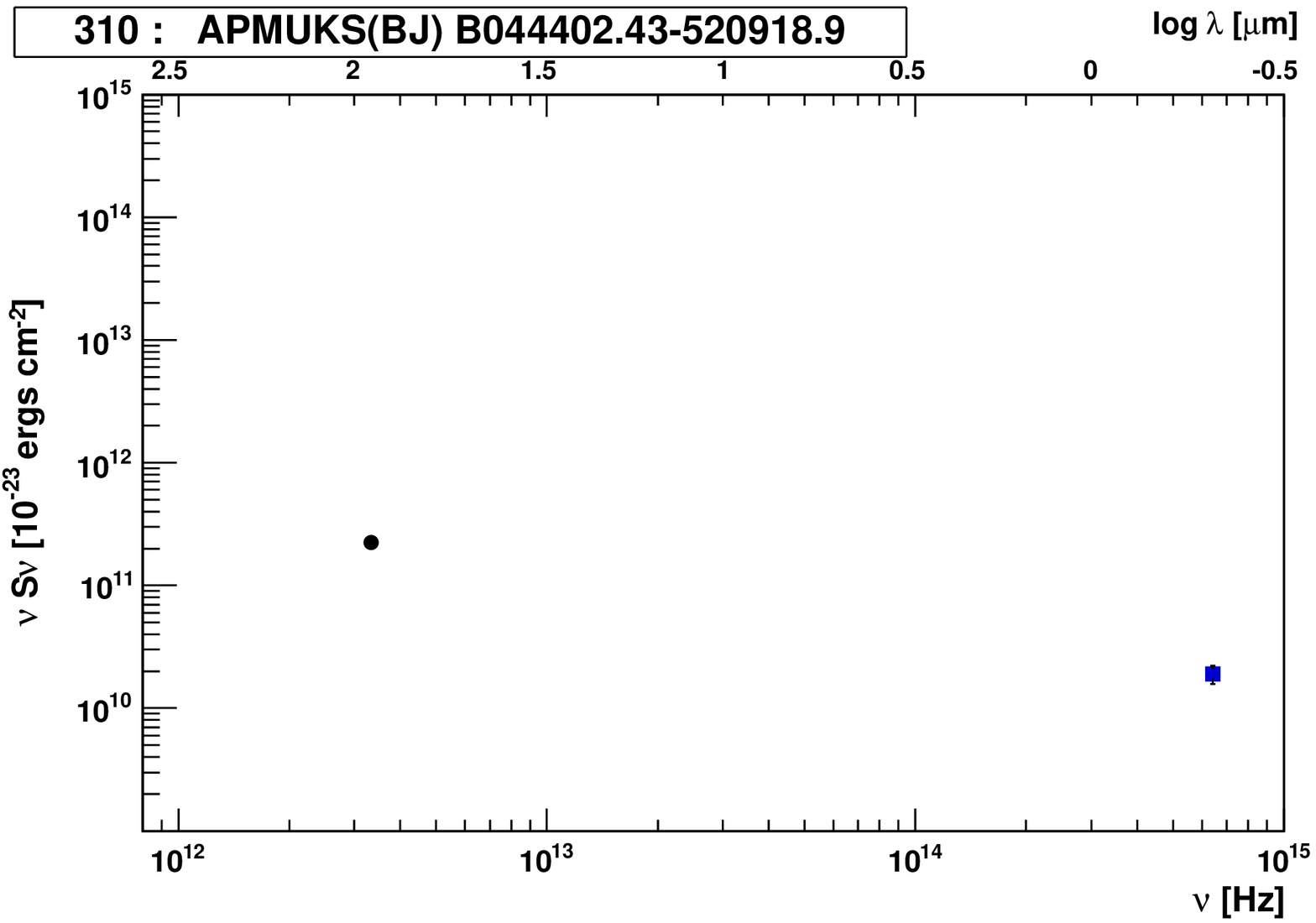}
\includegraphics[width=4cm]{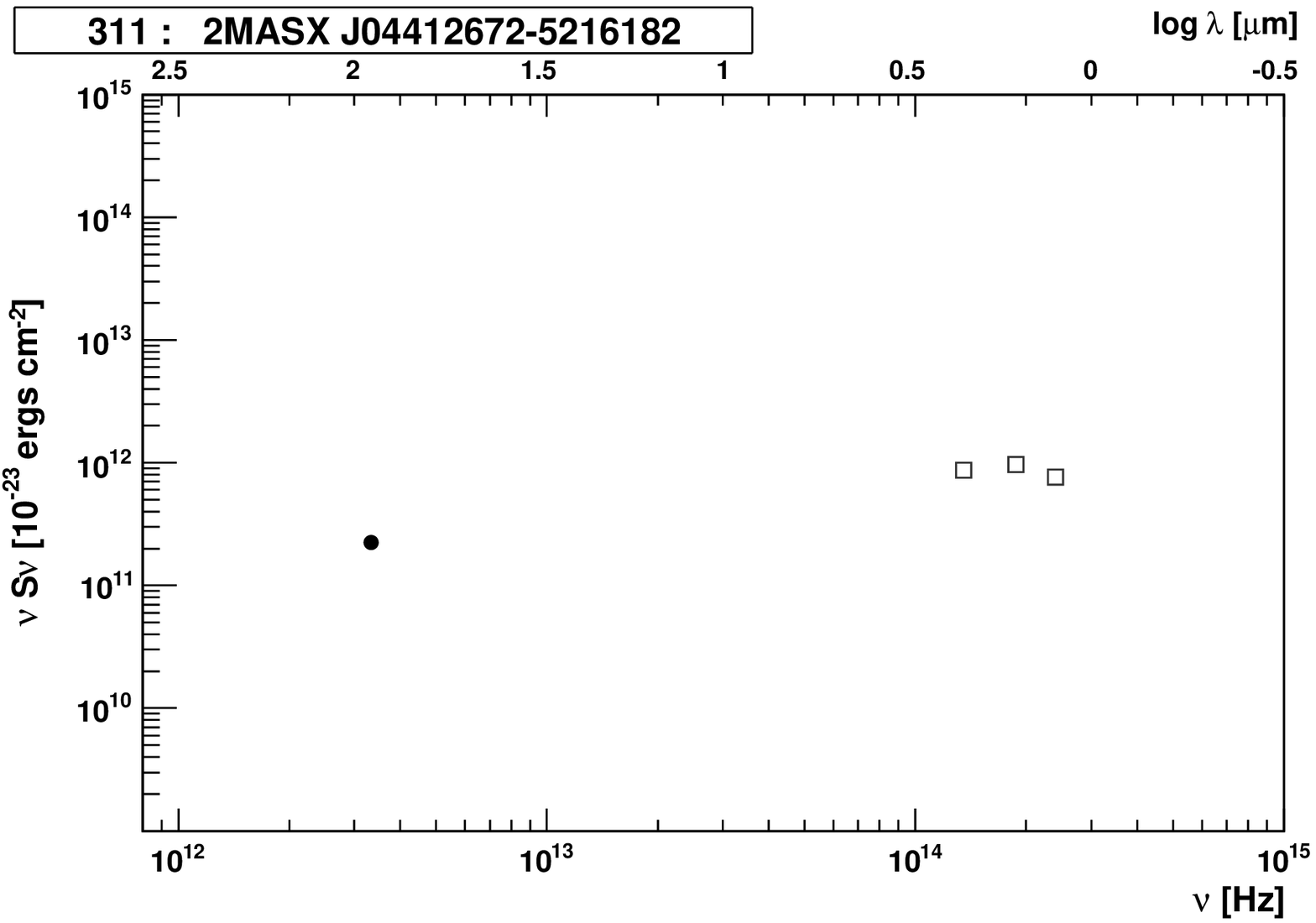}
\includegraphics[width=4cm]{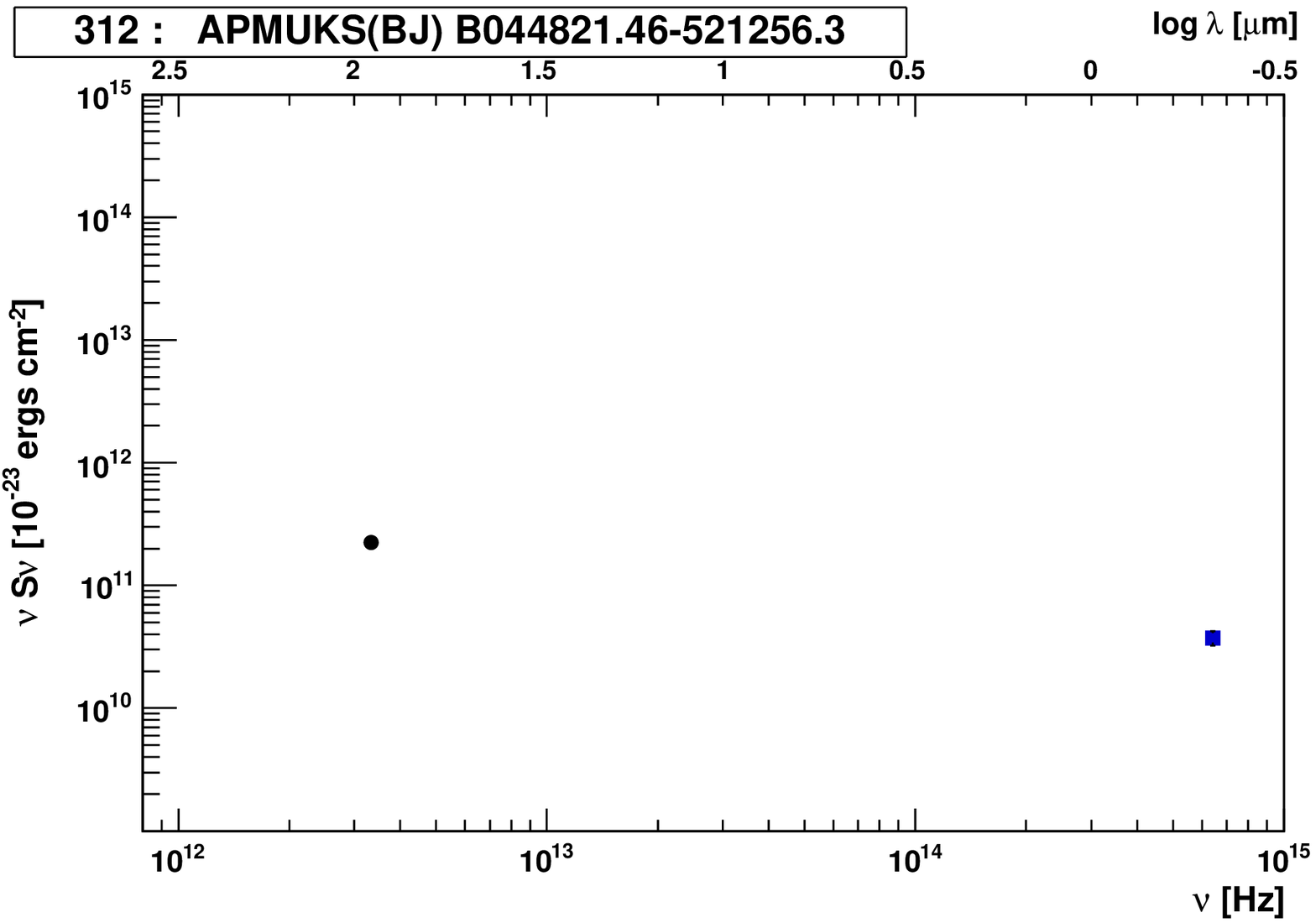}
\includegraphics[width=4cm]{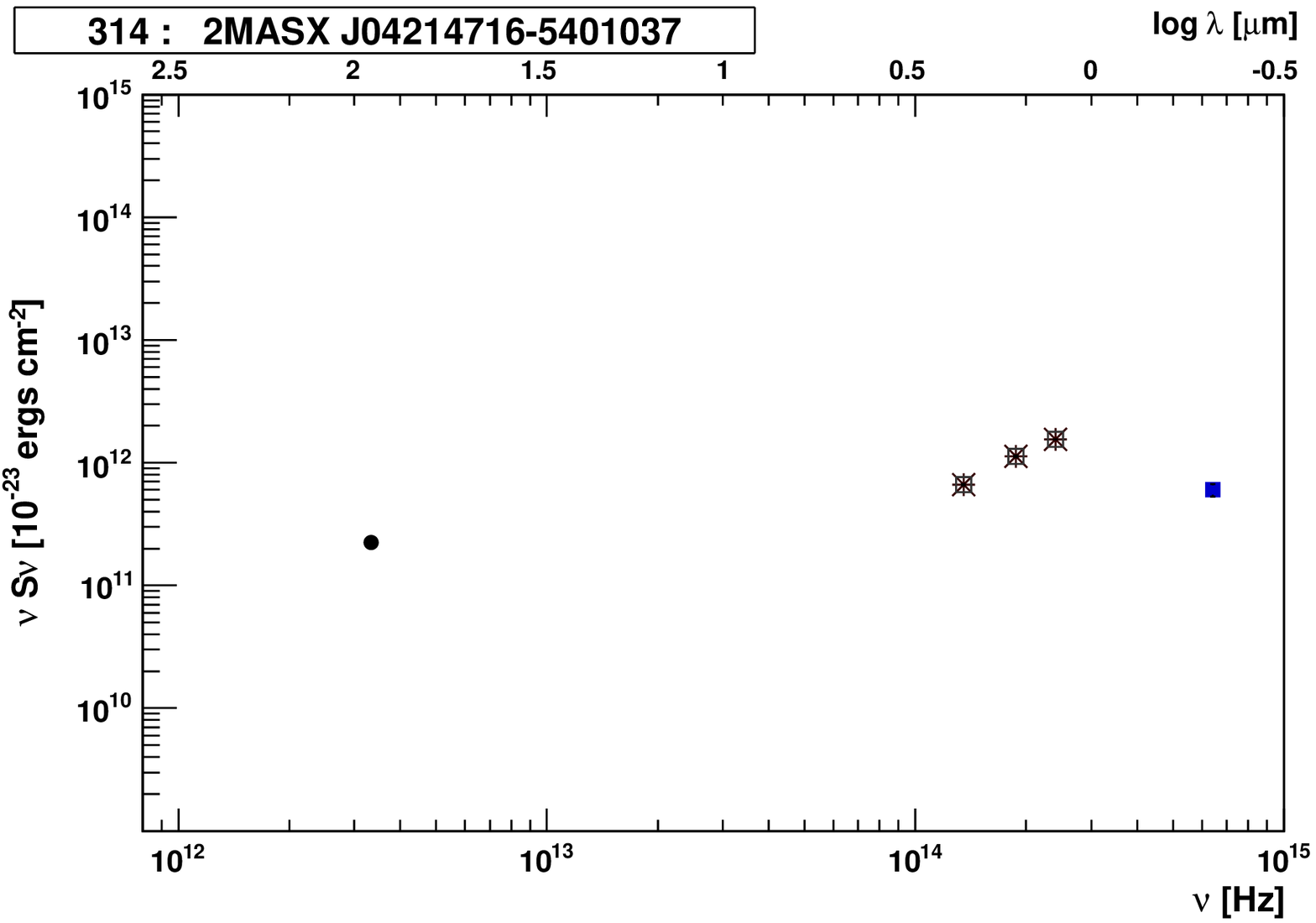}
\includegraphics[width=4cm]{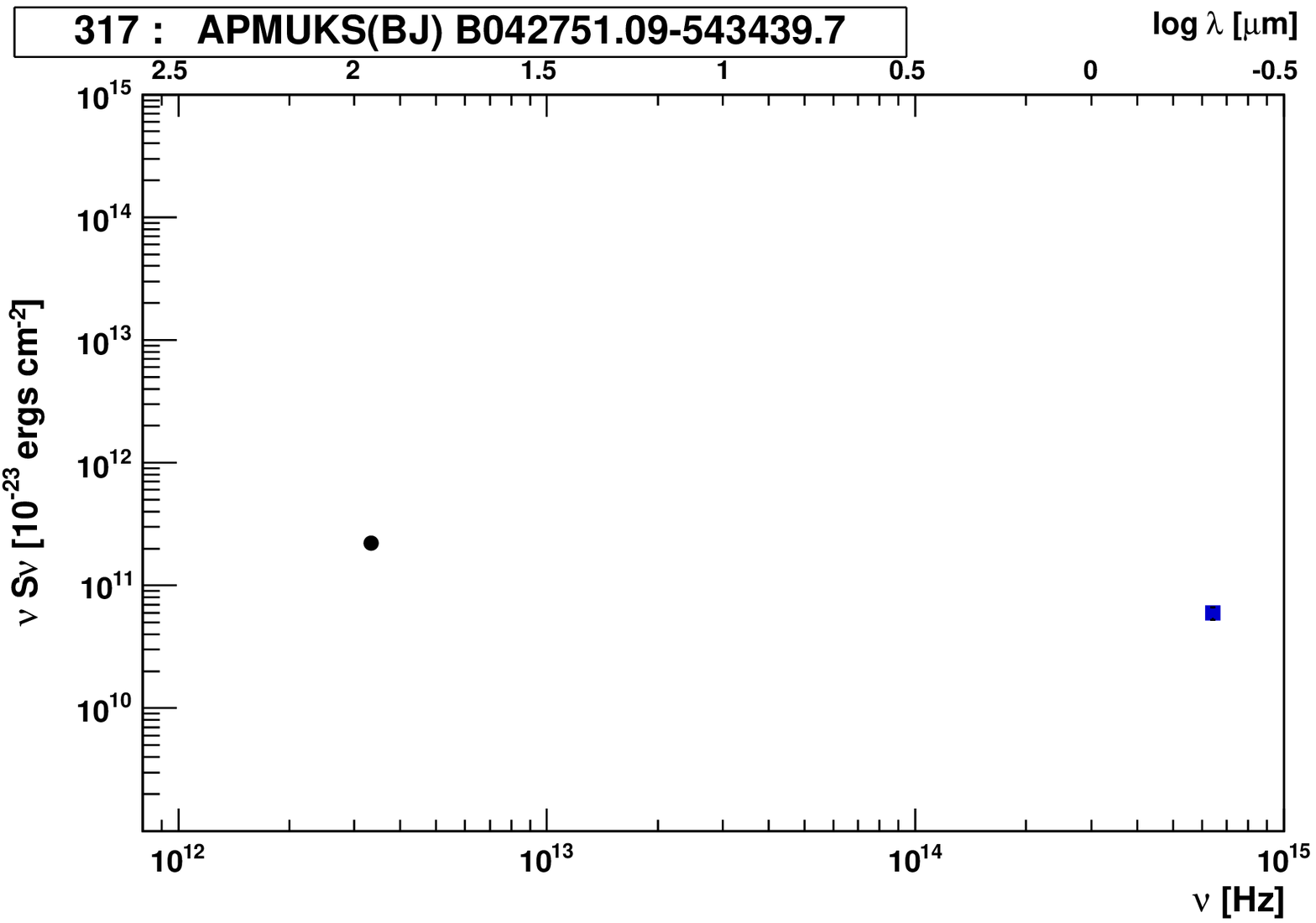}
\includegraphics[width=4cm]{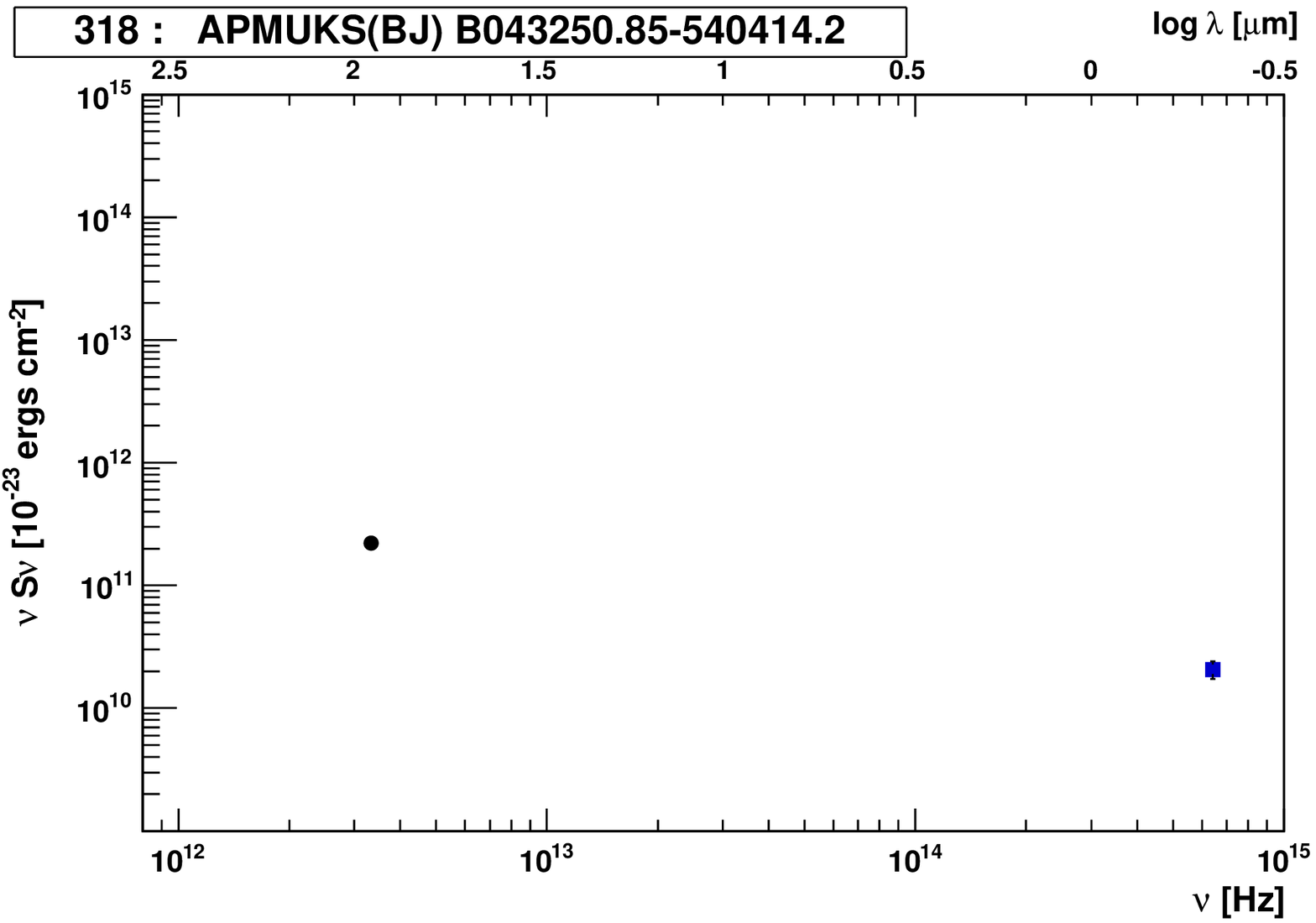}
\includegraphics[width=4cm]{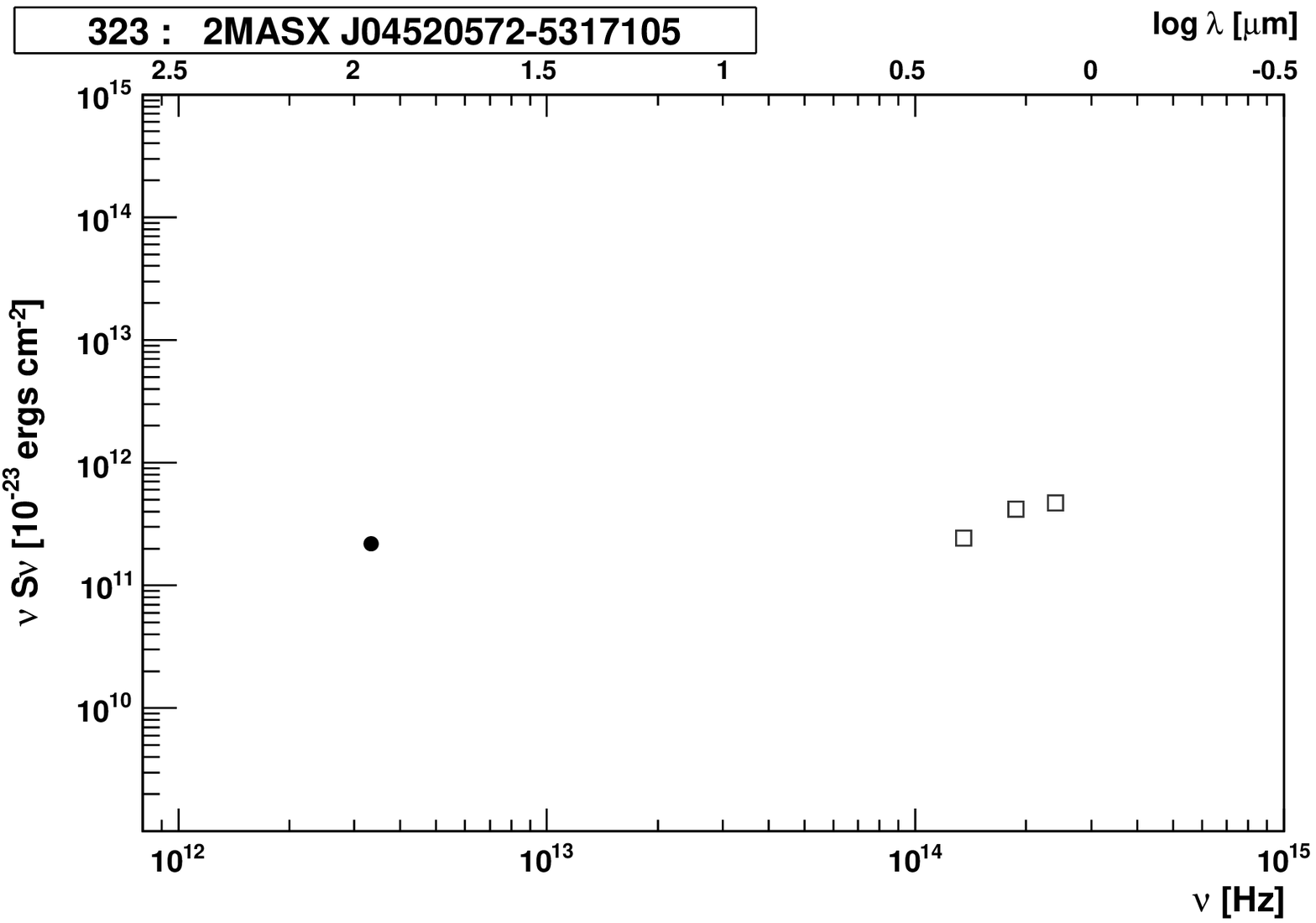}
\includegraphics[width=4cm]{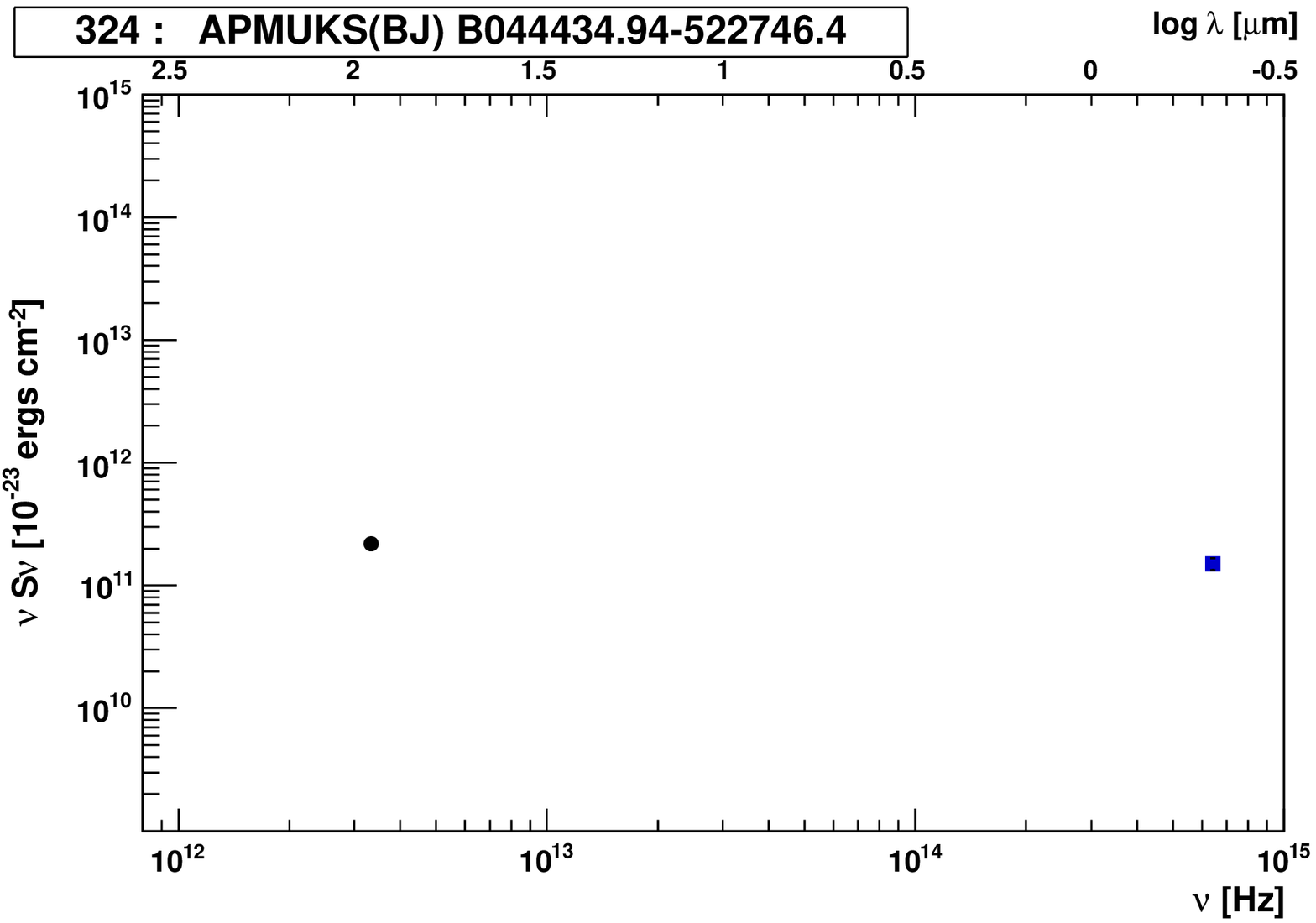}
\includegraphics[width=4cm]{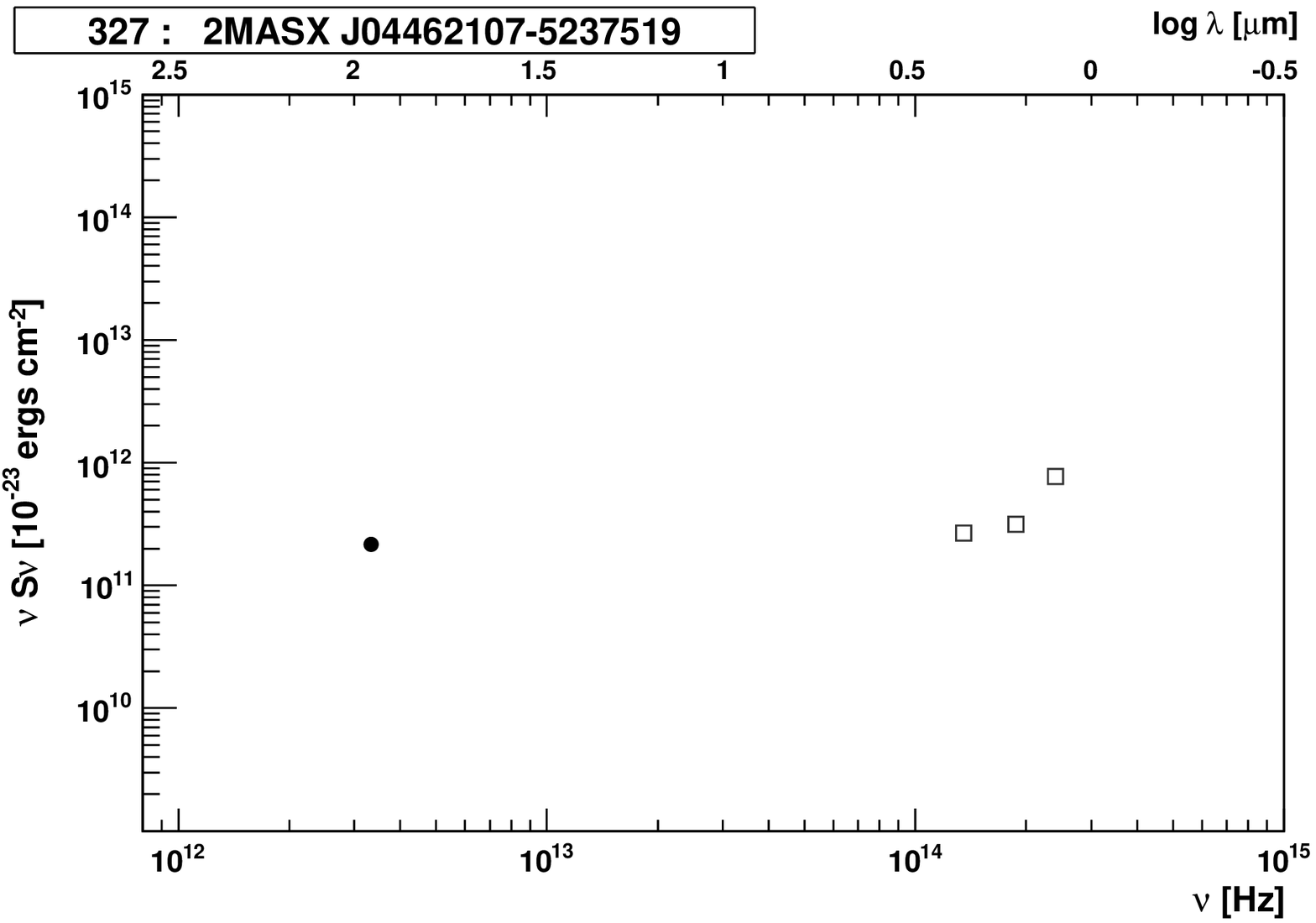}
\includegraphics[width=4cm]{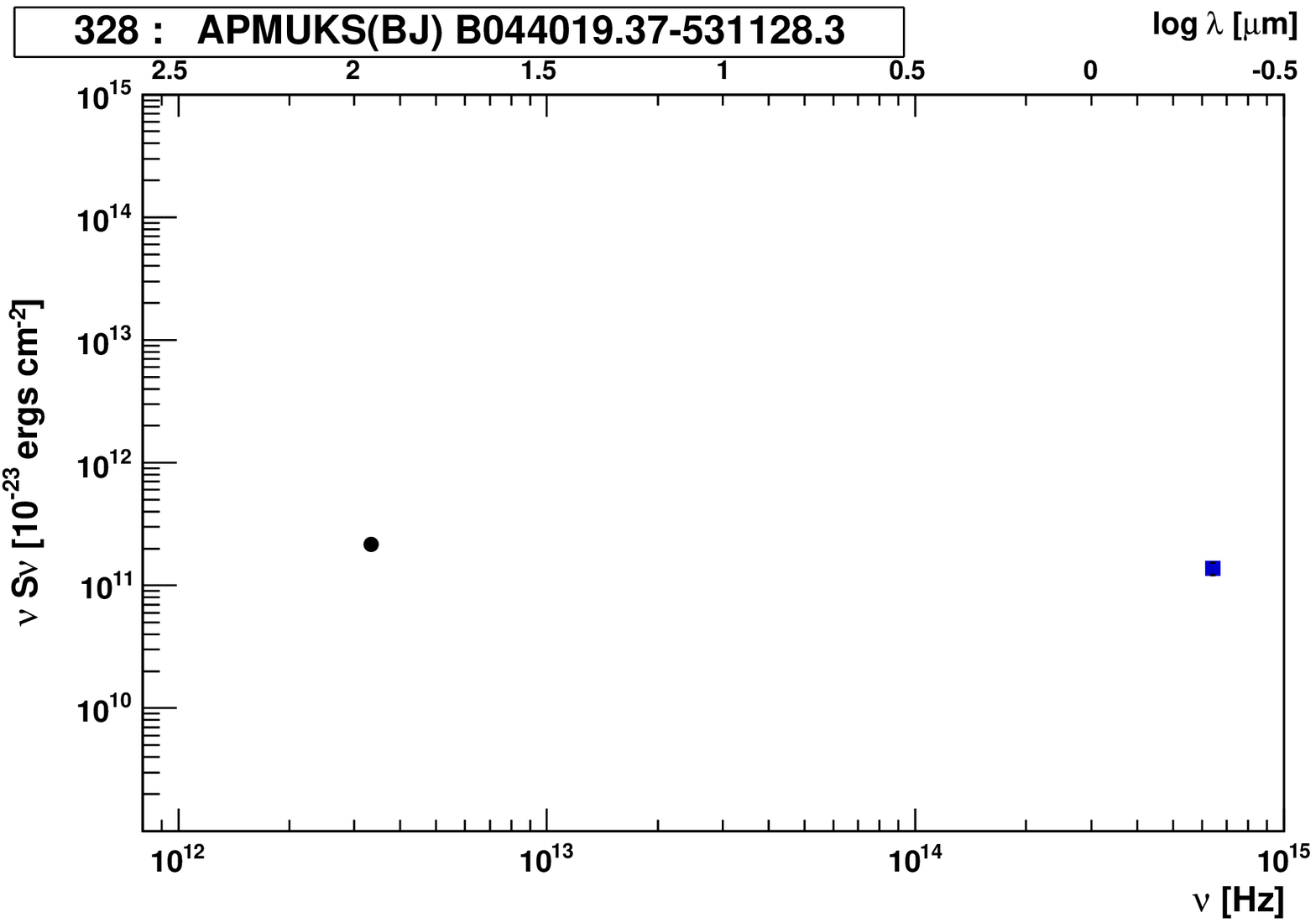}
\includegraphics[width=4cm]{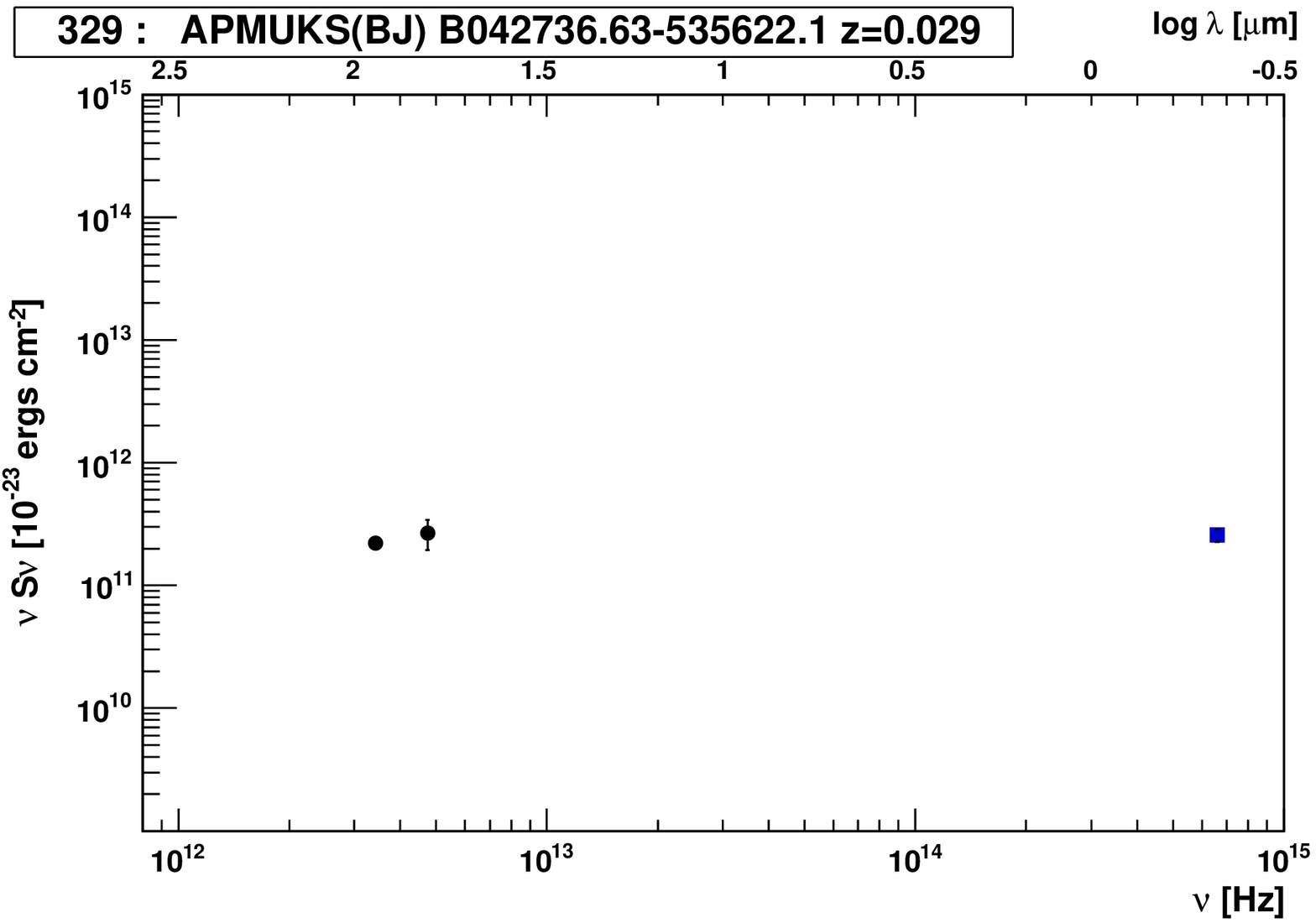}
\includegraphics[width=4cm]{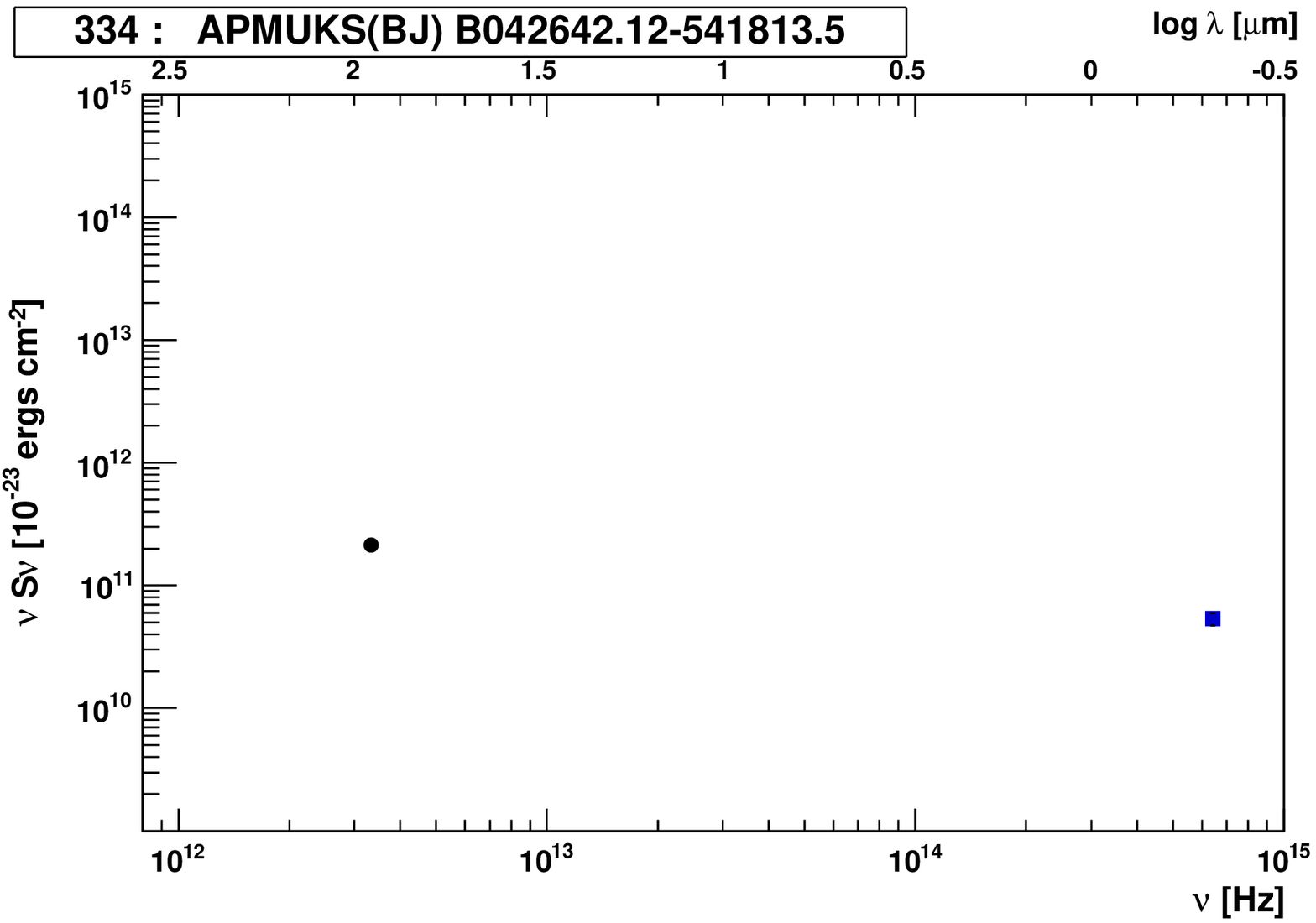}
\includegraphics[width=4cm]{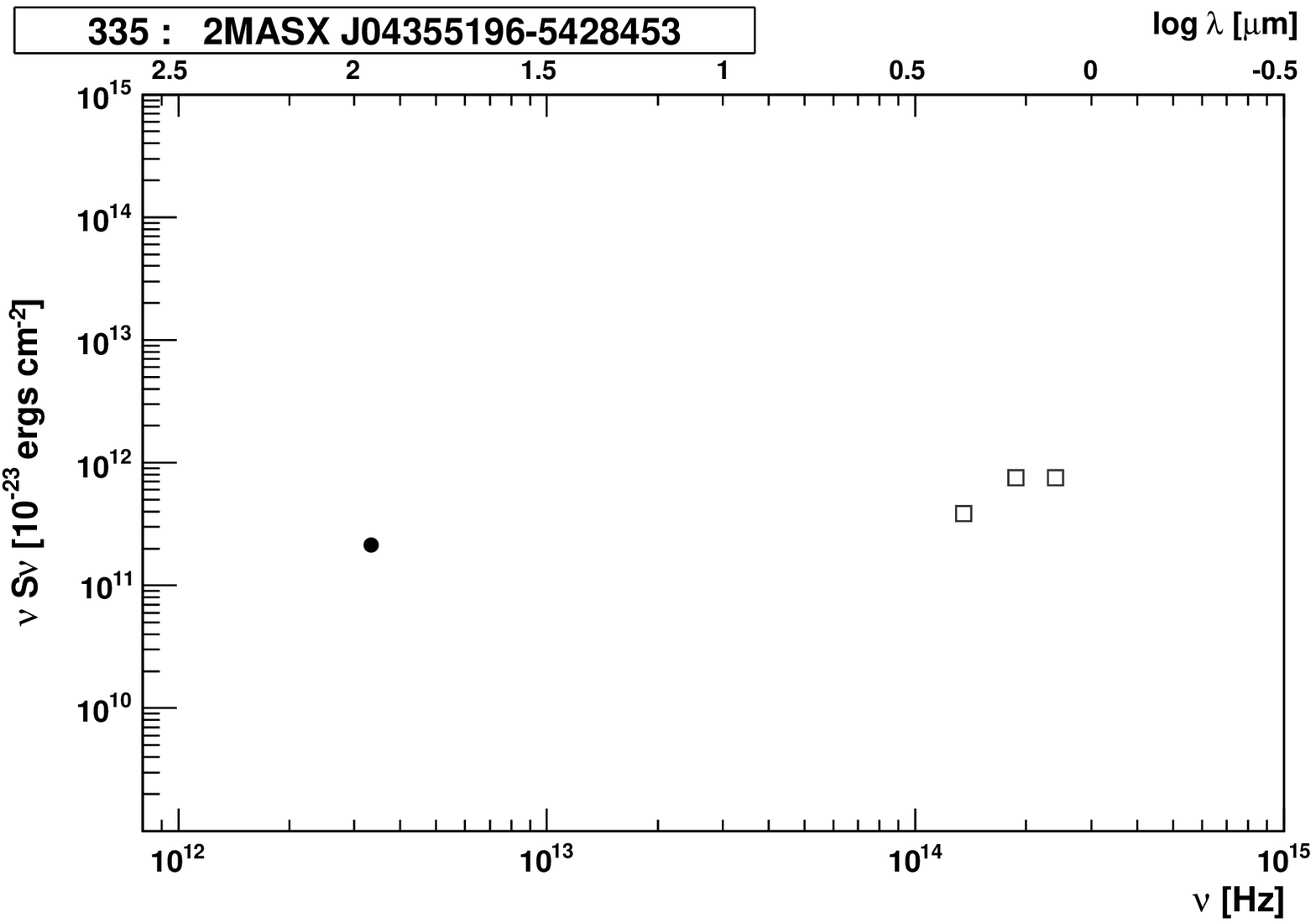}
\includegraphics[width=4cm]{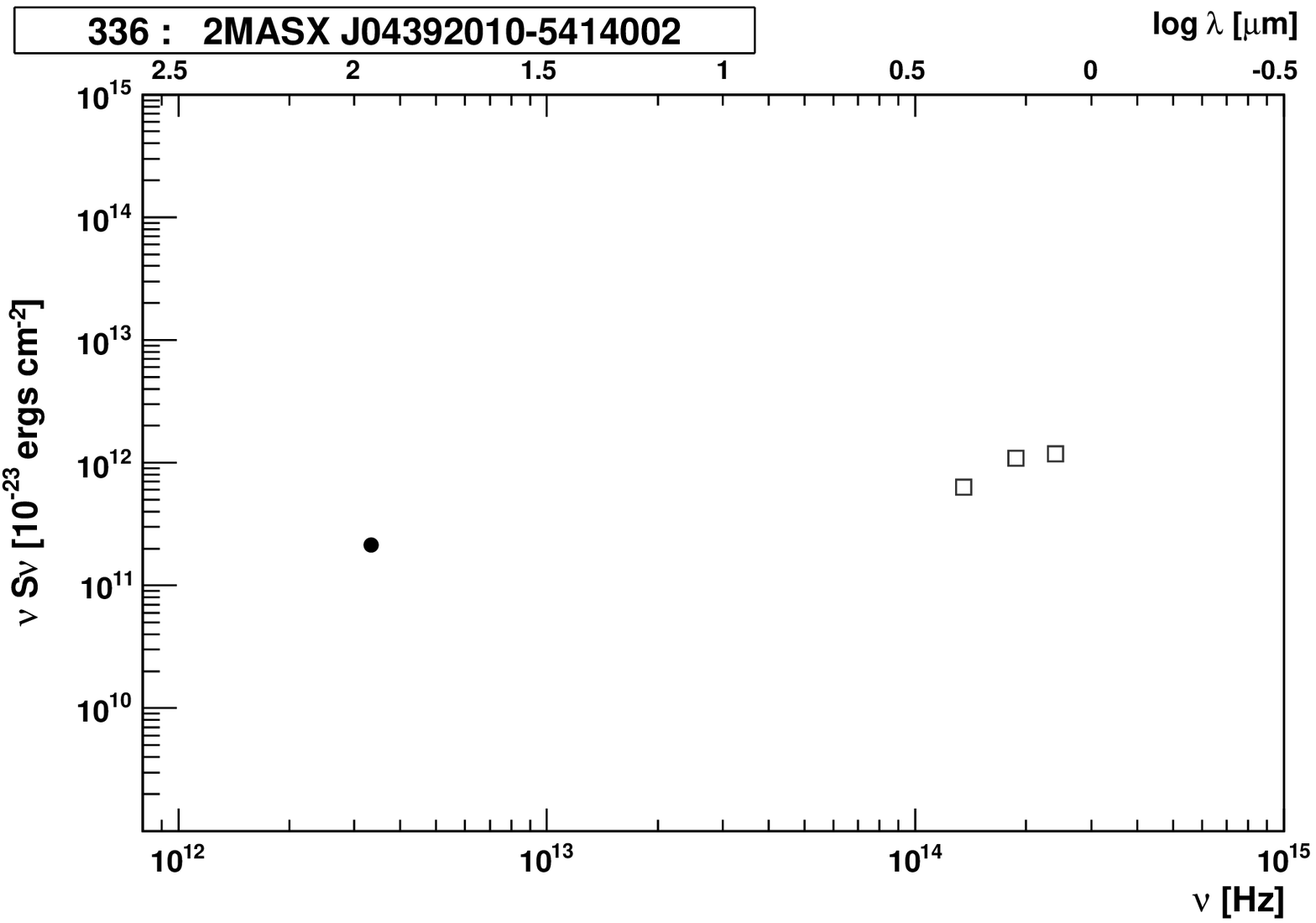}
\includegraphics[width=4cm]{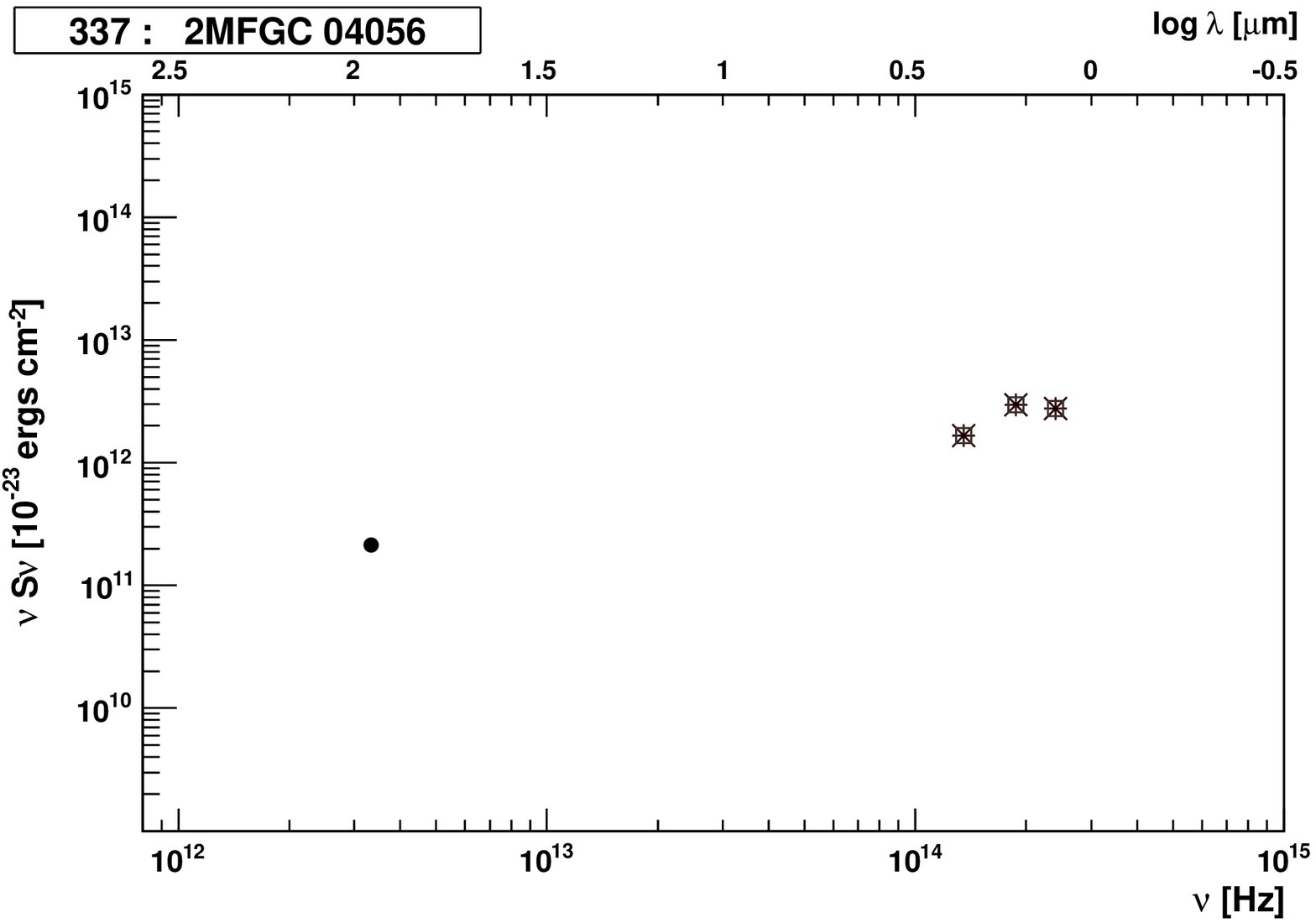}
\includegraphics[width=4cm]{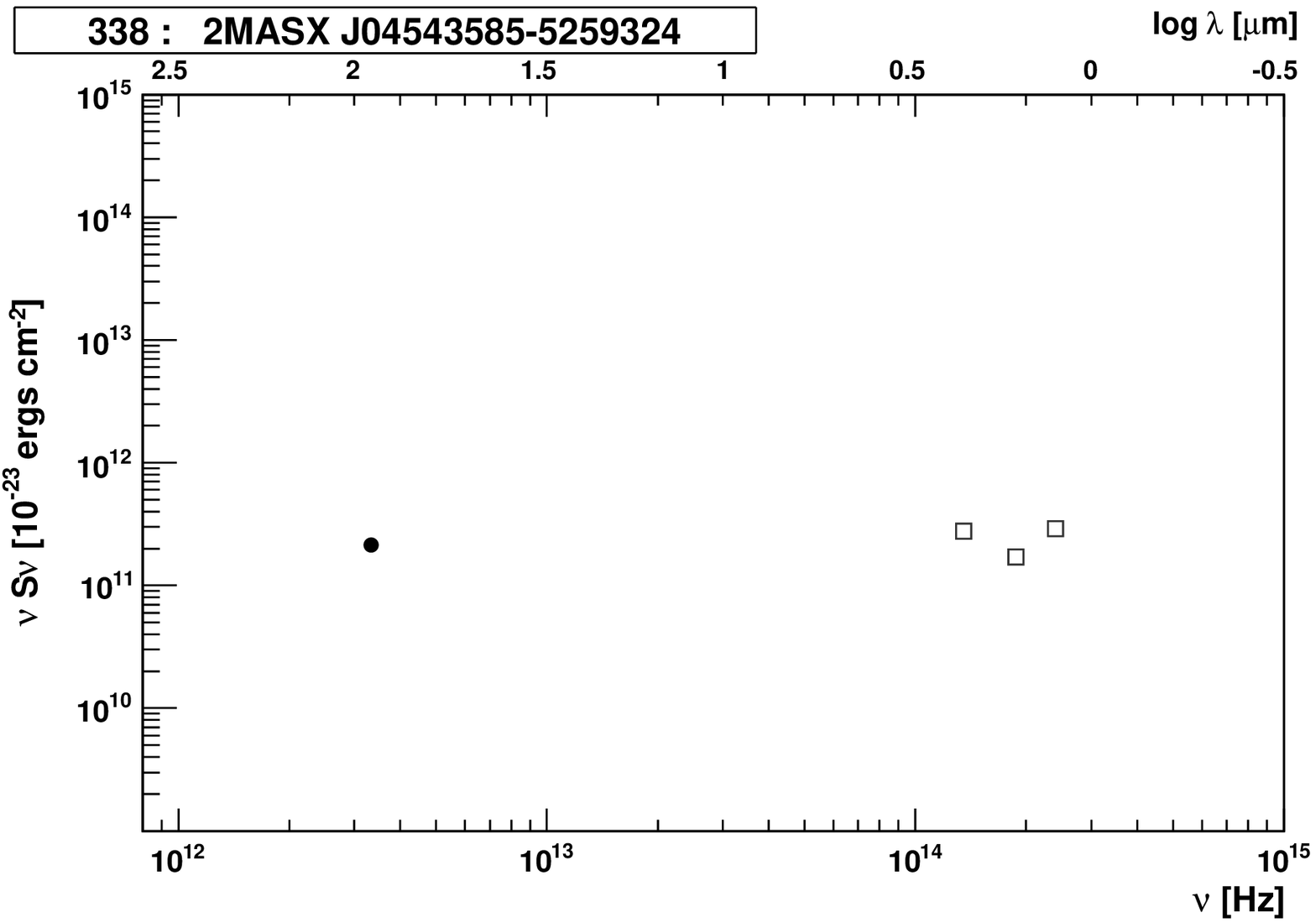}
\includegraphics[width=4cm]{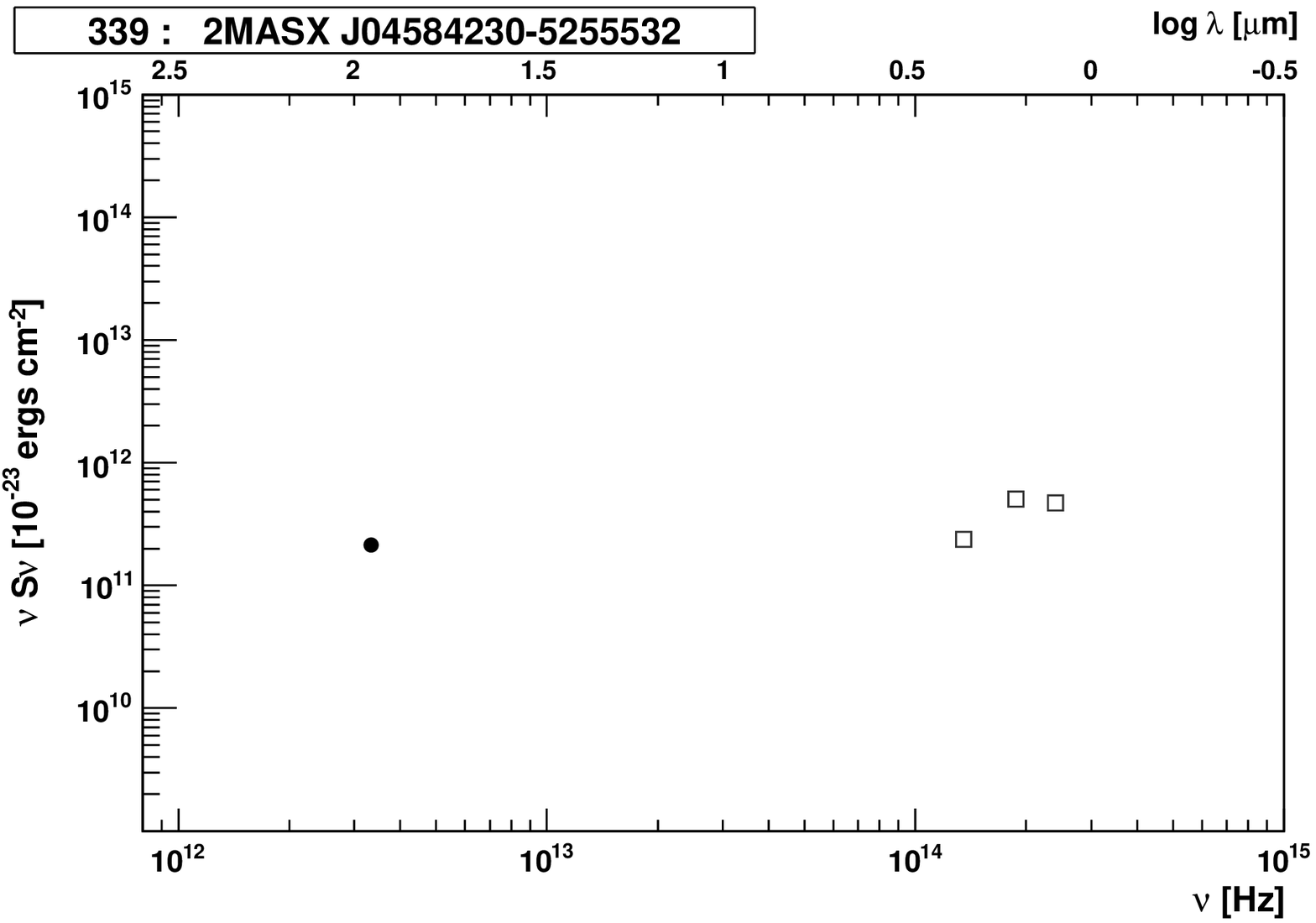}
\includegraphics[width=4cm]{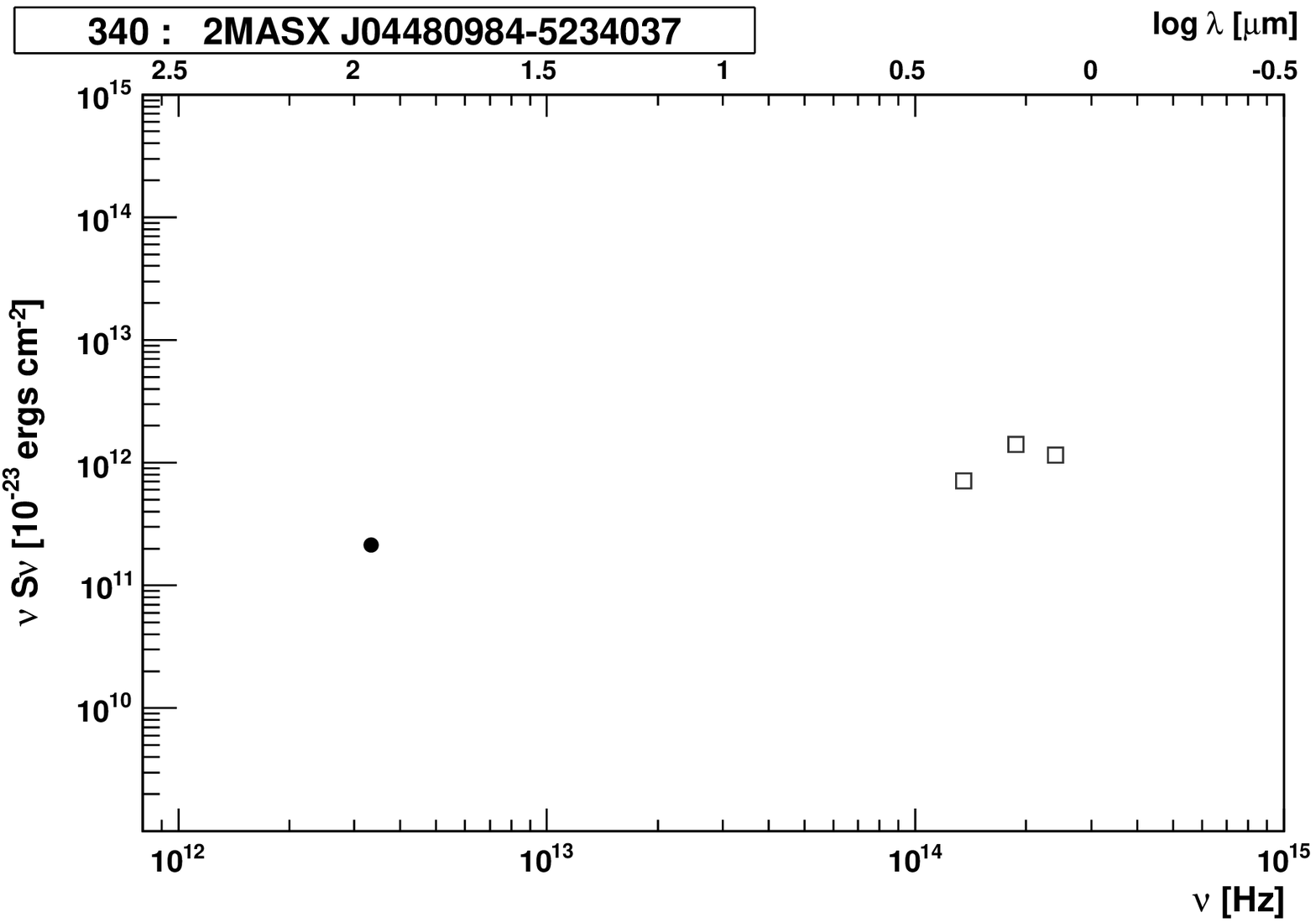}
\includegraphics[width=4cm]{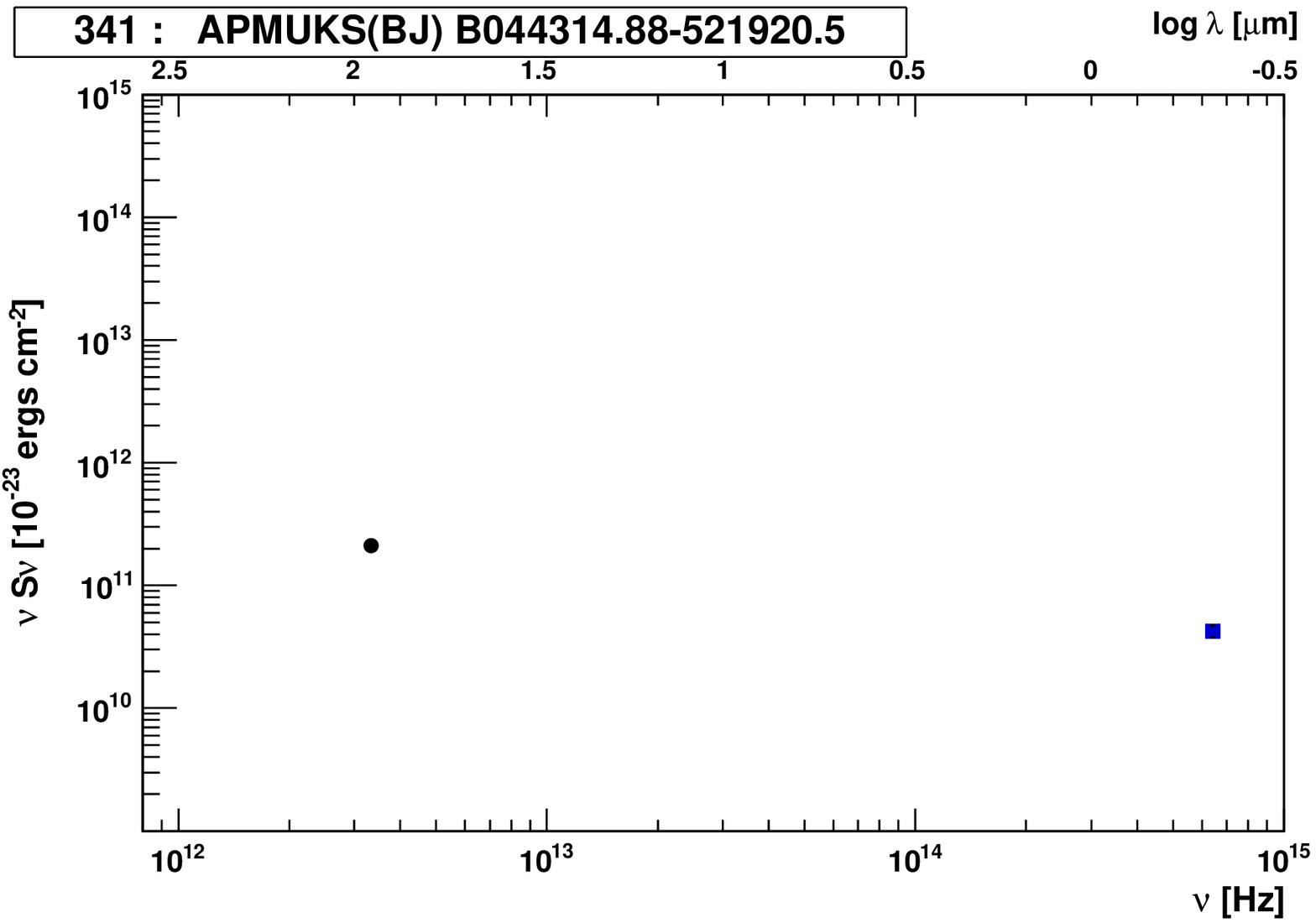}
\includegraphics[width=4cm]{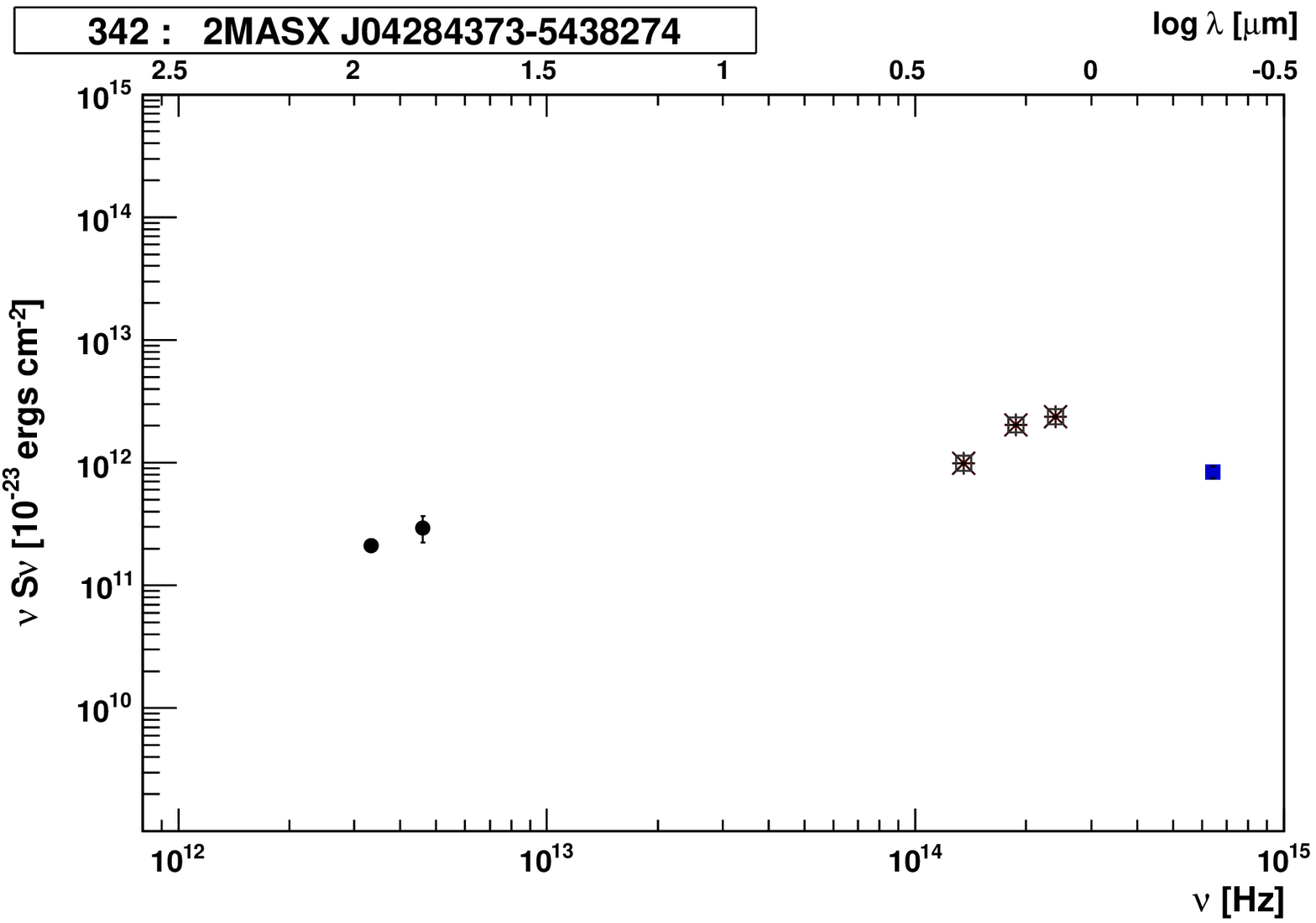}
\includegraphics[width=4cm]{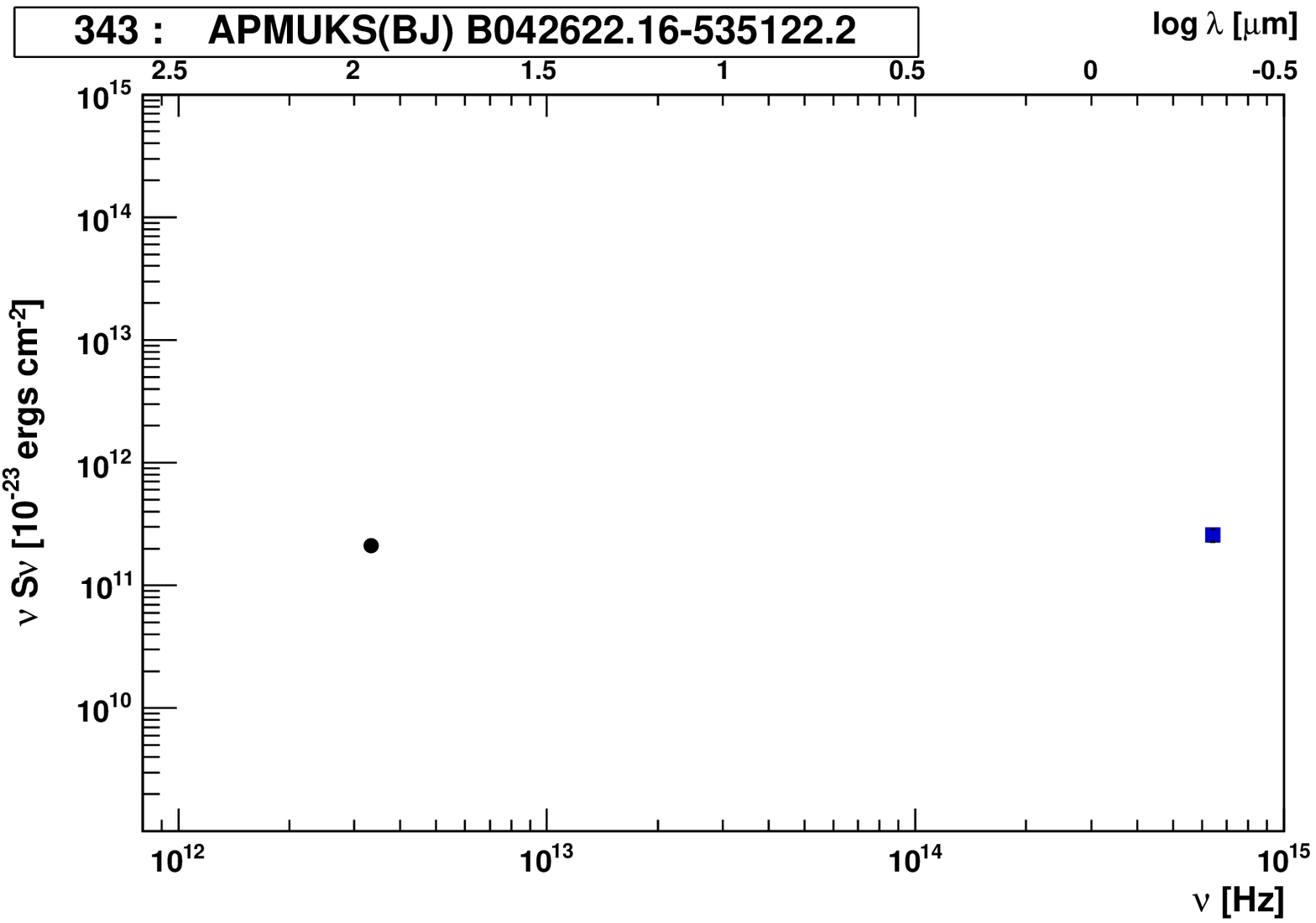}
\includegraphics[width=4cm]{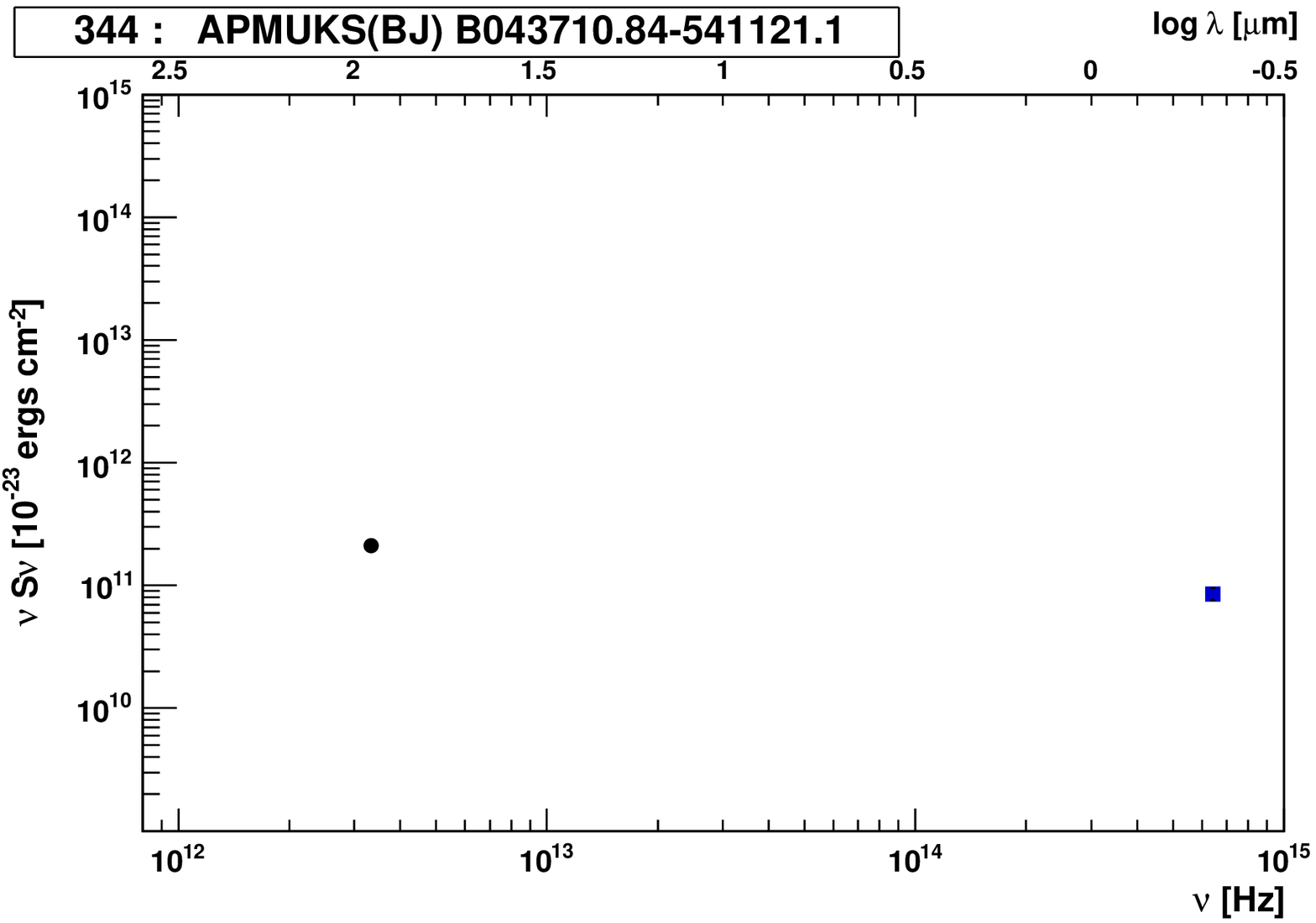}
\includegraphics[width=4cm]{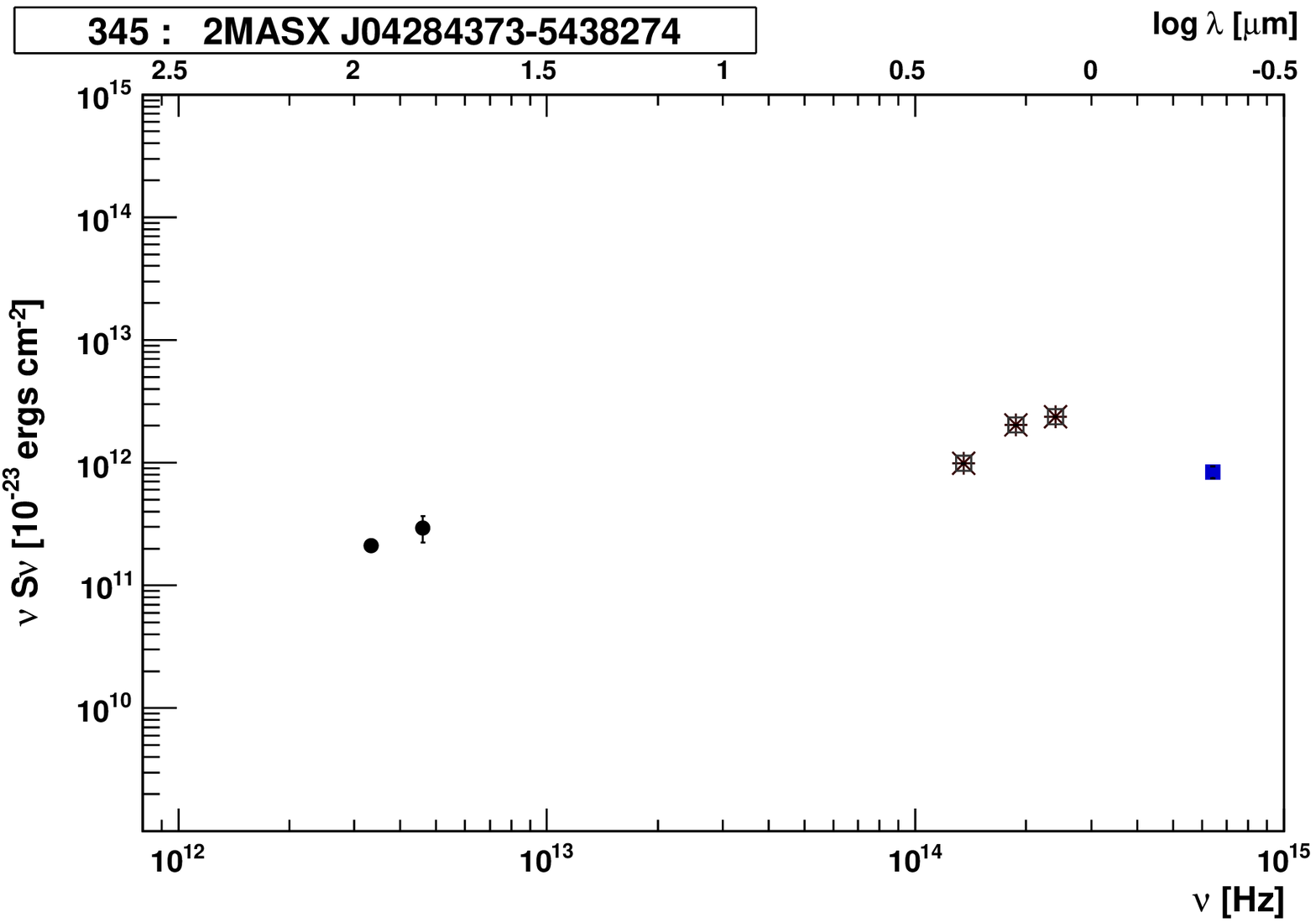}
\includegraphics[width=4cm]{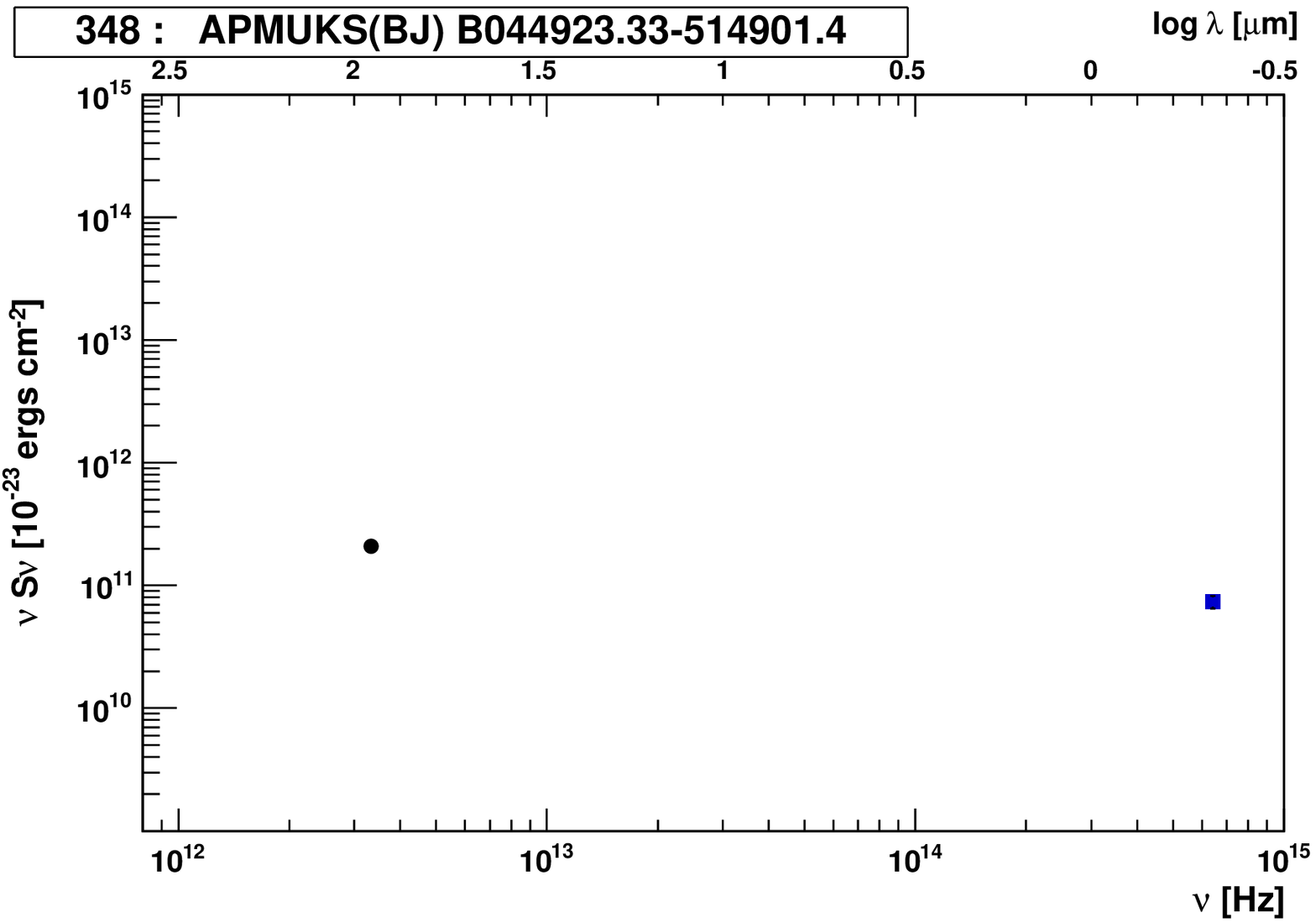}

\label{points7}
\caption {SEDs for the next 36 ADF-S identified sources, with symbols as in Figure~\ref{points1}.}
\end{figure*}
}

\clearpage

\onlfig{8}{
\begin{figure*}[t]
\centering
\includegraphics[width=4cm]{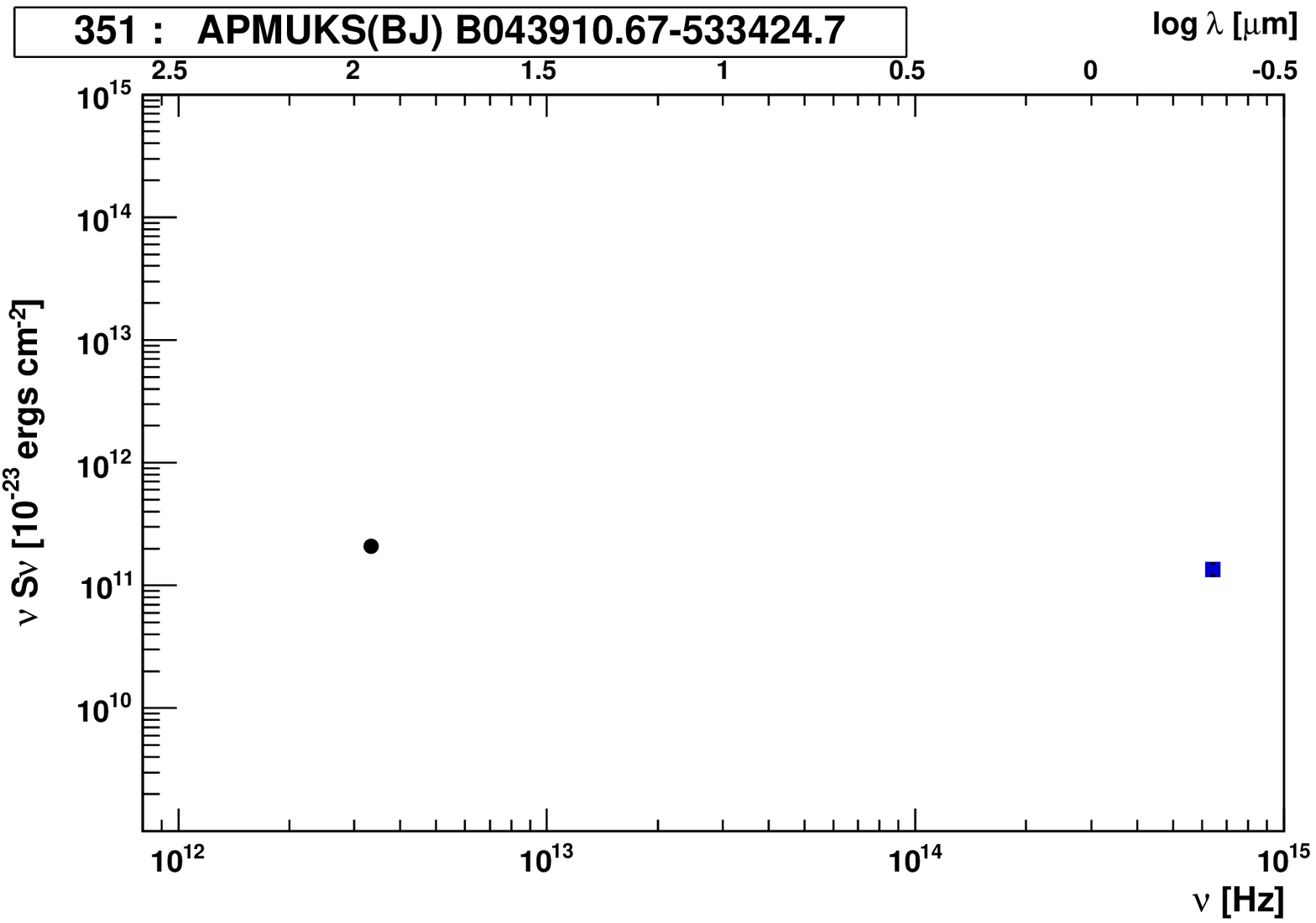}
\includegraphics[width=4cm]{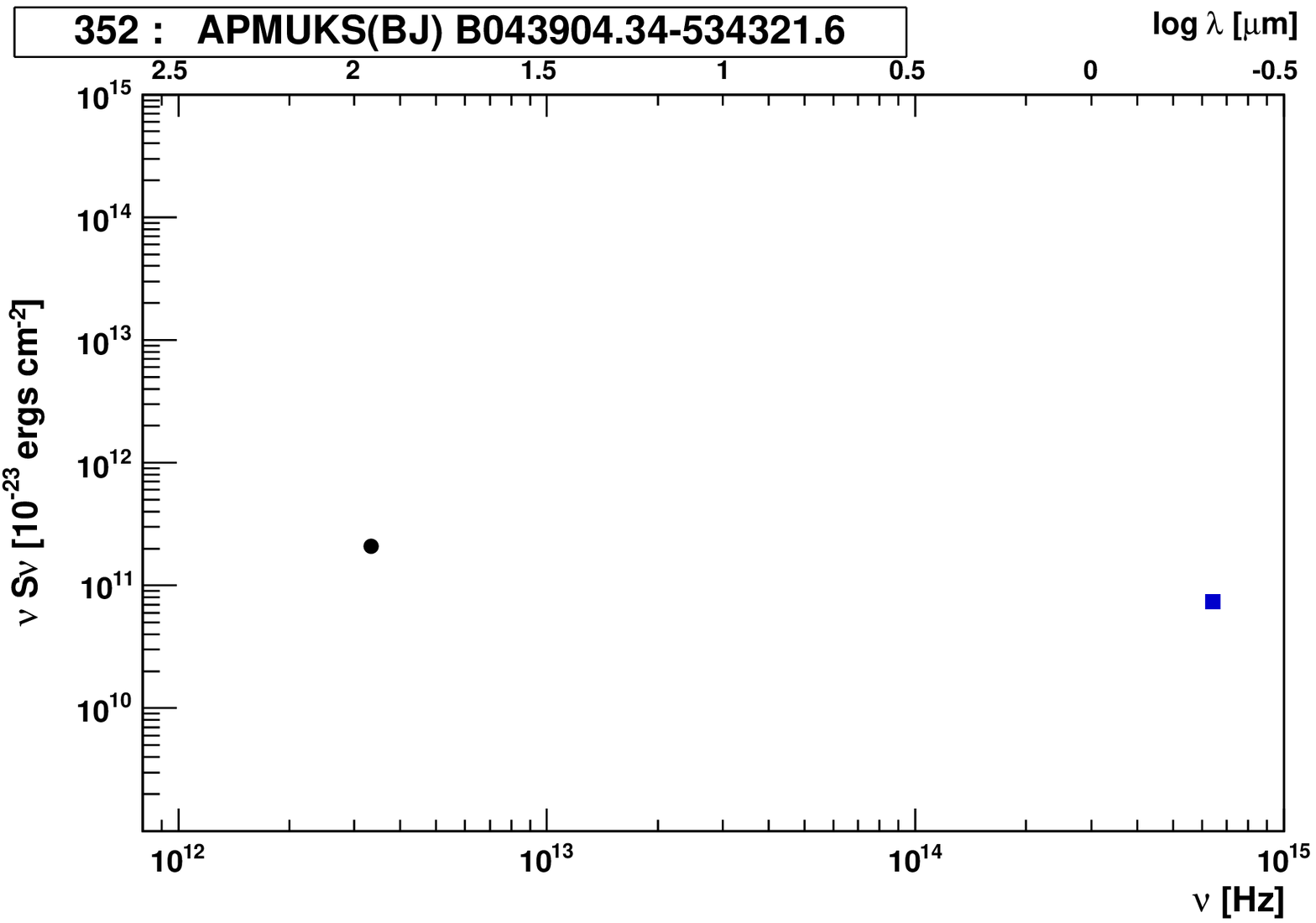}
\includegraphics[width=4cm]{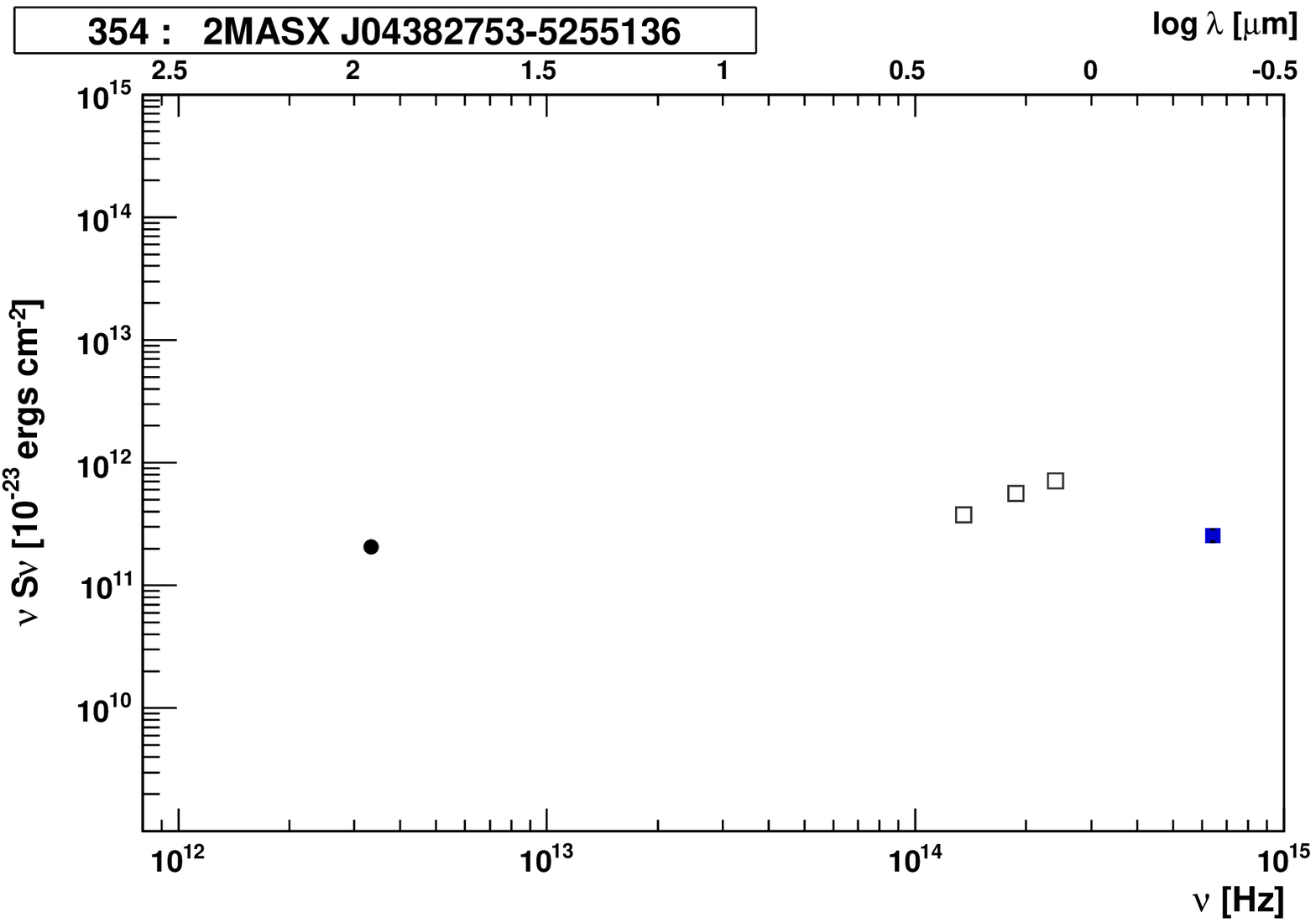}
\includegraphics[width=4cm]{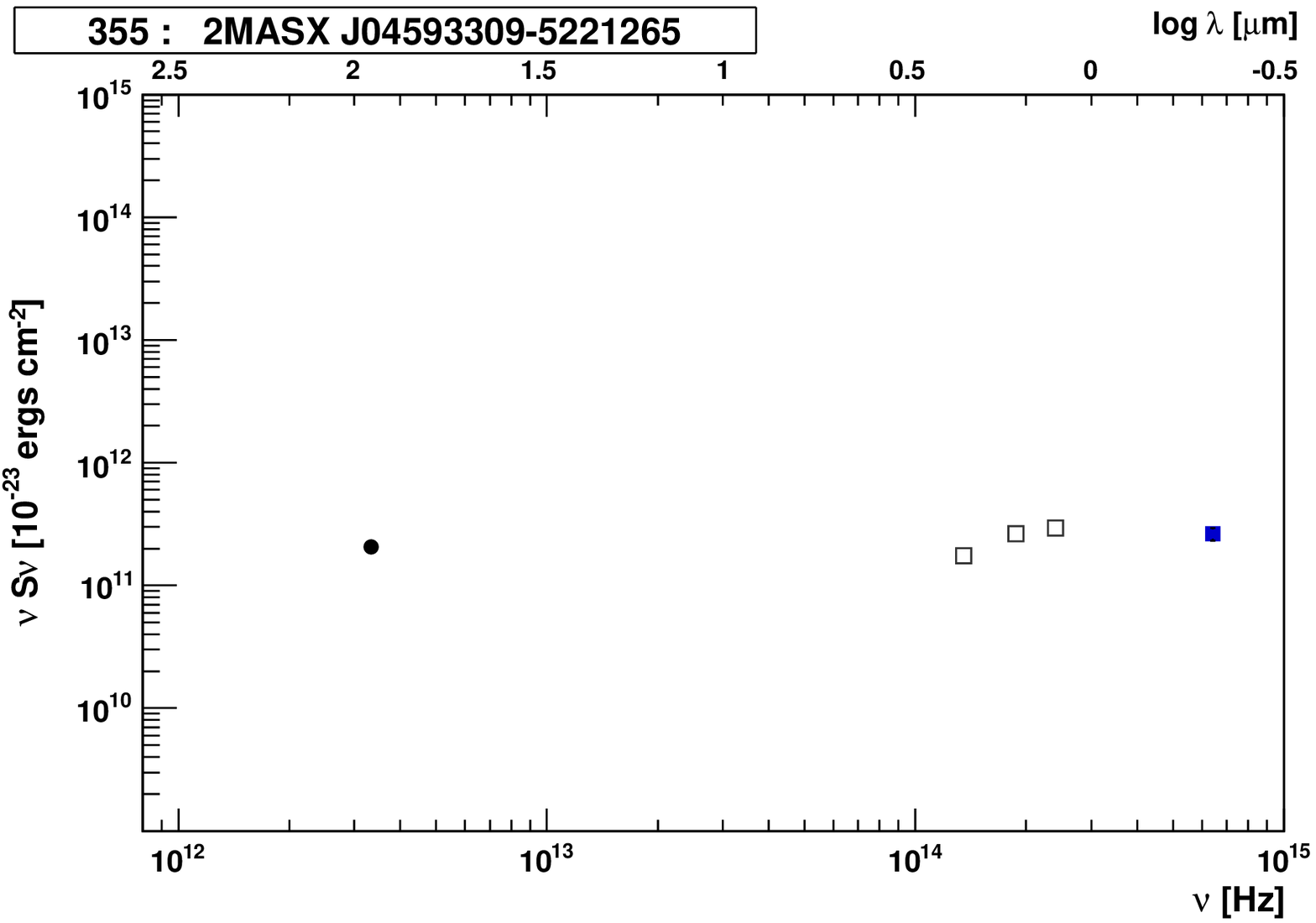}
\includegraphics[width=4cm]{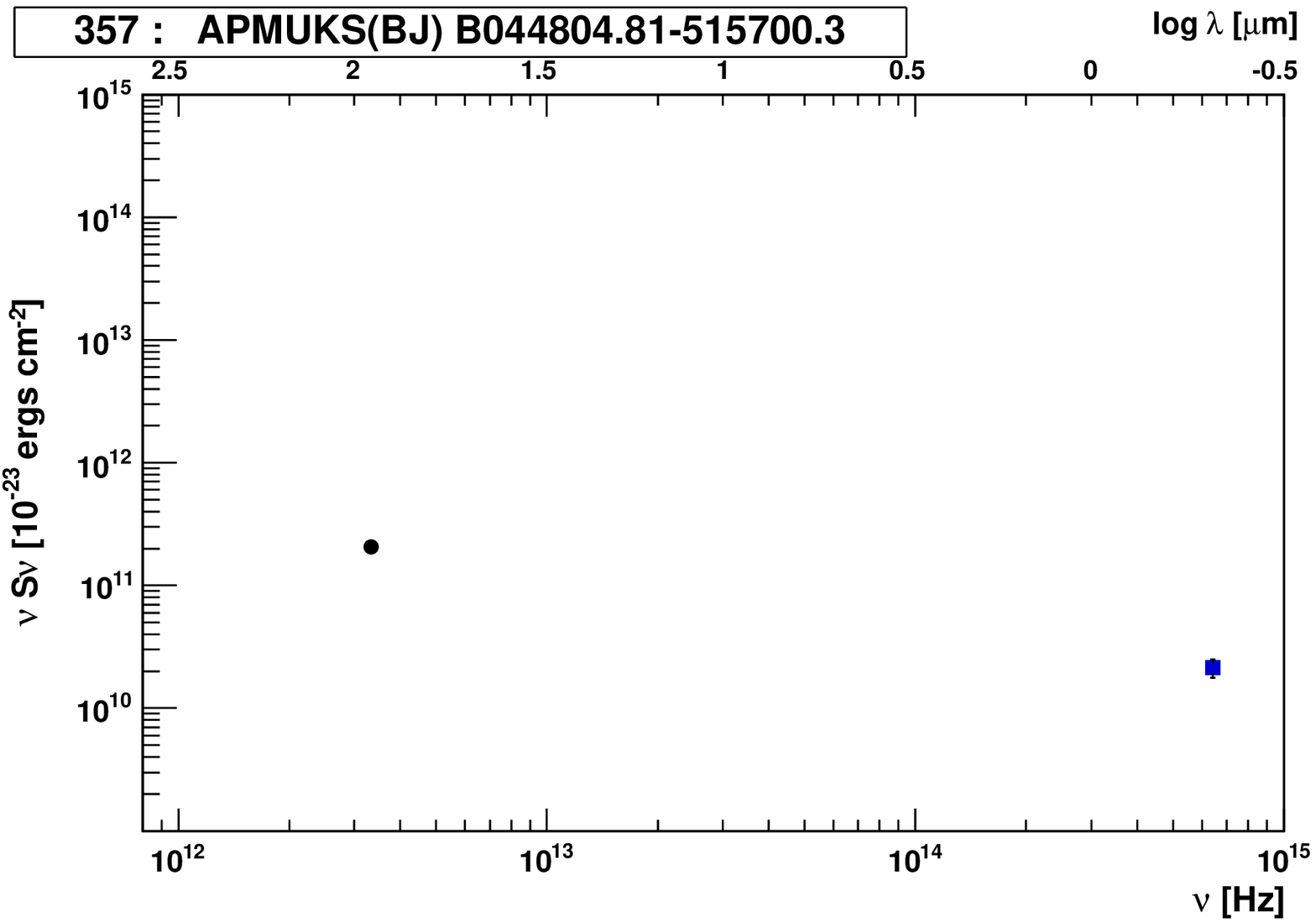}
\includegraphics[width=4cm]{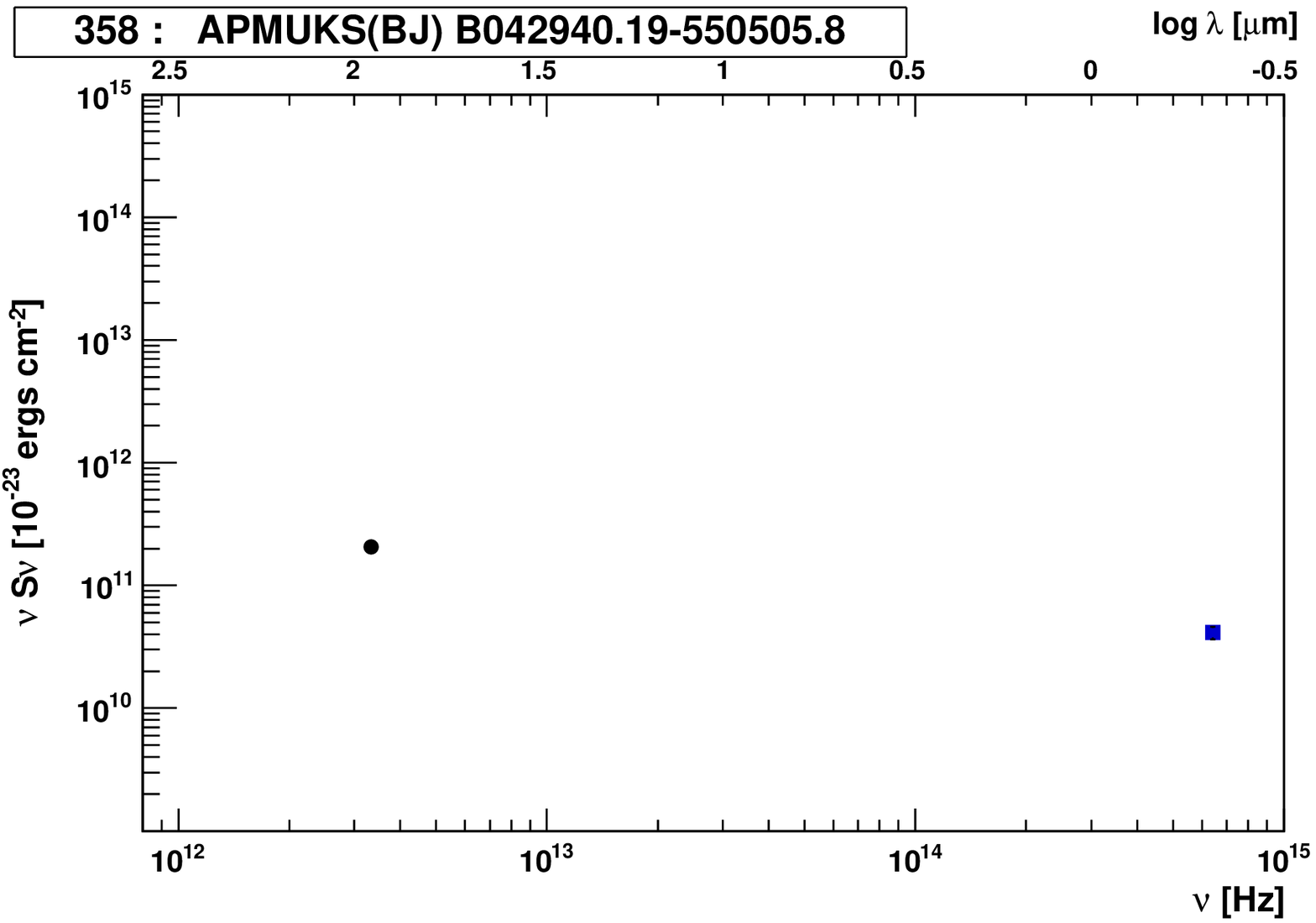}
\includegraphics[width=4cm]{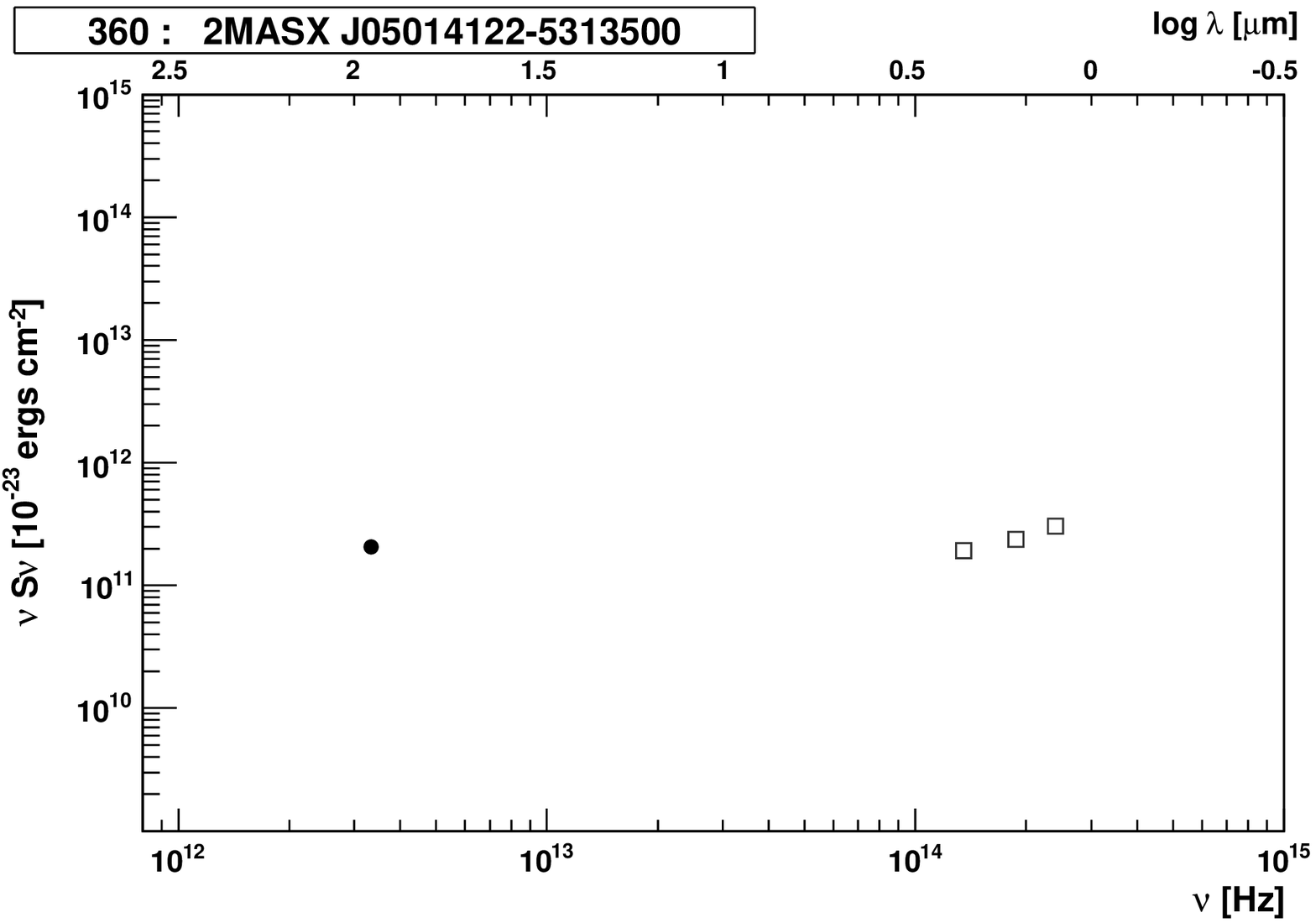}
\includegraphics[width=4cm]{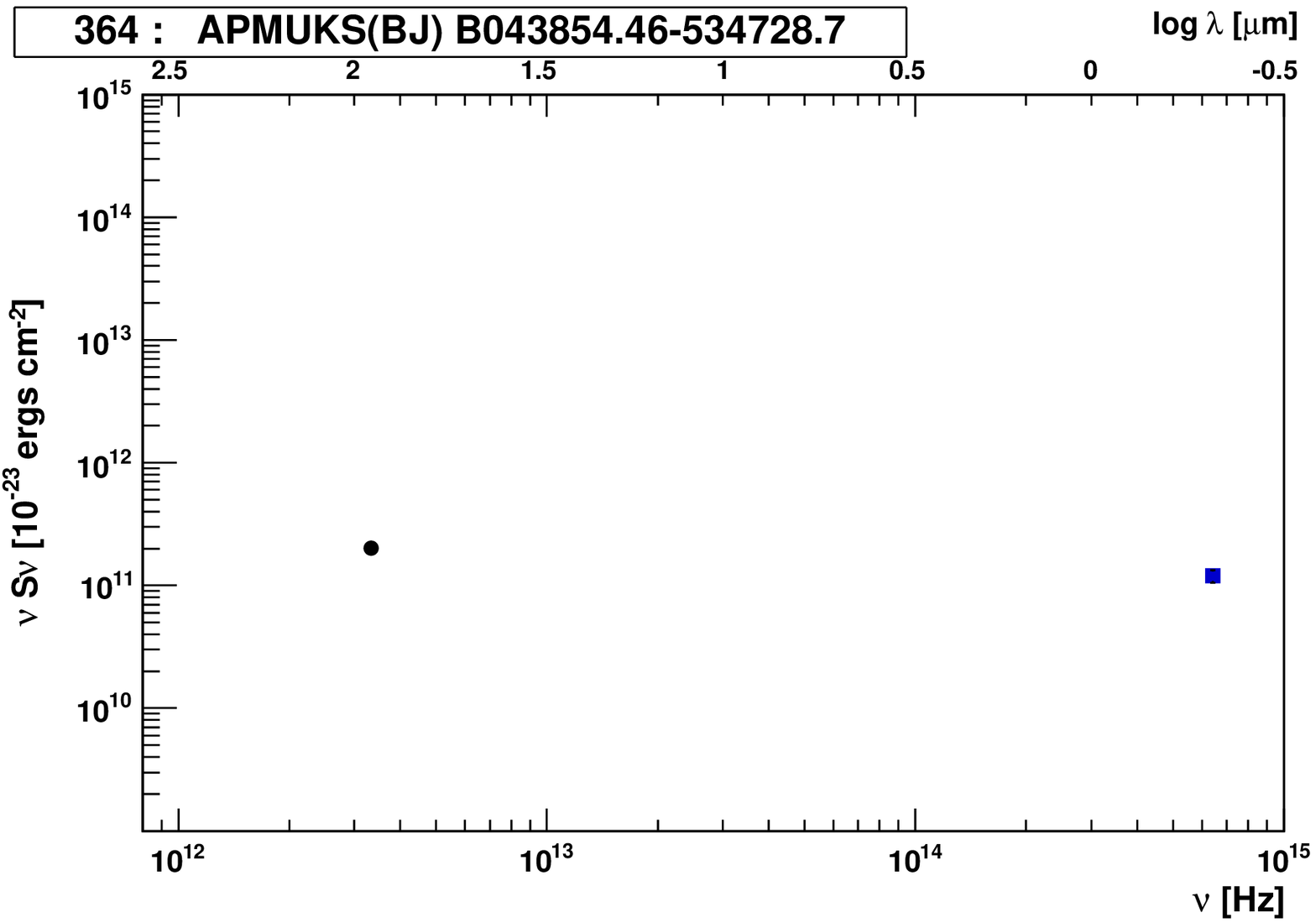}
\includegraphics[width=4cm]{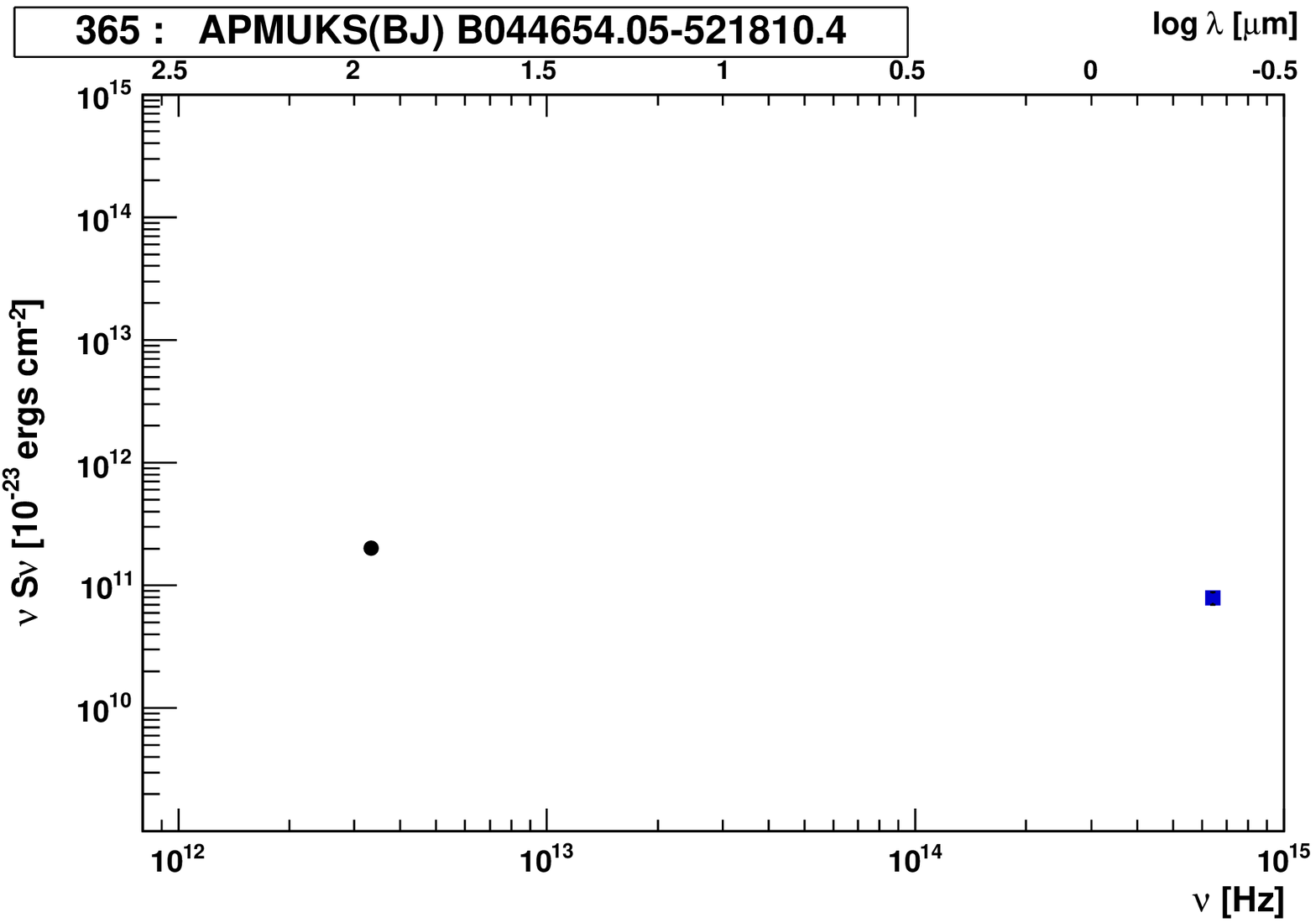}
\includegraphics[width=4cm]{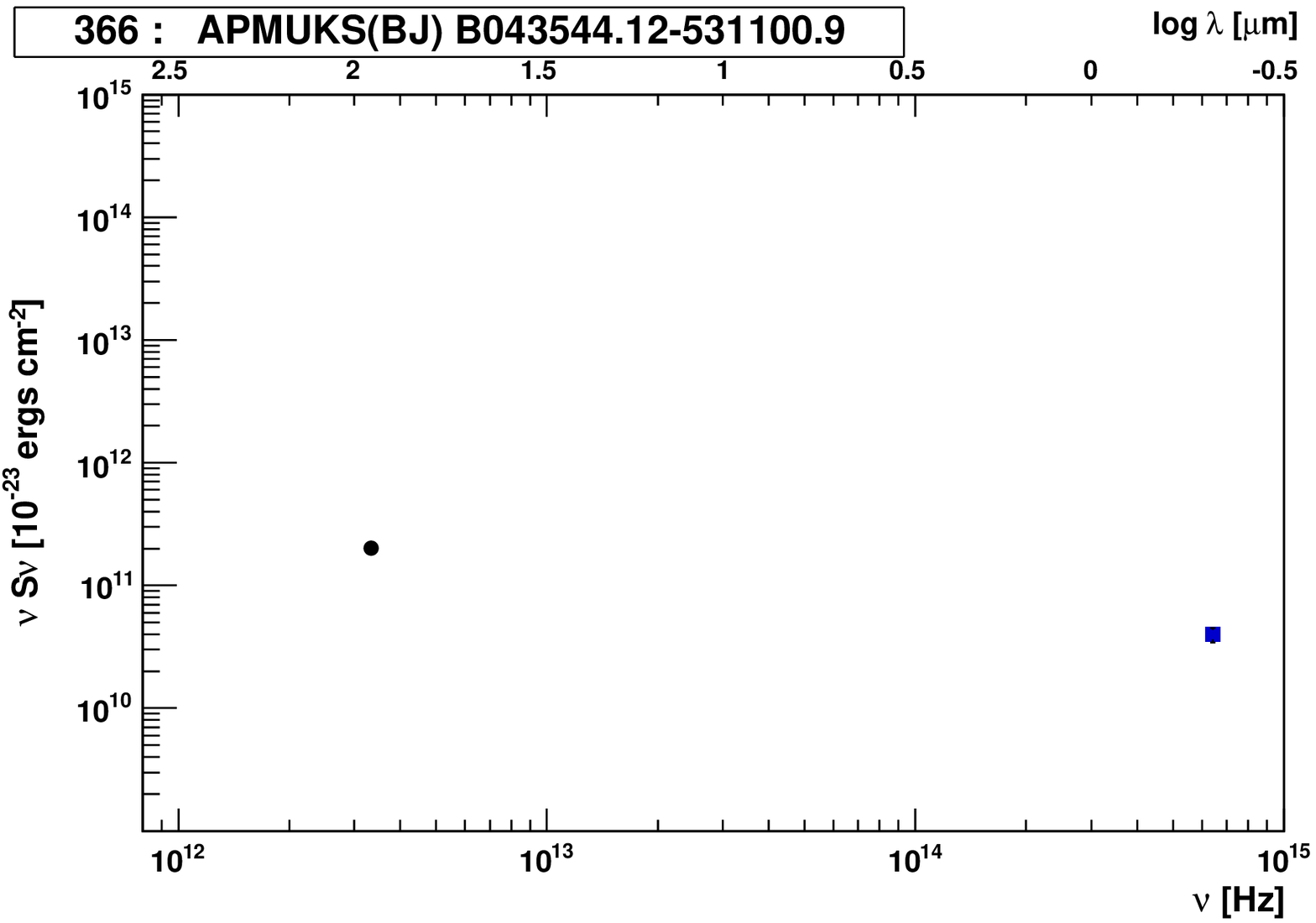}
\includegraphics[width=4cm]{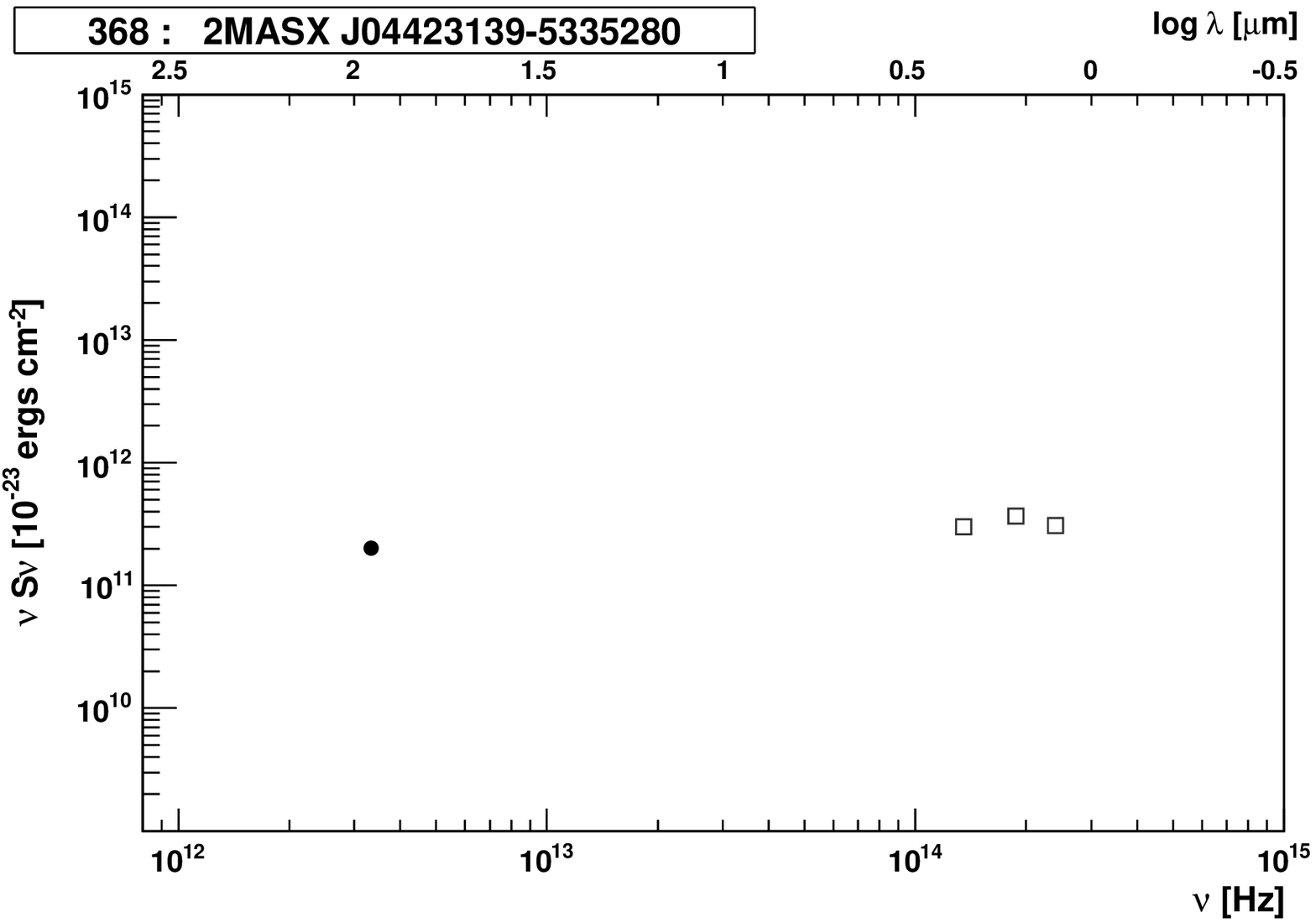}
\includegraphics[width=4cm]{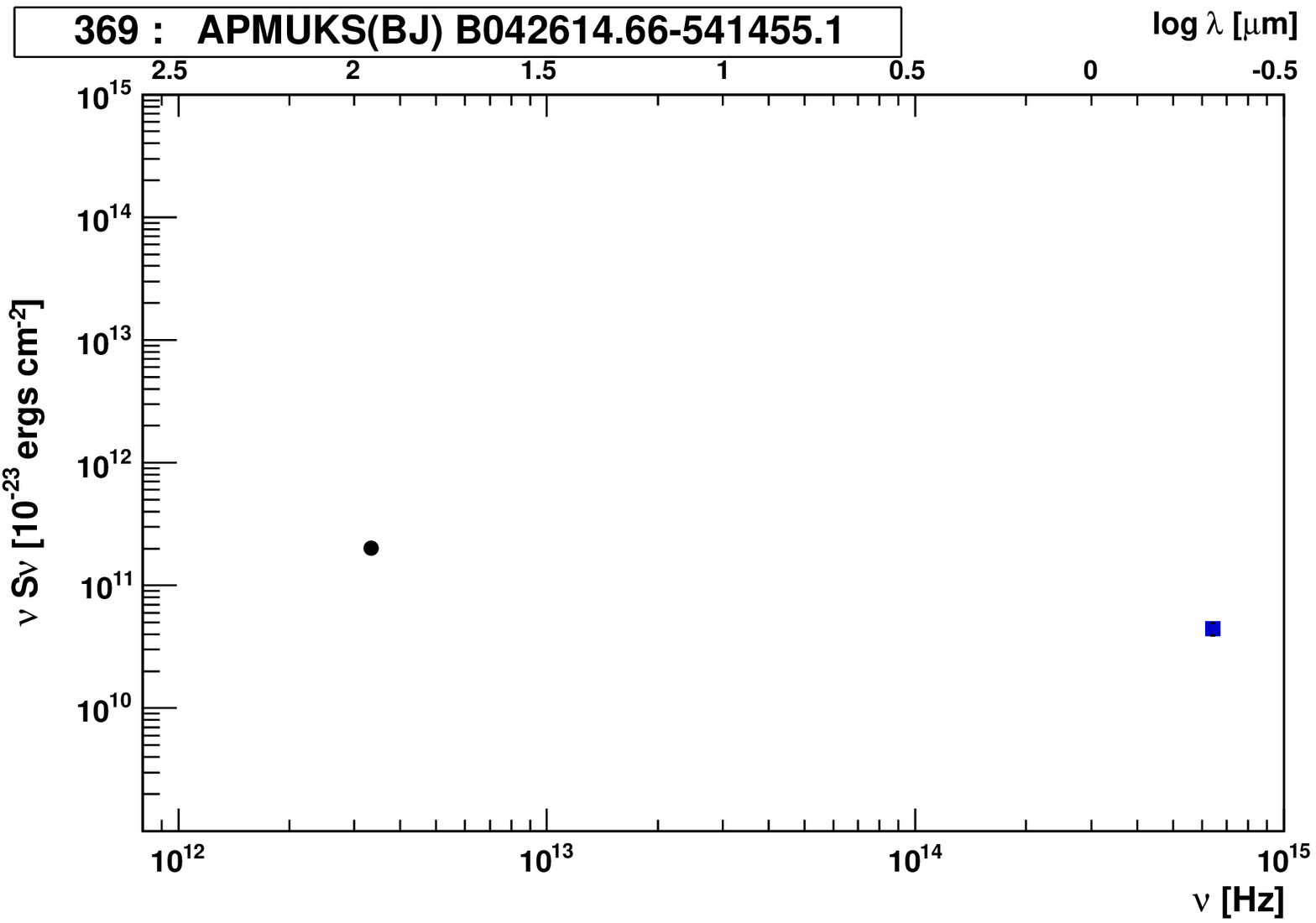}
\includegraphics[width=4cm]{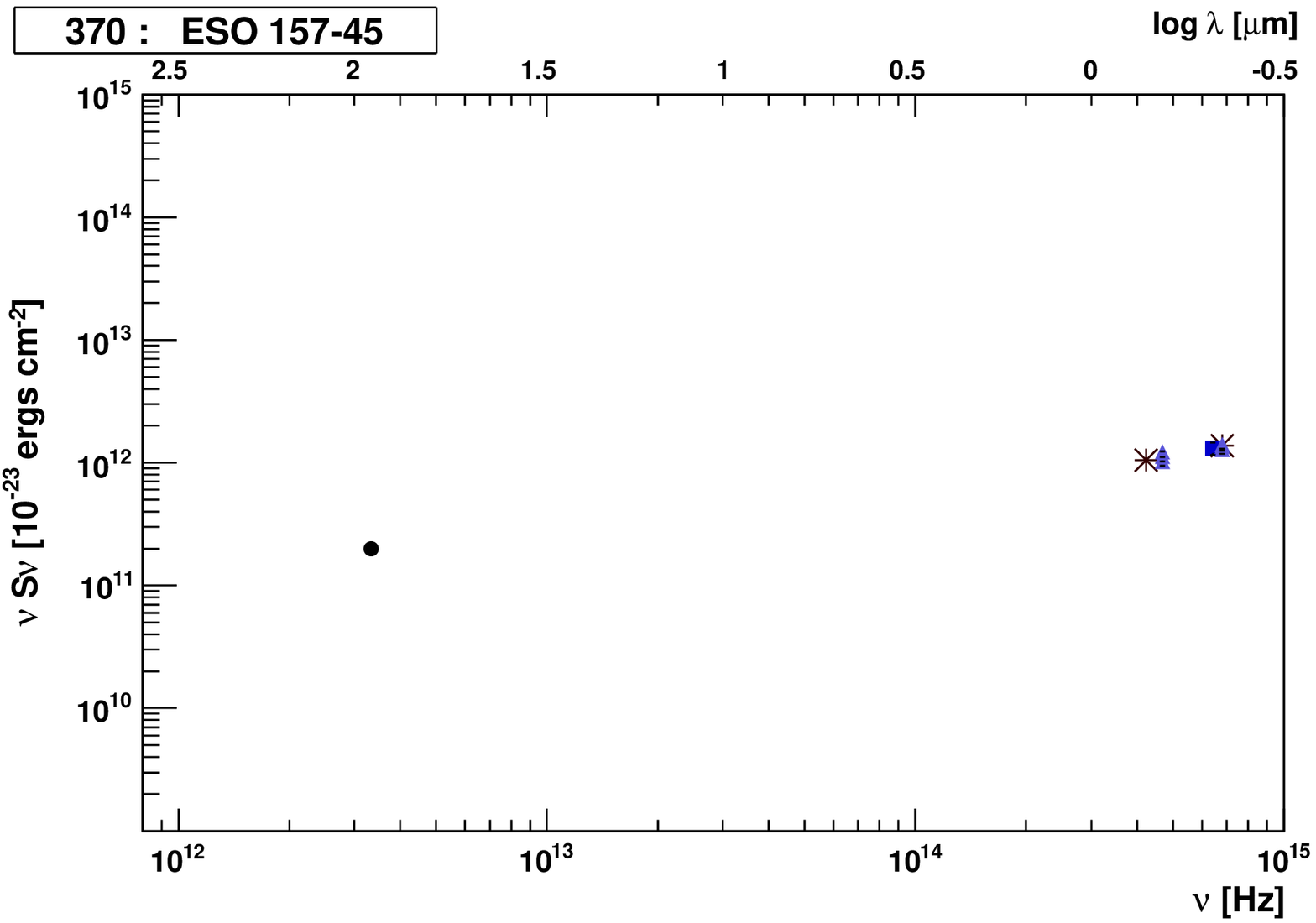}
\includegraphics[width=4cm]{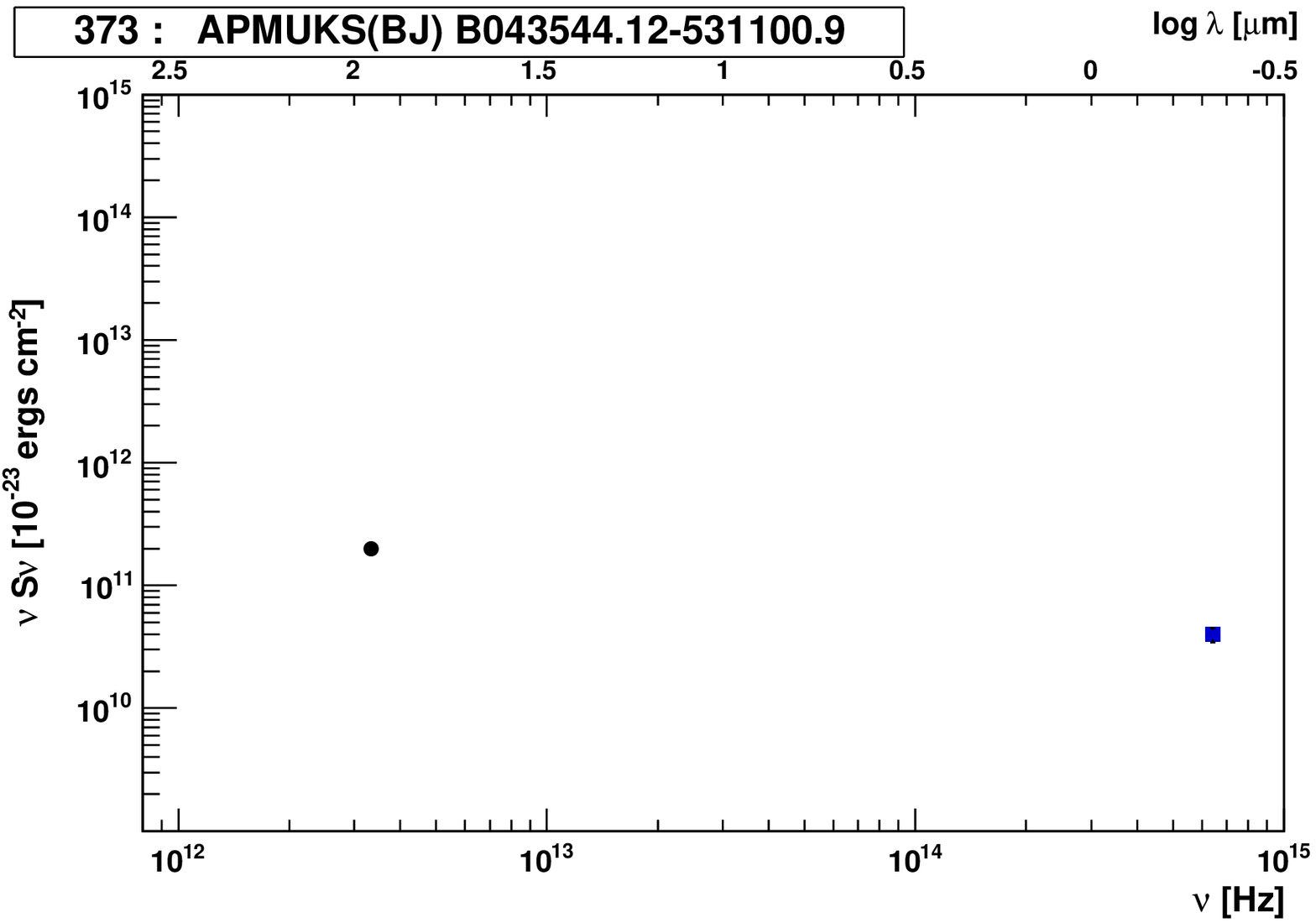}
\includegraphics[width=4cm]{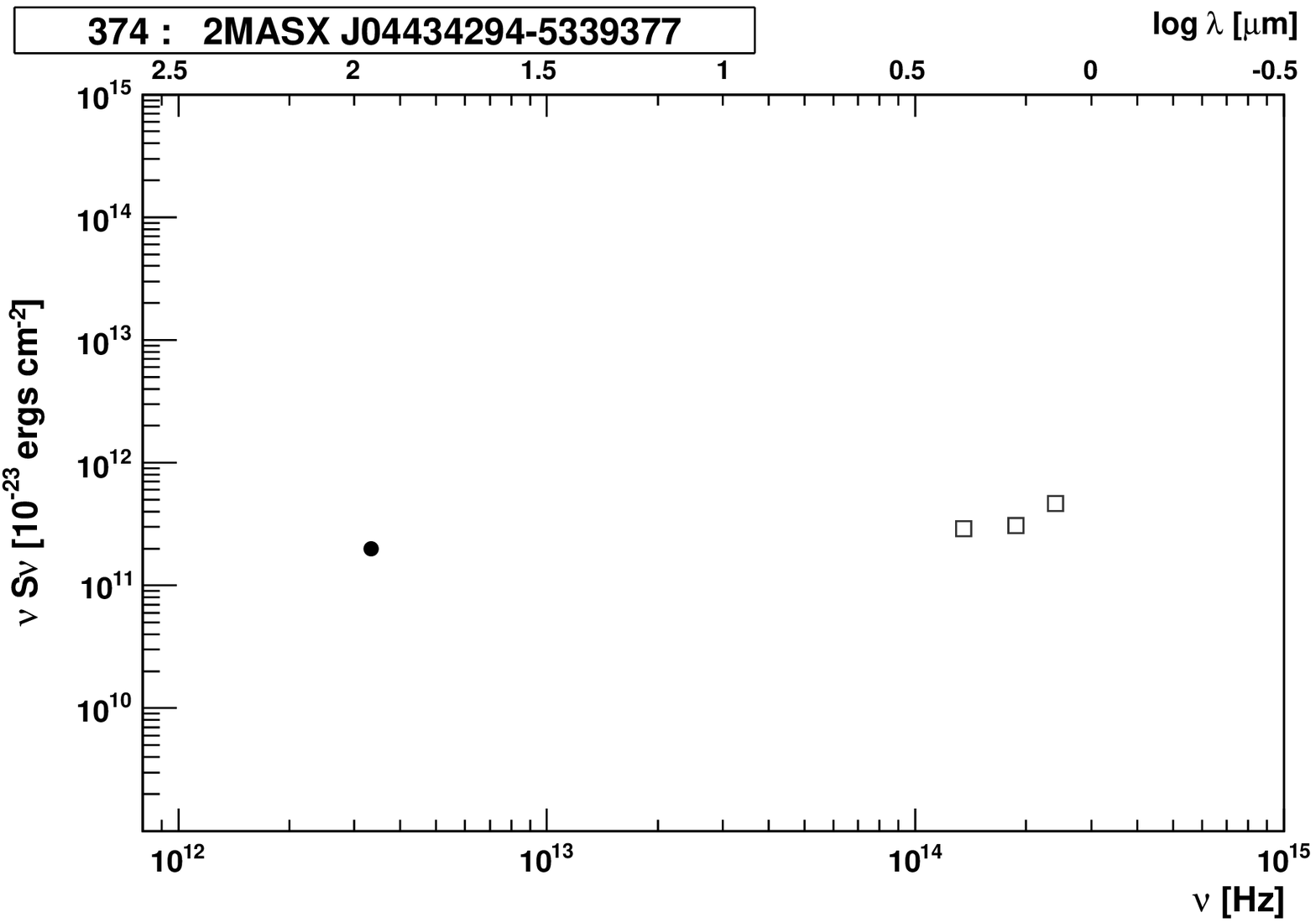}
\includegraphics[width=4cm]{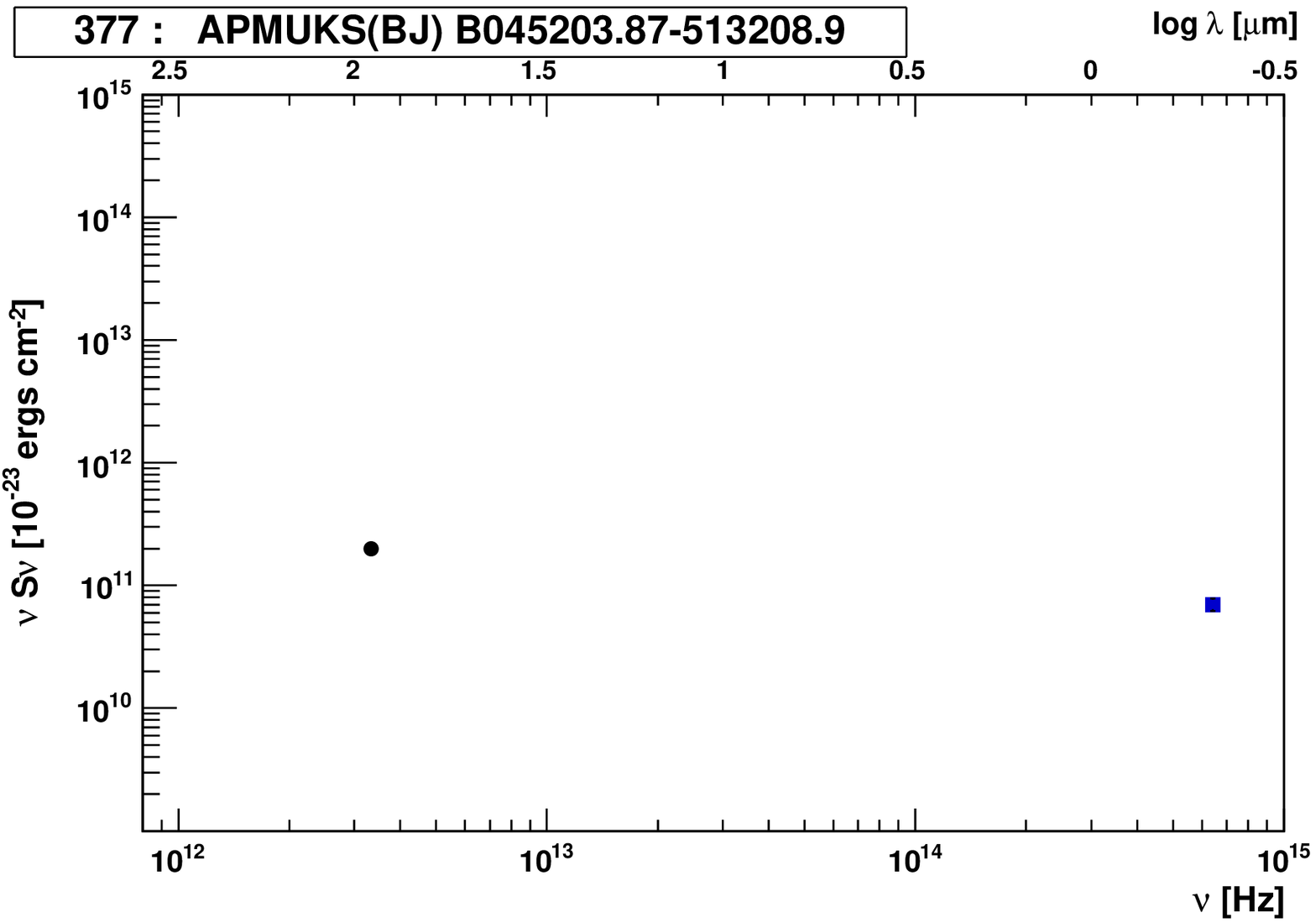}
\includegraphics[width=4cm]{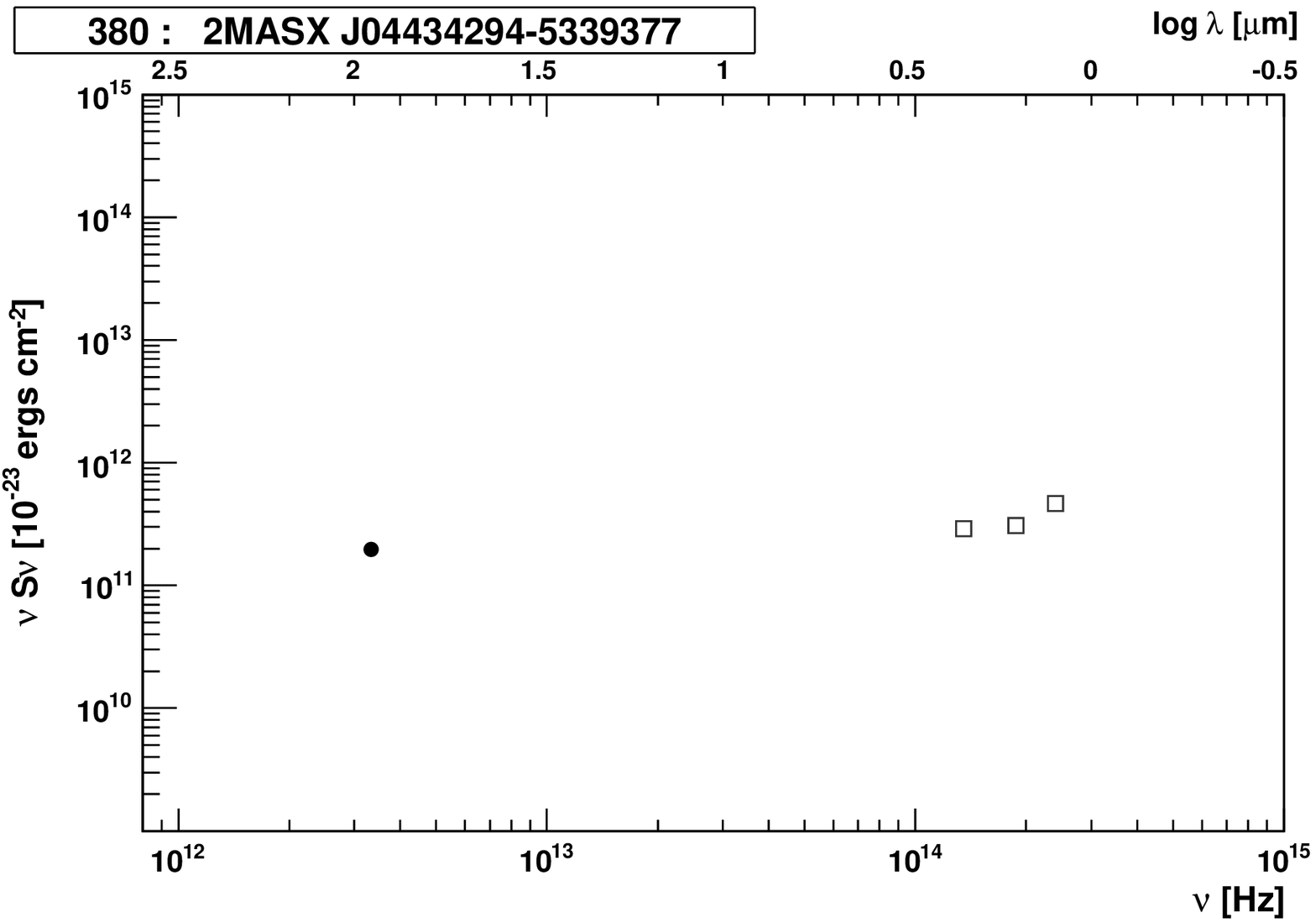}
\includegraphics[width=4cm]{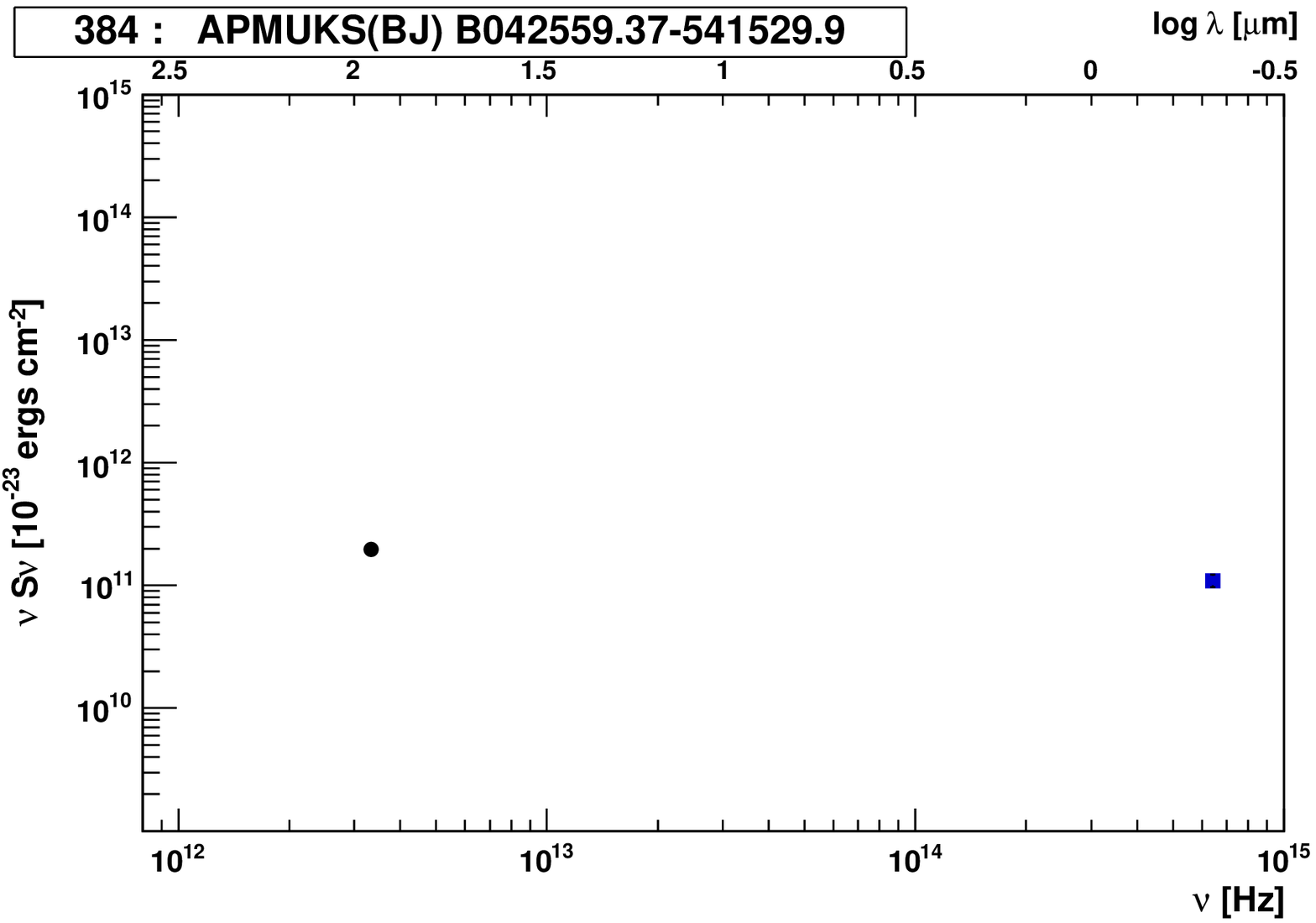}
\includegraphics[width=4cm]{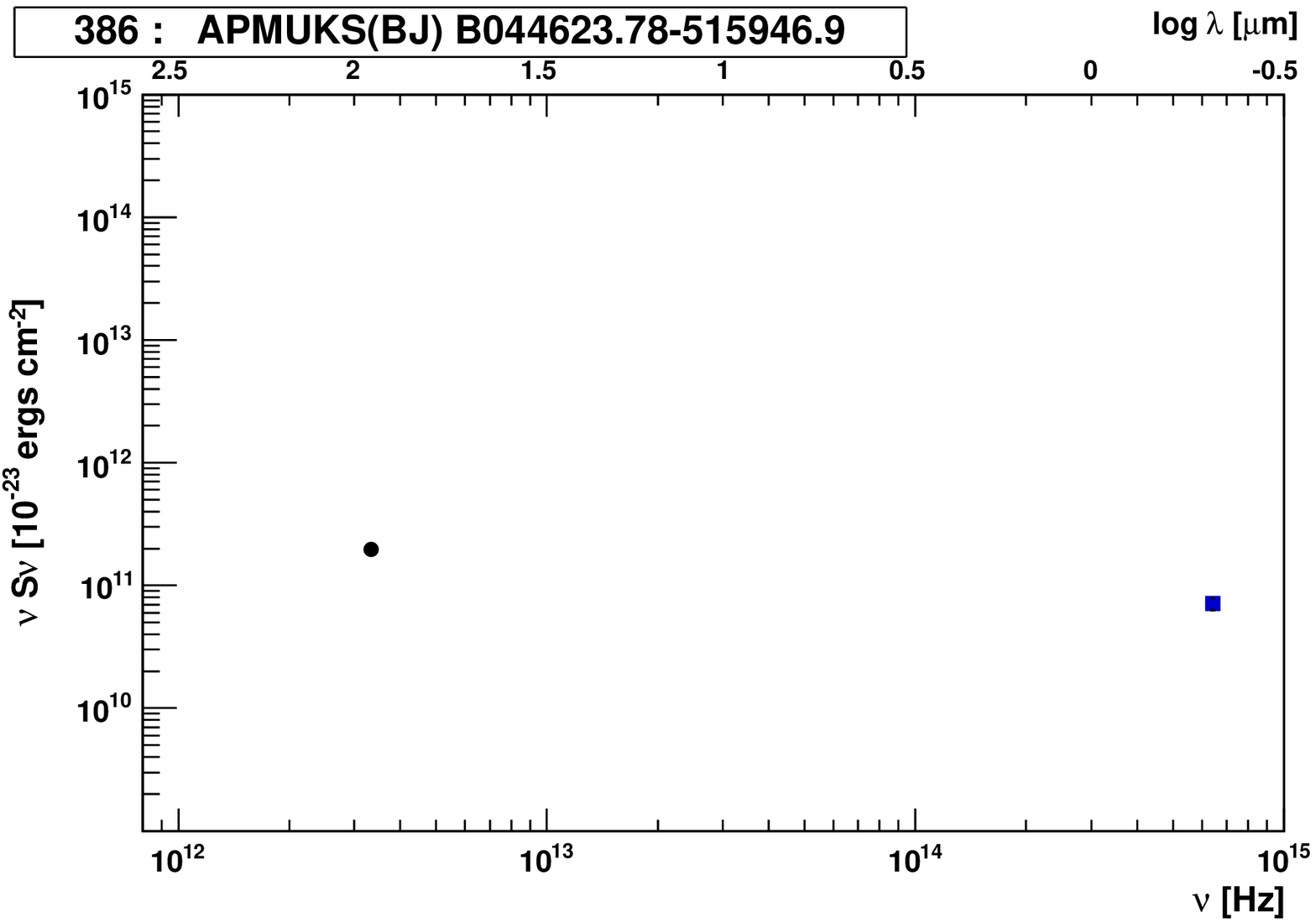}
\includegraphics[width=4cm]{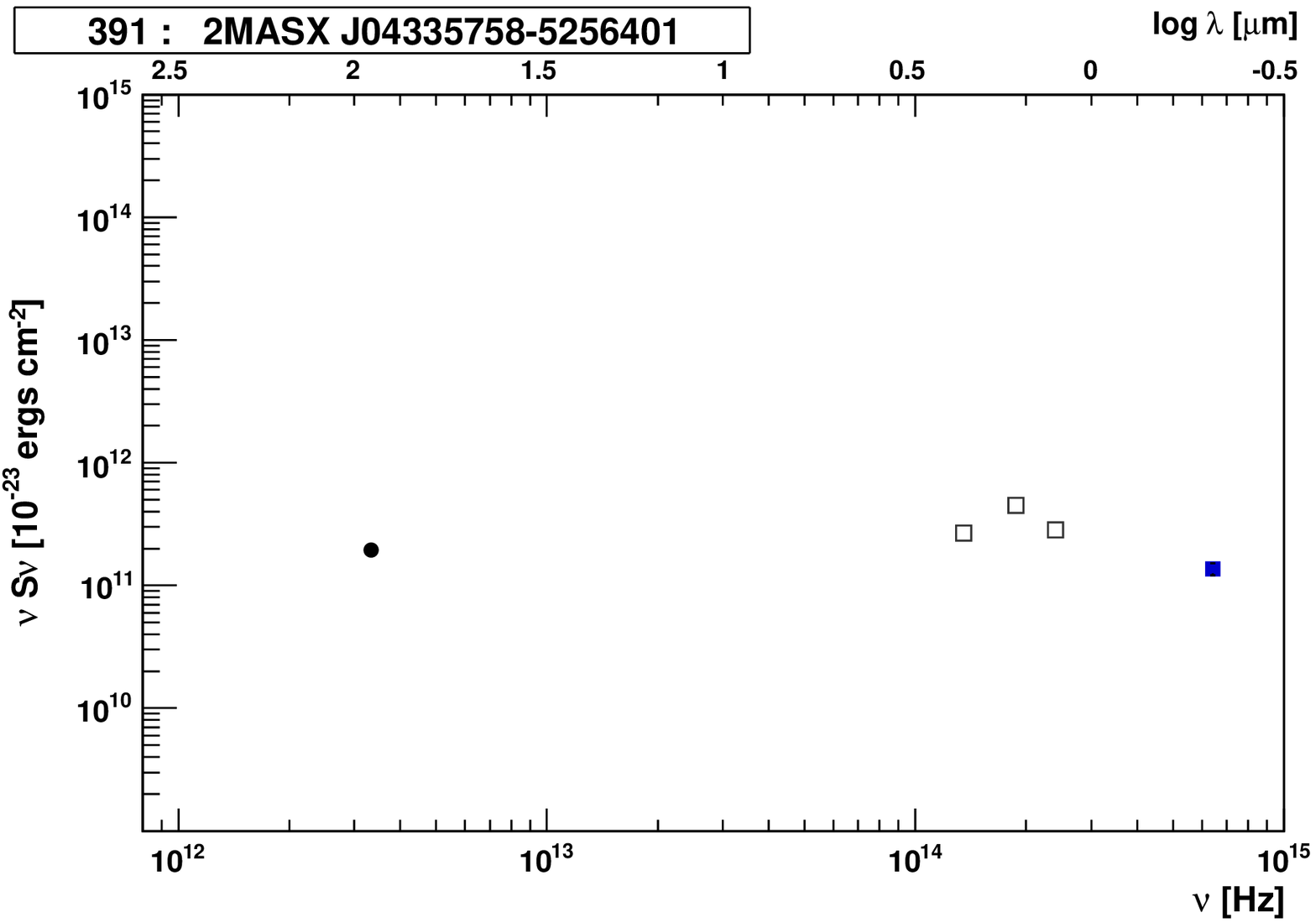}
\includegraphics[width=4cm]{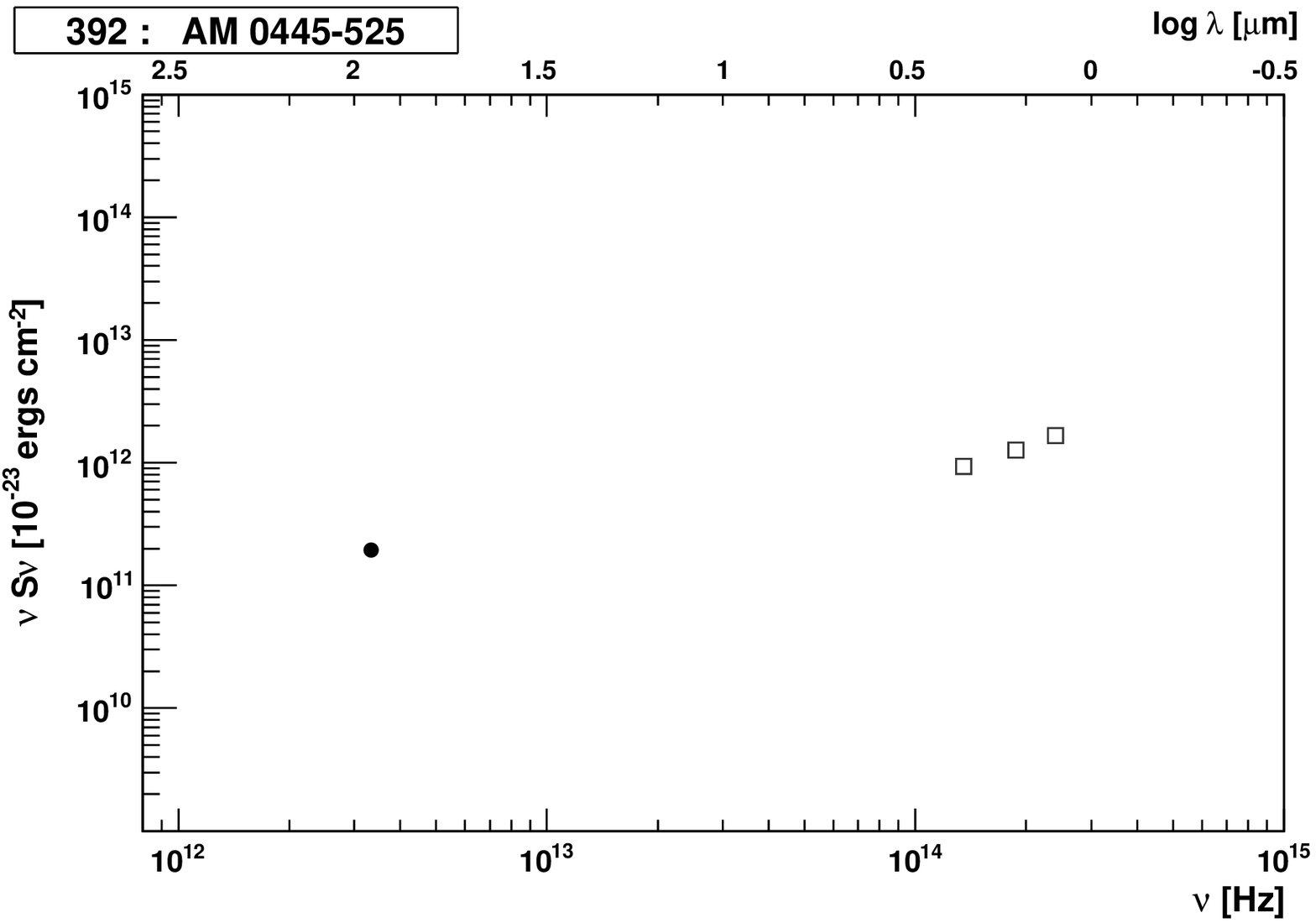}
\includegraphics[width=4cm]{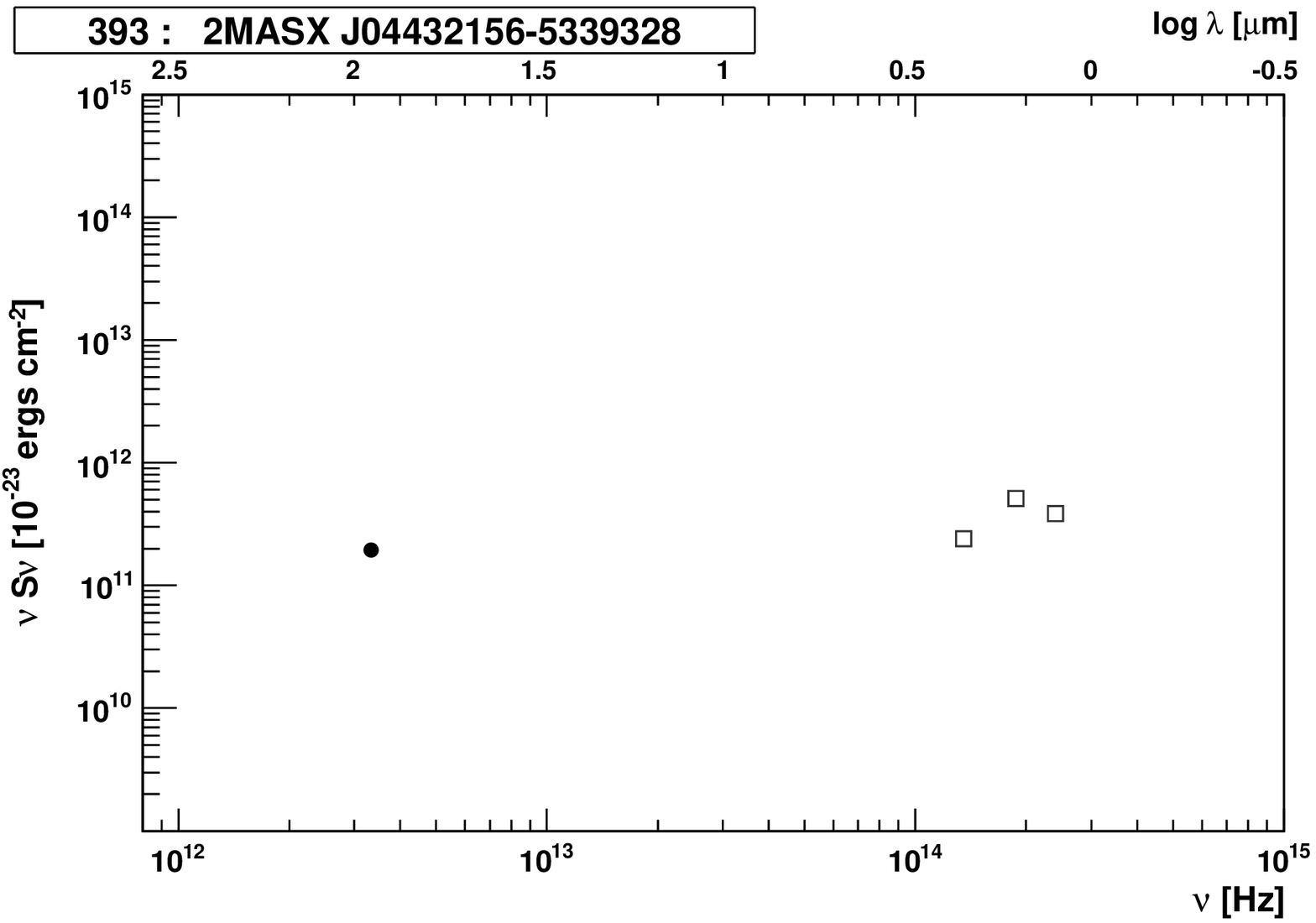}
\includegraphics[width=4cm]{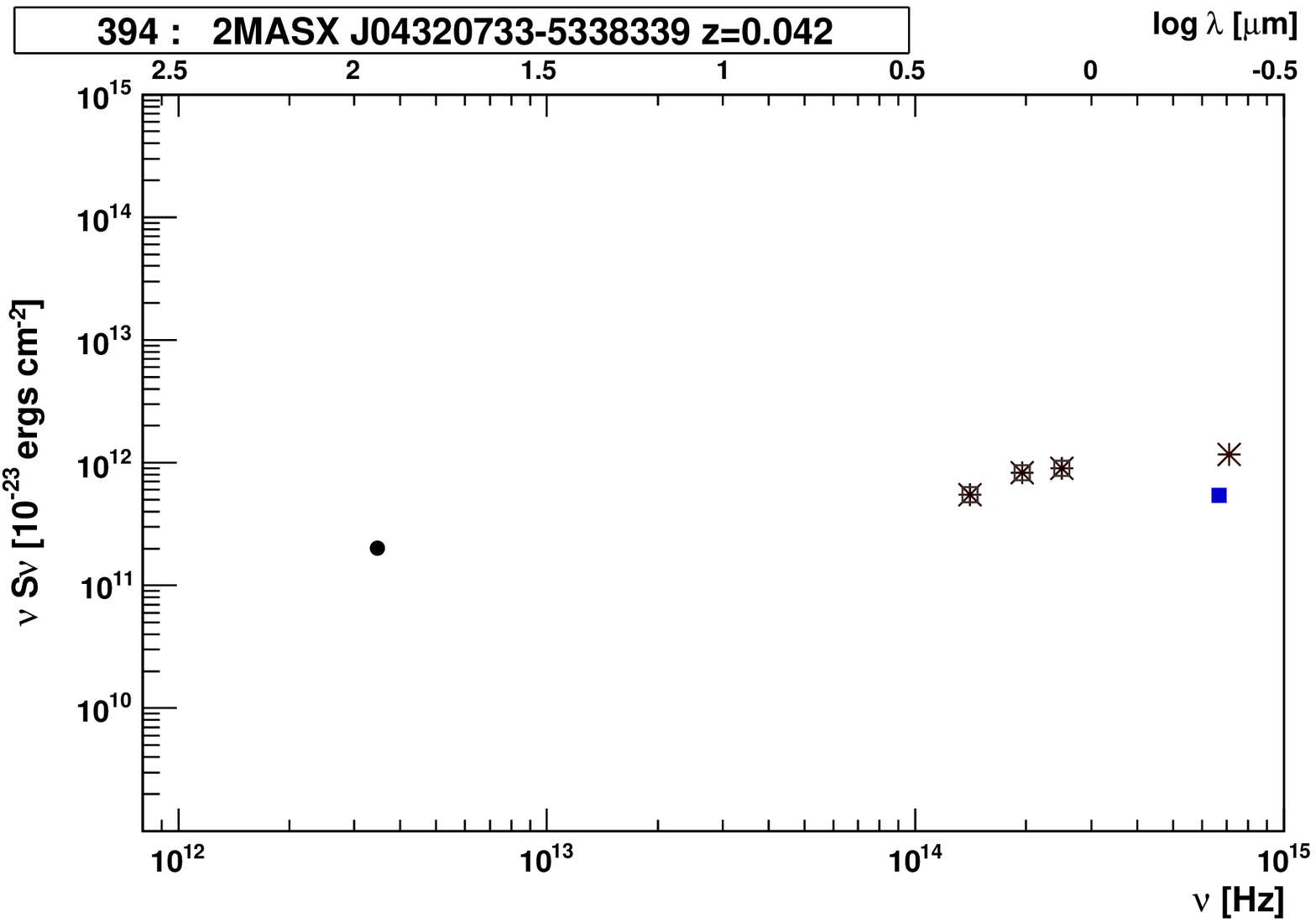}
\includegraphics[width=4cm]{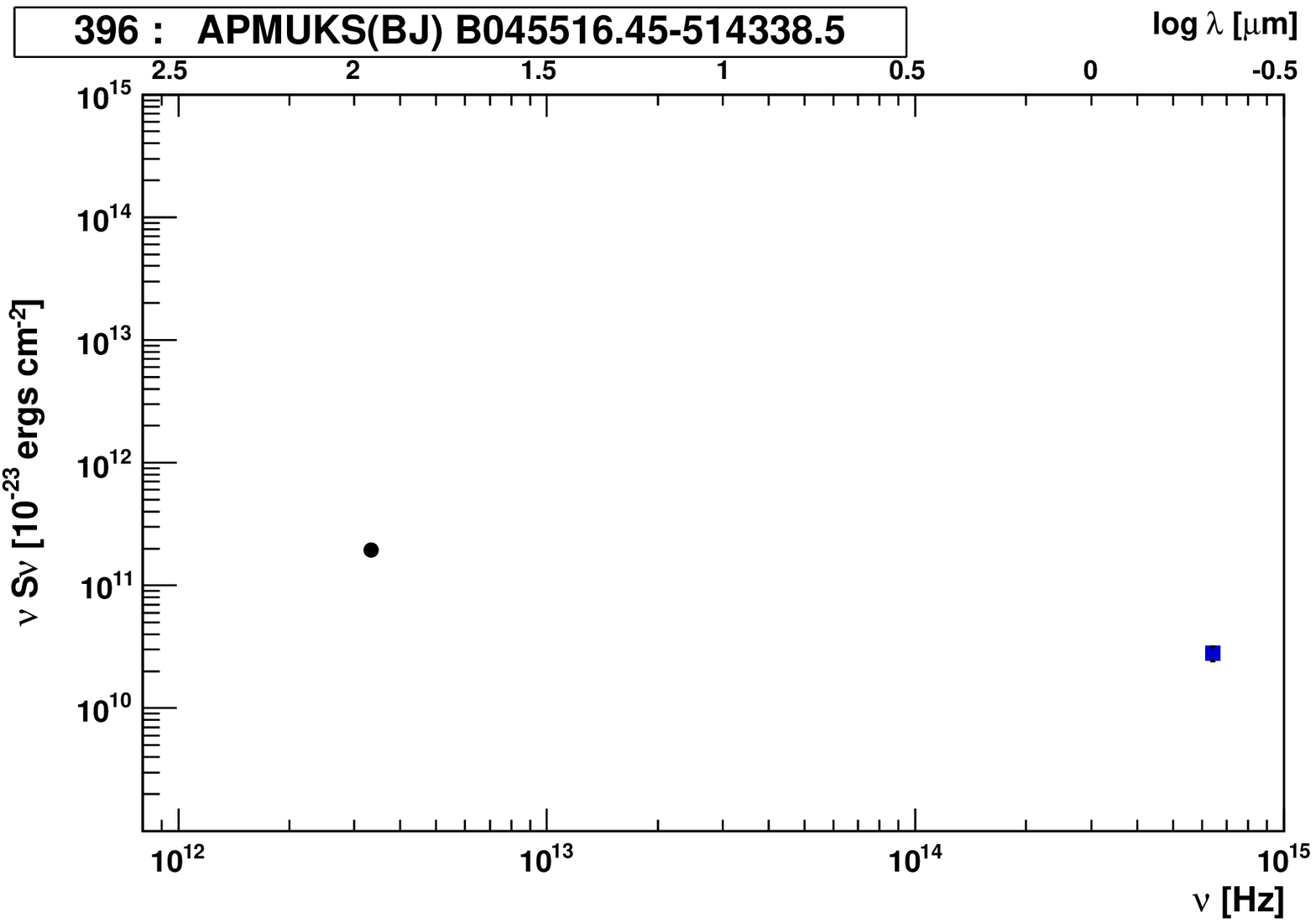}
\includegraphics[width=4cm]{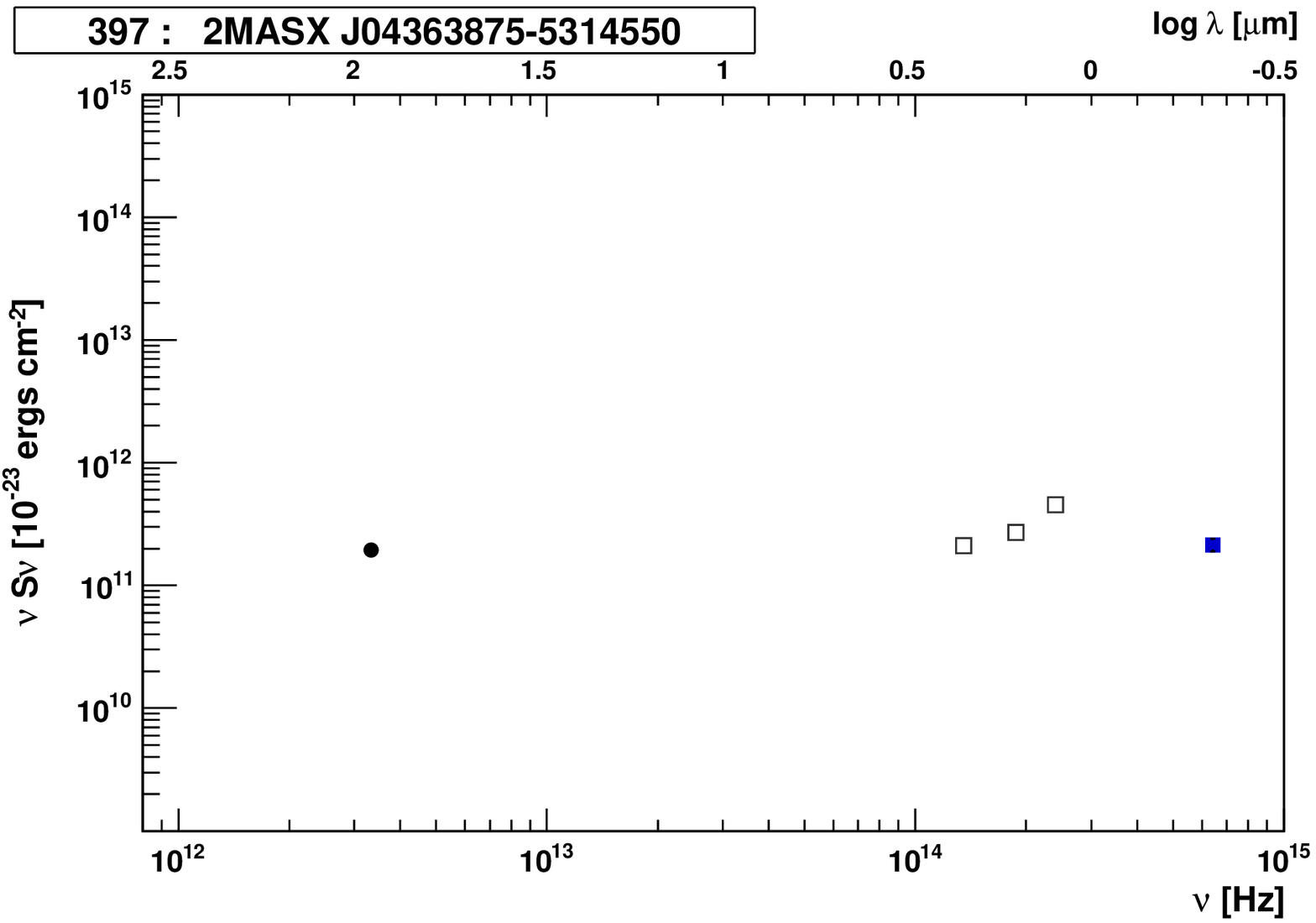}
\includegraphics[width=4cm]{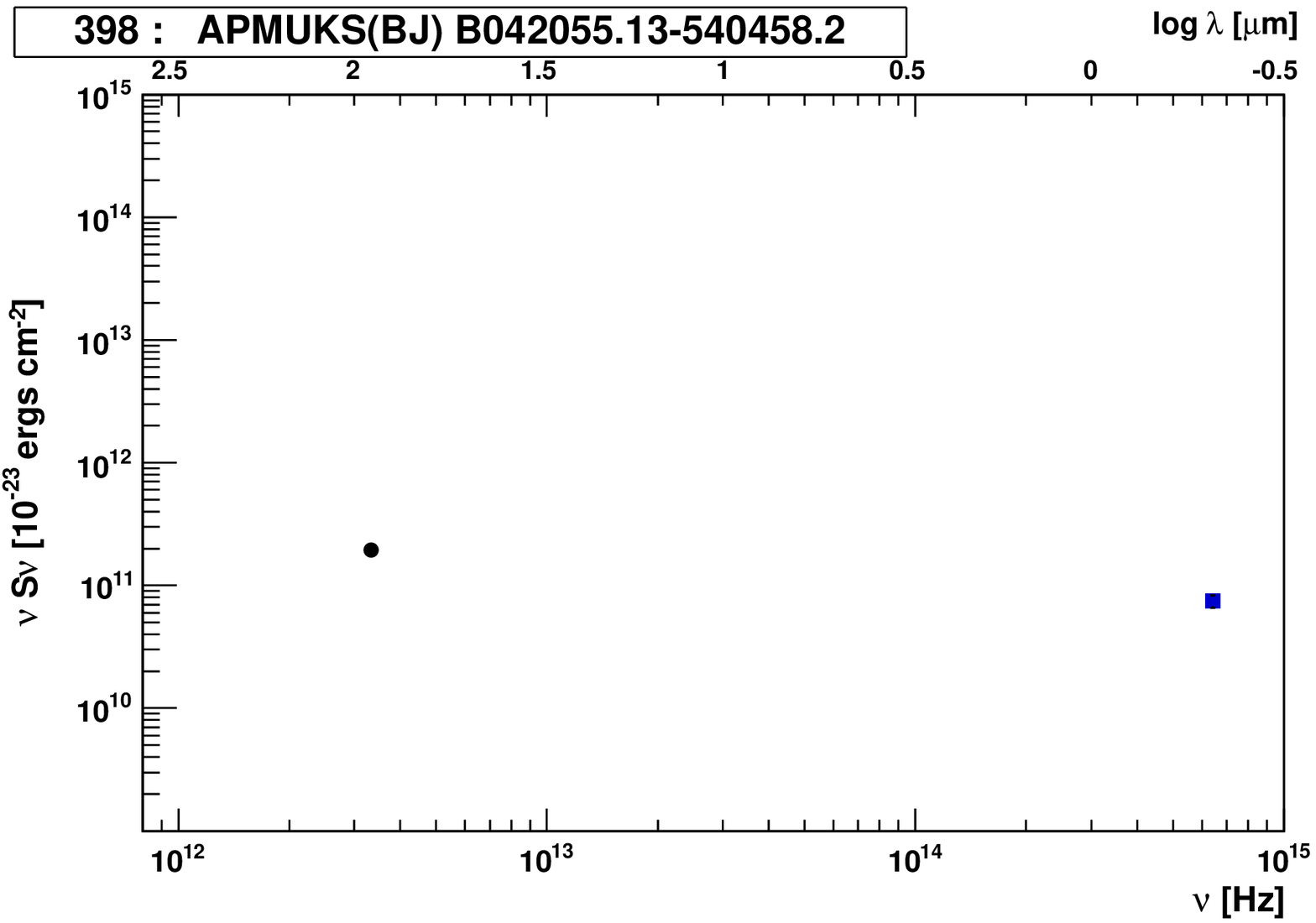}
\includegraphics[width=4cm]{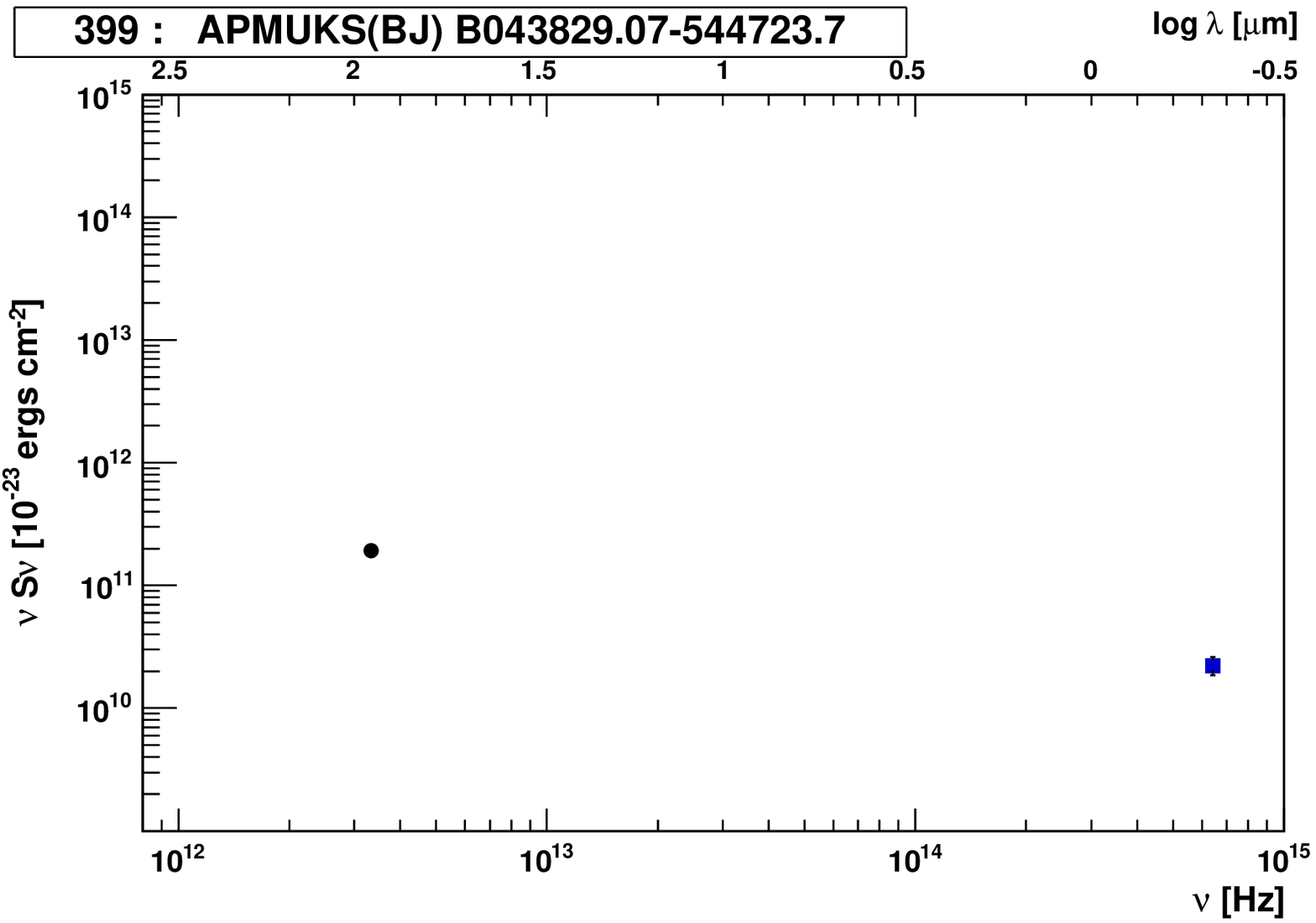}
\includegraphics[width=4cm]{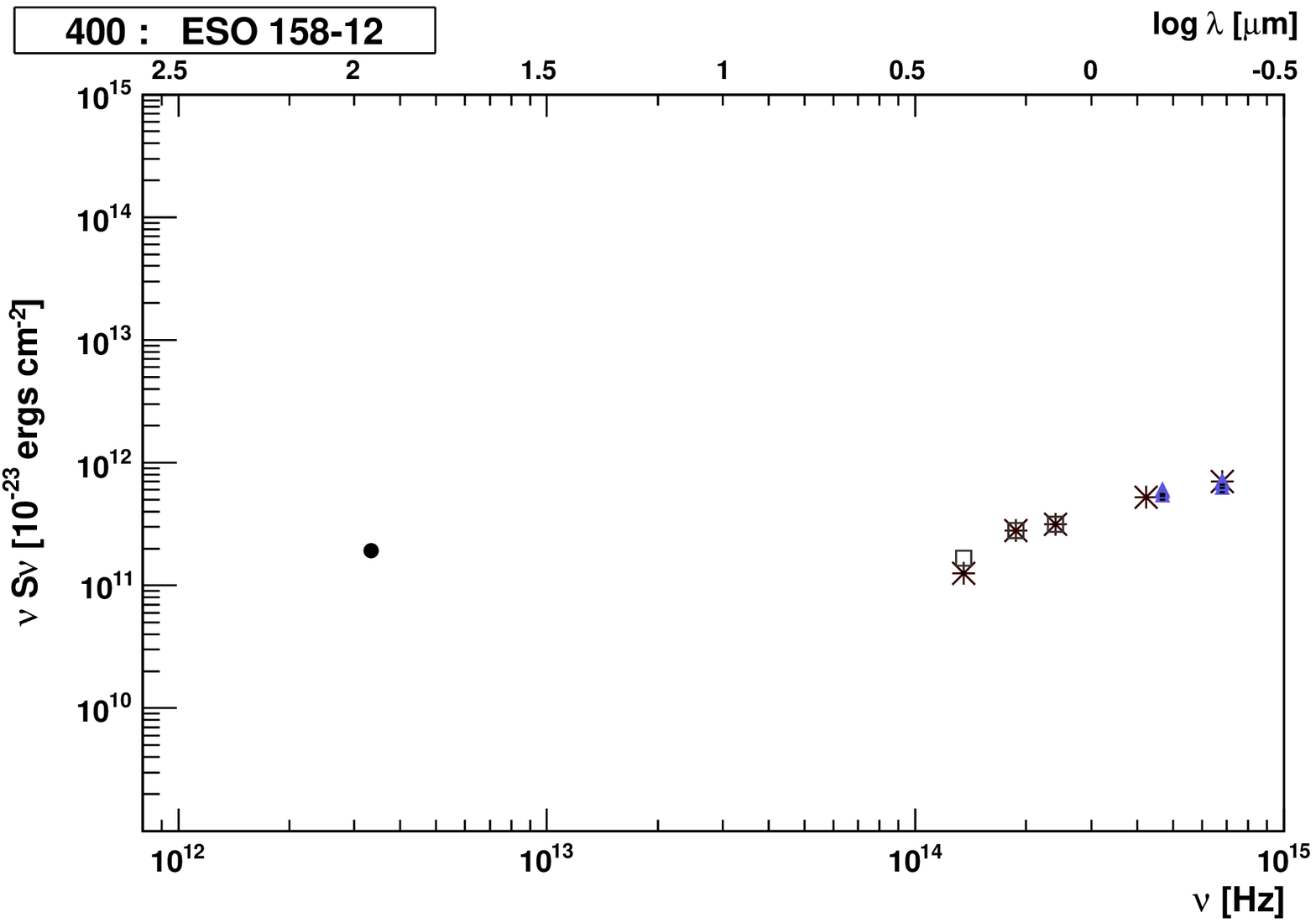}
\includegraphics[width=4cm]{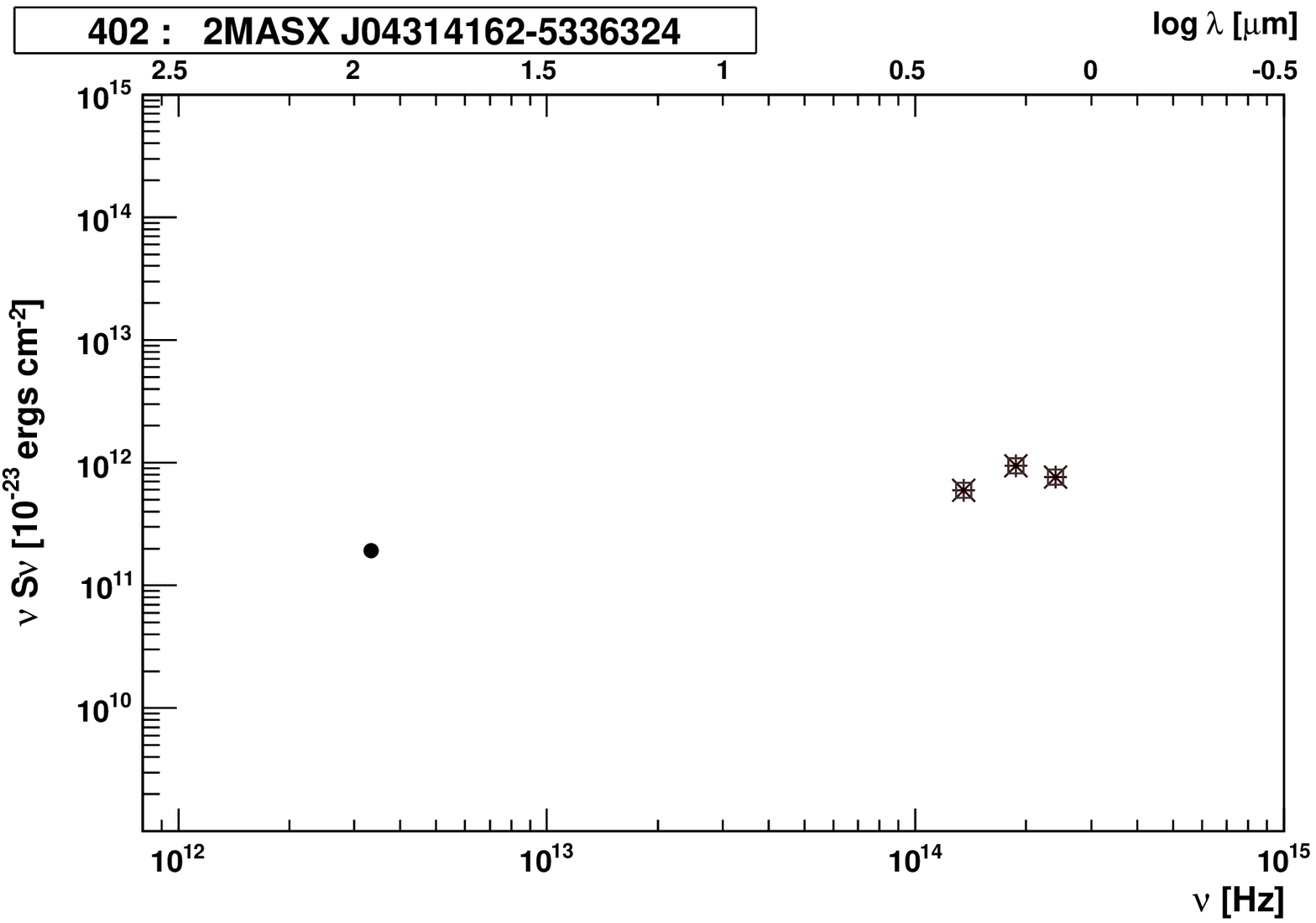}
\includegraphics[width=4cm]{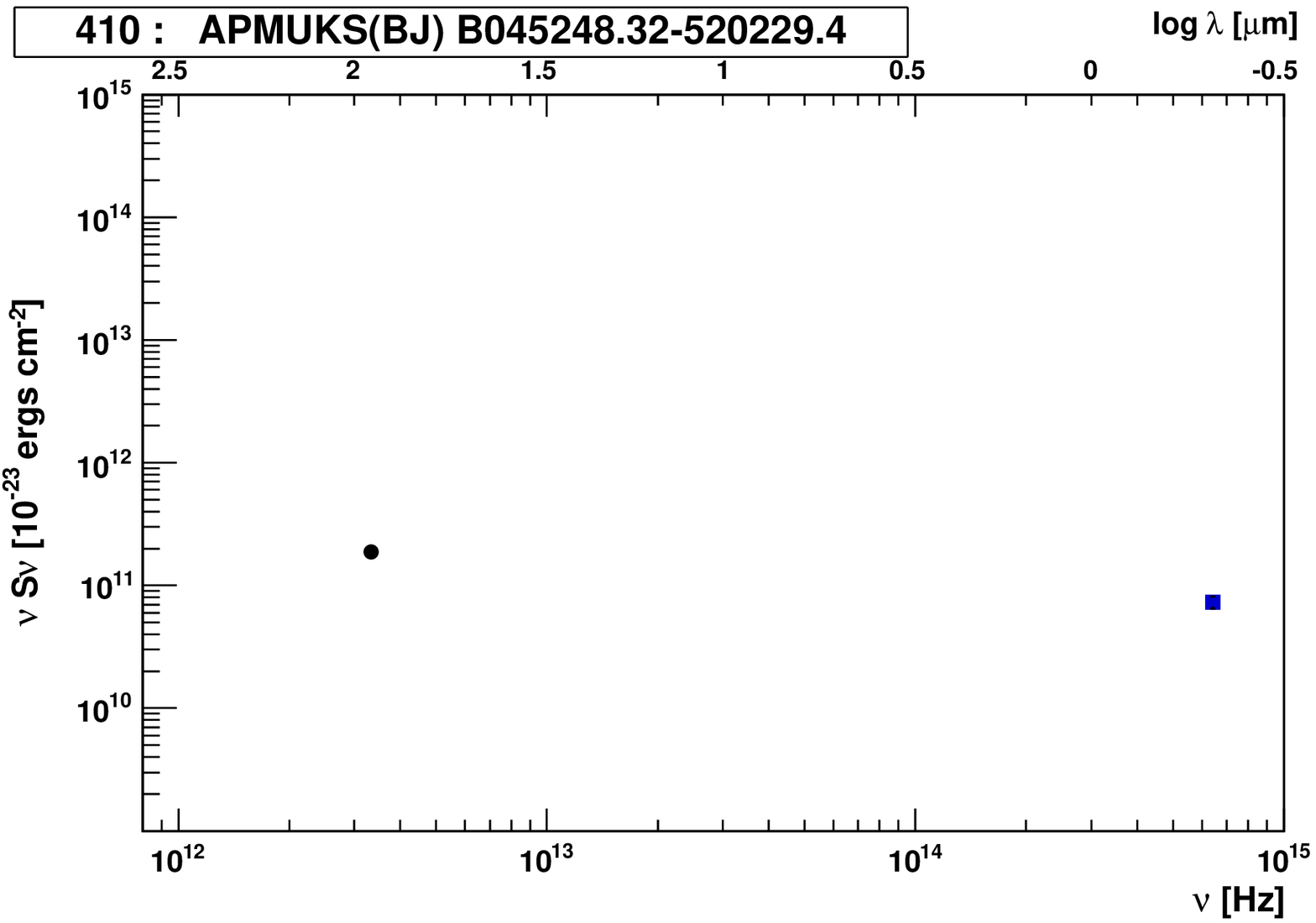}
\includegraphics[width=4cm]{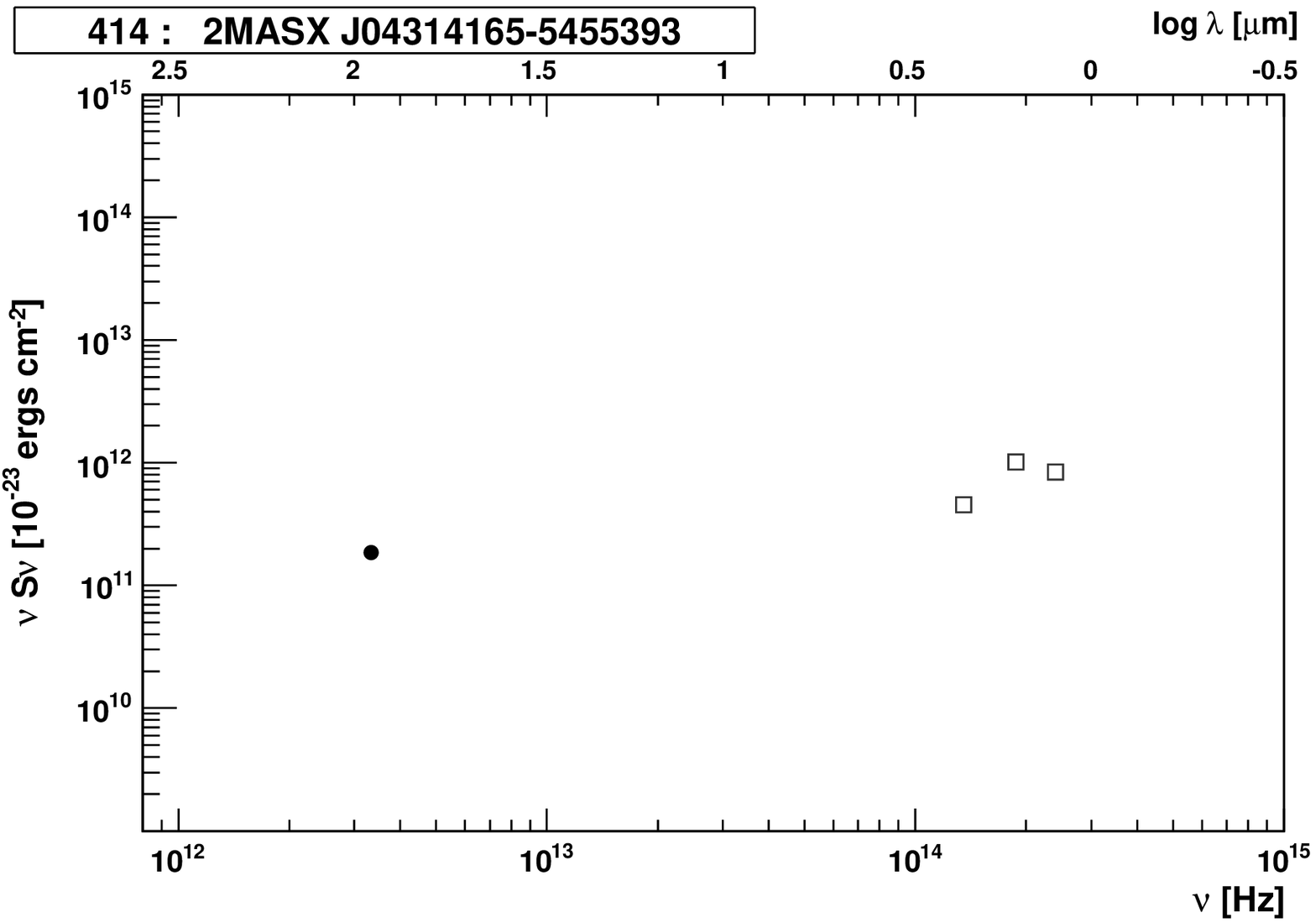}
\includegraphics[width=4cm]{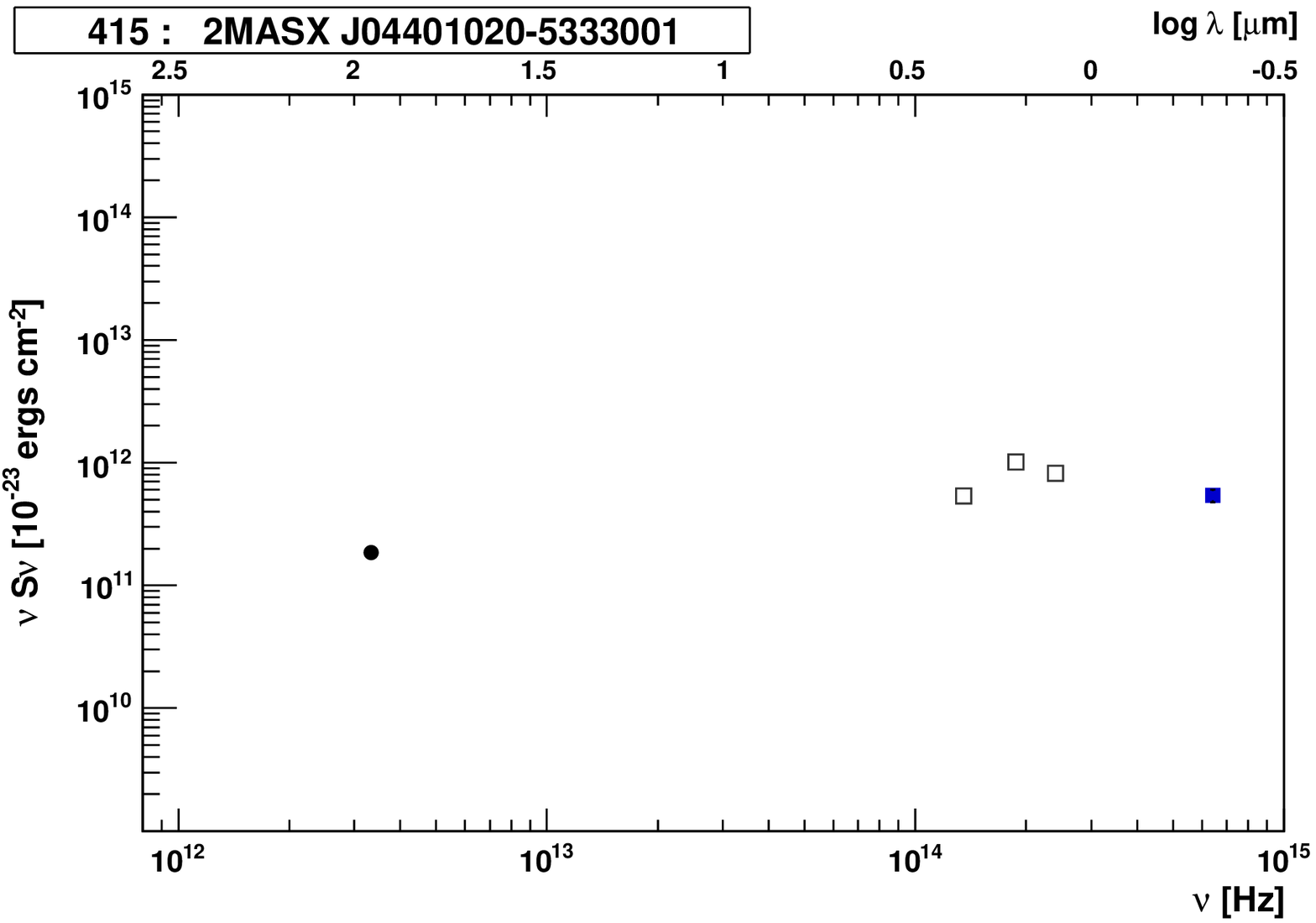}
\includegraphics[width=4cm]{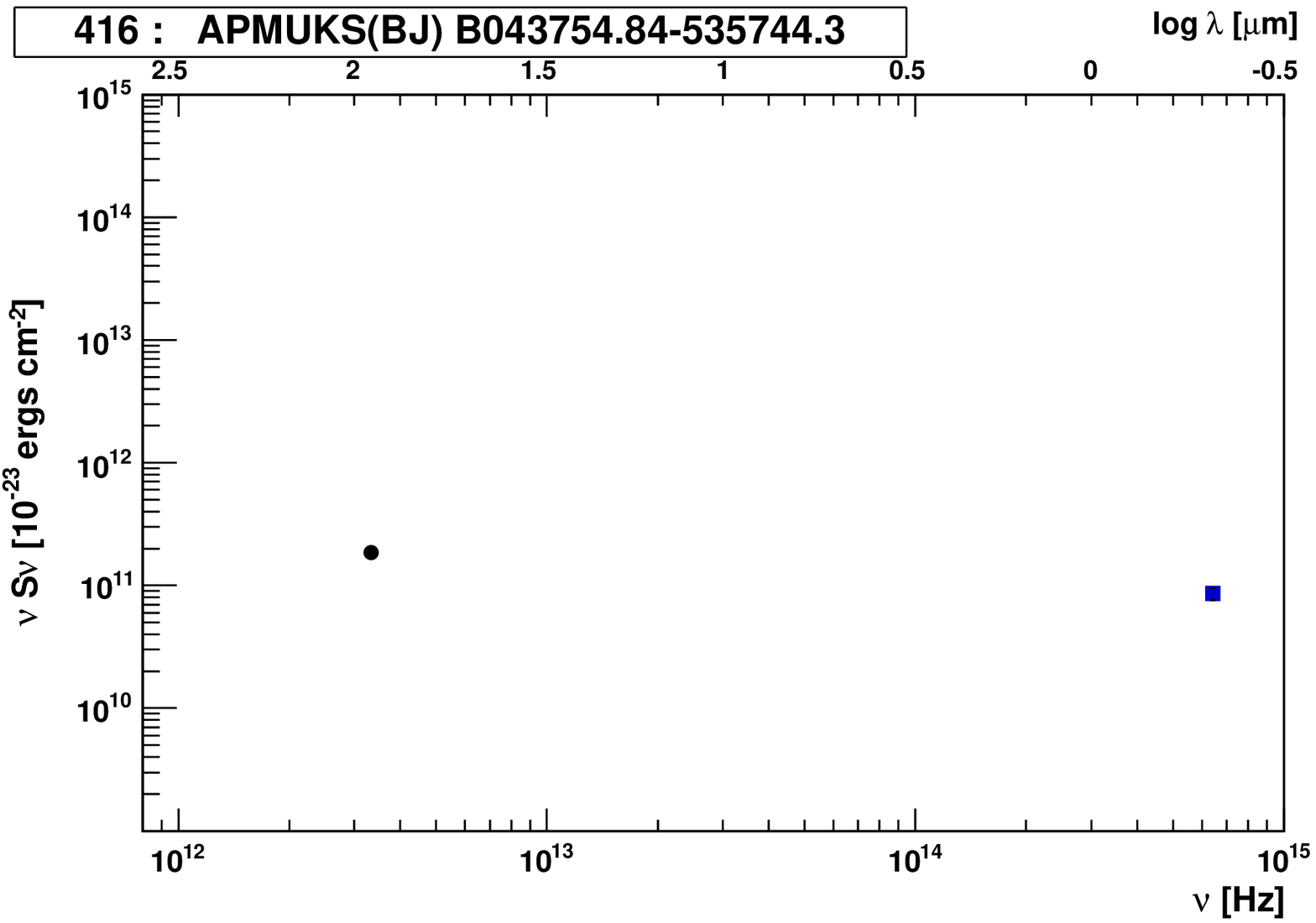}
\includegraphics[width=4cm]{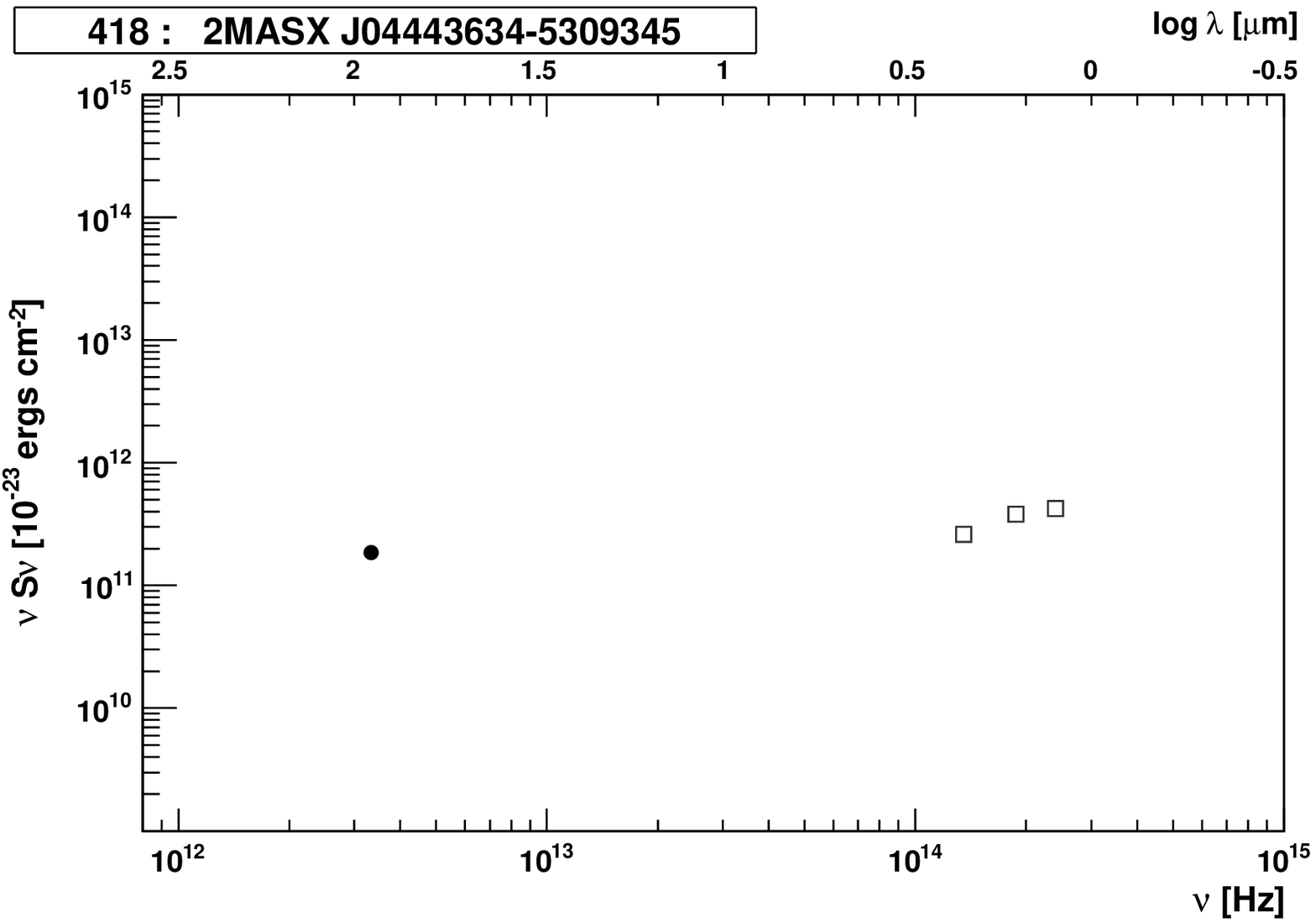}
\includegraphics[width=4cm]{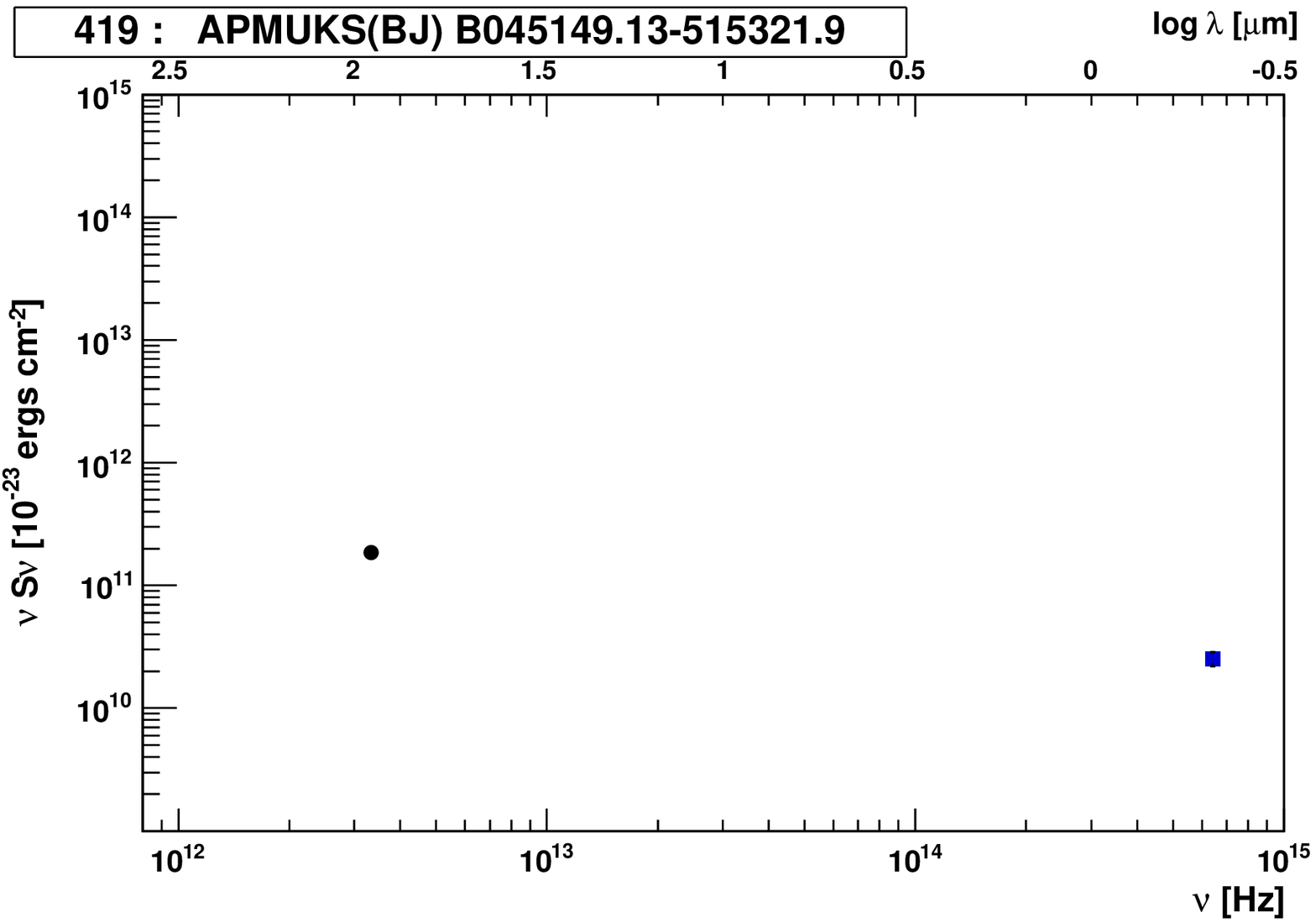}
\includegraphics[width=4cm]{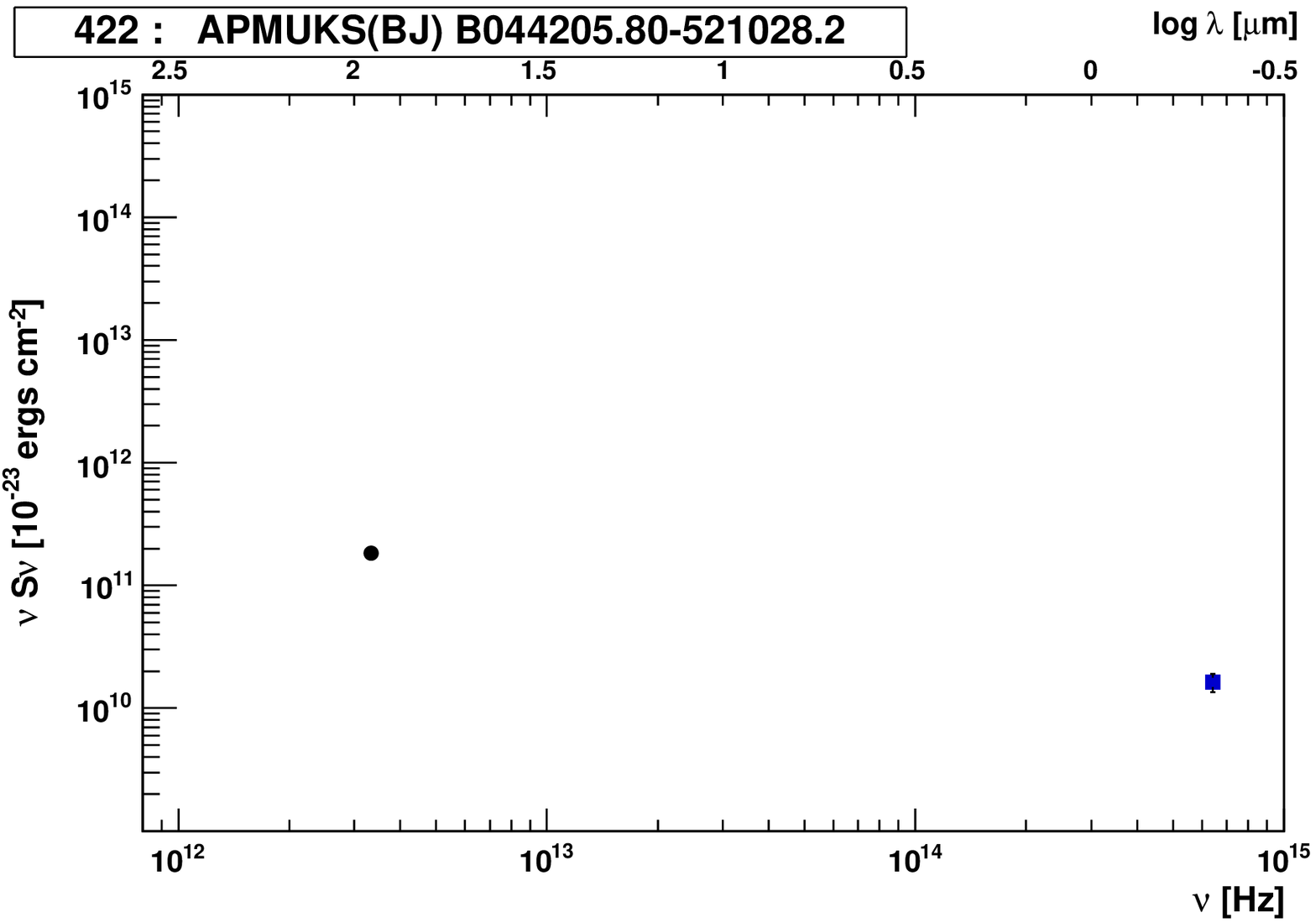}

\label{points8}
\caption {SEDs for the next 36 ADF-S identified sources, with symbols as in Figure~\ref{points1}.}
\end{figure*}
}

\clearpage

\onlfig{9}{
\begin{figure*}[t]
\centering
\includegraphics[width=4cm]{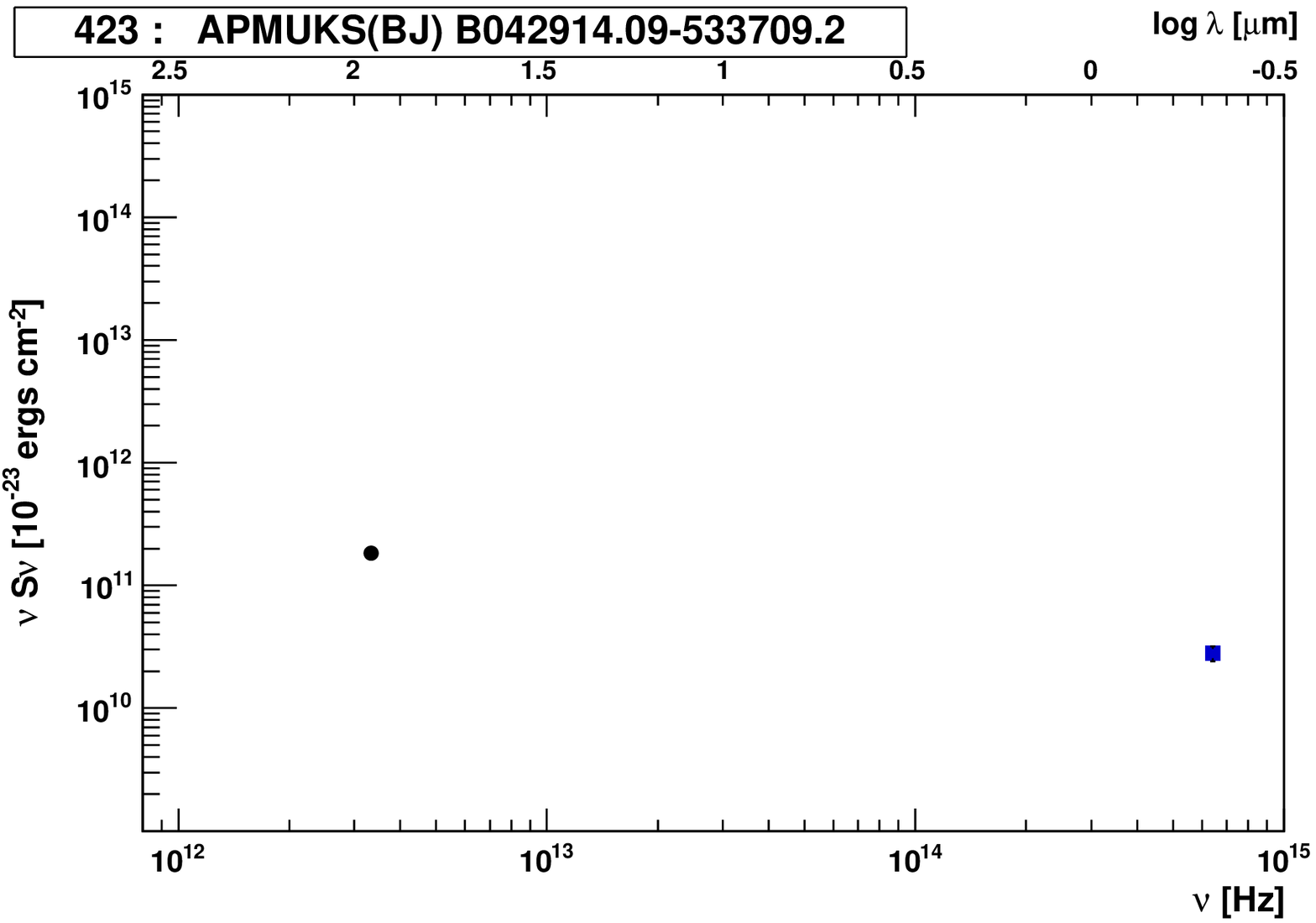}
\includegraphics[width=4cm]{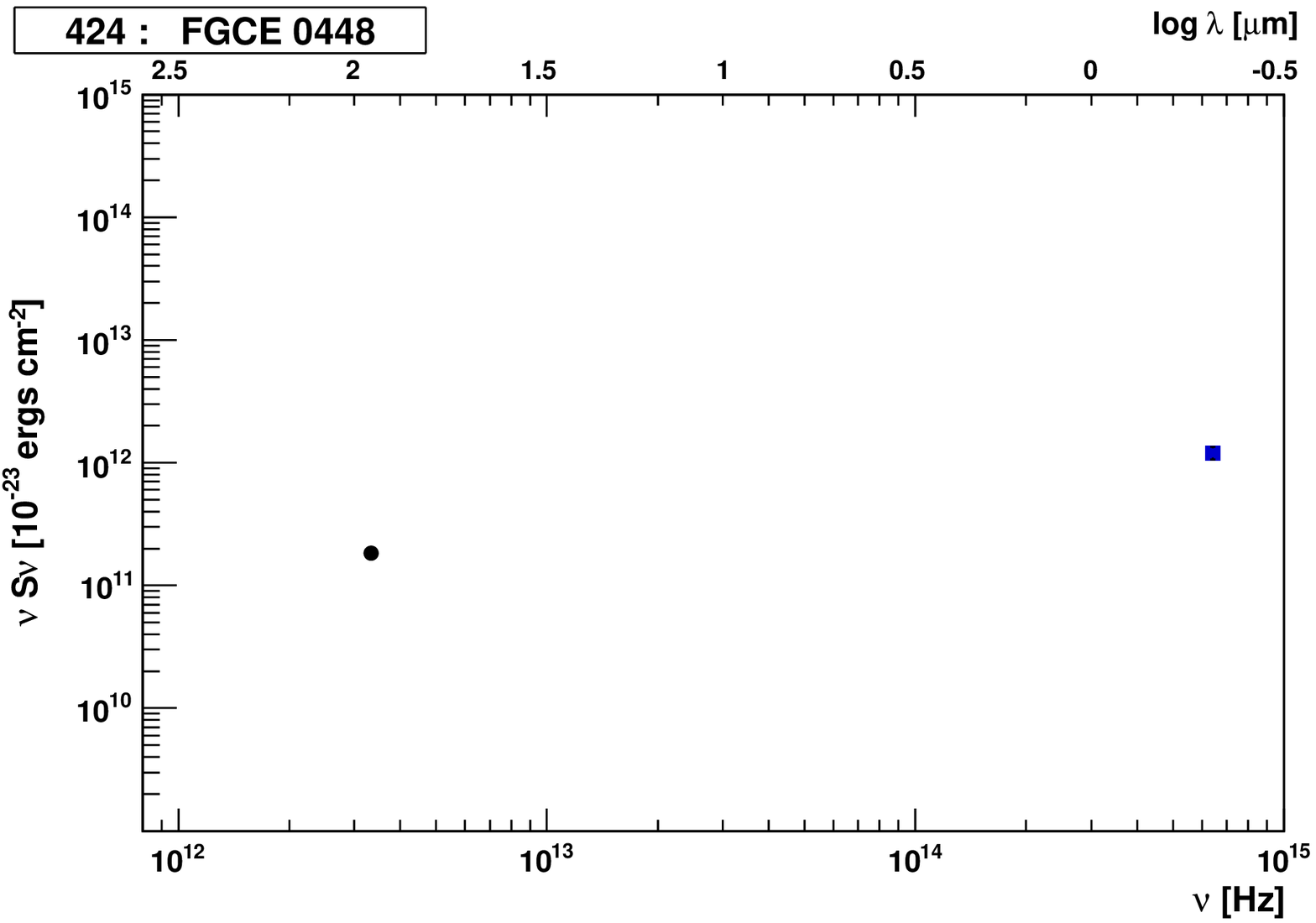}
\includegraphics[width=4cm]{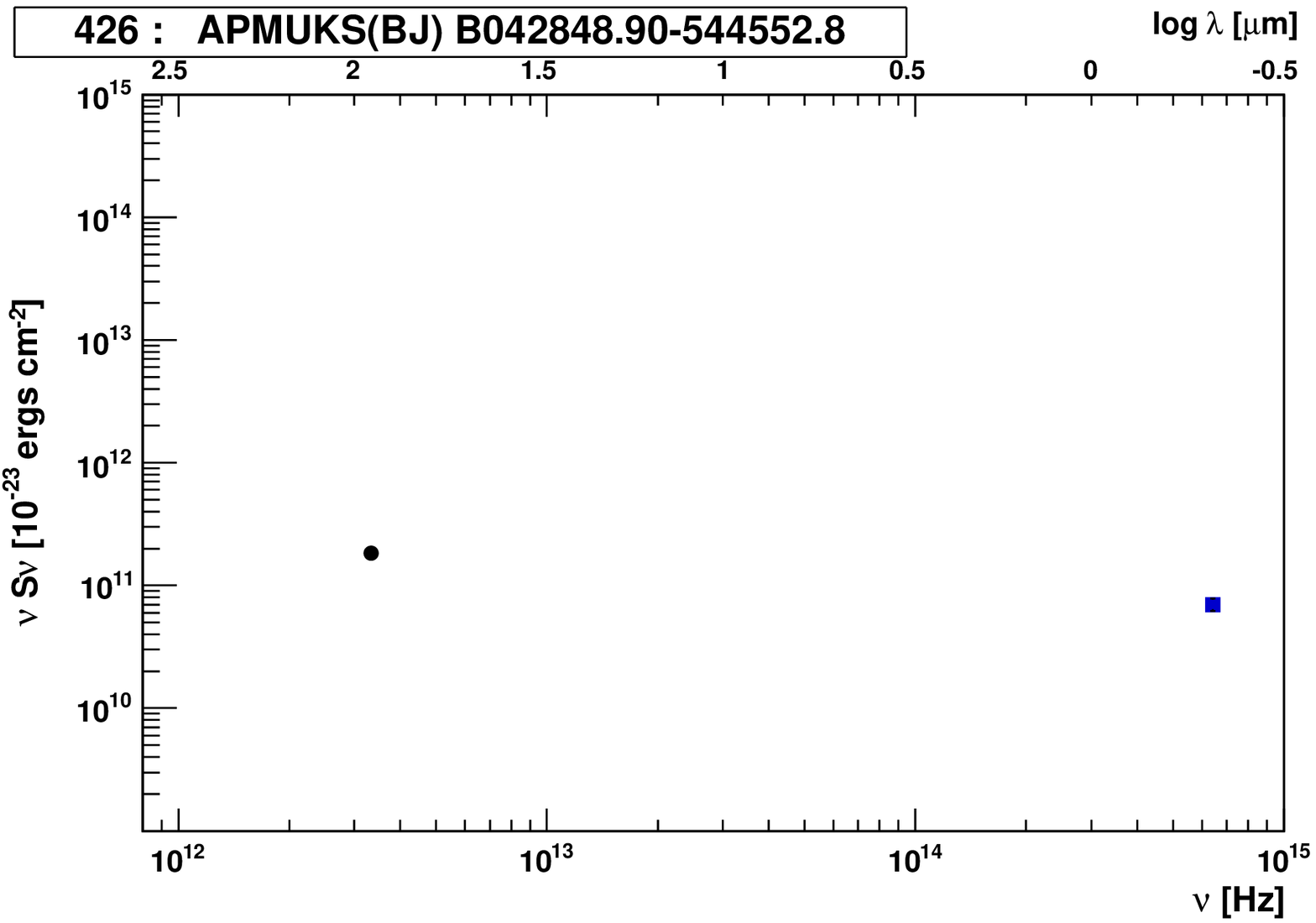}
\includegraphics[width=4cm]{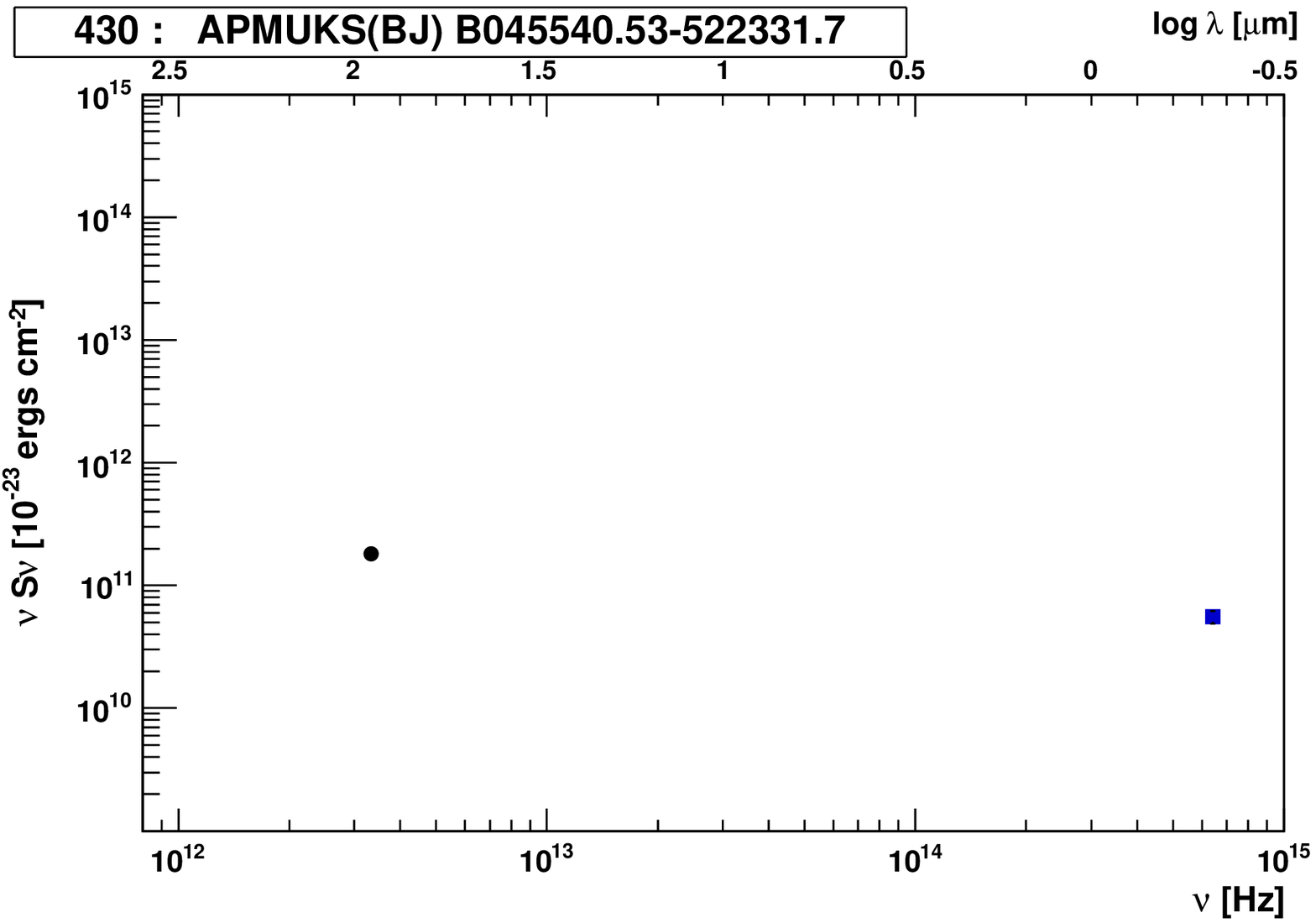}
\includegraphics[width=4cm]{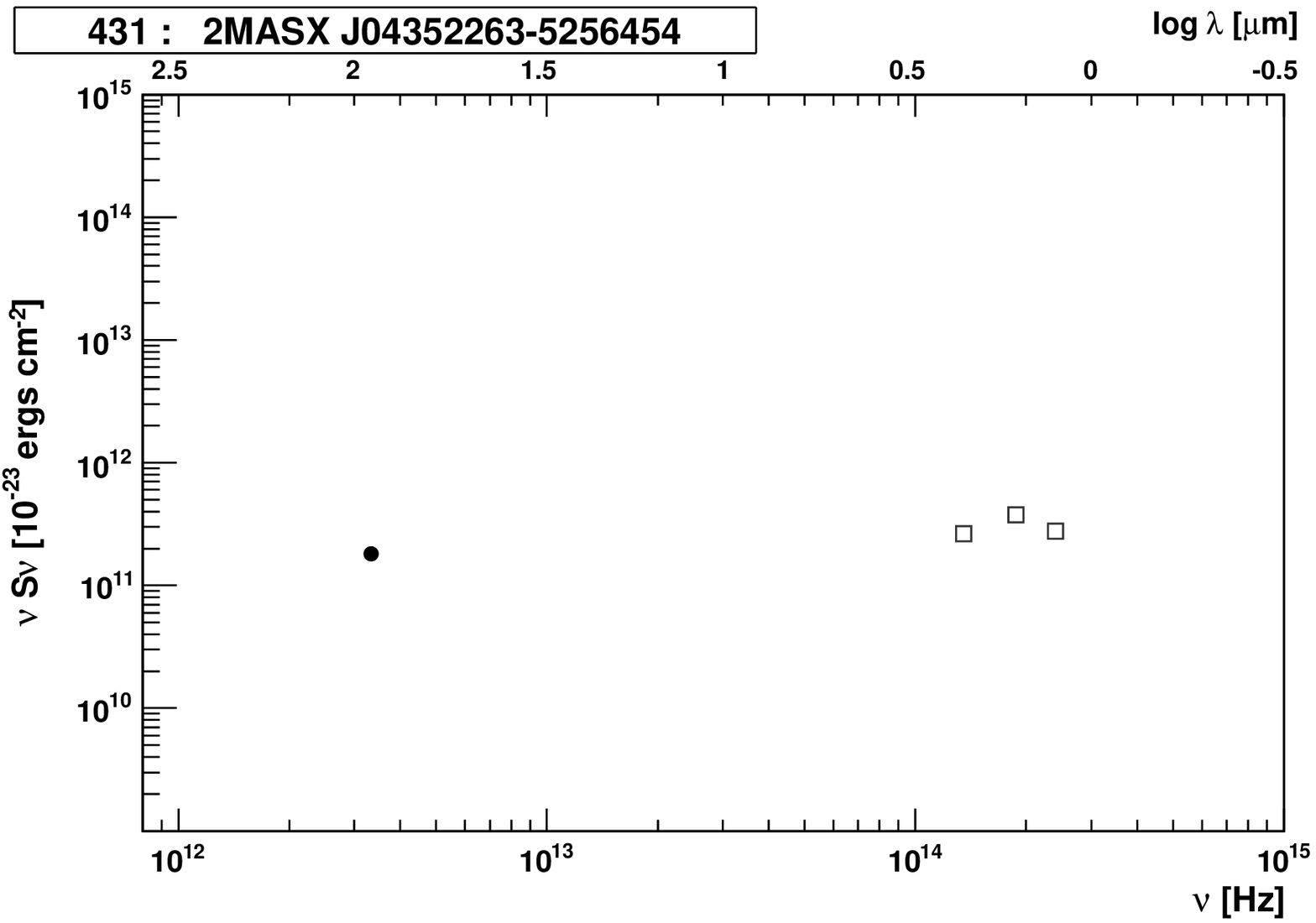}
\includegraphics[width=4cm]{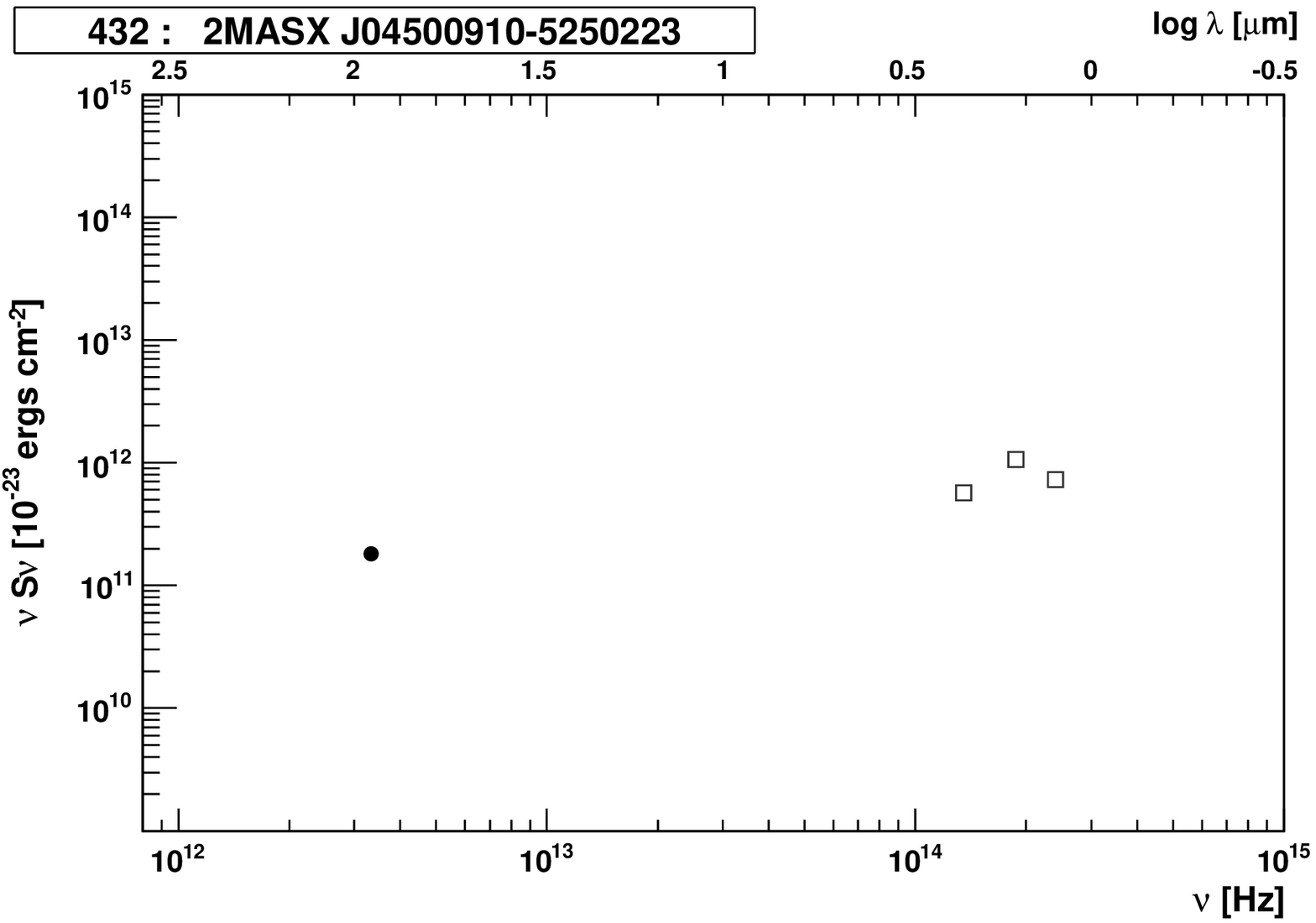}
\includegraphics[width=4cm]{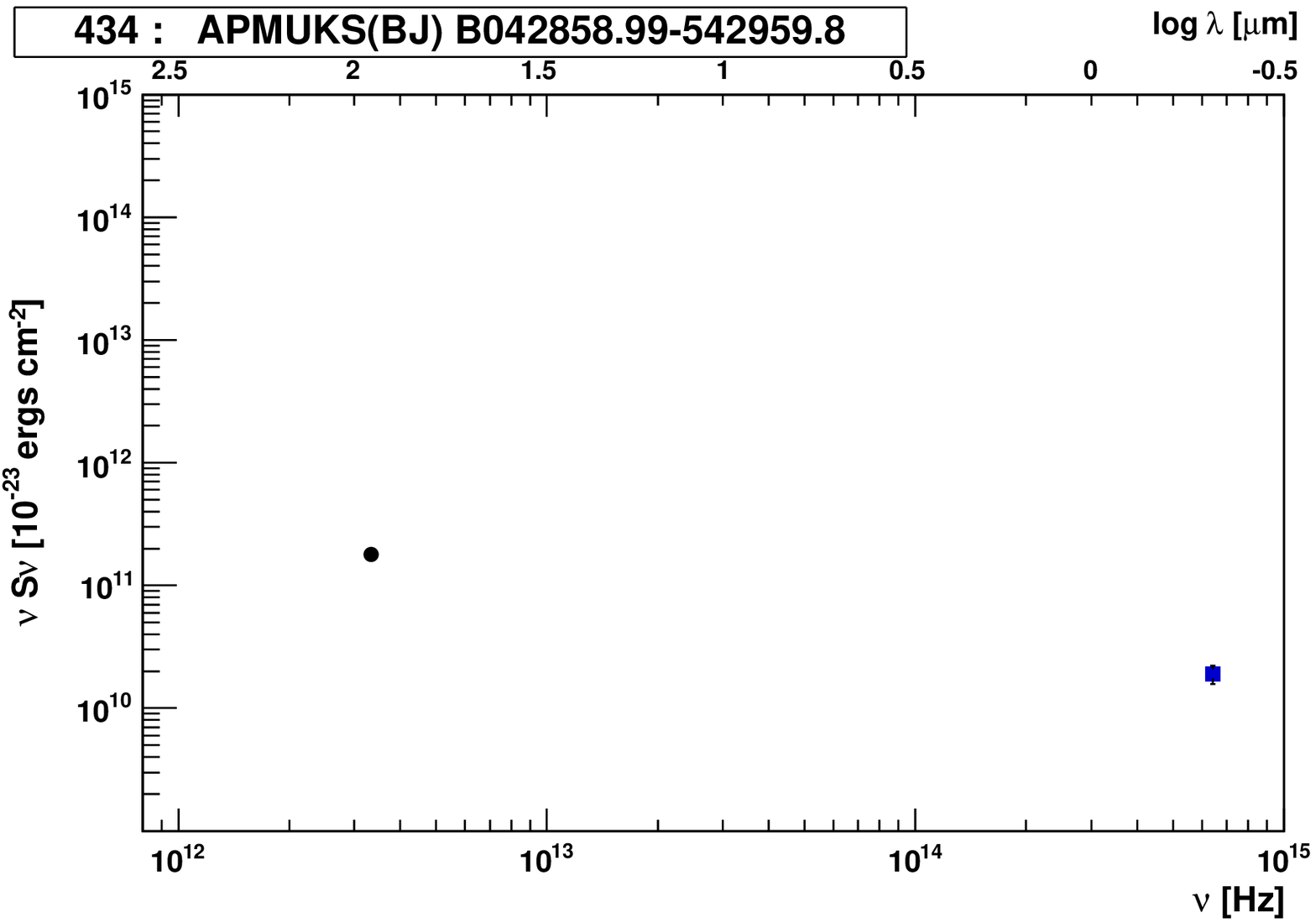}
\includegraphics[width=4cm]{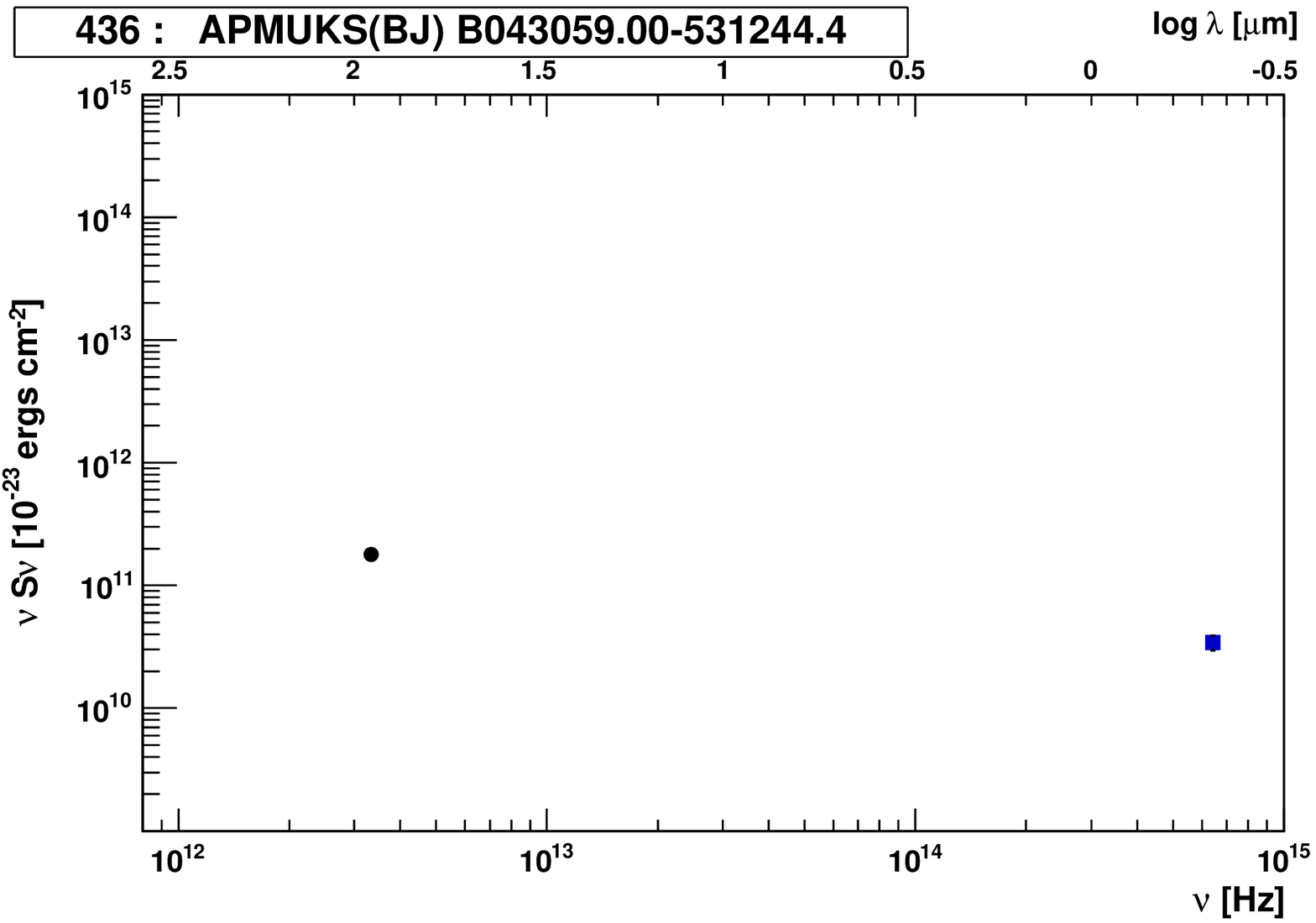}
\includegraphics[width=4cm]{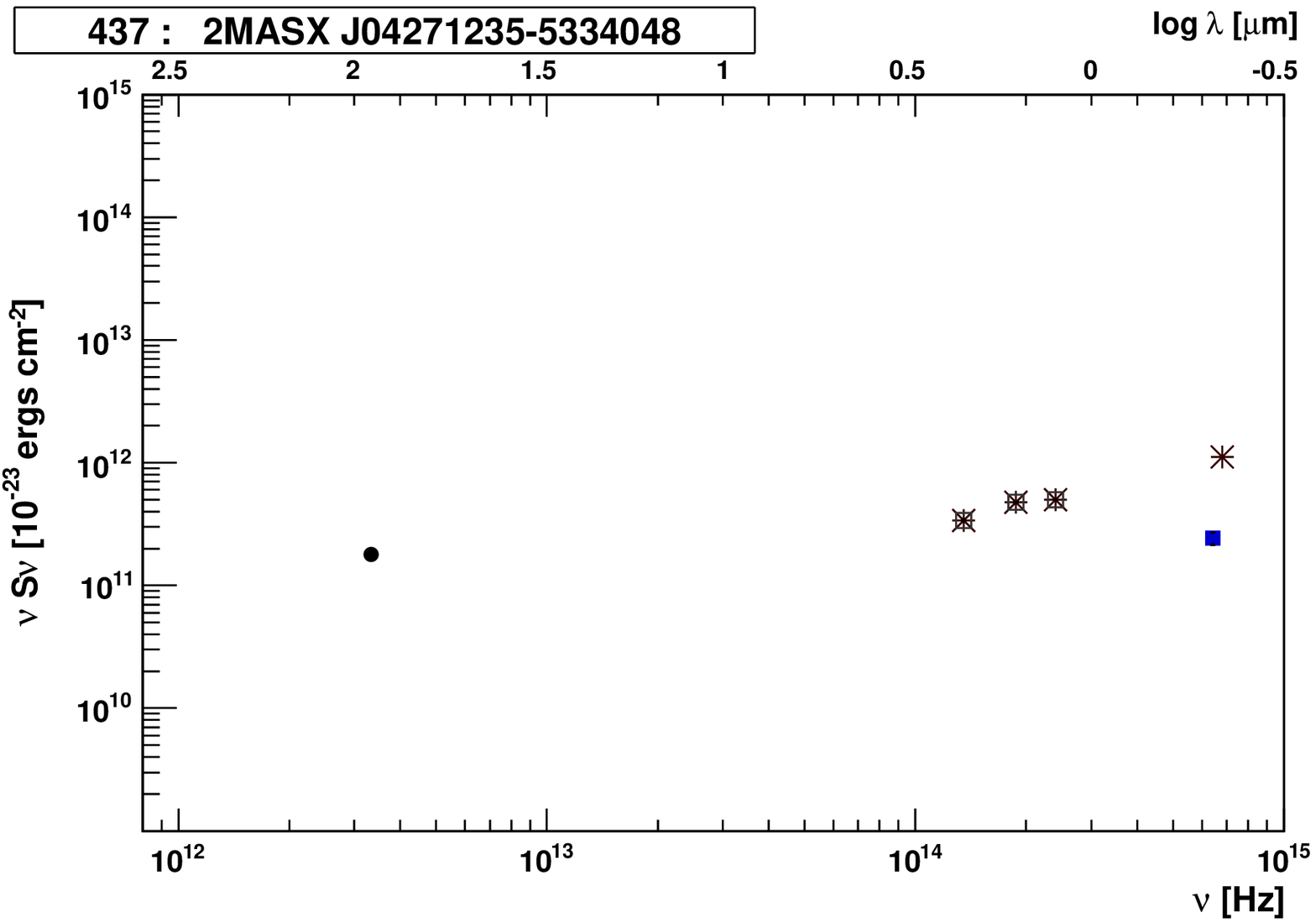}
\includegraphics[width=4cm]{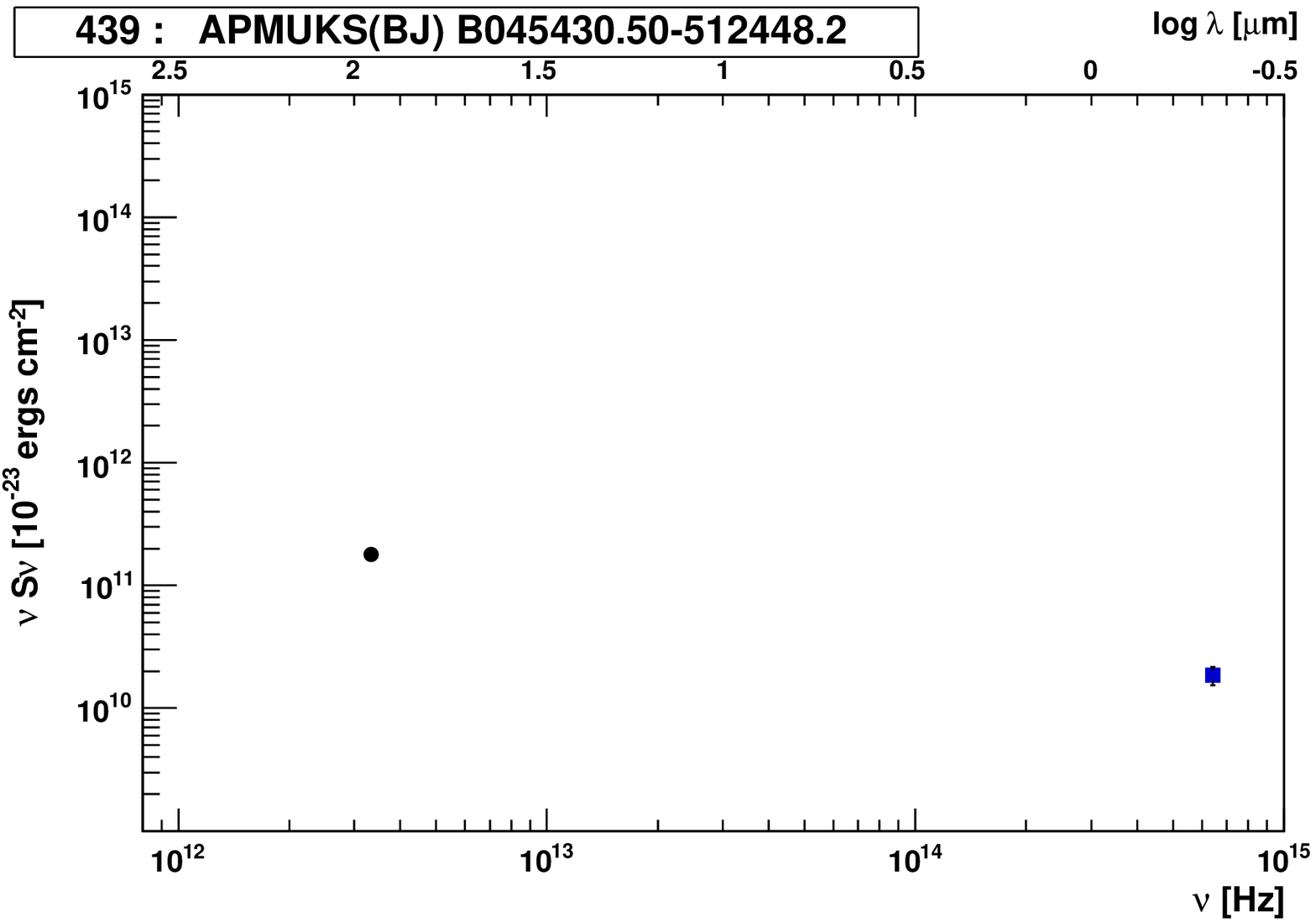}
\includegraphics[width=4cm]{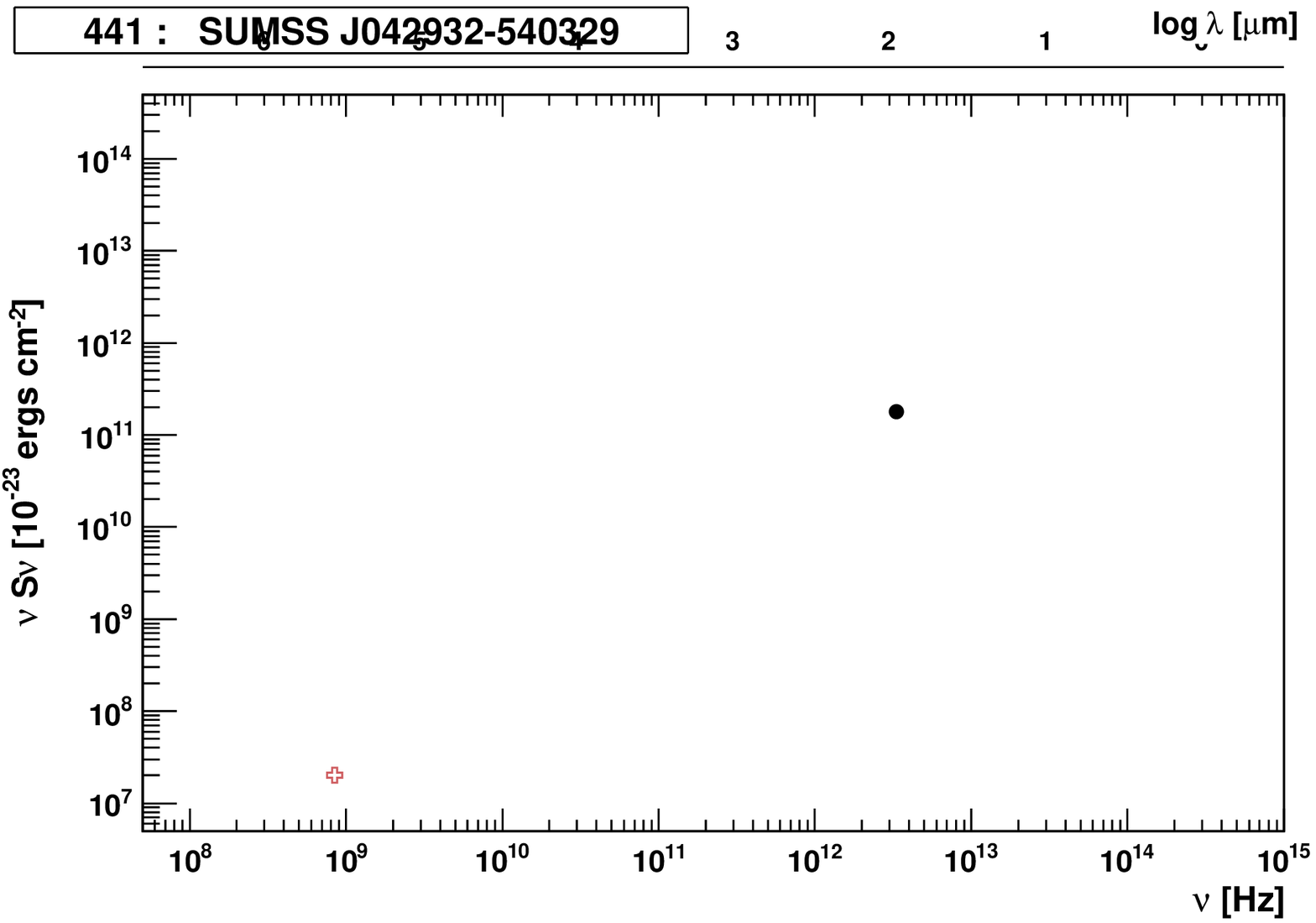}
\includegraphics[width=4cm]{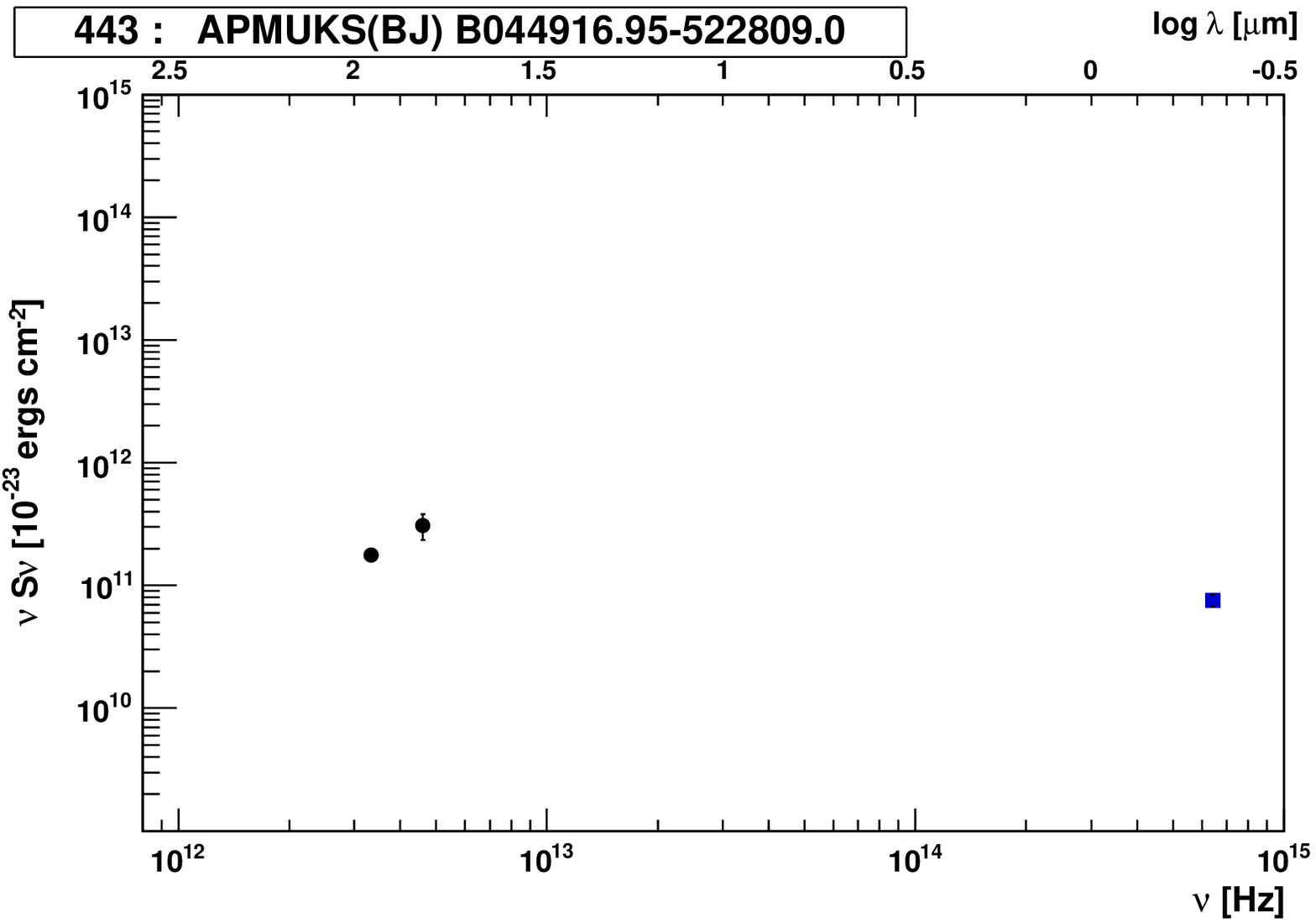}
\includegraphics[width=4cm]{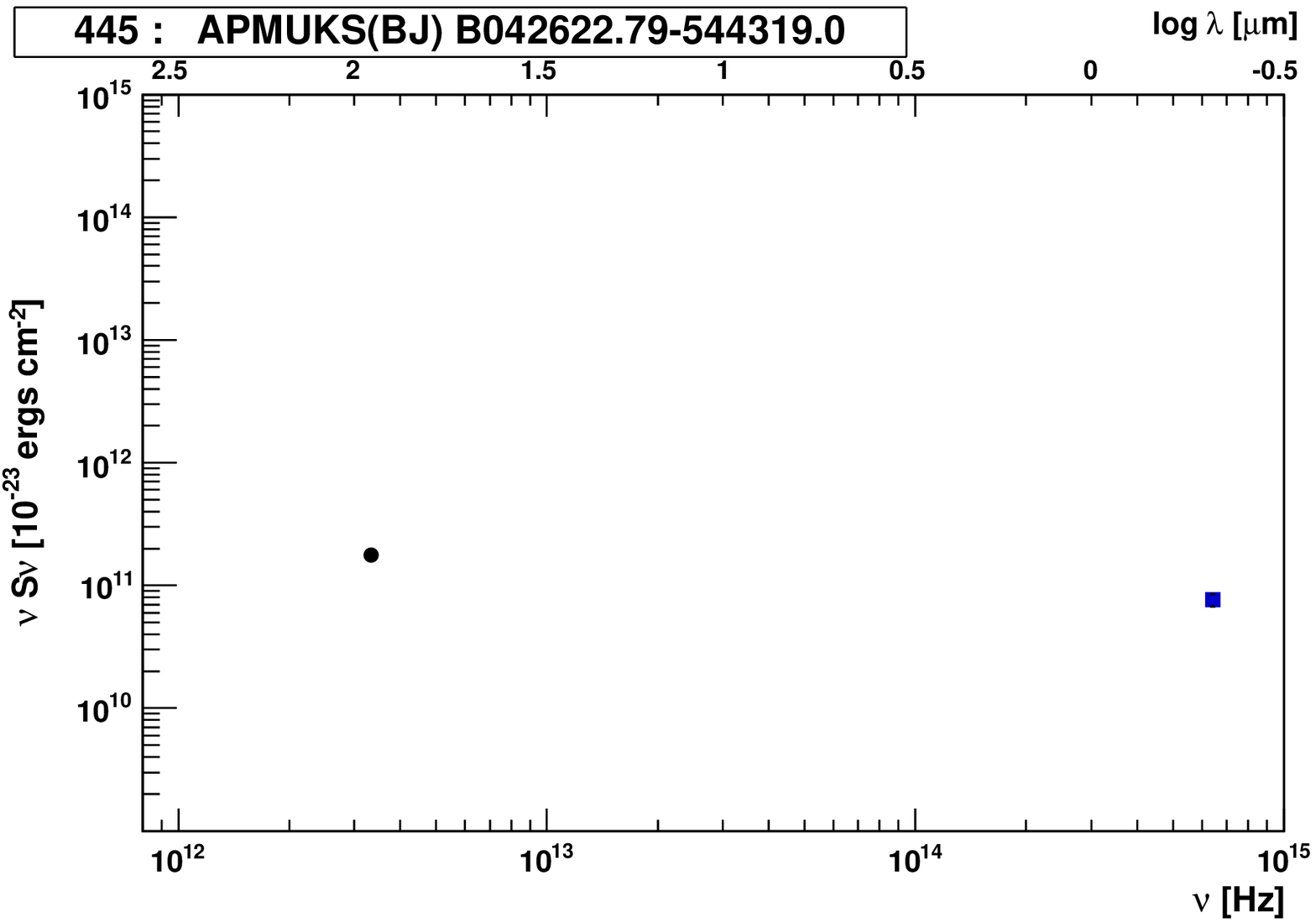}
\includegraphics[width=4cm]{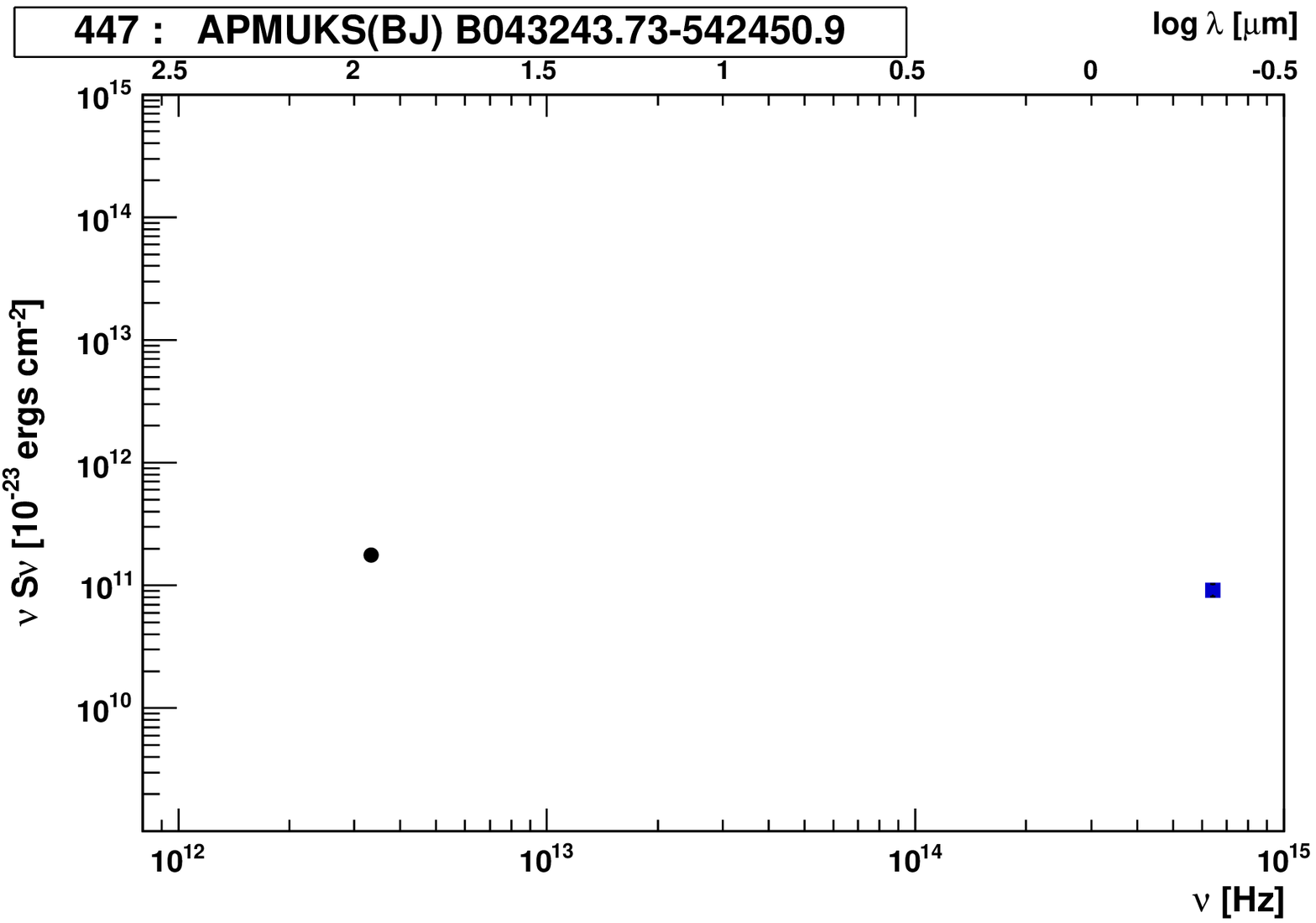}
\includegraphics[width=4cm]{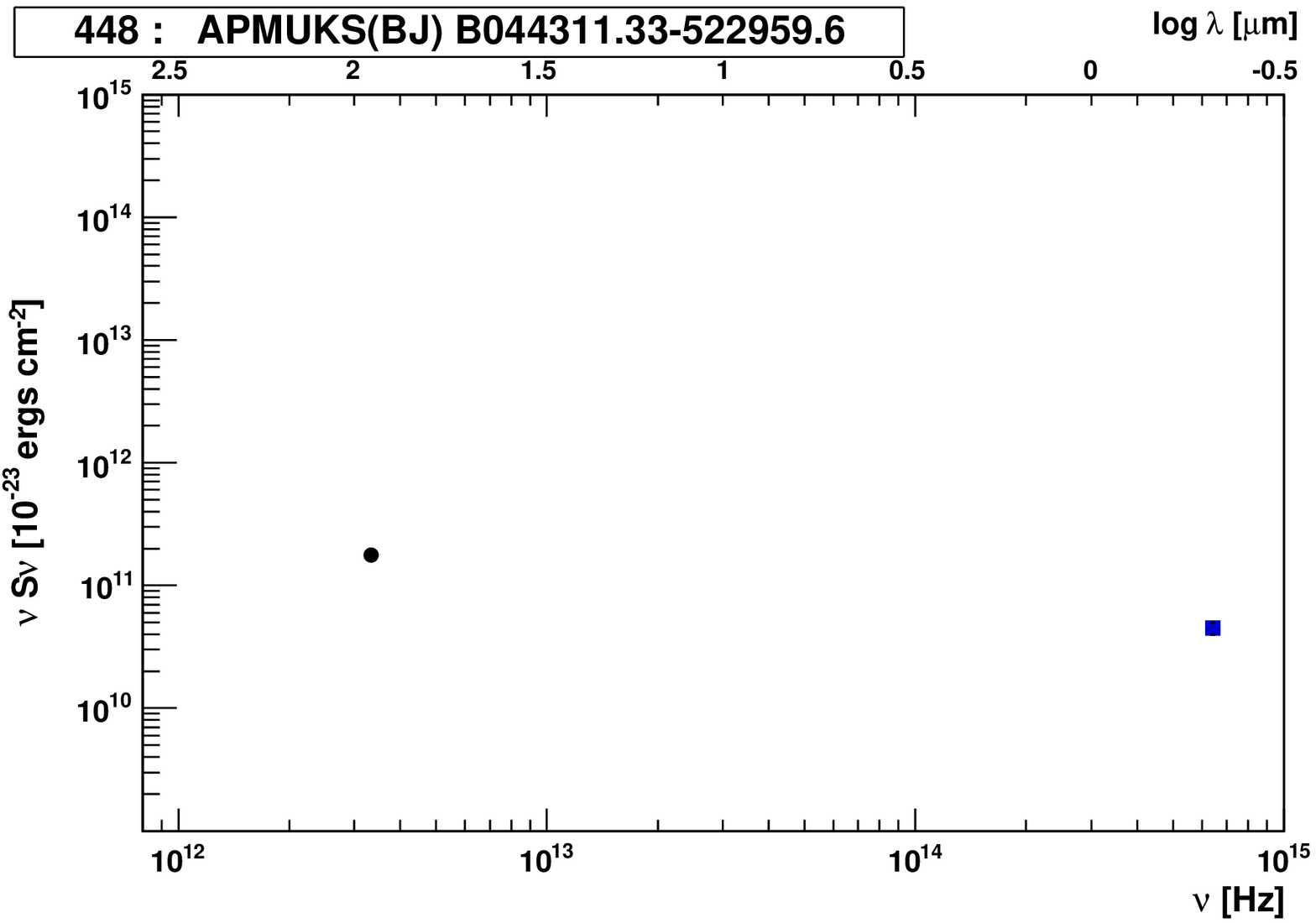}
\includegraphics[width=4cm]{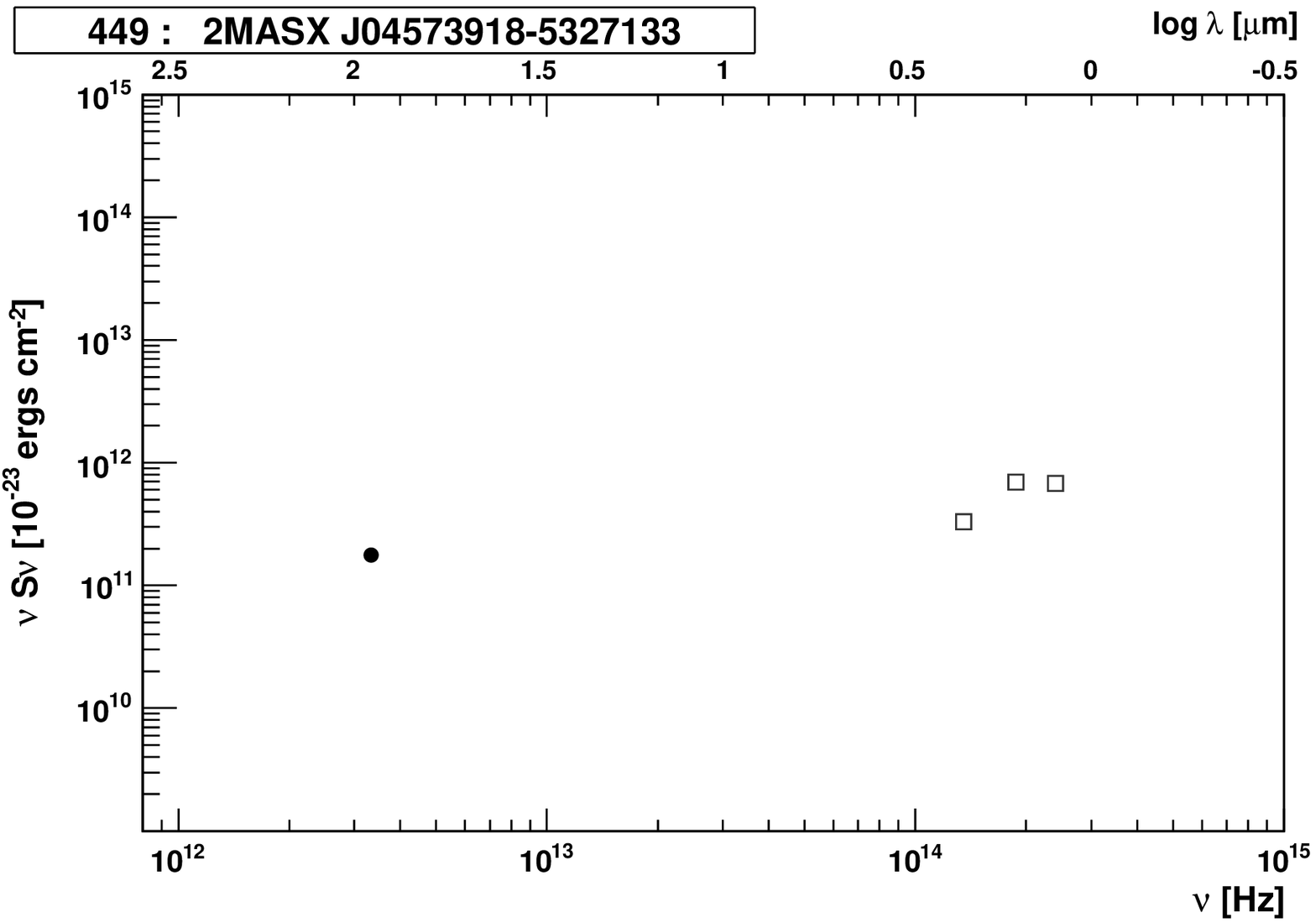}
\includegraphics[width=4cm]{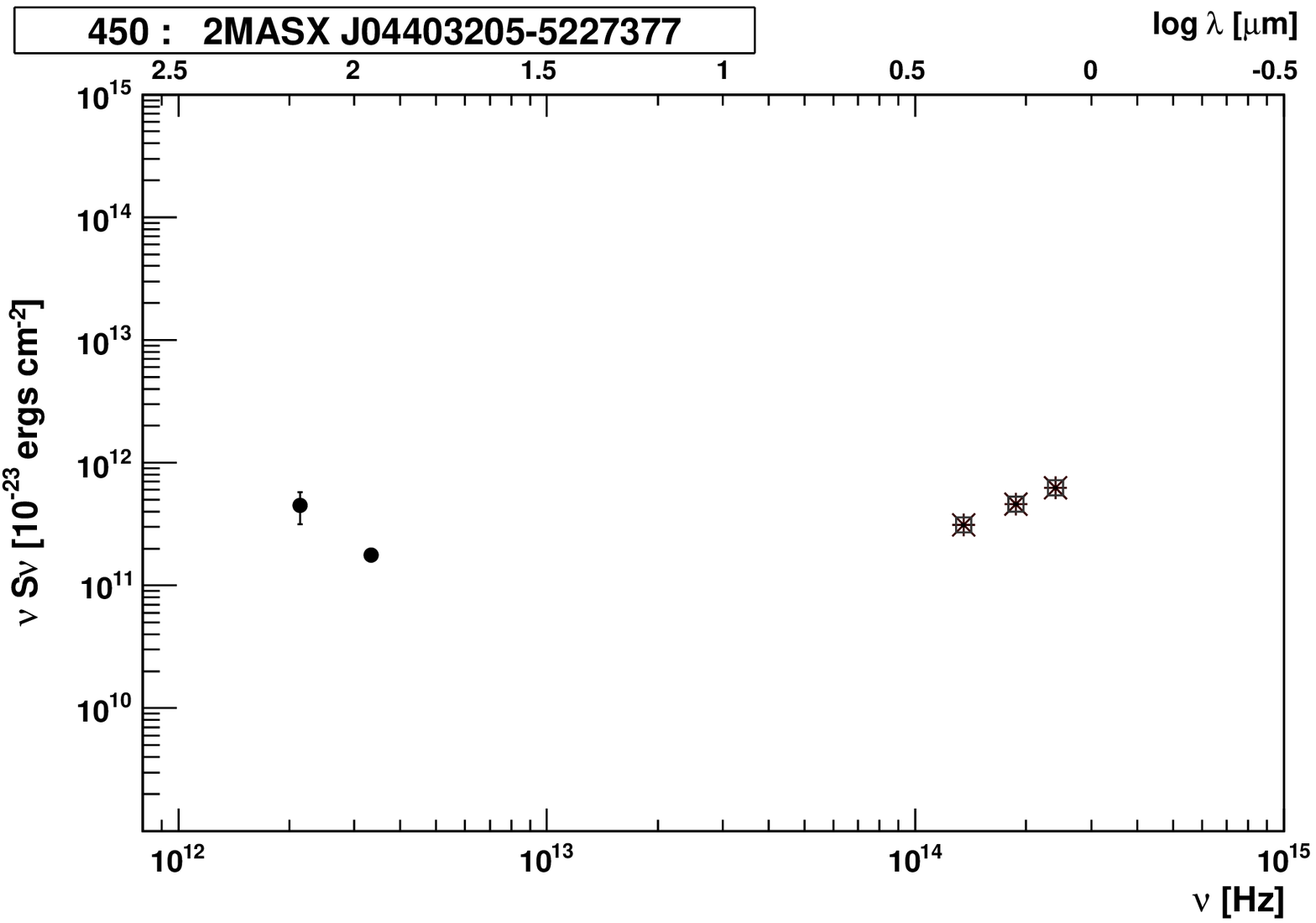}
\includegraphics[width=4cm]{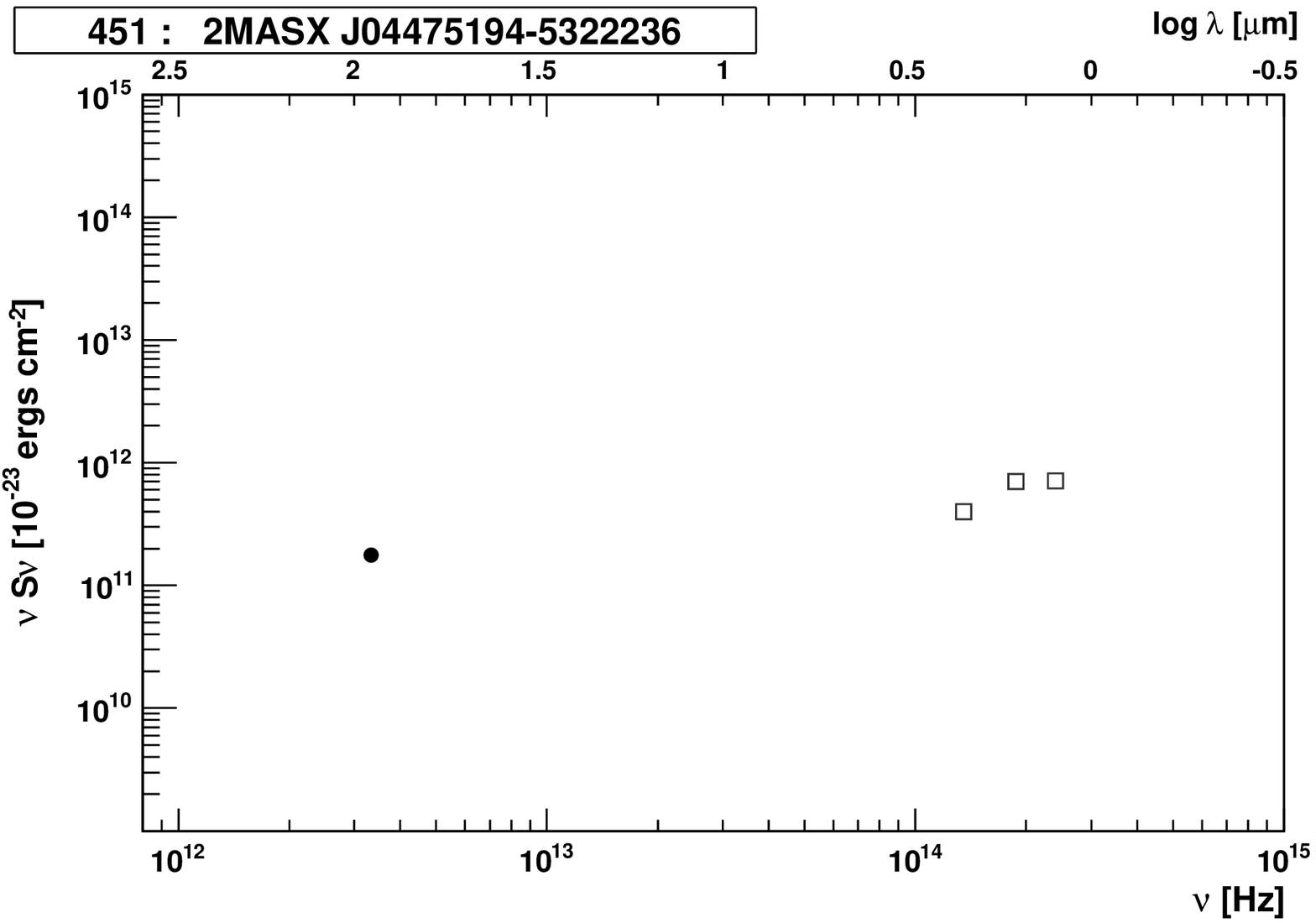}
\includegraphics[width=4cm]{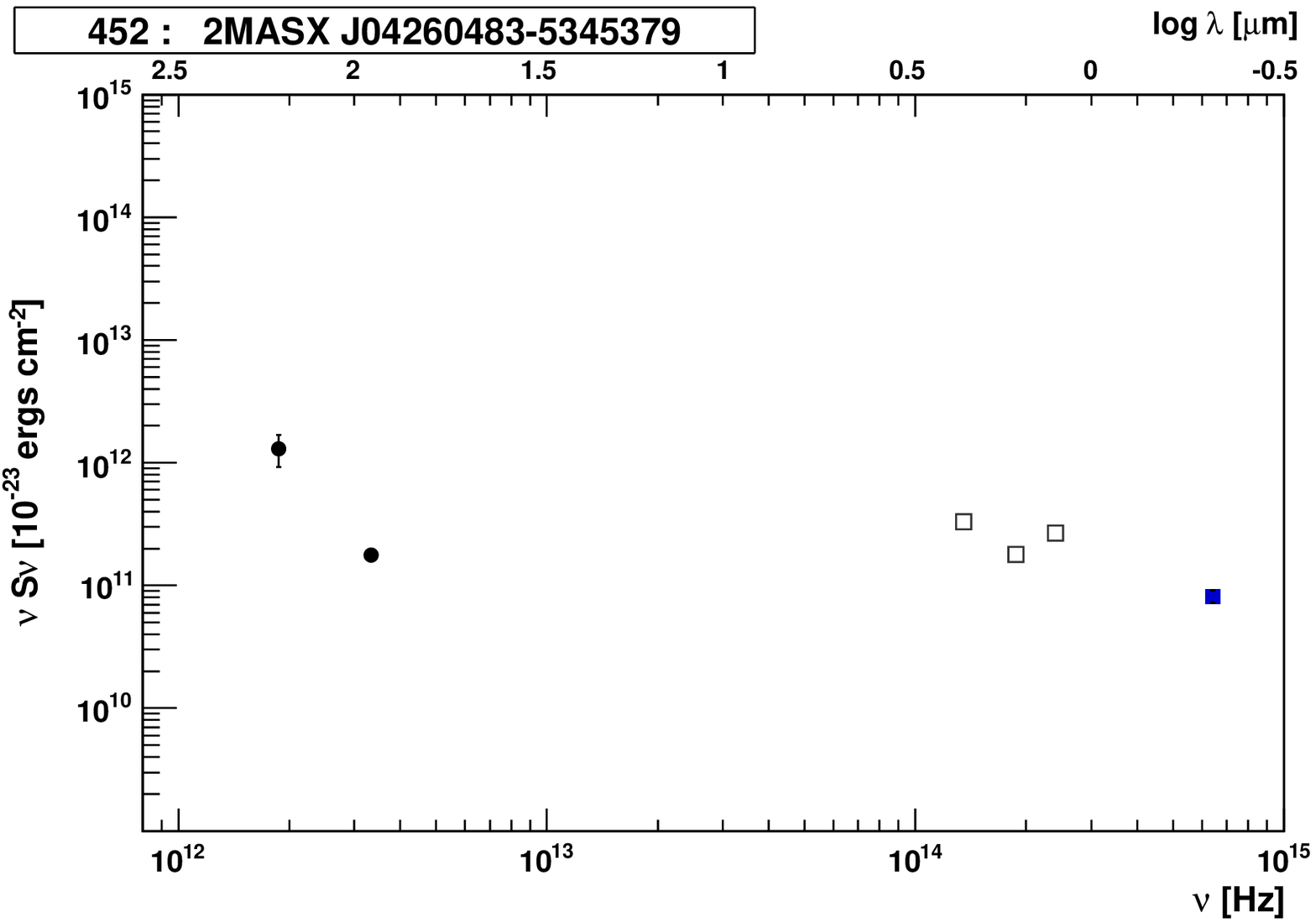}
\includegraphics[width=4cm]{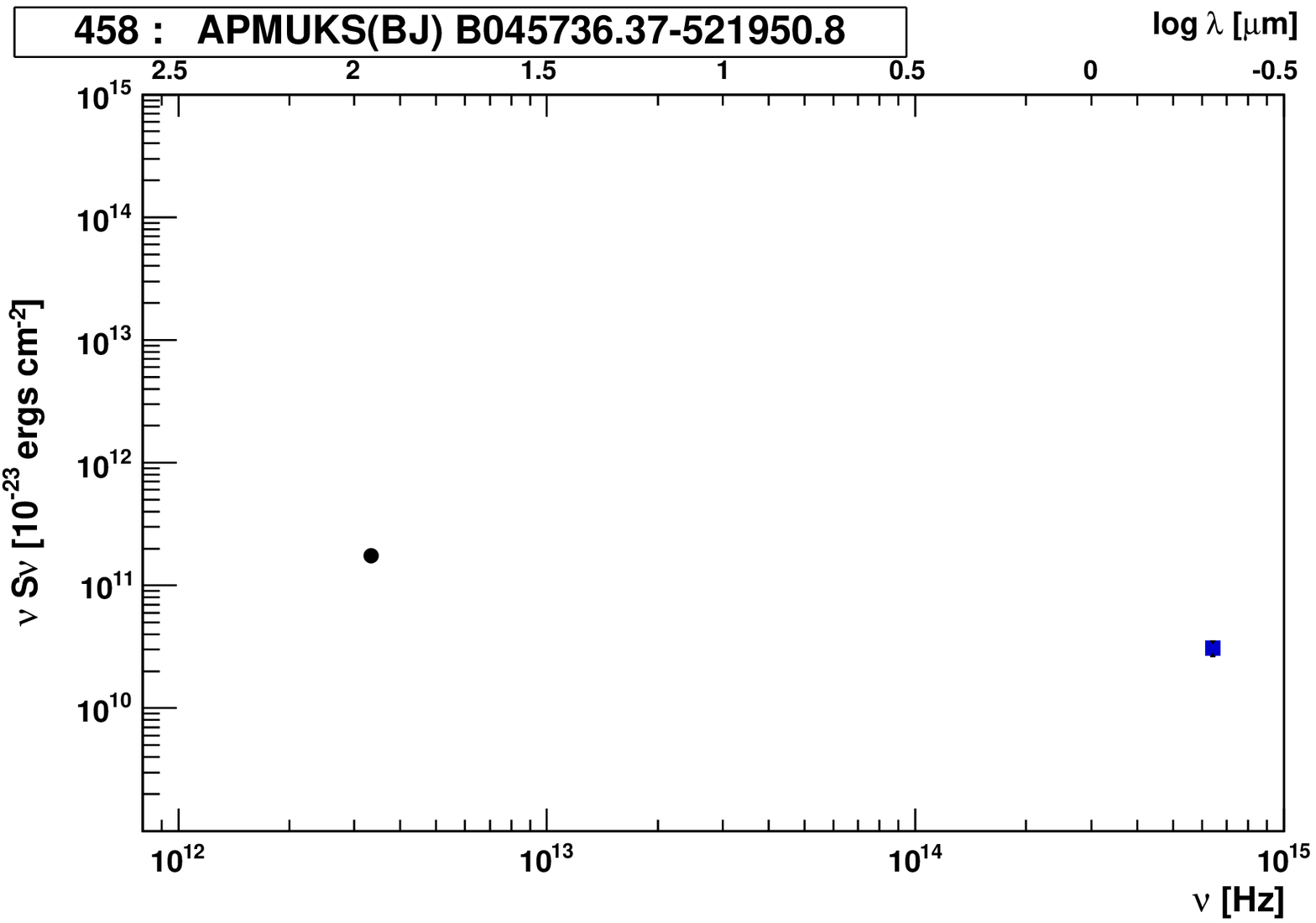}
\includegraphics[width=4cm]{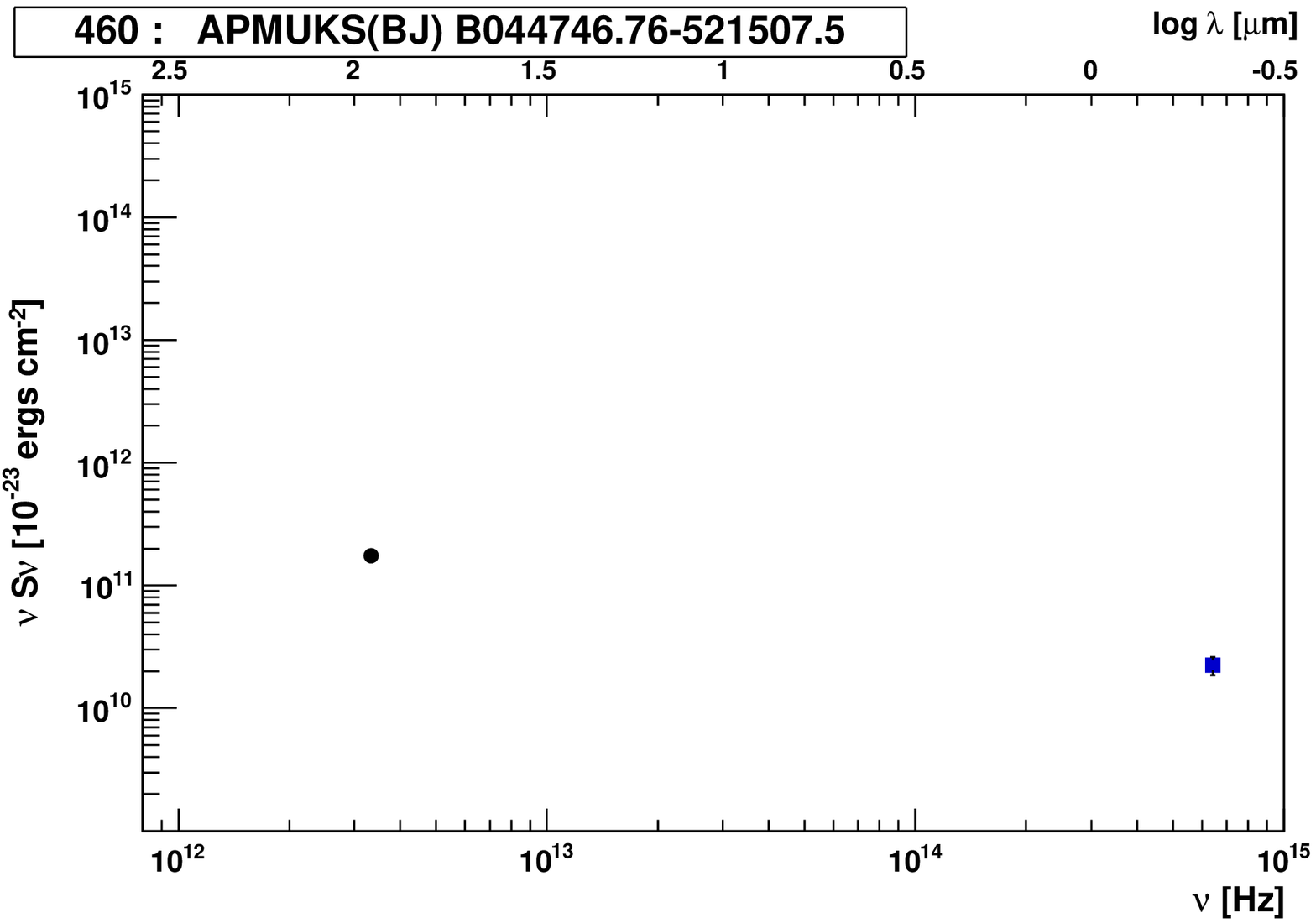}
\includegraphics[width=4cm]{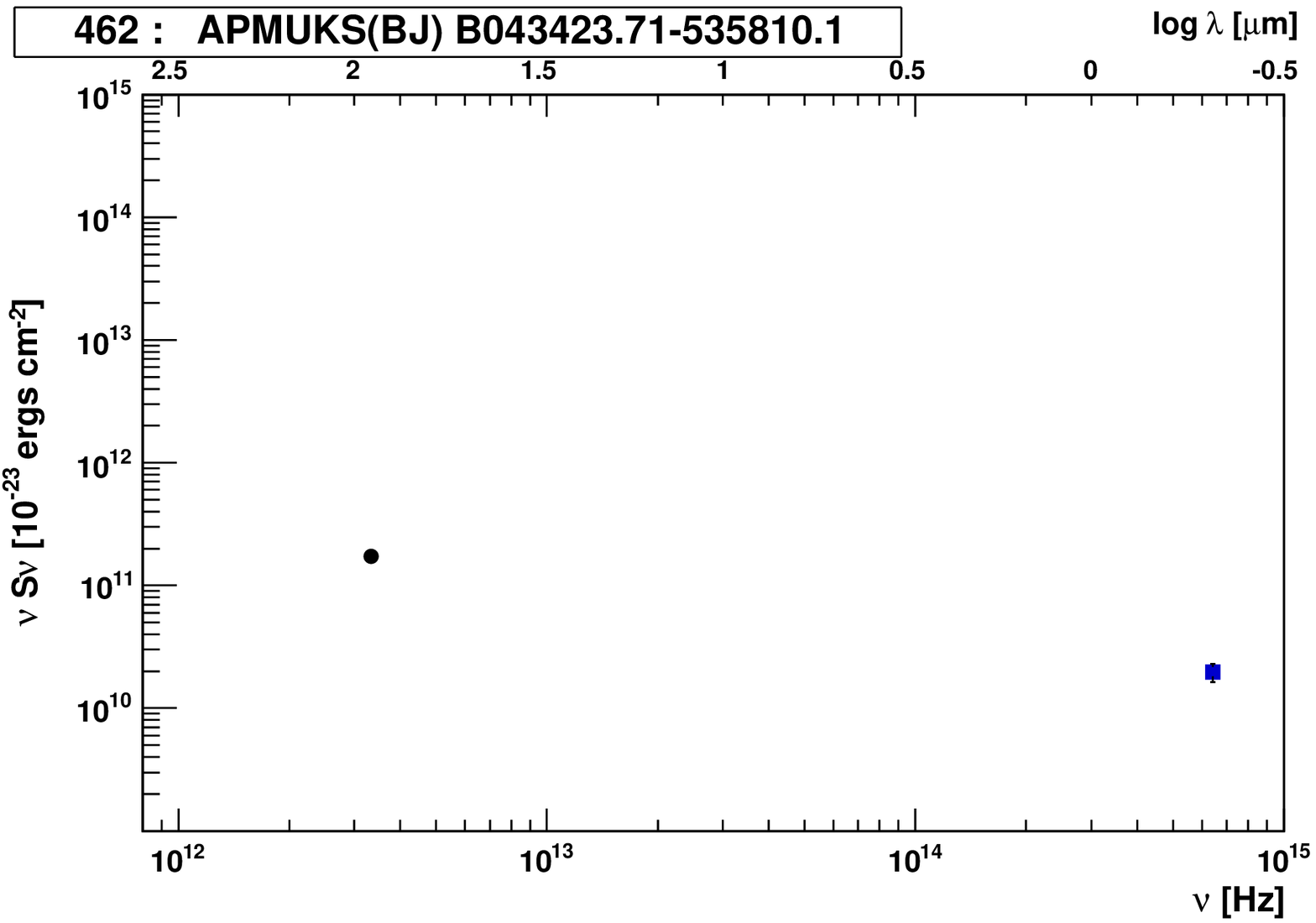}
\includegraphics[width=4cm]{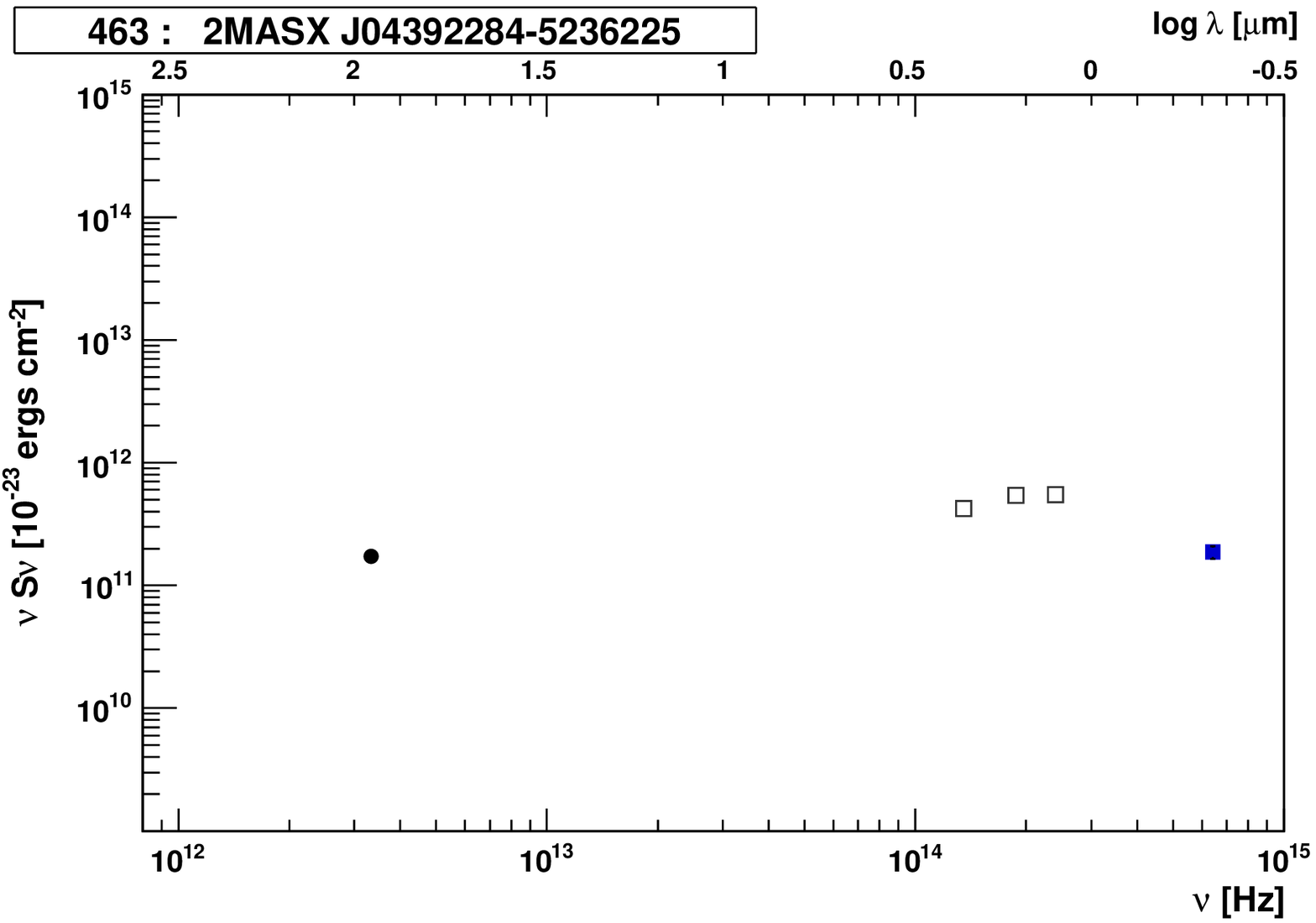}
\includegraphics[width=4cm]{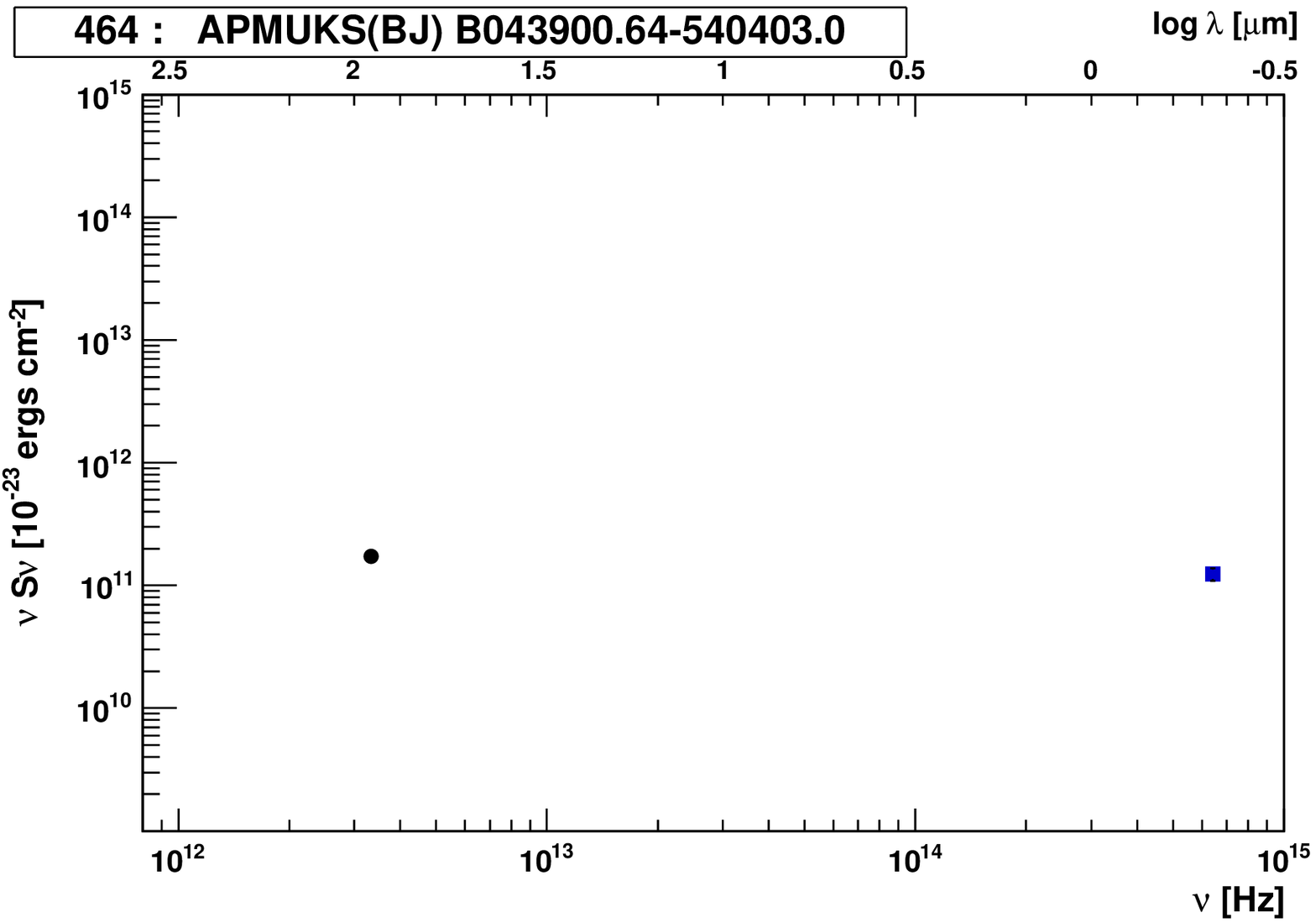}
\includegraphics[width=4cm]{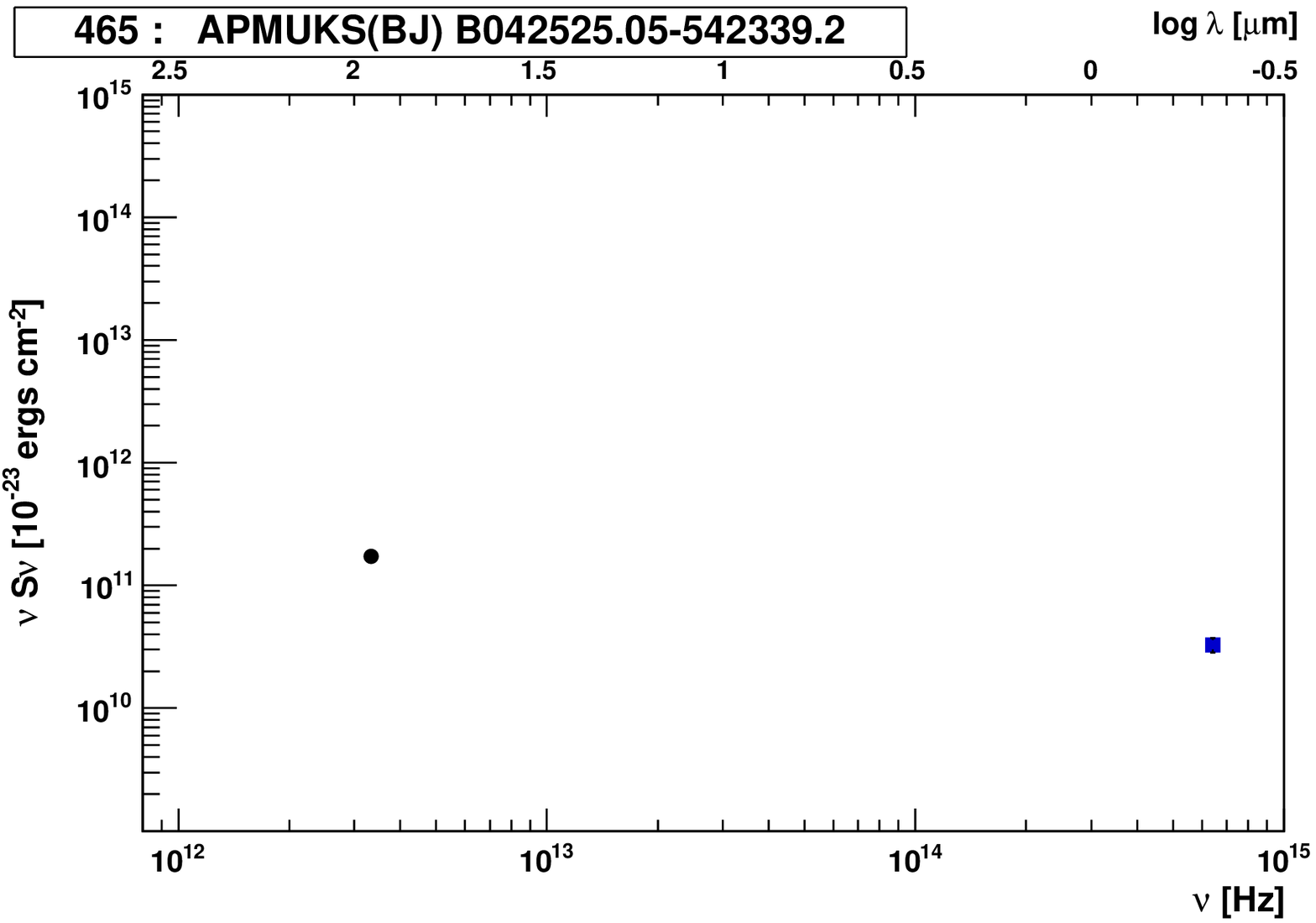}
\includegraphics[width=4cm]{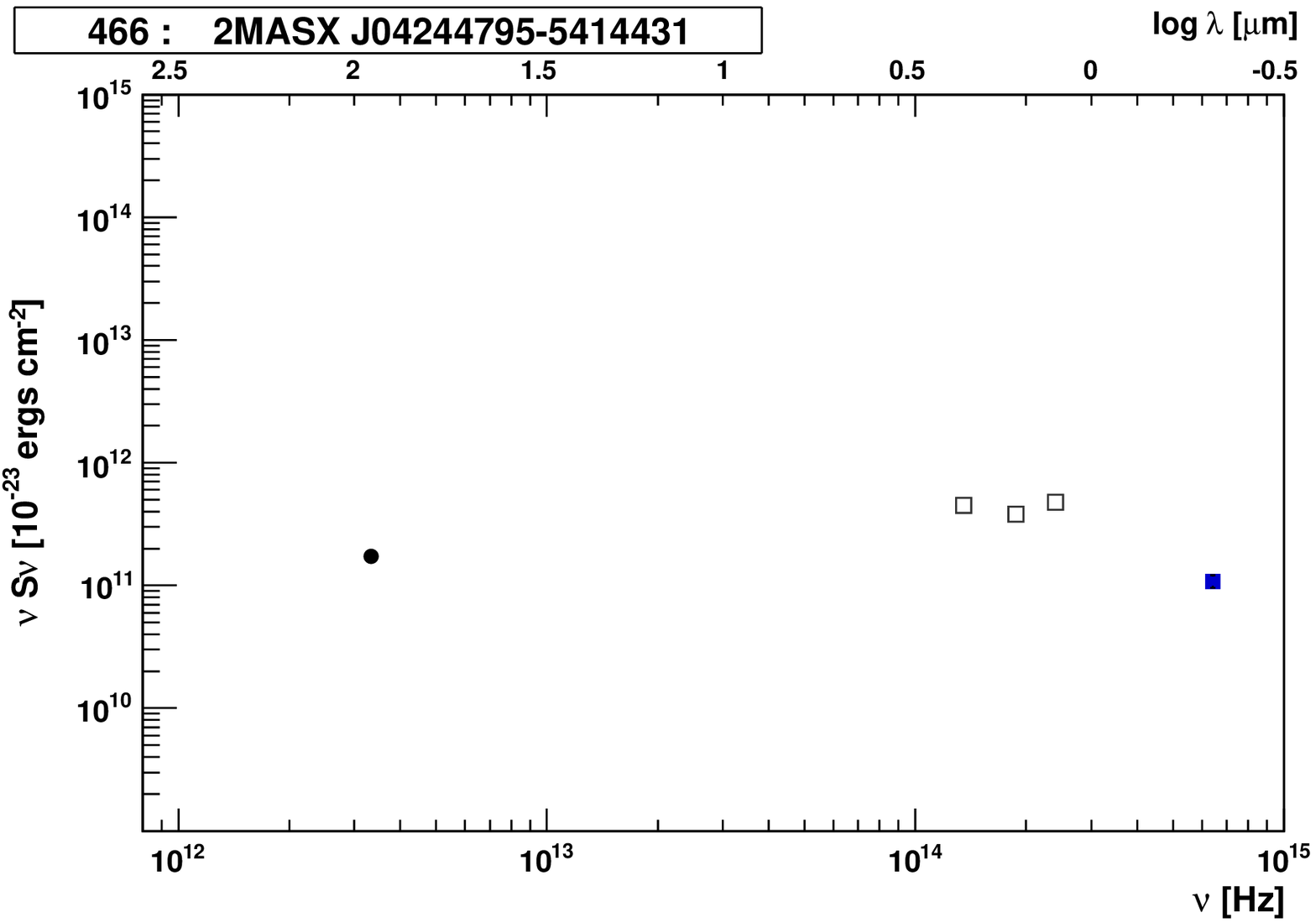}
\includegraphics[width=4cm]{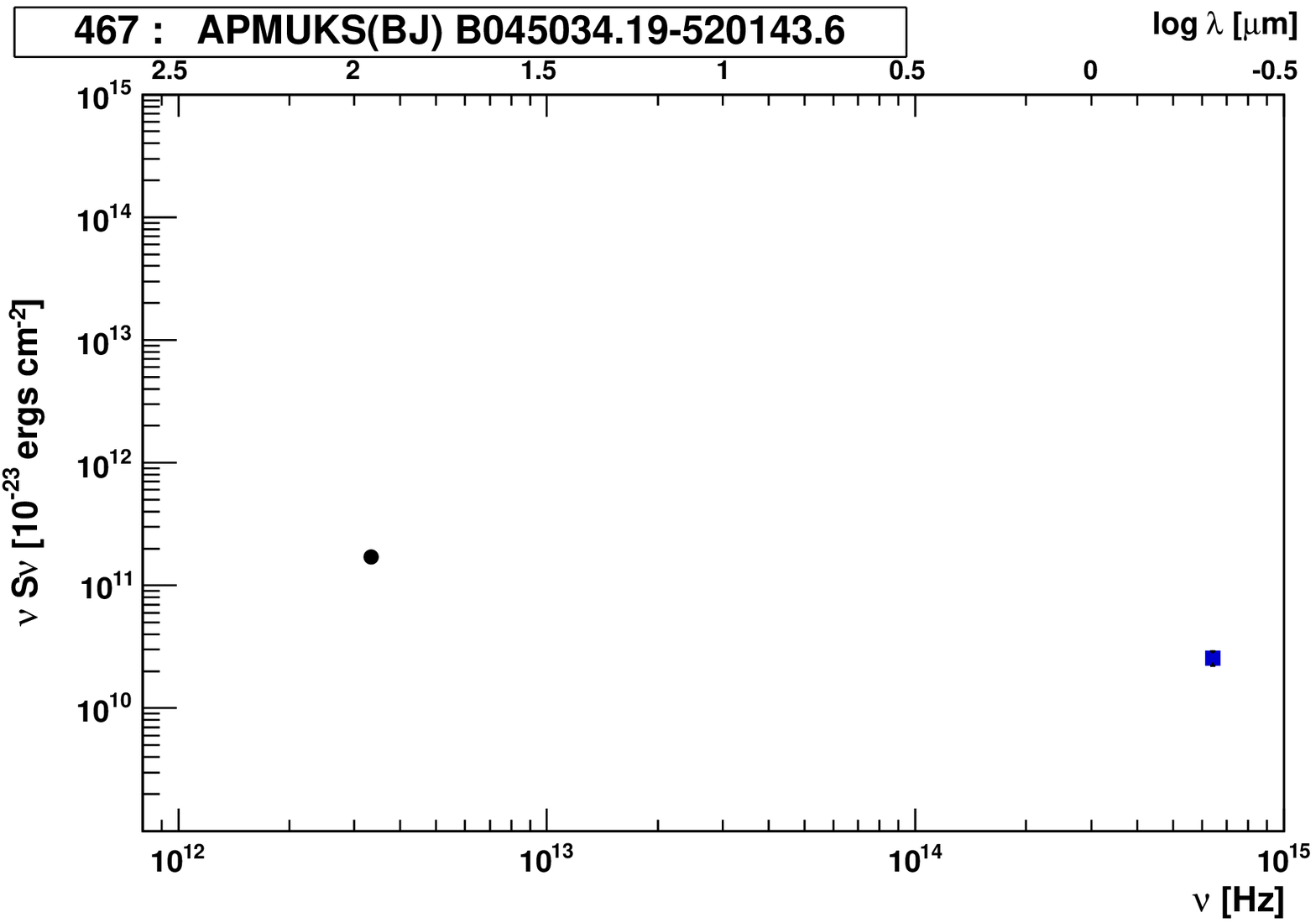}
\includegraphics[width=4cm]{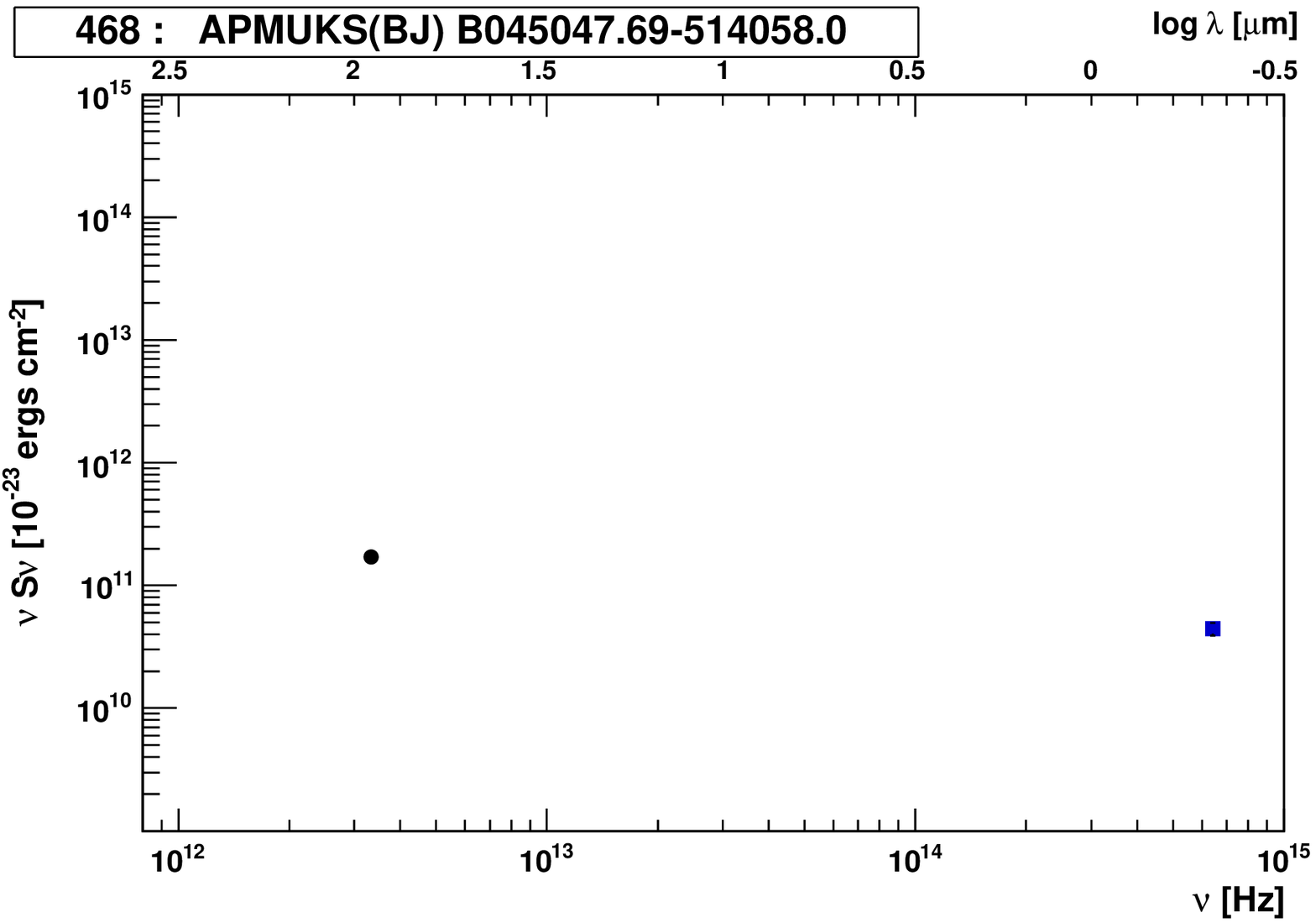}
\includegraphics[width=4cm]{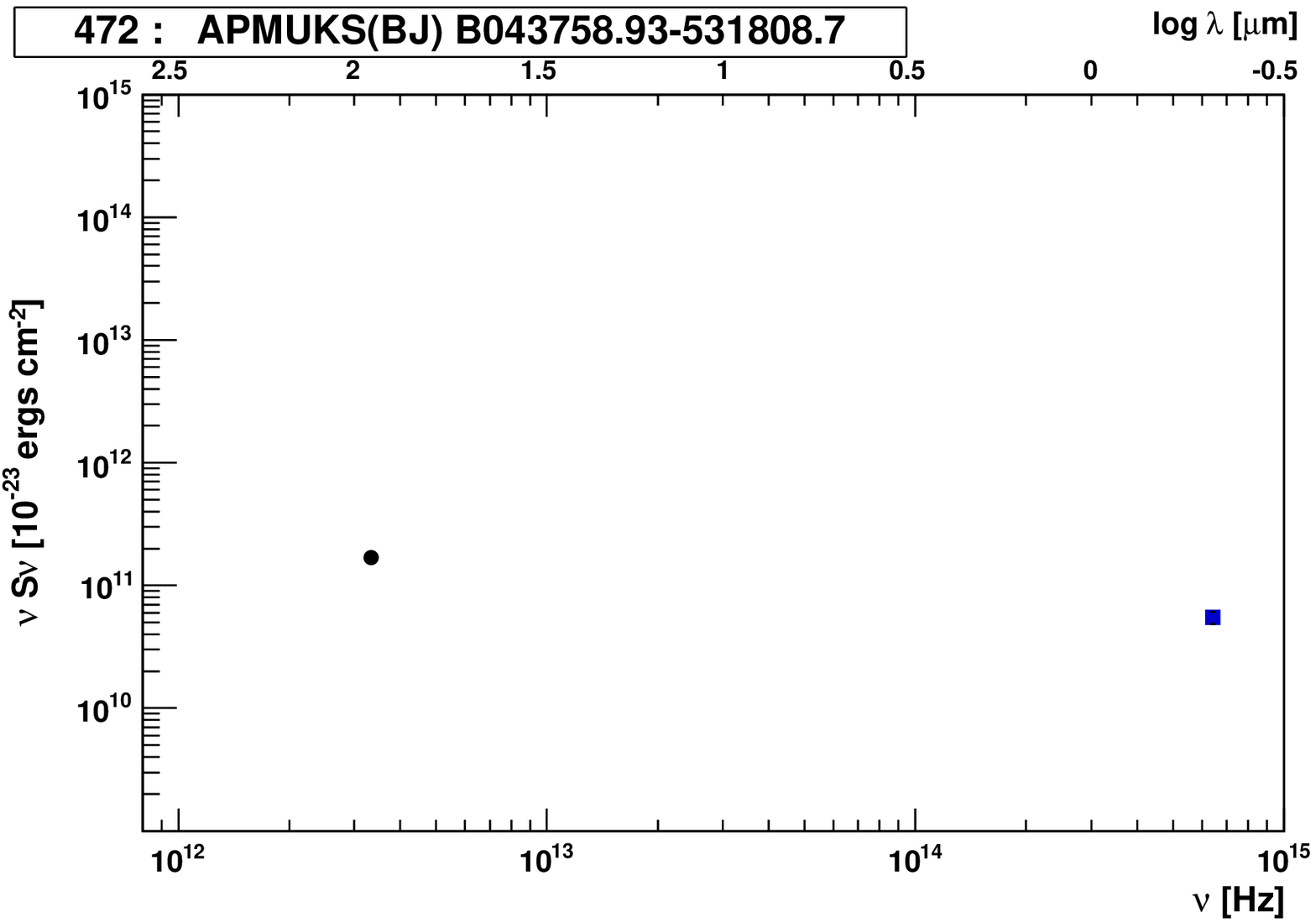}
\includegraphics[width=4cm]{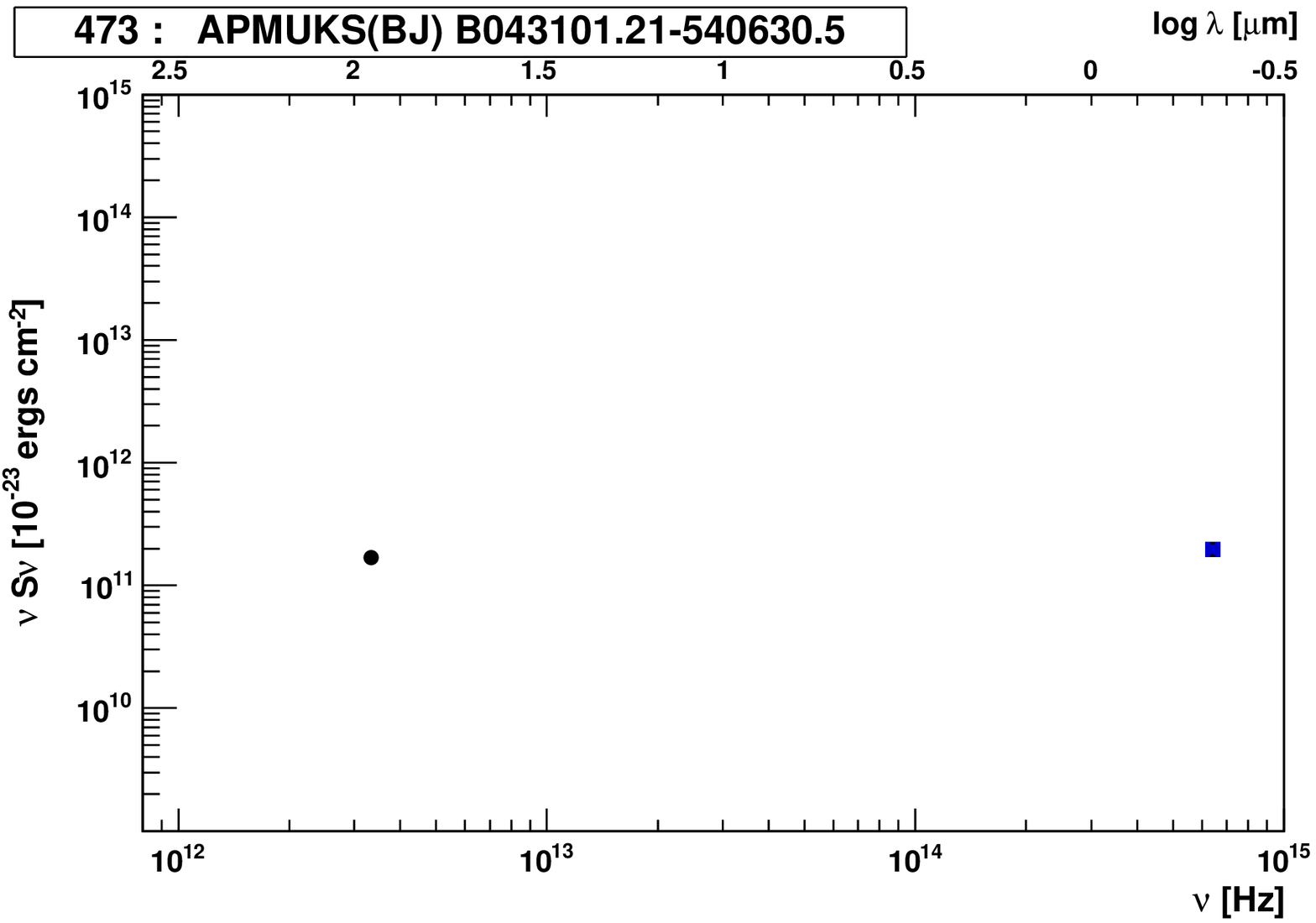}
\includegraphics[width=4cm]{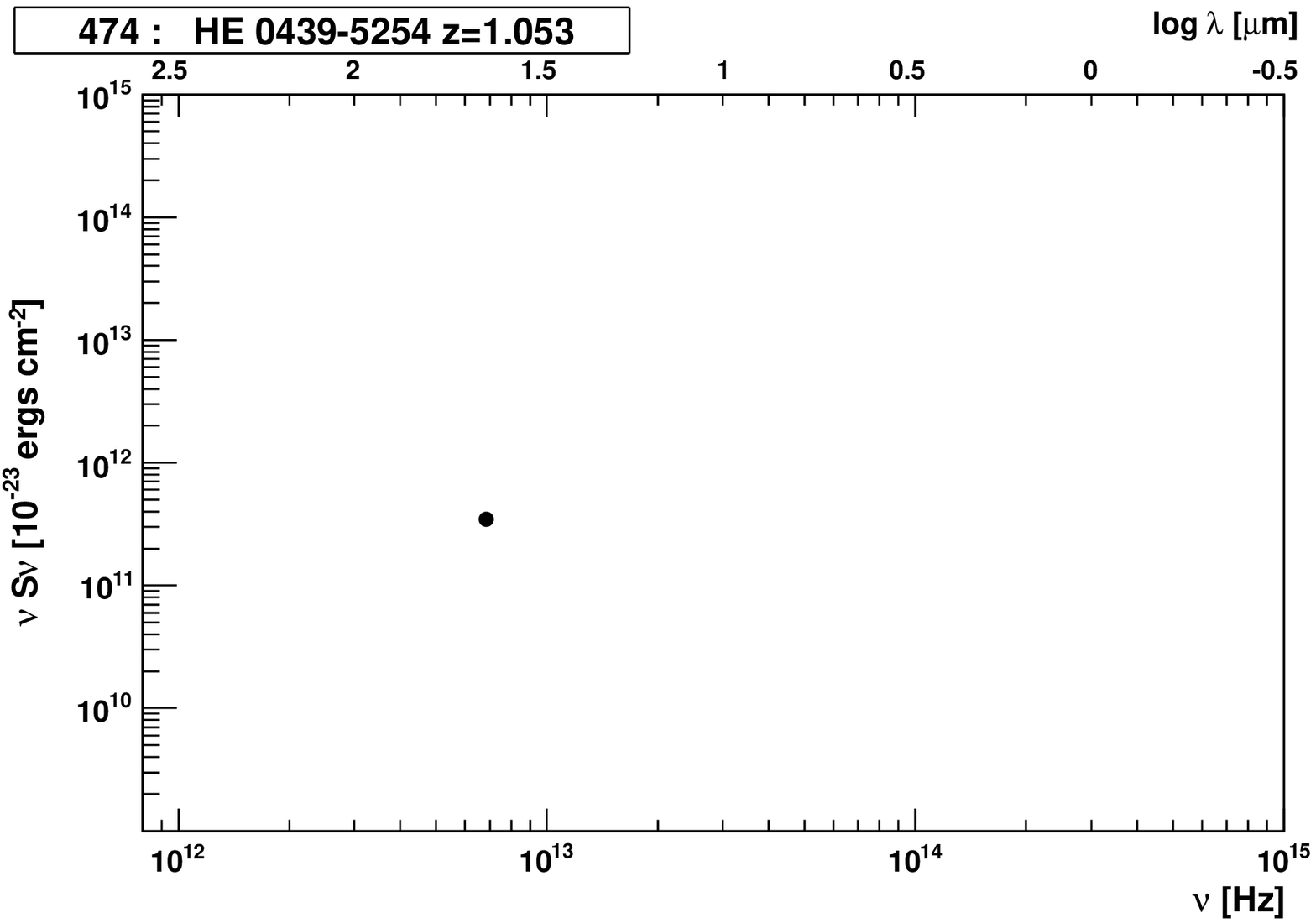}
\includegraphics[width=4cm]{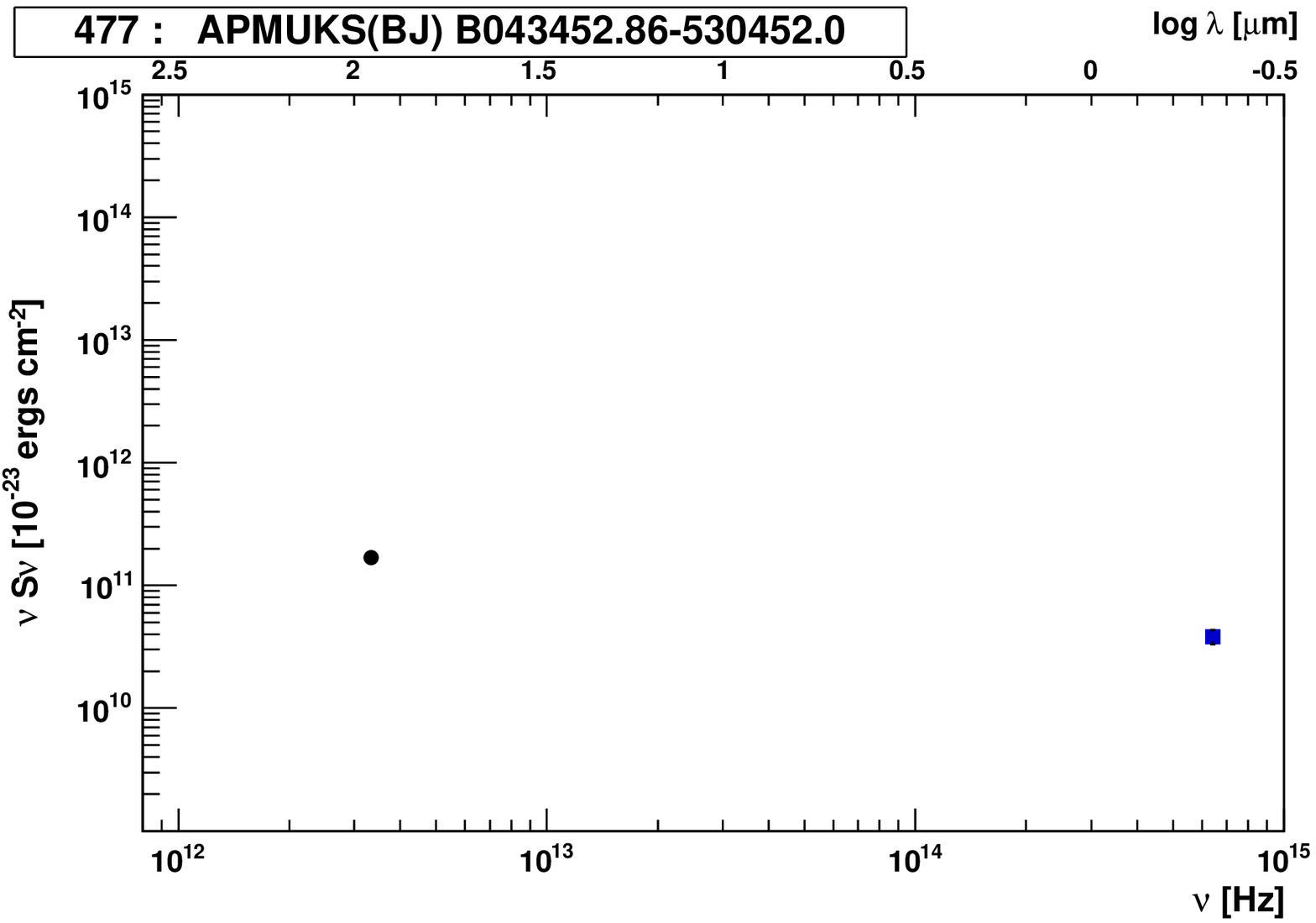}
\includegraphics[width=4cm]{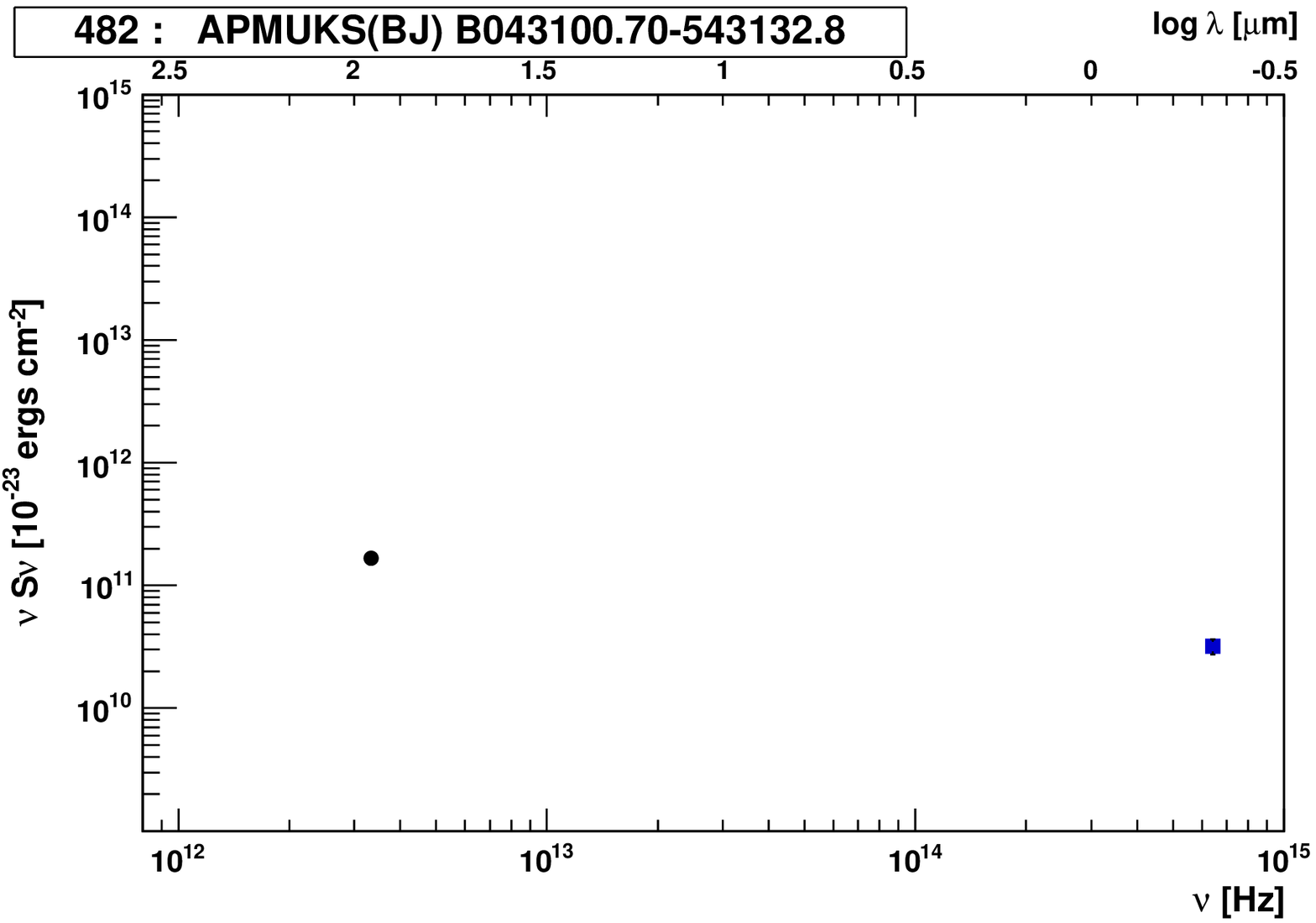}
\includegraphics[width=4cm]{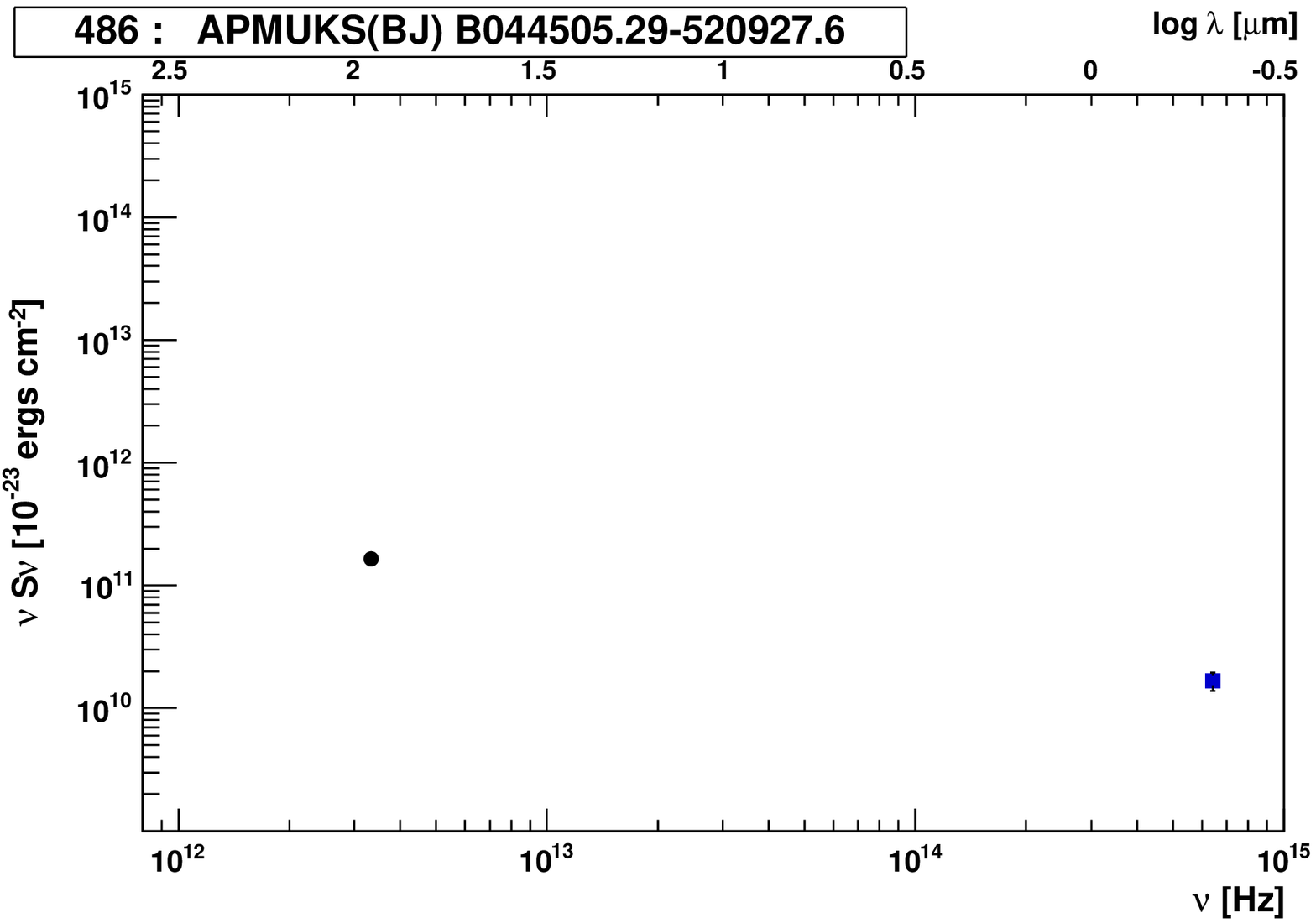}
\includegraphics[width=4cm]{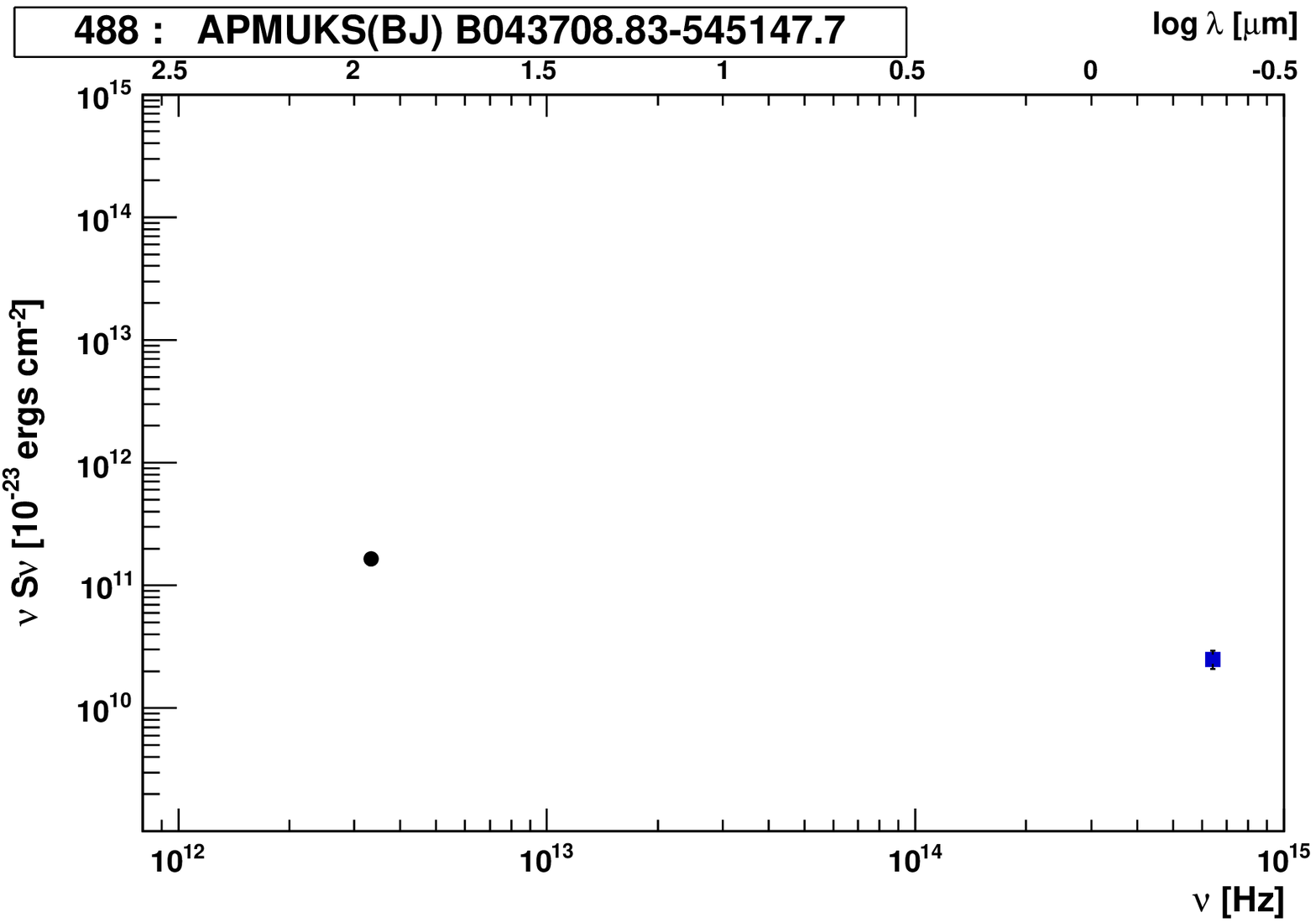}
\includegraphics[width=4cm]{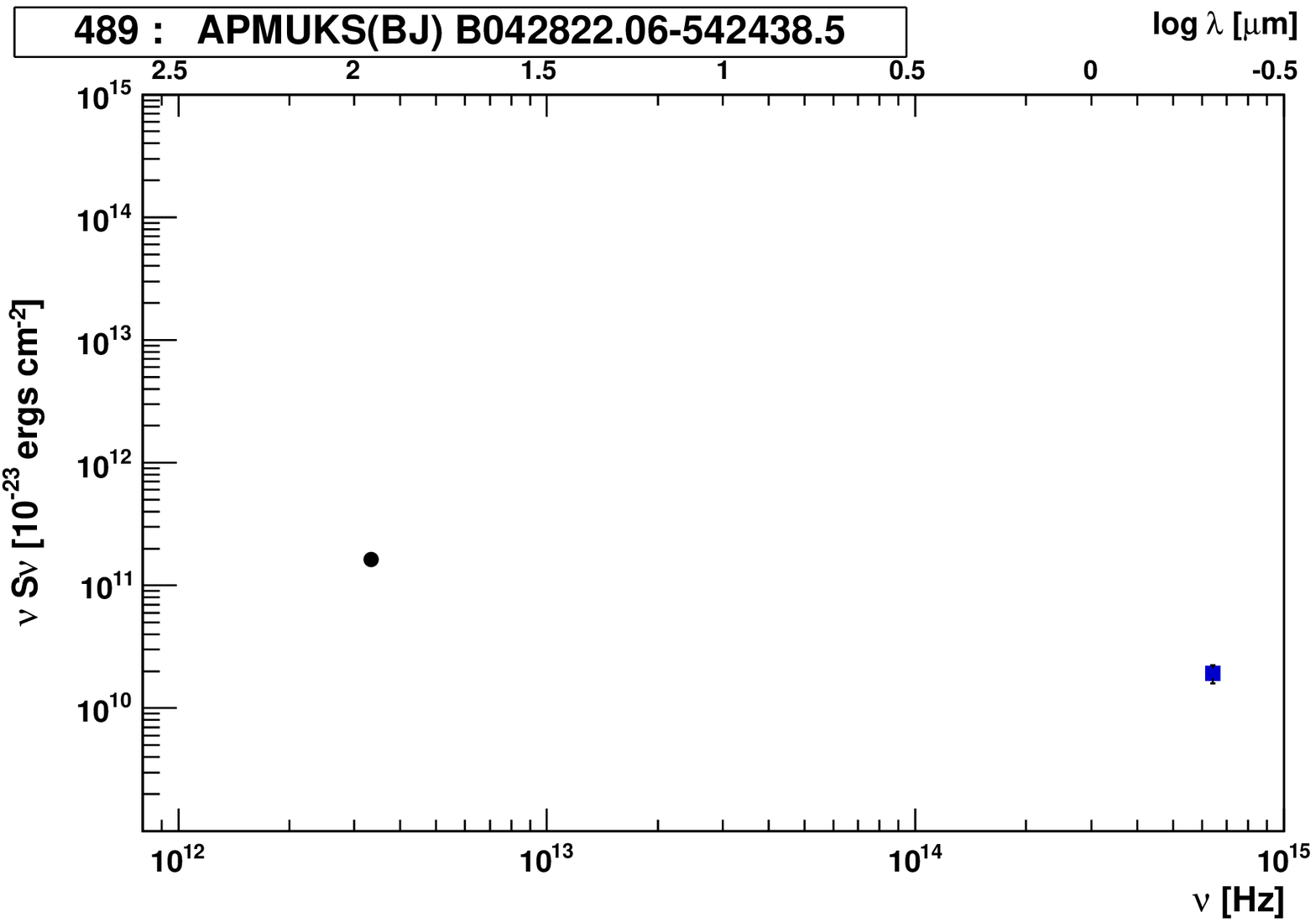}
\label{points9}
\caption {SEDs for the next 36 ADF-S identified sources, with symbols as in Figure~\ref{points1}.}
\end{figure*}
}

\clearpage

\onlfig{10}{
\begin{figure*}[t]
\centering

\includegraphics[width=4cm]{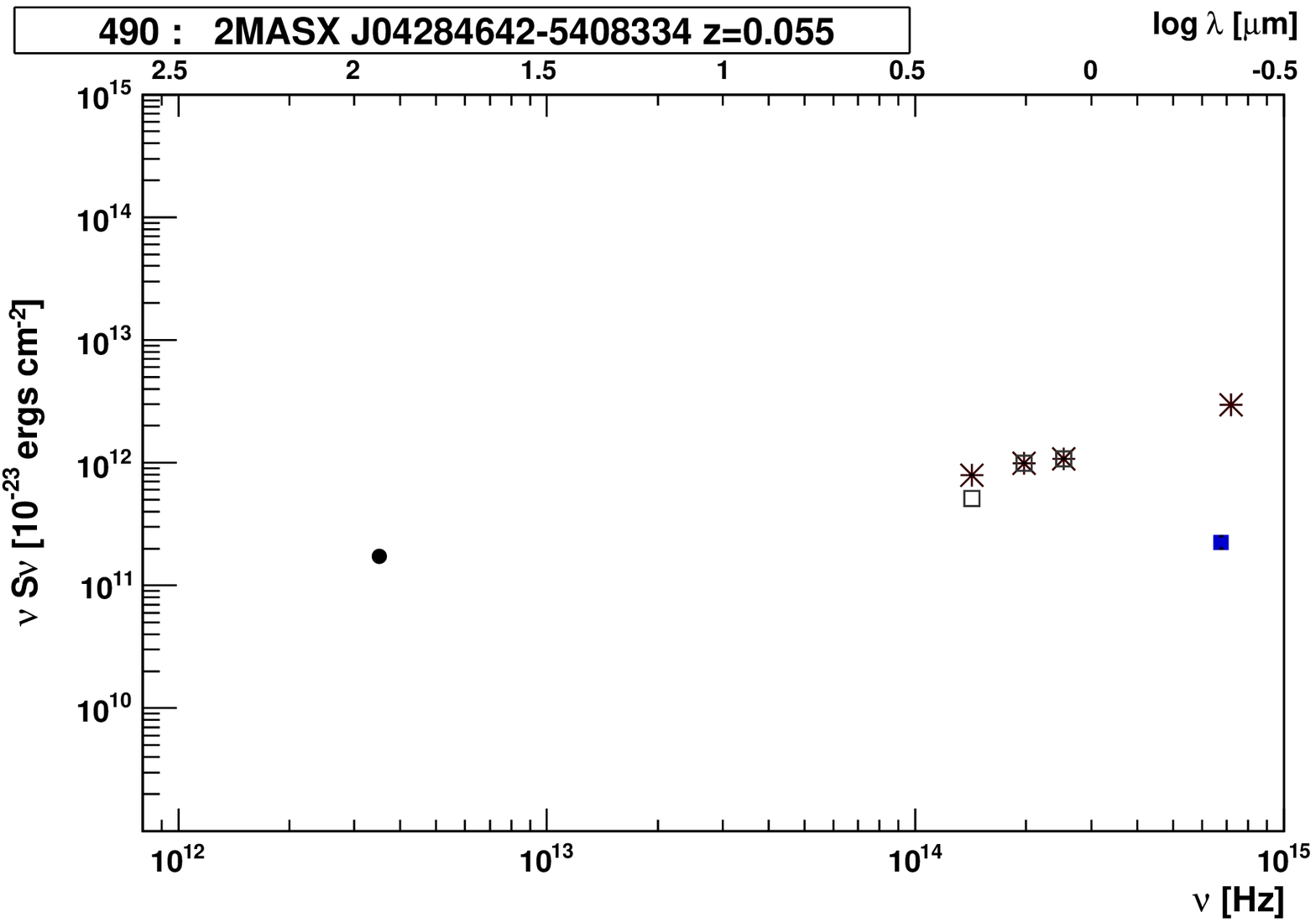}
\includegraphics[width=4cm]{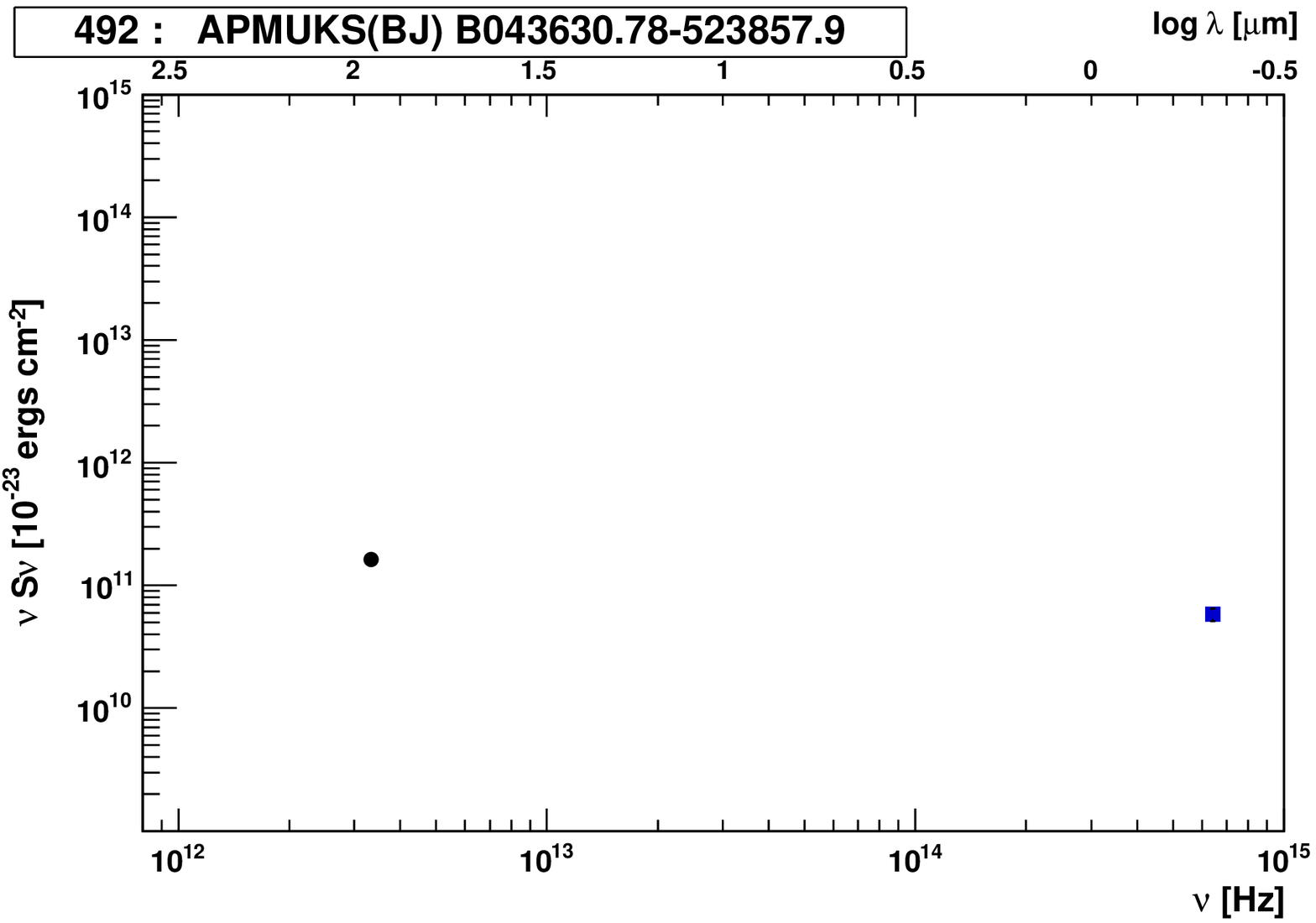}
\includegraphics[width=4cm]{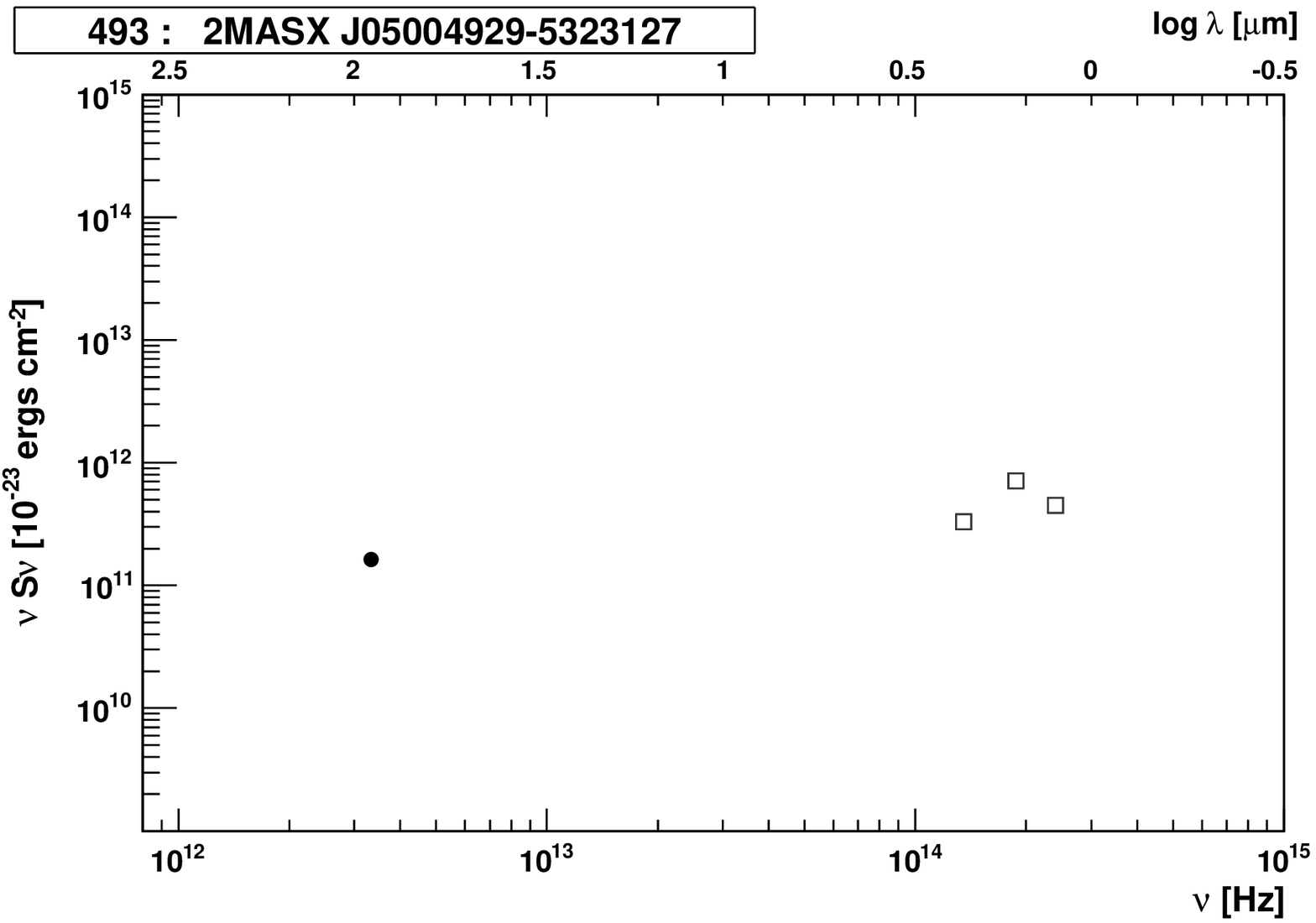}
\includegraphics[width=4cm]{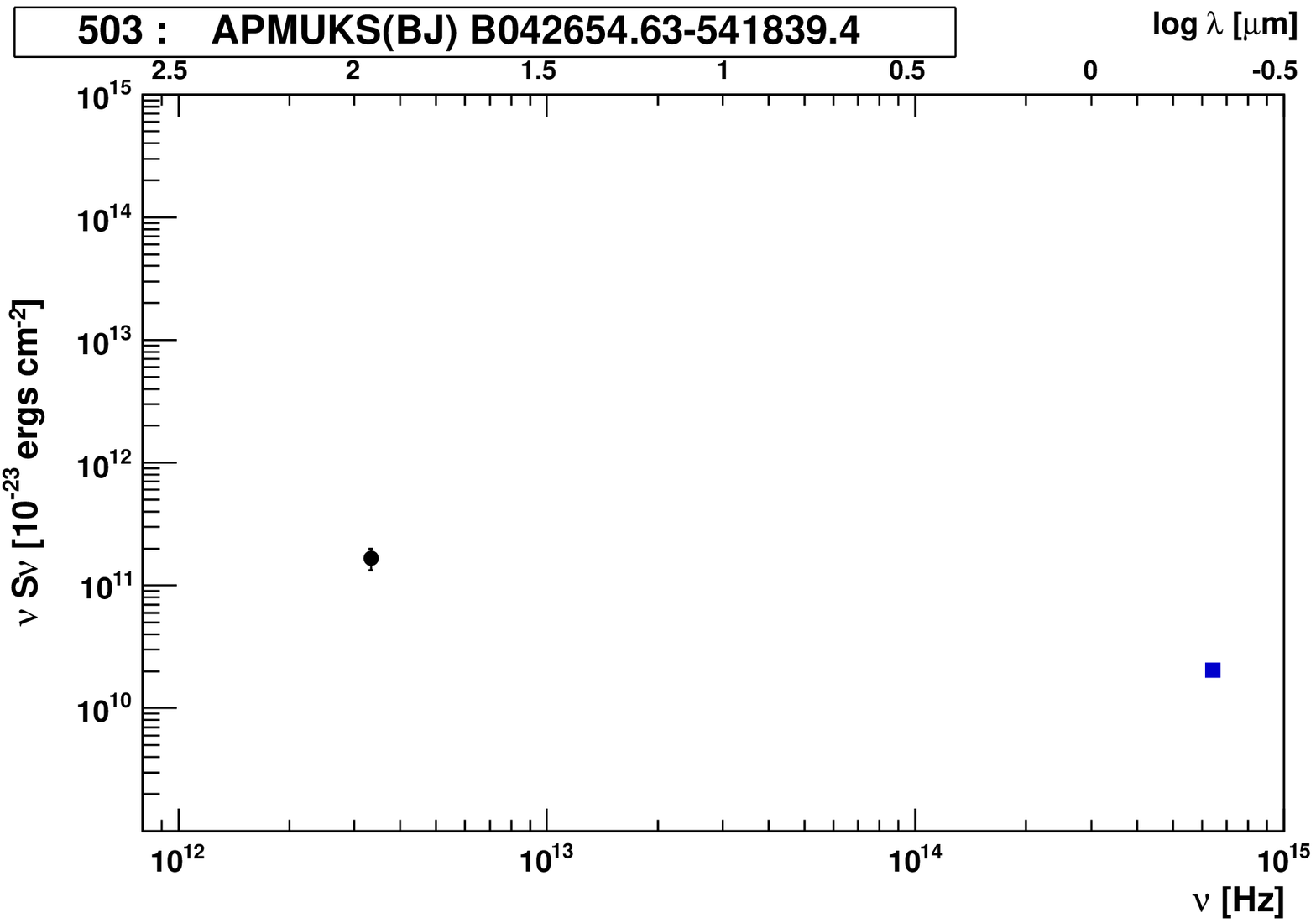}
\includegraphics[width=4cm]{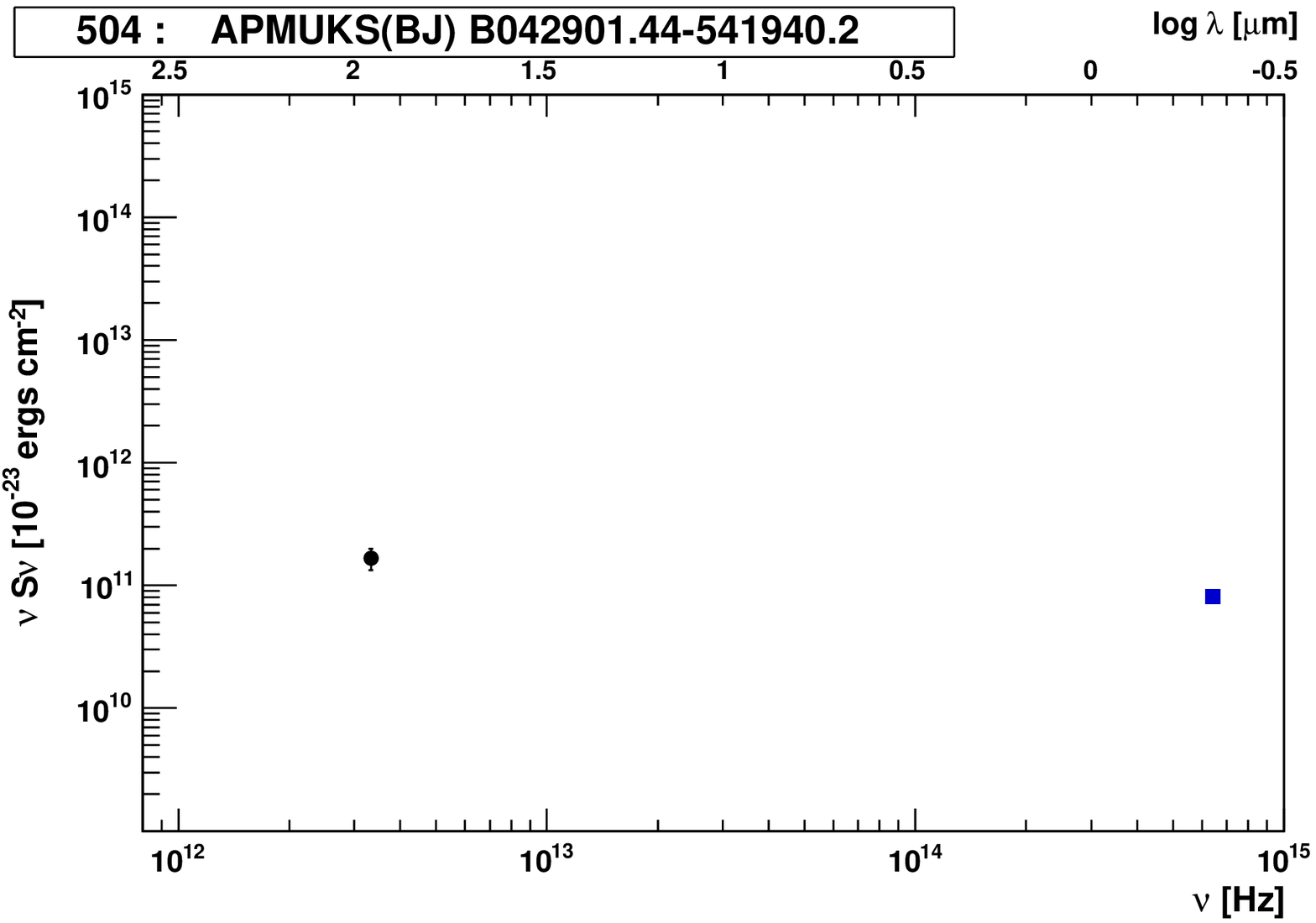}
\includegraphics[width=4cm]{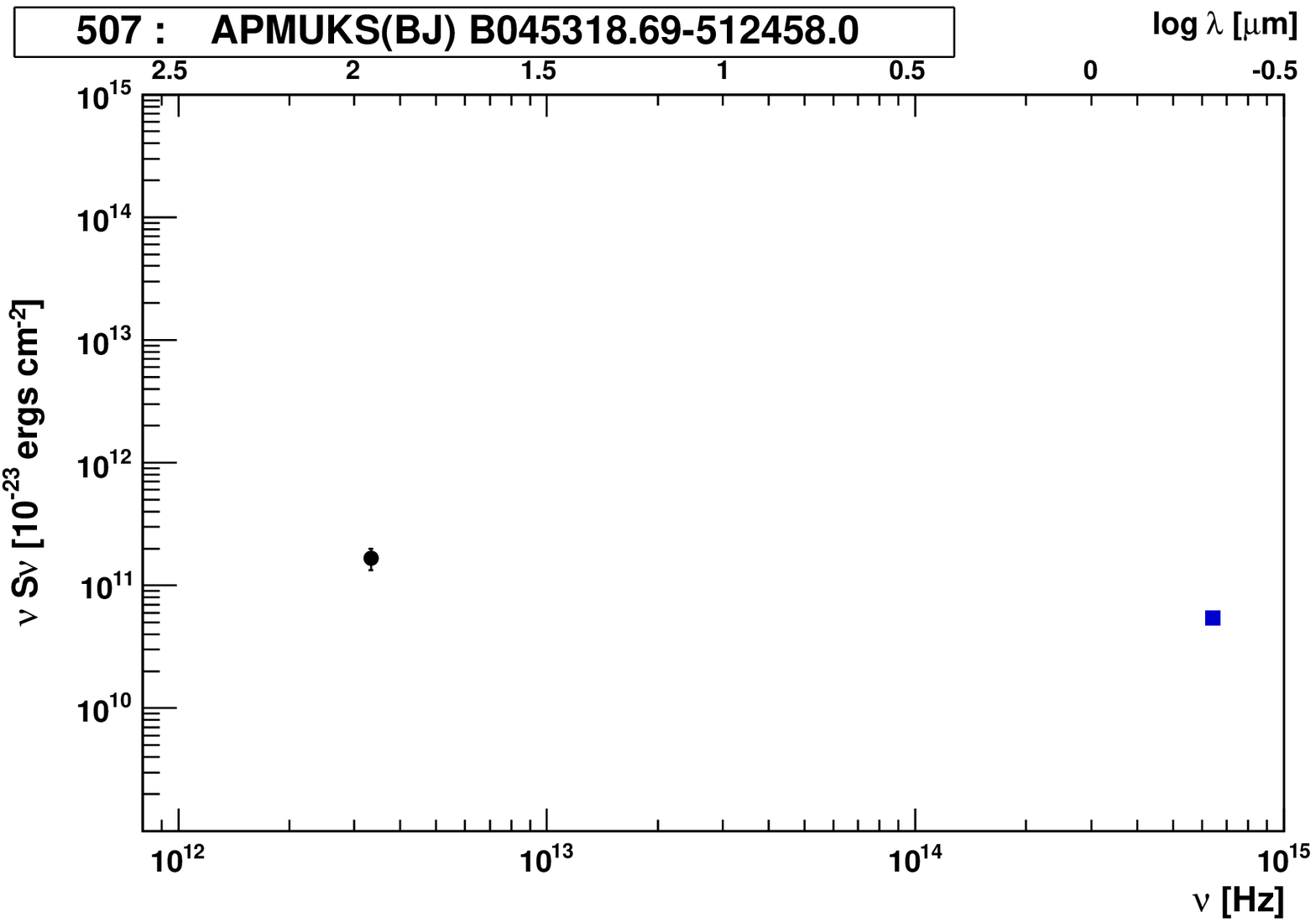}
\includegraphics[width=4cm]{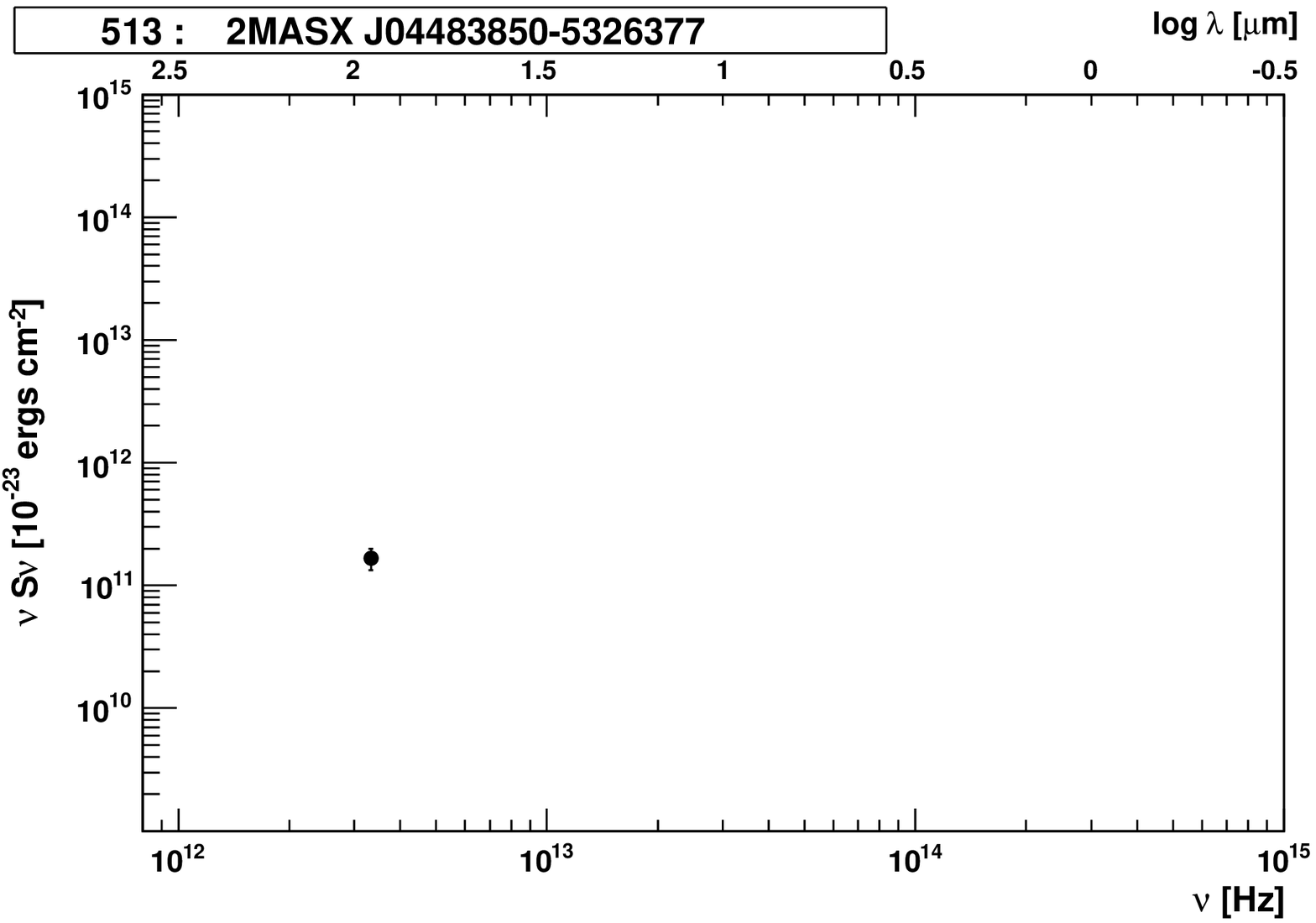}
\includegraphics[width=4cm]{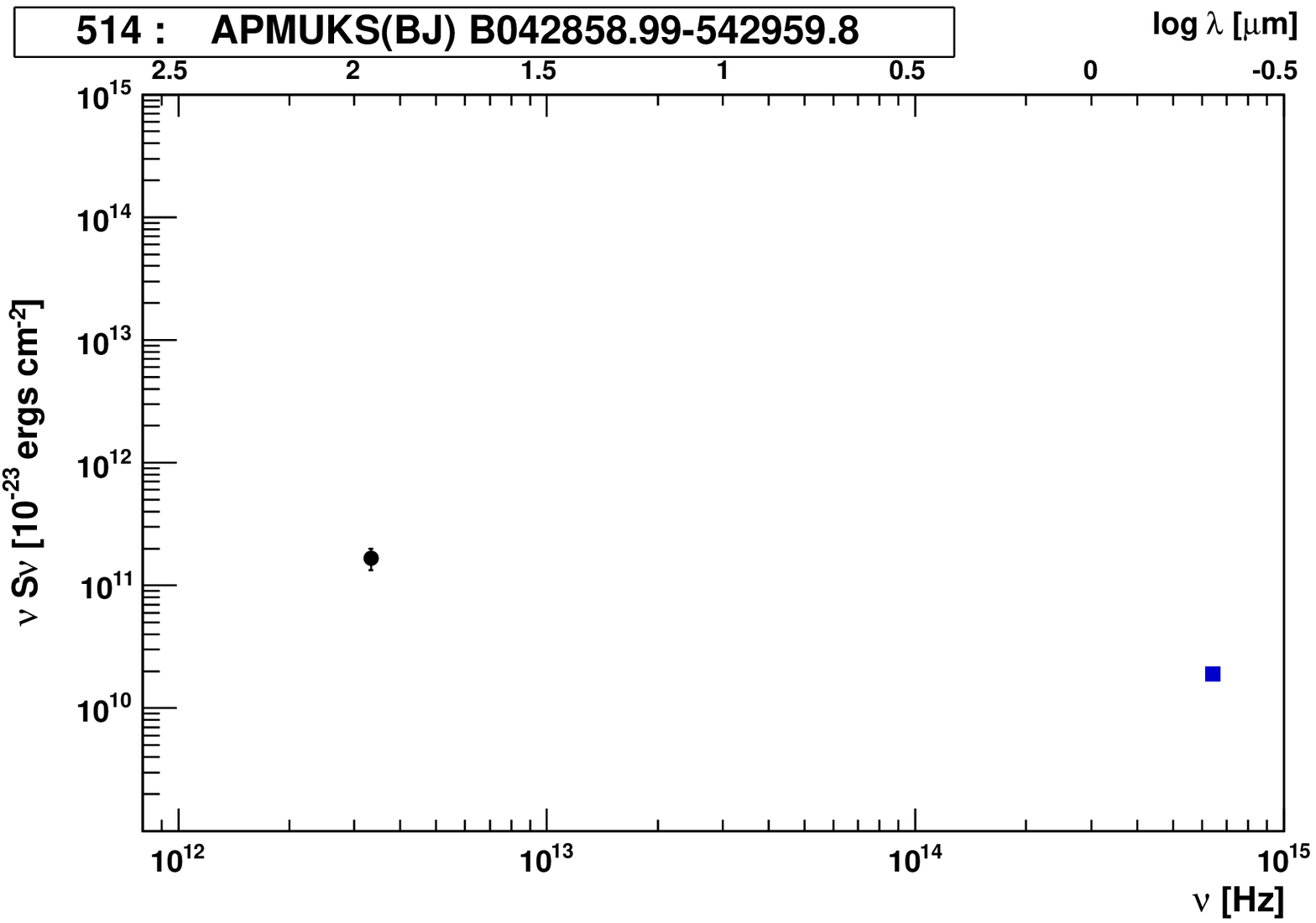}
\includegraphics[width=4cm]{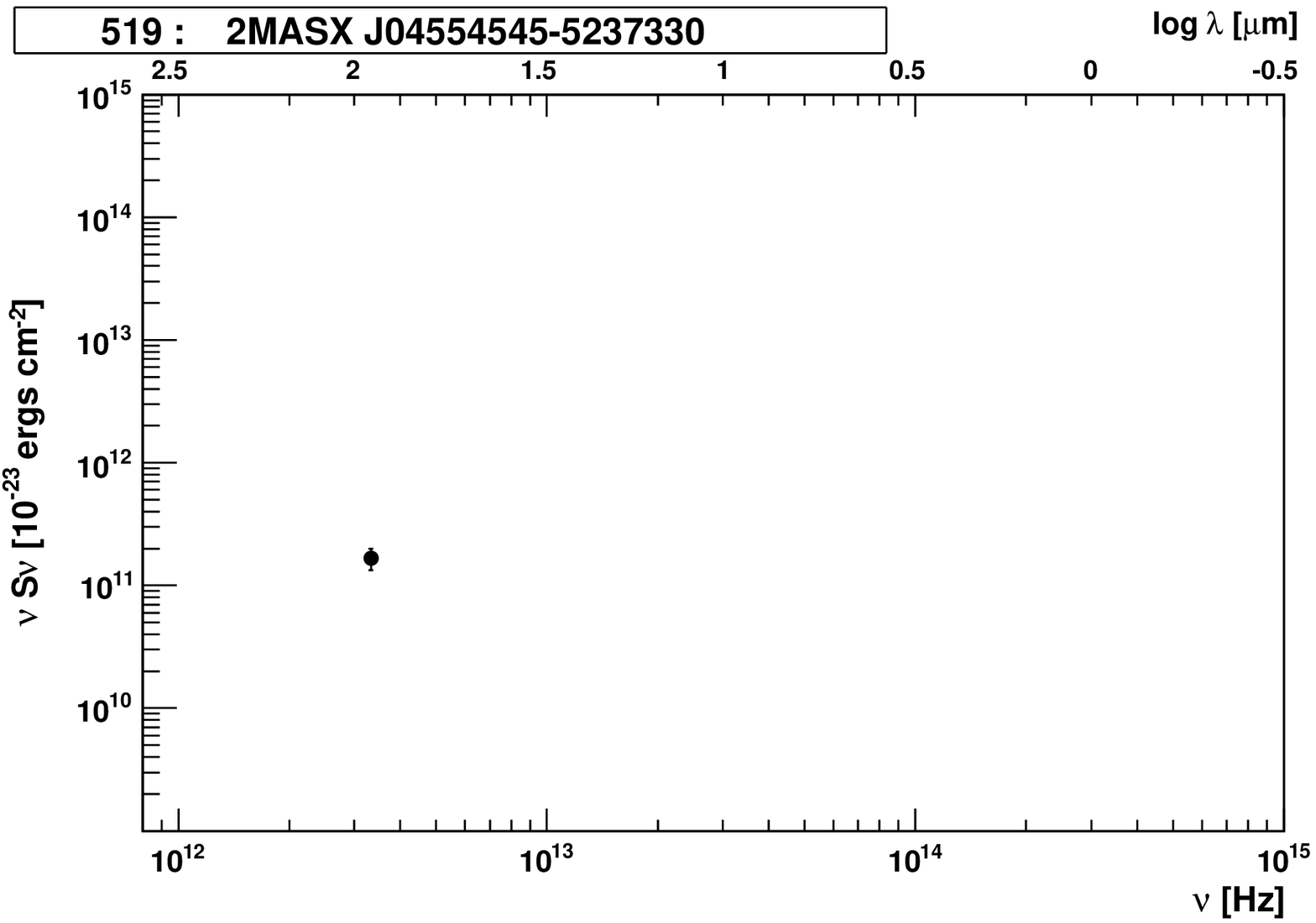}
\includegraphics[width=4cm]{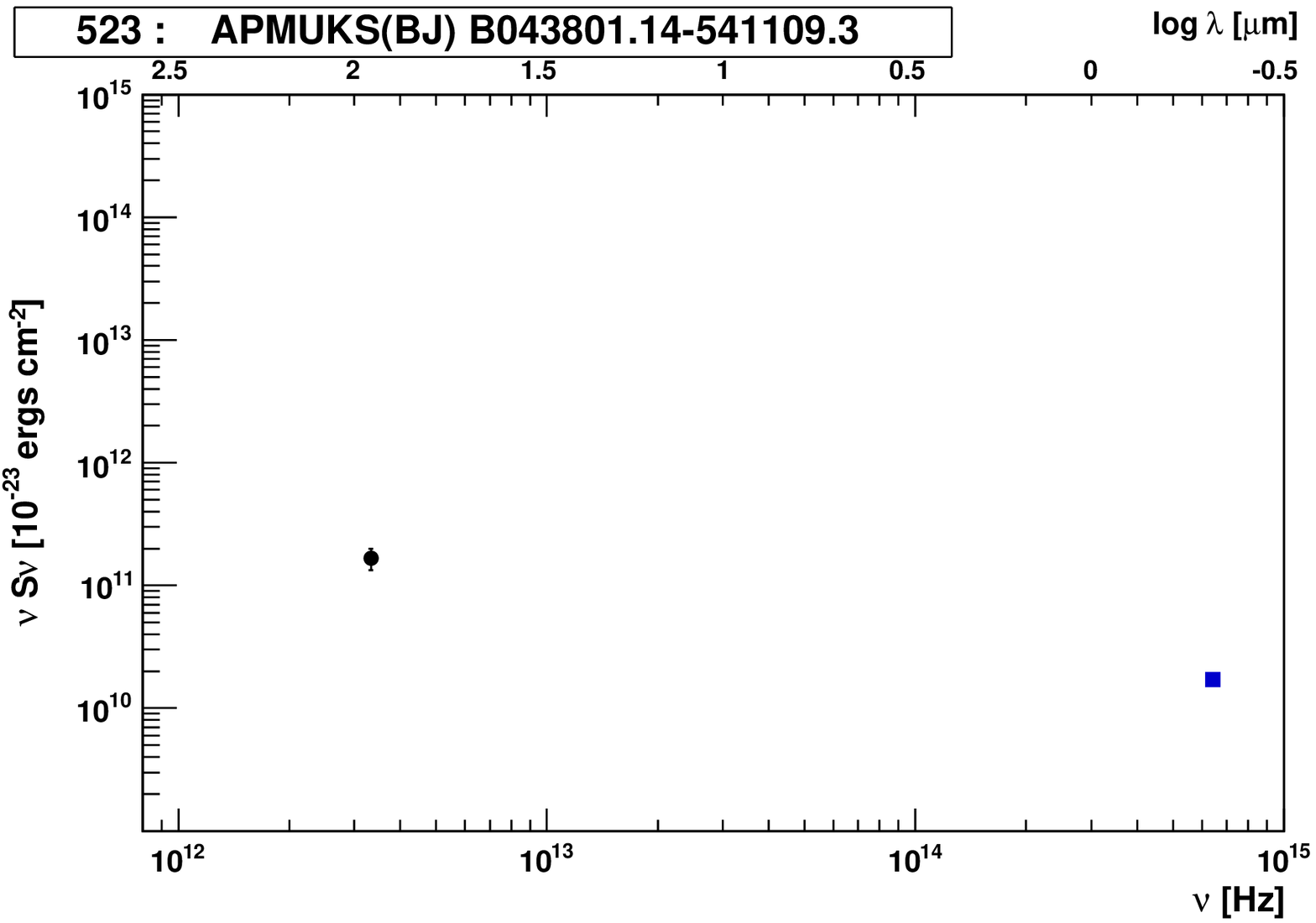}
\includegraphics[width=4cm]{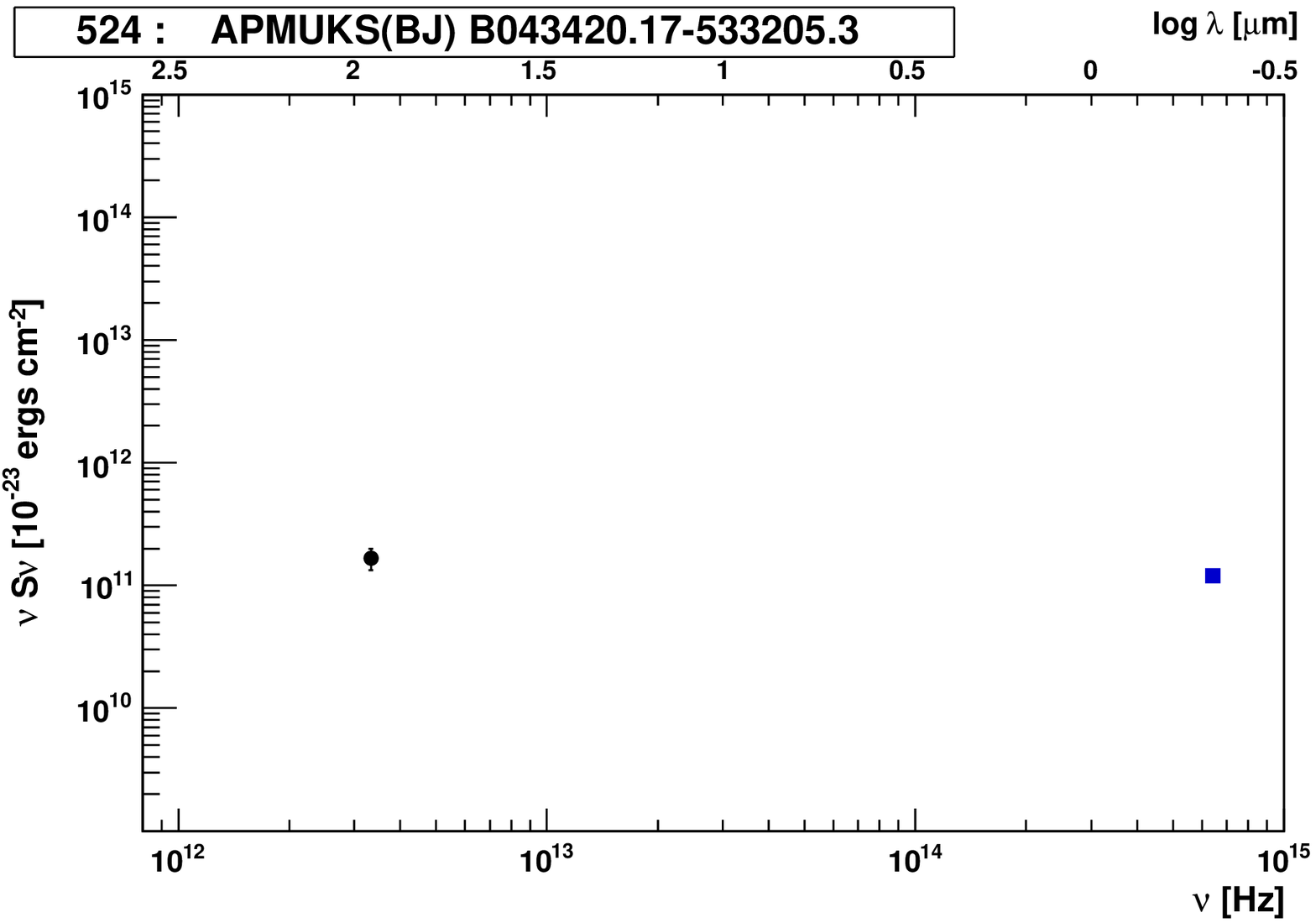}
\includegraphics[width=4cm]{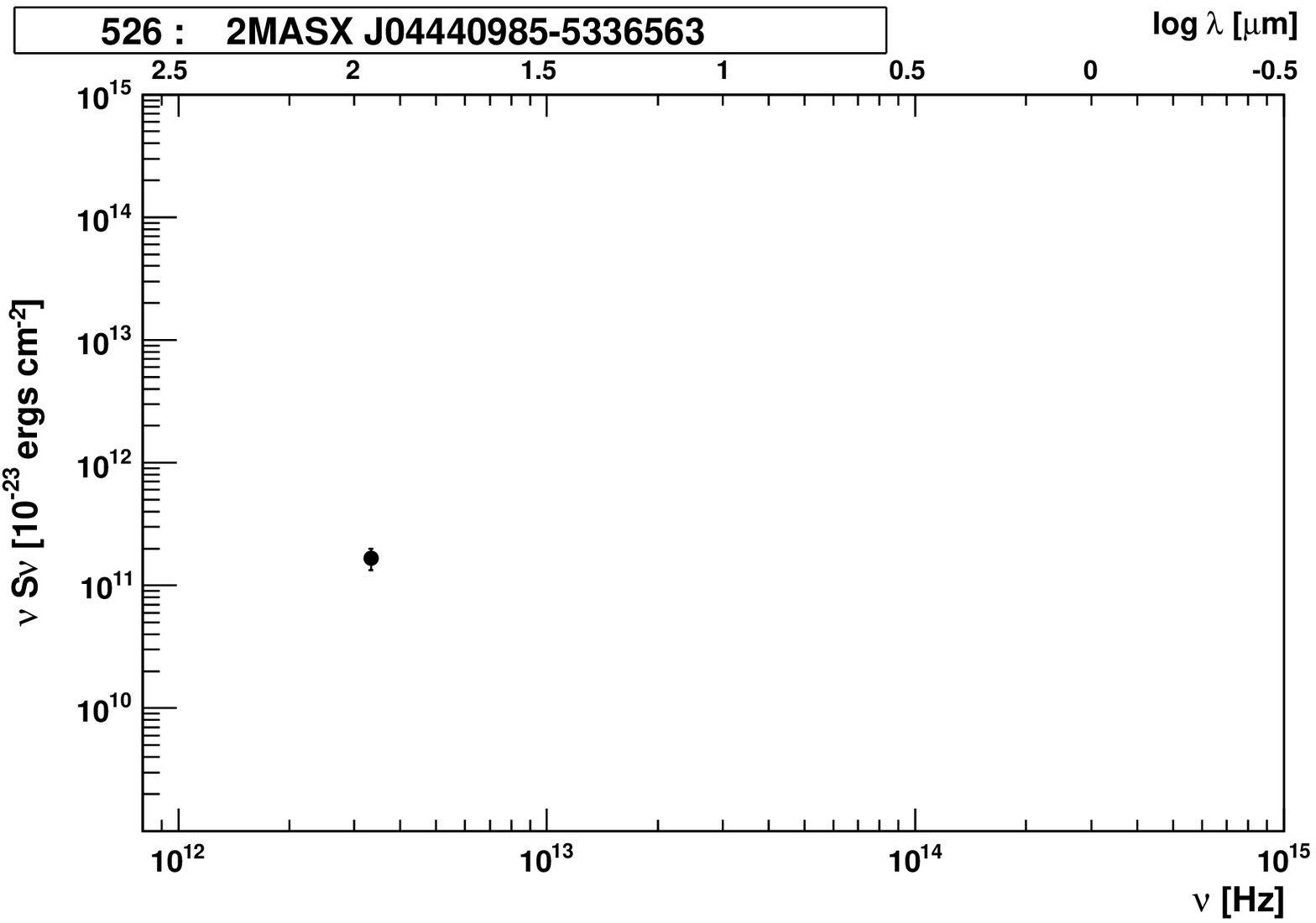}
\includegraphics[width=4cm]{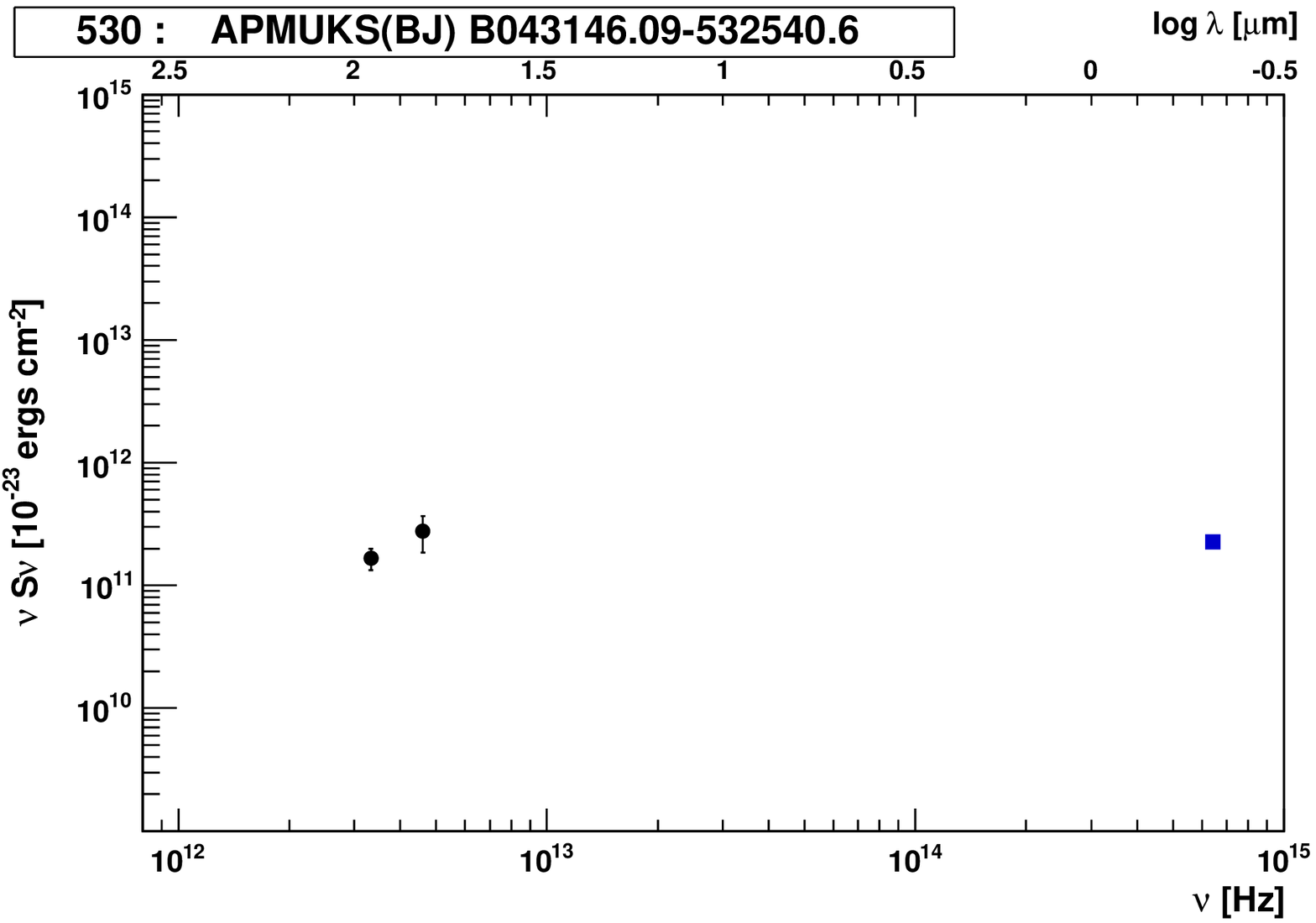}
\includegraphics[width=4cm]{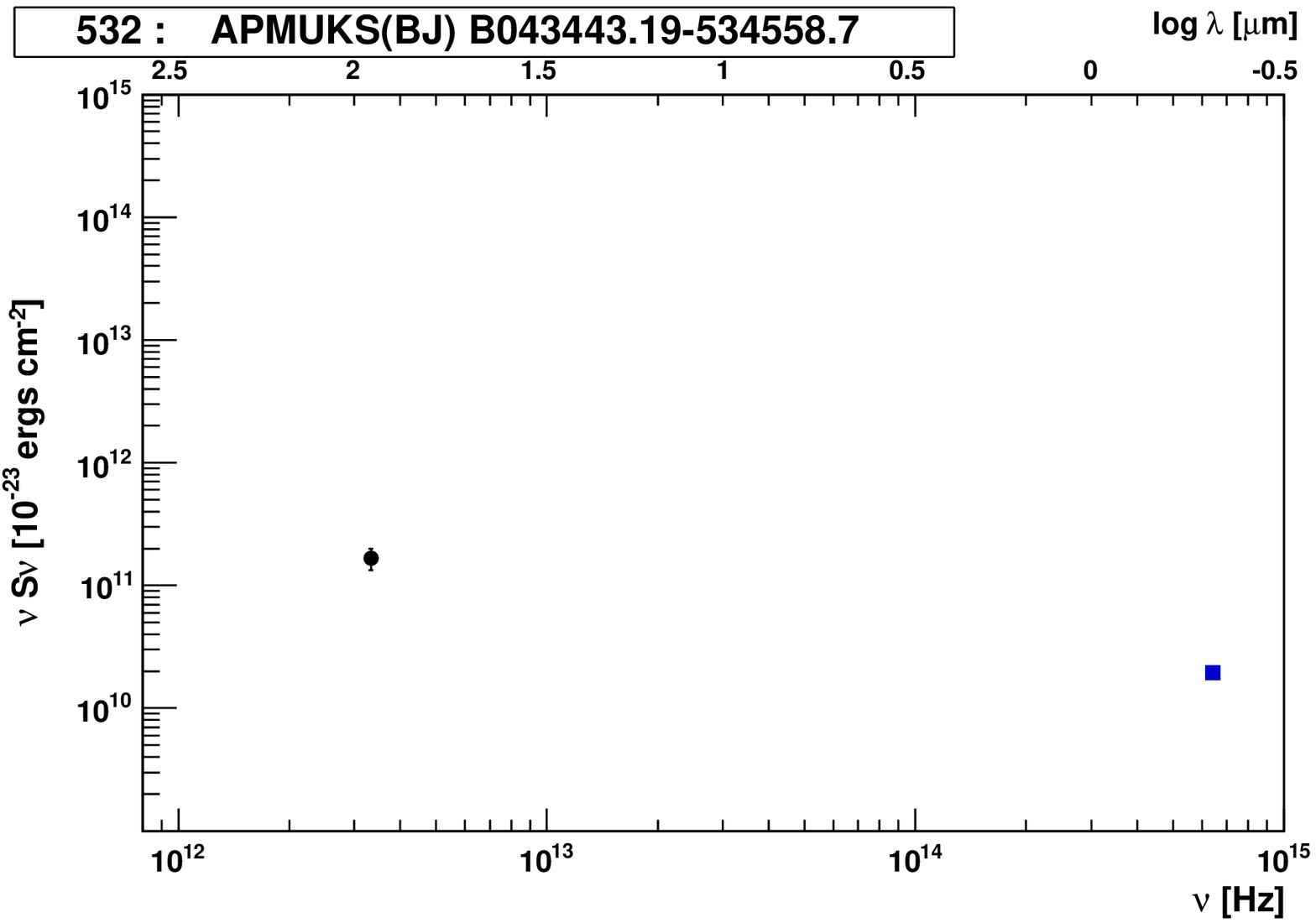}
\includegraphics[width=4cm]{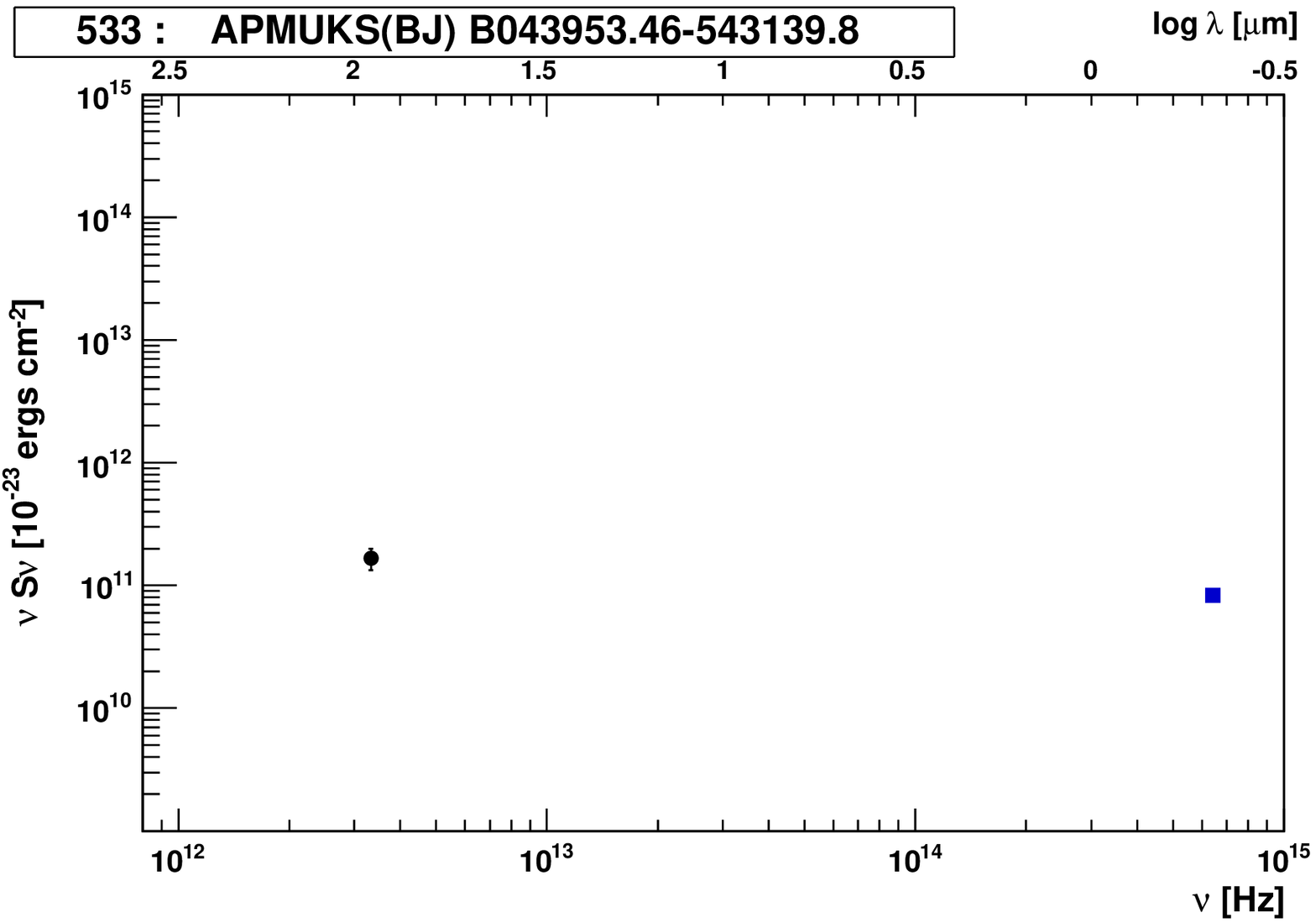}
\includegraphics[width=4cm]{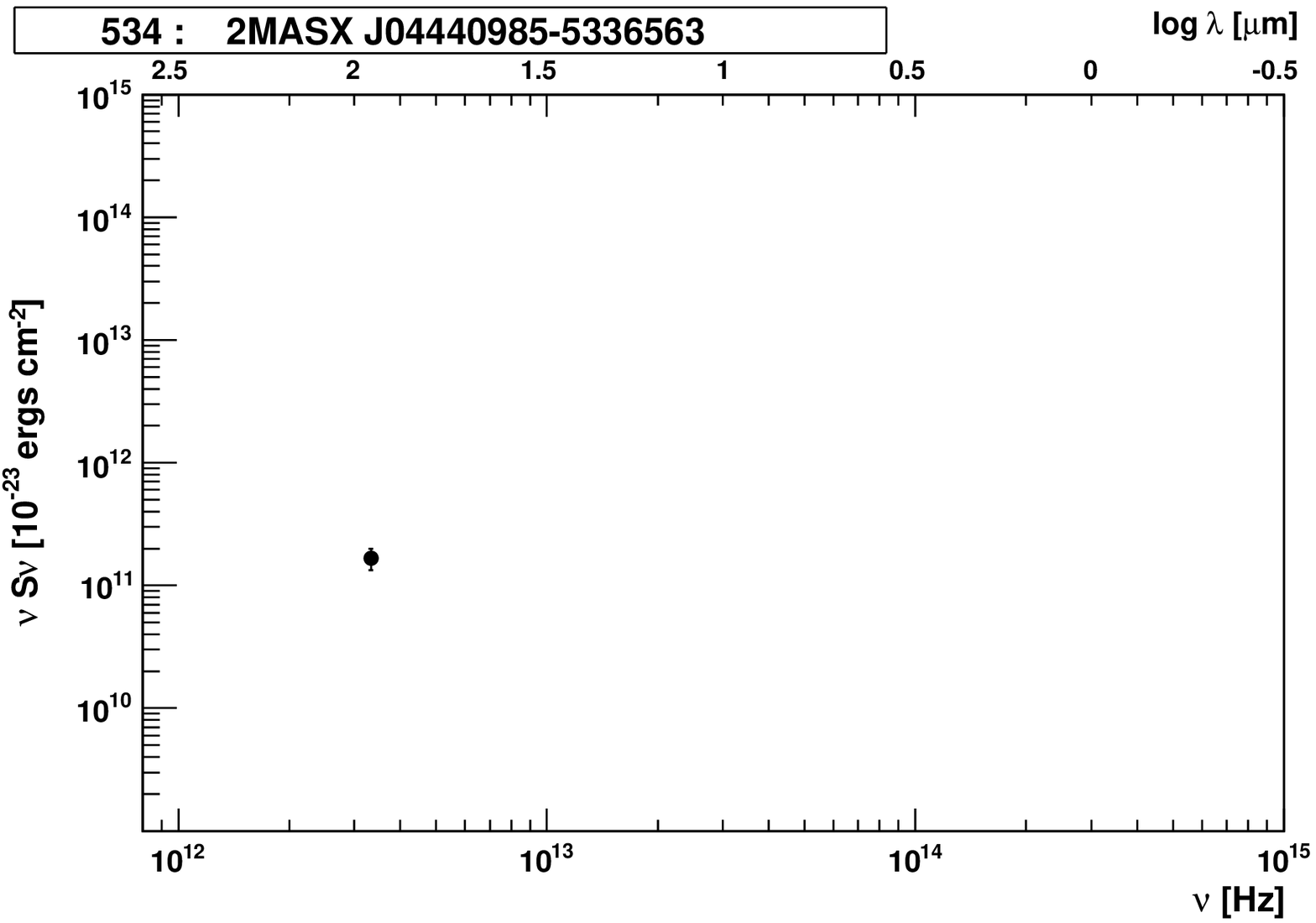}
\includegraphics[width=4cm]{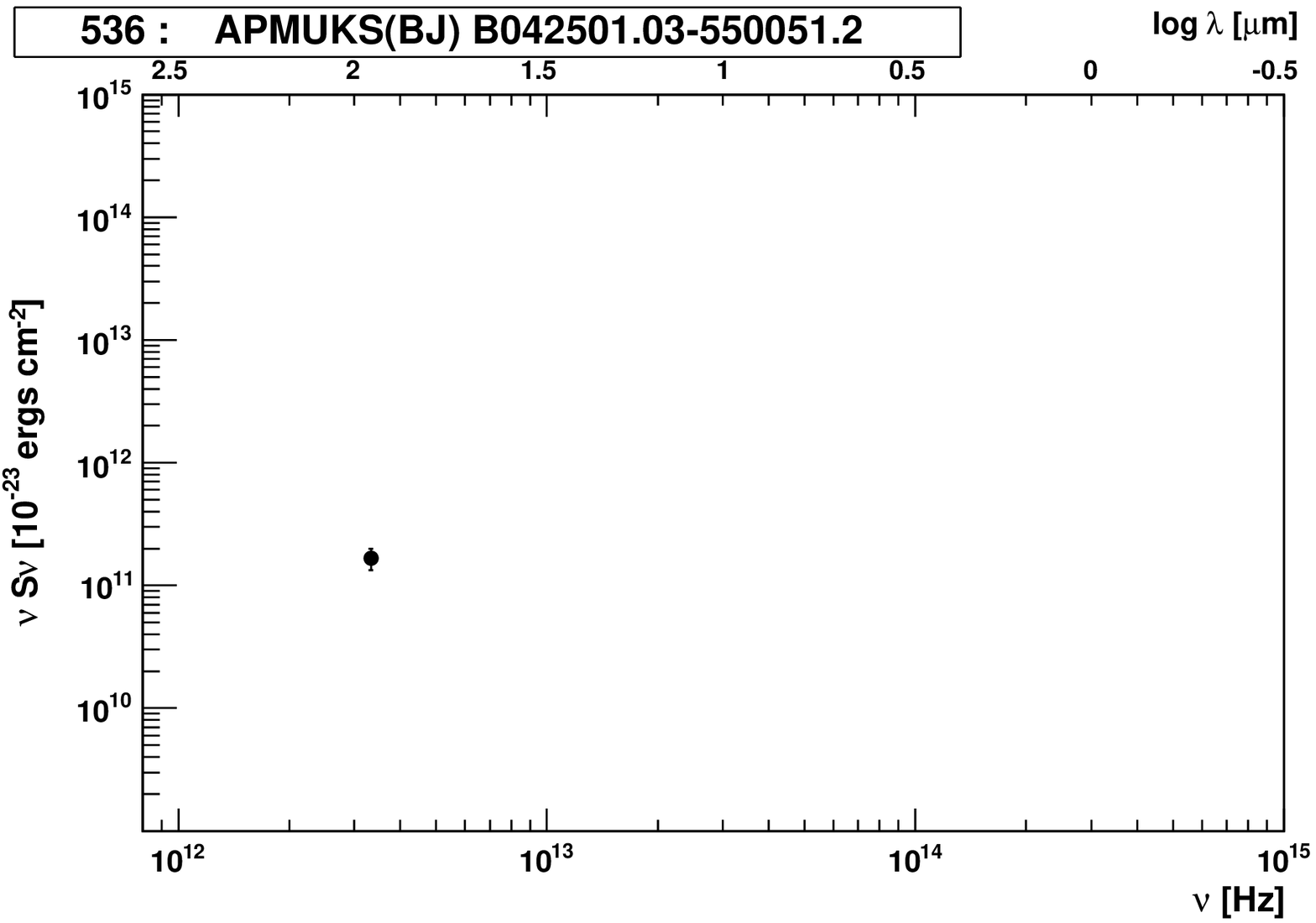}
\includegraphics[width=4cm]{points/kmalek_514.eps}
\includegraphics[width=4cm]{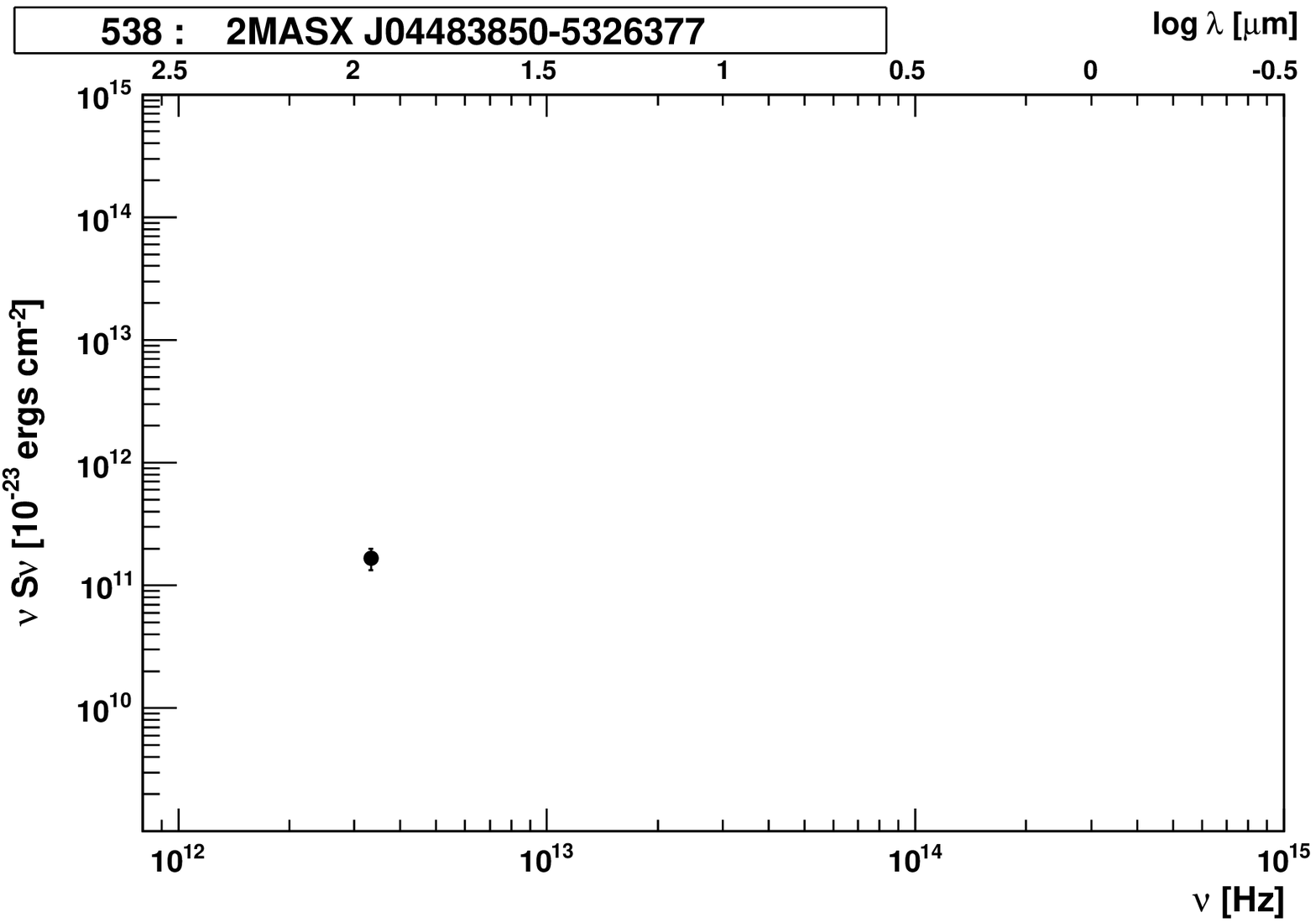}
\includegraphics[width=4cm]{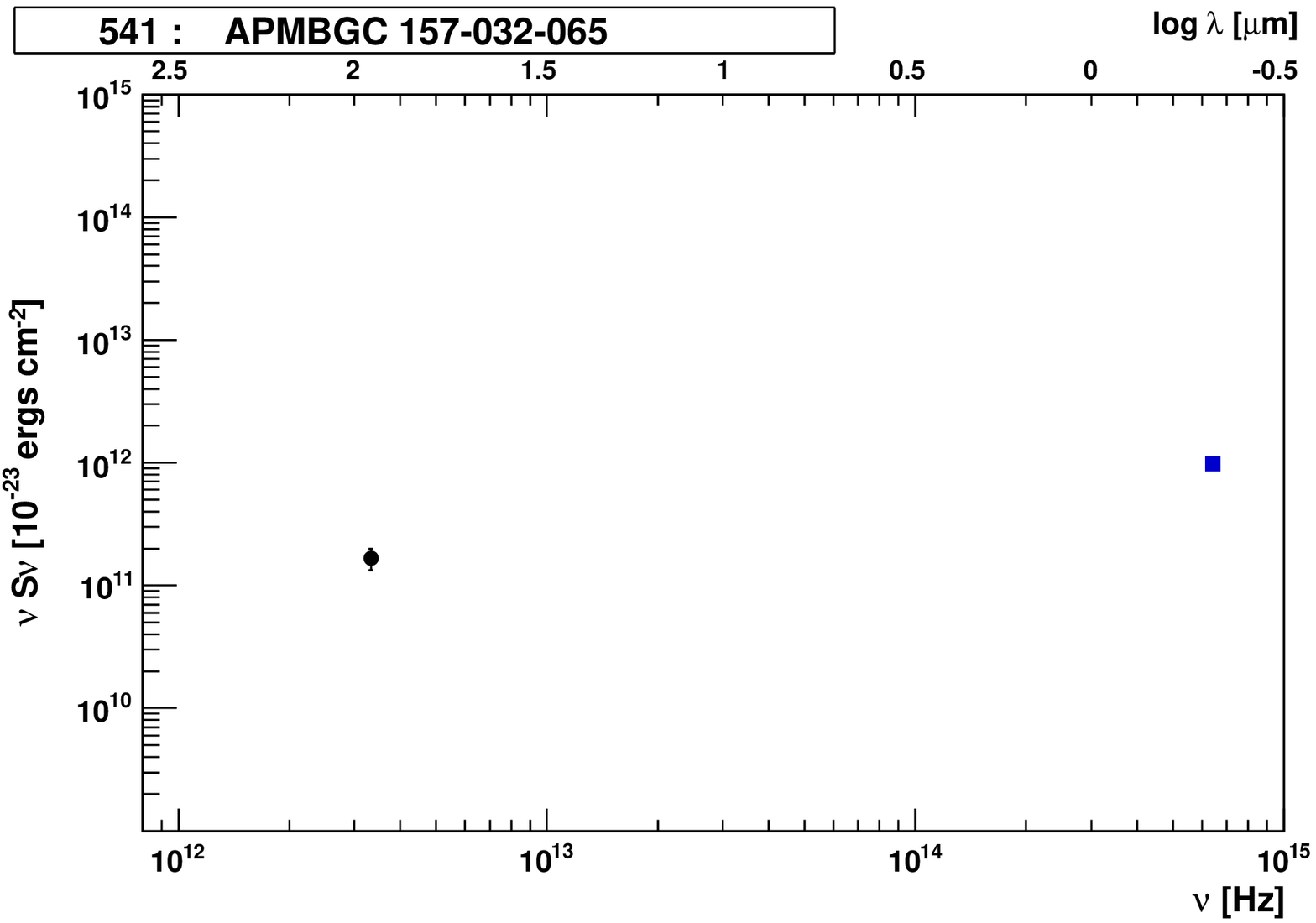}
\includegraphics[width=4cm]{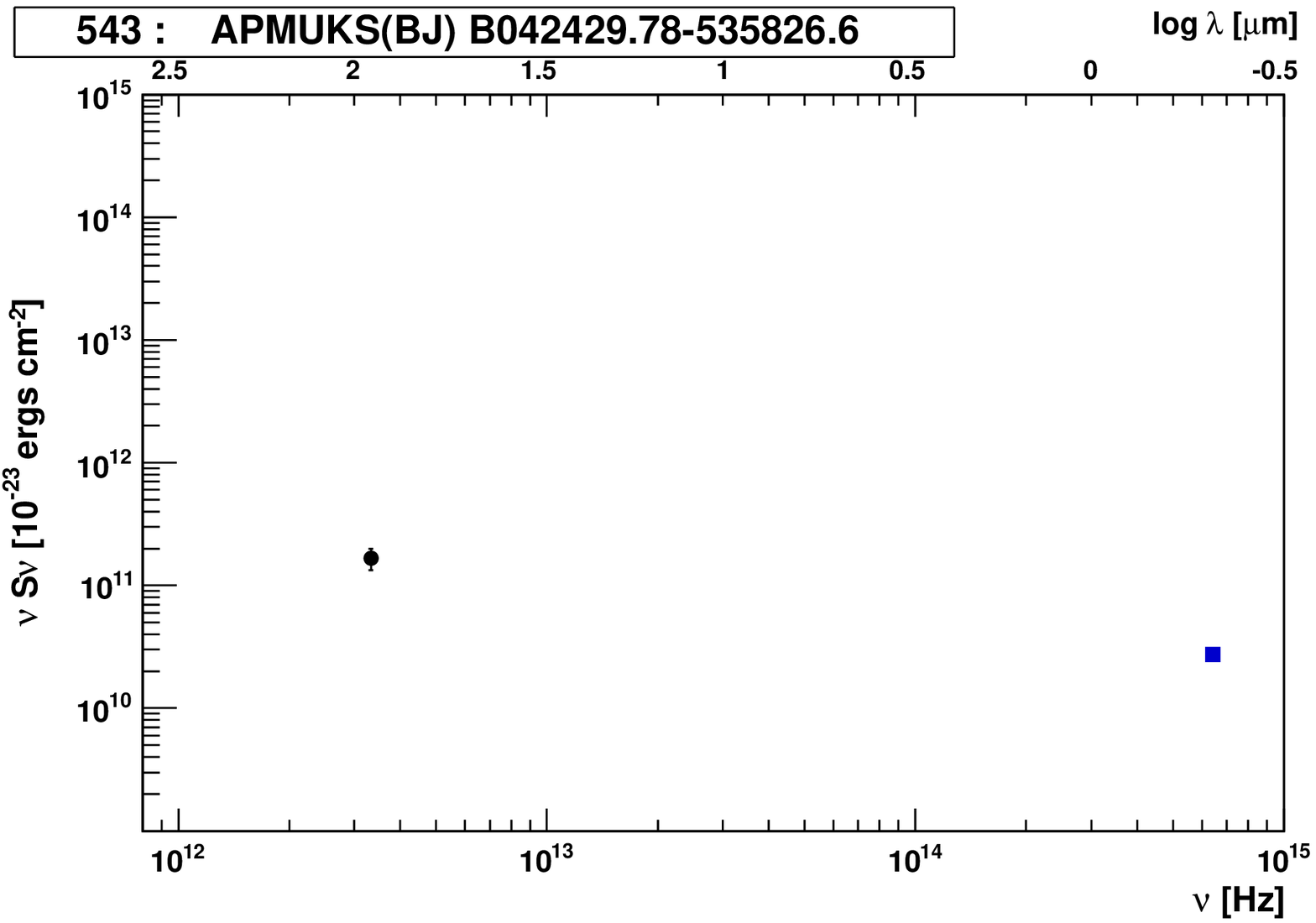}
\includegraphics[width=4cm]{points/kmalek_513.eps}
\includegraphics[width=4cm]{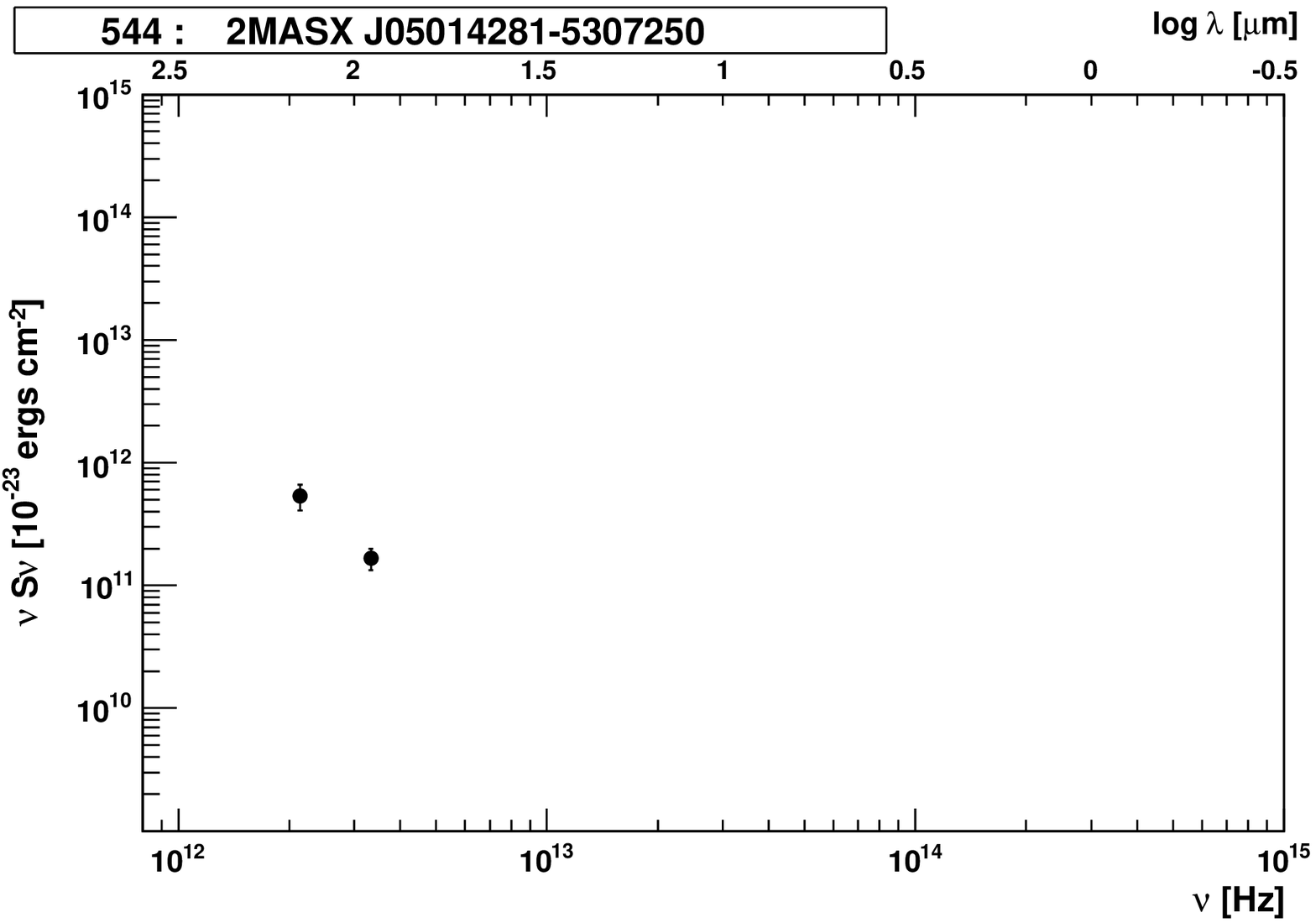}
\includegraphics[width=4cm]{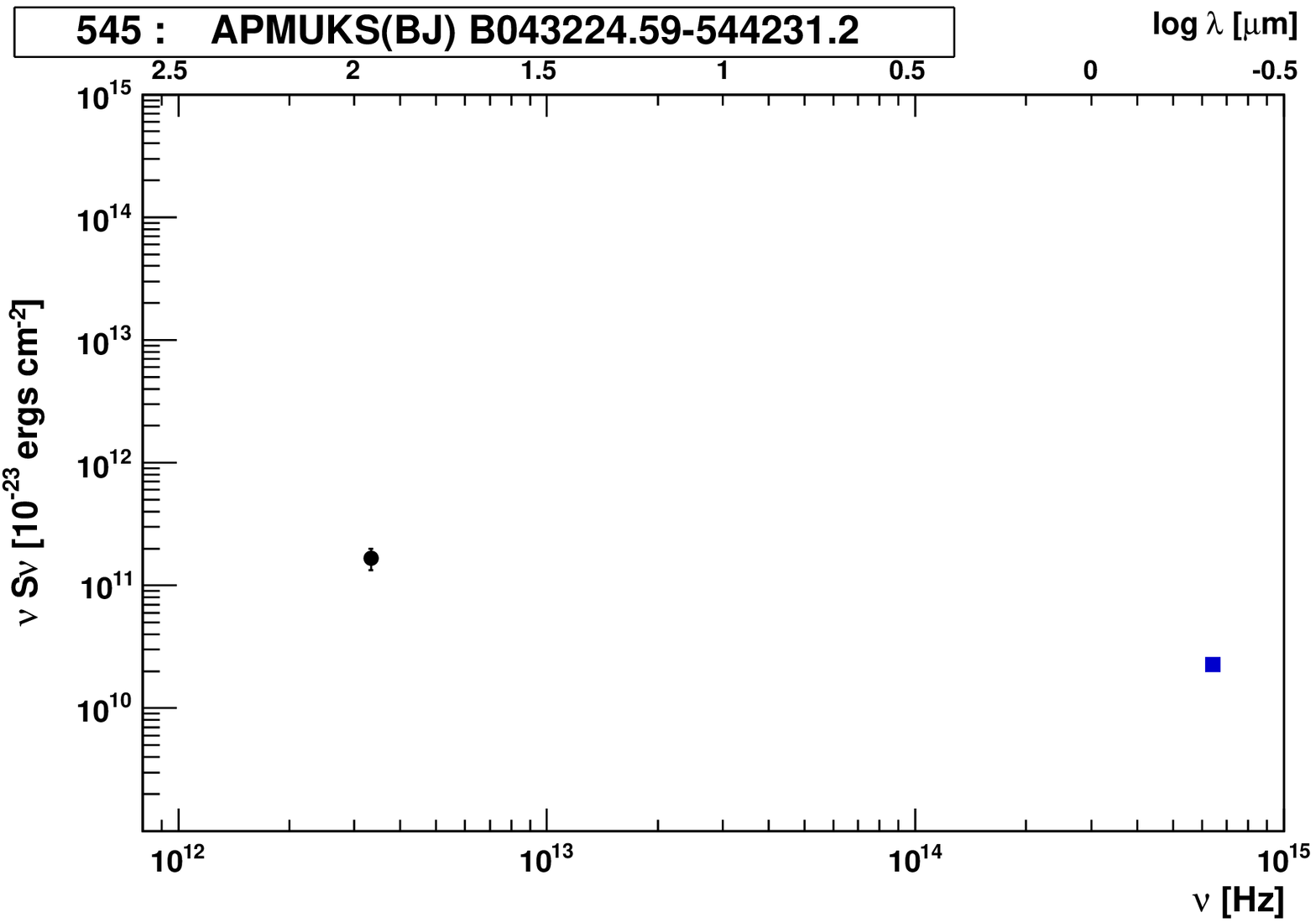}
\includegraphics[width=4cm]{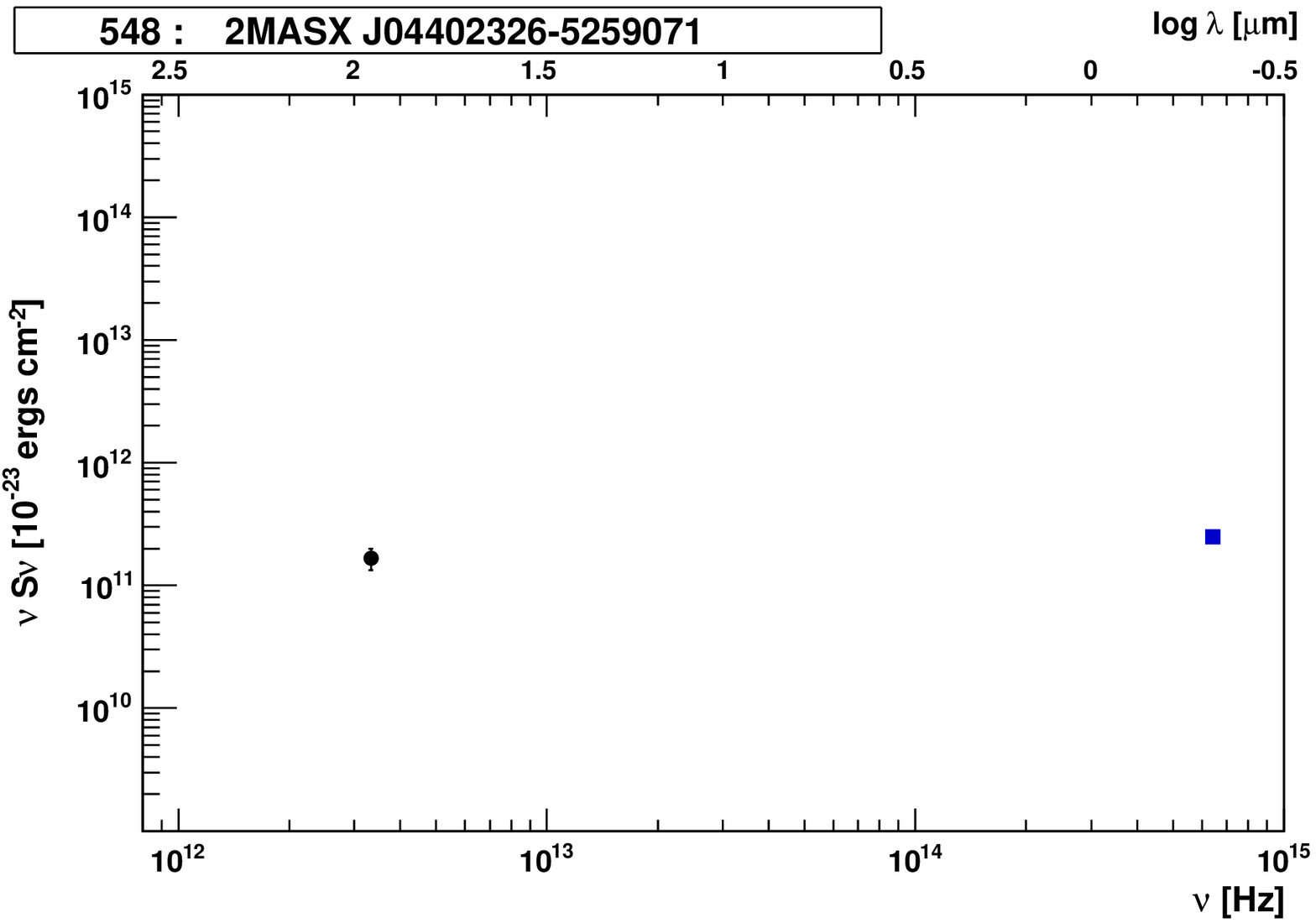}
\includegraphics[width=4cm]{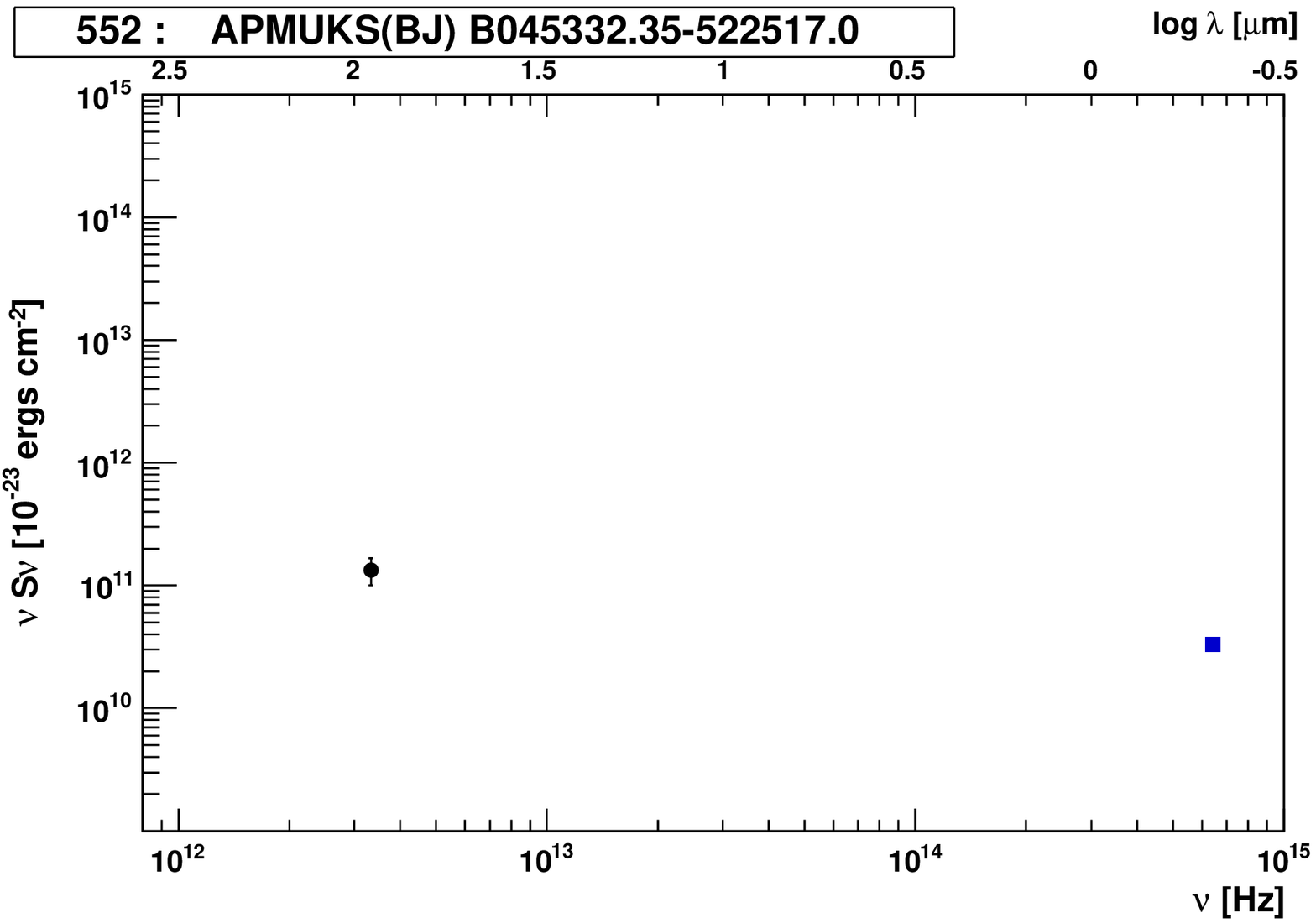}
\includegraphics[width=4cm]{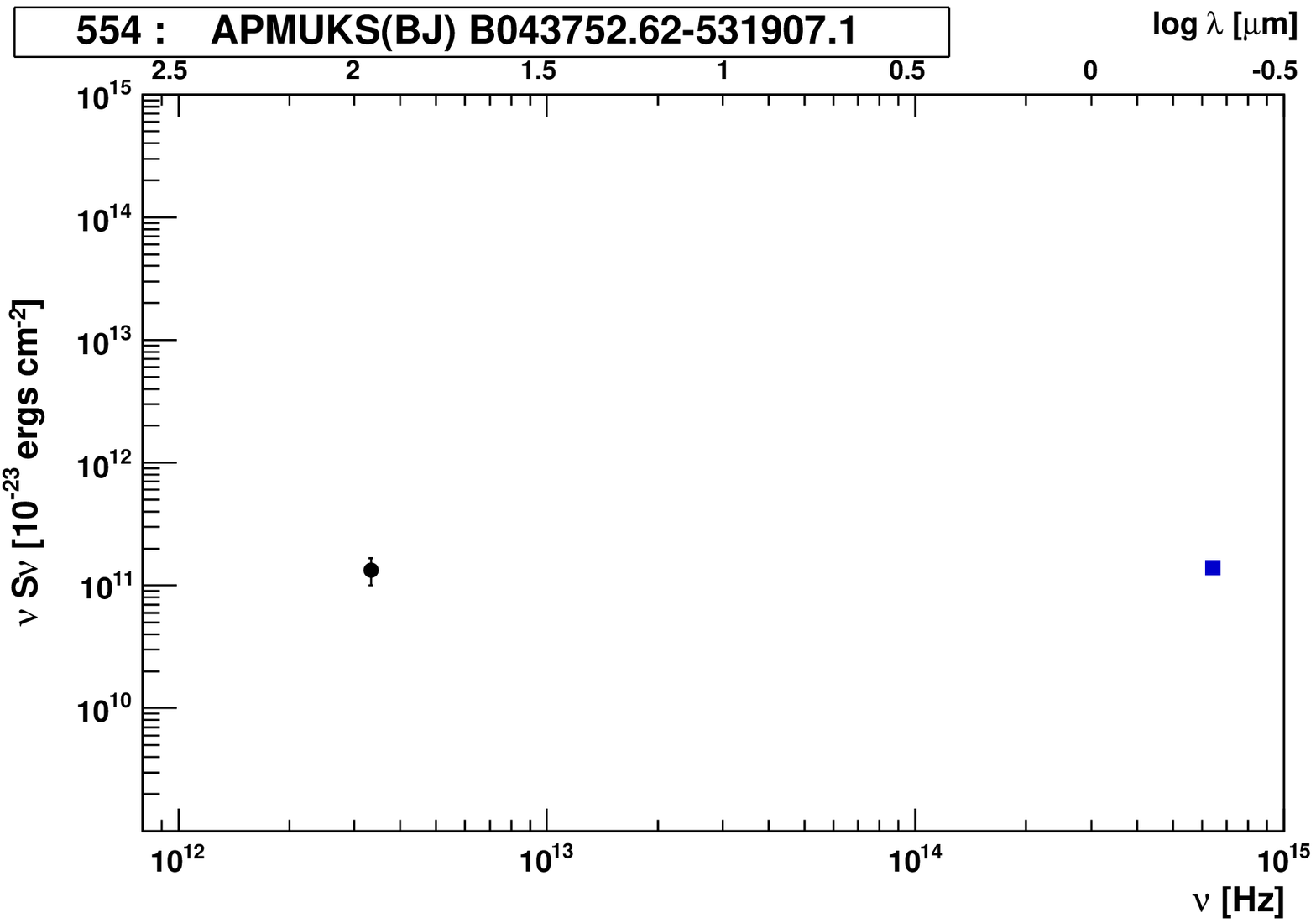}
\includegraphics[width=4cm]{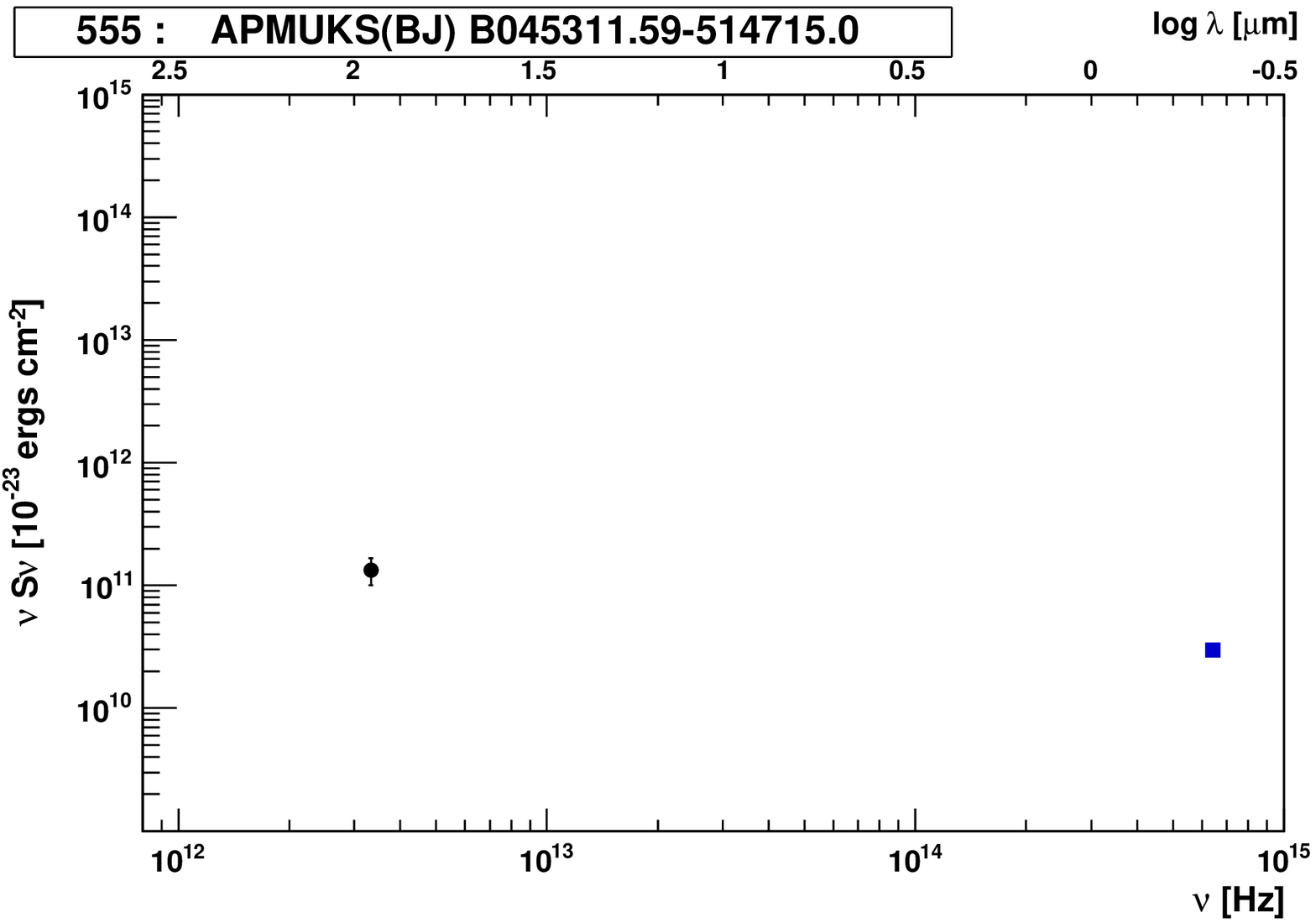}
\includegraphics[width=4cm]{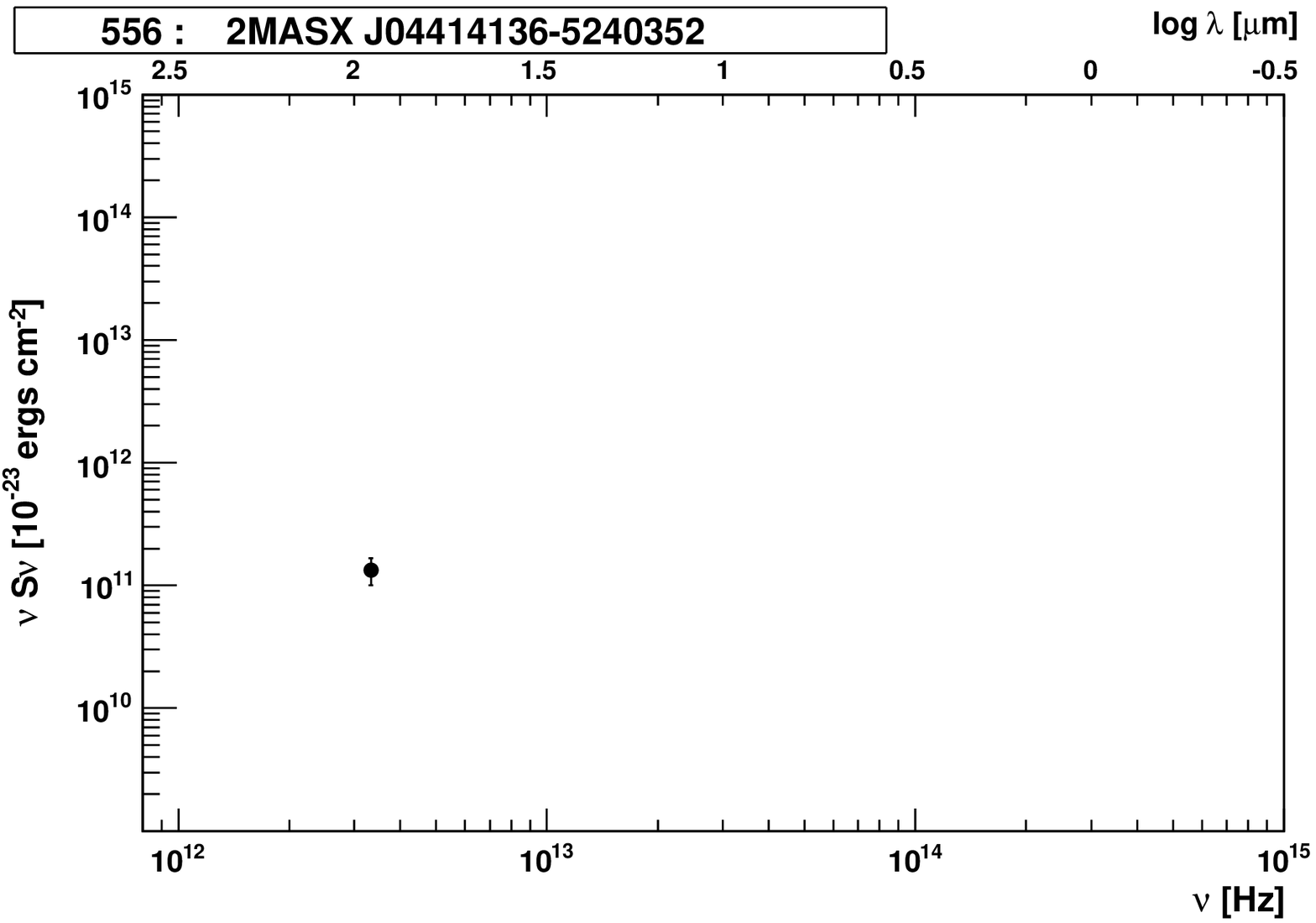}
\includegraphics[width=4cm]{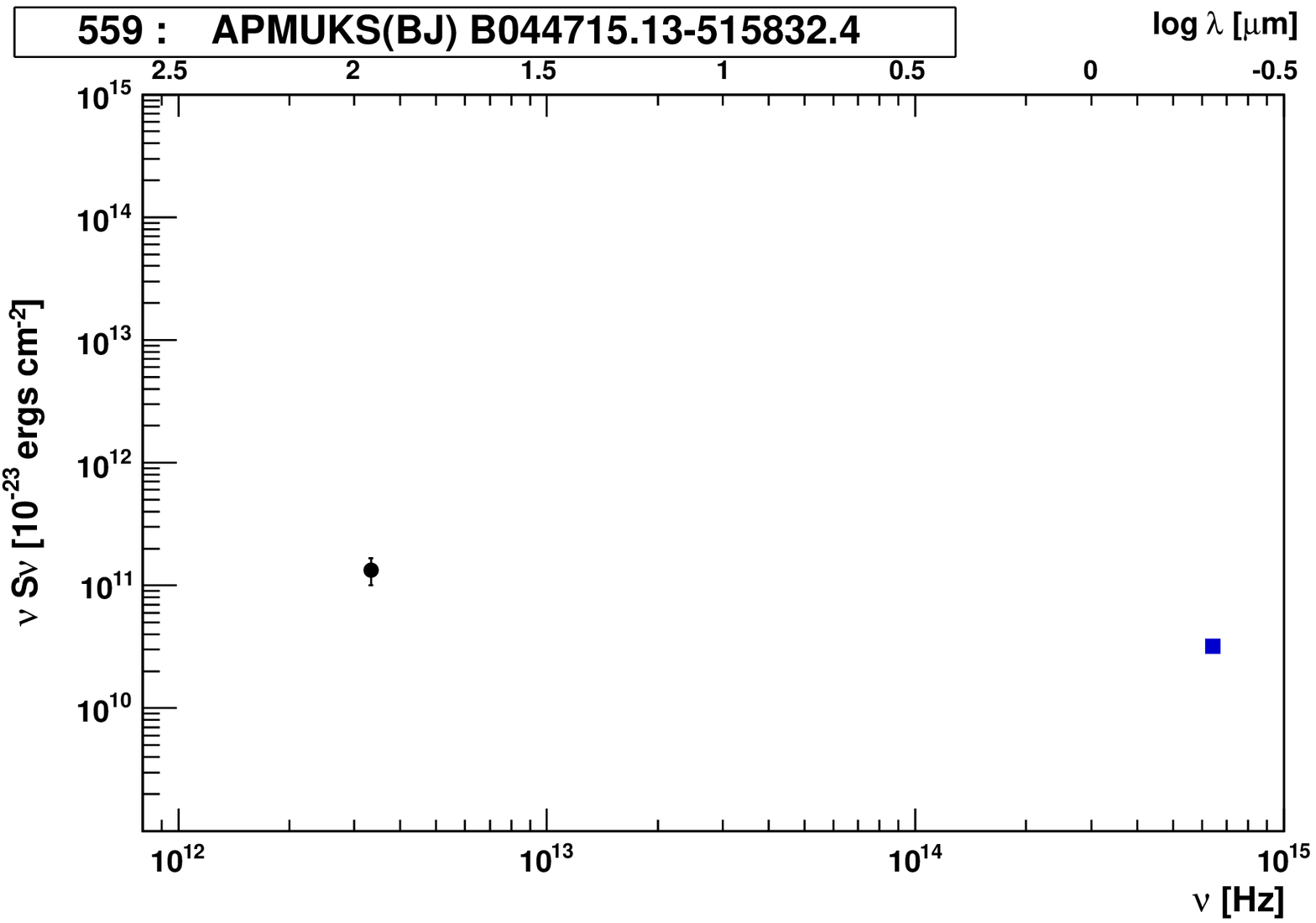}
\includegraphics[width=4cm]{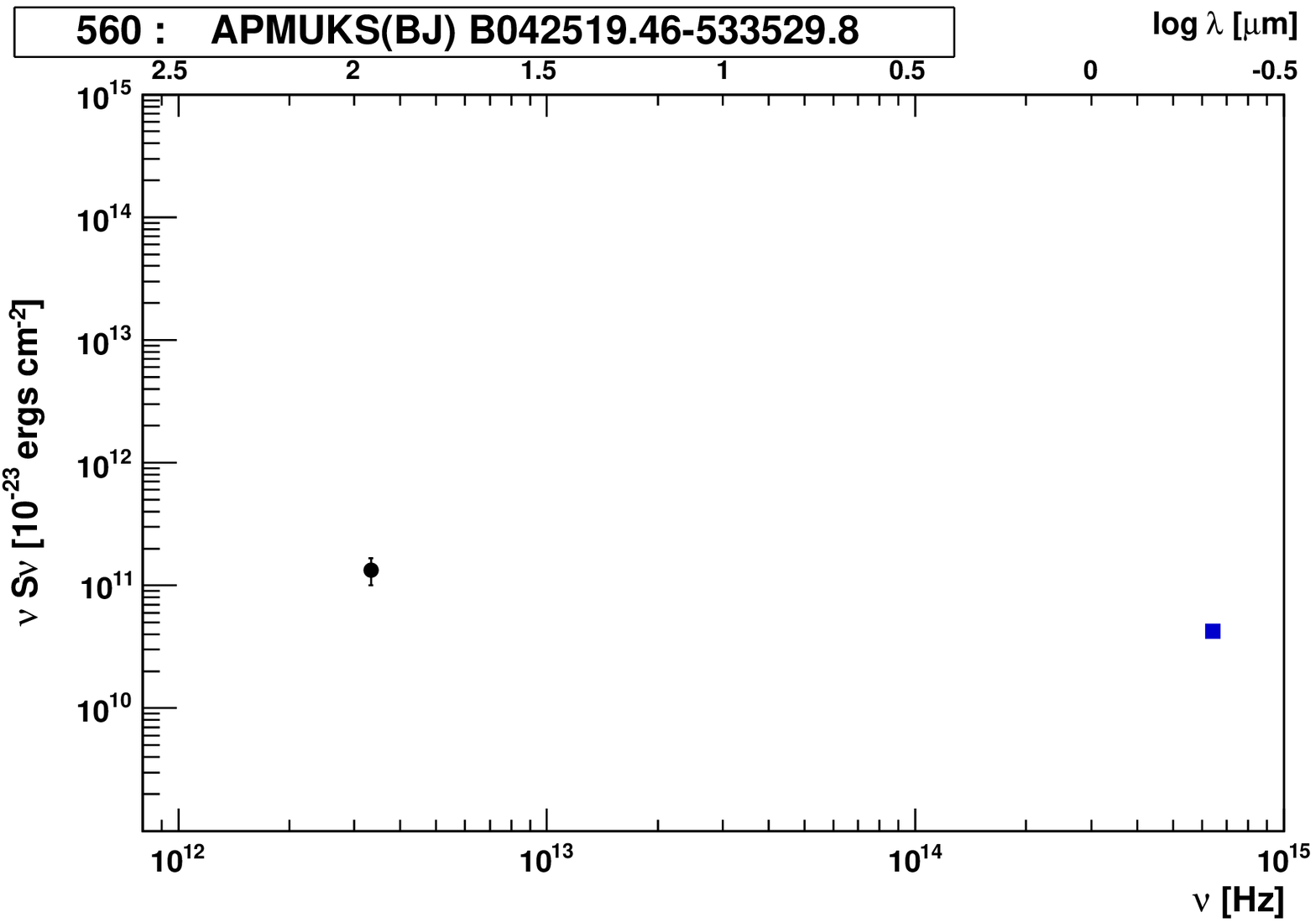}
\includegraphics[width=4cm]{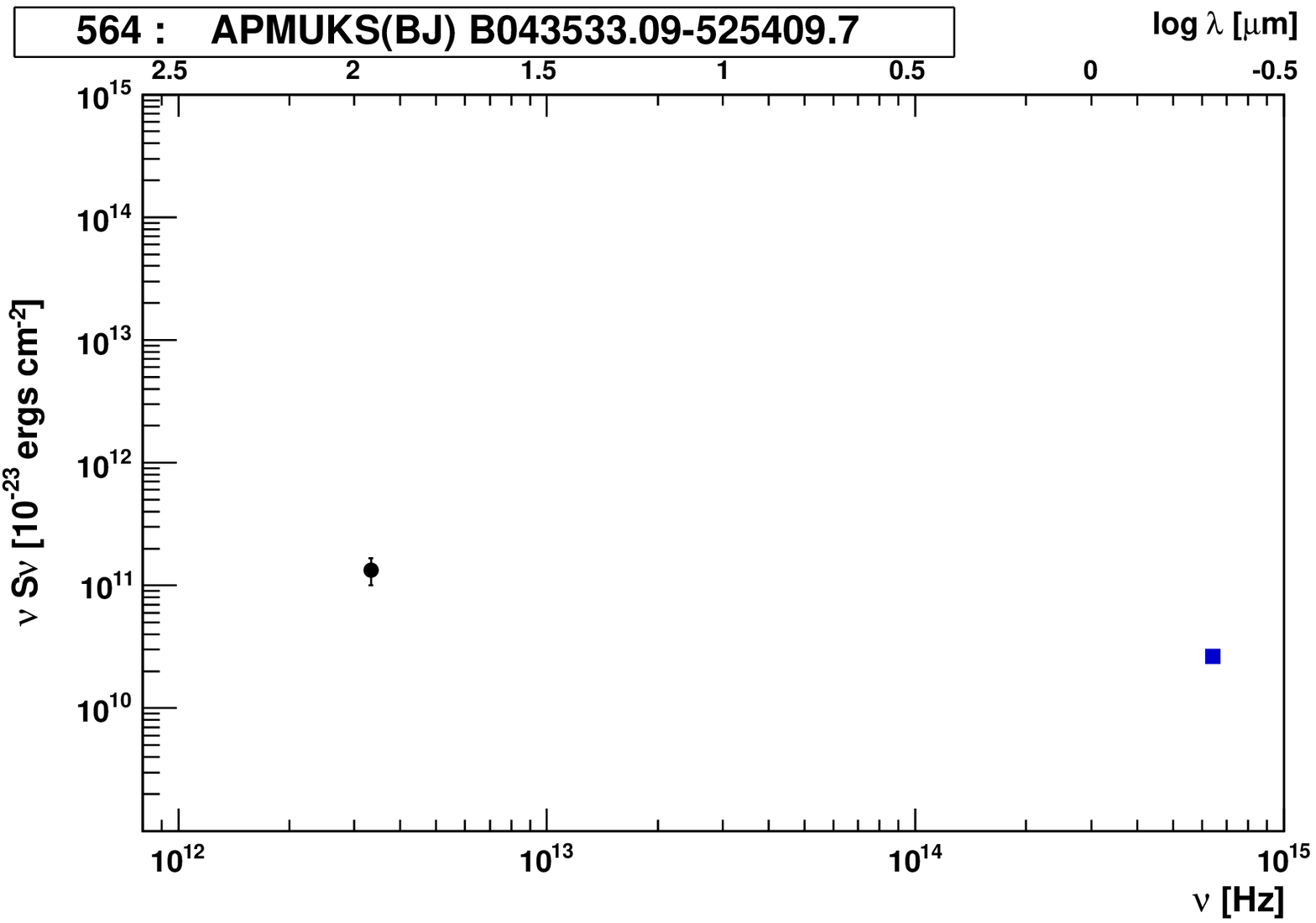}
\includegraphics[width=4cm]{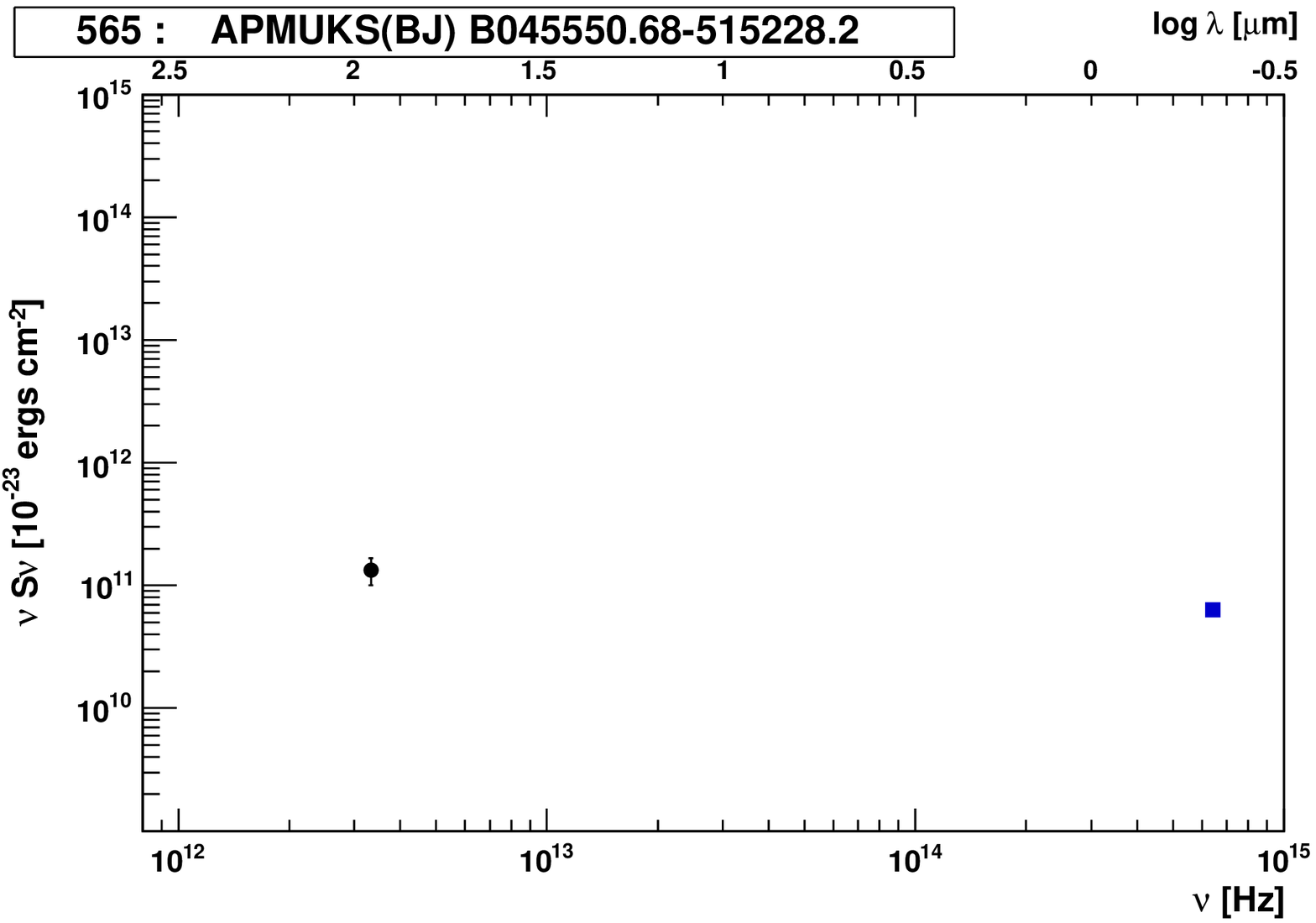}
\includegraphics[width=4cm]{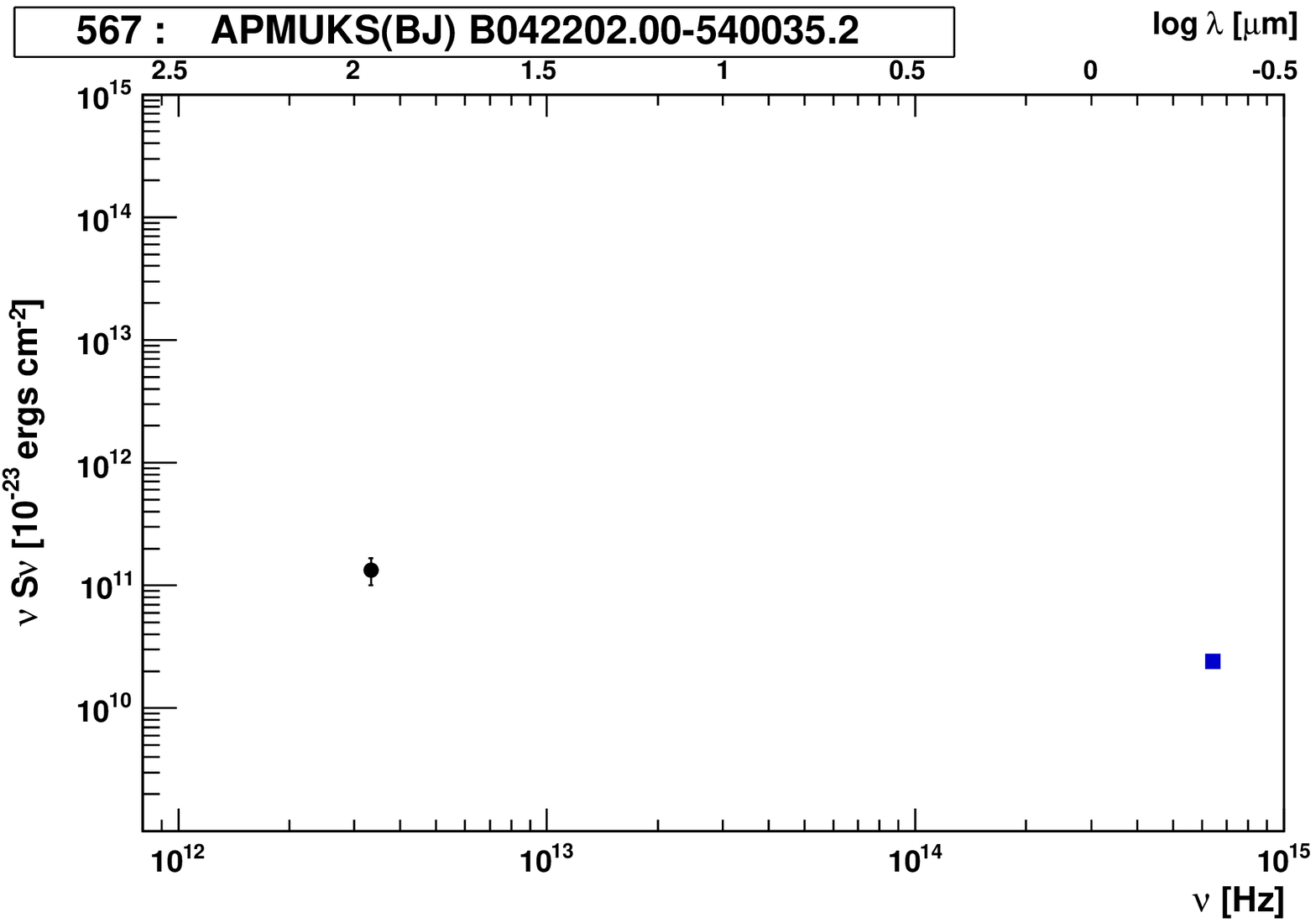}
\includegraphics[width=4cm]{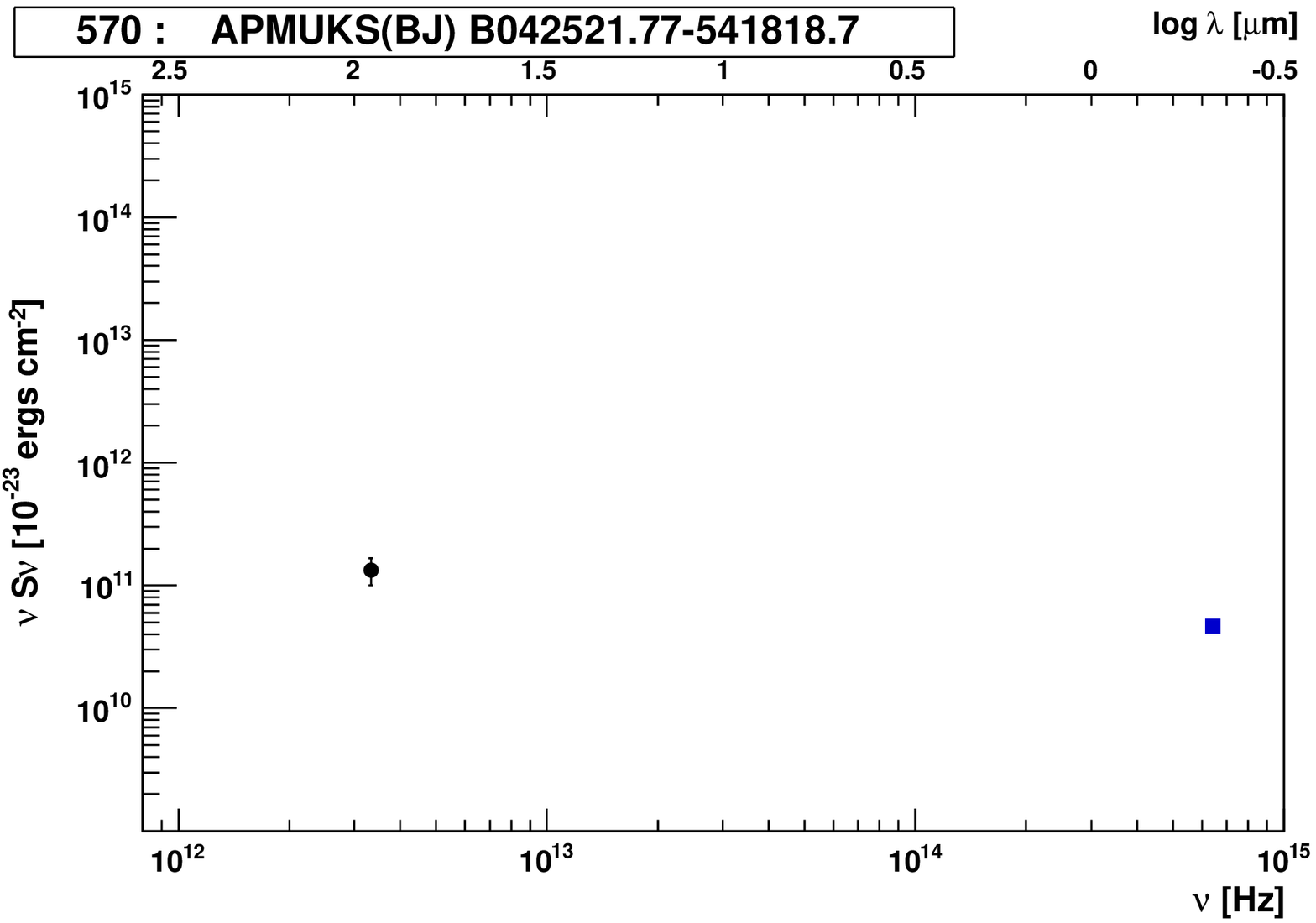}
\includegraphics[width=4cm]{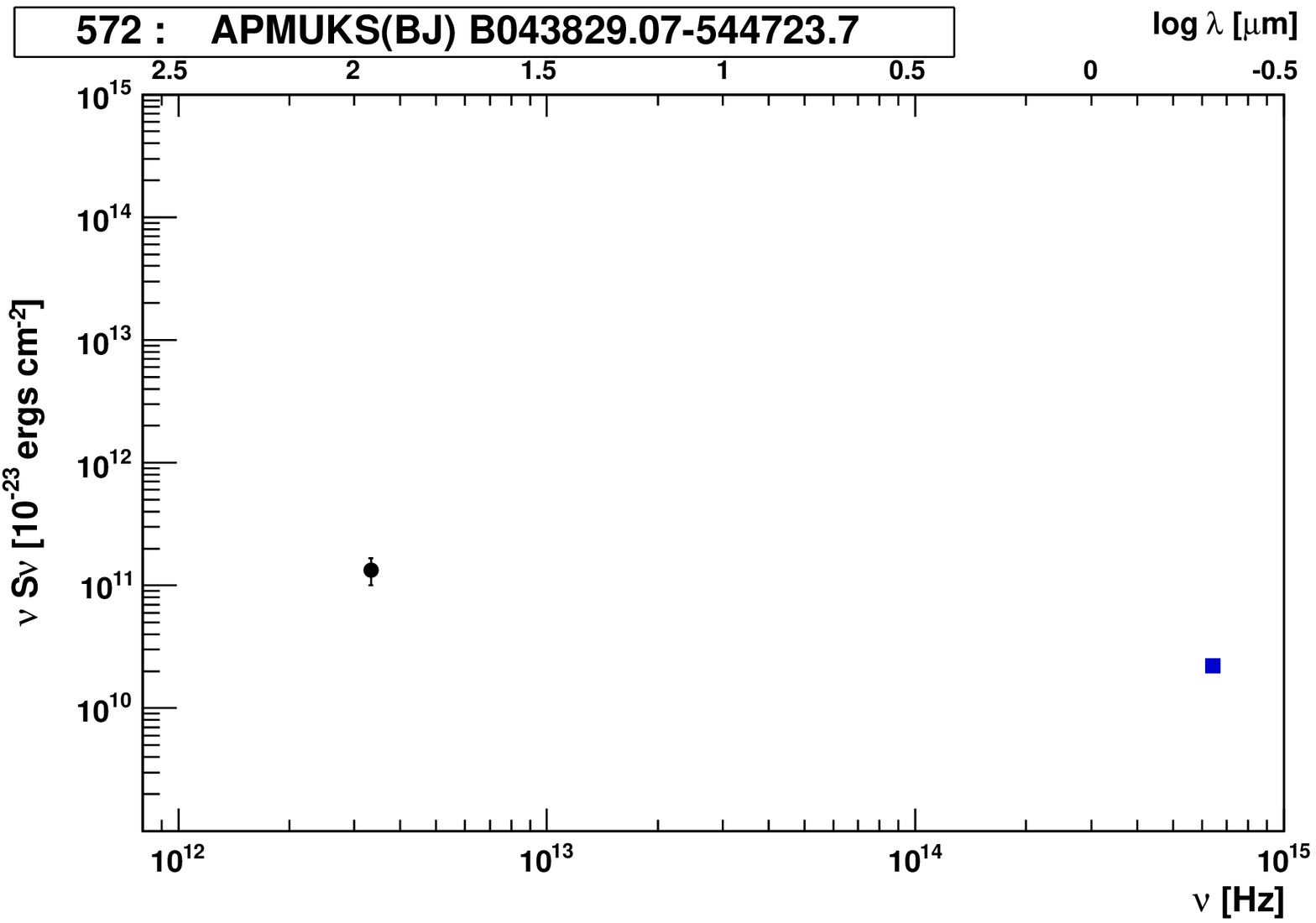}
\label{points10}
\caption {SEDs for the next 36 ADF-S identified sources, with symbols as in Figure~\ref{points1}.}
\end{figure*}
}

\clearpage

\onlfig{11}{
\begin{figure*}[t]
\centering

\includegraphics[width=4cm]{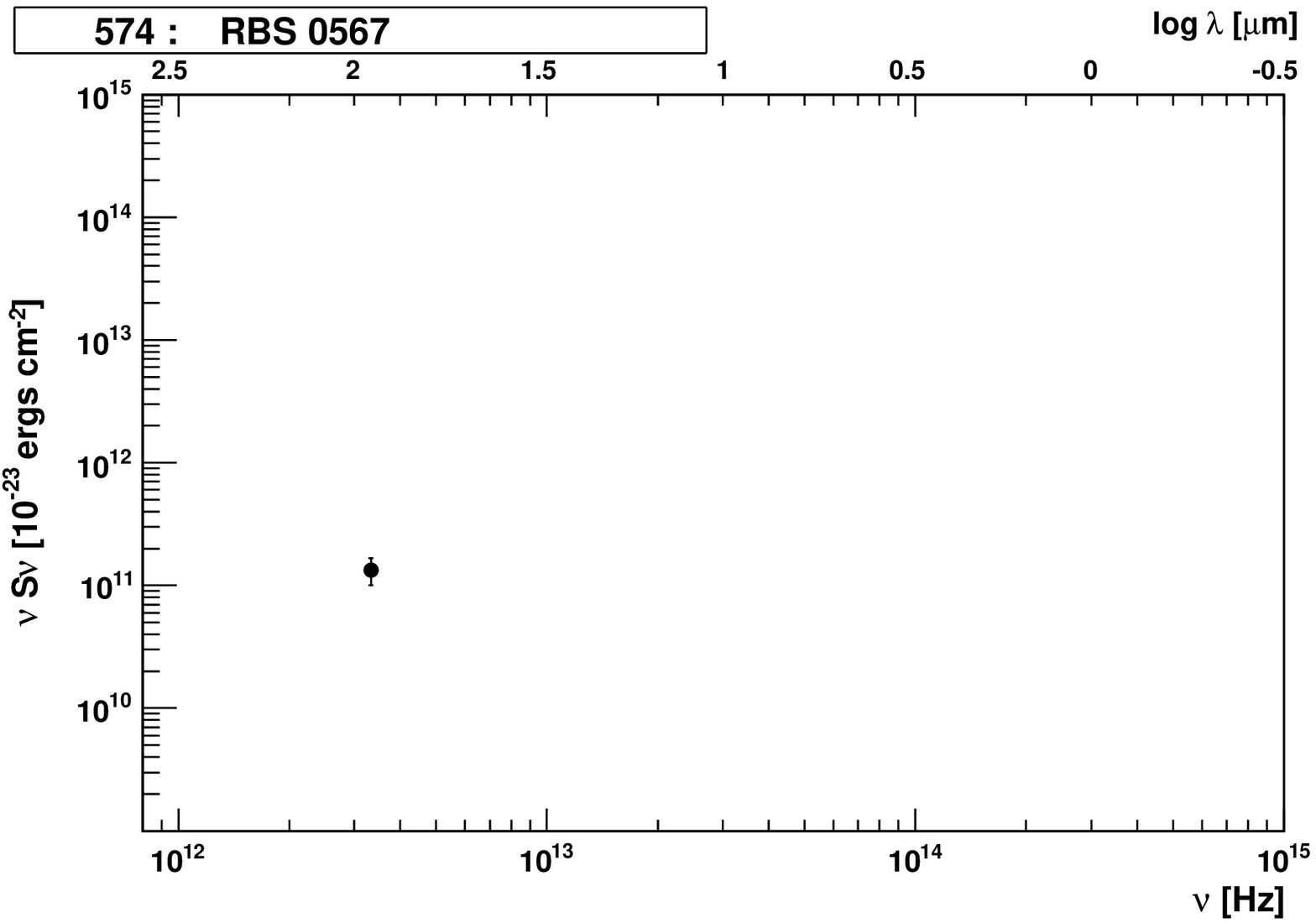}
\includegraphics[width=4cm]{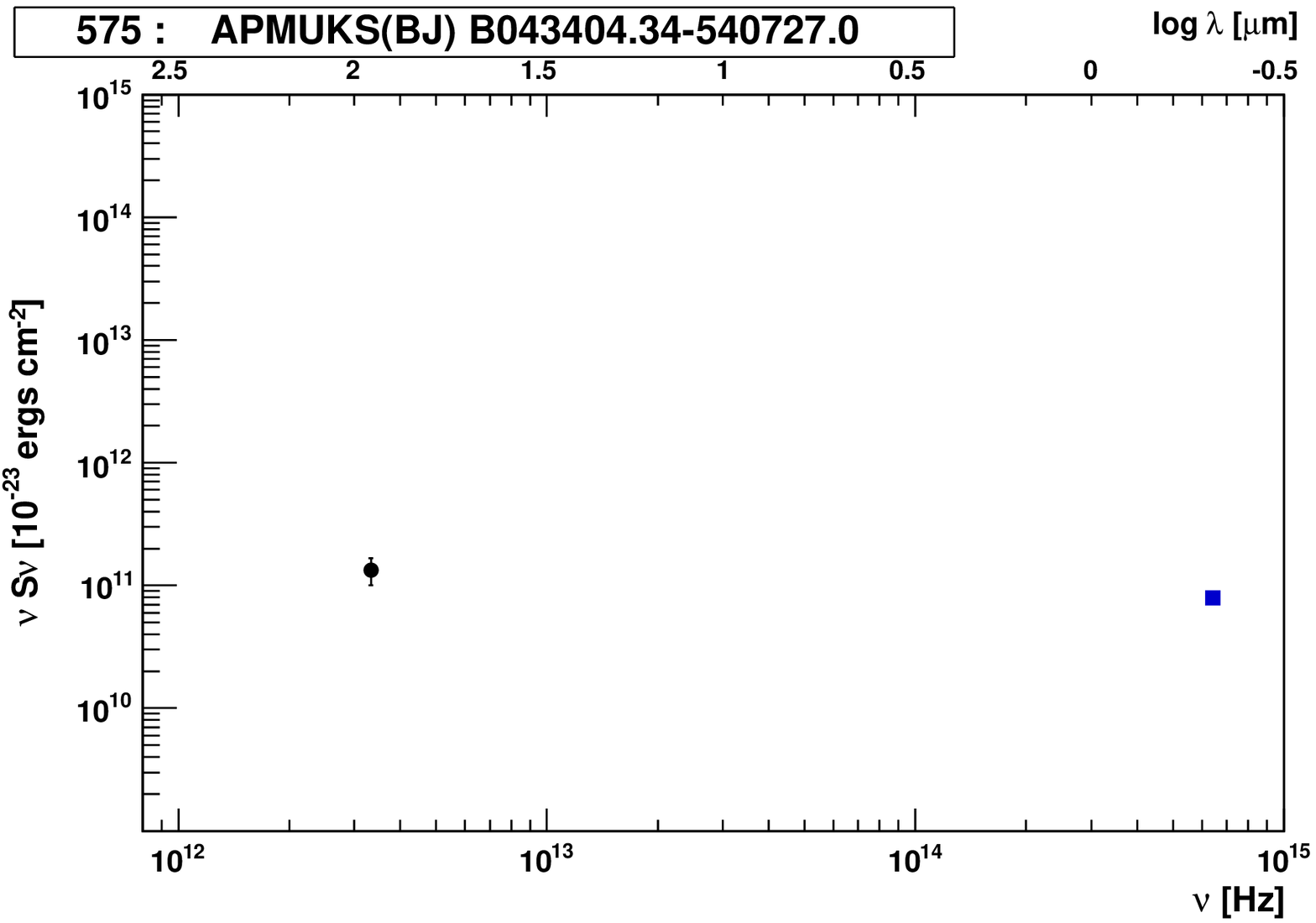}
\includegraphics[width=4cm]{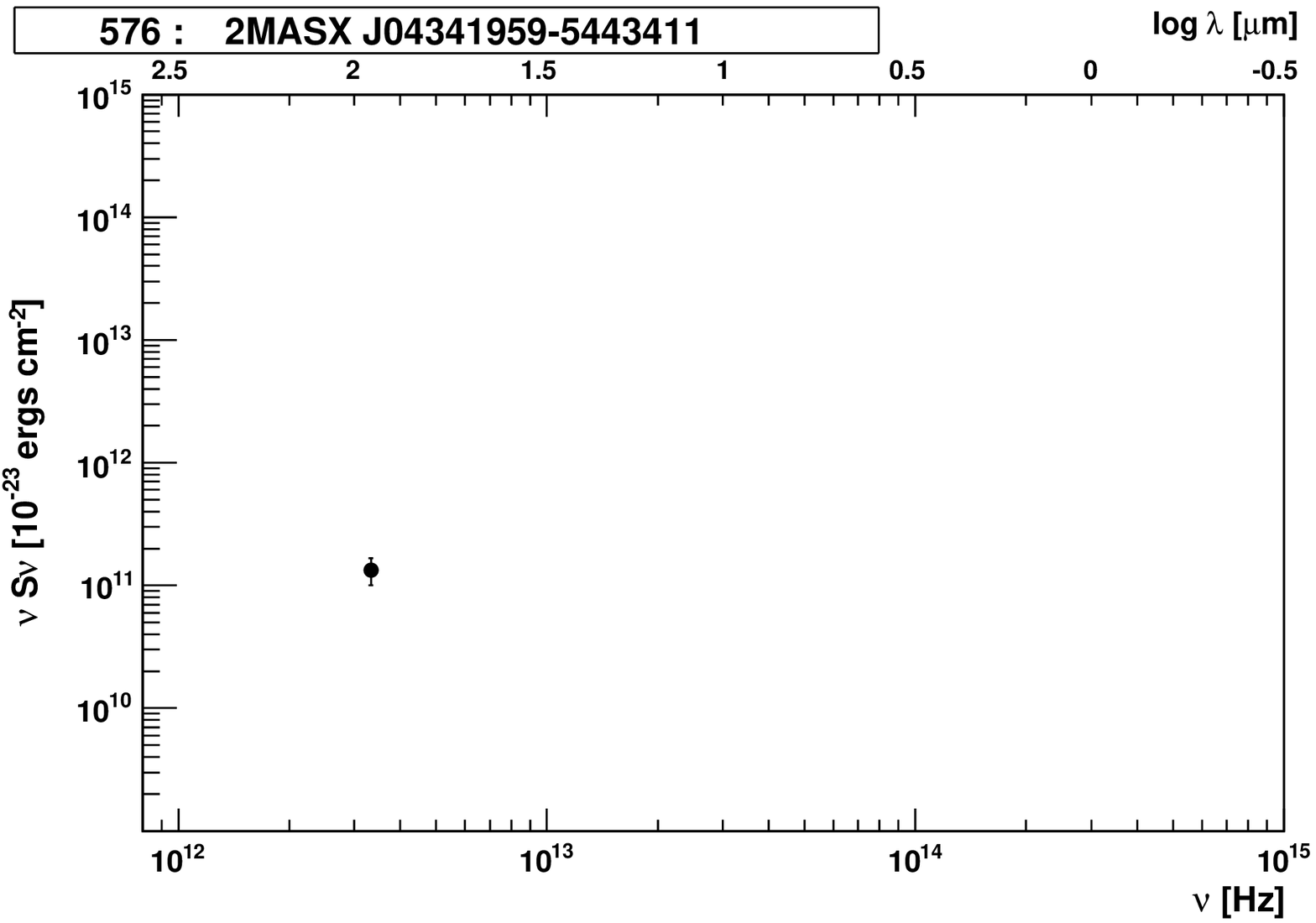}
\includegraphics[width=4cm]{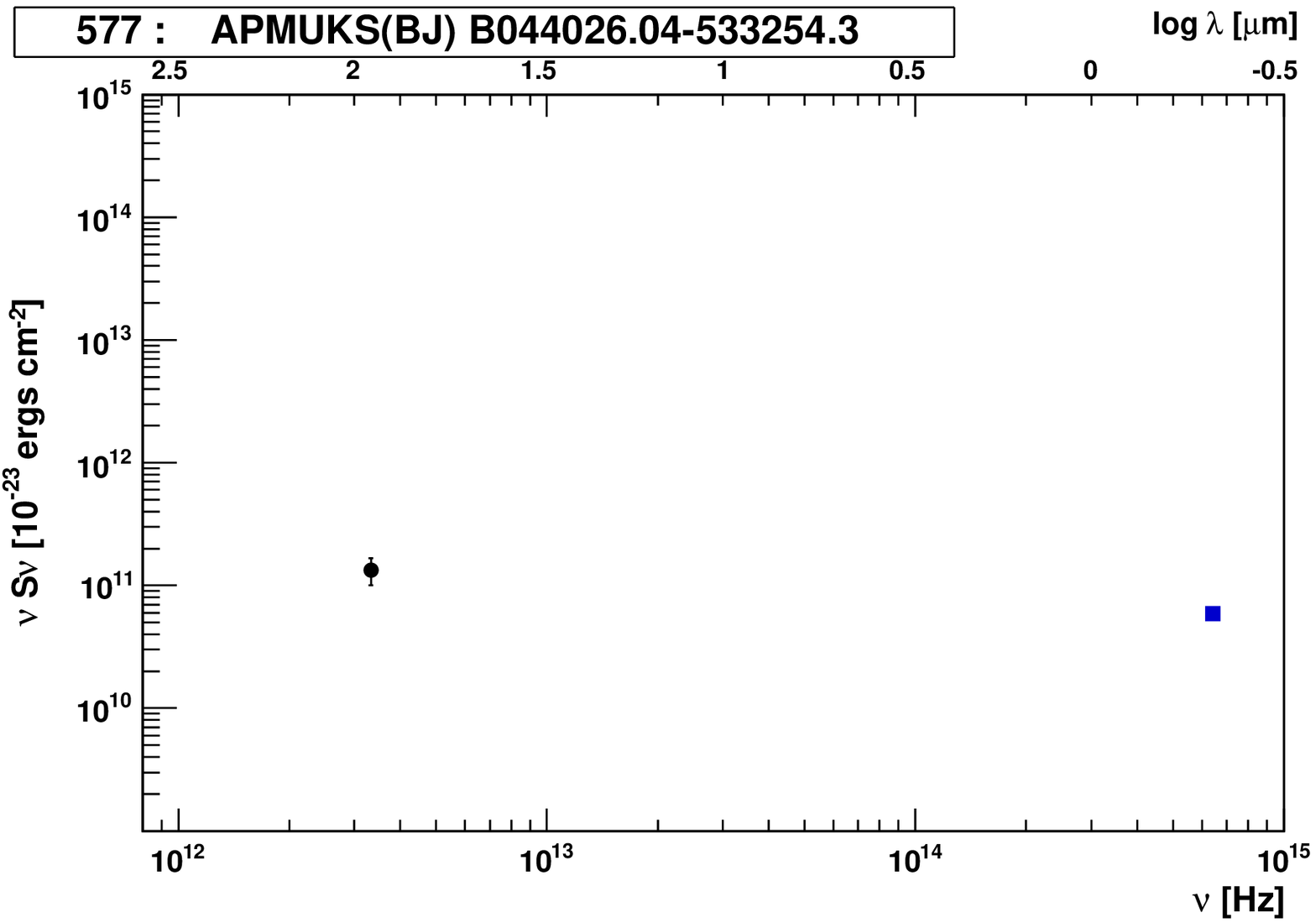}
\includegraphics[width=4cm]{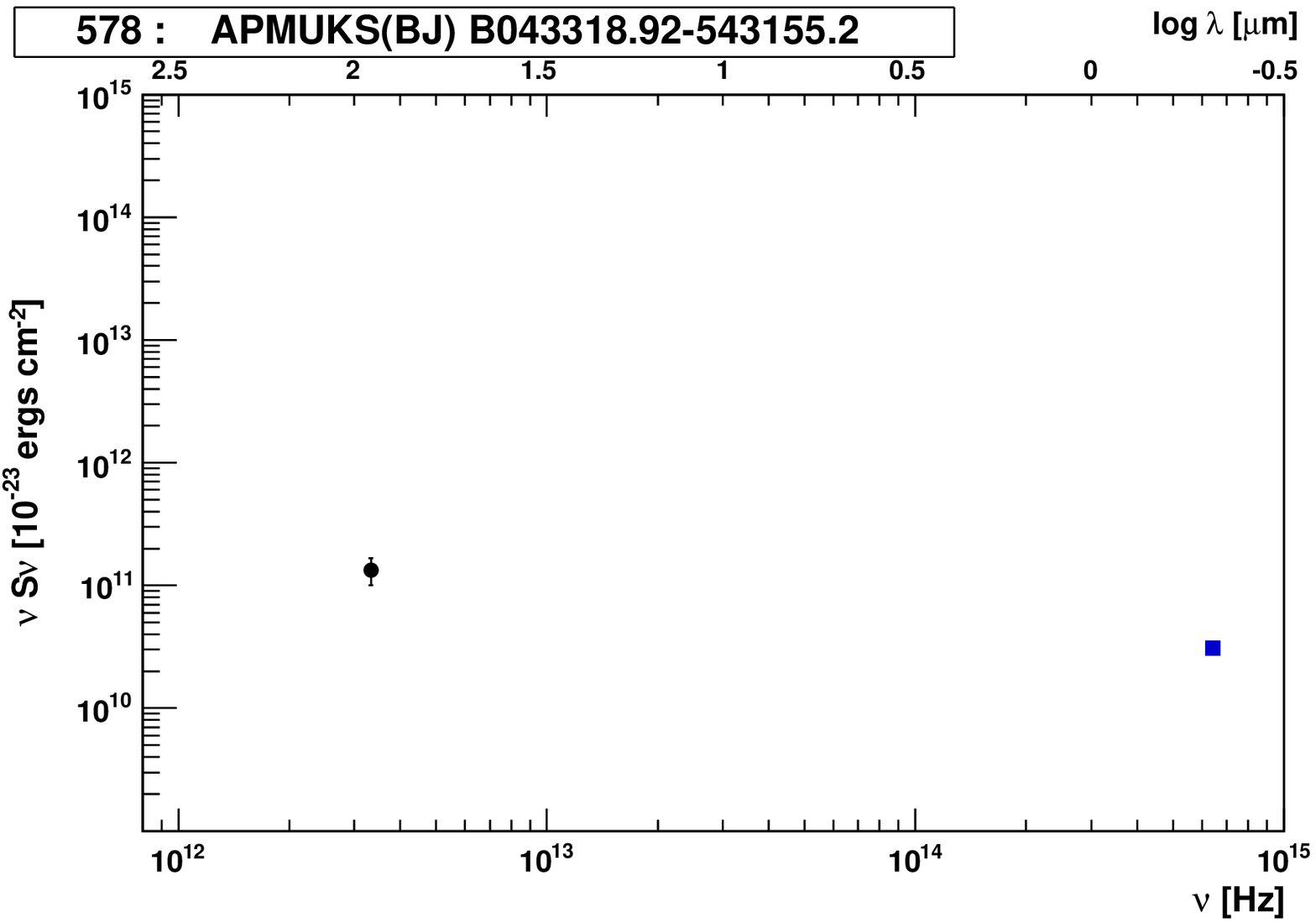}
\includegraphics[width=4cm]{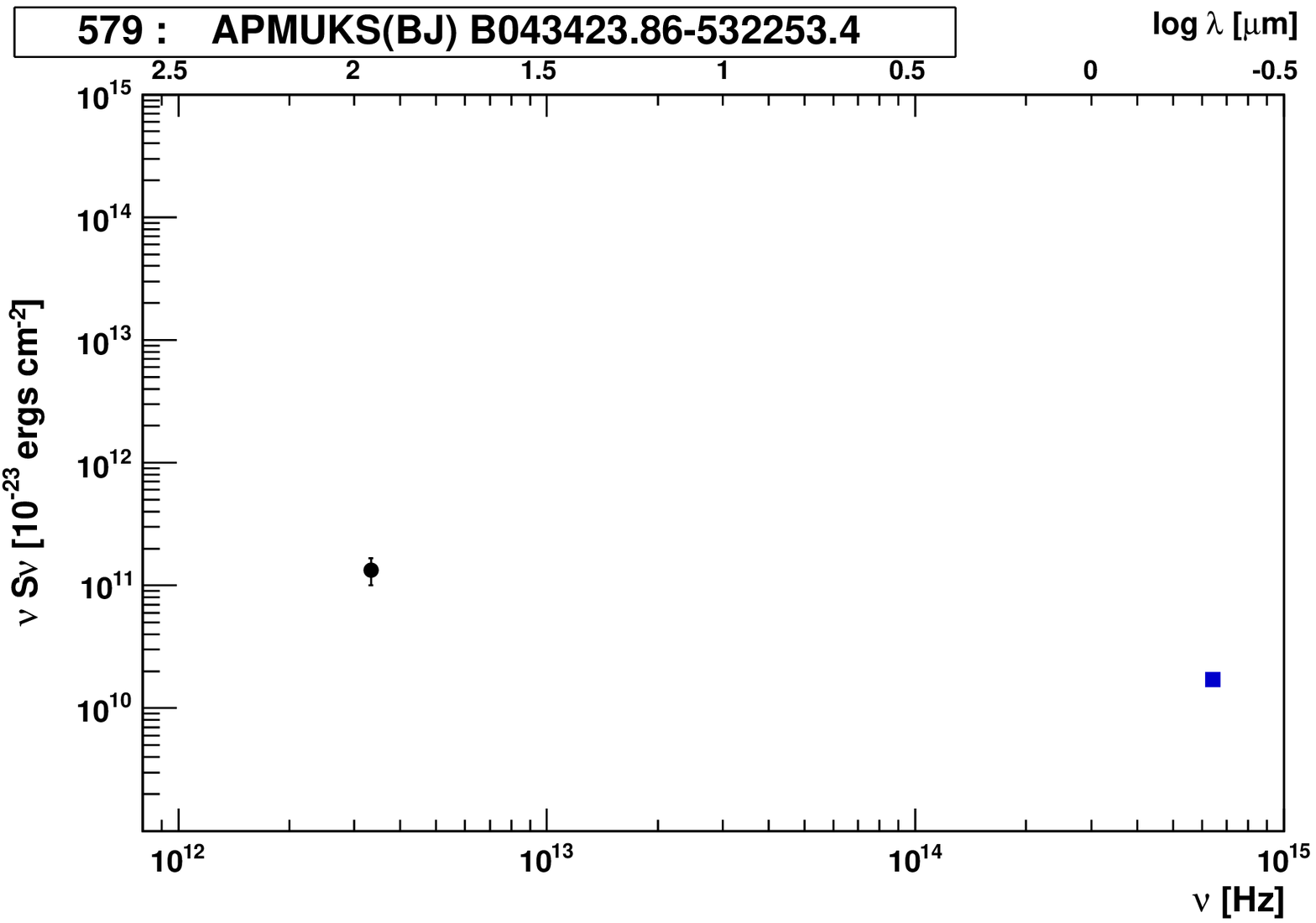}
\includegraphics[width=4cm]{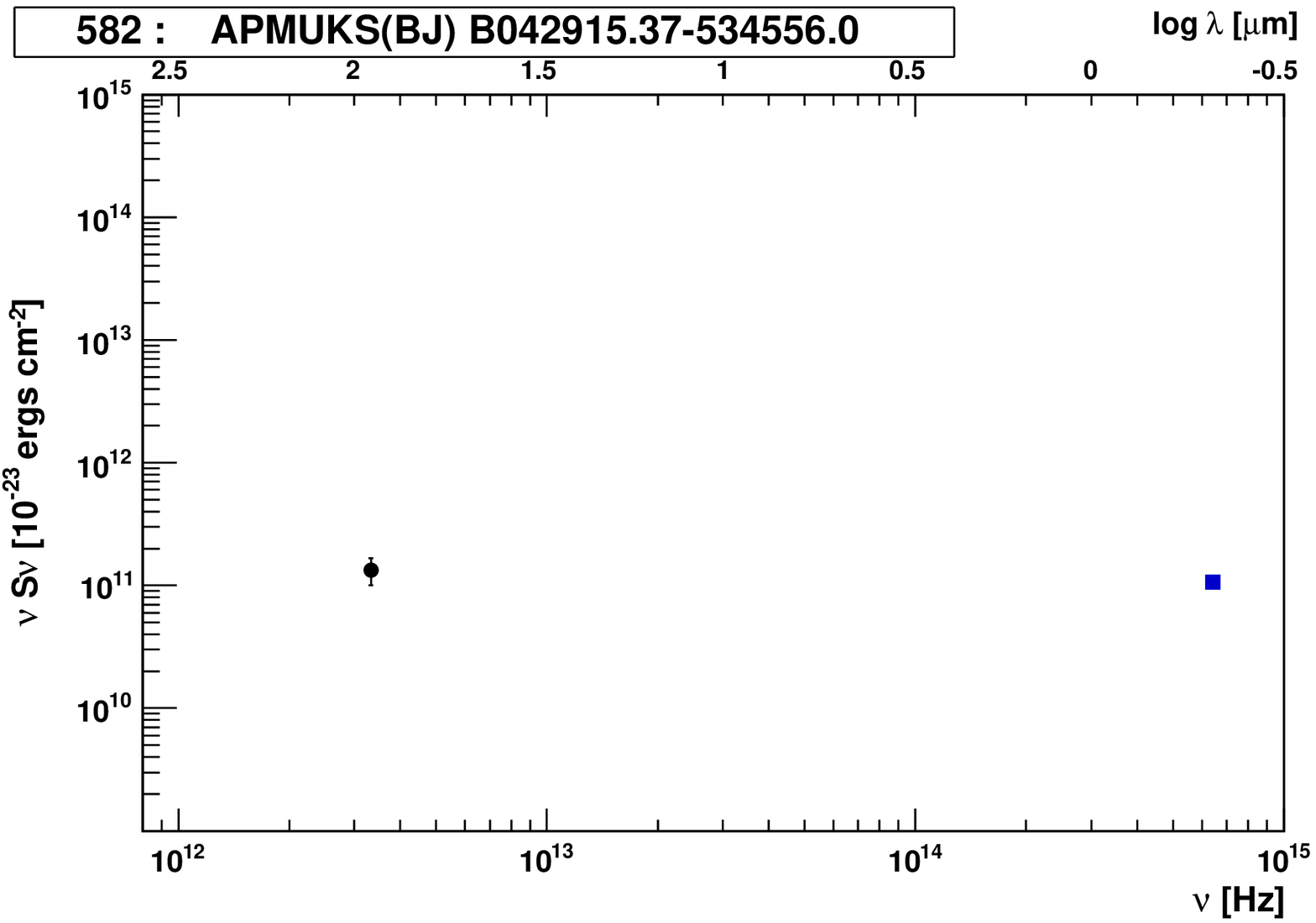}
\includegraphics[width=4cm]{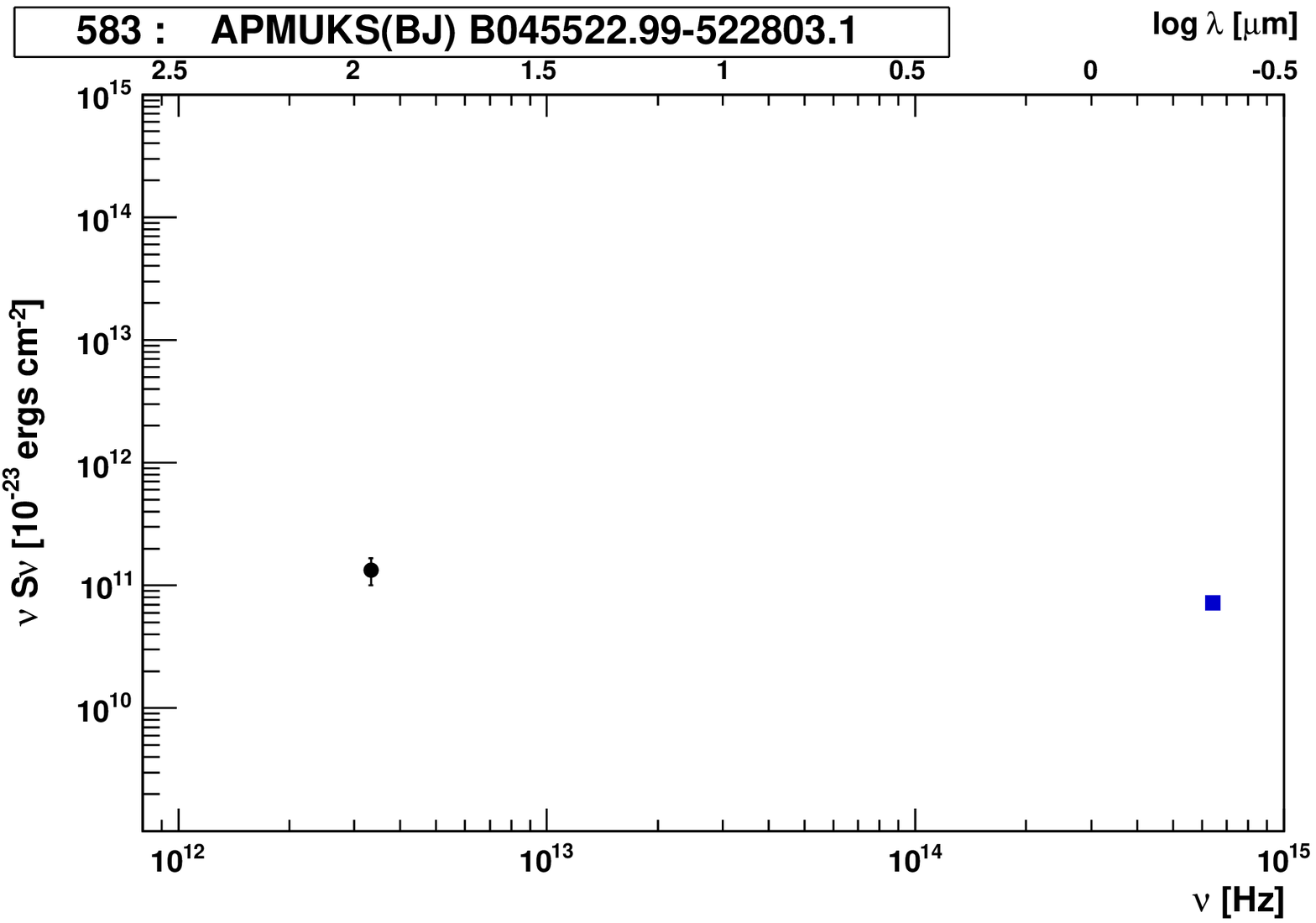}
\includegraphics[width=4cm]{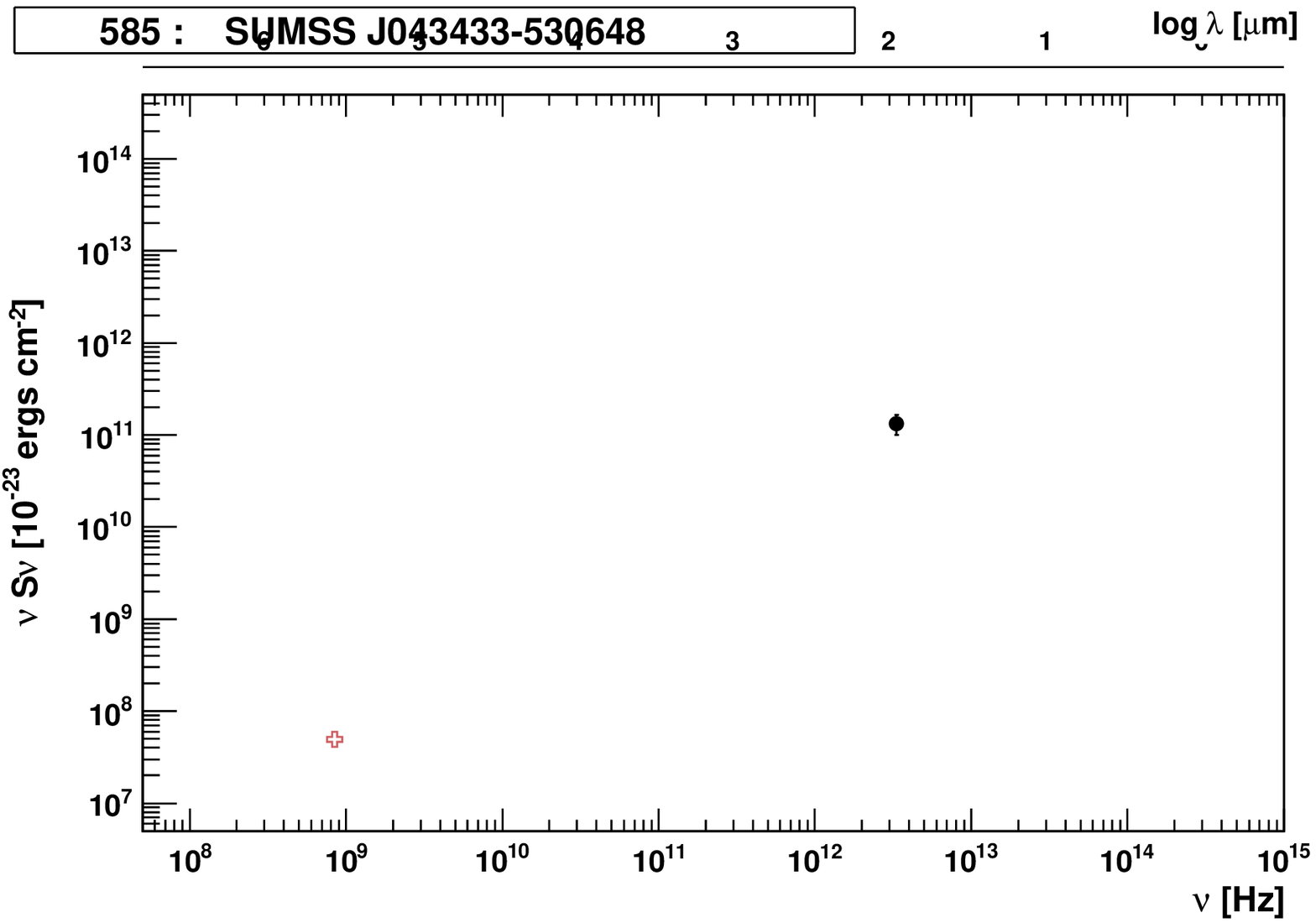}
\includegraphics[width=4cm]{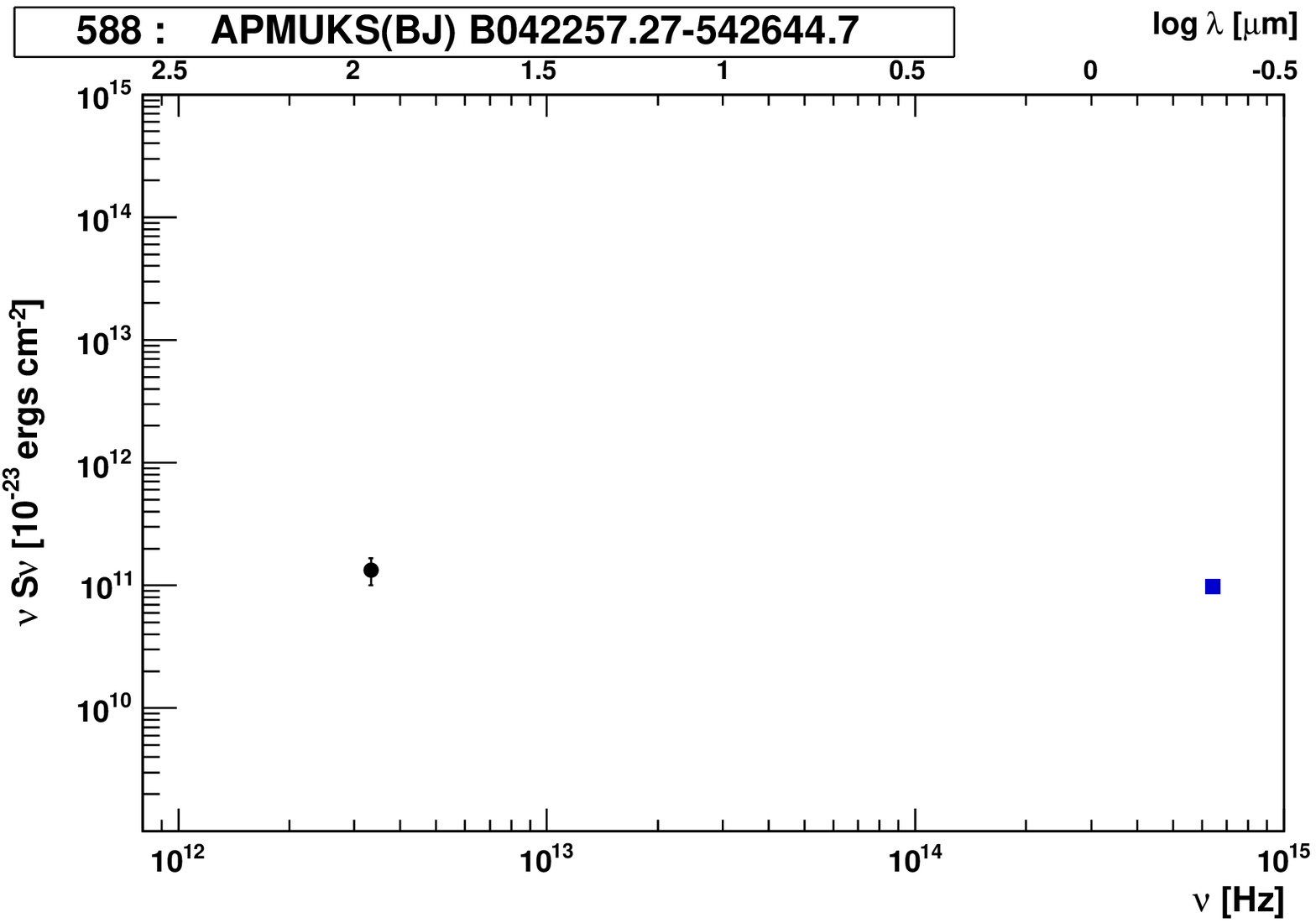}
\includegraphics[width=4cm]{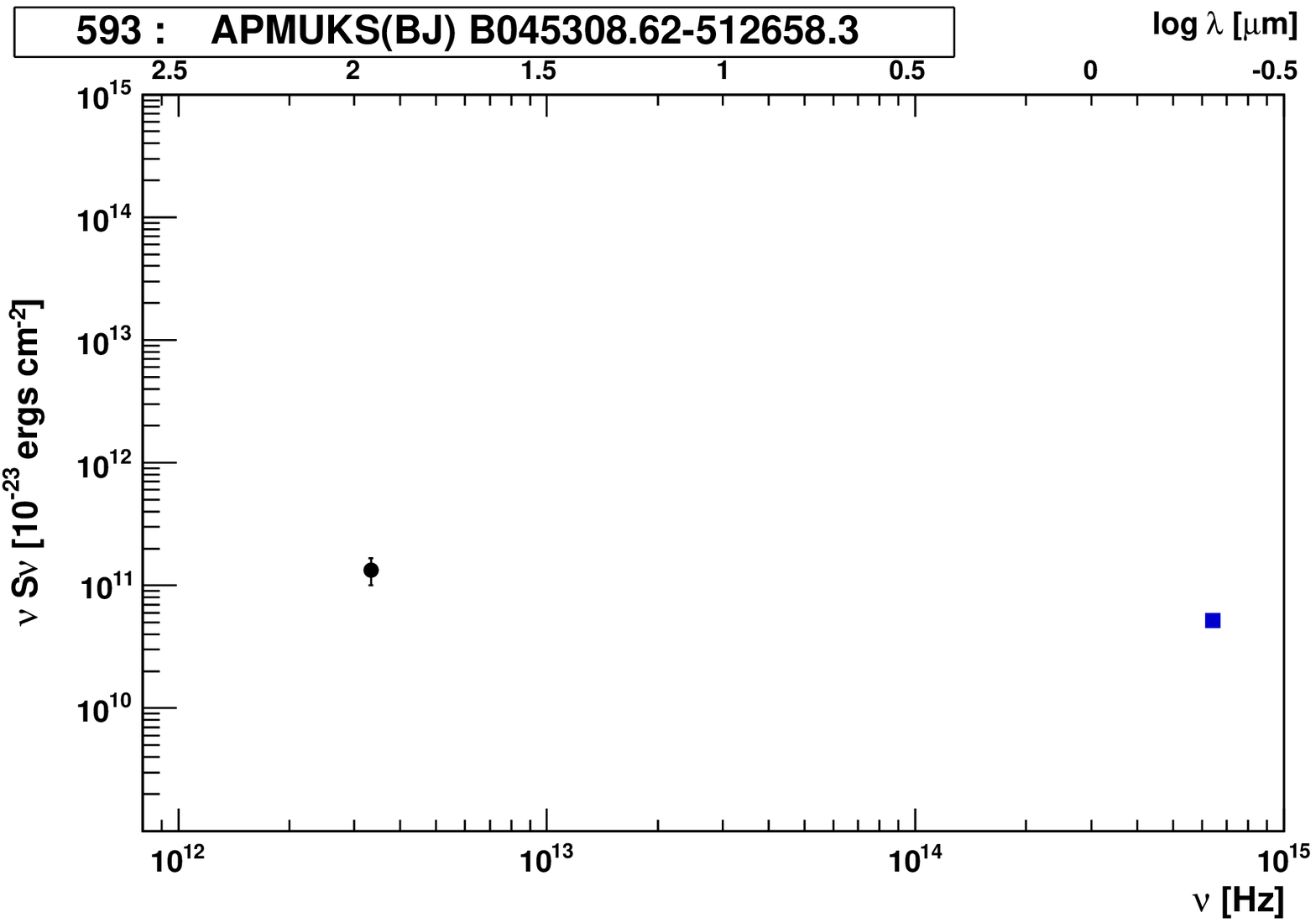}
\includegraphics[width=4cm]{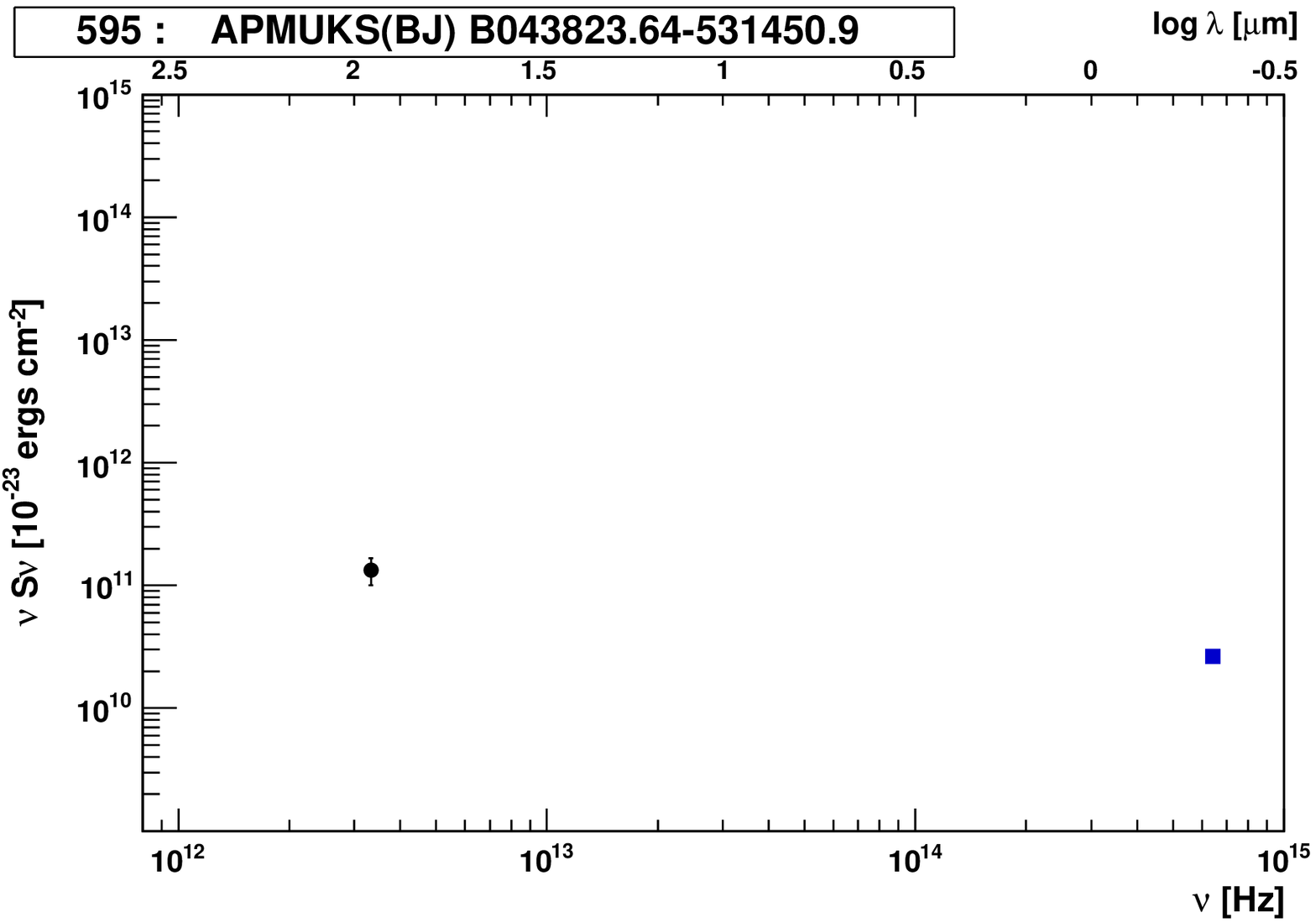}
\includegraphics[width=4cm]{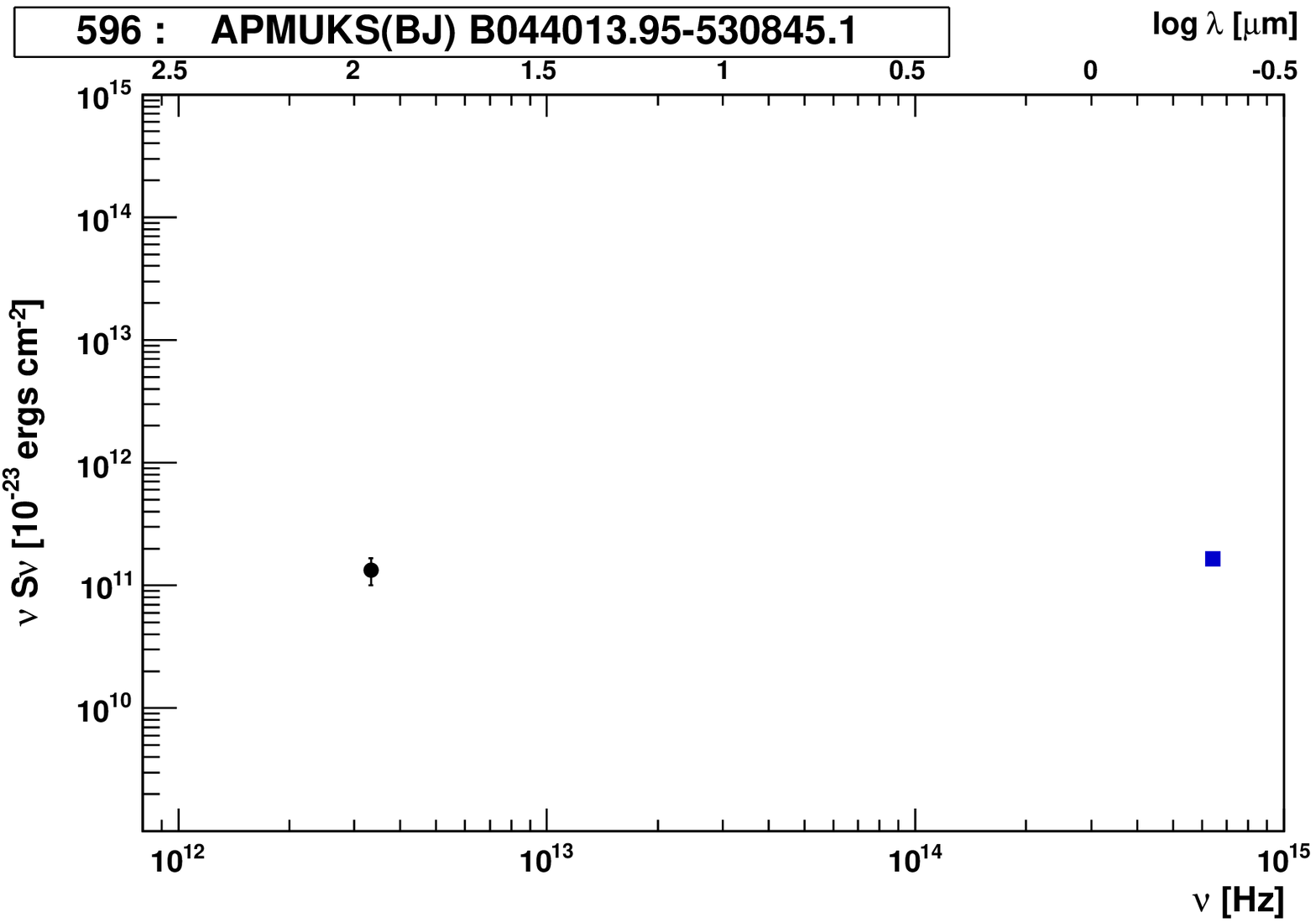}
\includegraphics[width=4cm]{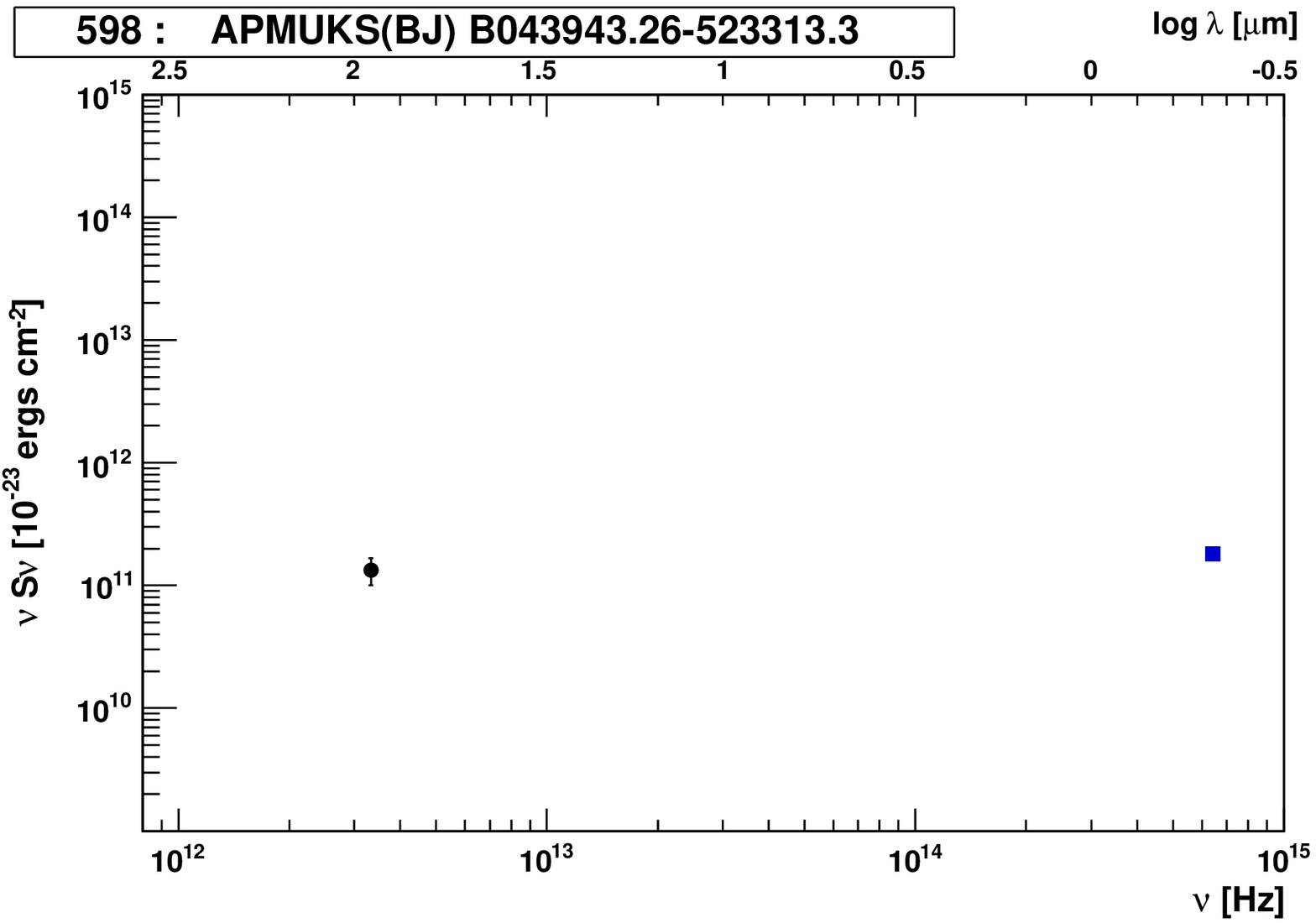}
\includegraphics[width=4cm]{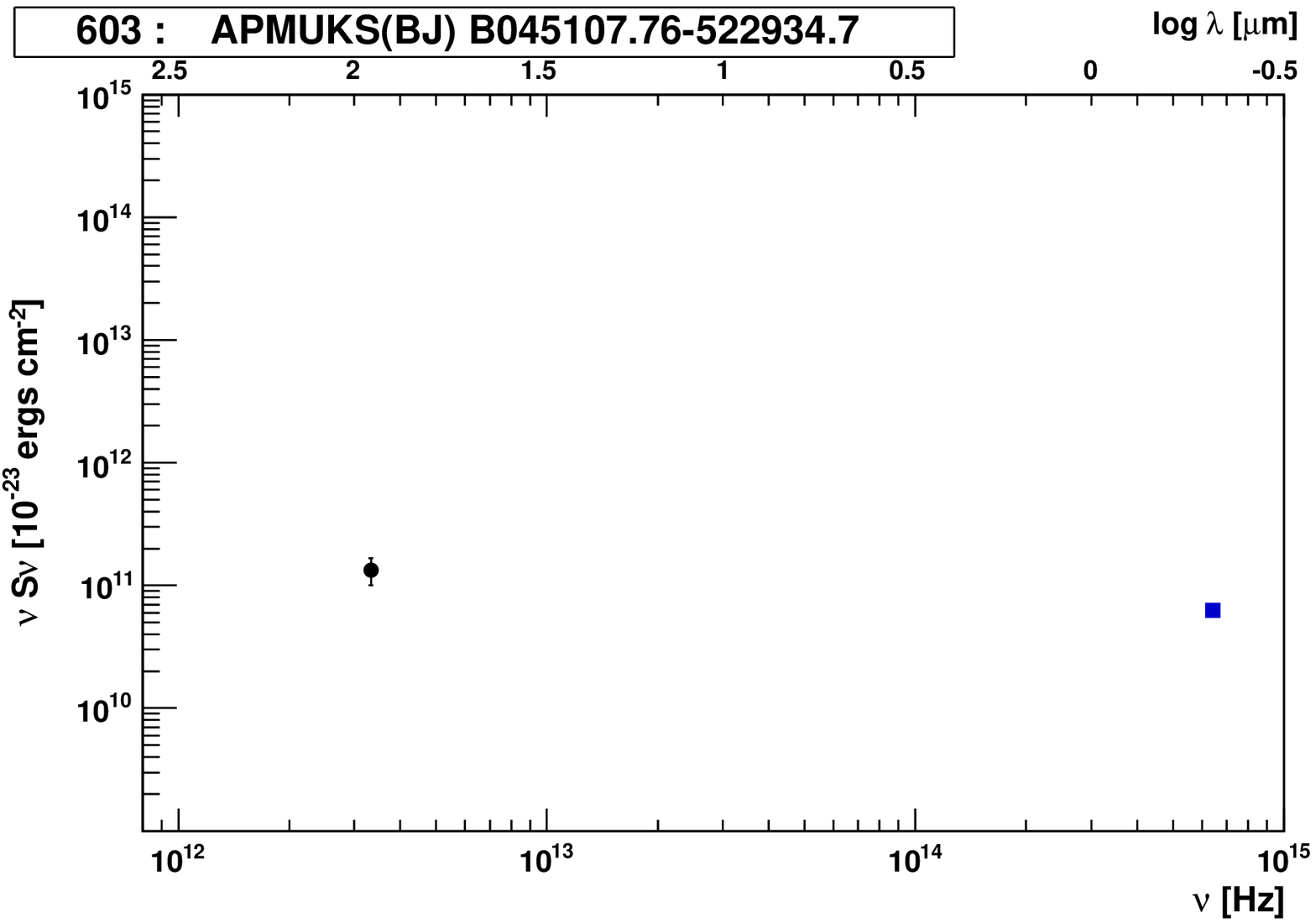}
\includegraphics[width=4cm]{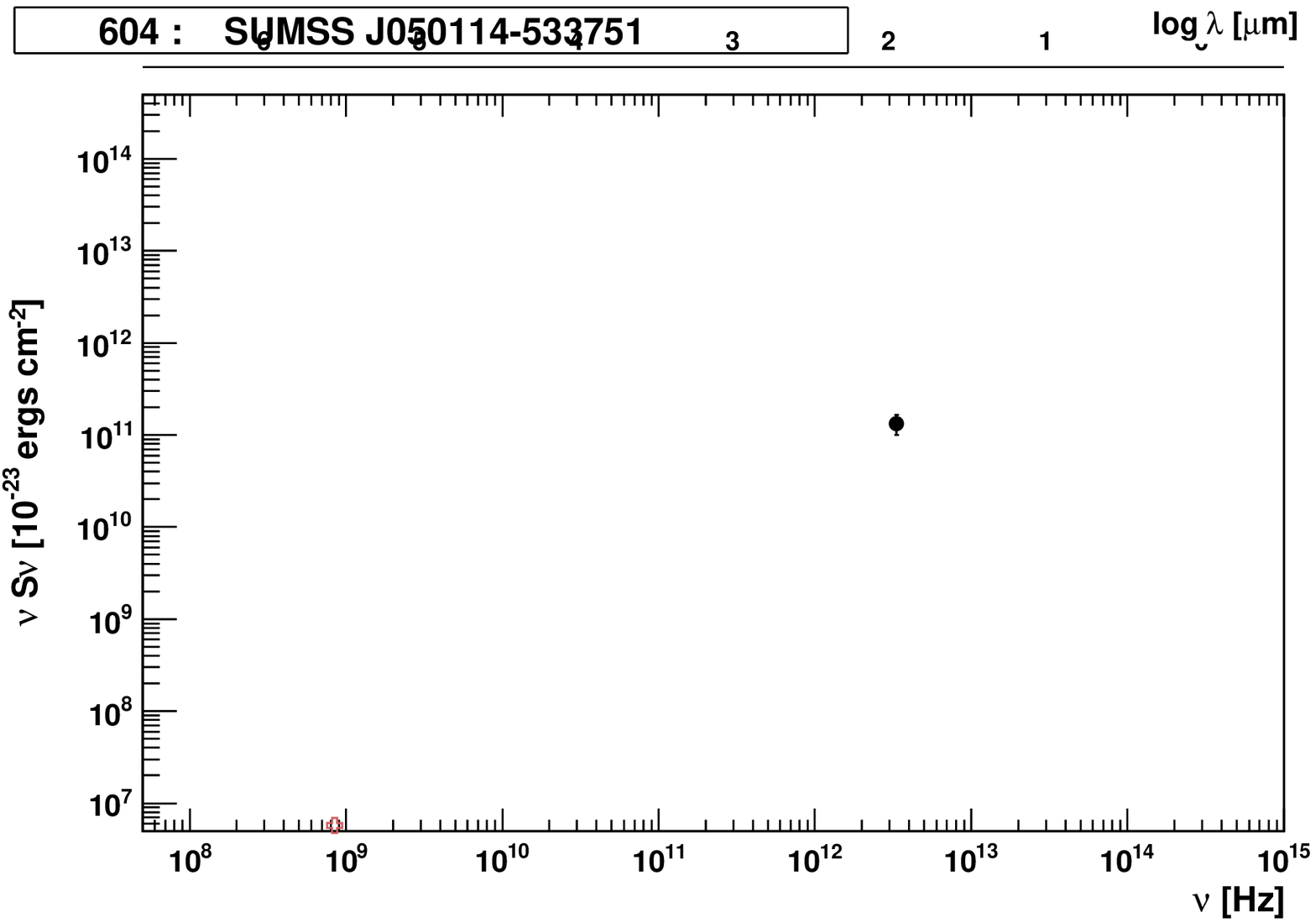}
\includegraphics[width=4cm]{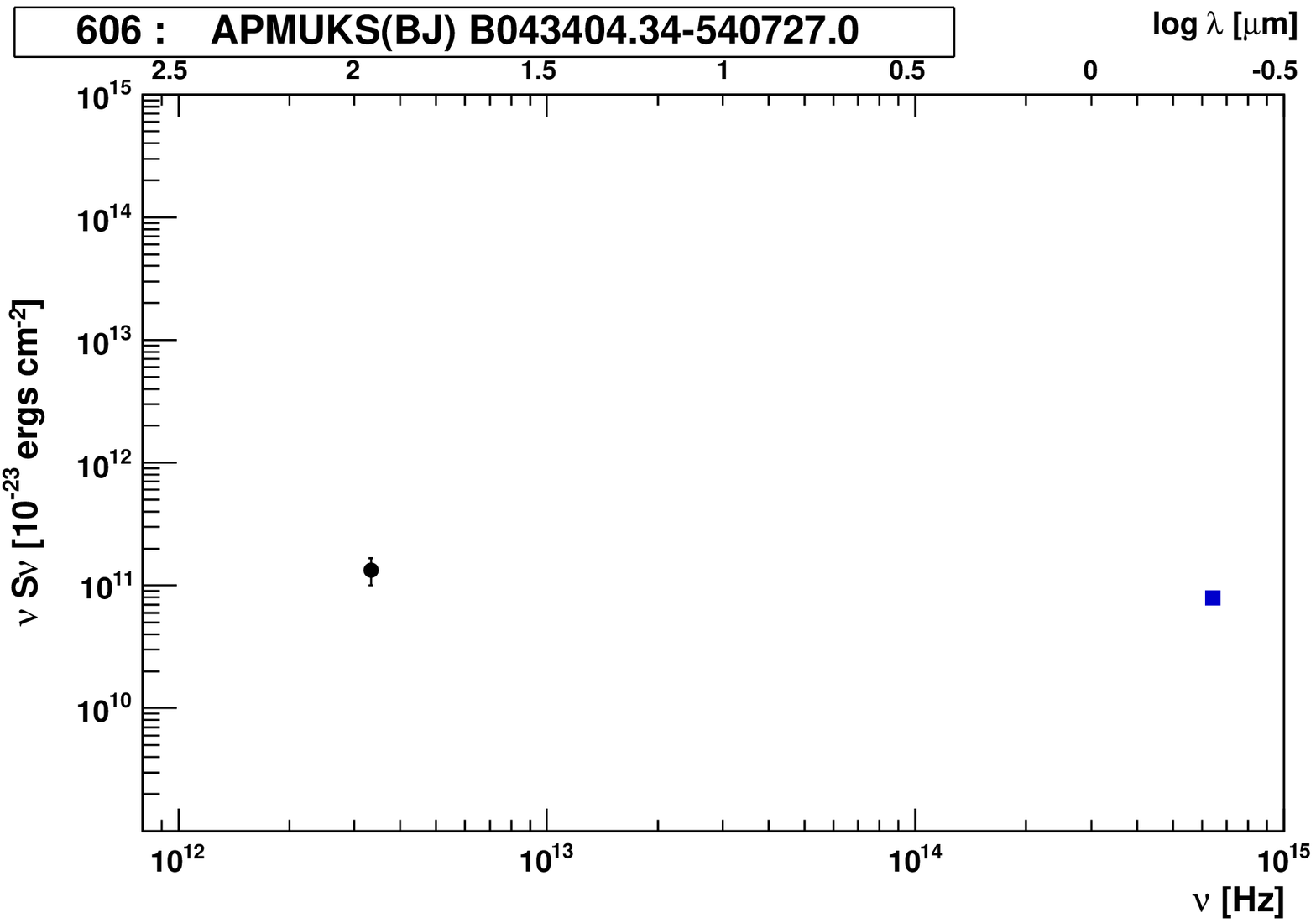}
\includegraphics[width=4cm]{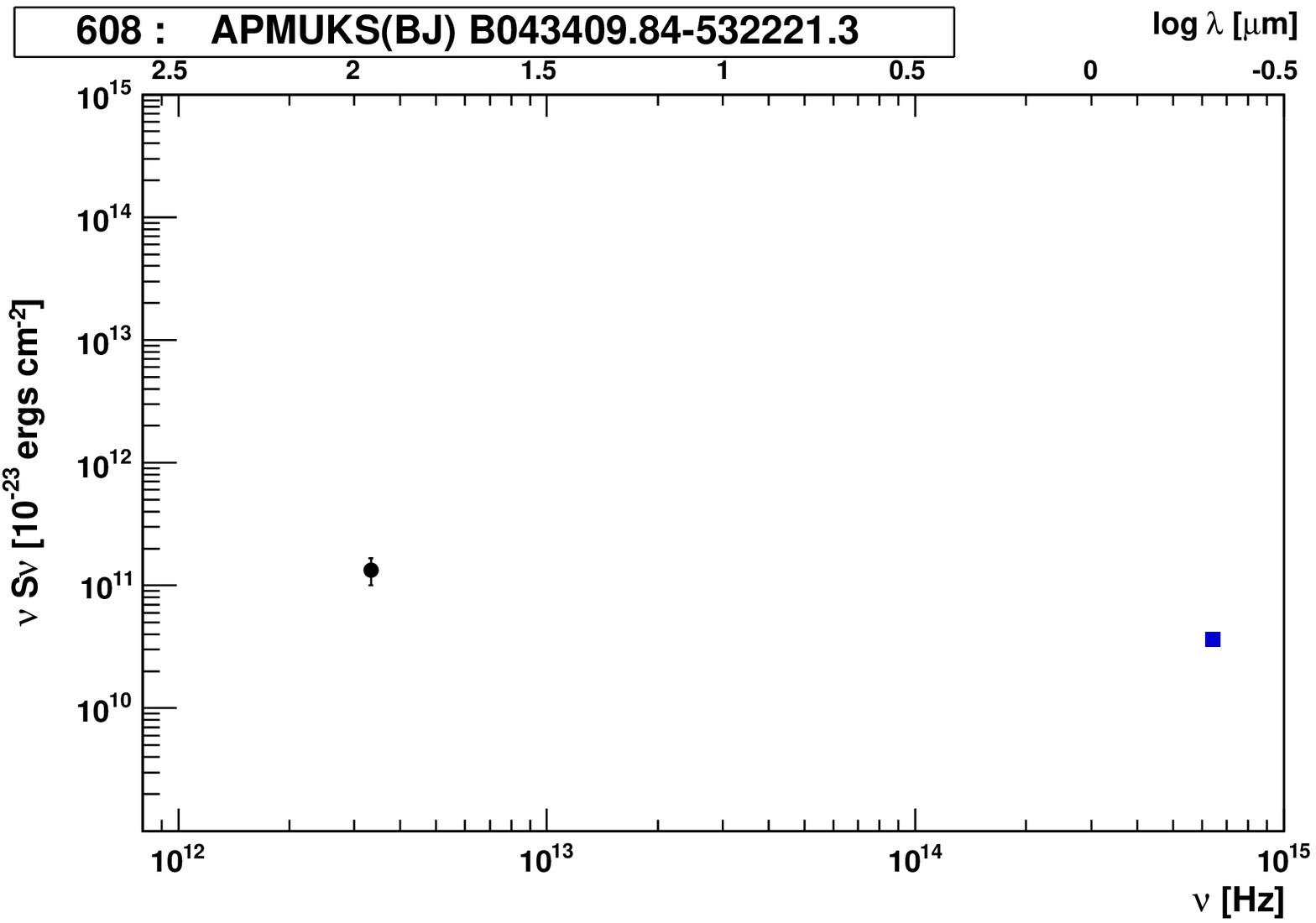}
\includegraphics[width=4cm]{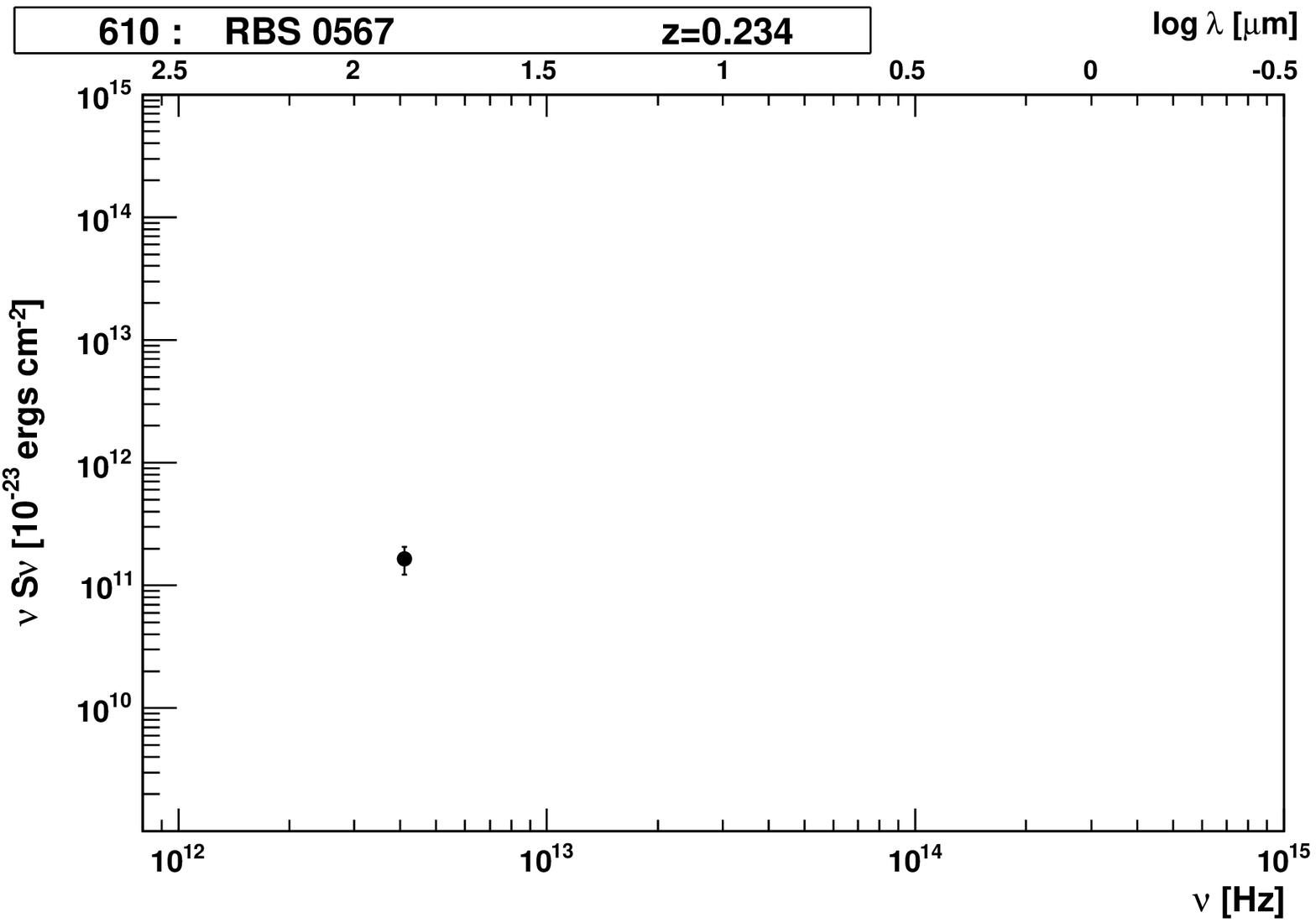}
\includegraphics[width=4cm]{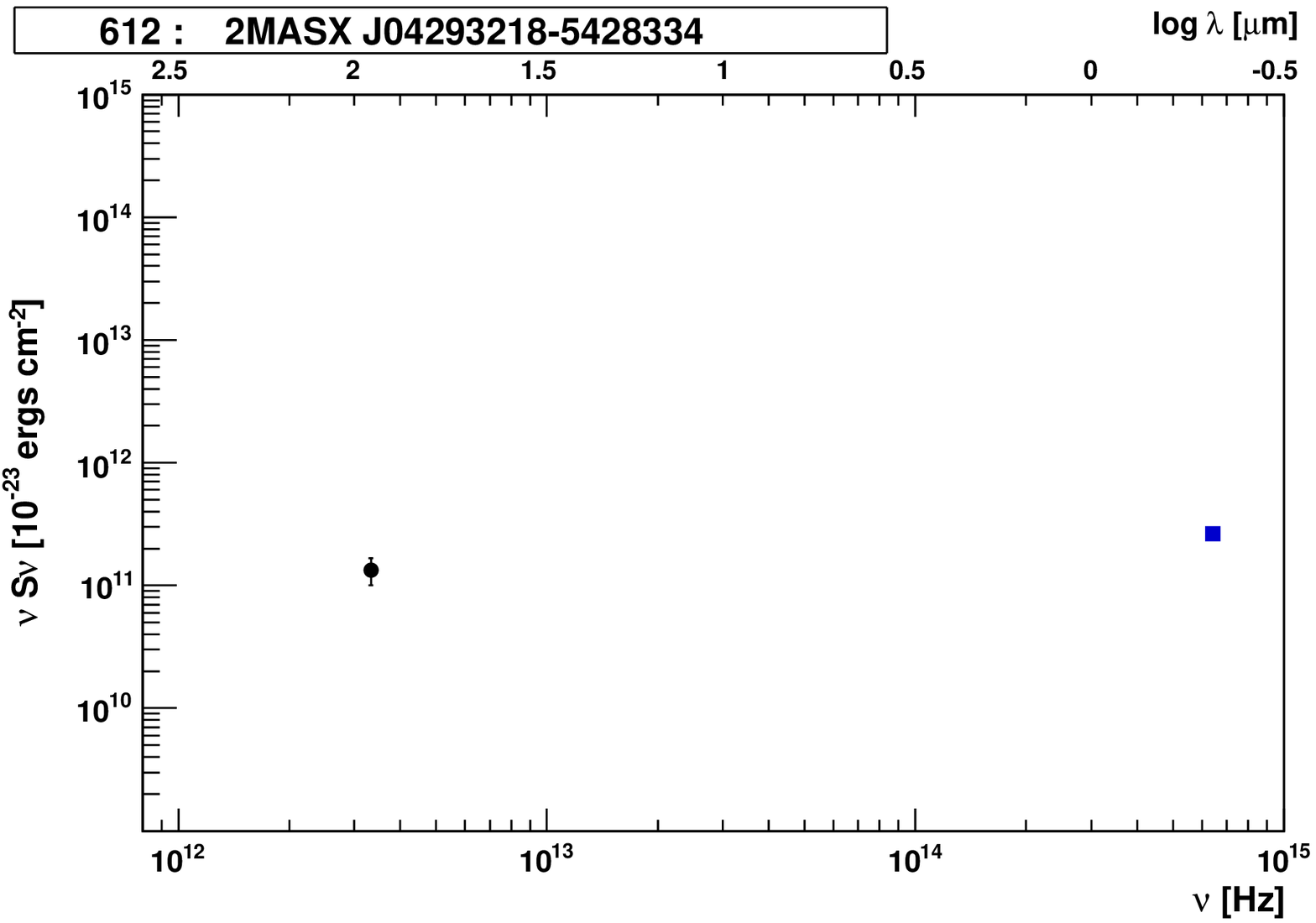}
\includegraphics[width=4cm]{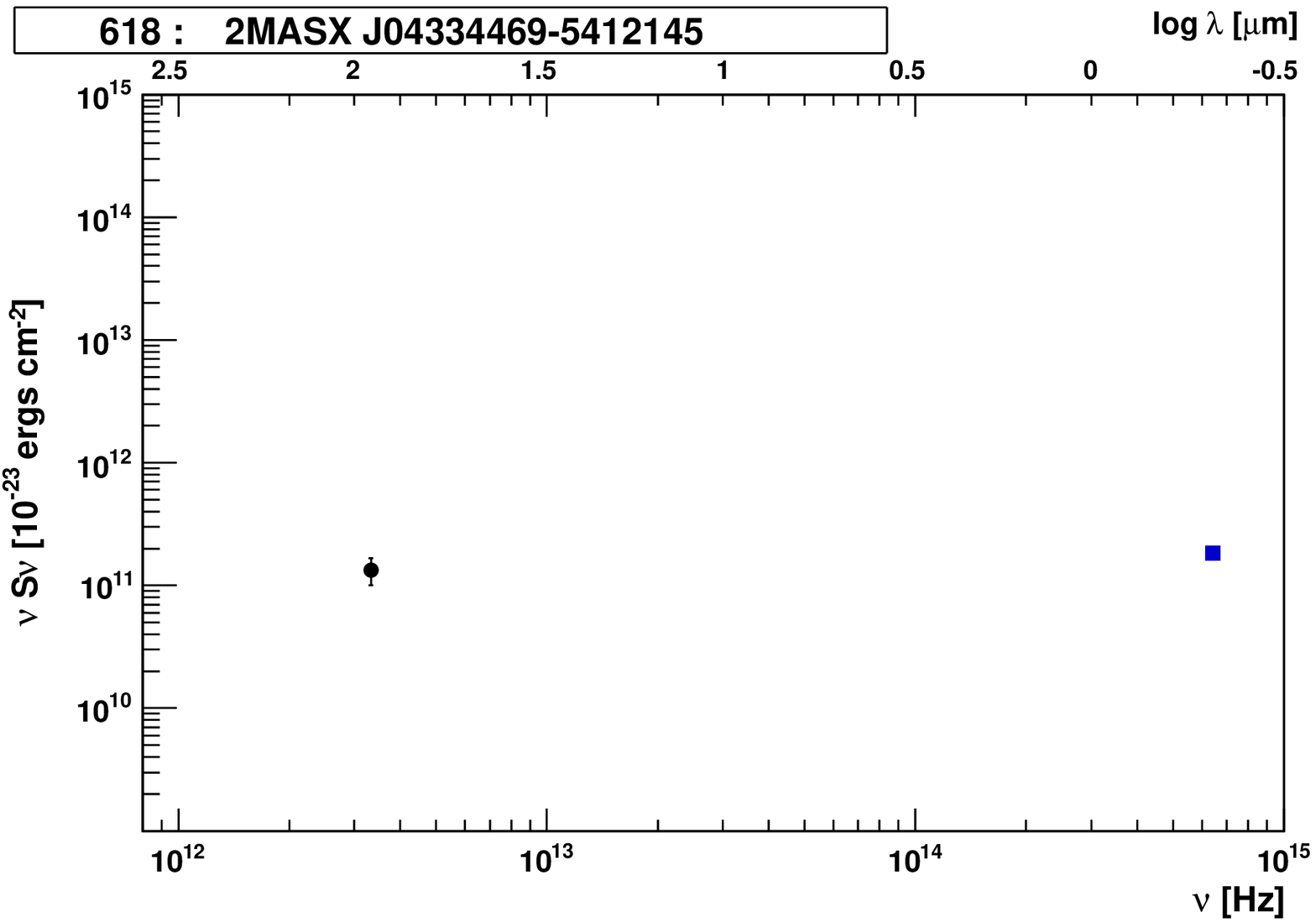}
\includegraphics[width=4cm]{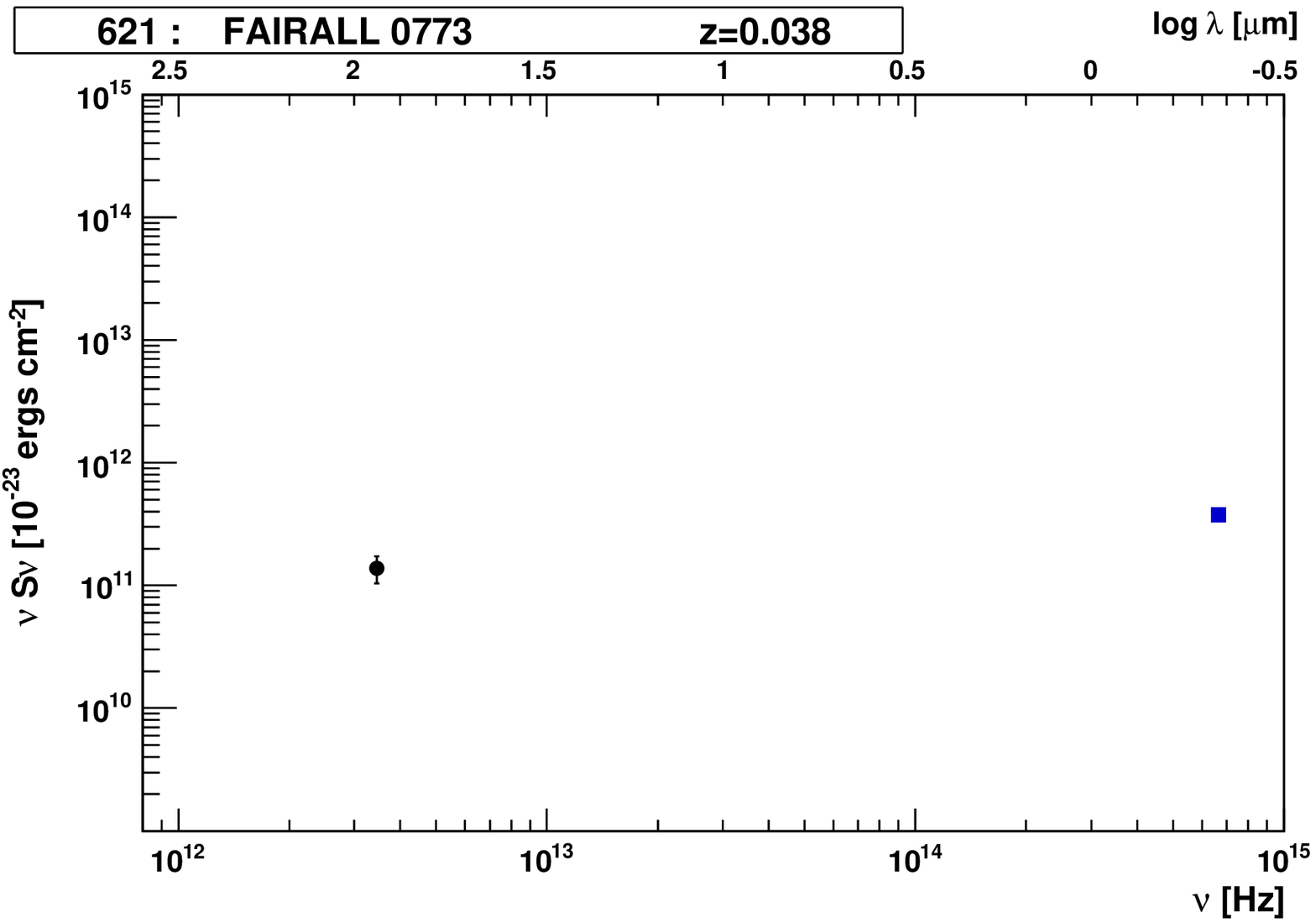}
\includegraphics[width=4cm]{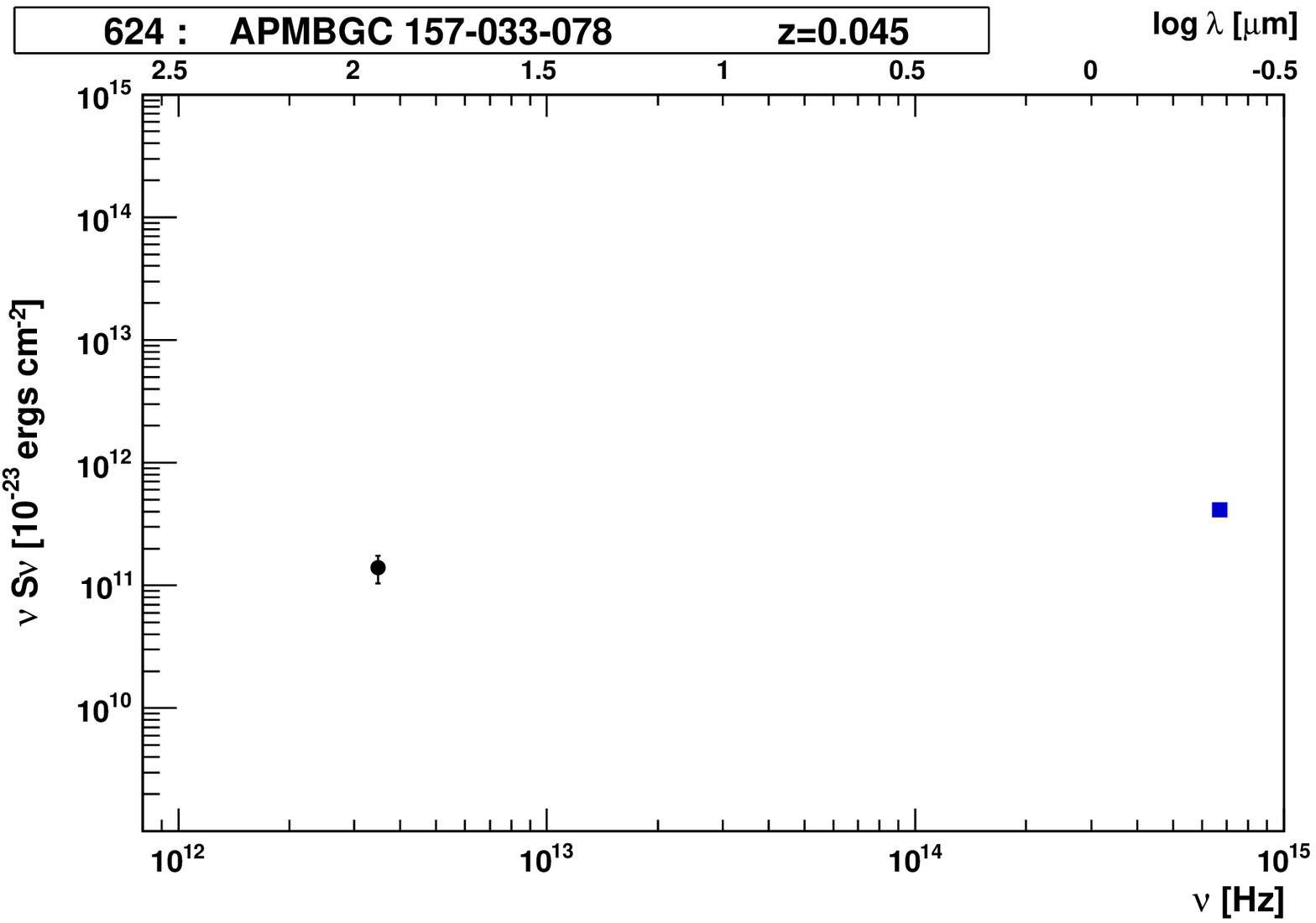}
\includegraphics[width=4cm]{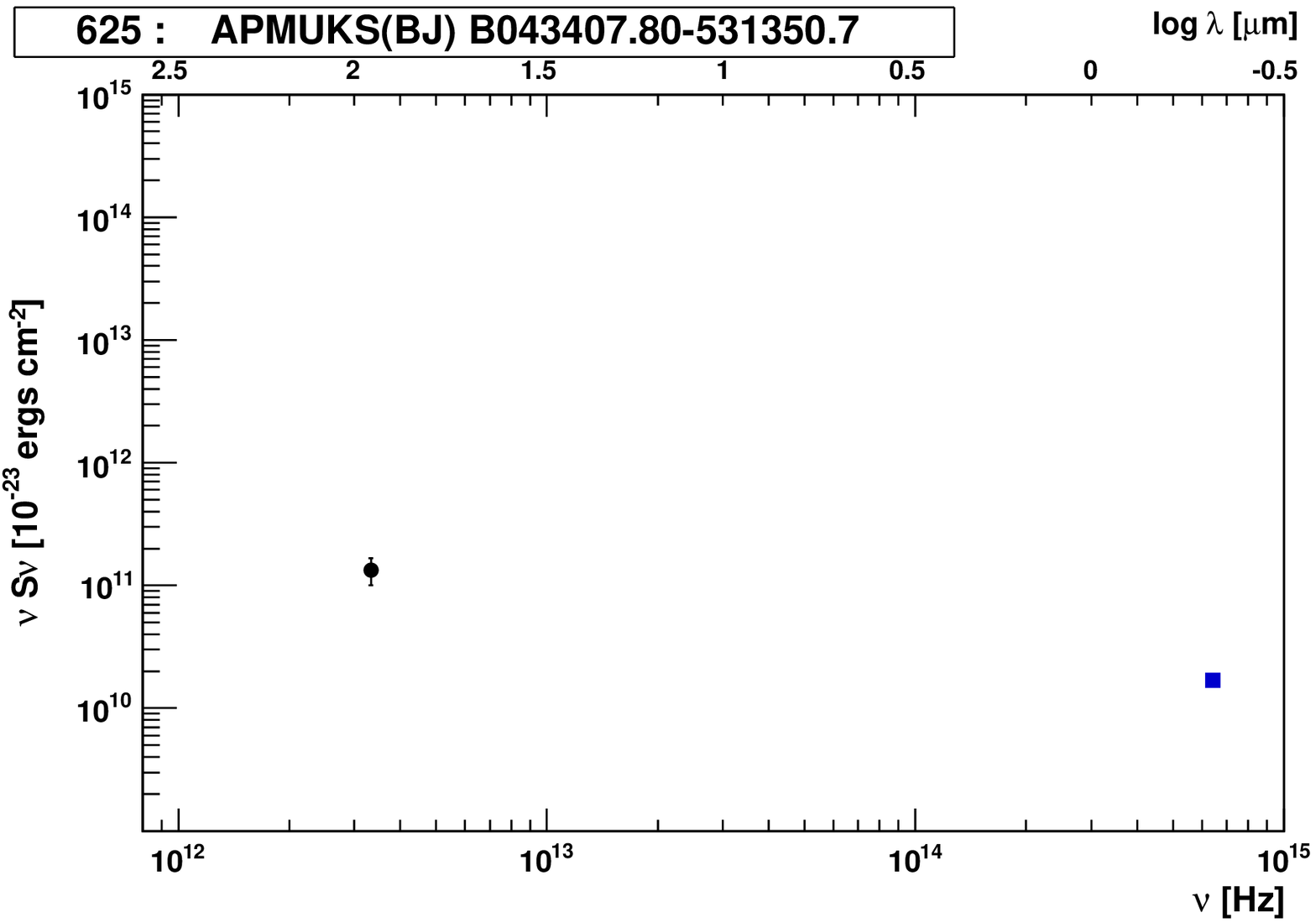}
\includegraphics[width=4cm]{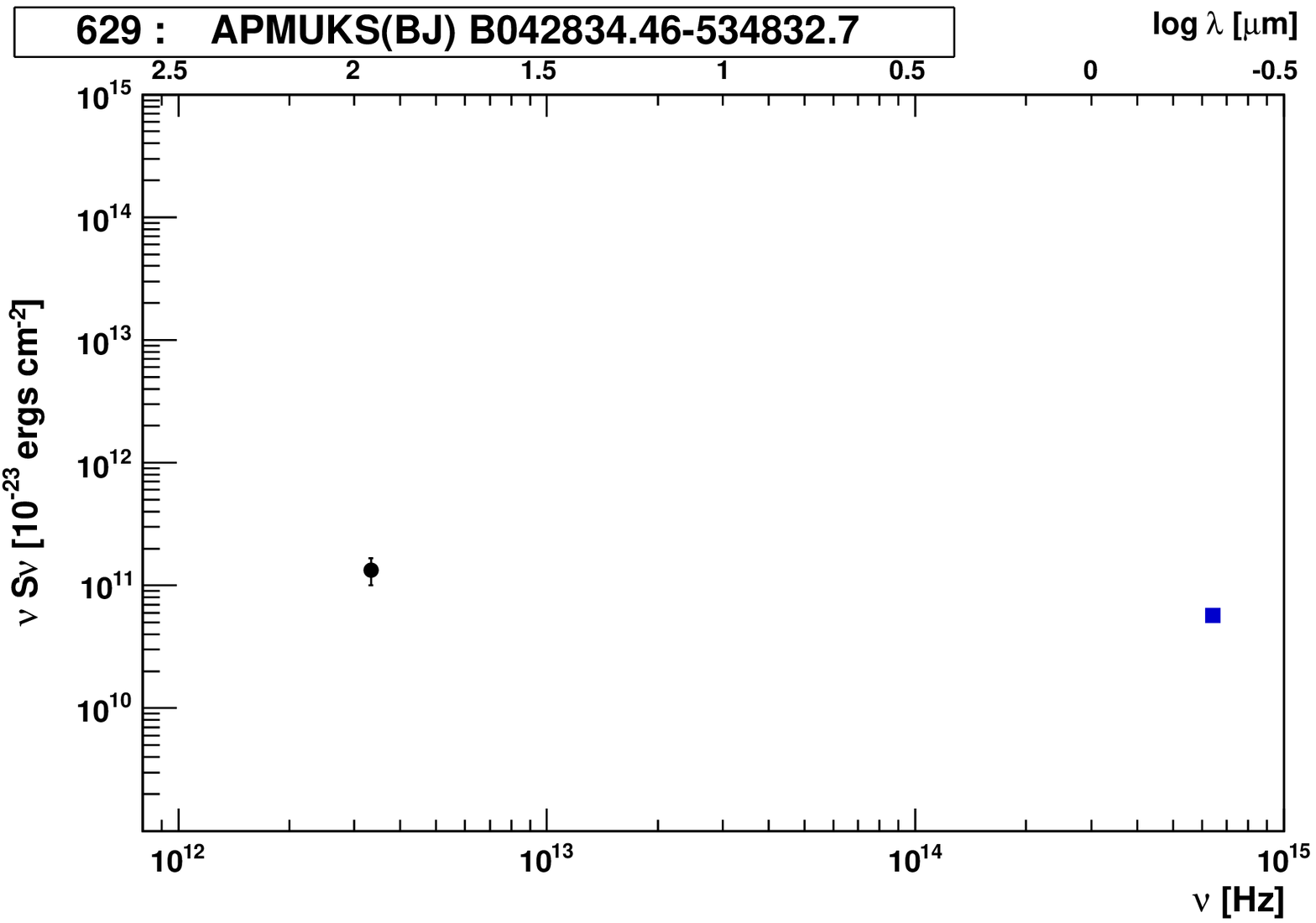}
\includegraphics[width=4cm]{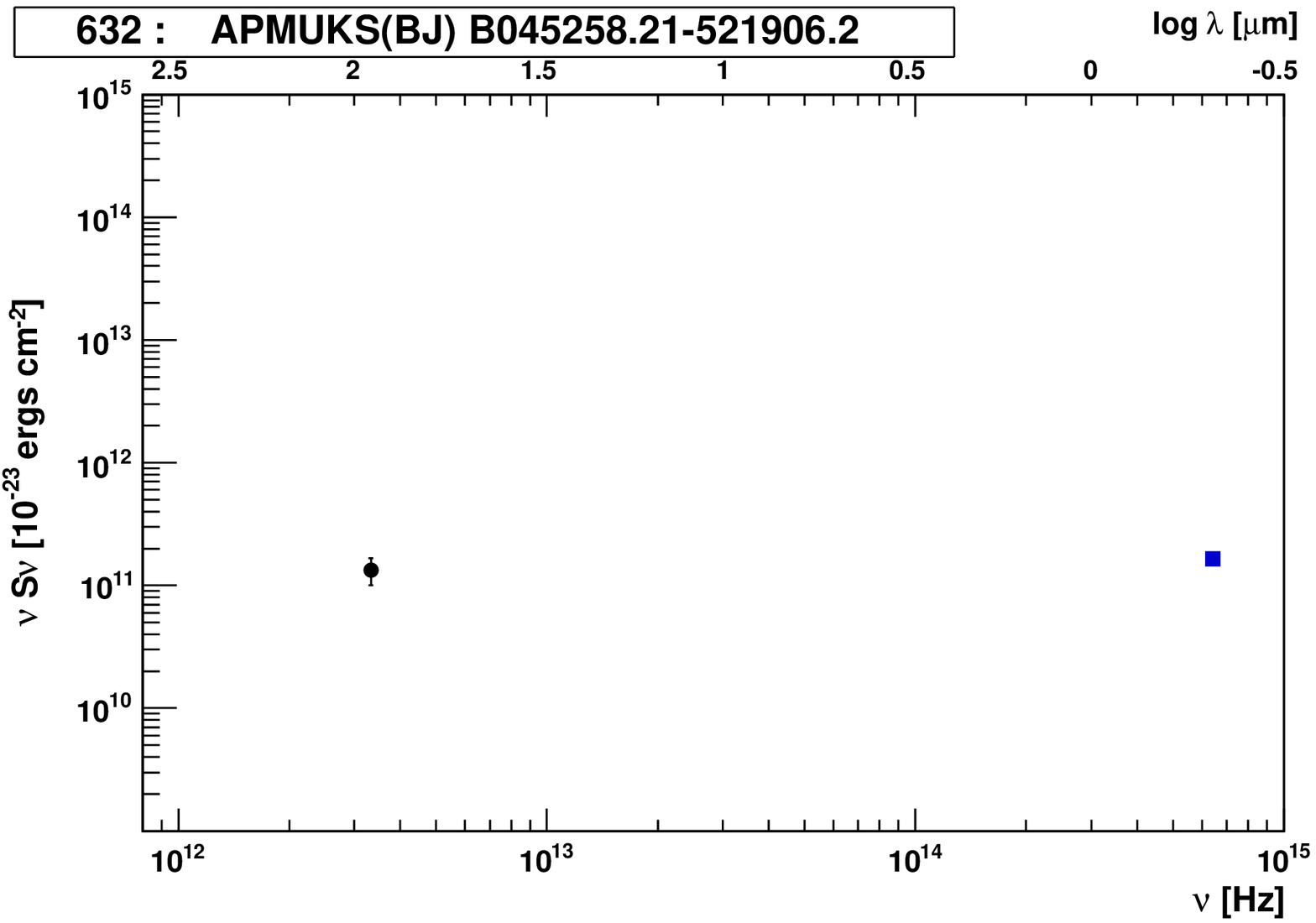}
\includegraphics[width=4cm]{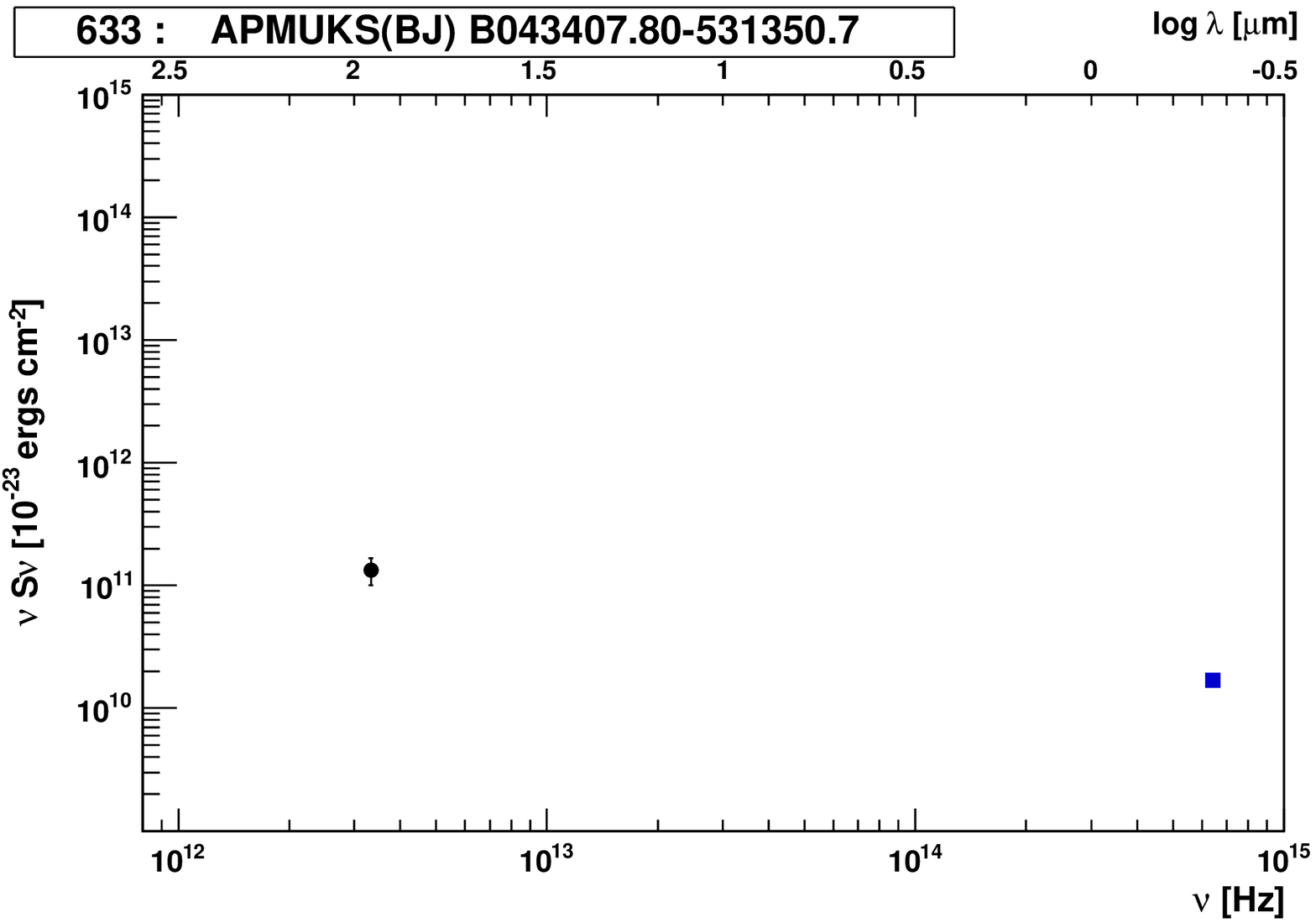}
\includegraphics[width=4cm]{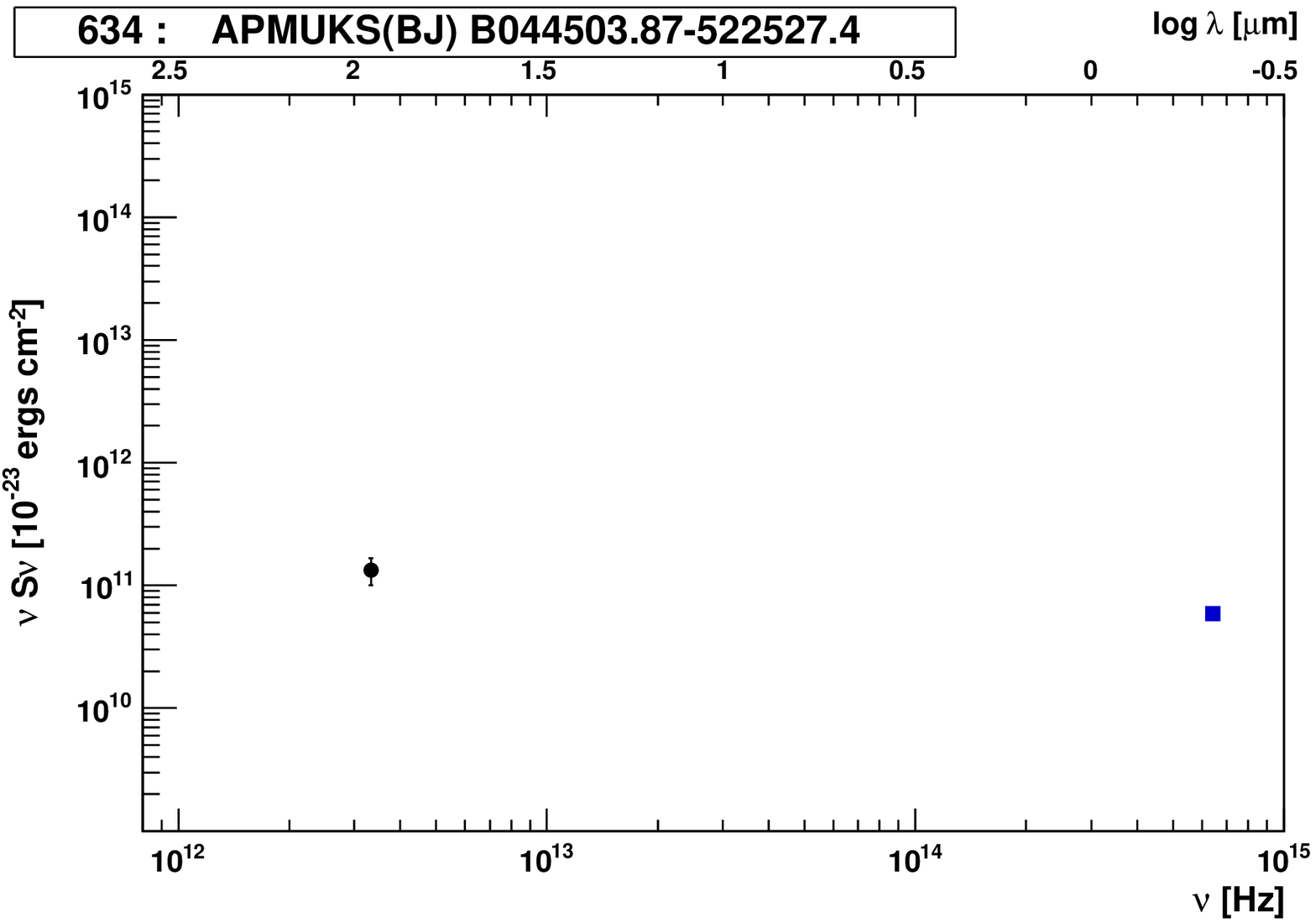}
\includegraphics[width=4cm]{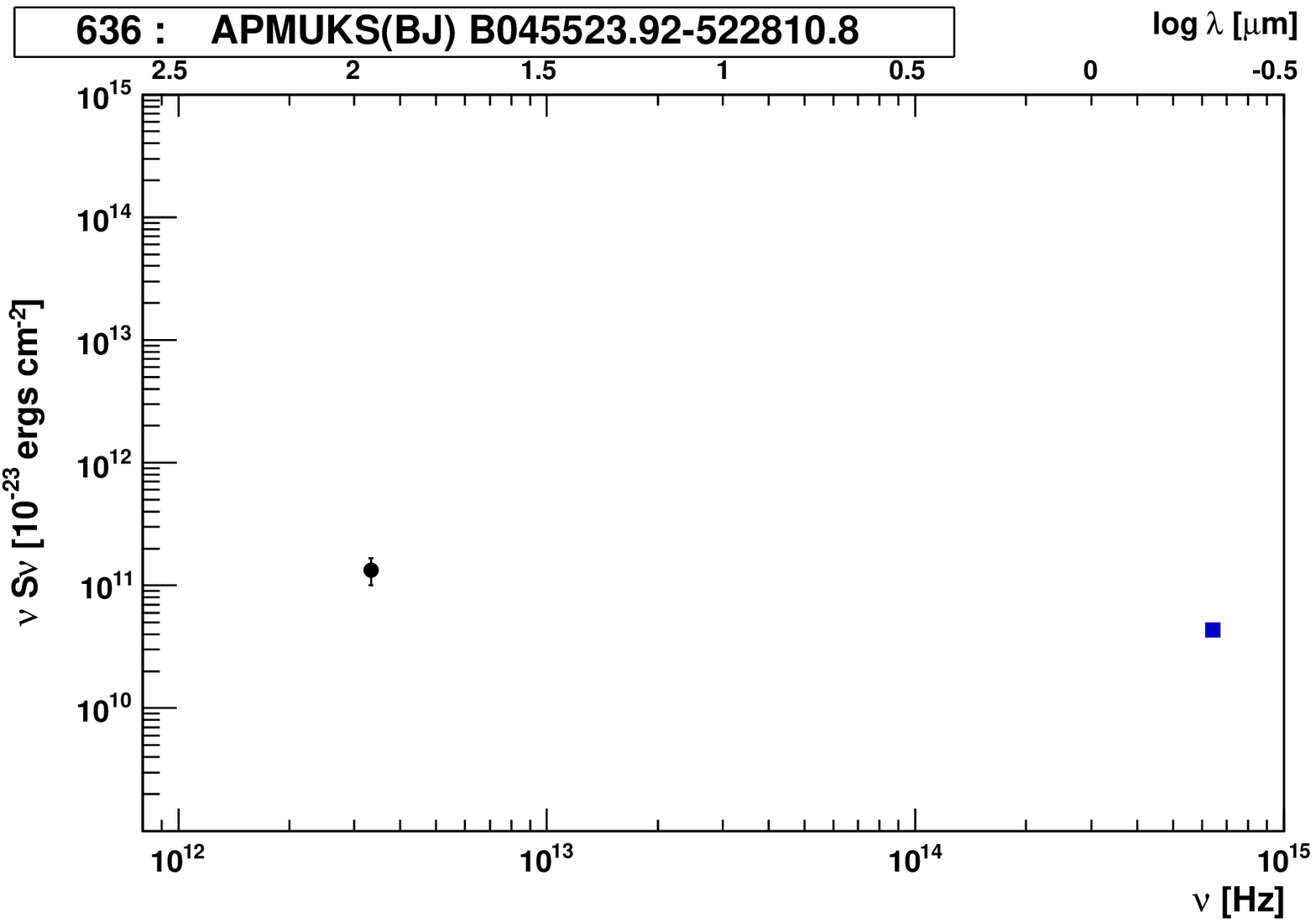}
\includegraphics[width=4cm]{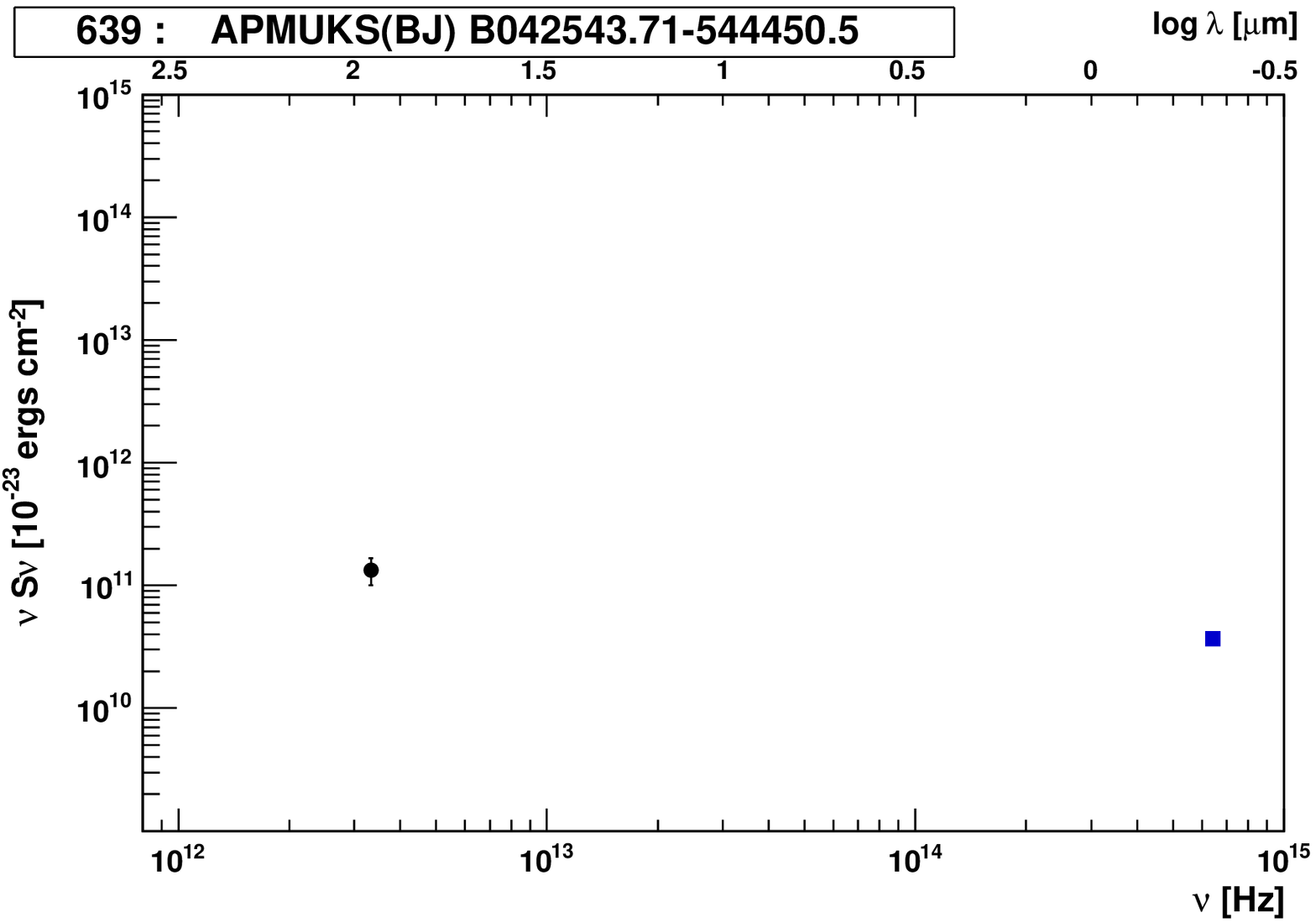}
\includegraphics[width=4cm]{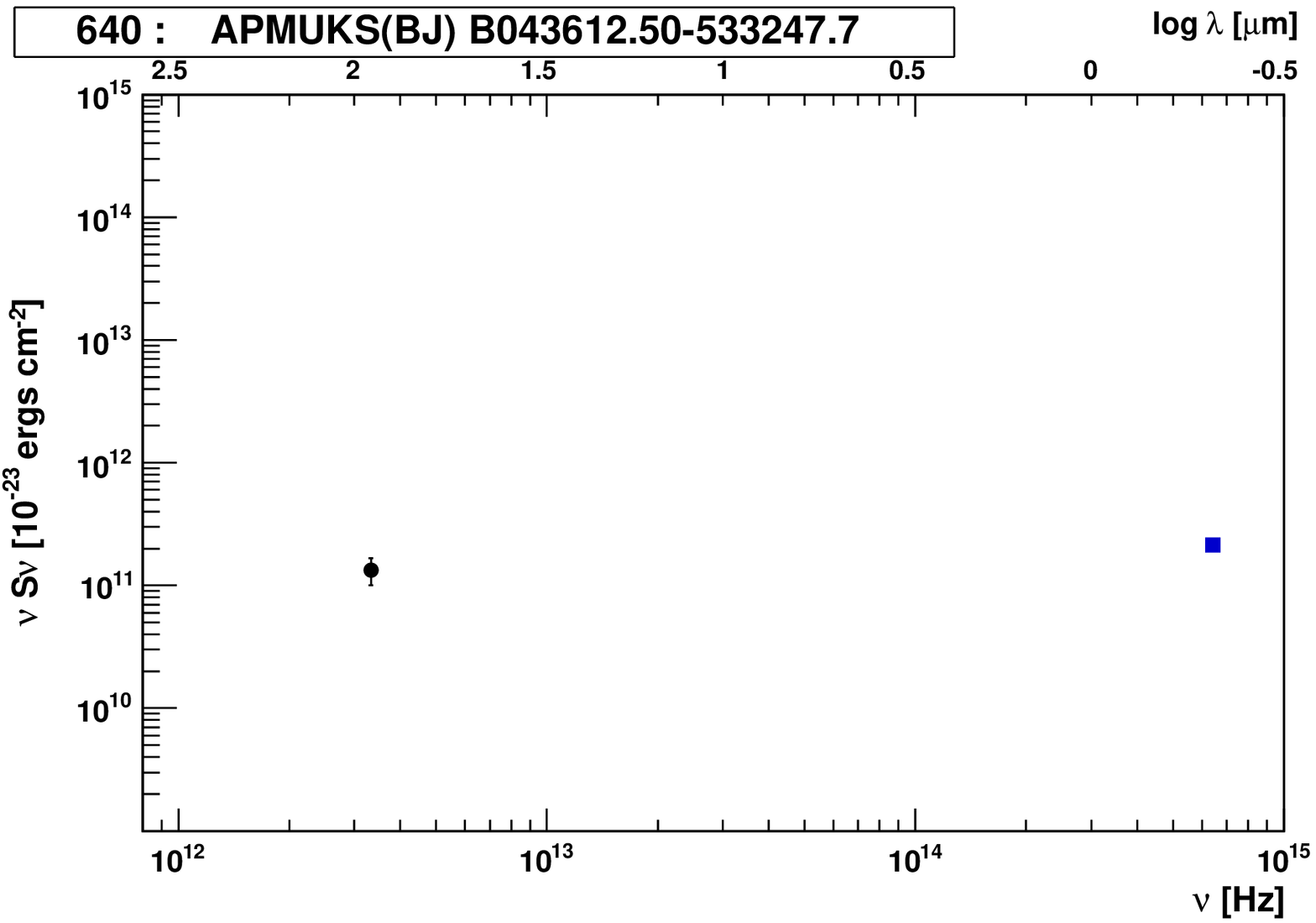}
\includegraphics[width=4cm]{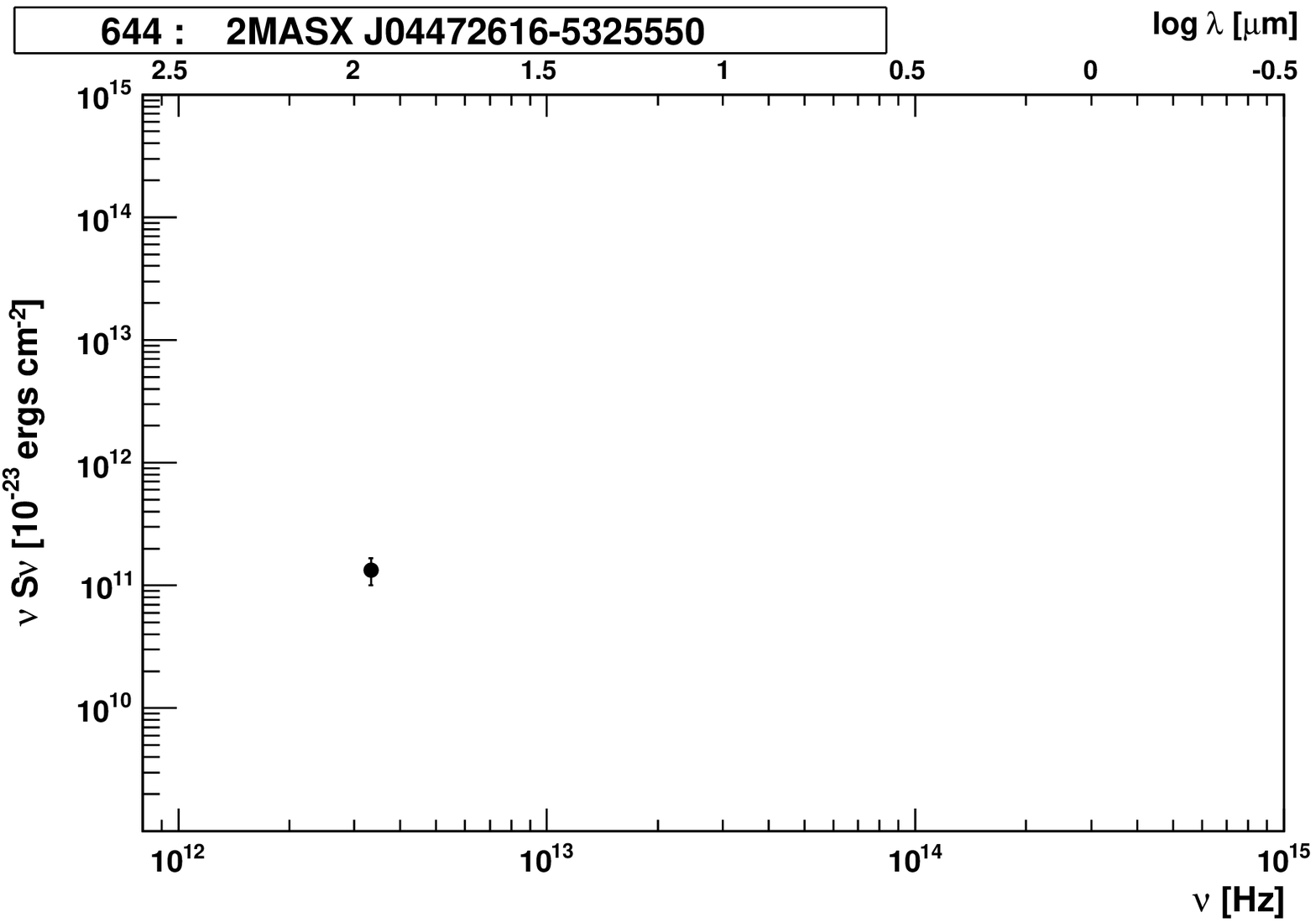}
\includegraphics[width=4cm]{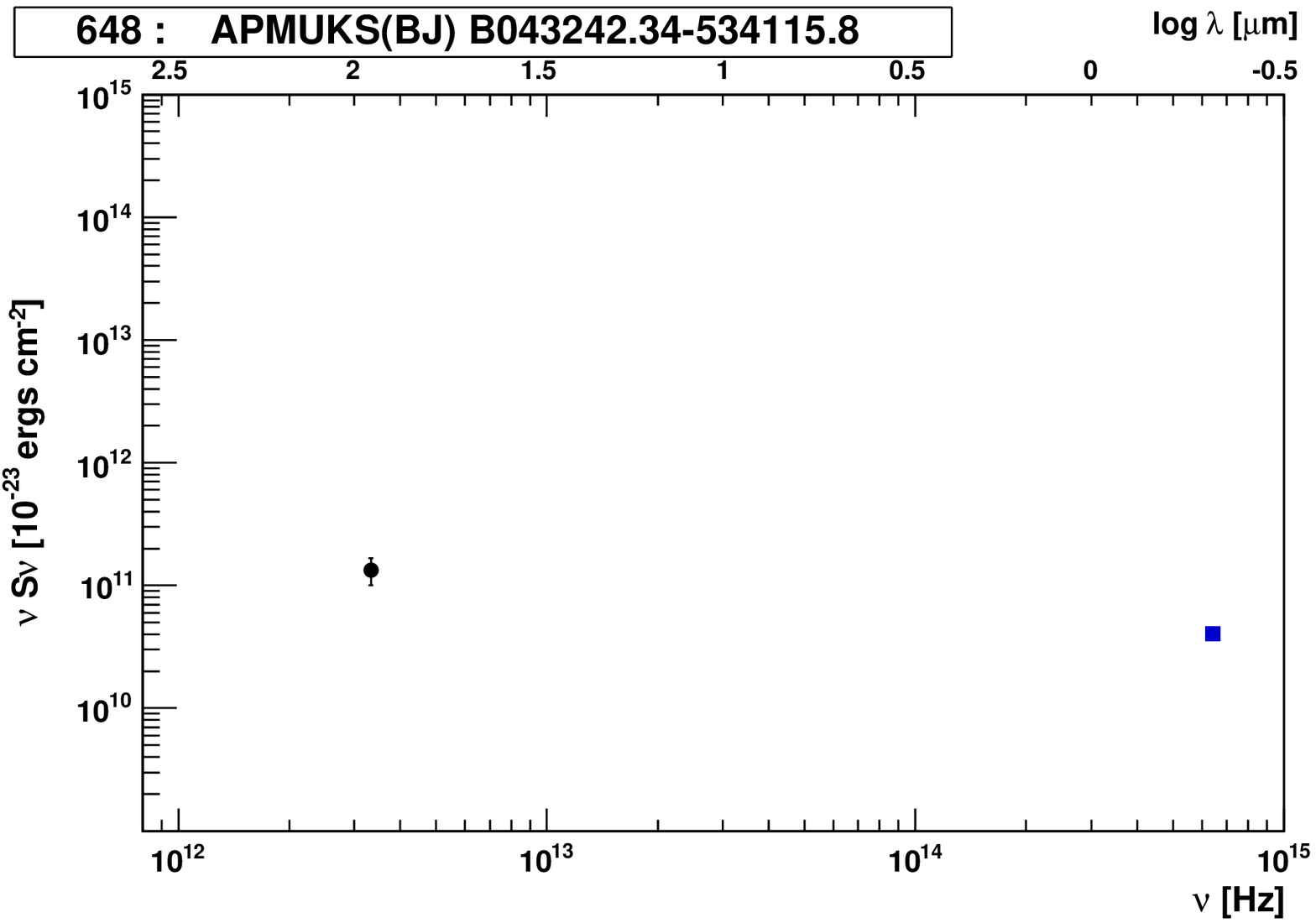}
\includegraphics[width=4cm]{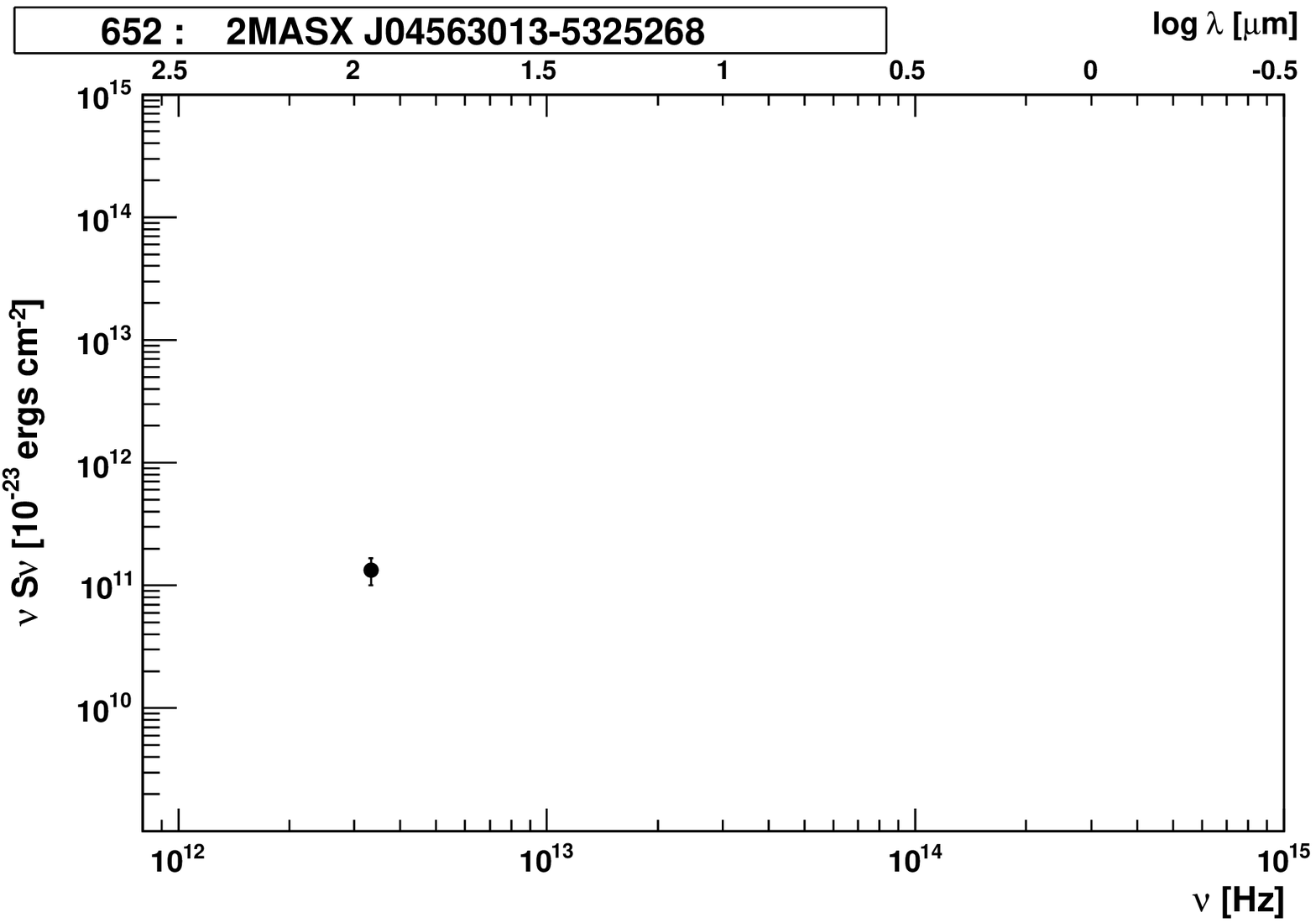}
\includegraphics[width=4cm]{points/kmalek_606.eps}
\includegraphics[width=4cm]{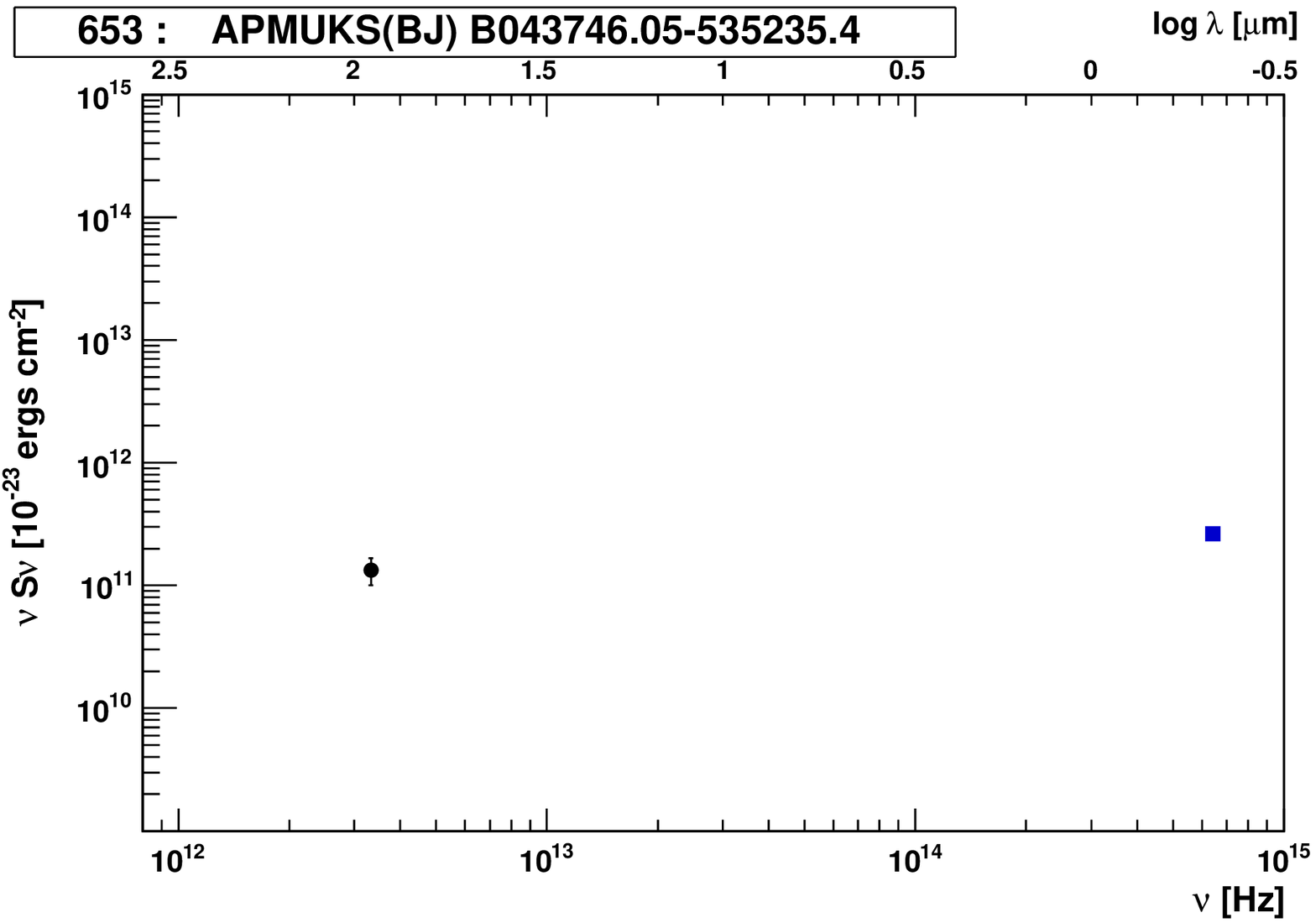}
\label{points11}
\caption {SEDs for the next 18 ADF-S identified sources, with symbols as in Figure~\ref{points1}.}
\end{figure*}
}

\clearpage

\onlfig{12}{
\begin{figure*}[t]
\centering

\includegraphics[width=4cm]{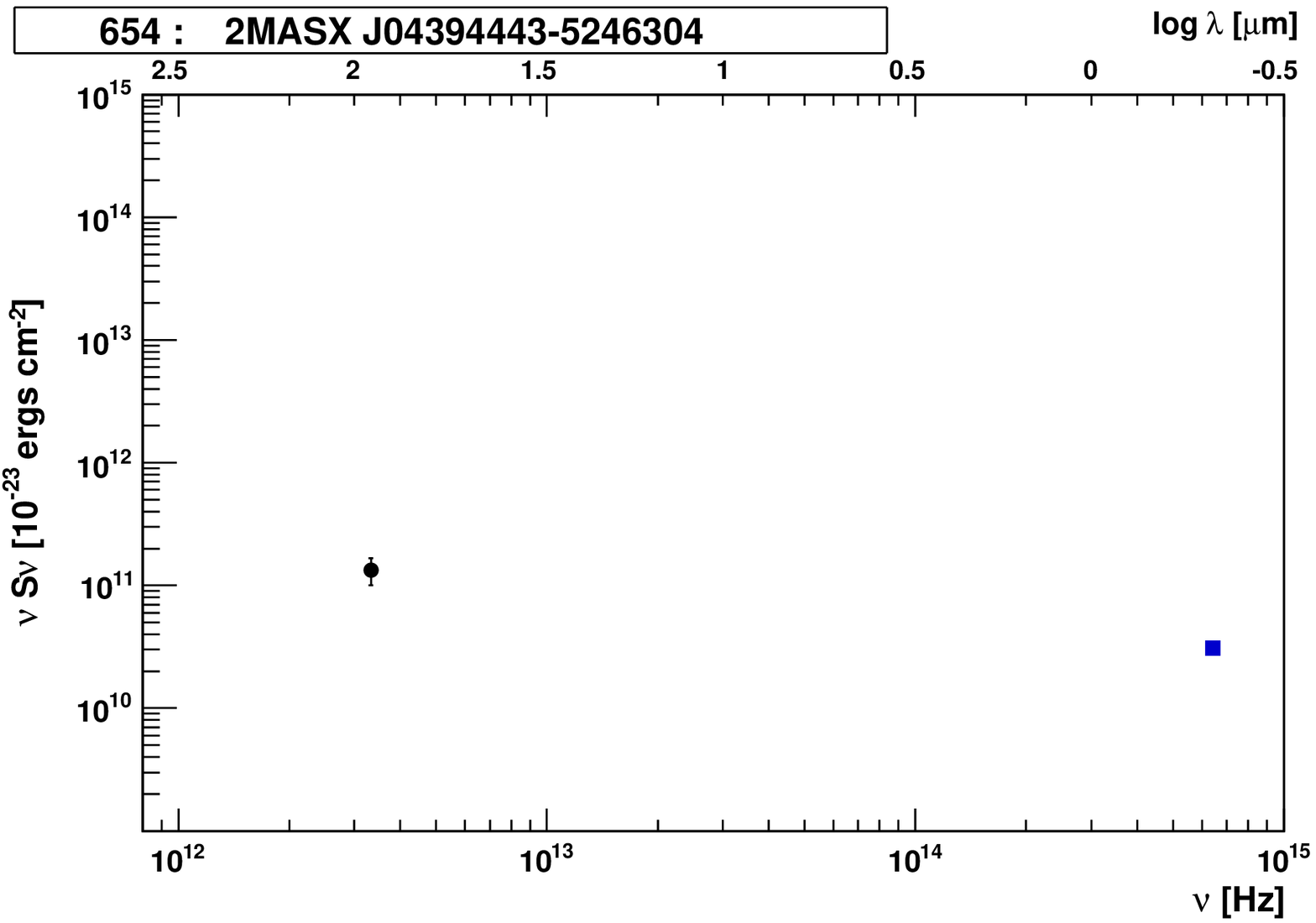}
\includegraphics[width=4cm]{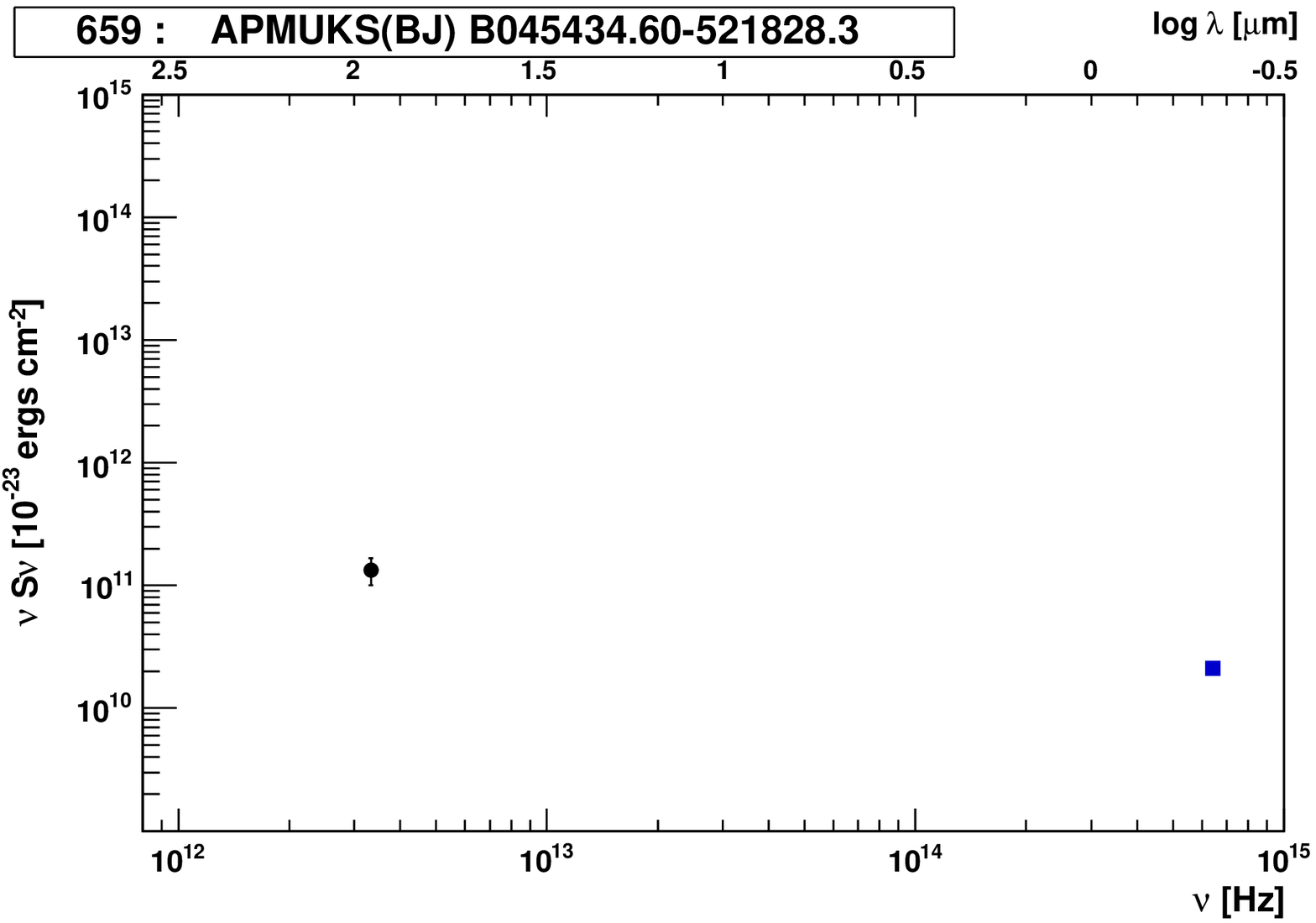}
\includegraphics[width=4cm]{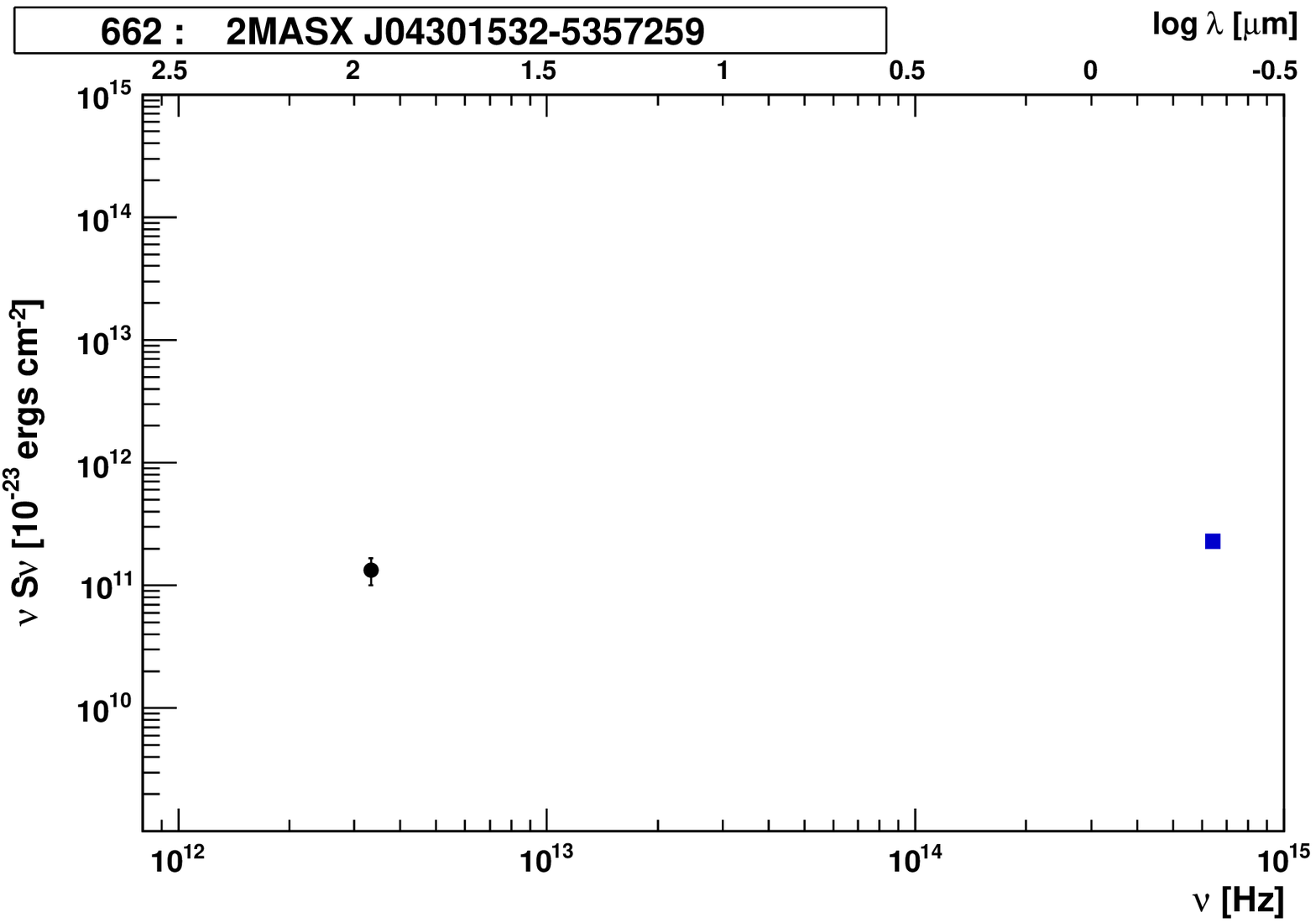}
\includegraphics[width=4cm]{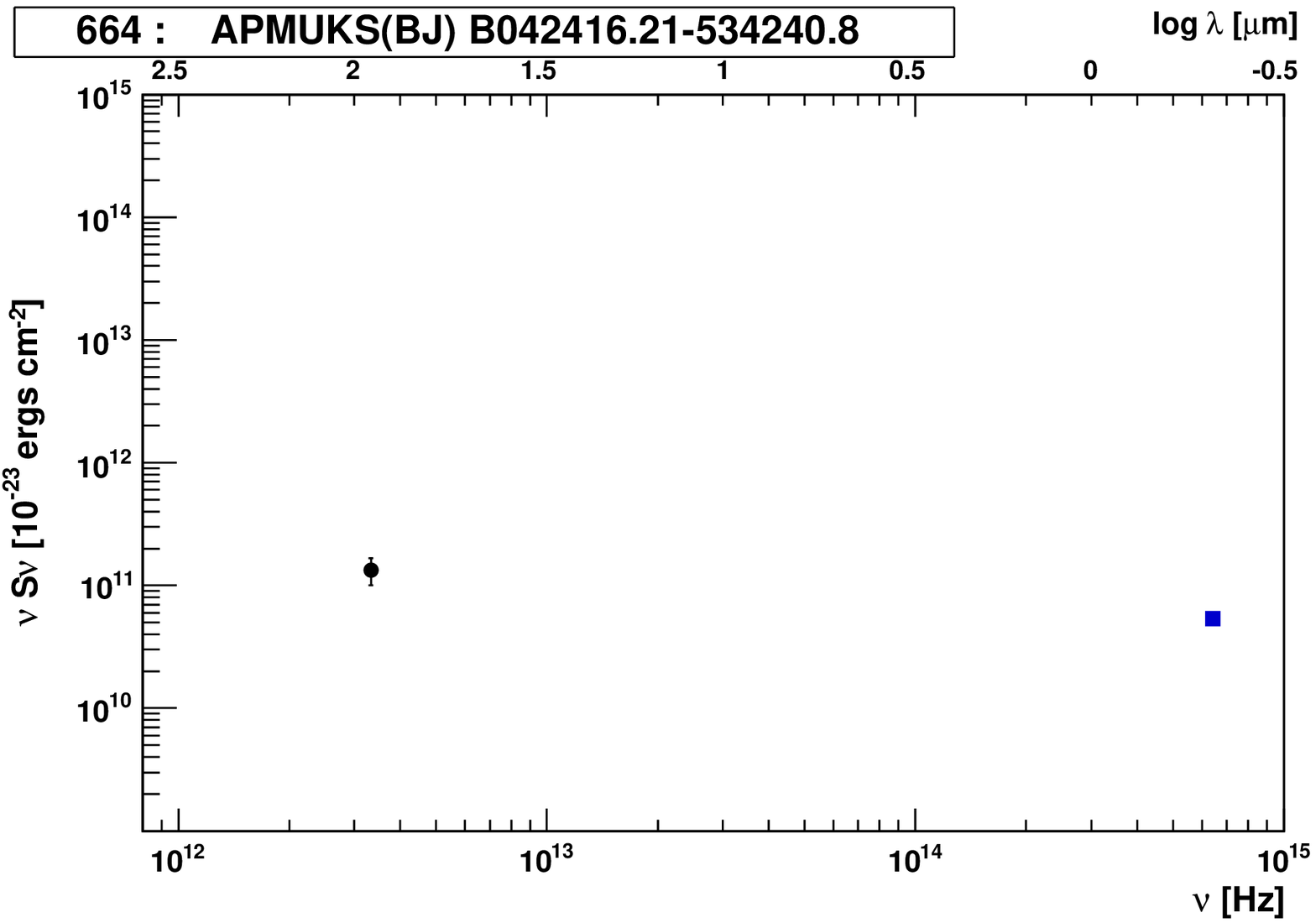}
\includegraphics[width=4cm]{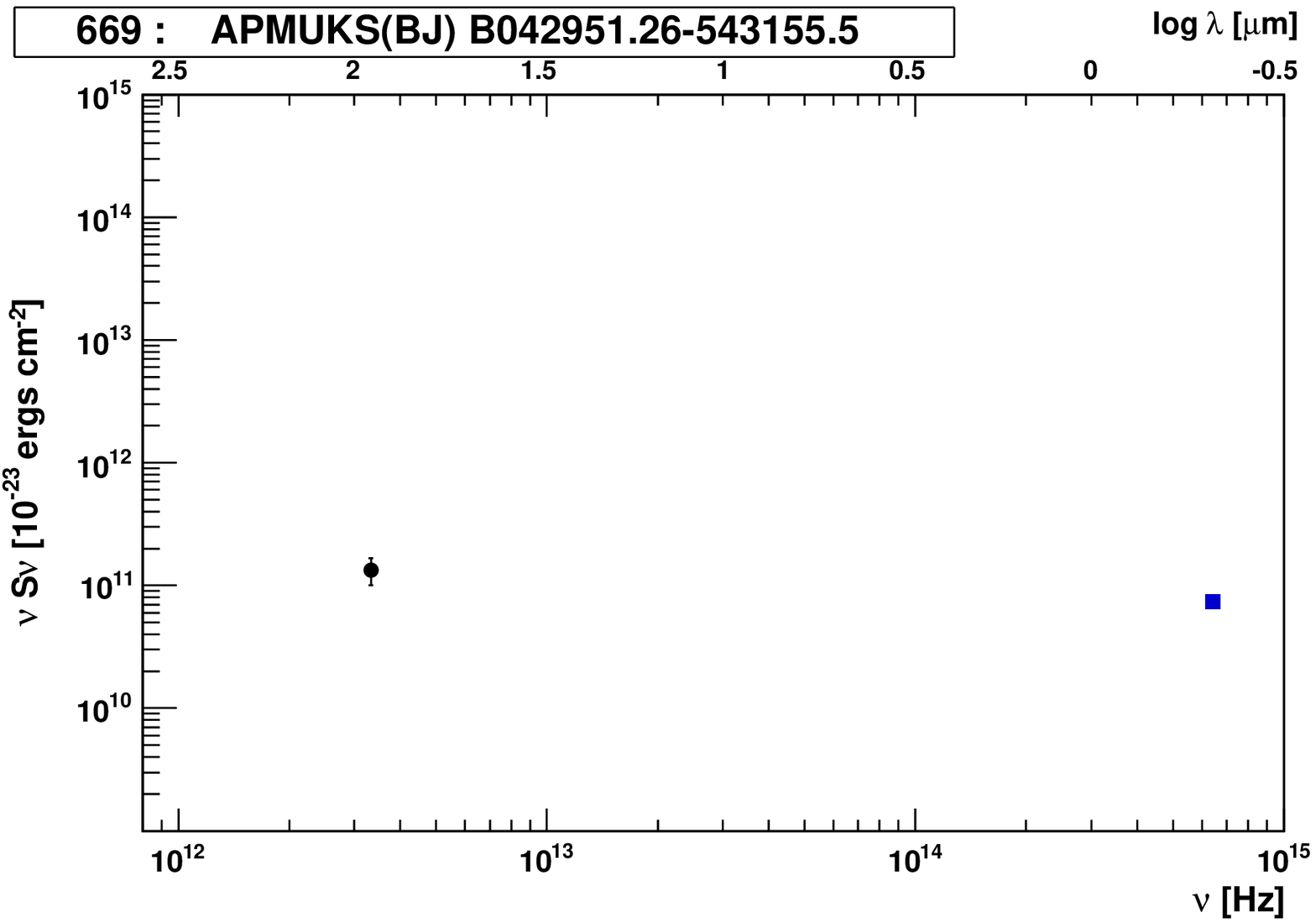}
\includegraphics[width=4cm]{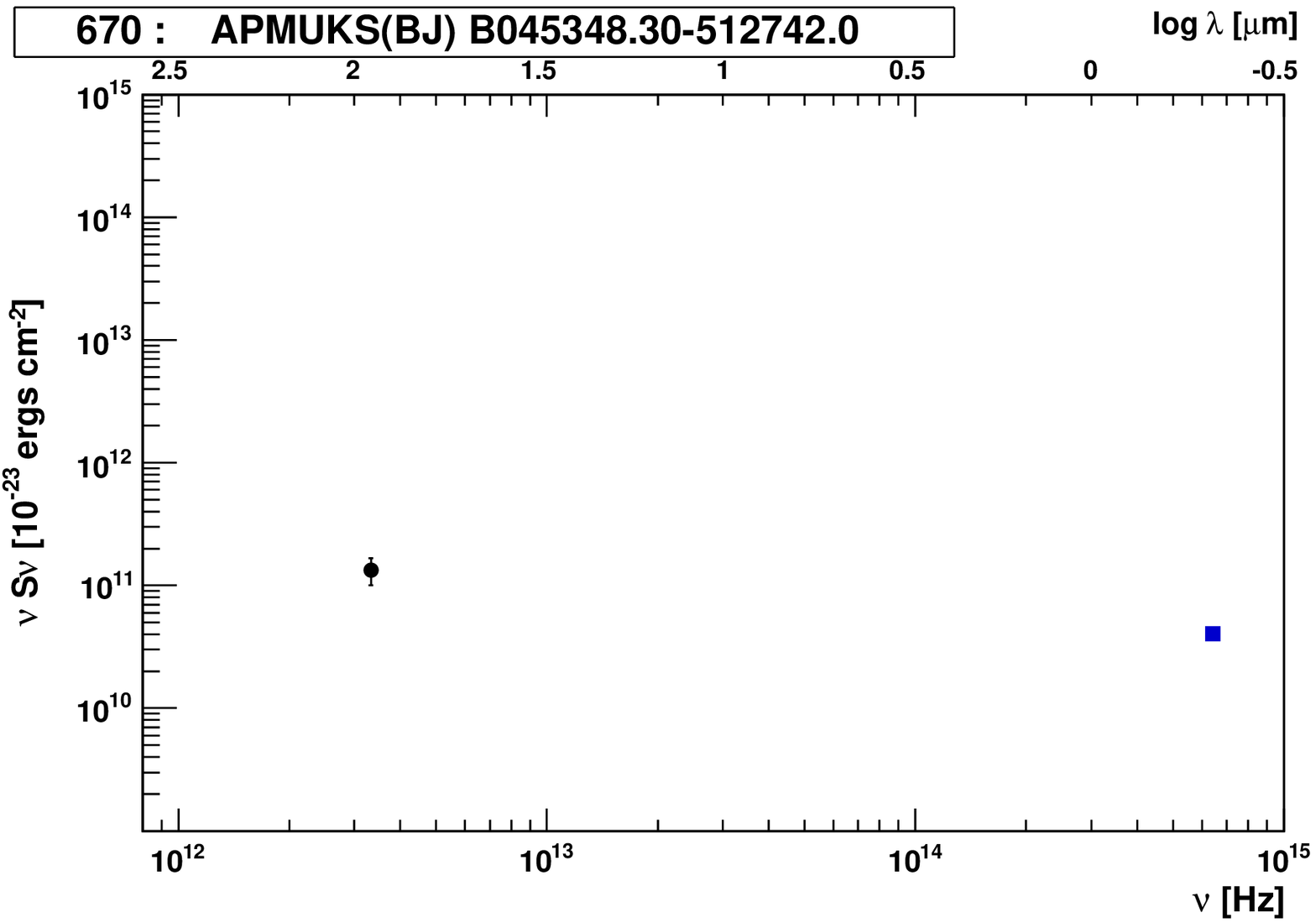}
\includegraphics[width=4cm]{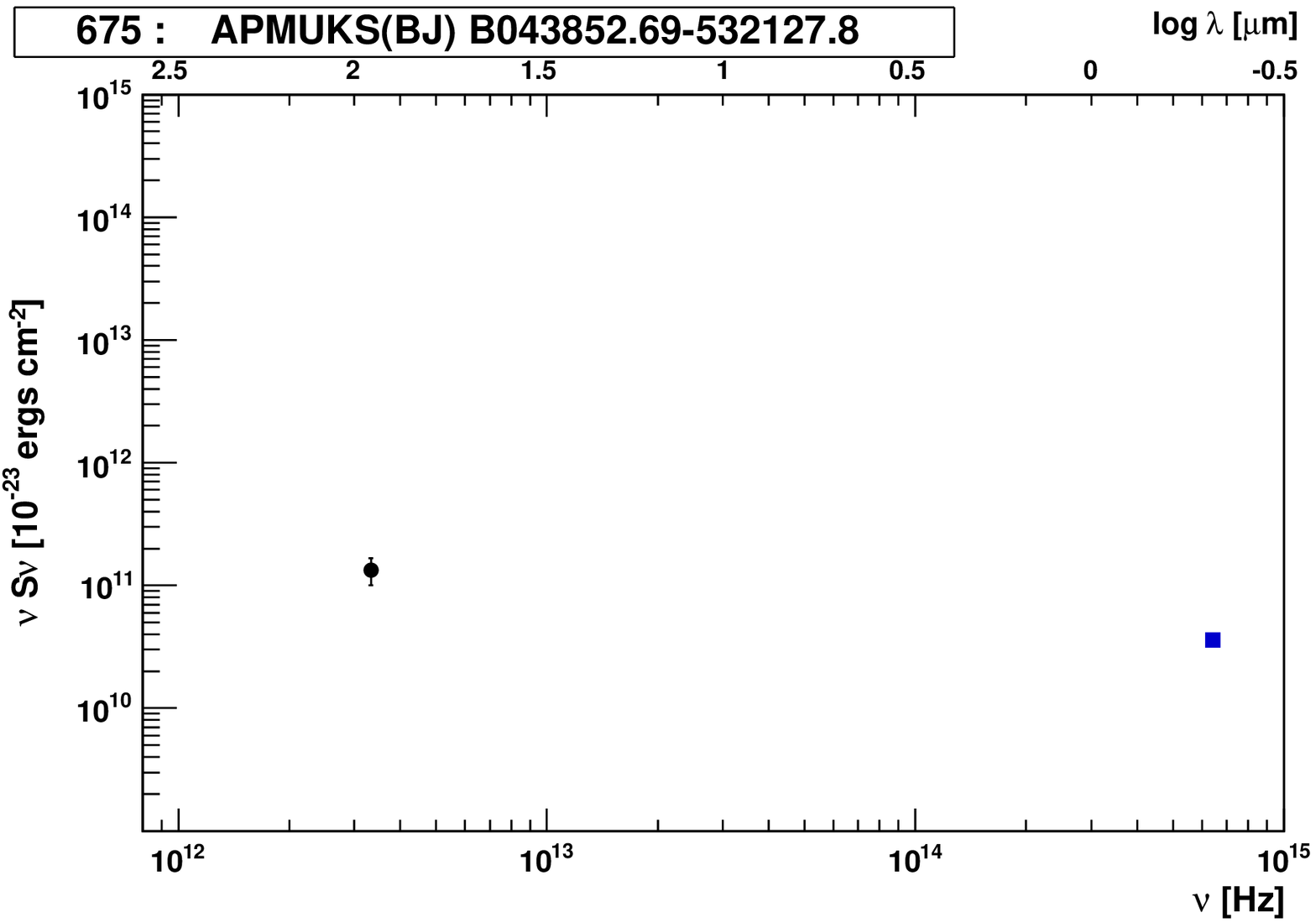}
\includegraphics[width=4cm]{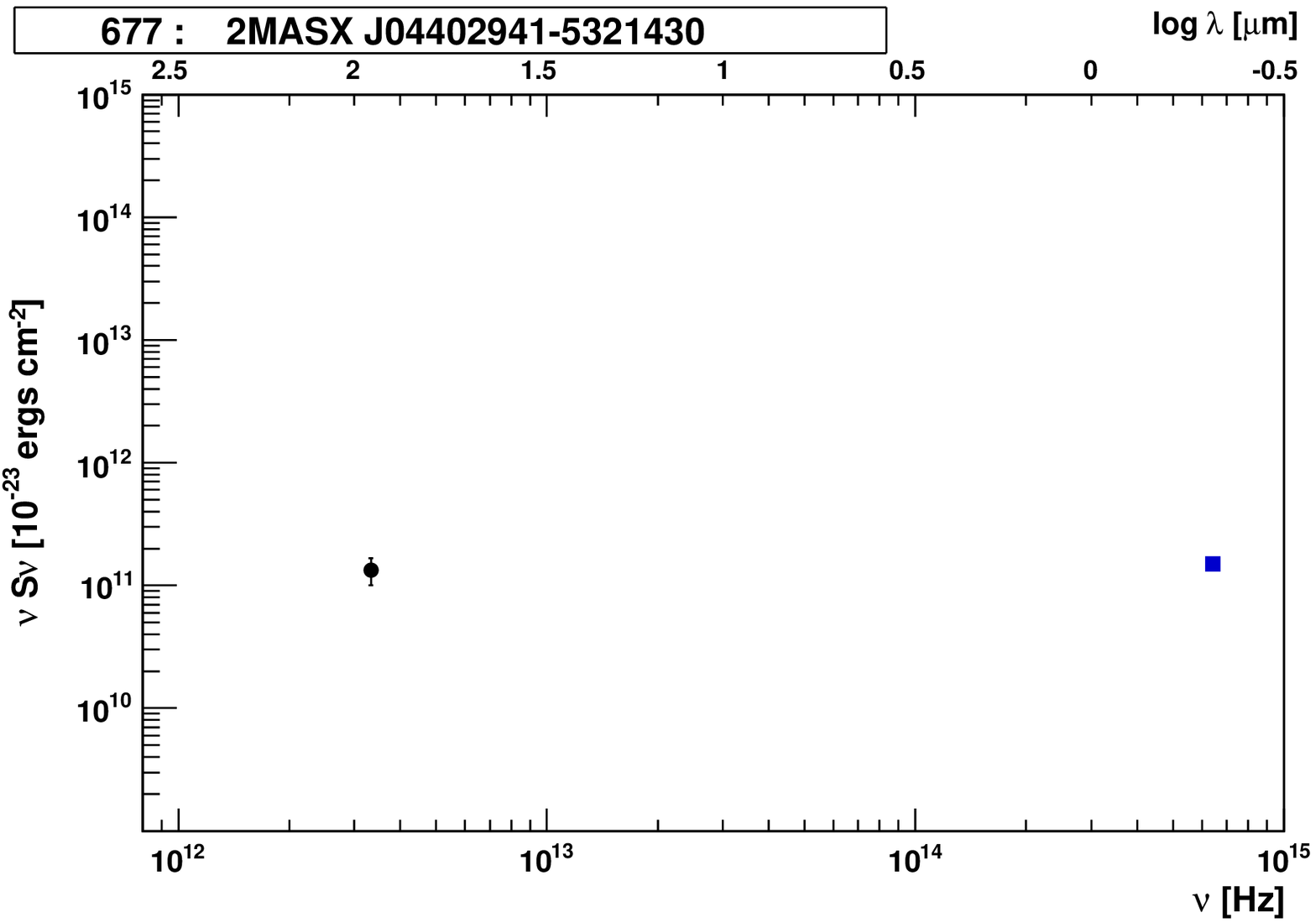}

\includegraphics[width=4cm]{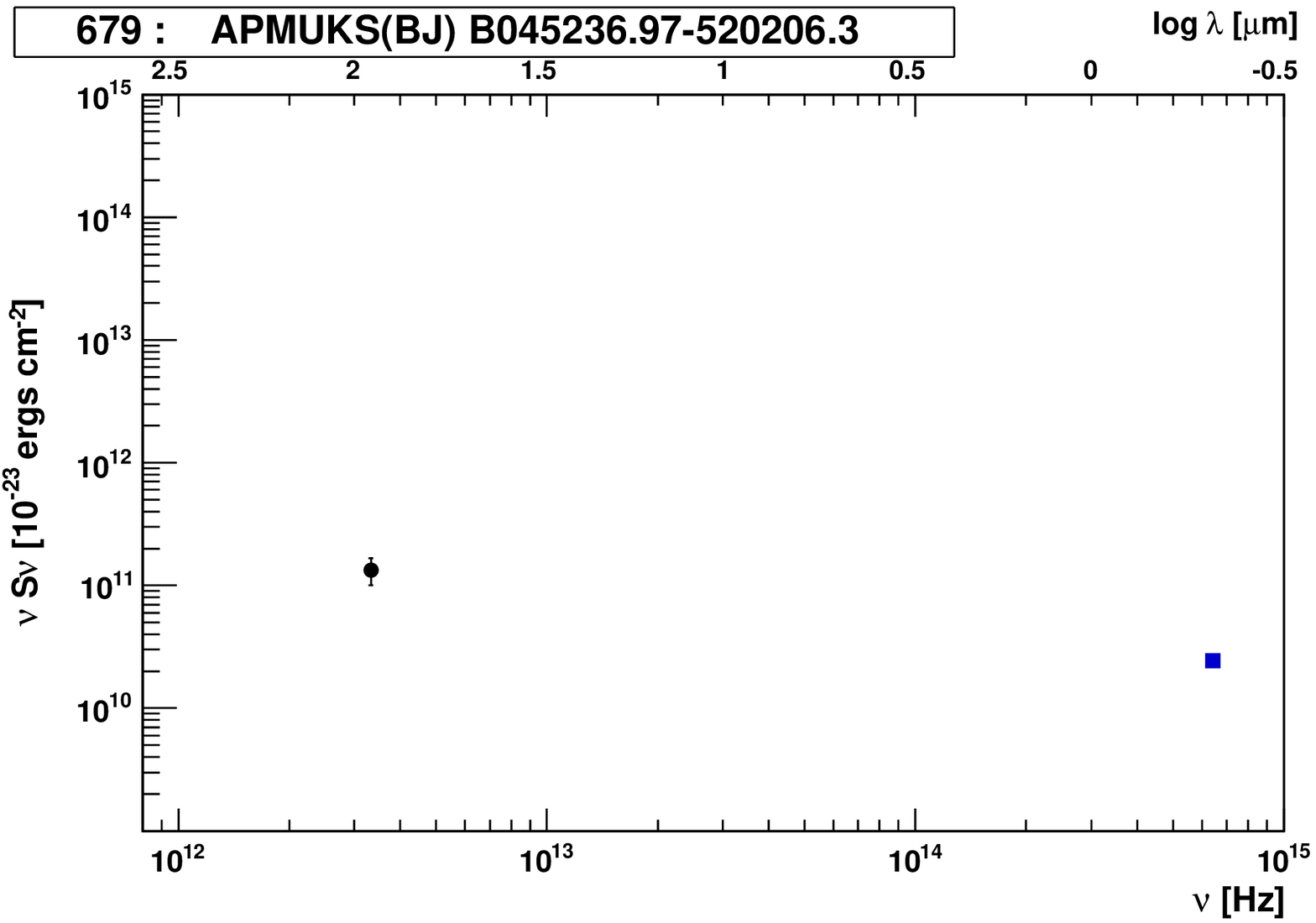}
\includegraphics[width=4cm]{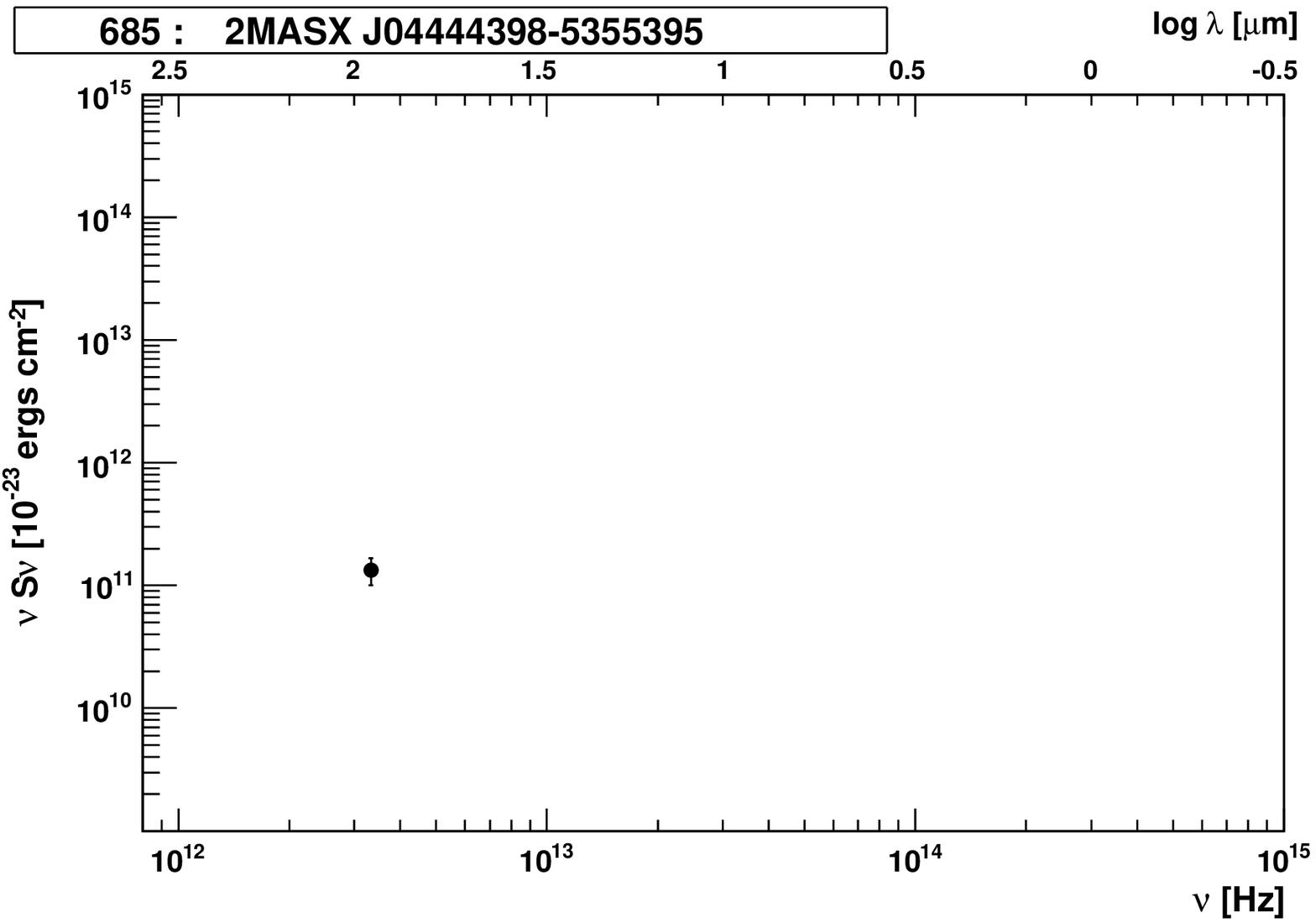}
\includegraphics[width=4cm]{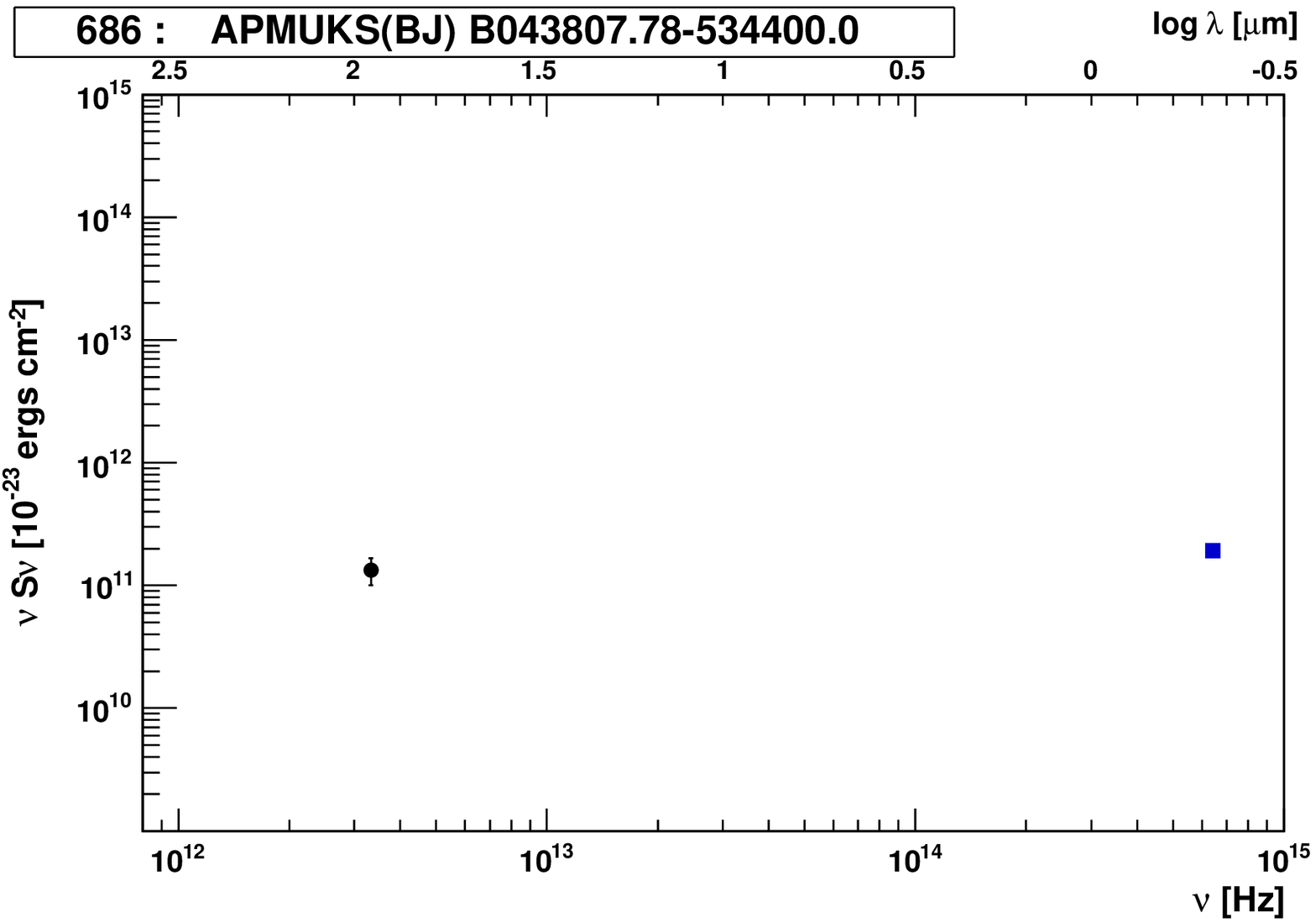}
\includegraphics[width=4cm]{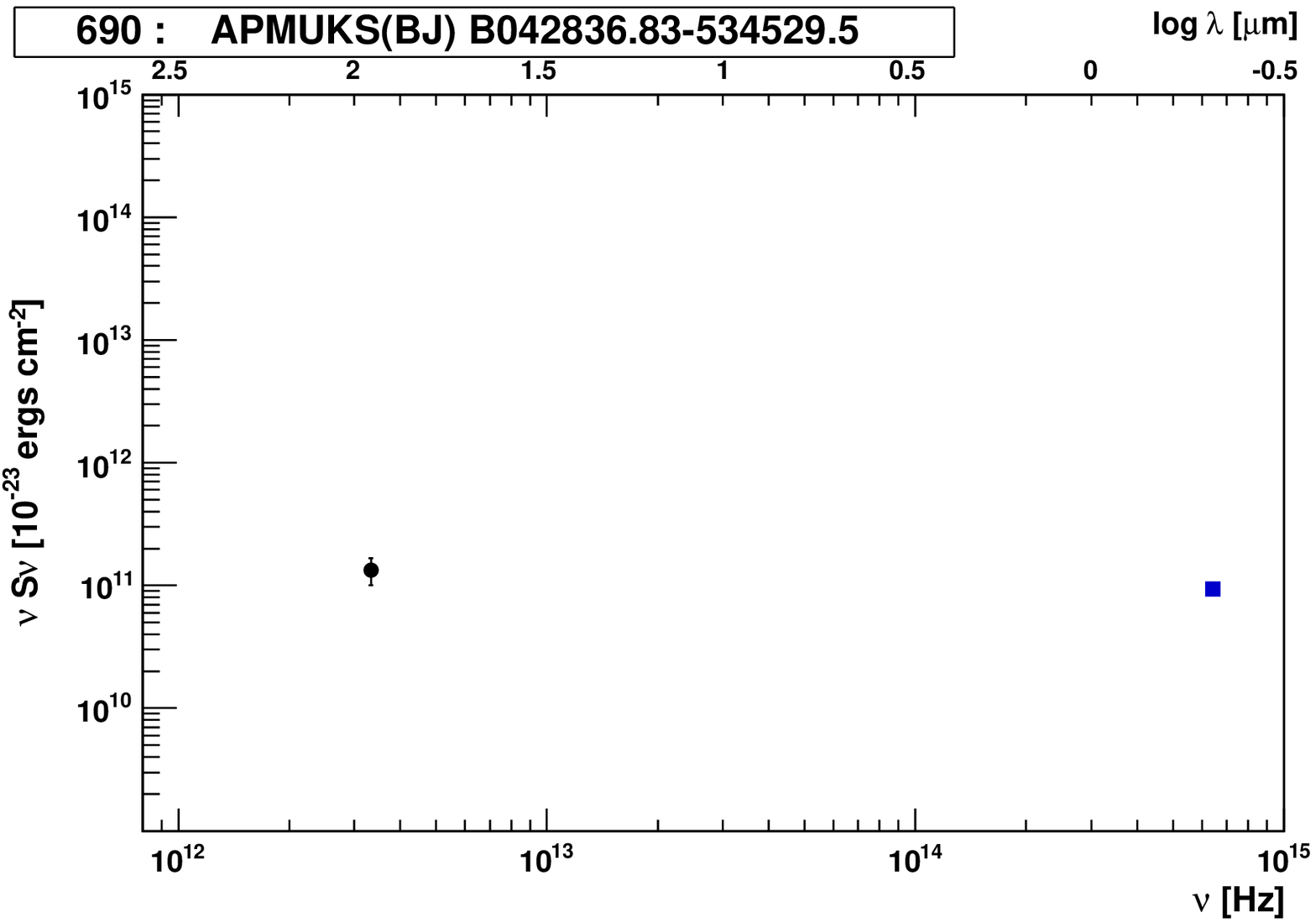}
\includegraphics[width=4cm]{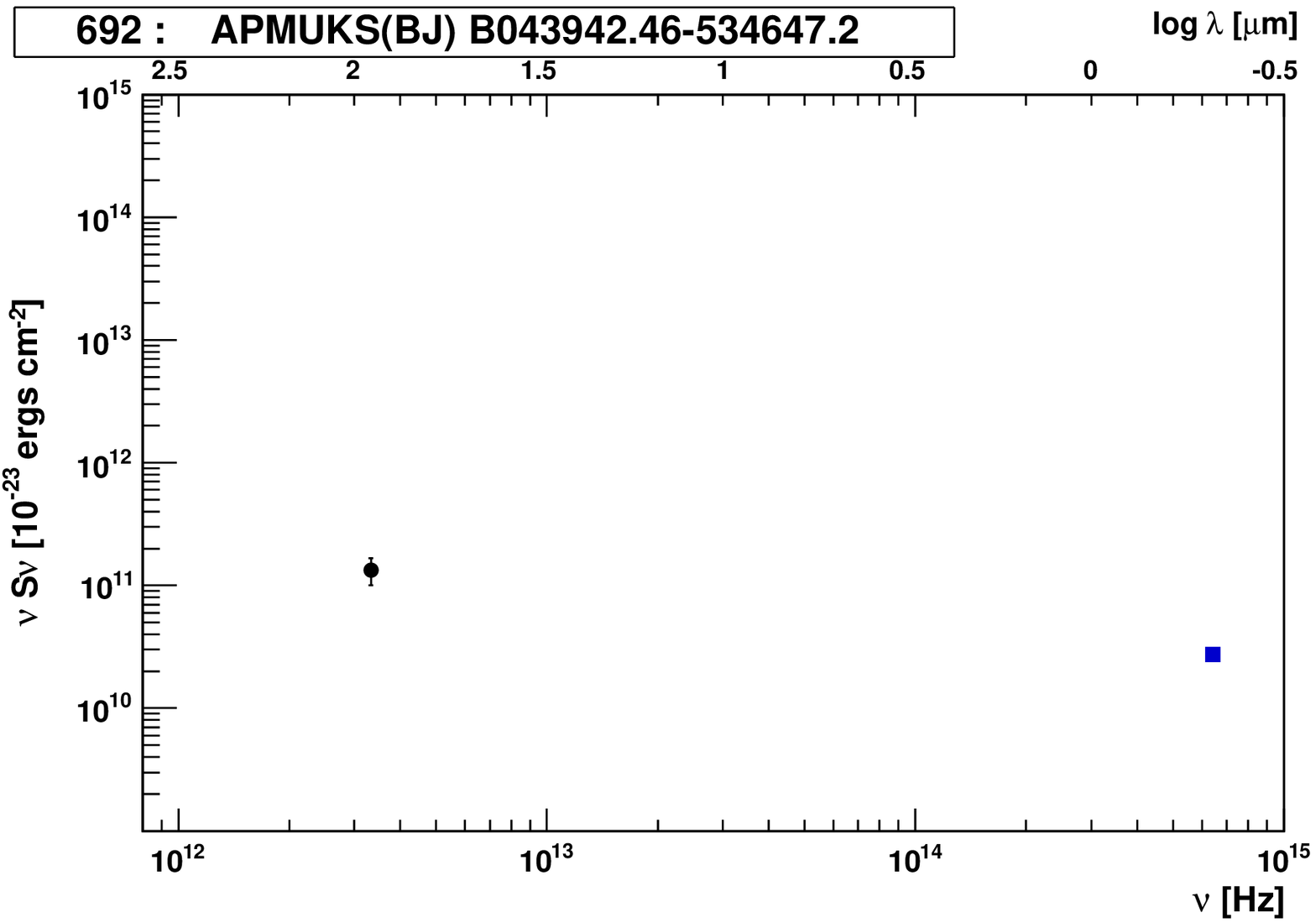}
\includegraphics[width=4cm]{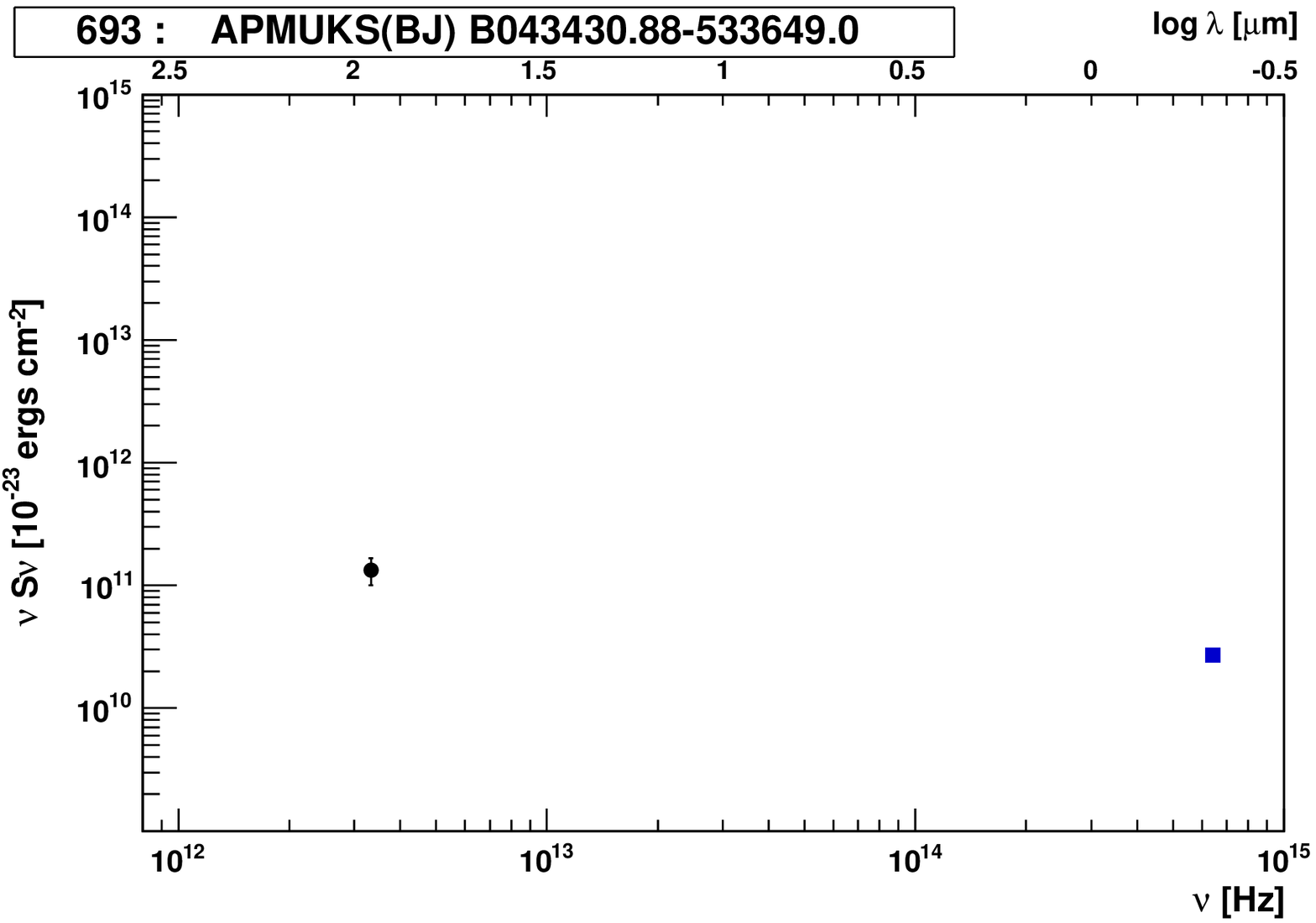}
\includegraphics[width=4cm]{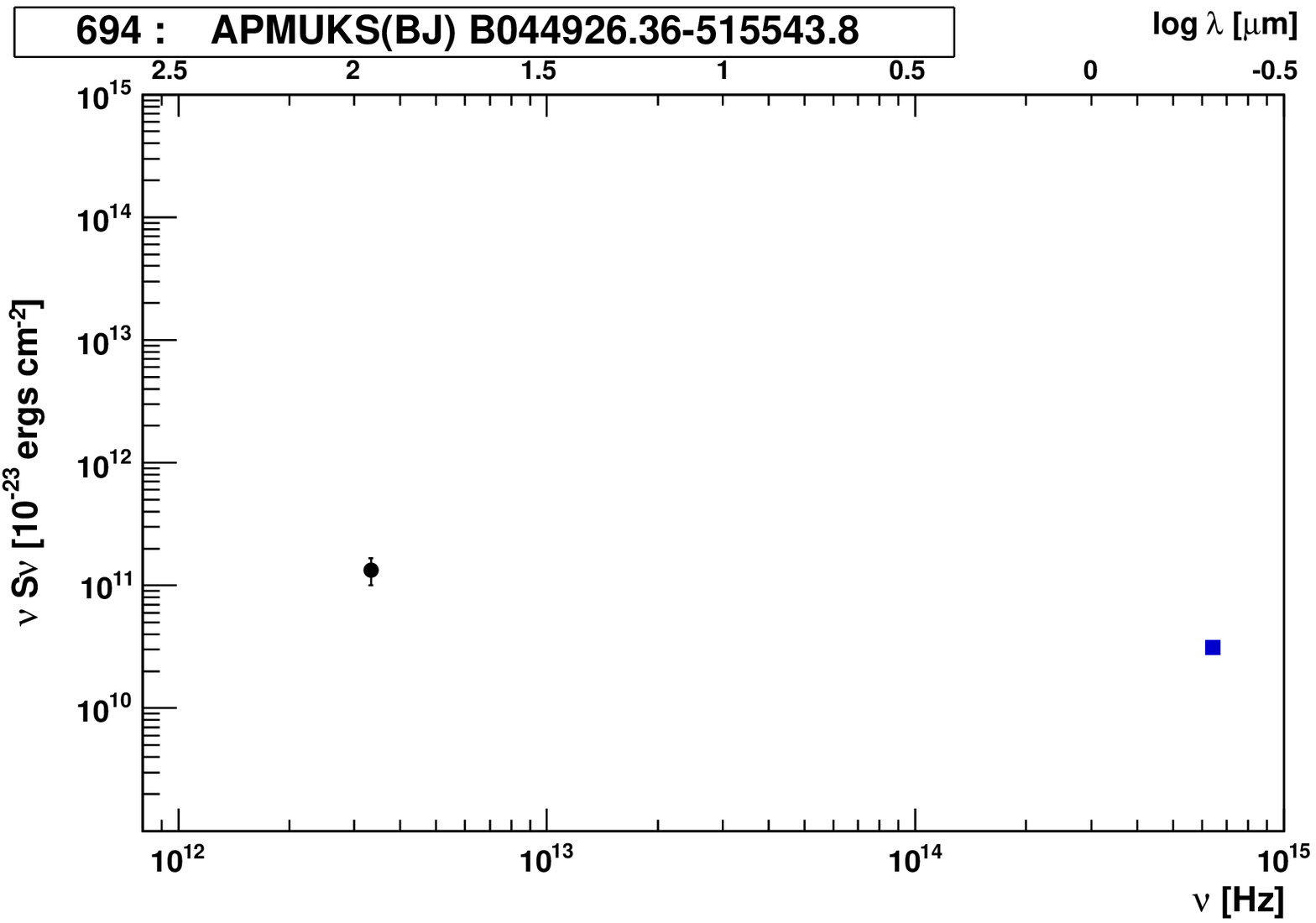}
\includegraphics[width=4cm]{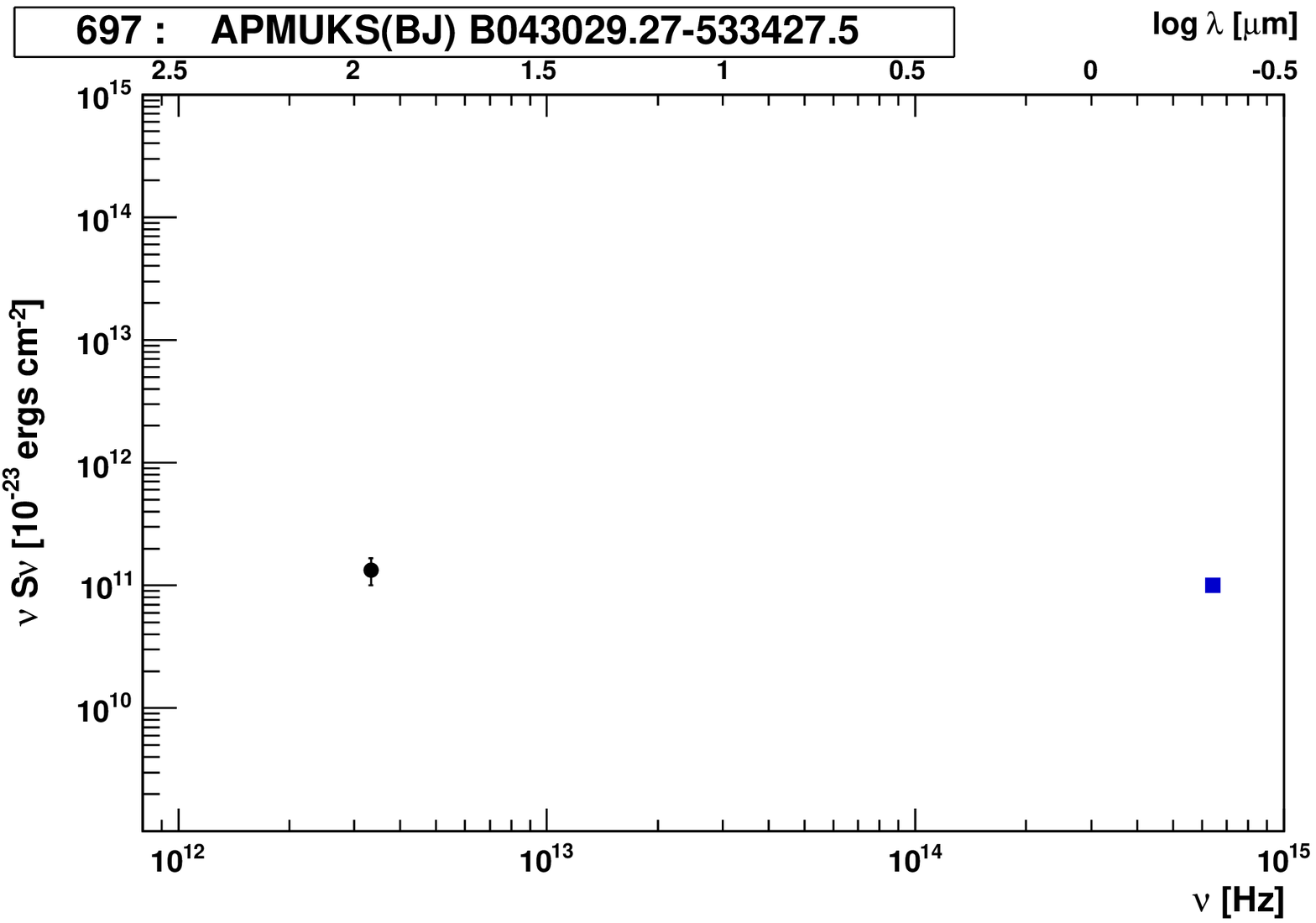}
\includegraphics[width=4cm]{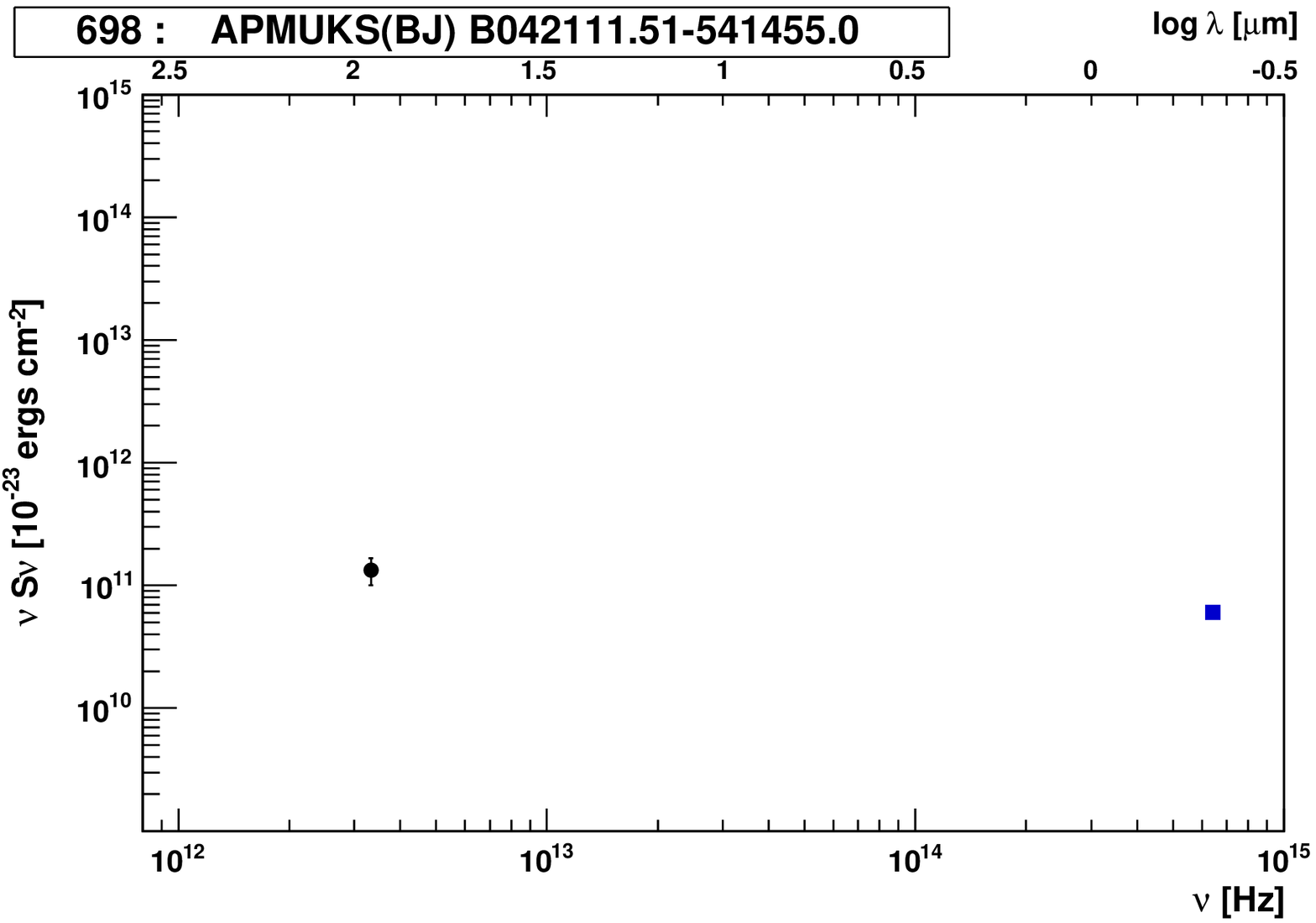}
\includegraphics[width=4cm]{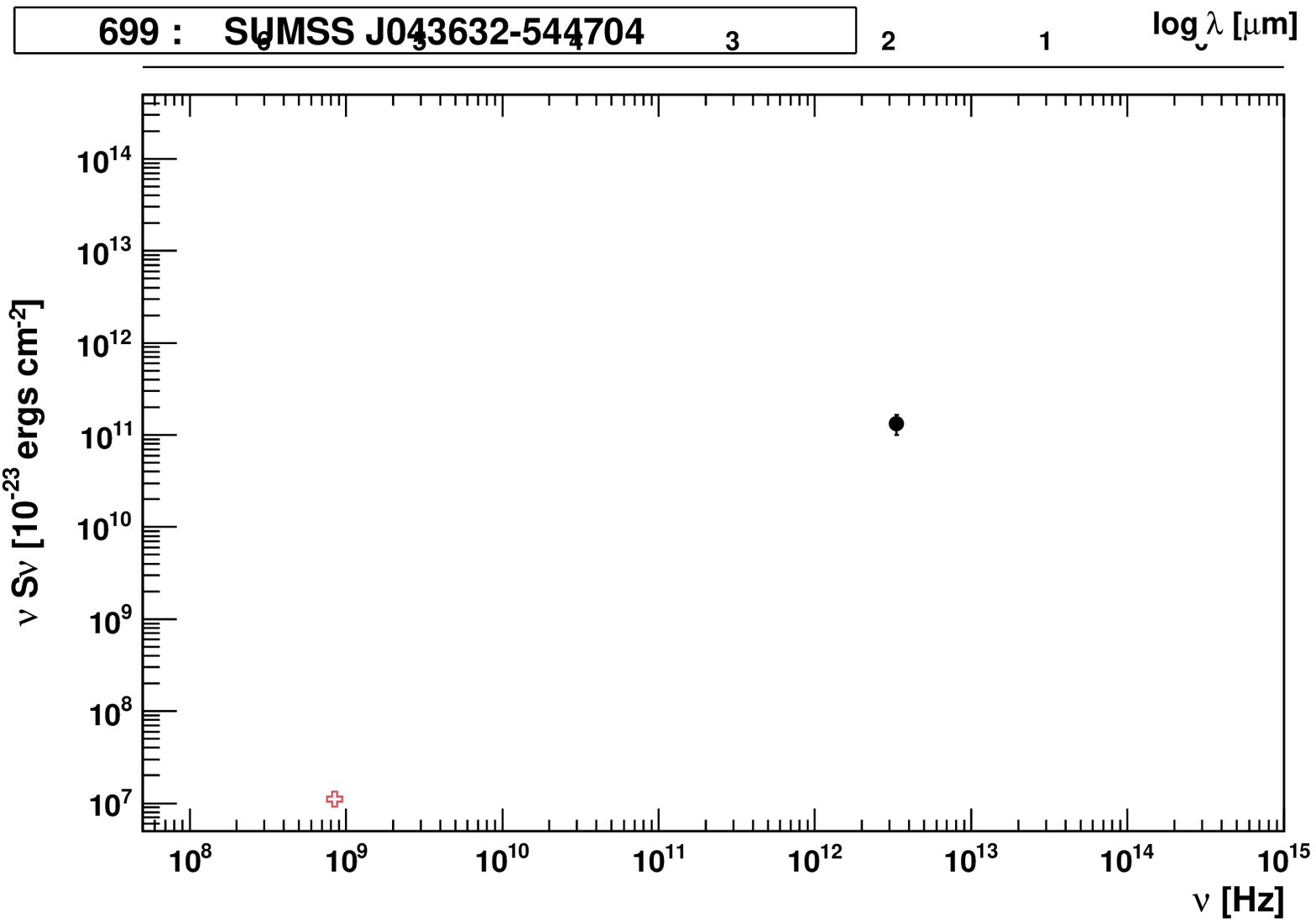}
\includegraphics[width=4cm]{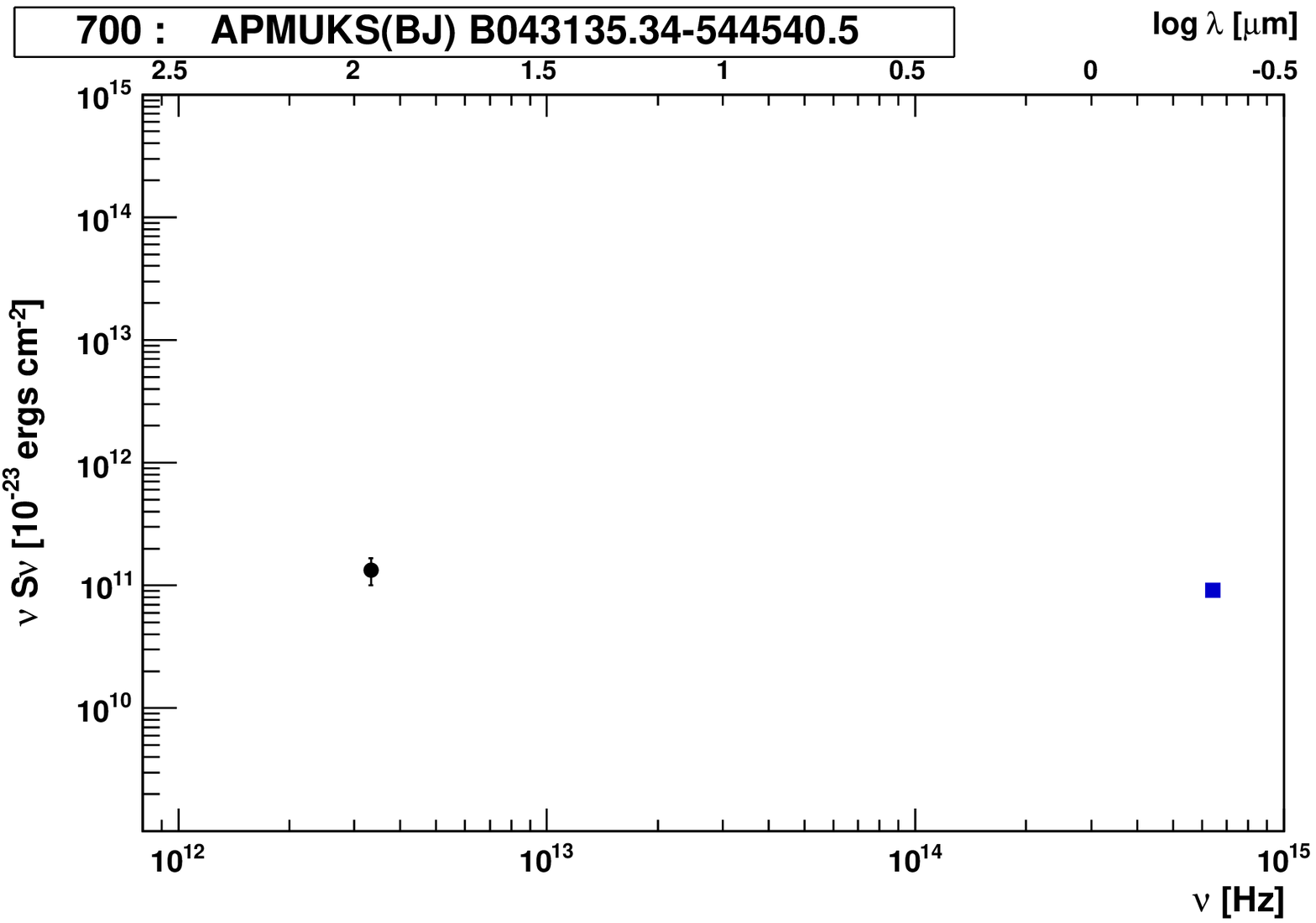}
\includegraphics[width=4cm]{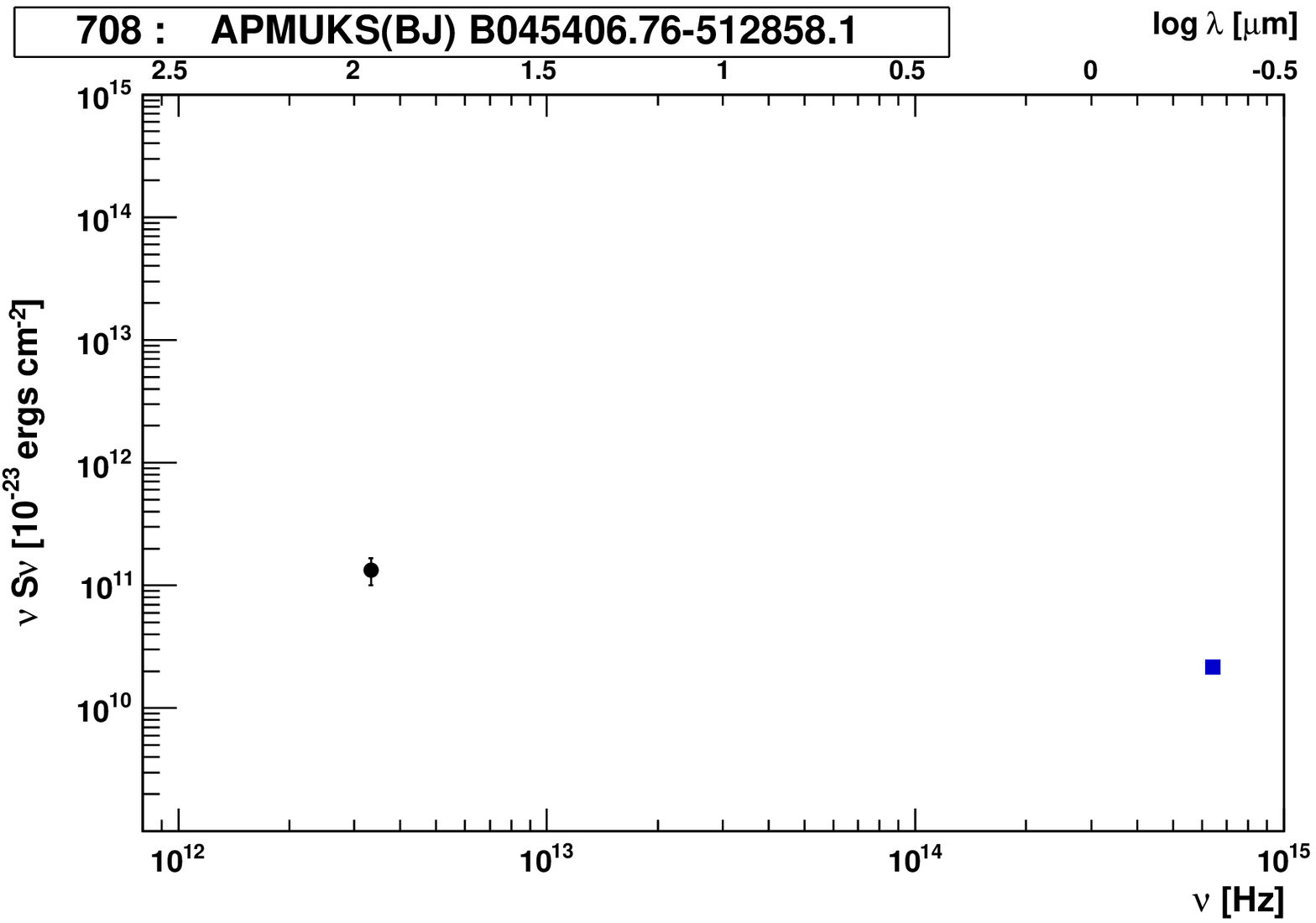}
\includegraphics[width=4cm]{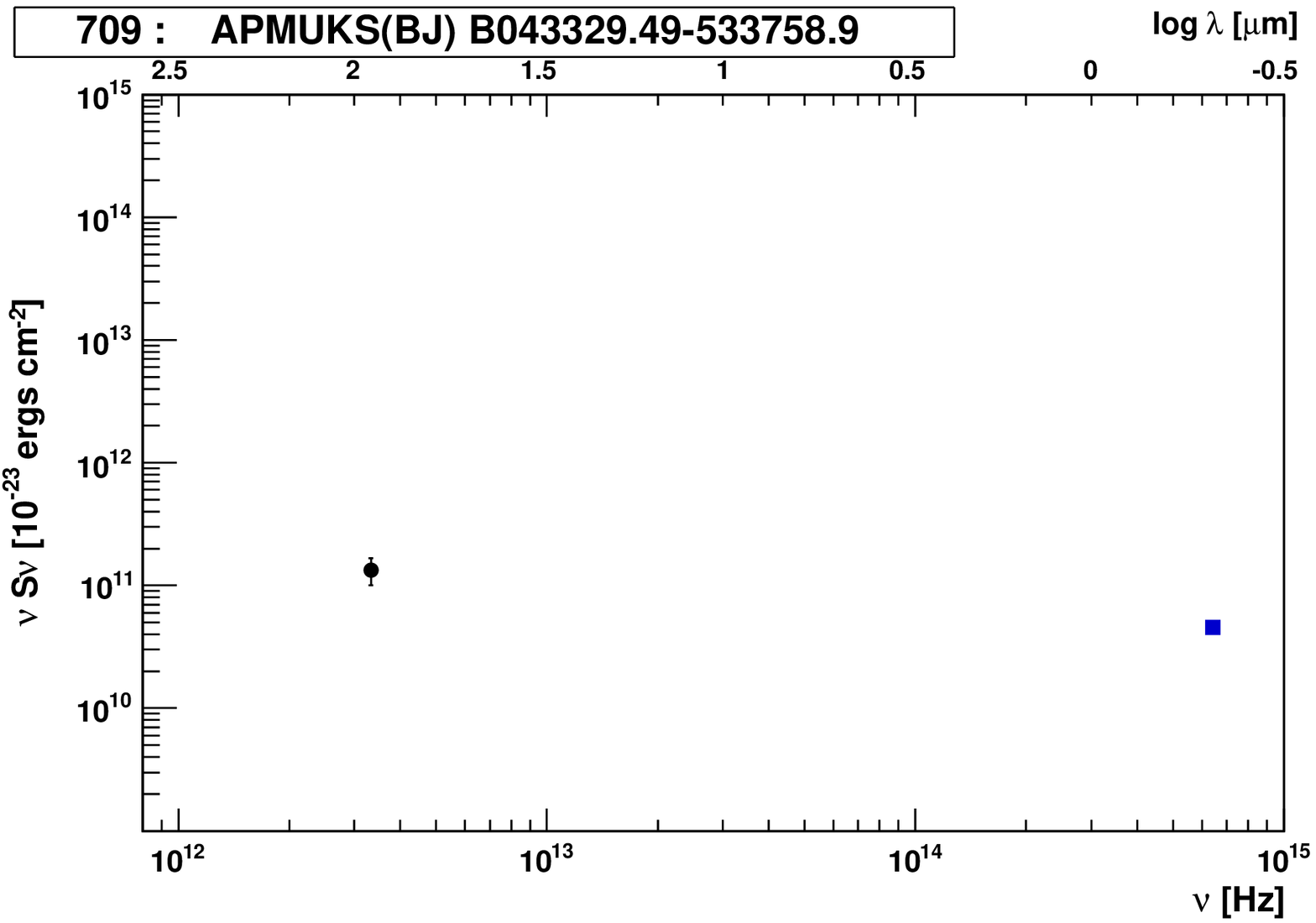}
\includegraphics[width=4cm]{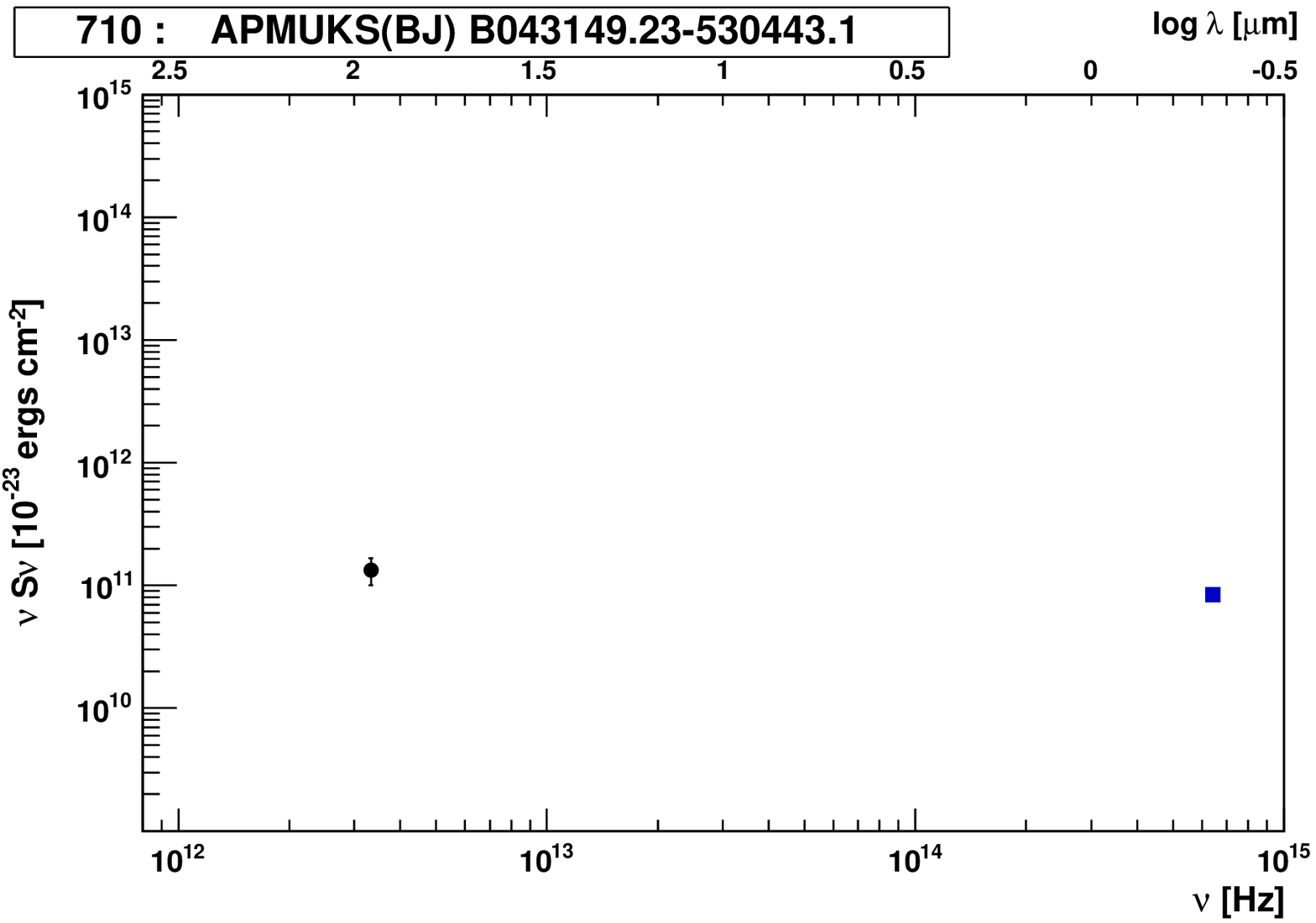}
\includegraphics[width=4cm]{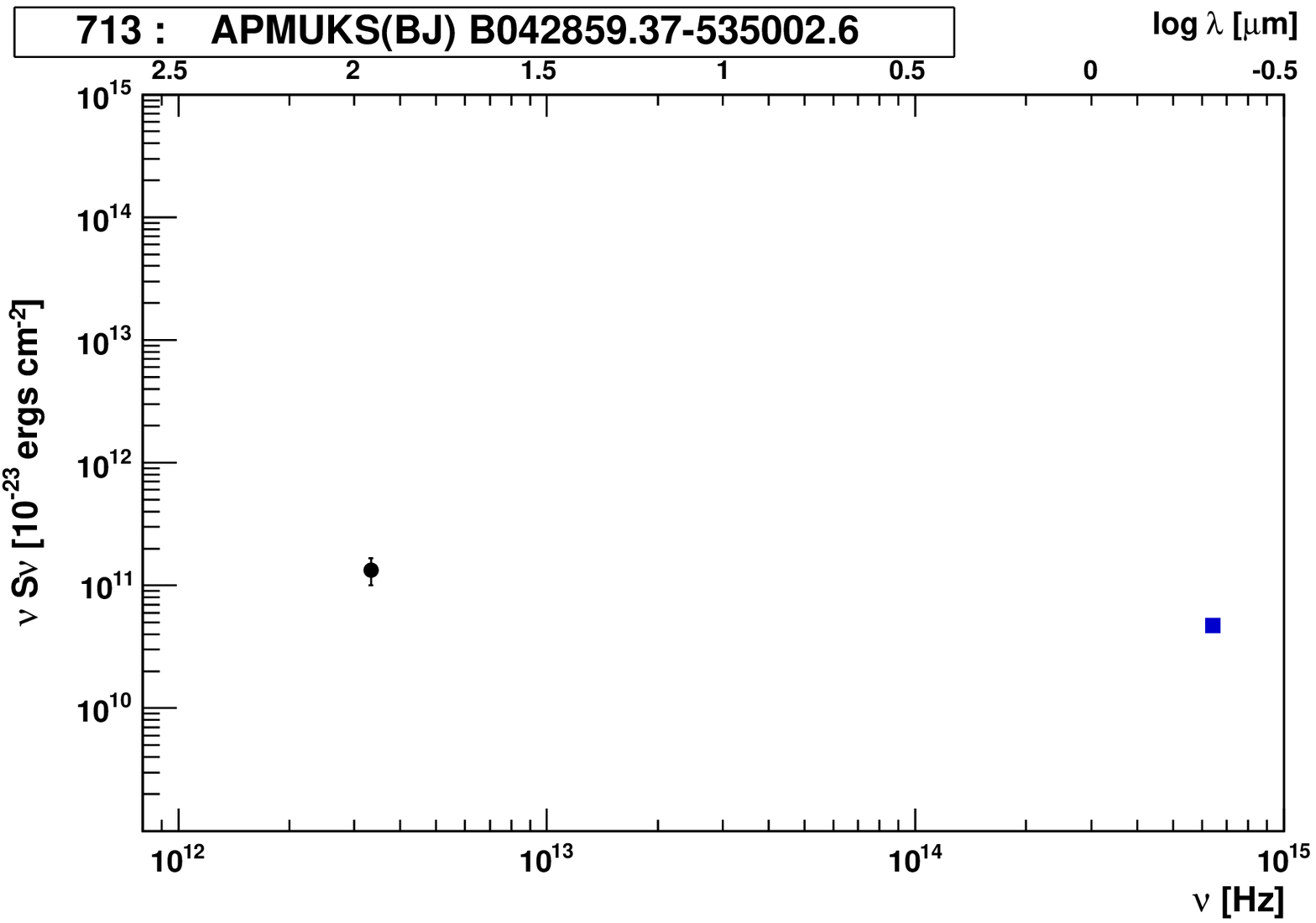}
\includegraphics[width=4cm]{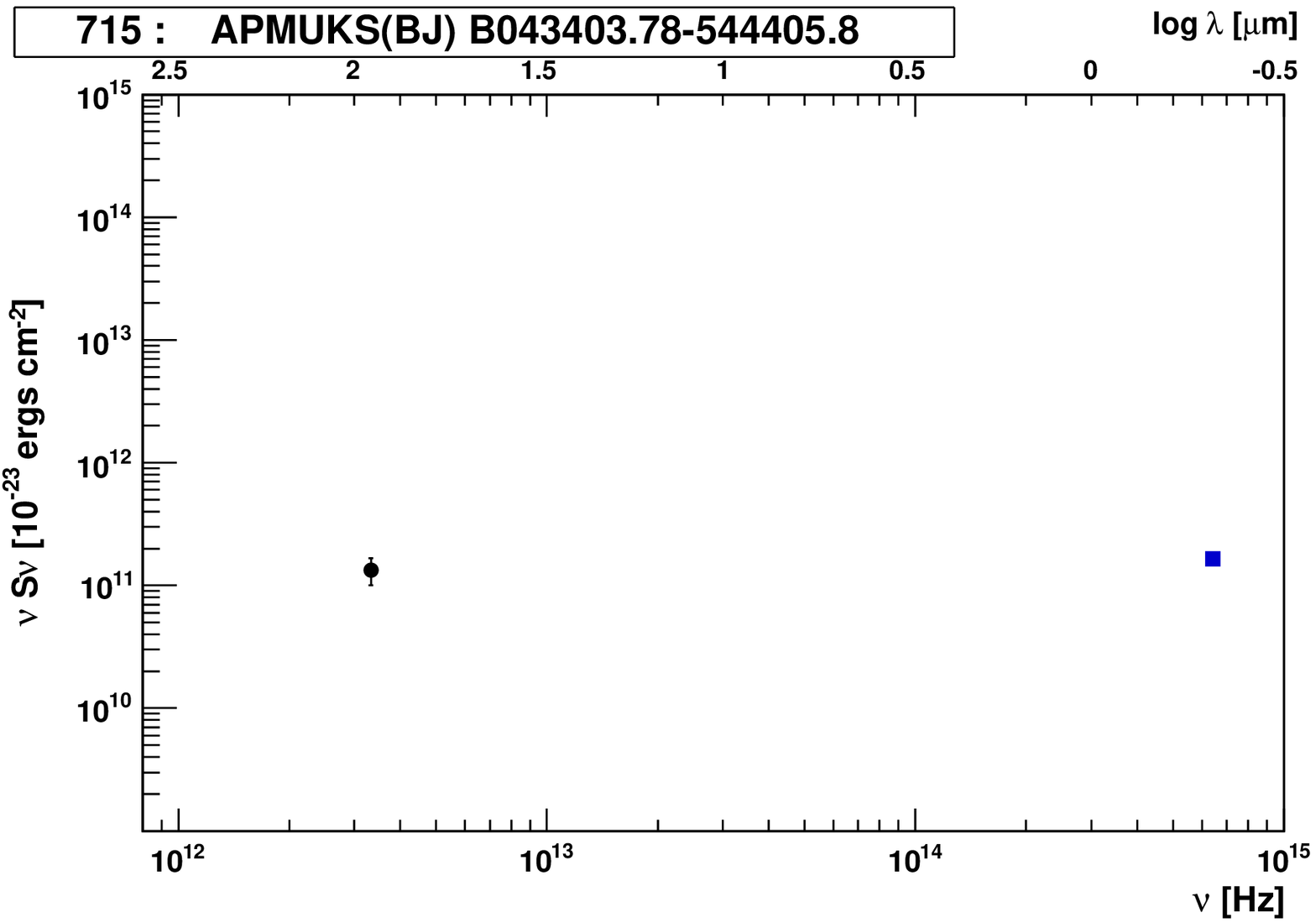}
\includegraphics[width=4cm]{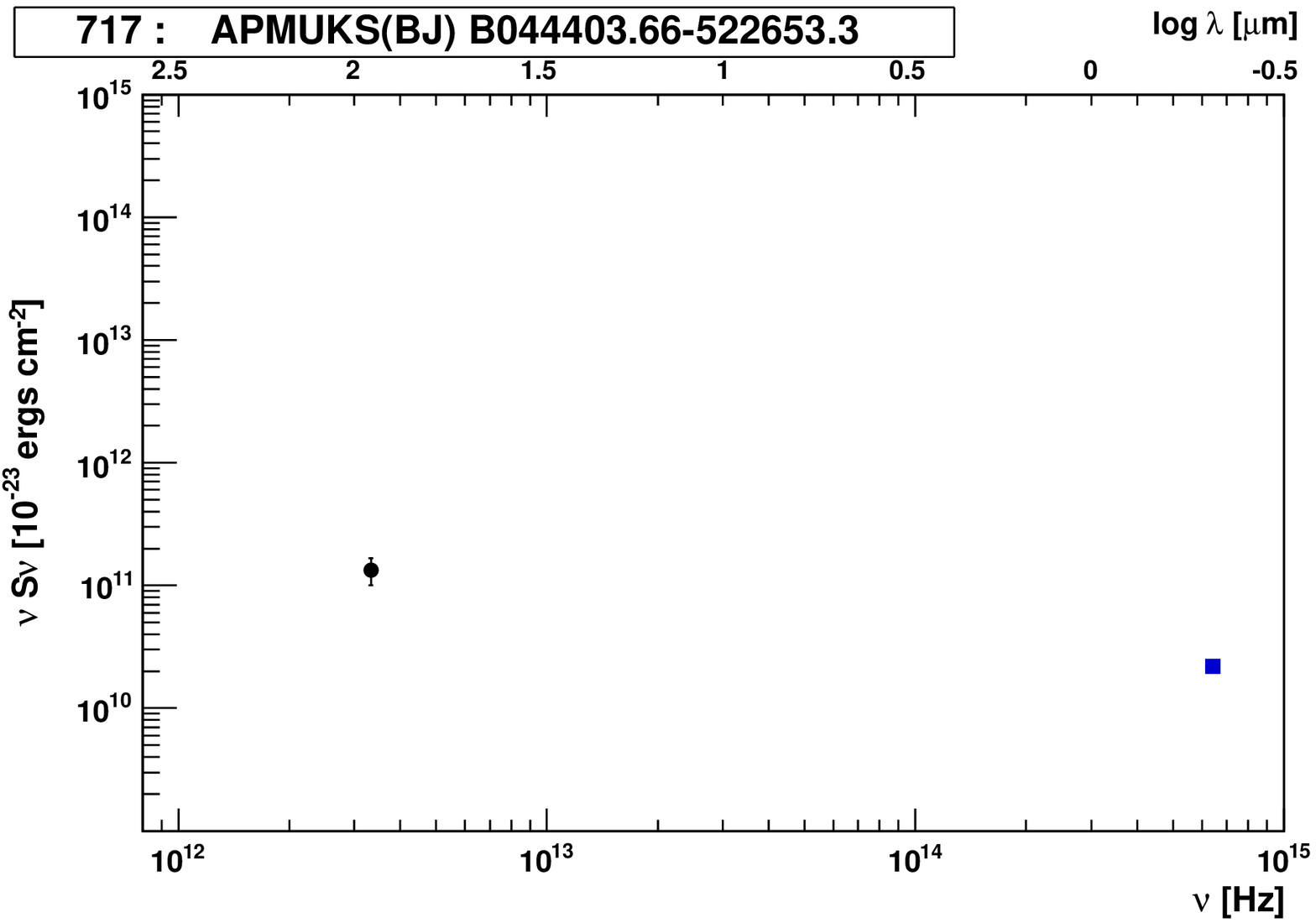}
\includegraphics[width=4cm]{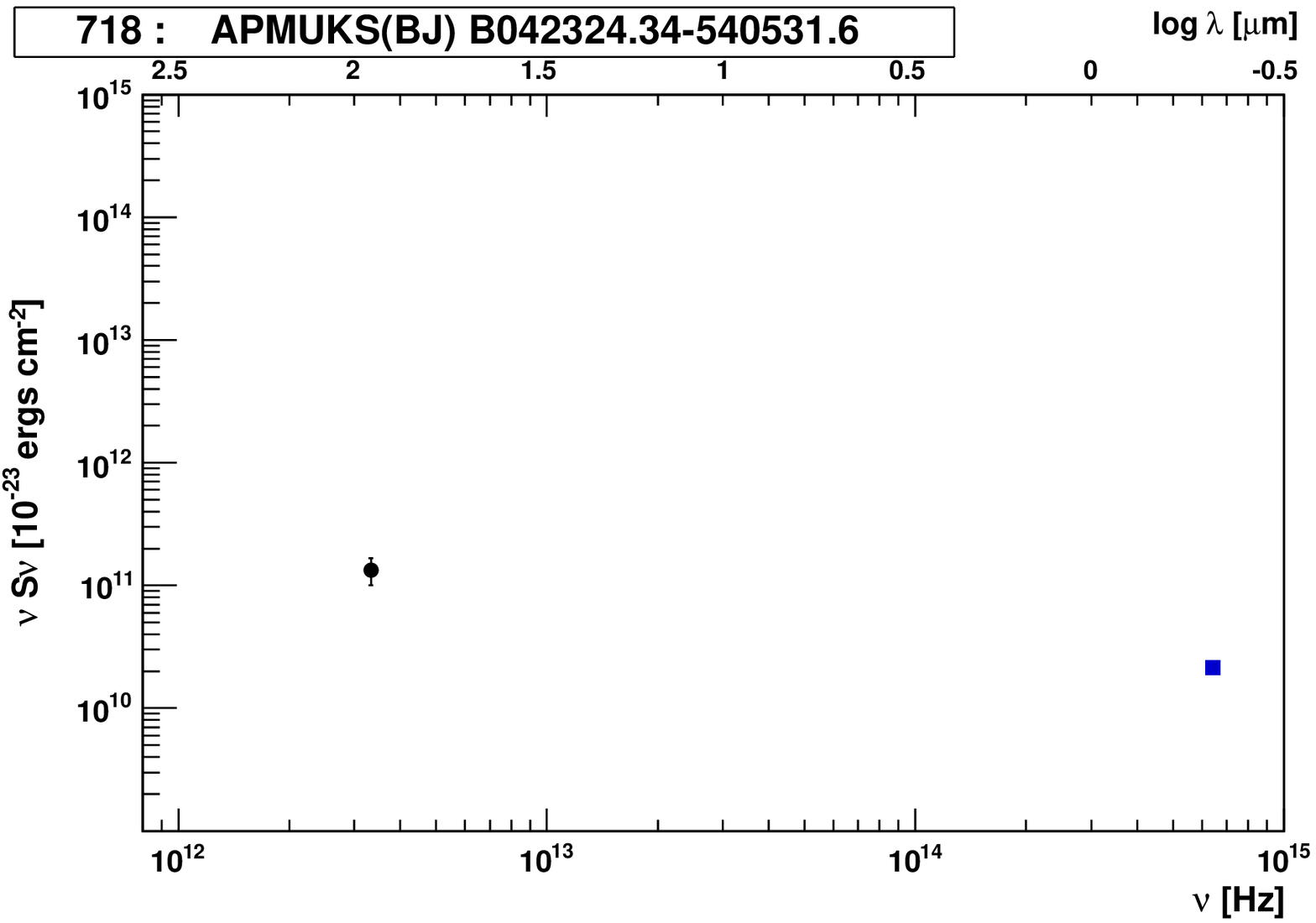}
\includegraphics[width=4cm]{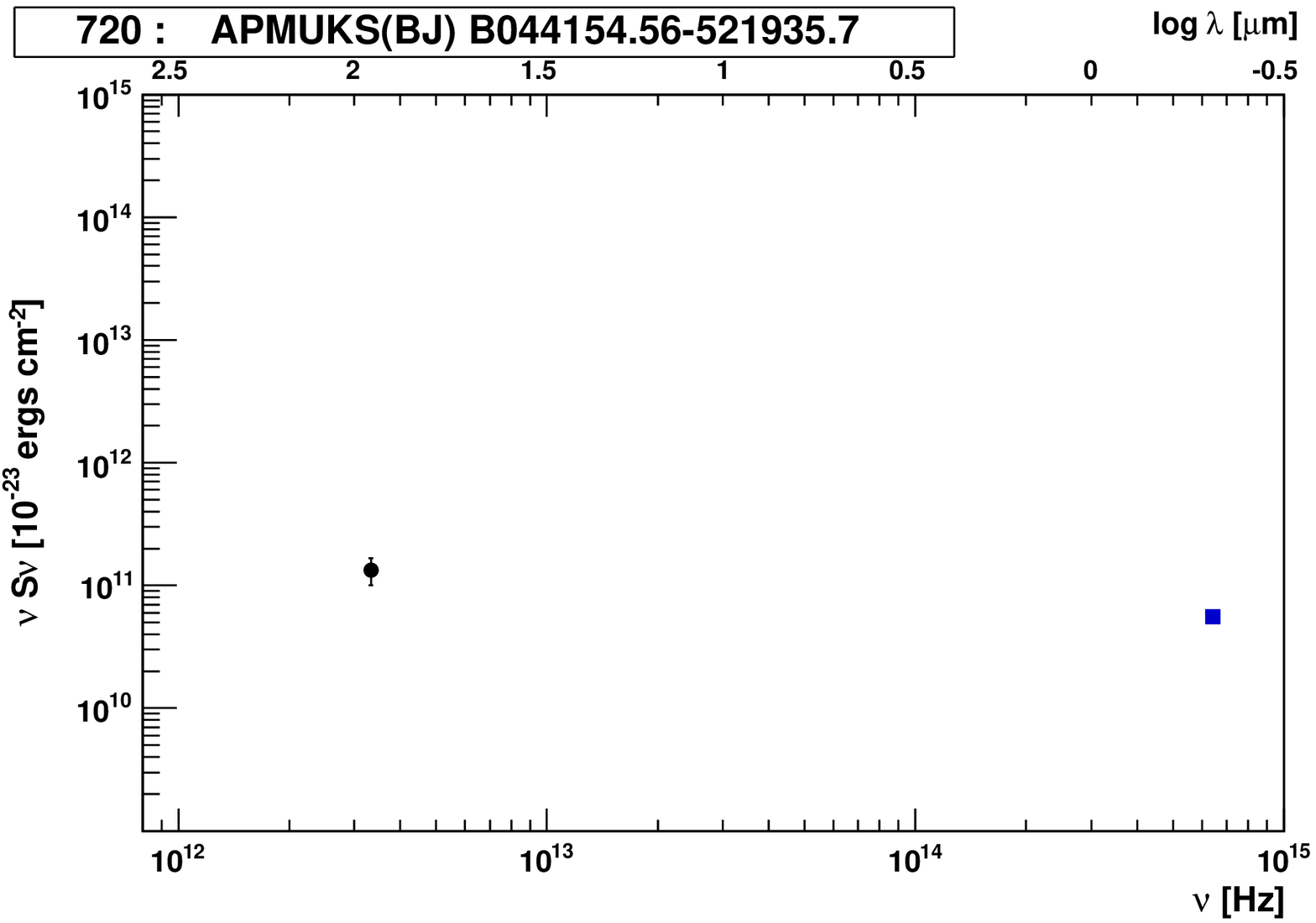}
\includegraphics[width=4cm]{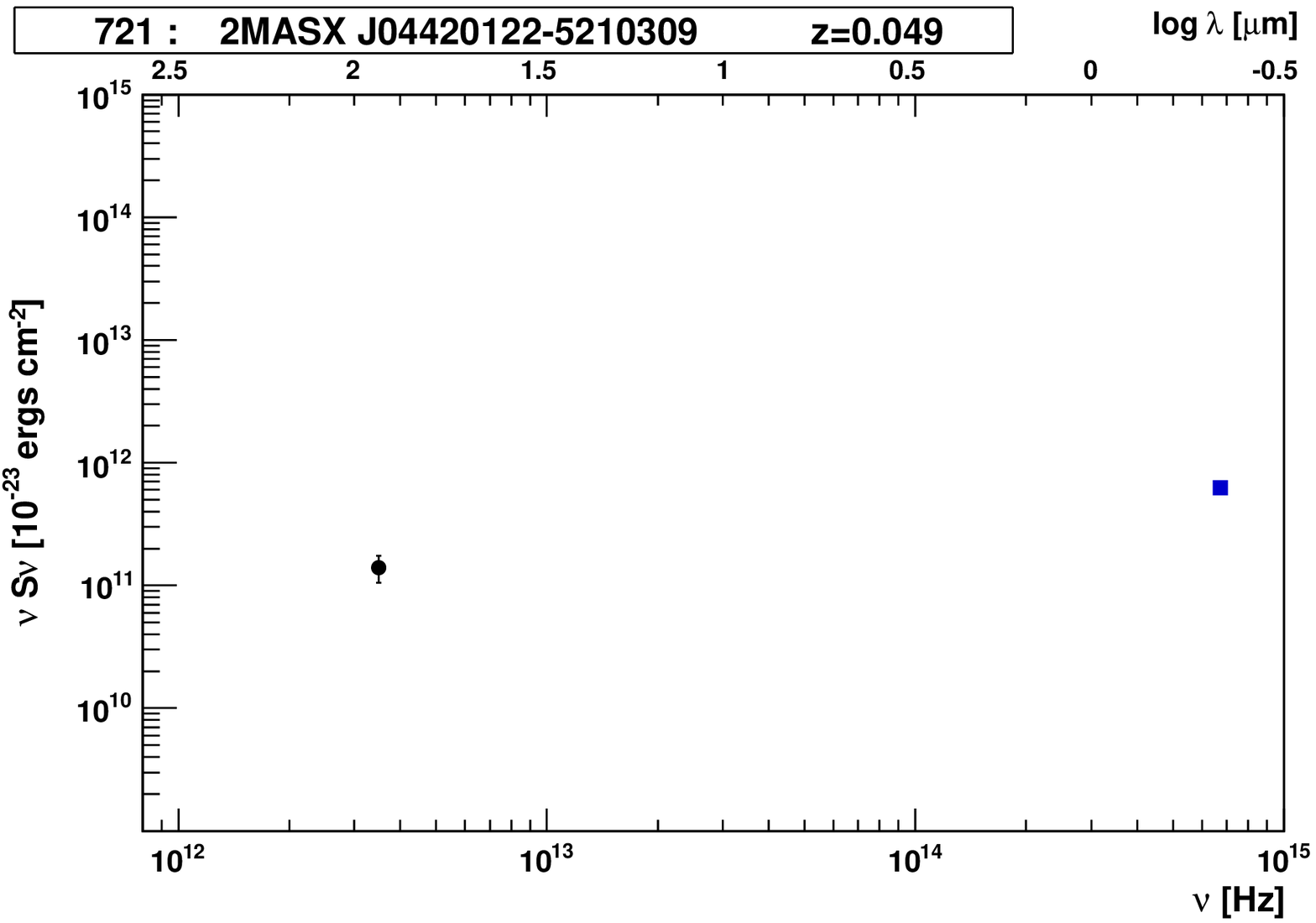}
\includegraphics[width=4cm]{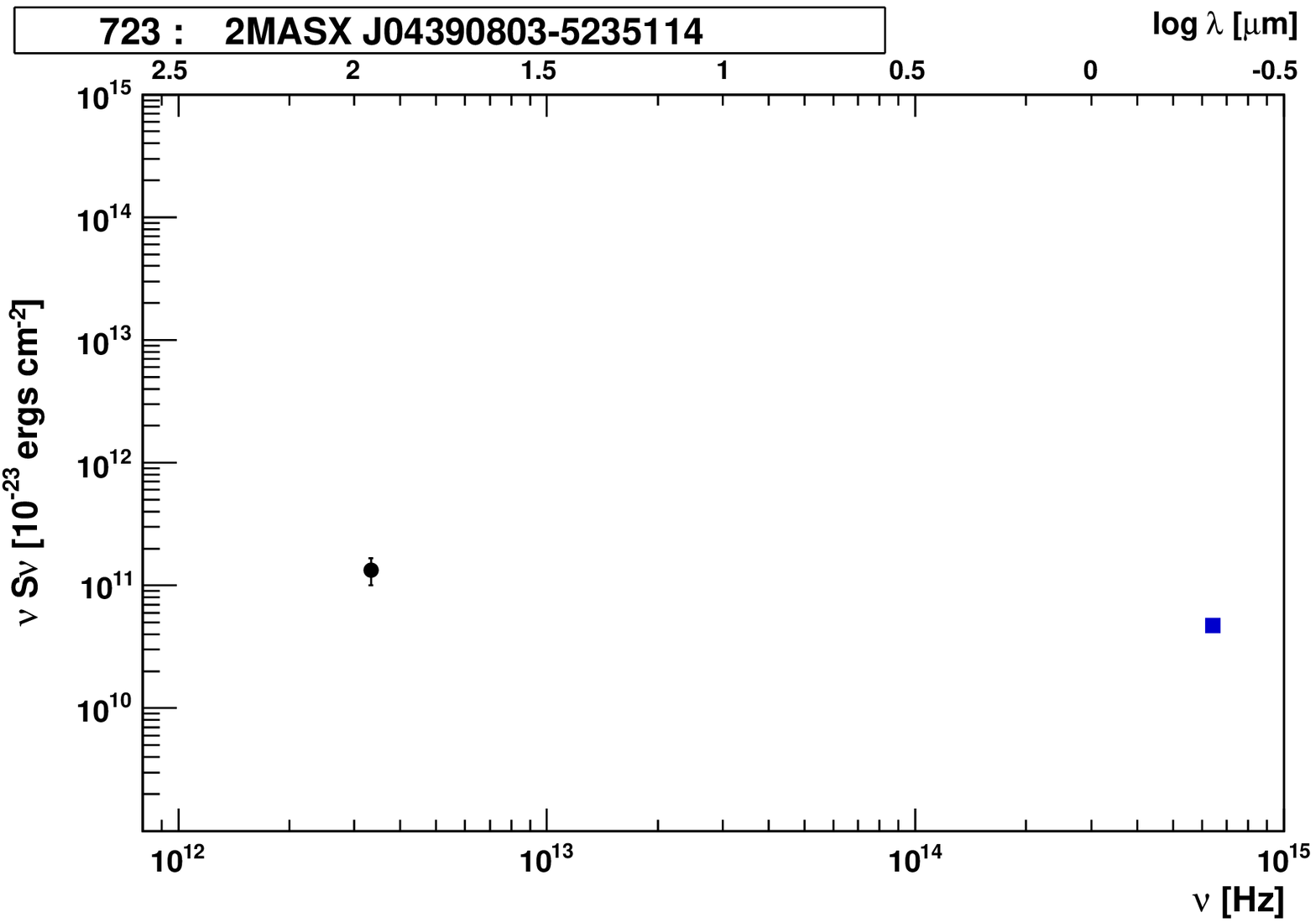}
\includegraphics[width=4cm]{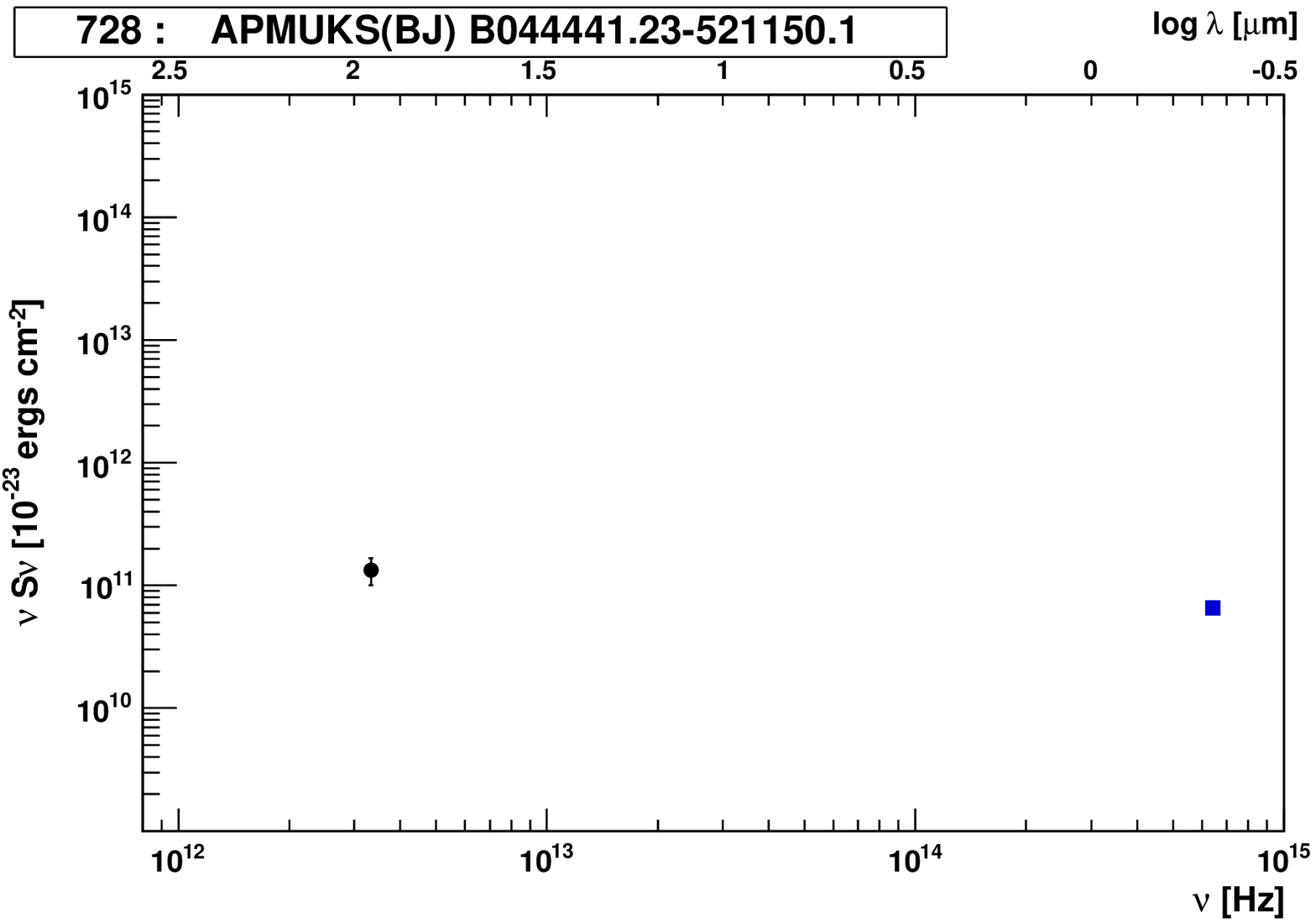}
\includegraphics[width=4cm]{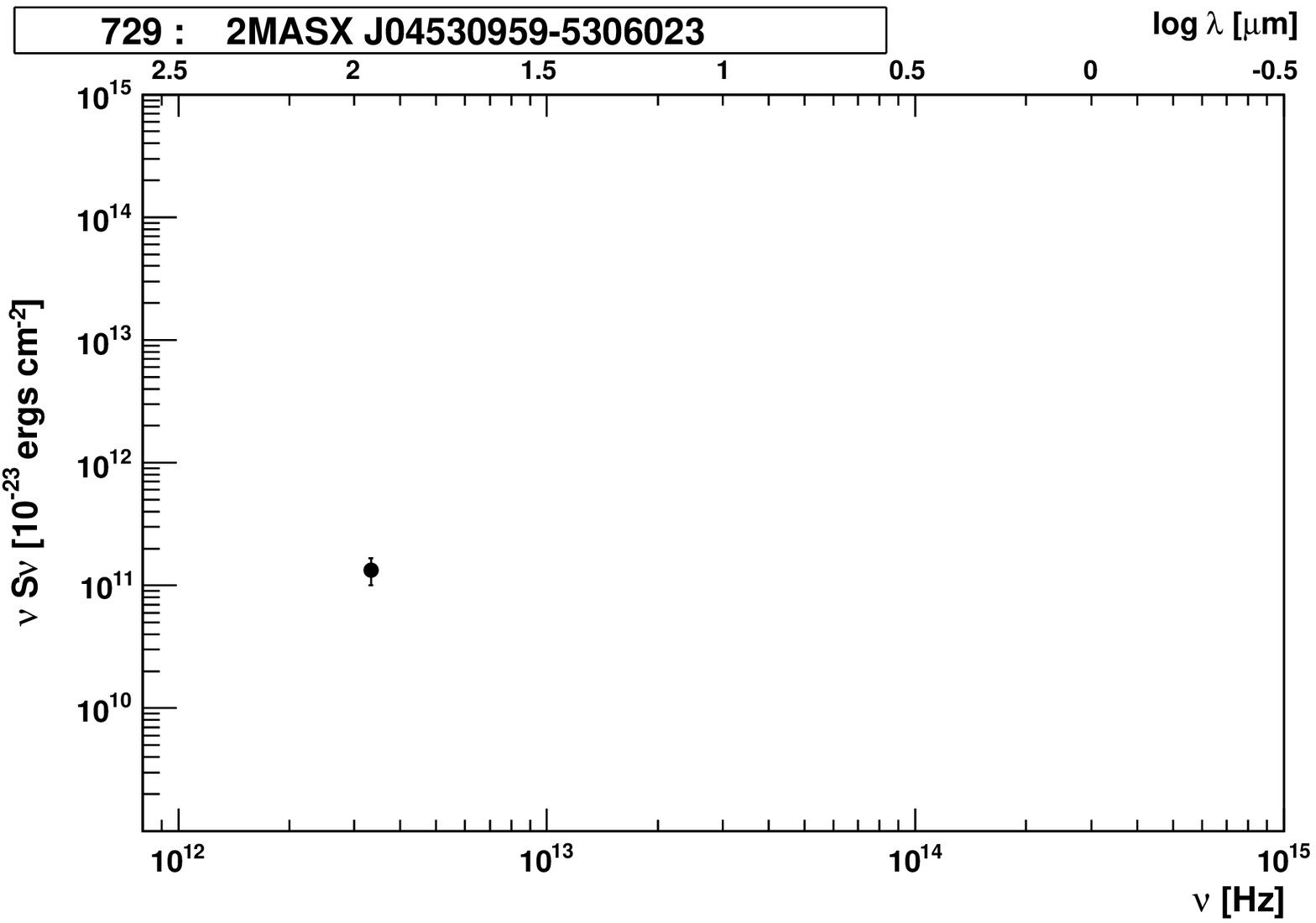}
\includegraphics[width=4cm]{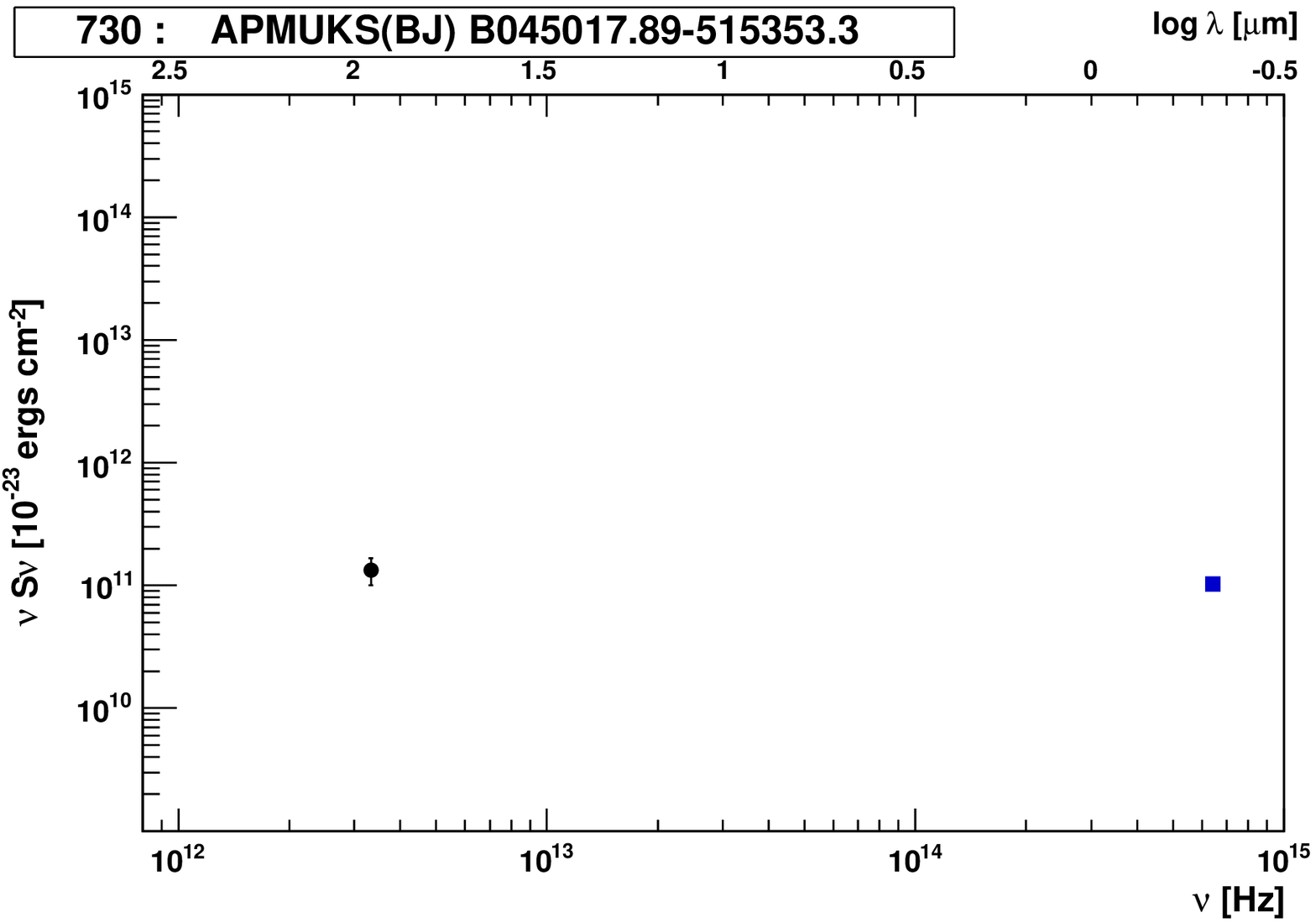}
\includegraphics[width=4cm]{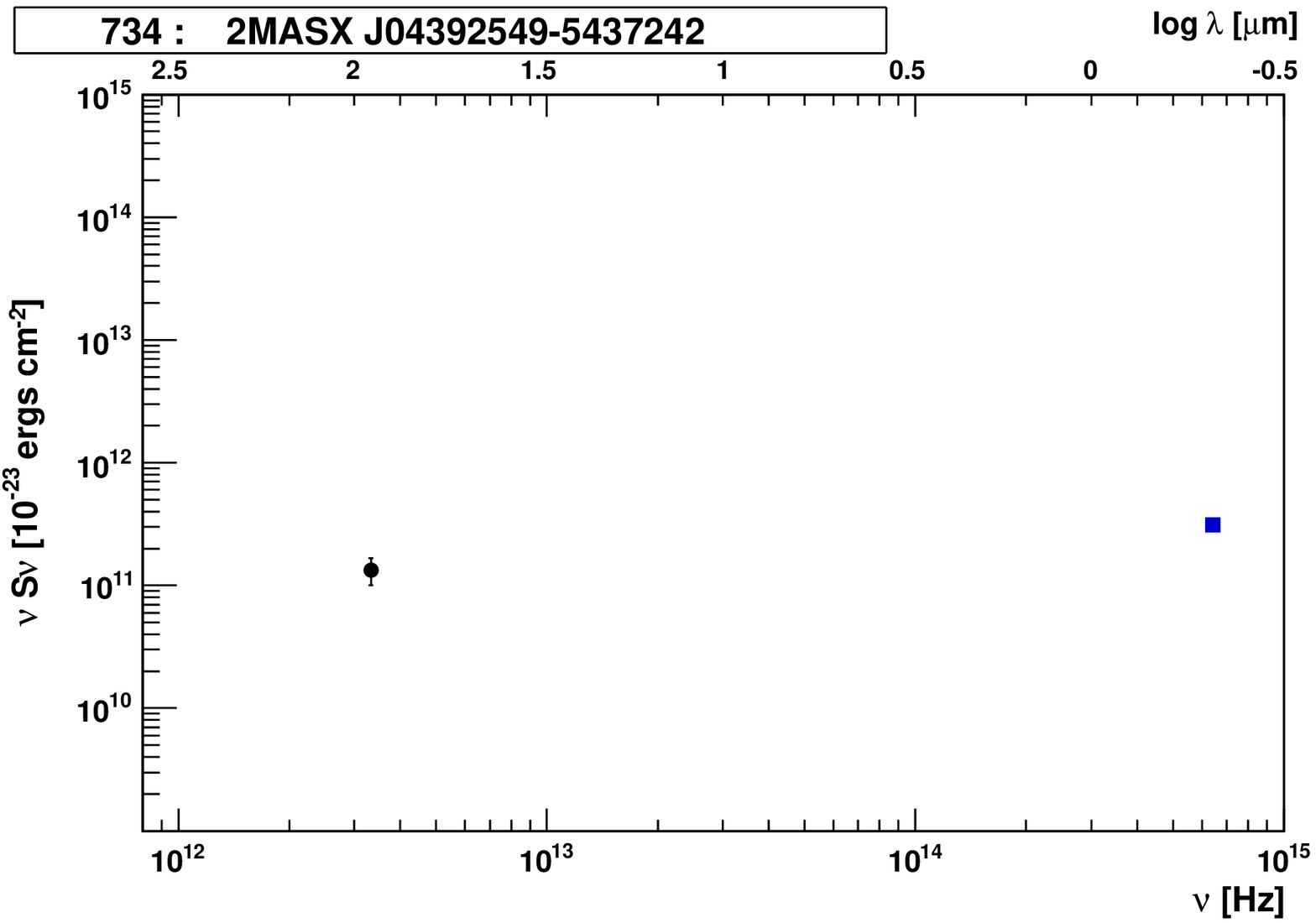}
\includegraphics[width=4cm]{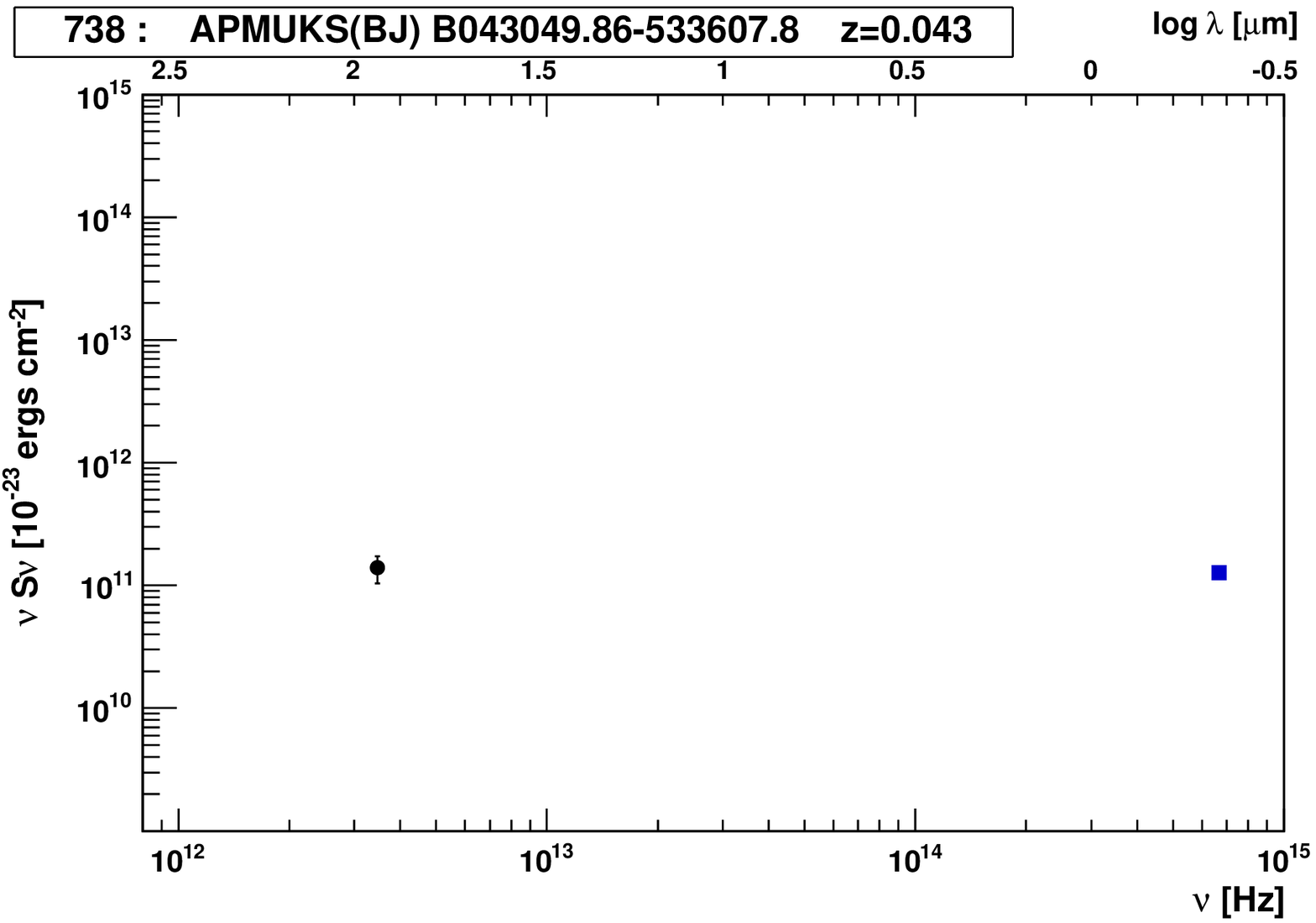}
\includegraphics[width=4cm]{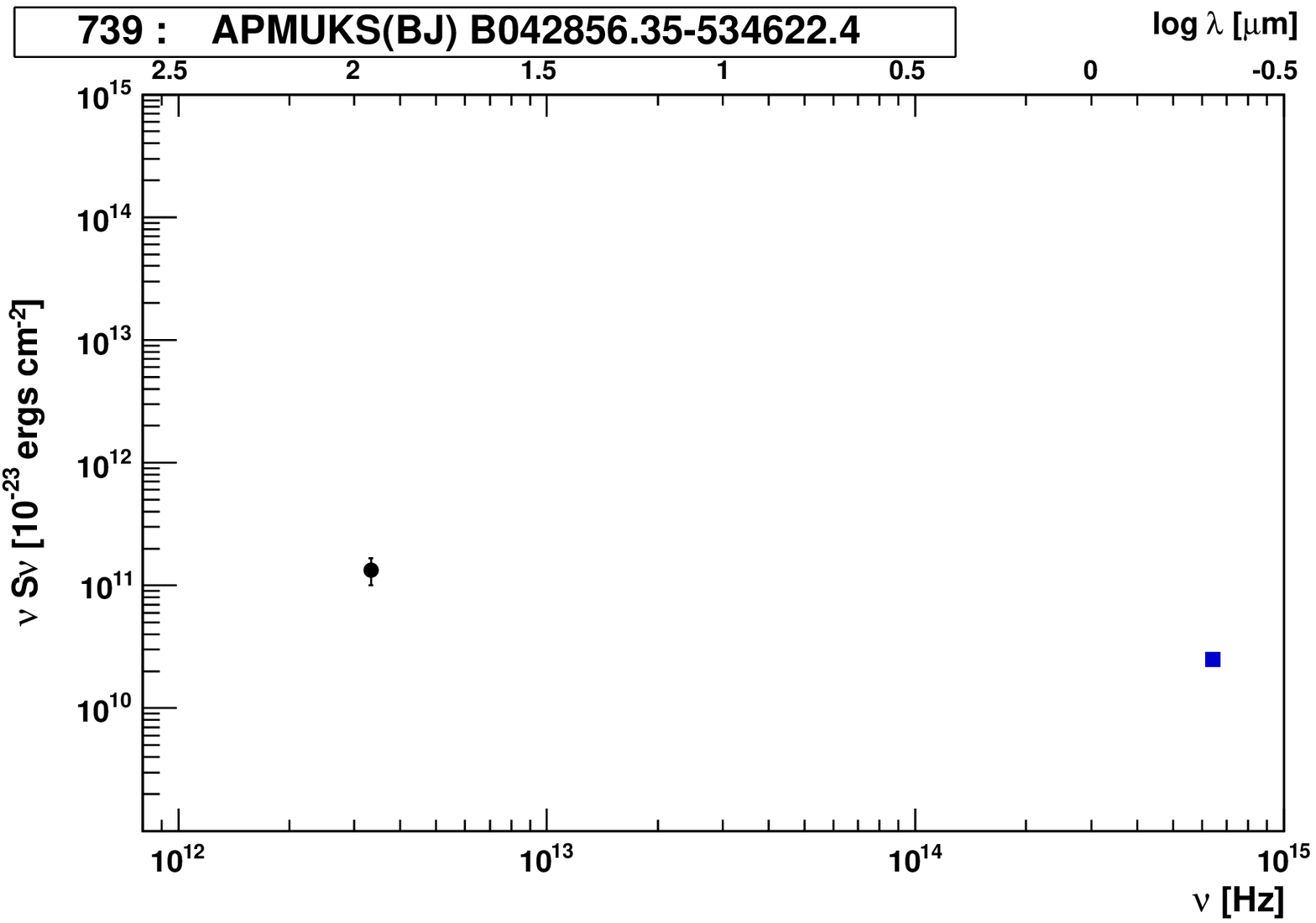}
\includegraphics[width=4cm]{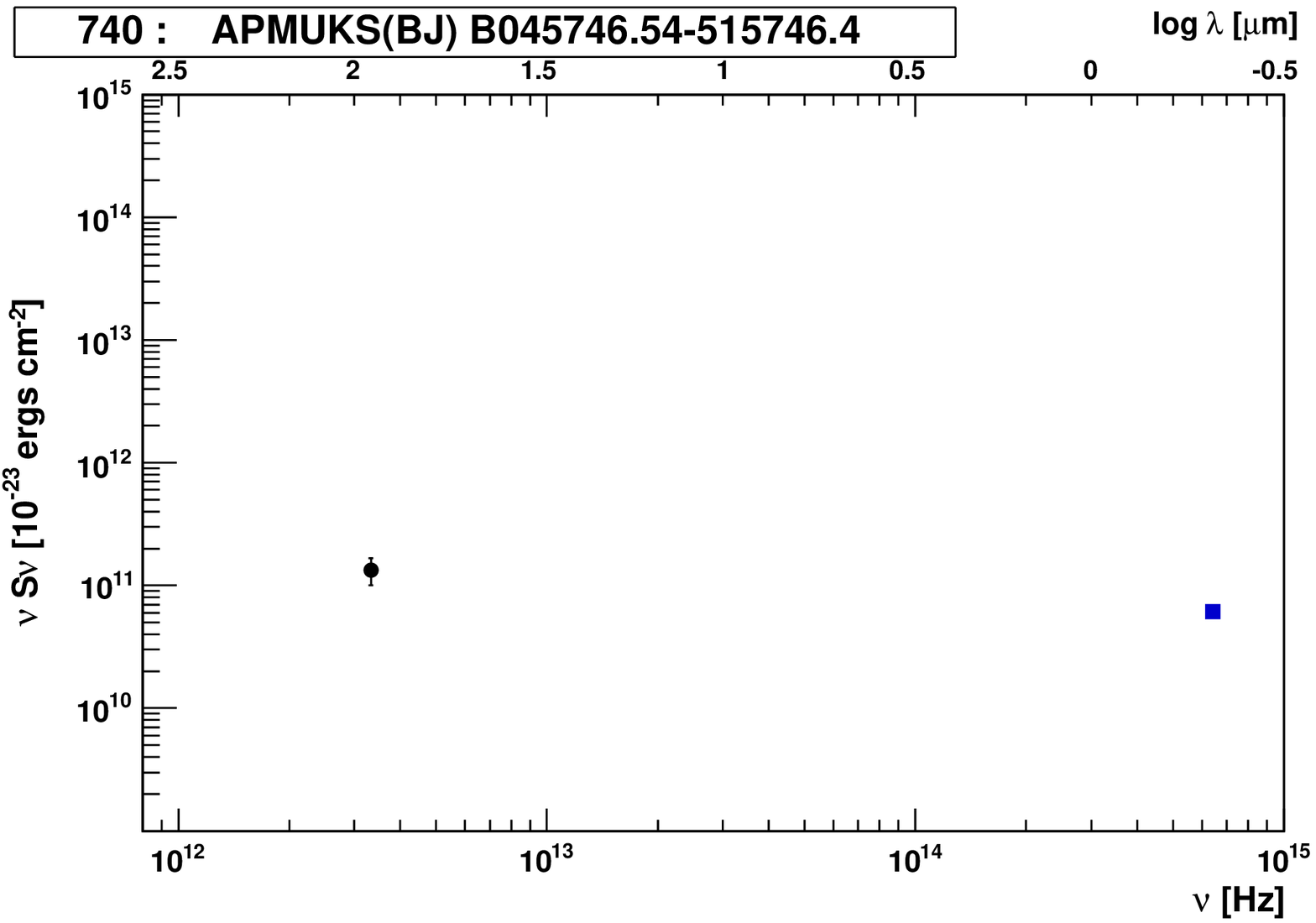}
\label{points12}
\caption {SEDs for the next 36 ADF-S identified sources, with symbols as in Figure~\ref{points1}.}
\end{figure*}
}

\clearpage

\onlfig{13}{
\begin{figure*}[t]
\centering

\includegraphics[width=4cm]{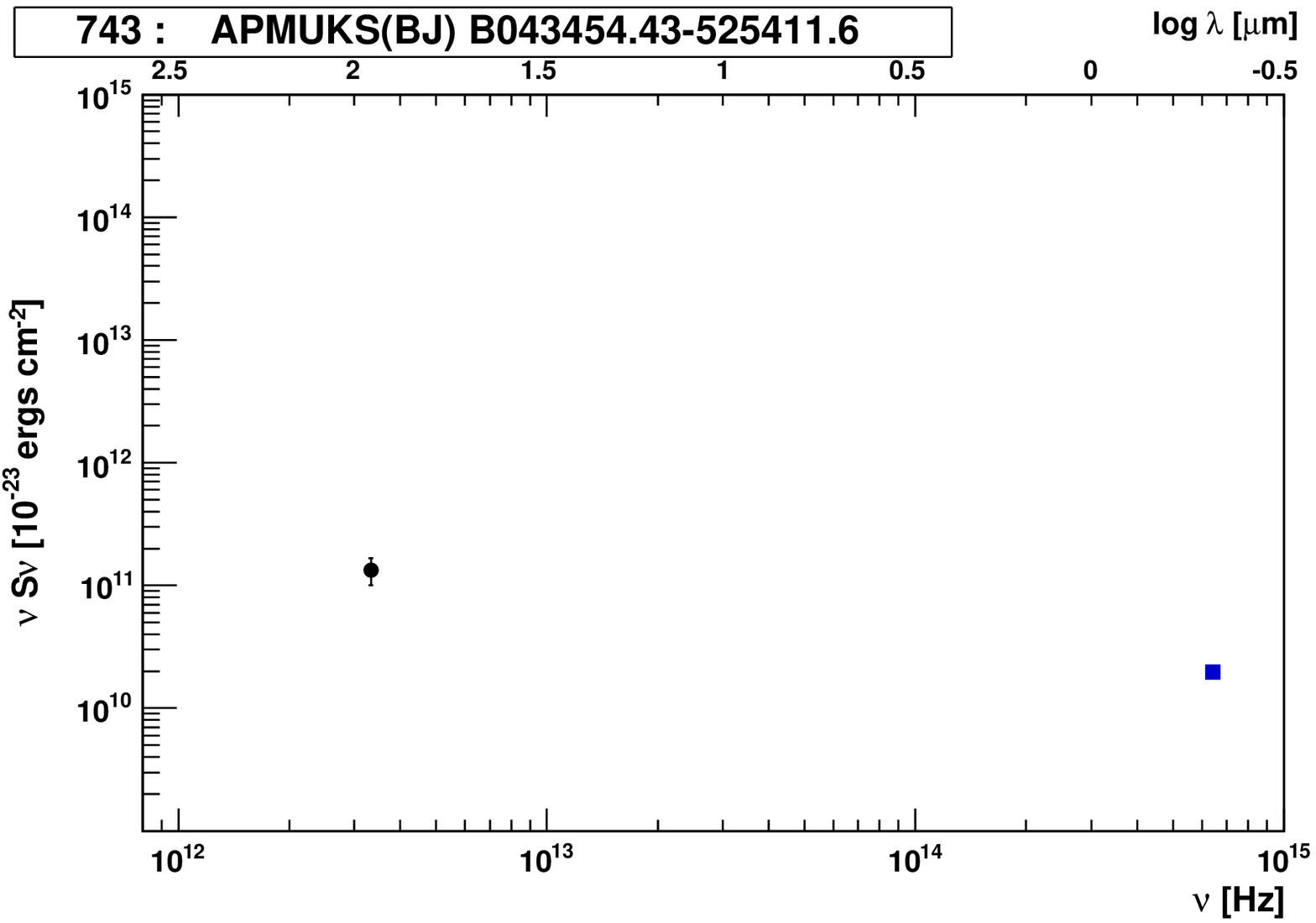}
\includegraphics[width=4cm]{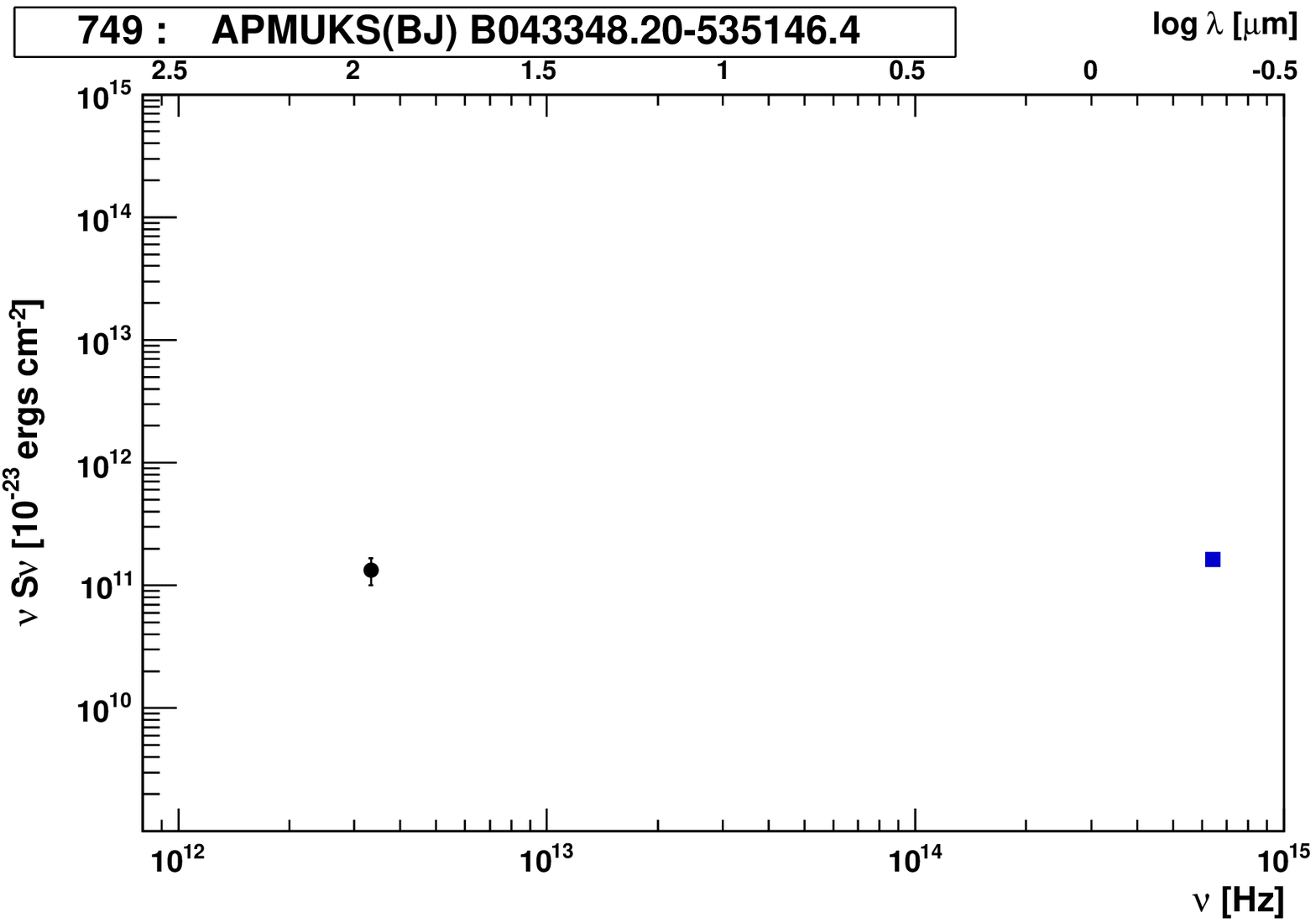}
\includegraphics[width=4cm]{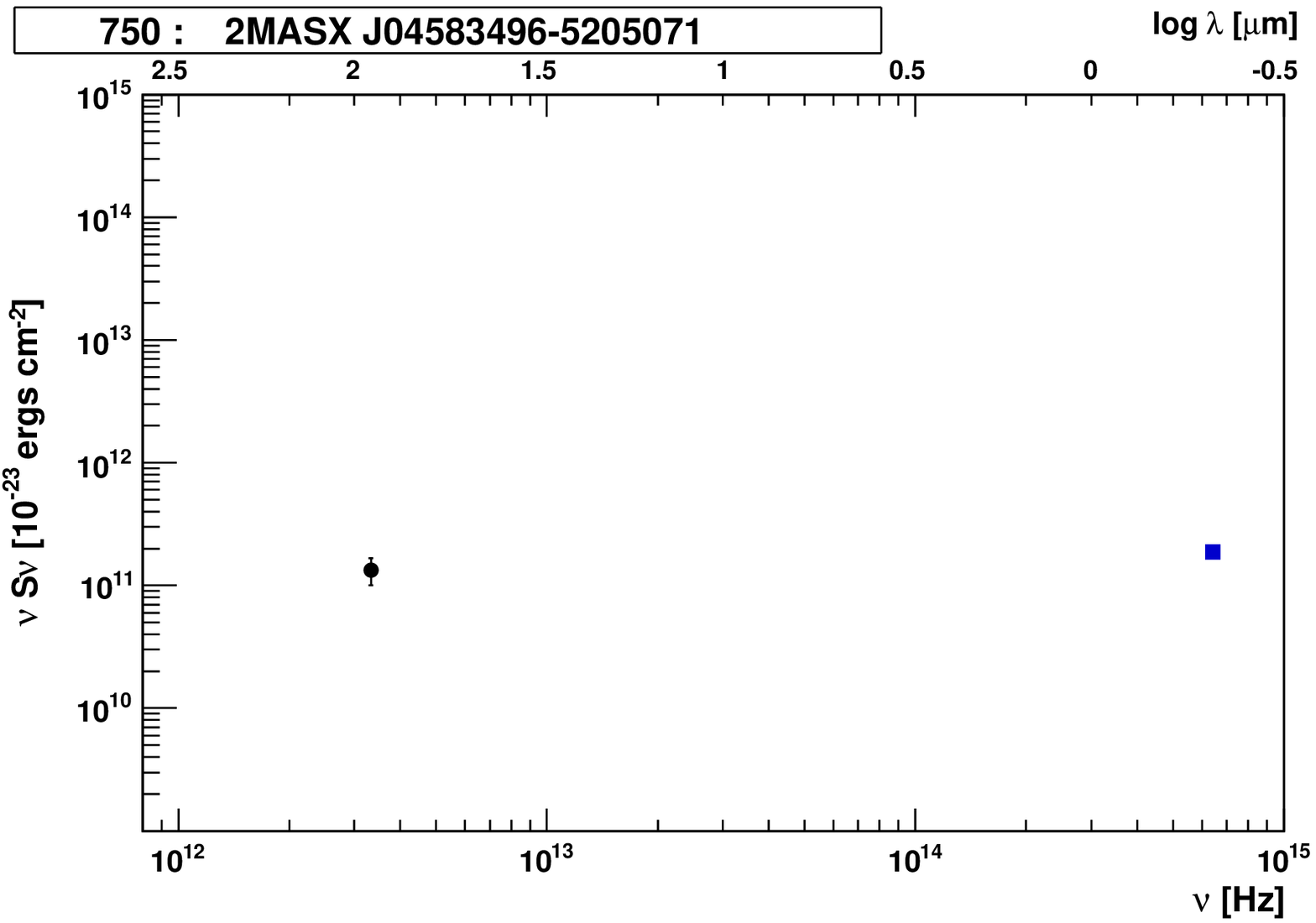}
\includegraphics[width=4cm]{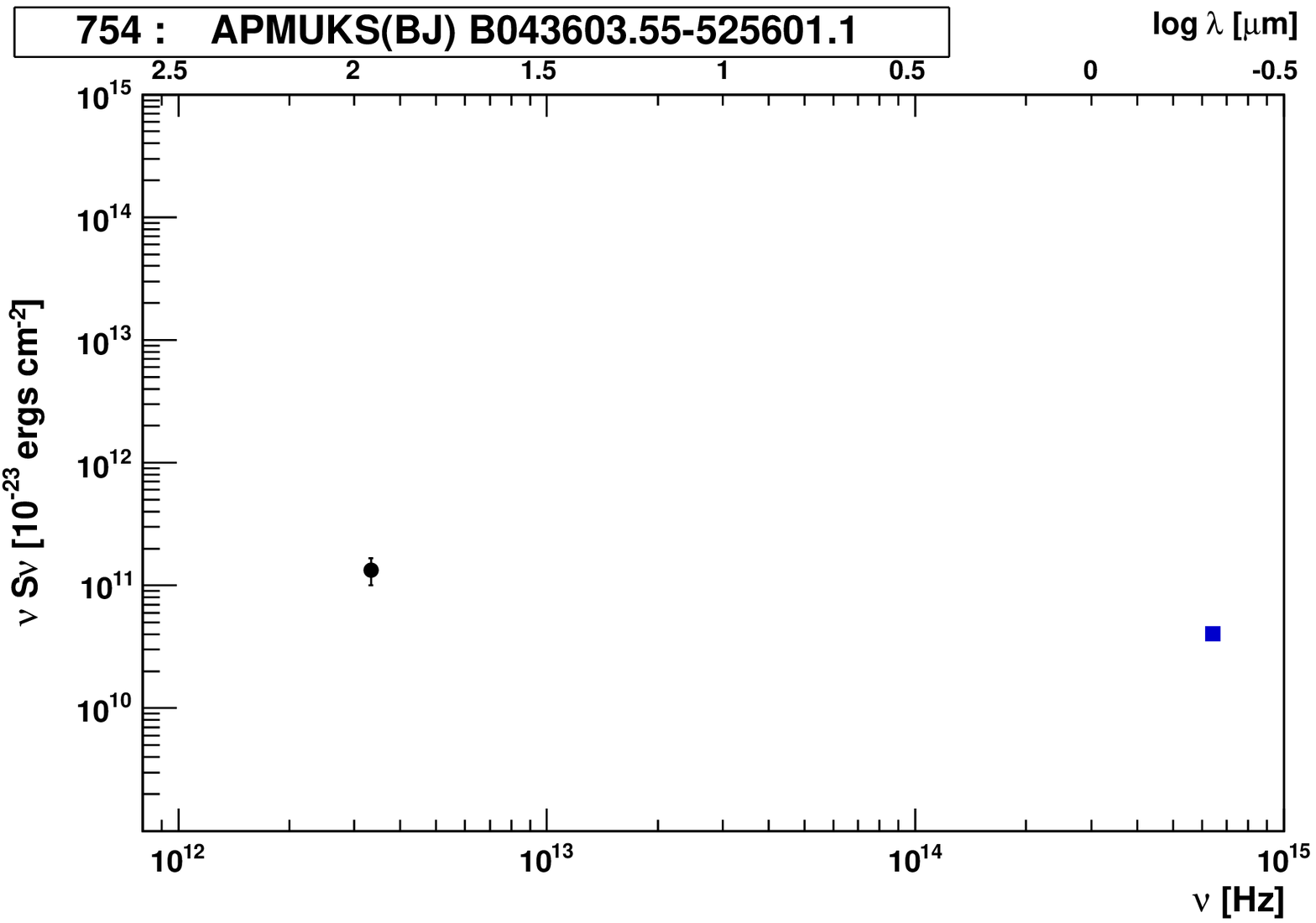}
\includegraphics[width=4cm]{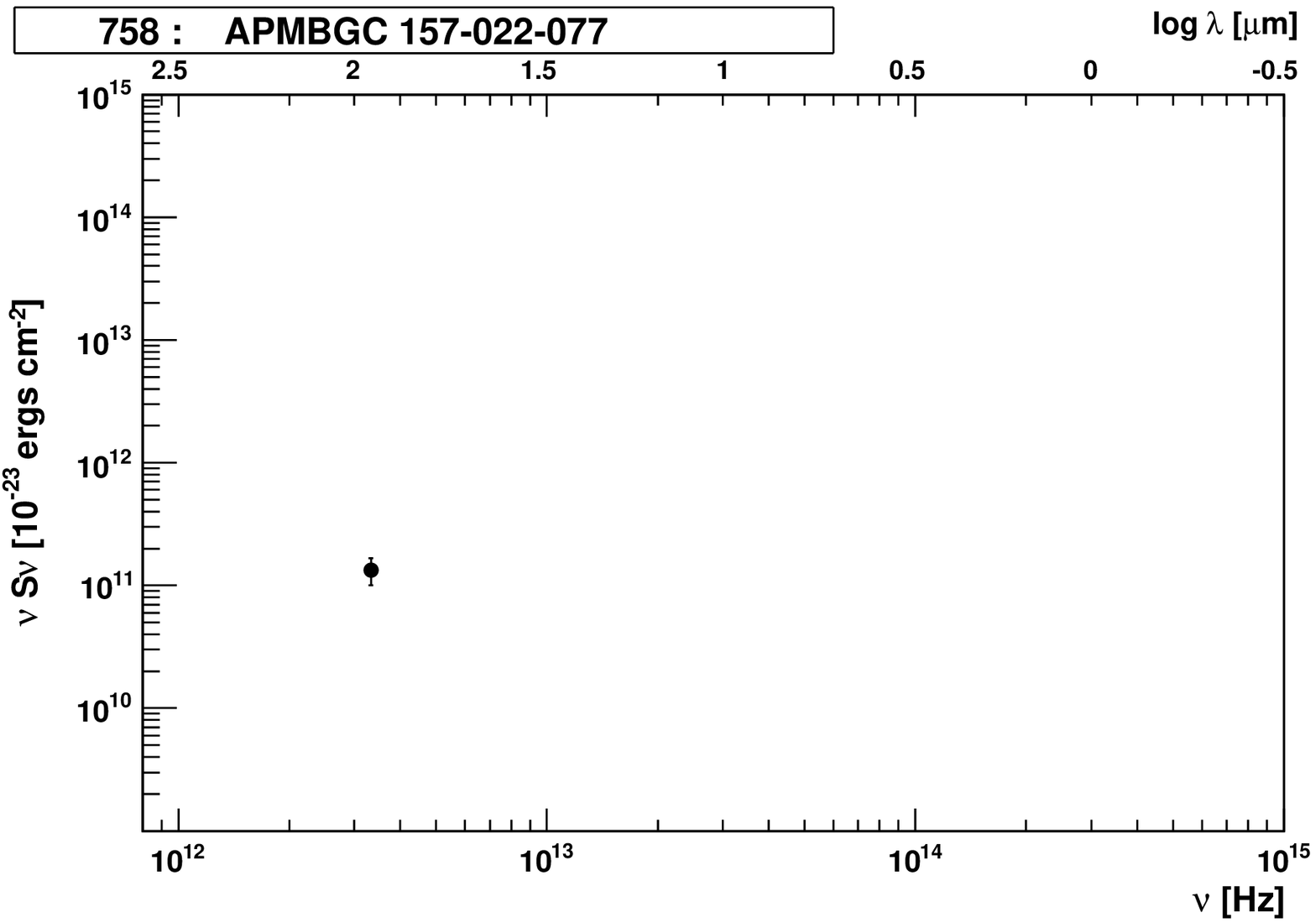}
\includegraphics[width=4cm]{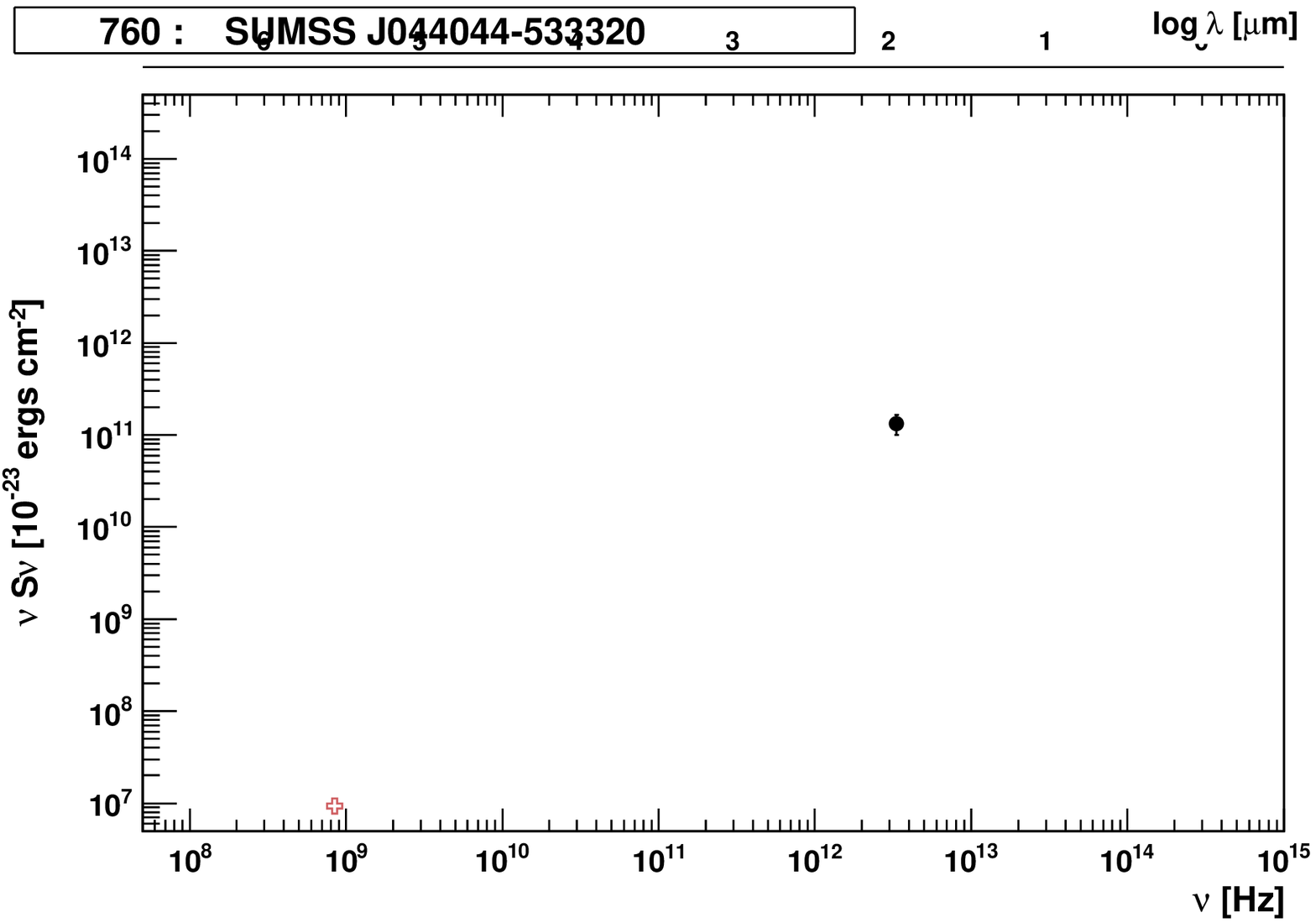}
\includegraphics[width=4cm]{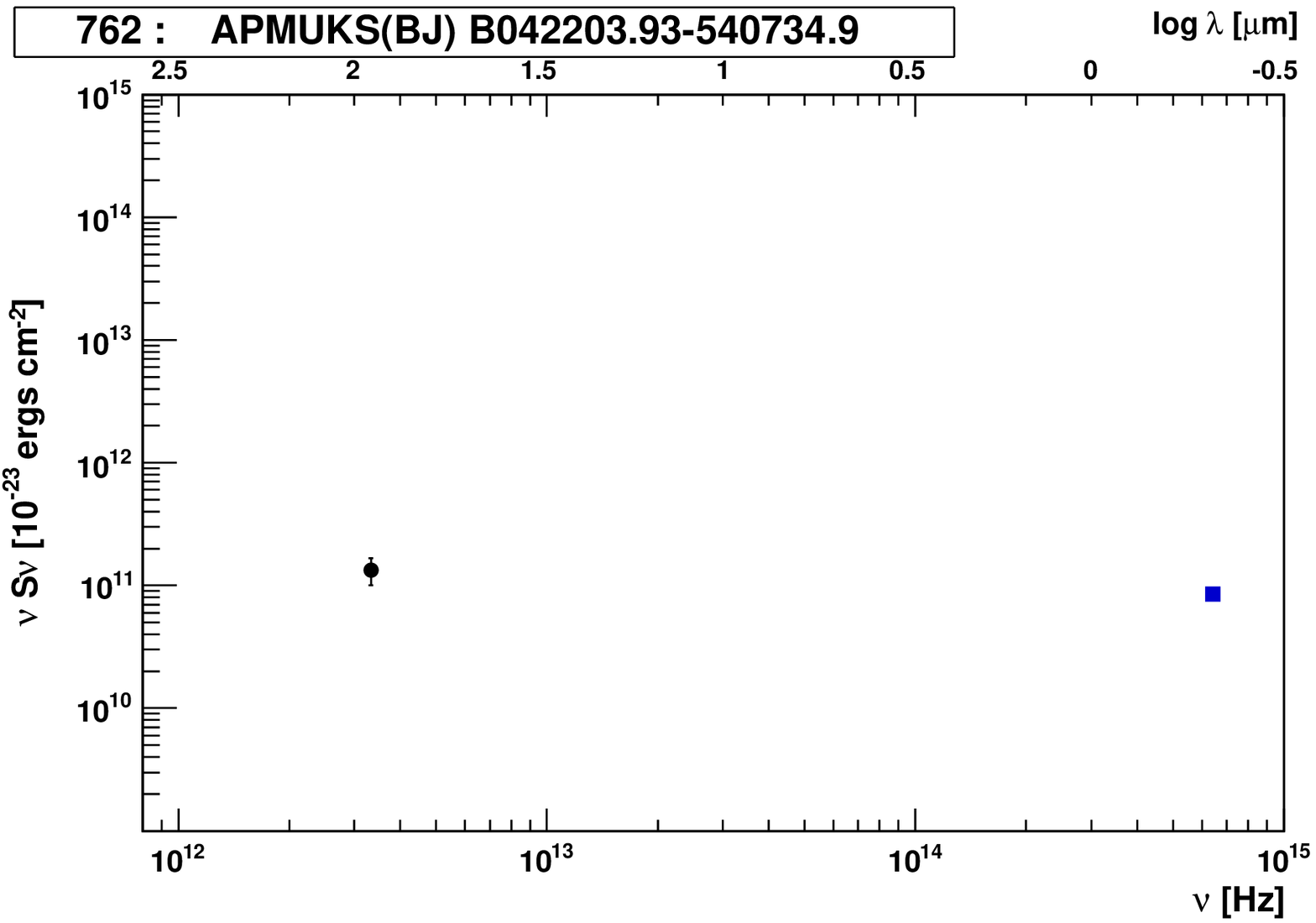}
\includegraphics[width=4cm]{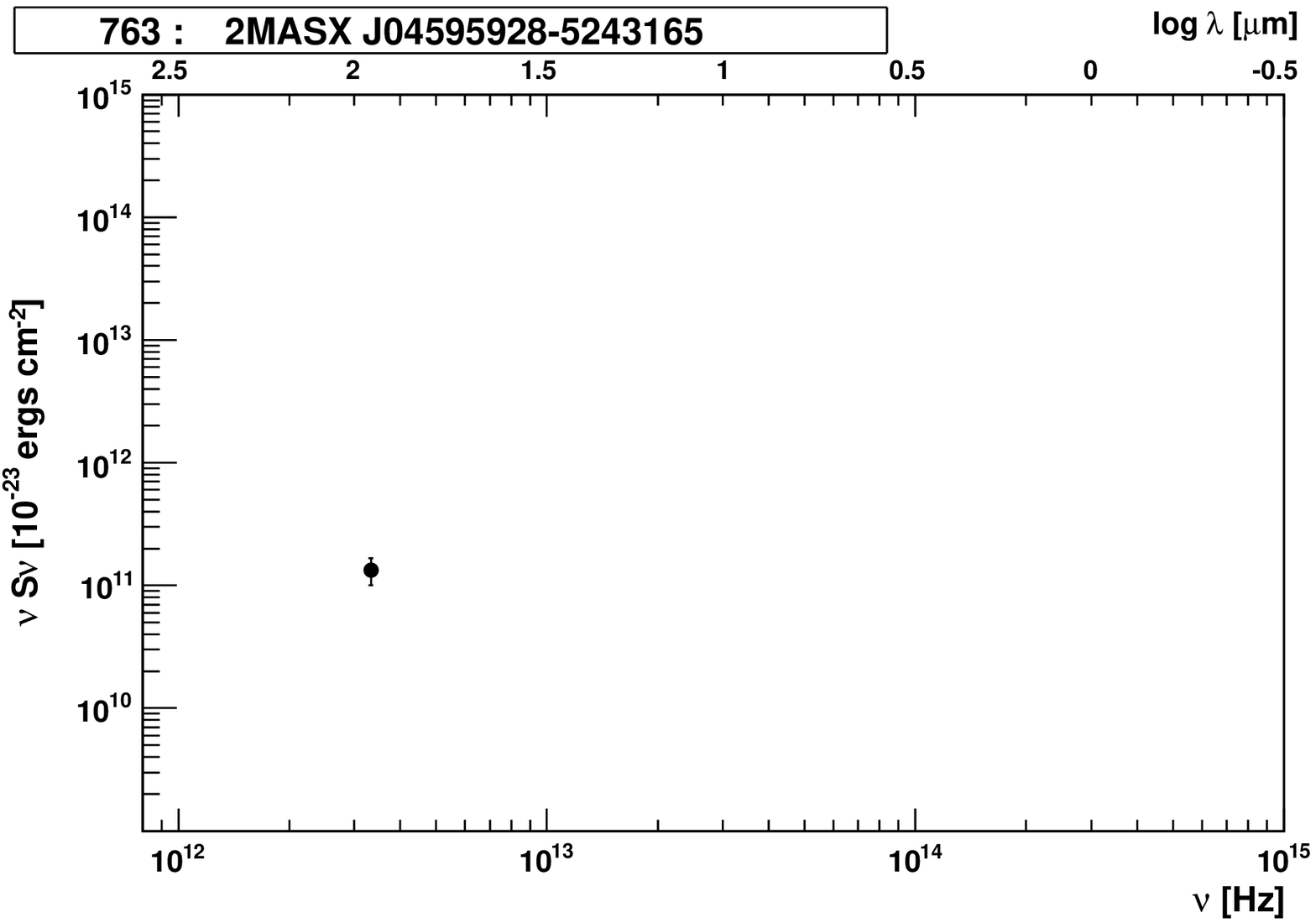}
\includegraphics[width=4cm]{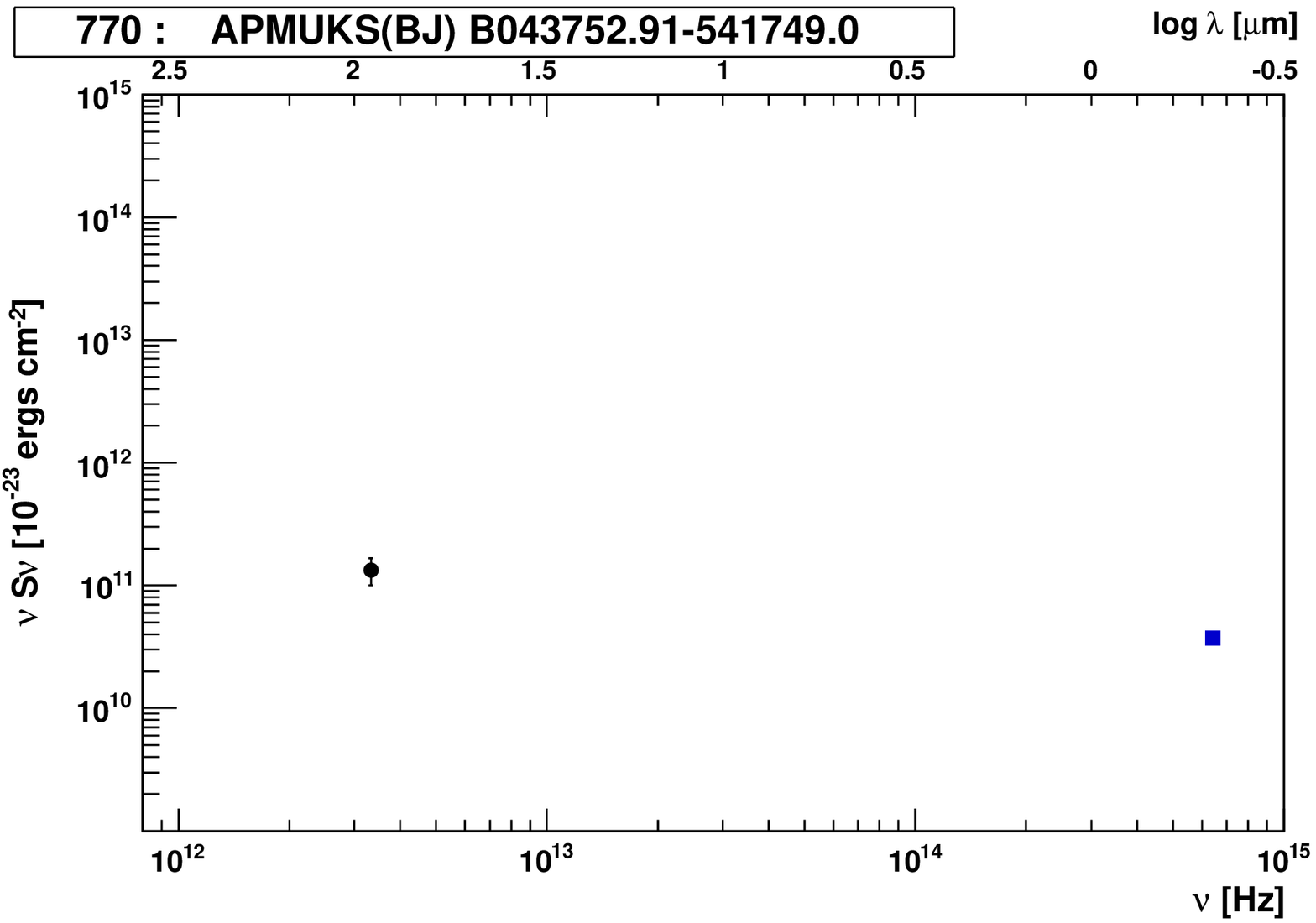}
\includegraphics[width=4cm]{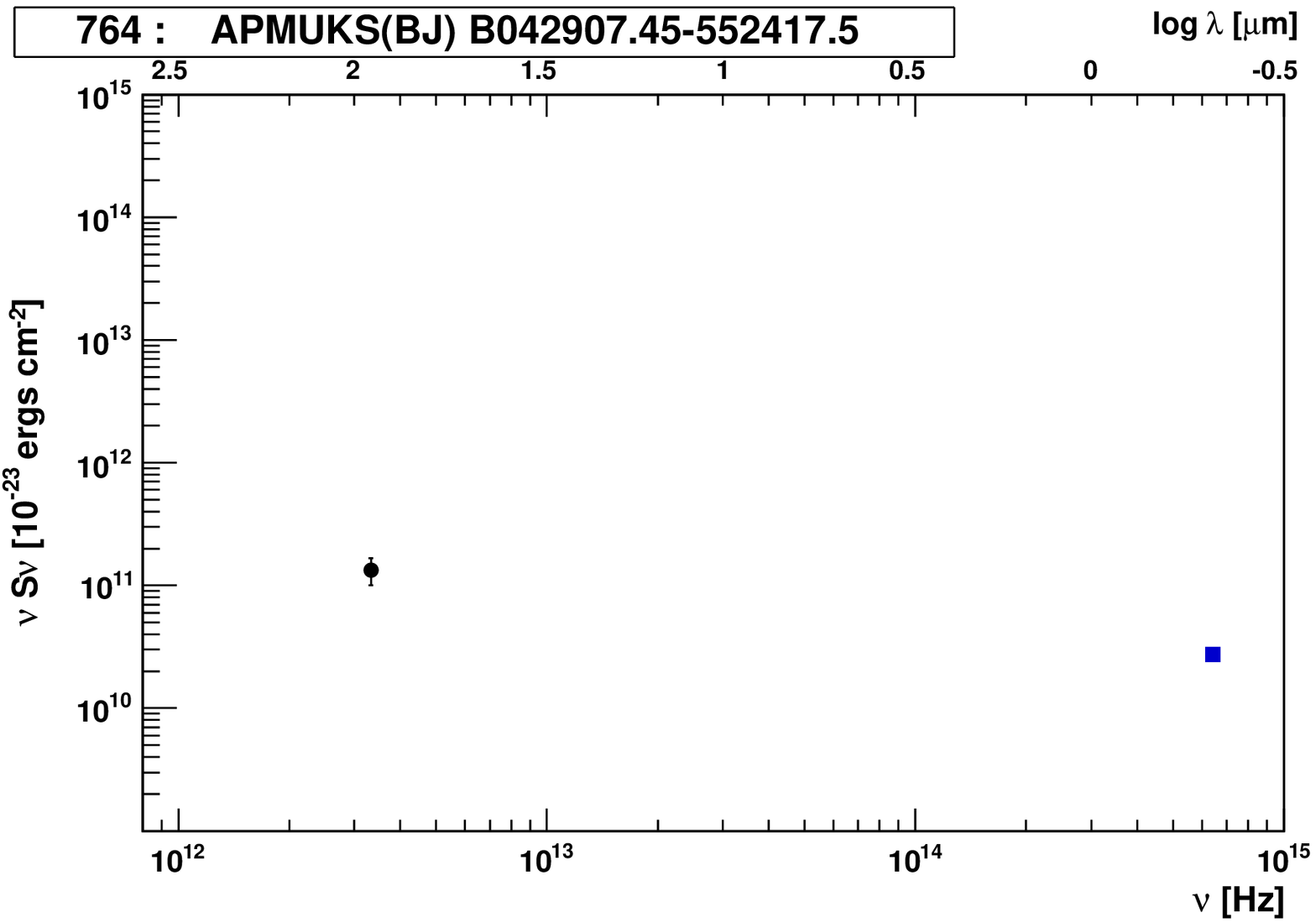}
\includegraphics[width=4cm]{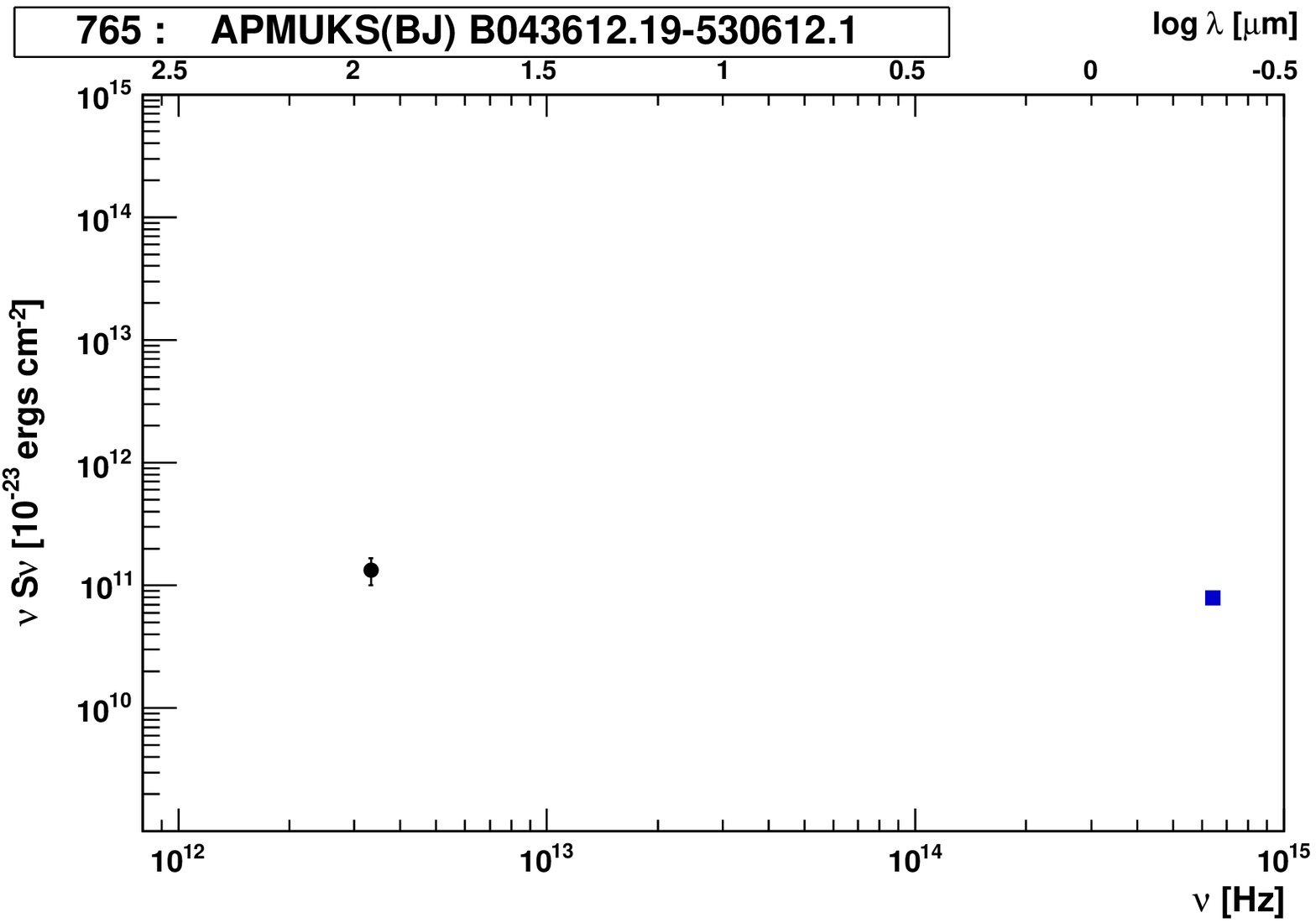}
\includegraphics[width=4cm]{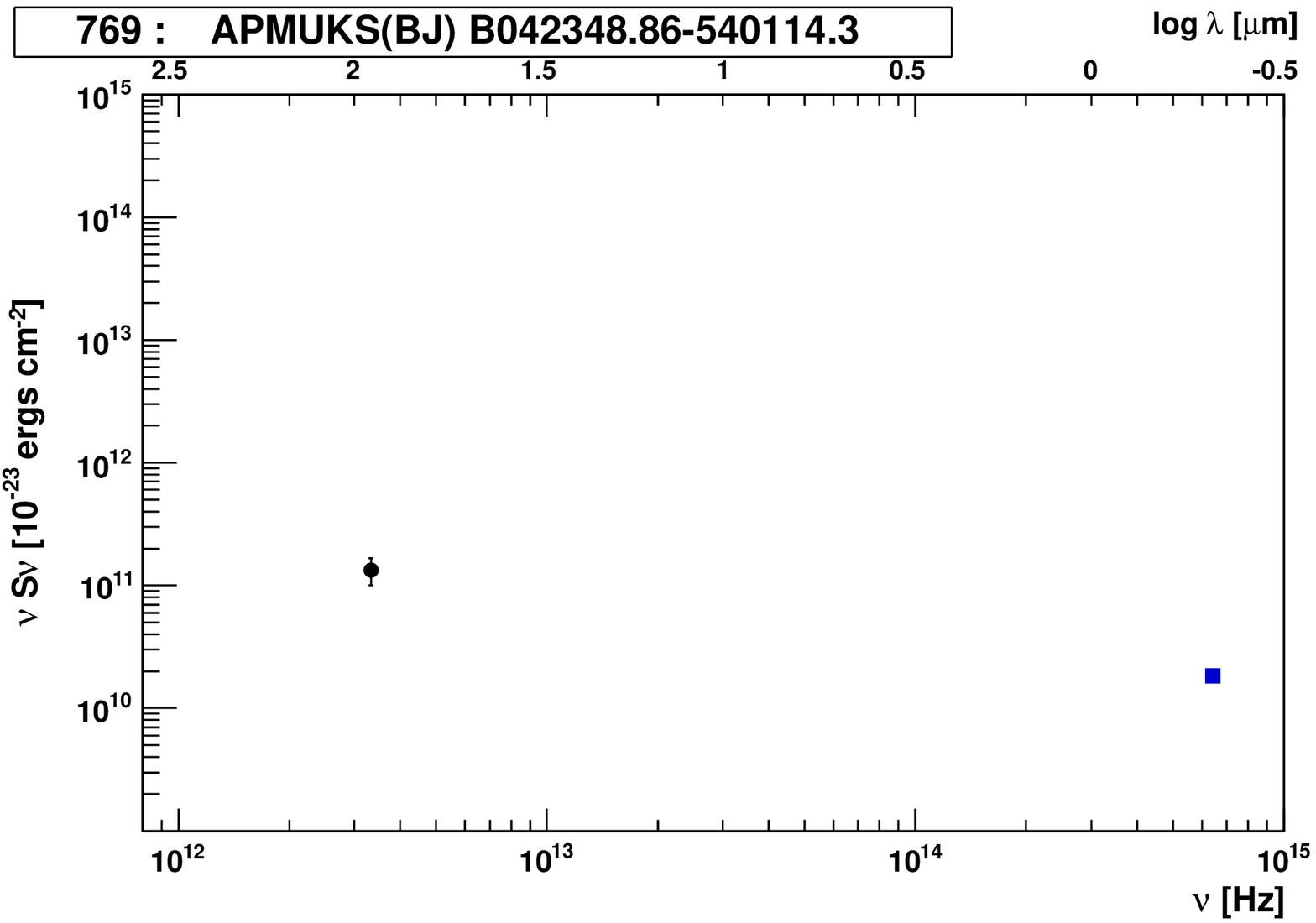}
\includegraphics[width=4cm]{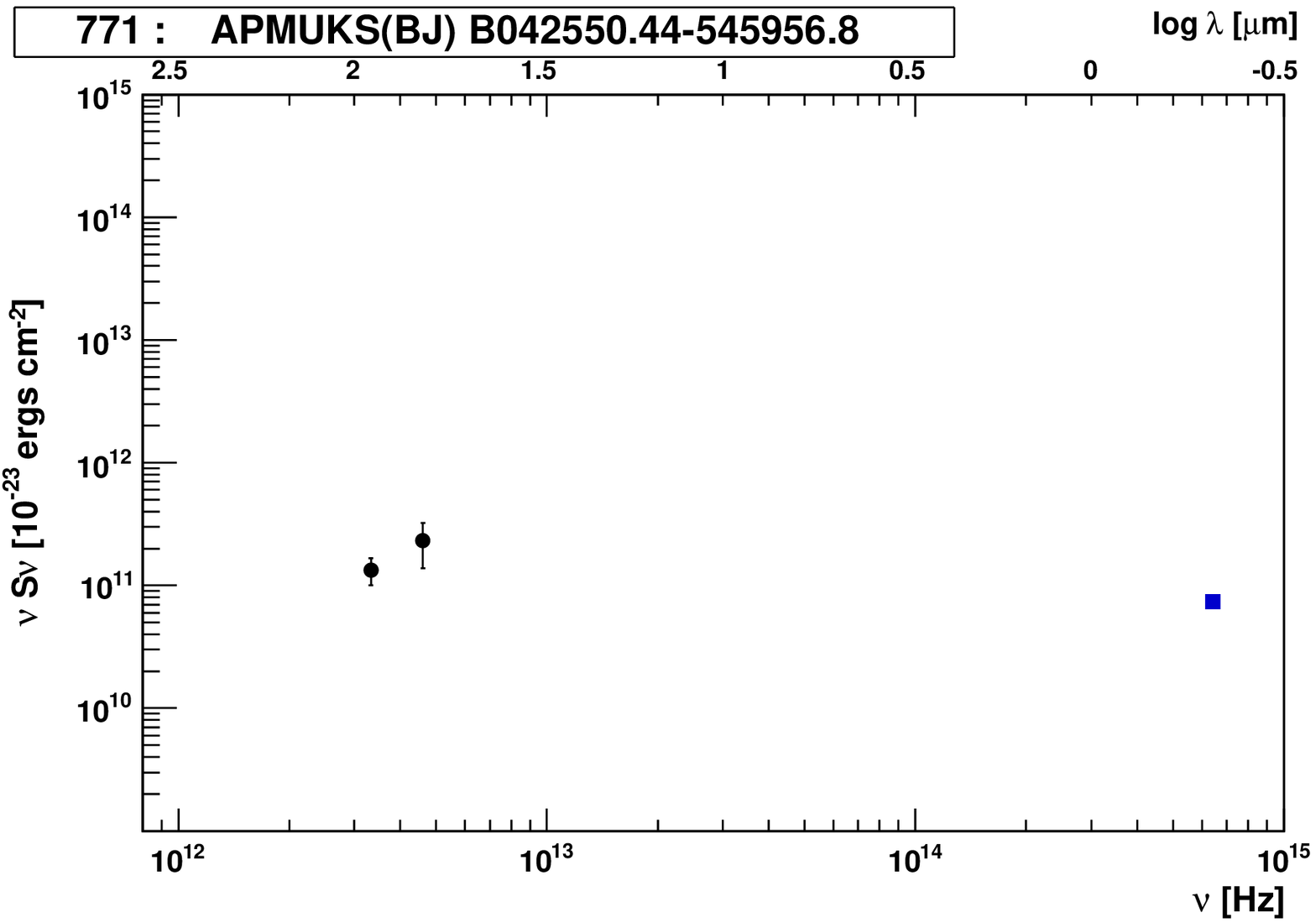}
\includegraphics[width=4cm]{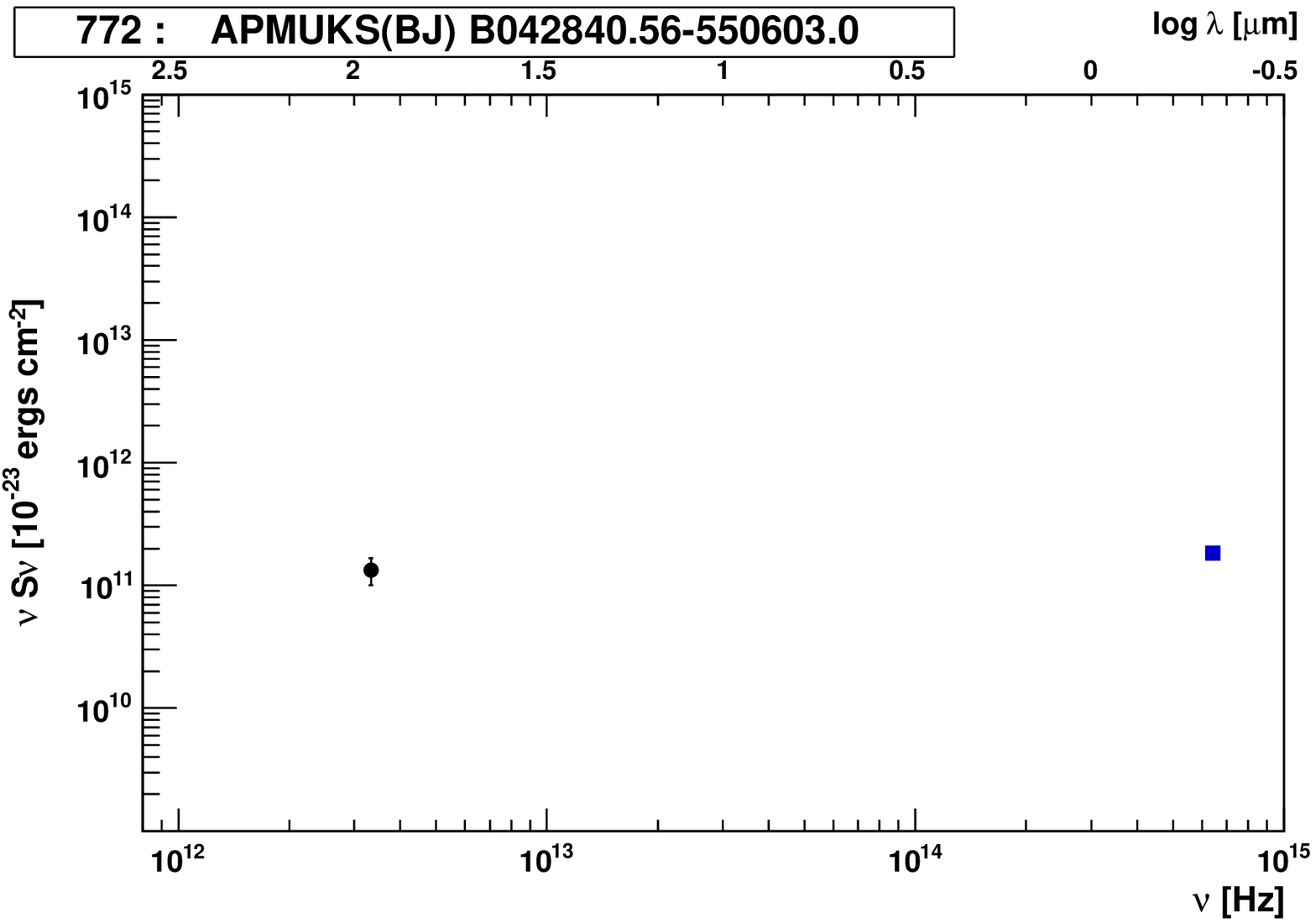}
\includegraphics[width=4cm]{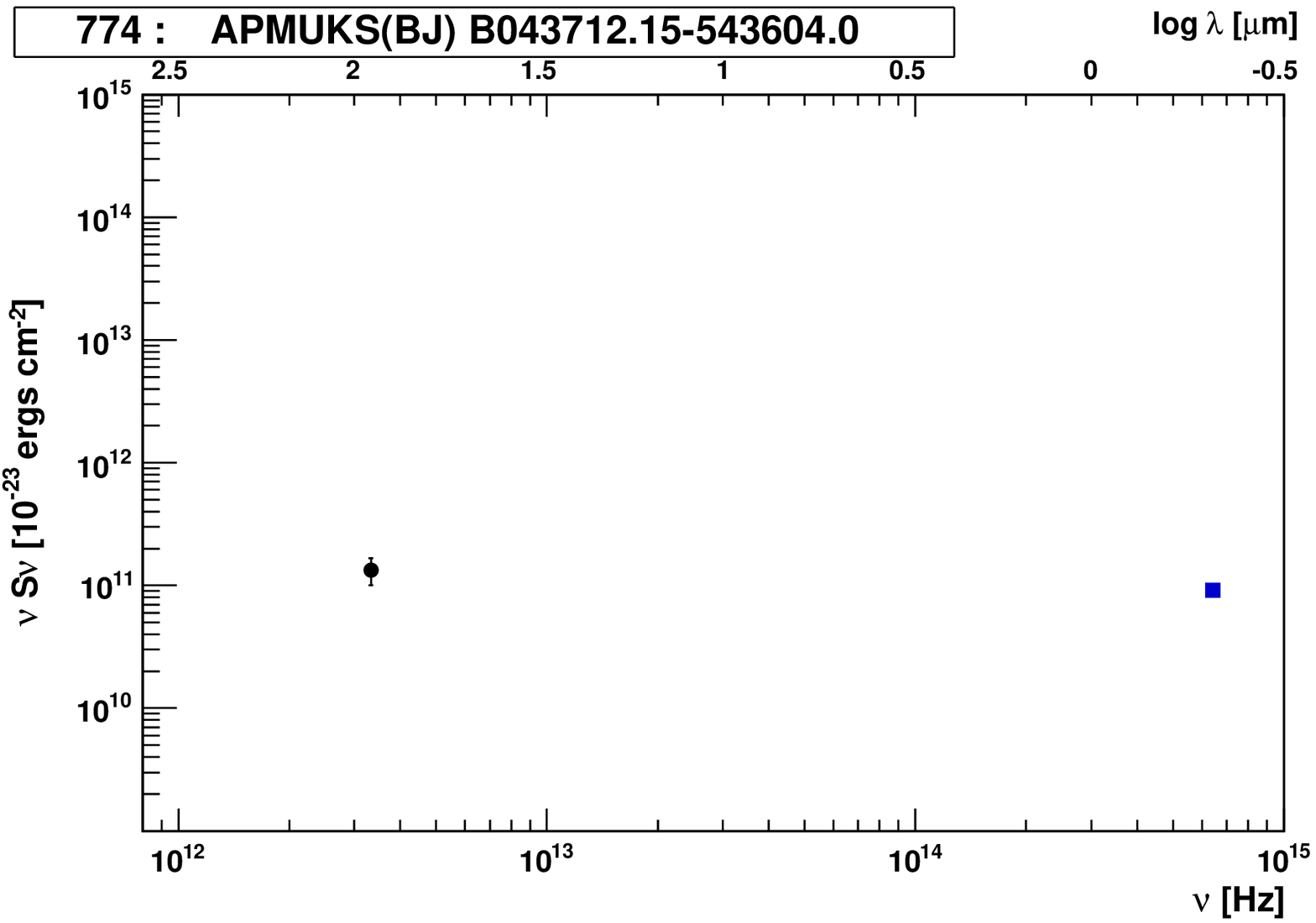}
\includegraphics[width=4cm]{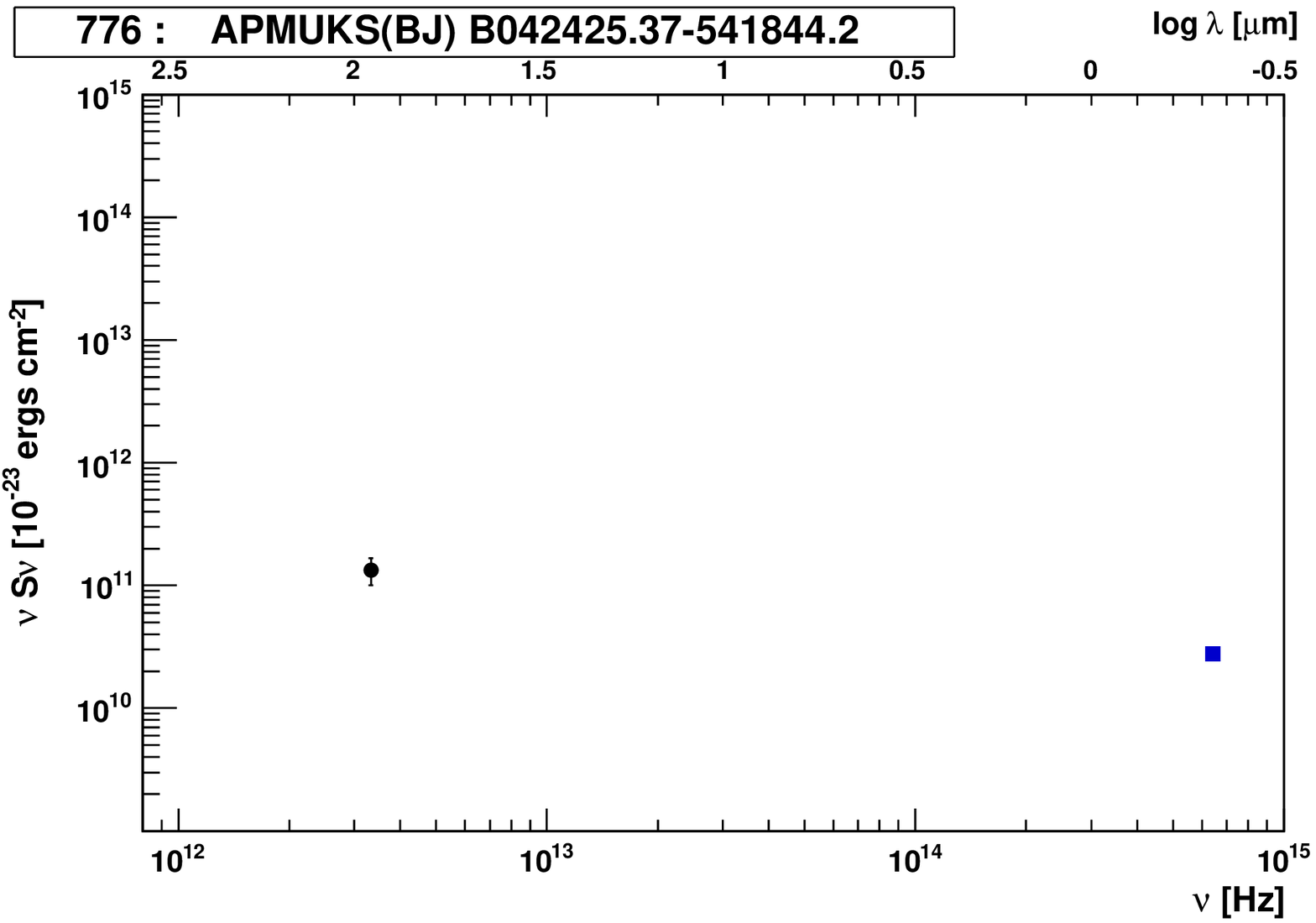}
\includegraphics[width=4cm]{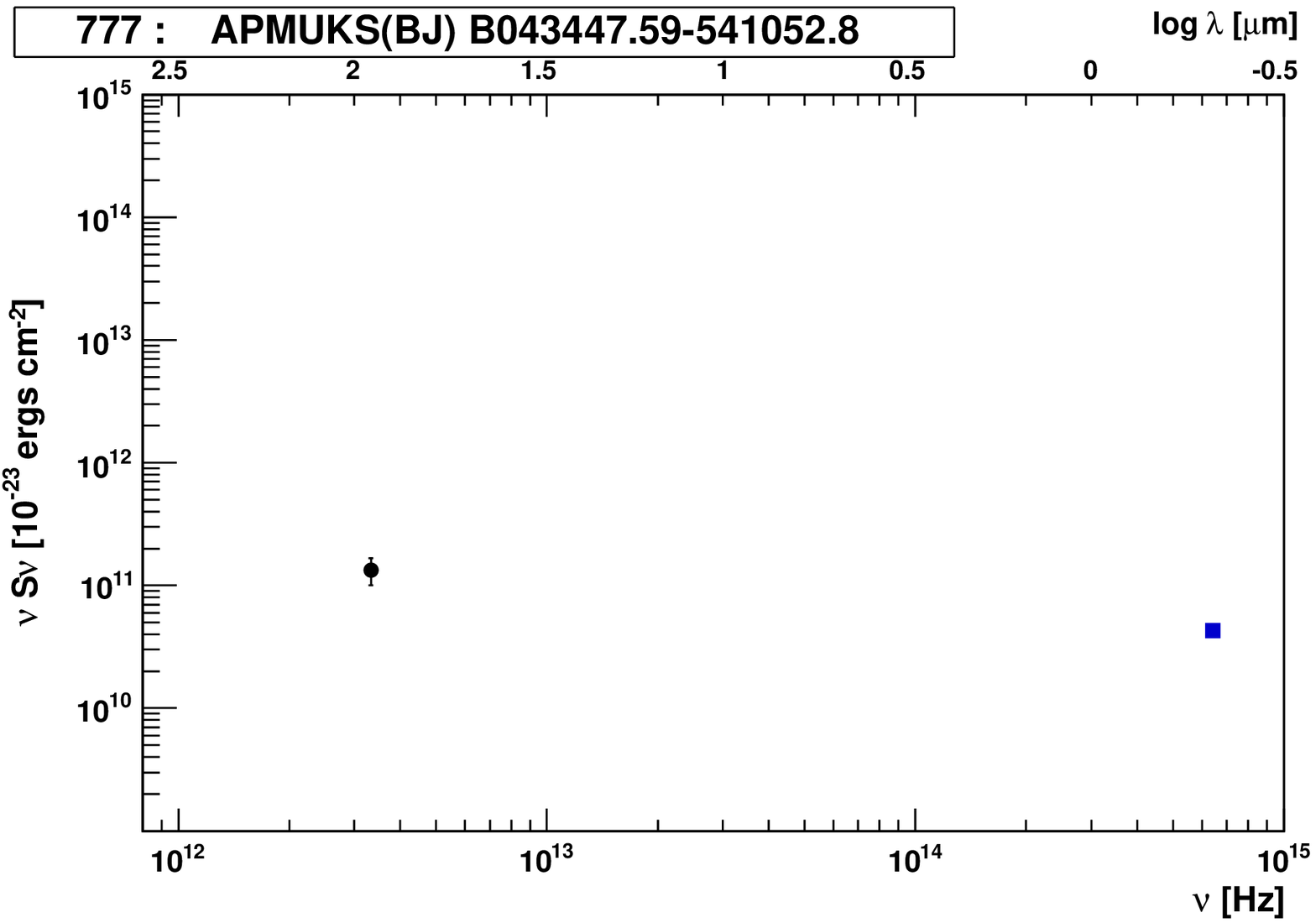}
\includegraphics[width=4cm]{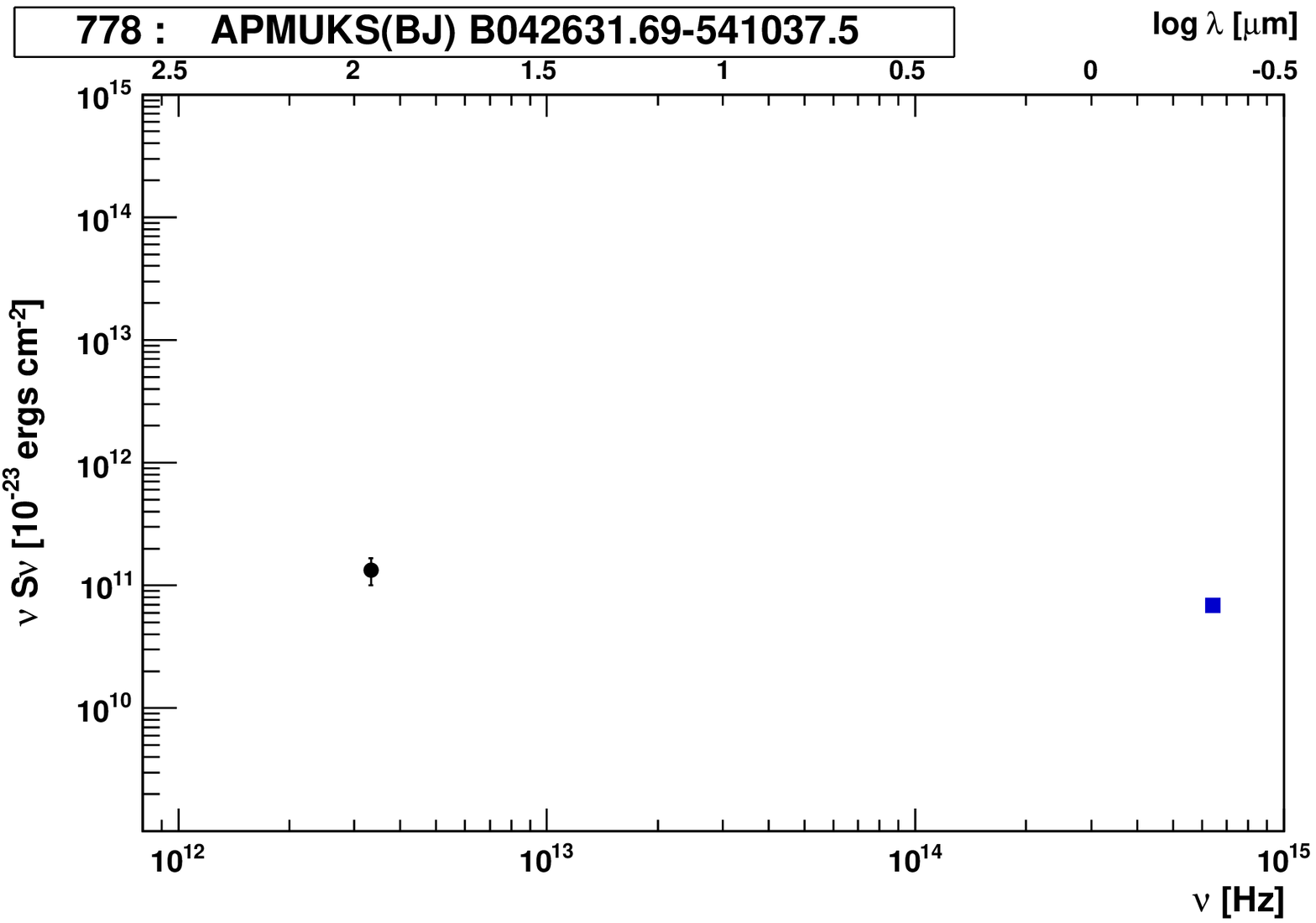}
\includegraphics[width=4cm]{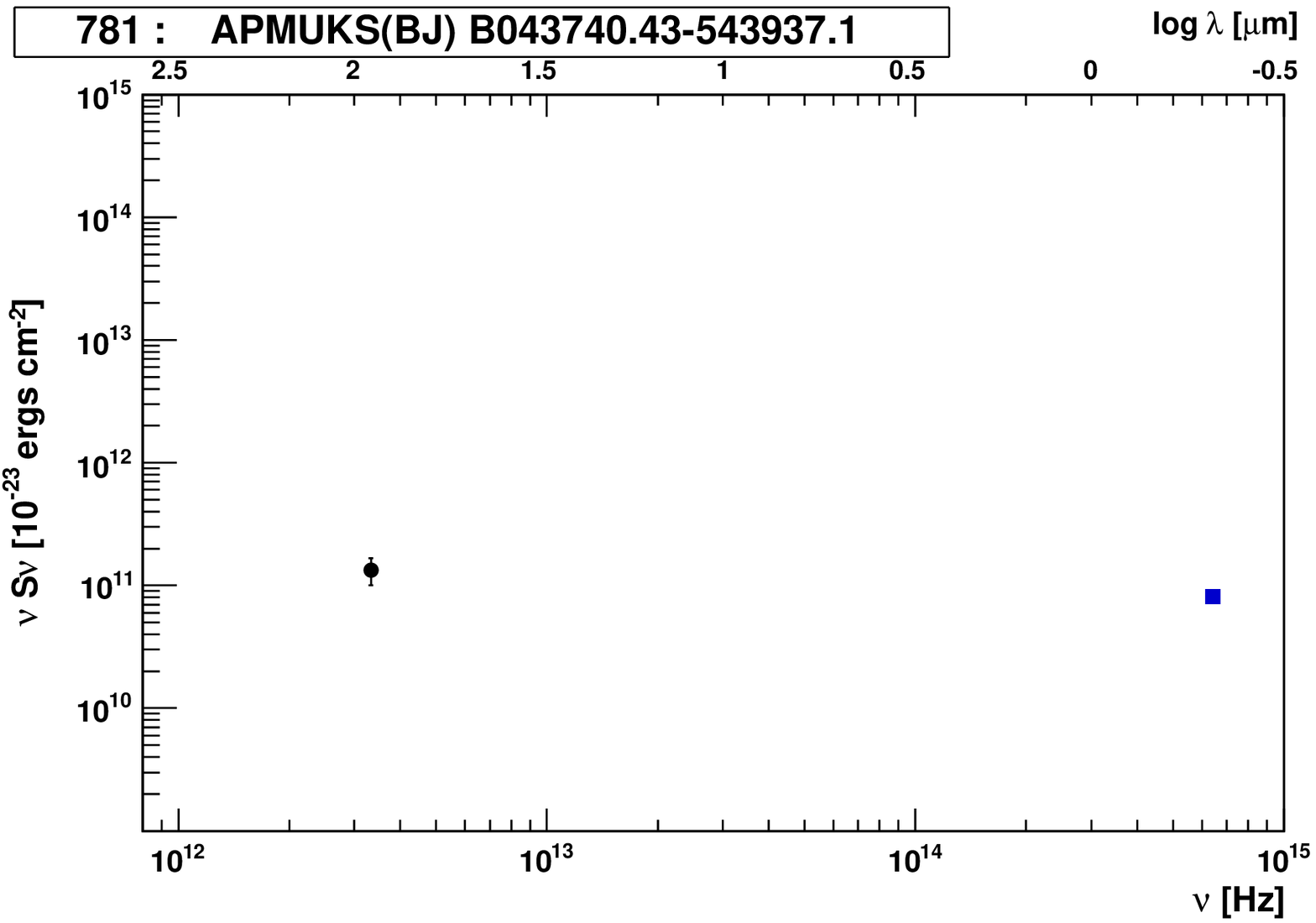}
\includegraphics[width=4cm]{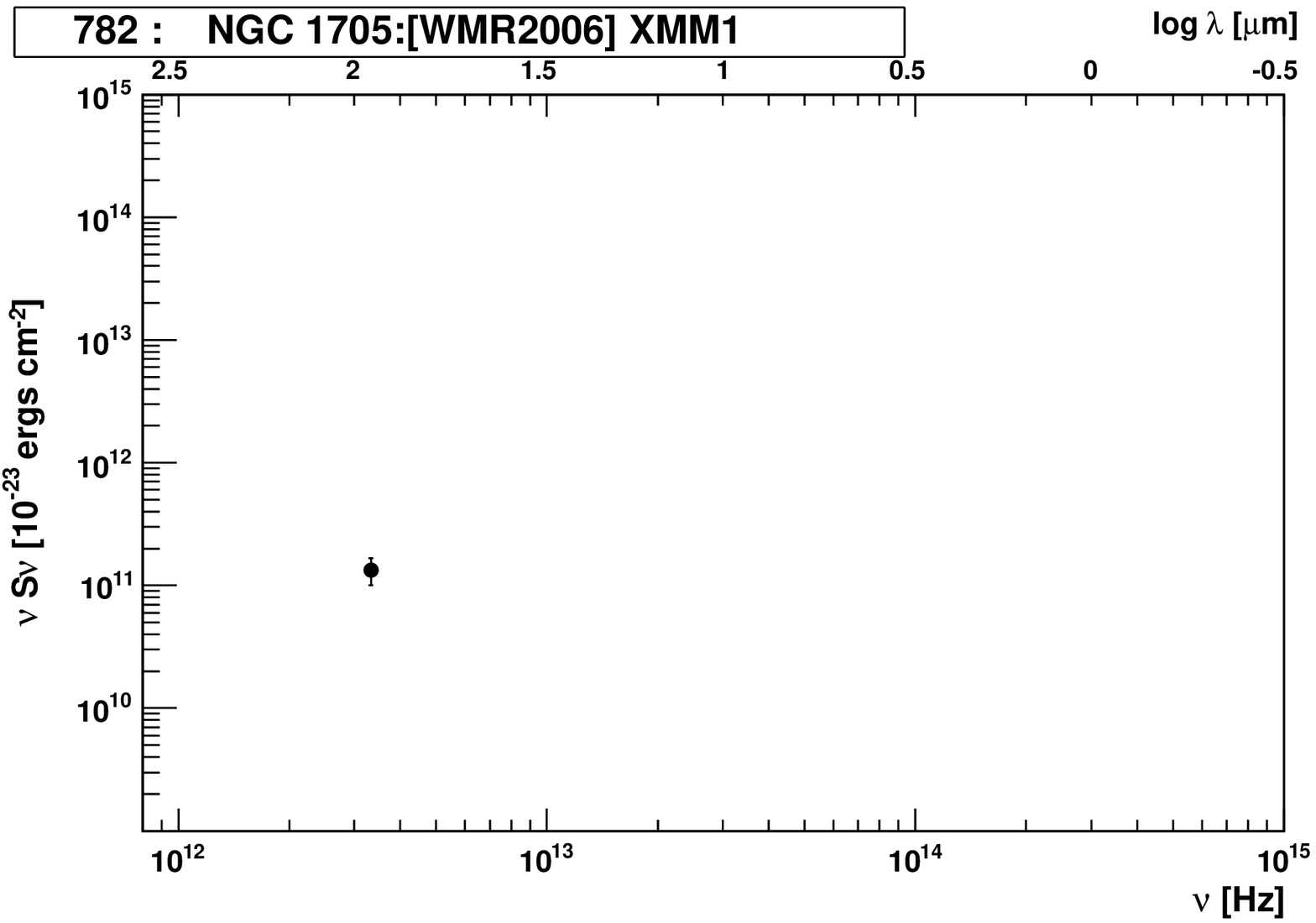}
\includegraphics[width=4cm]{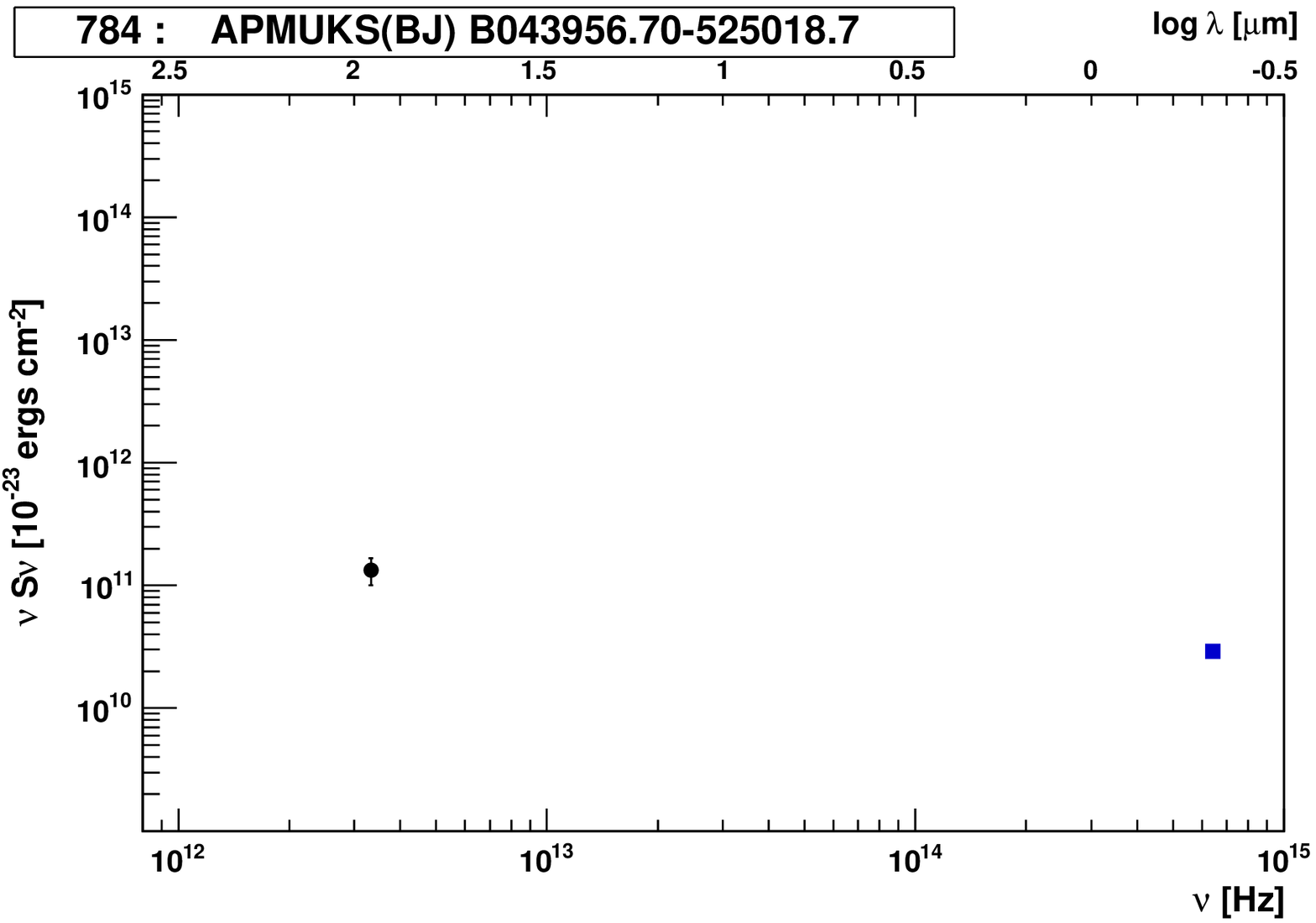}
\includegraphics[width=4cm]{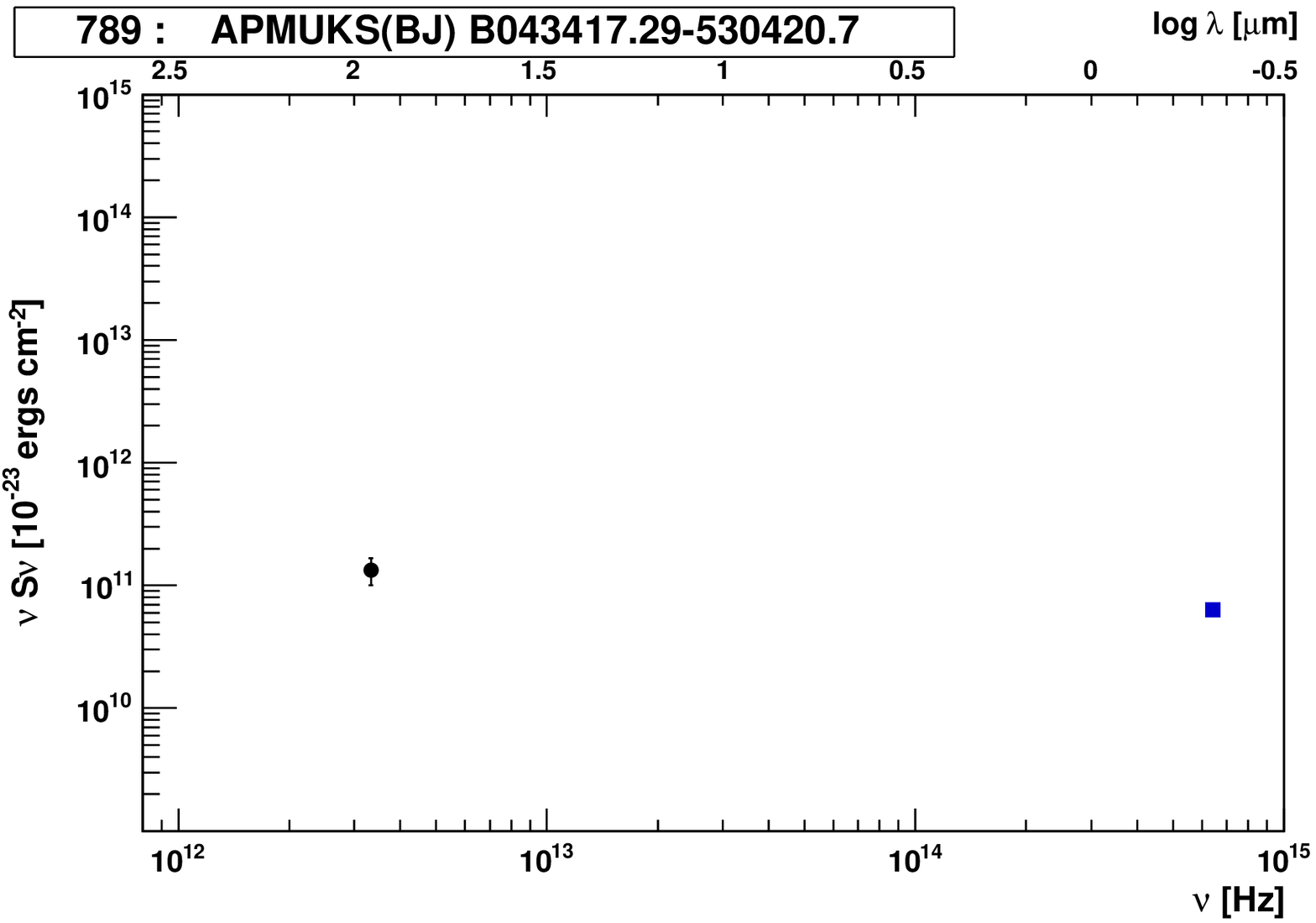}
\includegraphics[width=4cm]{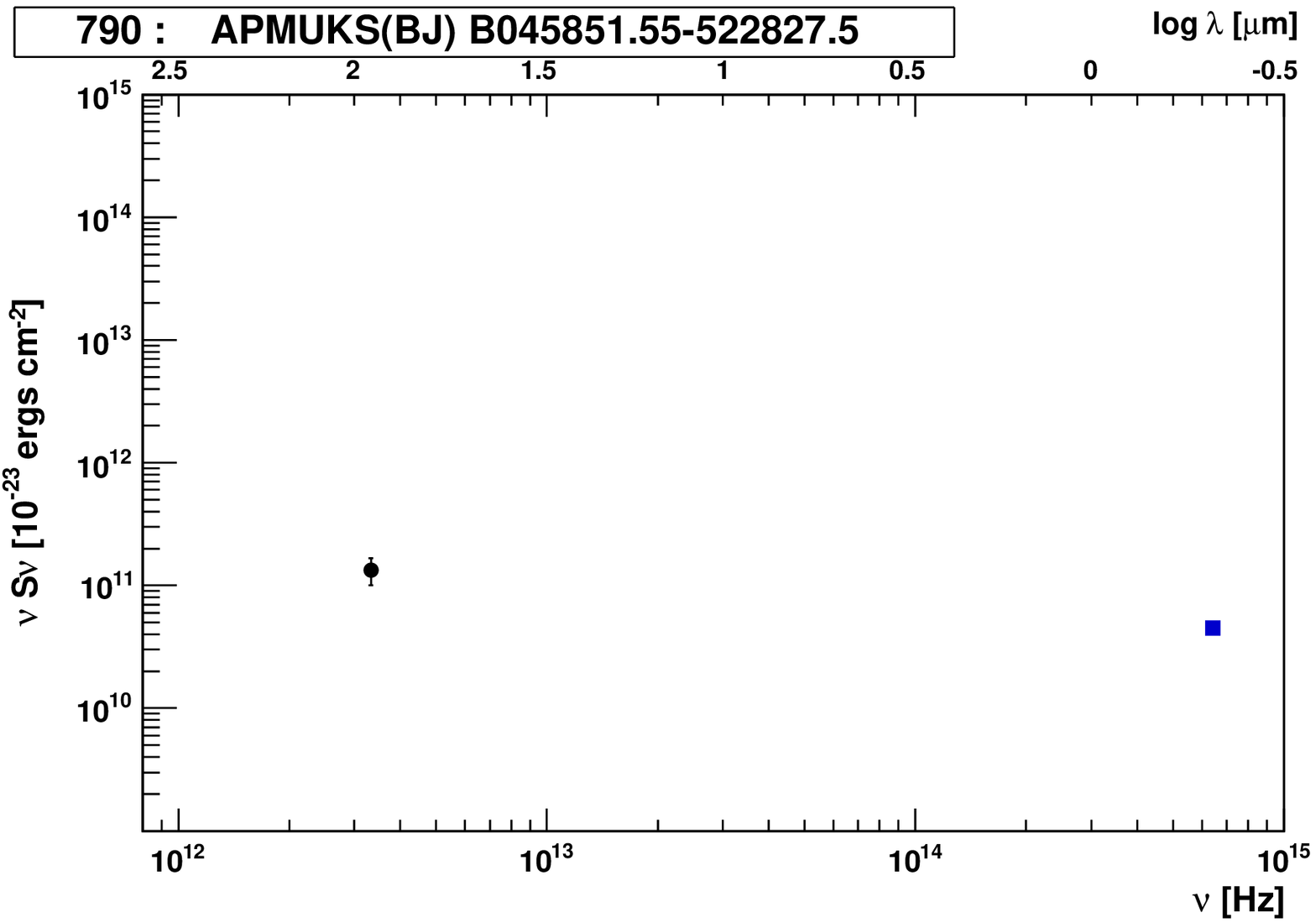}
\includegraphics[width=4cm]{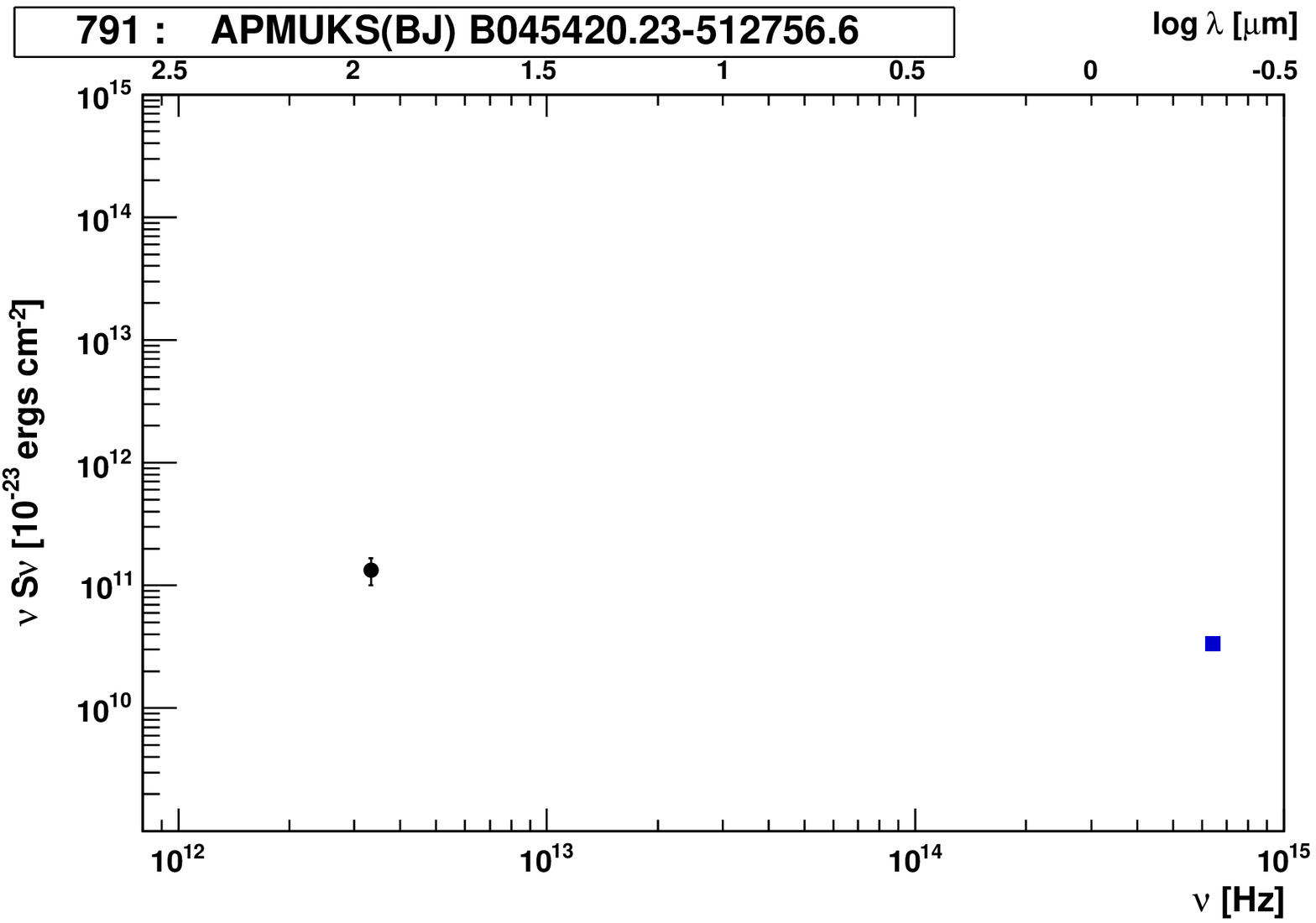}
\includegraphics[width=4cm]{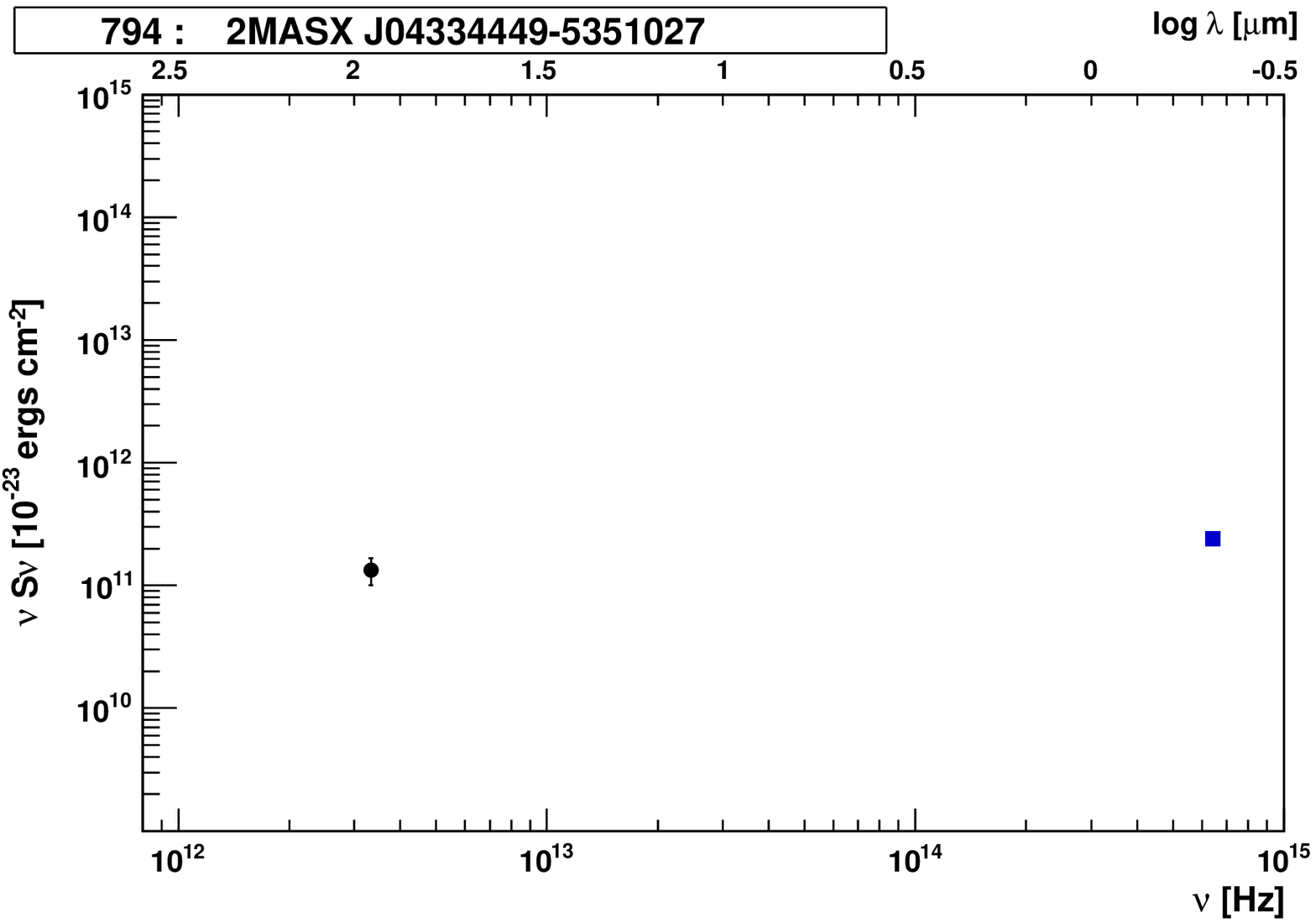}
\includegraphics[width=4cm]{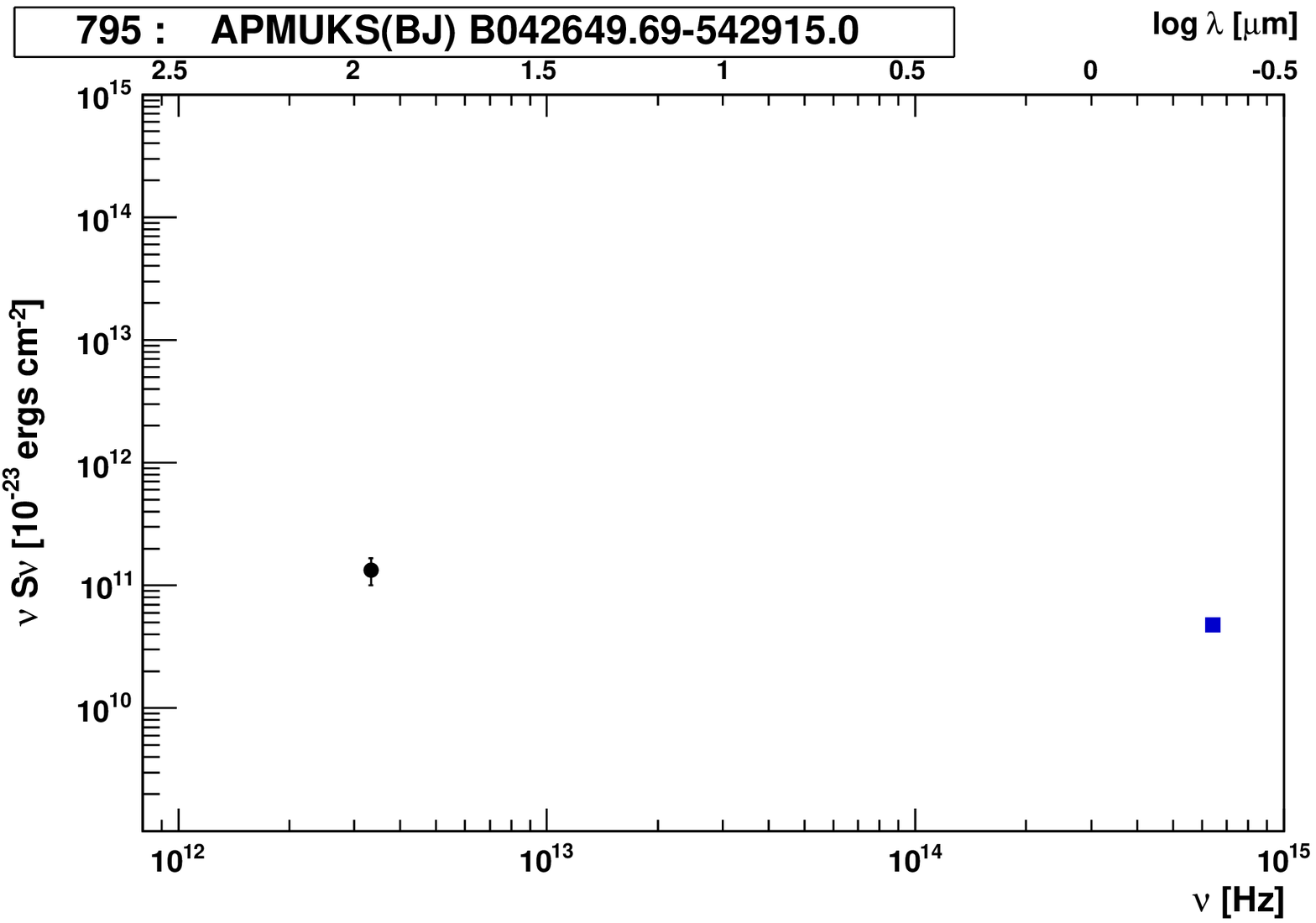}
\includegraphics[width=4cm]{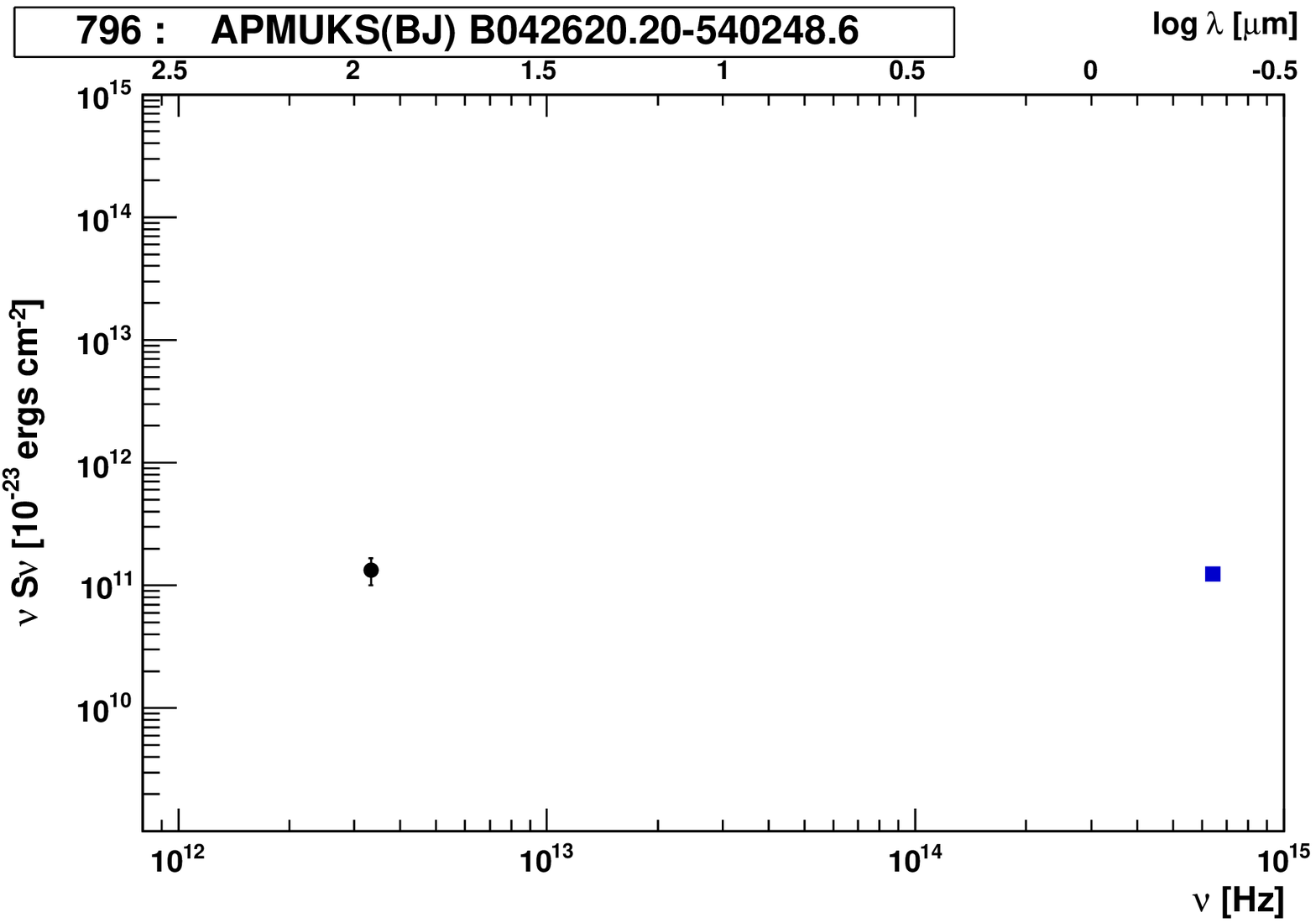}
\includegraphics[width=4cm]{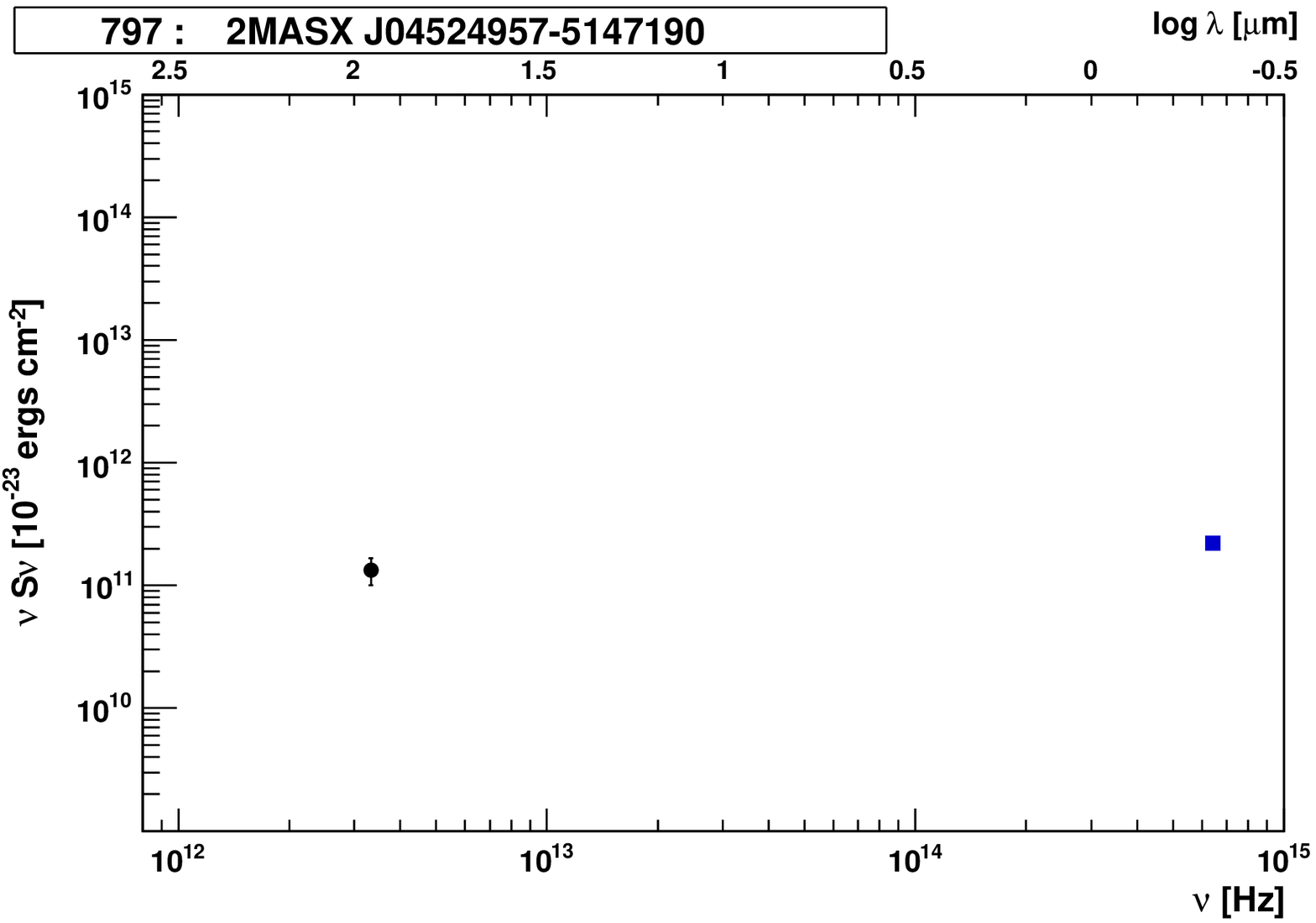}
\includegraphics[width=4cm]{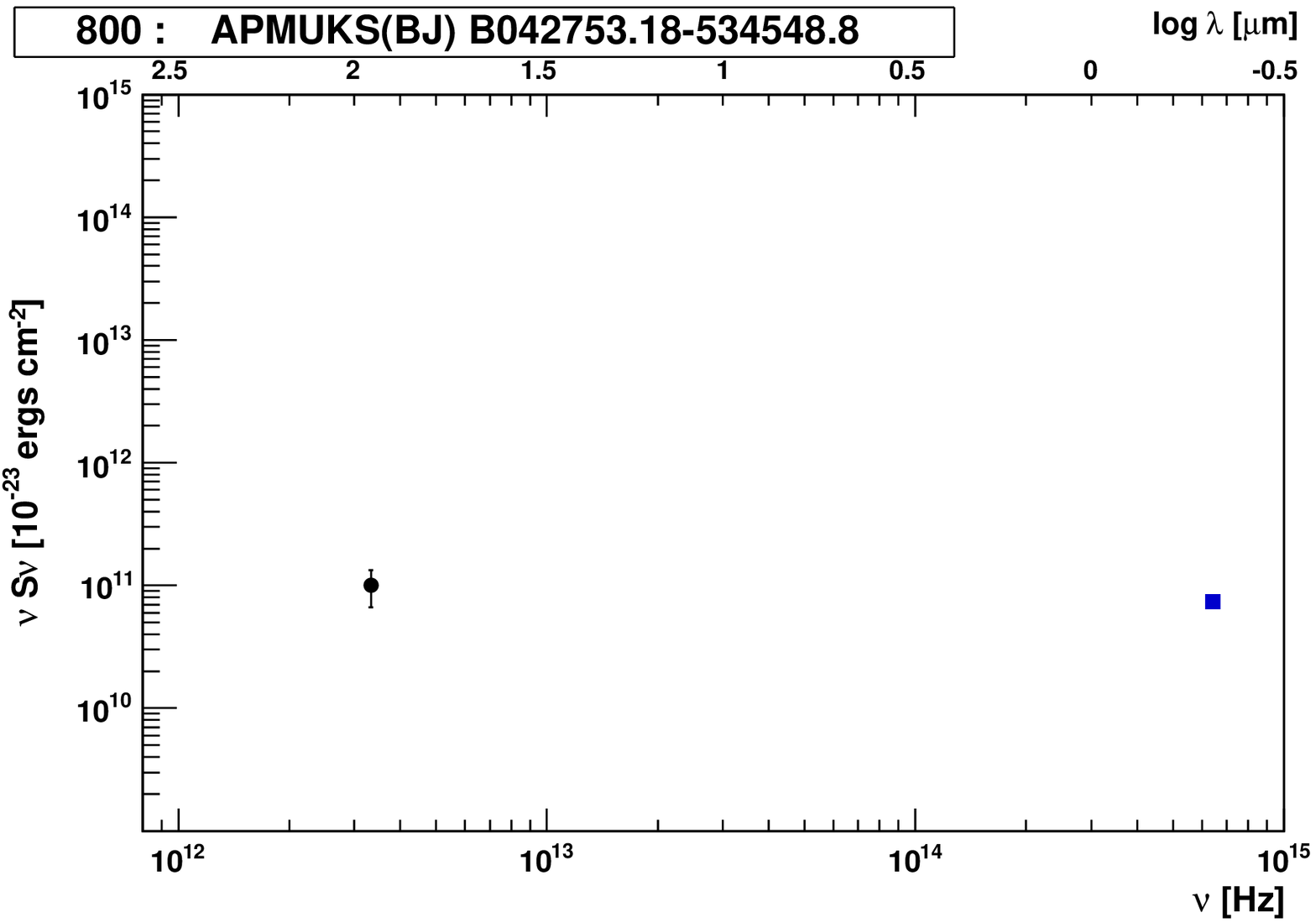}
\includegraphics[width=4cm]{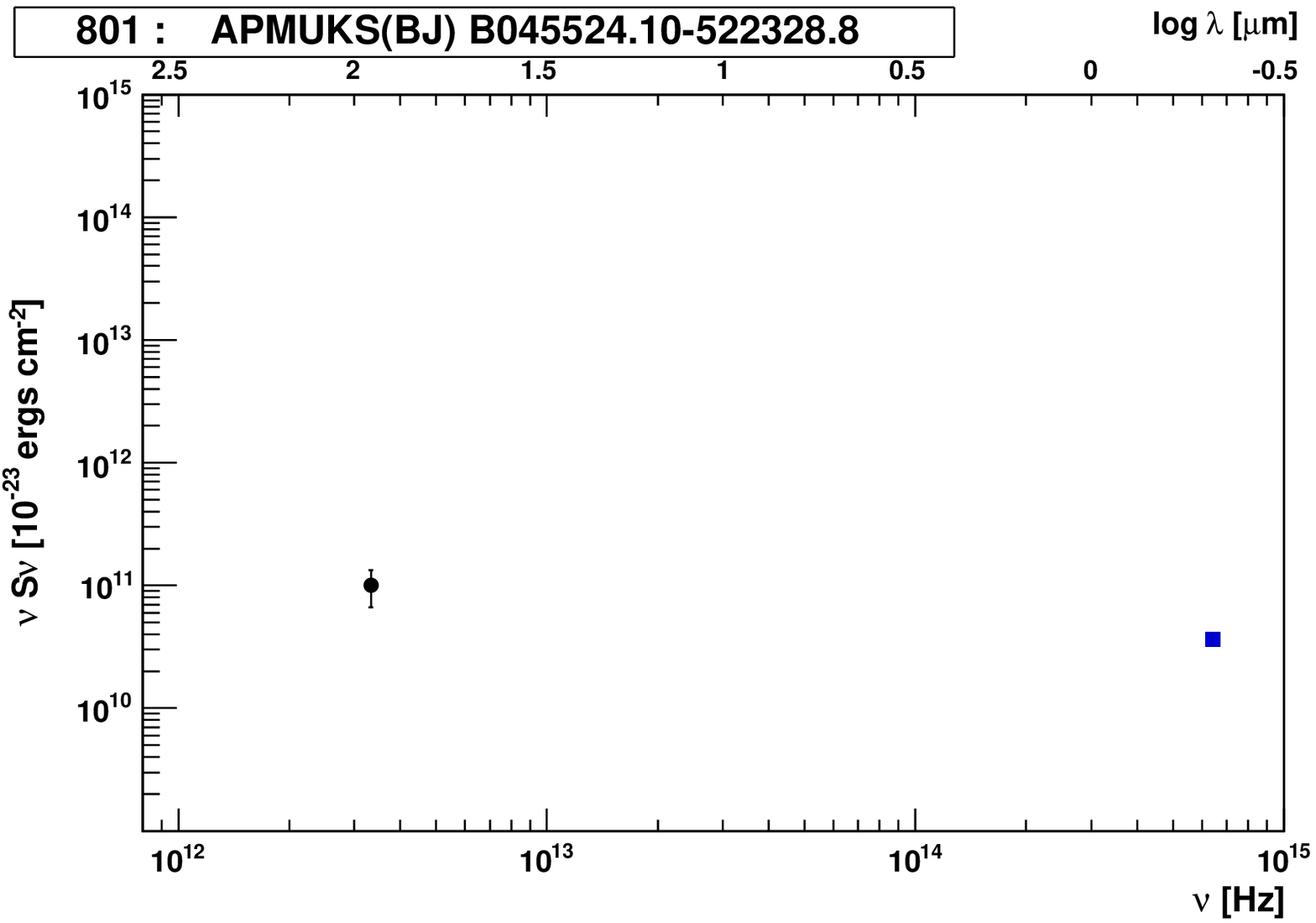}
\includegraphics[width=4cm]{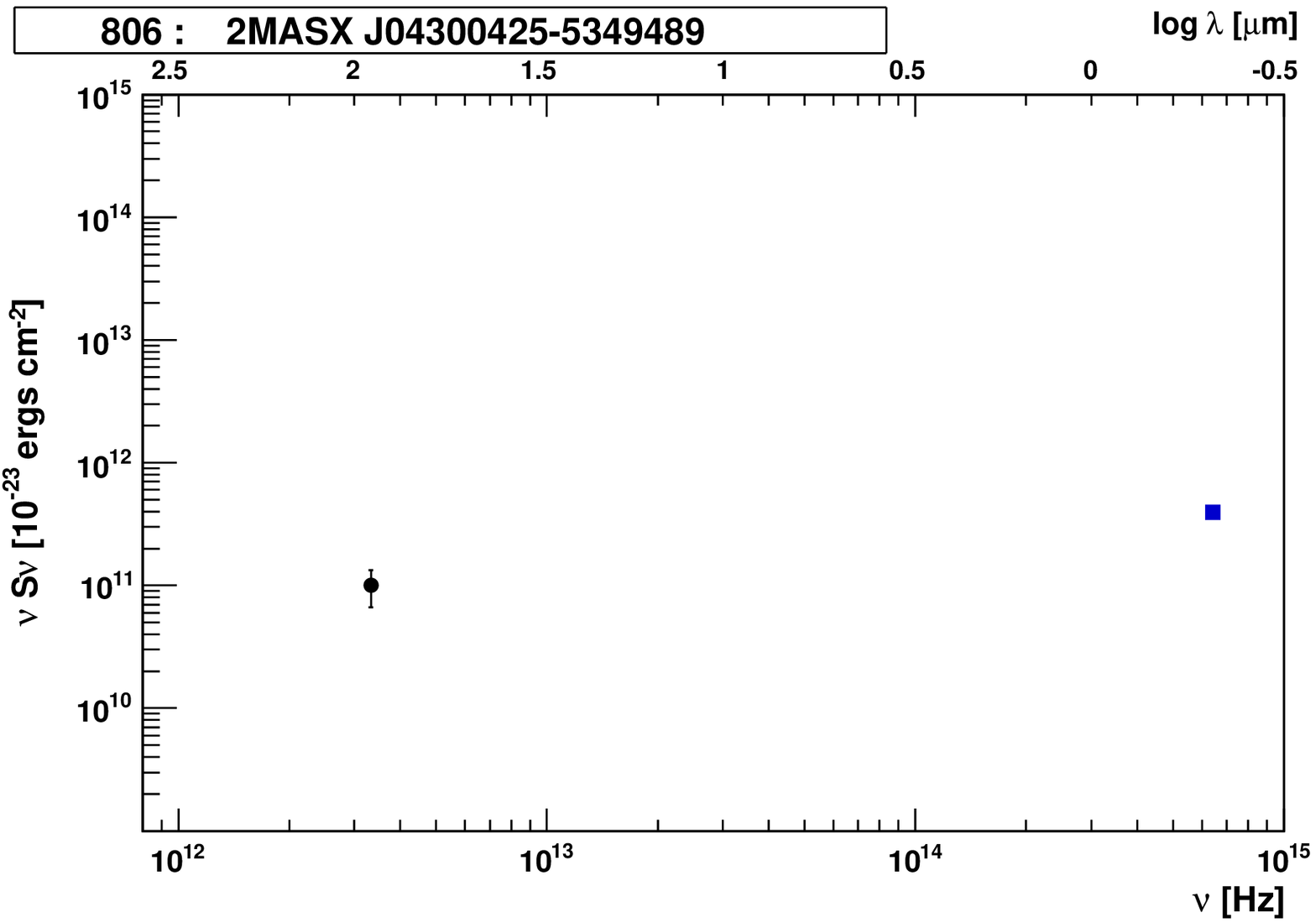}
\includegraphics[width=4cm]{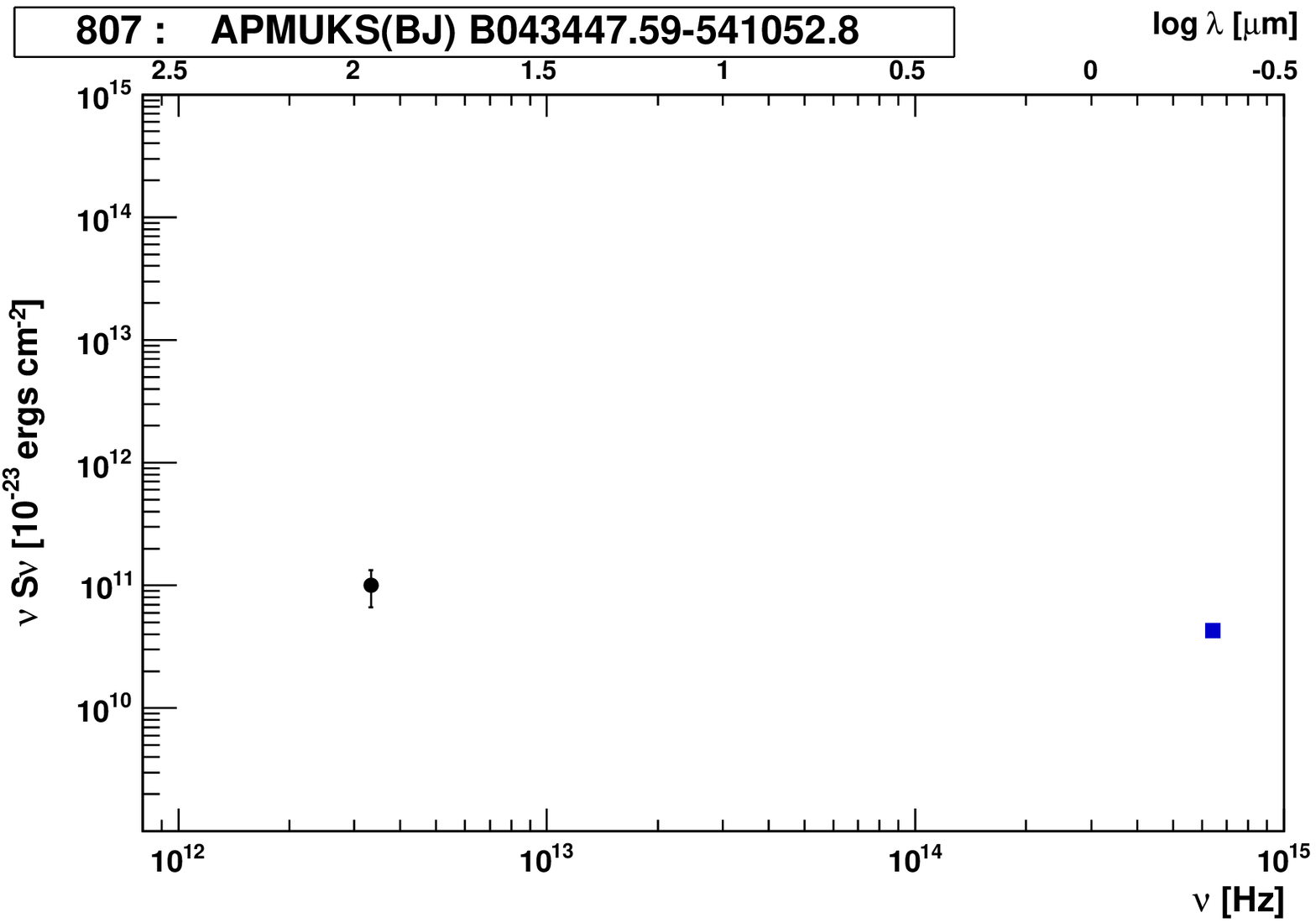}
\includegraphics[width=4cm]{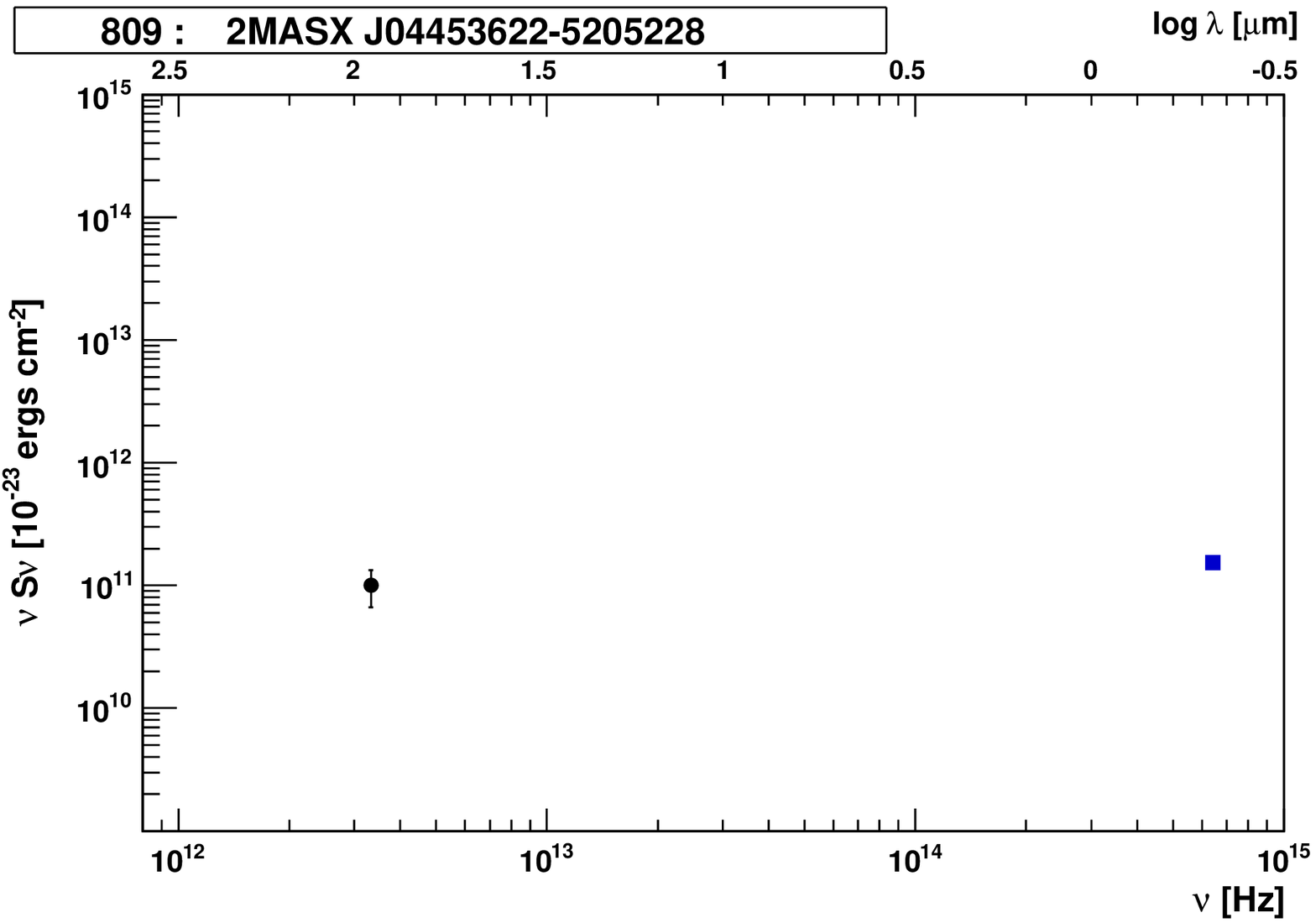}
\includegraphics[width=4cm]{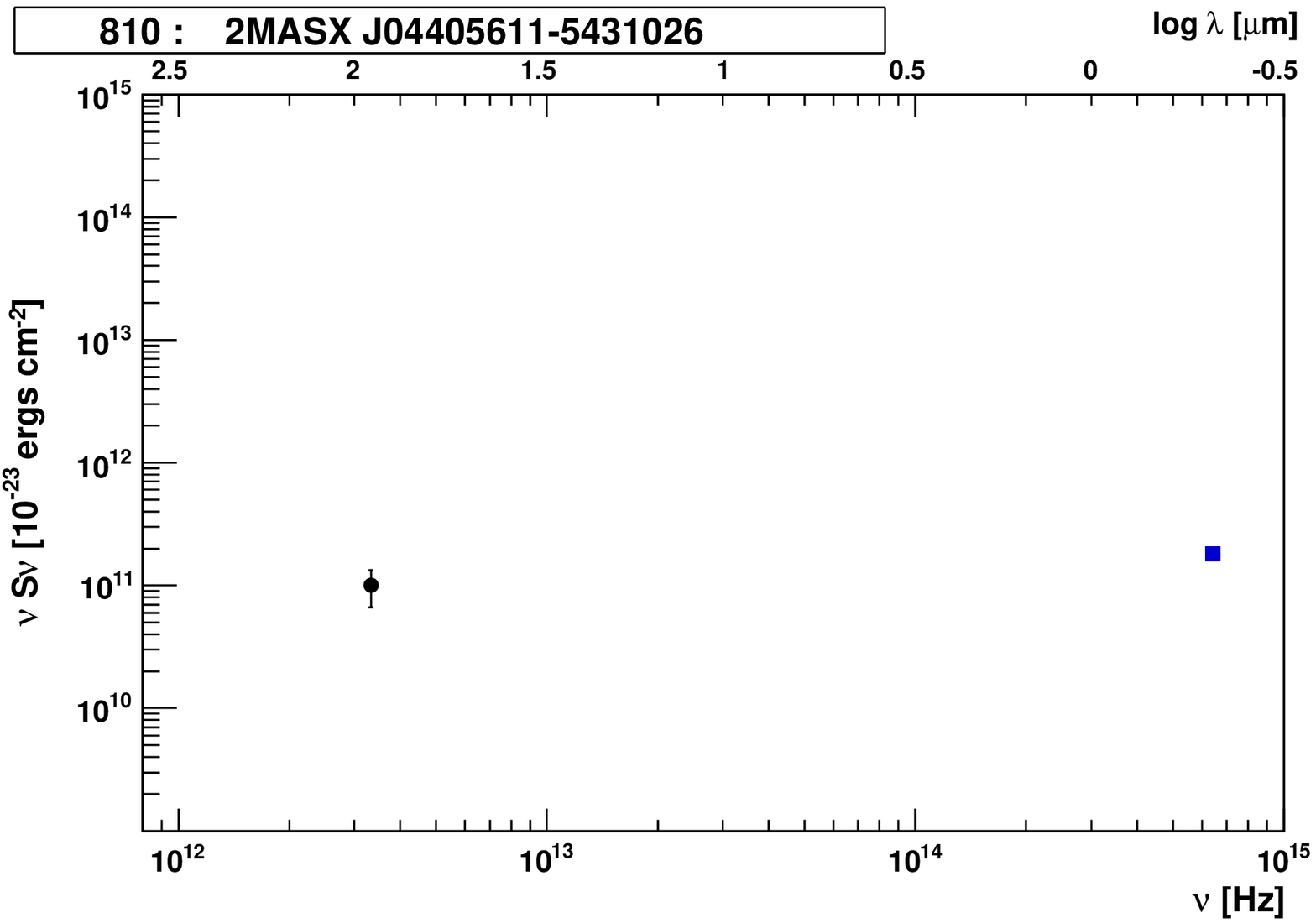}
\includegraphics[width=4cm]{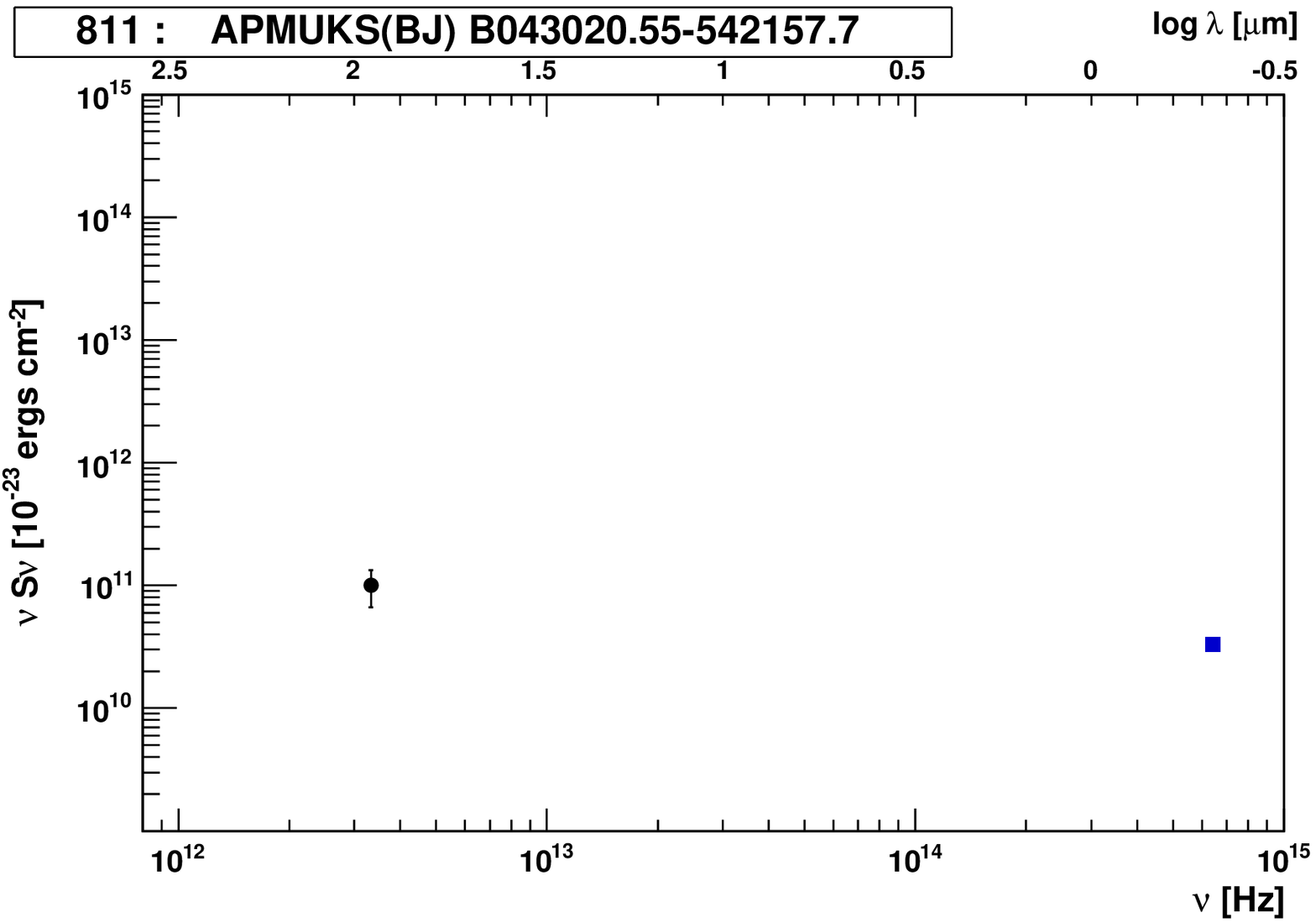}
\includegraphics[width=4cm]{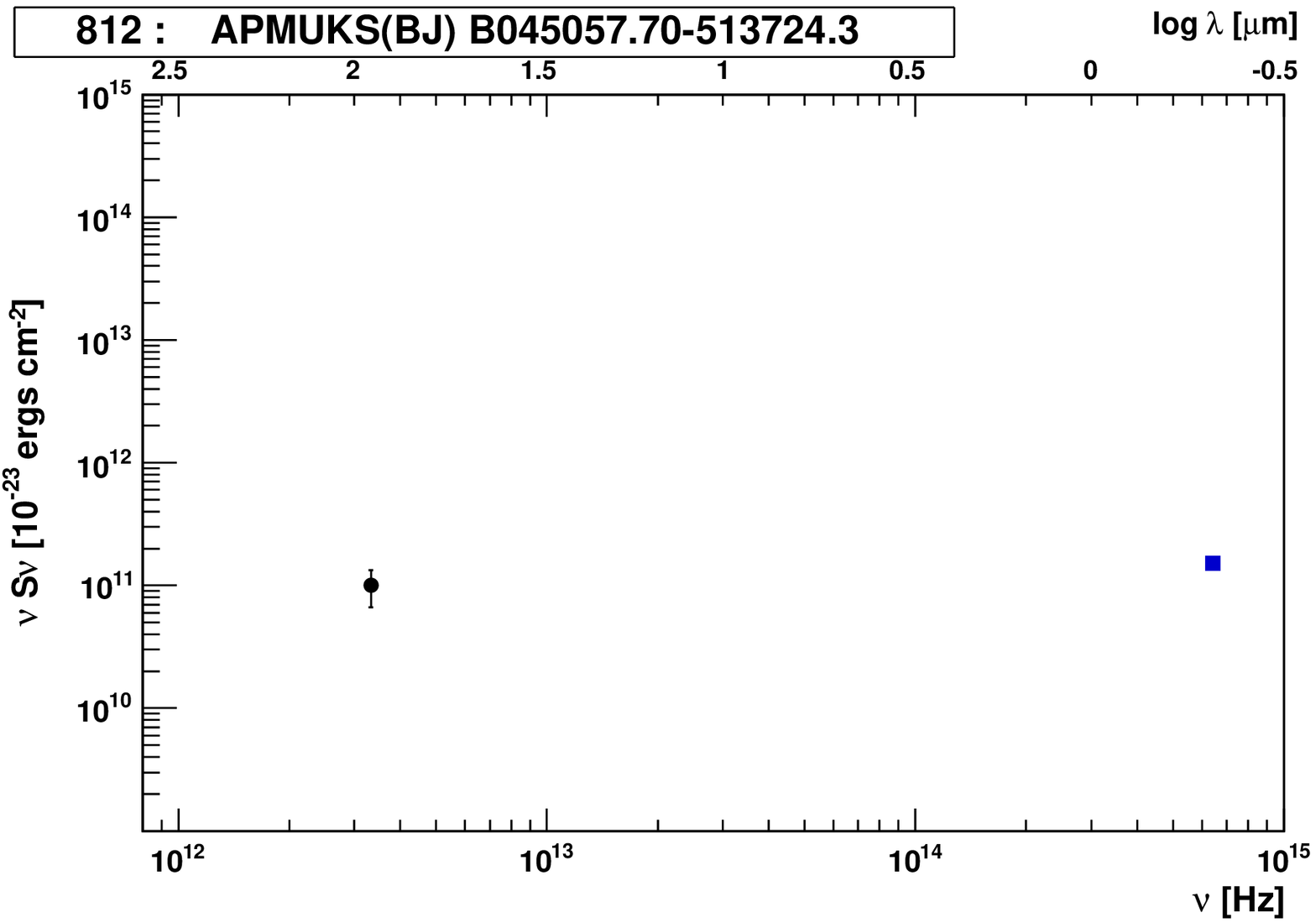}
\label{points13}
\caption {SEDs for the next 36 ADF-S identified sources, with symbols as in Figure~\ref{points1}.}
\end{figure*}
}

\clearpage

\onlfig{14}{
\begin{figure*}[t]
\centering

\includegraphics[width=4cm]{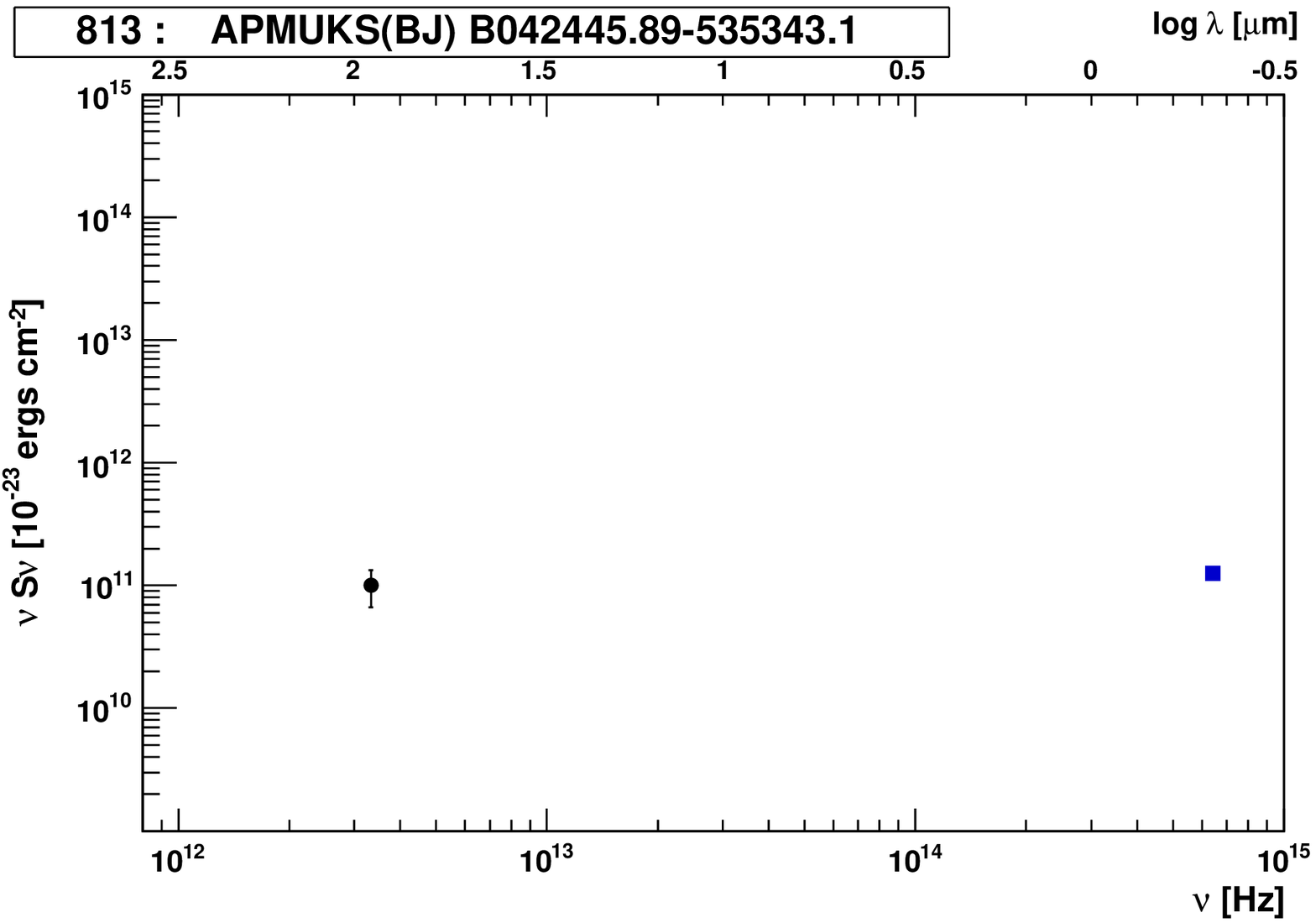}
\includegraphics[width=4cm]{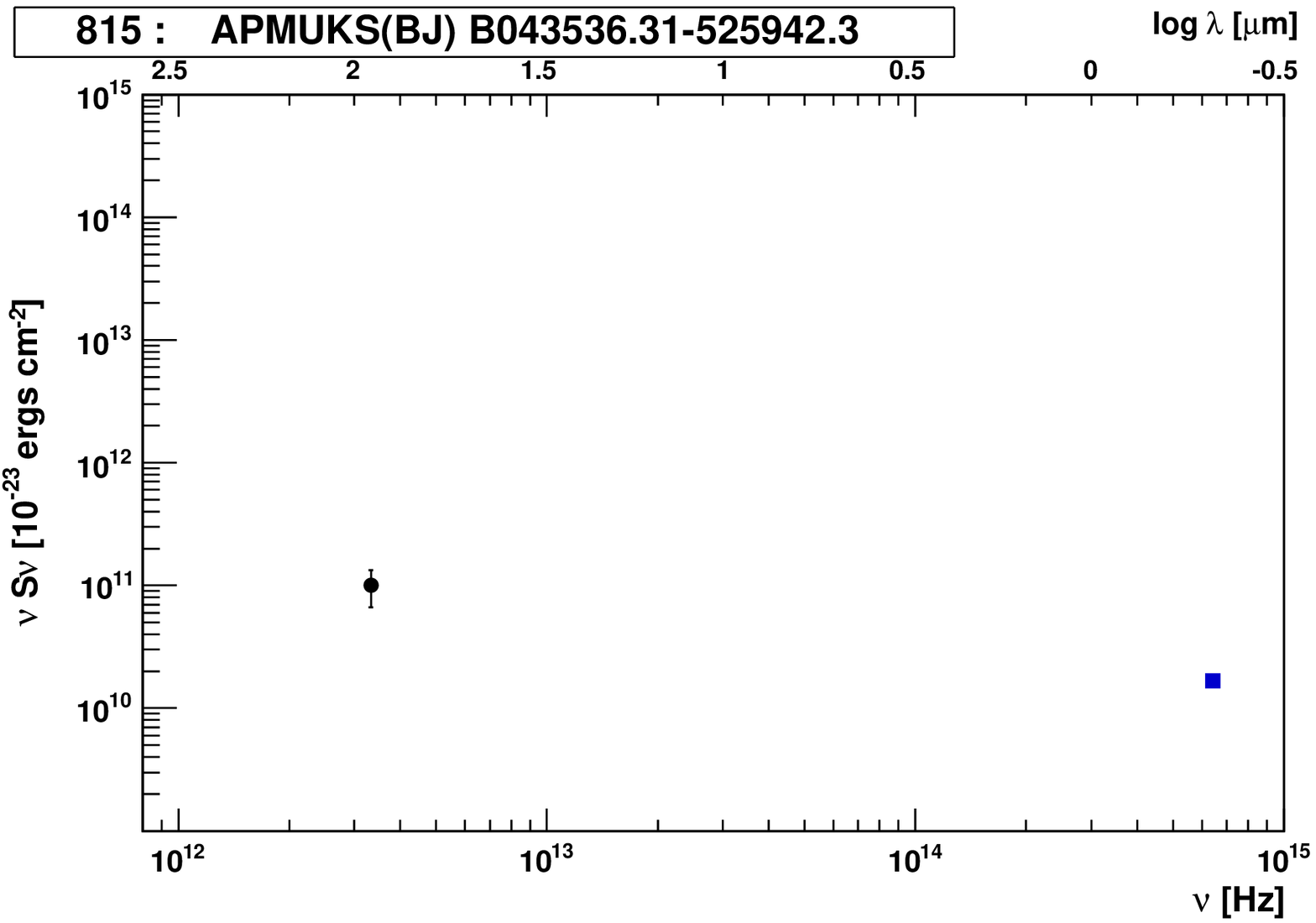}
\includegraphics[width=4cm]{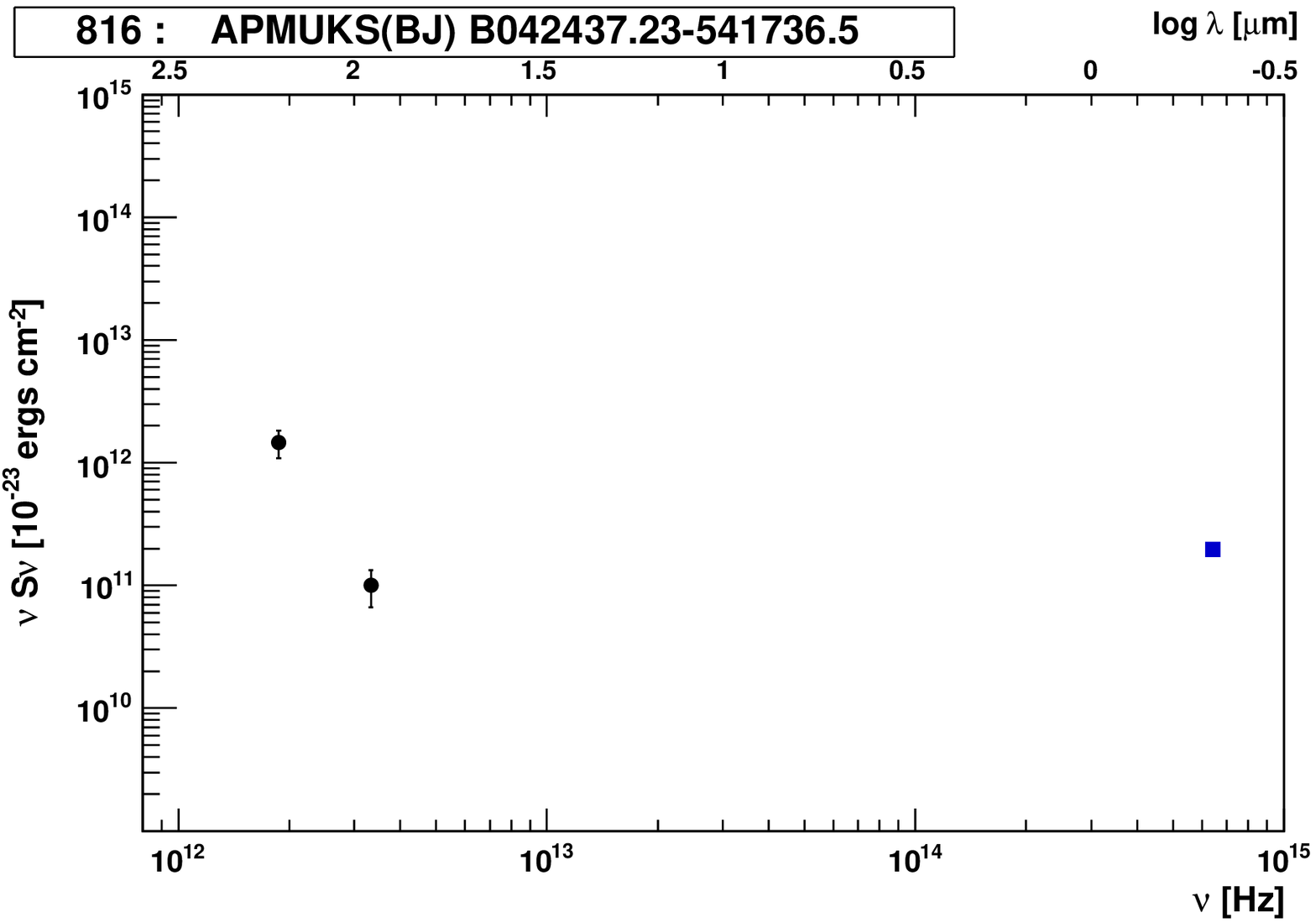}
\includegraphics[width=4cm]{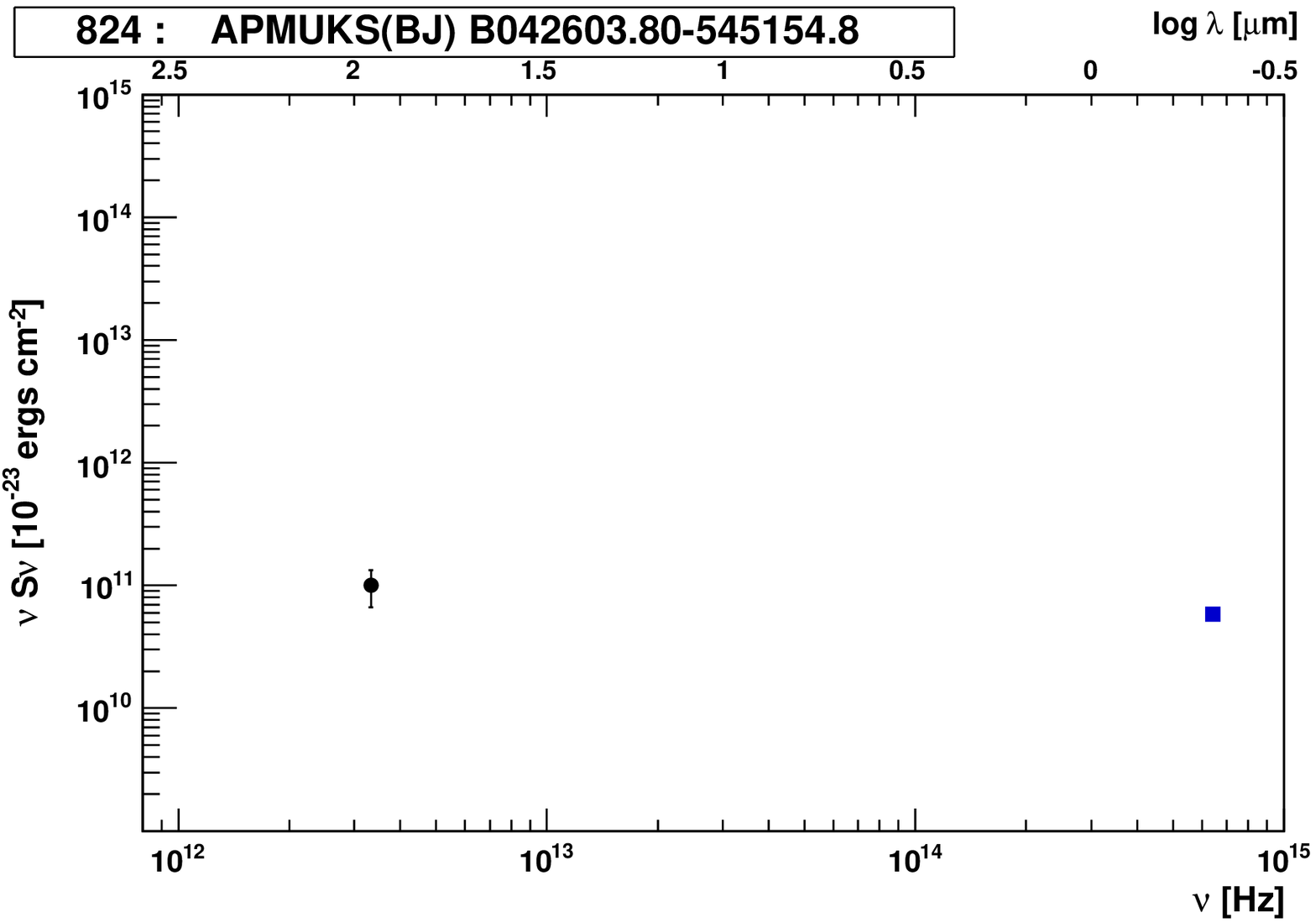}
\includegraphics[width=4cm]{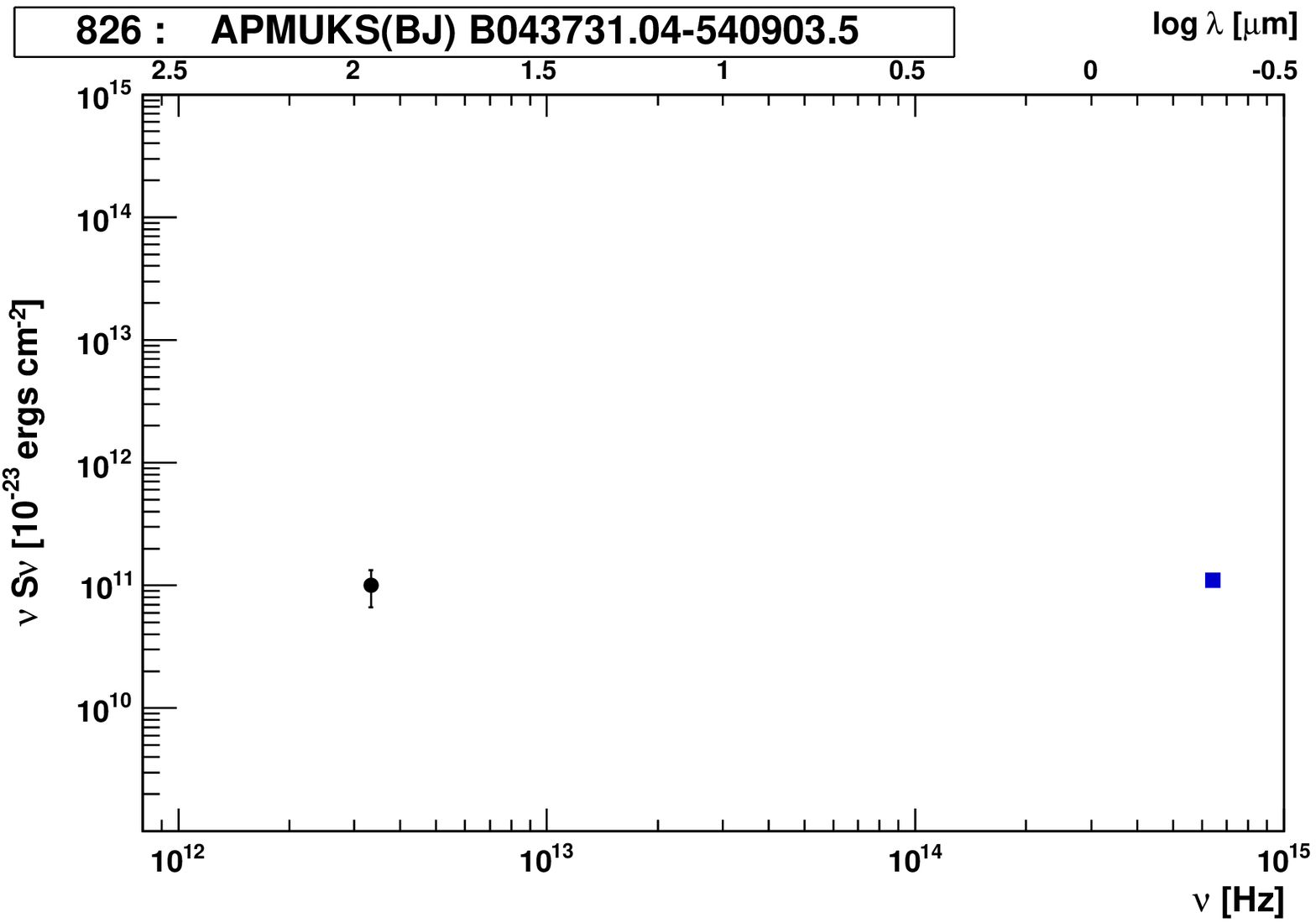}
\includegraphics[width=4cm]{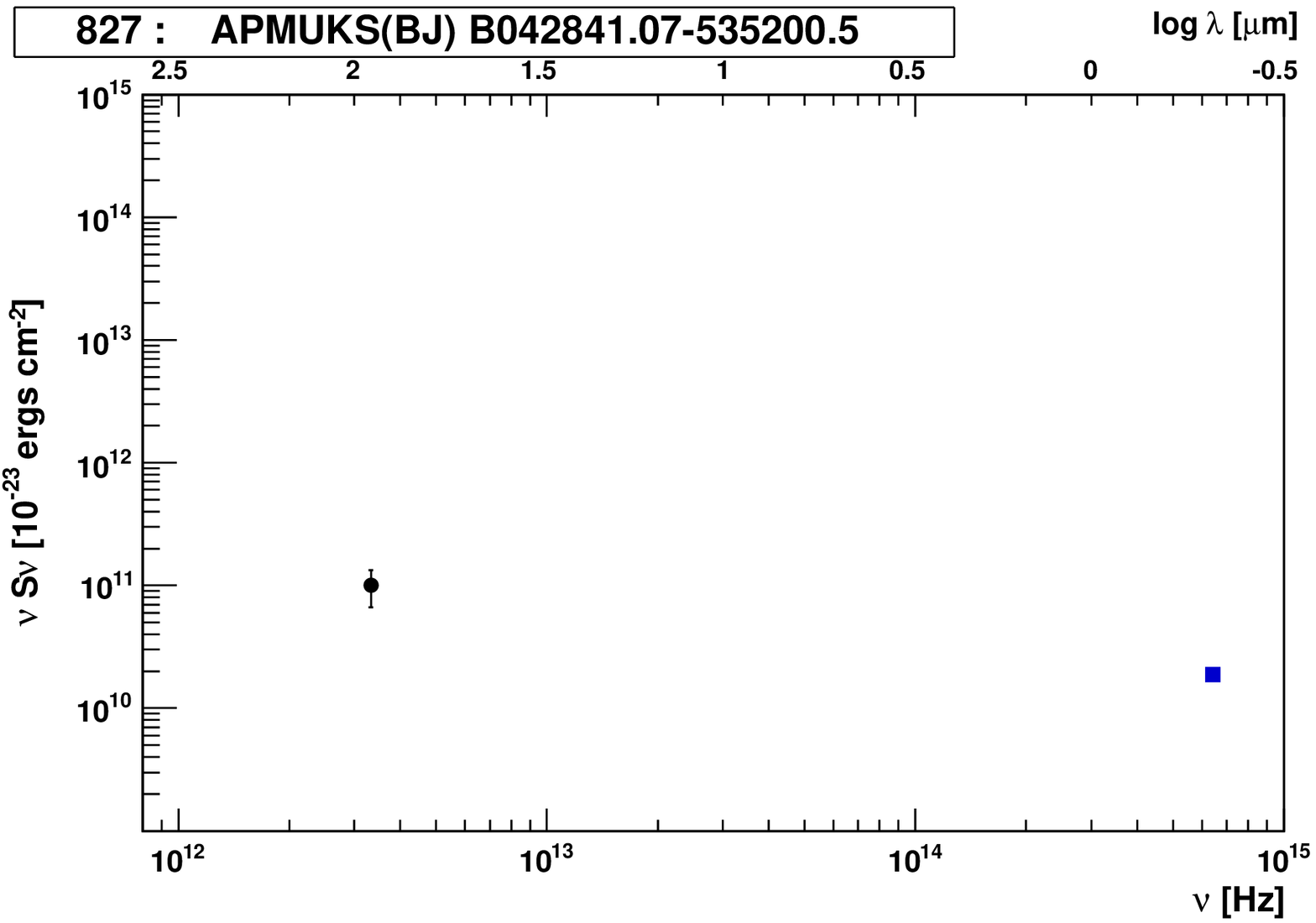}
\includegraphics[width=4cm]{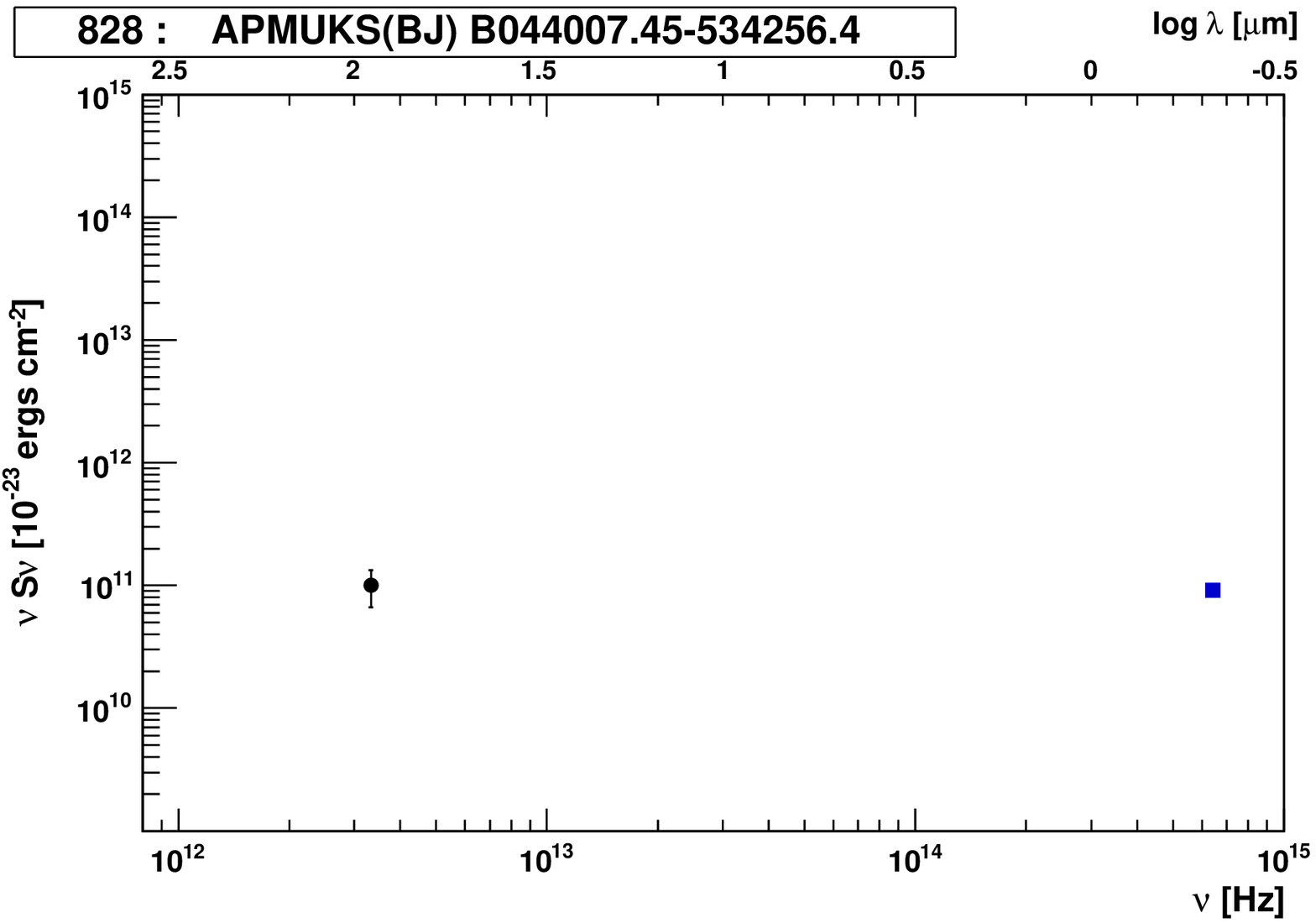}
\includegraphics[width=4cm]{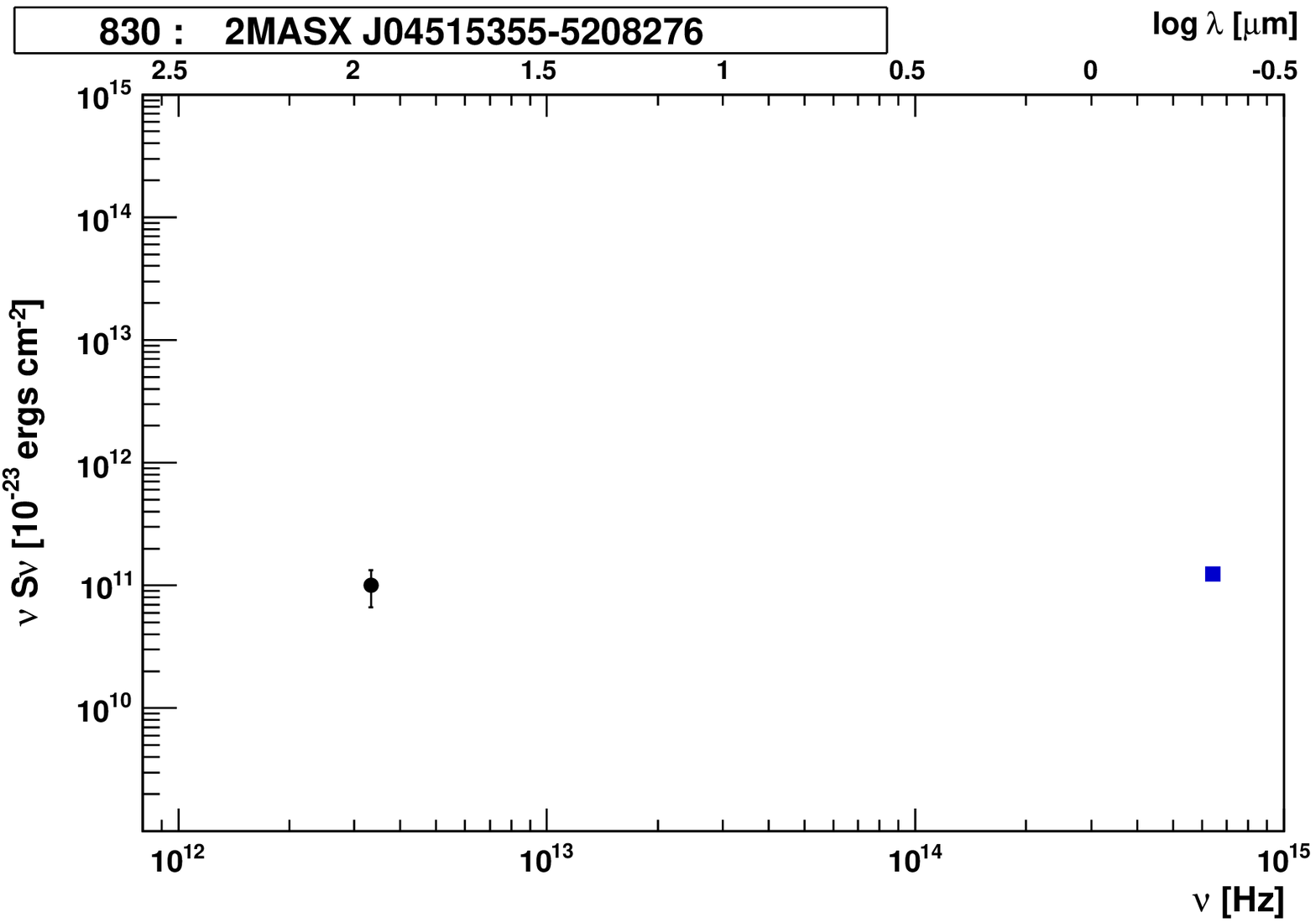}
\includegraphics[width=4cm]{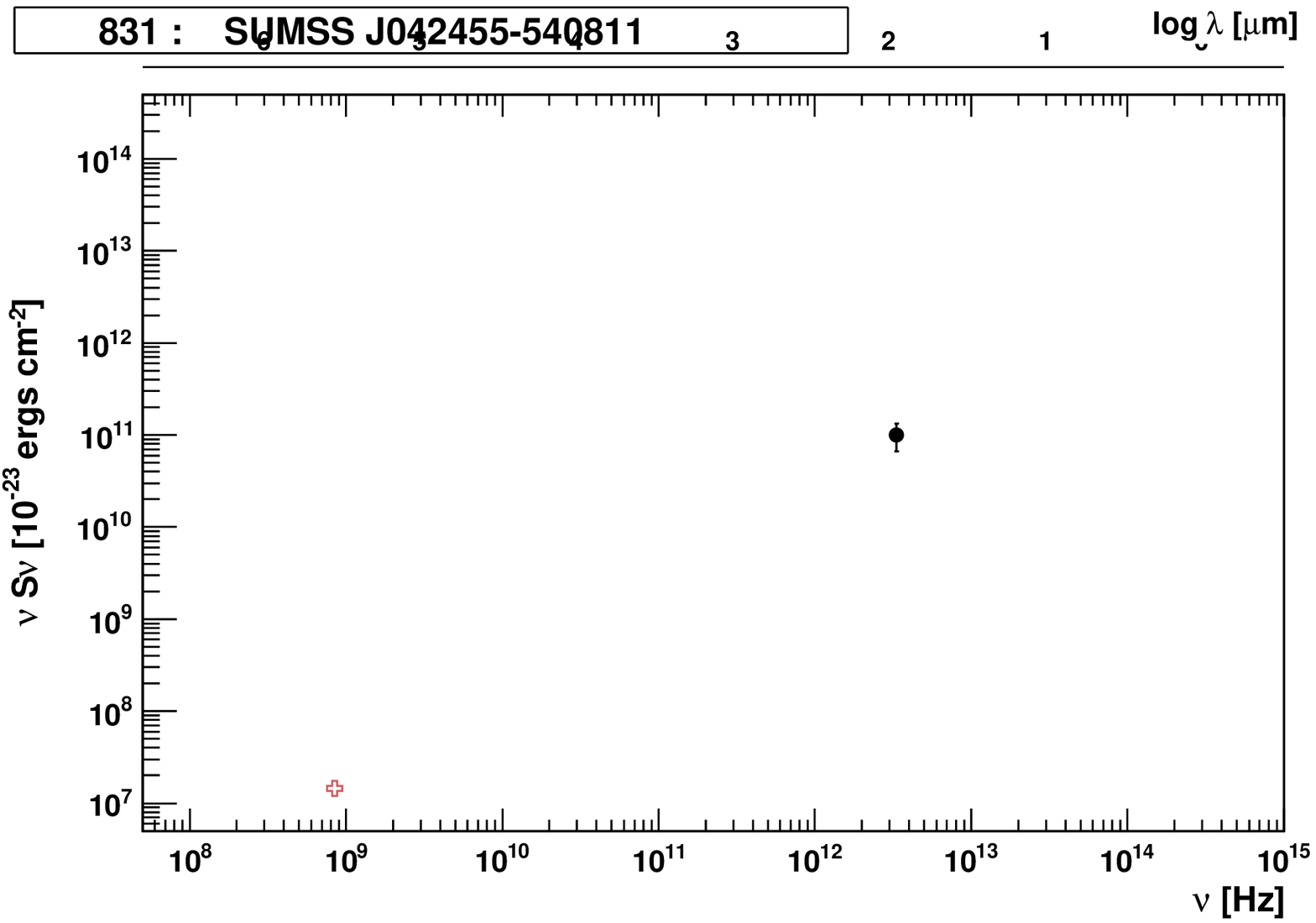}
\includegraphics[width=4cm]{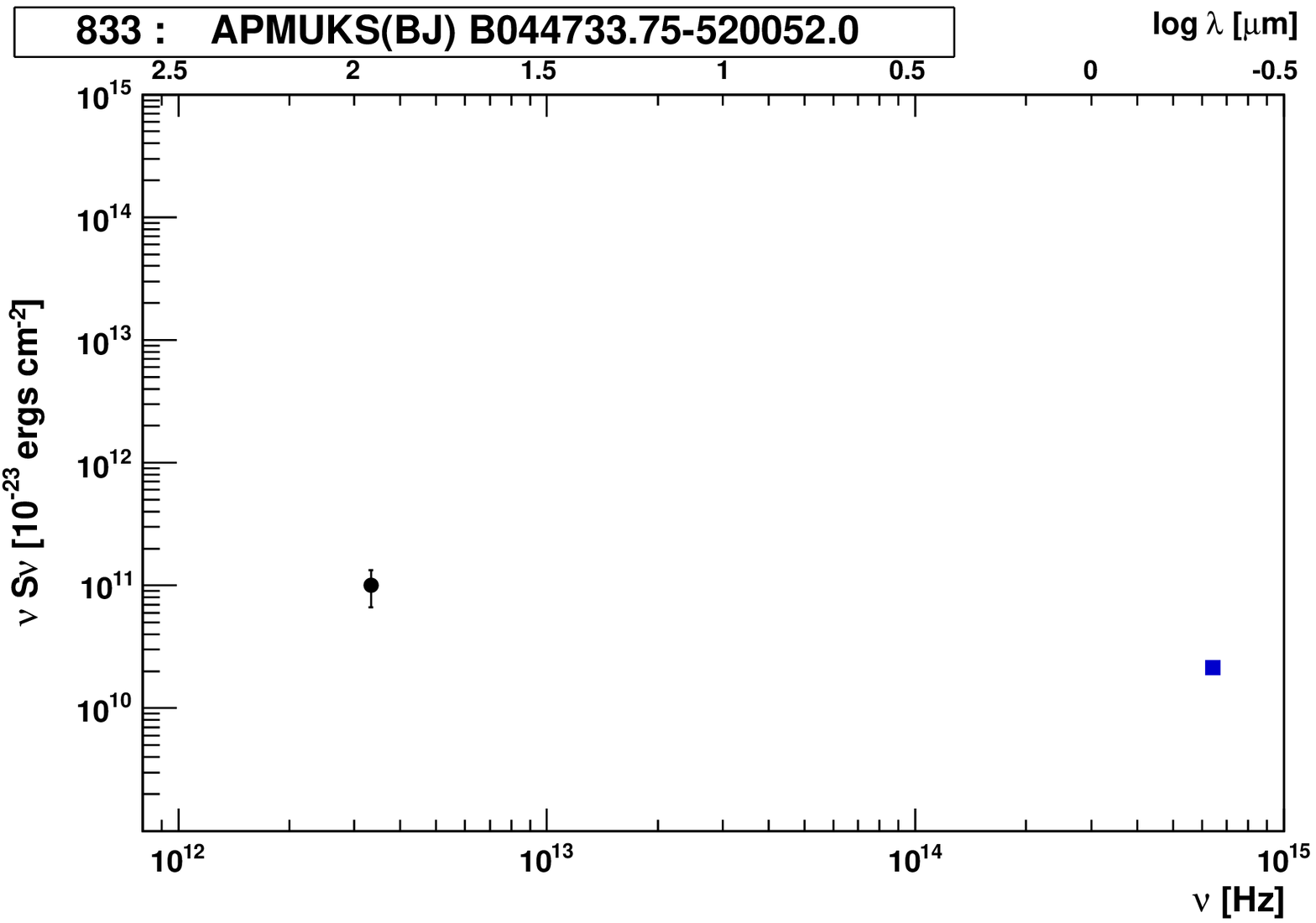}
\includegraphics[width=4cm]{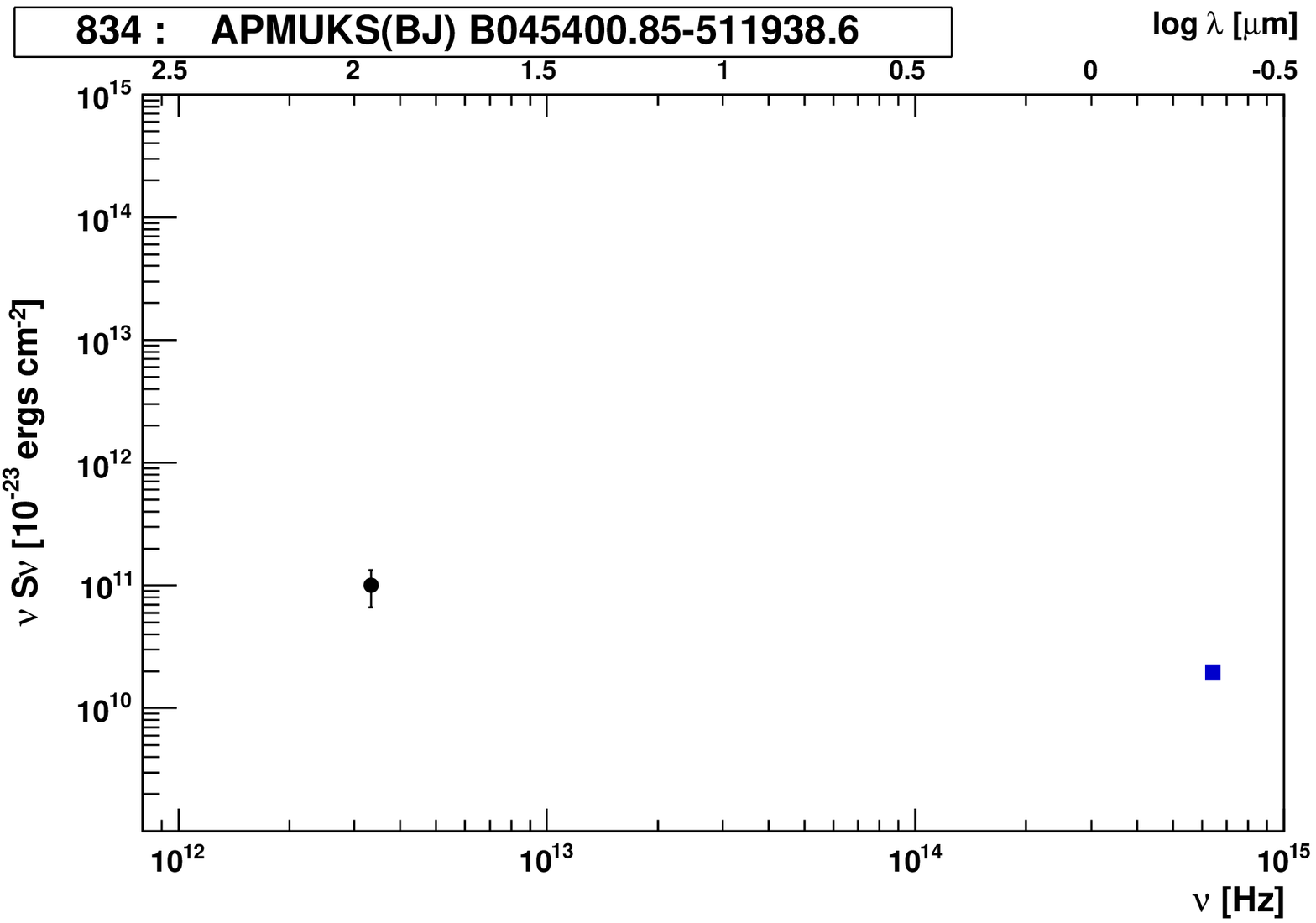}
\includegraphics[width=4cm]{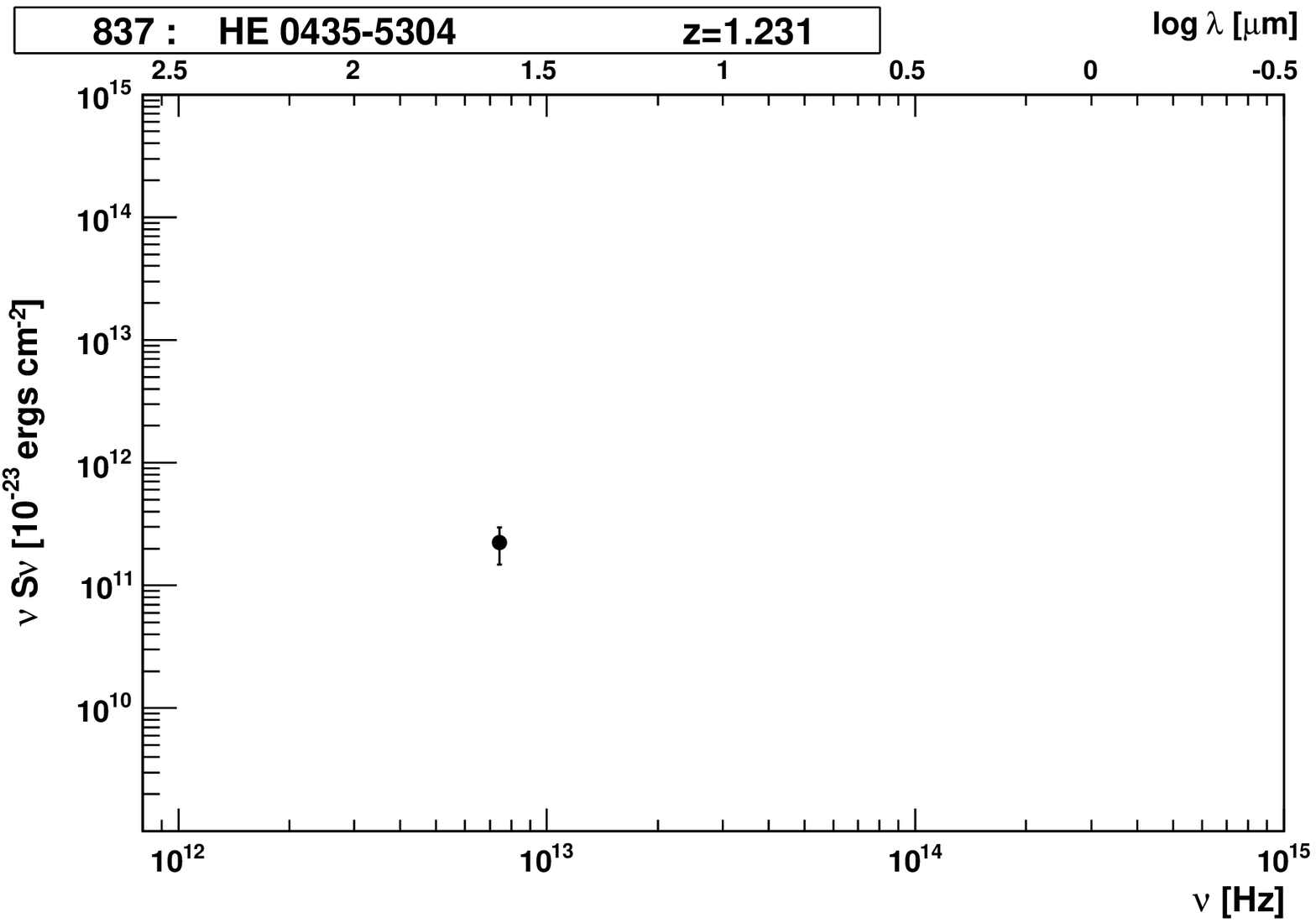}
\includegraphics[width=4cm]{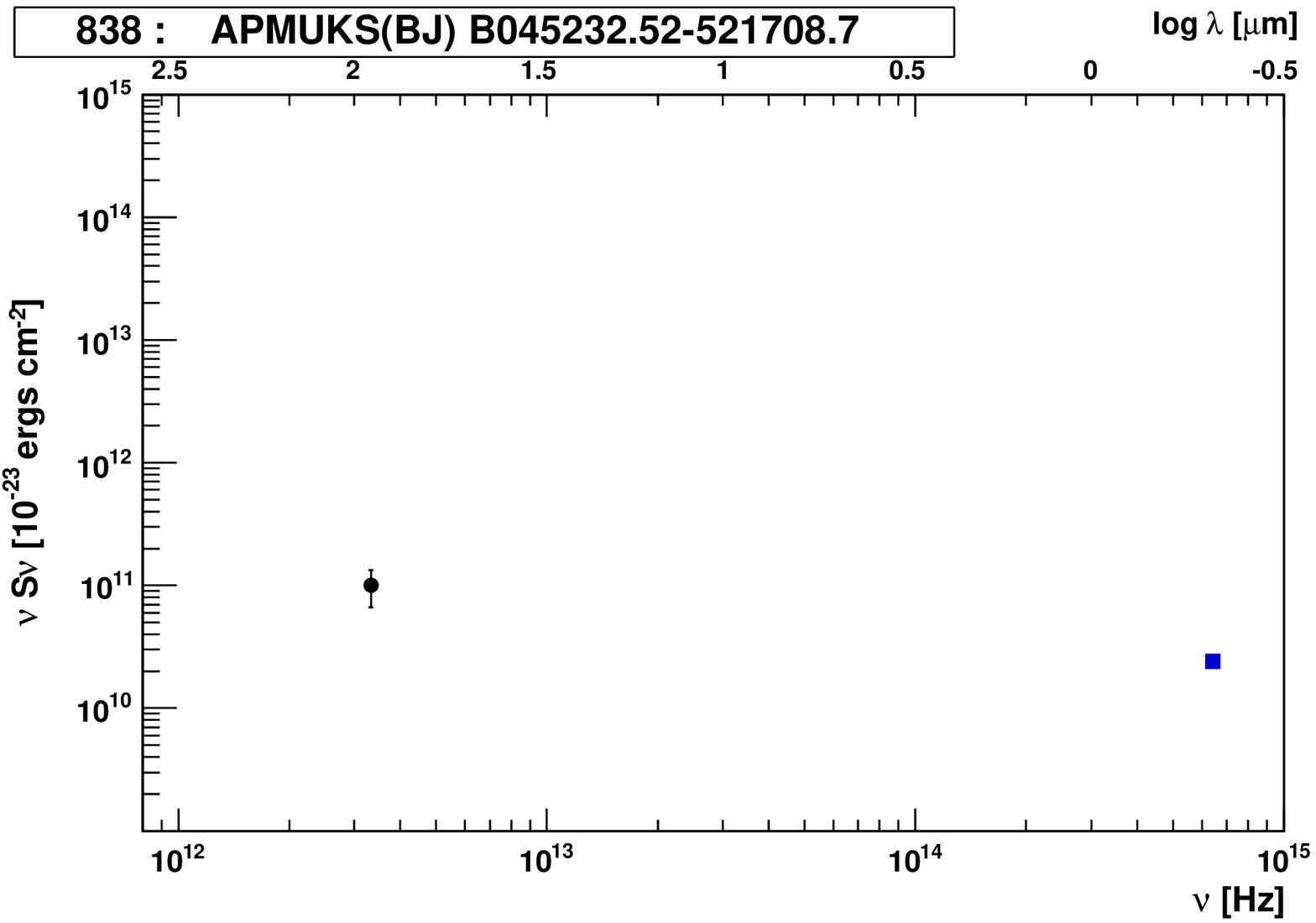}
\includegraphics[width=4cm]{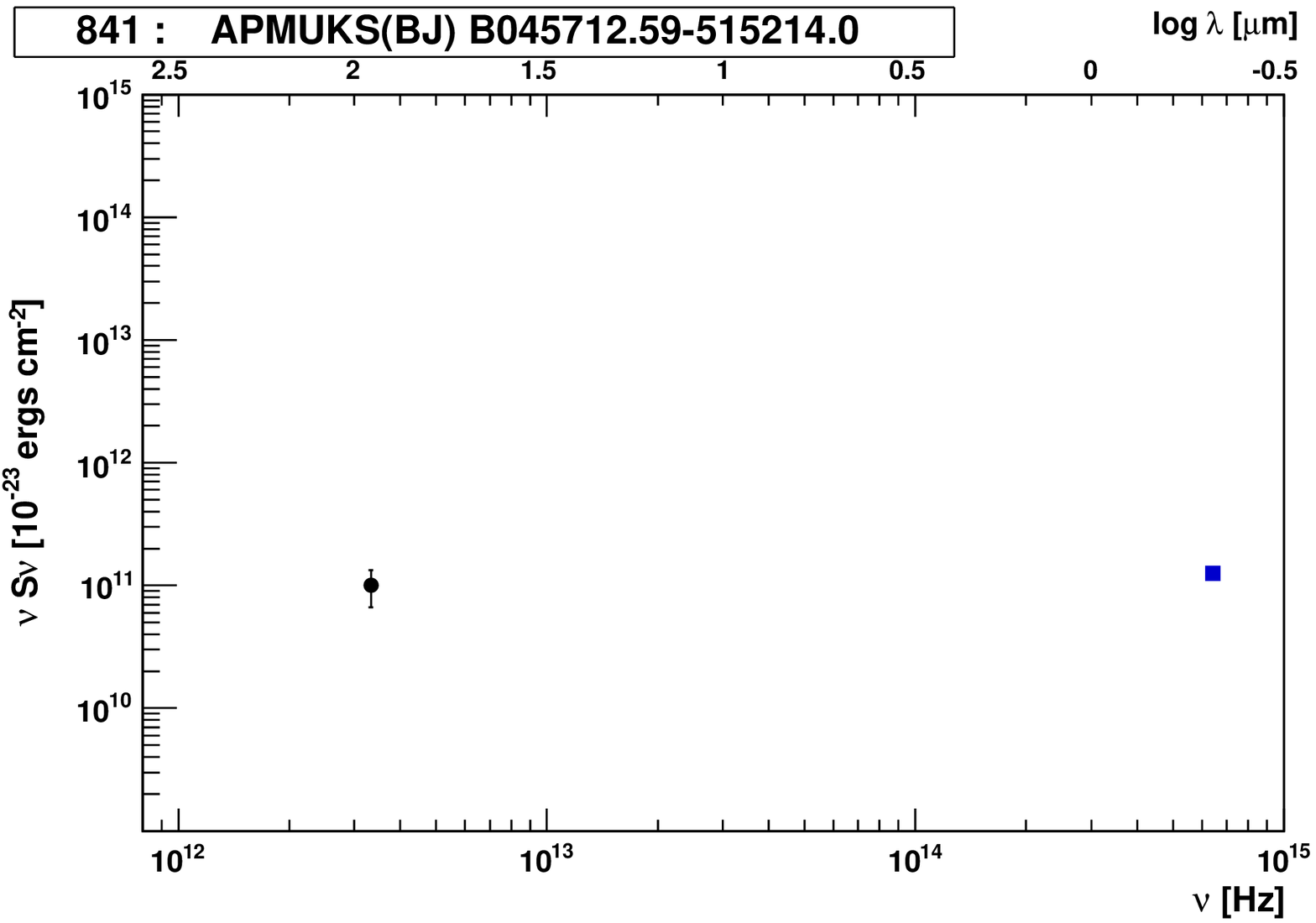}
\includegraphics[width=4cm]{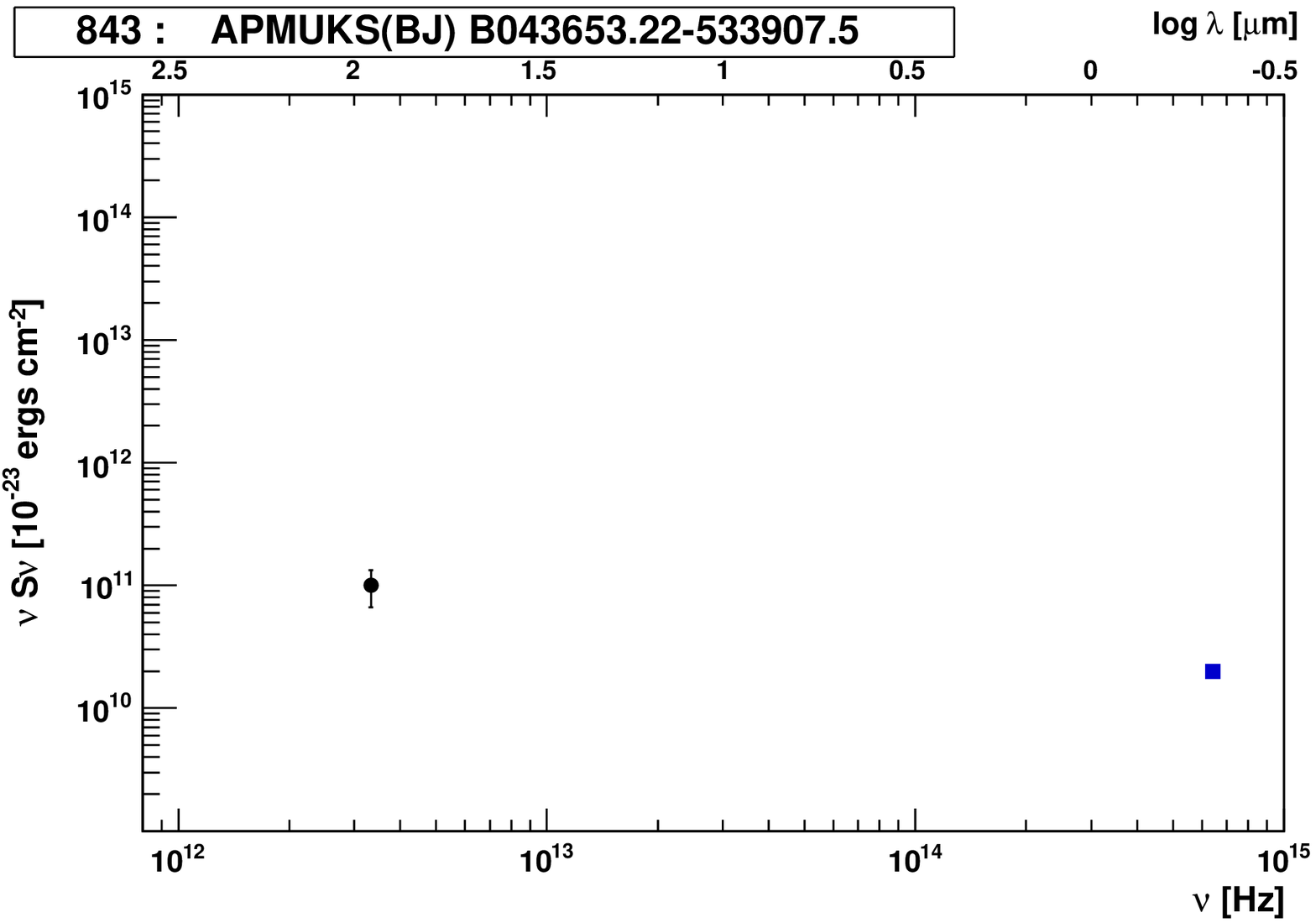}
\includegraphics[width=4cm]{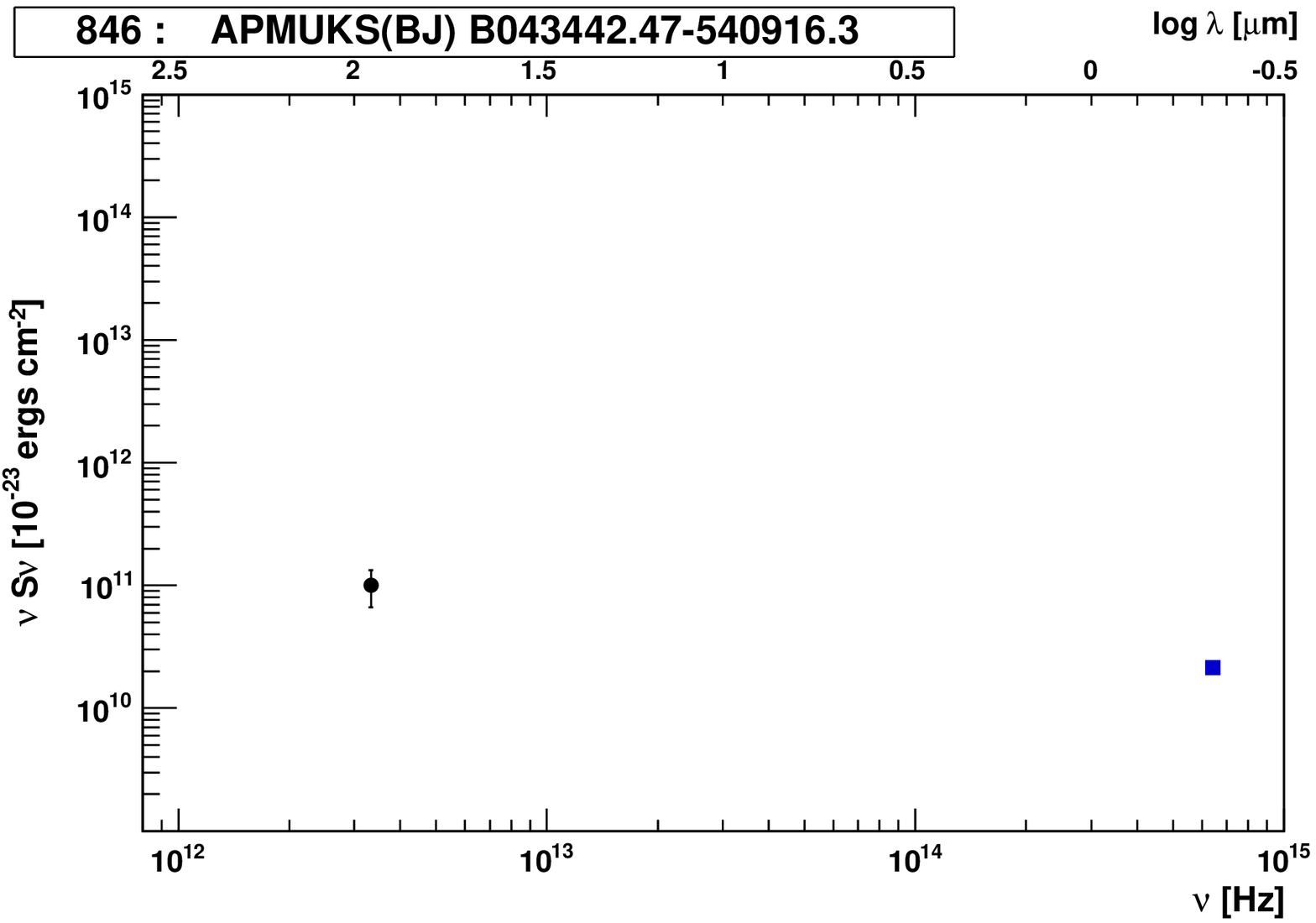}
\includegraphics[width=4cm]{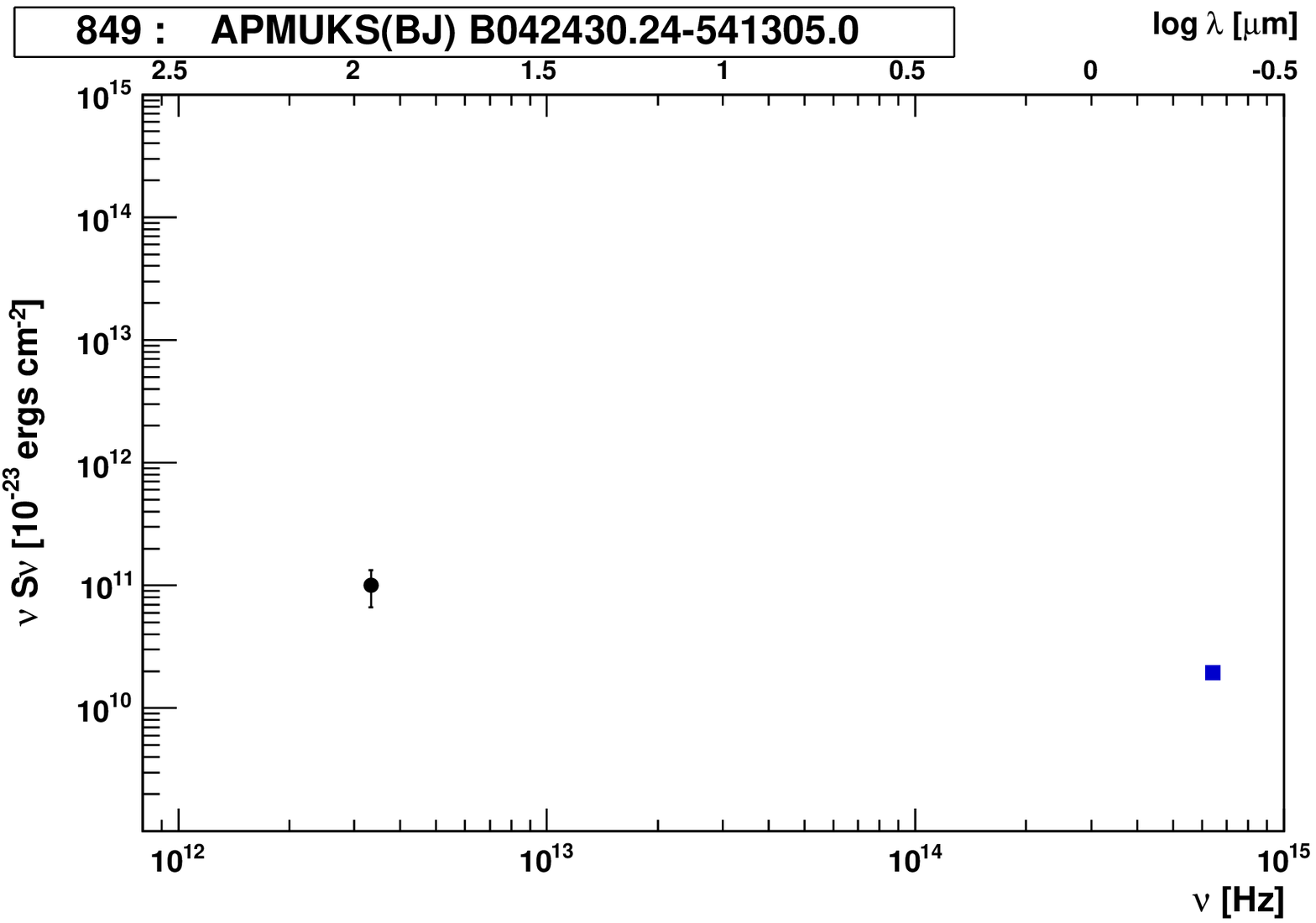}
\includegraphics[width=4cm]{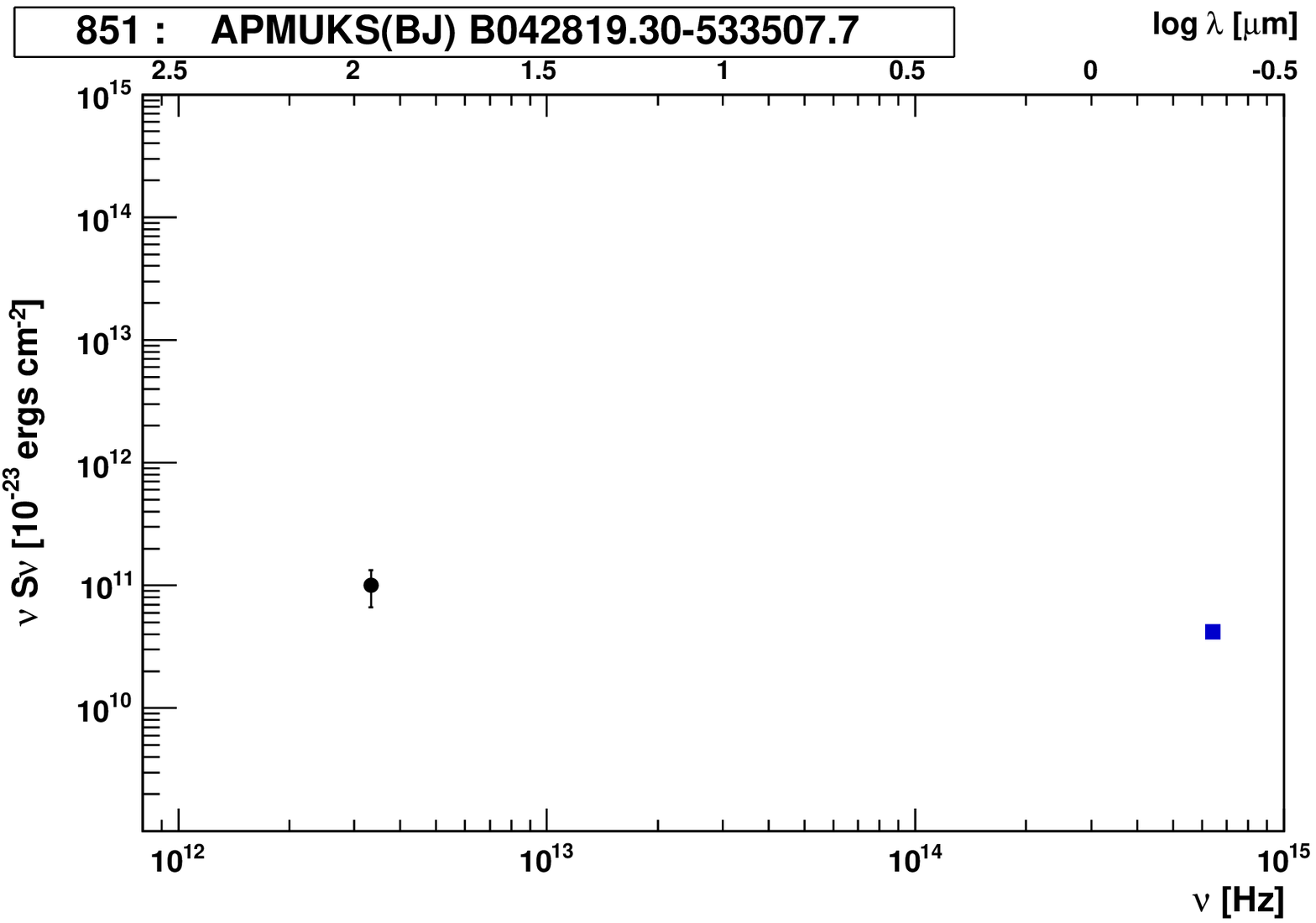}
\includegraphics[width=4cm]{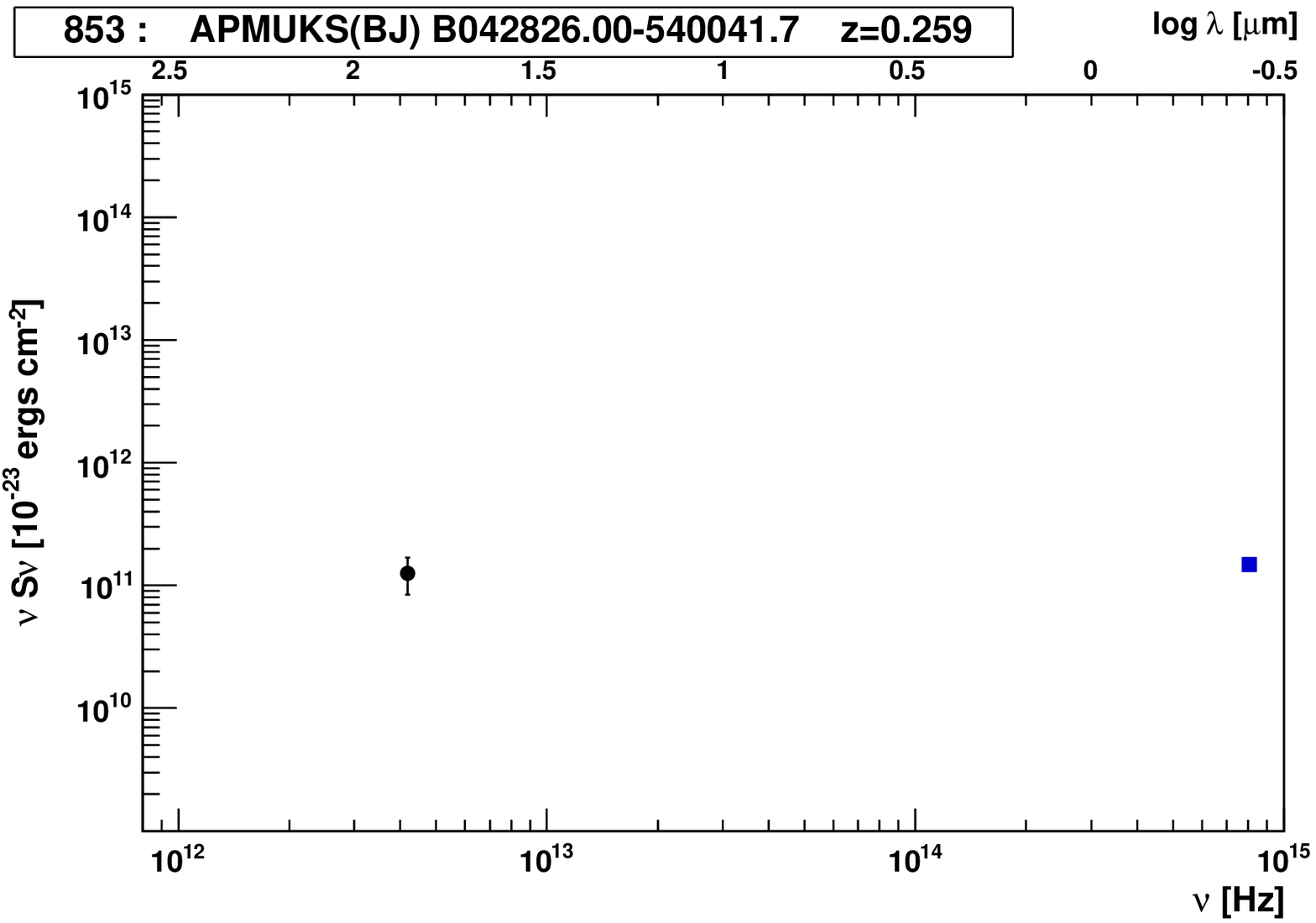}
\includegraphics[width=4cm]{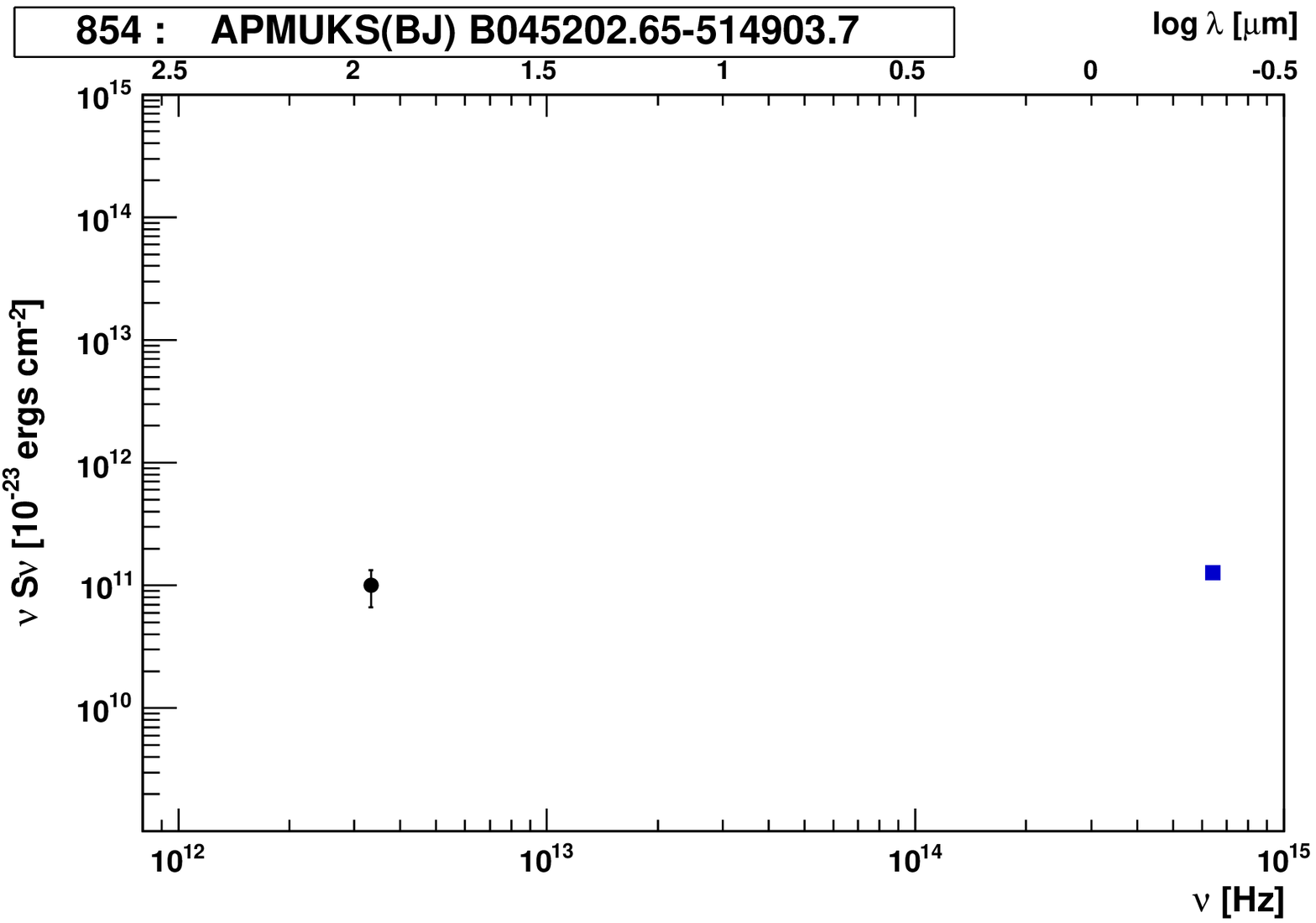}
\includegraphics[width=4cm]{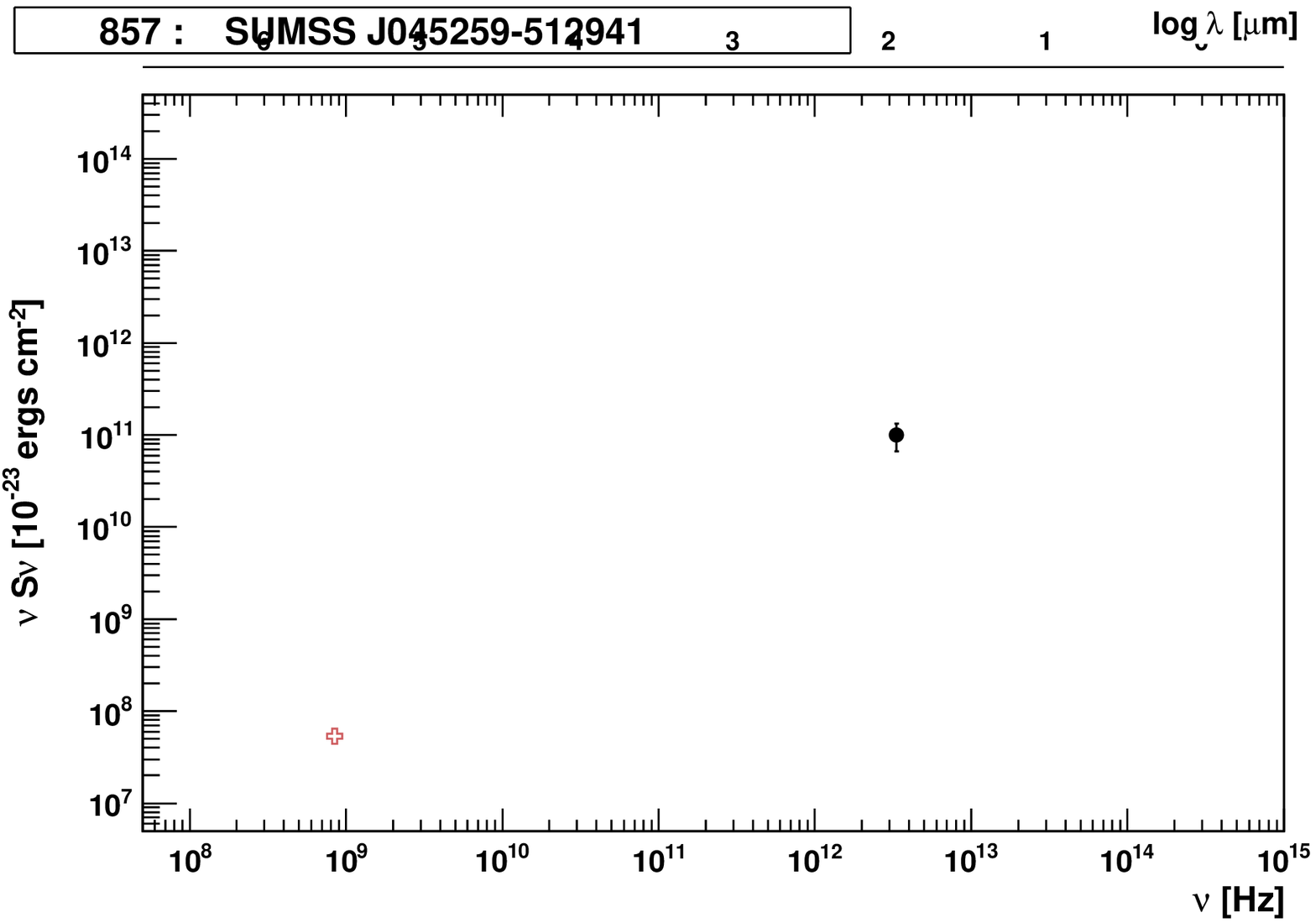}
\includegraphics[width=4cm]{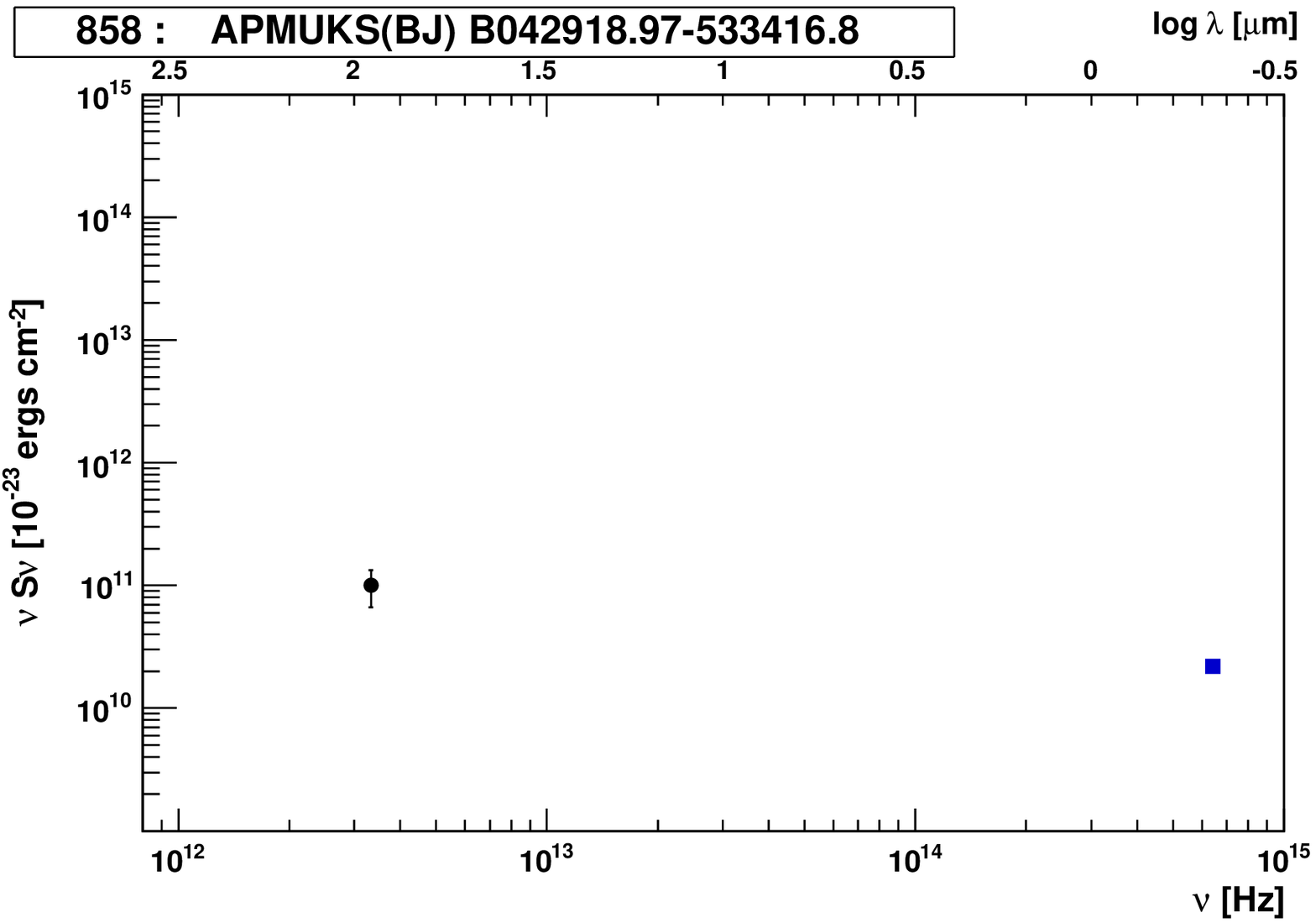}
\includegraphics[width=4cm]{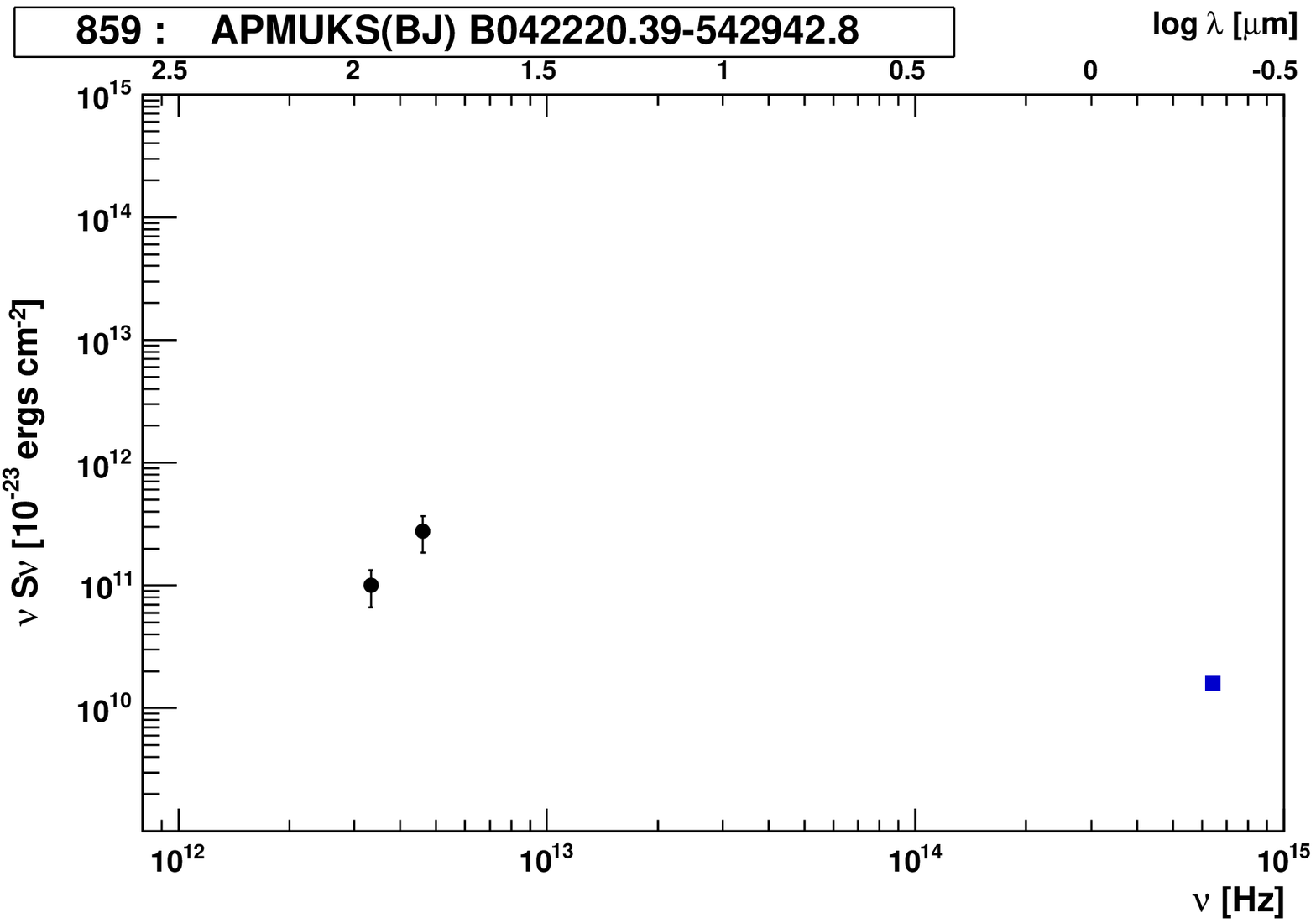}
\includegraphics[width=4cm]{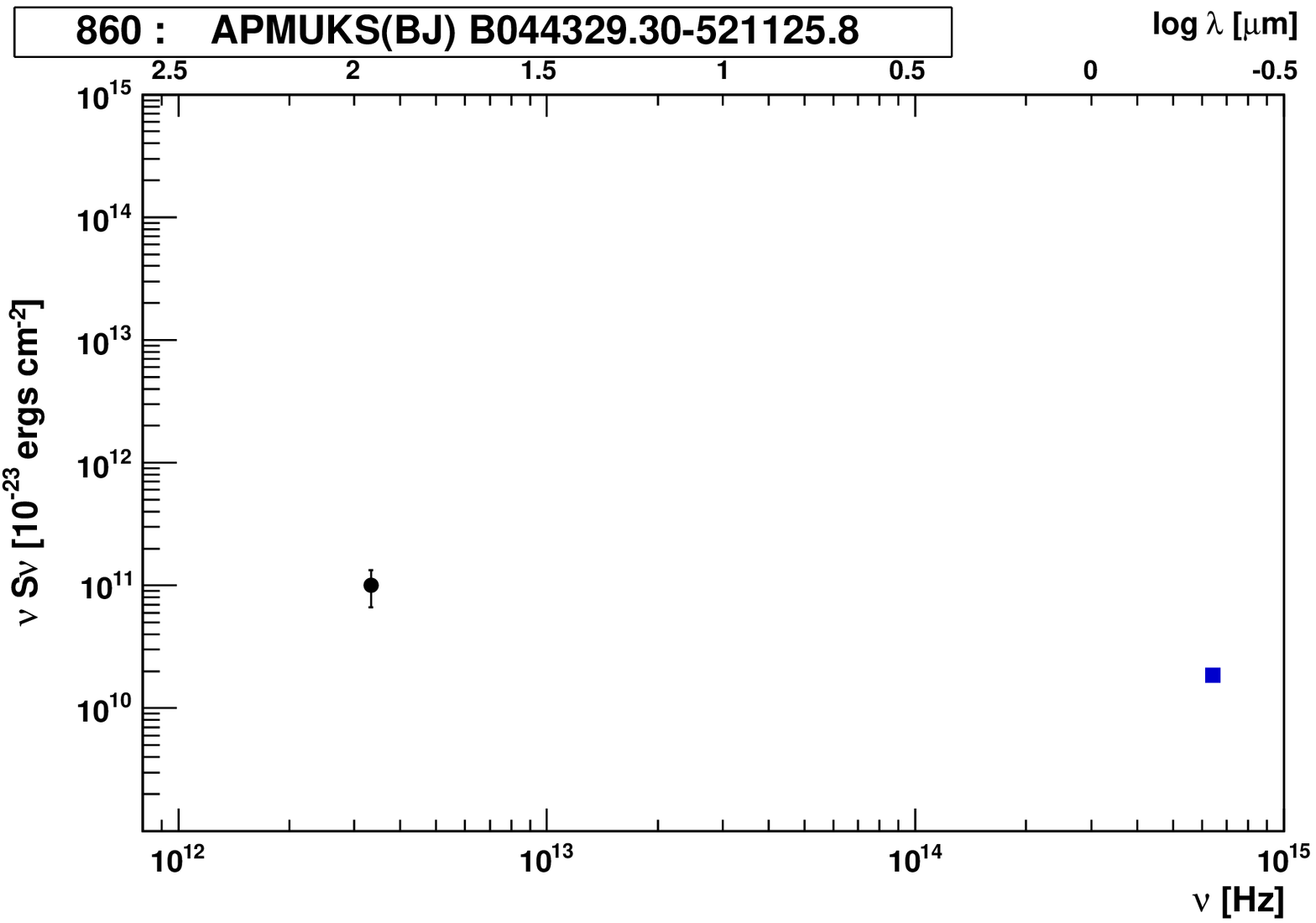}
\includegraphics[width=4cm]{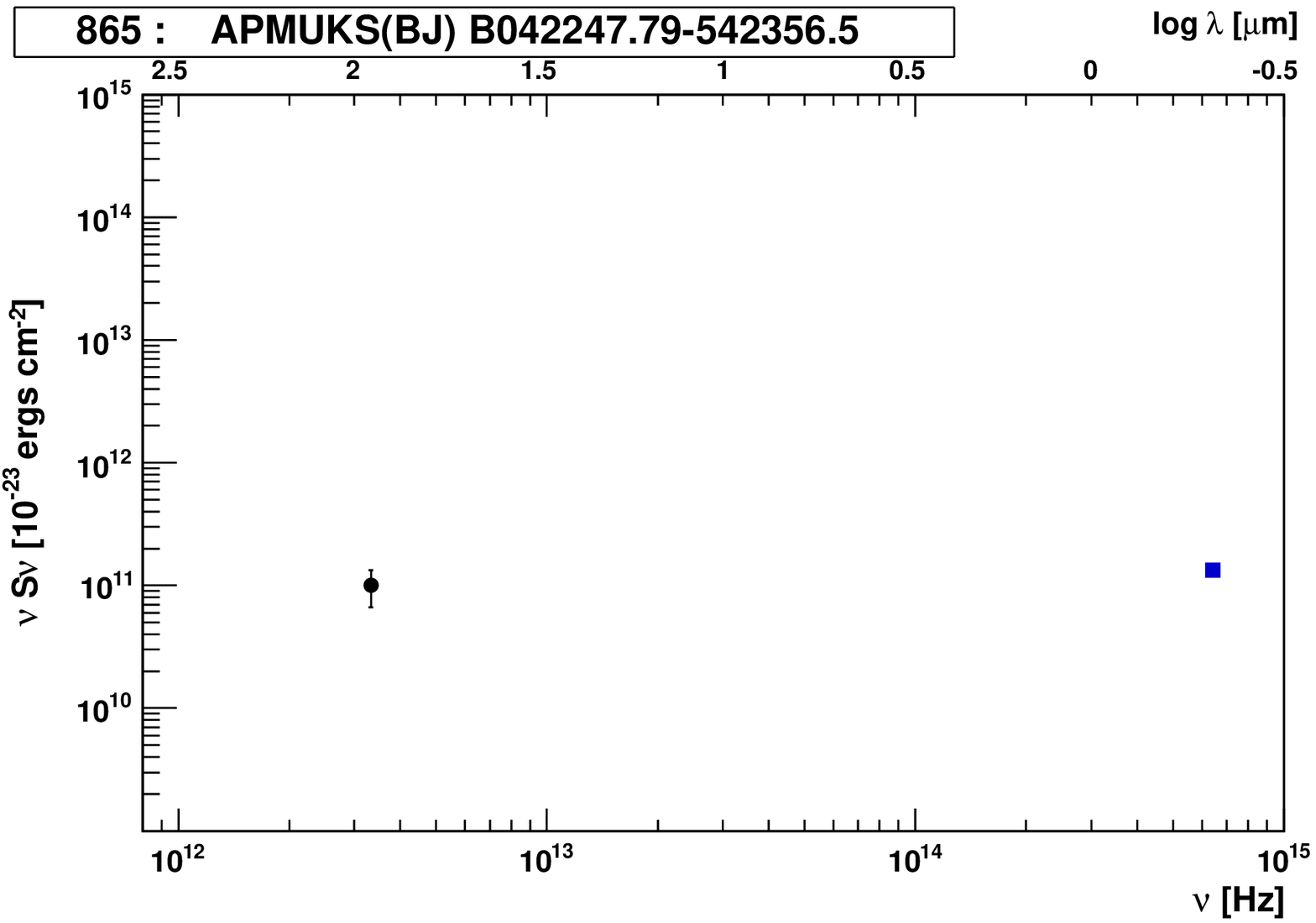}
\includegraphics[width=4cm]{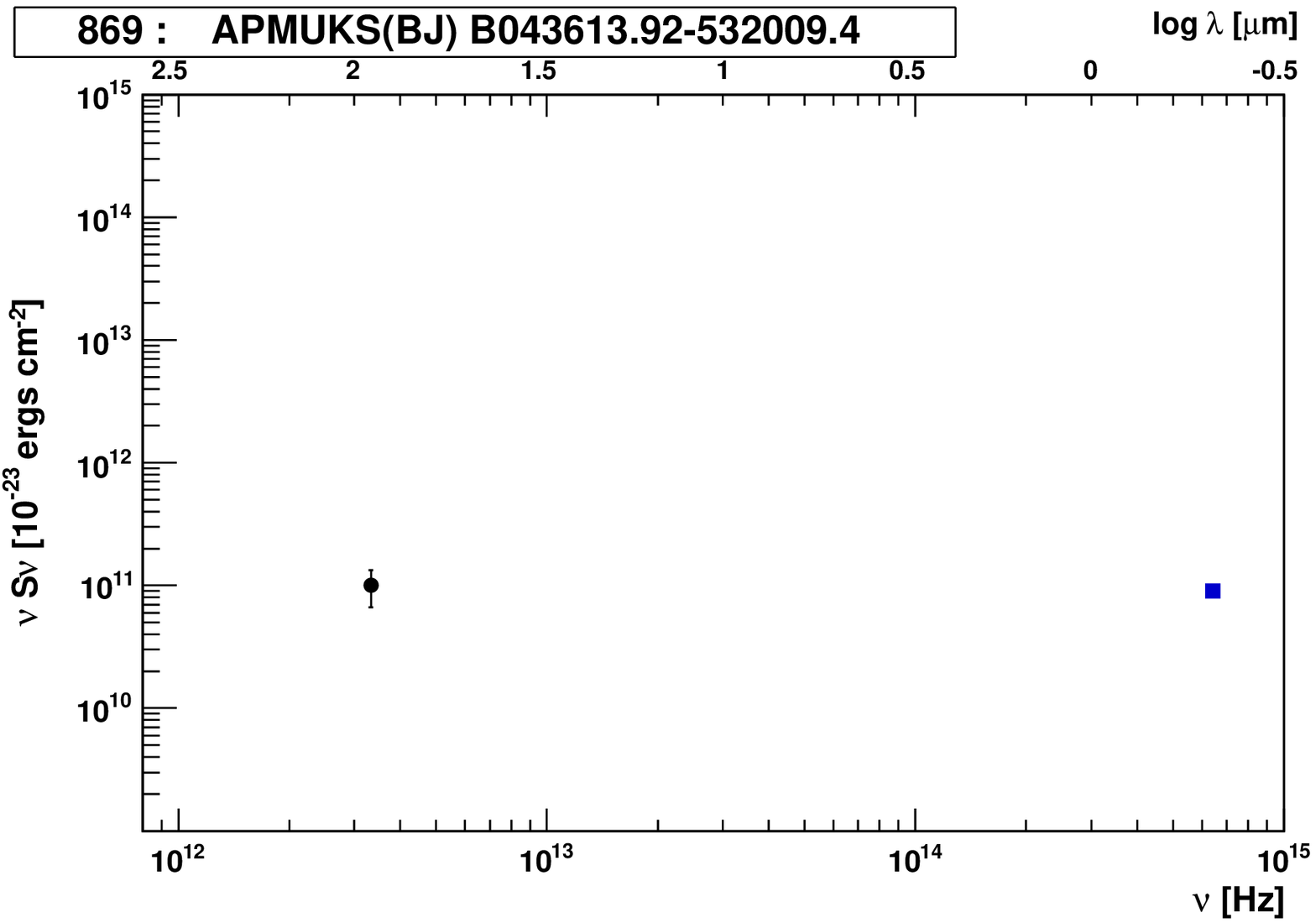}
\includegraphics[width=4cm]{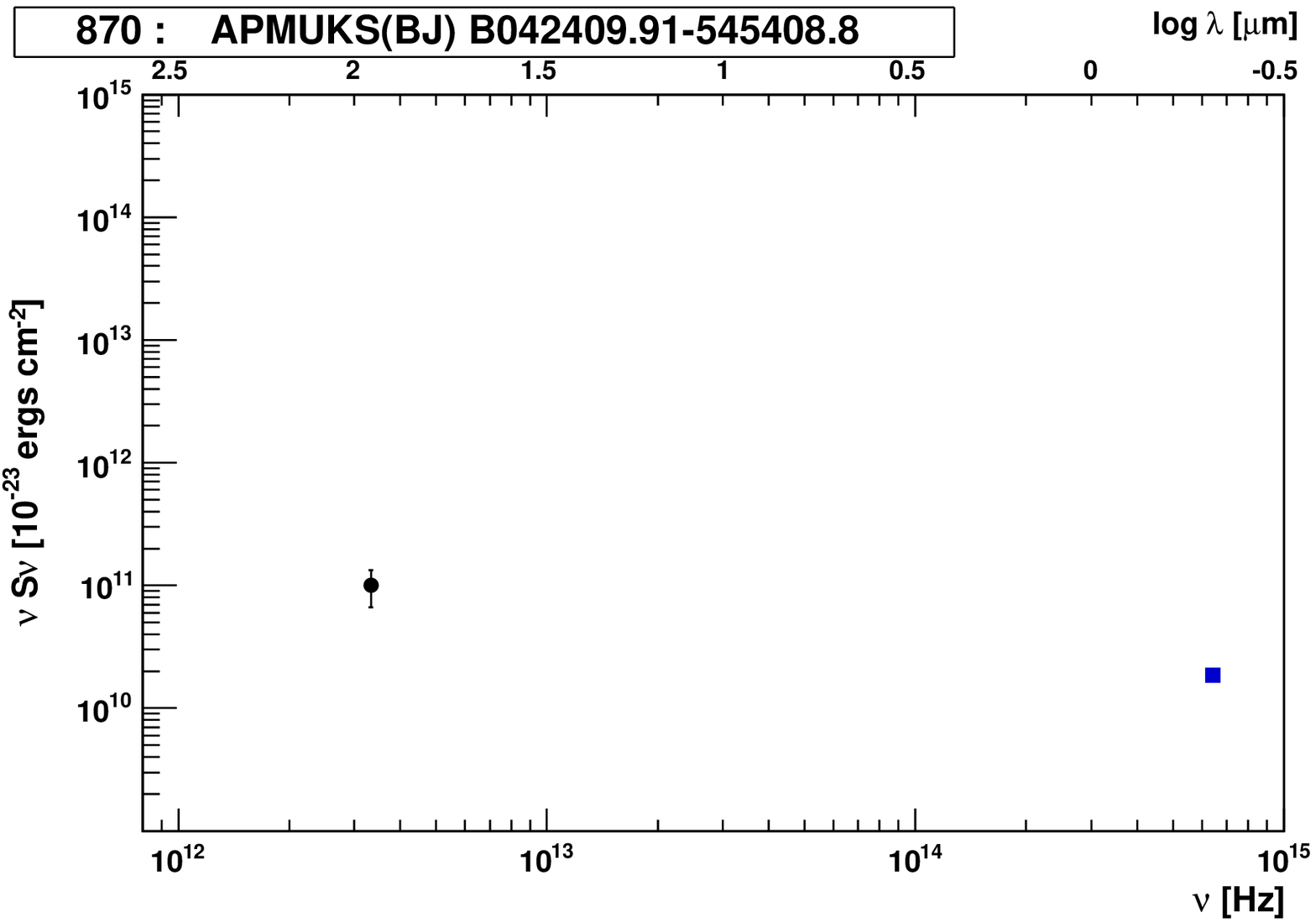}
\includegraphics[width=4cm]{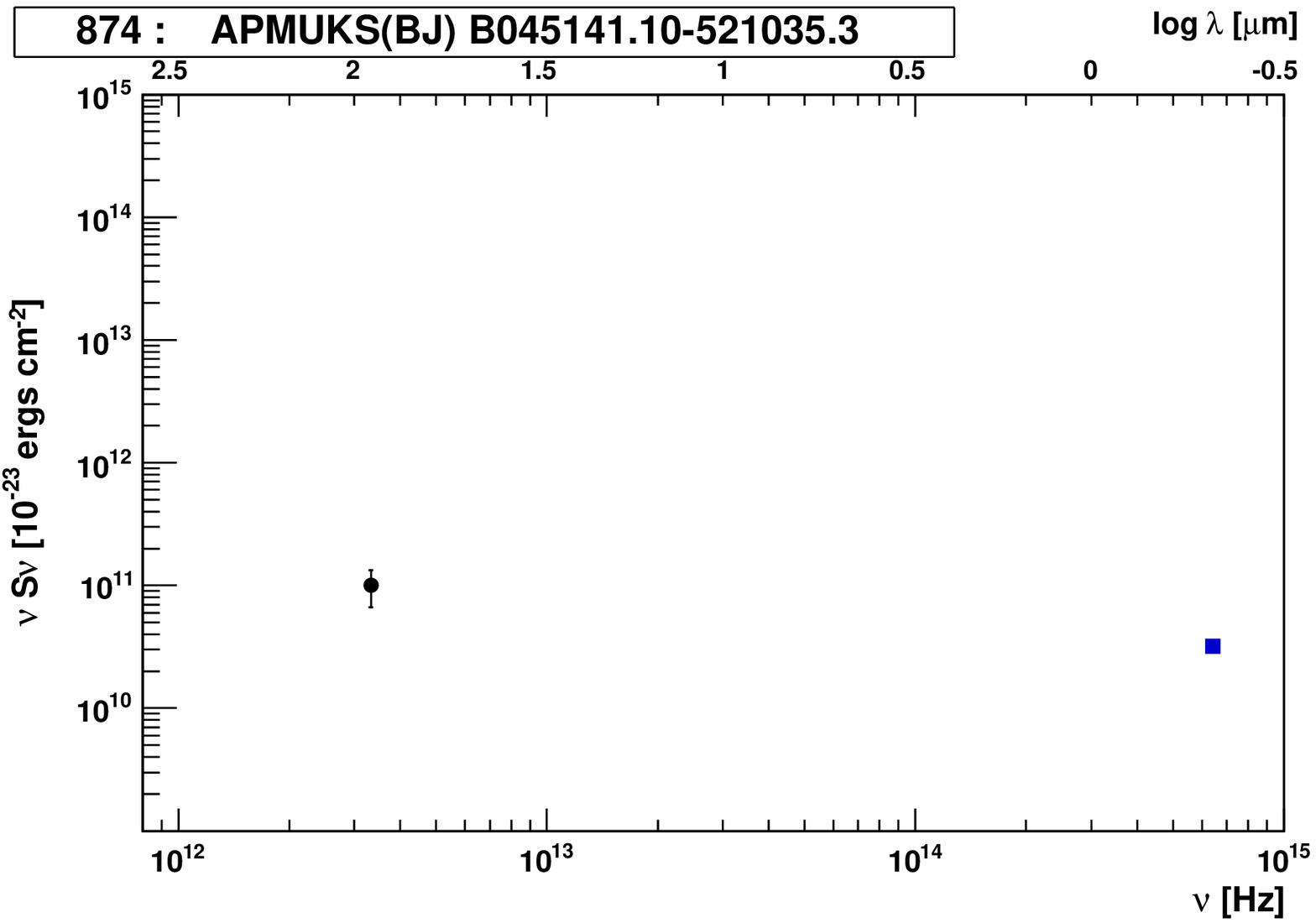}
\includegraphics[width=4cm]{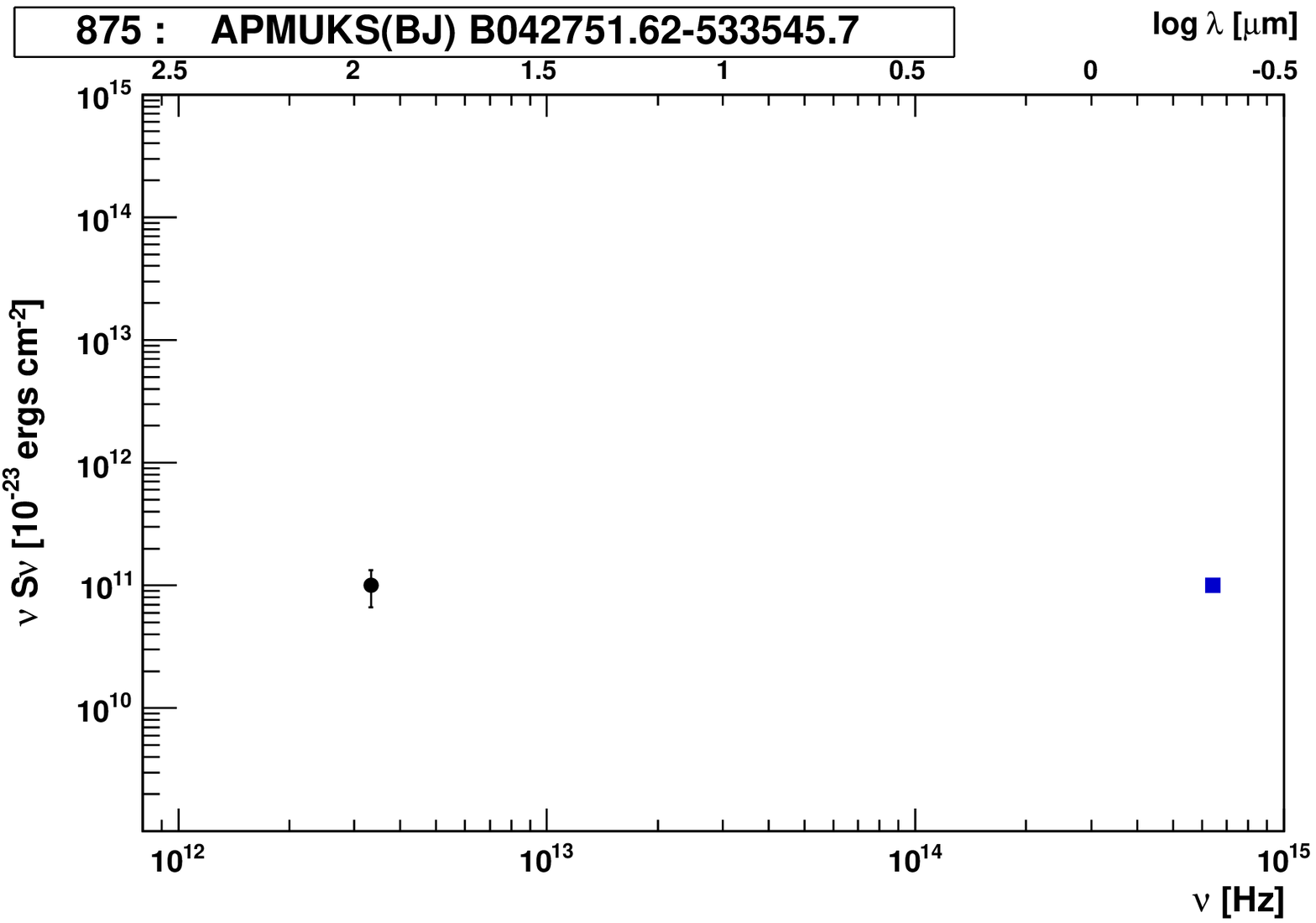}
\includegraphics[width=4cm]{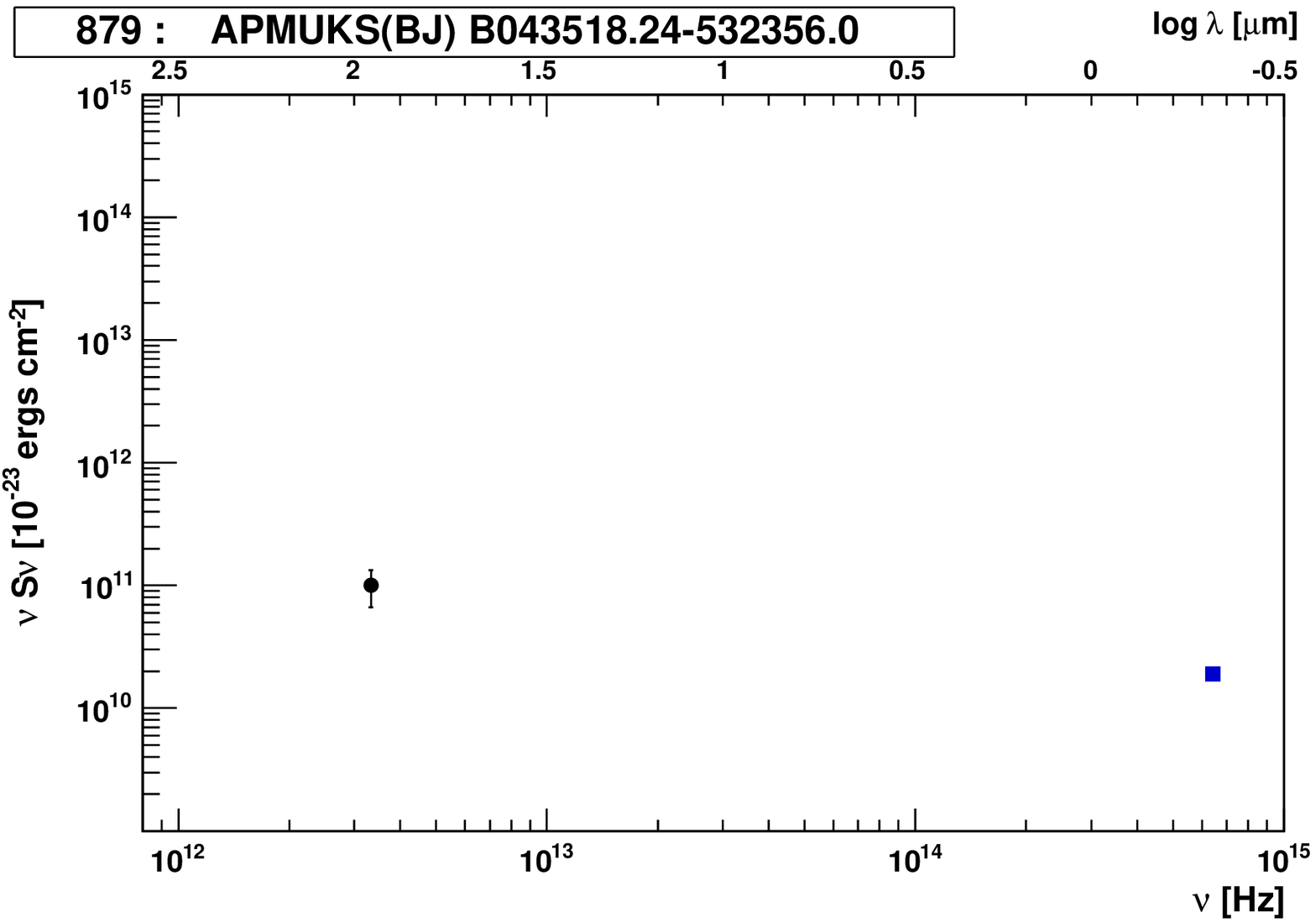}
\includegraphics[width=4cm]{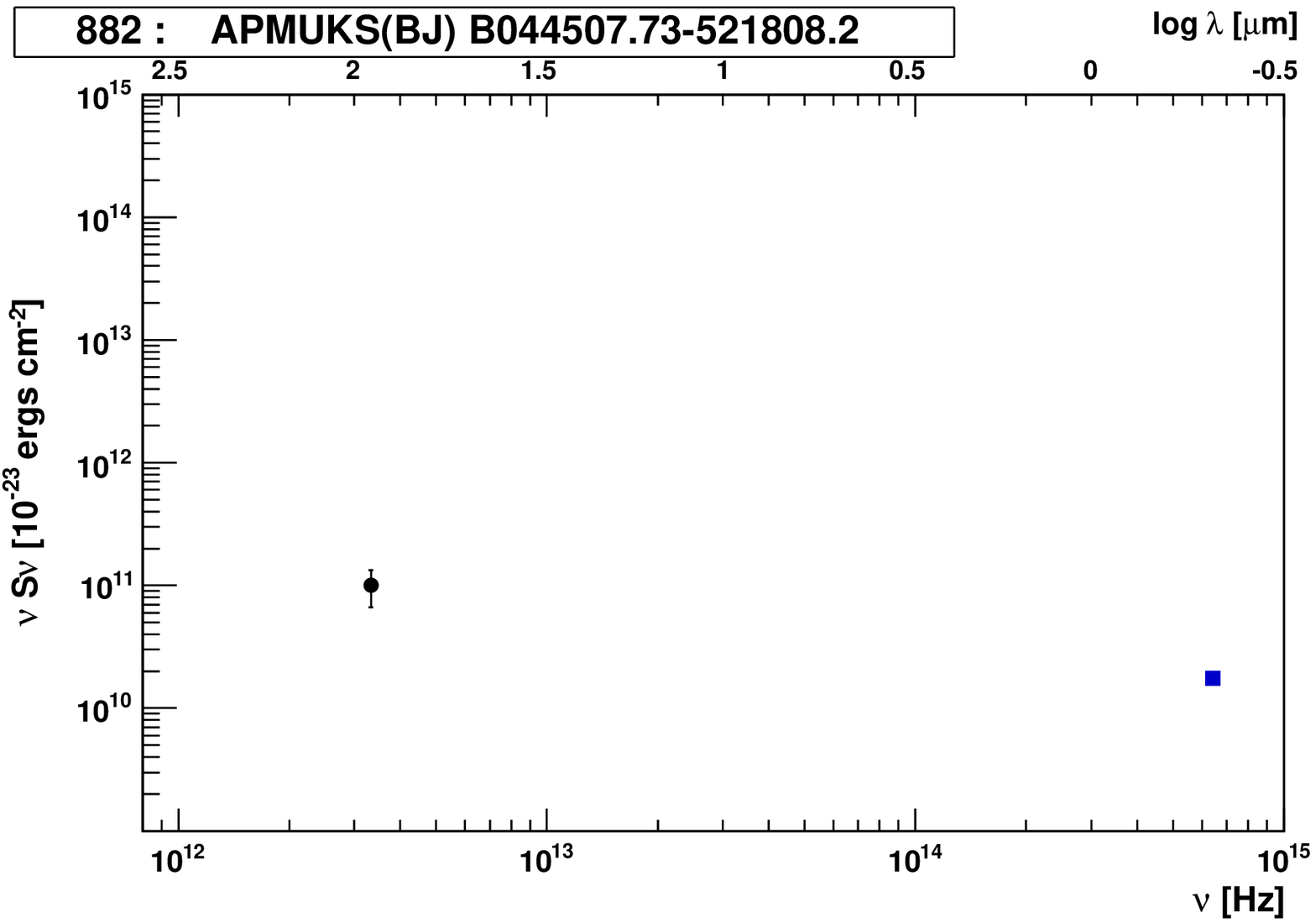}
\includegraphics[width=4cm]{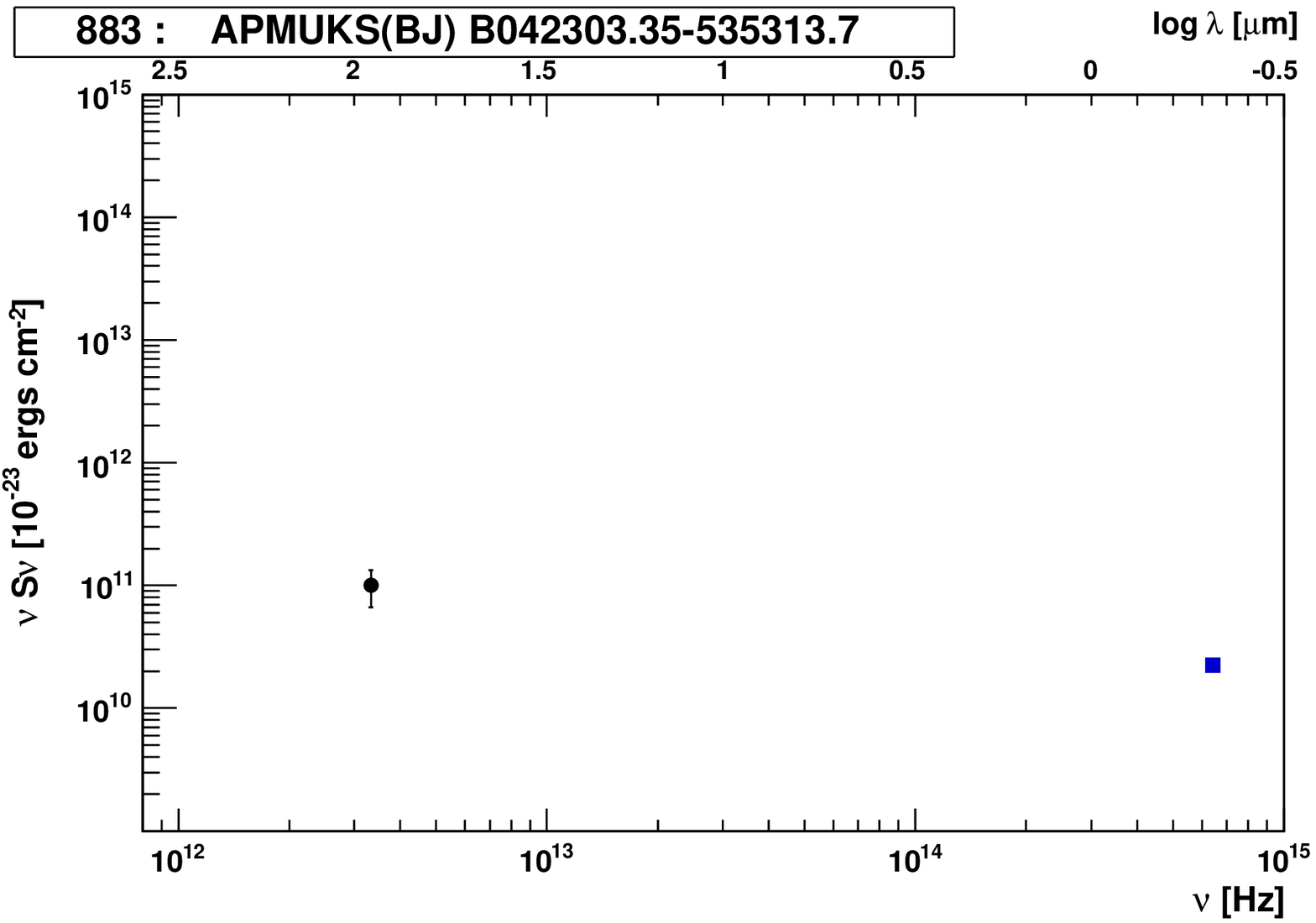}
\includegraphics[width=4cm]{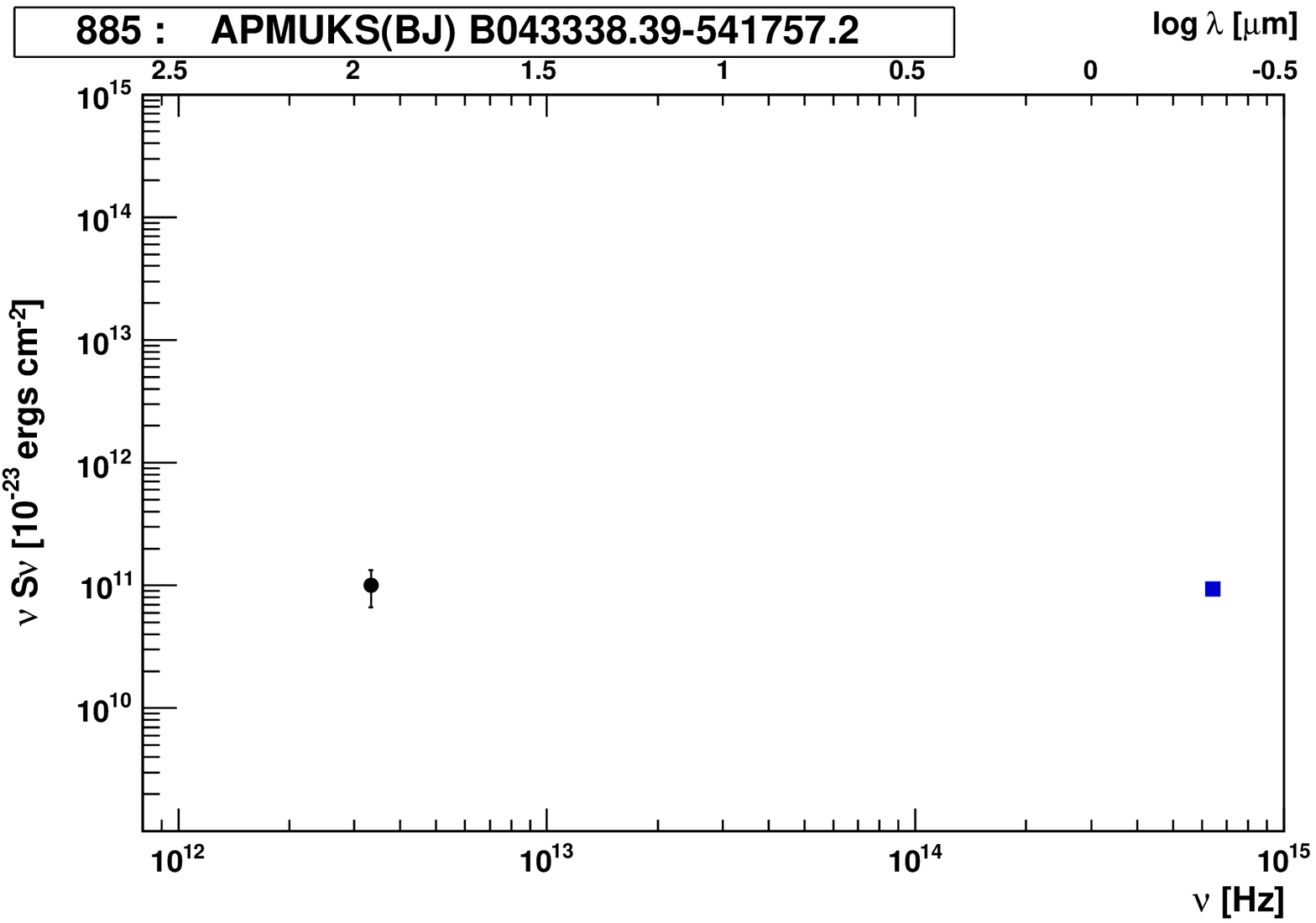}
\includegraphics[width=4cm]{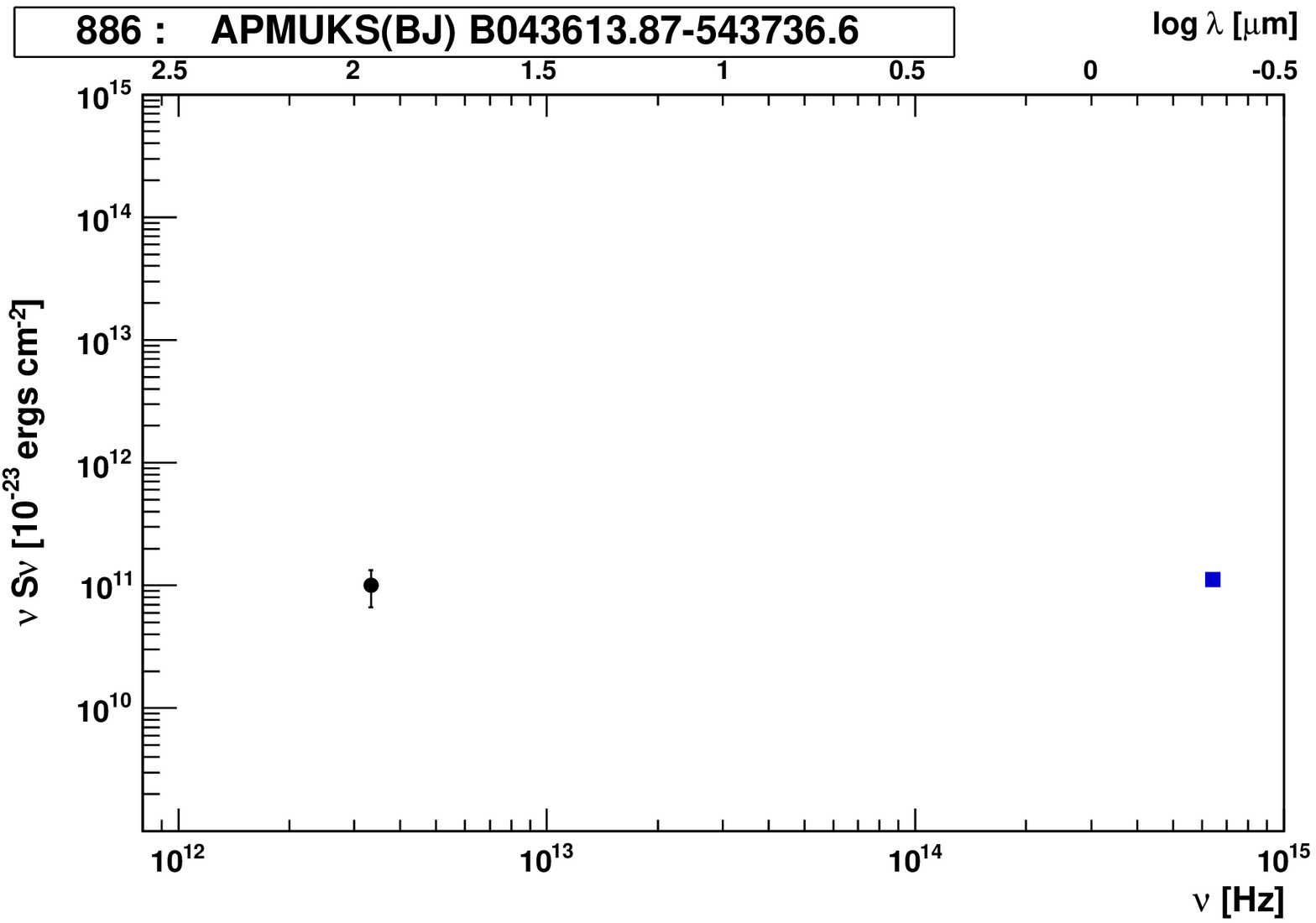}
\includegraphics[width=4cm]{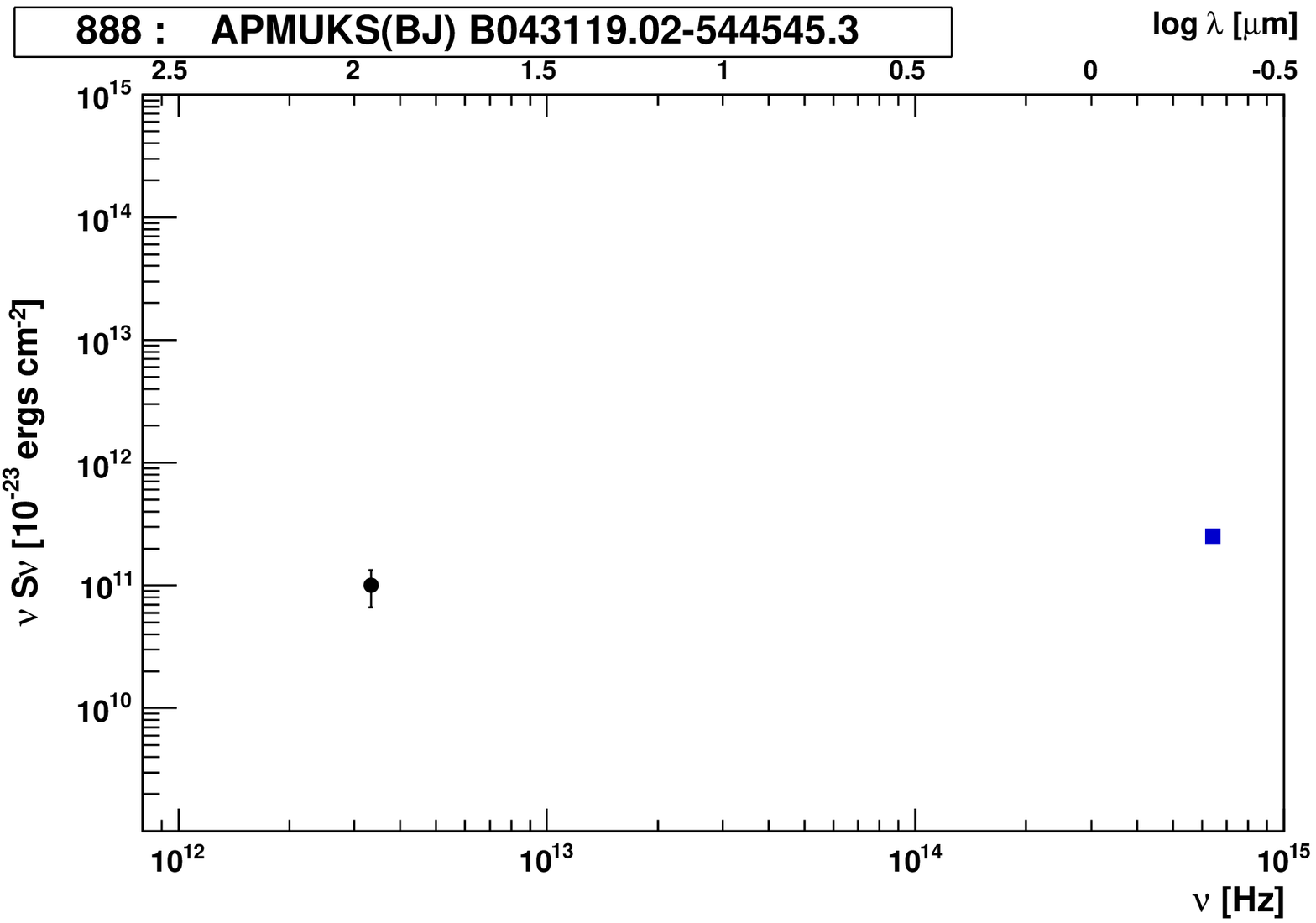}
\includegraphics[width=4cm]{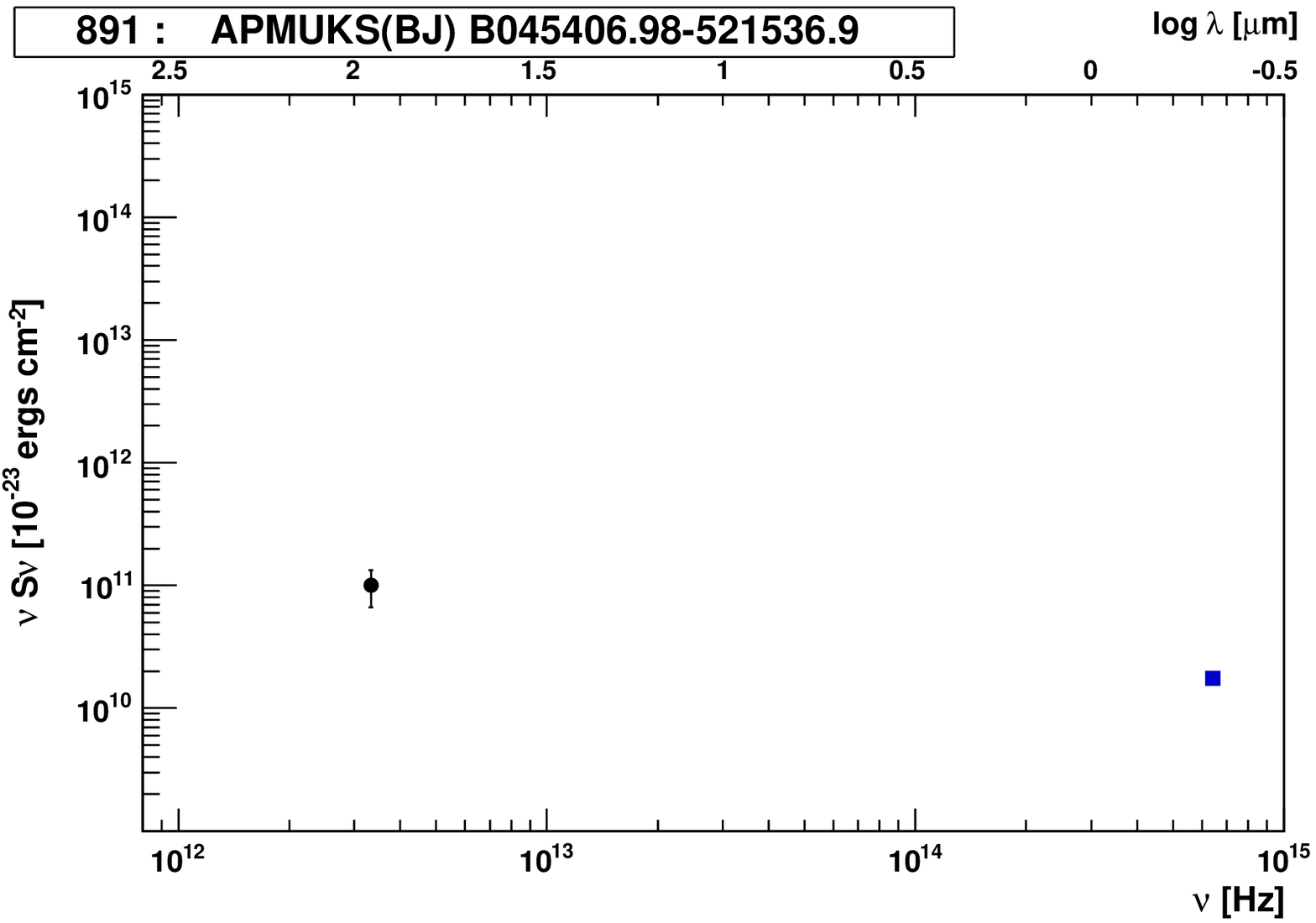}
\label{points14}
\caption {SEDs for the next 36 ADF-S identified sources, with symbols as in Figure~\ref{points1}.}
\end{figure*}
}

\clearpage

\onlfig{15}{
\begin{figure*}[t]
\centering

\includegraphics[width=4cm]{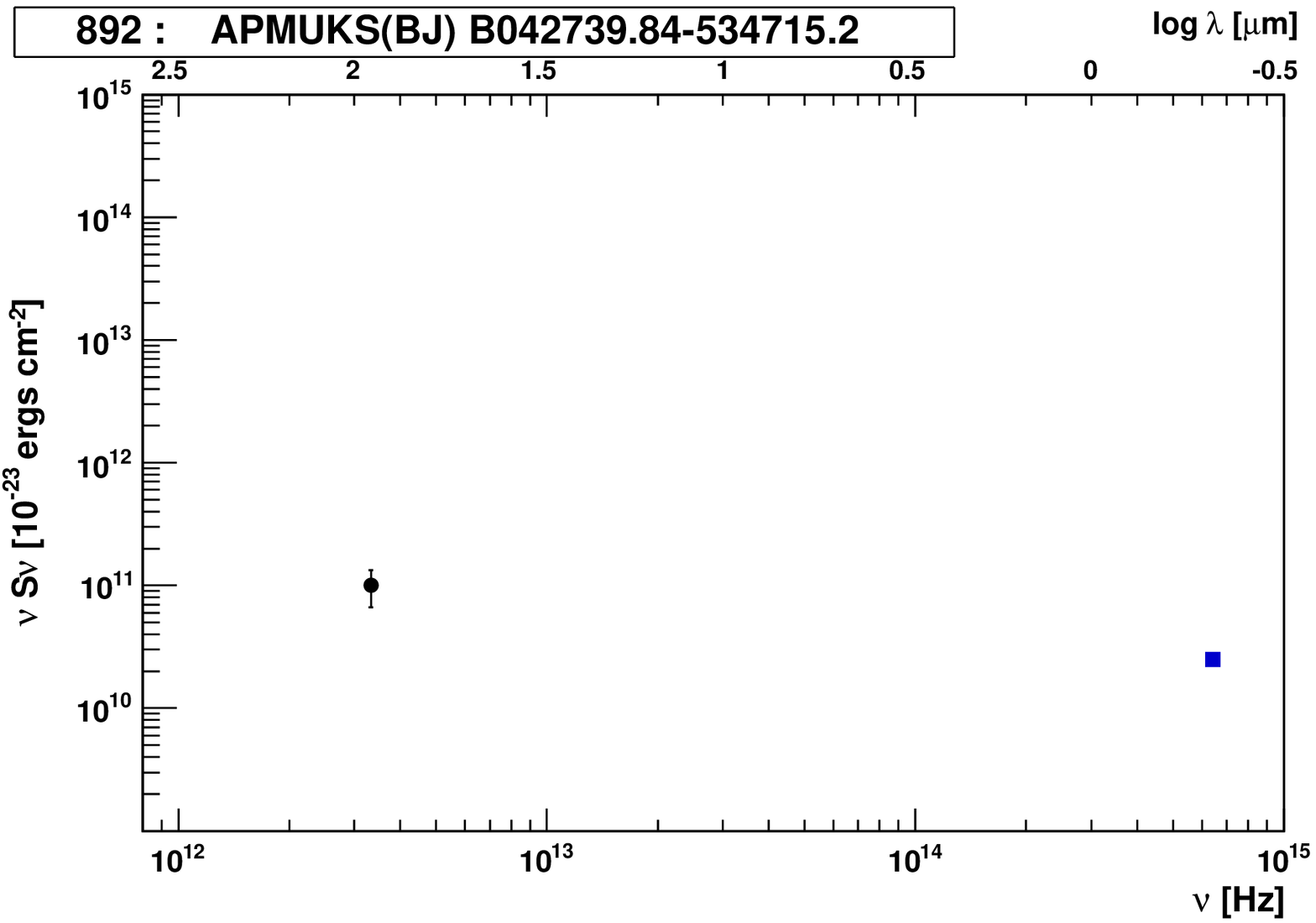}
\includegraphics[width=4cm]{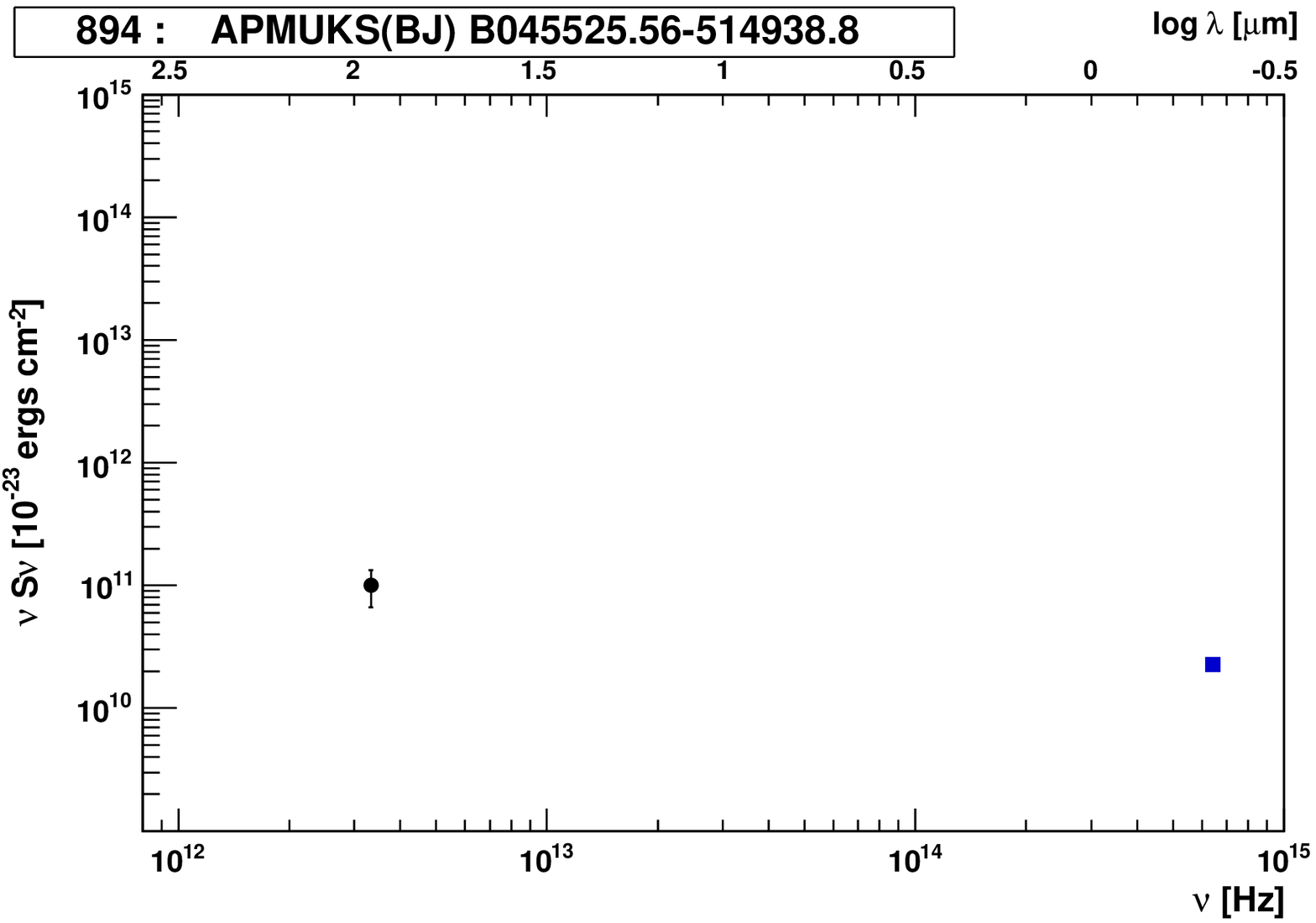}
\includegraphics[width=4cm]{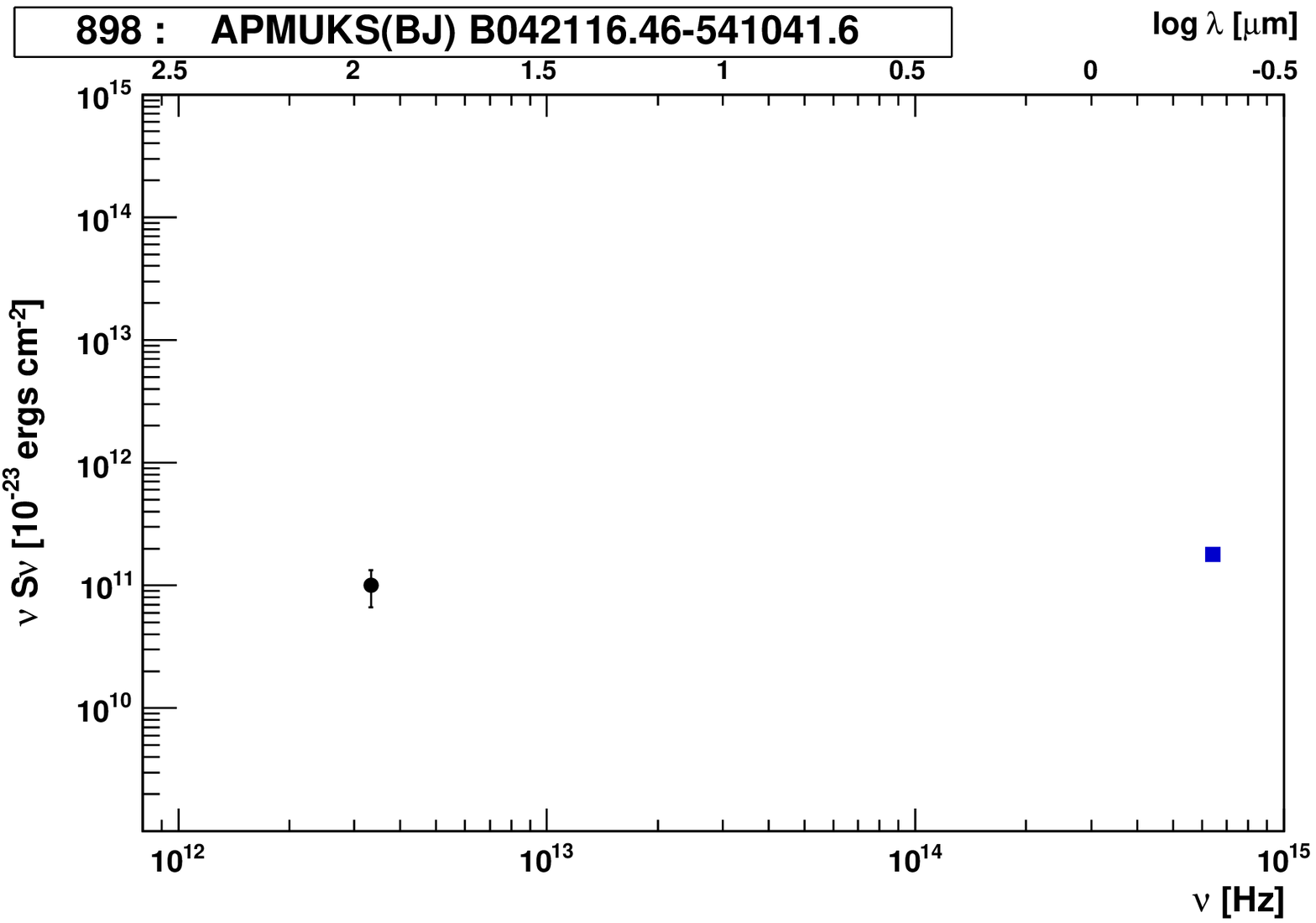}
\includegraphics[width=4cm]{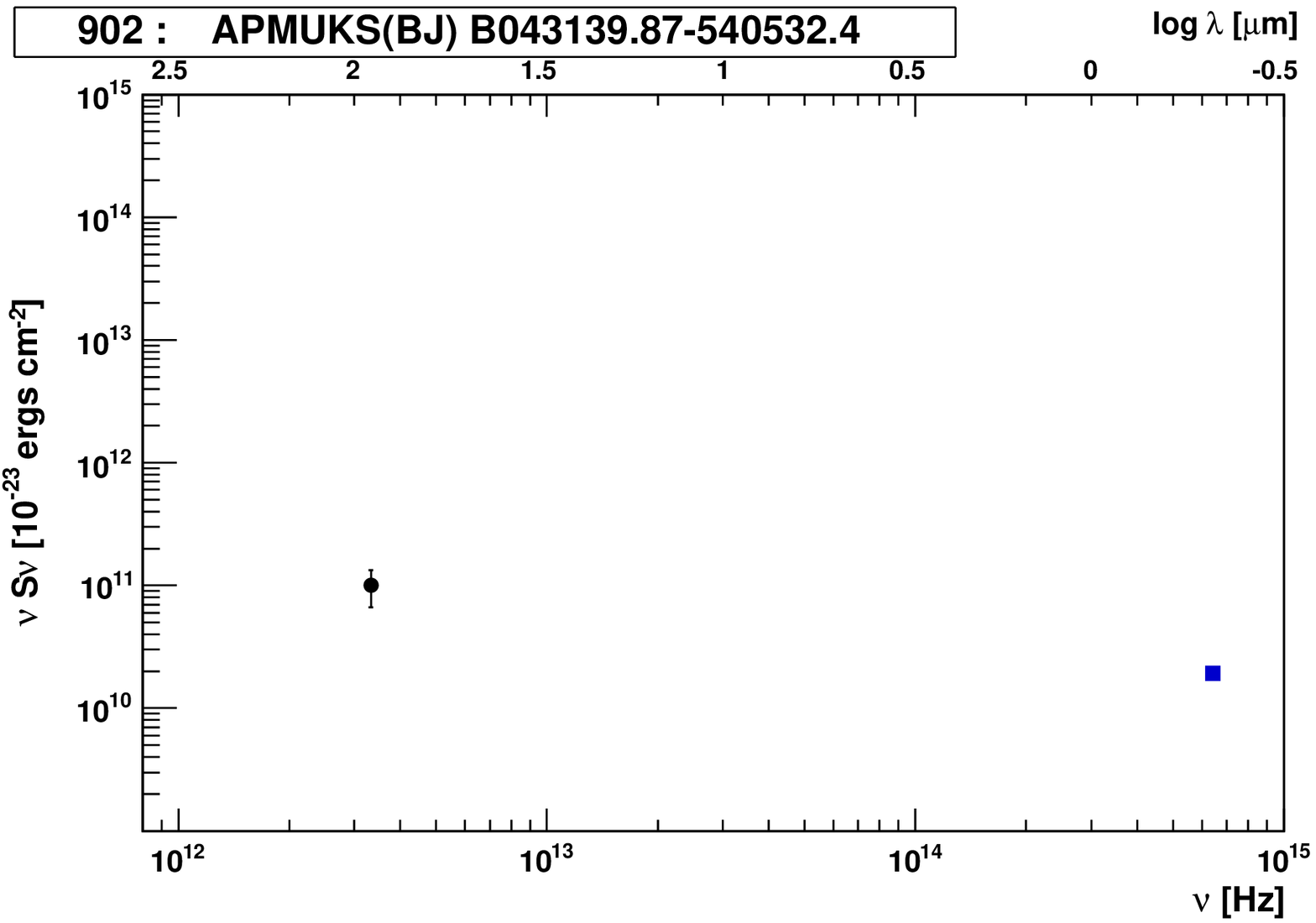}
\includegraphics[width=4cm]{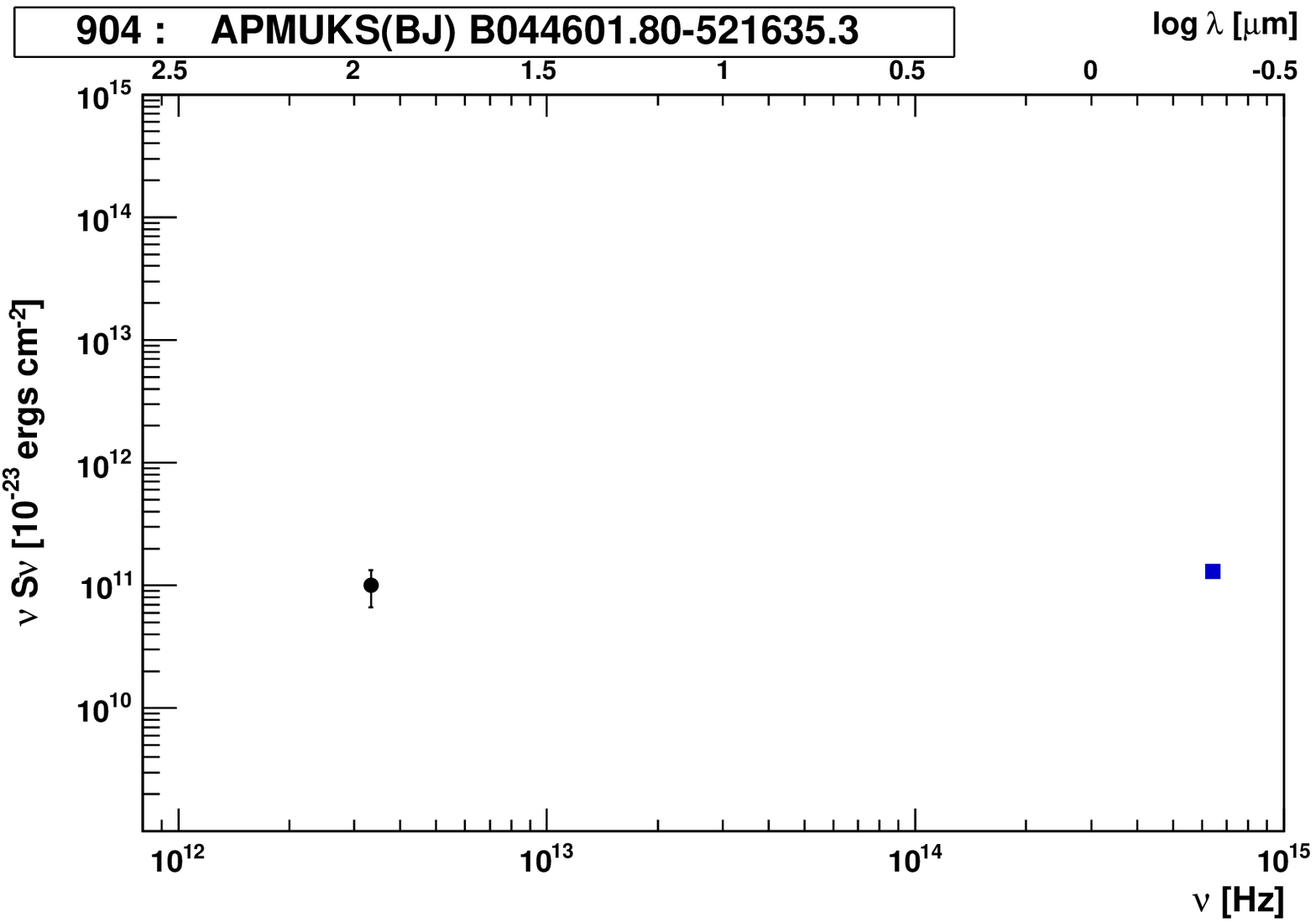}
\includegraphics[width=4cm]{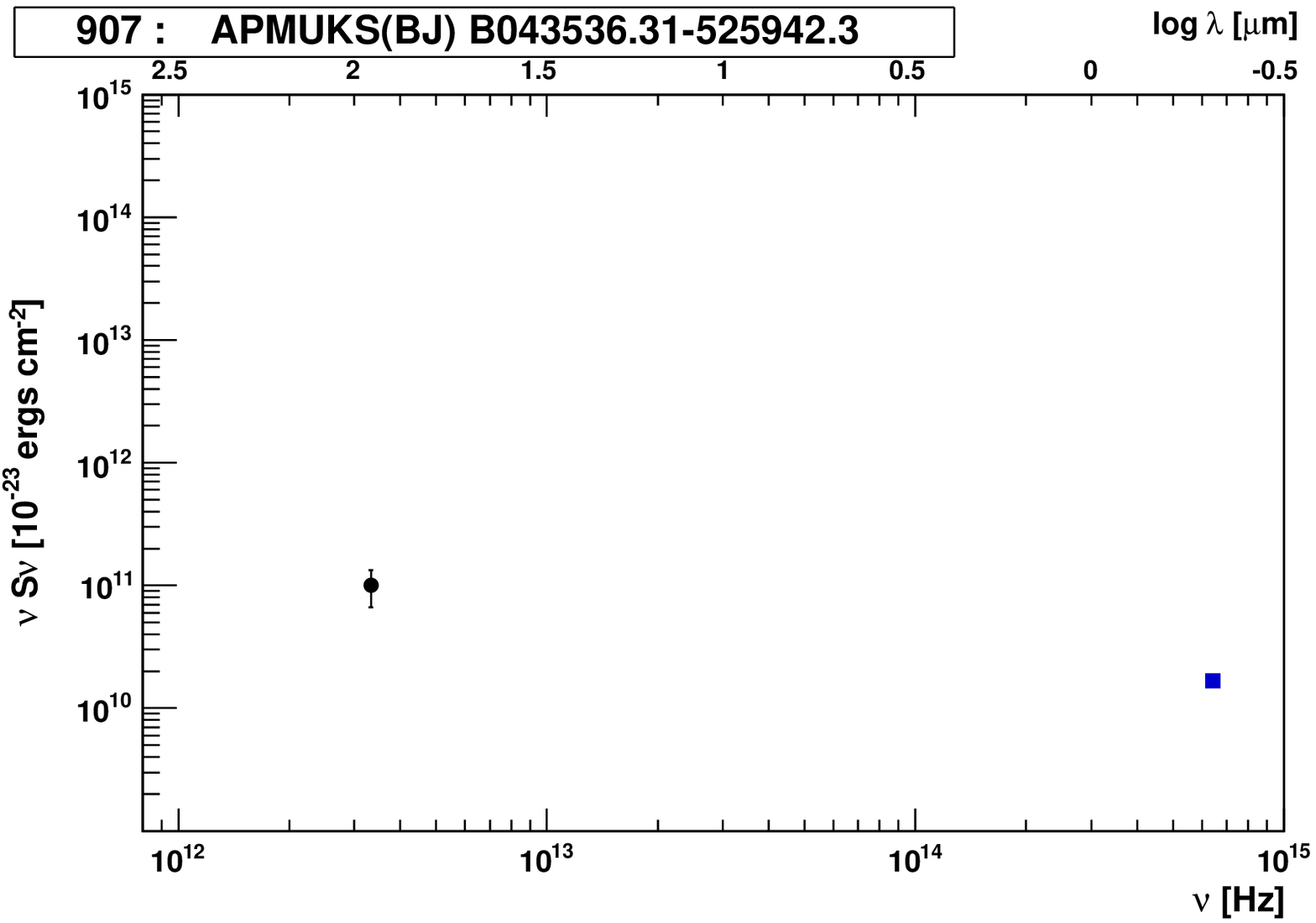}
\includegraphics[width=4cm]{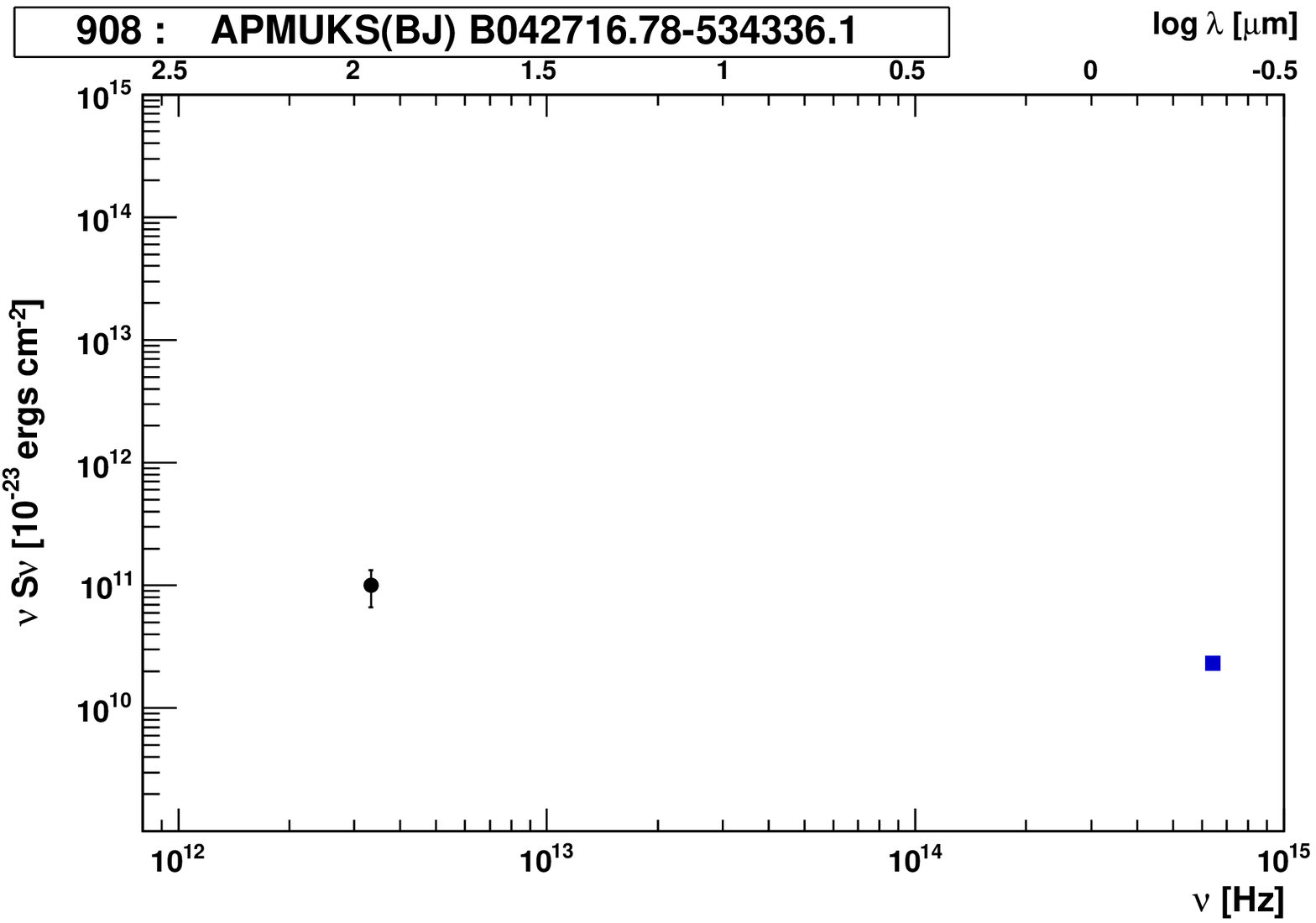}
\includegraphics[width=4cm]{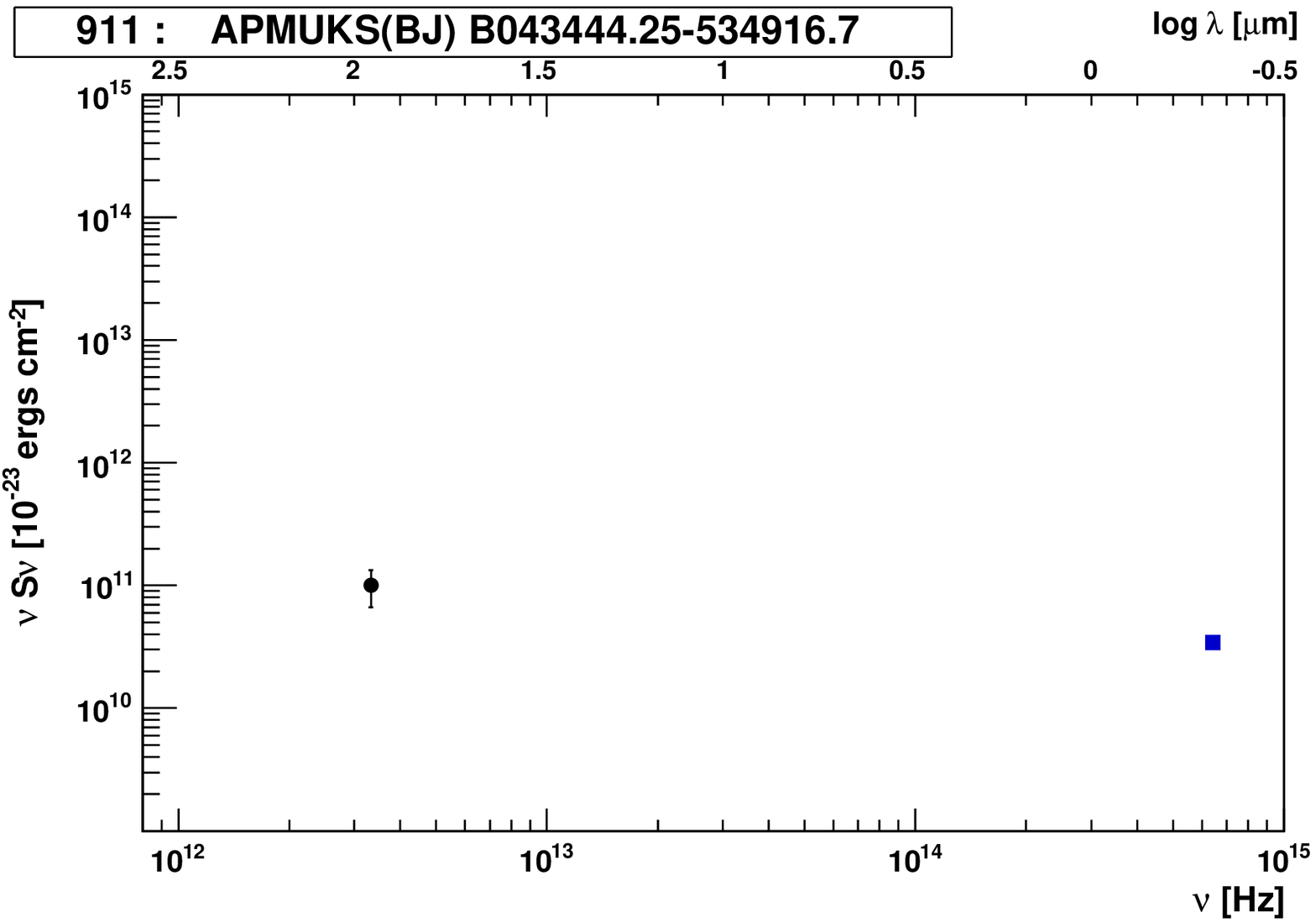}
\includegraphics[width=4cm]{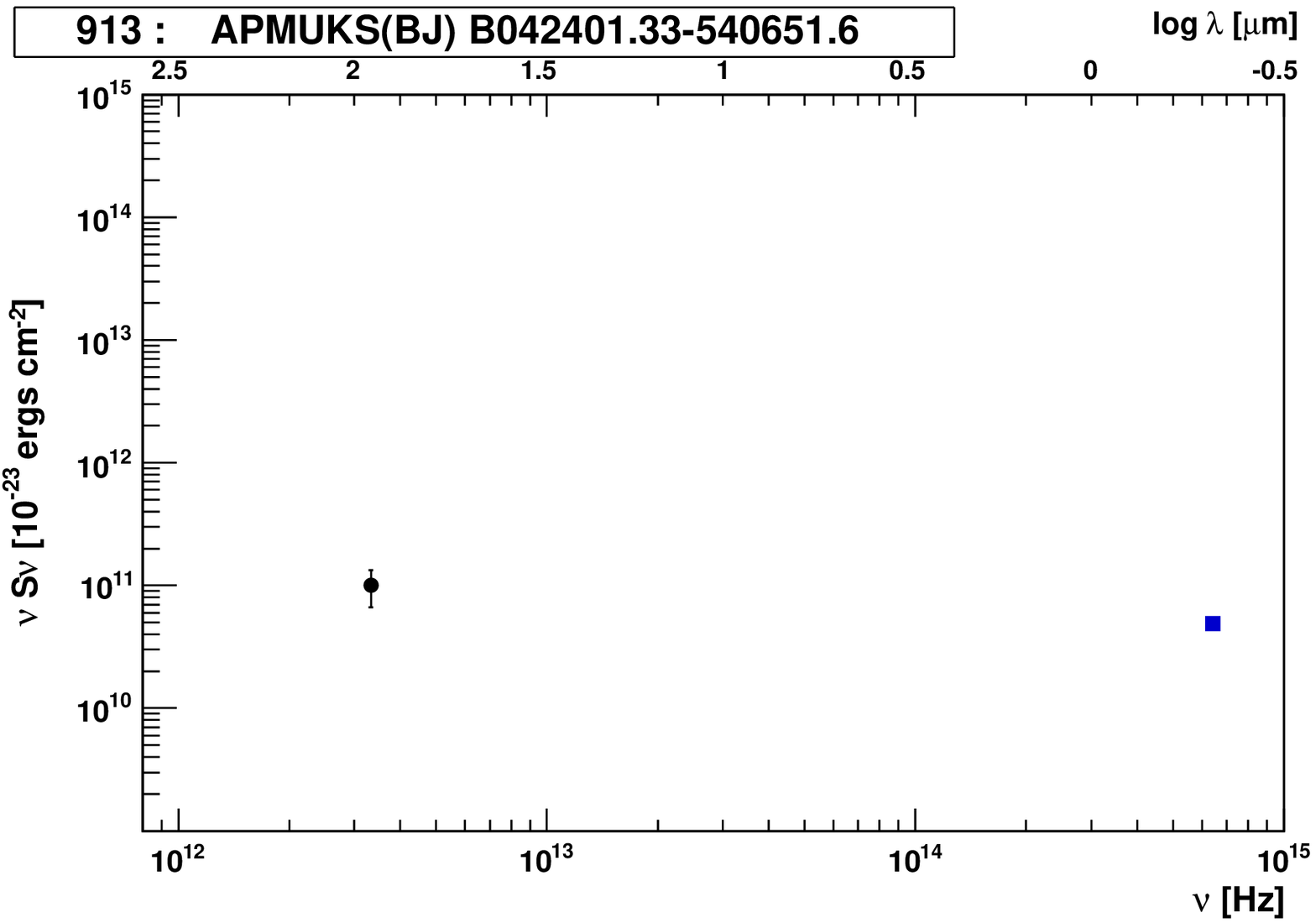}
\includegraphics[width=4cm]{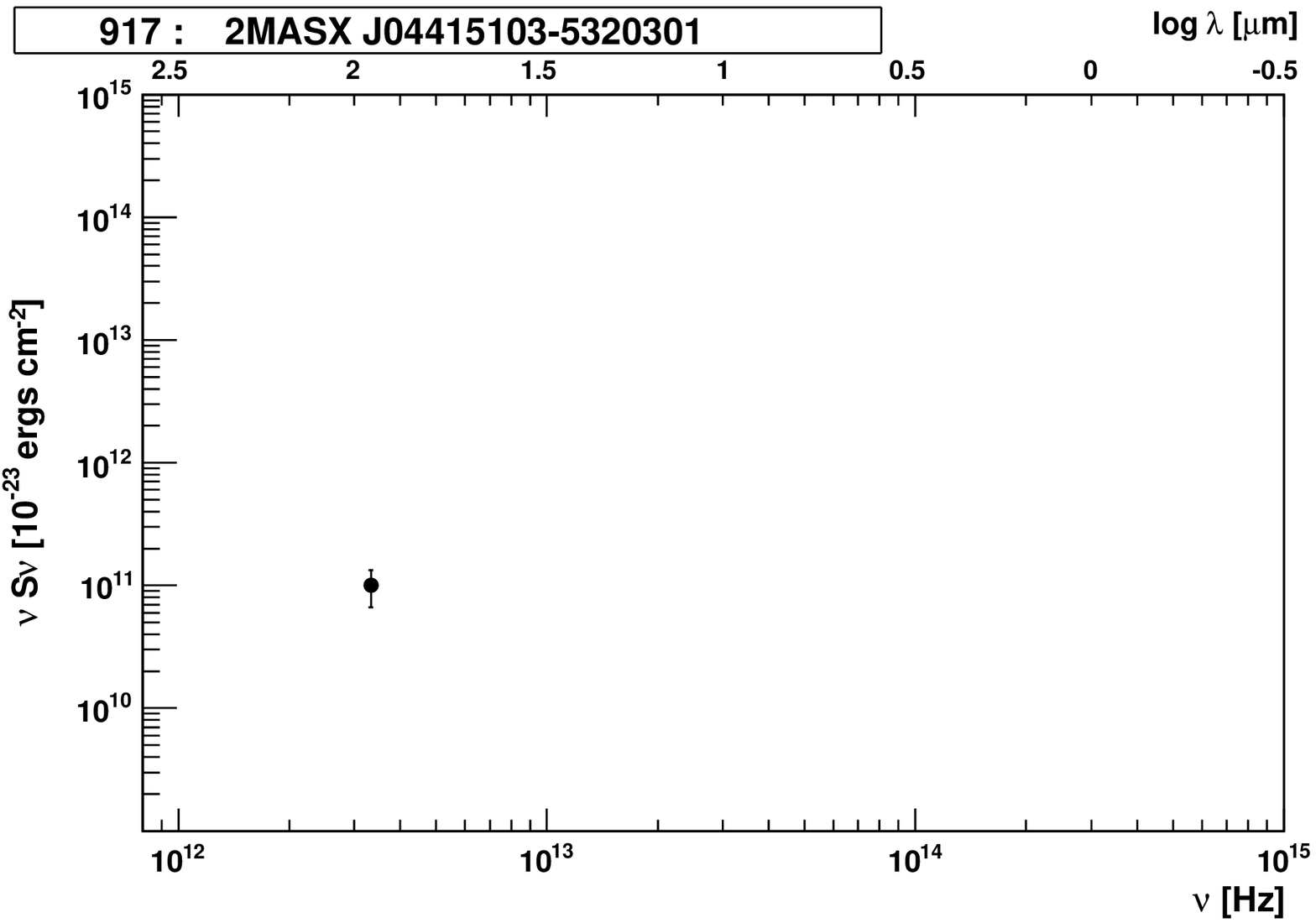}
\includegraphics[width=4cm]{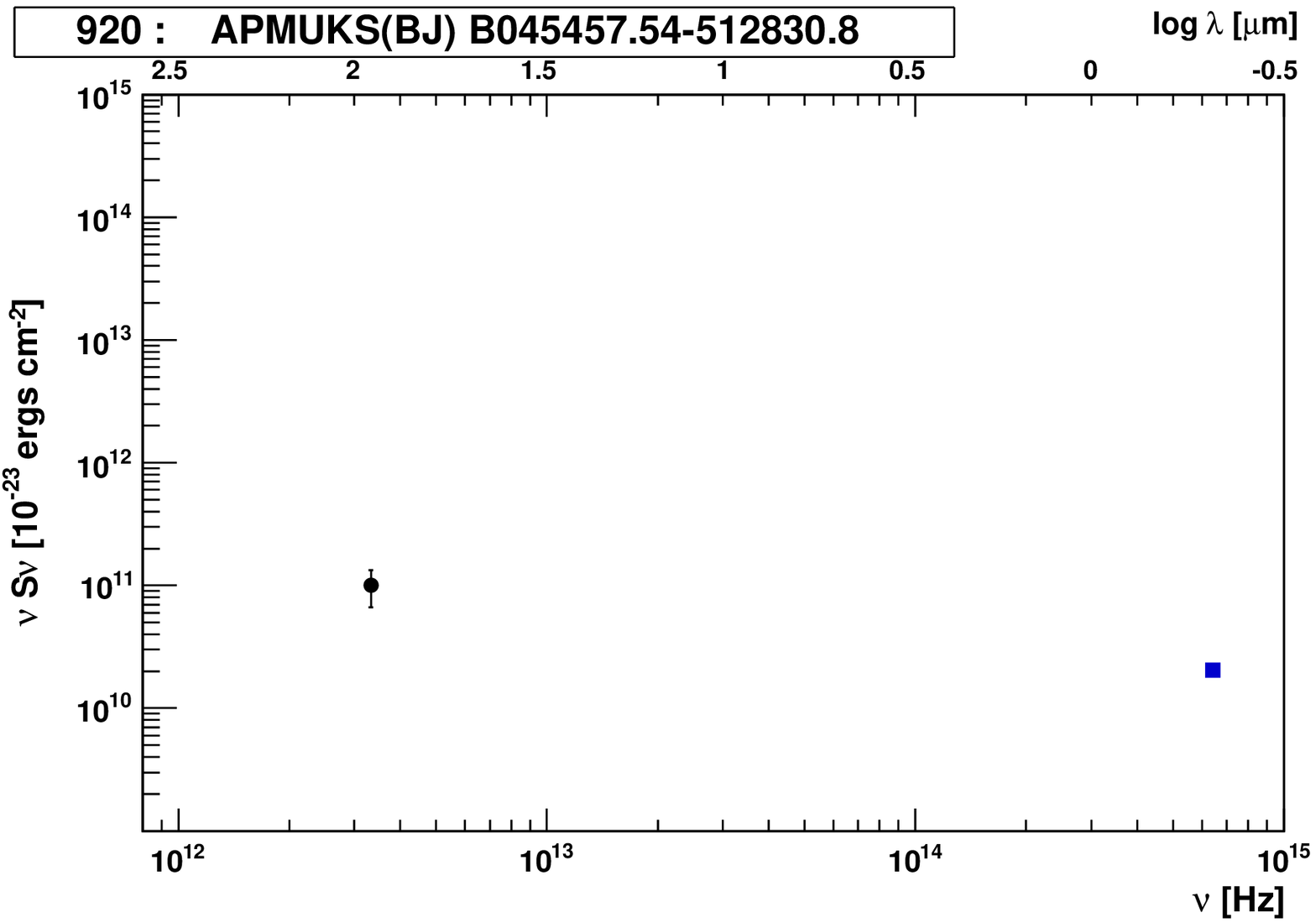}
\includegraphics[width=4cm]{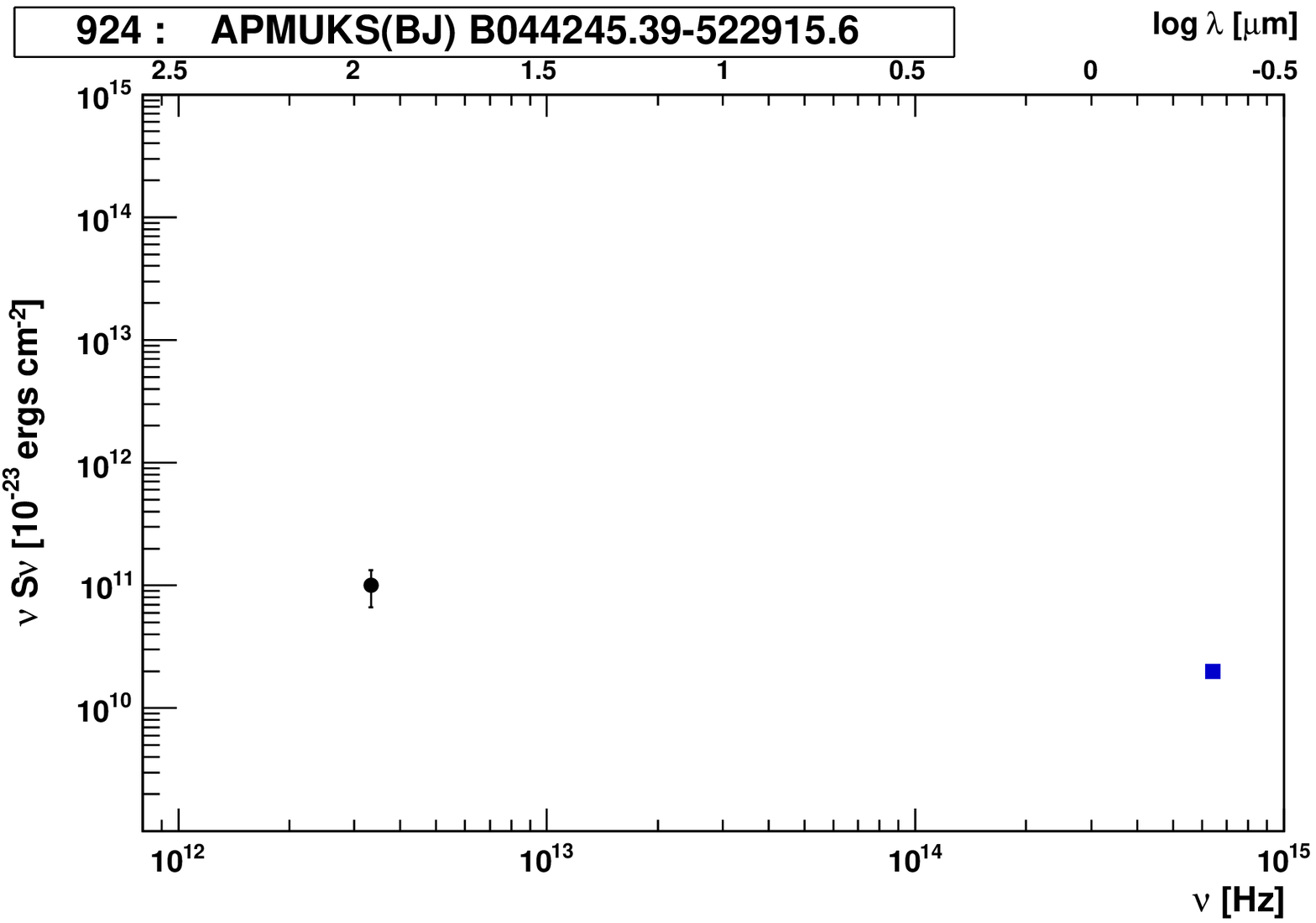}
\includegraphics[width=4cm]{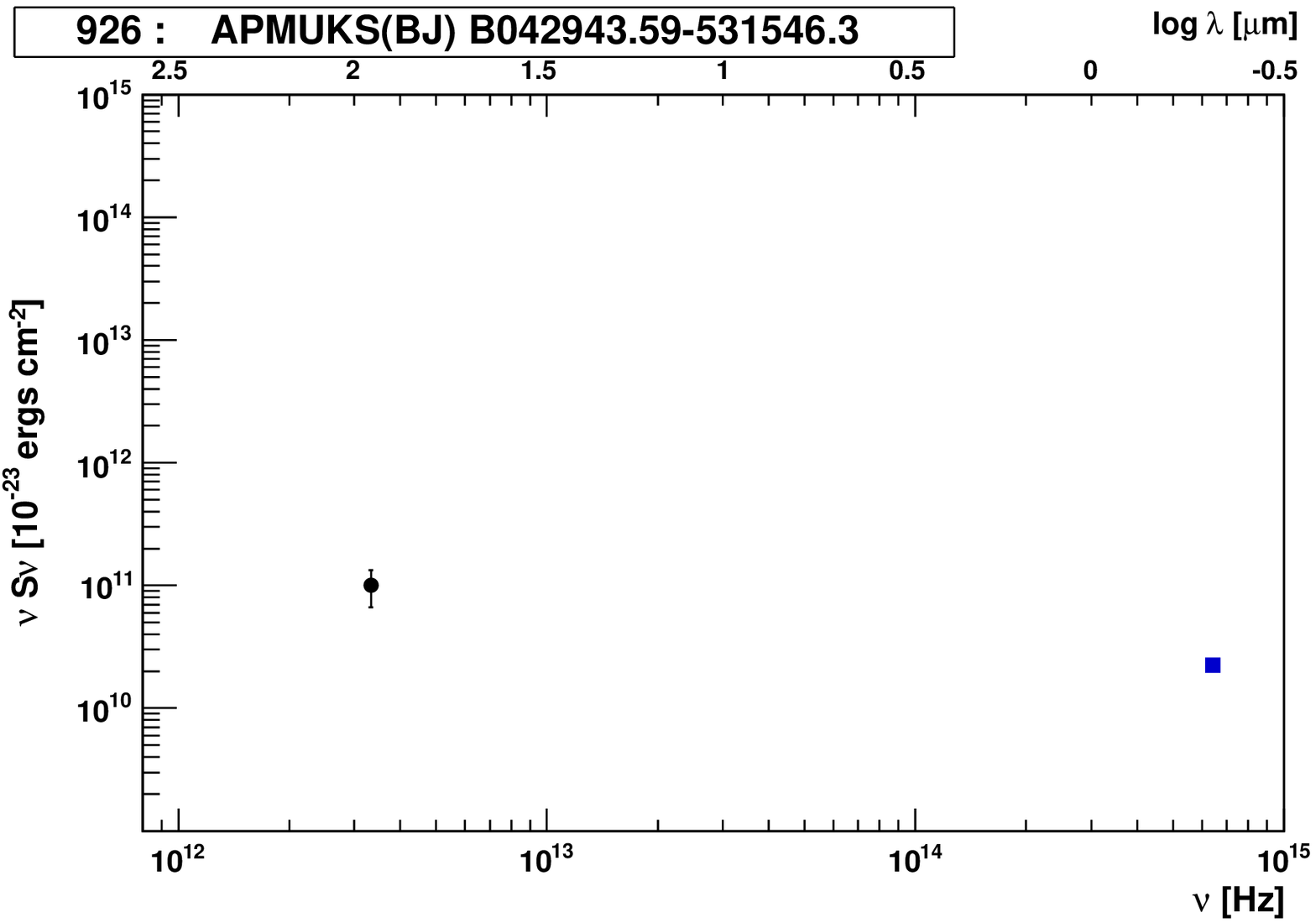}
\includegraphics[width=4cm]{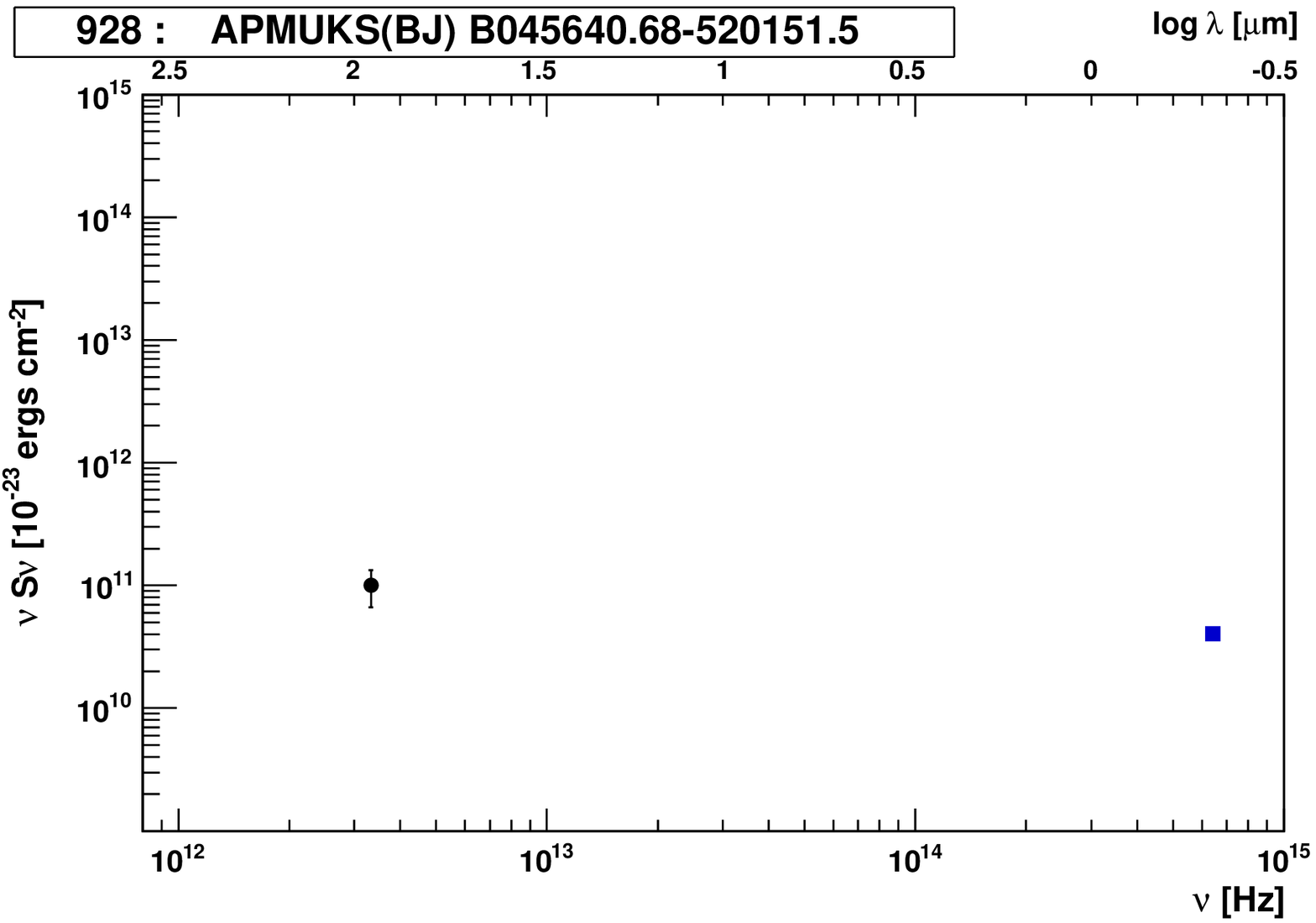}
\includegraphics[width=4cm]{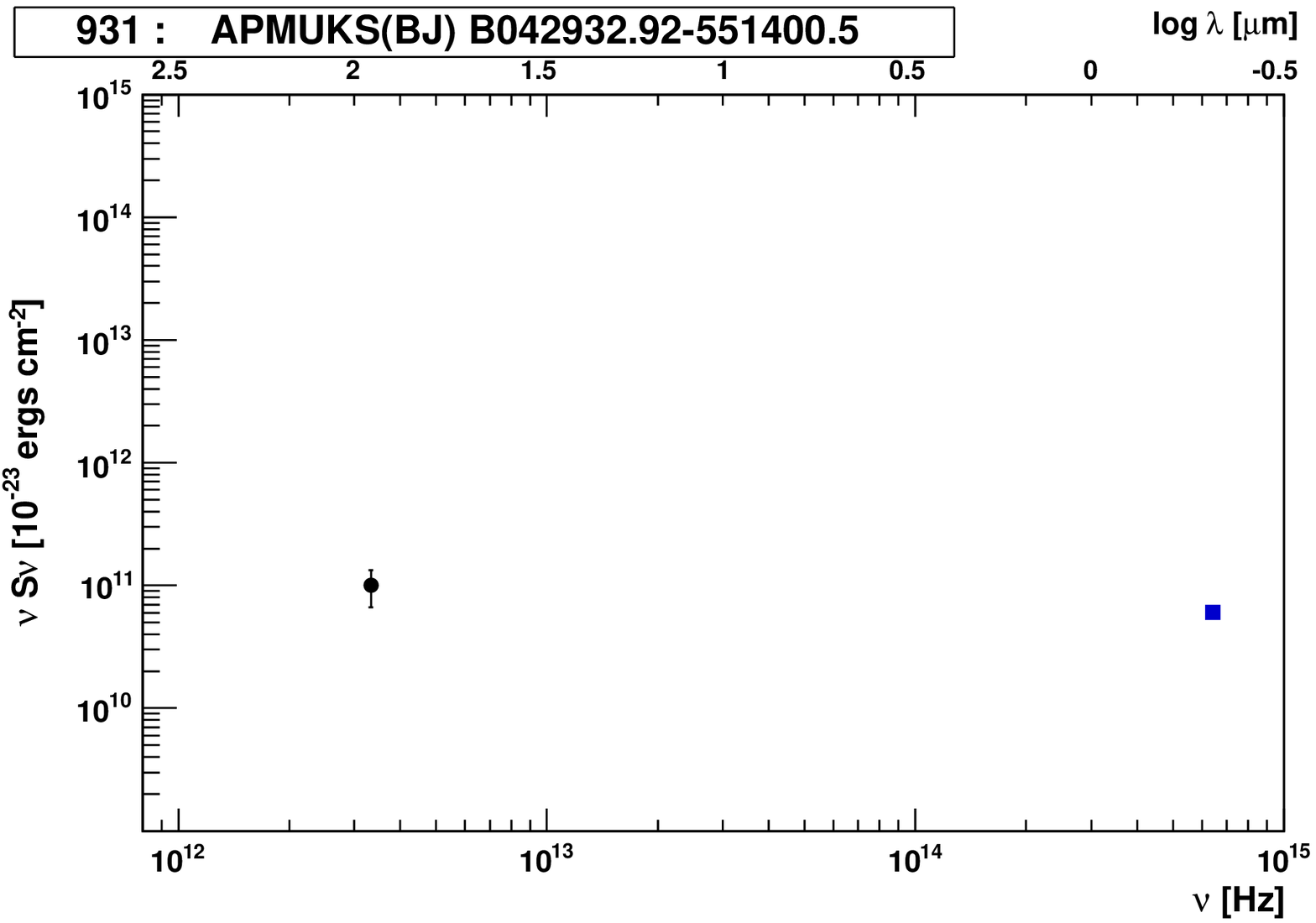}
\includegraphics[width=4cm]{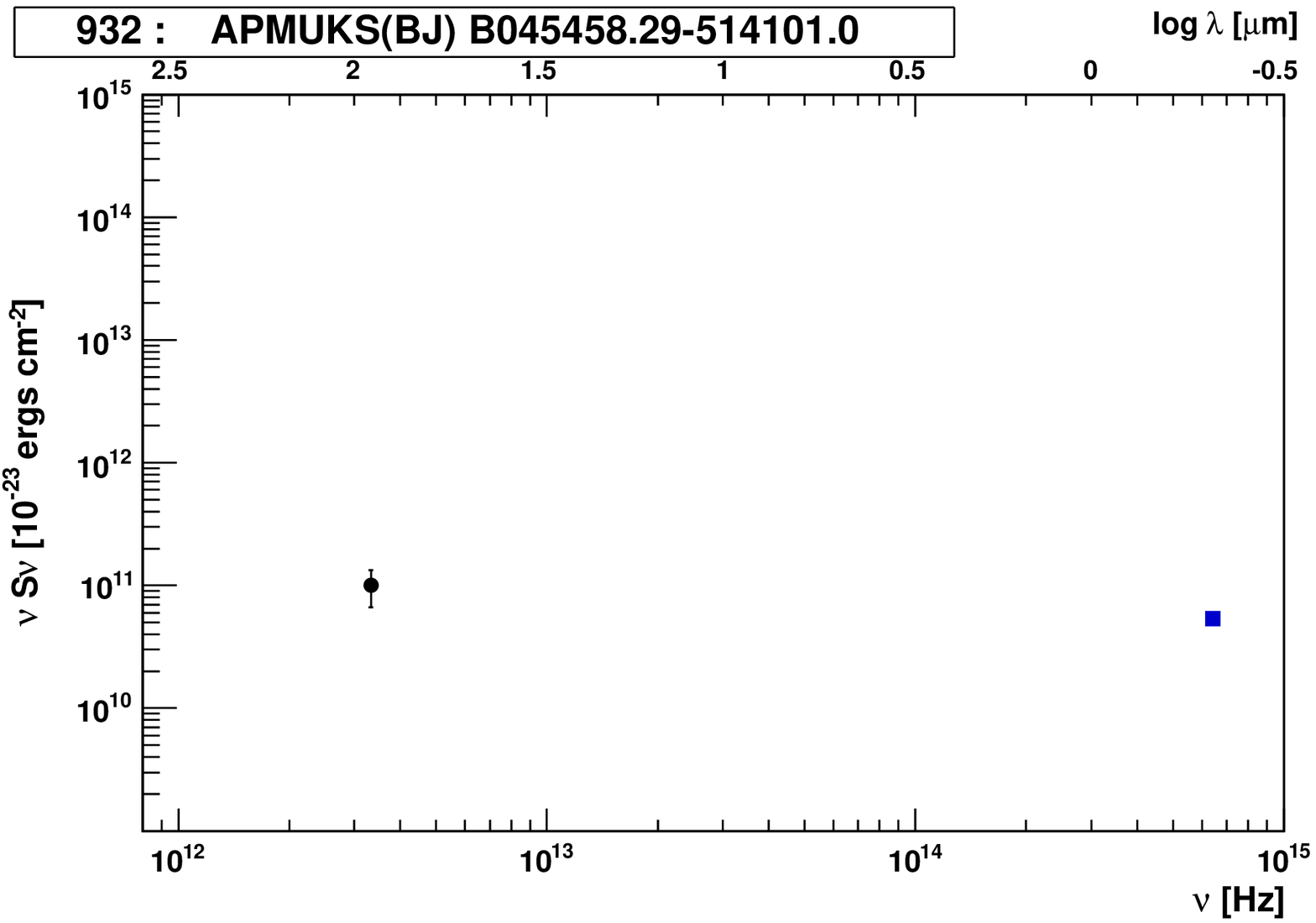}
\includegraphics[width=4cm]{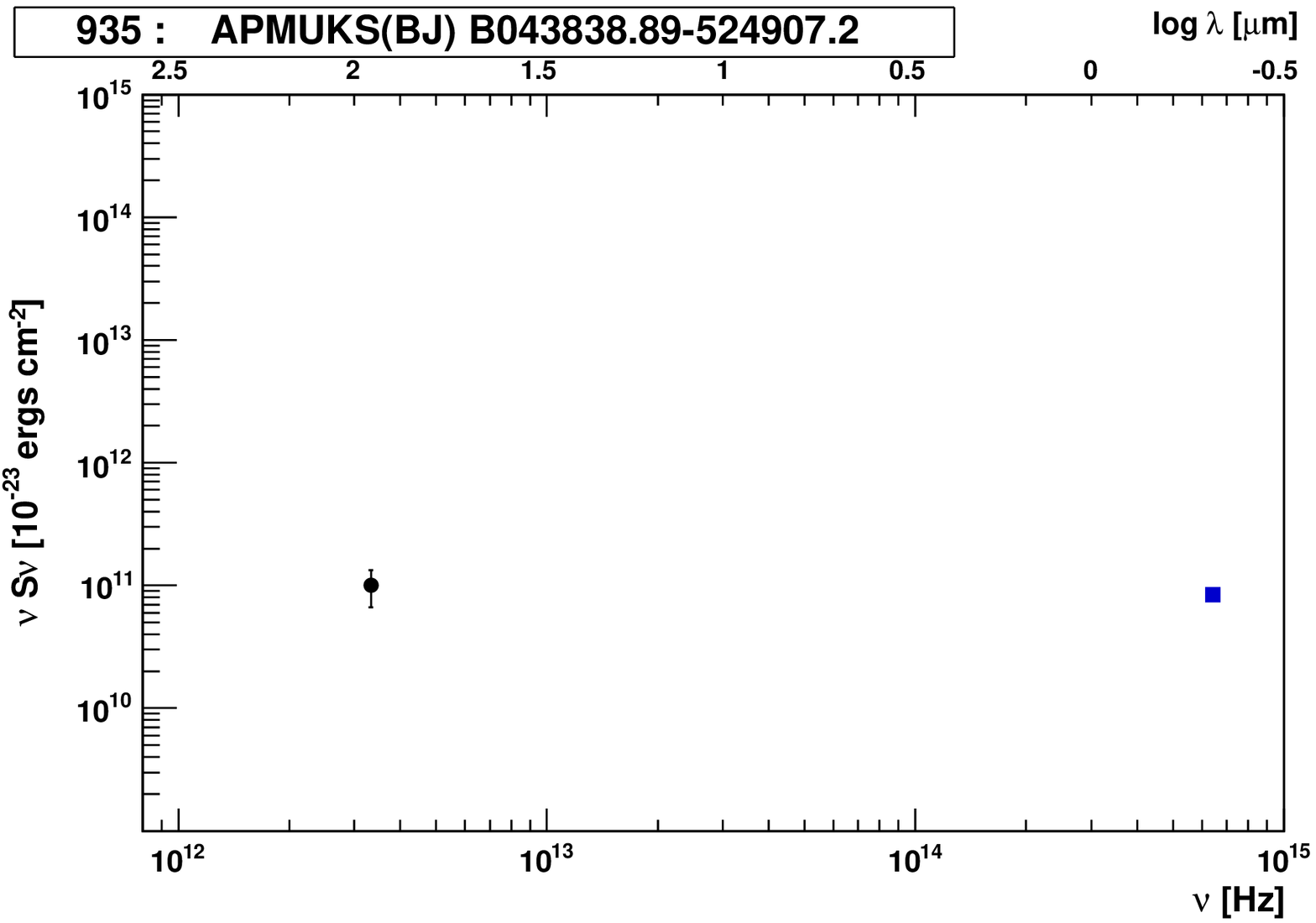}
\includegraphics[width=4cm]{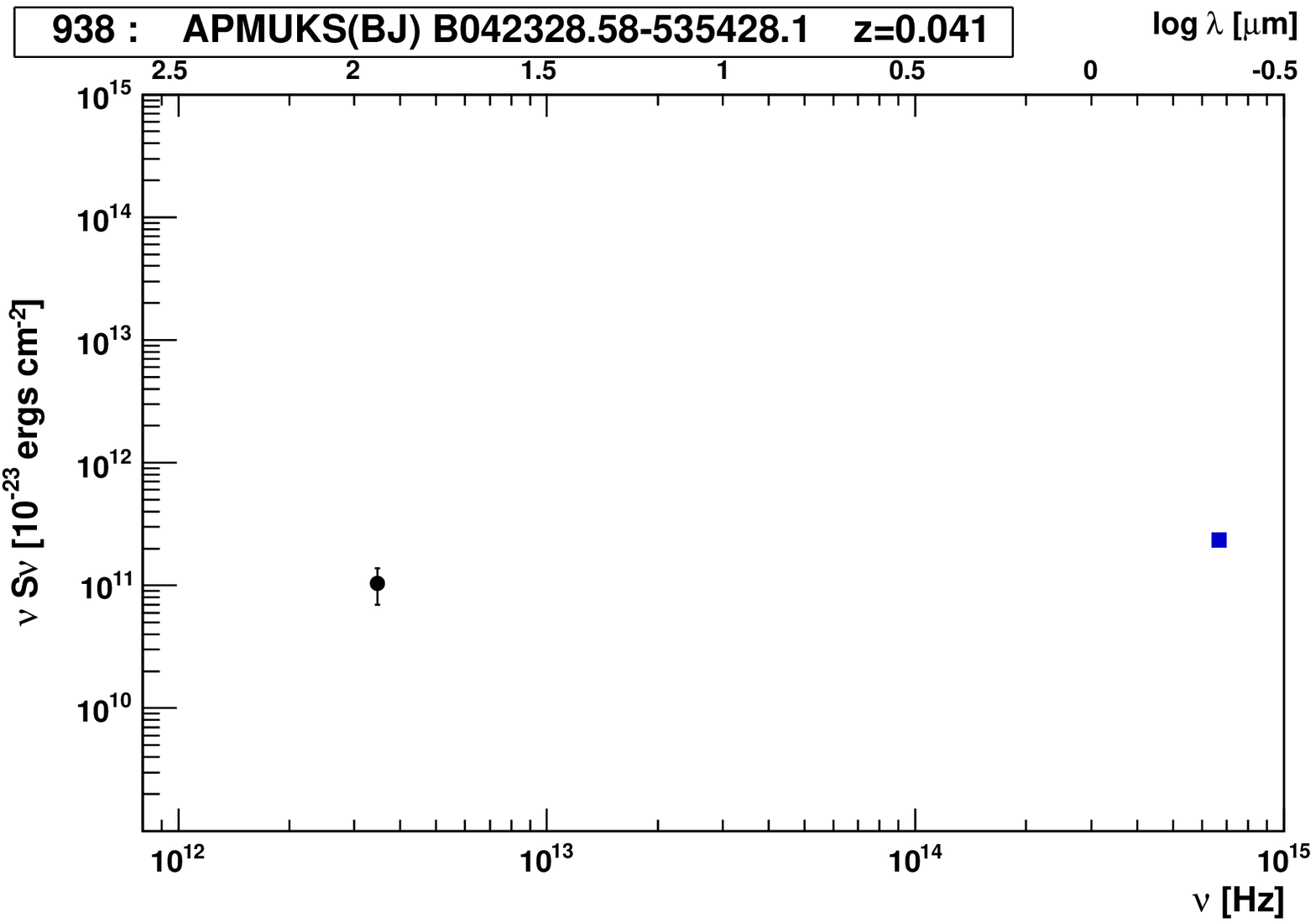}
\includegraphics[width=4cm]{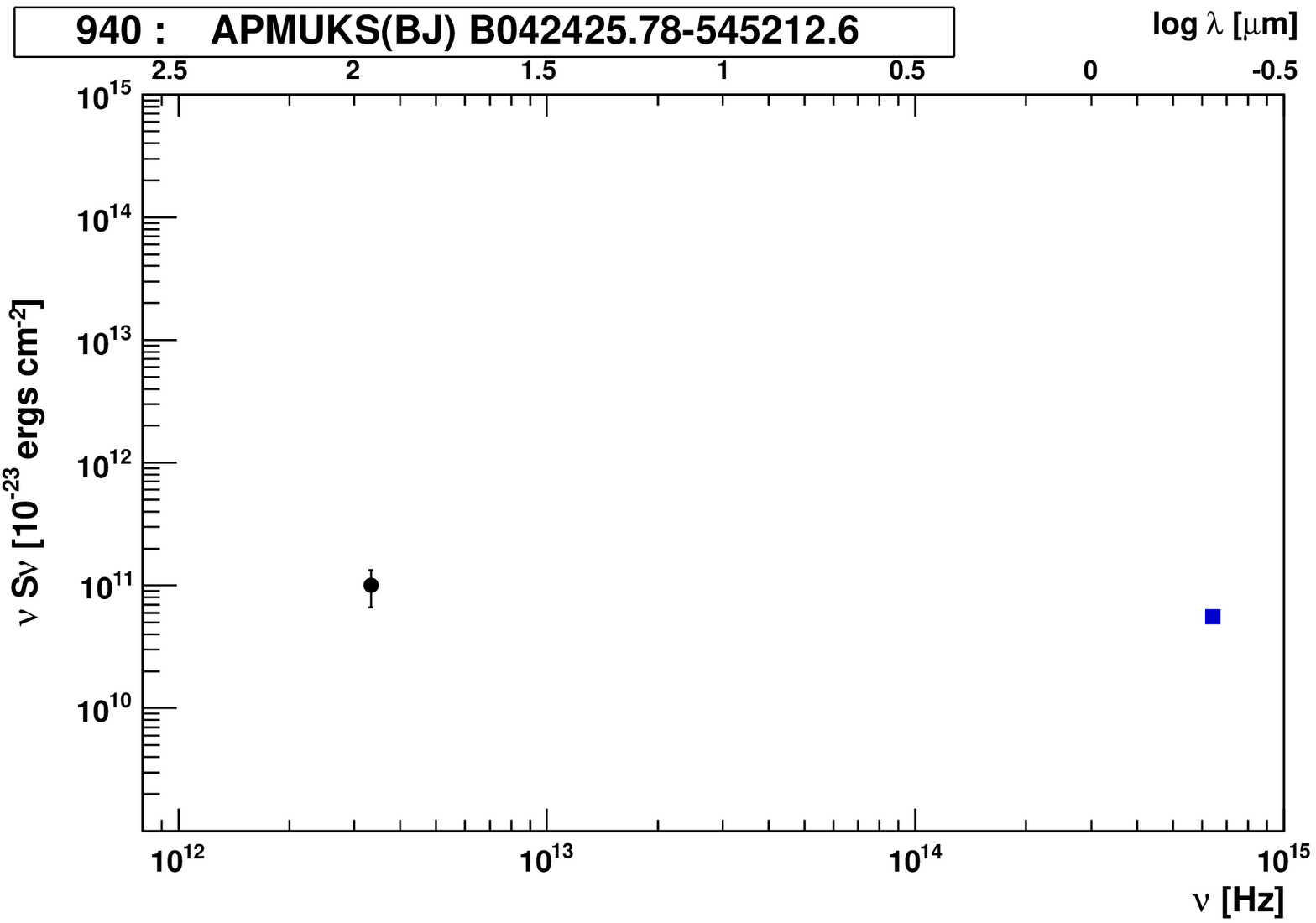}
\includegraphics[width=4cm]{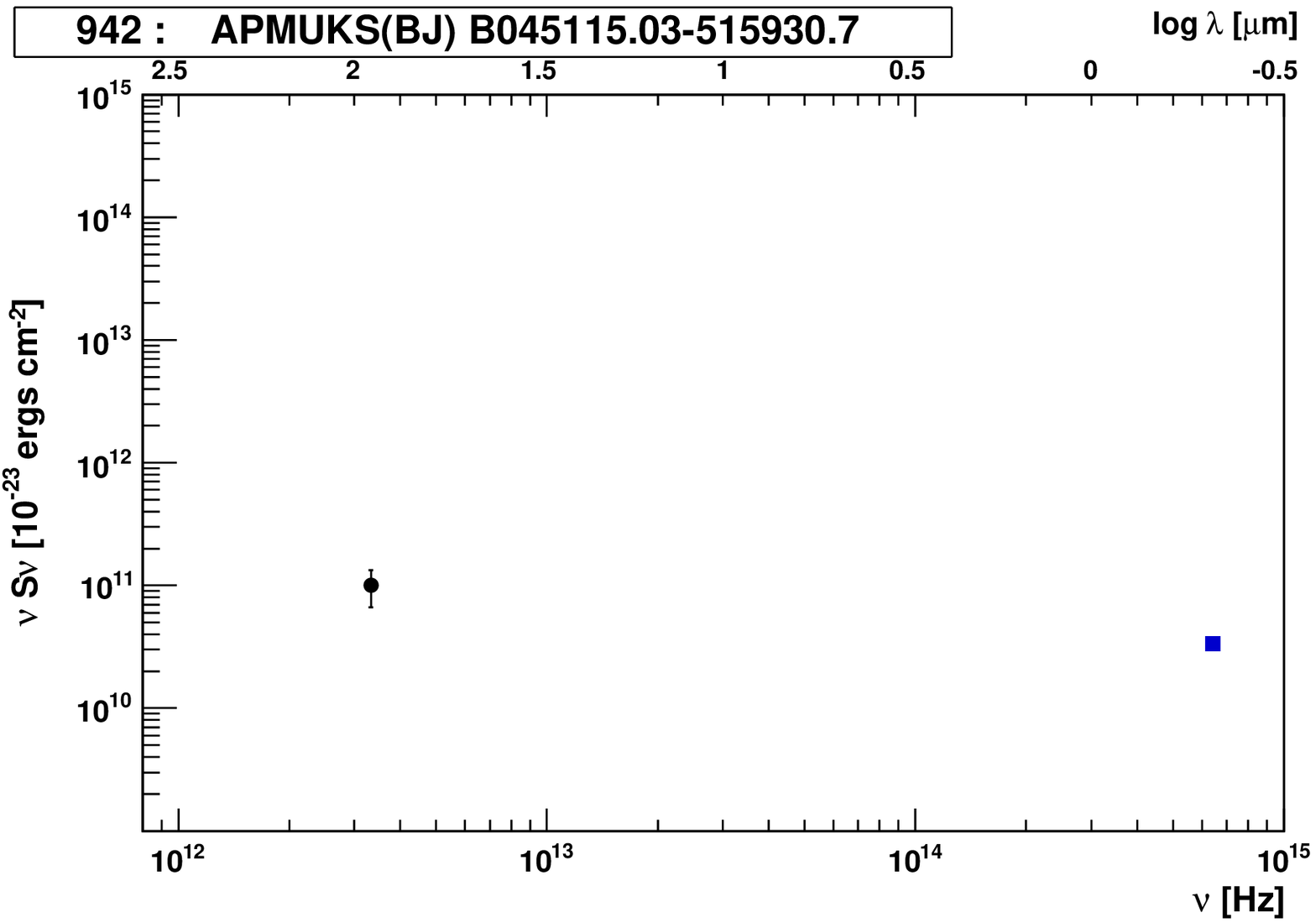}
\includegraphics[width=4cm]{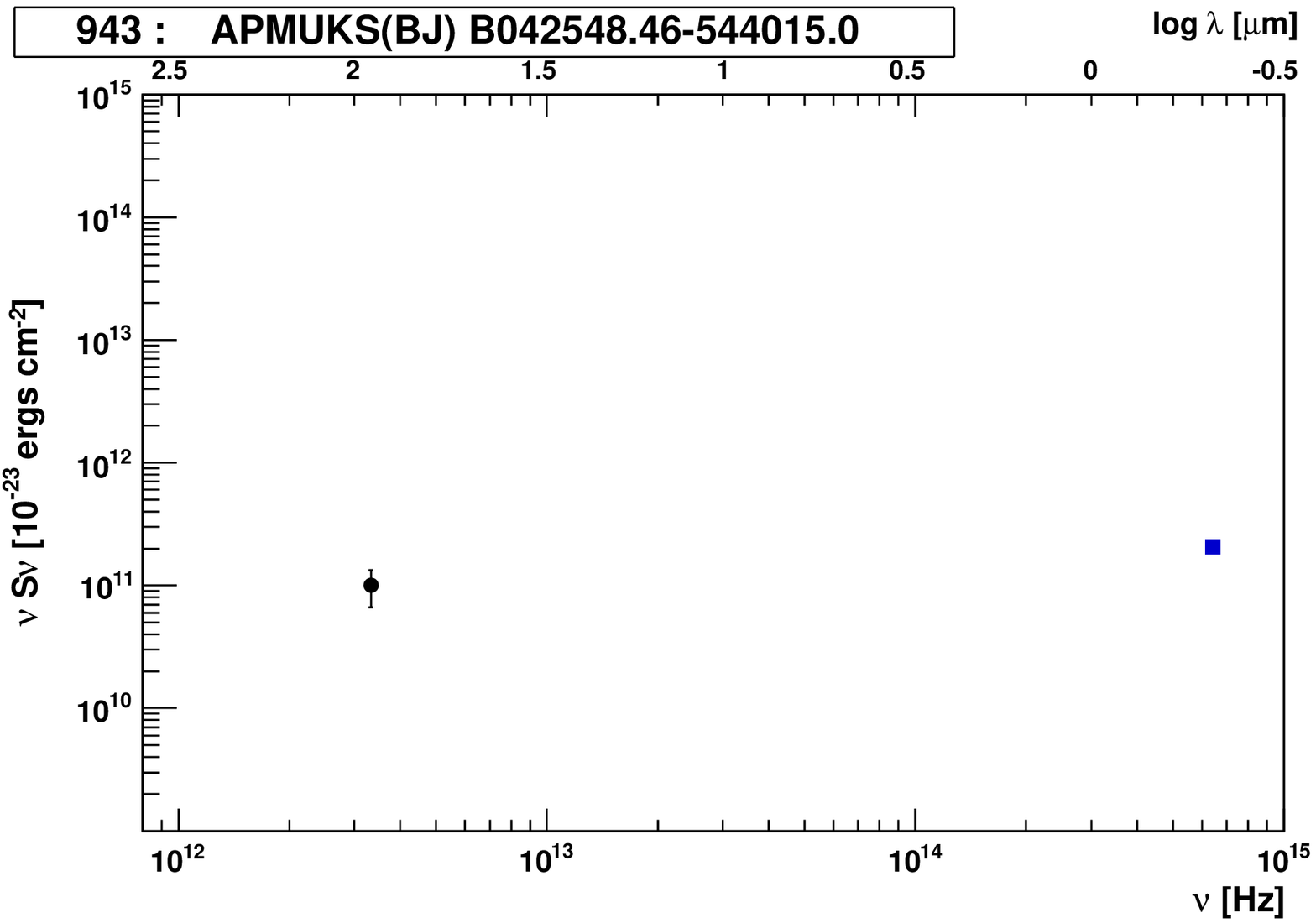}
\includegraphics[width=4cm]{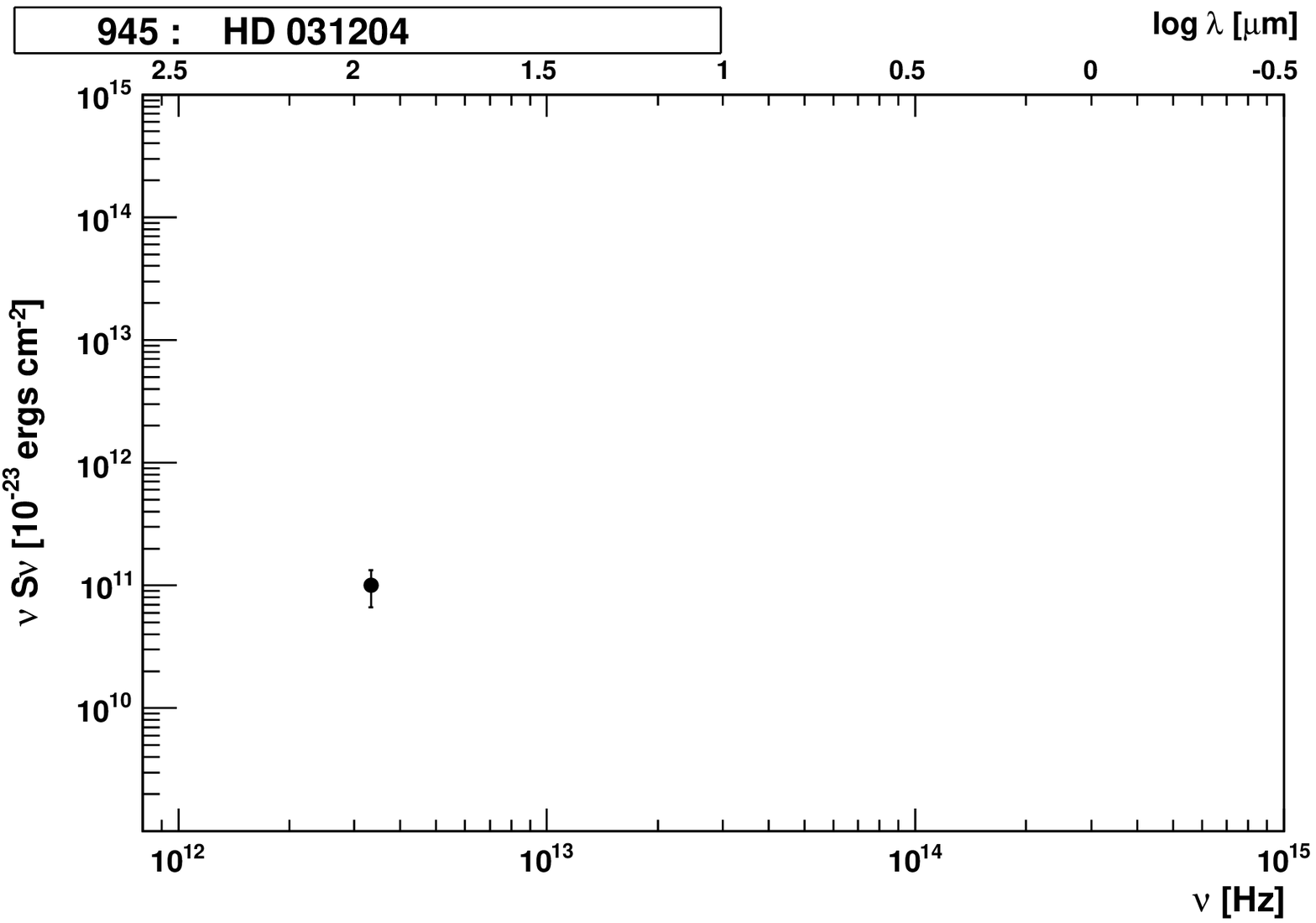}
\includegraphics[width=4cm]{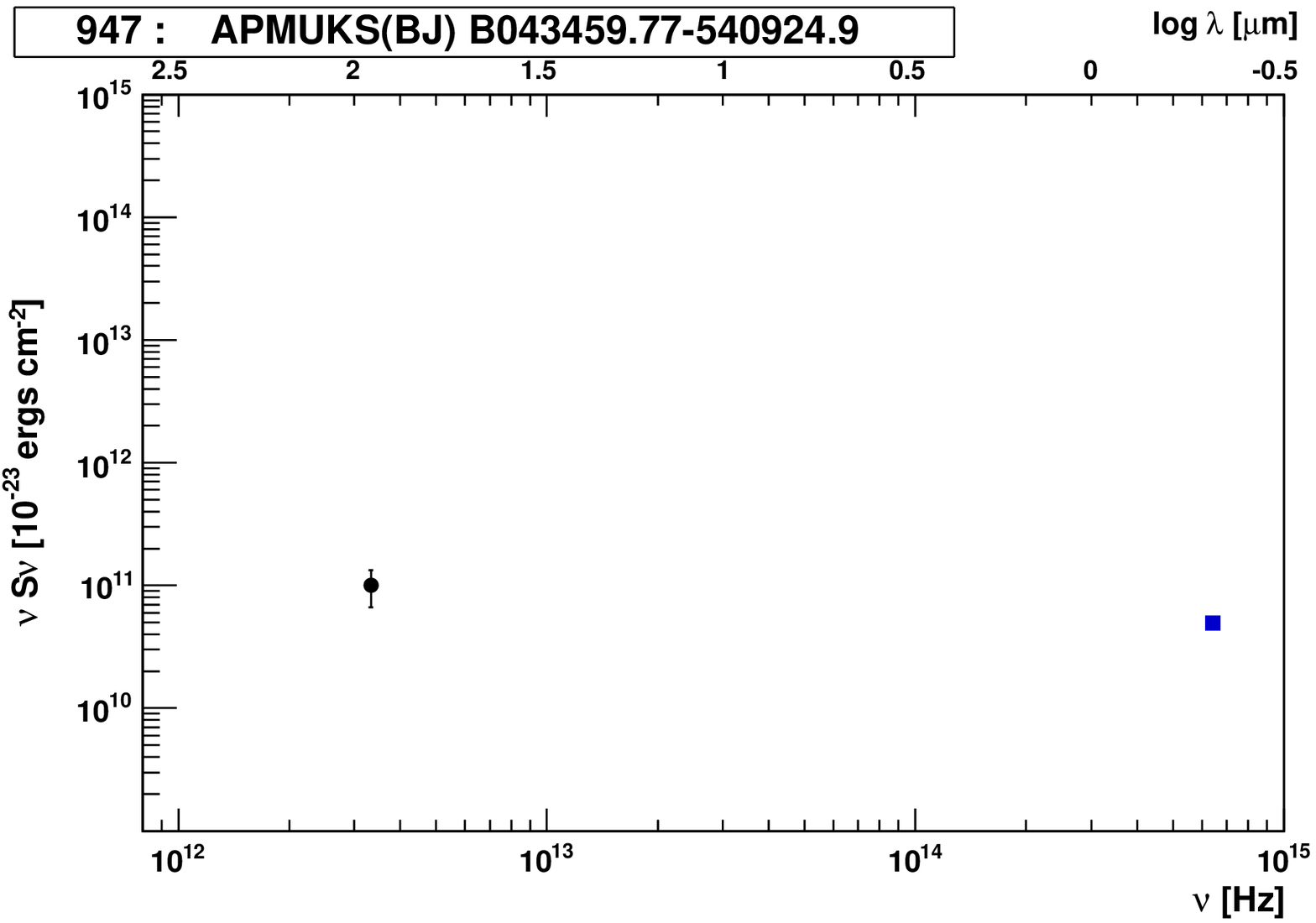}
\includegraphics[width=4cm]{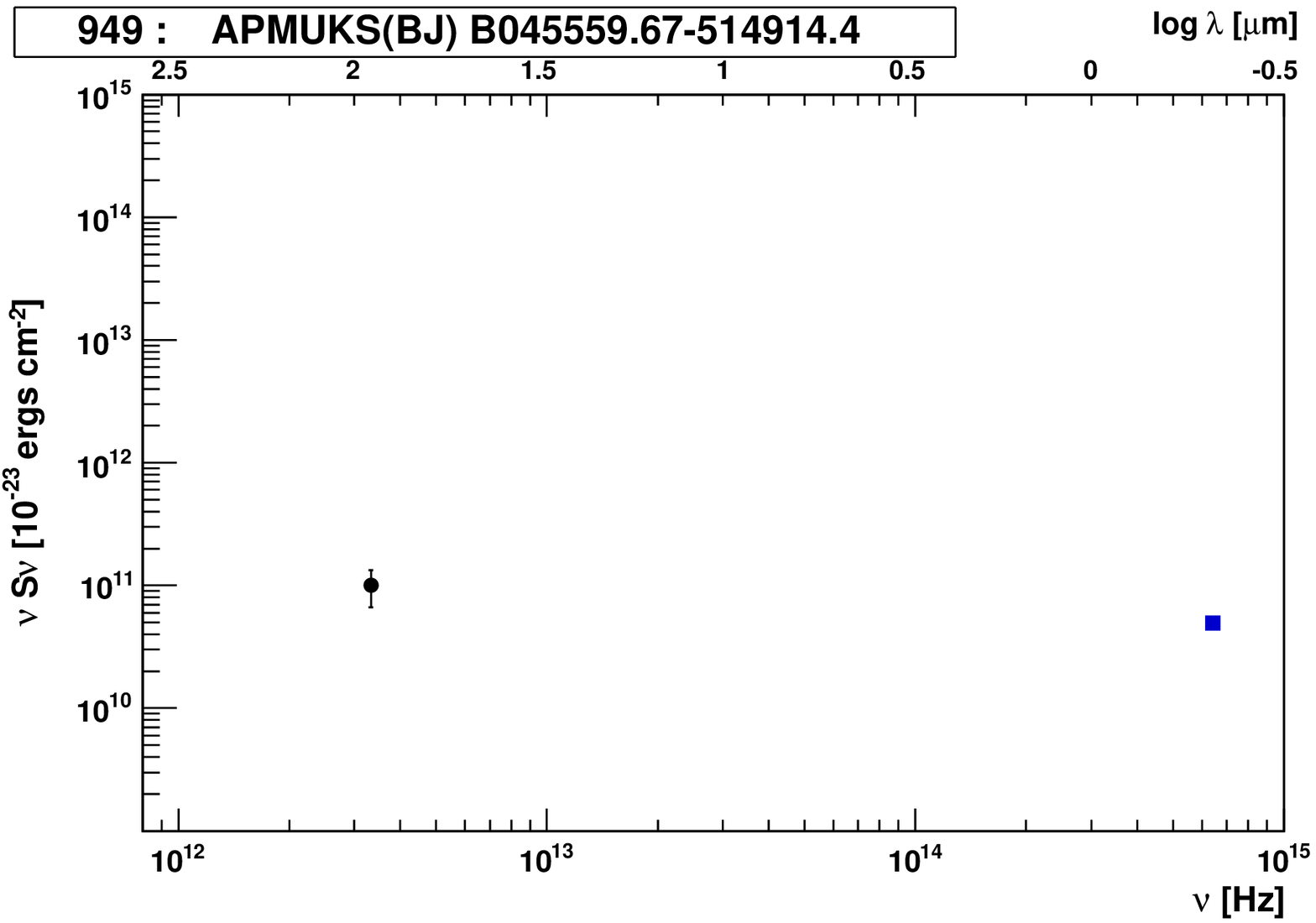}
\includegraphics[width=4cm]{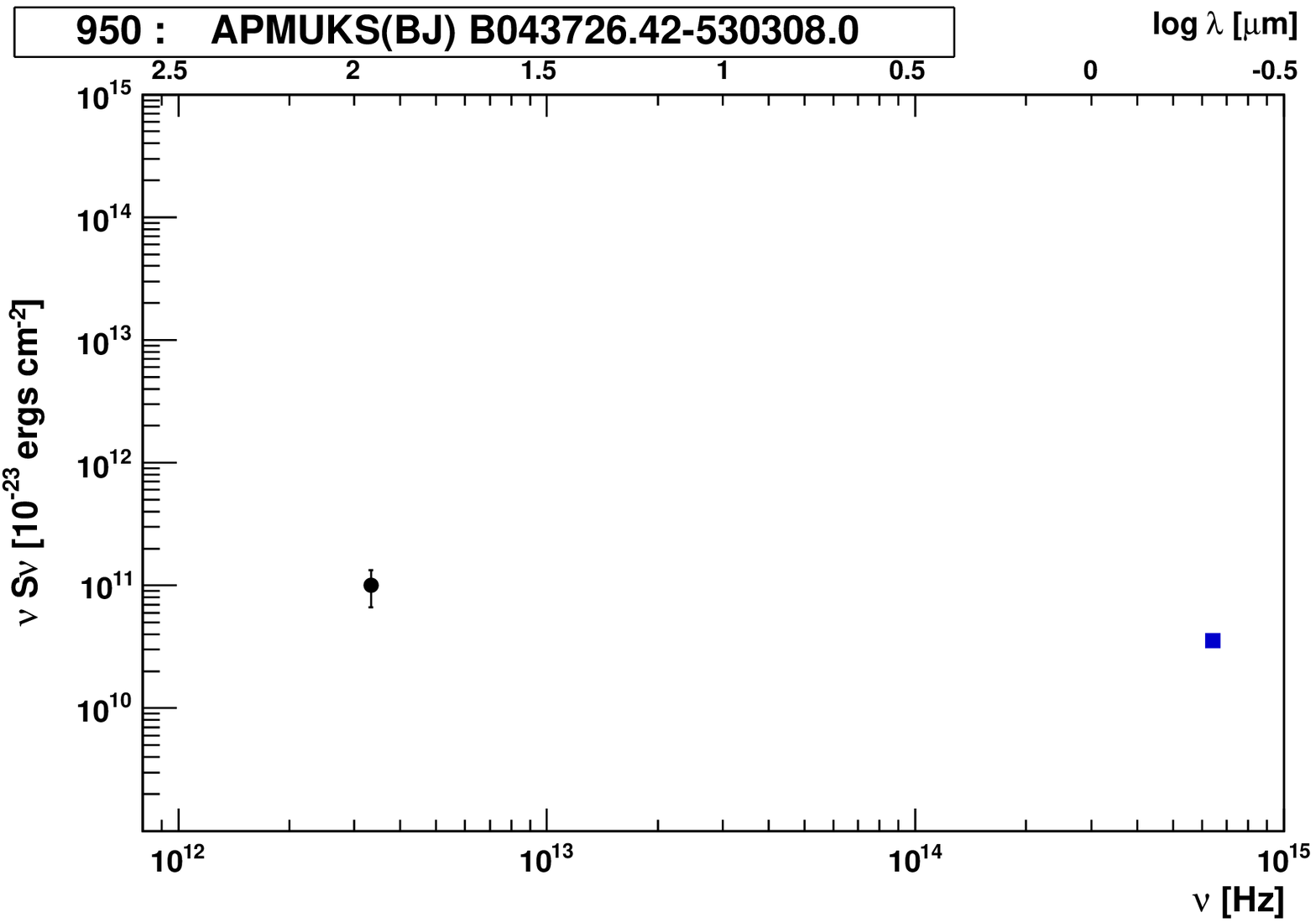}
\includegraphics[width=4cm]{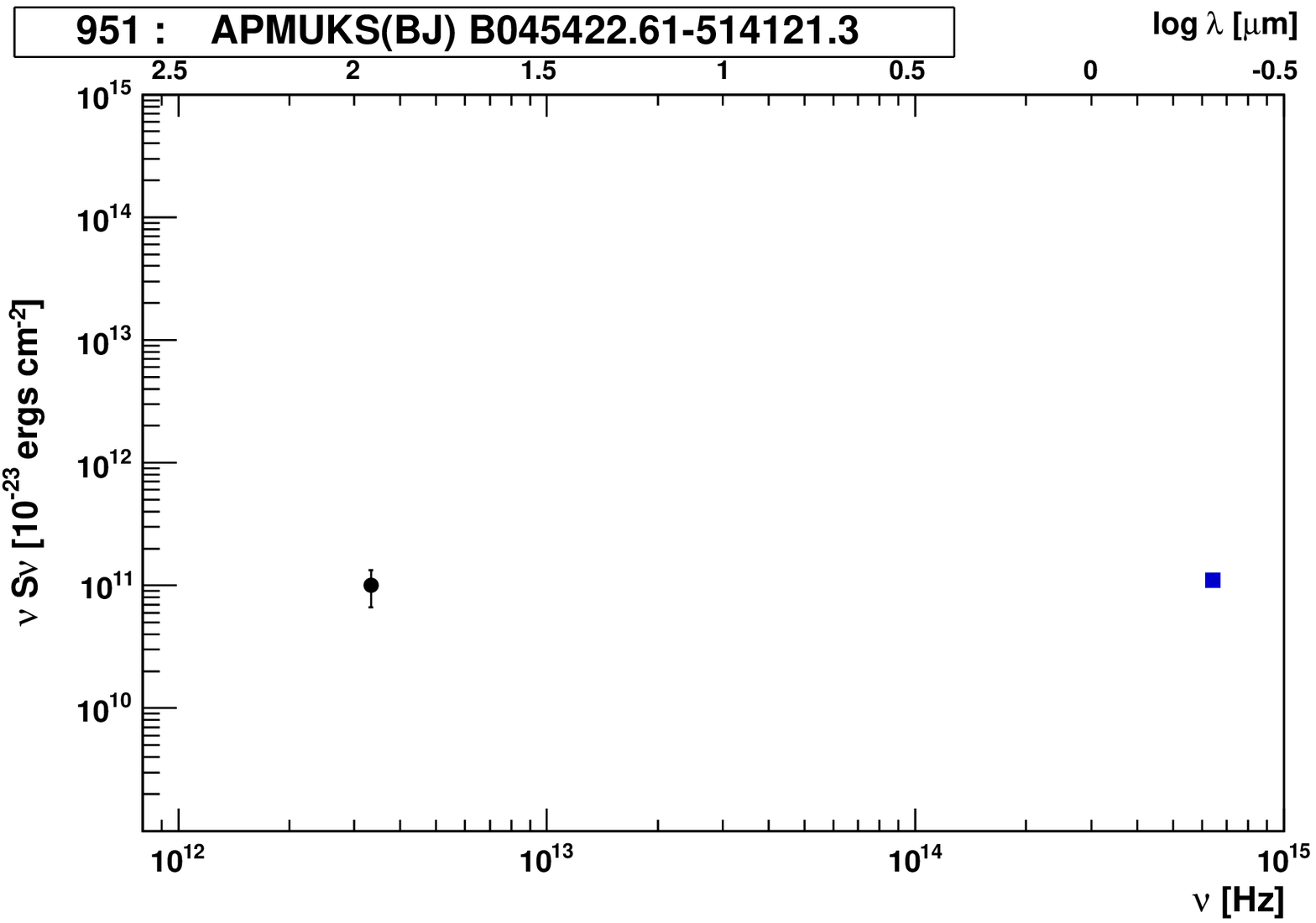}
\includegraphics[width=4cm]{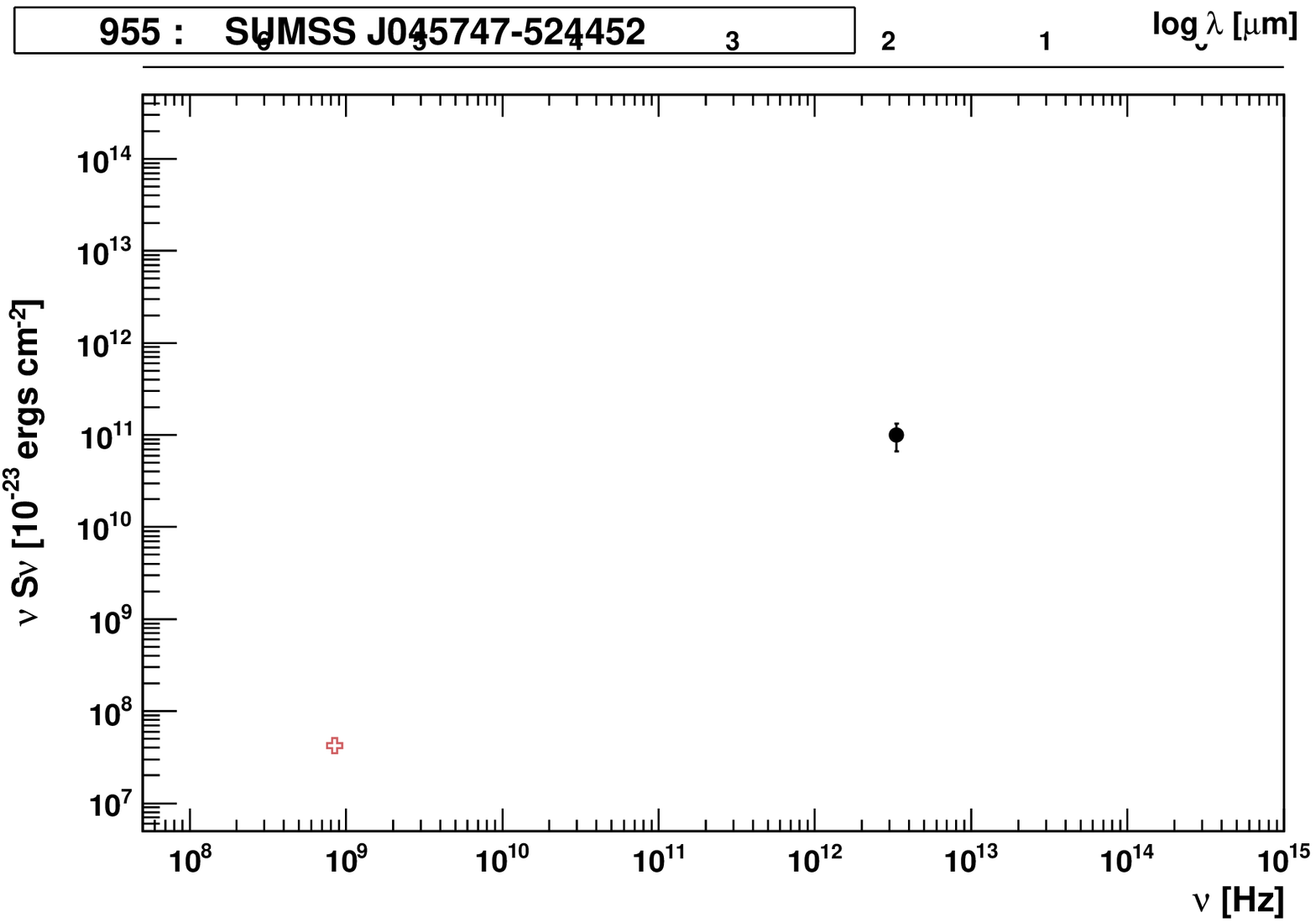}
\includegraphics[width=4cm]{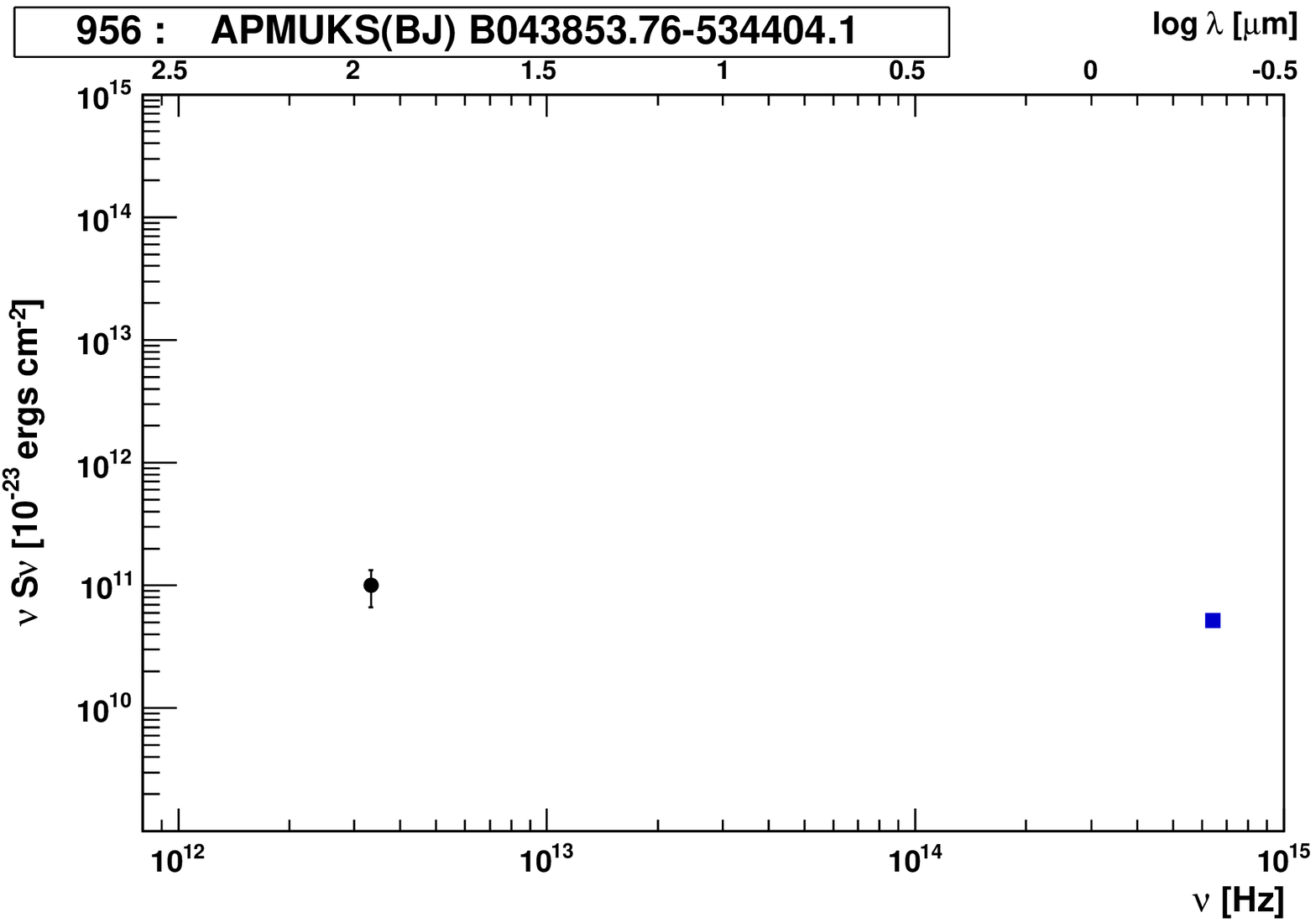}
\includegraphics[width=4cm]{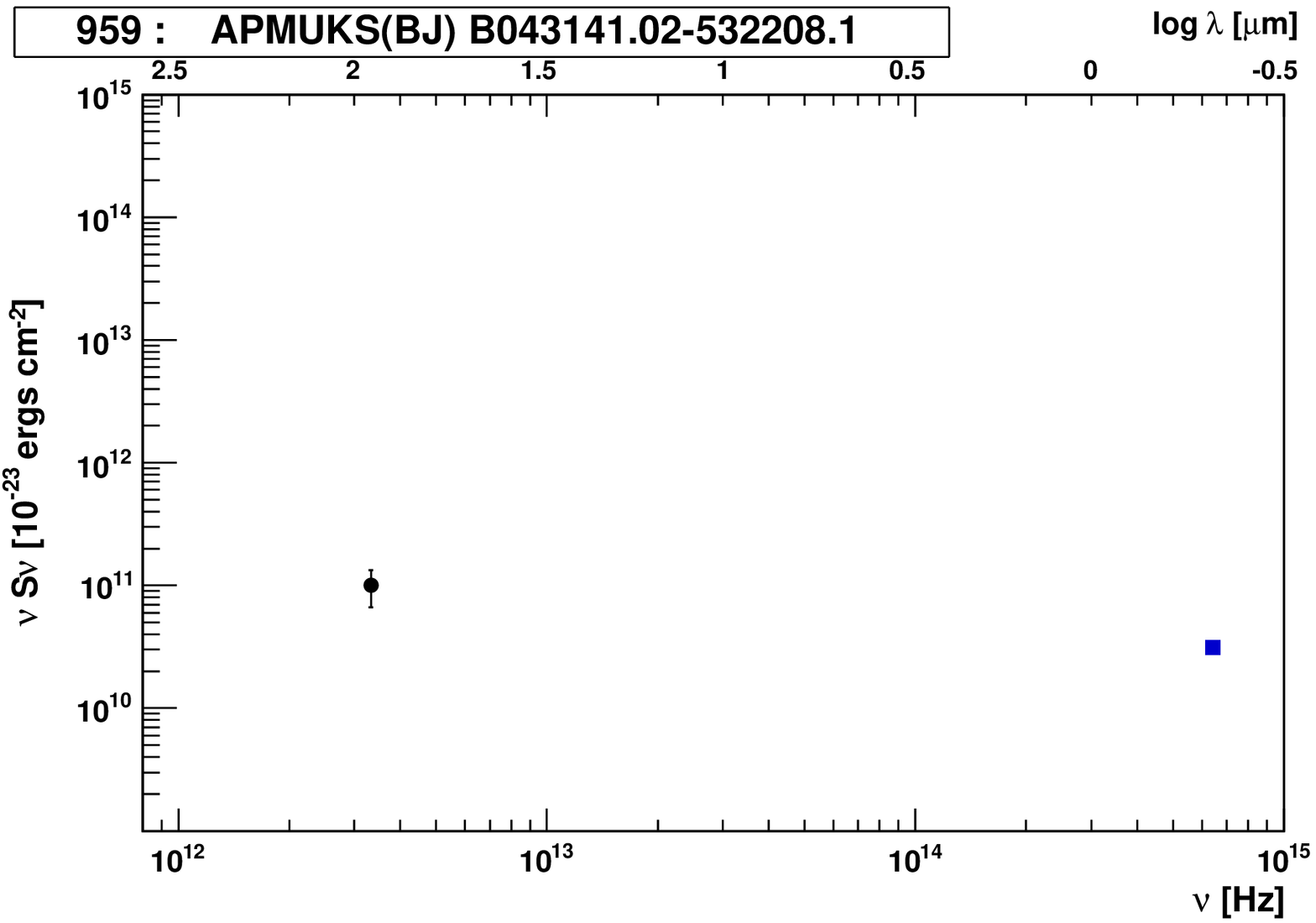}
\includegraphics[width=4cm]{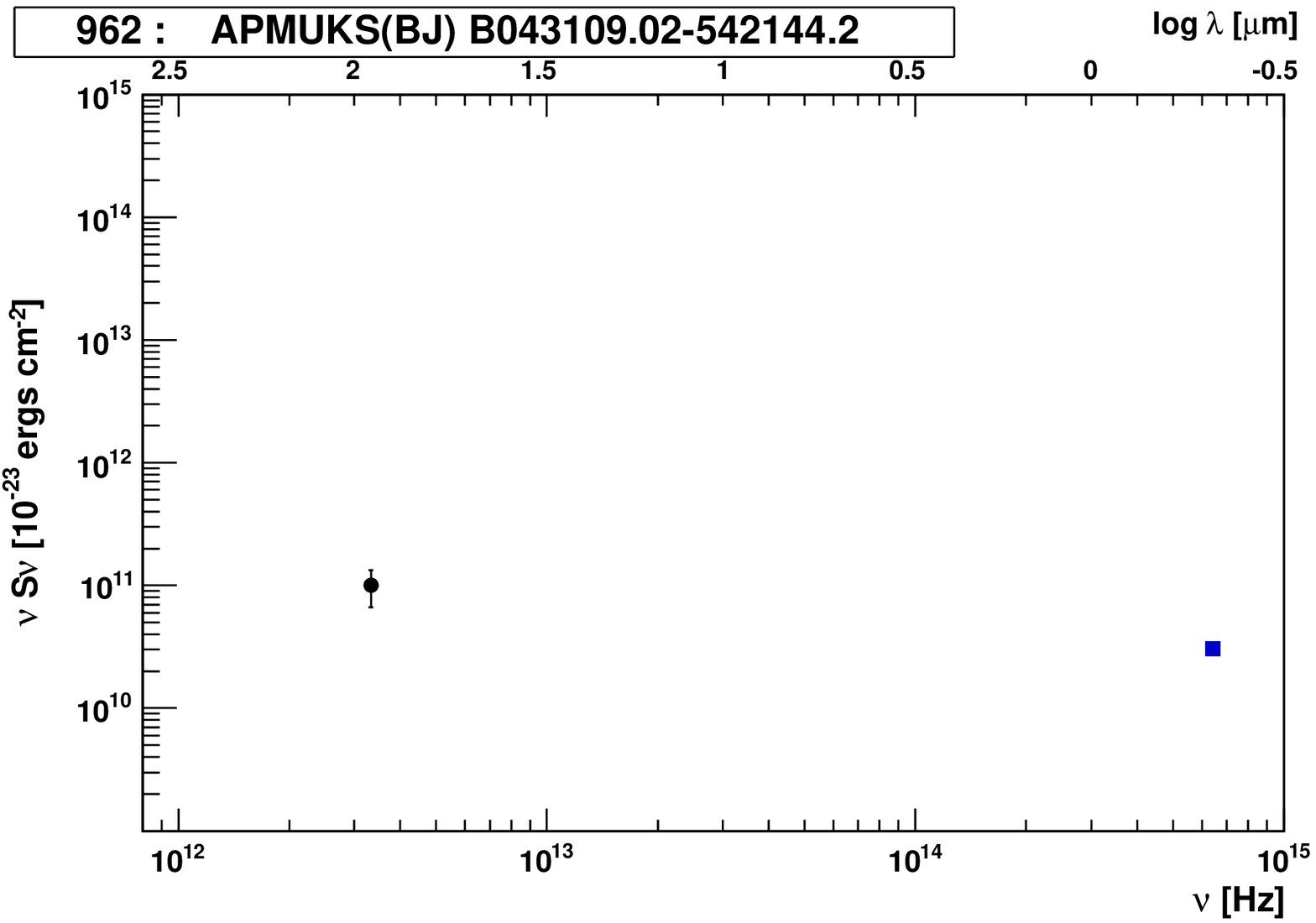}
\includegraphics[width=4cm]{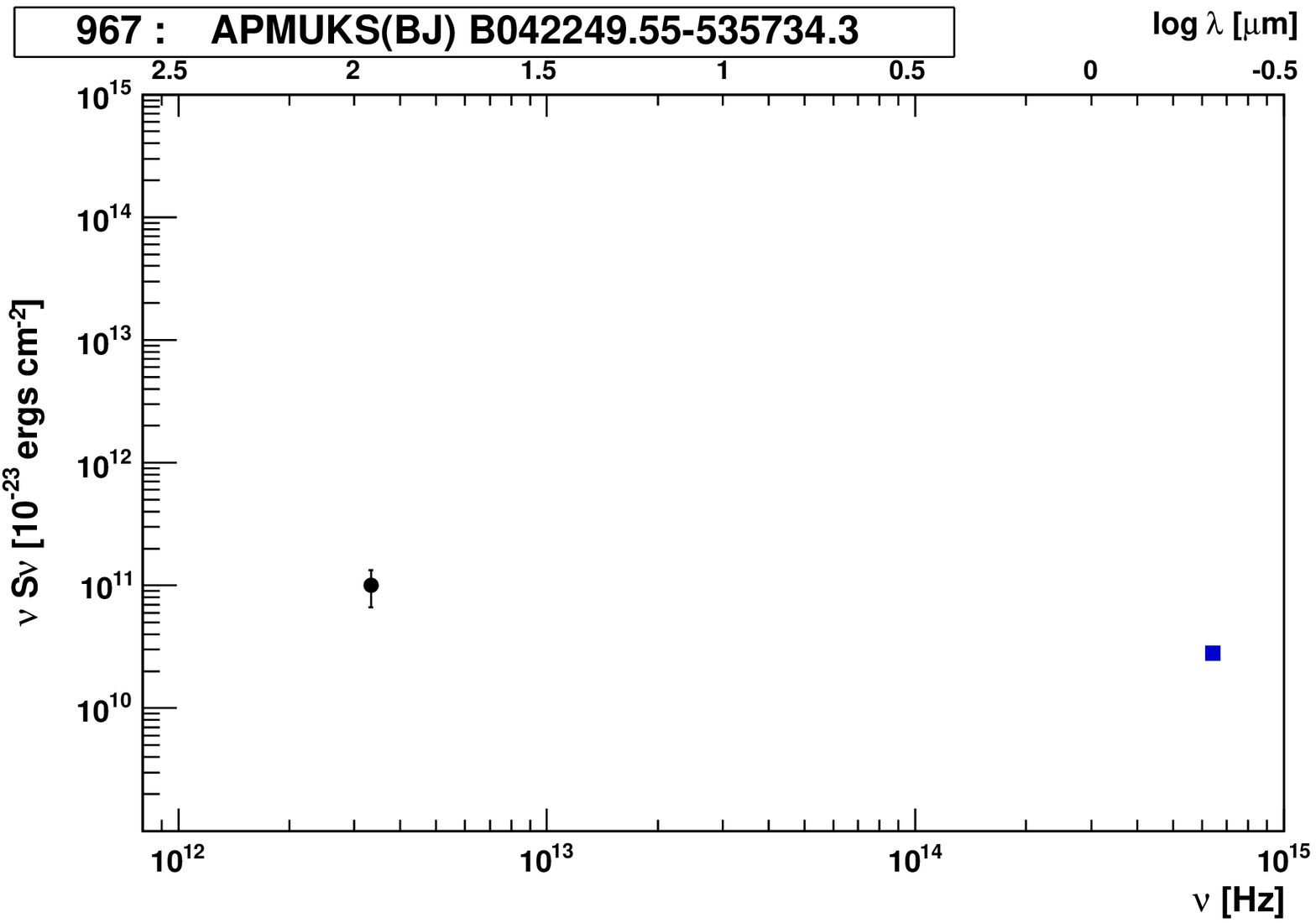}
\includegraphics[width=4cm]{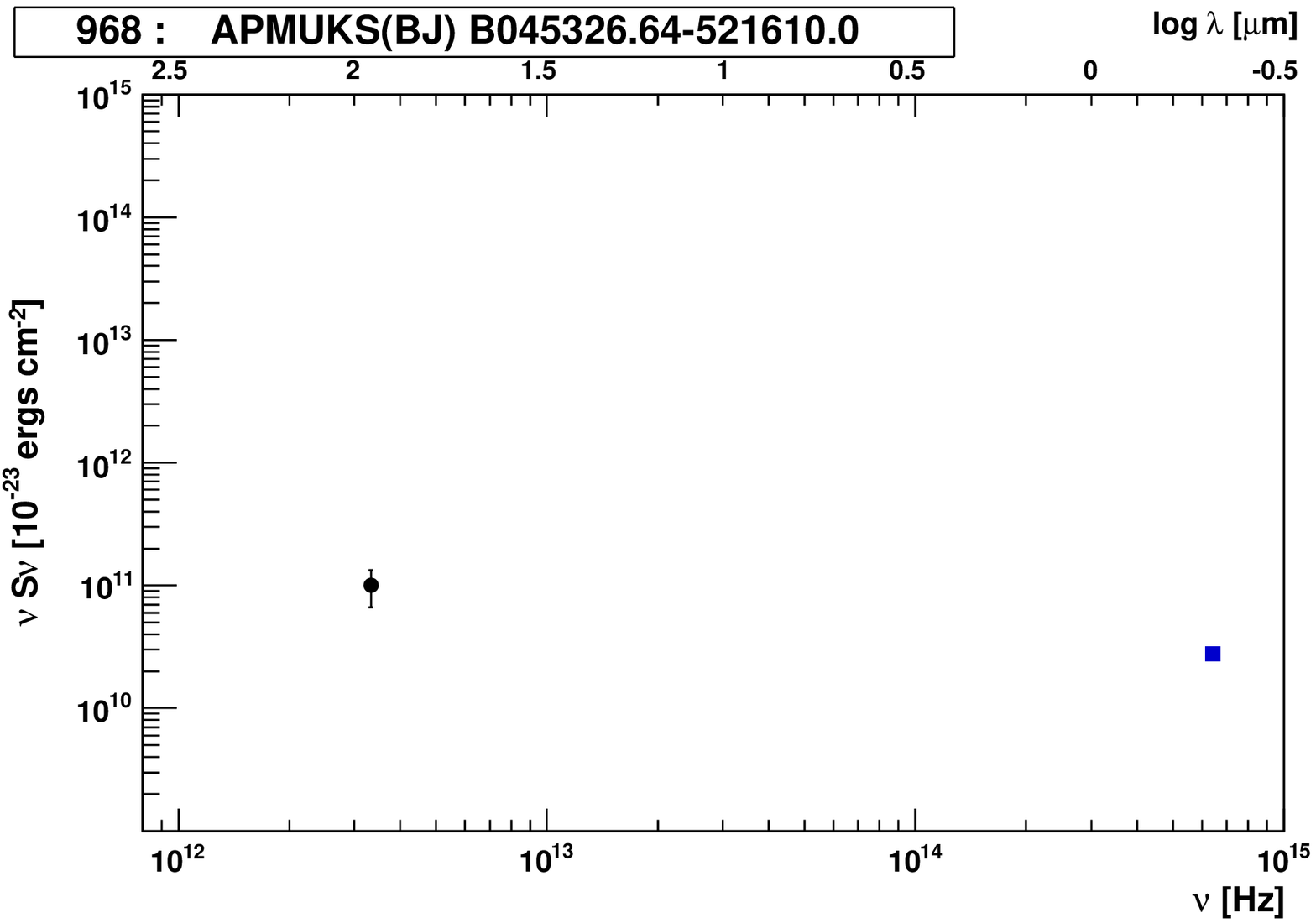}
\includegraphics[width=4cm]{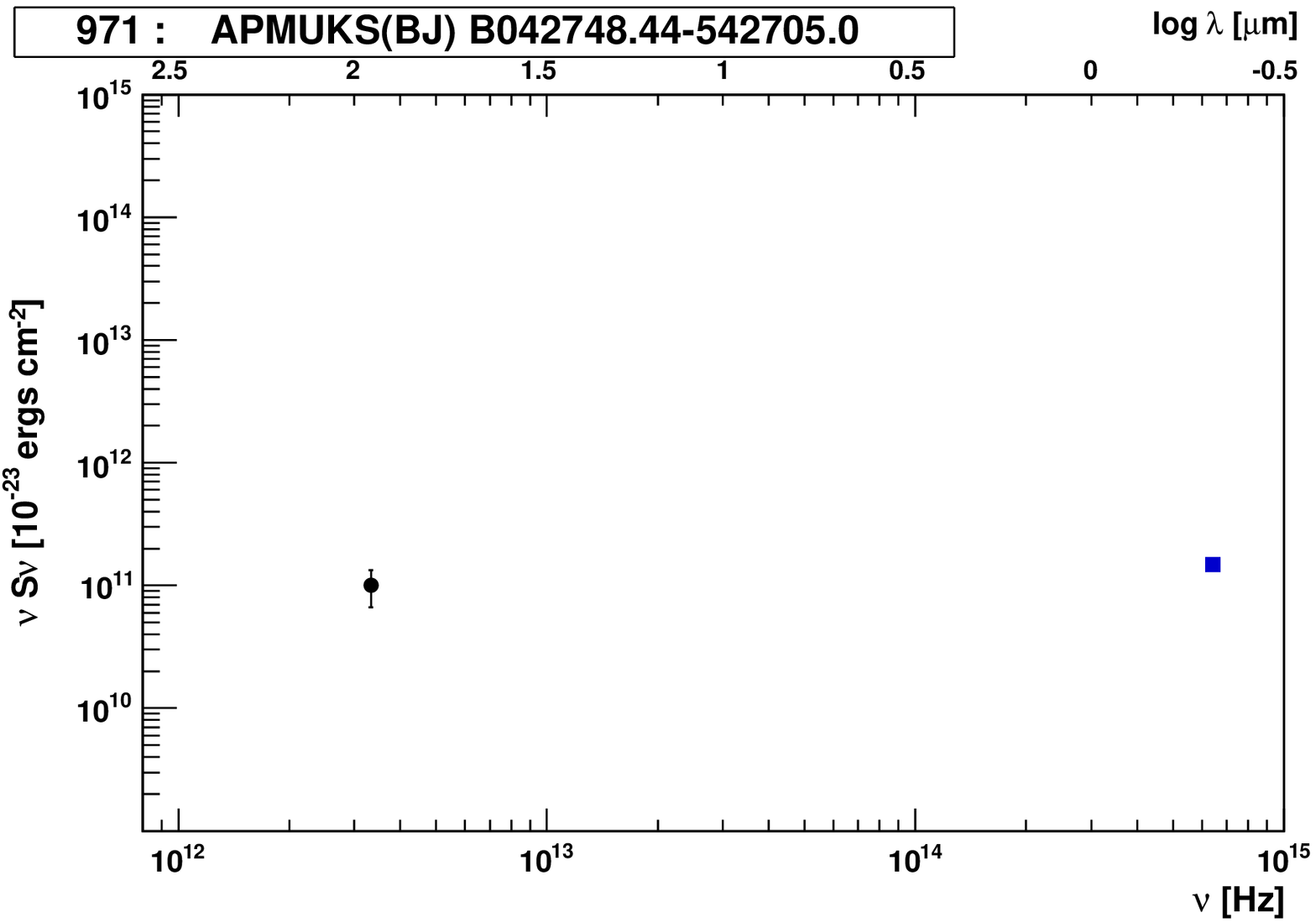}
\includegraphics[width=4cm]{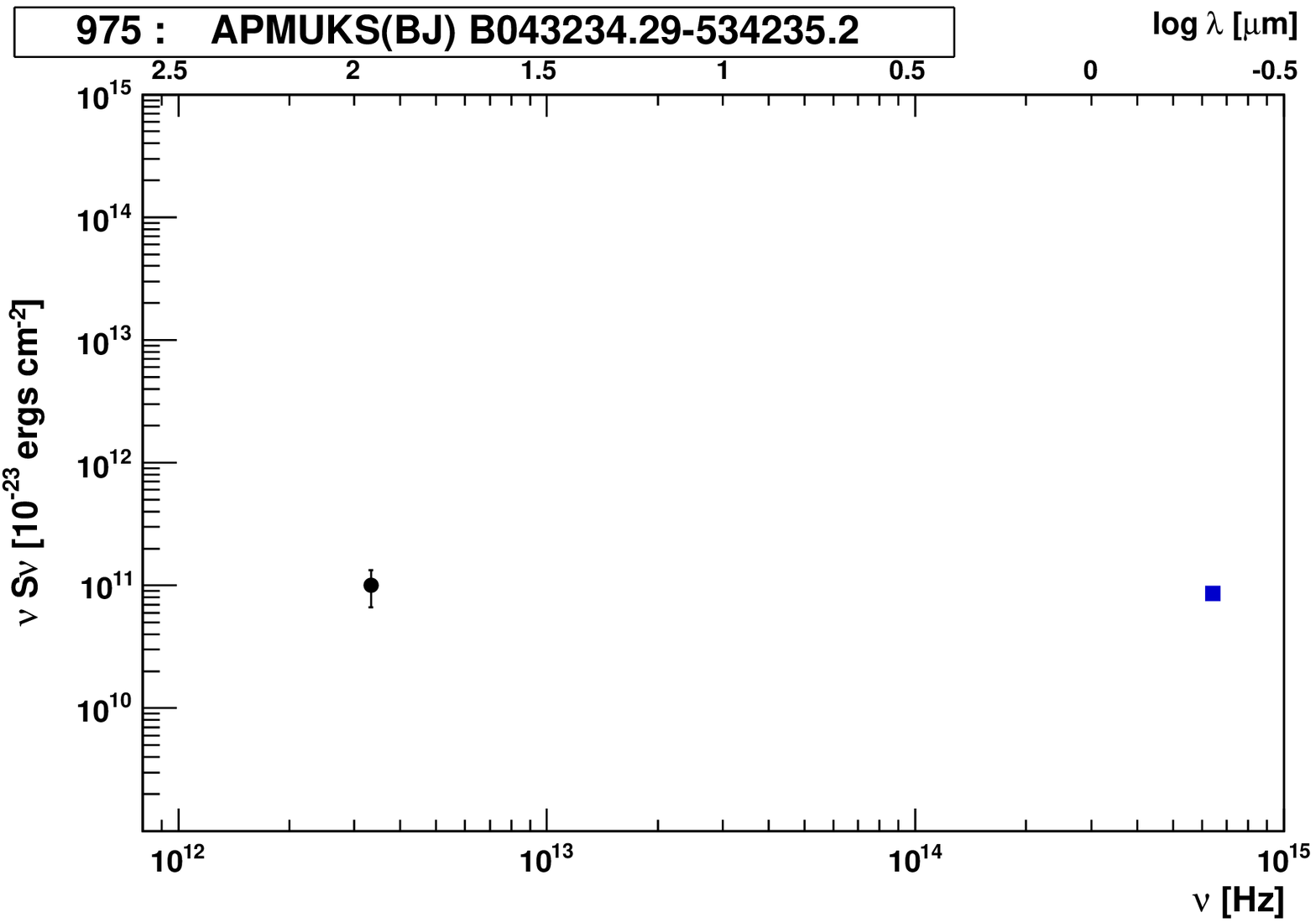}
\includegraphics[width=4cm]{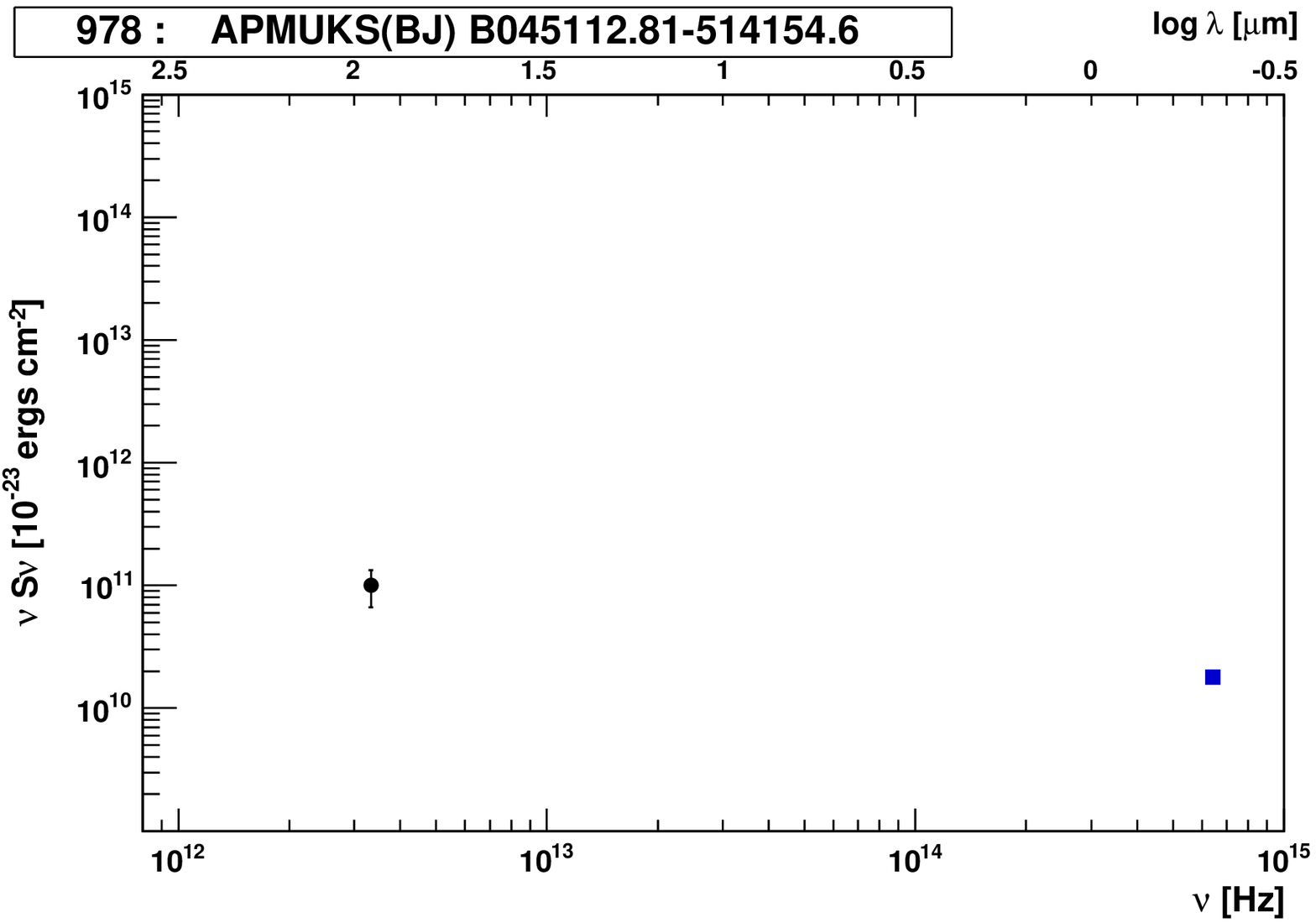}
\includegraphics[width=4cm]{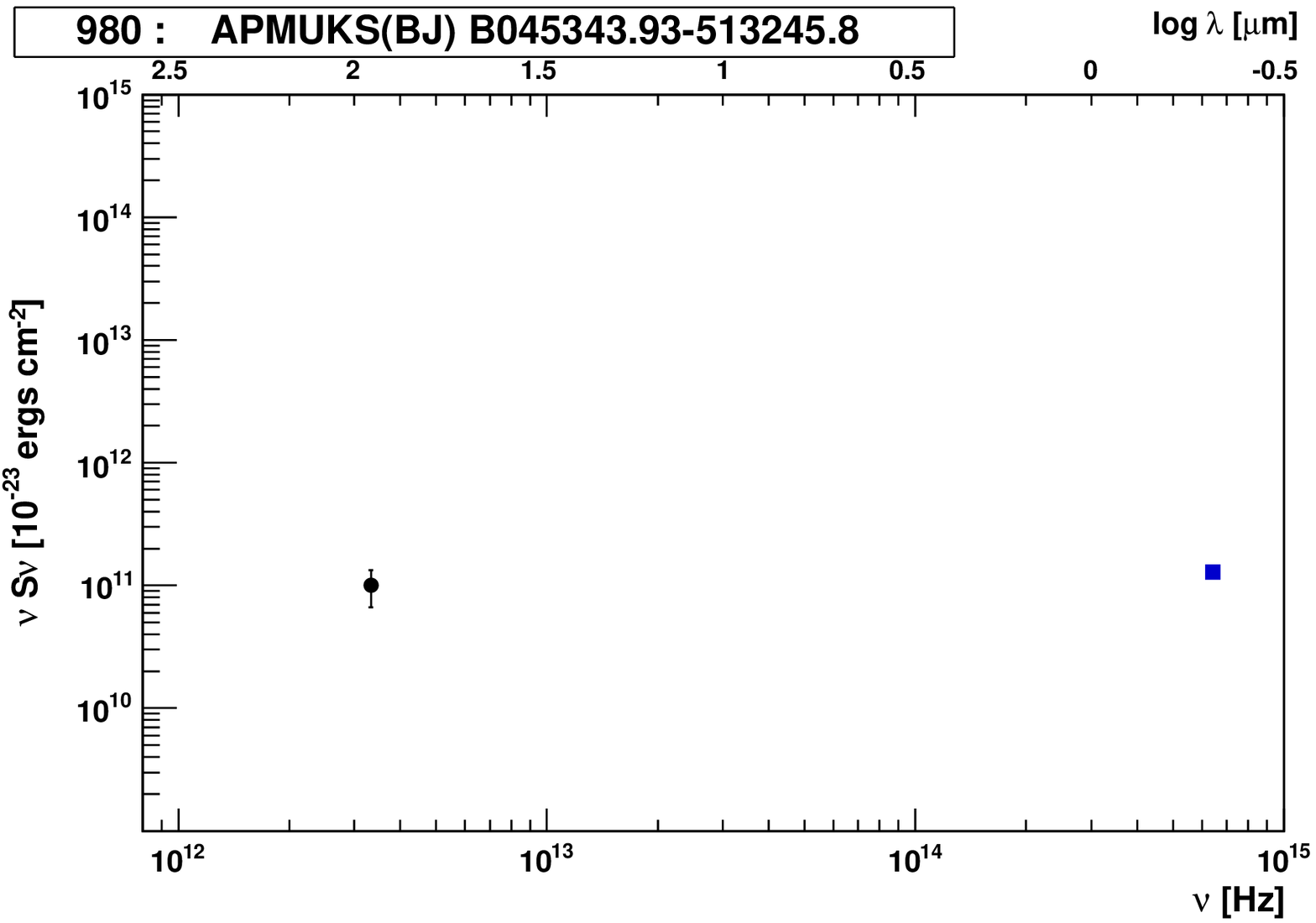}
\label{points15}
\caption {SEDs for the next 5 ADF-S identified sources, with symbols as in Figure~\ref{points1}.}
\end{figure*}
}

\clearpage

\onlfig{16}{
\begin{figure*}[t]
\centering
\includegraphics[width=4cm]{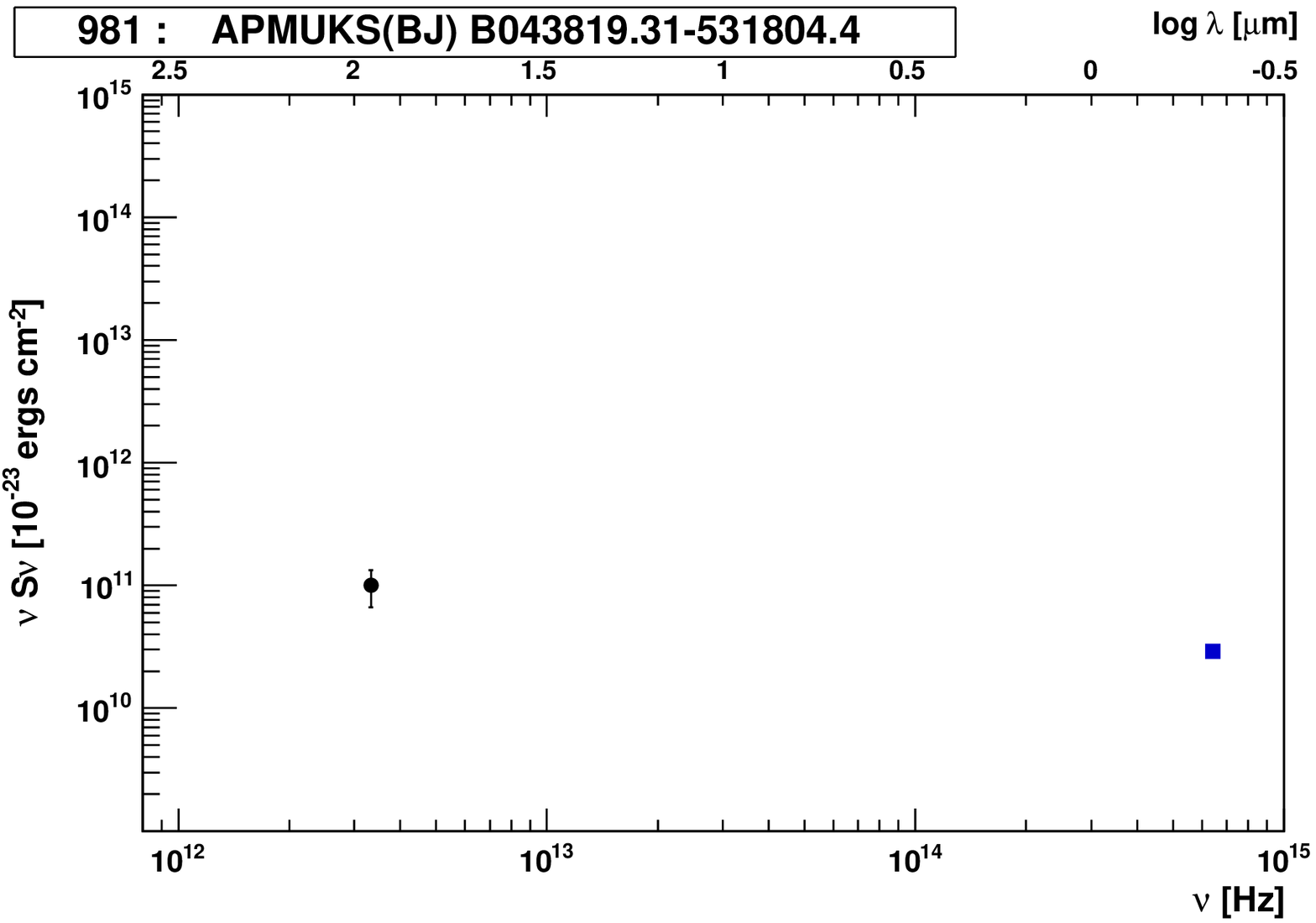}
\includegraphics[width=4cm]{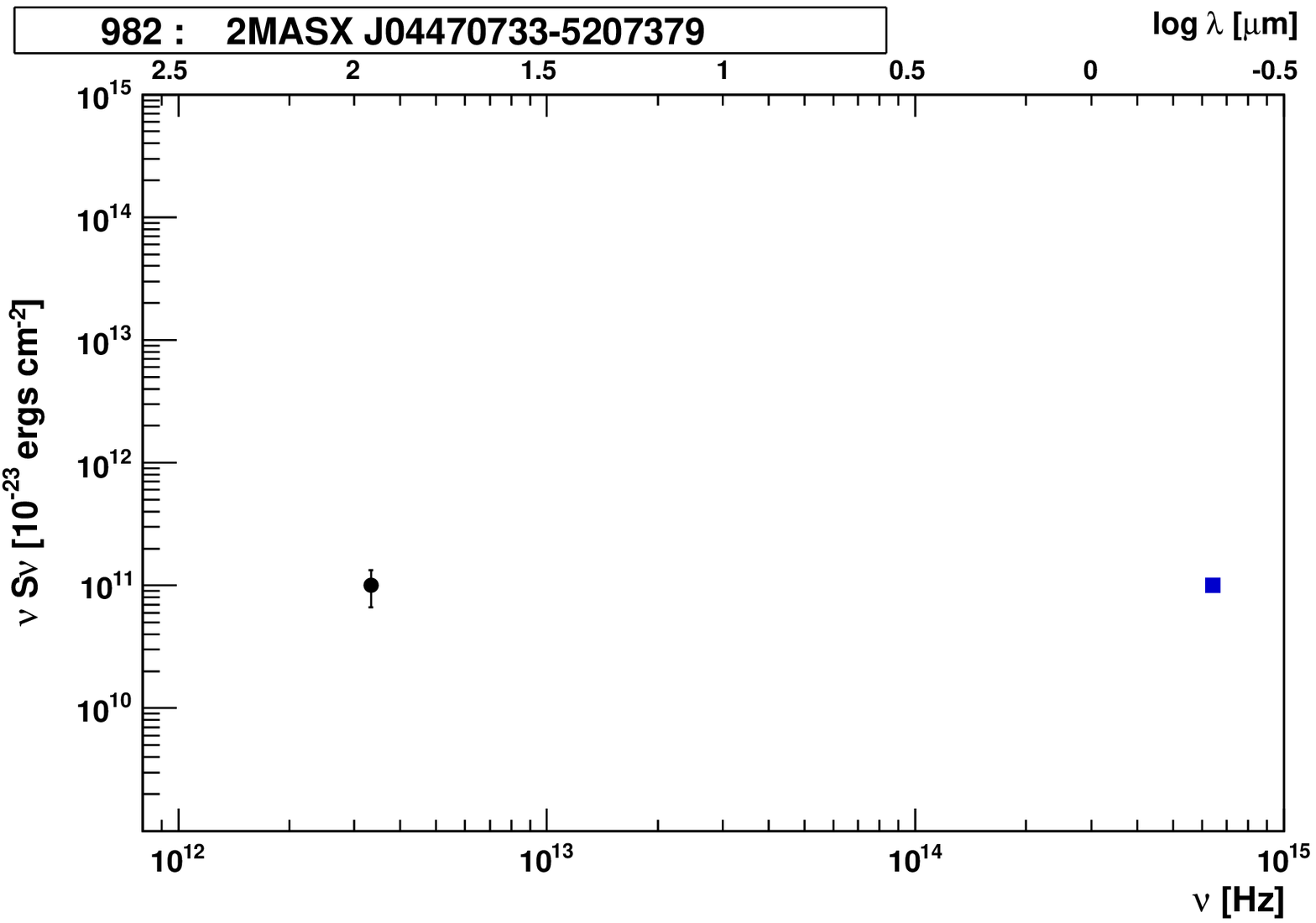}
\includegraphics[width=4cm]{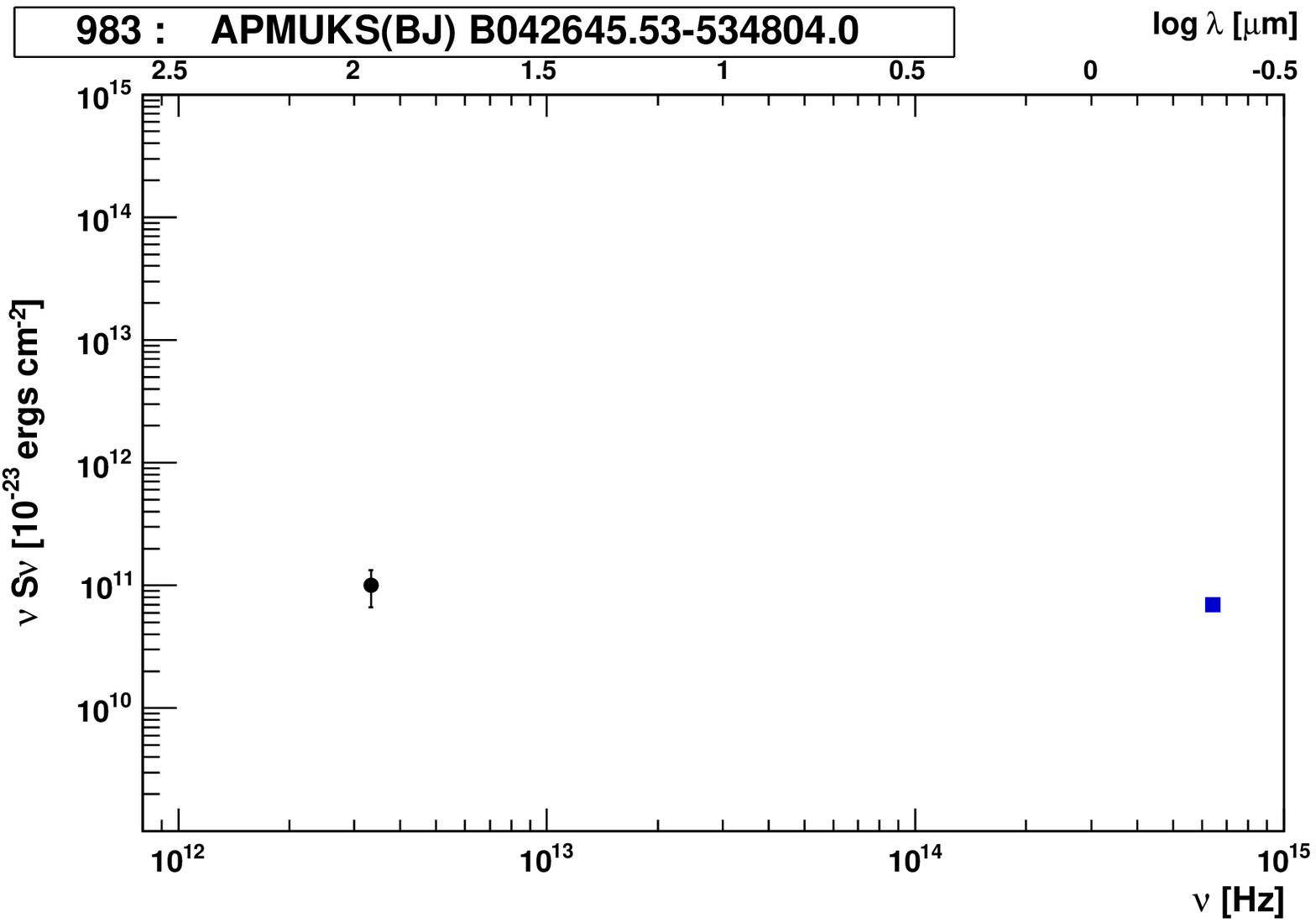}
\includegraphics[width=4cm]{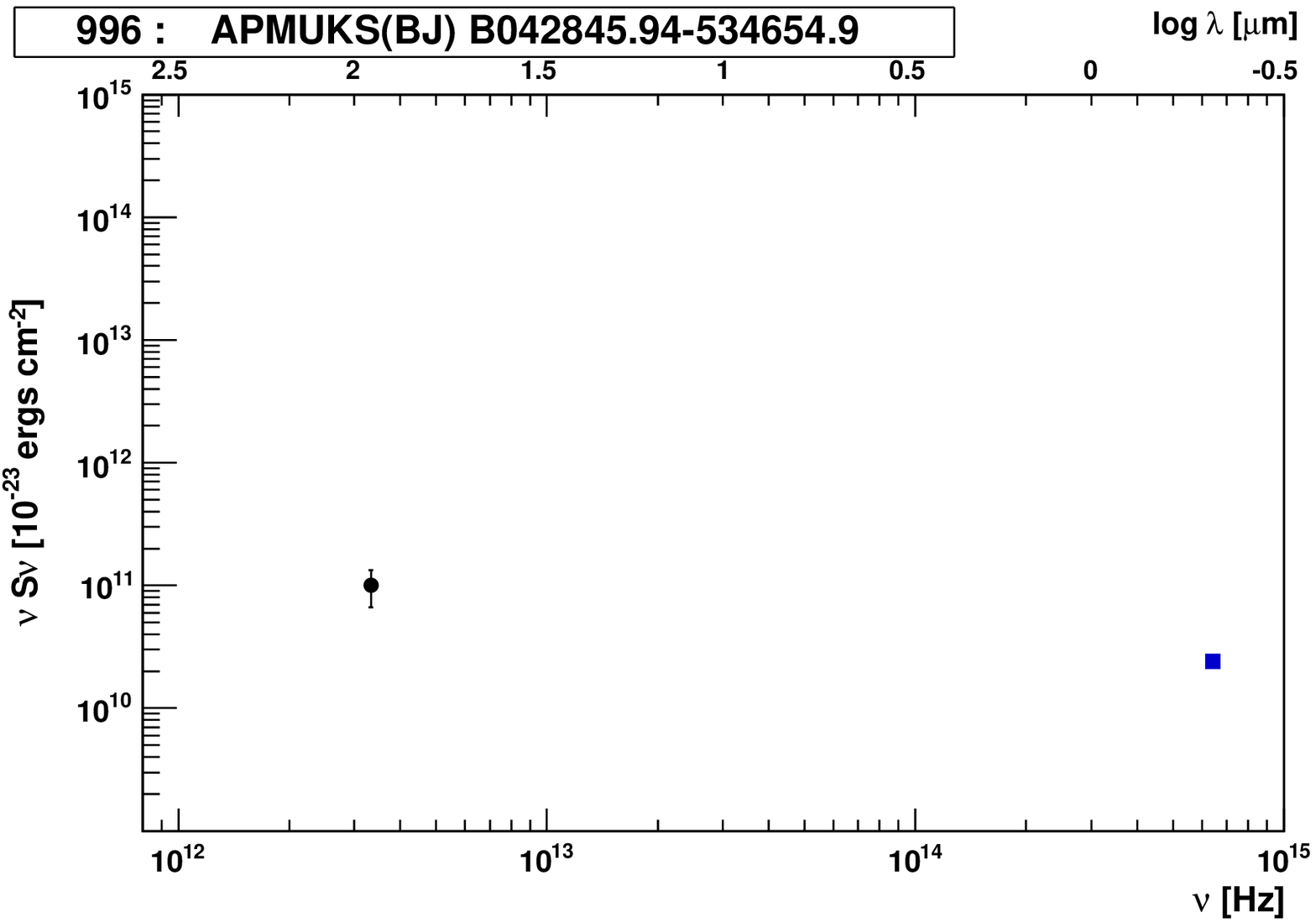}
\includegraphics[width=4cm]{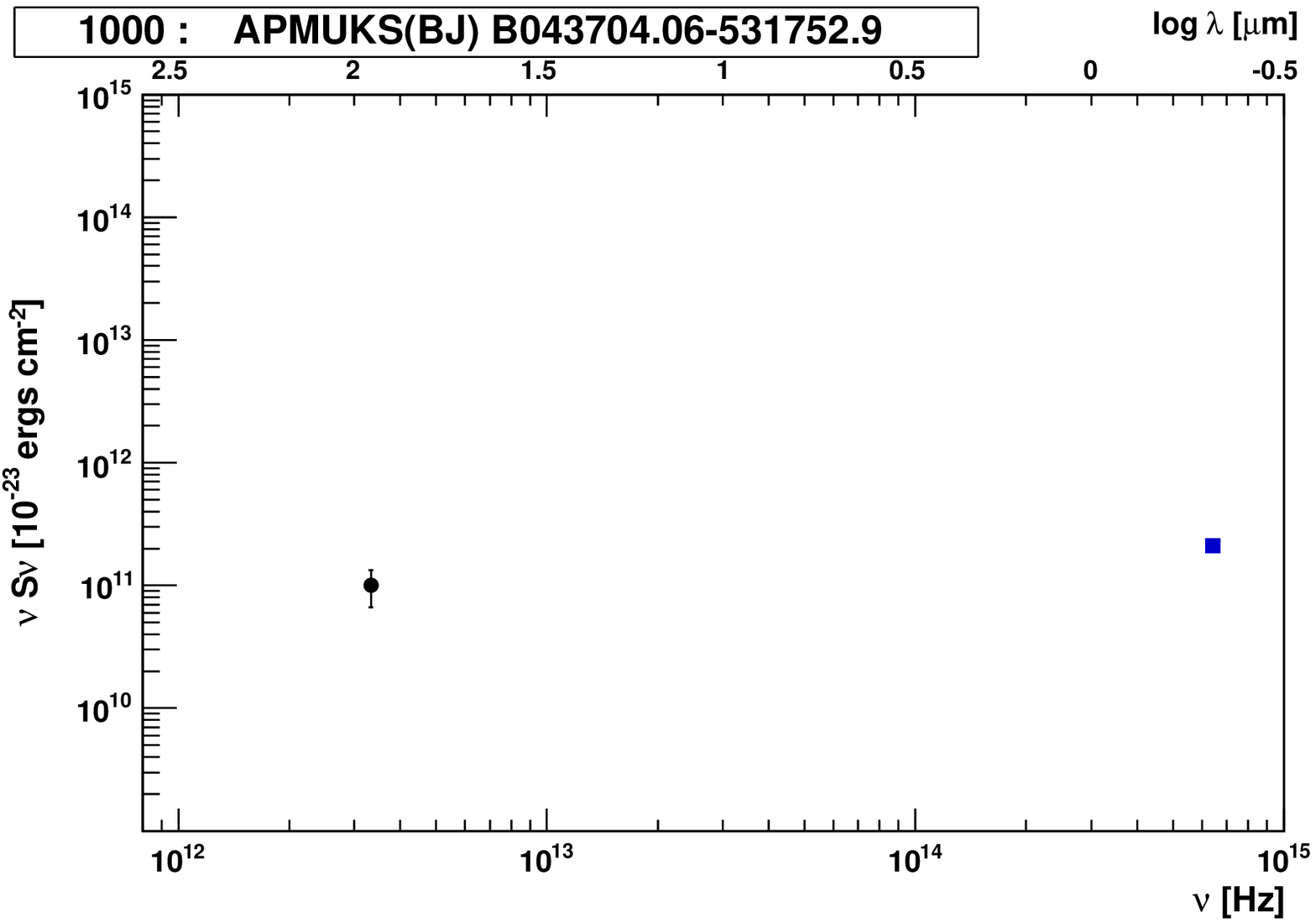}
\label{points16}
\caption {SEDs for the next 5 ADF-S identified sources, with symbols as in Figure~\ref{points1}.}
\end{figure*}
}

\end{document}